\documentclass[10pt]{book}
\usepackage[utf8]{inputenc}
\usepackage[T1]{fontenc}
\usepackage{etex}
\usepackage{amsmath}
\usepackage{amsfonts}
\usepackage{amssymb}
\usepackage[version=4]{mhchem}
\usepackage{stmaryrd}
\usepackage{graphicx}
\usepackage[export]{adjustbox}
\graphicspath{ {./images/} }
\usepackage{caption}
\usepackage{comment}
\usepackage{kbordermatrix} \renewcommand{\kbldelim}{(}
\renewcommand{\kbrdelim}{)}
\usepackage[a4paper,margin=2cm]{geometry}
\usepackage{ucharclasses}
\usepackage{index}
\usepackage{hyperref}
\hypersetup{colorlinks=true, linkcolor=blue, filecolor=magenta, urlcolor=cyan,}
\usepackage{graphicx}
\usepackage[export]{adjustbox}
\graphicspath{ {./images/} }
\usepackage{mathrsfs}
\usepackage{multirow}
\usepackage{diagbox}
\usepackage{bbold}
\usepackage{mathtools}
\usepackage{polyglossia}
\setdefaultlanguage{english}
\usepackage{fontspec}
\usepackage[backend=bibtex,style=numeric,sorting=none,defernumbers=true]{biblatex}
\newindex{default}{idx}{ind}{Subject Index}
\newindex{sym}{sdx}{snd}{Index of Symbols}

\usepackage{etoolbox}
\robustify{\widehat}
\robustify{\widetilde}

\newcommand{\isym} [1]{\index[sym]{Alatin #1}}%Latin letters
\newcommand{\isymg}[1]{\index[sym]{Bgreek #1}}%Greek letters
\newcommand{\isymo}[1]{\index[sym]{Cother #1}}%operators, brackets, symbols

\newcommand{\symgroupname}[1]{%
  \ifx A#1Latin letters%
  \else\ifx B#1Greek letters%
  \else\ifx C#1Special symbols and operators%
  \else #1\fi\fi\fi}

\usepackage{newunicodechar}
\usepackage{xcolor}
\usepackage{listings}
\lstdefinestyle{vmkcode}{%
  basicstyle=\ttfamily\footnotesize,
  keywordstyle=\color{blue!55!black},
  commentstyle=\color{green!40!black}\itshape,
  stringstyle=\color{purple!65!black},
  showstringspaces=false,
  columns=fullflexible,
  breaklines=true,
  frame=single,
  rulecolor=\color{black!25},
  xleftmargin=1em,
  aboveskip=0.7em,belowskip=0.4em,
}
\let\svthefootnote\thefootnote
\newcommand\blfootnotetext[1]{%
  \let\thefootnote\relax\footnote{#1}%
  \addtocounter{footnote}{-1}%
  \let\thefootnote\svthefootnote%
}

\let\svfootnotetext\footnotetext
\renewcommand\footnotetext[2][?]{%
  \if\relax#1\relax%
    \ifnum\value{footnote}=0\blfootnotetext{#2}\else\svfootnotetext{#2}\fi%
  \else%
    \if?#1\ifnum\value{footnote}=0\blfootnotetext{#2}\else\svfootnotetext{#2}\fi%
    \else\svfootnotetext[#1]{#2}\fi%
  \fi
}
\numberwithin{equation}{section}

\numberwithin{figure}{chapter}
\numberwithin{table}{chapter}

\newcommand{\del}{\boldsymbol{\nabla}}
\newcommand{\delop}{\widehat{\del}}
\newcommand{\clebsch}[6]{C_{#1\,#2\,#3\,#4}^{#5\,#6}}

\newcommand{\ket}[1]{\bigl| #1 \bigr\rangle}
\newcommand{\rket}[1]{\bigl| #1 \bigr)}
\newcommand{\bra}[1]{\bigl\langle #1 \bigr|}
\newcommand{\rbra}[1]{\bigl( #1 \bigr|}
\newcommand{\expect}[1]{\bigl\langle #1 \bigr\rangle}
\newcommand{\braket}[2]{\bigl\langle #1 \big| #2 \bigr\rangle}
\newcommand{\rbraket}[2]{\bigl( #1 \big| #2 \bigr)}
\newcommand{\braOket}[3]{\bigl\langle #1 \big| #2 \big| #3 \bigr\rangle}

\newcommand{\braOketred}[3]{\bigl\langle #1 \big\Vert #2 \big\Vert #3 \bigr\rangle}
\newcommand{\rbraOketred}[3]{\bigl( #1 \big\Vert #2 \big\Vert #3 \bigr)}

\newcommand{\vect}[1]{\mathbf{#1}}
\newcommand{\mat}[1]{\boldsymbol{#1}}
\newcommand{\threej}[6]{%
\renewcommand{\arraystretch}{1}\arraycolsep=0.2em\left(\hspace*{-0.2em}\begin{array}{rrr} #1 & #2 & #3 \\ #4 & #5 & #6 \end{array}\hspace*{-0.2em}\right)}
\newcommand{\sixj}[6]{%
{\renewcommand{\arraystretch}{1}\arraycolsep=0.2em\left\{\hspace*{-0.2em}%
\begin{array}{ccc} #1 & #2 & #3 \\ #4 & #5 & #6 \end{array}\hspace*{-0.2em}\right\}}}

\newcommand{\ninej}[9]{{\renewcommand{\arraystretch}{1}\arraycolsep=0.2em\left\{\hspace*{-0.2em}%
\begin{array}{ccc} #1 & #2 & #3 \\ #4 & #5 & #6 \\ #7 & #8 & #9 \end{array}\hspace*{-0.2em}\right\}}}

\RequirePackage{newlist}
\newlistcommand{\twelvejI}{%
{\renewcommand{\arraystretch}{1}\left\{\begin{array}{llll} 
 \argx{1}     &\argx{2}     &\argx{3}     &\argx{4} \\
\quad \argx{5}     &\quad \argx{6}     &\quad \argx{7}     &\quad \argx{8} \\
 \argx{9}     &\argx{10}     &\argx{11}     &\argx{12}  \end{array}\right\}}
}

\newlistcommand{\twelvejII}{%
{\renewcommand{\arraystretch}{1}\left\{\begin{array}{cccc} 
 -  &\argx{1}     &\argx{2}     &\argx{3} \\
 \argx{4} & - & \argx{5}     &\argx{6} \\
 \argx{7} & \argx{8}     & - & \argx{9} \\
 \argx{10} & \argx{11}     & \argx{12}     & - \end{array}\right\}}
}

\usepackage{enumitem}

\usepackage{titlesec}
\renewcommand{\thesubsubsection}{\alph{subsubsection}}
\titleformat{\subsubsection}
{\normalfont\itshape}
{(\thesubsubsection)}{0.5em}{}

\usepackage{relsize}
\newcommand{\Fact}[1]{{\mathlarger{\Pi}_{#1}}}
\newcommand{\Fhg}[2]{\prescript{}{#1}{F}_{#2}^{}}
\newcommand{\Fnm}[6][l]{\prescript{}{#2}{F}_{#3}^{}%
\if#1l \left(#4 ; #5 ; #6\right)
\else\left[\begin{array}{c} #4\\#5\end{array}\middle|#6\right]
\fi
}

\title{Quantum Theory Angular Momentum\\Electronic version 2026}

\author{originally by:\\
D. A. Varshalovich\\
A. N. Moskalev\\
V. K. Khersonskii\\
This version:\\
 Niels R. Walet}
\date{}
\begin{document}
\frontmatter
\maketitle
\tableofcontents
\mainmatter
%\captionsetup{singlelinecheck=false}
\hspace*{\fill}

\section*{Originally published by}
World Scientific Publishing Co. Pte. Ltd.\\
5 Toh Tuck Link, Singapore 596224\\
USA office: 27 Warren Street, Suite 401-402, Hackensack, NJ 07601\\
UK office: 57 Shelton Street, Covent Garden, London WC2H 9HE

\section*{Open Access}
This project is open access and is licensed under the terms of the Creative Commons Attribution 4.0 International License (\href{http://creativecommons.org/licenses/by/4.0/}{http://creativecommons.org/licenses/by/4.0/}), which permits use, sharing, adaptation, distribution and reproduction in any medium or format, as long as you give appropriate credit to the original author(s) and the source, provide a link to the Creative Commons license and indicate if changes were made.

It is based on the original work published by World Scientific Publishing Co. Pte. Ltd., Singapore, 1988, ISBN 978-9971-5-0107-5 (hardcover), ISBN 978-9971-5-0996-5 (paperback), ISBN 978-981-4415-49-1 (ebook for institutions), ISBN 978-981-4578-28-8 (ebook for individuals).

\section*{Open Access funded by SCOAP (Sponsoring Consortium for Open Access Publishing in Particle Physics)}
The original ebook was converted to Open Access in 2021 through the sponsorship of SCOAP , licensed under the terms of the Creative Commons Attribution 4.0 International License (\href{http://creativecommons.org/licenses/by/4.0/}{http://creativecommons.org/licenses/by/4.0/}), which permits use, sharing, adaptation, distribution and reproduction in any medium or format, as long as you give appropriate credit to the original author(s) and the source, provide a link to the Creative Commons license and indicate if changes were made.
Original copyright © 1988 by World Scientific Publishing Co. Pte. Ltd.

\newpage

\section*{Preface}
In the years since its publication, this book has become the standard reference on the quantum theory of angular momentum.  The book provides a comprehensive treatment of the subject, covering both the theoretical foundations and practical applications of angular momentum in quantum mechanics. As an open access publication, it is freely available to researchers and students worldwide, ensuring that the knowledge contained within its pages can be accessed by anyone interested in the field. The book has been widely cited and used in research and teaching, and it continues to be a valuable resource for those studying quantum mechanics and related areas of physics.

However, there are some drawbacks to the form in which it was provided. The scanned version of the book is not searchable, and the quality of the scanned images is not always optimal. This can make it difficult to find specific information or to read the text clearly. Additionally, the scanned version may not be compatible with all devices or software, which can limit its accessibility. Also, there is no verification of the accuracy of the scanned text, which can lead to errors or misinterpretations of the content. Finally, the scanned version may not be easily editable or adaptable for different purposes, such as creating new editions or translations.

Thus this attempt to provide a more accessible and user-friendly version of the book, with improved searchability, image quality, and compatibility with different devices and software. The text has been carefully reviewed and verified for accuracy, and it is provided in a format that can be easily edited or adapted for different purposes. We hope that this version of the book will be a valuable resource for researchers, students, and anyone interested in the quantum theory of angular momentum. The author would like to thank the original publisher, World Scientific Publishing Co. Pte. Ltd., for their support in making this open access version available to the public.

Clearly in  the era of AI, I have made substantial use of AI tools to assist in the preparation of this version of the book. I have used AI tools to help with tasks such as text recognition, formatting, and editing. These tools have helped to improve the accuracy and quality of the text, as well as to streamline the process generating code to verify equations. However, I have also carefully reviewed and verified the output of these tools to ensure that the final version of the book is as accurate and reliable. Overall, I believe that the use of AI tools has been a valuable aid in the preparation of this new open access version of the book.

Of course, the author would like to acknowledge that the original work was done by D. A. Varshalovich, A. N. Moskalev, and V. K. Khersonskii, and that this new version is based on their work. The author has made every effort to ensure that the content of this new version is faithful to the original work, while also improving its accessibility and usability for modern readers.

I do no see this as a static resource--there are undoubtedly errors in the text, and I would be grateful for any corrections or suggestions for improvement. Please post these at the GitHub repository for this project at \href{}. The intention is to improve the accessibility and usability of this important work, and I welcome any feedback, and contributions, that can help achieve that goal.

Manchester, Uk, September 2026 

\section*{Original Preface}
This book deals with one of the basic topics of quantum mechanics: the theory of angular momentum and irreducible tensors. Being rather versatile, the mathematical apparatus of this theory is widely used in atomic and molecular physics, in nuclear physics and elementary particle theory. It enables one to calculate atomic, molecular and nuclear structures, energies of ground and excited states, fine and hyperfine splittings, etc. The apparatus is also very handy for evaluating the probabilities of radiative transitions, cross sections of various processes such as elastic and nonelastic scattering, different decays and reactions (both chemical and nuclear) and for studying angular distributions and polarizations of particles.

Today this apparatus is finding ever increasing use in solving practical problems relating to quantum chemistry, kinetics, plasma physics, quantum optics, radiophysics and astrophysics.

The basic ideas of the theory of angular momentum were first put forward by M. Born, P. Dirac, W. Heisenberg and W. Pauli. However, the modern version of its mathematical apparatus was developed mainly in the works of E. Wigner, J. Racah, L. Biedenharn and others who applied group theoretical methods to problems in quantum mechanics.

At present a number of good books on the theory of angular momentum have been already published. The general principles and results of the theory may be found in the books by M. Rose [31], A. Edmonds [16], U. Fano and G. Racah [18], A. P. Yutsis, I. B. Levinson and V. V. Vanagas [44], A. P. Yutsis and A. A. Bandzaitis [45], D. Brink and G. Satcher [9]. Nevertheless, many formulas and relationships essential for practical calculations have escaped these books and are either scattered in various editions, or included as appendices in papers discussing somewhat disparate topics, making them generally inaccessible. Even greater difficulties arise when one tries to use the results, as each author employs his own phase conventions, initial definitions and symbols.

The authors of this book aimed at collecting and compiling ample material on the quantum theory of angular momentum within the framework of a single system of phases and definitions. This is why, in addition to the basic theoretical results, the book also includes a great number of formulas and relationships essential for practical applications.

This edition is the translated version of our book published in the USSR in 1975. In the course of its preparation we have tried to comply with a number of suggestions from our readers. For instance, each chapter opens with a comprehensive listing of its contents to ease the search for information needed. We also included some new results relating to different aspects of angular momentum theory which have recently appeared in journals. Unfortunately the limited volume of the present book prevented us from covering all the aforementioned results. We offer sincere apologies to the authors whose results we failed to include.

The monograph is a kind of handbook. Consequently the material is presented in concise form. Most of the formulas and relationships are given without proof. Their full derivation may be found in the literature\\
listed at the end of the text. Some results which have become generally known are given without references, and for this we also apologize.

The sequence adopted is as follows: chapter, section and subsection. Many chapters are self-contained and can be read independently of the others. Sections have double numbering: the first figure denotes the number of the chapter, the second, the number of the section. Equations are numbered within the confines of the section they are included in. When referring to an equation from the same section only the number of the equation is given, e.g., (3), (27); when reference is made to an equation from another section the numbers of the chapter, section and equation are given, e.g., Eq. 4.2.(17). A similar system is adopted when referring to individual subsections, e.g., Sec. 1.2.5. For convenience the book also contains a glossary of all symbols used in the text with references to the pages where their corresponding definitions are given. The list of references is divided into parts: the first part lists books and reviews; the second, papers on different subjects; the third, tables; the fourth, references added during translation.

The authors hope that many specialists will find in the book some fresh and interesting information. The material is prepared and arranged so as to make it useful to those less familiar with theory and for students of physics. These readers can effectively use the monograph as a supplementary text to their main courses.

For those who wish to thoroughly familiarize themselves with the fundamentals of angular momentum theory we recommend the excellent new book by L. Biedenharn and J. Louck [132] Angular Momentum in Quantum Physics. Theory and Applications.

The authors wish to express their deep appreciation to D. G. Yakovlev who took the trouble of reading the English translation of the book and gave some valuable suggestions on its preparation.

Leningrad
\chapter*{INTRODUCTION: BASIC CONCEPTS}\index{angular momentum!quantum theory of}
The evaluation of many physical quantities, such as expectation values of energy, electric and magnetic multipole moments, transition probabilities, etc., is considerably simplified by making use of the transformation properties of these quantities under coordinate rotation and inversion in three-dimensional space.\index{multipole moments}\index{transition probability}\index{expectation value}\index{rotation!of coordinate system}\index{coordinate inversion}

The transformation properties of physical quantities with respect to rotations reveal themselves either through rotations of the given physical system relative to some fixed reference frame or through rotations of the coordinate axes relative to the physical system.

Inversion characterizes the behavior of a physical quantity under transformation from a right-handed coordinate system to a left-handed one, or vice versa.

Many physical quantities, by their nature, are invariants under coordinate rotations. In particular, the properties of any closed physical system should be independent of rotations, as follows from the isotropy of space.\index{isotropy of space}\index{invariance!under rotations} As a consequence of this fundamental property of space, the total angular momentum of such a system is an integral of motion.\index{integral of motion}\index{angular momentum!conservation of}

A similar situation occurs with regard to coordinate inversion. Excluding phenomena connected with the weak interaction, a wealth of atomic, molecular and nuclear processes look alike in right-handed and left-handed coordinate systems. As a result of this "mirror" symmetry, quantum states of atoms, molecules, nuclei and elementary particles may be characterized by definite parity.\index{mirror symmetry}\index{parity}

Strictly speaking, a quantum mechanical wave function $\Psi(r)$ of any closed physical system may be characterized by four quantum numbers ( $\varepsilon, \pi, j, m$ ) which are the eigenvalues of four commuting operators: the Hamiltonian\index{wave function}\index{quantum numbers}\index{operators!commuting}\index{Hamiltonian}\isym{H-hat@$\widehat{H}$, Hamiltonian}\isym{Pr-hat@$\widehat{P}_{r}$, coordinate-inversion operator}\isym{J-hat@$\widehat{\vect{J}}$, total angular momentum operator} $\widehat{H}$, the parity operator $\widehat{P}_{\tau}$, the operator $\widehat{\vect{J}}^{2}$ of the square of the angular momentum and the operator $\widehat{J}_{z}$ of the projection of this momentum onto a quantization axis. Thus $\Psi(r) \equiv \Psi_{\varepsilon \pi \alpha j m}(r)$ obeys the equations

\[
\begin{gathered}
\widehat{H} \Psi_{\varepsilon \pi \alpha j m}(r)=\varepsilon \Psi_{\varepsilon \pi \alpha j m}(r), \\
\widehat{P}_{r} \Psi_{\varepsilon \pi \alpha j m}(r)=\pi \Psi_{e \pi \alpha j m}(r), \\
\widehat{\vect{J}}^{2} \Psi_{e \pi \alpha j m}(r)=j(j+1) \Psi_{\varepsilon \pi \alpha j m}(r), \\
\widehat{J}_{z} \Psi_{\varepsilon \pi \alpha j m}(r)=m \Psi_{\varepsilon \pi \alpha j m}(r),
\end{gathered}
\]

where $\alpha$ denotes all other quantum numbers (if available) and $r$ represents a variety of arguments.\\
$\varepsilon, \pi, j$ and $m$ are the integrals of motion not only for a closed system, but also for any system exposed to the action of an external spherically-symmetric field. Moreover, the ( $\varepsilon \pi j m$ )-representation appears to be very convenient for practical calculation even if the external field is not spherically-symmetric.\index{representation!$\varepsilon \pi j m$}\index{spherically-symmetric field}

For a given $j$, there exist $2 j+1$ wave functions which correspond to different $m$. These functions describe the quantum states of the system which differ only by the orientation of the angular momentum in space.

Under coordinate rotations these functions undergo linear mutual transformations which do not involve the functions of other quantum numbers ( $\varepsilon, \pi, \alpha, j$ ) and may be written as

\[
\Psi_{c \pi \alpha j m^{\prime}}\left(r^{\prime}\right)=\sum_{m} \Psi_{c \pi \alpha j m}(r) D_{m m^{\prime}}^{j}(\alpha, \beta, \gamma)
\]

where the transformations coefficients $D_{m m^{\prime}}^{j}(\alpha, \beta, \gamma)$ are called Wigner $D$-functions.\index{Wigner D-function@Wigner $D$-function}\index{rotation matrix}\index{Euler angles}\isym{DJMM@$D_{M M^{\prime}}^{J}$, Wigner $D$-function} They are the elements of the finite rotation matrix in the $j$-representation and depend on the Euler angles $\alpha, \beta, \gamma$ which determine rotation.

Let a quantum-mechanical system which possesses some fixed angular momentum $j$ and its projection $m$ consist of two subsystems, with certain angular momenta $j_{1}$ and $j_{2}$, respectively. In this case the wave function $\Psi_{j_{1} j_{2} j m}\left(r_{1}, r_{2}\right)$ may be constructed from the wave functions of subsystems according to the relation

\[
\Psi_{j_{1} j_{2} j m}\left(r_{1}, r_{2}\right)=\sum_{m_{1} m_{2}} \clebsch{j_{1}}{m_{1}}{j_{2}}{m_{2}}{j}{m} \Psi_{j_{1} m_{1}}\left(r_{1}\right) \Psi_{j_{2} m_{2}}\left(r_{2}\right)
\]

The quantities $\clebsch{j_{1}}{m_{1}}{j_{2}}{m_{2}}{j}{m}$ are called the Clebsch-Gordan coefficients.\index{Clebsch-Gordan coefficient}\index{addition of angular momenta}\isym{C-clebsch@$\clebsch{a}{\alpha}{b}{\beta}{c}{\gamma}$, Clebsch-Gordan coefficient} They are very important in quantum mechanics because they allow one to construct wave functions for various complex systems (nuclei, atoms, molecules, etc.). The addition of many angular momenta into some resultant total angular momentum may be performed in different ways, or, in other words, in accordance with different coupling schemes of the momenta in question.\index{coupling schemes} Each scheme may be associated with a certain representation of these angular momenta. The unitary transformations which relate various representations and describe the recoupling of angular momenta are realized by matrices expressed in terms of the $6 j$ - $9 j$-symbols and others $3 n j$-symbols of higher order.\index{recoupling}\index{recoupling coefficients}\index{6j symbol@$6j$ symbol}\index{9j symbol@$9j$ symbol}\index{3nj symbols@$3nj$ symbols}

One of the principal concepts of the quantum theory of angular momentum is that of an irreducible tensor.\index{irreducible tensor}\index{rank!of an irreducible tensor} By definition, an irreducible tensor of rank $\lambda$ has $2 \lambda+1$ components which transform under coordinate rotation as

\[
\mathfrak{M}_{\lambda \mu}=\sum_{\mu^{\prime}} \mathfrak{M}_{\lambda \mu^{\prime}} D_{\mu^{\prime} \mu}^{\lambda}(\alpha, \beta, \gamma)
\]

In particular, the wave functions $\Psi_{j m}$ with fixed $j$ but different $m$ constitute an irreducible tensor of rank $j$. Moreover, irreducible tensors may be constructed from various physical quantities. For instance, energy is a tensor of zero rank (scalar), spin and magnetic moments are tensors of first rank (vectors), the quadrupole moment is a tensor of second rank, etc.\index{scalar}\index{magnetic moment}\index{quadrupole moment} Generally, any physical quantity or operator which corresponds to these quantities may be represented as a linear combination of irreducible tensors.

The introduction of irreducible tensors into the analysis of physical quantities or corresponding operators substantially simplifies the evaluation of matrix elements of irreducible tensor operators,\index{matrix elements}\index{irreducible tensor operator}\isym{Mfrak-hat@$\widehat{\mathfrak{M}}_{k \kappa}$, irreducible tensor operator}

\[
\braOket{\varepsilon^{\prime} \pi^{\prime} \alpha^{\prime} j^{\prime} m^{\prime}}{\widehat{\mathfrak{M}}_{\lambda \mu}}{\varepsilon \pi \alpha j m}=\int \Psi_{c^{\prime} \pi^{\prime} \alpha^{\prime} j^{\prime} m^{\prime}}^{*}(r) \widehat{\mathfrak{M}}_{\lambda \mu} \Psi_{c \pi \alpha j m}(r) d r
\]

where $\int \ldots d r$ denotes integration over continuous variables and summation over discrete ones. These matrix elements determine expectation values of physical quantities in definite quantum states, the probabilities of quantum transitions between various states, etc.

According to the Wigner-Eckart theorem, all the dependence of matrix elements on the orientation of coordinate axes\index{Wigner-Eckart theorem}, i.e., on quantum numbers $m, m^{\prime}, \mu$ which determine the projections of angular momenta, is entirely contained in the Clebsch-Gordan coefficient

\[
\braOket{\varepsilon^{\prime} \pi^{\prime} \alpha^{\prime} j^{\prime} m^{\prime}}{\widehat{\mathfrak{M}}_{\lambda \mu}}{\varepsilon \pi \alpha j m}=\clebsch{j}{m}{\lambda}{\mu}{j^{\prime}}{m^{\prime}}\braOketred{\varepsilon^{\prime} \pi^{\prime} \alpha^{\prime} j^{\prime}}{\widehat{\mathfrak{M}}_{\lambda}}{\varepsilon \pi \alpha j} .
\]

In this case $\braOketred{\varepsilon^{\prime} \pi^{\prime} \alpha^{\prime} j^{\prime}}{\widehat{\mathfrak{M}}_{\lambda}}{\varepsilon \pi \alpha j}$ is an invariant factor called a reduced matrix element.\index{reduced matrix element}\index{matrix elements!reduced}\isym{reduced-me@$\braOketred{n^{\prime} j^{\prime}}{\widehat{\mathfrak{M}}_{k}}{n j}$, reduced matrix element} The Wigner-Eckart theorem allows one to reduce evaluation of transition probabilities, angular distributions, polarizations, etc., to the calculation of standard sums of the vector addition coefficients and recoupling coefficients.\index{vector-addition coefficients}\index{angular distribution}\index{polarized state}

All the facts mentioned above reveal that the quantum theory of angular momentum provides the formalism which is universal and extremely convenient for various practical calculations.

\chapter{Elements of Vector and Tensor Theory}\label{chap:1}
The theory of angular momenta and irreducible tensors represents, in principle, a development of the classical theory of vectors and tensors. In this chapter only the basic definitions and relations of the vector and tensor theory are represented which will be used throughout. For more detailed analysis see corresponding monographs (e.g., Refs. \cite{ref011, ref034, ref035}).

\section{Coordinate Systems. Basis Vectors}\label{sec:1.1}
In the quantum theory of angular momentum cartesian, polar and spherical coordinate systems are widely used.

\subsection{Cartesian Coordinate System}\label{sec:1.1.1}
\begin{figure}[tbh]
\begin{center}
\includegraphics{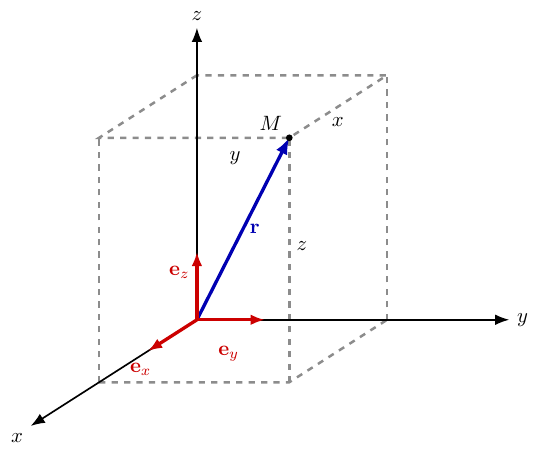}
\caption{Cartesian coordinate system.}
\label{chap1:fig:1}
\end{center}
\end{figure}

\index{Cartesian coordinate system}\index{coordinate system!Cartesian}\index{basis vectors!Cartesian}
In a rectangular cartesian coordinate system the position of a point is specified by three real numbers \(x\), \(y, z\) which represent the distances between the point and coordinate planes (Fig.~\ref{chap1:fig:1}). The position vector (radius vector) of a point \(\vect{r}\) may be written as\index{position vector}\index{radius vector|see{position vector}}
\begin{equation}
\vect{r}=x \vect{e}_{x}+y \vect{e}_{y}+z \vect{e}_{z} .\label{eq:1:1:1}
\end{equation}

The covariant cartesian basis (base) vectors \(\vect{e}_{x}, \vect{e}_{y}, \vect{e}_{z}\) form a real orthonormal basis
\begin{align}
\vect{e}_{i} \vect{e}_{k}=\delta_{i k}, & (i, k=x, y, z),  \label{eq:1:1:2}\\
\vect{e}_{i}^{*}=\vect{e}_{i}, & (i=x, y, z) . \label{eq:1:1:3}
\end{align}
The contravariant cartesian basis (base) vectors \(\vect{e}^{i}(i=x, y, z)\) coincide with the covariant ones
\begin{equation}
\vect{e}^{i}=\vect{e}_{i} . \label{eq:1:1:4}
\end{equation}
Throughout this book the right-handed coordinate system will be used. In this system
\begin{equation}
\left[\vect{e}_{i} \times \vect{e}_{k}\right]=\varepsilon_{i k l} \vect{e}_{l}, \quad(i, k, l=x, y, z), \label{eq:1:1:5}
\end{equation}
\begin{equation}
\varepsilon_{i k l}=\left[\vect{e}_{i} \times \vect{e}_{k}\right] \vect{e}_{l} . \label{eq:1:1:6}
\end{equation}
A detailed form of \eqref{eq:1:1:5} is
\begin{equation}
\left[\vect{e}_{x} \times \vect{e}_{y}\right]=\vect{e}_{z}, \quad\left[\vect{e}_{y} \times \vect{e}_{z}\right]=\vect{e}_{x}, \quad\left[\vect{e}_{z} \times \vect{e}_{x}\right]=\vect{e}_{y} . \label{eq:1:1:7}
\end{equation}
\subsection{Polar Coordinate System}\label{sec:1.1.2}
\begin{figure}[tbh]
\begin{center}
\includegraphics{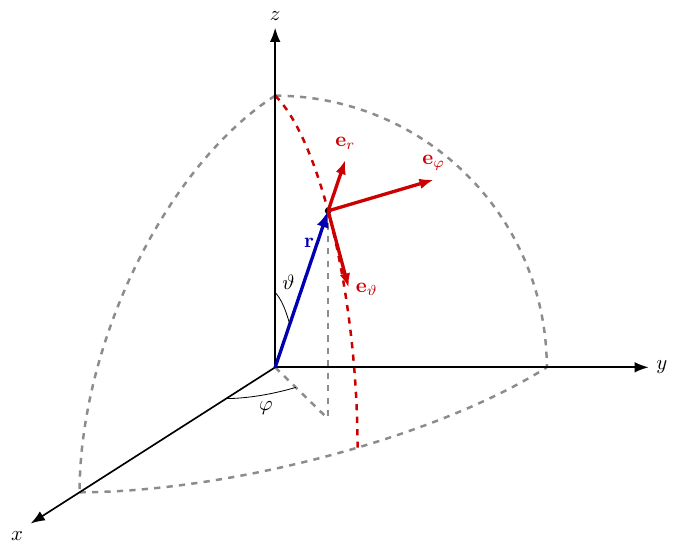}
\caption{Polar coordinate system.}
\label{chap1:fig:2}
\end{center}
\end{figure}

\index{polar coordinate system}\index{coordinate system!polar}\index{basis vectors!polar}\index{polar angles}\index{colatitude}\index{longitude}
In a polar coordinate system \({ }^{1}\) the position of a point is determined by \(r, \vartheta, \varphi\), where \(r\) is the position vector length, \(\vartheta\) is the colatitude, and \(\varphi\) is the longitude (Fig.~\ref{chap1:fig:2}). The angles \(\vartheta\) and \(\varphi\) are called the polar angles of vector \(\vect{r}\). The relations between cartesian and polar coordinates are

\begin{equation}
\begin{array}{lll}
x=r \sin \vartheta \cos \varphi, & r=\sqrt{x^{2}+y^{2}+z^{2}}, & 0 \leq r<\infty, \\
y=r \sin \vartheta \sin \varphi, & \vartheta=\arccos \frac{z}{\sqrt{x^{2}+y^{2}+z^{2}}}, & 0 \leq \vartheta \leq \pi,  \label{eq:1:1:8}\\
z=r \cos \vartheta, & \varphi=\arccos \frac{x}{\sqrt{x^{2}+y^{2}}},\left(\tan \varphi=\frac{y}{x}\right), & 0 \leq \varphi<2 \pi .
\end{array}
\end{equation}
The position vector \(r\) may be written as
\begin{equation}
\vect{r}=r \vect{e}_{r} . \label{eq:1:1:9}
\end{equation}
\footnotetext{\({ }^{1}\) Note that this coordinate system is often called ``spherical''. To avoid misunderstanding we prefer to call it the ``polar'' system reserving the name ``spherical'' only for the coordinate system considered in the next section.
}The covariant polar basis vectors \(\vect{e}_{r}, \vect{e}_{\vartheta}, \vect{e}_{\varphi}\) are shown in Fig.~\ref{chap1:fig:2}. They form a real orthonormal basis
\begin{align}
\vect{e}_{\alpha} \vect{e}_{\beta} & =\delta_{\alpha \beta}, \quad(\alpha, \beta=r, \vartheta, \varphi) , \label{eq:1:1:10}\\
\vect{e}_{\alpha}^{*} & =\vect{e}_{\alpha}, \quad(\alpha=r, \vartheta, \varphi) .\label{eq:1:1:11}
\end{align}
The contravariant polar basis vectors \(\vect{e}^{r}, \vect{e}^{\vartheta}, \vect{e}^{\varphi}\) coincide with the covariant ones
\begin{equation}
\vect{e}^{\alpha}=\vect{e}_{\alpha} \quad(\alpha=r, \vartheta, \varphi) . \label{eq:1:1:12}
\end{equation}
The unit vectors \(\vect{e}_{r}, \vect{e}_{\vartheta}, \vect{e}_{\varphi}\) form the right-handed basis
\begin{equation}
\left[\vect{e}_{r} \times \vect{e}_{\vartheta}\right]=\vect{e}_{\varphi}, \quad\left[\vect{e}_{\vartheta} \times \vect{e}_{\varphi}\right]=\vect{e}_{r}, \quad\left[\vect{e}_{\varphi} \times \vect{e}_{r}\right]=\vect{e}_{\vartheta} . \label{eq:1:1:13}
\end{equation}
The polar basis vectors \(\vect{e}_{r}, \vect{e}_{\vartheta}, \vect{e}_{\varphi}\), contrary to the cartesian ones, depend on the angles \(\vartheta, \varphi\). This should be taken into account when evaluating the derivatives
\begin{equation}
\begin{aligned}
&\frac{\partial}{\partial r} \vect{e}_{r}=0, && \frac{\partial}{\partial r} \vect{e}_{\vartheta}=0, && \frac{\partial}{\partial r} \vect{e}_{\varphi}=0, \\
&\frac{\partial}{\partial \vartheta} \vect{e}_{r}=\vect{e}_{\vartheta}, && \frac{\partial}{\partial \vartheta} \vect{e}_{\vartheta}=-\vect{e}_{r}, && \frac{\partial}{\partial \vartheta} \vect{e}_{\varphi}=0, \\
&\frac{\partial}{\partial \varphi} \vect{e}_{r}=\vect{e}_{\varphi} \sin \vartheta, && \frac{\partial}{\partial \varphi} \vect{e}_{\vartheta}=\vect{e}_{\varphi} \cos \vartheta, && \frac{\partial}{\partial \varphi} \vect{e}_{\varphi}=-\vect{e}_{r} \sin \vartheta-\vect{e}_{\vartheta} \cos \vartheta . \label{eq:1:1:14}
\end{aligned}
\end{equation}
The results of applying the \(\boldsymbol{\nabla}\) operator (see Sec. 1.3) to the polar basis vectors are presented in the form
\begin{gather}
\left(\boldsymbol{\nabla} \cdot \vect{e}_{r}\right)=\frac{2}{r}, \quad\left(\boldsymbol{\nabla} \cdot \vect{e}_{\vartheta}\right)=\frac{1}{r} \cot \vartheta, \quad\left(\boldsymbol{\nabla} \cdot \vect{e}_{\varphi}\right)=0,  \label{eq:1:1:15}\\
{\left[\boldsymbol{\nabla} \times \vect{e}_{r}\right]=0, \quad\left[\boldsymbol{\nabla} \times \vect{e}_{\vartheta}\right]=\frac{1}{r} \vect{e}_{\varphi}, \quad\left[\boldsymbol{\nabla} \times \vect{e}_{\varphi}\right]=\frac{1}{r} \cot \vartheta \vect{e}_{r}-\frac{1}{r} \vect{e}_{\vartheta}} .\label{eq:1:1:16}
\end{gather}
\subsection{Spherical Coordinate System}\label{sec:1.1.3}\index{spherical coordinate system}\index{coordinate system!spherical}\index{basis vectors!spherical}\index{spherical components}\index{components!covariant}\index{components!contravariant}
Spherical coordinates are widely used in the angular momentum theory.\\
The covariant spherical coordinates \(x_{\mu}\) (with \(\mu= \pm 1,0\) ) are defined by the relations
\begin{align}
x_{+1} & =-\frac{1}{\sqrt{2}}(x+i y)=-\frac{1}{\sqrt{2}} r \sin \vartheta e^{i \varphi},  \notag\\
x_{0} & =z=r \cos \vartheta , \label{eq:1:1:17}\\
x_{-1} & =\frac{1}{\sqrt{2}}(x-i y)=\frac{1}{\sqrt{2}} r \sin \vartheta e^{-i \varphi}. \notag
\end{align}
The contravariant spherical coordinates \(x^{\mu}\) (where \(\mu= \pm 1,0\) ) are given by
\begin{align}
x^{+1} & =-\frac{1}{\sqrt{2}}(x-i y)=-\frac{1}{\sqrt{2}} r \sin \vartheta e^{-i \varphi} , \notag\\
x^{0} & =z=r \cos \vartheta , \label{eq:1:1:18}\\
x^{-1} & =\frac{1}{\sqrt{2}}(x+i y)=\frac{1}{\sqrt{2}} r \sin \vartheta e^{i \varphi} .\notag
\end{align}
The relations between covariant and contravariant spherical coordinates are as follows
\begin{equation}
\begin{aligned}
&x^{\mu}=(-1)^{\mu} x_{-\mu}, & & x_{\mu}=(-1)^{\mu} x^{-\mu}, \\
&x^{\mu}=x_{\mu}^{*}, & & x_{\mu}=x^{\mu *}, 
\end{aligned} \quad(\mu= \pm 1,0) .\label{eq:1:1:19}
\end{equation}
The covariant spherical basis vectors \(\vect{e}_{\mu}(\mu= \pm 1,0)\) are defined by
\begin{align}
\vect{e}_{+1} & =-\frac{1}{\sqrt{2}}\left(\vect{e}_{x}+i \vect{e}_{y}\right),  \notag\\
\vect{e}_{0} & =\vect{e}_{z},  \label{eq:1:1:20}\\
\vect{e}_{-1} & =\frac{1}{\sqrt{2}}\left(\vect{e}_{x}-i \vect{e}_{y}\right) . \notag
\end{align}
The contravariant spherical basis vectors \(\vect{e}^{\mu}(\mu= \pm 1,0)\) are given by
\begin{align}
\vect{e}^{+1} & =-\frac{1}{\sqrt{2}}\left(\vect{e}_{x}-i \vect{e}_{y}\right) , \notag\\
\vect{e}^{0} & =\vect{e}_{z},  \label{eq:1:1:21}\\
\vect{e}^{-1} & =\frac{1}{\sqrt{2}}\left(\vect{e}_{x}+i \vect{e}_{y}\right) .\notag
\end{align}
Relations between the covariant and contravariant spherical basis vectors read
\begin{equation}
\begin{array}{lll}
\vect{e}^{\mu}=(-1)^{\mu} \vect{e}_{-\mu}, & \vect{e}_{\mu}=(-1)^{\mu} \vect{e}^{-\mu}, \\
\vect{e}^{\mu}=\vect{e}_{\mu}^{*}, & \vect{e}_{\mu}=\vect{e}^{\mu *}, & (\mu= \pm 1,0). \label{eq:1:1:22}
\end{array}
\end{equation}
The spherical basis vectors form a complex orthonormal basis
\begin{equation}
\vect{e}_{\mu} \vect{e}^{\nu}=\vect{e}_{\mu} \vect{e}_{\nu}^{*}=\delta_{\mu \nu}, \quad(\mu, \nu= \pm 1,0) .\label{eq:1:1:23}
\end{equation}
Vector products of spherical basis vectors may be written with the use of the Clebsch-Gordan coefficients (see Chap. 8) in the form\index{Clebsch-Gordan coefficient}
\begin{equation}
  \begin{aligned}
& \vect{e}_{\mu} \times \vect{e}_{\nu}=i \sqrt{2} \clebsch{1}{\mu}{1}{\nu}{1}{\lambda} \vect{e}_{\lambda}, \\
& \vect{e}^{\mu} \times \vect{e}^{\nu}=-i \sqrt{2} \clebsch{1}{\mu}{1}{\nu}{1}{\lambda} \vect{e}^{\lambda},
\end{aligned}
 \quad(\mu, \nu, \lambda= \pm 1,0).
\end{equation}
One may also rewrite these formulas in a form similar to \eqref{eq:1:1:5}
\begin{equation}
\begin{aligned}
& \vect{e}_{\mu} \times \vect{e}_{\nu}=-i \varepsilon_{\mu \nu \lambda} \vect{e}^{\lambda}, \quad(\mu, \nu, \lambda= \pm 1,0) \\
& \vect{e}^{\mu} \times \vect{e}^{\nu}=i \varepsilon_{\mu \nu \lambda} \vect{e}_{\lambda}, 
\end{aligned}
 \quad(\mu, \nu, \lambda= \pm 1,0),
\label{eq:1:1:25}
\end{equation}
where \(\varepsilon_{\mu \nu \lambda}=+1\) if the combination of indices \(\mu, \nu, \lambda\) is obtained by an even permutation of \(+1,0,-1, \varepsilon_{\mu \nu \lambda}=-1\) for an odd permutation, and \(\varepsilon_{\mu \nu \lambda}=0\) if at least two indices among \(\mu, \nu, \lambda\) are equal.

A detailed form of \eqref{eq:1:1:25} is as follows:
\begin{equation}
\begin{aligned}
&\vect{e}_{+1} \times \vect{e}_{0}=i \vect{e}_{+1}, && \vect{e}_{0} \times \vect{e}_{-1}=i \vect{e}_{-1}, \\
&\vect{e}^{0} \times \vect{e}_{+1}^{+1}=i \vect{e}_{-1}^{+1}=i \vect{e}_{0}^{+} && \vect{e}^{-1} \times \vect{e}^{0}=i \vect{e}^{-1},\\
&\vect{e}^{-1} \times \vect{e}^{+1}=i \vect{e}^{0}.
\end{aligned}
  \label{eq:1:1:26}
\end{equation}
Covariant and contravariant spherical components (see Sec. 1.2) of the basis vectors \(\vect{e}_{\mu}\) and \(\vect{e}^{\mu}\) are given by
\begin{equation}
\begin{aligned}
&{\left[\vect{e}_{\mu}\right]_{\sigma}=(-1)^{\sigma} \delta_{\sigma-\mu},} && {\left[\vect{e}_{\mu}\right]^{\sigma}=\delta_{\sigma \mu}} \\
&{\left[\vect{e}^{\mu}\right]_{\sigma}=\delta_{\sigma \mu},} && {\left[\vect{e}^{\mu}\right]^{\sigma}=(-1)^{\sigma} \delta_{\sigma-\mu}}. \label{eq:1:1:27}
\end{aligned}
\end{equation}
\subsection{Helicity Basis Vectors}\label{sec:1.1.4}\index{helicity basis vectors}\index{basis vectors!helicity}
By analogy with Eqs. \eqref{eq:1:1:20}, \eqref{eq:1:1:21} one may construct the following combinations of the polar basis vectors \(\vect{e}_{r}, \vect{e}_{\vartheta}, \vect{e}_{\varphi}\)
\begin{align}
\vect{e}_{+1}^{\prime} & =-\frac{1}{\sqrt{2}}\left(\vect{e}_{\vartheta}+i \vect{e}_{\varphi}\right), & \vect{e}^{\prime+1} & =-\frac{1}{\sqrt{2}}\left(\vect{e}_{\vartheta}-i \vect{e}_{\varphi}\right),  \notag\\
\vect{e}_{0}^{\prime} & =\vect{e}_{r}, & \vect{e}^{\prime 0} & =\vect{e}_{r},  \label{eq:1:1:28}\\
\vect{e}_{-1}^{\prime} & =\frac{1}{\sqrt{2}}\left(\vect{e}_{\vartheta}-i \vect{e}_{\varphi}\right), & \vect{e}^{\prime-1} & =\frac{1}{\sqrt{2}}\left(\vect{e}_{\vartheta}+i \vect{e}_{\varphi}\right) . \notag
\end{align}
The vectors \(\vect{e}_{\mu}^{\prime}(\mu= \pm 1,0)\) are called the covariant helicity basis vectors and \(\vect{e}^{\prime \mu}(\mu= \pm 1,0)\) are called the contravariant helicity basis vectors (the explanation of the term ``helicity'' is given below in Sec. 6.3.6).

The helicity basis vectors \(\vect{e}_{\mu}^{\prime}\) and \(\vect{e}^{\prime \mu}\) satisfy the same relations \eqref{eq:1:1:22}-\eqref{eq:1:1:26} as the spherical basis vectors \(\vect{e}_{\mu}\) and \(\vect{e}^{\mu}\).

\subsection{Relations Between Different Basis Vectors}\label{sec:1.1.5}\index{basis vectors!relations between}
(a) Cartesian and Polar Basis Vectors
\begin{align}
& \vect{e}_{x}=\vect{e}_{r} \sin \vartheta \cos \varphi+\vect{e}_{\vartheta} \cos \vartheta \cos \varphi-\vect{e}_{\varphi} \sin \varphi,  \notag\\
& \vect{e}_{y}=\vect{e}_{r} \sin \vartheta \sin \varphi+\vect{e}_{\vartheta} \cos \vartheta \sin \varphi+\vect{e}_{\varphi} \cos \varphi,  \label{eq:1:1:29}\\
& \vect{e}_{z}=\vect{e}_{r} \cos \vartheta-\vect{e}_{\vartheta} \sin \vartheta . \notag
\end{align}
Cartesian and Spherical Basis Vectors
\begin{align}
& \vect{e}_{x}=\frac{1}{\sqrt{2}}\left(\vect{e}_{-1}-\vect{e}_{+1}\right)=\frac{1}{\sqrt{2}}\left(\vect{e}^{-1}-\vect{e}^{+1}\right),  \notag\\
& \vect{e}_{y}=\frac{i}{\sqrt{2}}\left(\vect{e}_{-1}+\vect{e}_{+1}\right)=-\frac{i}{\sqrt{2}}\left(\vect{e}^{-1}+\vect{e}^{+1}\right),  \label{eq:1:1:30}\\
& \vect{e}_{z}=\vect{e}_{0}=\vect{e}^{0} . \notag
\end{align}
Cartesian and Helicity Covariant Basis Vectors
\begin{align}
& \vect{e}_{x}=-\vect{e}_{+1}^{\prime} \frac{1}{\sqrt{2}}(\cos \vartheta \cos \varphi+i \sin \varphi)+\vect{e}_{0}^{\prime} \sin \vartheta \cos \varphi+\vect{e}_{-1}^{\prime} \frac{1}{\sqrt{2}}(\cos \vartheta \cos \varphi-i \sin \varphi),  \notag\\
& \vect{e}_{y}=-\vect{e}_{+1}^{\prime} \frac{1}{\sqrt{2}}(\cos \vartheta \sin \varphi-i \cos \varphi)+\vect{e}_{0}^{\prime} \sin \vartheta \sin \varphi+\vect{e}_{-1}^{\prime} \frac{1}{\sqrt{2}}(\cos \vartheta \sin \varphi+i \cos \varphi),  \label{eq:1:1:31}\\
& \vect{e}_{x}=\vect{e}_{+1}^{\prime} \frac{1}{\sqrt{2}} \sin \vartheta+\vect{e}_{0}^{\prime} \cos \vartheta-\vect{e}_{-1}^{\prime} \frac{1}{\sqrt{2}} \sin \vartheta . \notag
\end{align}
Cartesian and Helicity Contravariant Basis Vectors
\begin{align}
& \vect{e}_{x}=-\vect{e}^{\prime+1} \frac{1}{\sqrt{2}}(\cos \vartheta \cos \varphi-i \sin \varphi)+\vect{e}^{\prime 0} \sin \vartheta \cos \varphi+\vect{e}^{\prime-1} \frac{1}{\sqrt{2}}(\cos \vartheta \cos \varphi+i \sin \varphi),  \notag\\
& \vect{e}_{y}=-\vect{e}^{\prime+1} \frac{1}{\sqrt{2}}(\cos \vartheta \sin \varphi+i \cos \varphi)+\vect{e}^{\prime 0} \sin \vartheta \sin \varphi+\vect{e}^{\prime-1} \frac{1}{\sqrt{2}}(\cos \vartheta \sin \varphi-i \cos \varphi),  \label{eq:1:1:32}\\
& \vect{e}_{z}=\vect{e}^{\prime+1} \frac{1}{\sqrt{2}} \sin \vartheta+\vect{e}^{\prime 0} \cos \vartheta-\vect{e}^{\prime-1} \frac{1}{\sqrt{2}} \sin \vartheta . \notag
\end{align}
(b) Polar and Cartesian Basis Vectors
\begin{align}
\vect{e}_{r} & =\vect{e}_{x} \sin \vartheta \cos \varphi+\vect{e}_{y} \sin \vartheta \sin \varphi+\vect{e}_{z} \cos \vartheta,  \notag\\
\vect{e}_{\vartheta} & =\vect{e}_{x} \cos \vartheta \cos \varphi+\vect{e}_{y} \cos \vartheta \sin \varphi-\vect{e}_{z} \sin \vartheta,  \label{eq:1:1:33}\\
\vect{e}_{\varphi} & =-\vect{e}_{x} \sin \varphi+\vect{e}_{y} \cos \varphi . \notag
\end{align}
Polar and Spherical Covariant Basis Vectors
\begin{align}
& \vect{e}_{r}=-\vect{e}_{+1} \frac{1}{\sqrt{2}} \sin \vartheta e^{-i \varphi}+\vect{e}_{0} \cos \vartheta+\vect{e}_{-1} \frac{1}{\sqrt{2}} \sin \vartheta e^{i \varphi},  \notag\\
& \vect{e}_{\vartheta}=-\vect{e}_{+1} \frac{1}{\sqrt{2}} \cos \vartheta e^{-i \varphi}-\vect{e}_{0} \sin \vartheta+\vect{e}_{-1} \frac{1}{\sqrt{2}} \cos \vartheta e^{i \varphi} ;  \label{eq:1:1:34}\\
& \vect{e}_{\varphi}=\vect{e}_{+1} \frac{i}{\sqrt{2}} e^{-i \varphi}+\vect{e}_{-1} \frac{i}{\sqrt{2}} e^{i \varphi} . \notag
\end{align}
Polar and Spherical Contravariant Basis Vectors
\begin{align}
& \vect{e}_{r}=-\vect{e}^{+1} \frac{1}{\sqrt{2}} \sin \vartheta e^{i \varphi}+\vect{e}^{0} \cos \vartheta+e^{-1} \frac{1}{\sqrt{2}} \sin \vartheta e^{-i \varphi},  \notag\\
& \vect{e}_{\vartheta}=-\vect{e}^{+1} \frac{1}{\sqrt{2}} \cos \vartheta e^{i \varphi}-\vect{e}^{0} \sin \vartheta+\vect{e}^{-1} \frac{1}{\sqrt{2}} \cos \vartheta e^{-i \varphi},  \label{eq:1:1:35}\\
& \vect{e}_{\varphi}=-\vect{e}^{+1} \frac{i}{\sqrt{2}} e^{i \varphi}-\vect{e}^{-1} \frac{i}{\sqrt{2}} e^{-i \varphi} . \notag
\end{align}
Polar and Helicity Basis Vectors
\begin{align}
& \vect{e}_{r}=\vect{e}_{0}^{\prime}=\vect{e}^{\prime 0},  \notag\\
& \vect{e}_{\vartheta}=\frac{1}{\sqrt{2}}\left(\vect{e}_{-1}^{\prime}-\vect{e}_{+1}^{\prime}\right)=\frac{1}{\sqrt{2}}\left(\vect{e}^{\prime-1}-\vect{e}^{\prime+1}\right),  \label{eq:1:1:36}\\
& \vect{e}_{\varphi}=\frac{i}{\sqrt{2}}\left(\vect{e}_{-1}^{\prime}+\vect{e}_{+1}^{\prime}\right)=-\frac{i}{\sqrt{2}}\left(\vect{e}^{\prime-1}+\vect{e}^{\prime+1}\right) . \notag
\end{align}
(c) Spherical and Cartesian Basis Vectors
\begin{align}
\vect{e}_{+1} & =-\frac{1}{\sqrt{2}}\left(\vect{e}_{x}+i \vect{e}_{y}\right), & \vect{e}^{+1} & =-\frac{1}{\sqrt{2}}\left(\vect{e}_{x}-i \vect{e}_{y}\right)  \notag\\
\vect{e}_{0} & =\vect{e}_{z}, & \vect{e}^{0} & =\vect{e}_{z},  \label{eq:1:1:37}\\
\vect{e}_{-1} & =\frac{1}{\sqrt{2}}\left(\vect{e}_{x}-i \vect{e}_{y}\right), & \vect{e}^{-1} & =\frac{1}{\sqrt{2}}\left(\vect{e}_{x}+i \vect{e}_{y}\right). \notag
\end{align}
Spherical Covariant and Polar Basis Vectors
\begin{align}
\vect{e}_{+1} & =-\vect{e}_{r} \frac{1}{\sqrt{2}} \sin \vartheta e^{i \varphi}-\vect{e}_{\vartheta} \frac{1}{\sqrt{2}} \cos \vartheta e^{i \varphi}-\vect{e}_{\varphi} \frac{i}{\sqrt{2}} e^{i \varphi},  \notag\\
\vect{e}_{0} & =\vect{e}_{r} \cos \vartheta-\vect{e}_{\vartheta} \sin \vartheta,  \label{eq:1:1:38}\\
\vect{e}_{-1} & =\vect{e}_{r} \frac{i}{\sqrt{2}} \sin \vartheta e^{-i \varphi}+\vect{e}_{\vartheta} \frac{1}{\sqrt{2}} \cos \vartheta e^{-i \varphi}-\vect{e}_{\varphi} \frac{i}{\sqrt{2}} e^{-i \varphi} . \notag
\end{align}
Spherical Contravariant and Polar Basis Vectors
\begin{align}
\vect{e}^{+1} & =-\vect{e}_{r} \frac{1}{\sqrt{2}} \sin \vartheta e^{-i \varphi}-\vect{e}_{\vartheta} \frac{1}{\sqrt{2}} \cos \vartheta e^{-i \varphi}+\vect{e}_{\varphi} \frac{i}{\sqrt{2}} e^{-i \varphi},  \notag\\
\vect{e}^{0} & =\vect{e}_{r} \cos \vartheta-\vect{e}_{\vartheta} \sin \vartheta,  \label{eq:1:1:39}\\
\vect{e}^{-1} & =\vect{e}_{r} \frac{1}{\sqrt{2}} \sin \vartheta e^{i \varphi}+\vect{e}_{\vartheta} \frac{1}{\sqrt{2}} \cos \vartheta e^{i \varphi}+\vect{e}_{\varphi} \frac{i}{\sqrt{2}} e^{i \varphi} .\notag
\end{align}
Spherical Covariant and Spherical Contravariant Basis Vectors
\begin{align}
\vect{e}_{+1} & =-\vect{e}^{-1}, & \vect{e}^{+1} & =-\vect{e}_{-1},  \notag\\
\vect{e}_{0} & =\vect{e}^{0}, & \vect{e}^{0} & =\vect{e}_{0} , \label{eq:1:1:40}\\
\vect{e}_{-1} & =-\vect{e}^{+1}, & \vect{e}^{-1} & =-\vect{e}_{+1}. \notag
\end{align}
Spherical Covariant and Helicity Covariant Basis Vectors
\begin{align}
\vect{e}_{+1} & =\vect{e}_{+1}^{\prime} \frac{1+\cos \vartheta}{2} e^{i \varphi}-\vect{e}_{0}^{\prime} \frac{\sin \vartheta}{\sqrt{2}} e^{i \varphi}+\vect{e}_{-1}^{\prime} \frac{1-\cos \vartheta}{2} e^{i \varphi},  \notag\\
\vect{e}_{0} & =\vect{e}_{+1}^{\prime} \frac{\sin \vartheta}{\sqrt{2}}+\vect{e}_{0}^{\prime} \cos \vartheta-\vect{e}_{-1}^{\prime} \frac{\sin \vartheta}{\sqrt{2}},  \label{eq:1:1:41}\\
\vect{e}_{-1} & =\vect{e}_{+1}^{\prime} \frac{1-\cos \vartheta}{2} e^{-i \varphi}+\vect{e}_{0}^{\prime} \frac{\sin \vartheta}{\sqrt{2}} \vect{e}^{-i \varphi}+\vect{e}_{-1}^{\prime} \frac{1+\cos \vartheta}{2} e^{-i \varphi} . \notag
\end{align}
Spherical Contravariant and Helicity Covariant Basis Vectors
\begin{align}
\vect{e}^{+1} & =-\vect{e}_{+1}^{\prime} \frac{1-\cos \vartheta}{2} e^{-i \varphi}-\vect{e}_{0}^{\prime} \frac{\sin \vartheta}{\sqrt{2}} \vect{e}^{-i \varphi}-\vect{e}_{-1}^{\prime} \frac{1+\cos \vartheta}{2} e^{-i \varphi} , \notag\\
\vect{e}^{0} & =\vect{e}_{+1}^{\prime} \frac{\sin \vartheta}{\sqrt{2}}+\vect{e}_{0}^{\prime} \cos \vartheta-\vect{e}_{-1}^{\prime} \frac{\sin \vartheta}{\sqrt{2}}  ,\label{eq:1:1:42}\\
\vect{e}^{-1} & =-\vect{e}_{+1}^{\prime} \frac{1+\cos \vartheta}{2} e^{i \varphi}+\vect{e}_{0}^{\prime} \frac{\sin \vartheta}{\sqrt{2}} e^{i \varphi}-\vect{e}_{-1}^{\prime} \frac{1-\cos \vartheta}{2} e^{i \varphi} \notag.
\end{align}
Spherical Covariant and Helicity Contravariant Basis Vectors
\begin{align}
\vect{e}_{+1} & =-\vect{e}^{\prime+1} \frac{1-\cos \vartheta}{2} e^{i \varphi}-\vect{e}^{\prime 0} \frac{\sin \vartheta}{\sqrt{2}} e^{i \varphi}-\vect{e}^{\prime-1} \frac{1+\cos \vartheta}{2} e^{i \varphi},  \notag\\
\vect{e}_{0} & =\vect{e}^{\prime+1} \frac{\sin \vartheta}{\sqrt{2}}+\vect{e}^{\prime 0} \cos \vartheta-\vect{e}^{\prime-1} \frac{\sin \vartheta}{\sqrt{2}},  \label{eq:1:1:43}\\
\vect{e}_{-1} & =-\vect{e}^{\prime+1} \frac{1+\cos \vartheta}{2} e^{-i \varphi}+\vect{e}^{\prime 0} \frac{\sin \vartheta}{\sqrt{2}} e^{-i \varphi}-\vect{e}^{\prime-1} \frac{1-\cos \vartheta}{2} e^{-i \varphi} . \notag
\end{align}
Spherical Contravariant and Helicity Contravariant Basis Vectors
\begin{align}
\vect{e}^{+1} & =\vect{e}^{\prime+1} \frac{1+\cos \vartheta}{2} e^{-i \varphi}-\vect{e}^{\prime 0} \frac{\sin \vartheta}{\sqrt{2}} e^{-i \varphi}+\vect{e}^{\prime-1} \frac{1-\cos \vartheta}{2} e^{-i \varphi},  \notag\\
\vect{e}^{0} & =\vect{e}^{\prime+1} \frac{\sin \vartheta}{\sqrt{2}}+\vect{e}^{\prime 0} \cos \vartheta-\vect{e}^{\prime-1} \frac{\sin \vartheta}{\sqrt{2}},  \label{eq:1:1:44}\\
\vect{e}^{-1} & =\vect{e}^{\prime+1} \frac{1-\cos \vartheta}{2} e^{i \varphi}+\vect{e}^{\prime 0} \frac{\sin \vartheta}{\sqrt{2}} e^{i \varphi}+\vect{e}^{\prime-1} \frac{1+\cos \vartheta}{2} e^{i \varphi} . \notag
\end{align}
Equations \eqref{eq:1:1:41}-\eqref{eq:1:1:44} may be written in a more compact form using the Wigner \(D\)-functions (see Chap. 4).
\begin{gather}
\vect{e}_{\mu}=\sum_{\nu} D_{-\nu-\mu}^{1}(0, \vartheta, \varphi) \vect{e}_{\nu}^{\prime}=\sum_{\nu}(-1)^{\nu} D_{\nu-\mu}^{1}(0, \vartheta, \varphi) \vect{e}^{\prime \nu},  \notag\\
\vect{e}^{\mu}=(-1)^{\mu} \sum_{\nu} D_{-\nu \mu}^{1}(0, \vartheta, \varphi) \vect{e}_{\nu}^{\prime}=\sum_{\nu}(-1)^{\mu+\nu} D_{\nu \mu}^{1}(0, \vartheta, \varphi) \vect{e}^{\prime \nu} , \label{eq:1:1:45}\\
(\mu, \nu= \pm 1,0) .\notag
\end{gather}
(d) Helicity Covariant and Cartesian Basis Vectors
\begin{align}
\vect{e}_{+1}^{\prime} & =-\vect{e}_{x} \frac{1}{\sqrt{2}}(\cos \vartheta \cos \varphi-i \sin \varphi)-\vect{e}_{y} \frac{1}{\sqrt{2}}(\cos \vartheta \sin \varphi+i \cos \varphi)+\vect{e}_{z} \frac{1}{\sqrt{2}} \sin \vartheta , \notag\\
\vect{e}_{0}^{\prime} & =\vect{e}_{x} \sin \vartheta \cos \varphi+\vect{e}_{y} \sin \vartheta \sin \varphi+\vect{e}_{z} \cos \vartheta , \label{eq:1:1:46}\\
\vect{e}_{-1}^{\prime} & =\vect{e}_{x} \frac{1}{\sqrt{2}}(\cos \vartheta \cos \varphi+i \sin \varphi)+\vect{e}_{y} \frac{1}{\sqrt{2}}(\cos \vartheta \sin \varphi-i \cos \varphi)-\vect{e}_{z} \frac{1}{\sqrt{2}} \sin \vartheta. \notag
\end{align}
Helicity Contravariant and Cartesian Basis Vectors
\begin{align}
\vect{e}^{\prime+1} & =-\vect{e}_{x} \frac{1}{\sqrt{2}}(\cos \vartheta \cos \varphi+i \sin \varphi)-\vect{e}_{y} \frac{1}{\sqrt{2}}(\cos \vartheta \sin \varphi-i \cos \varphi)+\vect{e}_{z} \frac{1}{\sqrt{2}} \sin \vartheta  ,\notag\\
\vect{e}^{\prime 0} & =\vect{e}_{x} \sin \vartheta \cos \varphi+\vect{e}_{y} \sin \vartheta \sin \varphi+\vect{e}_{z} \cos \vartheta  ,\label{eq:1:1:47}\\
\vect{e}^{\prime-1} & =\vect{e}_{x} \frac{1}{\sqrt{2}}(\cos \vartheta \cos \varphi-i \sin \varphi)+\vect{e}_{y} \frac{1}{\sqrt{2}}(\cos \vartheta \sin \varphi+i \cos \varphi)-\vect{e}_{z} \frac{1}{\sqrt{2}} \sin \vartheta .\notag
\end{align}
Helicity and Polar Basis Vectors
\begin{align}
\vect{e}_{+1}^{\prime} & =-\frac{1}{\sqrt{2}}\left(\vect{e}_{\vartheta}+i \vect{e}_{\varphi}\right), & \vect{e}^{\prime+1} & =-\frac{1}{\sqrt{2}}\left(\vect{e}_{\vartheta}-i \vect{e}_{\varphi}\right),  \notag\\
\vect{e}_{0}^{\prime} & =\vect{e}_{r}, & \vect{e}^{\prime 0} & =\vect{e}_{r},  \label{eq:1:1:48}\\
\vect{e}_{-1}^{\prime} & =\frac{1}{\sqrt{2}}\left(\vect{e}_{\vartheta}-i \vect{e}_{\varphi}\right), & \vect{e}^{\prime-1} & =\frac{1}{\sqrt{2}}\left(\vect{e}_{\vartheta}+i \vect{e}_{\varphi}\right) . \notag
\end{align}
Helicity Covariant and Spherical Covariant Basis Vectors
\begin{align}
\vect{e}_{+1}^{\prime} & =\vect{e}_{+1} \frac{1+\cos \vartheta}{2} e^{-i \varphi}+\vect{e}_{0} \frac{\sin \vartheta}{\sqrt{2}}+\vect{e}_{-1} \frac{1-\cos \vartheta}{2} e^{i \varphi},  \notag\\
\vect{e}_{0}^{\prime} & =-\vect{e}_{+1} \frac{\sin \vartheta}{\sqrt{2}} e^{-i \varphi}+\vect{e}_{0} \cos \vartheta+\vect{e}_{-1} \frac{\sin \vartheta}{\sqrt{2}} e^{i \varphi},  \label{eq:1:1:49}\\
\vect{e}_{-1}^{\prime} & =\vect{e}_{+1} \frac{1-\cos \vartheta}{2} e^{-i \varphi}-\vect{e}_{0} \frac{\sin \vartheta}{\sqrt{2}}+\vect{e}_{-1} \frac{1+\cos \vartheta}{2} e^{i \varphi} . \notag
\end{align}
Helicity Contravariant and Spherical Covariant Basis Vectors
\begin{align}
\vect{e}^{\prime+1} & =-\vect{e}_{+1} \frac{1-\cos \vartheta}{2} e^{-i \varphi}+\vect{e}_{0} \frac{\sin \vartheta}{\sqrt{2}}-\vect{e}_{-1} \frac{1+\cos \vartheta}{2} e^{i \varphi},  \notag\\
\vect{e}^{\prime 0} & =-\vect{e}_{+1} \frac{\sin \vartheta}{\sqrt{2}} e^{-i \varphi}+\vect{e}_{0} \cos \vartheta+\vect{e}_{-1} \frac{\sin \vartheta}{\sqrt{2}} e^{i \varphi},  \label{eq:1:1:50}\\
\vect{e}^{\prime-1} & =-\vect{e}_{+1} \frac{1+\cos \vartheta}{2} e^{-i \varphi}-\vect{e}_{0} \frac{\sin \vartheta}{\sqrt{2}}-\vect{e}_{-1} \frac{1-\cos \vartheta}{2} e^{i \varphi} . \notag
\end{align}
Helicity Covariant and Spherical Contravariant Basis Vectors
\begin{align}
\vect{e}_{+1}^{\prime} & =-\vect{e}^{+1} \frac{1-\cos \vartheta}{2} e^{i \varphi}+\vect{e}^{0} \frac{\sin \vartheta}{\sqrt{2}}-\vect{e}^{-1} \frac{1+\cos \vartheta}{2} e^{-i \varphi},  \notag\\
\vect{e}_{0}^{\prime} & =-\vect{e}^{+1} \frac{\sin \vartheta}{\sqrt{2}} e^{i \varphi}+\vect{e}^{0} \cos \vartheta+\vect{e}^{-1} \frac{\sin \vartheta}{\sqrt{2}} e^{-i \varphi},  \label{eq:1:1:51}\\
\vect{e}_{-1}^{\prime} & =-\vect{e}^{+1} \frac{1+\cos \vartheta}{2} e^{i \varphi}-\vect{e}^{0} \frac{\sin \vartheta}{\sqrt{2}}-\vect{e}^{-1} \frac{1-\cos \vartheta}{2} e^{-i \varphi} . \notag
\end{align}
Helicity Contravariant and Spherical Contravariant Basis Vectors
\begin{align}
\vect{e}^{\prime+1} & =\vect{e}^{+1} \frac{1+\cos \vartheta}{2} e^{i \varphi}+\vect{e}^{0} \frac{\sin \vartheta}{\sqrt{2}}+\vect{e}^{-1} \frac{1-\cos \vartheta}{2} e^{-i \varphi},  \notag\\
\vect{e}^{\prime 0} & =-\vect{e}^{+1} \frac{\sin \vartheta}{\sqrt{2}} e^{i \varphi}+\vect{e}^{0} \cos \vartheta+\vect{e}^{-1} \frac{\sin \vartheta}{\sqrt{2}} e^{-i \varphi},  \label{eq:1:1:52}\\
\vect{e}^{\prime-1} & =\vect{e}^{+1} \frac{1-\cos \vartheta}{2} e^{i \varphi}-\vect{e}^{0} \frac{\sin \vartheta}{\sqrt{2}}+\vect{e}^{-1} \frac{1+\cos \vartheta}{2} e^{-i \varphi} . \notag
\end{align}
Equations \eqref{eq:1:1:49}-\eqref{eq:1:1:52} may be written in a more compact form using the Wigner \(D\)-functions (see Chap. 4).
\begin{gather}
\vect{e}_{\mu}^{\prime}=\sum_{\nu} D_{\nu \mu}^{1}(\varphi, \vartheta, 0) \vect{e}_{\nu}=\sum_{\nu}(-1)^{\nu} D_{-\nu \mu}^{1}(\varphi, \vartheta, 0) \vect{e}^{\nu},  \notag\\
\vect{e}^{\prime \mu}=\sum_{\nu}(-1)^{\mu} D_{\nu-\mu}^{1}(\varphi, \vartheta, 0) \vect{e}_{\nu}=\sum_{\nu}(-1)^{\mu+\nu} D_{-\nu-\mu}^{1}(\varphi, \vartheta, 0) \vect{e}^{\nu},  \label{eq:1:1:53}\\
(\mu, \nu= \pm 1,0) . \notag
\end{gather}
Helicity Covariant and Helicity Contravariant Basis Vectors
\begin{align}
\vect{e}_{+1}^{\prime} & =-\vect{e}^{\prime-1}, & \vect{e}^{\prime+1} & =-\vect{e}_{-1}^{\prime}  ,\notag\\
\vect{e}_{0}^{\prime} & =\vect{e}^{\prime 0}, & \vect{e}^{\prime 0} & =\vect{e}_{0}^{\prime}  ,\label{eq:1:1:54}\\
\vect{e}_{-1}^{\prime} & =-\vect{e}^{\prime+1}, & \vect{e}^{\prime-1} & =-\vect{e}_{-1}^{\prime}. \notag
\end{align}
\section{Vectors. Tensors}\label{sec:1.2}\index{vector}\index{tensor}
Vectors and tensors are usually defined by transformation properties of their components under rotations of coordinate systems. The transformation rule for cartesian components of vectors and tensors is given below in Sec. 1.4 (Eqs. \eqref{eq:1:4:46}-\eqref{eq:1:4:51}). The transformation properties of spherical components of vectors and irreducible tensors are discussed in Chap. 3.

\subsection{Vector Components}\label{sec:1.2.1}\index{vector!components}\index{components!covariant}\index{components!contravariant}
Any vector can be expanded in terms of basis vectors, i.e., written as
\begin{equation}
\vect{A}=\sum_{\alpha} A^{\alpha} \vect{e}_{\alpha}=\sum_{\alpha} A_{\alpha} \vect{e}^{\alpha} .\label{eq:1:2:1}
\end{equation}
The expansion coefficients \(A_{\alpha}\) are called the covariant components of the vector, and \(A^{\alpha}\) are the contravariant vector components
\begin{equation}
A_{\alpha}=\vect{A} \cdot \vect{e}_{\alpha}, \quad A^{\alpha}=\vect{A} \cdot \vect{e}^{\alpha} . \label{eq:1:2:2}
\end{equation}
In a cartesian coordinate system one has
\begin{equation}
\vect{A}=A_{x} \vect{e}_{x}+A_{y} \vect{e}_{y}+A_{z} \vect{e}_{z}=A^{x} \vect{e}_{x}+A^{y} \vect{e}_{y}+A^{z} \vect{e}_{z} .\label{eq:1:2:3}
\end{equation}
The covariant cartesian components of a vector coincide with the contravariant ones.\\
In a polar coordinate system
\begin{equation}
\vect{A}=A_{r} \vect{e}_{r}+A_{\vartheta} \vect{e}_{\vartheta}+A_{\varphi} \vect{e}_{\varphi}=A^{r} \vect{e}_{r}+A^{\vartheta} \vect{e}_{\vartheta}+A^{\varphi} \vect{e}_{\varphi} . \label{eq:1:2:4}
\end{equation}
The covariant polar components coincide with the contravariant ones.\\
For a spherical coordinate system
\begin{equation}
\vect{A}=A^{+1} \vect{e}_{+1}+A^{0} \vect{e}_{0}+A^{-1} \vect{e}_{-1}=A_{+1} \vect{e}^{+1}+A_{0} \vect{e}^{0}+A_{-1} \vect{e}^{-1} \label{eq:1:2:5}
\end{equation}
The relations between covariant and contravariant spherical components are given by
\begin{equation}
A_{\mu}=(-1)^{\mu} A^{-\mu}, \quad A^{\mu}=(-1)^{\mu} A_{-\mu}, \quad(\mu= \pm 1,0) . \label{eq:1:2:6}
\end{equation}
If \(\vect{A}\) is a real vector, i.e., if \(\vect{A}^{*}=\vect{A}\), then
\begin{equation}
A_{\mu}^{*}=A^{\mu}, \quad A^{\mu *}=A_{\mu}, \quad(\mu= \pm 1,0) . \label{eq:1:2:7}
\end{equation}
If \(\vect{A}\) is a complex vector, then
\begin{equation}
A_{\mu}^{*}=\left(\vect{A}^{*}\right)^{\mu}, \quad A^{\mu *}=\left(\vect{A}^{*}\right)_{\mu}, \quad(\mu= \pm 1,0) . \label{eq:1:2:8}
\end{equation}
An expansion of a real vector \(\vect{A}\) in terms of spherical basis vectors is written as
\begin{align}
\vect{A} & =\sum_{\mu} A_{\mu} \vect{e}^{\mu}=\sum_{\mu} A^{\mu} \vect{e}_{\mu}=\sum_{\mu} A_{\mu}^{*} \vect{e}^{\mu *}=\sum_{\mu} A^{\mu *} \vect{e}_{\mu}^{*}  \notag\\
& =\sum_{\mu} A_{\mu} \vect{e}_{\mu}^{*}=\sum_{\mu} A_{\mu}^{*} \vect{e}_{\mu}=\sum_{\mu} A^{\mu} \vect{e}^{\mu *}=\sum_{\mu} A^{\mu *} \vect{e}^{\mu}  \notag\\
& =\sum_{\mu}(-1)^{\mu} A_{-\mu} \vect{e}_{\mu}=\sum_{\mu}(-1)^{\mu} A^{-\mu} \vect{e}^{\mu} . \label{eq:1:2:9}
\end{align}
An expansion of an arbitrary vector \(\vect{A}\) in terms of helicity basis vectors is given by
\begin{equation}
\vect{A}=A^{\prime+1} \vect{e}_{+1}^{\prime}+A^{\prime 0} \vect{e}_{0}^{\prime}+A^{\prime-1} \vect{e}_{-1}^{\prime}=A_{+1}^{\prime} \vect{e}^{\prime+1}+A_{0}^{\prime} \vect{e}^{\prime 0}+A_{-1}^{\prime} \vect{e}^{\prime-1} . \label{eq:1:2:10}
\end{equation}
The helicity components of a vector satisfy the same relations \eqref{eq:1:2:6}-\eqref{eq:1:2:9} as the spherical components.\\
The relations between vector components in different bases are the same as the relations between basis vectors. These relations are given by Eqs. \eqref{eq:1:1:29}-\eqref{eq:1:1:54} in which one should replace \(\vect{e}_{\alpha} \rightarrow A_{\alpha}\) and \(\vect{e}^{\alpha} \rightarrow A^{\alpha}\). In particular,
\begin{align}
A_{+1} & =-A^{-1}=-\frac{1}{\sqrt{2}}\left(A_{x}+i A_{y}\right), & A_{x}=\frac{1}{\sqrt{2}}\left(A_{-1}-A_{+1}\right)=\frac{1}{\sqrt{2}}\left(A^{-1}-A^{+1}\right),  \notag\\
A_{0} & =A^{0}=A_{z}, & A_{y}=\frac{i}{\sqrt{2}}\left(A_{-1}+A_{+1}\right)=\frac{-i}{\sqrt{2}}\left(A^{-1}+A^{+1}\right),  \label{eq:1:2:11}\\
A_{-1} & =-A^{+1}=\frac{1}{\sqrt{2}}\left(A_{x}-i A_{y}\right), & A_{z}=A_{0}=A^{0} . \notag
\end{align}
The matrices of transformations between cartesian, contravariant spherical and polar components of vectors are given in Tables 1.1 and 1.2.

Spherical components of a real vector \(\vect{A}\) which contains no derivatives and is independent of spin variables are
\begin{align}
A_{ \pm 1} & =\mp|\vect{A}| \frac{\sin \vartheta}{\sqrt{2}} e^{ \pm i \varphi}, & A^{ \pm 1} & =\mp|\vect{A}| \frac{\sin \vartheta}{\sqrt{2}} e^{\mp i \varphi},  \notag\\
A_{0} & =|\vect{A}| \cos \vartheta, & A^{0} & =|\vect{A}| \cos \vartheta, \label{eq:1:2:12}
\end{align}
where \(\vartheta, \varphi\) are the polar angles of the vector \(\vect{A}\).\\
Equation \eqref{eq:1:2:12} may be written in terms of spherical harmonics (see Chap. 5) as
\begin{align}
A_{\mu} & =\sqrt{\frac{4 \pi}{3}}|\vect{A}| Y_{1 \mu}(\vartheta, \varphi)  \notag\\
A^{\mu} & =\sqrt{\frac{4 \pi}{3}}|\vect{A}| Y_{1 \mu}^{*}(\vartheta, \varphi), \quad(\mu= \pm 1,0) \label{eq:1:2:13}
\end{align}
The expressions for cartesian components of \(\vect{A}\) in terms of spherical harmonics read
\begin{align}
A_{x} & =\sqrt{\frac{2 \pi}{3}}|\vect{A}|\left\{Y_{1-1}(\vartheta, \varphi)-Y_{1+1}(\vartheta, \varphi)\right\},  \notag\\
A_{y} & =i \sqrt{\frac{2 \pi}{3}}|\vect{A}|\left\{Y_{1-1}(\vartheta, \varphi)+Y_{1+1}(\vartheta, \varphi)\right\},  \label{eq:1:2:14}\\
A_{z} & =\sqrt{\frac{4 \pi}{3}}|\vect{A}| Y_{10}(\vartheta, \varphi) .\notag
\end{align}
\begin{table}[tbh]
  \scriptsize
\begin{center}
\caption{Matrix form of the transformations for vector components in different bases.}
\label{chap1:tab:1}
\begin{tabular}{c|c|c}
\hline
\hline
Cartesian coordinates & Spherical coordinates & Polar coordinates \\
\hline
\(\vect{A}=A_{x} \vect{e}_{x}+A_{y} \vect{e}_{y}+A_{z} \vect{e}_{z}\) & \(\vect{A}=A^{+1} \vect{e}_{+1}+A^{0} \vect{e}_{0}+A^{-1} \vect{e}_{-1}\) & \(\vect{A}=A_{r} \vect{e}_{r}+A_{\vartheta} \vect{e}_{\vartheta}+A_{\varphi} \vect{e}_{\varphi}\) \\
In terms of spherical components & In terms of cartesian components & In terms of cartesian components \\
\(\begin{pmatrix}A_{x} \\ A_{y} \\ A_{z}\end{pmatrix}=M(x, y, z \leftarrow+1,0,-1)\begin{pmatrix}A^{+1} \\ A^{0} \\ A^{-1}\end{pmatrix}\) & \(\begin{pmatrix}A^{+1} \\ A^{0} \\ A^{-1}\end{pmatrix}=M(+1,0,-1 \leftarrow x, y, z)\begin{pmatrix}A_{x} \\ A_{y} \\ A_{z}\end{pmatrix}\) & \(\begin{pmatrix}A_{r} \\ A_{\vartheta} \\ A_{\varphi}\end{pmatrix}=M(r, \vartheta, \varphi \leftarrow x, y, z)\begin{pmatrix}A_{x} \\ A_{y} \\ A_{z}\end{pmatrix}\) \\
In terms of polar components & In terms of polar components & In terms of spherical components \\
\(\begin{pmatrix}A_{x} \\ A_{y} \\ A_{z}\end{pmatrix}=M(x, y, z \leftarrow r, \vartheta, \varphi)\begin{pmatrix}A_{r} \\ A_{\vartheta} \\ A_{\varphi}\end{pmatrix}\) & \(\begin{pmatrix}A^{+1} \\ A^{0} \\ A^{-1}\end{pmatrix}=M(+1,0,-1 \leftarrow r, \vartheta, \varphi)\begin{pmatrix}A_{r} \\ A_{\vartheta} \\ A_{\varphi}\end{pmatrix}\) & \(\begin{pmatrix}A_{r} \\ A_{\vartheta} \\ A_{\varphi}\end{pmatrix}=M(r, \vartheta, \varphi \leftarrow+1,0,-1)\begin{pmatrix}A^{+1} \\ A^{0} \\ A^{-1}\end{pmatrix}\) \\
\hline
\hline
\end{tabular}
\end{center}
\end{table}

\begin{table}[ptbh]
\begin{center}
  \renewcommand{\arraystretch}{1.5}
\caption{Matrices of transformations between cartesian, spherical contravariant and polar components of vectors.}
\label{chap1:tab:2}
%\scriptsize
\begin{tabular}{cc}
\( M(x, y, z \leftarrow+1,0,-1) \) & \( M(+1,0,-1\leftarrow x, y, z ) \)\\
\(
\kbordermatrix{~& +1 & 0 & -1\cr
x & -\frac{1}{\sqrt{2}} & 0 & \frac{1}{\sqrt{2}} \cr
y & -\frac{i}{\sqrt{2}} & 0 & -\frac{i}{\sqrt{2}}\cr
z & 0                   & 1 & 0}
\) 
&
\(
\kbordermatrix{~& x & y & z\cr
-1 &-\frac{1}{\sqrt{2}} & -\frac{i}{\sqrt{2}} & 0 \cr
 0 & 0                  & 0                   & 1 \cr
 1 & \frac{1}{\sqrt{2}} & -\frac{i}{\sqrt{2}} & 0}
\)\\[10mm]
\(M(x, y, z \leftarrow r, \vartheta, \varphi)\) &  \(M(r, \vartheta, \varphi \leftarrow x, y, z)\) \\
\(
\kbordermatrix{~& r &\vartheta & \varphi \cr
x & \sin \vartheta \cos \varphi & \cos \vartheta \cos \varphi & -\sin\varphi\cr
y & \sin \vartheta \sin \varphi & \cos \vartheta \sin \varphi & \cos\varphi \cr
z & \cos \vartheta              & -\sin\vartheta & 0}
\) 
&
\(
\kbordermatrix{~& x & y & z\cr
r         & \sin \vartheta \cos \varphi & \sin \vartheta \sin \varphi & \cos \vartheta \cr
\vartheta & \cos \vartheta \cos \varphi & \cos \vartheta \sin \varphi & -\sin\vartheta\cr
 \varphi  & -\sin\varphi & \cos\varphi & 0 }
\)\\[10mm]
\( M(+1,0,-1 \longleftarrow r, \vartheta, \varphi)\) 
& \(M(r, \vartheta, \varphi \leftarrow+1,0,-1)\) \\
\(\kbordermatrix{ & r &\vartheta &\varphi \cr
 +1&-\frac{\sin \vartheta}{\sqrt{2}} e^{-i \varphi} & -\frac{\cos \vartheta}{\sqrt{2}} e^{-i \varphi} & \frac{i}{\sqrt{2}} e^{-i \varphi} \cr
0&\cos \vartheta & -\sin \vartheta & 0 \cr
-1&\frac{\sin \vartheta}{\sqrt{2}} e^{i \varphi} & \frac{\cos \vartheta}{\sqrt{2}} e^{i \varphi} & \frac{i}{\sqrt{2}} e^{i \varphi}}\)
&
-\(\kbordermatrix{ &+1&0&-1\cr
r & -\frac{\sin \vartheta}{\sqrt{2}} e^{i \varphi} & \cos \vartheta & \frac{\sin \vartheta}{\sqrt{2}} e^{-i \varphi} \cr
\vartheta&-\frac{\cos \vartheta}{\sqrt{2}} e^{i \varphi} & -\sin \vartheta & \frac{\cos \vartheta}{\sqrt{2}} e^{-i \varphi} \\
\varphi & -\frac{i}{\sqrt{2}} e^{i \varphi} & 0 & -\frac{i}{\sqrt{2}} e^{-i \varphi}}\)
\end{tabular}
\end{center}
\end{table}

\subsection{Scalar Product of Vectors}\label{sec:1.2.2}\index{scalar product}\index{product!scalar}
The scalar product of vectors \(\vect{A}\) and \(\vect{B}\) in an arbitrary orthonormal basis is defined by
\begin{equation}
\vect{A} \cdot \vect{B}=\sum_{\alpha} A_{\alpha} B^{\alpha}=\sum_{\alpha} A^{\alpha} B_{\alpha}. \label{eq:1:2:15}
\end{equation}
In a cartesian coordinate system we have
\begin{equation}
\vect{A} \cdot \vect{B}=A_{x} B_{x}+A_{y} B_{y}+A_{z} B_{z}. \label{eq:1:2:16}
\end{equation}
For polar coordinates the scalar product is given by
\begin{equation}
\vect{A} \cdot \vect{B}=A_{r} B_{r}+A_{\vartheta} B_{\vartheta}+A_{\varphi} B_{\varphi} . \label{eq:1:2:17}
\end{equation}
Equation \eqref{eq:1:2:17} is valid only if \(\vect{A}\) and \(\vect{B}\) do not contain derivatives because the polar basis vectors depend on the polar angles \(\vartheta, \varphi\) (see Section 1.1). In spherical coordinates we have
\begin{equation}
\vect{A} \cdot \vect{B}=\sum_{\mu} A^{\mu} B_{\mu}=\sum_{\mu} A_{\mu} B^{\mu}=\sum_{\mu}(-1)^{\mu} A_{\mu} B_{-\mu}=\sum_{\mu}(-1)^{\mu} A^{\mu} B^{-\mu}, \quad(\mu= \pm 1,0) ,\label{eq:1:2:18}
\end{equation}
or, in a more detailed form,
\begin{equation}
\vect{A} \cdot \vect{B}=-A_{+1} B_{-1}+A_{0} B_{0}-A_{-1} B_{+1} ,\label{eq:1:2:19}
\end{equation}
The scalar product in terms of helicity components of vectors is similar to \eqref{eq:1:2:18}-\eqref{eq:1:2:19}. The scalar product of vectors is invariant with respect to rotations of the coordinate system.

\subsection{Vector Product of Vectors}\label{sec:1.2.3}\index{vector product}\index{product!vector}
In a cartesian coordinate system the vector product of vectors \(\vect{A}\) and \(\vect{B}\) is defined by
\begin{equation}
\vect{A} \times \vect{B}=\left|\begin{array}{ccc}
\vect{e}_{x} & \vect{e}_{y} & \vect{e}_{z}  \label{eq:1:2:20}\\
A_{x} & A_{y} & A_{z} \\
B_{x} & B_{y} & B_{z}
\end{array}\right|=\sum_{i=x, y, z}\left[\vect{A} \times \vect{B}\right]_{i} \vect{e}_{i},
\end{equation}
where
\begin{align}
& {[\vect{A} \times \vect{B}]_{x}=A_{y} B_{z}-A_{z} B_{y}} , \notag\\
& {[\vect{A} \times \vect{B}]_{y}=A_{z} B_{x}-A_{x} B_{z}} , \label{eq:1:2:21}\\
& {[\vect{A} \times \vect{B}]_{z}=A_{x} B_{y}-A_{y} B_{x}} .\notag
\end{align}
Equation \eqref{eq:1:2:21} may be written in a more compact form as
\begin{equation}
[\vect{A} \times \vect{B}]_{i}=\sum_{k l} \varepsilon_{i k l} A_{k} B_{l}, \quad(i, k, l=x, y, z). \label{eq:1:2:22}
\end{equation}
In the polar coordinate system
\begin{equation}
\vect{A} \times \vect{B}=\left|\begin{array}{ccc}
\vect{e}_{r} & \vect{e}_{\vartheta} & \vect{e}_{\varphi}  \label{eq:1:2:23}\\
A_{r} & A_{\vartheta} & A_{\varphi} \\
B_{r} & B_{\vartheta} & B_{\varphi}
\end{array}\right|=\sum_{\alpha=r, \vartheta, \varphi}[\vect{A} \times \vect{B}]_{\alpha} \vect{e}_{\alpha},
\end{equation}
where
\begin{align}
{[\vect{A} \times \vect{B}]_{r} } & =A_{\vartheta} B_{\varphi}-A_{\varphi} B_{\vartheta} , \notag\\
{[\vect{A} \times \vect{B}]_{\vartheta} } & =A_{\varphi} B_{r}-A_{r} B_{\varphi} , \label{eq:1:2:24}\\
{[\vect{A} \times \vect{B}]_{\varphi} } & =A_{r} B_{\vartheta}-A_{\vartheta} B_{r} \notag .
\end{align}
Equation \eqref{eq:1:2:24} is valid only if \(\vect{A}\) and \(\vect{B}\) are not differential operators.\\
In the spherical coordinate system
\begin{equation}
\vect{A} \times \vect{B}=i\left|\begin{array}{lll}
\vect{e}_{+1} & \vect{e}_{0} & \vect{e}_{-1}  \label{eq:1:2:25}\\
A_{+1} & A_{0} & A_{-1} \\
B_{+1} & B_{0} & B_{-1}
\end{array}\right|=-i\left|\begin{array}{lll}
\vect{e}^{+1} & \vect{e}^{0} & \vect{e}^{-1} \\
A^{+1} & A^{0} & A^{-1} \\
B^{+1} & B^{0} & B^{-1}
\end{array}\right|=\sum_{\mu= \pm 1,0}[\vect{A} \times \vect{B}]_{\mu} \vect{e}^{\mu}=\sum_{\mu= \pm 1,0}[\vect{A} \times \vect{B}]^{\mu} \vect{e}_{\mu},
\end{equation}
where
\begin{align}
{[\vect{A} \times \vect{B}]_{+1} } & =i\left(A_{0} B_{+1}-A_{+1} B_{0}\right)=i\left(A^{-1} B^{0}-A^{0} B^{-1}\right),  \notag\\
{[\vect{A} \times \vect{B}]_{0} } & =i\left(A_{-1} B_{+1}-A_{+1} B_{-1}\right)=i\left(A^{+1} B^{-1}-A^{-1} B^{+1}\right) , \label{eq:1:2:26}\\
{[\vect{A} \times \vect{B}]_{-1} } & =i\left(A_{-1} B_{0}-A_{0} B_{-1}\right)=i\left(A^{0} B^{+1}-A^{+1} B^{0}\right) . \notag
\end{align}
\begin{align}
{[\vect{A} \times \vect{B}]^{+1} } & =i\left(A_{0} B_{-1}-A_{-1} B_{0}\right)=i\left(A^{+1} B^{0}-A^{0} B^{+1}\right)  ,\notag\\
{[\vect{A} \times \vect{B}]^{0} } & =i\left(A_{-1} B_{+1}-A_{+1} B_{-1}\right)=i\left(A^{+1} B^{-1}-A^{-1} B^{+1}\right) , \label{eq:1:2:27}\\
{[\vect{A} \times \vect{B}]^{-1} } & =i\left(A_{+1} B_{0}-A_{0} B_{+1}\right)=i\left(A^{0} B^{-1}-A^{-1} B_{0}\right) .\notag
\end{align}
Equations \eqref{eq:1:2:26}-\eqref{eq:1:2:27} may be written in a more compact form using the Clebsch-Gordan coefficients (see Chap. 8)
\begin{equation}
\begin{aligned}
& {[\vect{A} \times \vect{B}]_{\mu}=-i \sqrt{2} \sum_{\nu \lambda} \clebsch{1}{\nu}{1}{\lambda}{1}{\mu} A_{\nu} B_{\lambda}} ,\\
& {[\vect{A} \times \vect{B}]^{\mu}=i \sqrt{2} \sum_{\nu \lambda} \clebsch{1}{\nu}{1}{\lambda}{1}{\mu} A^{\nu} B^{\lambda}},
\end{aligned}
 \quad(\mu, \nu, \lambda= \pm 1,0) .\label{eq:1:2:28}
\end{equation}
Helicity components of the vector product are given by equations analogous to \eqref{eq:1:2:25}-\eqref{eq:1:2:28}.

\subsection{Products Involving Three or More Vectors}\label{sec:1.2.4}\index{triple product}\index{product!triple (scalar and vector)}
\begin{gather}
\vect{A} \cdot[\vect{B} \times \vect{C}]=\vect{B} \cdot[\vect{C} \times \vect{A}]=\vect{C} \cdot[\vect{A} \times \vect{B}]=-\vect{A} \cdot[\vect{C} \times \vect{B}]=-\vect{B} \cdot[\vect{A} \times \vect{C}]=-\vect{C} \cdot[\vect{B} \times \vect{A}],  \label{eq:1:2:29}\end{gather}
\begin{gather}
\vect{A} \times[\vect{B} \times \vect{C}]=\vect{B}(\vect{A} \cdot \vect{C})-\vect{C}(\vect{A} \cdot \vect{B}),  \label{eq:1:2:30}\end{gather}
\begin{gather}
{[\vect{A} \times \vect{B}] \cdot[\vect{C} \times \vect{D}]=(\vect{A} \cdot \vect{C})(\vect{B} \cdot \vect{D})-(\vect{A} \cdot \vect{D})(\vect{B} \cdot \vect{C}),}  \label{eq:1:2:31}\end{gather}
\begin{gather}
{[\vect{A} \times \vect{B}] \times[\vect{C} \times \vect{D}]=\vect{B}(\vect{A} \cdot[\vect{C} \times \vect{D}])-\vect{A}(\vect{B} \cdot[\vect{C} \times \vect{D}])=\vect{C}(\vect{A} \cdot[\vect{B} \times \vect{D}])-\vect{D}(\vect{A} \cdot[\vect{B} \times \vect{C}]) .}  \label{eq:1:2:32}\end{gather}
\begin{gather}
(\vect{A} \cdot[\vect{B} \times \vect{C}])(\vect{a} \cdot[\vect{b} \times \vect{c}])=\left|\begin{array}{lll}
\vect{A} \cdot \vect{a} & \vect{A} \cdot \vect{b} & \vect{A} \cdot \vect{c} \\
\vect{B} \cdot \vect{a} & \vect{B} \cdot \vect{b} & \vect{B} \cdot \vect{c} \\
\vect{C} \cdot \vect{a} & \vect{C} \cdot \vect{b} & \vect{C} \cdot \vect{c}
\end{array}\right|  \notag\\
=(\vect{A} \cdot \vect{a})(\vect{B} \cdot \vect{b})(\vect{C} \cdot \vect{c})-(\vect{A} \cdot \vect{a})(\vect{B} \cdot \vect{c})(\vect{C} \cdot \vect{b})-(\vect{B} \cdot \vect{b})(\vect{A} \cdot \vect{c})(\vect{C} \cdot \vect{a})  \notag\\
-(\vect{C} \cdot \vect{c})(\vect{A} \cdot \vect{b})(\vect{B} \cdot \vect{a})+(\vect{A} \cdot \vect{b})(\vect{B} \cdot \vect{c})(\vect{C} \cdot \vect{a})+(\vect{A} \cdot \vect{c})(\vect{B} \cdot \vect{a})(\vect{C} \cdot \vect{b}) . \label{eq:1:2:33}
\end{gather}
\subsection{Tensors \(\delta_{i k}\) and \(\varepsilon_{i k l}\)}\label{sec:1.2.5}\index{Kronecker delta symbol}\index{Levi-Civita tensor}\index{unit tensor}\index{tensor!unit}
In a cartesian basis two basic tensors \(\delta_{i k}\) and \(\varepsilon_{i k l}\) are widely used. The first tensor, \(\delta_{i k}\), is the symmetric unit tensor of rank 2. The second tensor, \(\varepsilon_{i k l}\), is the totally antisymmetric unit tensor of rank 3.

The tensor \(\delta_{i k}\) is called the Kronecker \(\delta\)-symbol and is defined by
\begin{equation}
\delta_{i k}=\left\{\begin{array}{ll}
1, & i=k,  \label{eq:1:2:34}\\
0, & i \neq k,
\end{array} \quad(i, k=x, y, z)\right.
\end{equation}
The components \(\delta_{i k}\) are invariant with respect to rotations and inversion of coordinate systems.\\
The tensor (or, more precisely, the pseudo-tensor) \(\boldsymbol{\varepsilon}_{i k l}\) is often called the Levi-Civita tensor. It is antisymmetric with respect to permutations of any pair of indices. Thus, \(\varepsilon_{i k l}=0\) if at least two of the indices \(i, k, l\) are equal, and \(\varepsilon_{i k l} \neq 0\) only if all indices \(i, k, l\) are different. The components \(\varepsilon_{i k l}\) are given by
\begin{align}
\varepsilon_{i i i} & =0, \quad(i=x, y, z) \quad(3 \text { components }),  \notag\\
\varepsilon_{i i k} & =\varepsilon_{i k i}=\varepsilon_{k i i}=0, \quad(i, k=x, y, z) \quad(18 \text { components }),  \label{eq:1:2:35}\\
\varepsilon_{x y z} & =\varepsilon_{y z x}=\varepsilon_{z x y}=-\varepsilon_{x z y}=-\varepsilon_{y x z}=-\varepsilon_{z y x}=1 \quad(6 \text { components }) . \notag
\end{align}
The components \(\varepsilon_{i k l}\) are invariant with respect to rotations and inversion of coordinate systems.

The tensor \(\varepsilon_{i k l}\) has the following properties: The product of two tensors \(\varepsilon_{i k l}\) and \(\varepsilon_{r g t}\) may be written in the form of a determinant
\begin{equation}
\varepsilon_{i k l} \varepsilon_{r s t}=\left|\begin{array}{ccc}
\delta_{i r} & \delta_{i s} & \delta_{i t}  \\
\delta_{k r} & \delta_{k s} & \delta_{k t} \\
\delta_{l r} & \delta_{l s} & \delta_{l t}
\end{array}\right|=\delta_{i r} \delta_{k s} \delta_{l t}+\delta_{i s} \delta_{k t} \delta_{l r}+\delta_{i t} \delta_{k r} \delta_{l s}-\delta_{i r} \delta_{k t} \delta_{l s}-\delta_{i s} \delta_{k r} \delta_{l t}-\delta_{i t} \delta_{k s} \delta_{l r}. \label{eq:1:2:36}
\end{equation}
By summing over a pair of indices, one obtains
\begin{equation}
\sum_{i} \varepsilon_{i k l} \varepsilon_{i s t}=\left|\begin{array}{ll}
\delta_{k s} & \delta_{k t}  \\
\delta_{l s} & \delta_{l t}
\end{array}\right|=\delta_{k s} \delta_{l t}-\delta_{k t} \delta_{l s}. \label{eq:1:2:37}
\end{equation}
Summation over two pairs of indices yields
\begin{equation}
\sum_{i, k} \varepsilon_{i k l} \varepsilon_{i k t}=2 \delta_{l t} . \label{eq:1:2:38}
\end{equation}
Finally, the summation over three pairs of indices gives
\begin{equation}
\sum_{i, k, l} \varepsilon_{i k l} \varepsilon_{i k l}=6 . \label{eq:1:2:39}
\end{equation}
For an arbitrary \(3 \times 3\) matrix \(\left\|A_{i k}\right\|(i, k=x, y, z)\) the following relation holds
\begin{equation}
\sum_{i, k, l} A_{x i} A_{y k} A_{z l} \varepsilon_{i k l}=\det\left\|A_{i k}\right\|=\left|\begin{array}{lll}
A_{x x} & A_{x y} & A_{x z}  \label{eq:1:2:40}\\
A_{y x} & A_{y y} & A_{y z} \\
A_{z x} & A_{z y} & A_{z z}
\end{array}\right| .
\end{equation}
\section{Differential Operations}\label{sec:1.3}\index{differential operations}
\subsection{Operator $\nabla$}\label{sec:1.3.1}\index{nabla operator@nabla ($\nabla$) operator}\index{operator!nabla ($\nabla$)}
The operator \(\boldsymbol{\nabla}\) (nabla) is the basic vector differential operator.\\
Cartesian components of \(\boldsymbol{\nabla}\) are given by
\begin{equation}
\nabla_{x}=\frac{\partial}{\partial x}, \quad \nabla_{y}=\frac{\partial}{\partial y}, \quad \nabla_{z}=\frac{\partial}{\partial z} . \label{eq:1:3:1}
\end{equation}
These components may be expressed in terms of polar coordinates as
\begin{align}
& \nabla_{x}=\sin \vartheta \cos \varphi \frac{\partial}{\partial r}+\frac{\cos \vartheta \cos \varphi}{r} \, \frac{\partial}{\partial \vartheta}-\frac{\sin \varphi}{r \sin \vartheta} \, \frac{\partial}{\partial \varphi},  \notag\\
& \nabla_{y}=\sin \vartheta \sin \varphi \frac{\partial}{\partial r}+\frac{\cos \vartheta \sin \varphi}{r} \, \frac{\partial}{\partial \vartheta}+\frac{\cos \varphi}{r \sin \vartheta} \, \frac{\partial}{\partial \varphi},  \label{eq:1:3:2}\\
& \nabla_{z}=\cos \vartheta \frac{\partial}{\partial r}-\frac{\sin \vartheta}{r} \, \frac{\partial}{\partial \vartheta} . \notag
\end{align}
An expansion of the operator $\boldsymbol{\nabla}$ in terms of spherical basis vectors reads
\begin{equation}
\boldsymbol{\nabla}=\sum_{\mu}(-1)^{\mu} \vect{e}_{\mu} \nabla_{-\mu}=-\vect{e}_{+1} \nabla_{-1}+\vect{e}_{0} \nabla_{0}-\vect{e}_{-1} \nabla_{+1}, \label{eq:1:3:3}
\end{equation}
where spherical components of \(\boldsymbol{\nabla}\) are given by
\begin{align}
\nabla_{+1} & =-\frac{1}{\sqrt{2}}\left(\frac{\partial}{\partial x}+i \frac{\partial}{\partial y}\right) , \notag\\
\nabla_{0} & =\frac{\partial}{\partial z} , \label{eq:1:3:4}\\
\nabla_{-1} & =\frac{1}{\sqrt{2}}\left(\frac{\partial}{\partial x}-i \frac{\partial}{\partial y}\right) .\notag
\end{align}
Spherical components of \(\boldsymbol{\nabla}\) in a polar coordinate system have the form
\begin{align}
\nabla_{+1} & =-\frac{e^{i \varphi}}{\sqrt{2}}\left\{\sin \vartheta \frac{\partial}{\partial r}+\frac{\cos \vartheta}{r} \, \frac{\partial}{\partial \vartheta}+\frac{i}{r \sin \vartheta} \, \frac{\partial}{\partial \varphi}\right\},  \notag\\
\nabla_{0} & =\cos \vartheta \frac{\partial}{\partial r}-\frac{\sin \vartheta}{r} \, \frac{\partial}{\partial \vartheta},  \label{eq:1:3:5}\\
\nabla_{-1} & =\frac{e^{-i \varphi}}{\sqrt{2}}\left\{\sin \vartheta \frac{\partial}{\partial r}+\frac{\cos \vartheta}{r} \, \frac{\partial}{\partial \vartheta}-\frac{i}{r \sin \vartheta} \, \frac{\partial}{\partial \varphi}\right\} . \notag
\end{align}
An expansion of \(\boldsymbol{\nabla}\) in terms of polar basis vectors may be written as
\begin{equation}
\boldsymbol{\nabla}=\vect{e}_{r} \nabla_{r}+\vect{e}_{\vartheta} \nabla_{\vartheta}+\vect{e}_{\varphi} \nabla_{\varphi}, \label{eq:1:3:6}
\end{equation}
where
\begin{equation}
\nabla_{r}=\frac{\partial}{\partial r}, \quad \nabla_{\vartheta}=\frac{1}{r} \, \frac{\partial}{\partial \vartheta}, \quad \nabla_{\varphi}=\frac{1}{r \sin \vartheta} \, \frac{\partial}{\partial \varphi} . \label{eq:1:3:7}
\end{equation}
The order of operator components relative to the basis vectors in Eq. \eqref{eq:1:3:6} is essential because \(\vect{e}_{r}, \vect{e}_{\vartheta}, \vect{e}_{\varphi}\) depend on \(\vartheta, \varphi\).

The operator $\boldsymbol{\nabla}$ may be written in the form
\begin{equation}
\boldsymbol{\nabla}=\vect{n} \frac{\partial}{\partial r}+\frac{1}{r} \boldsymbol\nabla_{\Omega}, \label{eq:1:3:8}
\end{equation}
where \(\boldsymbol{\nabla}_{\Omega}\) is the angular part of \(\boldsymbol{\nabla}\), and \(\vect{n}=\vect{r} / r\) is the unit vector determined by angles \(\vartheta\) and \(\varphi\). The operator \(\boldsymbol{\nabla}_{\Omega}\) acts only on variables \(\vartheta\) and \(\varphi\). In the polar coordinate system it has only two components
\begin{equation}
\left(\nabla_{\Omega}\right)_{\vartheta}=\frac{\partial}{\partial \vartheta}, \quad\left(\nabla_{\Omega}\right)_{\varphi}=\frac{1}{\sin \vartheta} \, \frac{\partial}{\partial \varphi} . \label{eq:1:3:9}
\end{equation}
The operator \(\boldsymbol{\nabla}_{\Omega}\) may be written as
\begin{equation}
\boldsymbol{\nabla}_{\Omega}=-i \vect{n} \times \hat{\vect{L}}, \label{eq:1:3:10}
\end{equation}
where \(\hat{\vect{L}}\) is the orbital angular momentum operator (see Sec. 2.2).\index{angular momentum operator!orbital}\index{operator!angular part of $\nabla$}

\subsection{Laplace Operator}\label{sec:1.3.2}\index{Laplace operator}\index{operator!Laplace}
The Laplace operator (Laplacian) \(\Delta\) is a scalar differential operator
\begin{equation}
\Delta=\boldsymbol{\nabla}^{2} . \label{eq:1:3:11}
\end{equation}
In the cartesian coordinate system \(\Delta\) has the form
\begin{equation}
\Delta=\frac{\partial^{2}}{\partial x^{2}}+\frac{\partial^{2}}{\partial y^{2}}+\frac{\partial^{2}}{\partial z^{2}} .\label{eq:1:3:12}
\end{equation}
In the polar coordinate system it is given by
\begin{equation}
\Delta=\frac{1}{r^{2}} \, \frac{\partial}{\partial r}\left\{r^{2} \frac{\partial}{\partial r}\right\}+\frac{1}{r^{2} \sin \vartheta} \, \frac{\partial}{\partial \vartheta}\left\{\sin \vartheta \frac{\partial}{\partial \vartheta}\right\}+\frac{1}{r^{2} \sin ^{2} \vartheta} \, \frac{\partial^{2}}{\partial \varphi^{2}} . \label{eq:1:3:13}
\end{equation}
The operator \(\Delta\) may also be written as
\begin{equation}
\Delta=\frac{1}{r^{2}} \, \frac{\partial}{\partial r}\left\{r^{2} \frac{\partial}{\partial r}\right\}+\frac{1}{r^{2}} \Delta_{\Omega}, \label{eq:1:3:14}
\end{equation}
where \(\Delta_{\Omega}\) is the angular part of \(\Delta\)
\begin{equation}
\Delta_{\Omega}=\boldsymbol{\nabla}_{\Omega}^{2}=\frac{1}{\sin \vartheta} \, \frac{\partial}{\partial \vartheta}\left\{\sin \vartheta \frac{\partial}{\partial \vartheta}\right\}+\frac{1}{\sin ^{2} \vartheta} \, \frac{\partial^{2}}{\partial \varphi^{2}} \label{eq:1:3:15}
\end{equation}
expressed in terms of the orbital angular momentum operator \(\widehat{\vect{L}}\) as
\begin{equation}
\Delta_{\Omega}=-\hat{\vect{L}}^{2} \label{eq:1:3:16}
\end{equation}
\subsection{Differential Operations on Scalars and Vectors}\label{sec:1.3.3}\index{gradient}\index{divergence}\index{curl}\index{directional derivative}
The gradient of a scalar function \(\Phi(\vect{r})\) is the vector defined in terms of the operator $\boldsymbol{
abla}$ as
\begin{equation}
\operatorname{grad} \Phi(\vect{r})=\boldsymbol{\nabla} \Phi(\vect{r}) \label{eq:1:3:17}
\end{equation}
The components of grad \(\Phi\) may be obtained by use of Eqs. \eqref{eq:1:3:1}-\eqref{eq:1:3:7} for the components of \(\boldsymbol{\nabla}\). If \(\Phi\) depends only on \(r=|\vect{r}|\) (spherically-symmetric field), then
\begin{equation}
\boldsymbol{\nabla} \Phi(r)=\vect{n} \frac{d \Phi(r)}{d r} \label{eq:1:3:18}
\end{equation}
where \(\vect{n}=\vect{r} / \vect{r}\).\\
The directional derivative of a scalar function \(\Phi(\vect{r})\) in the direction specified by a unit vector \(\vect{u}\) is the scalar defined by
\begin{equation}
\frac{d}{d s} \Phi(\vect{r})=(\vect{u} \cdot \boldsymbol{\nabla}) \Phi(\vect{r}) .\label{eq:1:3:19}
\end{equation}
The divergence of a vector field \(\vect{A}(\vect{r})\) is the scalar product of $\boldsymbol{
abla}$ and \(\vect{A}\)
\begin{equation}
\operatorname{div} \vect{A}=\boldsymbol{\nabla} \cdot \vect{A} . \label{eq:1:3:20}
\end{equation}
The expression for \(\operatorname{div} \vect{A}\) in a cartesian coordinate system is
\begin{equation}
\operatorname{div} \vect{A}=\frac{\partial A_{x}}{\partial x}+\frac{\partial A_{y}}{\partial y}+\frac{\partial A_{z}}{\partial z}=\sum_{i=x, y, z} \frac{\partial A_{i}}{\partial x_{i}} . \label{eq:1:3:21}
\end{equation}
and in a spherical coordinate system it has the form
\begin{equation}
\operatorname{div} \vect{A}=-\nabla_{+1} A_{-1}+\nabla_{0} A_{0}-\nabla_{-1} A_{+1}=\sum_{\mu= \pm 1,0}(-1)^{\mu} \nabla_{\mu} A_{-\mu} .\label{eq:1:3:22}
\end{equation}
where the spherical components \(\nabla_{\mu}\) are given by Eqs. \eqref{eq:1:3:4}-\eqref{eq:1:3:5}. In the polar coordinate system we have
\begin{equation}
\operatorname{div} \vect{A}=\frac{1}{r^{2}} \, \frac{\partial}{\partial r}\left(r^{2} A_{r}\right)+\frac{1}{r \sin \vartheta} \, \frac{\partial}{\partial \vartheta}\left(\sin \vartheta A_{\vartheta}\right)+\frac{1}{r \sin \vartheta} \, \frac{\partial A_{\varphi}}{\partial \varphi} . \label{eq:1:3:23}
\end{equation}
The curl of a vector field \(\vect{A}(\vect{r})\) is the vector product of \(\boldsymbol{\nabla}\) and \(\vect{A}\)
\begin{equation}
\operatorname{curl} \vect{A}=\nabla \times \vect{A} .\label{eq:1:3:24}
\end{equation}
The cartesian components of curl \(\vect{A}\) are given by
\begin{align}
& {[\operatorname{curl} \vect{A}]_{x}=\frac{\partial A_{z}}{\partial y}-\frac{\partial A_{y}}{\partial z}}  ,\notag\\
& {[\operatorname{curl} \vect{A}]_{y}=\frac{\partial A_{x}}{\partial z}-\frac{\partial A_{z}}{\partial x}} , \label{eq:1:3:25}\\
& {[\operatorname{curl} \vect{A}]_{z}=\frac{\partial A_{y}}{\partial x}-\frac{\partial A_{x}}{\partial y}} ,\notag
\end{align}
or, in a more compact form,
\begin{equation}
[\operatorname{curl} \vect{A}]_{i}=\sum_{k l} \varepsilon_{i k l} \frac{\partial A_{l}}{\partial x_{k}}. \label{eq:1:3:26}
\end{equation}
Moreover, curl A may also be written in the form
\begin{equation}
\operatorname{curl} \vect{A}=\left|\begin{array}{lll}
\vect{e}_{x} & \vect{e}_{y} & \vect{e}_{z}  \label{eq:1:3:27}\\
\frac{\partial}{\partial x} & \frac{\partial}{\partial y} & \frac{\partial}{\partial z} \\
A_{x} & A_{y} & A_{z}
\end{array}\right|.
\end{equation}
The spherical components of curl \(\vect{A}\) are given by
\begin{align}
{[\operatorname{curl} \vect{A}]_{+1} } & =i\left(\nabla_{0} A_{+1}-\nabla_{+1} A_{0}\right),  \notag\\
{[\operatorname{curl} \vect{A}]_{0} } & =i\left(\nabla_{-1} A_{+1}-\nabla_{+1} A_{-1}\right),  \label{eq:1:3:28}\\
{[\operatorname{curl} \vect{A}]_{-1} } & =i\left(\nabla_{-1} A_{0}-\nabla_{0} A_{-1}\right), \notag
\end{align}
or in a more compact form involving the Clebsch-Gordan coefficients
\begin{equation}
[\operatorname{curl} \vect{A}]_{\mu}=-i \sqrt{2} \sum_{\nu \lambda} \clebsch{1}{\nu}{1}{\lambda}{1}{\mu} \nabla_{\nu} A_{\lambda} \quad(\mu, \nu, \lambda= \pm 1,0) \label{eq:1:3:29}
\end{equation}
The spherical components of curl \(\vect{A}\) may also be written as
\begin{equation}
\operatorname{curl} \vect{A}=i\left|\begin{array}{ccc}
\vect{e}_{+1} & \vect{e}_{0} & \vect{e}_{-1}  \label{eq:1:3:30}\\
\nabla_{+1} & \nabla_{0} & \nabla_{-1} \\
A_{+1} & A_{0} & A_{-1}
\end{array}\right|.
\end{equation}
The polar components of curl \(\vect{A}\) read
\begin{align}
& {[\operatorname{curl} \vect{A}]_{r}=\frac{1}{r \sin \vartheta} \frac{\partial}{\partial \vartheta}\left(\sin \vartheta A_{\varphi}\right)-\frac{1}{r \sin \vartheta} \frac{\partial A_{\vartheta}}{\partial \varphi}} , \notag\\
& {[\operatorname{curl} \vect{A}]_{\vartheta}=\frac{1}{r \sin \vartheta} \frac{\partial A_{r}}{\partial \varphi}-\frac{1}{r} \frac{\partial}{\partial r}\left(r A_{\varphi}\right)} , \label{eq:1:3:31}\\
& {[\operatorname{curl} \vect{A}]_{\varphi}=\frac{1}{r} \frac{\partial}{\partial r}\left(r A_{\vartheta}\right)-\frac{1}{r} \frac{\partial A_{r}}{\partial \vartheta}} .\notag
\end{align}
The above equations are summarized in Table~\ref{chap1:tab:3}.\\
Note also the following differential operations of the second order
\begin{equation}
\operatorname{div} \operatorname{grad} \Phi=\boldsymbol{\nabla} \cdot(\boldsymbol{\nabla} \Phi)=\Delta \Phi \label{eq:1:3:32}.
\end{equation}
\begin{table}[tbh]
\begin{center}
  \renewcommand{\arraystretch}{1.8}
\caption{Differential operations.}
\label{chap1:tab:3}
\small
\begin{tabular}{c|c|c|c}
\hline
\hline
 & Cartesian coordinates & Spherical coordinates & Polar coordinates \\
\hline
\(\vect{r}\) & \(\vect{e}_{x}  x+\vect{e}_{y}  y+\vect{e}_{z}  z\) &  \(-\vect{e}_{+1} x_{-1}+\vect{e}_{0} x_{0}-\vect{e}_{-1} x_{+1}\) & \(\vect{e}_r r\)\\
\(d \vect{r}\) & \(\vect{e}_{x} d x+\vect{e}_{y} d y+\vect{e}_{z} d z\) & \(-\vect{e}_{+1} d x_{-1}+\vect{e}_{0} d x_{0}-\vect{e}_{-1} d x_{+1}\) & \(\vect{e}_{r} d r+\vect{e}_{\vartheta} r d \vartheta+\vect{e}_{\varphi} r \sin \vartheta d \varphi\) \\
\(d s^{2}\) & \(d x^{2}+d y^{2}+d z^{2}\) & \(-2 d x_{+1} d x_{-1}+d x_{0} d x_{0}\) & \(d r^{2}+r^{2} d \vartheta^{2}+r^{2} \sin ^{2} \vartheta d \varphi^{2}\) \\
\(d V\) & \(dx dy dz\) & \(i d x_{+1} d x_{0} d x_{-1}\) & \(r^{2} \sin \vartheta d r d \vartheta d \varphi\) \\
\(\vect{\boldsymbol{\nabla}} {\Phi}\) 
  & \(\vect{e}_{x} \frac{\partial \Phi}{\partial x}+\vect{e}_{y} \frac{\partial \Phi}{\partial y}+\vect{e}_{z} \frac{\partial \Phi}{\partial z}\)
  & \(-\vect e_{+1} \nabla_{-1} \Phi+\vect e_{0} \nabla_{0} \Phi-\) 
  & \(\vect{e}_{r} \frac{\partial \Phi}{\partial r}+\vect{e}_{\vartheta} \frac{1}{r} \frac{\partial \Phi}{\partial \vartheta}+\vect{e}_{\varphi} \frac{1}{r \sin \vartheta} \, \frac{\partial \Phi}{\partial \varphi}\) \\
  &&\( -\vect e_{-1} \nabla_{+1} \Phi\)&\\
\((\vect{\boldsymbol{\nabla}} \cdot \vect{A})\)  
  & \(\frac{\partial A_{x}}{\partial x}+\frac{\partial A_{y}}{\partial y}+\frac{\partial A_{z}}{\partial z}\) 
  & \(-\nabla_{+1} A_{-1}+\nabla_{0} A_{0}-\nabla_{-1} A_{+1}\) 
  & \( \frac{1}{r^{2}}   \frac{\partial}{\partial r}\left(r^{2} A_{r}\right)+\frac{1}{r \sin \vartheta}   \frac{\partial}{\partial \vartheta}\left(A_{\vartheta} \sin \vartheta\right)+\frac{1}{r \sin \vartheta}   \frac{\partial A_{\varphi}}{\partial \varphi} \)  \\
\([\vect{\boldsymbol{\nabla}} \times \vect{A}]\)  
  & \(\left|\begin{array}{ccc}\vect{e}_{x} & \vect{e}_{y} & \vect{e}_{z} \\ 
    \frac{\partial}{\partial x} & \frac{\partial}{\partial y} & \frac{\partial}{\partial z} \\ 
    A_{x} & A_{y} & A_{z}
   \end{array}\right|\) 
  & \(i\left|\begin{array}{ccc}e_{+1} & e_{0} & e_{-1} \\ \nabla_{+1} & \nabla_{0} & \nabla_{-1} \\ A_{+1} & A_{0} & A_{-1}\end{array}\right|\) 
  & \( \begin{array}{c} \vect{e}_{r} \frac{1}{r \sin \vartheta}\left[\frac{\partial}{\partial \vartheta}\left(A_{\varphi} \sin \vartheta\right)-\frac{\partial A_{\vartheta}}{\partial \varphi}\right] \\+\vect{e}_{\vartheta} \frac{1}{r}\left[\frac{1}{\sin \vartheta} \frac{\partial A_{r}}{\partial \varphi}-\frac{\partial}{\partial r}\left(r A_{\varphi}\right)\right] \\ +\vect{e}_{\varphi} \frac{1}{r}\left[\frac{\partial}{\partial r}\left(r A_{\vartheta}\right)-\frac{\partial A_{r}}{\partial \vartheta}\right] \end{array} \) \\
\(\Delta \Phi\) 
  & \(\frac{\partial^{2} \Phi}{\partial x^{2}}+\frac{\partial^{2} \Phi}{\partial y^{2}}+\frac{\partial^{2} \Phi}{\partial z^{2}}\) 
  & \(-2 \nabla_{+1} \nabla_{-1} \Phi+\nabla_{0} \nabla_{0} \Phi\) 
  & \(\frac{1}{r^{2}}   \frac{\partial}{\partial r}\left(r^{2} \frac{\partial \Phi}{\partial r}\right)+\frac{1}{r^{2} \sin \vartheta}   \frac{\partial}{\partial \vartheta}\left(\sin \vartheta \frac{\partial \Phi}{\partial \vartheta}\right)+\frac{1}{r^{2} \sin ^{2} \vartheta}   \frac{\partial^{2} \Phi}{\partial \varphi^{2}}\) \\
\hline
\end{tabular}
\end{center}
\end{table}
\begin{align}
& \operatorname{curl} \operatorname{grad} \Phi=\boldsymbol{\nabla} \times(\boldsymbol{\nabla} \Phi)=0,  \label{eq:1:3:33}\\
& \operatorname{div} \operatorname{curl} \vect{A}=\boldsymbol{\nabla} \cdot[\boldsymbol{\nabla} \times \vect{A}]=0, \label{eq:1:3:34}
\end{align}
\begin{align}
\operatorname{curl} \operatorname{curl} \vect{A} & =\boldsymbol{\nabla} \times[\boldsymbol{\nabla} \times \vect{A}]=\boldsymbol{\nabla}(\boldsymbol{\nabla} \cdot \vect{A})-\Delta \vect{A}  \notag\\
& =\operatorname{grad} \operatorname{div} \vect{A}-\Delta \vect{A} . \label{eq:1:3:35}
\end{align}
\section{Rotations of Coordinate System}\label{sec:1.4}\index{rotation!of coordinate system}\index{coordinate system!rotation of}
An arbitrary rotation of a coordinate system about the origin is completely specified by three real parameters. The most useful description of rotation is that in terms of the Euler angles \(\alpha, \beta, \gamma\). Note that two other sets of parameters are also widely used to describe rotations:\\
- direction of the rotation axis \(n(\Theta, \Phi)\) (2 parameters) and the rotation angle \(\omega\) (1 parameter);\\
- the Cayley-Klein parameters.

\subsection{Description of Rotations in Terms of the Euler Angles}\label{sec:1.4.1}
\begin{figure}[tbh]
\begin{center}
  \includegraphics[alt={},max width=0.9\textwidth]{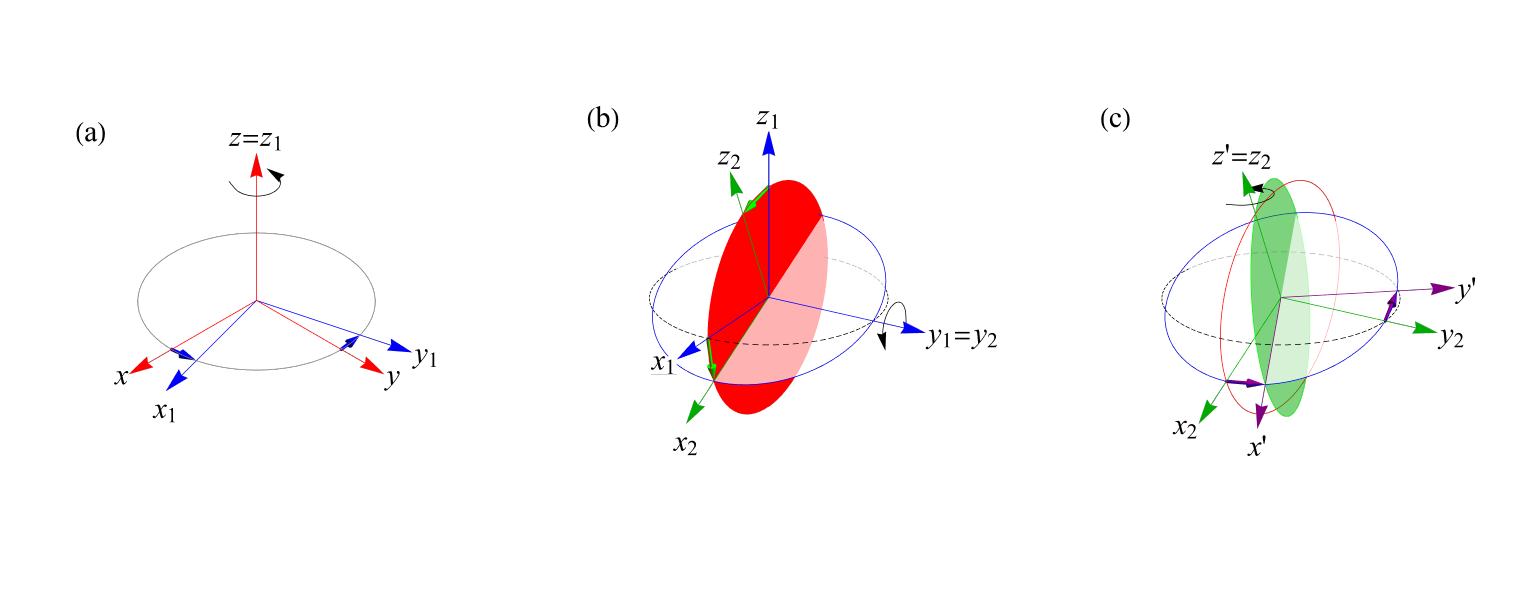}
\caption{Succession of rotations of a coordinate system according to scheme \(A\).}
\label{chap1:fig:3}
\end{center}
\end{figure}

\index{Euler angles}\index{rotation!in terms of Euler angles}
Any rotation of the coordinate system \(S\{x, y, z\} \rightarrow S^{\prime}\left\{x^{\prime}, y^{\prime}, z^{\prime}\right\}\) may be performed by three successive rotations about the coordinate axes (Fig.~\ref{chap1:fig:3})\\
\(\vect{A}\left\{\begin{array}{l}
  \text {(a) rotation about the \(z\)-axis through an angle \(\alpha(0 \leq \alpha<2 \pi\)),}\\ 
  \text {(b) rotation about the new \(y_{1}\)-axis through an angle \(\beta(0 \leq \beta \leq \pi)\),} \\ 
  \text {(c) rotation about the new axis \(z_{2}=z^{\prime}\) through an angle \(\gamma(0 \leq \gamma<2 \pi)\).}\end{array}\right.\)

\begin{figure}[tbh]
\begin{center}
  \includegraphics[alt={},max width=0.9\textwidth]{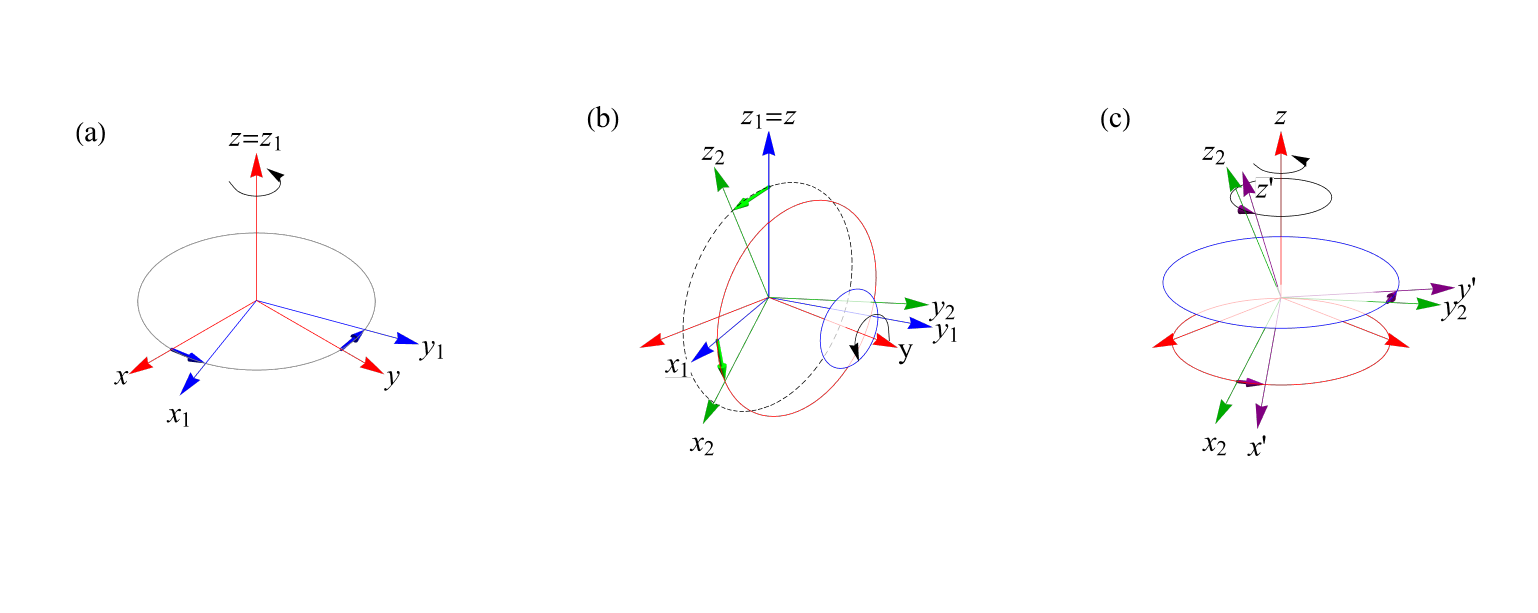}
\caption{Succession of rotations of a coordinate system according to scheme B.}
\label{chap1:fig:4}
\end{center}
\end{figure}

The same rotation \(S\{x, y, z\} \rightarrow S^{\prime}\left\{x^{\prime}, y^{\prime}, z^{\prime}\right\}\) may also be performed by another succession of rotations Fig.~\ref{chap1:fig:4}),

\begin{figure}[tbh]
\begin{center}
  \includegraphics[alt={},max width=0.4\textwidth]{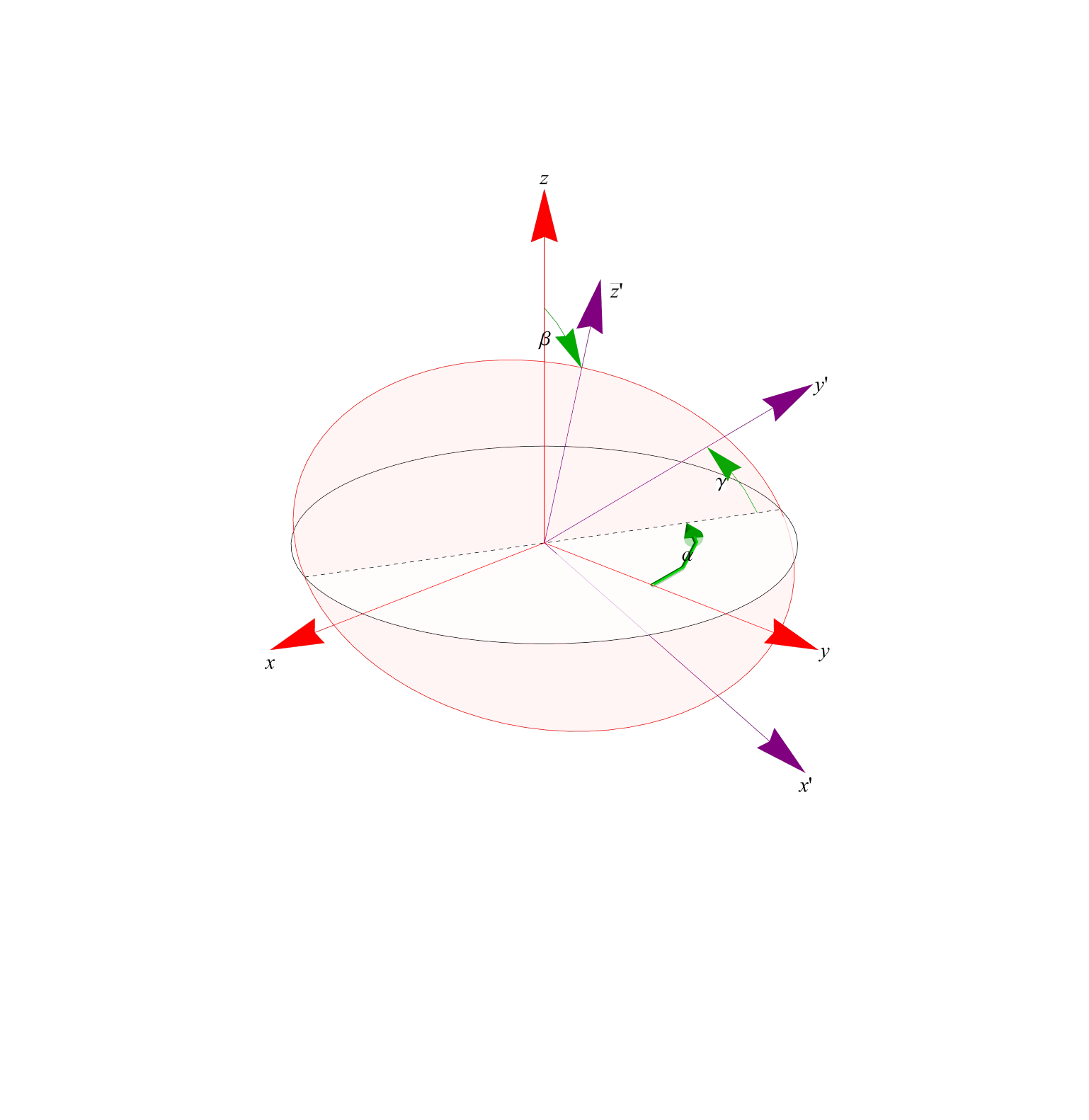}
\caption{The Euler angles \(\alpha, \beta, \gamma\).}
\label{chap1:fig:5}
\end{center}
\end{figure}

\(\vect{B}\left\{\begin{array}{l}
  \text {(a) rotation about the } z \text {-axis through an angle } \gamma(0 \leq \gamma<2 \pi), \\ 
  \text {(b) rotation about the initial } y \text {-axis through an angle } \beta(0 \leq \beta \leq \pi), \\ 
  \text {(c) rotation about the initial } z \text {-axis through an angle } \alpha(0 \leq \alpha<2 \pi) .\end{array}\right.\)\\
Here the angles \(\alpha, \beta, \gamma\) are the same as those in the first case.\\
The relative orientations of initial and final coordinate axes \(S\{x, y, z\}\) and \(S^{\prime}\left\{x^{\prime}, y^{\prime}, z^{\prime}\right\}\), obtained in both cases, \(\vect{A}\) and \(\vect{B}\), are shown in Fig.~\ref{chap1:fig:5}.

The angles \(\alpha, \beta, \gamma\) are called the Euler angles. They completely define the rotation of the coordinate system. The inverse rotation which returns the coordinate system \(S^{\prime}\left\{x^{\prime}, y^{\prime}, z^{\prime}\right\}\) back into \(S\{x, y, z\}\) is specified by the Euler angles \(-\gamma,-\beta,-\alpha\), or, equivalently, by the angles \(\pi-\gamma, \beta,-\pi-\alpha\).

Sometimes the following successive rotations are used to obtain the general rotation of the coordinate system:\\
(a) rotation about the \(z\)-axis through an angle \(\alpha^{\prime}\);\\
(b) rotation about the new \(x_{1}\)-axis through an angle \(\beta^{\prime}\);\\
(c) rotation about the new axis \(z_{2}=z^{\prime}\) through an angle \(\gamma^{\prime}\).

The angles \(\alpha^{\prime}, \beta^{\prime}, \gamma^{\prime}\) describe the same rotation of the coordinate system as the Euler angles \(\alpha, \beta, \gamma\) if
\begin{equation}
\alpha^{\prime}=\alpha+\frac{\pi}{2}, \quad \beta^{\prime}=\beta, \quad \gamma^{\prime}=\gamma-\frac{\pi}{2} . \label{eq:1:4:1}
\end{equation}
The absolute value of a vector is invariant with respect to rotations, but the polar angles \(\vartheta, \varphi\), which determine the vector direction, change. The relations between angles \(\vartheta, \varphi\) and \(\vartheta^{\prime}, \varphi^{\prime}\) which specify vector directions in the initial and final coordinate systems, \(S\{x, y, z\}\) and \(S^{\prime}\left\{x^{\prime}, y^{\prime}, z^{\prime}\right\}\), are given by
\begin{gather}
\cos \vartheta^{\prime}=\cos \vartheta \cos \beta+\sin \vartheta \sin \beta \cos (\varphi-\alpha) , \notag\\
\cot \left(\varphi^{\prime}+\gamma\right)=\cot (\varphi-\alpha) \cos \beta-\frac{\cot \vartheta \sin \beta}{\sin (\varphi-\alpha)}. \label{eq:1:4:2}
\end{gather}
The inverse relations are
\begin{gather}
\cos \vartheta=\cos \vartheta^{\prime} \cos \beta-\sin \vartheta^{\prime} \sin \beta \cos \left(\varphi^{\prime}+\gamma\right) , \notag\\
\cot (\varphi-\alpha)=\cot \left(\varphi^{\prime}+\gamma\right) \cos \beta+\frac{\cot \vartheta^{\prime} \sin \beta}{\sin \left(\varphi^{\prime}+\gamma\right)} . \label{eq:1:4:3}
\end{gather}
\subsection{Description of Rotations in Terms of Rotation Axis and Rotation Angle}\label{sec:1.4.2}
\begin{figure}[tbh]
\begin{center}
  \includegraphics[alt={},max width=0.4\textwidth]{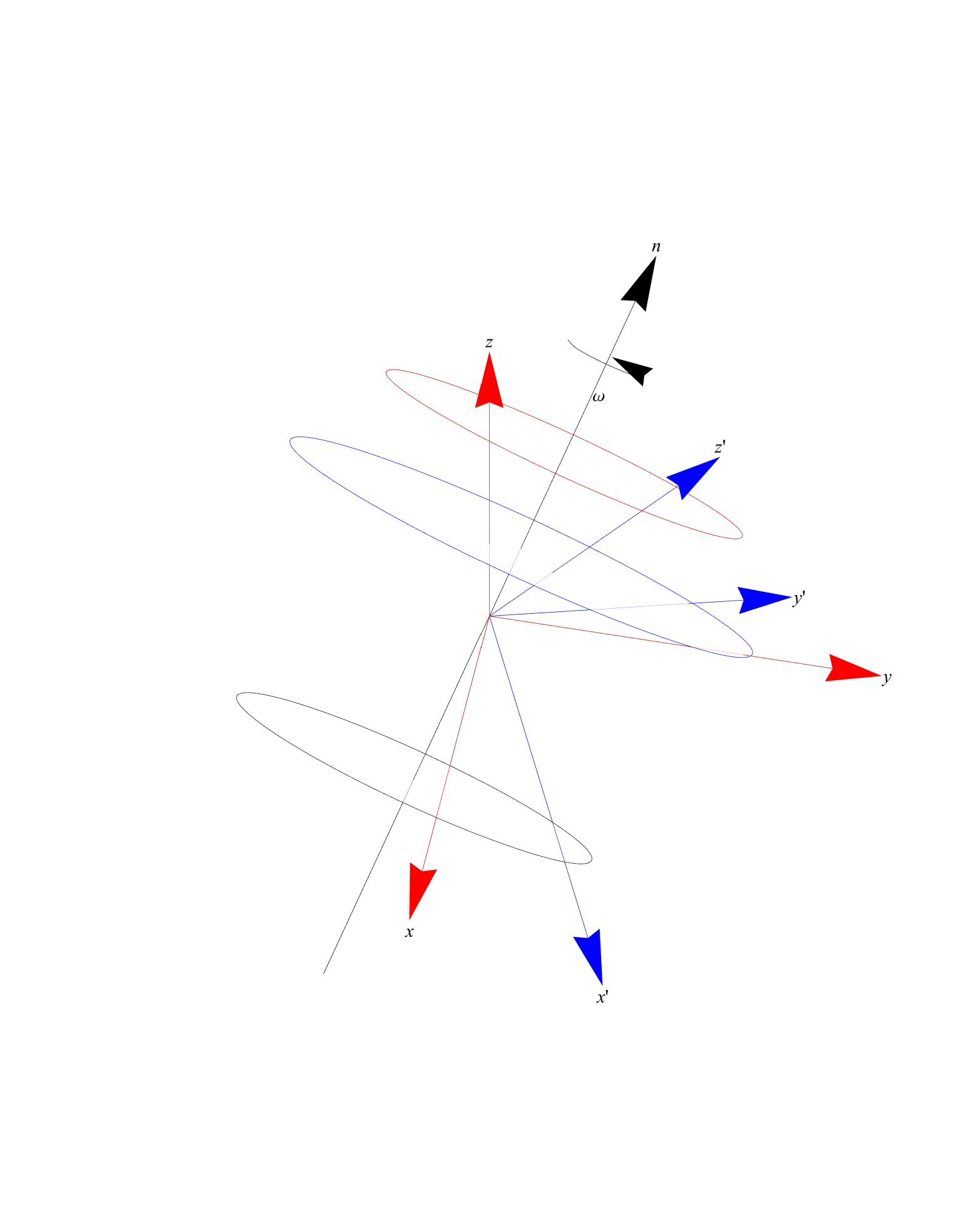}
\caption{Rotation of a coordinate system through an angle \(\omega\) about an axis \(\vect{n}(\Theta, \Phi)\).}
\label{chap1:fig:6}
\end{center}
\end{figure}

\index{rotation!axis and angle}\index{rotation axis}\index{rotation angle}
Any rotation of a coordinate system \(S\{x, y, z\} \rightarrow S^{\prime}\left\{x^{\prime}, y^{\prime}, z^{\prime}\right\}\) may be treated as one rotation through an angle \(\omega(0 \leq \omega \leq \pi)\) about some axis \(\vect{n}(\Theta, \Phi)\). The direction of this rotation axis \(\vect{n}\) is defined by the polar angles \(\Theta, \Phi(0 \leq \Theta \leq \pi, 0 \leq \Phi<2 \pi)\) which are the same in the initial coordinate system \(S\{x, y, z\}\) and in the final one \(S^{\prime}\left\{x^{\prime}, y^{\prime}, z^{\prime}\right\}\) (Fig.~\ref{chap1:fig:6}).

The angles \(\omega, \Theta, \Phi\) completely determine the relative orientation of the initial and final coordinate axes. The rotation defined by the angles \(-\omega, \pi-\Theta, \pi+\Phi\) is identical to the rotation defined by the angles \(\omega, \Theta, \Phi\). The inverse rotation \(S^{\prime}\left\{x^{\prime}, y^{\prime}, z^{\prime}\right\} \rightarrow S\{x, y, z\}\) is specified by the angles \(-\omega, \Theta, \Phi\) or, equivalently, by the angles \(\omega, \pi-\Theta, \pi+\Phi\).

The direction cosines of the unit vector \(\vect{n}(\Theta, \Phi)\) in the initial coordinate system \(S\{x, y, z\}\) coincide with those in the final coordinate system \(S^{\prime}\left\{x^{\prime}, y^{\prime}, z^{\prime}\right\}\).
\begin{align}
& \vect{n} \cdot \vect{e}_{x}=\vect{n} \cdot \vect{e}_{x}^{\prime}=\sin \Theta \cos \Phi , \notag\\
& \vect{n} \cdot \vect{e}_{y}=\vect{n} \cdot \vect{e}_{y}^{\prime}=\sin \Theta \sin \Phi , \label{eq:1:4:4}\\
& \vect{n} \cdot \vect{e}_{z}=\vect{n} \cdot \vect{e}_{z}^{\prime}=\cos \Theta \notag .
\end{align}
The polar angles \(\vartheta, \varphi\) of vectors not parallel to the $\vect{n}$-axis vary under coordinate rotations. The relations between the polar angles \(\vartheta, \varphi\) and \(\vartheta^{\prime}, \varphi^{\prime}\) which specify the direction of a vector with respect to \(S\{x, y, z\}\) and \(S^{\prime}\left\{x^{\prime}, y^{\prime}, z^{\prime}\right\}\), respectively, are given by
\begin{gather}
\cos \vartheta^{\prime}=\cos \vartheta\left(\cos \omega \sin ^{2} \Theta+\cos ^{2} \Theta\right)+\sin \vartheta \sin \Theta[(1-\cos \omega) \cos \Theta \cos (\varphi-\Phi)-\sin \omega \sin (\varphi-\Phi)]  ,\notag\\
\cot \left(\varphi^{\prime}-\Phi\right)=\frac{\cos (\varphi-\Phi)\left[\cos \omega \cos ^{2} \Theta+\sin ^{2} \Theta\right]+\sin (\varphi-\Phi) \sin \omega \cos \Theta-\cot \vartheta(\cos \omega-1) \sin \Theta \cos \Theta}{-\cos (\varphi-\Phi) \sin \omega \cos \Theta+\sin (\varphi-\Phi) \cos \omega+\cot \vartheta \sin \omega \sin \Theta} .\label{eq:1:4:5}
\end{gather}
The inverse relations are
\begin{gather}
\cos \vartheta=\cos \vartheta^{\prime}\left(\cos \omega \sin ^{2} \Theta+\cos ^{2} \Theta\right)+\sin \vartheta^{\prime} \sin \Theta\left[(1-\cos \omega) \cos \Theta \cos \left(\varphi^{\prime}-\Phi\right)+\sin \omega \sin \left(\varphi^{\prime}-\Phi\right)\right] , \notag\\
\cot (\varphi-\Phi)=\frac{\cos \left(\varphi^{\prime}-\Phi\right)\left[\cos \omega \cos ^{2} \Theta+\sin ^{2} \Theta\right]-\sin \left(\varphi^{\prime}-\Phi\right) \sin \omega \cos \Theta-\cot \vartheta^{\prime}(\cos \omega-1) \sin \Theta \cos \Theta}{\cos \left(\varphi^{\prime}-\Phi\right) \sin \omega \cos \Theta+\sin \left(\varphi^{\prime}-\Phi\right) \cos \omega-\cot \vartheta^{\prime} \sin \omega \sin \Theta} .\label{eq:1:4:6}
\end{gather}
Introducing parameters \(\omega, \vect{n}(\theta, \Phi)\) to describe rotations, we are able to write the transformation properties of components of the position vector \(r\) in compact vector form:
\begin{align}
& \vect{r}^{\prime}=\vect{r} \cos \omega+\vect{n}(\vect{n} \vect{r})(1-\cos \omega)+[\vect{n} \times \vect{r}] \sin \omega,  \notag\\
& \vect{r}=\vect{r}^{\prime} \cos \omega+\vect{n}\left(\vect{n} \vect{r}^{\prime}\right)(1-\cos \omega)-\left[\vect{n} \times \vect{r}^{\prime}\right] \sin \omega \label{eq:1:4:7} .
\end{align}
Equations \eqref{eq:1:4:5}-\eqref{eq:1:4:6} may be derived by projecting Eq. \eqref{eq:1:4:7} onto the coordinate axes.

\subsection{Description of Rotations in Terms of Unitary \(2 \times 2\) Matrices. Cayley-Klein Parameters.}\label{sec:1.4.3}\index{Cayley-Klein parameters}\index{rotation!Cayley-Klein parameters}\index{unitary matrix}\index{matrix!unitary $2\times2$}
The position vector of an arbitrary point \(\vect{r}=x \vect{e}_{x}+y \vect{e}_{y}+z \vect{e}_{z}\) may be represented by the following Hermitian \(2 \times 2\) matrix \(X\) :
\begin{equation}
X=X^{+}=\begin{pmatrix}
z & x+i y  \label{eq:1:4:8}\\
x-i y & -z
\end{pmatrix}=\sum_{i=x, y, z} x_{i} \tilde{\sigma}_{i},
\end{equation}
where \(\tilde{\sigma}_{i}(i=x, y, z)\) are the transposed Pauli matrices (Eq. \eqref{eq:2:5:4}).\index{Pauli matrices}\index{matrix!Pauli} Note that
\[
-\det X=\vect{r}^{2}=x^{2}+y^{2}+z^{2}
\]
Each rotation \(S\{x, y, z\} \rightarrow S^{\prime}\left\{x^{\prime}, y^{\prime}, z^{\prime}\right\}\) may be represented by a unitary transformation \(U\) of matrix \(X\) into \(\mat{X}^{\prime}\)
\begin{equation}
X^{\prime}=U X U^{-1} .\label{eq:1:4:9}
\end{equation}
Here \(U\) is the unitary unimodular \(2 \times 2\) matrix
\begin{equation}
U^{+}=U^{-1}, \quad \det U=1 .\label{eq:1:4:10}
\end{equation}
\begin{figure}[tbh]
\begin{center}
  \includegraphics[alt={},max width=0.4\textwidth]{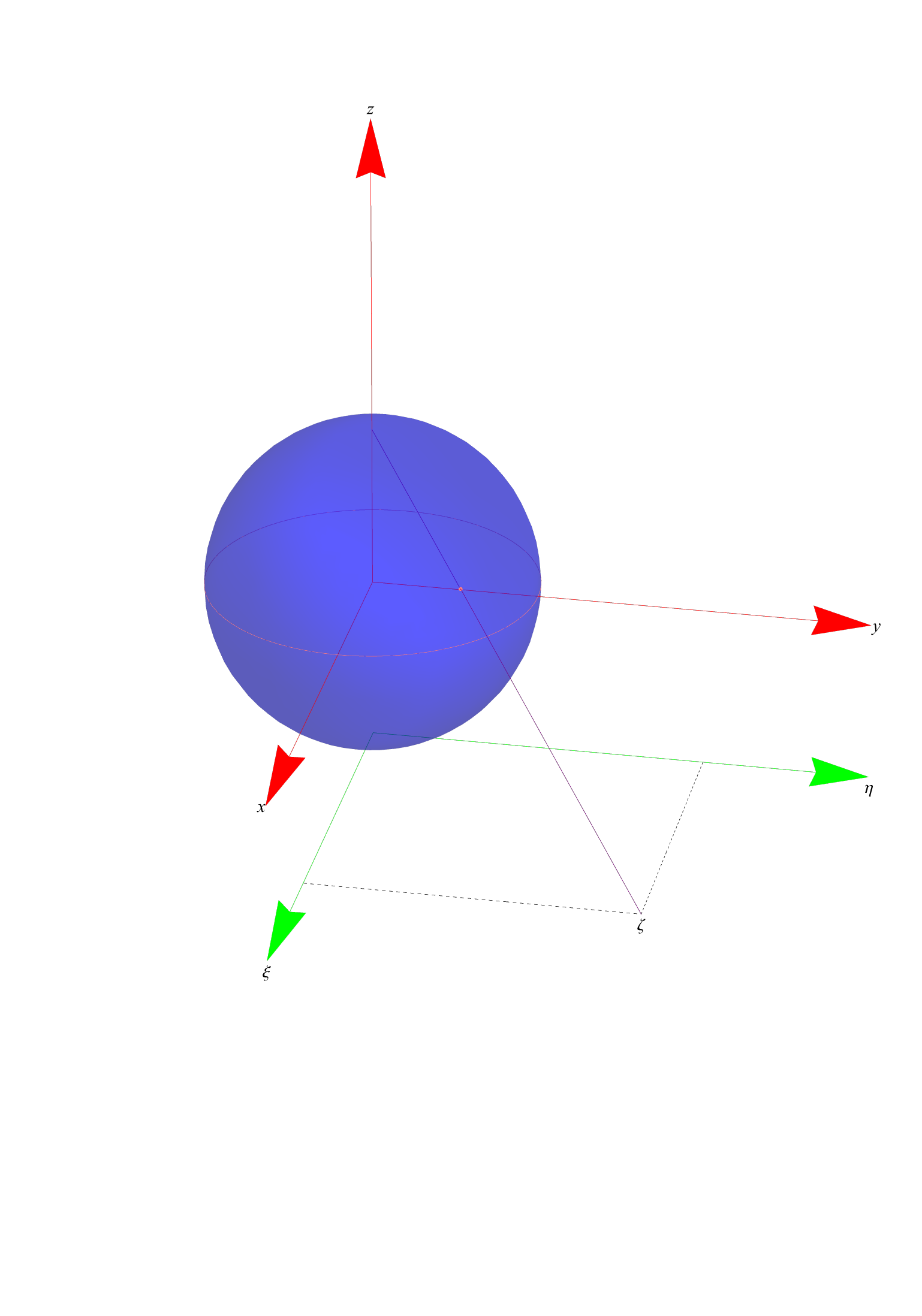}
\caption{Stereographic projection of a point on a sphere.}
\label{chap1:fig:7}
\end{center}
\end{figure}

Bearing in mind that \(\vect{r}\) is real, one can easily prove that \(U\), indeed, is unitary. The relation \(\det U=1\) ensures the invariance of the absolute value of the position vector under coordinate rotations.

Equations \eqref{eq:1:4:10} imply the following form of \(U\)
\begin{equation}
U=\begin{pmatrix}
a & b  \label{eq:1:4:11}\\
-b^{*} & a^{*}
\end{pmatrix}
\end{equation}
where \(a\) and \(b\) are complex numbers which satisfy the condition
\begin{equation}
|a|^{2}+|b|^{2}=1 \label{eq:1:4:12}
\end{equation}
Thus, the matrix \(U\) depends on three real independent parameters. The numbers \(a\) and \(b\) are called the Cayley-Klein parameters. They uniquely determine rotation of the coordinate system. The inverse statement is not true because the parameters \(-a\) and \(-b\) describe the same rotation as \(a\) and \(b\).

The inverse rotation \(S^{\prime}\left\{x^{\prime}, y^{\prime}, z^{\prime}\right\} \rightarrow S\{x, y, z\}\) corresponds to the matrix
\begin{equation}
U^{-1}=U^{+}=\begin{pmatrix}
a^{*} & -b  \label{eq:1:4:13}\\
b^{*} & a
\end{pmatrix}.
\end{equation}
The Cayley-Klein parameters permit us to propose an alternative interpretation of coordinate rotations. Let us consider a sphere (of unit diameter) about the origin. Each point of the sphere with coordinates \(x, y, z\left(x^{2}+y^{2}+z^{2}=1 / 4\right)\) corresponds to the point \(\varsigma=\xi+i \eta\) on the complex plane which is called the stereographic projection of the point \(x, y, z\) (Fig.~\ref{chap1:fig:7}).\index{stereographic projection} The complex number \(\varsigma\) is related to \(x, y, z\) by
\begin{equation}
\varsigma=\frac{x+i y}{\frac{1}{2}-z}=\frac{\frac{1}{2}+z}{x-i y} \label{eq:1:4:14}
\end{equation}
The rotation of the coordinate system which transforms the coordinates \(x, y, z\) of the point on the sphere into \(x^{\prime}, y^{\prime}, z^{\prime}\) generates the following complex-plane bilinear transformation
\begin{equation}
\varsigma^{\prime}=\frac{a \varsigma+b}{-b^{*} \varsigma+a^{*}} .\label{eq:1:4:15}
\end{equation}
The coefficients of this bilinear transformation are just the Cayley-Klein parameters \(a\) and \(b\).

\subsection{Relations Between Different Descriptions of Rotations}\label{sec:1.4.4}\index{rotation!descriptions, relations between}
(a) Relations Between Angles \(\omega, \Theta, \Phi\) and Euler Angles \(\alpha, \beta, \gamma\)

The angles \(\omega, \Theta, \Phi\) are expressed in terms of the Euler angles \(\alpha, \beta, \gamma\) by
\begin{gather}
\cos \frac{\omega}{2}=\cos \frac{\beta}{2} \cos \frac{\alpha+\gamma}{2}  \notag\\
\tan \Theta=\frac{\tan \frac{\beta}{2}}{\sin \frac{\alpha+\gamma}{2}}  \label{eq:1:4:16}\\
\Phi=\frac{\pi}{2}+\frac{\alpha-\gamma}{2} \notag
\end{gather}
The inverse relations are written as
\begin{gather}
\sin \frac{\beta}{2}=\sin \Theta \sin \frac{\omega}{2},  \notag\\
\tan \frac{\alpha+\gamma}{2}=\cos \Theta \tan \frac{\omega}{2} , \label{eq:1:4:17}\\
\frac{\alpha-\gamma}{2}=\Phi-\frac{\pi}{2}. \notag
\end{gather}
Note also the following useful relations between the angles \(\alpha, \beta, \gamma\) and \(\omega, \Theta, \Phi\).
\begin{equation}
\begin{aligned}
&\frac{\partial \alpha}{\partial \omega}=\frac{\partial \gamma}{\partial \omega}=\frac{\cos \Theta}{2 \cos ^{2} \frac{\beta}{2}}, && \frac{\partial \beta}{\partial \omega}=\frac{\sin ^{2} \Theta \sin \omega}{\sin \beta}, \\
&\frac{\partial \alpha}{\partial \Theta}=\frac{\partial \gamma}{\partial \Theta}=-\frac{1}{2} \tan \Theta \sin (\alpha+\gamma), && \frac{\partial \beta}{\partial \Theta}=\frac{2 \sin 2 \Theta \sin ^{2} \frac{\omega}{2}}{\sin \beta}, \\
&\frac{\partial \alpha}{\partial \Phi}=-\frac{\partial \gamma}{\partial \Phi}=1, && \frac{\partial \beta}{\partial \Phi}=0, 
\end{aligned}
\end{equation}
\begin{equation}
\begin{aligned}
&\frac{\partial \omega}{\partial \alpha}=\frac{\partial \omega}{\partial \gamma}=\cos \Theta, & &\frac{\partial \omega}{\partial \beta}=\frac{\tan \frac{\beta}{2}}{\tan \frac{\omega}{2}}, \\
&\frac{\partial \Theta}{\partial \alpha}=\frac{\partial \Theta}{\partial \gamma}=-\frac{\sin \Theta}{2 \tan \frac{\omega}{2}}, && \frac{\partial \Theta}{\partial \beta}=\frac{\cos \Theta \sin \Theta}{\sin \beta}, \\
&\frac{\partial \Phi}{\partial \alpha}=-\frac{\partial \Phi}{\partial \gamma}=\frac{1}{2}, && \frac{\partial \Phi}{\partial \beta}=0 .
\end{aligned}
\end{equation}
The Jacobian of the transformation is equal to
\begin{equation}
\left\|\frac{\partial(\alpha, \beta, \gamma)}{\partial(\omega, \Theta, \Phi)}\right\|=\left\|\frac{\partial(\omega, \Theta, \Phi)}{\partial(\alpha, \beta, \gamma)}\right\|^{-1}=\frac{4 \sin \Theta}{\sin \beta} \sin ^{2} \frac{\omega}{2} .\label{eq:1:4:20}
\end{equation}
A volume element of the three-dimensional rotation group is given by
\begin{equation}
d R \equiv \sin \beta d \alpha d \beta d \gamma=4 \sin ^{2} \frac{\omega}{2} d \omega \sin \Theta d \Theta d \Phi \label{eq:1:4:21}
\end{equation}
The total volume of the three-dimensional rotation group is equal to
\begin{equation}
\int d R=\int_{0}^{2 \pi} d \alpha \int_{0}^{\pi} \sin \beta d \beta \int_{0}^{2 \pi} d \gamma=4 \int_{0}^{\pi} \sin ^{2} \frac{\omega}{2} d \omega \int_{0}^{\pi} \sin \Theta d \Theta \int_{0}^{2 \pi} d \Phi=8 \pi^{2}. \label{eq:1:4:22}
\end{equation}
(b) Relations Between Cayley-Klein Parameters \(a, b\) and Euler Angles \(\alpha, \beta, \gamma\)

The parameters \(a\) and \(b\) are expressed in terms of the Euler angles \(\alpha, \beta, \gamma\) as
\begin{align}
& a=\cos \frac{\beta}{2} e^{-i \frac{\alpha+\gamma}{2}} , \notag\\
& b=\sin \frac{\beta}{2} e^{i \frac{\alpha-\gamma}{2}} .\label{eq:1:4:23}
\end{align}
while the inverse relations are
\begin{gather}
\cos \beta=|a|^{2}-|b|^{2}, \quad \cos \frac{\beta}{2}=|a|, \quad \sin \frac{\beta}{2}=|b|  \label{eq:1:4:24}\\
\cot \frac{\alpha+\gamma}{2}=-\frac{\operatorname{Re} a}{\operatorname{Im} a}, \quad \cot \frac{\alpha-\gamma}{2}=\frac{\operatorname{Re} b}{\operatorname{Im} b} \notag
\end{gather}
The parameters \(a\) and \(b\) may be expressed in terms of the Wigner \(D\)-functions (see Chap. 4)
\begin{equation}
a=D_{\frac{1}{2} \frac{1}{2}}^{\frac{1}{2}}(\alpha, \beta, \gamma), \quad b=D_{-\frac{1}{2} \frac{1}{2}}^{\frac{1}{2}}(\alpha, \beta, \gamma) . \label{eq:1:4:25}
\end{equation}
The unitary matrix \(U\), according to Eq.~\eqref{eq:1:4:11}, coincides with the transposed rotation matrix \(\widehat{D}^{\frac{1}{2}}(\alpha, \beta, \gamma)\) which transforms spin functions of particles of spin \(\frac{1}{2}\) (see Eq.~\eqref{eq:2:5:32}).\\
(c) Relations Between Cayley-Klein Parameters \(a, b\) and Angles \(\omega, \Theta, \Phi\)

The parameters \(a, b\) are expressed in terms of the angles \(\omega, \Theta, \Phi\) as
\begin{align}
& a=\cos \frac{\omega}{2}-i \sin \frac{\omega}{2} \cos \Theta , \label{eq:1:4:26}\\
& b=-i \sin \frac{\omega}{2} \sin \Theta e^{i \Phi} \notag.
\end{align}
and the inverse relations have the form
\begin{align}
\cos \omega & =2(\operatorname{Re} a)^{2}-1, \quad \cos \frac{\omega}{2}=\operatorname{Re} a  \notag\\
\cos \Theta & =-\frac{\operatorname{Im} a}{\sqrt{1-(\operatorname{Re} a)^{2}}}  ,\label{eq:1:4:27}\\
\cot \Phi & =-\frac{\operatorname{Im} b}{\operatorname{Re} b} \notag .
\end{align}
The unitary matrix \(U\), according to Eq.~\eqref{eq:1:4:11}, coincides with the transposed rotation matrix \(\widehat{U}^{\frac{1}{2}}(\omega ; \Theta, \Phi)\) which transforms spin functions of spin \(\frac{1}{2}\) (see Eq.~\eqref{eq:2:5:36}).

\subsection{Rotation Operator}\label{sec:1.4.5}\index{rotation operator}\index{operator!rotation}\index{angular momentum operator!total}\index{Wigner D-function@Wigner $D$-function}
Under rotations of coordinate systems quantum-mechanical quantities are transformed by the rotation operators \(\widehat{D}(\alpha, \beta, \gamma)\) or \(\widehat{U}(\omega ; \Theta, \Phi)\).

Wave functions (state vectors) \(\Psi^{\prime}\) and operators \(\widehat{O}^{\prime}\) in a rotated coordinate system are related to wave functions \(\Psi\) and operators \(\hat{O}\) in an initial coordinate system by
\begin{align}
\Psi^{\prime}=\widehat{D}(\alpha, \beta, \gamma) \Psi, & \widehat{O}=\widehat{D}(\alpha, \beta, \gamma) \widehat{O}[\widehat{D}(\alpha, \beta, \gamma)]^{-1}  ,\label{eq:1:4:28}\\
\Psi^{\prime}=\widehat{U}(\omega ; \Theta, \Phi) \Psi, & \widehat{O}=\widehat{U}(\omega ; \Theta, \Phi) \widehat{O}[\widehat{U}(\omega ; \Theta, \Phi)]^{-1} . \label{eq:1:4:29}
\end{align}
If the Euler angles \(\alpha, \beta, \gamma\) are chosen to describe the rotation, then the rotation operator \(\widehat{D}(\alpha, \beta, \gamma)\) may be written as
\begin{equation}
\widehat{D}(\alpha, \beta, \gamma)=e^{-i \gamma \hat{J}_{z^{\prime}}} e^{-i \beta \hat{J}_{y_{1}}} e^{-i \alpha \hat{J}_{z}} \label{eq:1:4:30}
\end{equation}
or, equivalently,
\begin{equation}
\hat{D}(\alpha, \beta, \gamma)=e^{-i \alpha \hat{J}_{z}} e^{-i \beta \hat{J}_{y}} e^{-i \gamma \hat{J}_{z}} .\label{eq:1:4:31}
\end{equation}
Here \(\widehat{J}_{i}\) is the projection of the total angular momentum operator (see Chap. 2) on an \(i\)-axis. The equivalence of Eqs. \eqref{eq:1:4:30} and \eqref{eq:1:4:31} follows from the fact that, according to \eqref{eq:1:4:28},
\begin{align}
& e^{-i \beta \hat{J}_{y_{1}}}=\hat{D}(\alpha, 0,0) e^{-i \beta \hat{J}_{y}}[\hat{D}(\alpha, 0,0)]^{-1}=e^{-i \alpha \hat{J}_{z}} e^{-i \beta \hat{J}_{y}} e^{i \alpha \hat{J}_{z}} , \label{eq:1:4:32}\\
& e^{-i \gamma \hat{J}_{z^{\prime}}}=\hat{D}(\alpha, \beta, 0) e^{-i \gamma \hat{J}_{z}}[\hat{D}(\alpha, \beta, 0)]^{-1}=e^{-i \alpha \hat{J}_{z}} e^{-i \beta \hat{J}_{y}} e^{-i \gamma \hat{J}_{z}} e^{i \beta \hat{J}_{y}} e^{i \alpha \hat{J}_{z}} .\notag
\end{align}
If the direction of the rotation axis \(n(\Theta, \Phi)\) and the rotation angle \(\omega\) are chosen to describe the coordinate rotation, the rotation operator \(\hat{U}\) may be written in the form
\begin{equation}
\hat{U}(\omega ; \Theta, \Phi)=e^{-i \omega n \cdot \hat{J}}, \label{eq:1:4:33}
\end{equation}
where \(\widehat{\vect{J}}\) is the total angular momentum operator (Chap. 2). Note that \(\widehat{D}(\alpha, \beta, \gamma) \equiv \widehat{U}(\omega ; \Theta, \Phi)\).\\
The rotation operator written in the forms \eqref{eq:1:4:30}, \eqref{eq:1:4:31} or \eqref{eq:1:4:33} is an unitary operator.
\begin{align}
\hat{D}^{+}(\alpha, \beta, \gamma) & =[\hat{D}(\alpha, \beta, \gamma)]^{-1}=\hat{D}(\pi-\gamma, \beta,-\pi-\alpha)=\hat{D}(-\gamma,-\beta,-\alpha) , \notag\\
\hat{U}^{+}(\omega ; \Theta, \Phi) & =[\hat{U}(\omega ; \Theta, \Phi)]^{-1}=\hat{U}(\omega ; \pi-\Theta, \pi+\Phi)=\hat{U}(-\omega ; \Theta, \Phi) . \label{eq:1:4:34}
\end{align}
Matrix elements of \(\widehat{D}\) between eigenstates of the operators \(\widehat{J}^{2}, \widehat{J}_{z}\) are the Wigner \(D\)-functions (see Chap. 4)
\begin{equation}
\braOket{J^{\prime} M^{\prime}}{\hat{D}(\alpha, \beta, \gamma)}{J M}=\delta_{J J^{\prime}} D_{M^{\prime} M}^{J}(\alpha, \beta, \gamma) .\label{eq:1:4:35}
\end{equation}
Matrix elements of \(\widehat{D}\) between states corresponding to the cartesian basis vectors \(e_{i}(i=x, y, z)\) coincide with elements of the rotation matrix \(a_{i k}\) (see Sec. 1.4.6)
\begin{equation}
\braOket{\vect{e}_{i}}{\hat{D}(\alpha, \beta, \gamma)}{\vect{e}_{k}}=a_{i k}, \quad(i, k=x, y, z) \label{eq:1:4:36}
\end{equation}
Effects of the rotation operator on various wave functions and quantum-mechanical operators are considered in Chaps. 3, 5-7.

\subsection{Transformation of Cartesian Vectors and Tensors Under Rotations of Coordinate Systems. Rotation Matrix $\vect{A}$}\label{sec:1.4.6}\index{rotation matrix}\index{matrix!rotation}\index{transformation!of Cartesian vectors and tensors}
An arbitrary vector \(\vect{A}\) may be written as a column
\begin{equation}
\vect{A}=\begin{pmatrix}
A_{x}  \label{eq:1:4:37}\\
A_{y} \\
A_{z}
\end{pmatrix},
\end{equation}
where \(A_{x}, A_{y}, A_{z}\) are cartesian components of A . In this representation the Cartesian basis vectors \(\vect{e}_{x}, \vect{e}_{y}, \vect{e}_{z}\) have the form
\begin{equation}
\vect{e}_{x}=\begin{pmatrix}
1  \label{eq:1:4:38}\\
0 \\
0
\end{pmatrix}, \quad \vect{e}_{y}=\begin{pmatrix}
0 \\
1 \\
0
\end{pmatrix}, \quad \vect{e}_{z}=\begin{pmatrix}
0 \\
0 \\
1
\end{pmatrix}.
\end{equation}
The effect of the rotation operator on the basis vectors written in such a form is equivalent to an action of some \(3 \times 3\) matrix \(a\) which may be regarded as one of the representations of the rotation operator
\begin{equation}
a=\begin{pmatrix}
a_{x x} & a_{x y} & a_{x z}  \label{eq:1:4:39}\\
a_{y x} & a_{y y} & a_{y z} \\
a_{z x} & a_{z y} & a_{z z}
\end{pmatrix} .
\end{equation}
The matrix \(a\) is real
\begin{equation}
a^{*}=a, \quad a_{i k}^{*}=a_{i k}, \quad(i, k=x, y, z) . \label{eq:1:4:40}
\end{equation}
and unitary
\begin{equation}
a^{+} a=a a^{+}=1. \label{eq:1:4:41}
\end{equation}
Equations \eqref{eq:1:4:40} and \eqref{eq:1:4:41} result in the orthogonality condition
\begin{equation}
\tilde{a} a=a \tilde{a}=1, \label{eq:1:4:42}
\end{equation}
where \(\tilde{a}\) is the transpose of \(a\). Equation \eqref{eq:1:4:42} written in a component form gives six independent relations for the elements \(a_{i k}\)
\begin{equation}
\sum_{i} a_{i k} a_{i l}=\delta_{k l}, \quad(i, k, l=x, y, z), \label{eq:1:4:43}
\end{equation}
or the equivalent relations
\begin{equation}
\sum_{k} a_{i k} a_{l k}=\delta_{i l}, \quad(i, k, l=x, y, z) . \label{eq:1:4:44}
\end{equation}
Relations \eqref{eq:1:4:43} or \eqref{eq:1:4:44} reveal that only three of the nine matrix elements \(a_{i k}\) are independent. This result is in agreement with the fact that any rotation of the coordinate system is completely determined by three real parameters.

The matrix \(a\) is unimodular, i.e.,
\begin{equation}
\det a=\left\|\begin{array}{lll}
a_{x x} & a_{x y} & a_{x z}  \label{eq:1:4:45}\\
a_{y x} & a_{y y} & a_{y z} \\
a_{z x} & a_{z y} & a_{z z}
\end{array}\right\|=1
\end{equation}
The relations between cartesian basis vectors \(e_{i}^{\prime}\) in a rotated coordinate system \(S^{\prime}\) and basis vectors \(e_{i}\) in an initial coordinate system \(S\) are given by
\begin{equation}
\vect{e}_{i}^{\prime}=a \vect{e}_{i}=\sum_{k} a_{k i} \vect{e}_{k} , \quad(i, k=x, y, z) .\label{eq:1:4:46}
\end{equation}
The transformation properties of cartesian vector components are given by
\begin{equation}
A_{i}^{\prime}=\sum_{k} a_{k i} A_{k}, \quad(i, k=x, y, z), \label{eq:1:4:47}
\end{equation}
where \(A_{k}\) are the components of \(\vect{A}\) in the initial coordinate system and \(A_{i}^{\prime}\) are the components of this vector in the rotated coordinate system. Equations \eqref{eq:1:4:43} and \eqref{eq:1:4:44} ensure that the absolute value of \(\vect{A}\) is unchanged by the rotation.

The transformation rule for cartesian components of a tensor of rank \(n\) ( \(n\) is integer) has the form
\begin{equation}
A_{i_{1} i_{2} \ldots i_{n}}^{\prime}=\sum_{k_{1} k_{2} \ldots k_{n}} a_{k_{1} i_{1}} a_{k_{2} i_{2}} \ldots a_{k_{n} i_{n}} A_{k_{1} k_{2} \ldots k_{n}} . \label{eq:1:4:48}
\end{equation}
The inverse transformation which corresponds to the rotation \(S^{\prime} \rightarrow S\) is performed by the transposed matrix \(\tilde{a}=a^{-1}\). The inverse relations are
\begin{gather}
\vect{e}_{k}=\sum_{i} a_{k i} \vect{e}_{i}^{\prime}, \quad(i, k=x, y, z) , \label{eq:1:4:49}\\
A_{k}=\sum_{i} a_{k i} A_{i}^{\prime}, \quad(i, k=x, y, z) , \label{eq:1:4:50}\\
A_{k_{1} k_{2} \ldots k_{n}}=\sum_{i_{1} i_{2} \ldots i_{n}} a_{k_{1} i_{1}} a_{k_{2} i_{2} \ldots a_{k_{n} i_{n}}} A_{i_{1} i_{2} \ldots i_{n}}^{\prime}, \quad\left(i_{1}, k_{1}, i_{2}, k_{2}, \ldots, i_{n}, k_{n}=x, y, z\right). \label{eq:1:4:51}
\end{gather}
The elements of the rotation matrix \(a_{i k}\) may be evaluated from
\begin{equation}
a_{i k}=\vect{e}_{i} \vect{e}_{k}^{\prime}, \quad(i, k=x, y, z), \label{eq:1:4:52}
\end{equation}
Thus, the elements \(a_{i k}\) are cosines of angles between the basis vectors in the initial ( \(S\) ) and rotated ( \(S^{\prime}\) ) coordinate systems. An equivalent definition of \(a_{i k}\) in terms of coordinates in \(S\) and \(S^{\prime}\) has the form
\begin{equation}
a_{i k}=\frac{\partial x_{i}^{\prime}}{\partial x_{k}}=\frac{\partial x_{k}}{\partial x_{i}^{\prime}}, \quad\left(x_{i}^{\prime}, x_{k}=x, y, z\right) \label{eq:1:4:53}
\end{equation}
The rotation matrix \(a\) is given in terms of the Euler angles by
\begin{equation}
a=\begin{pmatrix}
\cos \alpha \cos \beta \cos \gamma-\sin \alpha \sin \gamma & -\cos \alpha \cos \beta \sin \gamma-\sin \alpha \cos \gamma & \cos \alpha \sin \beta  \label{eq:1:4:54}\\
\sin \alpha \cos \beta \cos \gamma+\cos \alpha \sin \gamma & -\sin \alpha \cos \beta \sin \gamma+\cos \alpha \cos \gamma & \sin \alpha \sin \beta \\
-\sin \beta \cos \gamma & \sin \beta \sin \gamma & \cos \beta
\end{pmatrix}.
\end{equation}
The inverse matrix \(a^{-1}\) may be obtained from Eq. \eqref{eq:1:4:54} by transposing or, equivalently, by replacing the Euler angles \(\alpha, \beta, \gamma \rightarrow-\gamma,-\beta,-\alpha\).

The expression for the rotation matrix \(a\) in terms of the angles \(\Theta, \Phi\) which describe the direction of the rotation axis, and the rotation angle \(\omega\) has the form
\begin{align}
& a=\left(\begin{array}{cc}
(1-\cos \omega) \sin ^{2} \Theta \cos ^{2} \Phi+\cos \omega \\
(1-\cos \omega) \sin ^{2} \Theta \cos \Phi \sin \Phi+\sin \omega \cos \Theta \\
(1-\cos \omega) \sin \Theta \cos \Theta \cos \Phi-\sin \omega \sin \Theta \sin \Phi
\end{array}\right.  \notag\\
& \left.\begin{array}{ll}
(1-\cos \omega) \sin ^{2} \Theta \cos \Phi \sin \Phi-\sin \omega \cos \Theta & (1-\cos \omega) \sin \Theta \cos \Theta \cos \Phi+\sin \omega \sin \Theta \sin \Phi \\
(1-\cos \omega) \sin ^{2} \Theta \sin ^{2} \Phi+\cos \omega & (1-\cos \omega) \sin \Theta \cos \Theta \sin \Phi-\sin \omega \sin \Theta \cos \Phi \\
(1-\cos \omega) \sin \Theta \cos \Theta \sin \Phi+\sin \omega \sin \Theta \cos \Phi & (1-\cos \omega) \cos ^{2} \Theta+\cos \omega
\end{array}\right) . \label{eq:1:4:55}
\end{align}
Equation \eqref{eq:1:4:55} may be expanded into
\begin{equation}
a=\cos \omega\begin{pmatrix}
1 & 0 & 0  \label{eq:1:4:56}\\
0 & 1 & 0 \\
0 & 0 & 1
\end{pmatrix}+(1-\cos \omega)\begin{pmatrix}
n_{x}^{2} & n_{x} n_{y} & n_{x} n_{z} \\
n_{y} n_{x} & n_{y}^{2} & n_{y} n_{z} \\
n_{z} n_{x} & n_{z} n_{y} & n_{z}^{2}
\end{pmatrix}+\sin \omega\begin{pmatrix}
0 & -n_{z} & n_{y} \\
n_{z} & 0 & -n_{x} \\
-n_{y} & n_{x} & 0
\end{pmatrix}
\end{equation}
where \(n_{x}, n_{y}, n_{z}\) are components of the unit vector \(n\) which determines direction of the rotation axis. Using \eqref{eq:1:4:56} one can easily derive the following expressions for the matrix elements \(a_{i k}\)
\begin{equation}
a_{i k}=\cos \omega \delta_{i k}+(1-\cos \omega) n_{i} n_{k}-\sin \omega \varepsilon_{i k l} n_{l}, \quad(i, k, l=x, y, z) .\label{eq:1:4:57}
\end{equation}
The inverse matrix \(a^{-1}\) may be obtained from \eqref{eq:1:4:55} and \eqref{eq:1:4:56} by transposing or, equivalently, by replacing \(\omega, \Theta, \Phi \rightarrow-\omega, \Theta, \Phi\) or \(\omega, \Theta, \Phi \rightarrow \omega, \pi-\Theta, \pi+\Phi\).

The expressions for \(\omega, \Theta, \Phi\) in terms of the matrix elements \(a_{i k}\) read
\begin{align}
& \cos \omega=\frac{1}{2}[\operatorname{Tr} a-1]= \frac{1}{2}\left(a_{x x}+a_{y y}+a_{z z}-1\right) , \notag\\
& n_{x} \sin \omega \equiv \sin \omega \sin \Theta \cos \Phi= \frac{1}{2}\left(a_{z y}-a_{y z}\right) , \notag\\
& n_{y} \sin \omega \equiv \sin \omega \sin \Theta \sin \Phi= \frac{1}{2}\left(a_{x z}-a_{z x}\right) , \label{eq:1:4:58}\\
& n_{z} \sin \omega \equiv \sin \omega \cos \Theta= \frac{1}{2}\left(a_{y x}-a_{x y}\right)  ,\notag\\
& n_{i} \sin \omega=-\frac{1}{2} \sum_{k l} \varepsilon_{i k l} a_{k l}, \quad(i, k, l=x, y, z) .\notag
\end{align}
The rotation matrix \(a\) may be rewritten in terms of the Cayley-Klein parameters as
\begin{equation}
a=\begin{pmatrix}
\frac{1}{2}\left(a^{2}-b^{2}+a^{* 2}-b^{* 2}\right) & \frac{i}{2}\left(-a^{2}+b^{2}+a^{* 2}-b^{* 2}\right) & a b^{*}+a^{*} b  \label{eq:1:4:59}\\
\frac{i}{2}\left(a^{2}+b^{2}-a^{* 2}-b^{* 2}\right) & \frac{1}{2}\left(a^{2}+b^{2}+a^{* 2}+b^{* 2}\right) & i\left(a b^{*}-a^{*} b\right) \\
-\left(a b+a^{*} b^{*}\right) & i\left(a b-a^{*} b^{*}\right) & a a^{*}-b b^{*}
\end{pmatrix} .
\end{equation}
One can see that the parameters \(a, b\) and \(-a,-b\) correspond to the same rotation matrix.

\subsubsection{Particular Forms of Rotation Matrix}\label{sec:1.4.7}\index{rotation matrix!particular forms}
(a) Rotation through an angle \(\Psi\) about the \(x\)-axis:
\begin{equation}
a_{x}(\Psi)=\begin{pmatrix}
1 & 0 & 0  \label{eq:1:4:60}\\
0 & \cos \Psi & -\sin \Psi \\
0 & \sin \Psi & \cos \Psi
\end{pmatrix} .
\end{equation}
(b) Rotation through an angle \(\Psi\) about the \(y\)-axis:
\begin{equation}
a_{y}(\Psi)=\begin{pmatrix}
\cos \Psi & 0 & \sin \Psi  \label{eq:1:4:61}\\
0 & 1 & 0 \\
-\sin \Psi & 0 & \cos \Psi
\end{pmatrix} .
\end{equation}
(c) Rotation through an angle \(\Psi\) about the \(z\)-axis:
\begin{equation}
a_{z}(\Psi)=\begin{pmatrix}
\cos \Psi & -\sin \Psi & 0  \label{eq:1:4:62}\\
\sin \Psi & \cos \Psi & 0 \\
0 & 0 & 1
\end{pmatrix}.
\end{equation}
For an arbitrary rotation determined by the Euler angles \(\alpha, \beta, \gamma\) the rotation matrix, in accordance with Eq. \eqref{eq:1:4:31}, may be written in the form
\begin{equation}
a=a_{n}(\omega)=a_{z}(\alpha) a_{y}(\beta) a_{z}(\gamma). \label{eq:1:4:63}
\end{equation}
Equation \eqref{eq:1:4:63} represents a particular case of addition of coordinate rotations (see Sec. 1.4.7).

\subsection{Addition of Rotations}\label{sec:1.4.8}\index{rotation!addition of}\index{addition of rotations}
Let us consider two successive rotations of the coordinate system. Let the first rotation transform the coordinate system \(S\{x, y, z\}\) into \(S^{\prime}\left\{x^{\prime}, y^{\prime}, z^{\prime}\right\}\) and the second one transform \(S^{\prime}\left\{x^{\prime}, y^{\prime}, z^{\prime}\right\}\) into \(S^{\prime \prime}\left\{x^{\prime \prime}, y^{\prime \prime}, z^{\prime \prime}\right\}\).

Below the parameters describing the resultant rotation \(S\{x, y, z\} \rightarrow S^{\prime \prime}\left\{x^{\prime \prime}, y^{\prime \prime}, z^{\prime \prime}\right\}\) will be given in terms of the parameters specifying the rotations \(S\{x, y, z\} \rightarrow S^{\prime}\left\{x^{\prime}, y^{\prime}, z^{\prime}\right\}\) and \(S^{\prime}\left\{x^{\prime}, y^{\prime}, z^{\prime}\right\} \rightarrow S^{\prime \prime}\left\{x^{\prime \prime}, y^{\prime \prime}, z^{\prime \prime}\right\}\).

\subsubsection{(a) Description of rotations in terms of Euler angles}
Let both rotations be performed according to the scheme \(\vect{B}(\vect{p} .22)\). Let the first rotation \(S\{x, y, z\} \rightarrow S^{\prime}\left\{x^{\prime}, y^{\prime}, z^{\prime}\right\}\) be described by the Euler angles \(\alpha_{1}, \beta_{1}, \gamma_{1}\), the second one, \(S^{\prime}\left\{x^{\prime}, y^{\prime}, z^{\prime}\right\} \rightarrow S^{\prime \prime}\left\{x^{\prime \prime}, y^{\prime \prime}, z^{\prime \prime}\right\}\), by the Euler angles \(\alpha_{2}, \beta_{2}, \gamma_{2}\) and the resultant rotation, \(S\{x, y, z\} \rightarrow S^{\prime \prime}\left\{x^{\prime \prime}, y^{\prime \prime}, z^{\prime \prime}\right\}\) by the Euler angles \(\alpha, \beta, \gamma\). The Euler angles \(\alpha, \beta, \gamma, \alpha_{1}, \beta_{1}, \gamma_{1}\) and \(\alpha_{2}, \beta_{2}, \gamma_{2}\) are supposed to be defined with respect to an initial coordinate system \(S\{x, y, z\}\).

The operator of resultant rotation has the form
\begin{equation}
\widehat{D}(\alpha, \beta, \gamma)=\widehat{D}\left(\alpha_{2}, \beta_{2}, \gamma_{2}\right) \widehat{D}\left(\alpha_{1}, \beta_{1}, \gamma_{1}\right) ,\label{eq:1:4:64}
\end{equation}
or in more detail
\begin{equation}
e^{-i \alpha \hat{J}_{z}} e^{-i \beta \hat{J}_{y}} e^{-i \gamma \hat{J}_{z}}=e^{-i \alpha_{2} \hat{J}_{z}} e^{-i \beta_{2} \hat{J}_{y}} e^{-i \gamma_{2} \hat{J}_{z}} e^{-i \alpha_{1} \hat{J}_{z}} e^{-i \beta_{1} \hat{J}_{y}} e^{-i \gamma_{1} \hat{J}_{z}} . \label{eq:1:4:65}
\end{equation}
In Eq. \eqref{eq:1:4:65} \(\widehat{J}_{i}\) is the projection of the total angular momentum operator on an \(i\)-axis of the coordinate system \(S\{x, y, z\}\). The angles of the resultant rotation \(\alpha, \beta, \gamma\) are expressed in terms of \(\alpha_{1}, \beta_{1}, \gamma_{1}\) and \(\alpha_{2}, \beta_{2}, \gamma_{2}\) as
\begin{align}
\cot \left(\alpha-\alpha_{2}\right) & =\cos \beta_{2} \cot \left(\alpha_{1}+\gamma_{2}\right)+\cot \beta_{1} \frac{\sin \beta_{2}}{\sin \left(\alpha_{1}+\gamma_{2}\right)} , \notag\\
\cos \beta & =\cos \beta_{1} \cos \beta_{2}-\sin \beta_{1} \sin \beta_{2} \cos \left(\alpha_{1}+\gamma_{2}\right) , \label{eq:1:4:66}\\
\cot \left(\gamma-\gamma_{1}\right) & =\cos \beta_{1} \cot \left(\alpha_{1}+\gamma_{2}\right)+\cot \beta_{2} \frac{\sin \beta_{1}}{\sin \left(\alpha_{1}+\gamma_{2}\right)} .\notag
\end{align}
The following relations are useful for evaluation of \(\alpha, \beta, \gamma\).
\begin{gather}
\frac{\sin \left(\alpha-\alpha_{2}\right)}{\sin \beta_{1}}=\frac{\sin \left(\gamma-\gamma_{1}\right)}{\sin \beta_{2}}=\frac{\sin \left(\alpha_{1}+\gamma_{2}\right)}{\sin \beta},  \label{eq:1:4:67}\\
\cos \beta_{1}=\cos \beta \cos \beta_{2}+\sin \beta \sin \beta_{2} \cos \left(\alpha-\alpha_{2}\right),  \notag\\
\cos \beta_{2}=\cos \beta \cos \beta_{1}+\sin \beta \sin \beta_{1} \cos \left(\gamma-\gamma_{1}\right),  \label{eq:1:4:68}\\
\cos \beta=\cos \beta_{1} \cos \beta_{2}-\sin \beta_{1} \sin \beta_{2} \cos \left(\alpha_{1}+\gamma_{2}\right),  \notag\\
\cos \left(\gamma-\gamma_{1}\right)=\cos \left(\alpha_{1}+\gamma_{2}\right) \cos \left(\alpha-\alpha_{2}\right)+\sin \left(\alpha_{1}+\gamma_{2}\right) \sin \left(\alpha-\alpha_{2}\right) \cos \beta_{2},  \notag\\
\cos \left(\alpha-\alpha_{2}\right)=\cos \left(\alpha_{1}+\gamma_{2}\right) \cos \left(\gamma-\gamma_{1}\right)+\sin \left(\alpha_{1}+\gamma_{2}\right) \sin \left(\gamma-\gamma_{1}\right) \cos \beta_{1},  \label{eq:1:4:69}\\
\cos \left(\alpha_{1}+\gamma_{2}\right)=\cos \left(\gamma-\gamma_{1}\right) \cos \left(\alpha-\alpha_{2}\right)-\sin \left(\gamma-\gamma_{1}\right) \sin \left(\alpha-\alpha_{2}\right) \cos \beta,  \notag\\
\frac{\tan \frac{\beta-\beta_{1}}{2}}{\tan \frac{\beta+\beta_{1}}{2}}=\frac{\tan \frac{\alpha_{1}+\gamma_{2}+\alpha-\alpha_{2}}{2}}{\tan \frac{\alpha_{1}+\gamma_{2}-\alpha+\alpha_{2}}{2}},  \notag\\
\frac{\tan \frac{\beta-\beta_{2}}{2}}{\tan \frac{\beta+\beta_{2}}{2}}=\frac{\tan \frac{\alpha_{1}+\gamma_{2}+\gamma-\gamma_{1}}{2}}{\tan \frac{\alpha_{1}+\gamma_{2}-\gamma+\gamma_{1}}{2}},  \label{eq:1:4:70}\\
\frac{\tan \frac{\beta_{1}-\beta_{2}}{2}}{\tan \frac{\beta_{1}+\beta_{2}}{2}}=\frac{\tan \frac{\alpha-\alpha_{2}+\gamma-\gamma_{1}}{2}}{\tan \frac{\alpha-\alpha_{2}-\gamma+\gamma_{1}}{2}} . \notag
\end{gather}
\begin{figure}[tbh]
\begin{center}
  \caption{Addition of rotations in terms of the spherical geometry.}
\label{chap1:fig:8}
\end{center}
\end{figure}

Equations \eqref{eq:1:4:67}-\eqref{eq:1:4:70} may be easily interpreted in terms of the geometry on a sphere. Each rotation may be completely determined by a point of intersection of the \(z^{\prime}\)-axis with the spherical surface and by a unit vector in the direction of the \(x^{\prime}\)-axis which lies on a plane tangent to the surface at this point. In this case the determination of \(\alpha, \beta, \gamma\) is reduced to constructing the corresponding spherical triangle, (Fig.~\ref{chap1:fig:8}). Equations \eqref{eq:1:4:67}-\eqref{eq:1:4:70} represent the formulas of sines, cosines and tangents for the spherical triangle.

Another expression for the angles of the resultant rotation will be obtained if successive rotations are performed according to the scheme \(\vect{B}\) (p. 22) but the Euler angles \(\alpha_{2}, \beta_{2}, \gamma_{2}\) specifying the second rotation \(S^{\prime}\left\{x^{\prime}, y^{\prime}, z^{\prime}\right\} \rightarrow S^{\prime \prime}\left\{x^{\prime \prime}, y^{\prime \prime}, z^{\prime \prime}\right\}\) are defined with respect to the intermediate coordinate system \(S^{\prime}\left\{x^{\prime}, y^{\prime}, z^{\prime}\right\}\) rather than the initial system \(S\{x, y, z\}\). In this case the operator of the resultant rotation has the form
\begin{equation}
\widehat{D}(\alpha, \beta, \gamma)=\widehat{D}^{\prime}\left(\alpha_{2}, \beta_{2}, \gamma_{2}\right) \widehat{D}\left(\alpha_{1}, \beta_{1}, \gamma_{1}\right), \label{eq:1:4:71}
\end{equation}
where prime indicates that the operator of the second rotation is taken in the coordinate system \(S^{\prime}\left\{x^{\prime}, y^{\prime}, z^{\prime}\right\}\). According to \eqref{eq:1:4:28}, the operator \(\widehat{D}^{\prime}\) is related to the operator in the initial coordinate system by
\begin{equation}
\widehat{D}^{\prime}\left(\alpha_{2}, \beta_{2}, \gamma_{2}\right)=\widehat{D}\left(\alpha_{1}, \beta_{1}, \gamma_{1}\right) \widehat{D}\left(\alpha_{2}, \beta_{2}, \gamma_{2}\right)\left[\widehat{D}\left(\alpha_{1}, \beta_{1}, \gamma_{1}\right)\right]^{-1} . \label{eq:1:4:72}
\end{equation}
Substitution of this expression into Eq. \eqref{eq:1:4:71} yields
\begin{equation}
\widehat{D}(\alpha, \beta, \gamma)=\widehat{D}\left(\alpha_{1}, \beta_{1}, \gamma_{1}\right) \widehat{D}\left(\alpha_{2}, \beta_{2}, \gamma_{2}\right), \label{eq:1:4:73}
\end{equation}
i.e., the operator of the resultant rotation differs from \eqref{eq:1:4:64} in the order of operators of the first and second rotations. Thus, for such a description of the successive rotations the Euler angles \(\alpha, \beta, \gamma\) may be obtained from \eqref{eq:1:4:66}-\eqref{eq:1:4:70} by interchange of indices \(1 \rightleftharpoons 2\).

Finally, if successive rotations are performed according to the scheme \(\vect{A}\) (p. 21), i.e., if each rotation is made about the corresponding new axis, the operator of resultant rotation is given by Eq. \eqref{eq:1:4:73}. In this case the Euler angles \(\alpha, \beta, \gamma\) of the resultant rotation may also be derived from \eqref{eq:1:4:66}-\eqref{eq:1:4:70} by the interchange of indices \(1 \rightleftharpoons 2\).

\subsubsection{(b) Description of Rotations in Terms of Rotation Axis \(\vect{n}(\Theta, \Phi)\) and Rotation Angle \(\omega\)}
Let the first rotation, \(S\{x, y, z\} \rightarrow S^{\prime}\left\{x^{\prime}, y^{\prime}, z^{\prime}\right\}\), be performed about an axis \(\vect{n}_{1}\) through an angle \(\omega_{1}\), and the second one, \(S^{\prime}\left\{x^{\prime}, y^{\prime}, z^{\prime}\right\} \rightarrow S^{\prime \prime}\left\{x^{\prime \prime}, y^{\prime \prime}, z^{\prime \prime}\right\}\), be about an axis \(\vect{n}_{2}\) through an angle \(\omega_{2}\). The resultant rotation \(S\{x, y, z\} \rightarrow S^{\prime \prime}\left\{x^{\prime \prime}, y^{\prime \prime}, z^{\prime \prime}\right\}\) may be treated as a rotation about an axis \(n\) through an angle \(\omega\).

The operator of the resultant rotation has the form
\begin{equation}
e^{-i \omega n \cdot \hat{J}}=e^{-i \omega_{2} n_{2} \cdot \hat{\vect{J}}} e^{-i \omega_{1} n_{1} \cdot \hat{\vect{J}}} . \label{eq:1:4:74}
\end{equation}
The angle of the resultant rotation \(\omega\) and the axis of this rotation \(n\) are determined by
\begin{gather}
\cos \frac{\omega}{2}=\cos \frac{\omega_{1}}{2} \cos \frac{\omega_{2}}{2}-\left(n_{1} \cdot n_{2}\right) \sin \frac{\omega_{1}}{2} \sin \frac{\omega_{2}}{2},  \label{eq:1:4:75}\\
n \sin \frac{\omega}{2}=n_{1} \sin \frac{\omega_{1}}{2} \cos \frac{\omega_{2}}{2}+n_{2} \sin \frac{\omega_{2}}{2} \cos \frac{\omega_{1}}{2}-\left[n_{1} \times n_{2}\right] \sin \frac{\omega_{1}}{2} \sin \frac{\omega_{2}}{2} .\notag
\end{gather}
It follows from Eq. \eqref{eq:1:4:75} that the resultant rotation is independent of the order of successive rotations (i.e., the rotation operators commute) if and only if \(n_{1} \times n_{2}=0\), i.e., the axes of both rotations are parallel or antiparallel. In this case
\[
\omega=\omega_{1} \pm \omega_{2}
\]
If directions of the rotation axes \(\vect{n}_{1}, \vect{n}_{2}, \vect{n}\) are specified by the polar angles \(\Theta_{1}, \Phi_{1} ; \Theta_{2}, \Phi_{2}\) and \(\Theta, \Phi\), respectively, and the polar angles are defined with respect to the initial coordinate system \(S\{x, y, z\}\), then
\begin{gather}
\cos \frac{\omega}{2}=\cos \frac{\omega_{1}}{2} \cos \frac{\omega_{2}}{2}-\sin \frac{\omega_{1}}{2} \sin \frac{\omega_{2}}{2}\left[\cos \Theta_{1} \cos \Theta_{2}+\sin \Theta_{1} \sin \Theta_{2} \cos \left(\Phi_{1}-\Phi_{2}\right)\right]  \notag\\
\sin \frac{\omega}{2} \cos \Theta=\sin \frac{\omega_{1}}{2} \cos \frac{\omega_{2}}{2} \cos \Theta_{1}+\sin \frac{\omega_{2}}{2} \cos \frac{\omega_{1}}{2} \cos \Theta_{2}  \notag\\
+\sin \frac{\omega_{1}}{2} \sin \frac{\omega_{2}}{2} \sin \Theta_{1} \sin \Theta_{2} \sin \left(\Phi_{1}-\Phi_{2}\right)  \label{eq:1:4:76}\\
\cot \Phi=\frac{\sin \Theta_{1}\left(\cot \frac{\omega_{2}}{2} \cos \Phi_{1}-\cos \Theta_{2} \sin \Phi_{1}\right)+\sin \Theta_{2}\left(\cot \frac{\omega_{1}}{2} \cos \Phi_{2}+\cos \Theta_{1} \sin \Phi_{2}\right)}{\sin \Theta_{1}\left(\cot \frac{\omega_{2}}{2} \sin \Phi_{1}+\cos \Theta_{2} \cos \Phi_{1}\right)+\sin \Theta_{2}\left(\cot \frac{\omega_{1}}{2} \sin \Phi_{2}-\cos \Theta_{1} \cos \Phi_{2}\right)} . \notag
\end{gather}
If directions of the rotation axes \(n_{1}, n\) are defined by the polar angles \(\Theta_{1}, \Phi_{1}\) and \(\Theta, \Phi\) with respect to the initial coordinate system \(S\{x, y, z\}\) and the direction of the axis of the second rotation \(\vect{n}_{2}\) is defined by the polar angles \(\Theta_{2}, \Phi_{2}\) with respect to the intermediate coordinate system \(S^{\prime}\left\{x^{\prime}, y^{\prime}, z^{\prime}\right\}\) then the angles \(\omega, \Theta, \Phi\) of the resultant rotation will be given by Eqs. \eqref{eq:1:4:76} with interchanged indices \(1 \rightleftharpoons 2\). This situation is similar to the case when rotations are described by Euler angles.

\subsubsection{(c) Description of Rotation in Terms of Cayley-Klein Parameters}
Let the first rotation, \(S\{x, y, z\} \rightarrow S^{\prime}\left\{x^{\prime}, y^{\prime}, z^{\prime}\right\}\) be determined by the Cayley-Klein parameters \(a_{1}, b_{1}\) (1.4.3) and the second one, \(S^{\prime}\left\{x^{\prime}, y^{\prime}, z^{\prime}\right\} \rightarrow S^{\prime \prime}\left\{x^{\prime \prime}, y^{\prime \prime}, z^{\prime \prime}\right\}\), by the parameters \(a_{2}, b_{2}\). Then the resultant rotation \(S\{x, y, z\} \rightarrow S^{\prime \prime}\left\{x^{\prime \prime}, y^{\prime \prime}, z^{\prime \prime}\right\}\) will be determined by the parameters \(a, b\) such as
\begin{align}
& a=a_{1} a_{2}-b_{1}^{*} b_{2},  \notag\\
& b=a_{1}^{*} b_{2}+b_{1} a_{2} .\label{eq:1:4:77}
\end{align}
The matrix \(U\) (see Eq. \eqref{eq:1:4:11}) which describes the resultant rotation is a product of matrices corresponding to the first and second rotations
\begin{equation}
U(a, b)=U\left(a_{2}, b_{2}\right) U\left(a_{1}, b_{1}\right). \label{eq:1:4:78}
\end{equation}
In this case all the matrices are supposed to be given in an initial coordinate system.

\subsubsection{(d) Addition Theorem for Rotation Matrices a}\index{addition theorem!for rotation matrices}\index{rotation matrix!addition theorem}
Let us carry out two successive rotations \(S\{x, y, z\} \rightarrow S^{\prime}\left\{x^{\prime}, y^{\prime}, z^{\prime}\right\}\) and \(S^{\prime}\left\{x^{\prime}, y^{\prime}, z^{\prime}\right\} \rightarrow S^{\prime \prime}\left\{x^{\prime \prime}, y^{\prime \prime}, z^{\prime \prime}\right\}\). The matrix which transforms cartesian components of vectors and tensors under the resultant rotation \(S\{x, y, z\} \rightarrow S^{\prime \prime}\left\{x^{\prime \prime}, y^{\prime \prime}, z^{\prime \prime}\right\}\) represents a product of the matrices \(a(1)\) and \(a(2)\) corresponding to the first and second rotations. The order of these matrices in the product depends on the convention used for the rotation angles. If all angles which describe rotations are referred to the initial coordinate system, i.e., the operator of the resultant rotation is given by Eq. \eqref{eq:1:4:64}, then
\begin{equation}
a=a(2) a(1), \label{eq:1:4:79}
\end{equation}
or, in terms of matrix elements,
\begin{equation}
a_{i k}=\sum_{l} a_{i l}(2) a_{l k}(1), \quad(i, k, l=x, y, z). \label{eq:1:4:80}
\end{equation}
The rotation matrices in terms of the rotation angles are given by Eqs. \eqref{eq:1:4:54} and \eqref{eq:1:4:56}. The angles which determine the resultant rotation are related to the angles of the first and second rotations via Eqs. \eqref{eq:1:4:66} and \eqref{eq:1:4:76}.

If the angles which determine the first and resultant rotations, \(S \rightarrow S^{\prime}\) and \(S \rightarrow S^{\prime \prime}\), are defined with respect to the initial system \(S\{x, y, z\}\) but the angles of the second rotation \(S^{\prime} \rightarrow S^{\prime \prime}\) are defined with respect to the intermediate system \(S^{\prime}\left\{x^{\prime}, y^{\prime}, z^{\prime}\right\}\) (i.e., the operator of the resultant rotation is given by Eq. \eqref{eq:1:4:73}), then
\begin{equation}
a=a(1) a(2), \label{eq:1:4:81}
\end{equation}
or, equivalently,
\begin{equation}
a_{i k}=\sum_{l} a_{i l}(1) a_{l k}(2), \quad(i, k, l=x, y, z) .\label{eq:1:4:82}
\end{equation}
In this case the relationships between the angles of the resultant rotation and the angles of the first and second rotations may be derived from \eqref{eq:1:4:66} and \eqref{eq:1:4:76} by an interchange of indices \(1 \rightleftharpoons 2\).

\chapter{Angular momentum operators} \label{chap:2}
\section{Total angular momentum operator} \label{sec:2.1}\index{angular momentum operator!total}\index{total angular momentum operator}
\subsection{Definition}
In quantum mechanics the total angular momentum operator $\widehat{\vect{J}}$ is defined as an operator which generates transformations of wave functions (state vectors) and quantum operators under infinitesimal rotations of the coordinate system (see Eqs. \eqref{eq:2:1:1} and \eqref{eq:2:1:2}).

A transformation of an arbitrary wave function $\Psi$ under rotation of the coordinate system through an infinitesimal angle $\delta \omega$ about an axis $n$ may be written as
\begin{equation}
\Psi \rightarrow \Psi^{\prime}=(1-i \delta \omega n \cdot \hat{J}) \Psi . \label{eq:2:1:1}
\end{equation}
where $\widehat{\vect{J}}$ is the total angular momentum operator.\isym{J-hat@$\widehat{\vect{J}}$, total angular momentum operator}\\
A transformation of an arbitrary quantum operator $\widehat{O}$ under an infinitesimal rotation has the form
\begin{equation}
\widehat{O} \rightarrow \widehat{O}^{\prime}=\widehat{O}-i \delta \omega n \cdot[\widehat{\vect{J}}, \widehat{O}] . \label{eq:2:1:2}
\end{equation}
The finite rotation operator can also be written in terms of the total angular momentum operator (see Eqs. \ref{eq:1:4:29}, \ref{eq:1:4:30}, \ref{eq:1:4:32}). The total angular momentum operator is Hermitian,\index{Hermitian operator}
\begin{equation}
\widehat{\vect{J}}^{+}=\widehat{\vect{J}} . \label{eq:2:1:3}
\end{equation}
This property for cartesian and spherical components of the operator $\widehat{J}$ can be expressed as
\begin{equation}
\left(\widehat{J}_{i}\right)^{+}=\widehat{J}_{i}, \quad(i=x, y, z) ; \quad\left(\widehat{J}_{\mu}\right)^{+}=\widehat{J}^{\mu}=(-1)^{\mu} \widehat{J}_{-\mu}, \quad(\mu= \pm 1,0). \label{eq:2:1:4}
\end{equation}
The eigenfunctions of the operators $\widehat{\vect{J}}^{2}$ and $\widehat{J}_{z}$ represent well-known tensor spherical harmonics (see Chap.~\ref{chap:7}).

\subsection{Commutation Relations}\index{commutation relations}
Using the definition of the total angular momentum operator \eqref{eq:2:1:1} and equations for the rotation addition (see Sec.~\ref{sec:1.4.7}), one may obtain the commutation rules for the operator $\hat{\vect{J}}$. These rules may be written as
\begin{equation}
[(\vect{a} \cdot \widehat{\vect{J}}),(\vect{b} \cdot \widehat{\vect{J}})]=i \cdot[\vect{a} \times \vect{b}] \cdot \widehat{\vect{J}}, \label{eq:2:1:5}
\end{equation}
$a$ and $b$ being arbitrary constant vectors. The commutation relations for the total angular momentum operator may also be written symbolically as
\begin{equation}
[\widehat{\vect{J}} \times \widehat{\vect{J}}]=i \widehat{\vect{J}} .\label{eq:2:1:6}
\end{equation}
Cartesian components of $\widehat{\vect{J}}$ satisfy the commutation relations
\begin{equation}
\left[\widehat{J}_{i}, \widehat{J}_{k}\right]=i \varepsilon_{i k l} \widehat{J}_{l}, \quad\left[\widehat{\vect{J}}^{2}, \widehat{J}_{i}\right]=0 \quad(i, k, l=x, y, z). \label{eq:2:1:7}
\end{equation}
Equations \eqref{eq:2:1:7} can be derived from \eqref{eq:2:1:5} by substituting $\vect{a}=\vect{e}_{i}, \vect{b}=\vect{e}_{k}(i, k=x, y, z)$ where $\vect{e}_{i}$ and $\vect{e}_{k}$ are the cartesian basis vectors. In more detailed form Eq. \eqref{eq:2:1:7} reads
\begin{gather}
{\left[\hat{J}_{x}, \hat{J}_{x}\right]=\left[\hat{J}_{y}, \hat{J}_{y}\right]=\left[\hat{J}_{z}, \hat{J}_{z}\right]=0} , \notag\\
{\left[\hat{J}_{x}, \hat{J}_{y}\right]=-\left[\hat{J}_{y}, \hat{J}_{x}\right]=i \hat{J}_{z}, \quad\left[\hat{J}_{x}, \hat{J}_{z}\right]=-\left[\hat{J}_{z}, \hat{J}_{x}\right]=-i \hat{J}_{y}} , \notag\\
{\left[\hat{J}_{y}, \hat{J}_{z}\right]=-\left[\hat{J}_{z}, \hat{J}_{y}\right]=i \hat{J}_{x}}  ,\label{eq:2:1:8}\\
{\left[\hat{\vect{J}}^{2}, \hat{J}_{x}\right]=\left[\hat{\vect{J}}^{2}, \hat{J}_{y}\right]=\left[\hat{\vect{J}}^{2}, \hat{J}_{z}\right]=0} .\notag
\end{gather}
The square of the total angular momentum operator $\widehat{\vect{J}}^{2}$ may be expressed in terms of cartesian components $\widehat{J}_{i}(i=x, y, z)$ as
\begin{equation}
\widehat{\vect{J}}^{2}=\sum_{i} \widehat{J}_{i}^{2}=\widehat{J}_{x}^{2}+\widehat{J}_{y}^{2}+\widehat{J}_{z}^{2}.\label{eq:2:1:9}
\end{equation}
Covariant spherical components of $\widehat{\vect{J}}$ satisfy the following commutation relations
\begin{equation}
\left[\hat{J}_{\mu}, \hat{J}_{\nu}\right]=-\sqrt{2} \clebsch{1}{\mu}{1}{\nu}{1}{\lambda} \hat{J}_{\lambda}, \quad\left[\hat{J}^{2}, \hat{J}_{\mu}\right]=0, \quad(\mu, \nu, \lambda= \pm 1,0). \label{eq:2:1:10}
\end{equation}
These relations may be obtained from \eqref{eq:2:1:5} by putting $\vect{a}=\vect{e}_{\mu}, \vect{b}=\vect{e}_{\nu}(\mu, \nu= \pm 1,0), \vect{e}_{\mu}$ and $\vect{e}_{\nu}$ being covariant spherical basis vectors. A more detailed form of \eqref{eq:2:1:10} is
\begin{gather}
{\left[\widehat{J}_{+1}, \widehat{J}_{+1}\right]=\left[\widehat{J}_{0}, \widehat{J}_{0}\right]=\left[\widehat{J}_{-1}, \widehat{J}_{-1}\right]=0},  \notag\\
{\left[\widehat{J}_{+1}, \widehat{J}_{0}\right]=-\left[\widehat{J}_{0}, \widehat{J}_{+1}\right]=-\widehat{J}_{+1},\left[\widehat{J}_{+1}, \widehat{J}_{-1}\right]=-\left[\widehat{J}_{-1}, \widehat{J}_{+1}\right]=-\widehat{J}_{0}} , \notag\\
{\left[\widehat{J}_{0}, \widehat{J}_{-1}\right]=-\left[\widehat{J}_{-1}, \widehat{J}_{0}\right]=-\widehat{J}_{-1}} , \label{eq:2:1:11}\\
{\left[\widehat{J}^{2}, \widehat{J}_{+1}\right]=\left[\widehat{J}^{2}, \widehat{J}_{0}\right]=\left[\widehat{J}^{2}, \widehat{J}_{-1}\right]=0}. \notag
\end{gather}
For contravariant spherical components of $\widehat{J}$ the commutation relations may be written as
\begin{equation}
\left[\widehat{J}^{\mu}, \widehat{J}^{\nu}\right]=\sqrt{2} \clebsch{1}{\mu}{1}{\nu}{1}{\lambda} \widehat{J}^{\lambda}, \quad\left[\widehat{\vect{J}}^{2}, \widehat{J}^{\mu}\right]=0, \quad(\mu, \nu, \lambda= \pm 1,0) \label{eq:2:1:12}
\end{equation}
or, in a more detailed form
\begin{gather}
{\left[\widehat{J}^{+1}, \widehat{J}^{+1}\right]=\left[\widehat{J}^{0}, \widehat{J}^{0}\right]=\left[\widehat{J}^{-1}, \widehat{J}^{-1}\right]=0},  \notag\\
{\left[\widehat{J}^{+1}, \widehat{J}^{0}\right]=-\left[\widehat{J}^{0}, \widehat{J}^{+1}\right]=\widehat{J}^{+1},\left[\widehat{J}^{+1}, \widehat{J}^{-1}\right]=-\left[\widehat{J}^{-1}, \widehat{J}^{+1}\right]=\widehat{J}^{0}} , \label{eq:2:1:13}\\
{\left[\widehat{J}^{0}, \widehat{J}^{-1}\right]=-\left[\widehat{J}^{-1}, \widehat{J}^{0}\right]=\widehat{J}^{-1}}  ,\notag\\
{\left[\widehat{J}^{2}, \widehat{J}^{+1}\right]=\left[\widehat{J}^{2}, \widehat{J}^{0}\right]=\left[\widehat{J}^{2}, \widehat{J}^{-1}\right]=0} .\notag
\end{gather}
The operator $\widehat{\vect{J}}^{2}$ is expressed in terms of spherical components $\widehat{J}_{\mu}(\mu= \pm 1,0)$ as
\begin{equation}
\widehat{\vect{J}}^{2}=\sum_{\mu}(-1)^{\mu} \widehat{J}_{-\mu} \widehat{J}_{\mu}=-\widehat{J}_{+1} \widehat{J}_{-1}+\widehat{J}_{0} \widehat{J}_{0}-\widehat{J}_{-1} \widehat{J}_{+1}=\widehat{J}_{0}^{2}-\widehat{J}_{0}-2 \widehat{J}_{+1} \widehat{J}_{-1}=\widehat{J}_{0}^{2}+\widehat{J}_{0}-2 \widehat{J}_{-1} \widehat{J}_{+1}. \label{eq:2:1:14}
\end{equation}
\subsection{Coordinate Inversion. Time Reversal}\index{coordinate inversion}\index{time reversal}\index{axial vector}
The total angular momentum operator $\widehat{\vect{J}}$ is an axial vector, i.e., it is invariant with respect to coordinate inversion ( $\vect{r} \rightarrow-\vect{r}$ )
\begin{equation}
\begin{aligned}
\hat{P}_{r} \hat{J}_{i} \hat{P}_{r}^{-1}=\hat{J}_{i}, && (i=x, y, z),  \label{eq:2:1:15}\\
\hat{P}_{r} \hat{J}_{\mu} \hat{P}_{r}^{-1}=\hat{J}_{\mu}, && (\mu= \pm 1,0).
\end{aligned}
\end{equation}
Under time reversal $(t \rightarrow-t) \widehat{\vect{J}}$ changes its sign
\begin{equation}
\begin{aligned}
\hat{P}_{t} \hat{J}_{i} \hat{P}_{t}^{-1}=-\hat{J}_{i}, && (i=x, y, z)  \label{eq:2:1:16}\\
\hat{P}_{t} \hat{J}_{\mu} \hat{P}_{t}^{-1}=-\hat{J}_{\mu}, && (\mu= \pm 1,0)
\end{aligned}
\end{equation}
In Eqs. \eqref{eq:2:1:15} and \eqref{eq:2:1:16} $\widehat{P}_{r}$ and $\widehat{P}_{t}$ represent the operators of coordinate inversion and time reversal, respectively.\isym{Pr-hat@$\widehat{P}_{r}$, coordinate-inversion operator}\isym{Pt-hat@$\widehat{P}_{t}$, time-reversal operator}

\subsection{Total Angular Momentum of a System. Orbital and Spin Angular Momenta}\label{sec:2.1.4}\index{angular momentum!of a system}
The total angular momentum of some system consisting of $N$ subsystems with angular momenta $\widehat{\mathrm{j}}(1), \widehat{\mathrm{j}}(2)$, $\ldots, \widehat{\mathrm{j}}(N)$ is the vector sum
\begin{equation}
\widehat{\vect{J}}=\sum_{n=1}^{N} \widehat{\vect{j}}(n). \label{eq:2:1:17}
\end{equation}
For such a system the following commutation relations hold
\begin{gather}
{\left[\hat{j}_{i}(n), \hat{j}_{k}\left(n^{\prime}\right)\right]=i \delta_{n n^{\prime}} \varepsilon_{i k l} \hat{j}_{l}(n), \quad(i, k, l=x, y, z)},  \label{eq:2:1:18}\\
{\left[\hat{J}_{i}, \hat{j}_{k}(n)\right]=i \varepsilon_{i k l} \hat{j}_{l}(n),}  \notag
\end{gather}
\begin{gather}
{\left[\hat{j}_{\mu}(n), \hat{j}_{\nu}\left(n^{\prime}\right)\right]=-\sqrt{2} \delta_{n n^{\prime}} \clebsch{1}{\mu}{1}{\nu}{1}{\lambda} \hat{j}_{\lambda}(n), \quad(\mu, \nu, \lambda= \pm 1,0) .}  \label{eq:2:1:19} \\
{\left[\hat{J}_{\mu}, \hat{j}_{\nu}(n)\right]=-\sqrt{2} \clebsch{1}{\mu}{1}{\nu}{1}{\lambda} \hat{j}_{\lambda}(n), \quad} \notag
\end{gather}
In particular, Eq. \eqref{eq:2:1:17} determines the total angular momentum operator as the sum of the orbital angular momentum operator $\hat{\vect{L}}$ and the spin angular momentum operator $\widehat{\vect{S}}$.
\begin{equation}
\widehat{\vect{J}}=\widehat{\vect{L}}+\widehat{\vect{S}}. \label{eq:2:1:20}
\end{equation}
The properties of the operators $\hat{\vect{L}}$ and $\hat{\vect{S}}$ are considered below (Secs. 2.2 and 2.3). The commutation relations for $\hat{\vect{L}}$ and $\hat{\vect{J}}$ are
\begin{equation}
\begin{aligned}
{\left[\hat{J}_{i}, \hat{L}_{k}\right]=i \varepsilon_{i k l} \hat{L}_{l},} & (i, k, l=x, y, z),\\
{\left[\hat{J}_{\mu}, \hat{L}_{\nu}\right]=-\sqrt{2} \clebsch{1}{\mu}{1}{\nu}{1}{\lambda} \hat{L}_{\lambda},} & (\mu, \nu, \lambda= \pm 1,0). \label{eq:2:1:21}
\end{aligned}
\end{equation}
or, in more detail
\begin{gather}
{\left[\widehat{J}_{x}, \widehat{L}_{x}\right]=\left[\widehat{J}_{y}, \widehat{L}_{y}\right]=\left[\widehat{J}_{z}, \widehat{L}_{z}\right]=0}  ,\notag\\
{\left[\widehat{J}_{x}, \widehat{L}_{y}\right]=-\left[\widehat{J}_{y}, \widehat{L}_{x}\right]=i \widehat{L}_{z},\left[\widehat{J}_{x}, \widehat{L}_{z}\right]=-\left[\widehat{J}_{z}, \widehat{L}_{x}\right]=-i \widehat{L}_{y}}  ,\label{eq:2:1:22}\\
{\left[\widehat{J}_{y}, \widehat{L}_{z}\right]=-\left[\widehat{J}_{z}, \widehat{L}_{y}\right]=i \widehat{L}_{x}}  \notag\\
{\left[\widehat{J}_{+1}, \widehat{L}_{+1}\right]=\left[\widehat{J}_{0}, \widehat{L}_{0}\right]=\left[\widehat{J}_{-1}, \widehat{L}_{-1}\right]=0}  ,\notag\\
{\left[\widehat{J}_{+1}, \widehat{L}_{0}\right]=-\left[\widehat{J}_{0}, \widehat{L}_{+1}\right]=-\widehat{L}_{+1},\left[\widehat{J}_{+1}, \widehat{L}_{-1}\right]=-\left[\widehat{J}_{-1}, \widehat{L}_{+1}\right]=-\widehat{L}_{0}} , \label{eq:2:1:23}\\
{\left[\widehat{J}_{0}, \widehat{L}_{-1}\right]=-\left[\widehat{J}_{-1}, \widehat{L}_{0}\right]=-\widehat{L}_{-1}} .\notag
\end{gather}
Appropriate relations for the spin operator $\widehat{\vect{S}}$ and the total angular momentum operator $\widehat{\vect{J}}$ may be obtained from Eqs. \eqref{eq:2:1:21}-\eqref{eq:2:1:23} by replacing $\widehat{\vect{L}} \rightarrow \widehat{\vect{S}}$. The orbital angular momentum operator $\widehat{\vect{L}}$ commutes with the spin operator $\widehat{\vect{S}}$
\begin{equation}
\begin{aligned}
{\left[\widehat{L}_{i}, \widehat{S}_{k}\right]=0,} && (i, k=x, y, z) \\
{\left[\widehat{L}_{\mu}, \widehat{S}_{\nu}\right]=0,} && (\mu, \nu= \pm 1,0) \label{eq:2:1:24}
\end{aligned}
\end{equation}

\section{Orbital angular momentum operator}\label{sec:2.2}\index{angular momentum operator!orbital}\index{orbital angular momentum operator}
\subsection{Definition}
In classical mechanics the angular momentum $\vect{L}$ of a particle is defined as
\begin{equation}
\vect{L}=[\vect{r} \times \vect{p}],\label{eq:2:2:1}
\end{equation}
where $\vect{r}$ is the position vector of the particle and $\vect{p}$ is its linear momentum.\\
In the quantum mechanics the orbital angular momentum operator $\hat{\vect{L}}$ of a particle is obtained from Eq. \eqref{eq:2:2:1} by replacing $\vect{r} \rightarrow \widehat{\vect{r}}$ and $\vect{p} \rightarrow \widehat{\vect{p}}$, where $\widehat{\vect{r}}$ and $\widehat{\vect{p}}$ are the position and momentum operators, respectively.\isym{L-hat@$\widehat{\vect{L}}$, orbital angular momentum operator}\isym{r-hat@$\widehat{\vect{r}}$, position operator}\isym{p-hat@$\widehat{\vect{p}}$, momentum operator}

In the coordinate representation we have $\hat{\vect{r}}=\vect{r}, \hat{\vect{p}}=-i \nabla$,
\begin{equation}
\widehat{\vect{L}}=-i[\vect{r} \times \nabla] . \label{eq:2:2:2}
\end{equation}
In the momentum representation we have $\hat{\vect{r}}=i \vect{\nabla}_{\vect{p}}, \hat{\vect{p}}=\vect{p}$,
\begin{equation}
\widehat{\vect{L}}=-i\left[\vect{p} \times \vect{\nabla}_{\vect{p}}\right] . \label{eq:2:2:3}
\end{equation}
The formulas below are valid for both representations. The explicit form of $\hat{\vect{L}}$ (see Sec.~\ref{sec:2.2.3}) in the momentum representation is obtained by substituting $\vect{r} \rightarrow \vect{p}, \vect{\nabla} \equiv \partial / \partial \vect{r} \rightarrow \vect{\nabla}_{\vect{p}} \equiv \partial / \partial \vect{p}$.

The operator $\hat{\vect{L}}$ is Hermitian and purely imaginary.
\begin{equation}
\widehat{\vect{L}}^{+}=\widehat{\vect{L}}, \quad \widehat{\vect{L}}^{*}=-\widehat{\vect{L}} . \label{eq:2:2:4}
\end{equation}
For cartesian components of $\widehat{\vect{L}}$, Eqs. \eqref{eq:2:2:4} become
\begin{equation}
\widehat{L}_{i}^{+}=\widehat{L}_{i}, \quad \widehat{L}_{i}^{*}=-\widehat{L}_{i}, \quad(i=x, y, z) . \label{eq:2:2:5}
\end{equation}
and for spherical components Eqs. \eqref{eq:2:2:4} yield
\begin{equation}
\begin{aligned}
& \left(\widehat{L}_{\mu}\right)^{+}=(-1)^{\mu}\left(\widehat{L}^{+}\right)_{-\mu}=(-1)^{\mu} \widehat{L}_{-\mu}, \\
& \left(\widehat{L}_{\mu}\right)^{*}=(-1)^{\mu}\left(\widehat{L}^{*}\right)_{-\mu}=(-1)^{\mu+1} \widehat{L}_{-\mu}, \quad(\mu= \pm 1,0) .
\end{aligned}
\end{equation}
The orbital angular momentum operator $\hat{\vect{L}}$ generates transformations of scalar (spinless) wave functions under rotations of the coordinate system. A rotation through an infinitesimal angle $\delta \omega$ about the n -axis transforms the position vector $\vect{r}$ into $\vect{r}+\delta \vect{r}$, with $\delta \vect{r}=-\delta \omega[\vect{n} \times \vect{r}]$. The corresponding transformation of the scalar wave functions reads
\begin{equation}
\Psi(\vect{r}) \rightarrow \Psi(\vect{r}+\delta \vect{r})=(1+\delta \vect{r} \cdot \nabla) \Psi(\vect{r})=(1-i \delta \omega \vect{n} \cdot \hat{\vect{L}}) \Psi(\vect{r}) . \label{eq:2:2:7}
\end{equation}
In the case of a finite rotation through an angle $\omega$ about the axis $n$, scalar functions are transformed by the operator $e^{-i \omega n \cdot \widehat{\vect{L}}}$ which is the particular form of the rotation operator \ref{eq:1:4:32}. The eigenfunctions of the operators $\widehat{\vect{L}}^{2}$ and $\widehat{L}_{z}$ represent spherical harmonics $Y_{l m}(\vartheta, \varphi)$ (Chap.~\ref{chap:5}) which depend on the polar angles $\vartheta, \varphi$.

\subsection{Commutation Relations}\index{commutation relations}
The orbital angular momentum operator $\hat{\vect{L}}$ satisfies the same commutation relations as the total angular momentum operator $\widehat{\vect{J}}$. Appropriate expressions for the commutators are readily obtained from Eqs. \eqref{eq:2:1:5}-\eqref{eq:2:1:14} by substituting $\widehat{\vect{J}} \rightarrow \hat{\vect{L}}$.

The commutation rules for the operator $\hat{\vect{L}}$ with the position operator $\hat{\vect{r}}$ and the momentum operator $\hat{\vect{p}}$ are the following.

For cartesian components
\begin{equation}
\begin{aligned}
& {\left[\hat{L}_{i}, \hat{x}_{k}\right]=i \varepsilon_{i k l} x_{l},}\\
& {\left[\widehat{L}_{i}, \hat{p}_{k}\right]=i \varepsilon_{i k l} p_{l}, } 
\end{aligned} \quad(i, k, l=x, y, z). \label{eq:2:2:8}
\end{equation}
Note also that $\widehat{\vect{r}}$ and $\widehat{\vect{p}}$ obey the commutation relations
\begin{equation}
\left[\widehat{x}_{i}, \widehat{x}_{k}\right]=0, \quad\left[\widehat{p}_{i}, \widehat{p}_{k}\right]=0, \quad\left[\widehat{x}_{i}, \widehat{p}_{k}\right]=i \delta_{i k}, \quad(i, k=x, y, z) .\label{eq:2:2:9}
\end{equation}
For spherical components
\begin{equation}
\begin{aligned}
& {\left[\widehat{L}_{\mu}, \widehat{x}_{\nu}\right]=-\sqrt{2} \clebsch{1}{\mu}{1}{\nu}{1}{\lambda} \widehat{x}_{\lambda},} \\
& {\left[\widehat{L}_{\mu}, \widehat{p}_{\nu}\right]=-\sqrt{2} \clebsch{1}{\mu}{1}{\nu}{1}{\lambda} \widehat{p}_{\lambda},} 
\end{aligned}  \quad(\mu, \nu, \lambda= \pm 1,0) .\label{eq:2:2:10}
\end{equation}
The commutation relations for the spherical components of $\hat{\vect{r}}$ and $\hat{\vect{p}}$ are given by
\begin{equation}
\left[\widehat{x}_{\mu}, \widehat{x}_{\nu}\right]=0, \quad\left[\widehat{p}_{\mu}, \widehat{p}_{\nu}\right]=0, \quad\left[\widehat{x}_{\mu}, \widehat{p}_{\nu}\right]=i \delta_{\mu-\nu}(-1)^{\mu}, \quad(\mu, \nu= \pm 1,0) .\label{eq:2:2:11}
\end{equation}
Equation \eqref{eq:2:2:8} and \eqref{eq:2:2:10} show that the operators $\hat{\vect{r}}$ and $\hat{\vect{p}}$ are vectors.\\
The commutation relations for square of angular momentum operator $\widehat{\vect{L}}^{2}$ are
\begin{align}
{\left[\widehat{\vect{L}}^{2}, \widehat{\vect{r}}\right] } & =i[\widehat{\vect{r}} \times \widehat{\vect{L}}]-i[\widehat{\vect{L}} \times \widehat{\vect{r}}]  ,\notag\\
{\left[\widehat{\vect{L}}^{2}, \widehat{\vect{p}}\right] } & =i[\widehat{\vect{p}} \times \widehat{\vect{L}}]-i[\widehat{\vect{L}} \times \widehat{\vect{p}}] .\label{eq:2:2:12}
\end{align}
The operators $\widehat{\vect{r}}^{2}$ and $\widehat{\vect{p}}^{2}$ commute with $\widehat{\vect{L}}$ :
\[
\left[\widehat{\vect{L}}, \widehat{\vect{r}}^{2}\right]=0, \quad\left[\widehat{\vect{L}}, \widehat{\vect{p}}^{2}\right]=0
\]
The commutation rules of $\hat{\vect{L}}$ with the total angular momentum operator $\hat{\vect{J}}$ and the spin operator $\hat{\vect{S}}$ have been considered above (Eqs. \eqref{eq:2:1:21}-\eqref{eq:2:1:24}).

\subsection{Explicit Form}\label{sec:2.2.3}
The cartesian components of the operator $\hat{\vect{L}}$ are given by
\begin{equation}
\hat{L}_{x}=-i\left(y \frac{\partial}{\partial z}-z \frac{\partial}{\partial y}\right), \quad \hat{L}_{y}=-i\left(z \frac{\partial}{\partial x}-x \frac{\partial}{\partial z}\right), \quad \hat{L}_{z}=-i\left(x \frac{\partial}{\partial y}-y \frac{\partial}{\partial x}\right) . \label{eq:2:2:13}
\end{equation}
or in a more compact form
\begin{equation}
\hat{L}_{i}=-i \sum_{k, l} \varepsilon_{i k l} x_{k} \frac{\partial}{\partial x_{l}}, \quad(i, k, l=x, y, z) .\label{eq:2:2:14}
\end{equation}
The cartesian components of $\widehat{\vect{L}}$ are expressed in terms of polar angles $\vartheta, \varphi$ as
\begin{align}
& \hat{L}_{x}=i\left(\sin \varphi \frac{\partial}{\partial \vartheta}+\cot \vartheta \cos \varphi \frac{\partial}{\partial \varphi}\right)  \notag\\
& \hat{L}_{y}=i\left(-\cos \varphi \frac{\partial}{\partial \vartheta}+\cot \vartheta \sin \varphi \frac{\partial}{\partial \varphi}\right)  \label{eq:2:2:15}\\
& \hat{L}_{z}=-i \frac{\partial}{\partial \varphi} \notag
\end{align}
The spherical components of $\hat{\vect{L}}$ are written as
\begin{align}
\widehat{L}_{+1} & =x_{0} \nabla_{+1}-x_{+1} \nabla_{0} , \notag\\
\widehat{L}_{0} & =x_{-1} \nabla_{+1}-x_{+1} \nabla_{-1} , \label{eq:2:2:16}\\
\widehat{L}_{-1} & =x_{-1} \nabla_{0}-x_{0} \nabla_{-1} .\notag
\end{align}
or, in a more compact form
\begin{equation}
\widehat{L}_{\mu}=-\sqrt{2} \sum_{\nu, \lambda} \clebsch{1}{\nu}{1}{\lambda}{1}{\mu} x_{\nu} \nabla_{\lambda}, \quad(\mu, \nu, \lambda= \pm 1,0). \label{eq:2:2:17}
\end{equation}
The spherical components of $\hat{\vect{L}}$ may also be expressed in terms of polar angles $\vartheta, \varphi$ as
\begin{align}
\widehat{L}_{+1} & =-\frac{1}{\sqrt{2}} e^{i \varphi}\left\{\frac{\partial}{\partial \vartheta}+i \cot \vartheta \frac{\partial}{\partial \varphi}\right\}  ,\notag\\
\widehat{L}_{0} & =-i \frac{\partial}{\partial \varphi} , \label{eq:2:2:18}\\
\widehat{L}_{-1} & =-\frac{1}{\sqrt{2}} e^{-i \varphi}\left\{\frac{\partial}{\partial \vartheta}-i \cot \vartheta \frac{\partial}{\partial \varphi}\right\} . \notag
\end{align}
For polar components of $\widehat{\vect{L}}$ we have
\begin{equation}
\hat{L}_{r}=0, \quad \hat{L}_{\vartheta}=\frac{i}{\sin \vartheta} \frac{\partial}{\partial \varphi}, \quad \hat{L}_{\varphi}=-i \frac{\partial}{\partial \vartheta} . \label{eq:2:2:19}
\end{equation}
The square of the orbital angular momentum operator $\widehat{\vect{L}}^{2}$ is expressed in terms of its components as
\begin{gather}
\hat{\vect{L}}^{2}=\sum_{i} \hat{L}_{i}^{2}=\hat{L}_{x}^{2}+\widehat{L}_{y}^{2}+\hat{L}_{z}^{2}  ,\notag\\
\hat{\vect{L}}^{2}=\sum_{\mu}(-1)^{\mu} \hat{L}_{\mu} \hat{L}_{-\mu}=-\hat{L}_{+1} \hat{L}_{-1}+\hat{L}_{0} \hat{L}_{0}-\hat{L}_{-1} \hat{L}_{+1}. \label{eq:2:2:20}
\end{gather}
We may also write $\widehat{\vect{L}}^{2}$ in the form
\begin{equation}
\widehat{\vect{L}}^{2}=-\left[\frac{1}{\sin \vartheta} \cdot \frac{\partial}{\partial \vartheta}\left(\sin \vartheta \frac{\partial}{\partial \vartheta}\right)+\frac{1}{\sin ^{2} \vartheta} \cdot \frac{\partial^{2}}{\partial \varphi^{2}}\right] . \label{eq:2:2:21}
\end{equation}
This operator differs only in sign from the angular part of the Laplacian operator (Eq. \eqref{eq:1:3:15})
\begin{equation}
\widehat{\vect{L}}^{2}=-\Delta_{\Omega} . \label{eq:2:2:22}
\end{equation}
The operator $\widehat{\vect{L}}^{2}$ is Hermitian and real
\begin{equation}
\left(\widehat{\vect{L}}^{2}\right)^{+}=\left(\widehat{\vect{L}}^{2}\right)^{*}=\widehat{\vect{L}}^{2} .\label{eq:2:2:23}
\end{equation}
The representation of the orbital angular momentum $\hat{\vect{L}}$ in the form of a differential operator is only one of possible representations. Alternatively, $\hat{\vect{L}}$ may be represented by a set of three matrices (because it has three components). These matrix elements of $\hat{\vect{L}}$ will be given in Chap.~\ref{chap:13}.

\section{Spin angular momentum operator} \label{sec:2.3}\index{angular momentum operator!spin}\index{spin operator}
\subsection{Definition}
The spin angular momentum operator or briefly the spin operator $\widehat{\vect{S}}$\isym{S-hat@$\widehat{\vect{S}}$, spin operator} is usually represented by a set of three (since the vector $\widehat{\mathrm{S}}$ has three components) square $(2 S+1) \times(2 S+1)$ matrices, $S$ being the particle spin. These matrices act on the spin functions (see Chap.~\ref{chap:6}) and satisfy the same commutation relations as the components of the total angular momentum operator given in Sec.2.1. The spin operator $\widehat{\vect{S}}$ is Hermitian:
\begin{equation}
\widehat{\vect{S}}^{+}=\widehat{\vect{S}} \label{eq:2:3:1}
\end{equation}
For the cartesian components of $\widehat{\vect{S}}$ the Hermitian property \eqref{eq:2:3:1} has the form
\begin{equation}
\left(\widehat{S}_{i}\right)^{+}=\widehat{S}_{i}, \quad(i=x, y, z) \label{eq:2:3:2}
\end{equation}
and for spherical components
\begin{equation}
\left(\widehat{S}_{\mu}\right)^{+}=(-1)^{\mu} \widehat{S}_{-\mu}, \quad(\mu= \pm 1,0) .\label{eq:2:3:3}
\end{equation}
The behaviour of $\widehat{\vect{S}}$ under complex conjugation depends on the representation of $\widehat{\vect{S}}$ (see, e.g., Sec.~\ref{sec:2.6.2}). The eigenfunctions of the operators $\widehat{\vect{S}}^{2}$ and $\widehat{S}_{z}$ are spin functions $\chi_{S m}$ (see Chap.~\ref{chap:6}) which depend on the spin variable $\sigma$.\index{spin function}\isymg{chi-Sm@$\chi_{S m}$, spin function} These functions have $(2 S+1)$ components and describe polarization states of a particle.

\subsection{Commutation Relations}\index{commutation relations}
The commutation relations for the spin operator $\widehat{\mathrm{S}}$ are given by Eqs. \eqref{eq:2:1:5}-\eqref{eq:2:1:14} in which one must replace $\widehat{\vect{J}}$ by $\widehat{\vect{S}}$. The commutation rules for $\widehat{\vect{S}}$ and the orbital angular momentum operator $\widehat{\vect{L}}$ have already been considered in Sec.~\ref{sec:2.1.4}. The spin operator $\widehat{\vect{S}}$ commutes with the position operator $\widehat{\vect{r}}$ and the momentum operator $\widehat{\vect{p}}$,
\begin{align}
& {\left[\widehat{S}_{i}, \widehat{x}_{k}\right]=0, \quad\left[\widehat{S}_{i}, \widehat{p}_{k}\right]=0, \quad(i, k=x, y, z),}  \label{eq:2:3:4}\\
& {\left[\widehat{S}_{\mu}, \widehat{x}_{\nu}\right]=0, \quad\left[\widehat{S}_{\mu}, \widehat{p}_{\nu}\right]=0, \quad(\mu, \nu= \pm 1,0) .} \label{eq:2:3:5}
\end{align}
\subsection{Explicit Form}
The spherical components of the spin operator $\widehat{S}$ may be expressed in terms of the basis spin functions $\chi_{S_{m}}$ (Chap.~\ref{chap:6}) as
\begin{equation}
\widehat{S}_{\mu}=\sqrt{S(S+1)} \sum_{m, m^{\prime}} \clebsch{S}{m}{1}{\mu}{S}{m^{\prime}} \chi_{S_{m^{\prime}}} \chi_{S m}^{+}, \quad(\mu= \pm 1,0) .\label{eq:2:3:6}
\end{equation}
Cartesian components of $\widehat{\vect{S}}$ may be obtained from \eqref{eq:2:3:6}, using the relations
\begin{equation}
\widehat{S}_{x}=\frac{1}{\sqrt{2}}\left(\widehat{S}_{-1}-\widehat{S}_{+1}\right), \quad \widehat{S}_{y}=\frac{i}{\sqrt{2}}\left(\widehat{S}_{-1}+\widehat{S}_{+1}\right), \quad \widehat{S}_{z}=\widehat{S}_{0}. \label{eq:2:3:7}
\end{equation}
Equations \eqref{eq:2:3:6} and \eqref{eq:2:3:7} are independent of representation used for spin functions, while an explicit form of the spin matrices depends on this representation. For an arbitrary spin $S$ the simplest form of the spin matrices is that in the spherical basis representation. The basis spin functions $\chi_{S m}$ in this representation are given by
\begin{equation}
\chi_{S m}(\sigma)=\delta_{m \sigma}, \quad(m, \sigma=-S,-S+1, \ldots, S-1, S) \label{eq:2:3:8},
\end{equation}
and the matrix elements of the spin matrices are
\begin{equation}
\left(\widehat{S}_{\mu}\right)_{\sigma^{\prime} \sigma}=\sqrt{S(S+1)} \clebsch{S}{\sigma}{1}{\mu}{S}{\sigma^{\prime}}, \quad\left(\sigma, \sigma^{\prime}=-S,-S+1, \ldots, S-1, S\right) \label{eq:2:3:9}
\end{equation}
The elements are arranged in the matrices as follows
\begin{equation}
A=\begin{pmatrix}
A_{s s} \ldots \ldots A_{S-S}  \\
\ldots \ldots \ldots \ldots \ldots \\
\ldots \ldots \ldots \ldots \ldots \\
A_{-s s} \ldots A_{-S-S}
\end{pmatrix}.\label{eq:2:3:10}
\end{equation}
Explicit forms of the spin matrices for spin values $\frac{1}{2}$ and 1 are given below in Secs. 2.5 and 2.6.

\subsection{Traces of Products of Spin Matrices}\index{spin matrices!traces of}\index{trace!of spin matrices}
The traces of products of cartesian components of the spin matrices can be evaluated for an arbitrary spin $S$ by use of the following formulas (where $i, k, l$, etc. take the values $x, y, z$ )
\begin{align}
\operatorname{Tr}\left\{\widehat{S}_{i}\right\} & =0  \notag\\
\operatorname{Tr}\left\{\widehat{S}_{i} \widehat{S}_{k}\right\} & =\frac{S(S+1)(2 S+1)}{3} \delta_{i k} , \notag\\
\operatorname{Tr}\left\{\widehat{S}_{i} \widehat{S}_{k} \widehat{S}_{l}\right\} & =i \frac{S(S+1)(2 S+1)}{6} \varepsilon_{i k l} , \label{eq:2:3:11}\\
\operatorname{Tr}\left\{\widehat{S}_{i} \widehat{S}_{k} \widehat{S}_{l} \widehat{S}_{j}\right\} & =\frac{S(S+1)(2 S+1)}{15}\left\{\left[S(S+1)+\frac{1}{2}\right]\left(\delta_{i k} \delta_{l j}+\delta_{i j} \delta_{k l}\right)+[S(S+1)-2] \delta_{i l} \delta_{k j}\right\} .\notag
\end{align}
For spherical components of the spin operator $\widehat{S}$ the following relations hold $(\mu, \nu, \lambda$, etc. take the values $\pm 1,0$ )
\begin{align}
\operatorname{Tr}\left\{\widehat{S}_{\mu}\right\} & =0,  \notag\\
\operatorname{Tr}\left\{\widehat{S}_{\mu} \widehat{S}_{\nu}\right\} & =\frac{S(S+1)(2 S+1)}{3}(-1)^{\mu} \delta_{\mu-\nu},  \notag\\
\operatorname{Tr}\left\{\widehat{S}_{\mu} \widehat{S}_{\nu} \widehat{S}_{\lambda}\right\} & =-\frac{S(S+1)(2 S+1)}{\sqrt{6}}\begin{pmatrix}
1 & 1 & 1 \\
\mu & \nu & \lambda
\end{pmatrix}=\frac{S(S+1)(2 S+1)}{3 \sqrt{2}}(-1)^{1+\lambda} \clebsch{1}{\mu}{1}{\nu}{1}{-\lambda},  \label{eq:2:3:12}\\
\operatorname{Tr}\left\{\widehat{S}_{\mu} \widehat{S}_{\nu} \widehat{S}_{\lambda} \hat{S}_{\rho}\right\}= & \frac{S(S+1)(2 S+1)}{15}\left\{\left[S(S+1)+\frac{1}{2}\right](-1)^{\mu+\lambda}\left(\delta_{\mu-\nu} \delta_{\lambda-\rho}+\delta_{\mu-\rho} \delta_{\nu-\lambda}\right)\right.  \notag\\
& \left.\quad+[S(S+1)-2](-1)^{\mu+\nu} \delta_{\mu-\lambda} \delta_{\nu-\rho}\right\} . \notag
\end{align}
Some analytic expressions for traces of products of many spin matrices in the spherical basis are given in Refs. \cite{ref133, ref144, ref145}.

\section{Polarization operators}\label{sec:2.4}\index{polarization operator}
\subsection{Definition}
To describe a polarization (i.e., spin) state of a particle the so-called polarization operators are widely used. The polarization operators $\widehat{T}_{L M}(S)(M=-L,-L+1, \ldots L-1, L$ and $L=0,1, \ldots, 2 S ; L$ and $M$ being integers $)$ are $(2 S+1) \times(2 S+1)$ matrices which act on spin functions (see Chap.~\ref{chap:6}) and transform under rotations of the coordinate system according to the representation $D^{L}$. In other words, $\widehat{T}_{L M}(S)$ are irreducible tensors of rank $L$.\index{irreducible tensor}\index{tensor!irreducible} Such transformation properties of $\widehat{T}_{L M}(S)$ with respect to rotations of the coordinate system imply the following commutation relations with spherical components of the spin operator $\widehat{S}_{\mu}(\mu= \pm 1,0)$
\begin{equation}
\left[\widehat{S}_{\mu}, \widehat{T}_{L M}(S)\right]=\sqrt{L(L+1)} \clebsch{L}{M}{1}{\mu}{L}{M+\mu} \widehat{T}_{L M+\mu}(S) . \label{eq:2:4:1}
\end{equation}
Let us normalize the polarization operators by the condition
\begin{equation}
\operatorname{Tr}\left\{\widehat{T}_{L M}^{+}(S) \widehat{T}_{L^{\prime} M^{\prime}}(S)\right\}=\delta_{L L^{\prime}} \delta_{M M^{\prime}}, \label{eq:2:4:2}
\end{equation}
and choose the phase factors to satisfy the relations
\begin{equation}
\widehat{T}_{L M}^{+}(S)=(-1)^{M} \widehat{T}_{L-M}(S) . \label{eq:2:4:3}
\end{equation}
The conditions \eqref{eq:2:4:1}-\eqref{eq:2:4:3} completely determine the polarization operators. The full complement of polarization operators $\hat{T}_{L M}(S)$ with $-L \leq M \leq L, 0 \leq L \leq 2 S$ constitutes a complete set of $\sum_{L=0}^{2 S}(2 L+1)=(2 S+1)^{2}$ linearly independent square $(2 S+1) \times(2 S+1)$ matrices.

\subsection{Explicit Form}
The polarization operators $\widehat{T}_{L M}(S)$ may be constructed from products of the spin matrices as\isym{T-hat@$\widehat{T}_{L M}(S)$, polarization operator}
\begin{equation}
\hat{T}_{L M}(S)=N_{L}(S)(\widehat{\vect{S}} \cdot \nabla)^{L}\left\{r^{L} Y_{L M}(\vartheta, \varphi)\right\} ,\label{eq:2:4:4}
\end{equation}
where $\widehat{\mathrm{S}}$ is the spin operator, and $N_{L}(S)$ is the normalization factor given by\isym{NLS@$N_{L}(S)$, normalization factor}
\begin{equation}
N_{L}(S)=\frac{2^{L}}{L!}\left[\frac{4 \pi(2 S-L)!}{(2 S+L+1)!}\right]^{\frac{1}{2}} .\label{eq:2:4:5}
\end{equation}
Note that $\widehat{T}_{L M}(S)$ are actually independent of $r$, although the vector $r$ enters the right-hand side of Eq. \eqref{eq:2:4:4}. The operators $\widehat{T}_{L M}(S)$ may be expressed in terms of the basis spin functions $\chi_{S m}$ (Chap.~\ref{chap:6}) by
\begin{equation}
\hat{T}_{L M}(S)=\sqrt{\frac{2 L+1}{2 S+1}} \sum_{m, m^{\prime}} \clebsch{S}{m}{L}{M}{S}{m^{\prime}} \chi_{S_{m^{\prime}}} \chi_{S m}^{+} .\label{eq:2:4:6}
\end{equation}
The inverse relation is
\begin{equation}
\chi_{S_{m^{\prime}}} \chi_{S m}^{+}=\sum_{L} \sqrt{\frac{2 L+1}{2 S+1}} \clebsch{S}{m}{L}{M}{S}{m^{\prime}} \hat{T}_{L M}(S) .\label{eq:2:4:7}
\end{equation}
An explicit form of $\widehat{T}_{L M}(S)$ depends on the representation used for the spin functions. In particular, matrix elements of $\widehat{T}_{L M}(S)$ in the spherical basis representation are given by
\begin{equation}
\left[\hat{T}_{L M}(S)\right]_{\sigma^{\prime} \sigma}=\sqrt{\frac{2 L+1}{2 S+1}} \clebsch{S}{\sigma}{L}{M}{S}{\sigma^{\prime}}, \quad\left(\sigma, \sigma^{\prime}=-S,-S+1, \ldots, S\right) .\label{eq:2:4:8}
\end{equation}
For $L=0$ the operators $\widehat{T}_{L M}(S)$ are proportional to the unit $(2 S+1) \times(2 S+1)$ matrix
\begin{equation}
\widehat{T}_{00}(S)=\frac{1}{\sqrt{2 S+1}} \hat{I}. \label{eq:2:4:9}
\end{equation}
When $L=1$ the operators $\widehat{T}_{L M}(S)$ are proportional to the spherical components of the spin operator
\begin{equation}
\widehat{T}_{1 M}(S)=\frac{\sqrt{3}}{\sqrt{S(S+1)(2 S+1)}} \widehat{S}_{M}, \quad(M= \pm 1,0). \label{eq:2:4:10}
\end{equation}
The polarization operators for a spin value $S=1$ are considered in Sec.~\ref{sec:2.6}.\\
\subsection{Properties of $\widehat{T}_{L M}(S)$ under Transformations of the Coordinate System}
\subsubsection{Coordinate inversion ( $\vect{r} \rightarrow-\vect{r}$ )}

The polarization operators $\widehat{T}_{L M}(S)$ are invariant under coordinate inversion
\begin{equation}
\widehat{P}_{r} \widehat{T}_{L M}(S) \widehat{P}_{r}^{-1}=\widehat{T}_{L M}(S) . \label{eq:2:4:11}.
\end{equation}
\subsubsection{Rotation of coordinate system}

Under rotations specified by the Euler angles $\alpha, \beta, \gamma$ the polarization operators transform as
\begin{equation}
\widehat{T}_{L M^{\prime}}^{\prime}(S)=\widehat{D}(\alpha, \beta, \gamma) \widehat{T}_{L M^{\prime}}(S)[\widehat{D}(\alpha, \beta, \gamma)]^{-1}=\sum_{M} D_{M M^{\prime}}^{L}(\alpha, \beta, \gamma) \widehat{T}_{L M}(S) \label{eq:2:4:12}
\end{equation},
where $\widehat{D}(\alpha, \beta, \gamma)$ is the rotation operator (see Sec.~\ref{sec:1.4.5}), and $D_{M M^{\prime}}^{L}$ are the Wigner $D$-functions (Chap.~\ref{chap:4}).\index{Wigner D-function@Wigner $D$-function}

\subsection{Expansion Series of Polarization Operators}\index{polarization operator!expansion in series}
As has been mentioned above (Sec.2.4.1) the polarization operators $\hat{T}_{L M}(S)$ form a complete set of linearly independent matrices. An arbitrary square $(2 S+1) \times(2 S+1)$ matrix $\hat{A}$ (with $S$ integer or half-integer) may be expanded in a series of the polarization operators $\widehat{T}_{L M}(S)$, i.e., it may be written in the form
\begin{equation}
\widehat{A}=\sum_{L=0}^{2 S} \sum_{M=-L}^{L} A_{L M} \widehat{T}_{L M}(S) \label{eq:2:4:13},
\end{equation}
where the expansion coefficients $A_{L M}$ are given by
\begin{equation}
A_{L M}=\operatorname{Tr}\left\{\widehat{T}_{L M}^{+}(S) \hat{A}\right\} . \label{eq:2:4:14}.
\end{equation}
If the matrix $\widehat{A}$ is Hermitian, i.e., $\widehat{A}^{+}=\widehat{A}$, then
\begin{equation}
A_{L M}^{*}=(-1)^{M} A_{L-M} \label{eq:2:4:15}.
\end{equation}
Some examples of such expansions are given below.
\subsubsection{Clebsch-Gordan series for polarization operators}\index{Clebsch-Gordan series}\index{Clebsch-Gordan coefficient}

Products of two polarization operators $\widehat{T}_{L_{1} M_{1}}(S)$ and $\widehat{T}_{L_{2} M_{2}}(S)$ may be written in the form of a ClebschGordan series
\begin{equation}
\widehat{T}_{L_{1} M_{1}}(S) \widehat{T}_{L_{2} M_{2}}(S)=\sum_{L}(-1)^{2 S+L} \sqrt{\left(2 L_{1}+1\right)\left(2 L_{2}+1\right)}\sixj{L_{1} }{ L_{2} }{ L }{S }{ S }{ S} \clebsch{L_{1}}{M_{1}}{L_{2}}{M_{2}}{L}{M} \widehat{T}_{L M}(S).
\label{eq:2:4:16}
\end{equation}
\subsubsection{Expansion of rotation operator}\index{rotation operator!expansion of}
If one is interested only in transformation properties of spin functions and spin operators under rotations of the coordinate system, one may replace the total angular momentum operator $\widehat{\vect{J}}$ by the spin operator $\widehat{\vect{S}}$ in the general expressions for the rotation operator \eqref{eq:1:4:31} and \eqref{eq:1:4:33}. Such a modified operator will be labelled by a superscript $S$. If a rotation is defined by the Euler angles $\alpha, \beta, \gamma$, the rotation operator $\hat{D}^{S}(\alpha, \beta, \gamma)$ may be expressed as follows
\begin{equation}
\hat{D}^{S}(\alpha, \beta, \gamma) \equiv e^{-i \alpha \widehat{S}_{z}} e^{-i \beta \widehat{S}_{y}} e^{-i \gamma \widehat{S}_{z}}=\sum_{L, M, m, m^{\prime}} \frac{2 L+1}{2 S+1} \clebsch{S}{m^{\prime}}{L}{M}{S}{m} D_{m m^{\prime}}^{S}(\alpha, \beta, \gamma) \hat{T}_{L M}(S) .\label{eq:2:4:17}
\end{equation}
If a rotation is defined by the rotation axis $n(\Theta, \Phi)$ and the rotation angle $\omega$, the expansion of the rotation operator $\hat{U}^{S}(\omega ; \Theta, \Phi)$ is given by
\begin{equation}
\widehat{U}^{S}(\omega ; \Theta, \Phi) \equiv e^{-i \omega n \cdot \widehat{\vect{s}}}=2 \sqrt{\pi} \sum_{L=0}^{2 S} \frac{(-i)^{L}}{\sqrt{2 S+1}} \chi_{L}^{S}(\omega) \sum_{M=-L}^{L} Y_{L M}^{*}(\Theta, \Phi) \widehat{T}_{L M}(S) ,\label{eq:2:4:18}
\end{equation}
where the functions $\chi_{L}^{S}(\omega)$ are the generalized characters (see Sec.~\ref{sec:4.15}).\index{generalized characters}\isymg{chi-lambdaJ@$\chi_{\lambda}^{J}(\omega)$, generalized character}

\subsection{Commutators and Anticommutators}\index{commutator}\index{anticommutator}
The polarization operators satisfy the following commutation relations
\begin{align}
& {\left[\widehat{T}_{L_{1} M_{1}}(S), \widehat{T}_{L_{2} M_{2}}(S)\right]=}  \notag\\
& \quad \sqrt{\left(2 L_{1}+1\right)\left(2 L_{2}+1\right)} \sum_{L_{3}}(-1)^{2 S+L_{3}}\left[1-(-1)^{L_{1}+L_{2}+L_{3}}\right]\sixj{L_{1}}{L_{2}}{L_{3}} {S}{S}{S}
 \clebsch{L_{1}}{M_{1}}{L_{2}}{M_{2}}{L_{3}}{M_{3}} \widehat{T}_{L_{3} M_{3}}(S) , \label{eq:2:4:19}\\
& \left\{\widehat{T}_{L_{1} M_{1}}(S), \widehat{T}_{L_{2} M_{2}}(S)\right\}=  \notag\\
& \quad \sqrt{\left(2 L_{1}+1\right)\left(2 L_{2}+1\right)} \sum_{L_{3}}(-1)^{2 S+L_{3}}\left[1+(-1)^{L_{1}+L_{2}+L_{3}}\right]
\sixj{L_{1}}{L_{2}}{L_{3}} {S}{S}{S}
 \clebsch{L_{1}}{M_{1}}{L_{2}}{M_{2}}{L_{3}}{M_{3}}
 \widehat{T}_{L_{3} M_{3}}(S) .\label{eq:2:4:20}
\end{align}
Equation \eqref{eq:2:4:1} is a special case of \eqref{eq:2:4:19} for $L_{1}=1$.

\subsection{Traces of Products of Polarization Operators}\index{trace!of polarization operators}\index{polarization operator!traces of}
\begin{gather}
\operatorname{Tr}\left\{\widehat{T}_{L M}(S)\right\}=\sqrt{2 S+1} \delta_{L O} \delta_{M O},  \label{eq:2:4:21}\\
\operatorname{Tr}\left\{\widehat{T}_{L_{1} M_{1}}(S) \widehat{T}_{L_{2} M_{2}}(S)\right\}=(-1)^{M_{1}} \delta_{L_{1} L_{2}} \delta_{M_{1}-M_{2}},  \label{eq:2:4:22}\\
\operatorname{Tr}\left\{\widehat{T}_{L_{1} M_{1}}(S) \widehat{T}_{L_{2} M_{2}}(S) \widehat{T}_{L_{3} M_{3}}(S)\right\}=(-1)^{2 S+L_{3}+M_{3}} \sqrt{\left(2 L_{1}+1\right)\left(2 L_{2}+1\right)} \clebsch{L_{1}}{M_{1}}{L_{2}}{M_{2}}{L_{3}}{-M_{3}}\sixj{L_{1}}{L_{2}}{L_{3}} {S}{S}{S}
 . \label{eq:2:4:23}
\end{gather}
In general
\begin{align}
\operatorname{Tr}\left\{\hat{T}_{L_{1} M_{1}}(S)\right. & \left.\hat{T}_{L_{2} M_{2}}(S) \ldots \hat{T}_{L_{n} M_{n}}(S)\right\}=\frac{\left[\left(2 L_{1}+1\right)\left(2 L_{2}+1\right) \ldots\left(2 L_{n}+1\right)\right]^{\frac{1}{2}}}{(2 S+1)^{n / 2}}  ,\notag\\
& \times \delta_{M_{1}+M_{2}+M_{3}+\ldots+M_{n, 0}} \sum_{m} \clebsch{L_{1}}{M_{1}}{S}{m}{S}{m+\mu_{1}} \clebsch{L_{2}}{M_{2}}{S}{m+\mu_{1}}{S}{m+\mu_{2}} \ldots \clebsch{L_{n}}{M_{n}}{S}{m+\mu_{n-1}}{S}{m} ,\label{eq:2:4:24}
\end{align}
where
\[
\mu_{k}=\sum_{i=1}^{k} M_{i}
\]
\section{Spin matrices for $S=1 / 2$} \label{sec:2.5}\index{spin matrices!S=1/2@$S=1/2$}
\subsection{Explicit Form}
For the particular case $S=\frac{1}{2}$, the spin operator $\widehat{\mathrm{S}}$ is represented by a set of three square $2 \times 2$ matrices. Cartesian components of $\widehat{\vect{S}}$ in the spherical basis representation are given by
\begin{equation}
\hat{S}_{x}=\frac{1}{2}\begin{pmatrix}
0 & 1  \label{eq:2:5:1}\\
1 & 0
\end{pmatrix}, \quad \hat{S}_{y}=\frac{1}{2}\begin{pmatrix}
0 & -i \\
i & 0
\end{pmatrix}, \quad \hat{S}_{z}=\frac{1}{2}\begin{pmatrix}
1 & 0 \\
0 & -1
\end{pmatrix} .
\end{equation}
whereas spherical components are written as
\begin{equation}
\widehat{S}_{+1}=-\frac{1}{\sqrt{2}}\begin{pmatrix}
0 & 1  \label{eq:2:5:2}\\
0 & 0
\end{pmatrix}, \quad \widehat{S}_{0}=\frac{1}{2}\begin{pmatrix}
1 & 0 \\
0 & -1
\end{pmatrix}, \quad \widehat{S}_{-1}=\frac{1}{\sqrt{2}}\begin{pmatrix}
0 & 0 \\
1 & 0
\end{pmatrix} .
\end{equation}
In addition to the spin matrices, the unit $2 \times 2$ matrix $\hat{I}$ is usually introduced:\isym{I-hat@$\widehat{I}$, unit matrix}
\begin{equation}
\widehat{I}=\begin{pmatrix}
1 & 0  \label{eq:2:5:3}\\
0 & 1
\end{pmatrix}
\end{equation}
Note that the Pauli matrices $\hat{\sigma}$ are widely used instead of the spin operator $\widehat{\vect{S}}$.\index{Pauli matrices}\index{matrix!Pauli}\isymg{sigma-hat@$\widehat{\sigma}$, Pauli matrices} The Pauli matrices are proportional to the spin operator
\begin{equation}
\widehat{\mathrm{S}}=\frac{1}{2} \widehat{\sigma}. \label{eq:2:5:4}
\end{equation}
Equations \eqref{eq:2:5:1} and \eqref{eq:2:5:2} yield the following properties of the spin matrices
\begin{gather}
\widehat{S}_{i}^{+}=\widehat{S}_{i}, \quad(i=x, y, z),  \label{eq:2:5:5}\\
\widehat{S}_{\mu}^{+}=(-1)^{\mu} \widehat{S}_{-\mu}, \quad(\mu= \pm 1,0),  \label{eq:2:5:6}\\
\widehat{S}_{x}^{*}=\widehat{S}_{x}, \quad \widehat{S}_{y}^{*}=-\widehat{S}_{y}, \quad \widehat{S}_{z}^{*}=\widehat{S}_{z},  \label{eq:2:5:7}\\
\widehat{S}_{\mu}^{*}=\widehat{S}_{\mu}, \quad(\mu= \pm 1,0) . \label{eq:2:5:8}
\end{gather}
\subsection{Commutators and Anticommutators}\index{commutator}\index{anticommutator}
The cartesian components of the spin operator satisfy the following relations
\begin{equation}
\left[\widehat{S}_{i}, \widehat{S}_{k}\right]=i \varepsilon_{i k l} \widehat{S}_{l}, \quad\left\{\widehat{S}_{i}, \widehat{S}_{k}\right\}=\frac{1}{2} \delta_{i k} \widehat{I}, \quad(i, k, l=x, y, z) \label{eq:2:5:9}
\end{equation}
For spherical components these relations take the form
\begin{equation}
\left[\widehat{S}_{\mu}, \widehat{S}_{\nu}\right]=-\sqrt{2} \clebsch{1}{\mu}{1}{\nu}{1}{\lambda} S_{\lambda}, \quad\left\{\widehat{S}_{\mu}, \widehat{S}_{\nu}\right\}=\frac{1}{2}(-1)^{\mu} \delta_{\mu-\nu} \widehat{I}, \quad(\mu, \nu, \lambda= \pm 1,0). \label{eq:2:5:10}
\end{equation}
\subsection{Products of Spin Matrices}\index{spin matrices!products of}
The four matrices $\widehat{S}_{x}, \widehat{S}_{y}, \widehat{S}_{z}$ and $\widehat{I}$ (or $\widehat{S}_{+1}, \widehat{S}_{0}, \widehat{S}_{-1}$ and $\widehat{I}$ ) constitute a complete set of square $2 \times 2$ matrices. Any function of the spin operator ( $S=\frac{1}{2}$ ) may be expanded in a series in terms of these matrices. In particular, products of the spin matrices may also be written in such a form. Products of cartesian components (where the indices $i, k, l$, etc. take the values $x, y, z)$ are given by
\begin{align}
\widehat{S}_{i} \widehat{S}_{k}= & \frac{1}{4} \delta_{i k} \widehat{I}+\frac{i}{2} \varepsilon_{i k l} \widehat{S}_{l},  \label{eq:2:5:11}\\
\widehat{S}_{i} \widehat{S}_{k} \widehat{S}_{l}= & \frac{i}{8} \varepsilon_{i k l} \widehat{I}+\frac{1}{4}\left(\widehat{S}_{i} \delta_{k l}-\widehat{S}_{k} \delta_{i l}+\widehat{S}_{l} \delta_{i k}\right)  ,\label{eq:2:5:12}\\
\widehat{S}_{i} \widehat{S}_{k} \widehat{S}_{l} \widehat{S}_{j}= & \frac{1}{16}\left(\delta_{i k} \delta_{l j}-\delta_{i l} \delta_{k j}+\delta_{i j} \delta_{k l}\right) \widehat{I} , \notag\\
& +\frac{i}{8}\left(\delta_{i k} \varepsilon_{l j m}-\delta_{i l} \varepsilon_{k j m}+\delta_{i j} \varepsilon_{k l m}+\delta_{k l} \varepsilon_{i j m}-\delta_{k j} \varepsilon_{i l m}+\delta_{l j} \varepsilon_{i k m}\right) \widehat{S}_{m} .\label{eq:2:5:13}
\end{align}
In more detailed form, Eq. \eqref{eq:2:5:11} is written as
\begin{gather}
\widehat{S}_{x}^{2}=\widehat{S}_{y}^{2}=\widehat{S}_{z}^{2}=\frac{1}{4} \widehat{I} , \notag\\
\widehat{S}_{x} \widehat{S}_{y}=-\widehat{S}_{y} \widehat{S}_{x}=\frac{i}{2} \widehat{S}_{z}, \quad \widehat{S}_{y} \widehat{S}_{z}=-\widehat{S}_{z} \widehat{S}_{y}=\frac{i}{2} \widehat{S}_{x}  ,\label{eq:2:5:14}\\
\widehat{S}_{z} \hat{S}_{x}=-\widehat{S}_{x} \widehat{S}_{z}=\frac{i}{2} \hat{S}_{y}. \notag
\end{gather}
Products of spherical components are as follows:
\begin{gather}
\widehat{S}_{+1} \widehat{S}_{+1}=0, \quad \widehat{S}_{0} \widehat{S}_{0}=\frac{1}{4} \widehat{I}, \quad \widehat{S}_{-1} \widehat{S}_{-1}=0  ,\notag\\
\widehat{S}_{+1} \widehat{S}_{-1}=-\frac{1}{2}\begin{pmatrix}
1 & 0 \\
0 & 0
\end{pmatrix}=-\frac{1}{4} \widehat{I}-\frac{1}{2} \widehat{S}_{0} , \notag\\
\widehat{S}_{-1} \widehat{S}_{+1}=-\frac{1}{2}\begin{pmatrix}
0 & 0 \\
0 & 1
\end{pmatrix}=-\frac{1}{4} \widehat{I}+\frac{1}{2} \widehat{S}_{0} , \label{eq:2:5:15}\\
\widehat{S}_{+1} \widehat{S}_{0}=-\frac{1}{2} \widehat{S}_{+1}, \quad \widehat{S}_{-1} \widehat{S}_{0}=\frac{1}{2} \widehat{S}_{-1}  ,\notag\\
\widehat{S}_{0} \widehat{S}_{+1}=\frac{1}{2} \widehat{S}_{+1}, \quad \widehat{S}_{0} \widehat{S}_{-1}=-\frac{1}{2} \widehat{S}_{-1}  ,\notag\\
\widehat{S}_{\mu} \widehat{S}_{\nu}=\frac{1}{4}(-1)^{\mu} \delta_{\mu-\nu} \widehat{I}-\frac{1}{\sqrt{2}} \clebsch{1}{\mu}{1}{\nu}{1}{\lambda} \widehat{S}_{\lambda},(\mu, \nu, \lambda= \pm 1,0), \quad\left(\widehat{S}_{+1}\right)^{n}=0, \quad\left(\widehat{S}_{-1}\right)^{n}=0, \quad(n=2,3, \ldots),  \label{eq:2:5:16}\\
\left(\widehat{S}_{0}\right)^{2 n}=\left(\frac{1}{4}\right)^{n} \widehat{I},\left(\widehat{S}_{0}\right)^{2 n+1}=\left(\frac{1}{4}\right)^{n} \widehat{S}_{0}, \quad(n=0,1,2, \ldots),  \label{eq:2:5:17}\\
\widehat{\vect{S}} \cdot \widehat{\vect{S}}=\widehat{\vect{S}}^{2}=\widehat{S}_{x}^{2}+\widehat{S}_{y}^{2}+\widehat{S}_{z}^{2}=-\widehat{S}_{+1} \widehat{S}_{-1}+\widehat{S}_{0} \widehat{S}_{0}-\widehat{S}_{-1} \widehat{S}_{+1}=\frac{3}{4} \widehat{I}. \label{eq:2:5:18}
\end{gather}
Products of the operators $\widehat{\vect{S}}$ and any vectors $a$ and $b$ commuting with $\widehat{\vect{S}}$ may be written in the form
\begin{align}
\widehat{\vect{S}} \cdot(\widehat{\vect{S}} \cdot \vect{a}) & =\frac{1}{4} \widehat{I} \vect{a}-\frac{i}{2}[\widehat{\vect{S}} \times \vect{a}],  \label{eq:2:5:19}\\
(\widehat{\vect{S}} \cdot[\widehat{\vect{S}} \times \vect{a}]) & =([\widehat{\vect{S}} \times \widehat{\vect{S}}] \cdot \vect{a})=i(\widehat{\vect{S}} \cdot \vect{a}), \label{eq:2:5:20}
\end{align}
\begin{gather}
{[\widehat{\vect{S}} \times[\widehat{\vect{S}} \times \vect{a}]]=-\frac{1}{2} \widehat{I} \vect{a}+\frac{i}{2}[\widehat{\vect{S}} \times \vect{a}],}  \label{eq:2:5:21}\\
(\widehat{\vect{S}} \cdot \vect{a})(\widehat{\vect{S}} \cdot \vect{b})=\frac{1}{4}(\vect{a} \cdot \vect{b}) \widehat{I}+\frac{i}{2}(\widehat{\vect{S}} \cdot[\vect{a} \times \vect{b}]),  \label{eq:2:5:22}\\
([\widehat{\vect{S}} \times \vect{a}] \cdot[\widehat{\vect{S}} \times \vect{b}])=\frac{1}{2}(\vect{a} \cdot \vect{b}) \widehat{I}+\frac{i}{2}(\widehat{\vect{S}} \cdot[\vect{a} \times \vect{b}]),  \label{eq:2:5:23}\\
(\widehat{\vect{S}} \cdot \vect{a})[\widehat{\vect{S}} \times \vect{b}]=\frac{1}{4}[\vect{a} \times \vect{b}] \widehat{I}+\frac{i}{2}\{(\widehat{\vect{S}} \cdot \vect{b}) \vect{a}-(\vect{a} \cdot \vect{b}) \widehat{\vect{S}}\},  \label{eq:2:5:24}\\
{[[\widehat{\vect{S}} \times \vect{a}] \times[\widehat{\vect{S}} \times \vect{b}]]=\frac{1}{4}[\vect{a} \times \vect{b}] \widehat{I}+\frac{i}{2}\{(\widehat{\vect{S}} \cdot \vect{b}) \vect{a}+(\widehat{\vect{S}} \cdot \vect{a}) \vect{b}\} .} \label{eq:2:5:25}
\end{gather}
If $n$ is a unit vector ( $n^{2}=1$ ), then
\begin{equation}
(\widehat{\vect{S}} \cdot \vect{n})^{2 k}=\left(\frac{1}{4}\right)^{k} \widehat{I}, \quad(\widehat{\vect{S}} \cdot \vect{n})^{2 k+1}=\left(\frac{1}{4}\right)^{k}(\widehat{\vect{S}} \cdot \vect{n}), \quad k=0,1,2, \ldots \,.\label{eq:2:5:26}
\end{equation}
\subsection{Functions of Spin Matrices}\index{spin matrices!functions of}
\begin{gather}
e^{\alpha \widehat{S}_{i}}=\widehat{I} \cosh \frac{\alpha}{2}+2 \widehat{S}_{i} \sinh \frac{\alpha}{2} , \label{eq:2:5:27}\\
\cosh \left(\alpha \widehat{S}_{i}\right)=\widehat{I} \cosh \frac{\alpha}{2}, \quad \sinh \left(\alpha \hat{S}_{i}\right)=2 \widehat{S}_{i} \sinh \frac{\alpha}{2}, \label{eq:2:5:28}
\end{gather}
where $i=x, y, z$.
\begin{gather}
e^{\alpha \widehat{S}_{+1}}=\widehat{I}+\alpha \widehat{S}_{+1}, \quad e^{\alpha \widehat{S}_{0}}=\hat{I} \cosh \frac{\alpha}{2}+2 \widehat{S}_{0} \sinh \frac{\alpha}{2}, \quad e^{\alpha \widehat{S}_{-1}}=\widehat{I}+\alpha \widehat{S}_{-1}  ,\label{eq:2:5:29}\\
\cosh \left(\alpha \widehat{S}_{+1}\right)=\widehat{I}, \quad \cosh \left(\alpha \widehat{S}_{0}\right)=\widehat{I} \cosh \frac{\alpha}{2}, \quad \cosh \left(\alpha \widehat{S}_{-1}\right)=\widehat{I} , \label{eq:2:5:30}\\
\sinh \left(\alpha \widehat{S}_{+1}\right)=\alpha \widehat{S}_{+1}, \quad \sinh \left(\alpha \widehat{S}_{0}\right)=2 \widehat{S}_{0} \sinh \frac{\alpha}{2}, \quad \sinh \left(\alpha \widehat{S}_{-1}\right)=\alpha \widehat{S}_{-1} .\label{eq:2:5:31}
\end{gather}
\subsection{Rotation Operators}\index{rotation operator!for $S=1/2$}
\subsection{Euler angles}
 Under rotations described by the Euler angles $\alpha, \beta, \gamma$ spin functions and spin matrices for $S=\frac{1}{2}$ are transformed by a rotation operator $\widehat{D}^{\frac{1}{2}}(\alpha, \beta, \gamma)$ which is a special case of the operator given by Eq. \eqref{eq:1:4:31}
\begin{equation}
\hat{D}^{\frac{1}{2}}(\alpha, \beta, \gamma)=e^{-i \alpha \hat{S}_{z}} \cdot e^{-i \beta \hat{S}_{y}} \cdot e^{-i \gamma \hat{S}_{z}}=\begin{pmatrix}
\cos \frac{\beta}{2} e^{-i \frac{\alpha+\gamma}{2}} & -\sin \frac{\beta}{2} e^{-i \frac{\alpha-\gamma}{2}}  \label{eq:2:5:32}\\
\sin \frac{\beta}{2} e^{i \frac{\alpha-\gamma}{2}} & \cos \frac{\beta}{2} e^{i \frac{\alpha+\gamma}{2}}
\end{pmatrix}.
\end{equation}
The inverse matrix is
\begin{gather}
{\left[\hat{D}^{\frac{1}{2}}(\alpha, \beta, \gamma)\right]^{-1}=\left[\hat{D}^{\frac{1}{2}}(\alpha, \beta, \gamma)\right]^{+}=\hat{D}^{\frac{1}{2}}(\pi-\gamma, \beta,-\pi-\alpha)=\hat{D}^{\frac{1}{2}}(-\gamma,-\beta,-\alpha)} , \notag\\
=\begin{pmatrix}
\cos \frac{\beta}{2} e^{i \frac{\alpha+\gamma}{2}} & \sin \frac{\beta}{2} e^{-i \frac{\alpha-\gamma}{2}} \\
-\sin \frac{\beta}{2} e^{i \frac{\alpha-\gamma}{2}} & \cos \frac{\beta}{2} e^{-i \frac{\alpha+\gamma}{2}}
\end{pmatrix}. \label{eq:2:5:33}
\end{gather}
Matrix elements of the operator $\hat{D}^{\frac{1}{2}}(\alpha, \beta, \gamma)$ are the Wigner $D$-functions $D_{M M^{\prime}}^{\frac{1}{2}}(\alpha, \beta, \gamma)$ (Chap.~\ref{chap:4}). The expansion of $\widehat{D}^{\frac{1}{2}}(\alpha, \beta, \gamma)$ in terms of the spin matrices has the form
\begin{align}
\widehat{D}^{\frac{1}{2}}(\alpha, \beta, \gamma) & =\cos \frac{\beta}{2} \cos \frac{\alpha+\gamma}{2} \widehat{I}+2 i \sin \frac{\beta}{2} \sin \frac{\alpha-\gamma}{2} \widehat{S}_{x}-2 i \sin \frac{\beta}{2} \cos \frac{\alpha-\gamma}{2} \widehat{S}_{y}-2 i \cos \frac{\beta}{2} \sin \frac{\alpha+\gamma}{2} \widehat{S}_{z},  \label{eq:2:5:34}\\
\widehat{D}^{\frac{1}{2}}(\alpha, \beta, \gamma) & =\cos \frac{\beta}{2} \cos \frac{\alpha+\gamma}{2} \widehat{I}+\sqrt{2} \sin \frac{\beta}{2} e^{-i \frac{\alpha-\gamma}{2}} \widehat{S}_{+1}-2 i \cos \frac{\beta}{2} \sin \frac{\alpha+\gamma}{2} \widehat{S}_{0}+\sqrt{2} \sin \frac{\beta}{2} e^{i \frac{\alpha-\gamma}{2}} \widehat{S}_{-1} . \label{eq:2:5:35}
\end{align}
\subsubsection{Axis and angle}
 Under rotations of the coordinate system through an angle $\omega$ about an axis $\vect{n}(\boldsymbol{\Theta}, \Phi)$ spin functions and spin matrices for $S=\frac{1}{2}$ are transformed by the rotation operator $\hat{U}^{\frac{1}{2}}(\omega ; \Theta, \Phi)$ which is special case of the operator given by Eq. \eqref{eq:1:4:33}.
\begin{equation}
\hat{U}^{\frac{1}{2}}(\omega ; \Theta, \Phi)=e^{-i \omega \vect{n} \cdot \hat{\vect{s}}}=\begin{pmatrix}
\cos \frac{\omega}{2}-i \sin \frac{\omega}{2} \cos \Theta & -i \sin \frac{\omega}{2} \sin \Theta e^{-i \Phi}  \label{eq:2:5:36}\\
-i \sin \frac{\omega}{2} \sin \Theta e^{i \Phi} & \cos \frac{\omega}{2}+i \sin \frac{\omega}{2} \cos \Theta
\end{pmatrix} .
\end{equation}
The inverse matrix is
\begin{align}
{\left[\hat{U}^{\frac{1}{2}}(\omega ; \Theta, \Phi)\right]^{-1} } & =\left[\hat{U}^{\frac{1}{2}}(\omega ; \Theta, \Phi)\right]^{+}=\hat{U}^{\frac{1}{2}}(\omega ; \pi-\Theta, \pi+\Phi)=\hat{U}^{\frac{1}{2}}(-\omega ; \Theta, \Phi)  \notag\\
& =\begin{pmatrix}
\cos \frac{\omega}{2}+i \sin \frac{\omega}{2} \cos \Theta & i \sin \frac{\omega}{2} \sin \Theta e^{-i \Phi} \\
i \sin \frac{\omega}{2} \sin \Theta e^{i \Phi} & \cos \frac{\omega}{2}-i \sin \frac{\omega}{2} \cos \Theta
\end{pmatrix} \label{eq:2:5:37}
\end{align}
The expansion of the operator $\hat{U}^{\frac{1}{2}}(\omega ; \theta, \Phi)$ in terms of the spin matrices has the form
\begin{equation}
\widehat{U}^{\frac{1}{2}}(\omega ; \theta, \Phi)=e^{-i \omega n \cdot \widehat{\vect{S}}}=\widehat{I} \cos \frac{\omega}{2}-2 i(\mathrm{n} \cdot \widehat{\vect{S}}) \sin \frac{\omega}{2}. \label{eq:2:5:38}
\end{equation}
The relations between the angles $\omega, \Theta, \Phi$ and the Euler angles $\alpha, \beta, \gamma$ are given by Eqs. \eqref{eq:1:4:16}, \eqref{eq:1:4:17}. Note that
\begin{equation}
\hat{D}^{\frac{1}{2}}(\alpha, \beta, \gamma)=\hat{U}^{\frac{1}{2}}(\omega ; \Theta, \Phi) . \label{eq:2:5:39}
\end{equation}
\subsection{Result of rotation operator} The result of applying the rotation operator $\hat{D}^{\frac{1}{2}}(\alpha, \beta, \gamma)$ to the spin matrices may be presented as
\begin{equation}
\left.\left.\widehat{S}_{i}^{\prime} \equiv \widehat{D}^{\frac{1}{2}}(\alpha, \beta, \gamma) \widehat{S}_{i} \right\rvert\, \widehat{D}^{\frac{1}{2}}(\alpha, \beta, \gamma)\right]^{-1}=\sum_{k} a_{i k} \widehat{S}_{k}, \quad(i, k=x, y, z) .\label{eq:2:5:40}
\end{equation}
where $a_{i k}$ are elements of the rotation matrix (Sec.~\ref{sec:1.4.6}). For spherical components of the spin matrices we have
\begin{equation}
\widehat{S}_{\mu}^{\prime} \equiv \widehat{D}^{\frac{1}{2}}(\alpha, \beta, \gamma) \widehat{S}_{\mu}\left[\widehat{D}^{\frac{1}{2}}(\alpha, \beta, \gamma)\right]^{-1}=\sum_{\nu} D_{\nu \mu}^{1}(\alpha, \beta, \gamma) \widehat{S}_{\nu} ; \label{eq:2:5:41}
\end{equation}
where $D_{\nu \mu}^{1}$ are the Wigner D-functions (Chap.~\ref{chap:4}).

\subsection{Traces of Products of Spin Matrices ( $S=\frac{1}{2}$ )}
For cartesian components of the spin operator the following relations hold (where $i, k, l$ etc. take the values $x, y, z)$
\begin{gather}
\operatorname{Tr}\left\{\widehat{S}_{i}\right\}=0 ., \notag\\
\operatorname{Tr}\left\{\widehat{S}_{i} \widehat{S}_{k}\right\}=\frac{1}{2} \delta_{i k} , \notag\\
\operatorname{Tr}\left\{\widehat{S}_{i} \widehat{S}_{k} \widehat{S}_{l}\right\}=\frac{i}{4} \varepsilon_{i k l} , \label{eq:2:5:42}\\
\operatorname{Tr}\left\{\widehat{S}_{i} \widehat{S}_{k} \widehat{S}_{l} \widehat{S}_{j}\right\}=\frac{1}{8}\left(\delta_{i k} \delta_{l j}-\delta_{i l} \delta_{k j}+\delta_{i j} \delta_{k l}\right),  \notag\\
\operatorname{Tr}\left\{\widehat{S}_{i} \widehat{S}_{k} \widehat{S}_{l} \widehat{S}_{j} \widehat{S}_{m}\right\}=\frac{i}{16}\left(\delta_{i k} \varepsilon_{l j m}+\delta_{j l} \varepsilon_{i k m}+\delta_{m l} \varepsilon_{i j k}+\delta_{j m} \varepsilon_{i k l}\right) . \notag
\end{gather}
The evaluation of traces of products which contain many spin matrices can be facilitated by using the following recurrence relation: if we introduce the definition
\begin{equation}
T_{i_{1} i_{2} \ldots i_{n}} \equiv \operatorname{Tr}\left\{\widehat{S}_{i_{1}} \widehat{S}_{i_{2} \ldots} \widehat{S}_{i_{n}}\right\}, \quad(n \geq 3), \label{eq:2:5:43}
\end{equation}
then
\begin{equation}
T_{i_{1} i_{2} \ldots i_{n}}=\frac{1}{4} \delta_{i_{1} i_{2}} T_{i_{3} i_{4} \ldots i_{n}}+\frac{i}{2} \varepsilon_{i_{1} i_{2} i^{\prime}} T_{i^{\prime} i_{3} \ldots i_{n}}. \label{eq:2:5:44}
\end{equation}
If $n$ is even, one may also use the relation
\begin{equation}
T_{i_{1} i_{2} \ldots i_{n}}=\frac{1}{4}\left\{\delta_{i_{1} i_{2}} T_{i_{3} i_{4} \ldots i_{n}}-\delta_{i_{1} i_{3}} T_{i_{2} i_{4} \ldots i_{n}}+\ldots+\delta_{i_{1} i_{n}} T_{i_{2} i_{3} \ldots i_{n-1}}\right\} .\label{eq:2:5:45}
\end{equation}
For spherical components of the spin operator the following relations hold (where $\mu, \nu, \lambda$ etc. take the values $\pm 1,0$ )
\begin{gather}
\operatorname{Tr}\left\{\widehat{S}_{\mu}\right\}=0,  \notag\\
\operatorname{Tr}\left\{\widehat{S}_{\mu} \widehat{S}_{\nu}\right\}=\frac{1}{2}(-1)^{\mu} \delta_{\mu-\nu},  \notag\\
\operatorname{Tr}\left\{\widehat{S}_{\mu} \widehat{S}_{\nu} \widehat{S}_{\lambda}\right\}=-\frac{1}{2} \sqrt{\frac{3}{2}}\begin{pmatrix}
1 & 1 & 1 \\
\mu & \nu & \lambda
\end{pmatrix}=(-1)^{1+\lambda} \frac{1}{2 \sqrt{2}} \clebsch{1}{\mu}{1}{\nu}{1}{-\lambda},  \label{eq:2:5:46}\\
\operatorname{Tr}\left\{\widehat{S}_{\mu} \widehat{S}_{\nu} \widehat{S}_{\lambda} \widehat{S}_{\rho}\right\}=\frac{(-1)^{\mu}}{8}\left\{(-1)^{\lambda} \delta_{\mu-\nu} \delta_{\lambda-\rho}-(-1)^{\nu} \delta_{\mu-\lambda} \delta_{\nu-\rho}+(-1)^{\nu} \delta_{\mu-\rho} \delta_{\nu-\lambda}\right\} . \notag
\end{gather}
The recurrence relation for spherical components of the spin operator is written as
\begin{equation}
T_{\mu_{1} \mu_{2} \ldots \mu_{n}}=\frac{(-1)^{\mu_{1}}}{4} \delta_{\mu_{1}-\mu_{2}} T_{\mu_{3} \ldots \mu_{n}}-\frac{1}{\sqrt{2}} \clebsch{1}{\mu_{1}}{1}{\mu_{2}}{1}{\mu} T_{\mu \mu_{3} \ldots \mu_{n}} \label{eq:2:5:47}
\end{equation}
where
\begin{equation}
T_{\mu_{1} \mu_{2} \ldots \mu_{n}} \equiv \operatorname{Tr}\left\{\widehat{S}_{\mu_{1}} \widehat{S}_{\mu_{2} \ldots} \widehat{S}_{\mu_{n}}\right\}. \label{eq:2:5:48}
\end{equation}
If $n$ is even, one may use the formula
\begin{equation}
T_{\mu_{1} \mu_{2} \ldots \mu_{n}}=\frac{(-1)^{\mu_{1}}}{4}\left\{\delta_{\mu_{1}-\mu_{2}} T_{\mu_{3} \mu_{4} \ldots \mu_{n}}-\delta_{\mu_{1}-\mu_{3}} T_{\mu_{2} \mu_{4} \ldots \mu_{n}}+\ldots+\delta_{\mu_{1}-\mu_{n}} T_{\mu_{2} \mu_{3} \ldots \mu_{n-1}}\right\}. \label{eq:2:5:49}
\end{equation}
\section{Spin matrices and polarization operators for $S=1$}\label{sec:2.6}\index{spin matrices!S=1@$S=1$}\index{polarization operator!for $S=1$}
\subsection{Spin $S=1$}\label{sec:2.6.1}
If $S=1$, the spin operator $\widehat{S}$ and the polarization operators $\widehat{T}_{L M}(L=0,1,2 ;-L \leq M \leq L)$ are square $3 \times 3$ matrices. The polarization operator $\widehat{T}_{00}$ is proportional to the unit matrix
\begin{equation}
\widehat{T}_{00}=\frac{1}{\sqrt{3}} \widehat{I} ,\label{eq:2:6:1}
\end{equation}
where
\begin{equation}
\widehat{I}=\begin{pmatrix}
1 & 0 & 0  \\
0 & 1 & 0 \\
0 & 0 & 1
\end{pmatrix}.\label{eq:2:6:2}
\end{equation}
The polarization operators $\widehat{T}_{1 M}$ are proportional to spherical components of the spin matrices $\widehat{S}_{M}$ :
\begin{equation}
\widehat{T}_{1 M}=\frac{1}{\sqrt{2}} \widehat{S}_{M}, \quad(M=0, \pm 1). \label{eq:2:6:3}
\end{equation}
The operators $\widehat{T}_{2 M}(M=0, \pm 1, \pm 2)$ can be expressed in terms of spherical components of the spin matrices as
\begin{equation}
\widehat{T}_{2 M}=\sum_{\mu, \nu} \clebsch{1}{\mu}{1}{\nu}{2}{M} \widehat{S}_{\mu} \widehat{S}_{\nu} .\label{eq:2:6:4}
\end{equation}
The polarization operators $\widehat{T}_{2 M}$ are equivalent to some symmetric traceless cartesian tensor of second rank $\widehat{Q}_{i k}:$
\begin{gather}
\hat{Q}_{i k}=\frac{1}{2}\left(\widehat{S}_{i} \widehat{S}_{k}+\widehat{S}_{k} \widehat{S}_{i}-\frac{4}{3} \delta_{i k} \widehat{I}\right), \quad(i, k=x, y, z)  ,\label{eq:2:6:5}\\
\widehat{Q}_{i k}=\widehat{Q}_{k i}, \sum_{i} \hat{Q}_{i i}=0 .\label{eq:2:6:6}
\end{gather}
$\widehat{Q}_{i k}$ is called the quadrupole tensor. It has 5 linearly independent components.\index{quadrupole tensor}\index{tensor!quadrupole}\isym{Q-hat@$\widehat{Q}_{i k}$, quadrupole tensor} The relations between $\widehat{T}_{2 M}$ and $\widehat{Q}_{i k}$ are given by
\begin{gather}
\widehat{T}_{2 \pm 2}=\frac{1}{2}\left(\widehat{Q}_{x x}-\widehat{Q}_{y y} \pm 2 i \widehat{Q}_{x y}\right) , \notag\\
\widehat{T}_{2 \pm 1}=\mp\left(\widehat{Q}_{x z} \pm i \widehat{Q}_{y z}\right),  \label{eq:2:6:7}\\
\widehat{T}_{20}=\sqrt{\frac{3}{2}} \widehat{Q}_{z z} .\notag
\end{gather}
The inverse relations are
\begin{equation}
\begin{aligned}
\hat{Q}_{x x}=\frac{1}{2}\left(\hat{T}_{22}+\hat{T}_{2-2}\right)-\frac{1}{\sqrt{6}} \hat{T}_{20}, && \hat{Q}_{x y}=\hat{Q}_{y x}=\frac{i}{2}\left(\hat{T}_{2-2}-\hat{T}_{22}\right), \\
\hat{Q}_{y y}=-\frac{1}{2}\left(\hat{T}_{22}+\hat{T}_{2-2}\right)-\frac{1}{\sqrt{6}} \hat{T}_{20}, && \hat{Q}_{x z}=\hat{Q}_{z x}=\frac{1}{2}\left(\hat{T}_{2-1}-\hat{T}_{21}\right),  \label{eq:2:6:8}\\
\hat{Q}_{z z}=\sqrt{\frac{2}{3}} \hat{T}_{20}, && \hat{Q}_{y z}=\hat{Q}_{z y}=\frac{i}{2}\left(\hat{T}_{2-1}+\hat{T}_{21}\right).
\end{aligned}
\end{equation}
\subsection{Explicit Form}\label{sec:2.6.2}
Explicit forms of the spin matrices and the polarization operators depend on the representation used for basis spin functions (Chap.~\ref{chap:6}). We shall consider the spherical basis representation and the cartesian one. These representations are used very frequently.

\subsubsection{Spherical basis representation}\index{spherical basis representation}\index{representation!spherical basis}
Cartesian components of the spin operator $\widehat{\vect{S}}$ are given by
\begin{equation}
\widehat{S}_{x}=\frac{1}{\sqrt{2}}\begin{pmatrix}
0 & 1 & 0  \label{eq:2:6:9}\\
1 & 0 & 1 \\
0 & 1 & 0
\end{pmatrix}, \quad \widehat{S}_{y}=\frac{i}{\sqrt{2}}\begin{pmatrix}
0 & -1 & 0 \\
1 & 0 & -1 \\
0 & 1 & 0
\end{pmatrix}, \quad \widehat{S}_{z}=\begin{pmatrix}
1 & 0 & 0 \\
0 & 0 & 0 \\
0 & 0 & -1
\end{pmatrix}
\end{equation}
and the spherical components of $\hat{\vect{S}}$ are
\begin{equation}
\widehat{S}_{+1}=-\begin{pmatrix}
0 & 1 & 0  \label{eq:2:6:10}\\
0 & 0 & 1 \\
0 & 0 & 0
\end{pmatrix}, \quad \widehat{S}_{0}=\begin{pmatrix}
1 & 0 & 0 \\
0 & 0 & 0 \\
0 & 0 & -1
\end{pmatrix}, \quad \widehat{S}_{-1}=\begin{pmatrix}
0 & 0 & 0 \\
1 & 0 & 0 \\
0 & 1 & 0
\end{pmatrix}.
\end{equation}
The order of rows and columns in these matrices is shown in Eq. \eqref{eq:2:3:10}.\\
Matrix elements of the spin operators $\widehat{S}_{\mu}$ in the spherical basis representation are given by
\begin{equation}
\left(\widehat{S}_{\mu}\right)_{\sigma^{\prime} \sigma}=\sqrt{2} \clebsch{1}{\sigma}{1}{\mu}{1}{\sigma^{\prime}} \quad\left(\mu, \sigma, \sigma^{\prime}= \pm 1,0\right). \label{eq:2:6:11}
\end{equation}
The spin matrices in the spherical basis representation satisfy the relations
\begin{gather}
\widehat{S}_{i}^{+}=\widehat{S}_{i}, \quad(i=x, y, z) , \label{eq:2:6:12}\\
\widehat{S}_{x}^{*}=\widehat{S}_{x}, \quad \widehat{S}_{y}^{*}=-\widehat{S}_{y}, \quad \widehat{S}_{z}^{*}=\widehat{S}_{z}  ,\label{eq:2:6:13}\\
\widehat{S}_{\mu}^{*}=\widehat{S}_{\mu}, \quad \widehat{S}_{\mu}^{+}=(-1)^{\mu} \widehat{S}_{-\mu} \quad(\mu= \pm 1,0). \label{eq:2:6:14}
\end{gather}
The quadrupole tensor $\widehat{Q}_{i k}(i, k=x, y, z)$ in the spherical basis representation has the form
\begin{gather}
\hat{Q}_{x x}=\frac{1}{6}\begin{pmatrix}
-1 & 0 & 3 \\
0 & 2 & 0 \\
3 & 0 & -1
\end{pmatrix}, \quad \hat{Q}_{y y}=\frac{1}{6}\begin{pmatrix}
-1 & 0 & -3 \\
0 & 2 & 0 \\
-3 & 0 & -1
\end{pmatrix}, \quad \hat{Q}_{z z}=\frac{1}{3}\begin{pmatrix}
1 & 0 & 0 \\
0 & -2 & 0 \\
0 & 0 & 1
\end{pmatrix},  \label{eq:2:6:15}\\
\hat{Q}_{x y}=\hat{Q}_{y x}=\frac{i}{2}\begin{pmatrix}
0 & 0 & -1 \\
0 & 0 & 0 \\
1 & 0 & 0
\end{pmatrix}, \quad \hat{Q}_{x z}=\hat{Q}_{z x}=\frac{1}{2 \sqrt{2}}\begin{pmatrix}
0 & 1 & 0 \\
1 & 0 & -1 \\
0 & -1 & 0
\end{pmatrix}, \quad \hat{Q}_{y z}=\hat{Q}_{z y}=\frac{i}{2 \sqrt{2}}\begin{pmatrix}
0 & -1 & 0 \\
1 & 0 & 1 \\
0 & -1 & 0
\end{pmatrix} . \notag
\end{gather}
The components of the quadrupole tensor $\widehat{Q}_{i k}$ in the spherical basis representation have the following properties
\begin{equation}
\widehat{Q}_{i k}^{+}=\widehat{Q}_{i k}, \quad(i, k=x, y, z), \label{eq:2:6:16}
\end{equation}
The matrices $\hat{Q}_{x x}, \hat{Q}_{y y}, \hat{Q}_{z z}, \hat{Q}_{x z}, \hat{Q}_{z x}$ are real, whereas $\hat{Q}_{x y}, \hat{Q}_{y x}, \hat{Q}_{y z}, \hat{Q}_{z y}$ are purely imaginary. The polarization operators $\widehat{T}_{L M}$ in the spherical basis representation are given by
\begin{gather}
\hat{T}_{00}=\frac{1}{\sqrt{3}}\begin{pmatrix}
1 & 0 & 0 \\
0 & 1 & 0 \\
0 & 0 & 1
\end{pmatrix},  \label{eq:2:6:17}\\
\hat{T}_{1+1}=-\frac{1}{\sqrt{2}}\begin{pmatrix}
0 & 1 & 0 \\
0 & 0 & 1 \\
0 & 0 & 0
\end{pmatrix}, \quad \hat{T}_{10}=\frac{1}{\sqrt{2}}\begin{pmatrix}
1 & 0 & 0 \\
0 & 0 & 0 \\
0 & 0 & -1
\end{pmatrix}, \quad \hat{T}_{1-1}=\frac{1}{\sqrt{2}}\begin{pmatrix}
0 & 0 & 0 \\
1 & 0 & 0 \\
0 & 1 & 0
\end{pmatrix},  \label{eq:2:6:18}\\
\hat{T}_{2+2}=\begin{pmatrix}
0 & 0 & 1 \\
0 & 0 & 0 \\
0 & 0 & 0
\end{pmatrix}, \quad \hat{T}_{21}=\frac{1}{\sqrt{2}}\begin{pmatrix}
0 & -1 & 0 \\
0 & 0 & 1 \\
0 & 0 & 0
\end{pmatrix}, \quad \hat{T}_{20}=\frac{1}{\sqrt{6}}\begin{pmatrix}
1 & 0 & 0 \\
0 & -2 & 0 \\
0 & 0 & 1
\end{pmatrix},  \notag\\
\hat{T}_{2-1}=\frac{1}{\sqrt{2}}\begin{pmatrix}
0 & 0 & 0 \\
1 & 0 & 0 \\
0 & -1 & 0
\end{pmatrix}, \quad \hat{T}_{2-2}=\begin{pmatrix}
0 & 0 & 0 \\
0 & 0 & 0 \\
1 & 0 & 0
\end{pmatrix} . \label{eq:2:6:19}
\end{gather}
Matrix elements of the polarization operators in the spherical basis representation may be written as
\begin{equation}
\left(\widehat{T}_{L M}\right)_{\sigma^{\prime} \sigma}=\sqrt{\frac{2 L+1}{3}} \clebsch{1}{\sigma}{L}{M}{1}{\sigma^{\prime}}, \quad\left(\sigma, \sigma^{\prime}= \pm 1,0 ;-L \leq M \leq L\right) .\label{eq:2:6:20}
\end{equation}
The polarization operators $\widehat{T}_{L M}$ in the spherical basis representation are real, i.e.,
\begin{equation}
\widehat{T}_{L M}^{*}=\widehat{T}_{L M}. \label{eq:2:6:21}
\end{equation}
and satisfy the relations
\begin{equation}
\hat{T}_{L M}^{+}=(-1)^{M} \hat{T}_{L-M} .\label{eq:2:6:22}
\end{equation}
\subsubsection{Cartesian basis representation}\index{cartesian basis representation}\index{representation!cartesian basis}

Cartesian components of the spin operator $\widehat{S}$ in the cartesian basis representation are given by
\begin{equation}
\widehat{S}_{x}=\begin{pmatrix}
0 & 0 & 0  \label{eq:2:6:23}\\
0 & 0 & -i \\
0 & i & 0
\end{pmatrix}, \quad \widehat{S}_{y}=\begin{pmatrix}
0 & 0 & i \\
0 & 0 & 0 \\
-i & 0 & 0
\end{pmatrix}, \quad \widehat{S}_{z}=\begin{pmatrix}
0 & -i & 0 \\
i & 0 & 0 \\
0 & 0 & 0
\end{pmatrix} .
\end{equation}
The matrix elements of $\widehat{S}_{i}$ in this representation may be written as
\begin{equation}
\left(\widehat{S}_{i}\right)_{k l}=-i \varepsilon_{i k l}, \quad(i, k, l=x, y, z) . \label{eq:2:6:24}
\end{equation}
The spherical components of $\widehat{\vect{S}}$ are as follows
\begin{equation}
\widehat{S}_{+1}=\frac{1}{\sqrt{2}}\begin{pmatrix}
0 & 0 & 1  \label{eq:2:6:25}\\
0 & 0 & i \\
-1 & -i & 0
\end{pmatrix}, \quad \widehat{S}_{0}=\begin{pmatrix}
0 & -i & 0 \\
i & 0 & 0 \\
0 & 0 & 0
\end{pmatrix}, \quad \widehat{S}_{-1}=\frac{1}{\sqrt{2}}\begin{pmatrix}
0 & 0 & 1 \\
0 & 0 & -i \\
-1 & i & 0
\end{pmatrix} .
\end{equation}
The spin matrices in the cartesian basis representation satisfy the following relations
\begin{align}
\left(\widehat{S}_{\mu}\right)^{+}=(-1)^{\mu} \widehat{S}_{-\mu}, \quad \widehat{S}_{\mu}^{*}=(-1)^{1+\mu} \widehat{S}_{-\mu}, \quad(\mu= \pm 1,0) , \label{eq:2:6:26}\\
\left(\widehat{S}_{i}\right)^{+}=\widehat{S}_{i}, \quad \widehat{S}_{i}^{*}=-\widehat{S}_{i}, \quad(i=x, y, z) . \label{eq:2:6:27}
\end{align}
The quadrupole tensor $\widehat{Q}_{i k}(i, k=x, y, z)$ in the cartesian basis representation is given by
\begin{gather}
\hat{Q}_{x x}=\frac{1}{3}\begin{pmatrix}
-2 & 0 & 0 \\
0 & 1 & 0 \\
0 & 0 & 1
\end{pmatrix}, \quad \hat{Q}_{y y}=\frac{1}{3}\begin{pmatrix}
1 & 0 & 0 \\
0 & -2 & 0 \\
0 & 0 & 1
\end{pmatrix}, \quad \hat{Q}_{z z}=\frac{1}{3}\begin{pmatrix}
1 & 0 & 0 \\
0 & 1 & 0 \\
0 & 0 & -2
\end{pmatrix},  \notag\\
\hat{Q}_{x y}=\hat{Q}_{y x}=\frac{1}{2}\begin{pmatrix}
0 & -1 & 0 \\
-1 & 0 & 0 \\
0 & 0 & 0
\end{pmatrix}, \quad \hat{Q}_{x z}=\hat{Q}_{z x}=\frac{1}{2}\begin{pmatrix}
0 & 0 & -1 \\
0 & 0 & 0 \\
-1 & 0 & 0
\end{pmatrix}, \quad \hat{Q}_{y z}=\hat{Q}_{z y}=\frac{1}{2}\begin{pmatrix}
0 & 0 & 0 \\
0 & 0 & -1 \\
0 & -1 & 0
\end{pmatrix} . \label{eq:2:6:28}
\end{gather}
The matrix elements of $\widehat{Q}_{i k}$ in the cartesian basis representation may be written in the form
\begin{equation}
\left(\widehat{Q}_{i k}\right)_{l m}=-\frac{1}{2}\left(\delta_{i l} \delta_{k m}+\delta_{i m} \delta_{k l}-\frac{2}{3} \delta_{i k} \delta_{l m}\right), \quad(i, k, l, m=x, y, z). \label{eq:2:6:29}
\end{equation}
In the cartesian basis representation the matrices $\widehat{Q}_{i k}$ are Hermitian and real, i.e.,
\begin{equation}
\widehat{Q}_{i k}^{+}=\widehat{Q}_{i k}, \quad \widehat{Q}_{i k}^{*}=\widehat{Q}_{i k}, \quad(i, k=x, y, z) . \label{eq:2:6:30}
\end{equation}
The polarization operators $\widehat{T}_{L M}$ in the cartesian basis representation have the form
\begin{gather}
\widehat{T}_{00}=\frac{1}{\sqrt{3}}\begin{pmatrix}
1 & 0 & 0 \\
0 & 1 & 0 \\
0 & 0 & 1
\end{pmatrix},  \label{eq:2:6:31}\\
\widehat{T}_{1+1}=\frac{1}{2}\begin{pmatrix}
0 & 0 & 1 \\
0 & 0 & i \\
-1 & -i & 0
\end{pmatrix}, \quad \widehat{T}_{10}=\frac{1}{\sqrt{2}}\begin{pmatrix}
0 & -i & 0 \\
i & 0 & 0 \\
0 & 0 & 0
\end{pmatrix}, \quad \widehat{T}_{1-1}=\frac{1}{2}\begin{pmatrix}
0 & 0 & 1 \\
0 & 0 & -i \\
-1 & i & 0
\end{pmatrix},  \notag\\
\widehat{T}_{2+2}=\frac{1}{2}\begin{pmatrix}
-1 & -i & 0 \\
-i & 1 & 0 \\
0 & 0 & 0
\end{pmatrix}, \quad \widehat{T}_{2+1}=\frac{1}{2}\begin{pmatrix}
0 & 0 & 1 \\
0 & 0 & i \\
1 & i & 0
\end{pmatrix}, \quad \widehat{T}_{20}=\frac{1}{\sqrt{6}}\begin{pmatrix}
1 & 0 & 0 \\
0 & 1 & 0 \\
0 & 0 & -2
\end{pmatrix},  \label{eq:2:6:32}\\
\widehat{T}_{2-1}=\frac{1}{2}\begin{pmatrix}
0 & 0 & -1 \\
0 & 0 & i \\
-1 & i & 0
\end{pmatrix}, \quad \widehat{T}_{2-2}=\frac{1}{2}\begin{pmatrix}
-1 & i & 0 \\
i & 1 & 0 \\
0 & 0 & 0
\end{pmatrix} . \label{eq:2:6:33}
\end{gather}
The polarization operators $\widehat{T}_{L M}$ in the cartesian basis representation satisfy the relations
\begin{gather}
\widehat{T}_{L M}^{+}=(-1)^{M} \widehat{T}_{L-M} , \label{eq:2:6:34}\\
\widehat{T}_{L M}^{*}=(-1)^{L+M} \widehat{T}_{L-M}. \label{eq:2:6:35}
\end{gather}
One can easily transform any matrix in the cartesian basis representation into a matrix in the spherical basis representation and vice versa with the aid of the unitary matrix $U$,
\begin{align}
& \widehat{A}(\text { spherical basis })=U \widehat{A}(\text { cartesian basis }) U^{-1} , \notag\\
& \widehat{A}(\text { cartesian basis })=U^{-1} \widehat{A}(\text { spherical basis }) U .\label{eq:2:6:36}
\end{align}
In this case $\widehat{A}$ is any spin or polarization operator (i.e., $\widehat{S}_{i}, \widehat{S}_{\mu}, \widehat{Q}_{i k}, \widehat{T}_{L M}$ ),
\begin{equation}
U=\begin{pmatrix}
-\frac{1}{\sqrt{2}} & \frac{i}{\sqrt{2}} & 0  \label{eq:2:6:37}\\
0 & 0 & 1 \\
\frac{1}{\sqrt{2}} & \frac{i}{\sqrt{2}} & 0
\end{pmatrix}, \quad U^{-1}=U^{+}=\begin{pmatrix}
-\frac{1}{\sqrt{2}} & 0 & \frac{1}{\sqrt{2}} \\
-\frac{i}{\sqrt{2}} & 0 & -\frac{i}{\sqrt{2}} \\
0 & 1 & 0
\end{pmatrix} .
\end{equation}
The matrix $U$ coincides with $M(+1,0,-1 \leftarrow x, y, z)$ and $U^{-1}$ coincides with $M(x, y, z \leftarrow+1,0,-1)$ (Table 1.2).

The formulas given in Sec.~\ref{sec:2.6} (except those of Sec.~\ref{sec:2.6.2}) are independent of the representation unless the contrary is indicated.

\subsection{Products of Spin and Polarization Matrices}\index{polarization operator!products of}
For $S=1$, the nine matrices $\widehat{T}_{L M}(L=0,1,2 ;-L \leq M \leq L)$ or, equivalently, the matrices $\widehat{I}, \widehat{S}, \widehat{Q}_{i k}$ constitute a complete set of square $3 \times 3$ matrices for expanding any function of the spin and polarization operators. In particular, products of the spin and polarization operators may be expanded in a series in terms of these matrices.

Products of cartesian components of the matrices $\widehat{S}$ and $\widehat{Q}_{i k}$ may be expressed as follows $(i, k, l$, etc. take the values $x, y, z$ )
\subsubsection{Products of two cartesian components of spin matrices}
\begin{equation}
\widehat{S}_{i} \widehat{S}_{k}=\frac{2}{3} \delta_{i k} \widehat{I}+\frac{i}{2} \varepsilon_{i k l} \widehat{S}_{l}+\widehat{Q}_{i k}. \label{eq:2:6:38}
\end{equation}
In particular,
\begin{gather}
\widehat{\vect{S}}^{2}=\widehat{S}_{x}^{2}+\widehat{S}_{y}^{2}+\widehat{S}_{z}^{2}=2 \widehat{I} , \label{eq:2:6:39}\\
{\left[\widehat{S}_{i}, \widehat{S}_{k}\right] \equiv \widehat{S}_{i} \widehat{S}_{k}-\widehat{S}_{k} \widehat{S}_{i}=i \varepsilon_{i k l} \widehat{S}_{l}} , \label{eq:2:6:40}\\
\left\{\widehat{S}_{i}, \widehat{S}_{k}\right\} \equiv \widehat{S}_{i} \widehat{S}_{k}+\widehat{S}_{k} \widehat{S}_{i}=\frac{4}{3} \delta_{i k} \widehat{I}+2 \widehat{Q}_{i k}. \label{eq:2:6:41}
\end{gather}
\subsubsection{Products of  cartesian components of three spin matrices}
\begin{align}
\hat{S}_{i} \hat{S}_{k} \hat{S}_{l} & =\frac{i}{3} \varepsilon_{i k l} \hat{I}+\frac{1}{2}\left(\delta_{i k} \hat{S}_{l}+\delta_{k l} \hat{S}_{i}\right)+i \varepsilon_{i l m} \hat{Q}_{k m}  \notag\\
& =\frac{i}{3} \varepsilon_{i k l} \hat{I}+\frac{1}{2}\left(\delta_{i k} \hat{S}_{l}+\delta_{k l} \hat{S}_{i}\right)+\frac{i}{2}\left(\varepsilon_{i k m} \hat{Q}_{l m}+\varepsilon_{i l m} \hat{Q}_{k m}+\varepsilon_{k l m} \hat{Q}_{i m}\right) .\label{eq:2:6:42}
\end{align}
From Eq. \eqref{eq:2:6:42} one may obtain the Duffin-Kemmer relation\index{Duffin-Kemmer relation}
\begin{equation}
\widehat{S}_{i} \widehat{S}_{k} \widehat{S}_{l}+\widehat{S}_{l} \widehat{S}_{k} \widehat{S}_{i}=\delta_{i k} \widehat{S}_{l}+\delta_{k l} \widehat{S}_{i}. \label{eq:2:6:43}
\end{equation}
Other particular cases of \eqref{eq:2:6:42} are
\begin{gather}
\hat{S}_{i} \hat{S}_{k} \hat{S}_{i}=\delta_{i k} \hat{S}_{i} \quad \text { (no sum over $i$ ); } \quad \sum_{i} \hat{S}_{i} \hat{S}_{k} \hat{S}_{i}=\hat{S}_{k}  \label{eq:2:6:44}\\
\hat{S}_{i} \hat{S}_{k} \hat{S}_{k}+\hat{S}_{k} \hat{S}_{k} \hat{S}_{i}=\hat{S}_{i}+\delta_{i k} \hat{S}_{k} \quad \text { (no sum over $k$) } , \label{eq:2:6:45}\\
\hat{S}_{x} \hat{S}_{y} \hat{S}_{z}+\hat{S}_{y} \hat{S}_{z} \hat{S}_{x}+\hat{S}_{z} \hat{S}_{x} \hat{S}_{y}=i \hat{I},  \notag\\
\hat{S}_{x} \hat{S}_{z} \hat{S}_{y}+\hat{S}_{z} \hat{S}_{y} \hat{S}_{x}+\hat{S}_{y} \hat{S}_{x} \hat{S}_{z}=-i \hat{I} ,\label{eq:2:6:46}\\
\hat{S}_{x} \hat{S}_{z} \hat{S}_{z}=\hat{S}_{y} \hat{S}_{y} \hat{S}_{x}=\frac{1}{2} \hat{S}_{x}-i \hat{Q}_{y z}  ,\label{eq:2:6:47}\\
\hat{S}_{x} \hat{S}_{y} \hat{S}_{y}=\hat{S}_{z} \hat{S}_{z} \hat{S}_{x}=\frac{1}{2} \hat{S}_{x}+i \hat{Q}_{y z},  \notag\\
\hat{S}_{y} \hat{S}_{x} \hat{S}_{x}=\hat{S}_{z} \hat{S}_{z} \hat{S}_{y}=\frac{1}{2} \hat{S}_{y}-i \hat{Q}_{x z} , \notag\\
\hat{S}_{y} \hat{S}_{z} \hat{S}_{z}=\hat{S}_{x} \hat{S}_{x} \hat{S}_{y}=\frac{1}{2} \hat{S}_{y}+i \hat{Q}_{x z} , \label{eq:2:6:48}\\
\hat{S}_{z} \hat{S}_{y} \hat{S}_{y}=\hat{S}_{x} \hat{S}_{x} \hat{S}_{z}=\frac{1}{2} \hat{S}_{z}-i \hat{Q}_{x y} , \label{eq:2:6:49}\\
\hat{S}_{z} \hat{S}_{x} \hat{S}_{x}=\hat{S}_{y} \hat{S}_{y} \hat{S}_{z}=\frac{1}{2} \hat{S}_{z}+i \hat{Q}_{x y}. \notag 
\end{gather}
\subsubsection{General powers of  cartesian components of spin matrices}
\begin{equation}
\widehat{S}_{i}^{2 n}=\frac{2}{3} \widehat{I}+\widehat{Q}_{i i},(n=1,2, \ldots), \quad \widehat{S}_{i}^{2 n+1}=\widehat{S}_{i},(n=0,1,2, \ldots). \label{eq:2:6:50}
\end{equation}
In particular, if $n$ is the unit vector, then
\begin{gather}
(\widehat{\vect{S}} \cdot \vect{n})^{2 k}=\frac{2}{3} \widehat{I}+\sum_{i, l} n_{i} n_{l} \widehat{Q}_{i l}, \quad(k=1,2,3, \ldots) , \label{eq:2:6:51}\\
(\widehat{\vect{S}} \cdot \vect{n})^{2 k+1}=(\widehat{\vect{S}} \cdot \vect{n}), \quad(k=0,1,2, \ldots) .\notag
\end{gather}
\subsubsection{Products of  cartesian components of quadrupole tensors and spin matrices}
\begin{align}
& \widehat{Q}_{i k} \widehat{S}_{l}=\frac{1}{4}\left(\delta_{i l} \widehat{S}_{k}+\delta_{k l} \widehat{S}_{i}-\frac{2}{3} \delta_{i k} \widehat{S}_{l}\right)+\frac{i}{2}\left(\varepsilon_{i l m} \widehat{Q}_{k m}+\varepsilon_{k l m} \widehat{Q}_{i m}\right) , \label{eq:2:6:52}\\
& \widehat{S}_{i} \widehat{Q}_{k l}=\frac{1}{4}\left(\delta_{i k} \widehat{S}_{l}+\delta_{i l} \widehat{S}_{k}-\frac{2}{3} \delta_{k l} \widehat{S}_{i}\right)+\frac{i}{2}\left(\varepsilon_{i k m} \widehat{Q}_{l m}+\varepsilon_{i l m} \widehat{Q}_{k m}\right) .\label{eq:2:6:53}
\end{align}
\subsubsection{Products of  cartesian components of quadrupole tensors}
\begin{align}
& \hat{Q}_{i k} \hat{Q}_{l m}=\frac{1}{6}\left(\delta_{i l} \delta_{k m}+\delta_{i m} \delta_{k l}-\frac{2}{3} \delta_{i k} \delta_{l m}\right) \hat{I}-\frac{1}{4}\left(\delta_{i l} \hat{Q}_{k m}+\delta_{i m} \hat{Q}_{k l}+\delta_{k m} \hat{Q}_{i l}+\delta_{k l} \hat{Q}_{i m}-\right.  \notag\\
& \left.-\frac{4}{3} \delta_{i k} \hat{Q}_{l m}-\frac{4}{3} \delta_{l m} \hat{Q}_{i k}\right)+\frac{i}{8}\left(\delta_{i l} \varepsilon_{k m p} \hat{S}_{p}+\delta_{i m} \varepsilon_{k l p} \hat{S}_{p}+\delta_{k l} \varepsilon_{i m p} \hat{S}_{p}+\delta_{k m} \varepsilon_{i l p} \hat{S}_{p}\right) \label{eq:2:6:54}
\end{align}
Products involving spherical components of the spin operator $\widehat{\vect{S}}$ and the polarization operators $\widehat{T}_{\mathrm{LM}}$ may be expanded in a series as given below ( $\mu, \nu= \pm 1,0$ )\\
\subsubsection{Products of spherical components spin matrices}
\begin{equation}
\widehat{S}_{\mu} \widehat{S}_{\nu}=\frac{2}{3}(-1)^{\mu} \delta_{\mu-\nu} \widehat{I}-\frac{1}{\sqrt{2}} \clebsch{1}{\mu}{1}{\nu}{1}{\lambda} \widehat{S}_{\lambda}+\clebsch{1}{\mu}{1}{\nu}{2}{M} \widehat{T}_{2 M}. \label{eq:2:6:55}
\end{equation}
In particular
\begin{gather}
\widehat{S}^{2}=-\widehat{S}_{+1} \widehat{S}_{-1}+\widehat{S}_{0} \widehat{S}_{0}-\widehat{S}_{-1} \widehat{S}_{+1}=2 \widehat{I}  \label{eq:2:6:56}\\
{\left[\widehat{S}_{\mu}, \widehat{S}_{\nu}\right] \equiv \widehat{S}_{\mu} \widehat{S}_{\nu}-\widehat{S}_{\nu} \widehat{S}_{\mu}=-\sqrt{2} \clebsch{1}{\mu}{1}{\nu}{1}{\lambda} \widehat{S}_{\lambda}}  \label{eq:2:6:57}\\
\left\{\widehat{S}_{\mu}, \widehat{S}_{\nu}\right\} \equiv \widehat{S}_{\mu} \widehat{S}_{\nu}+\widehat{S}_{\nu} \widehat{S}_{\mu}=\frac{4}{3}(-1)^{\mu} \delta_{\mu-\nu}+2 \clebsch{1}{\mu}{1}{\nu}{2}{M} \widehat{T}_{2 M} \label{eq:2:6:58}
\end{gather}
\subsubsection{Products of spherical components spin matrices and polarisation operators}
\begin{gather}
\widehat{S}_{ \pm 1}^{2}=\widehat{T}_{2 \pm 2}, \widehat{S}_{ \pm 1}^{3}=\widehat{S}_{ \pm 1}^{4}=\widehat{S}_{ \pm 1}^{5}=\ldots=0 . \label{eq:2:6:59}\\
\widehat{S}_{0}^{2}=\frac{2}{3} \widehat{I}+\sqrt{\frac{2}{3}} \widehat{T}_{20}  \label{eq:2:6:60}\\
\widehat{S}_{0}^{2 n}=\widehat{S}_{0}^{2},(n=1,2, \ldots), \quad \widehat{S}_{0}^{2 n+1}=\widehat{S}_{0}, \quad(n=0,1,2, \ldots). \label{eq:2:6:61}
\end{gather}
\subsubsection{Products of  polarisation operators}
\begin{equation}
\hat{T}_{L_{1} M_{1}} \hat{T}_{L_{2} M_{2}}=\sqrt{\left(2 L_{1}+1\right)\left(2 L_{2}+1\right)} \sum_{L}(-1)^{L}\sixj{L_{1} }{ L_{2} }{ L }{1 }{ 1 }{ 1} \clebsch{L_{1}}{M_{1}}{L_{2}}{M_{2}}{L}{M} \hat{T}_{L M}. \label{eq:2:6:62}
\end{equation}
Particularly,
\begin{gather}
\widehat{T}_{L M} \widehat{T}_{00}=\widehat{T}_{00} \widehat{T}_{L M}=\frac{1}{\sqrt{3}} \widehat{T}_{L M}, \quad(L=0,1,2 ;-L \leq M \leq L),  \label{eq:2:6:63}\\
\widehat{S}_{\mu} \widehat{T}_{2 M}=-\frac{1}{2} \sqrt{\frac{5}{3}} \clebsch{1}{\mu}{2}{M}{1}{\nu} \widehat{S}_{\nu}-\sqrt{\frac{3}{2}} \clebsch{1}{\mu}{2}{M}{2}{N} \widehat{T}_{2 N},  \label{eq:2:6:64}\\
\widehat{T}_{2 M} \widehat{S}_{\mu}=-\frac{1}{2} \sqrt{\frac{5}{3}} \clebsch{1}{\mu}{2}{M}{1}{\nu} \widehat{S}_{\nu}+\sqrt{\frac{3}{2}} \clebsch{1}{\mu}{2}{M}{2}{N} \widehat{T}_{2 N},  \label{eq:2:6:65}\\
\widehat{T}_{2 M} \widehat{T}_{2 N}=(-1)^{M} \cdot \frac{1}{3} \delta_{M-N} \widehat{I}+\frac{1}{2} \sqrt{\frac{5}{2}} \clebsch{2}{M}{2}{N}{1}{\mu} \widehat{S}_{\mu}+\frac{1}{2} \sqrt{\frac{7}{3}} \clebsch{2}{M}{2}{N}{2}{A} \widehat{T}_{2 A} . \label{eq:2:6:66}
\end{gather}
\subsection{Functions of Spin Matrices}\index{spin matrices!functions of}
\begin{align}
e^{\alpha \widehat{S}_{i}}=\frac{1}{3}(1+2 \cosh \alpha) \widehat{I}+ & \left.\sinh \alpha \widehat{S}_{i}+(\cosh \alpha-1) \widehat{Q}_{i i} \quad \text { (no sum over } i\right) , \label{eq:2:6:67}\\
\cosh \left(\alpha \widehat{S}_{i}\right)= & \frac{1}{3}(1+2 \cosh \alpha) \widehat{I}+(\cosh \alpha-1) \widehat{Q}_{i i} , \label{eq:2:6:68}\\
& \sinh \left(\alpha \widehat{S}_{i}\right)=\sinh \alpha \widehat{S}_{i} ,\label{eq:2:6:69}
\end{align}
where $i=x, y, z$.
\begin{gather}
e^{\alpha \widehat{S}_{+1}}=\widehat{I}+\alpha \widehat{S}_{+1}+\frac{\alpha^{2}}{2} \widehat{T}_{22} , \label{eq:2:6:70}\\
e^{\alpha \widehat{S}_{0}}=\frac{1}{3}(1+2 \cosh \alpha) \widehat{I}+\sinh \alpha \widehat{S}_{0}+\sqrt{\frac{2}{3}}(\cosh \alpha-1) \widehat{T}_{20} , \label{eq:2:6:71}\\
e^{\alpha \widehat{S}_{-1}}=\widehat{I}+\alpha \widehat{S}_{-1}+\frac{\alpha^{2}}{2} \widehat{T}_{2-2} , \label{eq:2:6:72}\\
\sinh \left(\alpha \widehat{S}_{+1}\right)=\alpha \widehat{S}_{+1}, \quad \sinh \left(\alpha \widehat{S}_{0}\right)=\sinh \alpha \widehat{S}_{0}, \quad \sinh \left(\alpha \widehat{S}_{-1}\right)=\alpha \widehat{S}_{-1} .\label{eq:2:6:73}
\end{gather}
\begin{gather}
\cosh \left(\alpha \widehat{S}_{+1}\right)=\widehat{I}+\frac{\alpha^{2}}{2} \widehat{T}_{22} , \notag\\
\cosh \left(\alpha \widehat{S}_{0}\right)=\frac{1}{3}(1+2 \cosh \alpha) \widehat{I}+\sqrt{\frac{2}{3}}(\cosh \alpha-1) \widehat{T}_{20} , \label{eq:2:6:74}\\
\cosh \left(\alpha \widehat{S}_{-1}\right)=\widehat{I}+\frac{\alpha^{2}}{2} \widehat{T}_{2-2}. \notag
\end{gather}
\subsection{Operators of Coordinate Rotations}\index{rotation operator!for $S=1$}
\subsubsection{Coordinaterotations defined be Euler angles}

Under rotations of coordinate systems defined by the Euler angles $\alpha, \beta, \gamma$ spin functions and spin matrices for $S=1$ are transformed by the rotation operator $\hat{D}^{1}(\alpha, \beta, \gamma)$ which is a special case of the operator given by Eq. \eqref{eq:1:4:31}
\begin{equation}
\widehat{D}^{1}(\alpha, \beta, \gamma)=e^{-i \alpha \widehat{S}_{x}} e^{-i \beta \widehat{S}_{y}} e^{-i \gamma \widehat{S}_{z}} \label{eq:2:6:75}.
\end{equation}
In the spherical basis representation $\hat{D}^{1}(\alpha, \beta, \gamma)$ has the form
\begin{equation}
\widehat{D}^{1}(\alpha, \beta, \gamma)=\begin{pmatrix}
\frac{1+\cos \beta}{2} e^{-i(\alpha+\gamma)} & -\frac{\sin \beta}{\sqrt{2}} e^{-i \alpha} & \frac{1-\cos \beta}{2} e^{i(\gamma-\alpha)}  \label{eq:2:6:76}\\
\frac{\sin \beta}{\sqrt{2}} e^{-i \gamma} & \cos \beta & -\frac{\sin \beta}{\sqrt{2}} e^{i \gamma} \\
\frac{1-\cos \beta}{2} e^{i(\alpha-\gamma)} & \frac{\sin \beta}{\sqrt{2}} e^{i \alpha} & \frac{1+\cos \beta}{2} e^{i(\gamma+\alpha)}
\end{pmatrix} .
\end{equation}
The matrix elements of $\widehat{D}^{1}(\alpha, \beta, \gamma)$ in this representation are the Wigner $D$-functions $D_{M M^{\prime}}^{1}(\alpha, \beta, \gamma)$ (see Chap.~\ref{chap:4}).

In the cartesian basis representation we have
\begin{equation}
\hat{D}^{1}(\alpha, \beta, \gamma)=\begin{pmatrix}
\cos \beta \cos \alpha \cos \gamma-\sin \alpha \sin \gamma & -\cos \beta \cos \alpha \sin \gamma-\sin \alpha \cos \gamma & \sin \beta \cos \alpha  \label{eq:2:6:77}\\
\cos \beta \sin \alpha \cos \gamma+\cos \alpha \sin \gamma & -\cos \beta \sin \alpha \sin \gamma+\cos \alpha \cos \gamma & \sin \beta \sin \alpha \\
-\sin \beta \cos \gamma & \sin \beta \sin \gamma & \cos \beta
\end{pmatrix} .
\end{equation}
The matrix elements of $\widehat{D}^{1}(\alpha, \beta, \gamma)$ in the cartesian basis representation coincide with the elements of the rotation matrix $a_{i k}$ (see Sec.~\ref{sec:1.4.6}).

The inverse rotation of the coordinate system is performed by the matrix
\begin{equation}
\left[\widehat{D}^{1}(\alpha, \beta, \gamma)\right]^{-1}=\left[\widehat{D}^{1}(\alpha, \beta, \gamma)\right]^{+}=\widehat{D}^{1}(-\gamma,-\beta,-\alpha)=\widehat{D}^{1}(\pi-\gamma ; \beta,-\pi-\alpha) . \label{eq:2:6:78}
\end{equation}
\subsubsection{Coordinate rotations defined be an axis and angle of rotation}
Under rotation of the coordinate system through an angle $\omega$ about an axis $\vect{n}(\Theta, \Phi)$ the spin functions and spin matrices for $S=1$ are transformed by the rotation operator $\hat{U}^{1}(\omega ; \Theta, \Phi)$ which is a special case of the operator given by Eq. \eqref{eq:1:4:33}
\begin{equation}
\widehat{U}^{1}(\omega ; \Theta, \Phi)=e^{-i \omega \vect{n} \cdot \widehat{\vect{s}}} . \label{eq:2:6:79}
\end{equation}
In the spherical basis representation this operator may be written as
\begin{align}
& \hat{U}^{1}(\omega ; \Theta, \Phi)=\left(\begin{array}{l}
\frac{1}{2} \cos \omega\left(1+\cos ^{2} \Theta\right)-i \sin \omega \cos \Theta+\frac{1}{2} \sin ^{2} \Theta \\
\frac{1}{\sqrt{2}} \sin \Theta e^{i \Phi}[(\cos \omega-1) \cos \Theta-i \sin \omega] \\
\frac{1}{2}(\cos \omega-1) \sin ^{2} \Theta e^{i 2 \Phi}
\end{array}\right. \notag\\
& \left.\begin{array}{cc}
\frac{1}{\sqrt{2}} \sin \Theta e^{-i \Phi}[(\cos \omega-1) \cos \Theta-i \sin \omega] &
\frac{1}{2}(\cos \omega-1) \sin ^{2} \Theta e^{-i 2 \Phi} \\
\cos \omega \sin ^{2} \Theta+\cos ^{2} \Theta &
-\frac{1}{\sqrt{2}} \sin \Theta e^{i \Phi}[(\cos \omega-1) \cos \Theta+i \sin \omega] \\
-\frac{1}{\sqrt{2}} \sin \Theta e^{i \Phi}[(\cos \omega-1) \cos \Theta+i \sin \omega] & \frac{1}{2} \cos \omega\left(1+\cos ^{2} \Theta\right)+i \sin \omega \cos \Theta+\frac{1}{2} \sin ^{2} \Theta
\end{array}\right) . \label{eq:2:6:80}
\end{align}
In the cartesian basis representation $\hat{U}^{1}(\omega ; \Theta, \Phi)$ has the form
\begin{align}
& \hat{U}^{1}(\omega ; \Theta, \Phi)=\left(\begin{array}{l}
(1-\cos \omega) \sin ^{2} \Theta \cos ^{2} \Phi+\cos \omega \\
(1-\cos \omega) \sin ^{2} \Theta \cos \Phi \sin \Phi+\sin \omega \cos \Theta \\
(1-\cos \omega) \sin \Theta \cos \Theta \cos \Phi-\sin \omega \sin \Theta \sin \Phi
\end{array}\right.  \notag\\
& \left.\begin{array}{ll}
(1-\cos \omega) \sin ^{2} \Theta \cos \Phi \sin \Phi-\sin \omega \cos \Theta & (1-\cos \omega) \sin \Theta \cos \Theta \cos \Phi+\sin \omega \sin \Theta \sin \Phi \\
(1-\cos \omega) \sin ^{2} \Theta \sin ^{2} \Phi+\cos \omega & (1-\cos \omega) \sin \Theta \cos \Theta \sin \Phi-\sin \omega \sin \Theta \cos \Phi \\
(1-\cos \omega) \sin \Theta \cos \Theta \sin \Phi+\sin \omega \sin \Theta \cos \Phi & (1-\cos \omega) \cos ^{2} \Theta+\cos \omega
\end{array}\right) . \label{eq:2:6:81}
\end{align}
The inverse rotation of the coordinate system is performed by the matrix
\begin{equation}
\left[\hat{U}^{1}(\omega ; \Theta, \Phi)\right]^{-1}=\left[\hat{U}^{1}(\omega ; \Theta, \Phi)\right]^{+}=\hat{U}^{1}(-\omega ; \Theta, \Phi)=\hat{U}^{1}(\omega ; \pi-\Theta, \pi+\Phi) \label{eq:2:6:82}
\end{equation}
The expansion of the rotation operator $\hat{U}^{1}(\omega ; \Theta, \Phi)$ in terms of the spin matrices and the polarization operators has the form
\begin{equation}
\hat{U}^{1}(\omega ; \Theta, \Phi)=\frac{1}{3}(2 \cos \omega+1) \hat{I}-i \sin \omega(\mathrm{n} \cdot \widehat{\mathrm{~S}})+(\cos \omega-1) \sum_{i, k} n_{i} n_{k} \hat{Q}_{i k} .\label{eq:2:6:83}
\end{equation}
(See also Eqs. \eqref{eq:2:4:17} and \eqref{eq:2:4:18}.)\\
The relations between the angles $\omega, \Theta, \Phi$ and the Euler angles $\alpha, \beta, \gamma$ are given by Eqs. \eqref{eq:1:4:16}, \eqref{eq:1:4:17}. Note that
\begin{equation}
\widehat{D}^{1}(\alpha, \beta, \gamma)=\widehat{U}^{1}(\omega ; \Theta, \Phi) .\label{eq:2:6:84}
\end{equation}
\subsubsection{Effect of a rotation operator on a spin matrix}

Applying the rotation operator $\widehat{D}^{1}(\alpha, \beta, \gamma)$ to the spin matrices yields $(i, k, j, l=x, y, z)$
\begin{gather}
\widehat{S}_{i}^{\prime} \equiv \widehat{D}^{1}(\alpha, \beta, \gamma) \widehat{S}_{i}\left[\widehat{D}^{1}(\alpha, \beta, \gamma)\right]^{-1}=\sum_{k} a_{i k} \widehat{S}_{k} , \label{eq:2:6:85}\\
\widehat{Q}_{i k}^{\prime} \equiv \widehat{D}^{1}(\alpha, \beta, \gamma) \widehat{Q}_{i k}\left[\widehat{D}^{1}(\alpha, \beta, \gamma)\right]^{-1}=\sum_{j, l} a_{i j} a_{k l} \widehat{Q}_{j l} ,\label{eq:2:6:86}
\end{gather}
where $a_{i k}$ is the rotation matrix (Sec.~\ref{sec:1.4.6})
\begin{gather}
\widehat{S}_{\mu}^{\prime} \equiv \widehat{D}^{1}(\alpha, \beta, \gamma) \widehat{S}_{\mu}\left[\widehat{D}^{1}(\alpha, \beta, \gamma)\right]^{-1}=\sum_{\nu} D_{\nu \mu}^{1}(\alpha, \beta, \gamma) \widehat{S}_{\nu}, \quad(\mu, \nu= \pm 1,0) , \label{eq:2:6:87}\\
\widehat{T}_{L M}^{\prime} \equiv \widehat{D}^{1}(\alpha, \beta, \gamma) \widehat{T}_{L M}\left[\widehat{D}^{1}(\alpha, \beta, \gamma)\right]^{-1}=\sum_{M^{\prime}} D_{M^{\prime} M}^{L}(\alpha, \beta, \gamma) \widehat{T}_{L M^{\prime}} \quad\left(-L \leq M, M^{\prime} \leq L\right). \label{eq:2:6:88}
\end{gather}
$D_{M M^{\prime}}^{L}(\alpha, \beta, \gamma)$ being the Wigner $D$-functions (see Chap.~\ref{chap:4}).

\subsection{Traces of Products of Spin Matrices}\index{spin matrices!traces of}\index{trace!of spin matrices}
For cartesian components of the spin matrices $\widehat{S}_{i}$ and $\widehat{Q}_{i k}$ the following relations hold $(i, k, l$, etc. take the values $x, y, z)$.
\begin{gather}
\operatorname{Tr}\left\{\widehat{S}_{i}\right\}=0, \quad \operatorname{Tr}\left\{\widehat{Q}_{i k}\right\}=0,  \notag\\
\operatorname{Tr}\left\{\widehat{S}_{i} \widehat{S}_{k}\right\}=2 \delta_{i k}, \quad \operatorname{Tr}\left\{\widehat{S}_{i} \widehat{Q}_{k l}\right\}=0, \quad \operatorname{Tr}\left\{\widehat{Q}_{i k} \widehat{S}_{l}\right\}=0,  \notag\\
\operatorname{Tr}\left\{\widehat{S}_{i} \widehat{S}_{k} \widehat{S}_{l}\right\}=i \varepsilon_{i k l},  \label{eq:2:6:89}\\
\operatorname{Tr}\left\{\widehat{Q}_{i k} \widehat{Q}_{l m}\right\}=\frac{1}{2}\left(\delta_{i l} \delta_{k m}+\delta_{i m} \delta_{k l}-\frac{2}{3} \delta_{i k} \delta_{l m}\right), \operatorname{Tr}\left\{\widehat{S}_{i} \widehat{S}_{k} \widehat{S}_{l} \widehat{S}_{m}\right\}=\delta_{i k} \delta_{l m}+\delta_{i m} \delta_{k l}. \notag
\end{gather}
For spherical components of the spin operator $\widehat{S}_{\mu}$ we have ( $\mu, \nu, \lambda$, etc. take the values $\pm 1,0$ )
\begin{gather}
\operatorname{Tr}\left\{\widehat{S}_{\mu}\right\}=0 , \notag\\
\operatorname{Tr}\left\{\widehat{S}_{\mu} \widehat{S}_{\nu}\right\}=(-1)^{\mu} 2 \delta_{\mu-\nu} , \notag\\
\operatorname{Tr}\left\{\widehat{S}_{\mu} \widehat{S}_{\nu} \widehat{S}_{\lambda}\right\}=-\sqrt{6}\begin{pmatrix}
1 & 1 & 1 \\
\mu & \nu & \lambda
\end{pmatrix}=(-1)^{1+\lambda} \sqrt{2} \clebsch{1}{\mu}{1}{\nu}{1}{-\lambda},  \label{eq:2:6:90}\\
\operatorname{Tr}\left\{\widehat{S}_{\mu} \widehat{S}_{\nu} \widehat{S}_{\lambda} \widehat{S}_{\rho}\right\}=(-1)^{\mu+\lambda}\left(\delta_{\mu-\nu} \delta_{\lambda-\rho}+\delta_{\mu-\rho} \delta_{\nu-\lambda}\right),  \notag\\
\operatorname{Tr}\left\{\widehat{S}_{\mu_{1}} \widehat{S}_{\mu_{2}} \ldots \widehat{S}_{\mu_{n}}\right\}=0 \quad \text { if } \mu_{1}+\mu_{2}+\ldots+\mu_{n} \neq 0. \label{eq:2:6:91}
\end{gather}
Traces of products of the polarization operators $\widehat{T}_{L M}(L=0,1,2 ;-L \leq M \leq L)$ can be evaluated by using Eqs. \eqref{eq:2:4:21}-\eqref{eq:2:4:24} at $S=1$.

\chapter{Irreducible tensors}\label{chap:3}
\section{Definition and properties of irreducible tensors}\label{sec:3.1}\index{irreducible tensor}\index{tensor!irreducible}
\subsection{Definition}\label{sec:3.1.1}
Irreducible tensors occupy a central position in angular momentum theory. Under rotations of coordinate systems these tensors transform in the same manner as eigenfunctions of the angular momentum operator. The use of this property permits us to develop very effective methods for calculating matrix elements of different quantum-mechanical operators.

An irreducible tensor $\mathfrak{M}_{J}$ of rank $J$ (with $J$ integer or half-integer) is defined as a set of $2 J+1$ functions (components) $\mathfrak{M}_{J M}$ (where $M=-J,-J+1, \ldots, J-1, J$) which satisfy the following commutation rules with spherical components of the angular momentum operator\isym{Mfrak@$\mathfrak{M}_{J}$, irreducible tensor}\isym{MfrakJM@$\mathfrak{M}_{J M}$, components of irreducible tensor}\index{rank!of an irreducible tensor}\index{commutation relations!with angular momentum operator}
\begin{gather}
{\left[\hat{J}_{ \pm 1}, \mathfrak{M}_{J M}\right]=\mp \frac{1}{\sqrt{2}} e^{ \pm i \delta} \sqrt{J(J+1)-M(M \pm 1)} \mathfrak{M}_{J M \pm 1}}  ,\label{eq:3:1:1}\\
{\left[\hat{J}_{0}, \mathfrak{M}_{J M}\right]=M \mathfrak{M}_{J M}}. \notag
\end{gather}
In compact form
\begin{equation}
\left[\widehat{J}_{\mu}, \mathfrak{M}_{J M}\right]=e^{i \mu \delta} \sqrt{J(J+1)} \clebsch{J}{M}{1}{\mu}{J}{M+\mu} \mathfrak{M}_{J M+\mu} .\label{eq:3:1:2}
\end{equation}
From these relations it follows that
\begin{equation}
\left[\hat{\vect{J}}^{2}, \mathfrak{M}_{J M}\right]=J(J+1) \mathfrak{M}_{J M}. \label{eq:3:1:3}
\end{equation}
The quantity $\delta$ in Eq. \eqref{eq:3:1:1} which determines relative phases of different $\mathfrak{M}_{J M}$ components is arbitrary.\index{phase!of irreducible tensor}\isymg{delta@$\delta$, phase of irreducible tensor} Let us adopt $\delta=0$, i.e., $e^{ \pm i \delta}=1$, and choose the positive sign of the square root. The linear equations \eqref{eq:3:1:1} define the components of the irreducible tensor $\mathfrak{M}_{J M}$ within an arbitrary scalar factor, which is the same for all the components. This factor can be a real or complex number, function or operator. In the case of integer rank $J$ the overall phase of the $\mathfrak{M}_{J M}$ components is usually defined in such manner that
\begin{equation}
\left(\mathfrak{M}_{J M}\right)^{*}=(-1)^{-M} \mathfrak{M}_{J-M} . \label{eq:3:1:4}
\end{equation}
This choice of the phase coincides with that for spherical harmonics (Chap. \ref{chap:5}). However, sometimes in quantum-mechanical applications it is convenient to redefine irreducible tensors as\index{irreducible tensor!modified phase convention}\isym{Mfrak-tilde@$\tilde{\mathfrak{M}}_{J}$, modified irreducible tensor}
\begin{equation}
\tilde{\mathfrak{M}}_{J}=i^{J} \mathfrak{M}_{J} . \label{eq:3:1:5}
\end{equation}
Then one has
\begin{equation}
\left(\tilde{\mathfrak{M}}_{J M}\right)^{*}=(-1)^{J-M} \tilde{\mathfrak{M}}_{J-M} . \label{eq:3:1:6}
\end{equation}
The choice of the phase \eqref{eq:3:1:6} can be used for tensors of integer as well as half-integer rank $J$. Making use of this phase convention for tensor operators as well as for wavefunctions describing initial $|a\rangle$ and final states $|b\rangle$ we get the following relations for matrix elements of Hermitian $\tilde{\mathfrak{M}}_{J}^{\dagger}=\tilde{\mathfrak{M}}_{J}$ operators
\begin{equation}
\braOket{b}{\tilde{\mathfrak{M}}_{J M}}{a}=\left(\braOket{a}{\left(\tilde{\mathfrak{M}}_{J M}\right)^{*}}{b}\right)^{*} . \label{eq:3:1:7}
\end{equation}
\subsection{Covariant and Contravariant Components}\label{sec:3.1.2}\index{covariant component}\index{contravariant component}
Any irreducible tensor $\mathfrak{M}_{J}$ of rank $J$ can be expanded in a series based on a complete set of the unit orthonormalized irreducible tensors $\vect{e}_{J M}$ of rank $J$\index{unit irreducible tensor}\isym{e-vec@$\vect{e}_{J M}$, unit irreducible tensor}
\begin{equation}
\vect{e}_{J}^{M} \cdot \vect{e}_{J^{\prime} M^{\prime}}=\delta_{J J^{\prime}} \delta_{M M^{\prime}}. \label{eq:3:1:8}
\end{equation}
The tensors $\vect{e}_{J M}$ can be composed, for example, of the basis spin functions. The expansion of $\mathfrak{M}_{J}$ is written as
\begin{equation}
\mathfrak{M}_{J}=\sum_{M} \mathrm{e}_{J}^{M} \cdot \mathfrak{M}_{J M}=\sum_{M} \mathrm{e}_{J M} \cdot \mathfrak{M}_{J}^{M} . \label{eq:3:1:9}
\end{equation}
$\mathfrak{M}_{J M}$ represents the covariant component of the tensor $\mathfrak{M}_{J}$ and $\mathfrak{M}_{J}^{M}$ denotes the contravariant component.\isym{MfrakM@$\mathfrak{M}_{J}^{M}$, contravariant component of irreducible tensor} These components are related by
\begin{gather}
\mathfrak{M}_{J}^{M}=\left(\mathfrak{M}_{J M}\right)^{*}=(-1)^{-M} \mathfrak{M}_{J-M},  \notag\\
\tilde{\mathfrak{M}}_{J}^{M}=\left(\tilde{\mathfrak{M}}_{J M}\right)^{*}=(-1)^{J-M} \tilde{\mathfrak{M}}_{J-M} . \label{eq:3:1:10}
\end{gather}
\subsection{Transformation of Irreducible Tensors Under a Rotation of the Coordinate System}\label{sec:3.1.3}\index{irreducible tensor!transformation under rotation}\index{rotation of coordinate system}
Under rotations of the coordinate system described by the Euler angles $\alpha, \beta, \gamma$, the components of irreducible tensors $\mathfrak{M}_{J M}$ and $\tilde{\mathfrak{M}}_{J M}$ undergo linear transformation. The coefficients of such transformation are the Wigner $D$-functions (Chap.~\ref{chap:4})\index{Wigner D-function@Wigner $D$-function}\isym{DJMM@$D_{M M^{\prime}}^{J}$, Wigner $D$-function}
\begin{align}
& \mathfrak{M}_{J M^{\prime}}\left(X^{\prime}\right)=\hat{D}(\alpha, \beta, \gamma) \mathfrak{M}_{J M^{\prime}}(X)[\hat{D}(\alpha, \beta, \gamma)]^{-1}=\sum_{M} \mathfrak{M}_{J M}(X) D_{M M^{\prime}}^{J}(\alpha, \beta, \gamma),  \notag\\
& \tilde{\mathfrak{M}}_{J M^{\prime}}\left(X^{\prime}\right)=\hat{D}(\alpha, \beta, \gamma) \tilde{\mathfrak{M}}_{J M^{\prime}}(X)[\hat{D}(\alpha, \beta, \gamma)]^{-1}=\sum_{M} \tilde{\mathfrak{M}}_{J M}(X) D_{M M^{\prime}}^{J}(\alpha, \beta, \gamma) . \label{eq:3:1:11}
\end{align}
Here $X$ and $X^{\prime}$ denote sets of all arguments of the tensor in the initial and final coordinate systems, respectively.

\subsection{Transformation of Irreducible Tensors Under Inversion of the Coordinate System}\label{sec:3.1.4}\index{irreducible tensor!transformation under inversion}\index{coordinate inversion}
The transformation properties of irreducible tensor components under inversion of the coordinate system, $\vect{r} \rightarrow-\vect{r}$, permit us generally to represent an irreducible tensor $\mathfrak{M}_{J}$ of integer rank $J$ as the sum of two tensors $\mathfrak{M}_{J}^{(+1)}$ and $\mathfrak{M}_{J}^{(-1)}$, i.e.,
\begin{equation}
\mathfrak{M}_{J}=\mathfrak{M}_{J}^{(+1)}+\mathfrak{M}_{J}^{(-1)}. \label{eq:3:1:12}
\end{equation}
Each of these tensors has definite parity.\\
Under inversion of the coordinate system the tensors $\mathfrak{M}_{J}^{( \pm 1)}$ transform in the following way
\begin{equation}
\widehat{P}_{r} \mathfrak{M}_{J}^{\left(\pi_{J}\right)} \widehat{P}_{r}^{-1}=\pi_{J} \mathfrak{M}_{J}^{\left(\pi_{J}\right)}, \quad\left(\pi_{J}= \pm 1\right) . \label{eq:3:1:13}
\end{equation}
The tensor $\mathfrak{M}_{J}^{\left(\pi_{J}\right)}$ with parity $\pi_{J}=(-1)^{J}$ is called the true (or polar) irreducible tensor of rank $J$.\index{polar tensor}\index{true tensor}\index{parity!of irreducible tensor}\isymg{pi-J@$\pi_{J}$, parity of irreducible tensor} Tensor $\mathfrak{M}_{J}^{\left(\pi_{J}\right)}$ with parity $\pi_{J}=(-1)^{J+1}$ is the pseudotensor (or axial tensor) of rank $J$.\index{pseudotensor}\index{axial tensor}\index{tensor!axial} The relations \eqref{eq:3:1:12} and \eqref{eq:3:1:13} are valid for the tensors $\tilde{\mathfrak{M}}_{J}$ as well.

\subsection{Double Tensors}\label{sec:3.1.5}\index{double tensor}\index{tensor!double}
A double tensor $W_{J_{1} J_{2}}(1,2)$ of ranks $J_{1}$ and $J_{2}$ has\isym{W-double@$W_{J_{1} J_{2}}(1,2)$, double tensor} $\left(2 J_{1}+1\right)\left(2 J_{2}+1\right)$ components and depends on variables of two different subsystems, 1 and 2 . The components of the double tensor satisfy the following commutation relations
\begin{align}
\left[\widehat{J}_{ \pm 1}(1), W_{J_{1} M_{1} J_{2} M_{2}}(1,2)\right]&=\mp \frac{1}{\sqrt{2}} \sqrt{J_{1}\left(J_{1}+1\right)-M_{1}\left(M_{1} \pm 1\right)} W_{J_{1} M_{1} \pm 1 J_{2} M_{2}}(1,2)  ,\label{eq:3:1:14}\\
\left[\widehat{J}_{0}(1), W_{J_{1} M_{1} J_{2} M_{2}}(1,2)\right]&=M_{1} W_{J_{1} M_{1} J_{2} M_{2}}(1,2)  ,\label{eq:3:1:15}\\
\left[\widehat{J}_{ \pm 1}(2), W_{J_{1} M_{1} J_{2} M_{2}}(1,2)\right]&=\mp \frac{1}{\sqrt{2}} \sqrt{J_{2}\left(J_{2}+1\right)-M_{2}\left(M_{2} \pm 1\right)} W_{J_{1} M_{1} J_{2} M_{2} \pm 1}(1,2) , \label{eq:3:1:16}\\
\left[\widehat{J}_{0}(2), W_{J_{1} M_{1} J_{2} M_{2}}(1,2)\right]&=M_{2} W_{J_{1} M_{1} J_{2} M_{2}}(1,2). \label{eq:3:1:17}
\end{align}
Here $\widehat{\vect{J}}(1)$ and $\widehat{\vect{J}}(2)$ are the operators of the total angular momenta of subsystems 1 and 2 , respectively.\\
Under rotation of subsystem 1 and 2 the double tensor $W_{J_{1} J_{2}}(1,2)$ transforms according to the representation $D^{J_{1}}$ or $D^{J_{2}}$, respectively.
\begin{align}
& W_{J_{1} M_{1}^{\prime} J_{2} M_{2}}\left(1^{\prime}, 2\right)=\sum_{M_{1}} W_{J_{1} M_{1} J_{2} M_{2}}(1,2) D_{M_{1} M_{1}^{\prime}}^{J_{1}}\left(\alpha_{1}, \beta_{1}, \gamma_{1}\right) , \label{eq:3:1:18}\\
& W_{J_{1} M_{1} J_{2} M_{2}^{\prime}}\left(1,2^{\prime}\right)=\sum_{M_{2}} W_{J_{1} M_{1} J_{2} M_{2}}(1,2) D_{M_{2} M_{2}^{\prime}}^{J_{2}}\left(\alpha_{2}, \beta_{2}, \gamma_{2}\right) .\label{eq:3:1:19}
\end{align}
\subsection{Examples of Irreducible Tensors}\label{sec:3.1.6}
In this section we present some examples of irreducible tensors considered in this book.
\begin{enumerate}[label=(\alph*)]
\item The operators of the angular momenta $\widehat{\vect{J}}, \widehat{\vect{L}}, \widehat{\vect{S}}$ (Chap.~\ref{chap:2}) are irreducible tensor operators of rank 1.
\item The polarization operator $\widehat{T}_{L M}(S)$ (Sec.~\ref{sec:2.4}) is an irreducible tensor operator of rank $L$.
\item Spherical harmonics $Y_{l m}(\vartheta, \varphi)$ (Chap.~\ref{chap:5}) are irreducible tensors of rank $l$.
\item The spin wave functions of a particle of spin $S$ (Chap.~\ref{chap:6}) are irreducible tensors of rank $S$.
\item Tensor spherical harmonics $Y_{J M}^{L S}(\vartheta, \varphi)$ (Chap.~\ref{chap:7}) are irreducible tensors of rank $J$.
\end{enumerate}
We may apply the results given in this chapter for all the tensors mentioned above.

\subsection{Direct and Irreducible Tensor Products. Commutators of Tensor Products}\label{sec:3.1.7}\index{irreducible tensor product}\index{tensor product!irreducible}\index{direct product}\index{tensor product!direct}
An irreducible tensor product $\mathfrak{L}_{J}$ of two irreducible tensors $\mathfrak{M}_{J_{1}}$ and $\mathfrak{N}_{J_{2}}$ of ranks $J_{1}$ and $J_{2}$ is defined as the tensor of rank $J$ whose components $\mathfrak{L}_{J M}$ can be expressed in terms of $\mathfrak{M}_{J_{1} M_{1}}$ and $\mathfrak{N}_{J_{2} M_{2}}$ according to
\begin{equation}
\mathfrak{L}_{J M}=\sum_{M_{1} M_{2}} \clebsch{J_{1}}{M_{1}}{J_{2}}{M_{2}}{J}{M} \mathfrak{M}_{J_{1} M_{1}} \mathfrak{N}_{J_{2} M_{2}} \label{eq:3:1:20}
\end{equation}
Irreducible tensor product is denoted as\isym{Lfrak@$\mathfrak{L}_{J}=\{\mathfrak{M}_{J_1}\otimes\mathfrak{N}_{J_2}\}_{J}$, irreducible tensor product}
\begin{equation}
\mathfrak{L}_{J} \equiv\left\{\mathfrak{M}_{J_{1}} \otimes \mathfrak{N}_{J_{2}}\right\}_{J} . \label{eq:3:1:21}
\end{equation}
The direct product of two irreducible tensors $\mathfrak{M}_{J_{1}}$ and $\mathfrak{N}_{J_{2}}$ is defined as a set of $\left(2 J_{1}+1\right)\left(2 J_{2}+1\right)$ components $\mathfrak{M}_{J_{1} M_{1}} \mathfrak{N}_{J_{2} M_{2}}$ This tensor is generally reducible and can be decomposed into parts which themselves transform\\
independently under a rotation of the coordinate system. In other words, the direct product can be represented as a sum of irreducible tensors $\mathfrak{L}_{J M}$
\begin{equation}
\mathfrak{M}_{J_{1} M_{1}} \mathfrak{N}_{J_{2} M_{2}}=\sum_{J=\left|J_{1}-J_{2}\right|}^{J_{1}+J_{2}} \clebsch{J_{1}}{M_{1}}{J_{2}}{M_{2}}{J}{M} \mathfrak{L}_{J M} .\label{eq:3:1:22}
\end{equation}
It is essential that the irreducible tensor product $\tilde{\mathfrak{L}}_{J M}$ satisfies the same equality \eqref{eq:3:1:6} as $\tilde{\mathfrak{M}}_{J_{1} M_{1}}$ and $\tilde{\mathfrak{N}}_{J_{2} M_{2}}$
\begin{equation}
\left(\tilde{\mathfrak{L}}_{J M}\right)^{*}=(-1)^{J-M} \tilde{\mathfrak{L}}_{J-M}\left(\tilde{\mathfrak{L}}_{J M}=\left\{\tilde{\mathfrak{M}}_{J_{1}} \otimes \tilde{\mathfrak{N}}_{J_{2}}\right\}_{J M}\right) \label{eq:3:1:23}
\end{equation}
This property is specific for tensors of the $\tilde{\mathfrak{M}}_{J M}$-type. The tensor product $\mathfrak{L}_{J M}$ does not satisfy the relation \eqref{eq:3:1:4} although $\mathfrak{M}_{J_{1} M_{1}}$ and $\mathfrak{N}_{J_{2} M_{2}}$ satisfy this relation.

Let us introduce some definitions and notations which will be widely used in the consideration of products of non-commuting tensor operators.

The commutator of components of two irreducible tensors is defined by\index{commutator!of irreducible tensors}\isym{Rfrak@$\mathfrak{R}_{J_{1} M_{1} J_{2} M_{2}}$, commutator of irreducible tensors}
\begin{equation}
\mathfrak{R}_{J_{1} M_{1} J_{2} M_{2}} \equiv\left[\mathfrak{M}_{J_{1} M_{1}}, \mathfrak{N}_{J_{2} M_{2}}\right] \equiv \mathfrak{M}_{J_{1} M_{1}} \mathfrak{N}_{J_{2} M_{2}}-\mathfrak{N}_{J_{2} M_{2}} \mathfrak{M}_{J_{1} M_{1}} .\label{eq:3:1:24}
\end{equation}
The commutator of an irreducible tensor product is written as
\begin{equation}
\mathfrak{R}_{J M}^{J_{1} J_{2}} \equiv\left\{\mathfrak{M}_{J_{1}} \otimes \mathfrak{N}_{J_{2}}\right\}_{J M}-(-1)^{J_{1}+J_{2}-J}\left\{\mathfrak{N}_{J_{2}} \otimes \mathfrak{M}_{J_{1}}\right\}_{J M} . \label{eq:3:1:25}
\end{equation}
The functions $\mathfrak{R}_{J M}^{J_{1} J_{2}}$ are the components of some irreducible tensor. They may be expressed in the form
\begin{equation}
\mathfrak{R}_{J M}^{J_{1} J_{2}}=\sum_{M_{1} M_{2}} \clebsch{J_{1}}{M_{1}}{J_{2}}{M_{2}}{J}{M} \mathfrak{R}_{J_{1} M_{1} J_{2} M_{2}} \label{eq:3:1:26}
\end{equation}
For commuting tensors we get
\begin{equation}
\left\{\mathfrak{M}_{J_{1}} \otimes \mathfrak{N}_{J_{2}}\right\}_{J M}=(-1)^{J_{1}+J_{2}-J}\left\{\mathfrak{N}_{J_{2}} \otimes \mathfrak{M}_{J_{1}}\right\}_{J M} .\label{eq:3:1:27}
\end{equation}
On the other hand, for non-commuting tensors we have
\begin{equation}
\left\{\mathfrak{M}_{J_{1}} \otimes \mathfrak{N}_{J_{2}}\right\}_{J M}=(-1)^{J_{1}+J_{2}-J}\left\{\mathfrak{N}_{J_{2}} \otimes \mathfrak{M}_{J_{1}}\right\}_{J M}+\mathfrak{R}_{J M}^{J_{1} J_{2}} .\label{eq:3:1:28}
\end{equation}
In particular, from these equations one can see that an irreducible (rank-I) tensor product of two identical tensors $\mathfrak{M}_{J}$ is equal to zero, if $I=2 J-1,2 J-3, \ldots$ and the tensor components commute
\begin{equation}
\left\{\mathfrak{M}_{J} \otimes \mathfrak{M}_{J}\right\}_{I}=0 \label{eq:3:1:29}
\end{equation}
\subsection{Scalar Products of Irreducible Tensors}\label{sec:3.1.8}\index{scalar product!of irreducible tensors}
The scalar product of two irreducible tensors $\mathfrak{M}_{J}$ and $\mathfrak{N}_{J}$ of the same rank is defined as
\begin{equation}
\left(\mathfrak{M}_{J} \cdot \mathfrak{N}_{J}\right)=\sum_{M}(-1)^{-M} \mathfrak{M}_{J M} \mathfrak{N}_{J-M}=\sum_{M} \mathfrak{M}_{J M} \mathfrak{N}_{J M}^{*}=\sum_{M} \mathfrak{M}_{J M} \mathfrak{N}_{J}^{M} .\label{eq:3:1:30}
\end{equation}
Similarly, the scalar product of two irreducible tensors $\tilde{\mathfrak{N}}_{J}$ and $\tilde{\mathfrak{R}}_{J}$ is given by
\begin{equation*}
\left(\tilde{\mathfrak{M}}_{J} \cdot \tilde{\mathfrak{N}}_{J}\right)=\sum_{M}(-1)^{J-M} \tilde{\mathfrak{M}}_{J M} \tilde{\mathfrak{N}}_{J-M}=\sum_{M} \tilde{\mathfrak{M}}_{J M} \tilde{\mathfrak{N}}_{J M}^{*}=\sum_{M} \tilde{\mathfrak{M}}_{J M} \tilde{\mathfrak{N}}_{J}^{M} . 
\end{equation*}
Note that
\begin{equation}
\left(\mathfrak{M}_{J} \cdot \mathfrak{N}_{J}\right)=\left(\tilde{\mathfrak{M}}_{J} \cdot \tilde{\mathfrak{N}}_{J}\right) . \label{eq:3:1:31}
\end{equation}
Scalar products differ from irreducible tensor products of zero rank only by some numerical factor. The latter tensor products read
\begin{align}
& \left\{\mathfrak{M}_{J} \otimes \mathfrak{N}_{J}\right\}_{00}=\sum_{M_{1} M_{2}} \clebsch{J}{M_{1}}{J}{M_{2}}{0}{0} \mathfrak{M}_{J M_{1}} \mathfrak{N}_{J M_{2}}=\frac{1}{\sqrt{2 J+1}} \sum_{M}(-1)^{J-M} \mathfrak{M}_{J M} \mathfrak{N}_{J-M},  \label{eq:3:1:32}\\
& \left\{\tilde{\mathfrak{M}}{ }_{J} \otimes \tilde{\mathfrak{N}}_{J}\right\}_{00}=\sum_{M_{1} M_{2}} \clebsch{J}{M_{1}}{J}{M_{2}}{0}{0} \tilde{\mathfrak{M}}_{J M_{1}} \tilde{\mathfrak{N}}_{J M_{2}}=\frac{1}{\sqrt{2 J+1}} \sum_{M}(-1)^{J-M} \tilde{\mathfrak{M}}_{J M} \tilde{\mathfrak{N}}_{J-M} . \label{eq:3:1:33}
\end{align}
Hence, the relations between scalar products and tensor products of zero rank are determined by
\begin{gather}
\left(\mathfrak{M}_{J} \cdot \mathfrak{N}_{J}\right)=(-1)^{-J} \sqrt{2 J+1}\left\{\mathfrak{M}_{J} \otimes \mathfrak{N}_{J}\right\}_{00},  \notag\\
\left(\tilde{\mathfrak{M}}_{J} \cdot \tilde{\mathfrak{N}}_{J}\right)=\sqrt{2 J+1}\left\{\tilde{\mathfrak{M}}_{J} \otimes \tilde{\mathfrak{N}}_{J}\right\}_{00}. \label{eq:3:1:34}
\end{gather}
For the scalar product of double tensors we have
\begin{equation}
\left(\widetilde{W}_{J_{1} J_{2}}(1,2) \cdot \widetilde{U}_{J_{1} J_{2}}(1,2)\right)=\sum_{M_{1} M_{2}}(-1)^{J_{1}-M_{1}+J_{2}-M_{2}} \widetilde{W}_{J_{1} M_{1} J_{2} M_{2}}(1,2) \widetilde{U}_{J_{1}-M_{1} J_{2}-M_{2}}(1,2) . \label{eq:3:1:35}
\end{equation}
\section{Relation between the irreducible tensor algebra and vector and tensor theory}\label{sec:3.2}
\subsection{Vectors and Irreducible Tensors}\label{sec:3.2.1}\index{vector!as irreducible tensor}
An arbitrary vector $\vect{A}$ is an irreducible tensor of rank one. Its spherical components may be treated as components of an irreducible tensor $\vect{A}_{1}$, of rank one, for which\index{spherical components!of a vector}\isym{A1-vec@$\vect{A}_{1}$, vector as rank-one irreducible tensor} $A_{1}^{\mu}=(-1)^{\mu} A_{1-\mu}$.
\begin{equation}
A_{1 \mu} \equiv A_{\mu}, \quad A^{1 \mu} \equiv A^{\mu}. \label{eq:3:2:1}
\end{equation}
A polar vector is a true tensor of rank one.\index{polar vector}\index{vector!polar} Under inversion of the coordinate system its components change their sign. An axial vector is a pseudotensor of rank one.\index{axial vector}\index{vector!axial} Under coordinate inversion its components remain unchanged.

Using two vectors $\vect{A}=\vect{A}_{1}$ and $\vect{B}=\vect{B}_{1}$ one can construct three irreducible tensor products of ranks $0,1,2$,
\[
\left\{\vect{A}_{1} \otimes \vect{B}_{1}\right\}_{00}, \quad\left\{\vect{A}_{1} \otimes \vect{B}_{1}\right\}_{1 \mu}, \quad\left\{\vect{A}_{1} \otimes \vect{B}_{1}\right\}_{2 \mu} .
\]
If one of the vectors $\vect{A}$ or $\vect{B}$ is polar and the other axial, then $\left\{\vect{A}_{1} \otimes \vect{B}_{1}\right\}_{00}$ is a pseudoscalar, $\left\{\vect{A}_{1} \otimes \vect{B}_{1}\right\}_{1 \mu}$ is a polar vector, $\left\{\vect{A}_{1} \otimes \vect{B}_{1}\right\}_{2 \mu}$ is a pseudotensor of rank two. If both vectors $\vect{A}$ and $\vect{B}$ are either polar or axial, then $\left\{\vect{A}_{1} \otimes \vect{B}_{1}\right\}_{00}$ is a scalar, $\left\{\vect{A}_{1} \otimes \vect{B}_{1}\right\}_{1 \mu}$ is an axial vector and $\left\{\vect{A}_{1} \otimes \vect{B}_{1}\right\}_{2 \mu}$ is a true tensor of rank two.\\
\subsubsection{Rank zero tensor and scalar product}\index{scalar product!of vectors}
 The irreducible tensor product of rank zero differs from the scalar product of vectors $\vect{A}$ and $\vect{B}$ by a numerical factor:
\begin{equation}
\left\{\vect{A}_{1} \otimes \vect{B}_{1}\right\}_{00}=-\frac{1}{\sqrt{3}}(\vect{A} \cdot \vect{B}) \label{eq:3:2:2}
\end{equation}
The scalar product of irreducible tensors $\vect{A}_{1}$ and $\vect{B}_{1}$ introduced in Sec.~\ref{sec:3.1.8} coincides with the scalar product of vectors,
\begin{equation}
\left(\vect{A}_{1} \cdot \vect{B}_{1}\right)=(\vect{A} \cdot \vect{B}) . \label{eq:3:2:3}
\end{equation}
\subsubsection{Rank one tensor and vector product}\index{vector product}
 The irreducible tensor product of rank one is related to the vector product of vectors $\vect{A}$ and $\vect{B}$ by
\begin{equation}
\left\{\vect{A}_{1} \otimes \vect{B}_{1}\right\}_{1}=\frac{i}{\sqrt{2}}[\vect{A} \times \vect{B}] .\label{eq:3:2:4}
\end{equation}
The components of the tensor product of rank one can be expressed in terms of products of spherical components of vectors $\vect{A}$ and $\vect{B}$ as
\begin{equation}
\left\{\vect{A}_{1} \otimes \vect{B}_{1}\right\}_{1 M}=\frac{i}{\sqrt{2}}[\vect{A} \times \vect{B}]_{M}=\sum_{\mu \nu} \clebsch{1}{\mu}{1}{\nu}{1}{M} A_{\mu} B_{\nu} .\label{eq:3:2:5}
\end{equation}
\subsubsection{Rank two tensor in terms of vector components}
The components of a tensor product of rank two are also related to products of spherical components of $\vect{A}$ and $\vect{B}$ :
\begin{equation}
\left\{\vect{A}_{1} \otimes \vect{B}_{1}\right\}_{2 M}=\sum_{\mu, \nu} \clebsch{1}{\mu}{1}{\nu}{2}{M} A_{\mu} B_{\nu}=\sqrt{\frac{3|M|-2}{14|M|-12}} \sum_{\substack{\mu+\nu=M \\ \mu \geq \nu}}\left(A_{\mu} B_{\nu}+A_{\nu} B_{\mu}\right) .\label{eq:3:2:6}
\end{equation}
In a more detailed form Eq. \eqref{eq:3:2:6} may be written as
\begin{align}
\left\{\vect{A}_{1} \otimes \vect{B}_{1}\right\}_{2+2}&=A_{+1} B_{+1},  \notag\\
\left\{\vect{A}_{1} \otimes \vect{B}_{1}\right\}_{2+1}&=\frac{1}{\sqrt{2}}\left(A_{+1} B_{0}+A_{0} B_{+1}\right),  \notag\\
\left\{\vect{A}_{1} \otimes \vect{B}_{1}\right\}_{20}&=\frac{1}{\sqrt{6}}\left(A_{+1} B_{-1}+2 A_{0} B_{0}+A_{-1} B_{+1}\right),  \label{eq:3:2:7}\\
\left\{\vect{A}_{1} \otimes \vect{B}_{1}\right\}_{2-1}&=\frac{1}{\sqrt{2}}\left(A_{-1} B_{0}+A_{0} B_{-1}\right),  \notag\\
\left\{\vect{A}_{1} \otimes \vect{B}_{1}\right\}_{2-2}&=A_{-1} B_{-1} . \notag
\end{align}
Given three commuting vectors $\vect{A}_{1}, \vect{B}_{1}, \vect{C}_{1}$ we can compose the following irreducible tensor products of ranks 0 and 1:
\begin{align}
\left\{\left\{\vect{A}_{1} \otimes \vect{B}_{1}\right\}_{0} \otimes \vect{C}_{1}\right\}_{1}&=-\frac{1}{\sqrt{3}}(\vect{A} \cdot \vect{B}) \cdot \vect{C},  \label{eq:3:2:8}\\
\left\{\left\{\vect{A}_{1} \otimes \vect{B}_{1}\right\}_{1} \otimes \vect{C}_{1}\right\}_{0}&=-\frac{i}{\sqrt{6}}[\vect{A} \times \vect{B}] \cdot \vect{C},  \label{eq:3:2:9}\\
\left\{\left\{\vect{A}_{1} \otimes \vect{B}_{1}\right\}_{1} \otimes \vect{C}_{1}\right\}_{1}&=-\frac{1}{2}\left[[ \vect{A} \times \vect{B}] \times \vect{C}\right]=\frac{1}{2} \vect{A}(\vect{B} \cdot \vect{C})-\frac{1}{2} \vect{B}(\vect{A} \cdot \vect{C}),  \label{eq:3:2:10}\\
\left\{\left\{\vect{A}_{1} \otimes \vect{B}_{1}\right\}_{2} \otimes \vect{C}_{1}\right\}_{1}&=\sqrt{\frac{3}{5}}\left\{\frac{1}{3} \vect{C}(\vect{A} \cdot \vect{B})-\frac{1}{2} \vect{B}(\vect{A} \cdot \vect{C})-\frac{1}{2} \vect{A}(\vect{B} \cdot \vect{C})\right\} . \label{eq:3:2:11}
\end{align}
In addition to the foregoing formulas, there exist products which differ from Eqs. \eqref{eq:3:2:8}-\eqref{eq:3:2:11} by vector coupling schemes. The problems concerned with recoupling in tensor products are considered below (Sec.~\ref{sec:3.3}).

Given four commuting vectors $\vect{A}_{1}, \vect{B}_{1}, \vect{C}_{1}$ and $\vect{D}_{1}$ we can compose the following irreducible tensor products of ranks 0 and 1 :
\begin{align}
\left\{\left\{\vect{A}_{1} \otimes \vect{B}_{1}\right\}_{0} \otimes\left\{\vect{C}_{1} \otimes \vect{D}_{1}\right\}_{0}\right\}_{0}&=\frac{1}{3}(\vect{A} \cdot \vect{B})(\vect{D} \cdot \vect{C}),  \label{eq:3:2:12}\\
\left\{\left\{\vect{A}_{1} \otimes \vect{B}_{1}\right\}_{1} \otimes\left\{\vect{C}_{1} \otimes \vect{D}_{1}\right\}_{0}\right\}_{1}&=-\frac{i}{\sqrt{6}}[\vect{A} \times \vect{B}](\vect{C} \cdot \vect{D}),  \label{eq:3:2:13}\\
\left\{\left\{\vect{A}_{1} \otimes \vect{B}_{1}\right\}_{0} \otimes\left\{\vect{C}_{1} \otimes \vect{D}_{1}\right\}_{1}\right\}_{1}&=-\frac{i}{\sqrt{6}}(\vect{A} \cdot \vect{B})[\vect{C} \times \vect{D}],  \label{eq:3:2:14}\\
\left\{\left\{\vect{A}_{1} \otimes \vect{B}_{1}\right\}_{1} \otimes\left\{\vect{C}_{1} \otimes \vect{D}_{1}\right\}_{1}\right\}_{0}&=\frac{1}{2 \sqrt{3}}\{(\vect{A} \cdot \vect{C})(\vect{B} \cdot \vect{D})-(\vect{A} \cdot \vect{D})(\vect{B} \cdot \vect{C})\}, \label{eq:3:2:15}
\end{align}
\begin{align}
\left\{\left\{\vect{A}_{1} \otimes \vect{B}_{1}\right\}_{1} \otimes\left\{\vect{C}_{1} \otimes \vect{D}_{1}\right\}_{1}\right\}_{1}&=-\frac{i}{2 \sqrt{2}}\{\vect{C}(\vect{D} \cdot[\vect{A} \times \vect{B}])-\vect{D}(\vect{C} \cdot[\vect{A} \times \vect{B}])\}  \notag\\
&=-\frac{i}{2 \sqrt{2}}\{\vect{B}(\vect{A} \cdot[\vect{C} \times \vect{D}])-\vect{A}(\vect{B} \cdot[\vect{C} \times \vect{D}])\},  \label{eq:3:2:16}\\
\left\{\left\{\vect{A}_{1} \otimes \vect{B}_{1}\right\}_{2} \otimes\left\{\vect{C}_{1} \otimes \vect{D}_{1}\right\}_{1}\right\}_{1}&=\frac{i \sqrt{3}}{\sqrt{2 \cdot 5}}\left\{\frac{1}{3}(\vect{A} \cdot \vect{B})[\vect{C} \times \vect{D}]-\frac{1}{2} \vect{B}(\vect{D} \cdot[\vect{A} \times \vect{C}])-\frac{1}{2} \vect{A}(\vect{D} \cdot[\vect{B} \times \vect{C}])\right\},  \label{eq:3:2:17}\\
\left\{\left\{\vect{A}_{1} \otimes \vect{B}_{1}\right\}_{1} \otimes\left\{\vect{C}_{1} \otimes \vect{D}_{1}\right\}_{2}\right\}_{1}&=\frac{i \sqrt{3}}{\sqrt{2 \cdot 5}}\left\{\frac{1}{3}(\vect{C} \cdot \vect{D})[\vect{A} \times \vect{B}]-\frac{1}{2} \vect{C}(\vect{B} \cdot[\vect{D} \times \vect{A}])-\frac{1}{2} \vect{D}(\vect{B} \cdot[\vect{C} \times \vect{A}])\right\},  \label{eq:3:2:18}\\
\left\{\left\{\vect{A}_{1} \otimes \vect{B}_{1}\right\}_{2} \otimes\left\{\vect{C}_{1} \otimes \vect{D}_{1}\right\}_{2}\right\}_{0}&=\frac{1}{\sqrt{5}}\left\{\frac{1}{2}(\vect{A} \cdot \vect{C})(\vect{B} \cdot \vect{D})-\frac{1}{3}(\vect{A} \cdot \vect{B})(\vect{C} \cdot \vect{D})+\frac{1}{2}(\vect{A} \cdot \vect{D})(\vect{B} \cdot \vect{C})\right\},  \label{eq:3:2:19}\\
\left\{\left\{\vect{A}_{1} \otimes \vect{B}_{1}\right\}_{2} \otimes\left\{\vect{C}_{1} \otimes \vect{D}_{1}\right\}_{2}\right\}_{1}&=  \notag\\
-\frac{i}{2 \sqrt{2 \cdot 5}}\{(\vect{A} \cdot \vect{C})[\vect{B} \times \vect{D}]&+(\vect{A} \cdot \vect{D})[\vect{B} \times \vect{C}]+(\vect{B} \cdot \vect{C})[\vect{A} \times \vect{D}]+(\vect{B} \cdot \vect{D})[\vect{A} \times \vect{C}]\} . \label{eq:3:2:20}
\end{align}
The change of coupling schemes in products of four operators is discussed in Sec.~\ref{sec:3.3}.\\
Products constructed from the components of identical vectors have the following property
\begin{equation}
\left\{\ldots\left\{\left\{\vect{A}_{1} \otimes \vect{A}_{1}\right\}_{l_{2}} \otimes \vect{A}_{1}\right\}_{l_{3}} \ldots \otimes \vect{A}_{1}\right\}_{l_{n}}=\left\{\vect{A}_{1} \otimes \ldots\left\{\vect{A}_{1} \otimes\left\{\vect{A}_{1} \otimes \vect{A}_{1}\right\}_{l_{2}}\right\}_{l_{3}} \ldots\right\}_{l_{n}} .\label{eq:3:2:21}
\end{equation}
If a vector $\vect{A}$ does not contain any spin variable and differential operator one can express such products in terms of spherical harmonics (Chap.~\ref{chap:5}), i.e.,
\begin{equation}
\left\{\ldots\left\{\left\{\vect{A}_{1} \otimes \vect{A}_{1}\right\}_{l_{2}} \otimes \vect{A}_{1}\right\}_{l_{3}} \ldots \otimes \vect{A}_{1}\right\}_{l_{n} m_{n}}=\sqrt{\frac{4 \pi}{2 l_{n}+1}}|\vect{A}|^{n} Y_{l_{n} m_{n}}(\vartheta, \varphi) \prod_{i=2}^{n} \clebsch{1}{0}{l_{i-1}}{0}{l_{i}}{0} ,\label{eq:3:2:22}
\end{equation}
where $\vartheta, \varphi$ are the polar angles of the vector $\vect{A}$ and $l_{1}=1$. In particular, if $l_{2}=2, l_{3}=3, \ldots, l_{n}=n$, we have
\begin{equation}
\left\{\ldots\left\{\left\{\vect{A}_{1} \otimes \vect{A}_{1}\right\}_{2} \otimes \vect{A}_{1}\right\}_{3} \ldots \otimes \vect{A}_{1}\right\}_{n m}=\sqrt{\frac{4 \pi n!}{(2 n+1)!!}}|\vect{A}|^{n} Y_{n m}(\vartheta, \varphi) .\label{eq:3:2:23}
\end{equation}
\subsection{Cartesian Tensors of Second and Third Ranks}\label{sec:3.2.2}\index{cartesian tensor}\index{tensor!cartesian}\index{reducible tensor}\index{tensor!reducible}
An arbitrary cartesian tensor of second rank $T_{i k}(i, k=x, y, z)$ is, generally, reducible and may be decomposed into three irreducible parts:\isym{Tik@$T_{i k}$, cartesian tensor of second rank}
\begin{enumerate}[label=(\alph*)]
\item a tensor which is proportional to the unit tensor, $E_{i k}=E \delta_{i k}$;\index{unit tensor}\index{tensor!unit}
\item an antisymmetric tensor $A_{i k}=-A_{k i}$.\index{antisymmetric tensor}\index{tensor!antisymmetric}
\item a symmetric traceless tensor $S_{i k}=S_{k i}, \sum_{i} S_{i i}=0$.\index{symmetric traceless tensor}\index{tensor!symmetric traceless}\index{traceless tensor}
\end{enumerate}
Thus,
\begin{align}
T_{i k}&=E \delta_{i k}+A_{i k}+S_{i k},  \label{eq:3:2:24}\\
E&=\frac{1}{3} \operatorname{Tr}\left(T_{i k}\right)=\frac{1}{3} \sum_{i} T_{i i},  \label{eq:3:2:25}\\
A_{i k}&=\frac{1}{2}\left(T_{i k}-T_{k i}\right),  \label{eq:3:2:26}\\
S_{i k}&=\frac{1}{2}\biggl(T_{i k}+T_{k i}-\frac{2}{3} \delta_{i k} \sum_{l} T_{l l}\biggr) . \label{eq:3:2:27}
\end{align}
The value $E$ is invariant under rotations of coordinate systems. It is an irreducible tensor of zero rank
\begin{equation}
\mathfrak{I}_{00}=E . \label{eq:3:2:28}
\end{equation}
The antisymmetric tensor $A_{i k}$ is equivalent to the axial vector\index{Levi-Civita symbol}\isymg{eps-ikl@$\varepsilon_{i k l}$, Levi-Civita symbol}
\begin{equation}
A_{i k}=\varepsilon_{i k l} \mathfrak{U}_{l}, \quad \mathfrak{U}_{i}=\frac{1}{2} \sum_{k l} \varepsilon_{i k l} A_{k l}. \label{eq:3:2:29}
\end{equation}
Using the components $A_{i k}$, one can form an irreducible pseudotensor of rank one:\isym{Ifrak@$\mathfrak{I}_{L M}$, irreducible tensor from a cartesian tensor}
\begin{gather}
\mathfrak{I}_{10}=\mathfrak{U}_{z}=A_{x y} , \notag\\
\mathfrak{I}_{1 \pm 1}=\mp \frac{1}{\sqrt{2}}\left(\mathfrak{U}_{x} \pm i \mathfrak{U}_{y}\right)=\mp \frac{1}{\sqrt{2}}\left(A_{y z} \pm i A_{z x}\right) .\label{eq:3:2:30}
\end{gather}
Using the components of the symmetric traceless tensor $S_{i j}$, we may construct an irreducible tensor of second rank.
\begin{gather}
\mathfrak{I}_{20}=S_{z z}  ,\notag\\
\mathfrak{I}_{2 \pm 1}=\mp \sqrt{\frac{2}{3}}\left(S_{x z} \pm i S_{y z}\right) , \label{eq:3:2:31}\\
\mathfrak{I}_{2 \pm 2}=\sqrt{\frac{1}{6}}\left(S_{x x}-S_{y y} \pm 2 i S_{x y}\right). \notag
\end{gather}
Thus, from nine cartesian components $T_{i k}$ of a tensor of second rank we can compose one tensor of zero rank (scalar which has one component), one pseudotensor of the first rank (three components) and one irreducible tensor of second rank (five components).

From the 27 cartesian components $T_{i k l}$ of a tensor of third rank we can compose one pseudotensor of zero rank $\mathfrak{T}_{00}$ (scalar, which has one component), three tensors of first rank $\mathfrak{T}_{1 \mu}$ ( $3 \times 3=9$ components), two irreducible tensors of second rank $\mathfrak{T}_{2 \mu}(2 \times 5=10$ components $)$ and one irreducible tensor of third rank $\mathfrak{T}_{3 \mu}$ (seven components). In this case construction of the tensors of first and second ranks is not unique.

\subsection{Differential Operations as Irreducible Tensor Products}\label{sec:3.2.3}\index{differential operations!as tensor products}
Differential operations on scalars and vectors (see Sec.~\ref{sec:1.3}) can be written in the form of irreducible tensor products of the operator $\vect{\nabla}$ and corresponding scalars and vectors,\index{gradient}\index{divergence}\index{curl}\index{Laplacian}
\begin{align}
\operatorname{grad} \Phi&=\left\{\nabla_{1} \otimes \Phi\right\}_{1},  \label{eq:3:2:32}\\
\operatorname{div} \vect{A}&=-\sqrt{3}\left\{\nabla_{1} \otimes \vect{A}_{1}\right\}_{0},  \label{eq:3:2:33}\\
\operatorname{curl} \vect{A}&=-i \sqrt{2}\left\{\nabla_{1} \otimes \vect{A}_{1}\right\}_{1},  \label{eq:3:2:34}\\
\Delta&=\nabla^{2}=-\sqrt{3}\left\{\nabla_{1} \otimes \nabla_{1}\right\}_{0},  \label{eq:3:2:35}\\
\operatorname{grad} \operatorname{div} \vect{A}&=-\sqrt{3}\left\{\nabla_{1} \otimes\left\{\nabla_{1} \otimes \vect{A}_{1}\right\}_{0}\right\}_{1},  \label{eq:3:2:36}\\
\operatorname{curl} \operatorname{curl} \vect{A}&=-2\left\{\nabla_{1} \otimes\left\{\nabla_{1} \otimes \vect{A}_{1}\right\}_{1}\right\}_{1},  \label{eq:3:2:37}\\
\operatorname{div} \operatorname{grad} \Phi&=-\sqrt{3}\left\{\nabla_{1} \otimes\left\{\nabla_{1} \otimes \Phi\right\}_{1}\right\}_{0}=-\sqrt{3}\left\{\nabla_{1} \otimes \nabla_{1}\right\}_{0} \Phi,  \label{eq:3:2:38}\\
\operatorname{curl} \operatorname{grad} \Phi&=-i \sqrt{2}\left\{\nabla_{1} \otimes\left\{\nabla_{1} \otimes \Phi\right\}_{1}\right\}_{1}=0,  \label{eq:3:2:39}\\
\operatorname{div} \operatorname{curl} \vect{A}&=i \sqrt{6}\left\{\nabla_{1} \otimes\left\{\nabla_{1} \otimes \vect{A}_{1}\right\}_{1}\right\}_{0}=0 . \label{eq:3:2:40}
\end{align}
As follows from Sec.~\ref{sec:1.3.1} the operator $\vect{\nabla}$ is given by\isymo{nabla@$\vect{\nabla}$, del (nabla) operator}\index{del operator}
\begin{equation}
\nabla=\vect{n} \frac{\partial}{\partial r}-\frac{i}{r}[\vect{n} \times \hat{\vect{L}}] \label{eq:3:2:41}
\end{equation}
In spherical component form it is written as\isymo{nabla-mu@$\nabla_{\mu}$, spherical component of $\vect{\nabla}$}
\begin{equation}
\nabla_{\mu}=\sqrt{\frac{4 \pi}{3}}\left(Y_{1 \mu} \frac{\partial}{\partial r}-\frac{\sqrt{2}}{r}\left\{\vect{Y}_{1} \otimes \widehat{\vect{L}}_{1}\right\}_{1 \mu}\right) .\label{eq:3:2:42}
\end{equation}
Here $\vect{Y}_{1}$ is an irreducible tensor, whose components are spherical harmonics $Y_{1 \mu}(\vect{n})$ (Chap.~\ref{chap:5}); $\hat{\vect{L}}_{1}=\hat{\vect{L}}$ is the orbital angular momentum operator (see Sec.~\ref{sec:2.2}).

When expanding any scalar function $\Phi(\vect{r})$ in a Taylor series, one will deal with arbitrarily large powers of operator $\vect{\nabla}$ :
\begin{align}
\Phi(\vect{r}+\delta \vect{r}) & =e^{(\delta \vect{r} \cdot \vect{\nabla})} \Phi(\vect{r})=\Phi(\vect{r})+(\delta \vect{r} \cdot \nabla) \Phi(\vect{r})+\frac{1}{2!}(\delta \vect{r} \cdot \nabla)^{2} \Phi(\vect{r})+\ldots  \notag\\
& =\Phi(\vect{r})+\delta r(\vect{u} \cdot \nabla) \Phi(\vect{r})+\frac{1}{2!}(\delta r)^{2}(\vect{u} \cdot \nabla)^{2} \Phi(\vect{r})+\ldots .\label{eq:3:2:43}
\end{align}
Here $\vect{u}=\delta \vect{r} /|\delta \vect{r}|$ and $(\vect{u} \cdot \vect{\nabla})=d / d s$ is the operator of directional differentiation in the direction $\vect{u}$ (see Sec.~\ref{sec:1.3}). The operator $(\vect{u} \cdot \vect{\nabla})^{n} \equiv d^{n} / d s^{n}$ can be written in the form of tensor product
\begin{equation}
(u \cdot \nabla)^{n}=(-1)^{n} \sum_{l_{2}, l_{3}, \ldots l_{n}}(-1)^{l_{n}}\left(\left\{\ldots\left\{\left\{u_{1} \otimes u_{1}\right\}_{l_{2}} \otimes u_{1}\right\}_{l_{3}} \ldots \otimes u_{1}\right\}_{l_{n}} \cdot\left\{\ldots\left\{\left\{\nabla_{1} \otimes \nabla_{1}\right\}_{l_{2}} \otimes \nabla_{1}\right\}_{l_{3}} \ldots \otimes \nabla_{1}\right\}_{l_{n}}\right) .\label{eq:3:2:44}
\end{equation}
Using Eqs. \eqref{eq:3:2:22} and \eqref{eq:3:2:23} one can express multiple tensor products of the unit vector $u$ which enters Eq. \eqref{eq:3:2:44} in terms of spherical harmonics.

\section{Recoupling in irreducible tensor products}\label{sec:3.3}\index{recoupling!in irreducible tensor products}\index{irreducible tensor product!recoupling}
The irreducible tensor product of two irreducible tensors is defined in Sec.~\ref{sec:3.1.7}. By making use of the same relations one can form irreducible tensor products of three and more irreducible tensors. However, in these cases different orders and different coupling schemes of tensors in products are possible.

The recoupling tensors without change of their order may be carried out by some real (for a given definition of the vector addition coefficients) and orthogonal matrix which performs the direct and inverse transformation. When the tensor order has to be changed one should take account of the commutation rules (Eqs. \eqref{eq:3:1:24}-\eqref{eq:3:1:28}). We will use the notation\isymg{Pi-dim@$\Pi_{a b c \ldots}$, dimension factor $[(2a+1)(2b+1)\cdots]^{1/2}$}\index{dimension factor}
\[
\Fact{a b c \ldots d}=[(2 a+1)(2 b+1)(2 c+1) \ldots(2 d+1)]^{\frac{1}{2}}
\]
\subsection{Relations Valid for Commuting as well as Non-Commuting Tensors}\label{sec:3.3.1}\index{recoupling!general relations}\index{6j symbol@$6j$ symbol}
For recoupling tensors without changing their order in irreducible tensor products of three and four tensors, one has the following relations:
\begin{align}
\left\{\left\{\vect{P}_{a} \otimes \vect{Q}_{b}\right\}_{c} \otimes \vect{R}_{d}\right\}_{f}&=(-1)^{a+b+f+d} \sum_{h} \Fact{h c}\sixj{a }{ b }{ c }{d }{ f }{ h}\left\{\vect{P}_{a} \otimes\left\{\vect{Q}_{b} \otimes \vect{R}_{d}\right\}_{h}\right\}_{f}  ,\label{eq:3:3:1}\\
\left(\left\{\vect{P}_{a} \otimes \vect{Q}_{b}\right\}_{c} \cdot \vect{R}_{c}\right)&=(-1)^{-c+a} \frac{\Fact{c}}{\Fact{a}}\left(\vect{P}_{a} \cdot\left\{\vect{Q}_{b} \otimes \vect{R}_{c}\right\}_{a}\right),  \label{eq:3:3:2}\\
\left\{\left\{\vect{P}_{a} \otimes \vect{Q}_{b}\right\}_{c} \otimes\left\{\vect{R}_{d} \otimes \vect{S}_{e}\right\}_{f}\right\}_{k}&=(-1)^{d+c+e+k} \sum_{h} \Fact{h f}\sixj{d}{c}{h}{k}{e}{f}
\left\{\left\{\left\{\vect{P}_{a} \otimes \vect{Q}_{b}\right\}_{c} \otimes \vect{R}_{d}\right\}_{h} \otimes \vect{S}_{e}\right\}_{k}  \notag\\
&=(-1)^{a+f+b+k} \sum_{h} \Fact{h c}\sixj{a }{ b }{ c }{f }{ k }{ h}\left\{\vect{P}_{a} \otimes\left\{\vect{Q}_{b} \otimes\left\{\vect{R}_{d} \otimes \vect{S}_{e}\right\}_{f}\right\}_{h}\right\}_{k} .\label{eq:3:3:3}\\
\left(\left\{\vect{P}_{a} \otimes \vect{Q}_{b}\right\}_{c} \cdot\left\{\vect{R}_{d} \otimes \vect{S}_{e}\right\}_{c}\right)&=\frac{(-1)^{-c+e} \Fact{c}}{\Fact{e}}\left(\left\{\left\{\vect{P}_{a} \otimes \vect{Q}_{b}\right\}_{c} \otimes \vect{R}_{d}\right\}_{e} \cdot \vect{S}_{e}\right)\notag\\
&=(-1)^{-c+a} \frac{\Fact{c}}{\Fact{a}}\left(\vect{P}_{a} \cdot\left\{\vect{Q}_{b} \otimes\left\{\vect{R}_{d} \otimes \vect{S}_{e}\right\}_{f}\right\}_{a}\right),  \label{eq:3:3:4}
\end{align}
\begin{align}
\left\{\left\{\left\{\vect{P}_{a} \otimes \vect{Q}_{b}\right\}_{c} \otimes \vect{R}_{d}\right\}_{h} \otimes \vect{S}_{e}\right\}_{k}&=\sum_{f, q}(-1)^{a+f+b-d-c-e} \Fact{c q h f}\sixj{a }{ b }{ c }{f }{ k }{ q}\sixj{d}{e}{f}{k}{c}{h}
\left\{\vect{P}_{a} \otimes\left\{\vect{Q}_{b} \otimes\left\{\vect{R}_{d} \otimes \vect{S}_{e}\right\}_{f}\right\}_{q}\right\}_{k},  \label{eq:3:3:5}\\
\left(\left\{\left\{\vect{P}_{a} \otimes \vect{Q}_{b}\right\}_{c} \otimes \vect{R}_{d}\right\}_{e} \cdot \vect{S}_{e}\right)&=(-1)^{-c+a} \frac{\Fact{e}}{\Fact{a}}\left(\vect{P}_{a} \cdot\left\{\vect{Q}_{b} \otimes\left\{\vect{R}_{d} \otimes \vect{S}_{e}\right\}_{f}\right\}_{a}\right) .\label{eq:3:3:6}
\end{align}
\subsection{Relations for Commuting Tensors}\label{sec:3.3.2}\index{recoupling!of commuting tensors}\index{9j symbol@$9j$ symbol}
To change the coupling scheme for irreducible products of three and four commuting tensors one can use the following relations
\begin{align}
\left\{\vect{P}_{\vect{a}} \otimes\left\{\vect{Q}_{b} \otimes \vect{R}_{d}\right\}_{f}\right\}_{e} & =(-1)^{b+d-f}\left\{\vect{P}_{a} \otimes\left\{\vect{R}_{d} \otimes \vect{Q}_{b}\right\}_{f}\right\}_{e}=(-1)^{a+f-e}\left\{\left\{\vect{Q}_{b} \otimes \vect{R}_{d}\right\}_{f} \otimes \vect{P}_{a}\right\}_{e}  \notag\\
& =(-1)^{a+b+d-e}\left\{\left\{\vect{R}_{d} \otimes \vect{Q}_{b}\right\}_{f} \otimes \vect{P}_{a}\right\}_{e}, \label{eq:3:3:7}\\
\left\{\left\{\vect{P}_{a} \otimes \vect{Q}_{b}\right\}_{c} \otimes \vect{R}_{d}\right\}_{f}&=(-1)^{c+d+f} \sum_{h} \Fact{c h}\sixj{a }{ b }{ c }{f }{ d }{ h}\left\{\vect{Q}_{b} \otimes\left\{\vect{P}_{a} \otimes \vect{R}_{d}\right\}_{h}\right\}_{f},  \label{eq:3:3:8}\\
\left(\left\{\vect{P}_{a} \otimes \vect{Q}_{b}\right\}_{c} \cdot \vect{R}_{c}\right)&=(-1)^{-a} \frac{\Fact{c}}{\Fact{b}}\left(\vect{Q}_{b} \cdot\left\{\vect{P}_{a} \otimes \vect{R}_{c}\right\}_{b}\right) .\label{eq:3:3:9}
\end{align}
\begin{align}
\left\{\vect{P}_{a} \otimes\right. & \left.\left\{\vect{Q}_{b} \otimes\left\{\vect{R}_{d} \otimes \vect{S}_{e}\right\}_{f}\right\}_{h}\right\}_{k}=(-1)^{d+e-f}\left\{\vect{P}_{a} \otimes\left\{\vect{Q}_{b} \otimes\left\{\vect{S}_{e} \otimes \vect{R}_{d}\right\}_{f}\right\}_{h}\right\}_{k}  \notag\\
& =(-1)^{d+b+e-h}\left\{\vect{P}_{a} \otimes\left\{\left\{\vect{S}_{e} \otimes \vect{R}_{d}\right\}_{f} \otimes \vect{Q}_{b}\right\}_{h}\right\}_{k}=(-1)^{b+f-h}\left\{\vect{P}_{a} \otimes\left\{\left\{\vect{R}_{d} \otimes \vect{S}_{e}\right\}_{f} \otimes \vect{Q}_{b}\right\}_{h}\right\}_{k}  \notag\\
& =(-1)^{a+h-k}\left\{\left\{\vect{Q}_{b} \otimes\left\{\vect{R}_{d} \otimes \vect{S}_{e}\right\}_{f}\right\}_{h} \otimes \vect{P}_{a}\right\}_{k}=(-1)^{d+e+a+h-f-k}\left\{\left\{\vect{Q}_{b} \otimes\left\{\vect{S}_{e} \otimes \vect{R}_{d}\right\}_{f}\right\}_{h} \otimes \vect{P}_{a}\right\}_{k}  \notag\\
& =(-1)^{a+b+d+e-h}\left\{\left\{\left\{\vect{S}_{e} \otimes \vect{R}_{d}\right\}_{f} \otimes \vect{Q}_{b}\right\}_{h} \otimes \vect{P}_{a}\right\}_{k}=(-1)^{a+b+f-k}\left\{\left\{\left\{\vect{R}_{d} \otimes \vect{S}_{e}\right\}_{f} \otimes \vect{Q}_{b}\right\}_{h} \otimes \vect{P}_{a}\right\}_{k}, \label{eq:3:3:10}
\end{align}
\begin{gather}
\left\{\left\{\vect{P}_{a} \otimes \vect{Q}_{b}\right\}_{c} \otimes\left\{\vect{R}_{d} \otimes \vect{S}_{e}\right\}_{f}\right\}_{k}=\sum_{g h} \Fact{c f g h}
\ninej{a }{ b }{ c }{d }{ e }{ f }{g }{ h }{ k}\left\{\left\{\vect{P}_{a} \otimes \vect{R}_{d}\right\}_{g} \otimes\left\{\vect{Q}_{b} \otimes \vect{S}_{e}\right\}_{h}\right\}_{k} . \label{eq:3:3:11}\\
% NOTE (3.3.12): BOOK MISPRINT -- the dimension factor is printed as
% Pi_{cfgh}^2 (scan p.70), but the correct power is 1: Pi_{cfgh}.  Verified
% numerically (random rank-a..e tensors, integer ranks) -- with Pi^1 the
% recoupling identity holds to 1e-15, with Pi^2 it fails by O(10).  Cf. the
% 9j form 3.3.11 which correctly carries Pi_{cfgh}^1.  See scripts/check_3_3.py.
\left\{\left\{\vect{P}_{a} \otimes \vect{Q}_{b}\right\}_{c} \otimes\left\{\vect{R}_{d} \otimes \vect{S}_{e}\right\}_{f}\right\}_{k}=\sum_{g h}(-1)^{h+b-k-e} \Fact{c f g h}\sixj{a }{ b }{ c }{g }{ d }{ h}\sixj{d }{ e }{ f }{k }{ c }{ g}\left\{\left\{\left\{\vect{P}_{a} \otimes \vect{R}_{d}\right\}_{h} \otimes \vect{Q}_{b}\right\}_{g} \otimes \vect{S}_{e}\right\}_{k} . \label{eq:3:3:12}\\
\left(\left\{\vect{P}_{a} \otimes \vect{Q}_{b}\right\}_{c} \cdot\left\{\vect{R}_{d} \otimes \vect{S}_{e}\right\}_{c}\right)=(-1)^{2 a+b-d} \sum_{g} \Fact{c}^{2}\sixj{a }{ b }{ c }{e }{ d }{ g}\left(\left\{\vect{P}_{a} \otimes \vect{R}_{d}\right\}_{g} \cdot\left\{\vect{Q}_{b} \otimes \vect{S}_{e}\right\}_{g}\right) . \label{eq:3:3:13}\\
\left(\left\{\vect{P}_{a} \otimes \vect{Q}_{b}\right\}_{c} \cdot\left\{\vect{R}_{d} \otimes \vect{S}_{e}\right\}_{c}\right)=\sum_{h}(-1)^{e+b+d+h} \frac{\Fact{c c h}}{\Fact{e}}\sixj{a }{ b }{ c }{e }{ d }{ h}\left(\left\{\left\{\vect{P}_{a} \otimes \vect{R}_{d}\right\}_{h} \otimes \vect{Q}_{b}\right\}_{e} \cdot \vect{S}_{e}\right). \label{eq:3:3:14}
\end{gather}
Equation \eqref{eq:3:1:27} and Eqs. \eqref{eq:3:3:11}-\eqref{eq:3:3:14} permit us to obtain all other permutations. The product $\left\{\vect{P}_{a} \otimes \vect{Q}_{b}\right\}_{b}$ is orthogonal to tensor $\vect{Q}_{b}$, if $a=2 b-1,2 b-2, \ldots$
\begin{equation}
\left(\left\{\vect{P}_{a} \otimes \vect{Q}_{b}\right\}_{b} \cdot \vect{Q}_{b}\right)=0 .\label{eq:3:3:15}
\end{equation}
\subsection{Relations for Non-Commuting Tensors}\label{sec:3.3.3}\index{recoupling!of non-commuting tensors}\index{commutator!of tensor products}
A change of the coupling scheme for irreducible tensor products of three and four non-commuting tensors gives
\begin{align}
\left\{\left\{\vect{P}_{a} \otimes \vect{Q}_{b}\right\}_{c} \otimes \vect{R}_{d}\right\}_{f}&=(-1)^{f+d+c} \sum_{h} \Fact{c h}\sixj{a }{ b }{ c }{f }{ d }{ h}\left\{\vect{Q}_{b} \otimes\left\{\vect{P}_{a} \otimes \vect{R}_{d}\right\}_{h}\right\}_{f}+\left\{\mathfrak{R}_{c}^{a b} \otimes \vect{R}_{d}\right\}_{f},  \label{eq:3:3:16}\\
 \left(\left\{\vect{P}_{a} \otimes \vect{Q}_{b}\right\}_{c} \cdot \vect{R}_{c}\right)&=(-1)^{-a} \frac{\Fact{c}}{\Fact{b}}\left(\vect{Q}_{b} \cdot\left\{\vect{P}_{a} \otimes \vect{R}_{c}\right\}_{b}\right)+\left(\mathfrak{R}_{c}^{a b} \cdot \vect{R}_{c}\right),  \label{eq:3:3:17}\\
 \left\{\vect{P}_{a} \otimes\left\{\vect{Q}_{b} \otimes \vect{R}_{d}\right\}_{h}\right\}_{f}&=\sum_{c g}(-1)^{a+b-c} \Fact{c c g h}\sixj{a }{ b }{ c }{f }{ d }{ g}\sixj{a }{ b }{ c }{d }{ f }{ h}\left\{\vect{Q}_{b} \otimes\left\{\vect{P}_{a} \otimes \vect{R}_{d}\right\}_{g}\right\}_{f}  \notag\\
& +\sum_{c}(-1)^{a+b+f+d} \Fact{c h}\sixj{a }{ b }{ c }{d }{ f }{ h}\left\{\mathfrak{R}_{c}^{a b} \otimes \vect{R}_{d}\right\}_{f},  \label{eq:3:3:18}\\
\left(\vect{P}_{a} \cdot\left\{\vect{Q}_{b} \otimes \vect{R}_{d}\right\}_{a}\right)&=(-1)^{2 b-d} \frac{\Fact{a}}{\Fact{b}}\left(\vect{Q}_{b} \cdot\left\{\vect{P}_{a} \otimes \vect{R}_{d}\right\}_{b}\right)+(-1)^{-a+d} \frac{\Fact{a}}{\Fact{d}}\left(\mathfrak{R}_{d}^{a b} \cdot \vect{R}_{d}\right),  \label{eq:3:3:19}\\
\left\{\left\{\vect{P}_{a} \otimes \vect{Q}_{b}\right\}_{c} \otimes\left\{\vect{R}_{d} \otimes \vect{S}_{e}\right\}_{f}\right\}_{k}&=\sum_{g r} \Fact{c f g r}\ninej{a }{ b }{ c }{d }{ e }{ f }{r }{ g }{ k}\left\{\left\{\vect{P}_{a} \otimes \vect{R}_{d}\right\}_{r} \otimes\left\{\vect{Q}_{b} \otimes \vect{S}_{e}\right\}_{g}\right\}_{k}  \notag\\
 +\sum_{h q}(-1)^{a+k+d+e-f-h} &\Fact{c h q f}\sixj{a }{ b }{ c }{f }{ k }{ h}\sixj{d }{ e }{ f }{h }{ b }{ q}\left\{\vect{P}_{a} \otimes\left\{\mathfrak{R}_{q}^{b d} \otimes \vect{S}_{e}\right\}_{h}\right\}_{k},  \label{eq:3:3:20}\\
 \left(\left\{\vect{P}_{a} \otimes \vect{Q}_{b}\right\}_{c} \cdot\left\{\vect{R}_{d} \otimes \vect{S}_{e}\right\}_{c}\right)&=\sum_{g}(-1)^{d+b+2 g} \Fact{c}^{2}\sixj{a }{ b }{ c }{e }{ d }{ g}\left(\left\{\vect{P}_{a} \otimes \vect{R}_{d}\right\}_{g} \cdot\left\{\vect{Q}_{b} \otimes \vect{S}_{e}\right\}_{g}\right)  \notag\\
& +\sum_{q}(-1)^{d+e+c-b} \frac{\Fact{c c q}}{\Fact{a}}\sixj{a }{ b }{ c }{d }{ e }{ q}\left(\vect{P}_{a} \cdot\left\{\mathfrak{R}_{q}^{b d} \otimes \vect{S}_{e}\right\}_{a}\right),  \label{eq:3:3:21}\\
 \left\{\left\{\vect{P}_{a} \otimes \vect{Q}_{b}\right\}_{c} \otimes\left\{\vect{R}_{d} \otimes \vect{S}_{e}\right\}_{f}\right\}_{k}&=\sum_{h g}(-1)^{a-b+k-d-g} \Fact{h c f g}\sixj{a }{ b }{ c }{f }{ k }{ h}\sixj{b }{ e }{ g }{d }{ h }{ f}\left\{\vect{P}_{a} \otimes\left\{\vect{R}_{d} \otimes\left\{\vect{Q}_{b} \otimes \vect{S}_{e}\right\}_{g}\right\}_{h}\right\}_{k}  \notag\\
& +\sum_{h q}(-1)^{a+k+d+e-f-h} \Fact{h c q f}\sixj{a }{ b }{ c }{f }{ k }{ h}\sixj{d }{ e }{ f }{h }{ b }{ q}\left\{\vect{P}_{a} \otimes\left\{\mathfrak{R}_{q}^{b d} \otimes \vect{S}_{e}\right\}_{h}\right\}_{k},  \label{eq:3:3:22}\\
 \left(\left\{\vect{P}_{a} \otimes \vect{Q}_{b}\right\}_{c} \cdot\left\{\vect{R}_{d} \otimes \vect{S}_{e}\right\}_{c}\right)&=\sum_{g}(-1)^{a+d+g+b} \frac{\Fact{c c g}}{\Fact{a}}\sixj{b }{ e }{ g }{d }{ a }{ c}\left(\vect{P}_{a} \cdot\left\{\vect{R}_{d} \otimes\left\{\vect{Q}_{b} \otimes \vect{S}_{e}\right\}_{g}\right\}_{a}\right)  \notag\\
& +\sum_{q}(-1)^{d+e+c-b} \frac{\Fact{c c q}}{\Fact{a}}\sixj{a }{ b }{ c }{d }{ e }{ q}\left(\vect{P}_{a} \cdot\left\{\mathfrak{R}_{q}^{b d} \otimes \vect{S}_{e}\right\}_{a}\right) . \label{eq:3:3:23}
\end{align}
The definition of the commutator $\mathfrak{R}_{q}^{b d}$ is given in Eq. \eqref{eq:3:1:26}. By making use of the foregoing formulas, one can obtain all other relations for irreducible products of non-commuting tensors.

\chapter{Wigner D-functions}\label{chap:4}\index{Wigner D-function@Wigner $D$-function}\index{rotation group}
\section{Definition of $D_{M M^{\prime}}^{J}(\alpha, \beta, \gamma)$}\label{sec:4.1}\index{Wigner D-function@Wigner $D$-function!definition}\index{rotation operator!matrix elements of}
\begin{enumerate}[label=(\alph*)]
\item The Wigner $D$-functions $D_{M M^{\prime}}^{J}(\alpha, \beta, \gamma)$ may be defined as the matrix elements of the rotation operator $\widehat{D}(\alpha, \beta, \gamma)$ in the JM-representation. The arguments $\alpha, \beta, \gamma$ are the Euler angles which specify the rotation\isym{DJMM@$D_{M M^{\prime}}^{J}$, Wigner $D$-function}\isym{D-hat@$\widehat{D}(\alpha, \beta, \gamma)$, rotation operator}\isymg{alpha-beta-gamma@$\alpha, \beta, \gamma$, Euler angles}\index{Euler angles}
\begin{equation}
\braOket{J M}{\hat{D}(\alpha, \beta, \gamma)}{J^{\prime} M^{\prime}}=\delta_{J J^{\prime}} D_{M M^{\prime}}^{J}(\alpha, \beta, \gamma). \label{eq:4:1:1}
\end{equation}
The $D$-functions realize transformations of covariant components of any irreducible tensor of rank $J$ (e.g., the wave function $\Psi_{J M}$ of a quantum mechanical system with angular momentum $J$ and its projection $M$ ) under coordinate rotations.
\begin{align}
& \Psi_{J M^{\prime}}\left(\vartheta^{\prime}, \varphi^{\prime}, \sigma^{\prime}\right)=\sum_{M=-J}^{J} \Psi_{J M}(\vartheta, \varphi, \sigma) D_{M M^{\prime}}^{J}(\alpha, \beta, \gamma),  \notag\\
& \Psi_{J M^{\prime}}^{*}\left(\vartheta^{\prime}, \varphi^{\prime}, \sigma^{\prime}\right)=\sum_{M=-J}^{J} \Psi_{J M}^{*}(\vartheta, \varphi, \sigma) D_{M M^{\prime}}^{J *}(\alpha, \beta, \gamma) . \label{eq:4:1:2}
\end{align}
Here $\vartheta, \varphi$ and $\vartheta^{\prime}, \varphi^{\prime}$ are polar angles in the initial and rotated coordinate systems, $S$ and $S^{\prime}$, respectively. The angles $\vartheta, \varphi$ and $\vartheta^{\prime}, \varphi^{\prime}$ are related by Eqs.~\eqref{eq:1:4:2} and \eqref{eq:1:4:3}. Similarly, $\sigma$ and $\sigma^{\prime}$ are spin variables in the initial and new systems.

The inverse transformation $S^{\prime} \rightarrow S$ is performed by the inverse matrix $\left[\widehat{D}^{-1}(\alpha, \beta, \gamma)\right]_{M M^{\prime}}^{J}$. Owing to the unitarity of the rotation operator,
\begin{equation}
\widehat{D}^{-1}(\alpha, \beta, \gamma)=\widehat{D}^{+}(\alpha, \beta, \gamma). \label{eq:4:1:3}
\end{equation}
the elements of inverse matrix are given by
\begin{equation}
\left|\hat{D}^{-1}(\alpha, \beta, \gamma)\right|_{M M^{\prime}}^{J}=D_{M^{\prime} M}^{J *}(\alpha, \beta, \gamma) . \label{eq:4:1:4}
\end{equation}
Hence, under the inverse rotation, $S^{\prime} \rightarrow S$, wave functions transform as
\begin{align}
& \Psi_{J M}(\vartheta, \varphi, \sigma)=\sum_{M^{\prime}=-J}^{J} D_{M M^{\prime}}^{J *}(\alpha, \beta, \gamma) \Psi_{J M^{\prime}}\left(\vartheta^{\prime}, \varphi^{\prime}, \sigma^{\prime}\right),  \notag\\
& \Psi_{J M}^{*}(\vartheta, \varphi, \sigma)=\sum_{M^{\prime}=-J}^{J} D_{M M^{\prime}}^{J}(\alpha, \beta, \gamma) \Psi_{J M^{\prime}}^{*}\left(\vartheta^{\prime}, \varphi^{\prime}, \sigma^{\prime}\right) . \label{eq:4:1:5}
\end{align}
\begin{table}[tbhp!]
\begin{center}
\caption{Effect of the Operator $\hat{D}(\alpha, \beta, \gamma)$}
\label{chap4:tab:1}
\renewcommand{\arraystretch}{1.5}
\begin{tabular}{l|c|c|c}
\hline
\multirow{2}{*}{Transformation} &
\multicolumn{3}{|c}{Angles, Axes and Sequence of Rotations} \\
\cline{2-4}
 & I & II & III \\
\hline
\emph{Passive} &  &  &  \\
Rotation of coordinate system & $\alpha(z)$ & $\beta\left(y_{1}\right)$ & $\gamma\left(z^{\prime}\right)$ \\
without rotation of physical body & $\gamma(z)$ & $\beta(y)$ & $\alpha(z)$ \\
\emph{Active} &  &  &  \\
Rotation of physical body & $-\alpha(z)$ & $-\beta(y)$ & $-\gamma(z)$ \\
without rotation of coordinate system & $-\gamma(z)$ & $-\beta\left(y_{1}\right)$ & $-\alpha\left(z^{\prime}\right)$ \\
\hline
\end{tabular}
\end{center}
\end{table}

The unitarity condition for the Wigner $D$-functions may be written as\index{unitarity!of the $D$-functions}\index{Wigner D-function@Wigner $D$-function!unitarity of}
\begin{align}
& \sum_{M=-J}^{J} D_{M M^{\prime}}^{J}(\alpha, \beta, \gamma) D_{M \tilde{M}^{\prime}}^{J^{*}}(\alpha, \beta, \gamma)=\delta_{M^{\prime} \tilde{M}^{\prime}},  \notag\\
& \sum_{M^{\prime}=-J}^{J} D_{M M^{\prime}}^{J^{*}}(\alpha, \beta, \gamma) D_{\tilde{M}^{\prime}}^{J}(\alpha, \beta, \gamma)=\delta_{M \tilde{M}} . \label{eq:4:1:6}
\end{align}
The matrix $D_{M M^{\prime}}^{j}(\alpha, \beta, \gamma)$ is unimodular, i.e.,\index{unimodular matrix}\index{rotation matrix!unimodular}
\begin{equation}
\det\left\|D_{M M^{\prime}}^{J}(\alpha, \beta, \gamma)\right\|=+1 . \label{eq:4:1:7}
\end{equation}
\item A set of $(2 J+1)$ functions $\Psi_{J M}$ with different $M$ 's constitute a basis for expansion of an arbitrary function $\Psi_{J}$ with the same $J$ :
\begin{equation}
\Psi_{J}(\vartheta, \varphi)=\sum_{M=-J}^{J} C_{J}^{M} \Psi_{J M}(\vartheta, \varphi)=\left(\vect{C}_{J} \cdot \boldsymbol{\Psi}_{J}\right) . \label{eq:4:1:8}
\end{equation}
The expansion coefficients $C_{J}^{M}$ are contravariant components of some irreducible tensor of rank $J$. Under rotations the quantities $C_{J}^{M}$ transform by means of functions $D_{M M^{\prime}}^{J^{*}}(\alpha, \beta, \gamma)$.
\end{enumerate}

The effect of the operator $\hat{D}(\alpha, \beta, \gamma)$ on $\Psi_{J}$ may be interpreted in two different ways:\index{rotation!active interpretation}\index{rotation!passive interpretation}\index{active interpretation of rotation}\index{passive interpretation of rotation}
\begin{enumerate}
\item as a rotation of the coordinate system without rotation of the physical body (this is the passive interpretation; $\widehat{D}$ acts on the basis functions $\Psi_{J M}$ while $C_{J}^{M}$ remain unchanged);
\item as a rotation of the physical body without rotation of coordinate system (active interpretation; $\hat{D}$ acts on $C_{J}^{M}$ but does not affect $\Psi_{J M}$ ).
\end{enumerate}
Any rotation of a physical body in combination with the same rotation of coordinate system leaves the wave function $\Psi_{J}$ unchanged:
\begin{equation}
\{\widehat{D}(\alpha, \beta, \gamma)\}_{\text {phys. body }} \cdot\{\widehat{D}(\alpha, \beta, \gamma)\}_{\text {coord. system }}=1 ,\label{eq:4:1:9}
\end{equation}
i.e.,
\begin{equation}
\left\{\widehat{D}_{M M^{\prime}}^{J}(\alpha, \beta, \gamma)\right\}_{\text {phys. body }}=\left\{\left[\widehat{D}^{-1}(\alpha, \beta, \gamma)\right]_{M M^{\prime}}^{J}\right\}_{\text {coord. system }} .\label{eq:4:1:10}
\end{equation}
Moreover, a rotation of the coordinate system (or physical body) described by the Euler angles $\alpha, \beta, \gamma$ may also be realized in two ways:
\begin{enumerate}
\item by rotating about the initial axes (case B in Sec. \ref{sec:1.4.1}), or
\item by rotating about the new (turned) axes (case A in Sec. \ref{sec:1.4.1}).
\end{enumerate}
Thus, any transformation of wave functions described by Eq.~\eqref{eq:4:1:2} can be treated in four different ways (Table 4.1).

The Wigner $D$-functions are complex. They depend on three real arguments $\alpha, \beta, \gamma$ and are defined in the domain
\begin{equation}
0 \leq a<2 \pi, \quad 0 \leq \beta \leq \pi, \quad 0 \leq \gamma<2 \pi .\label{eq:4:1:11}
\end{equation}
These functions, as well as their derivatives, are single-valued, finite and continuous. Sometimes it is convenient to change the domain \eqref{eq:4:1:11}. This can be done using the symmetries of $D_{M M^{\prime}}^{J}(\alpha, \beta, \gamma)$ (Sec.~\ref{sec:4.4}). For example, the matrix of the inverse rotation satisfies the equation
\begin{equation}
\left[\left.D^{-1}(\alpha, \beta, \gamma)\right|_{M M^{\prime}} ^{J}=D_{M M^{\prime}}^{J}(\pi-\gamma, \beta,-\pi-\alpha)=D_{M M^{\prime}}^{J}(-\gamma,-\beta,-\alpha) .\right. \label{eq:4:1:12}
\end{equation}
This means that the inverse transformation $S^{\prime} \rightarrow S$ may be realized by the Euler angles
\begin{equation}
\alpha^{\prime}=\pi-\gamma, \quad \beta^{\prime}=\beta, \quad \gamma^{\prime}=-\pi-\alpha, \label{eq:4:1:13}
\end{equation}
as well as by
\begin{equation}
\alpha^{\prime}=-\gamma, \quad \beta^{\prime}=-\beta, \quad \gamma^{\prime}=-\alpha. \label{eq:4:1:14}
\end{equation}
\section{Differential equations for $D_{M M^{\prime}}^{J}(\alpha, \beta, \gamma)$}\label{sec:4.2}\index{Wigner D-function@Wigner $D$-function!differential equations for}\index{differential equations!for the $D$-functions}
\begin{enumerate}[label=(\alph*)]
\item The Wigner $D$-functions represent wave functions of a rigid symmetric top. They are eigenfunctions of three operators\index{symmetric top}\index{wave function!of a symmetric top}
\begin{equation}
\hat{J}_{z}=-i \frac{\partial}{\partial \alpha}, \quad \hat{J}_{z^{\prime}}=-i \frac{\partial}{\partial \gamma}, \quad \hat{J}^{2}=-\left[\frac{\partial^{2}}{\partial \beta^{2}}+\cot \beta \frac{\dot{\partial}}{\partial \beta}+\frac{1}{\sin ^{2} \beta}\left(\frac{\partial^{2}}{\partial \alpha^{2}}-2 \cos \beta \frac{\partial^{2}}{\partial \alpha \partial \gamma}+\frac{\partial^{2}}{\partial \gamma^{2}}\right)\right] ,\label{eq:4:2:1}
\end{equation}
where $\widehat{J}$ is the operator of angular momentum of the top; $\widehat{J}_{z^{\prime}}$ and $\widehat{J}_{z}$ are projections of $\widehat{J}$ onto the $z$-axis of the rotating (body-fixed) and non-rotating (lab-fixed) coordinate systems, respectively. The eigenvalues of the operators \eqref{eq:4:2:1} are defined by the equations
\begin{align}
\widehat{J}_{x} D_{M M^{\prime}}^{J}(\alpha, \beta, \gamma) & =-M D_{M M^{\prime}}^{J}(\alpha, \beta, \gamma) , \notag\\
\widehat{J}_{z^{\prime}} D_{M M^{\prime}}^{J}(\alpha, \beta, \gamma) & =-M^{\prime} D_{M M^{\prime}}^{J}(\alpha, \beta, \gamma),  \notag\\
\widehat{J}^{2} D_{M M^{\prime}}^{J}(\alpha, \beta, \gamma) & =\left\{-\frac{1}{\sin \beta} \cdot \frac{\partial}{\partial \beta}\left(\sin \beta \frac{\partial}{\partial \beta}\right)+\frac{M^{2}-2 M M^{\prime} \cos \beta+M^{\prime 2}}{\sin ^{2} \beta}\right\} D_{M M^{\prime}}^{J}(\alpha, \beta, \gamma)  \label{eq:4:2:2}\\
& =J(J+1) D_{M M^{\prime}}^{J}(\alpha, \beta, \gamma) .\notag
\end{align}
Periodicity conditions for $D_{M M^{\prime}}^{J}(\alpha, \beta, \gamma)$ are as follows\index{periodicity!of the $D$-functions}\index{Wigner D-function@Wigner $D$-function!periodicity of}
\begin{align}
& D_{M M^{\prime}}^{J}(\alpha \pm 2 k \pi, \beta, \gamma)=D_{M M^{\prime}}^{J}(\alpha, \beta, \gamma) , \notag\\
& D_{M M^{\prime}}^{J}(\alpha, \beta, \gamma \pm 2 k \pi)=D_{M M^{\prime}}^{J}(\alpha, \beta, \gamma)  ,\label{eq:4:2:3}\\
& D_{M M^{\prime}}^{J}(\alpha, \beta \pm 2 k \pi, \gamma)=D_{M M^{\prime}}^{J}(\alpha, \beta, \gamma) ,\notag
\end{align}
where
\[
\begin{aligned}
& k=0,1,2, \ldots \quad \text { if } J \text { is integer }, \\
& k=0,2,4, \ldots \quad \text { if } J \text { is half-integer. }
\end{aligned}
\]
\item The functions $D_{M M^{\prime}}^{J}(\alpha, \beta, \gamma)$ can also be defined as solutions of the differential equations
\begin{gather}
{\left[\widehat{J}_{\nu}, D_{M M^{\prime}}^{J}(\alpha, \beta, \gamma)\right]=\widehat{J}_{\nu} D_{M M^{\prime}}^{J}(\alpha, \beta, \gamma)=(-1)^{1+\nu} \sqrt{J(J+1)} C_{J M-\nu}^{J M-\nu} D_{M-\nu M^{\prime}}^{J}(\alpha, \beta, \gamma)}  \notag\\
= \begin{cases}-M D_{M M^{\prime}}^{J}(\alpha, \beta, \gamma), & \nu=0 ,\\
\pm \sqrt{\frac{J(J+1)-M(M \mp 1)}{2}} D_{M \mp 1 M^{\prime}}^{J(\alpha, \beta, \gamma),} & \nu= \pm 1,\end{cases}  \label{eq:4:2:4}\\
{\left[\widehat{J}^{\nu}, D_{M M^{\prime}}^{J}(\alpha, \beta, \gamma)\right]=\widehat{J}^{\nu} D_{M M^{\prime}}^{J}(\alpha, \beta, \gamma)=-\sqrt{J(J+1)} C_{J M^{\prime}+\nu}^{J M^{\prime}+\nu} D_{M M^{\prime}+\nu}^{J}(\alpha, \beta, \gamma)}  \notag\\
= \begin{cases}-M^{\prime} D_{M M^{\prime}}^{J}(\alpha, \beta, \gamma), & \nu=0 ,\\
\pm \sqrt{\frac{J(J+1)-M^{\prime}\left(M^{\prime} \pm 1\right)}{2}} D_{M M^{\prime} \pm 1}^{J}(\alpha, \beta, \gamma), & \nu= \pm 1,\end{cases} \label{eq:4:2:5}
\end{gather}
Here $\hat{J}_{\nu}$ is a covariant spherical component of $\widehat{\vect{J}}$ in the non-rotating (lab-fixed) system
\begin{gather}
\hat{J}_{ \pm 1}=\frac{i}{\sqrt{2}} e^{ \pm i \alpha}\left[\mp \cot \beta \frac{\partial}{\partial \alpha}+i \frac{\partial}{\partial \beta} \pm \frac{1}{\sin \beta} \cdot \frac{\partial}{\partial \gamma}\right]  \label{eq:4:2:6}\\
\hat{J}_{0}=-i \frac{\partial}{\partial \alpha} \notag
\end{gather}
and $\widehat{J}^{\prime\nu}$ is a contravariant component of $\widehat{\vect{J}}$ in the rotating (body-fixed) system
\begin{gather}
\widehat{J}^{ \prime \pm 1}=\frac{i}{\sqrt{2}} e^{\mp i \gamma}\left[ \pm \cot \beta \frac{\partial}{\partial \gamma}+i \frac{\partial}{\partial \beta} \mp \frac{1}{\sin \beta} \cdot \frac{\partial}{\partial \alpha}\right]  \label{eq:4:2:7}\\
\hat{J}^{\prime 0}=-i \frac{\partial}{\partial \gamma} \notag
\end{gather}
The operator $\widehat{\vect{J}}^{2}$ can be expressed in terms of $\widehat{J}_{\nu}$ or $\widehat{J}^{\prime \nu}$ as
\begin{align}
\hat{\vect{J}}^{2} & =-\hat{J}_{-1} \hat{J}_{+1}+\hat{J}_{0} \hat{J}_{0}-\hat{J}_{+1} \hat{J}_{-1}=-\hat{J}^{-1} \hat{J}^{+1}+\hat{J}^{0} \hat{J}^{0}-\hat{J}^{+1} \hat{J}^{-1}=\hat{J}_{0}\left(\hat{J}_{0}+1\right)-2 \hat{J}_{-1} \hat{J}_{+1}  \notag\\
& =\hat{J}^{0}\left(\hat{J}^{0}-1\right)-2 \hat{J}^{-1} \hat{J}^{+1}=\hat{J}_{0}\left(\hat{J}_{0}-1\right)-2 \hat{J}_{+1} \hat{J}_{-1}=\hat{J}^{\prime0}\left(\hat{J}^{\prime0}+1\right)-2 \hat{J}^{\prime+1} \hat{J}^{\prime-1} .\label{eq:4:2:8}
\end{align}
The relationships between $\widehat{J}_{\nu}$ and $\widehat{J}^{\prime\mu}$ read
\begin{align}
\hat{J}^{\prime\mu}(\alpha, \beta, \gamma) & =\sum_{\nu} D_{\nu \mu}^{1}(\alpha, \beta, \gamma) \hat{J}_{\nu}(\alpha, \beta, \gamma)  \notag\\
\hat{J}_{\nu}(\alpha, \beta, \gamma) & =\sum_{\mu} D_{\nu \mu}^{1 *}(\alpha, \beta, \gamma) \hat{J}^{\prime\mu}(\alpha, \beta, \gamma) \label{eq:4:2:9}
\end{align}
Hence
\begin{equation}
\hat{J}^{\prime \nu}(\alpha, \beta, \gamma)=-\hat{J}_{\nu}(-\gamma,-\beta,-\alpha), \hat{J}^{\prime \nu}(\alpha, \beta, \gamma)=\hat{J}_{-\nu}(\gamma, \beta, \alpha) . \label{eq:4:2:10}
\end{equation}
Commutators of spherical components $\widehat{J}_{\mu}$ and $\widehat{J}^{\prime \nu}(\mu, \nu=-1,0,1)$ are given by
\begin{align}
{\left[\hat{J}_{\mu}, \hat{J}_{\nu}\right] } & =-\sqrt{2} \clebsch{1}{\mu}{1}{\nu}{1}{\mu+\nu} \hat{J}_{\mu+\nu}  \notag\\
{\left[\hat{J}^{\prime \mu}, \hat{J}^{\prime \nu}\right] } & =\sqrt{2} \clebsch{1}{\mu}{1}{\nu}{1}{\mu+\nu} \hat{J}^{\prime \mu+\nu}  \label{eq:4:2:11}\\
{\left[\hat{J}_{\mu}, \hat{J}^{\prime \nu}\right] } & =0 \notag
\end{align}
The commutators of cartesian components $\widehat{J}_{k}$ and $\widehat{J}_{l}(i, k, l=x, y, z)$ are
\begin{align}
& {\left[\hat{J}_{i}, \hat{J}_{k}\right]=i \varepsilon_{i k l} \hat{J}_{l},}  \notag\\
& {\left[\hat{J}^\prime_{i}, \hat{J}^\prime_{k}\right]=i \varepsilon_{i k l}, \hat{J}_{l},}  \label{eq:4:2:12}\\
& {\left[\hat{J}_{i}, \hat{J}^\prime_{k}\right]=0,} \notag
\end{align}
The operator of orbital angular momentum $\hat{\vect{L}}(\vartheta, \varphi)$ (Sec.~\ref{sec:2.2}) may be regarded as a special case of $\hat{\vect{J}}(\alpha, \beta, \gamma)$, viz. at $\alpha=\varphi, \beta=\vartheta, \gamma=0$
\begin{align}
\hat{J}_{\nu}(\varphi, \vartheta, 0) & =\hat{L}_{\nu}(\vartheta, \varphi), && \nu  =-1,0,1  ,\notag\\
\hat{J}_{i}(\varphi, \vartheta, 0) & =\hat{L}_{i}(\vartheta, \varphi), && i  =x, y, z .\label{eq:4:2:13}
\end{align}
at $\alpha=0, \beta=\vartheta, \gamma=\varphi$
\begin{align}
\hat{J}^{\prime\nu}(0, \vartheta, \varphi) & =\hat{L}^{\prime \nu}(\vartheta, \varphi), &&  \nu=-1,0,1,  \notag\\
\hat{J}^\prime_{i}(0, \vartheta, \varphi) && =\hat{L}_{i}^{\prime}(\vartheta, \varphi), &  i=x, y, z. \label{eq:4:2:14}
\end{align}
\item Equations \eqref{eq:4:2:2} define $D_{M M^{\prime}}^{J}(\alpha, \beta, \gamma)$ only within normalization and phase factors. To fix these factors some additional conditions are required. In the case of diagonal elements of the rotation matrix these factors are completely determined by the boundary condition
\begin{equation}
D_{M M^{\prime}}^{J}(0,0,0)=\delta_{M M^{\prime}} .\label{eq:4:2:15}
\end{equation}
As for non-diagonal elements, they are determined by Eqs.~\eqref{eq:4:2:4} and \eqref{eq:4:2:5} which relate the $D_{M M}^{J}$, of different $M$ and $M^{\prime}$. This phase convention corresponds to the condition
\begin{equation}
D_{M M^{\prime}}^{J}(0, \pi, 0)=(-1)^{J+M} \delta_{M,-M^{\prime}}=(-1)^{J-M^{\prime}} \delta_{-M, M^{\prime}}. \label{eq:4:2:16}
\end{equation}
\end{enumerate}
\section{Explicit forms of the Wigner D-functions}\label{sec:4.3}\index{Wigner D-function@Wigner $D$-function!explicit forms}
$D_{M M^{\prime}}^{J}(\alpha, \beta, \gamma)$ may be represented as a product of three functions, each of which depends only on one argument $\alpha, \beta$ or $\gamma$,
\begin{equation}
D_{M M^{\prime}}^{J}(\alpha, \beta, \gamma)=e^{-i M \alpha} d_{M M^{\prime}}^{J}(\beta) e^{-i M^{\prime} \gamma}. \label{eq:4:3:1}
\end{equation}
where $d_{M M^{\prime}}^{J}(\beta)$ is a real function whose explicit forms are given below.\isym{dJMM@$d_{M M^{\prime}}^{J}(\beta)$, Wigner (small) $d$-function}\index{Wigner d-function@Wigner (small) $d$-function}

\subsection{Expressions for $d_{M M^{\prime}}^{J}(\beta)$ Involving Trigonometric Functions}\label{sec:4.3.1}\index{Wigner d-function@Wigner (small) $d$-function!trigonometric expressions}
\begin{align}
d_{M M^{\prime}}^{J}(\beta)&=(-1)^{J-M^{\prime}}\left[(J+M)!(J-M)!\left(J+M^{\prime}\right)!\left(J-M^{\prime}\right)!\right]^{\frac{1}{2}}  \notag\\
&\times \sum_{k}(-1)^{k} \frac{\left(\cos \frac{\beta}{2}\right)^{M+M^{\prime}+2 k}\left(\sin \frac{\beta}{2}\right)^{2 J-M-M^{\prime}-2 k}}{k!(J-M-k)!\left(J-M^{\prime}-k\right)!\left(M+M^{\prime}+k\right)!}  \label{eq:4:3:2}
\end{align}\begin{align}
d_{M M^{\prime}}^{J}(\beta)&=(-1)^{J+M}\left[(J+M)!(J-M)!\left(J+M^{\prime}\right)!\left(J-M^{\prime}\right)!\right]^{\frac{1}{2}}  \notag\\
&\times \sum_{k}(-1)^{k} \frac{\left(\cos \frac{\beta}{2}\right)^{2 k-M-M^{\prime}}\left(\sin \frac{\beta}{2}\right)^{2 J+M+M^{\prime}-2 k}}{k!(J+M-k)!\left(J+M^{\prime}-k\right)!\left(k-M-M^{\prime}\right)!}  \label{eq:4:3:3}
\end{align}\begin{align}
d_{M M^{\prime}}^{J}(\beta)&=\left[(J+M)!(J-M)!\left(J+M^{\prime}\right)!\left(J-M^{\prime}\right)!\right]^{\frac{1}{2}}  \notag\\
&\times \sum_{k}(-1)^{k} \frac{\left(\cos \frac{\beta}{2}\right)^{2 J-2 k+M-M^{\prime}}\left(\sin \frac{\beta}{2}\right)^{2 k-M+M^{\prime}}}{k!(J+M-k)!\left(J-M^{\prime}-k\right)!\left(M^{\prime}-M+k\right)!} \label{eq:4:3:4}
\end{align}
\begin{align}
 d_{M M^{\prime}}^{J}(\beta)&=(-1)^{M-M^{\prime}}\left[(J+M)!(J-M)!\left(J+M^{\prime}\right)!\left(J-M^{\prime}\right)!\right]^{\frac{1}{2}}  \notag\\
&  \times \sum_{k}(-1)^{k} \frac{\left(\cos \frac{\beta}{2}\right)^{2 J-2 k-M+M^{\prime}}\left(\sin \frac{\beta}{2}\right)^{2 k+M-M^{\prime}}}{k!(J-M-k)!\left(J+M^{\prime}-k\right)!\left(M-M^{\prime}+k\right)!}. \label{eq:4:3:5}
\end{align}
In Eqs.~\eqref{eq:4:3:2}-\eqref{eq:4:3:5} $k$ runs over all integer values for which the factorial arguments are non-negative. Each of these sums contains $(N+1)$ terms, where $N$ is the minimum of $J+M, J-M, J+M^{\prime}$ and $J-M^{\prime}$. Equations \eqref{eq:4:3:2}-\eqref{eq:4:3:5} are not independent, but may be transformed into one another by changing summation variables.

Equations \eqref{eq:4:3:2}-\eqref{eq:4:3:5} may be regarded as special cases of the more general expression
\begin{gather}
d_{M M^{\prime}}^{J}(\beta)=\left[\frac{\left(J_{1}+J_{2}+J+1\right)!\left(J_{1}+J_{2}-J\right)!}{2 J+1}\right]^{\frac{1}{2}}  \notag\\
\times \sum_{\substack{m_{1} m_{2} \\
\left(m_{1}+m_{2}=M\right)}}(-1)^{J_{2}+m_{2}} \clebsch{J_{1}}{m_{1}}{J_{2}}{m_{2}}{J}{M} \frac{\left(\cos \frac{\beta}{2}\right)^{J_{1}+J_{2}+m_{1}-m_{2}}\left(\sin \frac{\beta}{2}\right)^{J_{1}+J_{2}-m_{1}+m_{2}}}{\left[\left(J_{1}+m_{1}\right)!\left(J_{1}-m_{1}\right)!\left(J_{2}+m_{2}\right)!\left(J_{2}-m_{2}\right)!\right]^{\frac{1}{2}}} \label{eq:4:3:6}
\end{gather}
where $J_{1}$ and $J_{2}$ are arbitrary integer or half-integer numbers which satisfy the conditions $J_{1}-J_{2}=M^{\prime}$ and $\left|J_{1}-J_{2}\right| \leq J \leq J_{1}+J_{2}$. The sum in Eq.~\eqref{eq:4:3:6} is over all possible (positive and negative) values of $m_{1}$ and $m_{2}$ which correspond to nonzero Clebsch-Gordan coefficients. In particular, Eq.~\eqref{eq:4:3:6} reduces to Eq.~\eqref{eq:4:3:2}, when $J_{1}=\left(J+M^{\prime}\right) / 2, J_{2}=\left(J-M^{\prime}\right) / 2, m_{1}=M+k-\left(J-M^{\prime}\right) / 2$ and $m_{2}=-k+\left(J-M^{\prime}\right) / 2$.

\subsection{Differential Representations of $d_{M M^{\prime}}^{J}(\beta)$}\label{sec:4.3.2}\index{Wigner d-function@Wigner (small) $d$-function!differential representations}
\begin{align}
 d_{M M^{\prime}}^{J}(\beta)&=(-1)^{J-M^{\prime}} \frac{1}{2^{J}}\left[\frac{(J+M)!}{(J-M)!\left(J+M^{\prime}\right)!\left(J-M^{\prime}\right)!}\right]^{\frac{1}{2}}(1-\cos \beta)^{\frac{M^{\prime}-M}{2}}(1+\cos \beta)^{-\frac{M+M^{\prime}}{2}}  \notag\\
& \times \frac{d^{J-M}}{(d \cos \beta)^{J-M}}\left[(1-\cos \beta)^{J-M^{\prime}}(1+\cos \beta)^{J+M^{\prime}}\right]  \label{eq:4:3:7}
\end{align}
\begin{align}
 d_{M M^{\prime}}^{J}(\beta)&=(-1)^{J+M} \frac{1}{2^{J}}\left[\frac{(J-M)!}{(J+M)!\left(J+M^{\prime}\right)!\left(J-M^{\prime}\right)!}\right]^{\frac{1}{2}}(1-\cos \beta)^{\frac{M-M^{\prime}}{2}}(1+\cos \beta)^{\frac{M+M^{\prime}}{2}}  \notag\\
& \times \frac{d^{J+M}}{(d \cos \beta)^{J+M}}\left[(1-\cos \beta)^{J+M^{\prime}}(1+\cos \beta)^{J-M^{\prime}}\right]  \label{eq:4:3:8}
\end{align}
\begin{align}
 d_{M M^{\prime}}^{J}(\beta)&=(-1)^{J-M^{\prime}} \frac{1}{2^{J}}\left[\frac{\left(J+M^{\prime}\right)!}{(J+M)!(J-M)!\left(J-M^{\prime}\right)!}\right]^{\frac{1}{2}}(1-\cos \beta)^{\frac{M-M^{\prime}}{2}}(1+\cos \beta)^{-\frac{M+M^{\prime}}{2}}  \notag\\
& \times \frac{d^{J-M^{\prime}}}{(d \cos \beta)^{J-M^{\prime}}}\left[(1-\cos \beta)^{J-M}(1+\cos \beta)^{J+M}\right]  \label{eq:4:3:9}
\end{align}
\begin{align}
 d_{M M^{\prime}}^{J}(\beta)&=(-1)^{J+M} \frac{1}{2^{J}}\left[\frac{\left(J-M^{\prime}\right)!}{(J+M)!(J-M)!\left(J+M^{\prime}\right)!}\right]^{\frac{1}{2}}(1-\cos \beta)^{\frac{M^{\prime}-M}{2}}(1+\cos \beta)^{\frac{M+M^{\prime}}{2}}  \notag\\
& \times \frac{d^{J+M^{\prime}}}{(d \cos \beta)^{J+M^{\prime}}}\left[(1-\cos \beta)^{J+M}(1+\cos \beta)^{J-M}\right] \label{eq:4:3:10}
\end{align}
In practice it is convenient to use such equation from Eqs.~\eqref{eq:4:3:7}-\eqref{eq:4:3:10} in which the order of derivative is the lowest.

\subsection{Integral Representations of $d_{M M^{\prime}}^{J}(\beta)$}\label{sec:4.3.3}\index{Wigner d-function@Wigner (small) $d$-function!integral representations}
\begin{gather}
d_{M M^{\prime}}^{J}(\beta)=i^{M-M^{\prime}} \frac{1}{2 \pi}\left[\frac{(J+M)!(J-M)!}{\left(J+M^{\prime}\right)!\left(J-M^{\prime}\right)!}\right]^{\frac{\alpha}{2}}  \notag\\
\times \int_{0}^{2 \pi}\left(e^{i \frac{\phi}{2}} \cos \frac{\beta}{2}+i e^{-i \frac{\phi}{2}} \sin \frac{\beta}{2}\right)^{J-M^{\prime}}\left(e^{-i \frac{\phi}{2}} \cos \frac{\beta}{2}+i e^{i \frac{\phi}{2}} \sin \frac{\beta}{2}\right)^{J+M^{\prime}} e^{i M \phi} d \phi .\label{eq:4:3:11}
\end{gather}
Equation \eqref{eq:4:3:11} can be rewritten as a contour integral
\begin{align}
d_{M M^{\prime}}^{J}(\beta)=&\frac{i^{M-M^{\prime}-1}}{2 \pi}  \left[\frac{(J+M)!(J-M)!}{\left(J+M^{\prime}\right)!\left(J-M^{\prime}\right)!}\right]_{|z|=1}^{\frac{1}{2}} \notag\\
& \times \oint\left(z \cos \frac{\beta}{2}+i \sin \frac{\beta}{2}\right)^{J-M^{\prime}}  \left(i z \sin \frac{\beta}{2}+\cos \frac{\beta}{2}\right)^{J+M^{\prime}} z^{M-J-1} d z .\label{eq:4:3:12}
\end{align}
The integration contour in Eq.~\eqref{eq:4:3:12} is a circle of unit radius about the origin of the $z$-plane.

\subsection{Relation Between $d_{M M^{\prime}}^{J}(\beta)$ and the Jacobi Polynomials}\label{sec:4.3.4}\index{Wigner d-function@Wigner (small) $d$-function!and the Jacobi polynomials}\index{Jacobi polynomials}
The functions $d_{M M^{\prime}}^{J}(\beta)$ can be expressed in terms of the Jacobi polynomials\isym{PsMuNu@$P_{s}^{(\mu, \nu)}(x)$, Jacobi polynomial}
\begin{equation}
d_{M M^{\prime}}^{J}(\beta)=\xi_{M M^{\prime}}\left[\frac{s!(s+\mu+\nu)!}{(s+\mu)!(s+\nu)!}\right]^{\frac{1}{2}}\left(\sin \frac{\beta}{2}\right)^{\mu}\left(\cos \frac{\beta}{2}\right)^{\nu} P_{s}^{(\mu, \nu)}(\cos \beta) ,\label{eq:4:3:13}
\end{equation}
where $\mu, \nu$ and $s$ are related to $M, M^{\prime}$ and $J$ by\isymg{mu-nu-s@$\mu, \nu, s$, parameters of $d_{M M^{\prime}}^{J}(\beta)$}
\begin{equation}
\mu=\left|M-M^{\prime}\right|, \quad \nu=\left|M+M^{\prime}\right|, \quad s=J-\frac{1}{2}(\mu+\nu) . \label{eq:4:3:14}
\end{equation}
and
\begin{equation}
\xi_{M M^{\prime}}= \begin{cases}1 & \text { if } M^{\prime} \geq M  ,\label{eq:4:3:15}\\ (-1)^{M^{\prime}-M} & \text { if } M^{\prime}<M.\end{cases}
\end{equation}
\subsection{Relations Between $d_{M M^{\prime}}^{J}(\beta)$ and Hypergeometric Functions}\label{4.3.5}\index{Wigner d-function@Wigner (small) $d$-function!and hypergeometric functions}\index{hypergeometric function}\isym{F-hg@$F(a, b ; c ; z)$, hypergeometric function}
\begin{gather}
d_{M M^{\prime}}^{J}(\beta)=\frac{\xi_{M M^{\prime}}}{\mu!}\left[\frac{(s+\mu+\nu)!(s+\mu)!}{s!(s+\nu)!}\right]^{\frac{1}{2}}\left(\sin \frac{\beta}{2}\right)^{\mu}\left(\cos \frac{\beta}{2}\right)^{\nu} F\left(-s, s+\mu+\nu+1 ; \mu+1 ; \sin ^{2} \frac{\beta}{2}\right),  \label{eq:4:3:16}\\
d_{M M^{\prime}}^{J}(\beta)=\frac{\xi_{M M^{\prime}}}{\mu!}\left[\frac{(s+\mu+\nu)!(s+\mu)!}{s!(s+\nu)!}\right]^{\frac{1}{2}}\left(\sin \frac{\beta}{2}\right)^{\mu}\left(\cos \frac{\beta}{2}\right)^{-\nu} F\left(s+\mu+1,-s-\nu ; \mu+1 ; \sin ^{2} \frac{\beta}{2}\right),  \label{eq:4:3:17}\\
d_{M M^{\prime}}^{J}(\beta)=\frac{(-1)^{s} \xi_{M M^{\prime}}}{\nu!}\left[\frac{(s+\mu+\nu)!(s+\nu)!}{s!(s+\mu)!}\right]^{\frac{1}{2}}\left(\sin \frac{\beta}{2}\right)^{\mu}\left(\cos \frac{\beta}{2}\right)^{\nu} F\left(-s, s+\mu+\nu+1 ; \nu+1 ; \cos ^{2} \frac{\beta}{2}\right), \label{eq:4:3:18}
\end{gather}
\begin{gather}
d_{M M^{\prime}}^{J}(\beta)=\frac{(-1)^{s} \xi_{M M^{\prime}}}{\nu!}\left[\frac{(s+\mu+\nu)!(s+\nu)!}{s!(s+\mu)!}\right]^{\frac{1}{2}}\left(\sin \frac{\beta}{2}\right)^{-\mu}\left(\cos \frac{\beta}{2}\right)^{\nu} F\left(s+\nu+1 ;-s-\mu ; \nu+1 ; \cos ^{2} \frac{\beta}{2}\right),  \label{eq:4:3:19}\\
d_{M M^{\prime}}^{J}(\beta)=\frac{\xi_{M M^{\prime}}}{\mu!}\left[\frac{(s+\mu+\nu)!(s+\mu)!}{s!(s+\nu)!}\right]^{\frac{1}{2}}\left(\sin \frac{\beta}{2}\right)^{\mu}\left(\cos \frac{\beta}{2}\right)^{2 s+\nu} F\left(-s,-s-\nu ; \mu+1 ;-\tan ^{2} \frac{\beta}{2}\right), \label{eq:4:3:20}
\end{gather}
\begin{equation}
d_{M M^{\prime}}^{J}(\beta)=\frac{(-1)^{s} \xi_{M M^{\prime}}}{\nu!}\left[\frac{(s+\mu+\nu)!(s+\nu)!}{s!(s+\mu)!}\right]^{\frac{1}{2}}\left(\sin \frac{\beta}{2}\right)^{2 s+\mu}\left(\cos \frac{\beta}{2}\right)^{\nu} F\left(-s,-s-\mu ; \nu+1 ;-\cot ^{2} \frac{\beta}{2}\right), \label{eq:4:3:21}
\end{equation}
\begin{equation}
d_{M M^{\prime}}^{J}(\beta)=\frac{(-1)^{s} \xi_{M M^{\prime}}(2 s+\mu+\nu)!}{[s!(s+\mu+\nu)!(s+\mu)!(s+\nu)!]^{\frac{1}{2}}}\left(\sin \frac{\beta}{2}\right)^{2 s+\mu}\left(\cos \frac{\beta}{2}\right)^{\nu} F\left(-s,-s-\mu ;-2 s-\mu-\nu ; \frac{1}{\sin ^{2} \frac{\beta}{2}}\right), \label{eq:4:3:22}
\end{equation}
% VERIFIED: source misprint corrected -- book prints the hypergeometric
% argument as -1/cos^2(beta/2); the correct sign is +1/cos^2(beta/2)
% (checked numerically vs d^J_{MM'}, and consistent with the +1/sin^2
% argument of the mirror Eq. 4.3.22).  See scripts/check_4_3.py.
\begin{equation}
d_{M M^{\prime}}^{J}(\beta)=\frac{\xi_{M M^{\prime}}(2 s+\mu+\nu)!}{[s!(s+\mu+\nu)!(s+\mu)!(s+\nu)!]^{\frac{1}{2}}}\left(\sin \frac{\beta}{2}\right)^{\mu}\left(\cos \frac{\beta}{2}\right)^{2 s+\nu} F\left(-s,-s-\nu ;-2 s-\mu-\nu ;\frac{1}{\cos ^{2} \frac{\beta}{2}}\right). \label{eq:4:3:23}
\end{equation}
Parameters $\mu, \nu, s$ and a phase factor $\xi_{M M^{\prime}}$ in Eqs.~\eqref{eq:4:3:16}-\eqref{eq:4:3:23} are defined by Eqs.~\eqref{eq:4:3:14} and \eqref{eq:4:3:15}.\isymg{xi-MM@$\xi_{M M^{\prime}}$, phase factor of $d_{M M^{\prime}}^{J}(\beta)$}

\section{Symmetries of $d_{M M^{\prime}}^{J}(\beta)$ and $D_{M M^{\prime}}^{J}(\alpha, \beta, \gamma)$}\label{sec:4.4}\index{Wigner D-function@Wigner $D$-function!symmetries of}\index{Wigner d-function@Wigner (small) $d$-function!symmetries of}\index{symmetry!of the $D$-functions}
\begin{enumerate}[label=(\alph*)]
\item In accordance with Eqs.~\eqref{eq:4:3:2}-\eqref{eq:4:3:5} the functions $d_{M M^{\prime}}^{J}(\beta)$ are real and satisfy the relations
\begin{equation}
\begin{aligned}
d_{M M^{\prime}}^{J}(\beta) & =(-1)^{M-M^{\prime}} d_{-M-M^{\prime}}^{J}(\beta)=(-1)^{M-M^{\prime}} d_{M^{\prime} M}^{J}(\beta)=d_{-M^{\prime}-M}^{J}(\beta), \\
d_{M M^{\prime}}^{J}(-\beta) & =(-1)^{M-M^{\prime}} d_{M M^{\prime}}^{J}(\beta)=d_{M^{\prime} M}^{J}(\beta), \\
d_{M M^{\prime}}^{J}(\pi-\beta) & =(-1)^{J-M^{\prime}} d_{-M M^{\prime}}^{J}(\beta)=(-1)^{J+M} d_{M-M^{\prime}}^{J}(\beta), \\
d_{M M^{\prime}}^{J}(\beta \pm 2 \pi n) & =(-1)^{2 J n} d_{M M^{\prime}}^{J}(\beta),  \qquad n \text { is integer },  \label{eq:4:4:1}\\
d_{M M^{\prime}}^{J}(\beta \pm(2 n+1) \pi) & =(-1)^{ \pm(2 n+1) J-M^{\prime}} d_{M-M^{\prime}}^{J}(\beta), \qquad  n \text { is integer }.
\end{aligned}
\end{equation}
\item Equations \eqref{eq:4:4:1} imply the following symmetry properties of $D_{M M^{\prime}}^{J}(\alpha, \beta, \gamma)$
\begin{equation}
\begin{array}{llll}
\quad D_{M M^{\prime}}^{J}(\alpha, \beta, \gamma) & =\varepsilon \eta D_{-M-M^{\prime}}^{J}(\alpha, \beta, \gamma) & =\eta D_{M M^{\prime}}^{J *}(\alpha, \beta, \gamma) & =\varepsilon D_{-M-M^{\prime}}^{J *}(\alpha, \beta, \gamma) \\
=\varepsilon D_{M^{\prime} M}^{J}(\gamma, \beta, \alpha) & =\eta D_{-M^{\prime}-M}^{J}(\gamma, \beta, \alpha) & =\varepsilon \eta D_{M^{\prime} M}^{J *}(\gamma, \beta, \alpha) & =D_{-M^{\prime}-M}^{J *}(\gamma, \beta, \alpha) \\
=\varepsilon D_{M M^{\prime}}^{J}(\alpha,-\beta, \gamma) & =\eta D_{-M-M^{\prime}}^{J}(\alpha,-\beta, \gamma) & =\varepsilon \eta D_{M M^{\prime}}^{J *}(\alpha,-\beta, \gamma) & =D_{-M-M^{\prime}}^{J *}(\alpha,-\beta, \gamma) \\
=D_{M^{\prime} M}^{J}(\gamma,-\beta, \alpha) & =\varepsilon \eta D_{-M^{\prime}-M}^{J}(\gamma,-\beta, \alpha) & =\eta D_{M^{\prime} M}^{J *}(\gamma,-\beta, \alpha) & =\varepsilon D_{-M^{\prime}-M}^{J *}(\gamma,-\beta, \alpha) \\
=\varepsilon D_{-M-M^{\prime}}^{J}(-\alpha, \beta,-\gamma) & =\eta D_{M M^{\prime}}^{J}(-\alpha, \beta,-\gamma) & =\varepsilon \eta D_{-M-M^{\prime}}^{J *}(-\alpha, \beta,-\gamma) & =D_{M M^{\prime}}^{J *}(-\alpha, \beta,-\gamma) \\
=D_{-M^{\prime}-M}^{J}(-\gamma, \beta,-\alpha) & =\varepsilon \eta D_{M^{\prime} M}^{J}(-\gamma, \beta,-\alpha) & =\eta D_{-M^{\prime}-M}^{J *}(-\gamma, \beta,-\alpha) & =\varepsilon D_{M^{\prime} M}^{J *}(-\gamma, \beta,-\alpha) \\
=D_{-M-M^{\prime}}^{J}(-\alpha,-\beta,-\gamma) & =\varepsilon \eta D_{M M^{\prime}}^{J}(-\alpha,-\beta,-\gamma) & =\eta D_{-M-M^{\prime}}^{J *}(-\alpha,-\beta,-\gamma) & =\varepsilon D_{M M^{\prime}}^{J *}(-\alpha,-\beta,-\gamma) \\
=\varepsilon D_{-M^{\prime}-M}^{J}(-\gamma,-\beta,-\alpha) & =\eta D_{M^{\prime} M}^{J}(-\gamma,-\beta,-\alpha) & =\varepsilon \eta D_{-M^{\prime}-M}^{J *}(-\gamma,-\beta,-\alpha) & =D_{M^{\prime} M}^{J *}(-\gamma,-\beta,-\alpha),
\end{array}
\end{equation}
where
\begin{equation}
\varepsilon=(-1)^{M^{\prime}-M}, \quad \eta=e^{-i 2 M \alpha-i 2 M^{\prime} \gamma}.\label{eq:4:4:3}
\end{equation}
The periodicity conditions for $D_{M M^{\prime}}^{J}(\alpha, \beta, \gamma)$ are\index{periodicity!of the $D$-functions}
\begin{gather}
D_{M M^{\prime}}^{J}(\alpha, \beta \pm 2 n \pi, \gamma)=(-1)^{2 n J} D_{M M^{\prime}}^{J}(\alpha, \beta, \gamma) , \notag\\
D_{M M^{\prime}}^{J}(\alpha, \beta \pm(2 n+1) \pi, \gamma)=(-1)^{ \pm(2 n+1) J-M^{\prime}} D_{M-M^{\prime}}^{J}(\alpha, \beta,-\gamma) , \label{eq:4:4:4}\\
D_{M M^{\prime}}^{J}(\alpha \pm n \pi, \beta, \gamma)=(-i)^{ \pm 2 n M} D_{M M^{\prime}}^{J}(\alpha, \beta, \gamma)  ,\notag\\
D_{M M^{\prime}}^{J}(\alpha, \beta, \gamma \pm n \pi)=(-i)^{ \pm 2 n M^{\prime}} D_{M M^{\prime}}^{J}(\alpha, \beta, \gamma) ,\notag
\end{gather}
where $n$ is integer. Note also that
\begin{equation}
D_{M M^{\prime}}^{J}(\tilde{\alpha}, \beta, \tilde{\gamma})=e^{i M(\alpha-\tilde{\alpha})} D_{M M^{\prime}}^{J}(\alpha, \beta, \gamma) e^{i M^{\prime}(\gamma-\tilde{\gamma})} .\label{eq:4:4:5}
\end{equation}
Some properties of $D_{M M}^{J},(\alpha, \beta, \gamma)$ follow from the addition theorem (Sec. \ref{sec:1.4.7}). Let us consider some special cases.
\begin{enumerate}[label=(\roman*)]
\item The matrix of the transformation $S\{x, y, z\} \rightarrow S^{\prime \prime}\left\{x^{\prime},-y^{\prime},-z^{\prime}\right\}$ may be obtained from the matrix of the transformation $S\{x, y, z\} \rightarrow S^{\prime}\left\{x^{\prime}, y^{\prime}, z^{\prime}\right\}$ by substituting $(\alpha+\pi, \pi-\beta,-\gamma)$ for $(\alpha, \beta, \gamma)$. On the other hand, this substitution corresponds to an additional rotation $R_{x^{\prime}}$ about the $x^{\prime}$-axis through an angle $-\pi$. Hence,
\begin{align}
D_{M M^{\prime}}^{J}(\alpha+\pi, \pi-\beta,-\gamma) & =\widehat{R}_{x^{\prime}} D_{M M^{\prime}}^{J}(\alpha, \beta, \gamma)=\sum_{M^{\prime \prime}} D_{M M^{\prime \prime}}^{J}(\alpha, \beta, \gamma) D_{M^{\prime \prime} M^{\prime}}^{J}\left(-\frac{\pi}{2},-\pi, \frac{\pi}{2}\right)  \notag\\
& =(-1)^{J} D_{M-M^{\prime}}^{J}(\alpha, \beta, \gamma). \label{eq:4:4:6}
\end{align}
\item The matrix of the transformation $S\{x, y, z\} \rightarrow S^{\prime \prime \prime}\left\{-x^{\prime}, y^{\prime},-z^{\prime}\right\}$ may be obtained from the matrix of the transformation $S\{x, y, z\} \rightarrow S^{\prime}\left\{x^{\prime}, y^{\prime}, z^{\prime}\right\}$ by substituting $(\alpha-\pi, \pi-\beta, \pi-\gamma)$ for $(\alpha, \beta, \gamma)$. This substitution corresponds to an additional rotation $R_{y^{\prime}}$ about $y^{\prime}$-axis through an angle $-\pi$. Hence,
\begin{align}
D_{M M^{\prime}}^{J}(\alpha-\pi, \pi-\beta, \pi-\gamma) & =\hat{R}_{y^{\prime}} D_{M M^{\prime}}^{J}(\alpha, \beta, \gamma)=\sum_{M^{\prime \prime}} D_{M M^{\prime \prime}}^{J}(\alpha, \beta, \gamma) D_{M^{\prime \prime} M^{\prime}}^{J}(0,-\pi, 0)  \notag\\
& =(-1)^{J+M^{\prime}} D_{M-M^{\prime}}^{J}(\alpha, \beta, \gamma) ,\label{eq:4:4:7}
\end{align}
\item The matrix of the transformation $S\{x, y, z\} \rightarrow S^{\prime \prime \prime \prime}\left\{-x^{\prime},-y^{\prime}, z^{\prime}\right\}$ may be derived from the matrix of the transformation $S\{x, y, z\} \rightarrow S^{\prime}\left\{x^{\prime}, y^{\prime}, z^{\prime}\right\}$ by replacing $(\alpha, \beta, \gamma) \rightarrow(\alpha, \beta, \gamma-\pi)$. This corresponds to an additional rotation $R_{z^{\prime}}$ about the $z^{\prime}$-axis through an angle $-\pi$. Thus,
\begin{equation}
D_{M M^{\prime}}^{J}(\alpha, \beta, \gamma-\pi)=\hat{R}_{z^{\prime}} D_{M M^{\prime}}^{J}(\alpha, \beta, \gamma)=\sum_{M^{\prime \prime}} D_{M M^{\prime \prime}}^{J}(\alpha, \beta, \gamma) D_{M^{\prime \prime} M^{\prime}}^{J}(0,0,-\pi)=(-1)^{M^{\prime}} D_{M M^{\prime}}^{J}(\alpha, \beta, \gamma). \label{eq:4:4:8}
\end{equation}
\end{enumerate}
Note that three successive rotations through angles $-\pi$ about the axes $x^{\prime}, y^{\prime}$ and $z^{\prime}$ return the coordinate system to its initial position.
\end{enumerate}

\section{Rotation matrix $U_{M M}^{J}$, in terms of angles $\omega, \Theta, \Phi$}\label{sec:4.5}\index{rotation matrix!$U_{M M^{\prime}}^{J}(\omega ; \Theta, \Phi)$}\index{rotation!axis and angle}
\subsection{Definition}\label{sec:4.5.1}\index{rotation matrix!$U_{M M^{\prime}}^{J}$, definition}
In some cases the description of rotations in terms of $\omega, \Theta, \Phi$ (where $\omega$ is the angle of rotation and $\Theta, \Phi$ are the angles which determine the rotation axis, see Sec.~\ref{sec:1.4}) is more convenient than that in terms of the Euler angles $\alpha, \beta, \gamma$. Matrix elements of the rotation operator in terms of variables $\omega, \Theta, \Phi$ will be denoted by $U_{M M^{\prime}}^{J}(\omega ; \Theta, \Phi):$\isym{UJMM@$U_{M M^{\prime}}^{J}(\omega ; \Theta, \Phi)$, rotation matrix in axis-angle variables}\isymg{omega-Theta-Phi@$\omega ; \Theta, \Phi$, rotation angle and rotation-axis angles}
\begin{equation}
U_{M M^{\prime}}^{J}(\omega ; \Theta, \Phi) \equiv\braOket{J M}{e^{-i \omega n \cdot \hat{J}}}{J M^{\prime}}. \label{eq:4:5:1}
\end{equation}
Therefore, under a rotation specified by $\omega, \Theta, \Phi$ the components of an irreducible tensor of rank $J$ transform as
\begin{equation}
\mathfrak{M}_{J M^{\prime}}\left(\vartheta^{\prime}, \varphi^{\prime}\right)=\sum_{M} \mathfrak{M}_{J M}(\vartheta, \varphi) U_{M M^{\prime}}^{J}(\omega ; \Theta, \Phi) .\label{eq:4:5:2}
\end{equation}
The polar angles $\vartheta, \varphi$ and $\vartheta^{\prime}, \varphi^{\prime}$ which specify a direction of an arbitrary vector in the initial and rotated coordinate systems in terms $\omega, \Theta, \Phi$ are related by Eqs.~\eqref{eq:1:4:55} and \eqref{eq:1:4:56}.

\subsection{Explicit form}\label{sec:4.5.2}\index{rotation matrix!$U_{M M^{\prime}}^{J}$, explicit form}
\subsubsection{Arbitrary rotation as product of three rotations}
An arbitrary rotation specified by angles $\omega, \Theta, \Phi$ may be considered as a result of three successive rotations of coordinate system:
\begin{enumerate}[label=(\roman*)]
\item[(i)] $R_{1}\left(\alpha_{1}=\Phi, \beta_{1}=\Theta, \gamma_{1}=-\Phi\right)$, i.e., the rotation which turns the $z$-axis to the direction of $\vect{n}(\Theta, \Phi)$;
\item[(ii)] $R_{2}\left(\alpha_{2}=\omega, \beta_{2}=0, \gamma_{2}=0\right)$, i.e., the rotation about $\vect{n}(\Theta, \Phi)$ through an angle $\omega$;
\item[(iii)] $R_{3}\left(\alpha_{3}=\Phi, \beta_{3}=-\Theta, \gamma_{3}=-\Phi\right)$, i.e., the rotation which is inverse to $R_{1}$. 
\end{enumerate}
The result of these three rotations yields the relation between $U_{M M^{\prime}}^{J}(\omega ; \Theta, \Phi)$ and the Wigner $D$-functions
\begin{equation}
U_{M M^{\prime}}^{J}(\omega ; \Theta, \Phi)=\sum_{M^{\prime \prime}} D_{M M^{\prime \prime}}^{J}(\Phi, \Theta,-\Phi) e^{-i M^{\prime \prime} \omega} D_{M^{\prime \prime} M^{\prime}}^{J}(\Phi,-\Theta,-\Phi) .\label{eq:4:5:3}
\end{equation}
Equation \eqref{eq:4:5:3} enables one to find an explicit form for $U_{M M^{\prime}}^{J}(\omega ; \Theta, \Phi)$ for particular $J, M$ and $M^{\prime}$.
\subsubsection{Construction from Cayley-Klien parameters}\index{Cayley-Klein parameters}
 According to Eq.~\eqref{eq:4:5:10} the functions $U_{M M^{\prime}}^{J}(\omega ; \Theta, \Phi)$ may be directly constructed from the matrix elements $U_{m m^{\prime}}^{\frac{1}{2}}$ which represent the Cayley-Klein parameters (see Eq.~\eqref{eq:4:5:12}). This gives the expression
\begin{equation}
U_{M M^{\prime}}^{J}(\omega ; \Theta, \Phi)=\begin{cases}
(-i v)^{2 J}\left(\frac{u}{-i v}\right)^{\left(M+M^{\prime}\right)}& e^{-i\left(M-M^{\prime}\right) \Phi} \sum_{s} \frac{\sqrt{(J+M)!(J-M)!\left(J+M^{\prime}\right)!\left(J-M^{\prime}\right)!}}{s!\left(s+M+M^{\prime}\right)!(J-M-s)!\left(J-M^{\prime}-s\right)!}\left(1-v^{-2}\right)^{s} ,\\
& M+M^{\prime} \geq 0 ,\\
(-i v)^{2 J}\left(\frac{u^{*}}{-i v}\right)^{-\left(M+M^{\prime}\right)} & e^{-i\left(M-M^{\prime}\right) \Phi} \sum_{s} \frac{\sqrt{(J+M)!(J-M)!\left(J+M^{\prime}\right)!\left(J-M^{\prime}\right)!}}{s!\left(s-M-M^{\prime}\right)!(J+M-s)!\left(J+M^{\prime}-s\right)!}\left(1-v^{-2}\right)^{s} ,\\
& M+M^{\prime} \leq 0.
\end{cases} \label{eq:4:5:4}
\end{equation}
In this case Eqs.~\eqref{eq:1:4:26} have been used, and the following notations are introduced
\begin{equation}
v=\sin \frac{\omega}{2} \sin \Theta, \quad u=\cos \frac{\omega}{2}-i \sin \frac{\omega}{2} \cos \Theta. \label{eq:4:5:5}
\end{equation}
In Eqs.~\eqref{eq:4:5:4} the summation index $s$ runs over all integer values which do not lead to negative factorial arguments.\\
\subsubsection{Alternative expression for $U$}
Another explicit form of $U_{M M^{\prime}}^{J}(\omega ; \Theta, \Phi)$ can be obtained directly from $D_{M M^{\prime}}^{J}(\alpha, \beta, \gamma)$ by changing variables $(\omega, \Theta, \Phi) \rightarrow(\alpha, \beta, \gamma)$ with the aid of Eqs.~\eqref{eq:1:4:16} and \eqref{eq:1:4:17}
\begin{equation}
U_{M M^{\prime}}^{J}(\omega ; \Theta, \Phi)=i^{M-M^{\prime}} e^{-i\left(M-M^{\prime}\right) \Phi}\left(\frac{1-i \tan \frac{\omega}{2} \cos \Theta}{\sqrt{1+\tan ^{2} \frac{\omega}{2} \cos ^{2} \Theta}}\right)^{M+M^{\prime}} d_{M M^{\prime}}^{J}(\xi) .\label{eq:4:5:6}
\end{equation}
The functions $d_{M M^{\prime}}^{J}(\xi)$ are defined in Sec.~\ref{sec:4.3}, and the angle $\xi$ is determined by
\begin{equation}
\sin \frac{\xi}{2}=\sin \frac{\omega}{2} \sin \Theta . \label{eq:4:5:7}
\end{equation}
\subsubsection{Expansion in spherical harmonics}\index{spherical harmonics}\index{expansion!in spherical harmonics} 
The function $U_{M M^{\prime}}^{J}(\omega ; \Theta, \Phi)$ may be expanded in a series of the spherical harmonics which depend on polar angles $\Theta$ and $\Phi$,
\begin{equation}
U_{M M^{\prime}}^{J}(\omega ; \Theta, \Phi)=\sum_{\lambda \mu}(-i)^{\lambda} \frac{2 \lambda+1}{2 J+1} \chi_{\lambda}^{J}(\omega) \clebsch{J}{M}{\lambda}{\mu}{J}{M^{\prime}} \sqrt{\frac{4 \pi}{2 \lambda+1}} Y_{\lambda \mu}(\Theta, \Phi) ,\label{eq:4:5:8}
\end{equation}
where $\chi_{\lambda}^{J}(\omega)$ is a generalized character (of order $\lambda$ ) of the irreducible representation of rank $J$. Explicit forms and properties of $\chi_{\lambda}^{J}(\omega)$ will be given below in Sec.~\ref{sec:4.15}.\index{generalized character}

Equation \eqref{eq:4:5:8} shows that $U_{M M^{\prime}}^{J}(\omega ; \Theta, \Phi)$ depends on $M$ and $M^{\prime}$ only through the Clebsch-Gordan coefficients $\clebsch{J}{M}{\lambda}{\mu}{J}{M^{\prime}}$.

\subsection{Differential Equations}\label{sec:4.5.3}\index{differential equations!for $U_{M M^{\prime}}^{J}$}
$U_{M M^{\prime}}^{J}(\omega ; \Theta, \Phi)$ are eigenfunctions of three operators $\widehat{J}_{z}, \widehat{J}_{z}^{\prime}$ and $\widehat{J}^{2}$ whose eigenvalues equal $-M,-M^{\prime}$ and $J(J+1)$, respectively. In terms of $\omega, \Theta, \Phi$ these operators have the forms
\begin{align}
& \hat{J}_{z}=-i\left[\cos \Theta \frac{\partial}{\partial \omega}-\frac{1}{2} \cot \frac{\omega}{2} \sin \Theta \frac{\partial}{\partial \Theta}+\frac{1}{2} \cdot \frac{\partial}{\partial \Phi}\right] , \label{eq:4:5:9}\\
& \hat{J}_{z}^{\prime}=-i\left[\cos \Theta \frac{\partial}{\partial \omega}-\frac{1}{2} \cot \frac{\omega}{2} \sin \Theta \frac{\partial}{\partial \Theta}-\frac{1}{2} \cdot \frac{\partial}{\partial \Phi}\right] , \label{eq:4:5:10}\\
& \hat{J}^{2}=-\left[\frac{\partial^{2}}{\partial \omega^{2}}+\cot \frac{\omega}{2} \frac{\partial}{\partial \omega}+\frac{1}{4 \sin ^{2} \frac{\omega}{2}}\left(\frac{\partial^{2}}{\partial \Theta^{2}}+\cot \Theta \frac{\partial}{\partial \Theta}+\frac{1}{\sin ^{2} \Theta} \frac{\partial^{2}}{\partial \Phi^{2}}\right)\right] .\label{eq:4:5:11}
\end{align}
Thus, $U_{M M^{\prime}}^{J}(\omega ; \Theta, \Phi)$ is a solution of the differential equations
\begin{align}
{\left[\widehat{\vect{J}}^{2}-J(J+1)\right] U_{M M^{\prime}}^{J}(\omega ; \Theta, \Phi) } & =0  \notag\\
{\left[\widehat{J}_{z}+M\right] U_{M M^{\prime}}^{J}(\omega ; \Theta, \Phi) } & =0  ,\label{eq:4:5:12}\\
{\left[\widehat{J}_{z}+M^{\prime}\right] U_{M M^{\prime}}^{J}(\omega ; \Theta, \Phi) } & =0 .\notag
\end{align}
with the boundary conditions
\begin{gather}
U_{M M^{\prime}}^{J}(0 ; \Theta, \Phi)=\delta_{M M^{\prime}}  \notag\\
\left.\frac{\partial}{\partial \Phi} U_{M M^{\prime}}^{J}(\omega ; \Theta, \Phi)\right|_{\Theta=0}=\left.\frac{\partial}{\partial \Phi} U_{M M^{\prime}}^{J}(\omega ; \Theta, \Phi)\right|_{\Theta=\pi}=0 . \label{eq:4:5:13}
\end{gather}
\subsection{Orthogonality and Completeness}\label{sec:4.5.4}\index{orthogonality!of $U_{M M^{\prime}}^{J}$}\index{completeness!of $U_{M M^{\prime}}^{J}$}
A collection of the functions $U_{M M^{\prime}}^{J}(\omega ; \Theta, \Phi)$ with all possible integer and half-integer $J \geq 0$ constitutes a complete set of orthogonal functions of three variables $\omega, \Theta, \Phi$ defined in the domain
\begin{equation}
0 \leq \theta \leq \pi, \quad 0 \leq \Phi<2 \pi, \quad 0 \leq \omega<2 \pi, \label{eq:4:5:14}
\end{equation}
whose total volume is equal to $16 \pi^{2}$.\\
\subsubsection{ Orthogonality and Normalization}
\begin{equation}
4 \int_{0}^{2 \pi} d \omega \sin ^{2} \frac{\omega}{2} \int_{0}^{\pi} d \Theta \sin \Theta \int_{0}^{2 \pi} d \Phi U_{M_{1} M_{1}^{\prime}}^{J_{1} *}(\omega ; \Theta, \Phi) U_{M_{2} M_{2}^{\prime}}^{J_{2}}(\omega ; \Theta, \Phi)=\frac{16 \pi^{2}}{2 J_{1}+1} \delta_{J_{1} J_{2}} \delta_{M_{1} M_{2}} \delta_{M_{1}^{\prime} M_{2}^{\prime}} .\label{eq:4:5:15}
\end{equation}
\subsubsection{ Completeness}
\begin{equation}
\sum_{J=0,1 / 2,1 \ldots}^{\infty} \frac{2 J+1}{16 \pi^{2}} \sum_{M M^{\prime}} U_{M M^{\prime}}^{J}(\omega ; \Theta, \Phi) U_{M M^{\prime}}^{J *}(\tilde{\omega} ; \tilde{\Theta}, \tilde{\Phi})=\frac{\delta(\Phi-\tilde{\Phi}) \delta(\Theta-\tilde{\Theta}) \delta(\omega-\tilde{\omega})}{4 \sin \Theta \sin ^{2} \frac{\omega}{2}} .\label{eq:4:5:16}
\end{equation}
\subsection{Principal Properties}\label{sec:4.5.5}\index{rotation matrix!$U_{M M^{\prime}}^{J}$, principal properties}
\subsubsection{ Inverse transformation}
\begin{equation}
\left[U^{-1}(\omega ; \Theta, \Phi)\right]_{M M^{\prime}}^{J}=U_{M M^{\prime}}^{J}(-\omega ; \Theta, \Phi)=U_{M M^{\prime}}^{J}(\omega ; \pi-\Theta, \pi+\Phi) .\label{eq:4:5:17}
\end{equation}
\subsubsection{ Complex conjugation}
\begin{equation}
U_{M M^{\prime}}^{J *}(\omega ; \Theta, \Phi)=U_{M^{\prime} M}^{J}(\omega ; \pi-\Theta, \pi+\Phi)=U_{M^{\prime} M}^{J}(-\omega ; \Theta, \Phi) \label{eq:4:5:18}.
\end{equation}
\subsubsection{ Reversal of argument signs}
\begin{gather}
U_{M M^{\prime}}^{J}(-\omega ; \Theta, \Phi)=(-1)^{M-M^{\prime}} U_{-M^{\prime}-M}^{J}(\omega ; \Theta, \Phi)  ,\notag\\
U_{M M^{\prime}}^{J}(\omega ;-\Theta, \Phi)=(-1)^{M-M^{\prime}} U_{M M^{\prime}}^{J}(\omega ; \Theta, \Phi) , \label{eq:4:5:19}\\
U_{M M^{\prime}}^{J}(\omega ; \Theta,-\Phi)=U_{M^{\prime} M}^{J}(\omega ; \Theta, \Phi) .\notag
\end{gather}
\subsubsection{ Periodicity}
\begin{gather}
U_{M M^{\prime}}^{J}(\omega+4 \pi ; \Theta, \Phi)=U_{M M^{\prime}}^{J}(\omega ; \Theta, \Phi) , \notag\\
U_{M M^{\prime}}^{J}(\omega ; \Theta+2 \pi, \Phi)=U_{M M^{\prime}}^{J}(\omega ; \Theta, \Phi)  ,\label{eq:4:5:20}\\
U_{M M^{\prime}}^{J}(\omega ; \Theta, \Phi+2 \pi)=U_{M M^{\prime}}^{J}(\omega ; \Theta, \Phi) .\notag
\end{gather}
\subsubsection{Half-period Shift of Arguments}
\begin{align}
U_{M M^{\prime}}^{J}(\omega+2 \pi ; \Theta, \Phi) & =(-1)^{2 J} U_{M M^{\prime}}^{J}(\omega ; \Theta, \Phi), & U_{M M^{\prime}}^{J}(2 \pi-\omega ; \Theta, \Phi) & =(-1)^{M+M^{\prime}} U_{-M^{\prime}-M}^{J}(\omega ; \Theta, \Phi) , \notag\\
U_{M M^{\prime}}^{J}(\omega ; \Theta+\pi, \Phi) & =(-1)^{M-M^{\prime}} U_{-M^{\prime}-M}^{J}(\omega ; \Theta, \Phi), & U_{M M^{\prime}}^{J}(\omega ; \pi-\Theta, \Phi) & =U_{-M^{\prime}-M}^{J}(\omega ; \Theta, \Phi)  ,\notag\\
U_{M M^{\prime}}^{J}(\omega ; \Theta, \Phi+\pi) & =(-1)^{M-M^{\prime}} U_{M M^{\prime}}^{J}(\omega ; \Theta, \Phi), & U_{M M^{\prime}}^{J}(\omega ; \Theta, \pi-\Phi) & =(-1)^{M-M^{\prime}} U_{M^{\prime} M}^{J}(\omega ; \Theta, \Phi). \label{eq:4:5:21}
\end{align}
\subsubsection{ Permutation $M\rightleftarrows M^{\prime}$ and reversal of signs of $M$ and $M^{\prime}$}
\begin{align}
U_{M^{\prime} M}^{J}(\omega ; \Theta, \Phi) & =U_{M M^{\prime}}^{J}(\omega ; \Theta,-\Phi)=U_{M M^{\prime}}^{J *}(\omega ; \pi-\Theta, \pi+\Phi) , \notag\\
% VERIFIED: source misprint corrected -- book prints (omega; pi-Theta, pi+Phi)
% on the RHS; the correct third argument is pi-Phi (checked numerically vs the
% validated U-function; pi+Phi would force U_{-M-M'}=U_{-M'-M}). See
% scripts/check_4_5.py.
U_{-M-M^{\prime}}^{J}(\omega ; \Theta, \Phi) & =(-1)^{M-M^{\prime}} U_{M M^{\prime}}^{J}(\omega ; \pi-\Theta, \pi-\Phi) , \label{eq:4:5:22}\\
U_{M M^{\prime}}^{J}(\omega ; \Theta, \Phi) & =(-1)^{J-M^{\prime}} U_{-M M^{\prime}}^{J}\left(\omega_{1} ; \Theta_{1}, \Phi_{1}\right)=(-1)^{J+M} U_{M-M^{\prime}}^{J}\left(\omega_{1} ; \Theta_{1}, \Phi_{1}\right) .\notag
\end{align}
where angles $\omega_{1}, \Theta_{1}, \Phi_{1}$ and $\omega, \Theta, \Phi$ are related by the equations
\begin{gather}
\cos \frac{\omega_{1}}{2}=\sin \frac{\omega}{2} \sin \Theta \sin \Phi  \notag\\
\sin \frac{\omega_{1}}{2} \cos \Theta_{1}=-\sin \frac{\omega}{2} \sin \Theta \cos \Phi , \label{eq:4:5:23}\\
\cot \Phi_{1}=-\tan \frac{\omega}{2} \cos \Theta  ,\notag
\end{gather}
\begin{gather}
\cos \frac{\omega}{2}=\sin \frac{\omega_{1}}{2} \sin \Theta_{1} \sin \Phi_{1}  ,\notag\\
\sin \frac{\omega}{2} \cos \Theta=-\sin \frac{\omega_{1}}{2} \sin \Theta_{1} \cos \Phi_{1}  ,\label{eq:4:5:24}\\
\cot \Phi=-\tan \frac{\omega_{1}}{2} \cos \Theta_{1} ,\notag
\end{gather}
\begin{equation}
\sin ^{2} \frac{\omega_{1}}{2} \sin ^{2} \Theta_{1}+\sin ^{2} \frac{\omega}{2} \sin ^{2} \Theta =1 .\label{eq:4:5:25}
\end{equation}

Some relations which are valid for $D_{M M^{\prime}}^{J}(\alpha, \beta, \gamma)$ remain valid for $U_{M M^{\prime}}^{J}(\omega ; \Theta, \Phi)$. In particular, the Clebsch-Gordan expansion
\begin{equation}
U_{M_{1} M_{1}^{\prime}}^{J_{1}}(\omega ; \Theta, \Phi) U_{M_{2} M_{2}^{\prime}}^{J_{2}}(\omega ; \Theta, \Phi)=\sum_{J M} \clebsch{J_{1}}{M_{1}}{J_{2}}{M_{2}}{J}{M} U_{M M^{\prime}}^{J}(\omega ; \Theta, \Phi) \clebsch{J_{1}}{M_{1}^{\prime}}{J_{2}}{M_{2}^{\prime}}{J}{M^{\prime}} ,\label{eq:4:5:26}
\end{equation}
is equivalent to Eq.~\eqref{eq:4:6:1}, and the addition theorem
\begin{equation}
\sum_{M^{\prime \prime}} U_{M M^{\prime \prime}}^{J}\left(\omega_{2} ; \Theta_{2}, \Phi_{2}\right) U_{M^{\prime \prime} M^{\prime}}^{J}\left(\omega_{1} ; \Theta_{1}, \Phi_{1}\right)=U_{M M^{\prime}}^{J}(\omega ; \Theta, \Phi) \label{eq:4:5:27}
\end{equation}
is equivalent to Eq.~\eqref{eq:4:7:1}. The angles $\omega_{1}, \Theta_{1}, \Phi_{1}, \omega_{2}, \Theta_{2}, \Phi_{2}$ and $\omega, \Theta, \Phi$ are related by Eqs.~\eqref{eq:1:4:76}.

\subsection{Special Cases}\label{sec:4.5.6}\index{rotation matrix!$U_{M M^{\prime}}^{J}$, special cases}
\subsubsection{ The rotation axis $\mathrm{n}(\theta, \Phi)$ coincides with one of the coordinate axes}
\begin{enumerate}[label=(\roman*)]
\item The $x$-axis $\left(\Theta=\frac{\pi}{2}, \Phi=0\right)$
\begin{equation}
% VERIFIED: source misprint corrected -- book prints
% D(pi/2, omega, -pi/2) = (-i)^{M-M'} d(omega), which equals exp(+i omega J_x),
% the inverse of U(omega; pi/2,0)=exp(-i omega J_x).  The correct x-axis form is
% D(-pi/2, omega, pi/2) = (i)^{M-M'} d(omega) (checked vs the validated
% U-function; the y- and z-axis cases 4.5.29/4.5.30 are unaffected).
U_{M M^{\prime}}^{J}\left(\omega ; \frac{\pi}{2}, 0\right)=D_{M M^{\prime}}^{J}\left(-\frac{\pi}{2}, \omega,\frac{\pi}{2}\right)=(i)^{M-M^{\prime}} d_{M M^{\prime}}^{J}(\omega). \label{eq:4:5:28}
\end{equation}
\item The $y$-axis ( $\Theta=\frac{\pi}{2}, \Phi=\frac{\pi}{2}$ )
\begin{equation}
U_{M M^{\prime}}^{J}\left(\omega ; \frac{\pi}{2}, \frac{\pi}{2}\right)=D_{M M^{\prime}}^{J}(0, \omega, 0)=d_{M M^{\prime}}^{J}(\omega) .\label{eq:4:5:29}
\end{equation}
\item The $z$-axis $(\Theta=0)$
\begin{equation}
U_{M M^{\prime}}^{J}(\omega ; 0, \Phi)=\delta_{M M^{\prime}} e^{-i M \omega}. \label{eq:4:5:30}
\end{equation}
\end{enumerate}
\subsubsection{ Rotation about $\mathrm{n}(\Theta, \Phi)$ through a small angle $\omega \ll \pi / 2$}
\begin{gather}
U_{M M^{\prime}}^{J}(\omega ; \Theta, \Phi)=\delta_{M M^{\prime}}(1-i \omega M \cos \Theta)-\frac{i \omega}{2} e^{-i \Phi} \sin \Theta \sqrt{\left(J-M^{\prime}\right)(J+M)} \delta_{M, M^{\prime}+1}  ,\notag\\
-\frac{i \omega}{2} e^{i \Phi} \sqrt{(J-M)\left(J+M^{\prime}\right)} \delta_{M M^{\prime}-1}. \label{eq:4:5:31}
\end{gather}
\subsubsection{ Explicit forms of $U_{M M^{\prime}}^{J}(\omega ; \Theta, \Phi)$ at $M= \pm J$ and/or $M^{\prime}= \pm J$}
\begin{gather}
U_{J J}^{J}(\omega ; \theta, \Phi)=\left(\cos \frac{\omega}{2}-i \sin \frac{\omega}{2} \cos \Theta\right)^{2 J}, U_{-J-J}^{J}(\omega ; \theta, \Phi)=\left(\cos \frac{\omega}{2}+i \sin \frac{\omega}{2} \cos \Theta\right)^{2 J}  ,\notag\\
U_{J-J}^{J}(\omega ; \Theta, \Phi)=\left(-i \sin \frac{\omega}{2} \sin \Theta e^{-i \Phi}\right)^{2 J}, U_{-J J}^{J}(\omega ; \Theta, \Phi)=\left(-i \sin \frac{\omega}{2} \sin \Theta e^{i \Phi}\right)^{2 J}. \label{eq:4:5:32}
\end{gather}
\subsubsection{ Explicit forms of $U_{M M^{\prime}}^{J}(\omega ; \theta, \Phi)$ for $J=\frac{1}{2}, 1, \frac{3}{2}, 2$ are given below in Tables 4.23-4.26.}

\section{Sums involving $D$-functions}\label{sec:4.6}\index{Wigner D-function@Wigner $D$-function!sums involving}\index{sums!of the $D$-functions}
\subsection{The Clebsch-Gordan Series}\label{sec:4.6.1}\index{Clebsch-Gordan series}
The product of two $D$-functions with the same arguments may be expanded in the following series
\begin{equation}
D_{M_{1} N_{1}}^{J_{1}}(\alpha, \beta, \gamma) D_{M_{2} N_{2}}^{J_{2}}(\alpha, \beta, \gamma)=\sum_{J=\left|J_{1}-J_{2}\right|}^{J_{1}+J_{2}} \sum_{M N} \clebsch{J_{1}}{M_{1}}{J_{2}}{M_{2}}{J}{M} D_{M N}^{J}(\alpha, \beta, \gamma) \clebsch{J_{1}}{N_{1}}{J_{2}}{N_{2}}{J}{N}, \label{eq:4:6:1}
\end{equation}
Here $\clebsch{J_{1}}{M_{1}}{J_{2}}{M_{2}}{J}{M}$ is a Clebsch-Gordan coefficient (Chap.~\ref{chap:8}).\index{Clebsch-Gordan coefficient} The sum in Eq.~\eqref{eq:4:6:1} has $2 j+1$ terms, where $j=\min \left(J_{1}, J_{2}\right)$. Equation \eqref{eq:4:6:1} may be regarded as a particular case of the expansion of an arbitrary function in a series of the $D$-functions (Sec.~\ref{sec:4.10}).

\subsection{Some Applications of the Clebsch-Gordan Expansion}\label{sec:4.6.2}\index{Clebsch-Gordan series!applications of}
The Clebsch-Gordan expansion, Eq.~\eqref{eq:4:6:1}, together with the orthogonality condition of the Clebsch-Gordan coefficients, Eq.~\eqref{eq:8:1:8}, enable one to calculate sums of products of the $D$-functions with identical arguments. Hereafter we introduce the $3 j$-symbol $\left\{j_{1} j_{2} j_{3}\right\}$ defined by\isymo{braces-jjj@$\{j_{1} j_{2} j_{3}\}$, triangle symbol}\index{triangle symbol}\index{triangle condition}
\begin{equation}
\left\{j_{1} j_{2} j_{3}\right\}= \begin{cases}1 & \text { if } j_{1}+j_{2}+j_{3} \text { is integer and }\left|j_{1}-j_{2}\right| \leq j_{3} \leq j_{1}+j_{2},  \label{eq:4:6:2}\\ 0 & \text { otherwise },\end{cases}
\end{equation}
$\left\{j_{1} j_{2} j_{3}\right\}$ is invariant with respect to permutations of $j_{1}, j_{2}, j_{3}$. Using Eqs.~\eqref{eq:4:6:1} and \eqref{eq:8:1:8}, one obtains
\begin{align}
& \sum_{\substack{M_{1} M_{2} \\
N_{1} N_{2}}} \clebsch{J_{1}}{M_{1}}{J_{2}}{M_{2}}{J}{M} D_{M_{1} N_{1}}^{J_{1}}(\alpha, \beta, \gamma) D_{M_{2} N_{2}}^{J_{2}}(\alpha, \beta, \gamma) \clebsch{J_{1}}{N_{1}}{J_{2}}{N_{2}}{J^{\prime}}{N}=\delta_{J J^{\prime}}\left\{J_{1} J_{2} J\right\} D_{M N}^{J}(\alpha, \beta, \gamma)  ,\label{eq:4:6:3}\\
&  \sum_{J=\left|J_{1}-J_{2}\right|}^{J_{1}+J_{2}} \sum_{M_{1} M} \clebsch{J_{1}}{M_{1}}{J_{2}}{M_{2}}{J}{M} D_{M_{1} N_{1}}^{J_{1} *}(\alpha, \beta, \gamma) D_{M N}^{J}(\alpha, \beta, \gamma) \clebsch{J_{1}}{N_{1}^{\prime}}{J_{2}}{N_{2}}{J}{N}=\delta_{N_{1} N_{1}^{\prime}} D_{M_{2} N_{2}}^{J_{2}}(\alpha, \beta, \gamma)  ,\label{eq:4:6:4}\\
& \sum_{N_{1} N_{2}} D_{M_{1} N_{1}}^{J_{1}}(\alpha, \beta, \gamma) D_{M_{2} N_{2}}^{J_{2}}(\alpha, \beta, \gamma) \clebsch{J_{1}}{N_{1}}{J_{2}}{N_{2}}{J}{N}=\clebsch{J_{1}}{M_{1}}{J_{2}}{M_{2}}{J}{M} D_{M N}^{J}(\alpha, \beta, \gamma) , \label{eq:4:6:5}\\
& \sum_{N_{1} N_{2} N} D_{M N}^{J *}(\alpha, \beta, \gamma) D_{M_{1} N_{1}}^{J_{1}}(\alpha, \beta, \gamma) D_{M_{2} N_{2}}^{J_{2}}(\alpha, \beta, \gamma) \clebsch{J_{1}}{N_{1}}{J_{2}}{N_{2}}{J}{N}=\clebsch{J_{1}}{M_{1}}{J_{2}}{M_{2}}{J}{M}, \label{eq:4:6:6}\\
&\sum_{\substack{M_1M_2M\\N_1N_2N}}
\clebsch{J_{1}}{M_{1}}{J_{2}}{M_{2}}{J}{M}
 D_{M N}^{J *}(\alpha, \beta, \gamma) D_{M_{1} N_{1}}^{J_{1}}(\alpha, \beta, \gamma) D_{M_{2} N_{2}}^{J_{2}}(\alpha, \beta, \gamma) \clebsch{J_{1}}{N_{1}}{J_{2}}{N_{2}}{J^{\prime}}{N^{\prime}}=\delta_{J J^{\prime}} \delta_{N N^{\prime}}\left\{J_{1} J_{2} J\right\}. \label{eq:4:6:7}
\end{align}
\subsection{Generalization of the Clebsch-Gordan Expansion}\label{sec:4.6.3}\index{Clebsch-Gordan series!generalization of}
The Clebsch-Gordan expansion can be generalized to the case of an arbitrary number of $D$-functions of identical arguments by successive use of Eq.~\eqref{eq:4:6:3}. For instance, the summation of products of three $D$-functions yields
\begin{gather}
\sum_{\substack{M_{1} M_{2} M_{3} \\
N_{1} N_{2} N_{3}}} \clebsch{J_{12}}{M_{12}}{J_{3}}{M_{3}}{J}{M} \clebsch{J_{1}}{M_{1}}{J_{2}}{M_{2}}{J_{12}}{M_{12}} D_{M_{1} N_{1}}^{J_{1}}(\alpha, \beta, \gamma) D_{M_{2} N_{2}}^{J_{2}}(\alpha, \beta, \gamma) D_{M_{3} N_{3}}^{J_{3}}(\alpha, \beta, \gamma) \clebsch{J_{1}}{N_{1}}{J_{2}}{N_{2}}{J_{12}^{\prime}}{N_{12}} \clebsch{J_{12}^{\prime}}{N_{12}}{J_{3}}{N_{3}}{J^{\prime}}{N}  ,\notag\\
=\delta_{J J^{\prime}} \delta_{J_{12} J_{12}^{\prime}}\left\{J_{1} J_{2} J_{12}\right\}\left\{J_{12} J_{3} J\right\} D_{M N}^{J}(\alpha, \beta, \gamma) . \label{eq:4:6:8}
\end{gather}
In particular, for the case when $J_{3}=0$, Eq.~\eqref{eq:4:6:8} is reduced to Eq.~\eqref{eq:4:6:3}.\\
In general, the sum of products of $k D$-functions is given by
\begin{equation}
\sum_{\substack{m_{1}, \ldots, m_{k} \\ n_{1}, \ldots, n_{k}}} \prod_{i=1}^{k} \clebsch{J_{i-1}}{M_{i-1}}{j_{i}}{m_{i}}{J_{i}}{M_{i}} D_{m_{i} n_{i}}^{j_{i}}(\alpha, \beta, \gamma) \clebsch{J_{i-1}^{\prime}}{N_{i-1}}{j_{i}}{n_{i}}{J_{i}^{\prime}}{N_{i}}=D_{M_{k} N_{k}}^{J_{k}}(\alpha, \beta, \gamma) \prod_{i=1}^{k} \delta_{J_{i} J_{i}^{\prime}}\left\{j_{i} J_{i-1} J_{i}\right\} ,\label{eq:4:6:9}
\end{equation}
where $J_{i}(i=1,2, \ldots, k)$ is any angular momentum consistent with the vector addition rule,
\[
\vect{J}_{i}=\vect{j}_{1}+\vect{j}_{2}+\cdots+\vect{j}_{i}, \quad M_{i}=m_{1}+m_{2}+\cdots+m_{i}, \quad N_{i}=n_{1}+n_{2}+\cdots+n_{i},
\]
It is assumed in Eq.~\eqref{eq:4:6:9} that $J_{0}=J_{0}^{\prime}=M_{0}=N_{0}=0$.\\
In particular, for $j_{1}=j_{2}=\ldots=j_{k}=\frac{1}{2}$ and $J_{i+1}=J_{i}+\frac{1}{2}$ one has
\begin{align}
\sum_{\substack{m_{1}+\ldots+m_{k}=M \\
n_{1}+\ldots+n_{k}=N}} & D_{m_{1} n_{1}}^{\frac{1}{2}}(\alpha, \beta, \gamma) D_{m_{2} n_{2}}^{\frac{1}{2}}(\alpha, \beta, \gamma) \ldots D_{m_{k} n_{k}}^{\frac{1}{2}}(\alpha, \beta, \gamma)  ,\notag\\
& =\frac{(2 J)!}{\sqrt{(J+M)!(J-M)!(J+N)!(J-N)!}} D_{M N}^{J}(\alpha, \beta, \gamma) ,\label{eq:4:6:10}
\end{align}
where $J \equiv J_{k}=k / 2$ is either integer or half-integer. Similarly, for $j_{1}=j_{2}=\ldots=j_{k}=1$ and $J_{i+1}=J_{i}+1$,
\begin{gather}
\sum_{\substack{m_{1}+\ldots+m_{k}=M \\
n_{1}+\ldots+n_{k}=N}} \sqrt{\prod_{i, j=1}^{k}\left(1+\delta_{m_{i} 0}\right)\left(1+\delta_{n_{i} 0}\right)} D_{m_{1} n_{1}}^{1}(\alpha, \beta, \gamma) D_{m_{2} n_{2}}^{1}(\alpha, \beta, \gamma) \ldots D_{m_{k} n_{k}}^{1}(\alpha, \beta, \gamma)  ,\notag\\
=\frac{(2 J)!}{\sqrt{(J+M)!(J-M)!(J+N)!(J-N)!}} D_{M N}^{J}(\alpha, \beta, \gamma) ,\label{eq:4:6:11}
\end{gather}
where $J \equiv J_{k}=k$ is integer.\\
Equations \eqref{eq:4:6:10} and \eqref{eq:4:6:11} are useful for evaluating the Wigner $D$-functions. For example, Eq.~\eqref{eq:4:6:10} gives an explicit form of the $D$-functions in terms of the Cayley-Klein parameters (Sec. \ref{sec:1.4.3}.) defined by\index{Cayley-Klein parameters}\isym{ab-CK@$a, b$, Cayley-Klein parameters}
\begin{equation}
\left.\begin{array}{lc}
D_{\frac{1}{2} \frac{1}{2}}^{\frac{1}{2}}(\alpha, \beta, \gamma) \equiv a, & D_{\frac{1}{2}-\frac{1}{2}}^{\frac{1}{2}}(\alpha, \beta, \gamma) \equiv-b^{*}  .\label{eq:4:6:12}\\
D_{-\frac{1}{2} \frac{1}{2}}^{\frac{1}{2}}(\alpha, \beta, \gamma) \equiv b, & D_{-\frac{1}{2}-\frac{1}{2}}^{\frac{1}{2}}(\alpha, \beta, \gamma) \equiv a^{*}
\end{array}\right\}
\end{equation}
The appropriate expression for the $D$-functions has the form
\begin{equation}
D_{M N}^{J}(\alpha, \beta, \gamma)=\sqrt{(J+M)!(J-M)!(J+N)!(J-N)!} \sum_{p, q, r, s} \frac{(a)^{p}(b)^{q}\left(a^{*}\right)^{r}\left(-b^{*}\right)^{s}}{p!q!r!s!} .\label{eq:4:6:13}
\end{equation}
Here the summation indices $p, q, r$ and $s$ run over all integer values consistent with the condition
\begin{align}
& p+q+r+s=2 J  ,\notag\\
& p-q-r+s=2 M  ,\label{eq:4:6:14}\\
& p+q-r-s=2 N .\notag
\end{align}
According to Eqs.~\eqref{eq:4:6:14}, only one parameter from $p, q, r$ and $s$ is independent, i.e., in fact, Eq.~\eqref{eq:4:6:13} represents a single sum. The independent summation index may be chosen in different ways. This yields different explicit forms for the $D$-functions, Eqs.~\eqref{eq:4:3:2}-\eqref{eq:4:3:5}. For example, if $r$ is taken to be independent, Eq.~\eqref{eq:4:6:13} reduces to Eq.~\eqref{eq:4:3:2}.

\subsection{Determinant of Matrix $D_{M M^{\prime}}^{J}$}\label{sec:4.6.4}\index{rotation matrix!determinant of}\index{determinant!of the rotation matrix}
The determinant $\left\|D_{M M^{\prime}}^{J}\right\|$ of the rotation matrix is an invariant sum of products of $2 J+1 D$-functions. According to Eq.~\eqref{eq:4:1:7}, this matrix is unimodular, i.e.,
\begin{equation}
\sum_{P}(-1)^{P} D_{-J M_{1}}^{J}(\alpha, \beta, \gamma) D_{-J+1 M_{2}}^{J}(\alpha, \beta, \gamma) \ldots D_{J M_{2 J+1}}^{J}(\alpha, \beta, \gamma)=1 ,\label{eq:4:6:15}
\end{equation}
where $M_{1}, M_{2}, \ldots, M_{2 J+1}$ represent all possible permutations $P$ of $J, J-1, J-2, \ldots,-J$. The phase factor $(-1)^{P}$ equals +1 for even permutations, and -1 for odd ones.

\section{Addition of rotations}\label{sec:4.7}\index{rotation!addition of}\index{addition of rotations}
\subsection{The Addition Theorem for $D_{M M^{\prime}}^{J}(\alpha, \beta, \gamma)$}\label{sec:4.7.1}\index{addition theorem!for the $D$-functions}\index{Wigner D-function@Wigner $D$-function!addition theorem for}
Let two successive rotations of the coordinate system, $S\{x, y, z\} \rightarrow S^{\prime}\left\{x^{\prime}, y^{\prime}, z^{\prime}\right\}$ and $S^{\prime}\left\{x^{\prime}, y^{\prime}, z^{\prime}\right\} \rightarrow S^{\prime \prime}\left\{x^{\prime \prime}, y^{\prime \prime}, z^{\prime \prime}\right\}$, be described by the Euler angles $\alpha_{1}, \beta_{1}, \gamma_{1}$ and $\alpha_{2}, \beta_{2}, \gamma_{2}$, respectively, and the resultant rotation $S\{x, y, z\} \rightarrow S^{\prime \prime}\left\{x^{\prime \prime}, y^{\prime \prime}, z^{\prime \prime}\right\}$ be described by the angles $\alpha, \beta, \gamma$. In accordance with Sec. \ref{sec:1.4.7}, there are two alternative forms of the rotation addition.
\begin{enumerate}[label=(\alph*)]
\item The operator of the resultant rotation, $\widehat{D}(\alpha, \beta, \gamma)$, is given by Eq.~\eqref{eq:1:4:64}, if all rotations are performed according to scheme B (Sec. \ref{sec:1.4.1}) and the Euler angles $\alpha_{2}, \beta_{2}, \gamma_{2} ; \alpha_{1}, \beta_{1}, \gamma_{1}$ and $\alpha, \beta, \gamma$ are defined with respect to the initial system $S\{x, y, z\}$. Then the addition theorem reads
\begin{equation}
\sum_{M^{\prime \prime}=-J}^{J} D_{M M^{\prime \prime}}^{J}\left(\alpha_{2}, \beta_{2}, \gamma_{2}\right) D_{M^{\prime \prime} M^{\prime}}^{J}\left(\alpha_{1}, \beta_{1}, \gamma_{1}\right)=D_{M M^{\prime}}^{J}(\alpha, \beta, \gamma) ,\label{eq:4:7:1}
\end{equation}
where $\alpha, \beta, \gamma$ are related to $\alpha_{1}, \beta_{1}, \gamma_{1}$ and $\alpha_{2}, \beta_{2}, \gamma_{2}$ by Eqs.~\eqref{eq:1:4:66}-\eqref{eq:1:4:70}.\\
\item The operator of the resultant rotation $\hat{D}(\alpha, \beta, \gamma)$ is given by Eq.~\eqref{eq:1:4:73}, if $\alpha_{1}, \beta_{1}, \gamma_{1}$ and $\alpha, \beta, \gamma$ are defined with respect to the initial system $S\{x, y, z\}$, but $\alpha_{2}, \beta_{2}, \gamma_{2}$ are defined with respect to the intermediate system $S^{\prime}\left\{x^{\prime}, y^{\prime}, z^{\prime}\right\}$ (scheme B), or if successive rotations are performed according to scheme A. In these cases the addition theorem reads
\begin{equation}
\sum_{M^{\prime \prime}=-J}^{J} D_{M M^{\prime \prime}}^{J}\left(\alpha_{1}, \beta_{1}, \gamma_{1}\right) D_{M^{\prime \prime} M^{\prime}}^{J}\left(\alpha_{2}, \beta_{2}, \gamma_{2}\right)=D_{M M^{\prime}}^{J}(\alpha, \beta, \gamma) ,\label{eq:4:7:2}
\end{equation}
and $\alpha, \beta, \gamma$ are related to $\alpha_{1}, \beta_{1}, \gamma_{1}$ and $\alpha_{2}, \beta_{2}, \gamma_{2}$ by the equations which may be obtained from Eqs.~\eqref{eq:1:4:66}--\eqref{eq:1:4:70} by replacing $\left(\alpha_{1}, \beta_{1}, \gamma_{1}\right) \rightleftarrows\left(\alpha_{2}, \beta_{2}, \gamma_{2}\right)$.
\end{enumerate}
In particular,
\begin{equation}
\sum_{M^{\prime \prime}=-J}^{J} D_{M M^{\prime \prime}}^{J}\left(\alpha, \beta_{1}, \varphi\right) D_{M^{\prime \prime} M^{\prime}}^{J}\left(-\varphi, \beta_{2}, \gamma\right)=D_{M M^{\prime}}^{J}\left(\alpha, \beta_{1}+\beta_{2}, \gamma\right), \label{eq:4:7:3}
\end{equation}
where $\varphi$ is arbitrary, and
\begin{align}
\sum_{M^{\prime \prime}=-J}^{J} D_{M M^{\prime \prime}}^{J}(\alpha, \beta, \gamma) D_{M^{\prime} M^{\prime \prime}}^{J *}(\alpha, \beta, \gamma) & =\sum_{M^{\prime \prime}=-J}^{J} D_{M M^{\prime \prime}}^{J}(\alpha, \beta, \gamma) D_{M^{\prime \prime} M^{\prime}}^{J}(-\gamma,-\beta,-\alpha)  \notag\\
& =D_{M M^{\prime}}^{J}(0,0,0)=\delta_{M M^{\prime}}. \label{eq:4:7:4}
\end{align}
\subsection{The Addition Theorem for $d_{M M^{\prime}}^{J}(\beta)$}\label{sec:4.7.2}\index{addition theorem!for the $d$-functions}\index{Wigner d-function@Wigner (small) $d$-function!addition theorem for}
Equation \eqref{eq:4:7:2} may be rewritten as
\begin{equation}
\sum_{M^{\prime \prime}=-J}^{J} d_{M M^{\prime \prime}}^{J}\left(\beta_{1}\right) d_{M^{\prime \prime} M^{\prime}}^{J}\left(\beta_{2}\right) e^{-i M^{\prime \prime} \varphi}=e^{-i M \alpha} d_{M M^{\prime}}^{J}(\beta) e^{-i M^{\prime} \gamma}. \label{eq:4:7:5}
\end{equation}
Here
\begin{align}
& \cot \alpha=\cos \beta_{1} \cot \varphi+\cot \beta_{2} \frac{\sin \beta_{1}}{\sin \varphi}  ,\notag\\
& \cos \beta=\cos \beta_{1} \cos \beta_{2}-\sin \beta_{1} \sin \beta_{2} \cos \varphi  ,\label{eq:4:7:6}\\
& \cot \gamma=\cos \beta_{2} \cot \varphi+\cot \beta_{1} \frac{\sin \beta_{2}}{\sin \varphi} .\notag
\end{align}
Equation \eqref{eq:4:7:5} is simplified in the following particular cases. If $\varphi=0$ and $\beta_{1}+\beta_{2} \leq \pi$, then $\alpha=0, \beta=\beta_{1}+\beta_{2}$ and $\gamma=0$,
\begin{equation}
\sum_{M^{\prime \prime}=-J}^{J} d_{M M^{\prime \prime}}^{J}\left(\beta_{1}\right) d_{M^{\prime \prime} M^{\prime}}^{J}\left(\beta_{2}\right)=d_{M M^{\prime}}^{J}\left(\beta_{1}+\beta_{2}\right). \label{eq:4:7:7}
\end{equation}
If $\varphi=0$ and $\beta_{1}+\beta_{2}>\pi$, then $\alpha=\pi, \beta=2 \pi-\beta_{1}-\beta_{2}$ and $\gamma=\pi$,
\begin{equation}
\sum_{M^{\prime \prime}=-J}^{J} d_{M M^{\prime \prime}}^{J}\left(\beta_{1}\right) d_{M^{\prime \prime} M^{\prime}}^{J}\left(\beta_{2}\right)=(-1)^{M+M^{\prime}} d_{M M^{\prime}}^{J}\left(2 \pi-\beta_{1}-\beta_{2}\right). \label{eq:4:7:8}
\end{equation}
If $\varphi=\pi$ and $\beta_{1} \geq \beta_{2}$, then $\alpha=0, \beta=\beta_{1}-\beta_{2}$ and $\gamma=\pi$,
\begin{equation}
\sum_{M^{\prime \prime}=-J}^{J}(-1)^{M^{\prime \prime}-M^{\prime}} d_{M M^{\prime \prime}}^{J}\left(\beta_{1}\right) d_{M^{\prime \prime} M^{\prime}}^{J}\left(\beta_{2}\right)=d_{M M^{\prime}}^{J}\left(\beta_{1}-\beta_{2}\right) \label{eq:4:7:9}
\end{equation}
In particular, for $\beta_{1}=\beta_{2}$,
\begin{equation}
\sum_{M^{\prime \prime}=-J}^{J}(-1)^{M^{\prime \prime}-M^{\prime}} d_{M M^{\prime \prime}}^{J}(\beta) d_{M^{\prime \prime} M^{\prime}}^{J}(\beta)=\delta_{M M^{\prime}}. \label{eq:4:7:10}
\end{equation}
If $\varphi=\pi / 2$, then
\begin{equation}
\sum_{M^{\prime \prime}=-J}^{J}(-i)^{M^{\prime \prime}} d_{M M^{\prime \prime}}^{J}\left(\beta_{1}\right) d_{M^{\prime \prime} M^{\prime}}^{J}\left(\beta_{2}\right)=e^{-i M \alpha} d_{M M^{\prime}}^{J}(\beta) e^{-i M^{\prime} \gamma}. \label{eq:4:7:11}
\end{equation}
where
\begin{align}
& \cot \alpha=\cot \beta_{2} \sin \beta_{1} , \notag\\
& \cos \beta=\cos \beta_{1} \cos \beta_{2}  ,\label{eq:4:7:12}\\
& \cot \gamma=\cot \beta_{1} \sin \beta_{2} .\notag
\end{align}
\subsection{Addition of Two Identical Rotations}\label{sec:4.7.3}\index{rotation!addition of two identical}
When $\alpha_{1}=\alpha_{2}, \beta_{1}=\beta_{2}$ and $\gamma_{1}=\gamma_{2}$ the addition theorem, Eq.~\eqref{eq:4:7:1}, together with the Clebsch-Gordan expansion, Eq.~\eqref{eq:4:6:1}, yield
\begin{equation}
\sum_{J=0,1, \ldots}^{2 j} \sum_{m^{\prime \prime}} \clebsch{j}{m}{j}{m^{\prime \prime}}{J}{m+m^{\prime \prime}} D_{m+m^{\prime \prime} m^{\prime \prime}+m^{\prime}}^{J}(\alpha, \beta, \gamma) \clebsch{j}{m^{\prime \prime}}{j}{m^{\prime}}{J}{m^{\prime \prime}+m^{\prime}}=D_{m m^{\prime}}^{j}(\bar{\alpha}, \bar{\beta}, \bar{\gamma}) ,\label{eq:4:7:13}
\end{equation}
where
\begin{gather}
\cos \bar{\beta}=\cos ^{2} \beta-\sin ^{2} \beta \cos (\alpha+\gamma)  ,\notag\\
\tan (\bar{\alpha}-\alpha)=\tan (\bar{\gamma}-\gamma)=\frac{\tan \frac{\alpha+\gamma}{2}}{\cos \beta} .\label{eq:4:7:14}
\end{gather}
In particular, if $\alpha=\gamma=0$,
\begin{equation}
\sum_{J=0,1, \ldots}^{2 j} \sum_{m^{\prime \prime}} \clebsch{j}{m}{j}{m^{\prime \prime}}{J}{m+m^{\prime \prime}} d_{m+m^{\prime \prime} m^{\prime \prime}+m^{\prime}}^{J}(\beta) \clebsch{j}{m^{\prime \prime}}{j}{m^{\prime}}{J}{m^{\prime \prime}+m^{\prime}}=d_{m m^{\prime}}^{j}(2 \beta) .\label{eq:4:7:15}
\end{equation}
\subsection{The Multiplication Theorem for $d_{M M^{\prime}}^{J}(\beta)$}\label{sec:4.7.4}\index{multiplication theorem!for the $d$-functions}\index{Wigner d-function@Wigner (small) $d$-function!multiplication theorem for}
Equation \eqref{eq:4:7:5} leads to a representation of products of two $d_{M M^{\prime}}^{J}(\beta)$ functions
\begin{equation}
d_{M M^{\prime \prime}}^{J}\left(\beta_{1}\right) d_{M^{\prime \prime} M^{\prime}}^{J}\left(\beta_{2}\right)=\frac{1}{4 \pi} \int_{-2 \pi}^{2 \pi} e^{i\left(M^{\prime \prime} \varphi-M \alpha-M^{\prime} \gamma\right)} d_{M M^{\prime}}^{J}(\beta) d \varphi .\label{eq:4:7:16}
\end{equation}
Here $\alpha, \beta, \gamma$ are related to $\beta_{1}, \beta_{2}$ and $\varphi$ by means of Eq.~\eqref{eq:4:7:6}.

\subsection{Sums Involving the D-Functions of Different Arguments}\label{sec:4.7.5}\index{Wigner D-function@Wigner $D$-function!sums with different arguments}
In addition to Eqs.~\eqref{eq:4:7:1} or \eqref{eq:4:7:2} one can derive the following invariant sums which are equivalent to the characters of irreducible representations of rotation group for two rotations (Sec.~\ref{sec:4.14}):
\begin{enumerate}[label=(\alph*)]
\item
\begin{equation}
\sum_{M M^{\prime}} D_{M M^{\prime}}^{J}\left(\alpha_{1}, \beta_{1}, \gamma_{1}\right) D_{M^{\prime} M}^{J}\left(\alpha_{2}, \beta_{2}, \gamma_{2}\right) \equiv \chi^{J}\left(R_{1} R_{2}\right)=\chi^{J}\left(R_{2} R_{1}\right)=\frac{\sin \left[(2 J+1) \frac{\omega}{2}\right]}{\sin \frac{\omega}{2}} ,\label{eq:4:7:17}
\end{equation}
where
\begin{align}
\cos \frac{\omega}{2} & =\cos \frac{\beta_{1}}{2} \cos \frac{\beta_{2}}{2} \cos \frac{\alpha_{1}+\gamma_{1}+\alpha_{2}+\gamma_{2}}{2}-\sin \frac{\beta_{1}}{2} \sin \frac{\beta_{2}}{2} \cos \frac{\alpha_{1}-\gamma_{1}-\alpha_{2}+\gamma_{2}}{2}  ,\notag\\
& =\cos \frac{\beta_{1}+\beta_{2}}{2} \cos \frac{\alpha_{1}+\gamma_{2}}{2} \cos \frac{\gamma_{1}+\alpha_{2}}{2}-\cos \frac{\beta_{1}-\beta_{2}}{2} \sin \frac{\alpha_{1}+\gamma_{2}}{2} \sin \frac{\gamma_{1}+\alpha_{2}}{2} .\label{eq:4:7:18}
\end{align}
\item
\begin{equation}
\sum_{M M^{\prime}} D_{M M^{\prime}}^{J}\left(\alpha_{1}, \beta_{1}, \gamma_{1}\right) D_{M M^{\prime}}^{J *}\left(\alpha_{2}, \beta_{2}, \gamma_{2}\right) \equiv \chi^{J}\left(R_{1} R_{2}^{-1}\right)=\chi^{J}\left(R_{1}^{-1} R_{2}\right)=\frac{\sin \left[(2 J+1) \frac{\omega^{\prime}}{2}\right]}{\sin \frac{\omega^{\prime}}{2}} \label{eq:4:7:19}
\end{equation}
where
\begin{align}
\cos \frac{\omega^{\prime}}{2} & =\cos \frac{\beta_{1}}{2} \cos \frac{\beta_{2}}{2} \cos \frac{\alpha_{1}+\gamma_{1}-\alpha_{2}-\gamma_{2}}{2}+\sin \frac{\beta_{1}}{2} \sin \frac{\beta_{2}}{2} \cos \frac{\alpha_{1}-\gamma_{1}-\alpha_{2}+\gamma_{2}}{2}  ,\notag\\
& =\cos \frac{\beta_{1}-\beta_{2}}{2} \cos \frac{\alpha_{1}-\alpha_{2}}{2} \cos \frac{\gamma_{1}-\gamma_{2}}{2}-\cos \frac{\beta_{1}+\beta_{2}}{2} \sin \frac{\alpha_{1}-\alpha_{2}}{2} \sin \frac{\gamma_{1}-\gamma_{2}}{2} .\label{eq:4:7:20}
\end{align}
Equation \eqref{eq:4:7:20} may be obtained from Eq.~\eqref{eq:4:7:18} by replacing $\left(\alpha_{2}, \beta_{2}, \gamma_{2}\right) \rightarrow\left(-\gamma_{2},-\beta_{2},-\alpha_{2}\right)$.
\end{enumerate}

Similarly, one can arrive at invariant sums which may be reduced to the characters involving three and more rotations, i.e., $\chi^{J}\left(R_{1} R_{2} R_{3}\right)$ etc.

\subsection{The Ponzano-Regge sum}\label{sec:4.7.6}\index{Ponzano-Regge sum}\index{sums!Ponzano-Regge}
Note the following sum involving products of three $d_{M M^{\prime}}^{J}(\beta)$ functions with the same $J$ but different arguments (Ref. \cite{ref089})
\begin{equation}
\sum_{J=J_{\mathrm{min}}}^{\infty}(2 J+1) d_{M_{1} M_{2}}^{J}\left(\beta_{3}\right) d_{M_{2} M_{3}}^{J}\left(\beta_{1}\right) d_{M_{3} M_{1}}^{J}\left(\beta_{2}\right)=\frac{2 \Theta(B)}{\pi \sqrt{B}} \cos \left(\sum_{i=1}^{3} M_{i} \delta_{i}\right) .\label{eq:4:7:21}
\end{equation}
Here the summation index $J$ runs from $J_{\text {min }}=\max \left\{\left|M_{1}\right|,\left|M_{2}\right|,\left|M_{3}\right|\right\}$ to infinity, $M_{1}, M_{2}, M_{3}$ being fixed, and
\begin{gather}
B=\left|\begin{array}{ccc}
1 & \cos \beta_{3} & \cos \beta_{2} \\
\cos \beta_{3} & 1 & \cos \beta_{1} \\
\cos \beta_{2} & \cos \beta_{1} & 1
\end{array}\right|,  \label{eq:4:7:22}\\
\Theta(x)=\left\{\begin{array}{cc}
1 & x>0 \\
0 & x<0
\end{array}\right. \label{eq:4:7:23}
\end{gather}
\begin{figure}[htbp!]
\begin{center}
\includegraphics{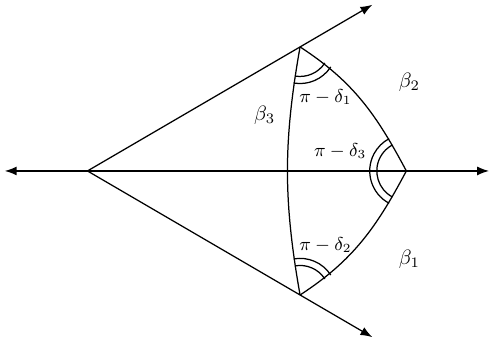}
\caption{Geometrical interpretation of the angles in Eq.~\eqref{eq:4:7:21}. The arguments $\beta_1,\beta_2,\beta_3$ are the sides of a spherical triangle, i.e.\ the angles they subtend at the centre of the sphere, whose three rays are shown emanating from the centre. The interior angle of the triangle at the vertex opposite the side $\beta_i$ equals $\pi-\delta_i$. The $\delta_i$ are thereby determined by $\beta_1,\beta_2,\beta_3$ through the spherical cosine and sine rules \eqref{eq:4:7:24}--\eqref{eq:4:7:25}.}
\label{chap4:fig:1}
\end{center}
\end{figure}

The relations between the angles $\delta_{1}, \delta_{2}, \delta_{3}$ and $\beta_{1}, \beta_{2}, \beta_{3}$ are
\begin{gather}
\cos \delta_{i}=\frac{\cos \beta_{j} \cos \beta_{k}-\cos \beta_{i}}{\sin \beta_{j} \sin \beta_{k}}, \quad(i \neq j \neq k),  \label{eq:4:7:24}\\
\frac{\sin \beta_{i}}{\sin \delta_{i}}=\frac{\sin \beta_{j}}{\sin \delta_{j}}, \quad(i, j=1,2,3) . \label{eq:4:7:25}
\end{gather}
These relations may be easily obtained by using spherical trigonometry (see Fig.~\ref{chap4:fig:1}).\index{spherical trigonometry}\index{spherical triangle}

\section{Recursion relations for $D_{M M^{\prime}}^{J}$}\label{sec:4.8}\index{recursion relations!for the $D$-functions}\index{Wigner D-function@Wigner $D$-function!recursion relations for}
\subsection{Relations between $D^{J}$ and $D^{J \pm 1}$}\label{sec:4.8.1}\index{recursion relations!in $J$}
\begin{align}
\cos \beta D_{M M^{\prime}}^{J}(\alpha, \beta, \gamma)= & \frac{\sqrt{\left(J^{2}-M^{2}\right)\left(J^{2}-M^{\prime 2}\right)}}{J(2 J+1)} D_{M M^{\prime}}^{J-1}(\alpha, \beta, \gamma)+\frac{M M^{\prime}}{J(J+1)} D_{M M^{\prime}}^{J}(\alpha, \beta, \gamma)  \notag\\
+ & \frac{\sqrt{\left((J+1)^{2}-M^{2}\right]\left[(J+1)^{2}-M^{\prime 2}\right]}}{(J+1)(2 J+1)} D_{M M^{\prime}}^{J+1}(\alpha, \beta, \gamma)  \label{eq:4:8:1}
\end{align}\begin{align}
\sin \beta e^{i \alpha} D_{M+1 M^{\prime}}^{J}(\alpha, \beta, \gamma)= & -\frac{\sqrt{(J+M)(J+M+1)\left(J^{2}-M^{\prime 2}\right)}}{J(2 J+1)} D_{M M^{\prime}}^{J-1}(\alpha, \beta, \gamma)  \notag\\
+ & \frac{M^{\prime} \sqrt{(J-M)(J+M+1)}}{J(J+1)} D_{M M^{\prime}}^{J}(\alpha, \beta, \gamma)  \notag\\
+ & \frac{\sqrt{(J-M)(J-M+1)\left[(J+1)^{2}-M^{\prime 2}\right]}}{(J+1)(2 J+1)} D_{M M^{\prime}}^{J+1}(\alpha, \beta, \gamma)  \label{eq:4:8:2}
\end{align}\begin{align}
\sin \beta e^{-i \alpha} D_{M-1 M^{\prime}}^{J}(\alpha, \beta, \gamma)= & \frac{\sqrt{(J-M)(J-M+1)\left(J^{2}-M^{\prime 2}\right)}}{J(2 J+1)} D_{M M^{\prime}}^{J-1}(\alpha, \beta, \gamma)  \notag\\
& +\frac{M^{\prime} \sqrt{(J+M)(J-M+1)}}{J(J+1)} D_{M M^{\prime}}^{J}(\alpha, \beta, \gamma)  \notag\\
& -\frac{\sqrt{(J+M)(J+M+1)\left[(J+1)^{2}-M^{\prime 2}\right]}}{(J+1)(2 J+1)} D_{M M^{\prime}}^{J+1}(\alpha, \beta, \gamma). \label{eq:4:8:3}
\end{align}
\begin{align}
 \sin \beta e^{i \gamma} D_{M M^{\prime}+1}^{J}(\alpha, \beta, \gamma)&= \frac{\sqrt{\left(J^{2}-M^{2}\right)\left(J+M^{\prime}\right)\left(J+M^{\prime}+1\right)}}{J(2 J+1)} D_{M M^{\prime}}^{J-1}(\alpha, \beta, \gamma)  \notag\\
&- \frac{M \sqrt{\left(J-M^{\prime}\right)\left(J+M^{\prime}+1\right)}}{J(J+1)} D_{M M^{\prime}}^{J}(\alpha, \beta, \gamma)  \notag\\
&- \frac{\sqrt{\left((J+1)^{2}-M^{2}\right)\left(J-M^{\prime}\right)\left(J-M^{\prime}+1\right)}}{(J+1)(2 J+1)} D_{M M^{\prime}}^{J+1}(\alpha, \beta, \gamma),  \label{eq:4:8:4}
\end{align}
\begin{align}
 \sin \beta e^{-i \gamma} D_{M M^{\prime}-1}^{J}(\alpha, \beta, \gamma)&=-\frac{\sqrt{\left(J^{2}-M^{2}\right)\left(J-M^{\prime}\right)\left(J-M^{\prime}+1\right)} D_{M M^{\prime}}^{J-1}(\alpha, \beta, \gamma)}{J(2 J+1)}  \notag\\
&- \frac{M \sqrt{\left(J+M^{\prime}\right)\left(J-M^{\prime}+1\right)} D_{M M^{\prime}}^{J}(\alpha, \beta, \gamma)}{J(J+1)}  \notag\\
&+ \frac{\sqrt{\left[(J+1)^{2}-M^{2}\right]\left(J+M^{\prime}\right)\left(J+M^{\prime}+1\right)} D_{M M^{\prime}}^{J+1}(\alpha, \beta, \gamma)}{(J+1)(2 J+1)} , \label{eq:4:8:5}
\end{align}
\begin{align}
(1+\cos \beta) e^{i(\alpha+\gamma)} D_{M+1 M^{\prime}+1}^{J}(\alpha, \beta, \gamma)&= \frac{\sqrt{(J+M+1)(J+M)\left(J+M^{\prime}+1\right)\left(J+M^{\prime}\right)}}{J(2 J+1)} D_{M M^{\prime}}^{J-1}(\alpha, \beta, \gamma)  \notag\\
&+\frac{\sqrt{(J-M)(J+M+1)\left(J-M^{\prime}\right)\left(J+M^{\prime}+1\right)}}{J(J+1)} D_{M M^{\prime}}^{J}(\alpha, \beta, \gamma)  \notag\\
&+\frac{\sqrt{(J-M)(J-M+1)\left(J-M^{\prime}\right)\left(J-M^{\prime}+1\right)}}{(J+1)(2 J+1)} D_{M M^{\prime}}^{J+1}(\alpha, \beta, \gamma), \label{eq:4:8:6}
\end{align}
\begin{align}
(1+\cos \beta) e^{-i(\alpha+\gamma)} D_{M-1 M^{\prime}-1}^{J}(\alpha, \beta, \gamma)&= \frac{\sqrt{(J-M)(J-M+1)\left(J-M^{\prime}\right)\left(J-M^{\prime}+1\right)}}{J(2 J+1)} D_{M M^{\prime}}^{J-1}(\alpha, \beta, \gamma)  \notag\\
&+\frac{\sqrt{(J+M)(J-M+1)\left(J+M^{\prime}\right)\left(J-M^{\prime}+1\right)}}{J(J+1)} D_{M M^{\prime}}^{J}(\alpha, \beta, \gamma)  \notag\\
&+\frac{\sqrt{(J+M)(J+M+1)\left(J+M^{\prime}\right)\left(J+M^{\prime}+1\right)}}{(J+1)(2 J+1)} D_{M M^{\prime}}^{J+1}(\alpha, \beta, \gamma),  \label{eq:4:8:7}
\end{align}\begin{align}
(1-\cos \beta) e^{i(\alpha-\gamma)} D_{M+1 M^{\prime}-1}^{J}(\alpha, \beta, \gamma) &= \frac{\sqrt{(J+M)(J+M+1)\left(J-M^{\prime}\right)\left(J-M^{\prime}+1\right)}}{J(2 J+1)} D_{M M^{\prime}}^{J-1}(\alpha, \beta, \gamma)  \notag\\
&-\frac{\sqrt{(J-M)(J+M+1)\left(J+M^{\prime}\right)\left(J-M^{\prime}+1\right)}}{J(J+1)} D_{M M^{\prime}}^{J}(\alpha, \beta, \gamma)  \notag\\
&+\frac{\sqrt{(J-M)(J-M+1)\left(J+M^{\prime}\right)\left(J+M^{\prime}+1\right)}}{(J+1)(2 J+1)} D_{M M^{\prime}}^{J+1}(\alpha, \beta, \gamma),  \label{eq:4:8:8}
\end{align}
\begin{align}
(1-\cos \beta) e^{-i(\alpha-\gamma)} D_{M-1 M^{\prime}+1}^{J}(\alpha, \beta, \gamma)&= \frac{\sqrt{(J-M)(J-M+1)\left(J+M^{\prime}\right)\left(J+M^{\prime}+1\right)}}{J(2 J+1)} D_{M M^{\prime}}^{J-1}(\alpha, \beta, \gamma)  \notag\\
&-\frac{\sqrt{(J+M)(J-M+1)\left(J-M^{\prime}\right)\left(J+M^{\prime}+1\right)}}{J(J+1)} D_{M M^{\prime}}^{J}(\alpha, \beta, \gamma)  \notag\\
&+\frac{\sqrt{(J+M)(J+M+1)\left(J-M^{\prime}\right)\left(J-M^{\prime}+1\right)}}{(J+1)(2 J+1)} D_{M M^{\prime}}^{J+1}(\alpha, \beta, \gamma)  .\label{eq:4:8:9}
\end{align}
Equations \eqref{eq:4:8:1}-\eqref{eq:4:8:9} may be obtained from the Clebsch-Gordan series, Eq.~\eqref{eq:4:6:1} with $J_{1}=1$.

\subsection{Relations Between $D^{J}$ and $D^{J \pm 1 / 2}$}\label{sec:4.8.2}\index{recursion relations!in half-integer steps of $J$}
\begin{align}
\cos \frac{\beta}{2} e^{i \frac{\alpha+\gamma}{2}} D_{M+\frac{1}{2} M^{\prime}+\frac{1}{2}}^{J}(\alpha, \beta, \gamma)= & \frac{\sqrt{\left(J+M+\frac{1}{2}\right)\left(J+M^{\prime}+\frac{1}{2}\right)}}{2 J+1} D_{M M^{\prime}}^{J-\frac{1}{2}}(\alpha, \beta, \gamma)  \notag\\
& +\frac{\sqrt{\left(J-M+\frac{1}{2}\right)\left(J-M^{\prime}+\frac{1}{2}\right)}}{2 J+1} D_{M M^{\prime}}^{J+\frac{1}{2}}(\alpha, \beta, \gamma),  \label{eq:4:8:10}
\end{align}\begin{align}
\sin \frac{\beta}{2} e^{i \frac{\alpha-\gamma}{2}} D_{M+\frac{1}{2} M^{\prime}-\frac{1}{2}}^{J}(\alpha, \beta, \gamma)= & -\frac{\sqrt{\left(J+M+\frac{1}{2}\right)\left(J-M^{\prime}+\frac{1}{2}\right)}}{2 J+1} D_{M M^{\prime}}^{J-\frac{1}{2}}(\alpha, \beta, \gamma)  \notag\\
& +\frac{\sqrt{\left(J-M+\frac{1}{2}\right)\left(J+M^{\prime}+\frac{1}{2}\right)}}{2 J+1} D_{M M^{\prime}}^{J+\frac{1}{2}}(\alpha, \beta, \gamma),  \label{eq:4:8:11}
\end{align}\begin{align}
\cos \frac{\beta}{2} e^{-i \frac{\alpha+\gamma}{2}} D_{M-\frac{1}{2} M^{\prime}-\frac{1}{2}}^{J}(\alpha, \beta, \gamma)= & \frac{\sqrt{\left(J-M+\frac{1}{2}\right)\left(J-M^{\prime}+\frac{1}{2}\right)}}{2 J+1} D_{M M^{\prime}}^{J-\frac{1}{2}}(\alpha, \beta, \gamma)  \notag\\
& +\frac{\sqrt{\left(J+M+\frac{1}{2}\right)\left(J+M^{\prime}+\frac{1}{2}\right)}}{2 J+1} D_{M M^{\prime}}^{J+\frac{1}{2}}(\alpha, \beta, \gamma),  \label{eq:4:8:12}
\end{align}\begin{align}
\sin \frac{\beta}{2} e^{-i \frac{\alpha-\gamma}{2}} D_{M-\frac{1}{2} M^{\prime}+\frac{1}{2}}^{J}(\alpha, \beta, \gamma)= & \frac{\sqrt{\left(J-M+\frac{1}{2}\right)\left(J+M^{\prime}+\frac{1}{2}\right)}}{2 J+1} D_{M M^{\prime}}^{J-\frac{1}{2}}(\alpha, \beta, \gamma)  \notag\\
& -\frac{\sqrt{\left(J+M+\frac{1}{2}\right)\left(J-M^{\prime}+\frac{1}{2}\right)}}{2 J+1} D_{M M^{\prime}}^{J+\frac{1}{2}}(\alpha, \beta, \gamma) . \label{eq:4:8:13}
\end{align}
Equations \eqref{eq:4:8:10}-\eqref{eq:4:8:13} may be obtained from Eq.~\eqref{eq:4:6:1} with $J_{1}=\frac{1}{2}$. They yield
\begin{align}
D_{M M^{\prime}}^{J}(\alpha, \beta, \gamma)= & \sqrt{\frac{J-M}{J-M^{\prime}}} \cos \frac{\beta}{2} e^{i \frac{\alpha+\gamma}{2}} D_{M+\frac{1}{2} M^{\prime}+\frac{1}{2}}^{J-\frac{1}{2}}(\alpha, \beta, \gamma)  \notag\\
& -\sqrt{\frac{J+M}{J-M^{\prime}}} \sin \frac{\beta}{2} e^{-i \frac{\alpha-\gamma}{2}} D_{M-\frac{1}{2} M^{\prime}+\frac{1}{2}}^{J-\frac{1}{2}}(\alpha, \beta, \gamma), \quad\left(M^{\prime} \neq J\right),  \label{eq:4:8:14}
\end{align}\begin{align}
D_{M M^{\prime}}^{J}(\alpha, \beta, \gamma)= & \sqrt{\frac{J-M}{J+M^{\prime}}} \sin \frac{\beta}{2} e^{i \frac{\alpha-\gamma}{2}} D_{M+\frac{1}{2} M^{\prime}-\frac{1}{2}}^{J-\frac{1}{2}}(\alpha, \beta, \gamma)  \notag\\
& +\sqrt{\frac{J+M}{J+M^{\prime}}} \cos \frac{\beta}{2} e^{-i \frac{\alpha+\gamma}{2}} D_{M-\frac{1}{2} M^{\prime}-\frac{1}{2}}^{J-\frac{1}{2}}(\alpha, \beta, \gamma), \quad\left(M^{\prime} \neq-J\right) . \label{eq:4:8:15}
\end{align}
\subsection{Relations Between $D_{M M^{\prime}}^{J}$ and $D_{M \pm 1 M^{\prime} \mp 1}^{J}$}\label{sec:4.8.3}\index{recursion relations!in $M$ and $M^{\prime}$}
\begin{align}
 \frac{-M+M^{\prime} \cos \beta}{\sin \beta} D_{M M^{\prime}}^{J}(\alpha, \beta, \gamma)=& \frac{1}{2} \sqrt{\left(J+M^{\prime}\right)\left(J-M^{\prime}+1\right)} D_{M M^{\prime}-1}^{J}(\alpha, \beta, \gamma) e^{-i \gamma} \notag\\
+&\frac{1}{2} \sqrt{\left(J-M^{\prime}\right)\left(J+M^{\prime}+1\right)} D_{M M^{\prime}+1}^{J}(\alpha, \beta, \gamma) e^{i \gamma}, \label{eq:4:8:16}
\end{align}\begin{align}
 \frac{M^{\prime}-M \cos \beta}{\sin \beta} D_{M M^{\prime}}^{J}(\alpha, \beta, \gamma)=& \frac{1}{2} \sqrt{(J+M)(J-M+1)} D_{M-1 M^{\prime}}^{J}(\alpha, \beta, \gamma) e^{-i \alpha} \notag\\
+&\frac{1}{2} \sqrt{(J-M)(J+M+1)} D_{M+1 M^{\prime}}^{J}(\alpha, \beta, \gamma) e^{i \alpha},  \label{eq:4:8:17}
\end{align}\begin{align}
D_{M+1 M^{\prime}}^{J}(\alpha, \beta, \gamma) e^{i \alpha}= & \sqrt{\frac{\left(J+M^{\prime}\right)\left(J-M^{\prime}+1\right)}{(J-M)(J+M+1)}} \frac{1+\cos \beta}{2} e^{-i \gamma} D_{M M^{\prime}-1}^{J}(\alpha, \beta, \gamma)\notag \\
+ & \frac{M^{\prime} \sin \beta}{\sqrt{(J-M)(J+M+1)}} D_{M M^{\prime}}^{J}(\alpha, \beta, \gamma) \notag\\
- & \sqrt{\frac{\left(J-M^{\prime}\right)\left(J+M^{\prime}+1\right)}{(J-M)(J+M+1)}} \frac{1-\cos \beta}{2} e^{i \gamma} D_{M M^{\prime}+1}^{J}(\alpha, \beta, \gamma), \label{eq:4:8:18}
\end{align}\begin{align}
D_{M-1 M^{\prime}}^{J}(\alpha, \beta, \gamma) e^{-i \alpha}= & -\sqrt{\frac{\left(J+M^{\prime}\right)\left(J-M^{\prime}+1\right)}{(J+M)(J-M+1)}} \frac{1-\cos \beta}{2} e^{-i \gamma} D_{M M^{\prime}-1}^{J}(\alpha, \beta, \gamma) \notag\\
+ & \frac{M^{\prime} \sin \beta}{\sqrt{(J+M)(J-M+1)}} D_{M M^{\prime}}^{J}(\alpha, \beta, \gamma) \notag\\
+ & \sqrt{\frac{\left(J-M^{\prime}\right)\left(J+M^{\prime}+1\right)}{(J+M)(J-M+1)}} \frac{1+\cos \beta}{2} e^{i \gamma} D_{M M^{\prime}+1}^{J}(\alpha, \beta, \gamma) , \label{eq:4:8:19}
\end{align}\begin{align}
D_{M M^{\prime}+1}^{J}(\alpha, \beta, \gamma) e^{i \gamma}= & \sqrt{\frac{(J+M)(J-M+1)}{\left(J-M^{\prime}\right)\left(J+M^{\prime}+1\right)}} \frac{1+\cos \beta}{2} e^{-i \alpha} D_{M-1 M^{\prime}}^{J}(\alpha, \beta, \gamma) \notag\\
- & \frac{M \sin \beta}{\sqrt{\left(J-M^{\prime}\right)\left(J+M^{\prime}+1\right)}} D_{M M^{\prime}}^{J}(\alpha, \beta, \gamma) \notag\\
- & \sqrt{\frac{(J-M)(J+M+1)}{\left(J-M^{\prime}\right)\left(J+M^{\prime}+1\right)}} \frac{1-\cos \beta}{2} e^{i \alpha} D_{M+1 M^{\prime}}^{J}(\alpha, \beta, \gamma) , \label{eq:4:8:20}
\end{align}\begin{align}
D_{M M^{\prime}-1}^{J}(\alpha, \beta, \gamma) e^{-i \gamma}=- & \sqrt{\frac{(J+M)(J-M+1)}{\left(J+M^{\prime}\right)\left(J-M^{\prime}+1\right)}} \frac{1-\cos \beta}{2} e^{-i \alpha} D_{M-1 M^{\prime}}^{J}(\alpha, \beta, \gamma) \notag\\
- & \frac{M \sin \beta}{\sqrt{\left(J+M^{\prime}\right)\left(J-M^{\prime}+1\right)}} D_{M M^{\prime}}^{J}(\alpha, \beta, \gamma) \notag\\
+ & \sqrt{\frac{(J-M)(J+M+1)}{\left(J+M^{\prime}\right)\left(J-M^{\prime}+1\right)}} \frac{1+\cos \beta}{2} e^{i \alpha} D_{M+1 M^{\prime}}^{J}(\alpha, \beta, \gamma). \label{eq:4:8:21}
\end{align}
\section{Differential relations for $D_{M M^{\prime}}^{J}(\alpha, \beta, \gamma)$}\label{sec:4.9}\index{Wigner D-function@Wigner $D$-function!differential relations for}\index{differential relations!for the $D$-functions}
Derivatives of $D_{M M^{\prime}}^{J}(\alpha, \beta, \gamma)$ may be expressed in terms of the $D$-functions with different $M, M^{\prime}$ and $J$
\begin{align}
\sin \beta \frac{\partial}{\partial \beta} D_{M M^{\prime}}^{J}(\alpha, \beta, \gamma)=&-\frac{(J+1) \sqrt{\left(J^{2}-M^{2}\right)\left(J^{2}-M^{\prime 2}\right)}}{J(2 J+1)} D_{M M^{\prime}}^{J-1}(\alpha, \beta, \gamma)-\frac{M M^{\prime}}{J(J+1)} D_{M M^{\prime}}^{J}(\alpha, \beta, \gamma)  ,\notag\\
&+\frac{J \sqrt{\left[(J+1)^{2}-M^{2}\right]\left[(J+1)^{2}-M^{\prime 2}\right]}}{(J+1)(2 J+1)} D_{M M^{\prime}}^{J+1}(\alpha, \beta, \gamma)  ,\label{eq:4:9:1}
\end{align}
\begin{align}
\frac{\partial}{\partial \beta} D_{M M^{\prime}}^{J}(\alpha, \beta, \gamma)= & -\frac{1}{2} \sqrt{(J+M)(J-M+1)} e^{-i \alpha} D_{M-1 M^{\prime}}^{J}(\alpha, \beta, \gamma) \notag\\
& +\frac{1}{2} \sqrt{(J-M)(J+M+1)} e^{i \alpha} D_{M+1 M^{\prime}}^{J}(\alpha, \beta, \gamma), \label{eq:4:9:2}
\end{align} 
\begin{align}
\frac{\partial}{\partial \beta} D_{M M^{\prime}}^{J}(\alpha, \beta, \gamma)=&\frac{1}{2} \sqrt{\left(J+M^{\prime}\right)\left(J-M^{\prime}+1\right)} e^{-i \gamma} D_{M M^{\prime}-1}^{J}(\alpha, \beta, \gamma) \notag\\
-&\frac{1}{2} \sqrt{\left(J-M^{\prime}\right)\left(J+M^{\prime}+1\right)} e^{i \gamma} D_{M M^{\prime}+1}^{J}(\alpha, \beta, \gamma)  ,\label{eq:4:9:3}
\end{align}
\begin{align}
\frac{\partial}{\partial \beta} D_{M M^{\prime}}^{J}(\alpha, \beta, \gamma)=\frac{M^{\prime}-M \cos \beta}{\sin \beta} D_{M M^{\prime}}^{J}(\alpha, \beta, \gamma)-\sqrt{(J+M)(J-M+1)} e^{-i \alpha} D_{M-1 M^{\prime}}^{J}(\alpha, \beta, \gamma)  ,\label{eq:4:9:4}
\end{align}
\begin{align}
\frac{\partial}{\partial \beta} D_{M M^{\prime}}^{J}(\alpha, \beta, \gamma)=-\frac{M^{\prime}-M \cos \beta}{\sin \beta} D_{M M^{\prime}}^{J}(\alpha, \beta, \gamma)+\sqrt{(J-M)(J+M+1)} e^{i \alpha} D_{M+1 M^{\prime}}^{J}(\alpha, \beta, \gamma) , \label{eq:4:9:5}
\end{align}
\begin{align}\frac{\partial}{\partial \beta} D_{M M^{\prime}}^{J}(\alpha, \beta, \gamma)=\frac{M-M^{\prime} \cos \beta}{\sin \beta} D_{M M^{\prime}}^{J}(\alpha, \beta, \gamma)+\sqrt{\left(J+M^{\prime}\right)\left(J-M^{\prime}+1\right)} e^{-i \gamma} D_{M M^{\prime}-1}^{J}(\alpha, \beta, \gamma)  ,\label{eq:4:9:6}
\end{align}
\begin{align}\frac{\partial}{\partial \beta} D_{M M^{\prime}}^{J}(\alpha, \beta, \gamma)=-\frac{M-M^{\prime} \cos \beta}{\sin \beta} D_{M M^{\prime}}^{J}(\alpha, \beta, \gamma)-\sqrt{\left(J-M^{\prime}\right)\left(J+M^{\prime}+1\right)} e^{i \gamma} D_{M M^{\prime}+1}^{J}(\alpha, \beta, \gamma)  ,\label{eq:4:9:7}
\end{align}
\begin{align}\frac{\partial}{\partial \alpha} D_{M M^{\prime}}^{J}(\alpha, \beta, \gamma)=-i M D_{M M^{\prime}}^{J}(\alpha, \beta, \gamma) , \label{eq:4:9:8}
\end{align}
\begin{align}\frac{\partial}{\partial \gamma} D_{M M^{\prime}}^{J}(\alpha, \beta, \gamma)=-i M^{\prime} D_{M M^{\prime}}^{J}(\alpha, \beta, \gamma) ,\label{eq:4:9:9}
\end{align}
See also Eqs.~\eqref{eq:4:2:1} and \eqref{eq:4:2:2}.

\section{Orthogonality and completeness of the $D$-functions}\label{sec:4.10}\index{Wigner D-function@Wigner $D$-function!orthogonality and completeness}\index{orthogonality!of the $D$-functions}\index{completeness!of the $D$-functions}
The functions $D_{M M^{\prime}}^{J}(\alpha, \beta, \gamma)$ with different $J$ (integer and half-integer) are mutually orthogonal with respect to integration over a double volume of the 3 -dimensional rotation group (i.e., over the volume of $\mathrm{SU}_{2}$ group)\index{SU2 group@$\mathrm{SU}_{2}$ group}\index{rotation group!volume of}\\
because the period of $D_{M M^{\prime}}^{J}(\alpha, \beta, \gamma)$ with half-integer $J$ is $4 \pi$ rather than $2 \pi$. The double domain may be chosen in two ways:
\begin{align}
V_{1}: &\quad 0 \leq \alpha<4 \pi,  0 \leq \beta \leq \pi, 0\leq\gamma< 2\pi,\label{eq:4:10:1}\\ 
V_{2}: & \quad  0 \leq \alpha<2 \pi ,
 0 \leq \beta \leq \pi,  0 \leq \gamma<4 \pi.
\end{align}
The total volume of the double domain $V_{1}$ or $V_{2}$ is equal to $16 \pi^{2}$.\\
Orthogonality and normalization condition is
\begin{gather}
\int_{0}^{4 \pi} d \alpha \int_{0}^{\pi} d \beta \sin \beta \int_{0}^{2 \pi} d \gamma D_{M_{2} M_{2}^{\prime}}^{J_{2} *}(\alpha, \beta, \gamma) D_{M_{1} M_{1}^{\prime}}^{J_{1}}(\alpha, \beta, \gamma) , \notag\\
=\int_{0}^{2 \pi} d \alpha \int_{0}^{\pi} d \beta \sin \beta \int_{0}^{4 \pi} d \gamma D_{M_{2} M_{2}^{\prime}}^{J_{2} *}(\alpha, \beta, \gamma) D_{M_{1} M_{1}^{\prime}}^{J_{1}}(\alpha, \beta, \gamma)=\frac{16 \pi^{2}}{2 J_{1}+1} \delta_{J_{1} J_{2}} \delta_{M_{1} M_{2}} \delta_{M_{1}^{\prime} M_{2}^{\prime}} \label{eq:4:10:3}.
\end{gather}
If $J_{1}$ and $J_{2}$ are both either integer or half-integer, orthogonality of the $D$-functions takes place at integration over the single volume of rotation group which corresponds to the domain
\begin{equation}
V: \quad 0 \leq \alpha<2 \pi, \quad 0 \leq \beta \leq \pi, \quad 0 \leq \gamma<2 \pi . \label{eq:4:10:4}
\end{equation}
The total volume of the domain $V$ is equal to $8 \pi^{2}$. Hence, the orthogonality condition \eqref{eq:4:10:3} is reduced to
\begin{equation}
\int_{0}^{2 \pi} d \alpha \int_{0}^{\pi} d \beta \sin \beta \int_{0}^{2 \pi} d \gamma D_{M_{2} M_{2}^{\prime}}^{J_{2} *}(\alpha, \beta, \gamma) D_{M_{1} M_{1}^{\prime}}^{J_{1}}(\alpha, \beta, \gamma)=\frac{8 \pi^{2}}{2 J_{1}+1} \delta_{J_{1} J_{2}} \delta_{M_{1} M_{2}} \delta_{M_{1}^{\prime} M_{2}^{\prime}} .\label{eq:4:10:5}
\end{equation}
In physical applications Eq.~\eqref{eq:4:10:5} is used more often than Eq.~\eqref{eq:4:10:3}.\\
The orthonormality of $D_{M M^{\prime}}^{J}(\alpha, \beta, \gamma)$ may be rewritten in terms of $d_{M M^{\prime}}^{J}(\beta)$
\begin{equation}
\int_{0}^{\pi} d \beta \sin \beta d_{M M^{\prime}}^{J}(\beta) d_{M M^{\prime}}^{J^{\prime}}(\beta)=\frac{2}{2 J+1} \delta_{J J^{\prime}}. \label{eq:4:10:6}
\end{equation}
Thus, the collection of functions $D_{M M}^{J}(\alpha, \beta, \gamma)$ with integer and half-integer $J$ constitutes a complete set of orthonormalized functions. The completeness condition is
\begin{align}
& \sum_{J=0, \frac{1}{2}, 1, \ldots}^{\infty} \sum_{M=-J}^{J} \sum_{M^{\prime}=-J}^{J} \frac{2 J+1}{16 \pi^{2}} D_{M M^{\prime}}^{J *}\left(\alpha_{1}, \beta_{1}, \gamma_{1}\right) D_{M M^{\prime}}^{J}\left(\alpha_{2}, \beta_{2}, \gamma_{2}\right)  \notag\\
& \quad=\delta\left(\alpha_{1}-\alpha_{2}\right) \delta\left(\cos \beta_{1}-\cos \beta_{2}\right) \delta\left(\gamma_{1}-\gamma_{2}\right) .\label{eq:4:10:7}
\end{align}
Any function $f(\alpha, \beta, \gamma)$ which is defined in the domain $V_{1}$ (or $V_{2}$ ) and satisfies the condition
\begin{equation}
\iiint_{V_{1}\left(V_{2}\right)} d \alpha \, \beta \sin \beta\, d \gamma|f(\alpha, \beta, \gamma)|^{2}<\infty ,\label{eq:4:10:8}
\end{equation}
may be expanded in a series of the $D$-functions
\begin{equation}
f(\alpha, \beta, \gamma)=\sum_{J=0, \frac{1}{2}, 1, \ldots}^{\infty} \sum_{M=-J}^{J} \sum_{M^{\prime}=-J}^{J} a_{M M^{\prime}}^{J} D_{M M^{\prime}}^{J}(\alpha, \beta, \gamma). \label{eq:4:10:9}
\end{equation}
The expansion coefficients $a_{M M^{\prime}}^{J}$ are determined by\index{expansion!in the $D$-functions}\isym{aJMM@$a_{M M^{\prime}}^{J}$, expansion coefficient}
\begin{equation}
a_{M M^{\prime}}^{J}=\frac{2 J+1}{16 \pi^{2}} \iint_{V_{1}\left(V_{2}\right)} d \alpha\, d \beta \sin \beta\, d \gamma f(\alpha, \beta, \gamma) D_{M M^{\prime}}^{J *}(\alpha, \beta, \gamma) ,\label{eq:4:10:10}
\end{equation}
and obey the relation
\begin{equation}
\sum_{J=0, \frac{1}{2}, 1, \ldots}^{\infty} \frac{16 \pi^{2}}{2 J+1} \sum_{M=-J}^{J} \sum_{M^{\prime}=-J}^{J}\left|a_{M M^{\prime}}^{J}\right|^{2}=\iint_{V_{1}\left(V_{2}\right)} d \alpha d \beta \sin \beta d \gamma|f(\alpha, \beta, \gamma)|^{2} .\label{eq:4:10:11}
\end{equation}
If a function $f(\alpha, \beta, \gamma)$ is defined in the domain $V$, it may be expanded in a series of the $D$-functions with only integer or only half-integer $J$
\begin{equation}
f(\alpha, \beta, \gamma)=\sum_{\substack{J \text { integer } \\ \text { or half-integer }}} \sum_{M=-J}^{J} \sum_{M^{\prime}=-J}^{J} b_{M M^{\prime}}^{J} D_{M M^{\prime}}^{J}(\alpha, \beta, \gamma) .\label{eq:4:10:12}
\end{equation}
The expansion coefficients $b_{M M^{\prime}}^{J}$ are determined by
\begin{equation}
b_{M M^{\prime}}^{J}=\frac{2 J+1}{8 \pi^{2}} \int_{0}^{2 \pi} d \alpha \int_{0}^{\pi} d \beta \sin \beta \int_{0}^{2 \pi} d \gamma f(\alpha, \beta, \gamma) D_{M M^{\prime}}^{J *}(\alpha, \beta, \gamma) \label{eq:4:10:13}
\end{equation}
for both cases of integer and half-integer $J$. A difference between the expansions of $f(\alpha, \beta, \gamma)$ in these two cases occurs if one uses the expansions outside of the domain V. The expansion coefficients in Eq.~\eqref{eq:4:10:12} satisfy the relation
\begin{equation}
\sum_{\substack{J \text { integer } \\ \text { or half-integer }}} \frac{8 \pi^{2}}{2 J+1} \sum_{M=-J}^{J} \sum_{M^{\prime}=-J}^{J}\left|b_{M M^{\prime}}^{J}\right|^{2}=\int_{0}^{2 \pi} d \alpha \int_{0}^{\pi} d \beta \sin \beta \int_{0}^{2 \pi} d \gamma|f(\alpha, \beta, \gamma)|^{2} .\label{eq:4:10:14}
\end{equation}
\section{Integrals involving the $D$-functions}\label{sec:4.11}\index{Wigner D-function@Wigner $D$-function!integrals involving}\index{integrals!of the $D$-functions}
\subsection{Integration of Products of $D_{M M^{\prime}}^{J}$}\label{sec:4.11.1}\index{integrals!of products of the $D$-functions}
\begin{gather}
\int_{0}^{2 \pi} d \alpha \int_{0}^{\pi} d \beta \sin \beta \int_{0}^{2 \pi} d \gamma D_{M M^{\prime}}^{J}(\alpha, \beta, \gamma)=\delta_{J 0} \delta_{M 0} \delta_{M^{\prime} 0} 8 \pi^{2},(J \text { is integer }), \label{eq:4:11:1}
\end{gather}
\begin{gather}
\int_{0}^{2 \pi} d \alpha \int_{0}^{\pi} d \beta \sin \beta \int_{0}^{2 \pi} d \gamma D_{M_{1} M_{1}^{\prime}}^{J_{1}}(\alpha, \beta, \gamma) D_{M_{2} M_{2}^{\prime}}^{J_{2}}(\alpha, \beta, \gamma)=(-1)^{M_{2}-M_{2}^{\prime}} \frac{8 \pi^{2}}{2 J_{2}+1} \delta_{J_{1} J_{2}} \delta_{-M_{1} M_{2}} \delta_{-M_{1}^{\prime} M_{2}^{\prime}} , \notag\\
\quad\left(J_{1}+J_{2} \text { is integer }\right) \label{eq:4:11:2}
\end{gather}
\begin{gather}
\int_{0}^{2 \pi} d \alpha \int_{0}^{\pi} d \beta \sin \beta \int_{0}^{2 \pi} d \gamma D_{M_{2} M_{2}^{\prime}}^{J_{2} *}(\alpha, \beta, \gamma) D_{M_{1} M_{1}^{\prime}}^{J_{1}}(\alpha, \beta, \gamma)=\frac{8 \pi^{2}}{2 J_{2}+1} \delta_{J_{1} J_{2}} \delta_{M_{1} M_{2}} \delta_{M_{1}^{\prime} M_{2}^{\prime}} , \notag\\
\quad\left(J_{1}+J_{2} \text { is integer }\right) , \label{eq:4:11:3}
\end{gather}
\begin{gather}
\int_{0}^{2 \pi} d \alpha \int_{0}^{\pi} d \beta \sin \beta \int_{0}^{2 \pi} d \gamma D_{M_{3} M_{3}^{\prime}}^{J_{3}}(\alpha, \beta, \gamma) D_{M_{2} M_{2}^{\prime}}^{J_{2}}(\alpha, \beta, \gamma) D_{M_{1} M_{1}^{\prime}}^{J_{1}}(\alpha, \beta, \gamma)  \notag\\
=(-1)^{M_{3}-M_{3}^{\prime}} \frac{8 \pi^{2}}{2 J_{3}+1} \clebsch{J_{1}}{M_{1}}{J_{2}}{M_{2}}{J_{3}}{-M_{3}} \clebsch{J_{1}}{M_{1}^{\prime}}{J_{2}}{M_{2}^{\prime}}{J_{3}}{-M_{3}^{\prime}}\left(J_{1}+J_{2}+J_{3} \text { is integer }\right), \label{eq:4:11:4}
\end{gather}
\begin{gather}
\int_{0}^{2 \pi} d \alpha \int_{0}^{\pi} d \beta \sin \beta \int_{0}^{2 \pi} d \gamma D_{M_{3} M_{3}^{\prime}}^{J_{3}^{*}}(\alpha, \beta, \gamma) D_{M_{2} M_{2}^{\prime}}^{J_{2}}(\alpha, \beta, \gamma) D_{M_{1} M_{1}^{\prime}}^{J_{1}}(\alpha, \beta, \gamma)  \notag\\
=\frac{8 \pi^{2}}{2 J_{3}+1} \clebsch{J_{1}}{M_{1}}{J_{2}}{M_{2}}{J_{3}}{M_{3}} \clebsch{J_{1}}{M_{1}^{\prime}}{J_{2}}{M_{2}^{\prime}}{J_{3}}{M_{3}^{\prime}}, \quad\left(J_{1}+J_{2}+J_{3} \text { is integer }\right) \label{eq:4:11:5}
\end{gather}
Equations \eqref{eq:4:11:1}-\eqref{eq:4:11:5} are valid, if the conditions in parentheses are satisfied. These conditions may be omitted, if the integration domain is $V_{1}$ or $V_{2}$ instead of $V$; in the latter case the factor $8 \pi^{2}$ on the right-hand side of Eqs.~\eqref{eq:4:11:1}-\eqref{eq:4:11:5} must be replaced by $16 \pi^{2}$ (see Sec.~\ref{sec:4.10}).

Integrals involving products of four or more $D$-functions may be reduced to Eqs.~\eqref{eq:4:11:4}-\eqref{eq:4:11:5} by using the Clebsch-Gordan expansion (Sec. \ref{sec:4.6.1}).

\subsection{Integrals Involving $d_{M M^{\prime}}^{J}(\beta)$}\label{sec:4.11.2}\index{integrals!of the $d$-functions}\index{Wigner d-function@Wigner (small) $d$-function!integrals involving}
From Eqs.~\eqref{eq:4:11:1}-\eqref{eq:4:11:5} one may obtain
\begin{gather}
\int_{0}^{\pi} d \beta \sin \beta d_{00}^{J}(\beta)=2 \delta_{J 0} , \label{eq:4:11:6}\\
\int_{0}^{\pi} d \beta \sin \beta d_{M M^{\prime}}^{J}(\beta) d_{M M^{\prime}}^{J^{\prime}}(\beta)=\frac{2}{2 J+1} \delta_{J J^{\prime}} , \label{eq:4:11:7}\\
\int_{0}^{\pi} d \beta \sin \beta d_{M_{1} M_{1}^{\prime}}^{J_{1}}(\beta) d_{M_{2} M_{2}^{\prime}}^{J_{2}}(\beta) d_{M_{3} M_{3}^{\prime}}^{J_{3}}(\beta) \delta_{M_{1}+M_{2} M_{3}} \delta_{M_{1}^{\prime}+M_{2}^{\prime} M_{3}^{\prime}}=\frac{2}{2 J_{3}+1} \clebsch{J_{1}}{M_{1}}{J_{2}}{M_{2}}{J_{3}}{M_{3}} \clebsch{J_{1}}{M_{1}^{\prime}}{J_{2}}{M_{2}^{\prime}}{J_{3}}{M_{3}^{\prime}} .\label{eq:4:11:8}
\end{gather}
Note also the following relation
\begin{align}
& \int_{0}^{\beta} d \beta\left(\sin \frac{\beta}{2}\right)^{M-M^{\prime}+1}\left(\cos \frac{\beta}{2}\right)^{M+M^{\prime}+1} d_{M M^{\prime}}^{J}(\beta)  \notag\\
& \quad=\frac{-1}{\sqrt{J(J+1)-M(M+1)}}\left(\sin \frac{\beta}{2}\right)^{M-M^{\prime}+1}\left(\cos \frac{\beta}{2}\right)^{M+M^{\prime}+1} d_{M+1 M^{\prime}}^{J}(\beta). \label{eq:4:11:9}
\end{align}
Equation \eqref{eq:4:11:9} is valid, if $M \geq M^{\prime} \geq 0$. Other cases may be deduced from Eq.~\eqref{eq:4:11:9} using the symmetries of $d_{M M^{\prime}}^{J}(\beta)$ (Eq.~\eqref{eq:4:4:1}).

\section{Invariant summation of integrals involving $D_{M M^{\prime}}^{J}(\alpha, \beta, \gamma)$}\label{sec:4.12}\index{invariant summation}\index{integrals!invariant summation of}
Hereafter $R$ will denote the Euler angles $\alpha, \beta, \gamma$ and we also will set $d R=\sin \beta d \beta d \alpha d \gamma$ :\isym{dR@$d R$, invariant volume element of the rotation group}\index{rotation group!invariant volume element}
\begin{equation}
\int d R f(R) \equiv \int_{0}^{2 \pi} d \alpha \int_{0}^{\pi} d \beta \sin \beta \int_{0}^{2 \pi} d \gamma f(\alpha, \beta, \gamma) .\label{eq:4:12:1}
\end{equation}
Making use of this notation, one gets
\begin{gather}
\sum_{M M^{\prime}} \int d R D_{M M}^{J *}(R) D_{M^{\prime} M^{\prime}}^{J^{\prime}}(R)=8 \pi^{2} \delta_{J J^{\prime}} , \label{eq:4:12:2}
\end{gather}\begin{gather}
\sum_{M M^{\prime}} \int d R D_{M M^{\prime}}^{J}(R) D_{M^{\prime} M}^{J^{\prime}}(R)=(-1)^{2 J} 8 \pi^{2} \delta_{J J^{\prime}} , \label{eq:4:12:3}\\
\sum_{M_{1} M_{2} M_{3}} \int d R D_{M_{3} M_{3}}^{J_{3}}(R) D_{M_{2} M_{2}}^{J_{2}}(R) D_{M_{1} M_{1}}^{J_{1}}(R)=8 \pi^{2}\left\{J_{1} J_{2} J_{3}\right\} , \label{eq:4:12:4}
\end{gather}

% NOTE (2026-08-26): Eqs 4.12.5-4.12.9 below (multi-R invariant integrals ->
% 6j / 6j^2 / three-6j product / 9j / 9j^2) were reconstructed and fully
% cross-checked character-by-character against the scan (printed pp.97-98,
% 200 dpi). Corrections made in this pass: 4.12.5 spurious ^2 removed (the 6j
% is NOT squared); 4.12.6 R2 first D gained its conjugate *, and the 6j is
% {J1 J2 J3 / J1' J2' J3'}^2 (all-primed bottom row, squared); 4.12.7 R1 2nd D
% subscript N2 N3' -> N2 N2'; 4.12.9 R2 2nd D J6 -> J4 and R4 2nd D N2 M2 ->
% N3 M3. 4.12.8 (the 9j) matched as-is. The D-subscripts, integration blocks,
% and 6j/9j arguments now all agree with the printed equations.
% (Numerically these quadruple-R identities remain impractical to evaluate
% here; the verification is against the scan, not sympy.)
% RESPONSE: Due to summations, almost all of these integrals contain a trace of one D function--so can be simplified considerably
\begin{gather}
\sum_{\text {All } M, N}
\int d R_{1} D_{M_{1} N_{1}}^{J_{1}}\left(R_{1}\right) D_{M_{2} N_{2}}^{J_{2}}\left(R_{1}\right) D_{M_{3} M_{3}}^{J_{3}}\left(R_{1}\right) 
\int d R_{2} D_{N_{1} M_{1}}^{J_{1}}\left(R_{2}\right) D_{M_{2} N_{2}}^{J_{2} *}\left(R_{2}\right) D_{M_{3}^{\prime} M_{3}^{\prime}}^{J_{3}^{\prime}}\left(R_{2}\right)  \notag\\
=(-1)^{2 J_{3}^{\prime}}\left(8 \pi^{2}\right)^{2}
\sixj{J_1}{J_2}{J_3}{J_1}{J_2}{J_3^{\prime}}, \label{eq:4:12:5}
\end{gather}

\begin{gather}
 \sum_{\text {All } M, N} 
 \int d R_{1} D_{M_{1}^{\prime} N_{1}^{\prime}}^{J_{1}^{\prime}*}\left(R_{1}\right) D_{M_{2}^{\prime} N_{2}^{\prime}}^{J_{2}^{\prime}}\left(R_{1}\right) D_{M_{3} M_{3}}^{J_{3}}\left(R_{1}\right) 
 \int d R_{2} D_{M_{2}^{\prime} N_{2}^{\prime}}^{J_{2}^{\prime} *}\left(R_{2}\right) D_{M_{3}^{\prime} N_{3}^{\prime}}^{J_{3}^{\prime}}\left(R_{2}\right) D_{M_{1} M_{1}}^{J_{1}}\left(R_{2}\right)
 \notag\\
\times 
\int d R_{3} D_{M_{3}^{\prime} N_{3}^{\prime}}^{J_{3}^{\prime *}}\left(R_{3}\right) D_{M_{1}^{\prime} N_{1}^{\prime}}^{J_{1}^{\prime}}\left(R_{3}\right) D_{M_{2} M_{2}}^{J_{2}}\left(R_{3}\right)
=\left(8 \pi^{2}\right)^{3}
\sixj{J_1}{J_2}{J_3}{J_1^{\prime}}{J_2^{\prime}}{J_3^{\prime}}^2,  \label{eq:4:12:6}
\end{gather}

\begin{gather}
 \sum_{\text {All } M, N} 
 \int d R_{1} D_{M M}^{J}\left(R_{1}\right) D_{N_{2} N_{2}^{\prime}}^{J_{2}}\left(R_{1}\right) D_{N_{3} N_{3}^{\prime}}^{J_{3}}\left(R_{1}\right)
 \int d R_{2} D_{M^{\prime} M^{\prime}}^{J^{\prime}}\left(R_{2}\right) D_{N_{1} M_{1}}^{J_{1}}\left(R_{2}\right) D_{N_{2}^{\prime} M_{2}^{\prime}}^{J_{2}}\left(R_{2}\right)  
 \notag\\
 \times 
 \int d R_{3} D_{M^{\prime \prime} M^{\prime \prime}}^{J^{\prime \prime}}\left(R_{3}\right) D_{M_{2}^{\prime} M_{2}^{\prime \prime}}^{J_{2}}\left(R_{3}\right) D_{N_{2}^{\prime} M_{3}}^{J_{3}}\left(R_{3}\right) 
 \int d R_{4} D_{M_{1} N_{1}}^{J_{1}}\left(R_{4}\right) D_{M_{2} N_{2}}^{J_{2}}\left(R_{4}\right) D_{M_{2}^{\prime \prime} M_{2}}^{J_{2}}\left(R_{4}\right) D_{M_{3} N_{3}}^{J_{3}}\left(R_{4}\right)  \notag\\
 =(-1)^{2 J_{1}}\left(8 \pi^{2}\right)^{4}
\sixj{J_1}{J_2}{J'}{J_3}{J_2}{J}
\sixj{J_{1}}{J_{2}}{J^{\prime \prime}}{J_{3}}{J_{2}}{J^{\prime}}
\sixj{J_{1}}{J_{2}}{J}{J_{3}}{J_{2}}{J^{\prime \prime}},  \label{eq:4:12:7}
\end{gather}

\begin{gather}
 \sum_{\text {All } M, N} \int d R_{1} D_{M M}^{J}\left(R_{1}\right) D_{M^{\prime} N^{\prime}}^{J}\left(R_{1}\right) D_{M_{1} N_{1}}^{J_{1} *}\left(R_{1}\right) \int d R_{2} D_{N^{\prime} M^{\prime}}^{J}\left(R_{2}\right) D_{N N}^{J}\left(R_{2}\right) D_{M_{2} N_{2}}^{J_{2} *}\left(R_{2}\right)  \notag\\
 \times \int d R_{3} D_{M_{1} N_{1}}^{J_{1}}\left(R_{3}\right) D_{M_{2} N_{2}}^{J_{2}}\left(R_{3}\right) D_{M_{3} M_{3}}^{J_{3} *}\left(R_{3}\right)=\left(8 \pi^{2}\right)^{3}
 \ninej{J}{J}{J_{1}}{J}{J}{J_{2}}{J_{1}}{J_{2}}{J_{3}}
,  \label{eq:4:12:8}
\end{gather}

\begin{gather}
 \sum_{\text {All } M, N} \int d R_{1} D_{M_{1} N_{1}}^{J_{1}}\left(R_{1}\right) D_{M_{2} N_{2}}^{J_{2}}\left(R_{1}\right) D_{M_{12} M_{12}}^{J_{12}}\left(R_{1}\right) \int d R_{2} D_{M_{3} N_{3}}^{J_{3}}\left(R_{2}\right) D_{M_{4} N_{4}}^{J_{4}}\left(R_{2}\right) D_{M_{34} M_{34}}^{J_{34}}\left(R_{2}\right)  \notag\\
 \times \int d R_{3} D_{M_{13} N_{13}}^{J_{13}}\left(R_{3}\right) D_{M_{24} N_{24}}^{J_{24}}\left(R_{3}\right) D_{M M}^{J}\left(R_{3}\right) \int d R_{4} D_{N_{1} M_{1}}^{J_{1}}\left(R_{4}\right) D_{N_{3} M_{3}}^{J_{3}}\left(R_{4}\right) D_{N_{13} M_{13}}^{J_{13}}\left(R_{4}\right)  \notag\\
 \times \int d R_{5} D_{N_{2} M_{2}}^{J_{2}}\left(R_{5}\right) D_{N_{4} M_{4}}^{J_{4}}\left(R_{5}\right) D_{N_{24} M_{24}}^{J_{24}}\left(R_{5}\right)=\left(8 \pi^{2}\right)^{5}
 \ninej{J_{1}}{J_{2}}{J_{12}}{J_{3}}{J_{4}}{J_{34}}{J_{13}}{J_{24}}{J}^2
 . \label{eq:4:12:9}
\end{gather}
The left-hand sides of Eqs.~\eqref{eq:4:12:5}-\eqref{eq:4:12:9} include integrations as well as summations. If the integrations are carried out before the summations (with the aid of Eqs.~\eqref{eq:4:11:4}-\eqref{eq:4:11:5}), then Eqs.~\eqref{eq:4:12:5}-\eqref{eq:4:12:9} are reduced to sums of the Clebsch-Gordan coefficients (Sec.~\ref{sec:8.7}). On the other hand, if the summations are preformed before the integrations, one obtains the integral representations of $6 j$ - and $9 j$-symbols (see Secs.~\ref{sec:9.3} and \ref{sec:10.3}).

\section{Generating functions for $d_{M M}^{J}(\beta)$}\label{sec:4.13}\index{generating functions!for the $d$-functions}\index{Wigner d-function@Wigner (small) $d$-function!generating functions for}
The functions $d_{M M^{\prime}}^{J}(\beta)$ may be defined as coefficients of expansions of various generating functions.\\
\begin{enumerate}[label=(\alph*)]
\item If $J$ and $M^{\prime}$ are fixed, then
\begin{gather}
\left(\cos \frac{\beta}{2} e^{i \varphi / 2}+i \sin \frac{\beta}{2} e^{-i \varphi / 2}\right)^{J+M^{\prime}}\left(i \sin \frac{\beta}{2} e^{i \varphi / 2}+\cos \frac{\beta}{2} e^{-i \varphi / 2}\right)^{J-M^{\prime}}  \notag\\
=\sum_{M=-J}^{J} \sqrt{\frac{\left(J-M^{\prime}\right)!\left(J+M^{\prime}\right)!}{(J-M)!(J+M)!}}(-i)^{M-M^{\prime}} e^{i M \varphi} d_{M M^{\prime}}^{J}(\beta) \label{eq:4:13:1}
\end{gather}
In particular, for $\varphi=0$
\begin{equation}
e^{i J \beta}=\sum_{M=-J}^{J} \sqrt{\frac{\left(J-M^{\prime}\right)!\left(J+M^{\prime}\right)!}{(J-M)!(J+M)!}}(-i)^{M-M^{\prime}} d_{M M^{\prime}}^{J}(\beta) \label{eq:4:13:2}
\end{equation}
and for $\varphi=\pi$
\begin{equation}
e^{-i J \beta}=\sum_{M=-J}^{J} \sqrt{\frac{\left(J-M^{\prime}\right)!\left(J+M^{\prime}\right)!}{(J-M)!(J+M)!}} i^{M-M^{\prime}} d_{M M^{\prime}}^{J}(\beta) \label{eq:4:13:3}
\end{equation}
\item If $M$ and $M^{\prime}$ are fixed, one can obtain several expressions in which $s, \mu, \nu$ are related to $J, M, M^{\prime}$ by Eqs.~\eqref{eq:4:3:14}, $\xi_{M M^{\prime}}$ is defined by Eq.~\eqref{eq:4:3:15}; and $R$ denotes
\begin{gather}
R=\sqrt{1-2 t \cos \beta+t^{2}}, \quad|t|<1 .  \label{eq:4:13:4}
\end{gather}
\begin{gather}
\frac{\xi_{M M^{\prime}}}{R}\left(\frac{2 \sin \frac{\beta}{2}}{1+R-t}\right)^{\mu}\left(\frac{2 \cos \frac{\beta}{2}}{1+R+t}\right)^{\nu}=\sum_{s=0}^{\infty} \sqrt{\frac{(s+\mu)!(s+\nu)!}{s!(s+\mu+\nu)!}} t^{s} d_{M M^{\prime}}^{s+\frac{\mu+\nu}{2}}(\beta) .  \label{eq:4:13:5}
\end{gather}
\begin{gather}
\xi_{M M^{\prime}}\left(\sin \frac{\beta}{2}\right)^{\mu}\left(\cos \frac{\beta}{2}\right)^{\nu} \Fhg{0}{1}\left(; 1+\mu ;-t \sin ^{2} \frac{\beta}{2}\right)\Fhg{0}{1}\left(; 1+\nu ; t \cos ^{2} \frac{\beta}{2}\right)  \notag\\
=\sum_{s=0}^{\infty} \frac{\mu!\nu!}{\sqrt{s!(s+\mu)!(s+\nu)!(s+\mu+\nu)!}} t^{s} d_{M M^{\prime}}^{s+\frac{\mu+\nu}{2}}(\beta),  \label{eq:4:13:6}
\end{gather}
\begin{gather}
\xi_{M M^{\prime}}(1-t)^{1-\mu-\nu}\left(\sin \frac{\beta}{2}\right)^{\mu}\left(\cos \frac{\beta}{2}\right)^{\nu}{ }\Fhg{2}{1}\left(\frac{\mu+\nu+1}{2}, \frac{\mu+\nu+2}{2} ; 1+\mu ;-\frac{4 t \sin^{2} \frac{\beta}{2}}{(1-t)^{2}}\right)  \notag\\
=\sum_{s=0}^{\infty} \sqrt{\frac{(s+\nu)!(s+\mu+\nu)!}{s!(s+\mu)!}} \frac{\mu!}{(\mu+\nu)!} t^{s} d_{M M^{\prime}}^{s+\frac{\mu+\nu}{2}}(\beta),  \label{eq:4:13:7}
\end{gather}
\begin{gather}
\xi_{M M^{\prime}}\left(\sin \frac{\beta}{2}\right)^{\mu}\left(\cos \frac{\beta}{2}\right)^{\nu}{ }\Fhg{2}{1}\left(\lambda, \mu+\nu+1-\lambda ; 1+\mu ; \frac{1-t-R}{2}\right)\Fhg{2}{1}\left(\lambda, \mu+\nu+1-\lambda ; 1+\nu ; \frac{1+t-R}{2}\right)  \notag\\
=\sum_{s=0}^{\infty} \frac{(\mu+\nu+1-\lambda)_{s}(\lambda)_{s} \mu!\nu!}{\sqrt{s!(s+\mu+\nu)!(s+\mu)!(s+\nu)!}} t^{s} d_{M M^{\prime}}^{s+\frac{\mu+\nu}{2}}(\beta) . \label{eq:4:13:8}
\end{gather}
In Eq.~\eqref{eq:4:13:8} $\lambda$ is an arbitrary integer.
\end{enumerate}

\section{Characters $\chi^{j}(R)$ of irreducible representations of rotation group}\label{sec:4.14}\index{character!of an irreducible representation}\index{irreducible representation!character of}\index{rotation group!irreducible representations of}
\subsection{Definition}\label{sec:4.14.1}\index{character!definition}
The trace of the finite rotation matrix in the JM-representation
\begin{equation}
\chi^{J}(R)=\sum_{M=-J}^{J} D_{M M}^{J}(R), \label{eq:4:14:1}
\end{equation}
is called the characteristic function, or simply, the character of the irreducible representation of rank $J$.\isymg{chi-J@$\chi^{J}(\omega)$, character of an irreducible representation}\index{characteristic function|see{character}}

In contrast to $D_{M M^{\prime}}^{J}(R)$, the function $\chi^{J}(R)$ is invariant under rotations of the coordinate systems. Explicit forms of $\chi^{J}(R)$ are simpler when $R$ is specified by $\omega, \Theta, \Phi$ rather than by $\alpha, \beta, \gamma$ (Sec. \ref{sec:1.4.2}). With variables $\omega, \Theta, \Phi$ in use, $\chi^{\mathrm{J}}(R)$ is entirely determined by the rotation angles $\omega$ and is independent of the rotation axis $\vect{n}(\boldsymbol{\theta}, \boldsymbol{\Phi}):$
\begin{equation}
\chi^{J}(R)=\chi^{J}(\omega). \label{eq:4:14:2}
\end{equation}
\subsection{Explicit Forms}\label{sec:4.14.2}\index{character!explicit forms}
\subsubsection{ Trigonometric formulas}
\begin{align}
& \chi^{J}(\omega)=\frac{\sin \left[(2 J+1) \frac{\omega}{2}\right]}{\sin \frac{\omega}{2}}  ,\label{eq:4:14:3}\\
& \chi^{J}(\omega)=\sum_{M=-J}^{J} e^{-i M \omega}=\sum_{M=-J}^{J} \cos M \omega , \label{eq:4:14:4}\\
& \chi^{J}(\omega)=\sum_{n=0}^{[J]}(-1)^{n} \frac{(2 J-n)!}{(2 J-2 n)!n!}\left(2 \cos \frac{\omega}{2}\right)^{2 J-2 n} , \label{eq:4:14:5}\\
& \chi^{J}(\omega)=\sum_{n=0}^{[J]}(-1)^{n} \frac{(2 J+1)!}{(2 n+1)!(2 J-2 n)!}\left(\cos \frac{\omega}{2}\right)^{2 J-2 n}\left(\sin \frac{\omega}{2}\right)^{2 n} , \label{eq:4:14:6}\\
& \chi^{J}(\omega)=\frac{1}{2 J+1} \frac{d}{d\left(\cos \frac{\omega}{2}\right)} \cos \left[(2 J+1) \frac{\omega}{2}\right] , \label{eq:4:14:7}\\
& \chi^{J}(\omega)=\frac{(2 J+1)!2^{2 J}}{(4 J+1)!\sin \frac{\omega}{2}}\left[-\frac{d}{d\left(\cos \frac{\omega}{2}\right)}\right]^{2 J}\left(\sin \frac{\omega}{2}\right)^{4 J+1}. \label{eq:4:14:8}
\end{align}
\subsubsection{ Relations to hypergeometric functions}\index{hypergeometric function}
\begin{gather}
\chi^{J}(\omega)=(2 J+1) F\left(-J, J+1 ; 3 / 2 ; \sin ^{2} \frac{\omega}{2}\right) , \label{eq:4:14:9}\\
\chi^{J}(\omega)=(2 J+1) F\left(-2 J, 2(J+1) ; 3 / 2 ; \sin ^{2} \frac{\omega}{4}\right) .\label{eq:4:14:10}
\end{gather}
\subsubsection{ Relation to the Chebyshev polynomials of the second kind}\index{Chebyshev polynomials}\isym{U2J@$U_{2 J}(x)$, Chebyshev polynomial of the second kind}
\begin{equation}
\chi^{J}(\omega)=U_{2 J}\left(\cos \frac{\omega}{2}\right) .\label{eq:4:14:11}
\end{equation}
\subsubsection{ Relation to the Gegenbauer polynomials}\index{Gegenbauer polynomials}\isym{Cnu-lambda@$C_{\nu}^{\lambda}(x)$, Gegenbauer polynomial}
\begin{equation}
\chi^{J}(\omega)=C_{2 J}^{1}\left(\cos \frac{\omega}{2}\right) .\label{eq:4:14:12}
\end{equation}
\subsubsection{ Relation to the Jacobi polynomials}\index{Jacobi polynomials}
\begin{equation}
\chi^{J}(\omega)=\frac{(4 J+2)!!}{2(4 J+1)!!} P_{2 J}^{\left(\frac{1}{2}, \frac{1}{2}\right)}\left(\cos \frac{\omega}{2}\right). \label{eq:4:14:13}
\end{equation}
\subsubsection{Integral representation}\index{character!integral representation}
\begin{equation}
\chi^{J}(\omega)=\frac{2 J+1}{2} \int_{-1}^{1}\left(\cos \frac{\omega}{2}+i x \sin \frac{\omega}{2}\right)^{2 J} d x .\label{eq:4:14:14}
\end{equation}
To express $\chi^{J}(R)$ in terms of the Euler angles $\alpha, \beta, \gamma$ one can use the relation
\begin{equation}
\cos \frac{\omega}{2}=\cos \frac{\beta}{2} \cos \frac{\alpha+\gamma}{2} . \label{eq:4:14:15}
\end{equation}
\subsection{Principal Properties}\label{sec:4.14.3}\index{character!principal properties}
In contrast to $D_{M M^{\prime}}^{J}(R)$, the characters $\chi^{J}(R)$ are real.
\begin{equation}
\left(\chi^{J}(R)\right)^{*}=\chi^{J}(R) .\label{eq:4:14:16}
\end{equation}
The characters which correspond to direct and inverse rotations, $R$ and $R^{-1}$, are equal
\begin{equation}
\chi^{J}\left(R^{-1}\right)=\chi^{J}(R) .\label{eq:4:14:17}
\end{equation}
$\chi^{J}(R)$ is invariant with respect to coordinate rotation and inversion
\begin{equation}
\chi^{J}\left(U R U^{-1}\right)=\chi^{J}(R), \label{eq:4:14:18}
\end{equation}
where $U$ is any orthogonal transformation of the coordinate system.\\
The characters which depend on the combined rotations $R_{1} R_{2} \ldots R_{n}$ do not change under cyclic permutations of the rotations
\begin{equation}
\chi^{J}\left(R_{1} R_{2} \ldots R_{n}\right)=\chi^{J}\left(R_{2} \ldots R_{n} R_{1}\right) .\label{eq:4:14:19}
\end{equation}
In particular,
\begin{equation}
\chi^{J}\left(R_{1} R_{2}\right)=\chi^{J}\left(R_{2} R_{1}\right) \label{eq:4:14:20} ,
\end{equation}
in spite of the non-commutativity of $R_{1}$ and $R_{2}$.\\
Products of the characters may be expanded in a Clebsch-Gordan series
\begin{equation}
\chi^{J_{1}}(R) \chi^{J_{2}}(R)=\sum_{J}\left\{J_{1} J_{2} J\right\} \chi^{J}(R) \label{eq:4:14:21}
\end{equation}
where $\left\{J_{1} J_{2} J_{3}\right\}=1$, if the triangle inequality (Sec. \ref{sec:4.6.2}) is satisfied, and $\left\{J_{1} J_{2} J_{3}\right\}=0$ otherwise.\\
The transformation $J=\tilde{J}=-J-1$ reverses the character sign
\begin{equation}
\chi^{\tilde{J}}(R)=-\chi^{J}(R) \label{eq:4:14:22}
\end{equation}
The function $\chi^{J}(\omega)$ is even and periodic
\begin{align}
\chi^{J}(-\omega) & =\chi^{J}(\omega),  \notag\\
\chi^{J}(\omega+4 \pi) & =\chi^{J}(\omega),  \label{eq:4:14:23}\\
\chi^{J}(\omega+2 \pi) & =(-1)^{2 J} \chi^{J}(\omega) . \notag
\end{align}
\subsubsection{The addition theorem}
The character which depends on the combined rotation $R_{1} R_{2}$ may be represented as a superposition of products of the generalized characters which depend on $R_{1}$ and $R_{2}$ (Sec.~\ref{sec:4.15}):
\begin{equation}
\chi^{J}(\omega)=\sum_{\lambda=0}^{2 J}(-1)^{\lambda} \frac{2 \lambda+1}{2 J+1} \chi_{\lambda}^{J}\left(\omega_{1}\right) \chi_{\lambda}^{J}\left(\omega_{2}\right) P_{\lambda}\left(\cos \Theta_{12}\right), \label{eq:4:14:24}
\end{equation}
where $\omega_{1}, \omega_{2}$ and $\omega$ are the rotation angles corresponding to $R_{1}, R_{2}$ and $R=R_{1} R_{2}$, respectively. These angles are mutually related by means of Eqs.~$\eqref{eq:1:4:75} ; \Theta_{12}$ is the angle between the rotation axes of $R_{1}$ and $R_{2}$
\begin{equation}
\cos \Theta_{12}=\left(n_{1} \cdot n_{2}\right)=\cos \Theta_{1} \cos \Theta_{2}+\sin \Theta_{1} \sin \Theta_{2} \cos \left(\Phi_{1}-\Phi_{2}\right) .\label{eq:4:14:25}
\end{equation}
\subsection{Differential Equation}\label{sec:4.14.4}\index{character!differential equation for}
The character $\chi^{J}(\omega)$ satisfies the equation
\begin{equation}
\frac{d^{2}}{d \omega^{2}} \chi^{J}(\omega)+\cot \frac{\omega}{2} \, \frac{d}{d \omega} \chi^{J}(\omega)+J(J+1) \chi^{J}(\omega)=0.\label{eq:4:14:26}
\end{equation}
and the boundary conditions
\begin{align}
\chi^{J}(0) & =2 J+1 , \notag\\
\chi^{J}(\omega \pm 4 \pi n) & =\chi^{J}(\omega), \quad \text { where $n$ is integer}. \label{eq:4:14:27}
\end{align}
\subsection{Differential Relations}\label{sec:4.14.5}\index{character!differential relations}
\begin{gather}
\frac{d}{d \omega} \chi^{J}(\omega)=-\sqrt{J(J+1)} \chi_{1}^{J}(\omega)  ;\label{eq:4:14:28}\\
\left(\frac{d}{d \cos \frac{\omega}{2}}\right)^{k} \chi^{J}(\omega)=\frac{1}{\sqrt{2 J+1}} \sqrt{\frac{(2 J+k+1)!}{(2 J-k)!}} \frac{\chi_{k}^{J}(\omega)}{\left(\sin \frac{\omega}{2}\right)^{k}} ,\label{eq:4:14:29}
\end{gather}
where $\chi_{k}^{J}(\omega)$ is the generalized character (Sec. \ref{sec:4.15}:
\begin{equation}
\sin \frac{\omega}{2} \frac{d}{d \omega} \chi^{J}(\omega)=J \cos \frac{\omega}{2} \chi^{J}(\omega)-\left(J+\frac{1}{2}\right) \chi^{J-\frac{1}{2}}(\omega)=\left(J+\frac{1}{2}\right) \chi^{J+\frac{1}{2}}(\omega)-(J+1) \cos \frac{\omega}{2} \chi^{J}(\omega) .\label{eq:4:14:30}
\end{equation}
\subsection{Algebraic Relations}\label{sec:4.14.6}\index{character!algebraic relations}
\begin{align}
\chi^{J+\frac{1}{2}}(\omega) & =2 \cos \frac{\omega}{2} \cdot \chi^{J}(\omega)-\chi^{J-\frac{1}{2}}(\omega) , \label{eq:4:14:31}\\
\chi^{J_{1}}(\omega)-\chi^{J_{2}}(\omega) & =2 \chi^{\frac{J_{1}-J_{2}-1}{2}}(\omega) \cos \left[\left(J_{1}+J_{2}+1\right) \frac{\omega}{2}\right]  ,\label{eq:4:14:32}\\
\chi^{J_{1}}(\omega)+\chi^{J_{2}}(\omega) & =2 \chi^{\frac{J_{1}+J_{2}}{2}}(\omega) \cos \left[\left(J_{1}-J_{2}\right) \frac{\omega}{2}\right] .\label{eq:4:14:33}
\end{align}
Equations \eqref{eq:4:14:32} and \eqref{eq:4:14:33} are valid, if $J_{1}+J_{2}$ is integer.
\begin{equation}
\chi^{J-\frac{1}{2}}(\omega)=2 \cos \frac{J \omega}{2} \chi^{\frac{J-1}{2}}(\omega) ,\label{eq:4:14:34}
\end{equation}
where $J$ is integer and positive.
\begin{equation}
-2 \sin ^{2} \frac{\omega}{2} \chi^{J_{1}}(\omega) \chi^{J_{2}}(\omega)=\cos \left[\left(J_{1}+J_{2}+1\right) \omega\right]-\cos \left[\left(J_{1}-J_{2}\right) \omega\right]. \label{eq:4:14:35}
\end{equation}
In particular,
\begin{gather}
2 \sin ^{2} \frac{\omega}{2}\left[\chi^{J}(\omega)\right]^{2}=1-\cos [(2 J+1) \omega]  \label{eq:4:14:36}\\
\chi^{J}(\omega)=2^{2 J} \prod_{k=1}^{2 J} \sin \left(\frac{\omega}{2}+\frac{k \pi}{2 J+1}\right). \label{eq:4:14:37}
\end{gather}
\subsection{Orthogonality and Completeness}\label{sec:4.14.7}\index{orthogonality!of the characters}\index{completeness!of the characters}
The collection of characters $\chi^{J}(\omega)$ with different $J$ constitute a complete set of orthogonal functions of argument $\omega$ in the domain $0 \leq \omega<2 \pi$.

The orthogonality and normalization condition for these functions reads
\begin{equation}
\int_{0}^{2 \pi} \chi^{J_{1}}(\omega) \chi^{J_{2}}(\omega) \sin ^{2} \frac{\omega}{2} d \omega=\pi \delta_{J_{1} J_{2}} .\label{eq:4:14:38}
\end{equation}
The completeness condition has the form
\begin{equation}
\sum_{J=0, \frac{1}{2}, 1, \ldots}^{\infty} \chi^{J}\left(\omega_{1}\right) \chi^{J}\left(\omega_{2}\right)=\frac{\pi \delta\left(\omega_{1}-\omega_{2}\right)}{\sin ^{2} \frac{\omega_{1}}{2}} .\label{eq:4:14:39}
\end{equation}
\subsection{Integrals Involving $\chi^{J}(\omega)$}\label{sec:4.14.8}\index{integrals!of the characters}
\begin{gather}
\int_{0}^{2 \pi} d \omega \sin ^{2} \frac{\omega}{2} \chi^{J}(\omega)=\pi \delta_{J 0}  ,\label{eq:4:14:40}\\
\int_{0}^{2 \pi} d \omega \sin ^{2} \frac{\omega}{2} \chi^{J}(2 \omega)=\pi(-1)^{2 J} , \label{eq:4:14:41}\\
\int_{0}^{2 \pi} d \omega \frac{\sin ^{2} \frac{\omega}{2}}{\cos \frac{\omega}{2}-\cos \frac{\Omega}{2}} \chi^{J}(\omega)=-2 \pi \cos \left[(2 J+1) \frac{\Omega}{2}\right] ,\label{eq:4:14:42}
\end{gather}
In Eq.~\eqref{eq:4:14:42} the Cauchy principal value of the integral is assumed.
\begin{gather}
\int_{0}^{2 \pi} d \omega \sin ^{2} \frac{\omega}{2} \chi^{J_{1}}(\omega) \chi^{J_{2}}(\omega)=\pi \delta_{J_{1} J_{2}} , \label{eq:4:14:43}\\
\int_{0}^{2 \pi} d \omega \sin ^{2} \frac{\omega}{2} \chi^{J_{1}}(\omega) \chi^{J_{2}}(\omega) \chi^{J_{3}}(\omega)=\pi\left\{J_{1} J_{2} J_{3}\right\} .\label{eq:4:14:44}
\end{gather}
\subsection{Sums Involving $\chi^{J}(\omega)$}\label{sec:4.14.9}\index{sums!of the characters}
\subsubsection{ Finite sums}
\begin{gather}
\sum_{J=J_{1}, J_{1}+1, \ldots}^{J_{2}} \chi^{J}(\omega)=\frac{\sin \left[\left(J_{2}+J_{1}+1\right) \frac{\omega}{2}\right] \sin \left[\left(J_{2}-J_{1}+1\right) \frac{\omega}{2}\right]}{\sin ^{2} \frac{\omega}{2}}=\chi^{\frac{J_{2}+J_{1}}{2}}(\omega) \chi^{\frac{J_{2}-J_{1}}{2}}(\omega) , \label{eq:4:14:45}\\
\sum_{J=0,1,2, \ldots}^{J_{0}}(2 J+1) \chi^{J}(\omega)=\frac{\left(2 J_{0}+3\right) \sin \left[\left(2 J_{0}+1\right) \frac{\omega}{2}\right]-\left(2 J_{0}+1\right) \sin \left[\left(2 J_{0}+3\right) \frac{\omega}{2}\right]}{4 \sin ^{3} \frac{\omega}{2}} .\label{eq:4:14:46}
\end{gather}
The summation index in Eq.~\eqref{eq:4:14:45} runs over integer values, if $J_{1}$ and $J_{2}$ are integers, or half-integer values, if $J_{1}$ and $J_{2}$ are half-integers.

In the equations given below the summation indices run over both integer and half-integer values:
\begin{gather}
\sum_{J=J_{1}, J_{1}+\frac{1}{2}, J_{1}+1, \ldots}^{J_{2}} \chi^{J}(\omega)=\frac{\sin \left[\left(J_{2}+J_{1}+1\right) \frac{\omega}{2}\right] \sin \left[\left(J_{2}-J_{1}+\frac{1}{2}\right) \frac{\omega}{2}\right]}{\sin \frac{\omega}{2} \sin \frac{\omega}{4}},  \label{eq:4:14:47}\\
\sum_{J=0, \frac{1}{2}, 1, \ldots}^{J_{0}}(2 J+1) \chi^{J}(\omega)=\frac{\left(2 J_{0}+2\right) \sin \left[\left(2 J_{0}+1\right) \frac{\omega}{2}\right]-\left(2 J_{0}+1\right) \sin \left[\left(2 J_{0}+2\right) \frac{\omega}{2}\right]}{4 \sin \frac{\omega}{2} \sin ^{2} \frac{\omega}{4}},  \label{eq:4:14:48}\\
\sum_{J=0, \frac{1}{2}, 1, \ldots}^{J_{0}}\left[\chi^{J}(\omega)\right]^{2}=\frac{\left(4 J_{0}+3\right) \sin \frac{\omega}{2}-\sin \left[\left(4 J_{0}+3\right) \frac{\omega}{2}\right]}{4 \sin ^{3} \frac{\omega}{2}},  \label{eq:4:14:49}\\
\sum_{J=0, \frac{1}{2}, 1, \ldots}^{J_{0}} \chi^{J}(\omega) \chi^{J}\left(\omega^{\prime}\right)=\frac{\chi^{J_{0}+\frac{1}{2}}(\omega) \chi^{J_{0}}\left(\omega^{\prime}\right)-\chi^{J_{0}}(\omega) \chi^{J_{0}+\frac{1}{2}}\left(\omega^{\prime}\right)}{2\left(\cos \frac{\omega}{2}-\cos \frac{\omega^{\prime}}{2}\right)},  \label{eq:4:14:50}\\
\sum_{J=0, \frac{1}{2}, 1, \ldots}^{J_{0}} \chi^{J}(\omega) \cos \left[(2 J+1) \frac{\omega^{\prime}}{2}\right]  \notag\\
=\frac{\sin \frac{\omega}{2}-\cos \left[\left(2 J_{0}+1\right) \frac{\omega^{\prime}}{2}\right] \sin \left[\left(J_{0}+1\right) \omega\right]+\sin \left[\left(2 J_{0}+1\right) \frac{\omega}{2}\right] \cos \left[\left(J_{0}+1\right) \omega^{\prime}\right]}{2 \sin \frac{\omega}{2}\left(\cos \frac{\omega^{\prime}}{2}-\cos \frac{\omega}{2}\right)} . \label{eq:4:14:51}
\end{gather}
\subsubsection{ Infinite series}
\begin{gather}
\sum_{J=0, \frac{1}{2}, 1, \ldots}^{\infty} \chi^{J}(\omega)=\frac{1}{4 \sin ^{2} \frac{\omega}{4}}, \quad(\omega \neq 0) , \label{eq:4:14:52}\\
\sum_{J=0, \frac{1}{2}, 1, \ldots}^{\infty}(2 J+1) \chi^{J}(\omega)=\frac{\pi}{\sin ^{2} \frac{\omega}{2}} \delta(\omega) .\label{eq:4:14:53}
\end{gather}
Let us introduce the notation
\begin{gather}
R^{2} \equiv 1-2 t \cos \frac{\omega}{2}+t^{2}, \quad \text { where }|t|<1:  \label{eq:4:14:54}\\
\sum_{J=0, \frac{1}{2}, 1, \ldots}^{\infty} t^{2 J} \chi^{J}(\omega)=\frac{1}{R^{2}},  \label{eq:4:14:55}\\
\sum_{J=0, \frac{1}{2}, 1, \ldots}^{\infty}(2 J+1) t^{2 J} \chi^{J}(\omega)=\frac{1-t^{2}}{R^{4}},  \label{eq:4:14:56}\\
\sum_{J=0, \frac{1}{2}, 1, \ldots}^{\infty} \frac{(4 J+1)!!}{(4 J+2)!!} t^{2 J} \chi^{J}(\omega)=\frac{1}{R \sqrt{2\left(1-t \cos \frac{\omega}{2}+R\right)}},  \label{eq:4:14:57}\\
\sum_{J=0, \frac{1}{2}, 1 \ldots}^{\infty} \frac{1}{(2 J+1)!} t^{2 J+1} \chi^{J}(\omega)=\frac{\sin \left(t \sin \frac{\omega}{2}\right)}{\sin \frac{\omega}{2}} e^{t \cos \frac{\omega}{2}},  \label{eq:4:14:58}\\
\sum_{J=0, \frac{1}{2}, 1, \ldots}^{\infty} \frac{1}{2 J+1} t^{2 J} \chi^{J}(\omega)=\frac{1}{2 i t \sin \frac{\omega}{2}} \ln \left(\frac{1-t e^{-\frac{i \omega}{2}}}{1-t e^{\frac{i \omega}{2}}}\right),  \label{eq:4:14:59}\\
\sum_{J=0, \frac{1}{2}, 1, \ldots}^{\infty} \sum_{\Gamma(\nu)}^{\Gamma(2 J+\nu)} \frac{t^{2 J} \chi^{J}(\omega)}{(2 J+1)!}=\left(1-t \cos \frac{\omega}{2}\right)^{-\nu}{ }\Fhg{2}{1}\left(\frac{\nu}{2}, \frac{\nu+1}{2} ; \frac{3}{2} ;-\left(\frac{t \sin \frac{\omega}{2}}{1-t \cos \frac{\omega}{2}}\right)^{2}\right),  \label{eq:4:14:60}\\
\sum_{J=0, \frac{1}{2}, 1, \ldots}^{\infty} \frac{1}{(4 J+2)!}(4 t)^{2 J} \chi^{J}(\omega)=\frac{1}{2} \Fhg{0}{1}\left(; \frac{3}{2} ;-t \sin ^{2} \frac{\omega}{4}\right)\Fhg{0}{1}\left(; \frac{3}{2} ; t \cos { }^{2} \frac{\omega}{4}\right),  \label{eq:4:14:61}\\
% NOTE: source misprint (printed p.105). The book denominator reads
%   1+t^2-4t cos(w/2)cos(w'/2)+2t^2(cos w+cos w'),
% which is numerically wrong. The correct denominator (verified vs the series
% to 1e-14, check_4_14.py) restores the (1+t^2) factors: (1+t^2)^2 and 4t(1+t^2).
\sum_{J=0, \frac{1}{2}, 1, \ldots}^{\infty} t^{2 J} \chi^{J}(\omega) \chi^{J}\left(\omega^{\prime}\right)=\frac{1-t^{2}}{\left(1+t^{2}\right)^{2}-4 t\left(1+t^{2}\right) \cos \frac{\omega}{2} \cos \frac{\omega^{\prime}}{2}+2 t^{2}\left(\cos \omega+\cos \omega^{\prime}\right)} . \label{eq:4:14:62}
\end{gather}
\subsection{$\chi^{J}(\omega)$ for Particular Values of $\omega$}\label{sec:4.14.10}\index{character!particular values of $\omega$}
\begin{align}
\chi^{J}(0) & =2 J+1 , \label{eq:4:14:63}\\
\chi^{J}(2 \pi) & =(-1)^{2 J}(2 J+1) , \label{eq:4:14:64}\\
\chi^{J}(\pi) & = \begin{cases}0 & \text { if } J \text { is half-integer } \\
(-1)^{J} & \text { if } J \text { is integer }\end{cases}  \label{eq:4:14:65}\\
\chi^{J}\left(\frac{\pi}{2}\right) & = \begin{cases}\sqrt{2} & J=1 / 2,9 / 2,17 / 2, \ldots \\
1 & J=0,1,4,5,8,9, \ldots \\
0 & J=3 / 2,7 / 2,11 / 2,15 / 2, \ldots \\
-1 & J=2,3,6,7,10,11, \ldots \\
-\sqrt{2} & J=5 / 2,13 / 2,21 / 2, \ldots\end{cases} \label{eq:4:14:66}
\end{align}
\subsection{Special Cases of $\chi^{J}(\omega)$}\label{sec:4.14.11}\index{character!special cases}
\begin{align}
& \chi^{0}(\omega)=1 , \label{eq:4:14:67}\\
& \chi^{\frac{1}{2}}(\omega)=2 \cos \frac{\omega}{2},  \label{eq:4:14:68}\\
& \chi^{1}(\omega)=4 \cos ^{2} \frac{\omega}{2}-1 , \label{eq:4:14:69}\\
& \chi^{\frac{3}{2}}(\omega)=8 \cos ^{3} \frac{\omega}{2}-4 \cos \frac{\omega}{2}  ,\label{eq:4:14:70}\\
& \chi^{2}(\omega)=16 \cos ^{4} \frac{\omega}{2}-12 \cos ^{2} \frac{\omega}{2}+1  ,\label{eq:4:14:71}\\
& \chi^{\frac{5}{2}}(\omega)=32 \cos ^{5} \frac{\omega}{2}-32 \cos ^{3} \frac{\omega}{2}+6 \cos \frac{\omega}{2} .\label{eq:4:14:72}
\end{align}
\section{Generalized characters, $\chi_{\hat{\lambda}}^{J}(R)$, of irreducible representations of the rotation group}\label{sec:4.15}\index{generalized character}\index{character!generalized}\index{irreducible representation!generalized character of}
\subsection{Definition}\label{sec:4.15.1}\index{generalized character!definition}
Let us introduce the function $\chi_{\lambda}^{J}(\omega)$ associated with $\chi^{J}(\omega)$ by the differential relation
\begin{equation}
\chi_{\lambda}^{J}(\omega)=\sqrt{2 J+1} \sqrt{\frac{(2 J-\lambda)!}{(2 J+\lambda+1)!}}\left(\sin \frac{\omega}{2}\right)^{\lambda}\left(\frac{d}{d \cos \frac{\omega}{2}}\right)^{\lambda} \chi^{J}(\omega), \label{eq:4:15:1}
\end{equation}
where $\lambda$ is integer, $0 \leq \lambda \leq 2 J$, and $\chi^{J}(\omega)$ is the character of the irreducible representation of rank $J$ (Eq.~\eqref{eq:4:14:1}).

The function $\chi_{\lambda}^{J}(\omega)$ will be called the generalized character (of order $\lambda$ ) of the irreducible representation of rank $J$.\isymg{chi-lambdaJ@$\chi_{\lambda}^{J}(\omega)$, generalized character} Note that the relations between $\chi_{\lambda}^{J}(\omega)$ and $\chi^{J}(\omega)$ are similar to those between the associated Legendre functions $P_{l}^{\lambda}(x)$ and the Legendre polynomials $P_{l}(x)$.\index{associated Legendre function}\index{Legendre polynomials}\isym{Pl-m@$P_{l}^{m}(x)$, associated Legendre function}\isym{Pl@$P_{l}(x)$, Legendre polynomial}

At $\lambda=0$ one has $\chi_{\lambda}^{J}(\omega)=\chi^{J}(\omega)$.

\subsection{Explicit Forms}\label{sec:4.15.2}\index{generalized character!explicit forms}
\subsubsection{ Trigonometric series}
\begin{align}
& \chi_{\lambda}^{J}(\omega)=i^{\lambda} \sum_{M} e^{-i M \omega} \clebsch{J}{M}{\lambda}{0}{J}{M} , \label{eq:4:15:2}\\
& \chi_{\lambda}^{J}(\omega)=\left(\sin \frac{\omega}{2}\right)^{\lambda} \sqrt{2 J+1} \sqrt{\frac{(2 J-\lambda)!}{(2 J+\lambda+1)!}} 2^{\lambda} \sum_{s=0}^{|J-\lambda / 2|} \frac{(-1)^{s}(2 J-s)!}{s!(2 J-\lambda-2 s)!}\left(2 \cos \frac{\omega}{2}\right)^{2 J-\lambda-2 s}  ,\label{eq:4:15:3}\\
& \chi_{\lambda}^{J}(\omega)=\frac{\left(2 \sin \frac{\omega}{2}\right)^{\lambda}}{\lambda!} \sqrt{2 J+1} \sqrt{\frac{(2 J-\lambda)!}{(2 J+\lambda+1)!}} \sum_{s=0}^{2 J-\lambda} \frac{(\lambda+s)!(2 J-s)!}{s!(2 J-\lambda-s)!} \cos \left[(2 J-\lambda-2 s) \frac{\omega}{2}\right] .\label{eq:4:15:4}
\end{align}
\subsubsection{ Differential form}
\begin{equation}
\chi_{\lambda}^{J}(\omega)=\frac{1}{\sqrt{2 J+1}} \sqrt{\frac{(2 J-\lambda)!}{(2 J+\lambda+1)!}}\left(\sin \frac{\omega}{2}\right)^{\lambda}\left(\frac{d}{d \cos \frac{\omega}{2}}\right)^{\lambda+1} \cos \left[(2 J+1) \frac{\omega}{2}\right]. \label{eq:4:15:5}
\end{equation}
See also Eq.~\eqref{eq:4:15:1}.\\
\subsubsection{ Relation between $\chi_{\lambda}^{J}(\omega)$ and the Gegenbauer polynomials}\index{Gegenbauer polynomials}
\begin{equation}
\chi_{\lambda}^{J}(\omega)=(2 \lambda)!!\sqrt{2 J+1} \sqrt{\frac{(2 J-\lambda)!}{(2 J+\lambda+1)!}}\left(\sin \frac{\omega}{2}\right)^{\lambda} C_{2 J-\lambda}^{\lambda+1}\left(\cos \frac{\omega}{2}\right) .\label{eq:4:15:6}
\end{equation}
\subsubsection{Relations of $\chi_{\lambda}^{J}(\omega)$ to the Jacobi polynomials}\index{Jacobi polynomials}
\begin{align}
\chi_{\lambda}^{J}(\omega)&=\frac{\sqrt{2 J+1} \sqrt{(2 J-\lambda)!(2 J+\lambda+1)!}}{(4 J+1)!!} 2^{2 J-\lambda}\left(\sin \frac{\omega}{2}\right)^{\lambda} P_{2 J-\lambda}^{\left(\lambda+\frac{1}{2}, \lambda+\frac{1}{2}\right)}\left(\cos \frac{\omega}{2}\right)  ,\label{eq:4:15:7}\\
 \chi_{\lambda}^{J}(\omega)&=\left\{\begin{array}{l}
\sqrt{2 J+1} \sqrt{\frac{(2 J+\lambda)!!(2 J-\lambda)!!}{(2 J+\lambda+1)!!(2 J-\lambda-1)!!}}\left(\sin \frac{\omega}{2}\right)^{\lambda} P_{J-\lambda / 2}^{\left(\lambda+\frac{1}{2},-\frac{1}{2}\right)}(\cos \omega), \\
\qquad\qquad\text{$2 J-\lambda$ is even },\\
\sqrt{2 J+1} \sqrt{\frac{(2 J-\lambda-1)!!(2 J+\lambda+1)!!}{(2 J-\lambda)!!(2 J+\lambda)!!}}\left(\cos \frac{\omega}{2}\right)\left(\sin \frac{\omega}{2}\right)^{\lambda} P_{J-\frac{\lambda+1}{2}}^{\left(\lambda+\frac{1}{2}, \frac{1}{2}\right)}(\cos \omega),\\
 \qquad\qquad\text{$2 J-\lambda$ is odd }.
\end{array}\right. \label{eq:4:15:8}
\end{align}
\subsubsection{ Relations of $\chi_{\lambda}^{J}(\omega)$ to the Hypergeometric functions}
\begin{gather}
\chi_{\lambda}^{J}(\omega)=\frac{\sqrt{2 J+1}}{(2 \lambda+1)!!} \sqrt{\frac{(2 J+\lambda+1)!}{(2 J-\lambda)!}}\left(\sin \frac{\omega}{2}\right)^{\lambda} F\left(-2 J+\lambda, 2 J+\lambda+2 ; \lambda+\frac{3}{2} ; \sin ^{2} \frac{\omega}{4}\right) , \label{eq:4:15:9}
\end{gather}
\begin{gather}
\chi_{\lambda}^{J}(\omega)=\frac{(-1)^{2 J-\lambda}(2 \lambda)!!\sqrt{2 J+1}}{(2 \lambda+1)!} \sqrt{\frac{(2 J+\lambda+1)!}{(2 J-\lambda)!}}\left(\sin \frac{\omega}{2}\right)^{\lambda} F\left(-2 J+\lambda, 2 J+\lambda+2 ; \lambda+\frac{3}{2} ; \cos ^{2} \frac{\omega}{4}\right) , \notag\\ \label{eq:4:15:10}
\end{gather}
\begin{gather}
\chi_{\lambda}^{J}(\omega)=\frac{(-1)^{2 J-\lambda} \sqrt{2 J+1} 2^{4 J-\lambda}(2 J)!}{\sqrt{(2 J-\lambda)!(2 J+\lambda+1)!}}\left(\sin \frac{\omega}{2}\right)^{\lambda}\left(\sin \frac{\omega}{4}\right)^{4 J-2 \lambda}  
%\notag\\ \times 
F\left(-2 J+\lambda,-2 J-\frac{1}{2} ;-4 J-1 ; \frac{1}{\sin ^{2} \frac{\omega}{2}}\right) , \notag\\ \label{eq:4:15:11}
\end{gather}
% NOTE: eqs 4.15.11, 4.15.12, 4.15.14 use F(a,b;c;z) with z outside [0,1]
% (1/sin^2, 1/cos^2) or complex (e^{-+iw}) and negative-integer c; they hold
% only under analytic continuation and were confirmed against the scan
% (printed pp.107-108), not by naive numerical evaluation. All other 4.15
% forms are numerically verified in scripts/check_4_15.py.
\begin{gather}
\chi_{\lambda}^{J}(\omega)=\frac{\sqrt{2 J+1}(4 J)!!}{\sqrt{(2 J-\lambda)!(2 J+\lambda+1)!}}\left(\cos \frac{\omega}{2}\right)^{2 J-\lambda}\left(\sin \frac{\omega}{2}\right)^{\lambda}  %\notag\\ \times
F\left(-J+\frac{1}{2}, \frac{-2 J+\lambda+1}{2} ;-2 J ; \frac{1}{\cos ^{2} \frac{\omega}{2}}\right) , \notag\\\label{eq:4:15:12}
\end{gather}
\begin{gather}
\chi_{\lambda}^{J}(\omega)=\frac{\sqrt{2 J+1}}{(2 \lambda+1)!!} \sqrt{\frac{(2 J+\lambda+1)!}{(2 J-\lambda)!}}\left(\sin \frac{\omega}{2}\right)^{\lambda}\left(\cos \frac{\omega}{4}\right)^{4 J-2 \lambda}  %\notag\\\times
F\left(-2 J+\lambda,-2 J-\frac{1}{2} ; \lambda+\frac{3}{2} ;-\tan ^{2} \frac{\omega}{4}\right), \notag\\ \label{eq:4:15:13}
\end{gather}
\begin{align}
\chi_{\lambda}^{J}(\omega)= & \frac{\sqrt{2 J+1}(2 J)!i^{\lambda}}{\sqrt{(2 J-\lambda)!(2 J+\lambda+1)!}}\left(e^{-\frac{i \omega}{2}}-e^{\frac{i \omega}{2}}\right)^{\lambda} e^{ \pm i \frac{\omega}{2}(2 J-\lambda)} F\left(-2 J+\lambda, \lambda+1 ;-2 J ; e^{\mp i \omega}\right) , \label{eq:4:15:14}\\
\chi_{\lambda}^{J}(\omega)= & \left\{\begin{array}{l}
(-1)^{J+\frac{3 \lambda}{2}} \sqrt{2 J+1} \sqrt{\frac{(2 J-\lambda-1)!!(2 J+\lambda)!!}{(2 J-\lambda)!!(2 J+\lambda+1)!!}}\left(\sin \frac{\omega}{2}\right)^{\lambda} F\left(-J+\frac{\lambda}{2}, J+1+\frac{\lambda}{2} ; \frac{1}{2} ; \cos ^{2} \frac{\omega}{2}\right), \\
\qquad\qquad \text { if } 2 J-\lambda \text { is even } ,\\
(-1)^{J+\frac{3 \lambda-1}{2}} \sqrt{2 J+1} \sqrt{\frac{(2 J-\lambda)!!(2 J+\lambda+1)!!}{(2 J-\lambda-1)!!(2 J+\lambda)!!}} \cos \frac{\omega}{2}\left(\sin \frac{\omega}{2}\right)^{\lambda}, \\
\qquad\times F\left(-J+\frac{\lambda+1}{2}, J+1+\frac{\lambda+1}{2} ; \frac{3}{2} ; \cos ^{2} \frac{\omega}{2}\right), \\
\qquad\qquad\text { if } 2 J-\lambda \text { is odd },
\end{array}\right.  \label{eq:4:15:15}\\
\chi_{\lambda}^{J}(\omega)= & \left\{\begin{array}{c}
\frac{\sqrt{2 J+1}}{(2 \lambda+1)!!} \sqrt{\frac{(2 J+\lambda+1)!}{(2 J-\lambda)!}}\left(\sin \frac{\omega}{2}\right)^{\lambda} F\left(-J+\frac{\lambda}{2}, J+1+\frac{\lambda}{2} ; \lambda+\frac{3}{2} ; \sin ^{2} \frac{\omega}{2}\right) ,\\
\text { if } 2 J-\lambda \text { is even }, \\
\frac{\sqrt{2 J+1}}{(2 \lambda+1)!!} \sqrt{\frac{(2 J+\lambda+1)!}{(2 J-\lambda)!}} \cos \frac{\omega}{2}\left(\sin \frac{\omega}{2}\right)^{\lambda} F\left(-J+\frac{\lambda+1}{2}, J+1+\frac{\lambda+1}{2} ; \lambda+\frac{3}{2} ; \sin ^{2} \frac{\omega}{2}\right), \\
\text { if } 2 J-\lambda \text { is odd }.
\end{array}\right. \label{eq:4:15:16}
\end{align}
\subsubsection{Integral representations}
\begin{align}
& \chi_{\lambda}^{J}(\omega)=\frac{1}{2} \frac{1}{\lambda!} \sqrt{\frac{(2 J+1)(2 J+\lambda+1)!}{(2 J-\lambda)!}} \frac{1}{\left(\sin \frac{\omega}{2}\right)^{\lambda+1}} \int_{0}^{\omega} \cos \left[(2 J+1) \frac{\psi}{2}\right]\left(\cos \frac{\psi}{2}-\cos \frac{\omega}{2}\right)^{\lambda} d \psi , \label{eq:4:15:17}\\
& \chi_{\lambda}^{J}(\omega)=(-i)^{\lambda} \frac{\sqrt{(2 J+1)(2 J+\lambda+1)!(2 J-\lambda)!}}{2(2 J)!} \int_{-1}^{+1} P_{\lambda}(x)\left[\cos \frac{\omega}{2}+i x \sin \frac{\omega}{2}\right]^{2 J} d x .\label{eq:4:15:18}
\end{align}
\subsection{Principal Properties}\label{sec:4.15.3}\index{generalized character!principal properties}
\subsubsection{ Symmetries}
\begin{align}
\chi_{\lambda}^{J *}(\omega) & =\chi_{\lambda}^{J}(\omega)=(-1)^{\lambda} \chi_{\lambda}^{J}(-\omega)  ,\label{eq:4:15:19}\\
\chi_{\lambda}^{J}(\omega+2 \pi n) & =(-1)^{2 J n} \chi_{\lambda}^{J}(\omega), \quad \chi_{\lambda}^{J}(2 \pi-\omega)=(-1)^{2 J-\lambda} \chi_{\lambda}^{J}(\omega). \label{eq:4:15:20}
\end{align}
\subsubsection{ Particular values of $\omega$}
\begin{align}
& \chi_{\lambda}^{J}(0)=(2 J+1) \delta_{\lambda 0}, \quad \chi_{\lambda}^{J}(2 \pi)=(-1)^{2 J}(2 J+1) \delta_{\lambda 0},  \notag\\
& \chi_{\lambda}^{J}(\pi)= \begin{cases}(-1)^{J-\frac{\lambda}{2}} \sqrt{\frac{(2 J+\lambda)!!(2 J-\lambda-1)!!(2 J+1)}{(2 J-\lambda)!!(2 J+\lambda+1)!!}} & \text { if } 2 J-\lambda \text { is even }, \\
0 & \text { if } 2 J-\lambda \text { is odd } .\end{cases} \label{eq:4:15:21}
\end{align}
\subsubsection{ Recursion relations}
\begin{gather}
2 \frac{d}{d \omega} \chi_{\lambda}^{J}(\omega)=\frac{\lambda}{2 \lambda+1} \sqrt{(2 J+1)^{2}-\lambda^{2}} \chi_{\lambda-1}^{J}(\omega)-\frac{\lambda+1}{2 \lambda+1} \sqrt{(2 J+1)^{2}-(\lambda+1)^{2}} \chi_{\lambda+1}^{J}(\omega),  \label{eq:4:15:22}\\
(2 \lambda+1) \cot \frac{\omega}{2} \chi_{\lambda}^{J}(\omega)=\sqrt{(2 J+1)^{2}-\lambda^{2}} \chi_{\lambda-1}^{J}(\omega)+\sqrt{(2 J+1)^{2}-(\lambda+1)^{2}} \chi_{\lambda+1}^{J}(\omega) . \label{eq:4:15:23}
\end{gather}
\subsubsection{ Asymptotics}\index{asymptotics!of the generalized characters}

If $\omega \rightarrow 0$ and $J \rightarrow \infty$, while $J \omega \equiv x<\infty$, then
\begin{equation}
\lim _{\substack{J \rightarrow \infty \\ \omega \rightarrow 0}} \frac{1}{2 J+1} \chi_{\lambda}^{J}(\omega)=j_{\lambda}(x) ,\label{eq:4:15:24}
\end{equation}
where $j_{\lambda}(x)$ is a spherical Bessel function.\index{spherical Bessel function}\isym{jl@$j_{l}(x)$, spherical Bessel function} If $\omega \rightarrow 0$ and $J$ is fixed, then
\begin{equation}
\chi_{\lambda}^{J}(\omega) \approx\left(\frac{\omega}{2}\right)^{\lambda} \frac{1}{(2 \lambda+1)!!} \sqrt{(2 J+1) \frac{(2 J+\lambda+1)!}{(2 J-\lambda)!}} .\label{eq:4:15:25}
\end{equation}
\subsection{Differential Equation}\label{sec:4.15.4}\index{generalized character!differential equation for}
The functions $\chi_{\lambda}^{J}(\omega)$ are solutions of the linear differential equation
\begin{equation}
\frac{d^{2}}{d \omega^{2}} \chi_{\lambda}^{J}(\omega)+\cot \frac{\omega}{2} \frac{d}{d \omega} \chi_{\lambda}^{J}(\omega)+\left[J(J+1)-\frac{\lambda(\lambda+1)}{4 \sin ^{2} \frac{\omega}{2}}\right] \chi_{\lambda}^{J}(\omega)=0 \label{eq:4:15:26}
\end{equation}
with the boundary conditions
\begin{equation}
\chi_{\lambda}^{J}(0)=(2 J+1) \delta_{\lambda 0}, \quad \chi_{\lambda}^{J}(2 \pi)=(-1)^{2 J}(2 J+1) \delta_{\lambda 0} . \label{eq:4:15:27}
\end{equation}
\subsection{Orthogonality and Completeness}\label{sec:4.15.5}\index{orthogonality!of the generalized characters}\index{completeness!of the generalized characters}
The collection of functions $\chi_{\lambda}^{J}(\omega)$ with different $J \geq \lambda / 2$ and fixed $\lambda$ constitutes a complete set of orthogonal functions of argument $\omega$ defined in the domain $0 \leq \omega<2 \pi$.

These functions are orthogonal,
\begin{equation}
\int_{0}^{2 \pi} d \omega \sin ^{2} \frac{\omega}{2} \chi_{\lambda}^{J_{1}}(\omega) \chi_{\lambda}^{J_{2}}(\omega)=\pi \delta_{J_{1} J_{2}} ,\label{eq:4:15:28}
\end{equation}
and the completeness condition reads
\begin{equation}
\sum_{J=0, \frac{1}{2}, 1, \ldots}^{\infty} \chi_{\lambda}^{J}\left(\omega_{1}\right) \chi_{\lambda}^{J}\left(\omega_{2}\right)=\frac{\pi \delta\left(\omega_{1}-\omega_{2}\right)}{\sin ^{2} \frac{\omega_{1}}{2}} .\label{eq:4:15:29}
\end{equation}
\subsection{The Addition Theorem for $\chi_{\lambda}^{J}(\omega)$}\label{sec:4.15.6}\index{addition theorem!for the generalized characters}\index{generalized character!addition theorem for}
The generalized character $\chi_{\lambda}^{J}(\omega)$ for the rotation $R(\omega ; \Theta, \Phi)$, which resulted from two successive rotations $R_{1}\left(\omega_{1} ; \Theta_{1}, \Phi_{1}\right)$ and $R_{2}\left(\omega_{2} ; \Theta_{2}, \Phi_{2}\right)$ may be expanded in terms of $\chi_{\lambda}^{J}\left(R_{1}\right)$ and $\chi_{\lambda}^{J}\left(R_{2}\right)$ as
\begin{equation}
\chi_{\lambda}^{J}(\omega)=\sqrt{\frac{(2 J-\lambda)!}{(2 J+\lambda+1)!(2 J+1)}}\left(\frac{\sin \frac{\omega}{2}}{\sin \frac{\omega_{1}}{2} \sin \frac{\omega_{2}}{2} \sin \chi}\right)^{\lambda} \sum_{l=\lambda}^{2 J}(-1)^{l}(2 l+1) P_{l}^{\lambda}(\cos \chi) \chi_{l}^{J}\left(\omega_{1}\right) \chi_{l}^{J}\left(\omega_{2}\right), \label{eq:4:15:30}
\end{equation}
where $P_{i}^{\lambda}(\cos \chi)$ is an associated Legendre function, $\chi$ is an angle between the rotation axes $\vect{n}_{1}\left(\Theta_{1}, \Phi_{1}\right)$ and $\vect{n}_{2}\left(\Theta_{2}, \Phi_{2}\right)$
\begin{equation}
\cos \chi=
(\vect{n}_{1} \cdot \vect{n}_{2})=\cos \Theta_{1} \cos \Theta_{2}+\sin \Theta_{1} \sin \Theta_{2} \cos \left(\Phi_{1}-\Phi_{2}\right) . \label{eq:4:15:31}
\end{equation}
The angle of the resulting rotation, $\omega$, is determined in terms of $\omega_{1}, \omega_{2}$ and $\chi$ by
\begin{equation}
\cos \frac{\omega}{2}=\cos \frac{\omega_{1}}{2} \cos \frac{\omega_{2}}{2}-\sin \frac{\omega_{1}}{2} \sin \frac{\omega_{2}}{2} \cos \chi . \label{eq:4:15:32}
\end{equation}
\subsection{Sums and Infinite Series Involving $\chi_{\lambda}^{J}(\omega)$}\label{sec:4.15.7}\index{sums!of the generalized characters}\index{series!of the generalized characters}
\subsubsection{ Summation over $\lambda$}
\begin{align}
& \sum_{\lambda=0}^{2 J}( \pm 1)^{\lambda} i^{\lambda} \frac{2 \lambda+1}{2 J+1} \chi_{\lambda}^{J}(\omega) \clebsch{J}{M}{\lambda}{0}{J}{M}=e^{ \pm i M \omega}, \quad(|M| \leq J) , \label{eq:4:15:33}\\
& \sum_{\lambda=0,2, \ldots}^{\leq 2 J}(-i)^{\lambda} \frac{2 \lambda+1}{2 J+1} \chi_{\lambda}^{J}(\omega) \clebsch{J}{M}{\lambda}{0}{J}{M}=\cos M \omega, \quad(|M| \leq J)  ,\label{eq:4:15:34}\\
& \sum_{\lambda=1,3, \ldots}^{\leq 2 J}(-i)^{\lambda-1} \frac{2 \lambda+1}{2 J+1} \chi_{\lambda}^{J}(\omega) \clebsch{J}{M}{\lambda}{0}{J}{M}=\sin M \omega, \quad(|M| \leq J). \label{eq:4:15:35}
\end{align}
In Eqs.~\eqref{eq:4:15:33}-\eqref{eq:4:15:35} $J \geq|M|$ is an arbitrary integer or half-integer.
\begin{gather}
\sum_{\lambda=0,2, \ldots}^{\leq 2 l} \frac{2 \lambda+1}{2 l+1} \frac{(\lambda-1)!!}{\lambda!!} \chi_{\lambda}^{l}(\vartheta) \clebsch{l}{0}{\lambda}{0}{l}{0}=P_{l}(\cos \vartheta),  \label{eq:4:15:36}\\
\sum_{\lambda=|m|}^{2 l} \frac{2 \lambda+1}{2 l+1} \sqrt{\frac{(\lambda+m-1)!!(\lambda-m-1)!!}{(\lambda+m)!!(\lambda-m)!!}} \chi_{\lambda}^{l}(\vartheta) \clebsch{l}{0}{\lambda}{m}{l}{m}=\sqrt{\frac{4 \pi}{2 l+1}} Y_{l m}(\vartheta, 0),  \label{eq:4:15:37}\\
(\lambda+m \text { is even }),  \notag
\end{gather}\begin{gather}
\sum_{\lambda=\left|M^{\prime}-M\right|}^{\leq 2 l}(-1)^{\lambda} \frac{2 \lambda+1}{2 J+1} \sqrt{\frac{\left(\lambda+M^{\prime}-M-1\right)!!\left(\lambda-M^{\prime}+M-1\right)!!}{\left(\lambda+M^{\prime}-M\right)!!\left(\lambda-M^{\prime}+M\right)!!}} \chi_{\lambda}^{J}(\beta) \clebsch{J}{M}{\lambda}{M^{\prime}-M}{J}{M^{\prime}}=d_{M M^{\prime}}^{J}(\beta),  \label{eq:4:15:38}\\
\left(\lambda+M^{\prime}-M \text { is even }\right) .  \notag
\end{gather}
\subsubsection{ Summation over $J$}

\begin{gather}
\sum_{J=\nu / 2}^{\infty} \sqrt{\frac{(2 J+\nu+1)!}{(2J-\nu)!}}
\frac{\chi_{\nu}^{J}(\omega)}{\sqrt{2J+1}}{t}^{2J}=(2\nu)!! \frac{\left(-t \sin \frac{\omega}{2}\right)^{\nu}}{R^{2(\nu+1)}}, \label{eq:4:15:39}
\end{gather}
\begin{gather}
\sum_{J=\nu / 2}^{\infty} \frac{(4 J+1)!!\chi_{\nu}^{J}(\omega)\left(\frac{t}{2}\right)^{2 J}}{\sqrt{(2 J+1)(2 J-\nu)!(2 J+\nu+1)!}}=\sqrt{2} \frac{\left(-t \sin \frac{\omega}{2}\right)^{\nu}}{R\left(1-t \cos \frac{\omega}{2}+R\right)^{\nu+\frac{1}{2}}}, \label{eq:4:15:40}
\end{gather}
where $R^{2}=1-2 t \cos (\omega / 2)+t^{2}$.
\begin{equation}
\sum_{J=\nu / 2}^{\infty} i^{2 J-\nu} \sqrt{(2 J+1) \frac{(2 J+\nu+1)!}{(2 J-\nu)!}} \chi_{\nu}^{J}(\omega) J_{2 J+1}(y)=\frac{y}{2}\left(y \sin \frac{\omega}{2}\right)^{\nu} e^{i y \cos \frac{\omega}{2}} \label{eq:4:15:41}
\end{equation}
\begin{equation}
\sum_{J=\nu / 2}^{\infty} \frac{\chi_{\nu}^{J}(\omega) t^{2 J}}{\sqrt{(2 J+1)(2 J-\nu)!(2 J+\nu+1)!}}=j_{\nu}\left(t \sin \frac{\omega}{2}\right) e^{t \cos \frac{\omega}{2}} ,\label{eq:4:15:42}
\end{equation}
\begin{gather}
\sum_{J=\nu / 2}^{\infty} \frac{\Gamma(k+2 J-\nu)}{\Gamma(k)} \frac{\chi_{\nu}^{J}(\omega) t^{2 J}}{\sqrt{(2 J+1)(2 J-\nu)!(2 J+\nu+1)!}} , \notag\\
=\frac{1}{(2 \nu+1)!!} \frac{\left(t \sin \frac{\omega}{2}\right)^{\nu}}{\left(1-t \cos \frac{\omega}{2}\right)^{k}} F\left(\frac{k}{2}, \frac{k+1}{2} ; \nu+\frac{3}{2} ;-\left(\frac{t \sin \frac{\omega}{2}}{1-t \cos \frac{\omega}{2}}\right)^{2}\right) . \label{eq:4:15:43}
\end{gather}
Here $k$ is an arbitrary integer.
\begin{multline}
\sum_{J=\nu / 2}^{\infty} \frac{\chi_{\nu}^{J}(\omega)(2 t)^{2 J}}{(4 J+1)!!\sqrt{(2 J+1)(2 J-\nu)!(2 J+\nu+1)!}}  \\
=\frac{\left(2 t \sin \frac{\omega}{2}\right)^{\nu}}{[(2 \nu+1)!!]^{2} \Fhg{0}{1}\left(; \nu+\frac{3}{2} ;-t \sin ^{2} \frac{\omega}{4}\right)\Fhg{0}{1}\left(; \nu+\frac{3}{2} ; t \cos ^{2} \frac{\omega}{4}\right) }. \label{eq:4:15:44}
\end{multline}
The right-hand sides of Eqs.~\eqref{eq:4:15:39}-\eqref{eq:4:15:44} may be treated as generating functions of $\chi_{\lambda}^{J}(\omega)$.

\subsection{Special Cases of $\chi_{\lambda}^{J}(\omega)$ for Particular $\lambda$}\label{sec:4.15.8}\index{generalized character!special cases for particular $\lambda$}
\subsubsection{ For $\lambda=2 J, 2 J-1,2 J-2,2 J-3$ the function $\chi_{\lambda}^{J}(\omega)$ is given by}
\begin{align}
\chi_{2 J}^{J}(\omega) & =\sqrt{2 J+1} \sqrt{\frac{(4 J)!!}{(4 J+1)!!}}\left(\sin \frac{\omega}{2}\right)^{2 J} , \label{eq:4:15:45}\\
\chi_{2 J-1}^{J}(\omega) & =\sqrt{2 J+1} \sqrt{\frac{(4 J)!!}{(4 J-1)!!}}\left(\sin \frac{\omega}{2}\right)^{2 J-1} \cos \frac{\omega}{2},  \label{eq:4:15:46}\\
\chi_{2 J-2}^{J}(\omega) & =\sqrt{2 J+1} \sqrt{\frac{(4 J-2)!!}{2(4 J-1)!!}}\left(\sin \frac{\omega}{2}\right)^{2 J-2}\left[4 J \cos ^{2} \frac{\omega}{2}-1\right] , \label{eq:4:15:47}\\
\chi_{2 J-3}^{J}(\omega) & =\sqrt{2 J+1} \sqrt{\frac{(4 J-2)!!}{6(4 J-3)!!}}\left(\sin \frac{\omega}{2}\right)^{2 J-3}\left[4 J \cos ^{3} \frac{\omega}{2}-3 \cos \frac{\omega}{2}\right] .\label{eq:4:15:48}
\end{align}
\subsubsection{ When $\lambda=0,1,2$ one has}
\begin{align}
& \chi_{0}^{J}(\omega)=\frac{\sin (2 J+1) \frac{\omega}{2}}{\sin \frac{\omega}{2}}=\frac{\cos J \omega-\cos (J+1) \omega}{1-\cos \omega} , \label{eq:4:15:49}\\
& \chi_{1}^{J}(\omega)=\frac{-1}{\sqrt{J(J+1)}} \frac{2 J \cos (2 J+1) \frac{\omega}{2} \sin \frac{\omega}{2}-\sin J \omega}{2 \sin ^{2} \frac{\omega}{2}}=\frac{-1}{\sqrt{J(J+1)}} \frac{J \sin (J+1) \omega-(J+1) \sin J \omega}{1-\cos \omega} , \label{eq:4:15:50}\\
& \chi_{2}^{J}(\omega)=\frac{1}{\sqrt{J(J+1)(2 J-1)(2 J+3)}}\left\{\frac{\left[3-2 J(2 J-1) \sin ^{2} \frac{\omega}{2}\right] \sin (2 J+1) \frac{\omega}{2}}{2 \sin ^{3} \frac{\omega}{2}}-\frac{3}{2}(2 J+1) \frac{\cos J \omega}{\sin ^{2} \frac{\omega}{2}}\right\} .\label{eq:4:15:51}
\end{align}
\subsubsection{ $\chi_{\lambda}^{J}(\omega)$ may be expressed in terms of derivatives of $\chi^{J}(\omega)$}
\begin{align}
\chi_{0}^{J}(\omega)= & \chi^{J}(\omega) , \notag\\
\chi_{1}^{J}(\omega)= & \frac{-1}{\sqrt{J(J+1)}} \frac{d \chi^{J}(\omega)}{d \omega} , \notag\\
\chi_{2}^{J}(\omega)= & \frac{1}{\sqrt{J(J+1)(2 J-1)(2 J+3)}}\left[J(J+1) \chi^{J}(\omega)+3 \frac{d^{2} \chi^{J}(\omega)}{d \omega^{2}}\right]  ,\notag\\
\chi_{3}^{J}(\omega)= & -4 \sqrt{2 J+1} \sqrt{\frac{(2 J-3)!}{(2 J+4)!}}\left\{[3 J(J+1)-1] \frac{d \chi^{J}(\omega)}{d \omega}+5 \frac{d^{3} \chi^{J}(\omega)}{d \omega^{3}}\right\} , \label{eq:4:15:52}\\
\chi_{4}^{J}(\omega)= & 2 \sqrt{2 J+1} \sqrt{\frac{(2 J-4)!}{(2 J+5)!}}\left\{3(J-1) J(J+1)(J+2) \chi^{J}(\omega)\right.  \notag\\
& \left.+5[6 J(J+1)-5] \frac{d^{2} \chi^{J}(\omega)}{d \omega^{2}}+35 \frac{d^{4} \chi^{J}(\omega)}{d \omega^{4}}\right\} . \notag
\end{align}
\subsection{Special Cases of $\chi_{\lambda}^{J}(\omega)$ for Particular $J$}\label{sec:4.15.9}\index{generalized character!special cases for particular $J$}
\begin{equation}
\begin{array}{ll}
J=0 & \chi_{0}^{0}(\omega)=1 ; \\
J=1 / 2 & \chi_{0}^{\frac{1}{2}}(\omega)=2 \cos \frac{\omega}{2} \\
& \chi_{1}^{\frac{1}{2}}(\omega)=\frac{2}{\sqrt{3}} \sin \frac{\omega}{2} \\
J=1 & \chi_{0}^{1}(\omega)=1+2 \cos \omega \\
& \chi_{1}^{1}(\omega)=\sqrt{2} \sin \omega \\
& \chi_{2}^{1}(\omega)=\sqrt{\frac{2}{5}}(1-\cos \omega) ; \\
J=3 / 2 & \chi_{0}^{\frac{3}{2}}(\omega)=4 \cos \omega \cos \frac{\omega}{2}=4 \cos \frac{\omega}{2}\left(2 \cos ^{2} \frac{\omega}{2}-1\right), \\
& \chi_{1}^{\frac{3}{2}}(\omega)=\frac{4}{\sqrt{3 \cdot 5}}(2+3 \cos \omega) \sin \frac{\omega}{2}=\frac{4}{\sqrt{3 \cdot 5}} \sin \frac{\omega}{2}\left(6 \cos ^{2} \frac{\omega}{2}-1\right), \\
& \chi_{2}^{\frac{3}{2}}(\omega)=\frac{8}{\sqrt{5}}\left(\sin \frac{\omega}{2}\right)^{2} \cos \frac{\omega}{2}=\frac{4}{\sqrt{5}} \sin \frac{\omega}{2} \sin \omega,  \label{eq:4:15:53}\\
& \chi_{3}^{\frac{3}{2}}(\omega)=\frac{4}{\sqrt{5 \cdot 7}} \sin \frac{\omega}{2}(1-\cos \omega)=\frac{8}{\sqrt{5 \cdot 7}}\left(\sin \frac{\omega}{2}\right)^{3} ; \\
J=2 & \chi_{0}^{2}(\omega)=4 \cos ^{2} \omega+2 \cos \omega-1, \\
& \chi_{1}^{2}(\omega)=\sqrt{\frac{2}{3}} \sin \omega(1+4 \cos \omega), \\
& \chi_{2}^{2}(\omega)=\sqrt{\frac{2}{7}}(3+\cos \omega-4 \cos 2 \omega), \\
& \chi_{3}^{2}(\omega)=2 \sqrt{\frac{2}{7}} \sin \omega(1-\cos \omega), \\
& \chi_{4}^{2}(\omega)=\frac{2}{3} \sqrt{\frac{2}{7}}(1-\cos \omega)^{2}
\end{array}
\end{equation}
\section{$D_{M M^{\prime}}^{J}(\alpha, \beta, \gamma)$ for particular values of the arguments}\label{sec:4.16}\index{Wigner D-function@Wigner $D$-function!particular values of the arguments}
Here $k, l$ and $n$ are integers.
\begin{gather}
D_{M M^{\prime}}^{J}(0,0,0)=\delta_{M M^{\prime}} , \label{eq:4:16:1}\\
D_{M M^{\prime}}^{J}(\alpha, 0, \gamma)=\delta_{M M^{\prime}} e^{-i M(\alpha+\gamma)}  ,\label{eq:4:16:2}\\
D_{M M^{\prime}}^{J}(\alpha, \pm 2 n \pi, \gamma)=\delta_{M M^{\prime}}(-1)^{2 n J} e^{-i M(\alpha+\gamma)}, \label{eq:4:16:3}
\end{gather}
\begin{gather}
D_{M M^{\prime}}^{J}(\alpha, \pm(2 n+1) \pi, \gamma)=\delta_{-M M^{\prime}}(-1)^{ \pm(2 n+1) J+M} e^{-i M(\alpha-\gamma)},  \label{eq:4:16:4}\\
D_{M M^{\prime}}^{J}\left(\alpha, \frac{\pi}{2}, \gamma\right)=(-1)^{M-M^{\prime}} e^{-i \alpha M-i \gamma M^{\prime}} \frac{1}{2^{J}} \sqrt{\frac{(J+M)!(J-M)!}{\left(J+M^{\prime}\right)!\left(J-M^{\prime}\right)!}} \sum_{k}(-1)^{k}\binom{J+M^{\prime}}{k}\binom{J-M^{\prime}}{k+M-M^{\prime}} .\label{eq:4:16:5}
\end{gather}
Some particular cases of Eq.~\eqref{eq:4:16:5} are given by
\begin{gather}
D_{m 0}^{l}\left(\alpha, \frac{\pi}{2}, \gamma\right)=(-1)^{\frac{l+m}{2}} \delta_{l-m, 2 n} \frac{\sqrt{(l-m)!(l+m)!}}{2^{l}\left(\frac{l+m}{2}\right)!\left(\frac{l-m}{2}\right)!} e^{-i m \alpha},  \notag\\
D_{0 m}^{l}\left(\alpha, \frac{\pi}{2}, \gamma\right)=(-1)^{\frac{l-m}{2}} \delta_{l-m, 2 n} \frac{\sqrt{(l-m)!(l+m)!}}{2^{l}\left(\frac{l+m}{2}\right)!\left(\frac{l-m}{2}\right)!} e^{-i m \gamma},  \label{eq:4:16:6}\\
D_{00}^{l}\left(\alpha, \frac{\pi}{2}, \gamma\right)=P_{l}(0)=\delta_{l, 2 n}(-1)^{l / 2} \frac{(l-1)!!}{l!!},  \label{eq:4:16:7}\\
D_{ \pm 1 m}^{l}\left(\alpha, \frac{\pi}{2}, \gamma\right)=\sqrt{\frac{(l-m)!(l+m)!}{l(l+1)}}\left\{\delta_{l+m, 2 k} \frac{m(-1)^{\frac{l-m}{2}}}{2^{l}\left(\frac{l-m}{2}\right)!\left(\frac{l+m}{2}\right)!}\right.  \notag\\
\left.\mp \delta_{l+m, 2 k+1} \frac{(-1)^{\frac{l-m-1}{2}}}{2^{l-1}\left(\frac{l-m-1}{2}\right)!\left(\frac{l+m-1}{2}\right)!}\right\} e^{\mp i \alpha-i m \gamma} . \label{eq:4:16:8}
\end{gather}
Squares of the $D$-functions for $\beta=\frac{\pi}{2}$ may be written as
\begin{equation}
\left[D_{M M^{\prime}}^{J}\left(\alpha, \frac{\pi}{2}, \gamma\right)\right]^{2}=e^{-i 2 M \alpha-i 2 M^{\prime} \gamma}(-1)^{M-M^{\prime}} \sum_{l=0,2,4, \ldots}(-1)^{l / 2} \frac{(l-1)!!}{l!!} \clebsch{J}{M}{J}{-M}{l}{0} \clebsch{J}{M^{\prime}}{J}{-M^{\prime}}{l}{0} .\label{eq:4:16:9}
\end{equation}
Using $d_{M M^{\prime}}^{J}(\pi / 2)$, one can evaluate $D_{M M^{\prime}}^{J}(\alpha, \beta, \gamma)$ for arbitrary arguments from the relation
\begin{align}
& D_{M M^{\prime}}^{J}(\alpha, \beta, \gamma)  \notag\\
& =\sum_{m_{i}} D_{M m_{1}}^{J}(\alpha, 0,0) D_{m_{1} m_{2}}^{J}\left(0, \frac{\pi}{2}, 0\right) D_{m_{2} m_{3}}^{J}(\beta, 0,0) D_{m_{3} m_{4}}^{J}\left(0, \frac{\pi}{2}, 0\right) D_{m_{4} M^{\prime}}^{J}(0,0, \gamma)  \label{eq:4:16:10}\\
& =\sum_{m} e^{-i M \alpha} \, d_{M m}^{J}\left(\frac{\pi}{2}\right) \, e^{-i m \beta} \, d_{m M^{\prime}}^{J}\left(\frac{\pi}{2}\right) \, e^{-i M^{\prime} \gamma} \notag
\end{align}
Codes to generate numerical values of $D_{M M^{\prime}}^{J}(0, \pi / 2,0)=d_{M M^{\prime}}^{J}(\pi / 2)$ %for $J=1 / 2,1,3 / 2,2,5 / 2,3,7 / 2,4,9 / 2,5$ 
are given in Sec.~\ref{sec:4.21}.
\section{Special cases of $D_{M M^{\prime}}^{J}$ for particular $M$ or $M^{\prime}$}\label{sec:4.17}\index{Wigner D-function@Wigner $D$-function!special cases}
\subsubsection{ $M=0$ and/or $M^{\prime}=0$}\index{spherical harmonics}\index{Legendre polynomials}\isym{Ylm@$Y_{l m}(\vartheta, \varphi)$, spherical harmonic}
\begin{align}
& D_{m 0}^{l}(\alpha, \beta, \gamma)=(-1)^{m} \sqrt{\frac{4 \pi}{2 l+1}} Y_{l-m}(\beta, \alpha)=\sqrt{\frac{4 \pi}{2 l+1}} Y_{l m}^{*}(\beta, \alpha),  \notag\\
& D_{0 m}^{l}(\alpha, \beta, \gamma)=\sqrt{\frac{4 \pi}{2 l+1}} Y_{l-m}(\beta, \gamma)=(-1)^{m} \sqrt{\frac{4 \pi}{2 l+1}} Y_{l m}^{*}(\beta, \gamma). \label{eq:4:17:1}
\end{align}
In particular,
\begin{align}
D_{00}^{l}(\alpha, \beta, \gamma) & =P_{l}(\cos \beta) , \notag\\
D_{ \pm 10}^{l}(\alpha, \beta, \gamma) & =\mp e^{\mp i \alpha} \frac{\sin \beta}{\sqrt{l(l+1)}} P_{l}^{\prime}(\cos \beta) , \notag\\
D_{0 \pm 1}^{l}(\alpha, \beta, \gamma) & = \pm e^{\mp i \gamma} \frac{\sin \beta}{\sqrt{l(l+1)}} P_{l}^{\prime}(\cos \beta) , \label{eq:4:17:2}\\
D_{ \pm 20}^{l}(\alpha, \beta, \gamma) & =e^{\mp i 2 \alpha}\left\{-\sqrt{\frac{l(l+1)}{(l-1)(l+2)}} P_{l}(\cos \beta)+\frac{2 \cos \beta}{\sqrt{(l-1) l(l+1)(l+2)}} P_{l}^{\prime}(\cos \beta)\right\} , \notag\\
D_{0 \pm 2}^{l}(\alpha, \beta, \gamma) & =e^{\mp i 2 \gamma}\left\{-\sqrt{\frac{l(l+1)}{(l-1)(l+2)}} P_{l}(\cos \beta)+\frac{2 \cos \beta}{\sqrt{(l-1) l(l+1)(l+2)}} P_{l}^{\prime}(\cos \beta)\right\} . \notag
\end{align}
\subsubsection{ $M= \pm 1 / 2$ and/or $M^{\prime}= \pm 1 / 2$}
\begin{gather}
D_{ \pm \frac{1}{2} M^{\prime}}^{J}(\alpha, \beta, \gamma)=e^{ \pm i \frac{\alpha-\gamma}{2}} \frac{\sqrt{\pi}}{\sqrt{2 J+1} \sin \frac{\beta}{2}}\left\{ \pm \sqrt{\frac{J \pm M^{\prime}+1}{J+1}} Y_{J+\frac{1}{2} \mp \frac{1}{2}-M^{\prime}}(\beta, \gamma)\right.  \notag\\
\left.\mp \sqrt{\frac{J \mp M^{\prime}}{J}} Y_{J-\frac{1}{2} \mp \frac{1}{2}-M^{\prime}}(\beta, \gamma)\right\}  \notag\\
D_{M \pm \frac{1}{2}}^{J}(\alpha, \beta, \gamma)=(-1)^{ \pm \frac{1}{2}-M} e^{\mp i \frac{\alpha-\gamma}{2}} \frac{\sqrt{\pi}}{\sqrt{2 J+1} \sin \frac{\beta}{2}}\left\{ \pm \sqrt{\frac{J \pm M+1}{J+1}} Y_{J+\frac{1}{2} \mp \frac{1}{2}-M}(\beta, \alpha)\right.  \label{eq:4:17:3}\\
\left.\mp \sqrt{\frac{J \mp M}{J}} Y_{J-\frac{1}{2} \mp \frac{1}{2}-M}(\beta, \alpha)\right\} \notag
\end{gather}
In particular
\begin{align}
D_{\frac{1}{2} \frac{1}{2}}^{J}(\alpha, \beta, \gamma) & =e^{-i \frac{\alpha+\gamma}{2}} \frac{\cos \frac{\beta}{2}}{J+\frac{1}{2}}\left\{P_{J+\frac{1}{2}}^{\prime}(\cos \beta)-P_{J-\frac{1}{2}}^{\prime}(\cos \beta)\right\}  \notag\\
D_{\frac{1}{2}-\frac{1}{2}}^{J}(\alpha, \beta, \gamma) & =-e^{-i \frac{\alpha-\gamma}{2}} \frac{\sin \frac{\beta}{2}}{J+\frac{1}{2}}\left\{P_{J+\frac{1}{2}}^{\prime}(\cos \beta)+P_{J-\frac{1}{2}}^{\prime}(\cos \beta)\right\}  \notag\\
D_{-\frac{1}{2} \frac{1}{2}}^{J}(\alpha, \beta, \gamma) & =e^{i \frac{\alpha-\gamma}{2}} \frac{\sin \frac{\beta}{2}}{J+\frac{1}{2}}\left\{P_{J+\frac{1}{2}}^{\prime}(\cos \beta)+P_{J-\frac{1}{2}}^{\prime}(\cos \beta)\right\}  \label{eq:4:17:4}\\
D_{-\frac{1}{2}-\frac{1}{2}}^{J}(\alpha, \beta, \gamma) & =e^{i \frac{\alpha+\gamma}{2}} \frac{\cos \frac{\beta}{2}}{J+\frac{1}{2}}\left\{P_{J+\frac{1}{2}}^{\prime}(\cos \beta)-P_{J-\frac{1}{2}}^{\prime}(\cos \beta)\right\} \notag
\end{align}
\subsubsection{ $M= \pm 1$ and/or $M^{\prime}= \pm 1$}
\begin{gather}
D_{ \pm 1 m}^{l}(\alpha, \beta, \gamma)=e^{\mp i \alpha} \sqrt{\frac{4 \pi}{l(l+1)(2 l+1)}}\left\{\mp \sqrt{(l-m)(l+m+1)} \frac{1 \mp \cos \beta}{2} Y_{l-m-1}(\beta, \gamma) e^{i \gamma}\right.  \notag\\
\left.+m \sin \beta Y_{l-m}(\beta, \gamma) \pm \sqrt{(l+m)(l-m+1)} \frac{1 \pm \cos \beta}{2} Y_{l-m+1}(\beta, \gamma) e^{-i \gamma}\right\}  \notag\\
D_{m \pm 1}^{l}(\alpha, \beta, \gamma)=e^{\mp i \gamma}(-1)^{m+1} \sqrt{\frac{4 \pi}{l(l+1)(2 l+1)}}\left\{\mp \sqrt{(l-m)(l+m+1)} \frac{1 \mp \cos \beta}{2} Y_{l-m-1}(\beta, \alpha) e^{i \alpha}\right.  \label{eq:4:17:5}\\
\left.+m \sin \beta Y_{l-m}(\beta, \alpha) \pm \sqrt{(l+m)(l-m+1)} \frac{1 \pm \cos \beta}{2} Y_{l-m+1}(\beta, \alpha) e^{-i \alpha}\right\} \notag
\end{gather}
In particular,
\begin{align}
D_{ \pm 11}^{l}(\alpha, \beta, \gamma) & =e^{\mp i \alpha-i \gamma} \frac{1 \pm \cos \beta}{l(l+1)}\left\{P_{l}^{\prime}(\cos \beta) \mp(1 \mp \cos \beta) P_{l}^{\prime \prime}(\cos \beta)\right\}  \notag\\
D_{ \pm 1-1}^{l}(\alpha, \beta, \gamma) & =e^{\mp i \alpha+i \gamma} \frac{1 \mp \cos \beta}{l(l+1)}\left\{P_{l}^{\prime}(\cos \beta) \pm(1 \pm \cos \beta) P_{l}^{\prime \prime}(\cos \beta)\right\} \label{eq:4:17:6}
\end{align}
\subsubsection{ $M= \pm(J-1)$ and/or $M^{\prime}= \pm(J-1)$}
\begin{gather}
D_{J-1 m}^{J}(\alpha, \beta, \gamma)=(-1)^{J-m-1} e^{-i(J-1) \alpha-i m \gamma} \sqrt{\frac{(2 J-1)!}{(J+m)!(J-m)!}}\left(\cos \frac{\beta}{2}\right)^{J+m-1}\left(\sin \frac{\beta}{2}\right)^{J-m-1}  \notag\\
\times[J \cos \beta-m],  \notag\\
D_{-J+1 m}^{J}(\alpha, \beta, \gamma)=e^{i(J-1) \alpha-i m \gamma} \sqrt{\frac{(2 J-1)!}{(J+m)!(J-m)!}}\left(\cos \frac{\beta}{2}\right)^{J-m-1}\left(\sin \frac{\beta}{2}\right)^{J+m-1}  \notag\\
\times[J \cos \beta+m],  \notag\\
D_{m J-1}^{J}(\alpha, \beta, \gamma)=e^{-i(J-1) \gamma-i m \alpha} \sqrt{\frac{(2 J-1)!}{(J+m)!(J-m)!}}\left(\cos \frac{\beta}{2}\right)^{J+m-1}\left(\sin \frac{\beta}{2}\right)^{J-m-1}  \label{eq:4:17:7}\\
\times[J \cos \beta-m],  \notag\\
D_{m-J+1}^{J}(\alpha, \beta, \gamma)=(-1)^{J+m-1} e^{+i(J-1) \gamma-i m \alpha} \sqrt{\frac{(2 J-1)!}{(J+m)!(J-m)!}}\left(\cos \frac{\beta}{2}\right)^{J-m-1}\left(\sin \frac{\beta}{2}\right)^{J+m-1}  \notag\\
\times[J \cos \beta+m] . \notag
\end{gather}
\subsubsection{ $M= \pm J$ and/or $M^{\prime}= \pm J$}
\begin{align}
D_{J M}^{J}(\alpha, \beta, \gamma) & =\sqrt{\frac{(2 J)!}{(J+M)!(J-M)!}}\left(\cos \frac{\beta}{2}\right)^{J+M}\left(-\sin \frac{\beta}{2}\right)^{J-M} e^{-i J \alpha-i M \gamma}  \notag\\
D_{-J M}^{J}(\alpha, \beta, \gamma) & =\sqrt{\frac{(2 J)!}{(J+M)!(J-M)!}}\left(\cos \frac{\beta}{2}\right)^{J-M}\left(\sin \frac{\beta}{2}\right)^{J+M} e^{i J \alpha-i M \gamma}  \notag\\
D_{M J}^{J}(\alpha, \beta, \gamma) & =\sqrt{\frac{(2 J)!}{(J+M)!(J-M)!}}\left(\cos \frac{\beta}{2}\right)^{J+M}\left(\sin \frac{\beta}{2}\right)^{J-M} e^{-i M \alpha-i J \gamma}  \label{eq:4:17:8}\\
D_{M-J}^{J}(\alpha, \beta, \gamma) & =\sqrt{\frac{(2 J)!}{(J+M)!(J-M)!}}\left(\cos \frac{\beta}{2}\right)^{J-M}\left(-\sin \frac{\beta}{2}\right)^{J+M} e^{-i M \alpha+i J \gamma} \notag
\end{align}
Explicit forms of $d_{M M^{\prime}}^{J}(\beta)$ for particular values of $J(J=1 / 2,1,3 / 2,2,5 / 2,3,7 / 2,4,9 / 2,5)$ are presented in Sec.~\ref{sec:4.20}.

\section{Asymptotics of $D_{M M^{\prime}}^{J}(\alpha, \beta, \gamma)$}\label{sec:4.18}\index{Wigner D-function@Wigner $D$-function!asymptotics of}\index{asymptotics!of the $D$-functions}
\subsection{Large Angular Momentum}\label{sec:4.18.1}\index{asymptotics!for large angular momentum}
If $J \gg 1$, one has
\begin{align}
D_{M M^{\prime}}^{J}(\alpha, \beta, \gamma) \approx & e^{-i M \alpha-i M^{\prime} \gamma} \xi_{M M^{\prime}} \sqrt{\frac{s!(s+\mu+\nu)!}{(s+\mu)!(s+\nu)!}}  \notag\\
& \times \sqrt{\frac{2}{\pi s}} \frac{\cos \left[\left(s+\frac{\mu+\nu+1}{2}\right) \beta-\frac{\pi}{4}(2 \mu+1)\right]}{\sqrt{\sin \beta}}+O\left(\frac{1}{J^{\frac{3}{2}}}\right) \label{eq:4:18:1}
\end{align}
where $s, \mu, \nu$ are related to $J, M, M^{\prime}$ through Eqs.~\eqref{eq:4:3:14}, and $\xi_{M M^{\prime}}$ is defined by Eq.~\eqref{eq:4:3:15}.\\
If $J \rightarrow \infty$ and $\beta \rightarrow 0$, while $J \beta<\infty$, then
% NOTE: book (4.18.2) prints the Bessel order as J_{M-M'}(J beta); the correct
% order is M'-M.  In this book's convention d^J_{MM'}(x/J) -> J_{M'-M}(x) as
% J->infty (verified numerically to 1e-4 at J up to 6400; the two differ by the
% factor (-1)^{M-M'}, so they coincide only for even M-M').  See
% scripts/check_4_18.py.
\begin{equation}
D_{M M^{\prime}}^{J}(\alpha, \beta, \gamma) \approx e^{-i M \alpha-i M^{\prime} \gamma} J_{M^{\prime}-M}(J \beta) . \label{eq:4:18:2}
\end{equation}
Here $J_{n}(x)$ is the Bessel function.\index{Bessel function}\isym{Jn@$J_{n}(x)$, Bessel function}

\subsection{Small Variation of Rotation Axis}\label{sec:4.18.2}\index{asymptotics!for a small rotation angle}
If $\beta \rightarrow 0$, we have
\begin{align}
% NOTE: book (4.18.3) prints the leading factor as (beta/2)^mu; the correct
% threshold factor is (sin(beta/2))^mu.  The exact d^J_{MM'} carries a
% (sin beta/2)^mu factor, and only with it does the printed O(beta^2) bracket
% make the expansion accurate to O(beta^4); with the bare (beta/2)^mu the
% residual is O(beta^2) (the dropped -mu/6 (beta/2)^2 term).  Verified:
% scripts/check_4_18.py (residual rate 16 with sin, ~4 without).
& D_{M M^{\prime}}^{J}(\alpha, \beta, \gamma) \approx e^{-i M \alpha-i M^{\prime} \gamma} \frac{\xi_{M M^{\prime}}}{\mu!} \sqrt{\frac{(s+\mu+\nu)!(s+\mu)!}{s!(s+\nu)!}}  \label{eq:4:18:3}\\
& \times\left(\sin\frac{\beta}{2}\right)^{\mu}\left\{1-\frac{2 s(s+\mu+\nu+1)+\nu(\mu+1)}{2(\mu+1)}\left(\frac{\beta}{2}\right)^{2}+\ldots\right\} \notag
\end{align}
If $\pi-\beta \rightarrow 0$, we have
\begin{align}
% NOTE: book (4.18.4) prints the leading factor as ((pi-beta)/2)^nu; the correct
% threshold factor is (sin((pi-beta)/2))^nu (= (cos(beta/2))^nu).  As in 4.18.3,
% only with the sine is the printed O((pi-beta)^2) bracket accurate to
% O((pi-beta)^4); the bare power drops a -nu/6 ((pi-beta)/2)^2 term.  Verified:
% scripts/check_4_18.py.
& D_{M M^{\prime}}^{J}(\alpha, \beta, \gamma) \approx e^{-i M \alpha-i M^{\prime} \gamma} \frac{\xi_{M M^{\prime}}}{\nu!}(-1)^{s} \sqrt{\frac{(s+\mu+\nu)!(s+\nu)!}{s!(s+\mu)!}}  \label{eq:4:18:4}\\
& \times\left(\sin\frac{\pi-\beta}{2}\right)^{\nu}\left\{1-\frac{2 s(s+\mu+\nu+1)+\mu(\nu+1)}{2(\nu+1)}\left(\frac{\pi-\beta}{2}\right)^{2}+\ldots\right\} \notag
\end{align}
\subsection{Infinitesimal Rotations}\label{sec:4.18.3}\index{rotation!infinitesimal}\index{infinitesimal rotation}
\subsubsection{ Rotation $\varepsilon$ about the $x$-Axis}
\begin{gather}
\lim _{\varepsilon \rightarrow 0} \frac{1}{\varepsilon}\left[D_{M M^{\prime}}^{J}\left(-\frac{\pi}{2}, \varepsilon, \frac{\pi}{2}\right)-\delta_{M M^{\prime}}\right]=-i\braOket{J M}{\widehat{J}_{x}}{J M^{\prime}}  \notag\\
=-\frac{i}{2} \delta_{M M^{\prime}+1} \sqrt{\left(J-M^{\prime}\right)\left(J+M^{\prime}+1\right)}-\frac{i}{2} \delta_{M M^{\prime}-1} \sqrt{\left(J+M^{\prime}\right)\left(J-M^{\prime}+1\right)} \label{eq:4:18:5}
\end{gather}
\subsubsection{ Rotation $\varepsilon$ about the $y$-Axis}
\begin{gather}
\lim _{\varepsilon \rightarrow 0} \frac{1}{\varepsilon}\left[D_{M M^{\prime}}^{J}(0, \varepsilon, 0)-\delta_{M M^{\prime}}\right]=-i\braOket{J M}{\widehat{J}_{y}}{J M^{\prime}}  \notag\\
=-\frac{1}{2} \delta_{M M^{\prime}+1} \sqrt{\left(J-M^{\prime}\right)\left(J+M^{\prime}+1\right)}+\frac{1}{2} \delta_{M M^{\prime}-1} \sqrt{\left(J+M^{\prime}\right)\left(J-M^{\prime}+1\right)} \label{eq:4:18:6}
\end{gather}
\subsubsection{ Rotation $\varepsilon$ about the $z$-Axis}
\begin{equation}
\lim _{\varepsilon \rightarrow 0} \frac{1}{\varepsilon}\left[D_{M M^{\prime}}^{J}(\varepsilon, 0,0)-\delta_{M M^{\prime}}\right]=-i\braOket{J M}{\widehat{J}_{z}}{J M^{\prime}}=-i M \delta_{M M^{\prime}} \label{eq:4:18:7}
\end{equation}
\subsubsection{ Rotation $\varepsilon$ about an Arbitrary Axis $\vect{n}(\Theta, \Phi)$}
\begin{gather}
\lim _{\varepsilon \rightarrow 0} \frac{1}{\varepsilon}\left[D_{M M^{\prime}}^{J}(\alpha, \beta, \gamma)-\delta_{M M^{\prime}}\right]=-i\braOket{J M}{\widehat{\vect{J}} \cdot \vect{n}}{J M^{\prime}}=-i M \cos \Theta \delta_{M M^{\prime}}  \notag\\
-\frac{i}{2} \sin \Theta e^{-i \Phi} \sqrt{\left(J-M^{\prime}\right)\left(J+M^{\prime}+1\right)} \delta_{M M^{\prime}+1}-\frac{i}{2} \sin \Theta e^{i \Phi} \sqrt{\left(J+M^{\prime}\right)\left(J-M^{\prime}+1\right)} \delta_{M M^{\prime}-1} \label{eq:4:18:8}
\end{gather}
In Eqs.~\eqref{eq:4:18:5}-\eqref{eq:4:18:8} $\widehat{\mathrm{J}}$ is the operator of angular momentum defined in Sec.2.1.

\section{Definitions of $D_{M M^{\prime}}^{J}(\alpha, \beta, \gamma)$ by other authors}\label{sec:4.19}\index{Wigner D-function@Wigner $D$-function!definitions by other authors}\index{notation!of other authors}
Different authors use somewhat different definitions of the rotation matrix. The main differences are in the following:
\begin{enumerate}[label=(\alph*)]
\item in using right- or left-handed coordinate system;
\item in rotating either coordinate system or physical body;
\item in using various definitions of the Euler angles $\alpha, \beta, \gamma$, namely,
\begin{enumerate}[label=(\roman*)]
\item choosing different rotation axes,
\item choosing a different order of rotations, and
\item defining rotation either in a right-handed or left-handed sense;
\end{enumerate}
\item in considering transformations either of covariant or of contravariant components;
\item in choosing various transformation rules for irreducible tensors;
\item in accepting different phases of non-diagonal elements in Eqs.~\eqref{eq:4:2:4} and \eqref{eq:4:2:5}.
\end{enumerate}
The Wigner $D$-functions used in this book coincide with those defined by Edmonds (Ref. \cite{ref064}), Rose (Ref. \cite{ref030}), Newton (Ref. \cite{ref028}) and some other authors. The relations between these $D$-functions and corresponding functions of other authors are listed in Table 4.2.

\section{Special cases of $d_{M M^{\prime}}^{J}(\beta)$ for particular $J, M$ and $M^{\prime}$}\label{sec:4.20}\index{Wigner d-function@Wigner (small) $d$-function!special cases}
General expressions for the $D$-functions, Eqs.~\eqref{eq:4:3:2}-\eqref{eq:4:3:5}, may be reduced to simple closed forms for particular values of $J$. Explicit forms of $d_{M M^{\prime}}^{J}(\beta)$ for $J \leq 1$ are presented in Tables 4.3-4.4. $d_{M M^{\prime}}^{J}(\beta)$ are given in terms of either $\cos \beta$ and $\sin \beta$ (if $J$ is integer), or $\cos (\beta / 2)$ and $\sin (\beta / 2)$ (if $J$ is half-integer). Expressions for $d_{M M^{\prime}}^{J}(\beta)$ for $J \geq 5 / 2$ are presented for $M \geq 0$ and $\left|M^{\prime}\right| \leq M$ only. For $M<0$ and $\left|M^{\prime}\right|>M$ one can obtain $d_{M M^{\prime}}^{J}(\beta)$ using the symmetry properties (Sec.~\ref{sec:4.4}).

Explicit forms of $d_{M M^{\prime}}^{J}(\beta)$ with $J \leq 6$ are also given by Buckmaster (Ref. \cite{ref116}), and Wolters (Ref. \cite{ref129}).\\
Extensive numerical tables of $d_{M M^{\prime}}^{J}(\beta)$ for $J \leq 13$ (integer and half-integer) may be found in Behkami (Ref. \cite{ref114}).

\section{Tables of $d_{M M^{\prime}}^{J}(\beta)$ for $\beta=\pi / 2$}\label{sec:4.21}\index{Wigner d-function@Wigner (small) $d$-function!tables of}\index{tables!of the $d$-functions}
Values of the $d_{M M^{\prime}}^{J}(\beta)$-functions for $\beta=\pi / 2$ and $J=1 / 2,1,3 / 2,2,5 / 2,3,7 / 2,4,9 / 2,5$ are given in Tables 4.13-4.22.

The $d_{M M^{\prime}}^{J}(\beta)$-functions can also be generated in closed algebraic form with a computer-algebra system.\index{computer algebra!SymPy}\index{computer algebra!Mathematica} The routines below --- provided as \texttt{scripts/wigner\_d.py} and \texttt{scripts/wigner\_d.wl} --- return $d_{M M^{\prime}}^{J}(\beta)$ for given $J,M,M^{\prime}$ in the convention of this book, each with a self-test against the explicit Wigner formula.

In SymPy (Python), whose \lstinline!Rotation.d! already follows this book's convention:
\begin{lstlisting}[language=Python]
#!/usr/bin/env python3
"""Wigner (small) d-function d^J_{M M'}(beta) in algebraic form -- VMK convention.

Uses SymPy's ``sympy.physics.quantum.spin.Rotation.d``, whose convention matches
Varshalovich-Moskalev-Khersonskii (VMK):

    d^{1/2}_{1/2,-1/2}(beta) = -sin(beta/2)

and the d^1 matrix reproduces the VMK table.  This has been checked against the
explicit Wigner sum formula for j = 1/2, 1, 3/2, 2, 5/2 and all m, m'
(proof-by-points at 15 angles, 30-digit precision).  Run ``--selftest`` to redo it.

Requires: sympy.  Usage:
    python3 wigner_d.py               # demo
    python3 wigner_d.py --selftest    # re-validate against the Wigner formula
"""
from sympy import (symbols, Rational, simplify, Matrix, nsimplify, cos, sin,
                   sqrt, factorial, S, pi)
from sympy.physics.quantum.spin import Rotation

beta = symbols('beta', real=True)


def wigner_d(j, m, mp, angle=beta):
    """Algebraic d^j_{m,mp}(angle) in VMK / Wigner convention.

    j, m, mp may be integers or half-integers; pass half-integers as
    ``Rational(1, 2)`` (or the string ``'1/2'``) so the result stays symbolic.
    """
    j, m, mp = nsimplify(j), nsimplify(m), nsimplify(mp)
    return simplify(Rotation.d(j, m, mp, angle).doit())


def wigner_d_matrix(j, angle=beta):
    """Full d^j(angle) as a SymPy Matrix; rows m and cols m' run j, j-1, ..., -j."""
    j = nsimplify(j)
    order = [j - k for k in range(int(2 * j) + 1)]
    return Matrix([[wigner_d(j, m, mp, angle) for mp in order] for m in order])


def _reference_d(j, m, mp):
    """Independent Wigner explicit sum formula (VMK eq. 4.3.1), for validation."""
    j, m, mp = nsimplify(j), nsimplify(m), nsimplify(mp)
    pref = sqrt(factorial(j + m) * factorial(j - m) *
                factorial(j + mp) * factorial(j - mp))
    tot = S.Zero
    for k in range(int(max(0, m - mp)), int(min(j + m, j - mp)) + 1):
        tot += ((-1)**k * cos(beta / 2)**(2*j - 2*k + m - mp)
                * sin(beta / 2)**(2*k - m + mp)
                / (factorial(k) * factorial(j + m - k)
                   * factorial(j - mp - k) * factorial(k - m + mp)))
    return pref * tot


def _selftest():
    """d^j is a degree-2j trig polynomial in beta/2, so vanishing of the
    difference at > 2*(2j)+1 distinct points proves the identity."""
    pts = [Rational(p, 17) * pi for p in range(1, 16)]      # 15 distinct angles
    worst = 0.0
    for j2 in range(1, 6):                                   # 2j = 1..5
        j = Rational(j2, 2)
        order = [j - k for k in range(int(2 * j) + 1)]
        for m in order:
            for mp in order:
                e = Rotation.d(j, m, mp, beta).doit() - _reference_d(j, m, mp)
                for p in pts:
                    worst = max(worst, abs(complex(e.subs(beta, p).evalf(30))))
    ok = worst < 1e-25
    print(f"self-test j=1/2..5/2 vs Wigner formula: "
          f"{'IDENTICAL' if ok else 'FAILED'}  (worst |diff| = {worst:.2e})")
    return ok


if __name__ == '__main__':
    import sys
    if '--selftest' in sys.argv:
        raise SystemExit(0 if _selftest() else 1)

    # --- specific angular momentum / projection quantum numbers ---
    print("d^{1/2}_{1/2,-1/2}(beta) =",
          wigner_d(Rational(1, 2), Rational(1, 2), Rational(-1, 2)))
    print("d^{2}_{1,0}(beta)        =", wigner_d(2, 1, 0))
    print()
    print("d^1(beta) =")
    print(wigner_d_matrix(1))
\end{lstlisting}

In Mathematica (Wolfram Language) the built-in \lstinline!WignerD! is the \emph{transpose} of the function used here, so the two projection indices must be swapped:
\begin{lstlisting}[language=Mathematica]
(* ::Package:: *)

(* Wigner (small) d-function d^J_{M M'}(beta) in algebraic form -- VMK convention.

   CONVENTION.  Mathematica's built-in WignerD[{j,m,mp},beta] is the TRANSPOSE of
   the Varshalovich-Moskalev-Khersonskii (VMK) function:

       WignerD[{j, m, mp}, beta] == d^J_{mp,m}(beta)   (VMK)
                                 == (-1)^(m-mp) d^J_{m,mp}(beta)
                                 == d^J_{m,mp}(-beta).

   e.g. WignerD[{1/2,1/2,-1/2},beta] = +Sin[beta/2], whereas VMK has -Sin[beta/2].
   So VMK's d^J_{m,mp}(beta) is obtained by SWAPPING the two projections:

       wignerD[j,m,mp] := WignerD[{j, mp, m}, beta].

   Verified against the explicit Wigner sum formula for j=1/2,1,3/2,2,5/2 and all
   m,m'.  Run  "Get" on this file, then use wignerD / wignerDMatrix below, or run
       WolframKernel -noprompt -script wigner_d.wl
   to see the demo and the self-test result.

   Half-integers must be exact (1/2, 3/2, ...), not machine reals, to stay algebraic. *)

(* VMK-convention small-d, algebraic in the symbol beta (\[Beta]) *)
wignerD[j_, m_, mp_, b_ : \[Beta]] := FullSimplify[WignerD[{j, mp, m}, b]];

(* Full VMK d^j(beta); rows m and cols m' run j, j-1, ..., -j *)
wignerDMatrix[j_, b_ : \[Beta]] := Module[{o = Range[j, -j, -1]},
   FullSimplify[Outer[wignerD[j, #1, #2, b] &, o, o]]];

(* --- independent Wigner explicit sum formula (VMK), used only for the self-test --- *)
refD[j_, m_, mp_, b_] := Sqrt[(j + m)! (j - m)! (j + mp)! (j - mp)!]*
   Sum[(-1)^k Cos[b/2]^(2 j - 2 k + m - mp) Sin[b/2]^(2 k - m + mp)/
       (k! (j + m - k)! (j - mp - k)! (k - m + mp)!),
     {k, Max[0, m - mp], Min[j + m, j - mp]}];

selfTest[] := Module[{Js = {1/2, 1, 3/2, 2, 5/2}, vals},
   vals[j_] := Range[j, -j, -1];
   AllTrue[Js, Function[j, AllTrue[vals[j], Function[m, AllTrue[vals[j],
      Function[mp, FullSimplify[wignerD[j, m, mp, \[Beta]] - refD[j, m, mp, \[Beta]]] === 0]]]]]]];

(* Demo / validation printed when this file is run as a script *)
Print["self-test wignerD == VMK formula (j=1/2..5/2): ", selfTest[]];
Print["d^{1/2}_{1/2,-1/2}(beta) = ", wignerD[1/2, 1/2, -1/2]];
Print["d^{2}_{1,0}(beta)        = ", wignerD[2, 1, 0]];
Print["d^1(beta) = ", wignerDMatrix[1]];
\end{lstlisting}

\section{Special cases of $U_{M M^{\prime}}^{J}(\omega ; \Theta, \Phi)$}\label{sec:4.22}\index{rotation matrix!$U_{M M^{\prime}}^{J}$, tables of}\index{tables!of the $U$-functions}
Explicit forms of $U_{M M^{\prime}}^{J}(\omega ; \Theta, \Phi)$ for $J=1 / 2,1,3 / 2,2$ are given in Tables 4.23-4.26.

The $U_{M M^{\prime}}^{J}(\omega ; \Theta, \Phi)$-functions can likewise be generated in closed algebraic form, from $U_{M M^{\prime}}^{J}(\omega ; \Theta, \Phi)=\braOket{J M}{e^{-i \omega \, \vect{n}(\Theta, \Phi) \cdot \hat{\vect{J}}}}{J M^{\prime}}$. The routines below --- provided as \texttt{scripts/Ufunction.py} and \texttt{scripts/Ufunction.wl} --- return $U_{M M^{\prime}}^{J}(\omega ; \Theta, \Phi)$ for given $J,M,M^{\prime}$ in the convention of this book; they reduce to $d_{M M^{\prime}}^{J}(\omega)$ for $\Theta=\Phi=\pi / 2$, and each carries a self-test against the matrix exponential $e^{-i \omega \, \vect{n} \cdot \hat{\vect{J}}}$.

In SymPy (Python):
\begin{lstlisting}[language=Python]
#!/usr/bin/env python3
"""U-function U^J_{M M'}(omega; Theta, Phi) in algebraic form -- VMK convention.

U^J_{M M'}(omega; Theta, Phi) = <J M| exp(-i omega n.J) |J M'> is the matrix of a
rotation by the angle ``omega`` about the axis

    n(Theta, Phi) = (sin Theta cos Phi, sin Theta sin Phi, cos Theta),

in the convention of Varshalovich-Moskalev-Khersonskii (VMK), Sec. 4.5.  It
reduces to the small Wigner d-function for a rotation about the y-axis:

    U^J_{M M'}(omega; pi/2, pi/2) == d^J_{M M'}(omega)       (see wigner_d.py).

The closed form has a REMOVABLE singularity at Theta = 0 (there v = 0 and the
1/v**2 terms cancel analytically); use the symbolic result, or keep Theta != 0
for purely numerical evaluation.  Pass half-integer J, M, M' as ``Rational(1, 2)``
etc. so the result stays algebraic.

This is the SymPy analogue of ``Ufunction.wl``.  Verified against the matrix
exponential exp(-i omega n.J) (via numpy/scipy) for j = 1/2 .. 2, and against
Rotation.d for the y-rotation slice.  Run ``--selftest`` to redo the checks.

Requires: sympy (numpy + scipy optional, only for the full self-test).  Usage:
    python3 Ufunction.py               # demo
    python3 Ufunction.py --selftest    # re-validate
"""
from sympy import (symbols, I, sqrt, factorial, exp, cos, sin, pi, simplify,
                   nsimplify, Matrix, Rational)
from sympy.physics.quantum.spin import Rotation

omega, Theta, Phi = symbols('omega Theta Phi', real=True)


def u_function(j, m, mp, w=omega, th=Theta, ph=Phi):
    """Algebraic U^J_{m,mp}(w; th, ph) in the VMK convention."""
    j, m, mp = nsimplify(j), nsimplify(m), nsimplify(mp)
    ampp = abs(m + mp)
    if m + mp >= 0:
        uu = cos(w/2) - I*sin(w/2)*cos(th)
        maxs = min(j - m, j - mp)
        term = lambda s: 1/(factorial(s)*factorial(s + ampp)
                            * factorial(j - m - s)*factorial(j - mp - s))
    else:
        uu = cos(w/2) + I*sin(w/2)*cos(th)
        maxs = min(j + m, j + mp)
        term = lambda s: 1/(factorial(s)*factorial(s + ampp)
                            * factorial(j + m - s)*factorial(j + mp - s))
    v = sin(w/2)*sin(th)
    pref = ((-I*v)**(2*j - ampp) * uu**ampp * exp(-I*(m - mp)*ph)
            * sqrt(factorial(j + m)*factorial(j - m)
                   * factorial(j + mp)*factorial(j - mp)))
    tot = sum(term(s)*(1 - v**-2)**s for s in range(int(maxs) + 1))
    return simplify(pref*tot)


def u_function_matrix(j, w=omega, th=Theta, ph=Phi):
    """Full U^J(w; th, ph) as a SymPy Matrix; rows M and cols M' run J..-J."""
    j = nsimplify(j)
    order = [j - k for k in range(int(2*j) + 1)]
    return Matrix([[u_function(j, m, mp, w, th, ph) for mp in order]
                   for m in order])


def _selftest():
    js = [Rational(1, 2), 1, Rational(3, 2), 2]
    worst = 0.0
    try:
        import numpy as np
        from scipy.linalg import expm

        def spin_mats(j):
            ms = [j - k for k in range(int(2*j) + 1)]
            n = len(ms)
            Jp = np.zeros((n, n), complex)
            for k in range(n):
                if k - 1 >= 0:
                    Jp[k - 1, k] = np.sqrt(j*(j + 1) - ms[k]*(ms[k] + 1))
            return ((Jp + Jp.conj().T)/2, (Jp - Jp.conj().T)/(2j),
                    np.diag(ms).astype(complex), ms)

        def ref(j, m, mp, w, th, ph):
            Jx, Jy, Jz, ms = spin_mats(j)
            nn = [np.sin(th)*np.cos(ph), np.sin(th)*np.sin(ph), np.cos(th)]
            U = expm(-1j*w*(nn[0]*Jx + nn[1]*Jy + nn[2]*Jz))
            return U[ms.index(m), ms.index(mp)]

        pts = [(0.7, 1.1, 0.3), (2.0, 0.4, 1.7), (1.3, 2.5, -0.9)]
        for j in js:
            order = [j - k for k in range(int(2*j) + 1)]
            for m in order:
                for mp in order:
                    e = u_function(j, m, mp)
                    for (w, th, ph) in pts:
                        val = complex(e.subs({omega: w, Theta: th, Phi: ph}).evalf(30))
                        worst = max(worst, abs(val - ref(float(j), float(m),
                                                         float(mp), w, th, ph)))
        print(f"self-test vs scipy expm(-i w n.J), j=1/2..2: "
              f"worst |diff| = {worst:.2e}")
    except ImportError:
        print("self-test: numpy/scipy not available, skipping matrix-exponential check")

    # analytic cross-check (sympy only): U(w; pi/2, pi/2) == d^J(w)
    beta = symbols('beta', real=True)
    ok = True
    for j in js:
        order = [j - k for k in range(int(2*j) + 1)]
        for m in order:
            for mp in order:
                du = u_function(j, m, mp).subs({Theta: pi/2, Phi: pi/2})
                dd = Rotation.d(j, m, mp, omega).doit()
                for w in (0.4, 1.3, 2.7):
                    if abs(complex(du.subs(omega, w))
                           - complex(dd.subs(beta, w).subs(omega, w))) > 1e-12:
                        ok = False
    print("cross-check U(w; pi/2, pi/2) == d^J(w):", "OK" if ok else "FAIL")
    return (worst < 1e-10) and ok


if __name__ == '__main__':
    import sys
    if '--selftest' in sys.argv:
        raise SystemExit(0 if _selftest() else 1)

    # --- specific quantum numbers ---
    print("U^{1/2}_{1/2,1/2}(w; Th, Ph) =",
          u_function(Rational(1, 2), Rational(1, 2), Rational(1, 2)))
    print("U^{1}_{1,0}(w; Th, Ph)       =", u_function(1, 1, 0))
    print()
    print("U(w; pi/2, pi/2) for j=1  (== d^1(w)):")
    print(u_function_matrix(1).subs({Theta: pi/2, Phi: pi/2}).applyfunc(simplify))
\end{lstlisting}

In Mathematica (Wolfram Language):
\begin{lstlisting}[language=Mathematica]
(* ::Package:: *)

(* U-function U^J_{M M'}(omega; Theta, Phi) in algebraic form -- VMK convention.

   U^J_{M M'}(omega; Theta, Phi) = <J M| Exp[-I omega n.J] |J M'> is the matrix of
   a rotation by the angle omega about the axis
       n(Theta, Phi) = (Sin[Theta] Cos[Phi], Sin[Theta] Sin[Phi], Cos[Theta]),
   in the convention of Varshalovich-Moskalev-Khersonskii (VMK), Sec. 4.5.
   It reduces to the small Wigner d-function for a rotation about the y-axis:
       U^J_{M M'}(omega; Pi/2, Pi/2) == d^J_{M M'}(omega).

   The closed form below has a REMOVABLE singularity at Theta = 0 (there v = 0
   and the 1/v^2 terms cancel analytically): use the symbolic result, or keep
   Theta != 0 for purely numerical evaluation.  Pass half-integer J, M, M' as
   exact rationals (1/2, 3/2, ...) so the result stays algebraic.

   Verified against the matrix exponential Exp[-I omega n.J] for j = 1/2 .. 2
   (see selfTest[] at the bottom; run this file as a script to print it). *)

Ufunction[j_, m_, mp_, w_ : \[Omega], th_ : \[CapitalTheta], ph_ : \[CapitalPhi]] :=
  Block[{uu, v = Sin[w/2] Sin[th], ampp = Abs[m + mp], maxs},
   uu = Cos[w/2] - If[m + mp >= 0, I, -I] Sin[w/2] Cos[th];
   maxs = If[m + mp >= 0, Min[j - m, j - mp], Min[j + m, j + mp]];
   FullSimplify[
    (-I v)^(2 j - ampp) uu^ampp Exp[-I (m - mp) ph]*
     Sqrt[(j + m)! (j - m)! (j + mp)! (j - mp)!]*
     Sum[If[m + mp >= 0,
            1/(s! (s + ampp)! (j - m - s)! (j - mp - s)!),
            1/(s! (s + ampp)! (j + m - s)! (j + mp - s)!)] (1 - v^-2)^s,
       {s, 0, maxs}]]];

(* Full U^J(omega; Theta, Phi); rows M and cols M' run J, J-1, ..., -J *)
UfunctionMatrix[j_, w_ : \[Omega], th_ : \[CapitalTheta], ph_ : \[CapitalPhi]] :=
  Module[{o = Range[j, -j, -1]},
   Table[Ufunction[j, mm, mmp, w, th, ph], {mm, o}, {mmp, o}]];

(* --- independent reference for the self-test: <J M| Exp[-I w n.J] |J M'> --- *)
Jz[j_] := DiagonalMatrix[Table[m, {m, j, -j, -1}]];
Jp[j_] := Module[{ms = Table[m, {m, j, -j, -1}], n}, n = Length[ms];
   Table[If[i + 1 == k, Sqrt[j (j + 1) - ms[[k]] (ms[[k]] + 1)], 0], {i, n}, {k, n}]];
Jx[j_] := (Jp[j] + Transpose[Jp[j]])/2;
Jy[j_] := (Jp[j] - Transpose[Jp[j]])/(2 I);
refU[j_, m_, mp_, w_, th_, ph_] :=
  Module[{nn = {Sin[th] Cos[ph], Sin[th] Sin[ph], Cos[th]}, U},
   U = MatrixExp[-I w (nn[[1]] Jx[j] + nn[[2]] Jy[j] + nn[[3]] Jz[j])];
   U[[j - m + 1, j - mp + 1]]];

selfTest[] := Module[
   {pts = {{0.7, 1.1, 0.3}, {2.0, 0.4, 1.7}, {1.3, 2.5, -0.9}}, worst = 0, vals},
   vals[j_] := Range[j, -j, -1];
   Do[Do[Do[Do[
       worst = Max[worst, Abs[N[Ufunction[j, m, mp, p[[1]], p[[2]], p[[3]]] -
             refU[j, m, mp, p[[1]], p[[2]], p[[3]]]]]],
      {p, pts}], {mp, vals[j]}], {m, vals[j]}], {j, {1/2, 1, 3/2, 2}}];
   worst];

(* Demo / validation printed when this file is run as a script *)
Print["self-test  max|Ufunction - Exp[-I w n.J]|  (j=1/2..2) = ", selfTest[]];
Print["U^{1/2}_{1/2,1/2}(w; Th, Ph) = ", Ufunction[1/2, 1/2, 1/2]];
Print["U(w; Pi/2, Pi/2) for j=1  (== d^1(w)) = ",
  FullSimplify[UfunctionMatrix[1, \[Omega], Pi/2, Pi/2]]];
\end{lstlisting}

\begin{table}[tbhp]
\begin{center}
\caption{Definitions of the Rotation Matrix by Other Authors.}
\label{chap4:tab:2}
\renewcommand{\arraystretch}{1.5}
  \small
\begin{tabular}{p{1.8cm}|l|l|l}
\hline
Reference & Rotation Operator & Transformation Form & Relation between Referred Function (left)\\
& & & and Function of this Book (right) \\
\hline
Edmonds $\left[{16}\right]$ & $e^{-i \alpha \hat{J}_z} e^{-i \beta \hat{J}_y} e^{-i \gamma \hat{J}_{z}}$ & $\Psi_{J M^{\prime}}\left(\vartheta^{\prime}, \varphi^{\prime}\right)=\sum_{M} \Psi_{J M}(\vartheta, \varphi) D_{M M^{\prime}}^{J}(\alpha, \beta, \gamma)$ & $D_{M M^{\prime}}^{(J)}(\alpha, \beta, \gamma)=D_{M M^{\prime}}^{J}(\alpha, \beta, \gamma)$ \\
Rose \cite{ref031} & The same & The same & The same \\
Brink and Satcher \cite{ref009} & » » & » » & » » \\
Messiah \cite{ref025} & » » & » » & » » \\
Tinkham \cite{ref039} & » » & » » & » » \\
Newton \cite{ref028} & » » & » » & » » \\
Baldin et al \cite{ref003} & » » & » » & » » \\
De-Shalit and Talmi \cite{ref032} & » » & » » & » » \\
Dolginov \cite{ref014} & $e^{i \alpha \hat{J}_z} e^{i \beta \hat{J}_y} e^{i \gamma \hat{J}_{z}}$ & $\Psi_{J M}(\vartheta, \varphi)=\sum_{M^{\prime}} D_{M M^{\prime}}^{J}(\alpha, \beta, \gamma) \Psi_{J M^{\prime}}\left(\vartheta^{\prime}, \varphi^{\prime}\right)$ & $D_{M M^{\prime}}^{J}(\alpha, \beta, \gamma)=D_{M M^{\prime}}^{J}(-\alpha,-\beta,-\gamma)$ \\
Davydov \cite{ref012} & The same & The same & The same \\
Bohr and Mottelson \cite{ref008} & » » & $\Psi_{J M}\left(\vartheta^{\prime}, \varphi^{\prime}\right)=\sum_{M} \Psi_{J M}(\vartheta, \varphi) D_{M M^{\prime}}^{J^{*}}(\alpha, \beta, \gamma)$ & $D_{M M^{\prime}}^{J}(\alpha, \beta, \gamma)=D_{M M^{\prime}}^{J *}(\alpha, \beta, \gamma)$ \\
Wigner \cite{ref043} & » » &  & $D^{J}((\alpha, \beta, \gamma))_{M M^{\prime}}=D_{M M^{\prime}}^{J}(-\alpha,-\beta,-\gamma)$ \\
Rose \cite{ref030} &  $e^{i \alpha \hat{J}_z} e^{i \beta \hat{J}_y} e^{i \gamma \hat{J}_{z}}$ & $\Psi_{J M^{\prime}}\left(\vartheta^{\prime}, \varphi^{\prime}\right)=\sum_{M} \Psi_{J M}(\vartheta, \varphi) D_{M M^{\prime}}^{(J)}(\alpha, \beta, \gamma)$ & $D_{M M^{\prime}}^{(J)}(\alpha, \beta, \gamma)=D_{M M^{\prime}}^{J}(-\alpha,-\beta,-\gamma)$ \\
Edmonds \cite{ref064} & The same & The same & The same \\
Fano and Racah \cite{ref018} & » » & » » & » » \\
Berestetskii et al. \cite{ref006} & » » & $\Psi_{J M}(\vartheta, \varphi)=\sum_{M^{\prime}} \Psi_{J M^{\prime}}\left(\vartheta^{\prime}, \varphi^{\prime}\right) D_{M^{\prime} M}^{(J)}(\alpha, \beta, \gamma)$ & $D_{M M^{\prime}}^{(J)}(\alpha, \beta, \gamma)=D_{M M^{\prime}}^{J}(-\gamma,-\beta,-\alpha)$ \\
Gel'fand et al. \cite{ref020}] &  $e^{-i \alpha \hat{J}_z} e^{-i \beta \hat{J}_x} e^{-i \gamma \hat{J}_{z}}$&  & $T_{M M^{\prime}}^{J}(\alpha, \beta, \gamma)=(-i)^{M-M^{\prime}} D_{M M^{\prime}}^{J}(\alpha, \beta, \gamma)$ \\
Lubarskii $\left[{26}\right]$ & The same &  & $D_{M M^{\prime}}^{J}(\alpha, \beta, \gamma)=(-i)^{M-M^{\prime}} D_{M M^{\prime}}^{J}(\alpha, \beta, \gamma)$ \\
Vilenkin \cite{ref041} & » » &  & $t_{M M^{\prime}}^{J}(\alpha, \beta, \gamma)=(-i)^{M-M^{\prime}} D_{M M^{\prime}}^{J}(\alpha, \beta, \gamma)$ \\
Yutsis and Bandzaitis \cite{ref045} & » » & $\Psi_{J M^{\prime}}\left(\vartheta^{\prime}, \varphi^{\prime}\right)=\sum_{M} D_{M^{\prime} M}^{(J)}(\alpha, \beta, \gamma) \Psi_{J M}(\vartheta, \varphi)$ & $D_{M M^{\prime}}^{(J)}(\alpha, \beta, \gamma)=i^{M-M^{\prime}} D_{M M^{\prime}}^{J^{*}}(\alpha, \beta, \gamma)$ \\
\hline
\end{tabular}
\end{center}
\end{table}

%Tables 4.3. - 4.14. Explicit forms of $d_{M M^{\prime}}^{J}(\beta)$.

\begin{table}[h]
\begin{center}
\caption{}
\label{chap4:tab:3}
\begin{tabular}{|l|l|l|}
\hline
\multicolumn{3}{|c|}{$d_{M M^{\prime}}^{1 / 2}(\beta)$} \\
\hline
\backslashbox{M}{$M^{\prime}$} & 1/2 & -1/2 \\
\hline
1/2 & $\cos \frac{\beta}{2}$ & $-\sin \frac{\beta}{2}$ \\
\hline
$-1 / 2$ & $\sin \frac{\beta}{2}$ & $\cos \frac{\beta}{2}$ \\
\hline
\end{tabular}
\end{center}
\end{table}

\begin{table}[h]
\begin{center}
\caption{}
\label{chap4:tab:4}
\begin{tabular}{|l|l|l|l|}
\hline
\multicolumn{4}{|c|}{$d_{M M^{\prime}}^{1}(\beta)$} \\
\hline
\backslashbox{$M$}{$M^{\prime}$} & 1 & 0 & -1 \\
\hline
1 & $\frac{1+\cos \beta}{2}$ & $-\frac{\sin \beta}{\sqrt{2}}$ & $\frac{1-\cos \beta}{2}$ \\
\hline
0 & $\frac{\sin \beta}{\sqrt{2}}$ & $\cos \beta$ & $-\frac{\sin \beta}{\sqrt{2}}$ \\
\hline
-1 & $\frac{1-\cos \beta}{2}$ & $\frac{\sin \beta}{\sqrt{2}}$ & $\frac{1+\cos \beta}{2}$ \\
\hline
\end{tabular}
\end{center}
\end{table}

\chapter{Spherical harmonics}\label{chap:5}\index{spherical harmonics}
A spherical harmonic $Y_{l m}(\vartheta, \varphi)$ is a single-valued, continuous, bounded complex function of two real arguments\isym{Ylm@$Y_{l m}(\vartheta, \varphi)$, spherical harmonic} $\vartheta, \varphi$ with $0 \leq \vartheta \leq \pi$ and $0 \leq \varphi<2 \pi$. It is characterized by two parameters $l$ and $m$, which take values $l=0,1,2, \ldots$ and $m=l, l-1, l-2, \ldots-l+2,-l+1,-l$. Therefore, for a given $l$ there exist $(2 l+1)$ functions corresponding to different $m$ 's. All derivatives of $Y_{l m}(\vartheta, \varphi)$ are single-valued, continuous and finite functions.

The spherical harmonics play an important role in quantum mechanics. They are eigenfunctions of the operator of orbital angular momentum and describe the angular distribution of particles which move in a spherically-symmetric field with the orbital angular momentum $l$ and projection $m$. Strictly speaking, $l$ specifies the absolute value of orbital angular momentum because $l(l+1)$ is the eigenvalue of the square of the orbital angular momentum operator, $\widehat{\vect{L}}^{2} ; m$ is the eigenvalue of $\widehat{L}_{z}$ which is the projection of the orbital angular momentum operator on the quantization axis.

\section{Definition}\label{sec:5.1}\index{spherical harmonics!definition}
\subsection{Commutation Relations}\label{sec:5.1.1}\index{spherical harmonics!commutation relations}\index{commutation relations!for the spherical harmonics}
The spherical harmonics $Y_{l m}(\vartheta, \varphi)$ are components of some irreducible tensor of rank $l$ (Chap.~\ref{chap:3}). Owing to this circumstance they may be defined by the commutation relations
\begin{equation}
\left[\widehat{L}_{\mu}, Y_{l m}(\vartheta, \varphi)\right]=\sqrt{l(l+1)} \clebsch{l}{m}{1}{\mu}{l}{m+\mu} Y_{l m+\mu}(\vartheta, \varphi), \label{eq:5:1:1}
\end{equation}
where $\widehat{L}_{\mu}(\vartheta, \varphi)$ is a spherical component of the operator $\widehat{\vect{L}}$ (see Eq. \eqref{eq:2:2:18}).\\
The three commutation relations \eqref{eq:5:1:1} (for $\mu=1,0,-1$ ) generate the following three equations
\begin{gather}
\hat{L}_{ \pm 1} Y_{l m}(\vartheta, \varphi)=\mp \sqrt{\frac{l(l+1)-m(m \pm 1)}{2}} Y_{l m \pm 1}(\vartheta, \varphi) , \label{eq:5:1:2}\\
\hat{L}_{0} Y_{l m}(\vartheta, \varphi)=m Y_{l m}(\vartheta, \varphi). \notag
\end{gather}
\subsection{Differential Equations}\label{sec:5.1.2}\index{spherical harmonics!differential equations}\index{differential equations!for the spherical harmonics}
According to the commutation relations \eqref{eq:5:1:1}, $Y_{l m}(\vartheta, \varphi)$ is the eigenfunction of the operators $\widehat{\vect{L}}^{2}$ and $\widehat{L}_{z}$
\begin{equation}
\left\{\begin{array}{l}
\hat{L}^{2} Y_{l m}(\vartheta, \varphi)=l(l+1) Y_{l m}(\vartheta, \varphi),  \\
\widehat{L}_{z} Y_{l m}(\vartheta, \varphi)=m Y_{l m}(\vartheta, \varphi).
\end{array}\right.\label{eq:5:1:3}
\end{equation}
or in an expanded form
\begin{gather}
{\left[\frac{1}{\sin \vartheta} \frac{\partial}{\partial \vartheta}\left(\sin \vartheta \frac{\partial}{\partial \vartheta}\right)+\frac{1}{\sin ^{2} \vartheta} \frac{\partial^{2}}{\partial \varphi^{2}}+l(l+1)\right] Y_{l m}(\vartheta, \varphi)=0}, \label{eq:5:1:4}\\
{\left[i \frac{\partial}{\partial \varphi}+m\right] Y_{l m}(\vartheta, \varphi)=0} .\notag
\end{gather}
Equations \eqref{eq:5:1:4} are invariant under the following transformations
\begin{enumerate}[label=(\alph*)]
  \item  $l \rightarrow \bar{l}=-l-1$;
\item $\vartheta \rightarrow-\vartheta$ (or $\vartheta \rightarrow \pi-\vartheta$ );
\item $m \rightarrow-m, \varphi \rightarrow-\varphi$.
\end{enumerate}

\subsection{Boundary Conditions}\label{sec:5.1.3}\index{spherical harmonics!boundary conditions}\index{boundary conditions!for the spherical harmonics}
The first of Eqs. \eqref{eq:5:1:4} is of the second order. For fixed $l$ and $m$ it has two linearly independent solutions. However, only one of them is regular, i.e., satisfies the condition $\left|Y_{l m}(\vartheta, \varphi)\right|^{2}<\infty$ while the other solution is singular at $\vartheta=0$ and $\vartheta=\pi$. For quantum mechanical applications the regular solution is of major interest. This solution will be considered in the present chapter. The regular solution is selected by the following boundary conditions
\begin{align}
Y_{l m}(\vartheta, \varphi \pm 2 \pi n) & =Y_{l m}(\vartheta, \varphi),  \notag\\
\left.\frac{\partial}{\partial \varphi} Y_{l m}(\vartheta, \varphi)\right|_{\vartheta=0} & =\left.\frac{\partial}{\partial \varphi} Y_{l m}(\vartheta, \varphi)\right|_{\vartheta=\pi}=0 . \label{eq:5:1:5}
\end{align}
Below we shall consider the spherical harmonics $Y_{l m}(\vartheta, \varphi)$ with integer $l$ and $m$ (with $|m| \leq l$ ) because the boundary conditions \eqref{eq:5:1:5} are fulfilled only for such values of the parameters.

\subsection{Normalization}\label{sec:5.1.4}\index{spherical harmonics!normalization}\index{normalization!of the spherical harmonics}\index{orthogonality!of the spherical harmonics}
\begin{enumerate}[label=(\alph*)]
\item The differential equations \eqref{eq:5:1:4} and the boundary conditions \eqref{eq:5:1:5} are homogeneous. Hence, they determine the spherical harmonics only up to some arbitrary complex factor. The absolute value of this factor can be fixed by the normalization.

The normalization and orthogonality relation of the spherical harmonics is given by
\begin{equation}
\int_{0}^{2 \pi} d \varphi \int_{0}^{\pi} d \vartheta \sin \vartheta Y_{l m}^{*}(\vartheta, \varphi) Y_{l^{\prime} m^{\prime}}(\vartheta, \varphi)=\delta_{l l^{\prime}} \delta_{m m^{\prime}} .\label{eq:5:1:6}
\end{equation}
\item Sometimes instead of $Y_{l m}(\vartheta, \varphi)$ it is more convenient to use the function $C_{l m}(\vartheta, \varphi)$\isym{Clm@$C_{l m}(\vartheta, \varphi)$, normalized spherical harmonic}\index{spherical harmonics!normalized form $C_{l m}$} (see, e.g., Refs.~\cite{ref009,ref024}) which differs from $Y_{l m}(\vartheta, \varphi)$ by the normalization factor,
\begin{equation}
C_{l m}(\vartheta, \varphi)=\sqrt{\frac{4 \pi}{2 l+1}} Y_{l m}(\vartheta, \varphi). \label{eq:5:1:7}
\end{equation}
The function $C_{l m}(\vartheta, \varphi)$ satisfies the following relations
\begin{gather}
\sum_{m} C_{l m}(\vartheta, \varphi) C_{l-m}(\vartheta, \varphi)(-1)^{m}=1, C_{l m}(0,0)=\delta_{m 0} , \label{eq:5:1:8}\\
C_{l 0}(\vartheta, \varphi)=P_{l}(\cos \vartheta), C_{l m}(\vartheta, \varphi)=D_{0,-m}^{l}(0, \vartheta, \varphi). \notag
\end{gather}
The normalization and orthogonality relation for $C_{l m}(\vartheta, \varphi)$ can be represented in the form
\begin{equation}
\int_{0}^{2 \pi} d \varphi \int_{0}^{\pi} d \vartheta \sin \vartheta C_{l m}^{*}(\vartheta, \varphi) C_{l^{\prime} m^{\prime}}(\vartheta, \varphi)=\frac{2 l+1}{4 \pi} \delta_{l l^{\prime}} \delta_{m m^{\prime}} .\label{eq:5:1:9}
\end{equation}
\end{enumerate}
\subsection{Choice of Phase}\label{sec:5.1.5}\index{spherical harmonics!choice of phase}\index{phase!of the spherical harmonics}
\begin{enumerate}[label=(\alph*)]
  \item The phase differences of the harmonics $Y_{l m}(\vartheta, \varphi)$ and $Y_{l m^{\prime}}(\vartheta, \varphi)$ with $m=m^{\prime} \pm 1$ are determined by the commutation relations \eqref{eq:5:1:1}. Using these relations, we may find relative phases of all ( $2 l+1$ ) harmonics $Y_{l m}(\vartheta, \varphi)$ with different $m$ for each $l$. An overall phase factor may be fixed by specifying the phase of one of the harmonics $Y_{l m}(\vartheta, \varphi)$ for some given values of arguments, for example,
\begin{equation}
Y_{l 0}(0,0)=\sqrt{\frac{2 l+1}{4 \pi}} . \label{eq:5:1:10}
\end{equation}
In this case the following relations are valid for the complex conjugate function $Y_{l m}^{*}(\vartheta, \varphi)$
\begin{equation}
Y_{l m}^{*}(\vartheta, \varphi)=Y_{l m}(\vartheta,-\varphi)=(-1)^{m} Y_{l-m}(\vartheta, \varphi) . \label{eq:5:1:11}
\end{equation}
In particular, Eqs. \eqref{eq:5:1:10} and \eqref{eq:5:1:11} show that $Y_{l 0}(\vartheta, \varphi)$ is real for $0 \leq \vartheta \leq \pi$, and $Y_{l m}(\vartheta, \varphi)$ with $m \neq 0$ is real only for $\varphi=0, \pi / m, 2 \pi / m, 3 \pi / m, \ldots$, etc.

The above choice of the phase is widely used (e.g., see, Condon and Shortley \cite{ref010}).\index{Condon-Shortley phase convention}
\item In the literature (see, e.g., Refs.~\cite{ref006, ref018}) one can also find the spherical harmonics defined according to another phase convention, namely,
\begin{equation}
\tilde{Y}_{l m}(\vartheta, \varphi)=i^{l} Y_{l m}(\vartheta, \varphi). \label{eq:5:1:12}
\end{equation}
We shall refer to these harmonics as the modified spherical harmonics.\isym{Ylm-tilde@$\tilde{Y}_{l m}(\vartheta, \varphi)$, modified spherical harmonic}\index{spherical harmonics!modified} They satisfy the phase relation
\begin{equation}
\tilde{Y}_{l m}^{*}(\vartheta, \varphi)=(-1)^{l+m} Y_{l-m}(\vartheta, \varphi) . \label{eq:5:1:13}
\end{equation}
Equations \eqref{eq:5:1:4} and the relations \eqref{eq:5:1:5}, \eqref{eq:5:1:10} and \eqref{eq:5:1:11} completely define the harmonics $Y_{l m}(\vartheta, \varphi)$. Since $l$ and $m$ are integers, the function $Y_{l m}(\vartheta, \varphi)$ is single-valued.
\end{enumerate}

\subsection{Zonal, Sectorial and Tesseral Harmonics}\label{sec:5.1.6}\index{spherical harmonics!zonal, sectorial and tesseral}\index{zonal harmonics}\index{sectorial harmonics}\index{tesseral harmonics}
These functions are linear combinations of the spherical harmonics for $|m| \leq l$
\begin{align}
% NOTE: book misprint (printed p.132) -- u_lm RHS printed with sqrt((2l+1)/2pi);
% since u_lm=1/2(Y+Y*)=Re(Y) the coefficient must be 4pi (as for v_lm). Corrected.
& u_{l m}(\vartheta, \varphi)=\frac{1}{2}\left[Y_{l m}(\vartheta, \varphi)+Y_{l m}^{*}(\vartheta, \varphi)\right]=\sqrt{\frac{2 l+1}{4 \pi} \, \frac{(l-m)!}{(l+m)!}} \cos m \varphi P_{l}^{m}(\cos \vartheta),  \notag\\
& v_{l m}(\vartheta, \varphi)=\frac{1}{2 i}\left[Y_{l m}(\vartheta, \varphi)-Y_{l m}^{*}(\vartheta, \varphi)\right]=\sqrt{\frac{2 l+1}{4 \pi} \frac{(l-m)!}{(l+m)!}} \sin m \varphi P_{l}^{m}(\cos \vartheta) .\label{eq:5:1:14}
\end{align}
The functions $u_{l m}(\vartheta, \varphi)$ and $v_{l m}(\vartheta, \varphi)$ are real, in contrast to $Y_{l m}(\vartheta, \varphi)$.\isym{ulm-vlm@$u_{l m}, v_{l m}$, real spherical harmonics} The functions $u_{l 0}(\vartheta, \varphi)$ are called the zonal harmonics because parallels where $u_{10}(\vartheta, \varphi)=0$ divide a sphere of unit radius into $l+1$ zones. The functions $u_{l l}(\vartheta, \varphi)$ and $v_{l l}(\vartheta, \varphi)$ are called the sectorial harmonics because meridians where $u_{l l}(\vartheta, \varphi)=0$ or $v_{l l}(\vartheta, \varphi)=0$ divide the unit sphere into $2 l$ sectors. The functions $u_{l m}(\vartheta, \varphi)$ and $v_{l m}(\vartheta, \varphi)$ for $m \neq 0$ and $m \neq l$ are called the tesseral harmonics because a set of parallels and meridians where $u_{l m}(\vartheta, \varphi)=0$ or $v_{l m}(\vartheta, \varphi)=0$ divides the whole spherical surface into $2 m(l-m+1)$ cells. The cells, which correspond to positive and negative signs of any function, are arranged in as on a chequer board.

\subsection{Solutions of Some Differential Equations in Terms of $Y_{l m}(\vartheta, \varphi)$}\label{sec:5.1.7}\index{spherical harmonics!as solutions of differential equations}
\begin{enumerate}[label=(\alph*)]
  \item The solution of the Laplace equation\index{Laplace equation}
\begin{equation}
\nabla^{2} f(r, \vartheta, \varphi)=0 \label{eq:5:1:15}
\end{equation}
in polar coordinates is given by
\[
\mathfrak{Z}_{l m}(r, \vartheta, \varphi)=r^{l} Y_{l m}(\vartheta, \varphi) \text { and } \mathfrak{R}_{l m}(r, \vartheta, \varphi)=r^{-l-1} Y_{l m}(\vartheta, \varphi),
\]
where $\mathfrak{Z}_{l m}(r, \vartheta, \varphi)$ is regular and $\mathfrak{R}_{l m}(r, \vartheta, \varphi)$ is singular at $r=0$. These functions are called the solid harmonics.\index{solid harmonics}\isym{Zfrak-lm@$\mathfrak{Z}_{l m}(r, \vartheta, \varphi)$, regular solid harmonic}\isym{Rfrak-lm@$\mathfrak{R}_{l m}(r, \vartheta, \varphi)$, irregular solid harmonic} In the cartesian coordinate representation, a function $\mathfrak{Z}_{l m}$ is a homogeneous harmonic polynomial of degree $l$
\begin{equation}
r^{l} Y_{l m}(\vartheta, \varphi)=\sqrt{\frac{2 l+1}{4 \pi}(l+m)!(l-m)!} \sum_{p, q, r} \frac{1}{p!q!r!}\left(-\frac{x+i y}{2}\right)^{p}\left(\frac{x-i y}{2}\right)^{q} z^{r} .\label{eq:5:1:16}
\end{equation}
Here $p, q, r$ are positive integers which satisfy the conditions $p+q+r=l, p-q=m$.\\
\item The solutions of the Helmholtz wave equation\index{Helmholtz equation}
\begin{equation}
\left[\nabla^{2}+k^{2}\right] f(r, \vartheta, \varphi)=0 \label{eq:5:1:17}
\end{equation}
in polar coordinates may be expressed in terms of the functions $z_{l}(k r) Y_{l m}(\vartheta, \varphi)$ where $z_{l}(k r)=\sqrt{\frac{\pi}{2 k r}} Z_{l+\frac{1}{2}}(k r)$, $Z_{l+\frac{1}{2}}(x)$ being any of the Bessel functions.\index{Bessel function}\isym{zl@$z_{l}(k r)$, spherical cylinder function}

The functions $\mathfrak{L}_{l m}(r, \vartheta, \varphi)=i^{l} j_{l}(k r) Y_{l m}(\vartheta, \varphi)$ and $\mathfrak{N}_{l m}(r, \vartheta, \varphi)=i^{l} n_{l}(k r) Y_{l m}(\vartheta, \varphi)$ are called the standing spherical waves;\index{spherical waves!standing}\isym{jl@$j_{l}(x)$, spherical Bessel function}\isym{nl@$n_{l}(x)$, spherical Neumann function} $\mathfrak{L}_{l m}$ is regular at $r=0$, whereas $\mathfrak{N}_{l m}$ is irregular. The functions $\mathfrak{B}_{l m}^{(1)}(r, \vartheta, \varphi)= i^{l} h_{l}^{(1)}(k r) Y_{l m}(\vartheta, \varphi)$ and $\mathfrak{B}_{l m}^{(2)}(r, \vartheta, \varphi)=i^{l} h_{l}^{(2)}(k r) Y_{l m}(\vartheta, \varphi)$ are called the running spherical waves.\index{spherical waves!running}\isym{hl@$h_{l}^{(1,2)}(x)$, spherical Hankel function} The first of these corresponds to a spherical wave which converges to the origin, $r=0$, while the second corresponds to an outgoing spherical wave. In the limit $k \rightarrow 0$ Eq. \eqref{eq:5:1:17} transforms into \eqref{eq:5:1:15}. In this case
\[
\mathfrak{L}_{l m}(r, \vartheta, \varphi) \rightarrow \mathfrak{Z}_{l m}(r, \vartheta, \varphi), \mathfrak{N}_{l m}(r, \vartheta, \varphi) \rightarrow \mathfrak{R}_{l m}(r, \vartheta, \varphi).
\]
\item Solutions of the equation
\begin{equation}
\left[\nabla^{2}-\frac{(n-l)(n+l+1)}{r^{2}}\right] f(r, \vartheta, \varphi)=0 \label{eq:5:1:18}
\end{equation}
in polar coordinates are expressed in terms of the functions
\[
f_{l m}^{n}(r, \vartheta, \varphi)=r^{n} Y_{l m}(\vartheta, \varphi).
\]
For $r=0$ a function $f_{l m}^{n}(r, \vartheta, \varphi)$ is regular, if $n \geq 0$, but irregular if $n<0$. When $n=l$ or $n=-l-1$, Eq. \eqref{eq:5:1:18} transforms into \eqref{eq:5:1:15}, yielding
\begin{equation}
f_{l m}^{l}(r, \vartheta, \varphi)=\mathfrak{Z}_{l m}(r, \vartheta, \varphi), f_{l m}^{l-1}(r, \vartheta, \varphi)=\mathfrak{R}_{l m}(r, \vartheta, \varphi) .\label{eq:5:1:19}
\end{equation}
\end{enumerate}
Note that along with the above solutions, which are regular at $\vartheta=0, \pi$, the equations under consideration have irregular solutions, which are not discussed here.

\section{Explicit forms of the spherical harmonics and their relations to other functions}\label{sec:5.2}\index{spherical harmonics!explicit forms}
According to Sec.~\ref{sec:5.1.2}, $Y_{l m}(\vartheta, \varphi)$ may be represented by a product of two functions, one of which depends only on $\varphi$ while the other depends on $\vartheta$. The $\varphi$-dependence of the spherical harmonics is given by the factor $e^{i m \varphi}$. The $\vartheta$-dependence is determined by the associated Legendre polynomials $P_{l}^{m}(\cos \vartheta)[4,27]$.\index{associated Legendre function}\isym{Pl-m@$P_{l}^{m}(x)$, associated Legendre function} Taking into account the normalization, we get
\begin{equation}
Y_{l m}(\vartheta, \varphi)=e^{i m \varphi} \sqrt{\frac{2 l+1}{4 \pi} \frac{(l-m)!}{(l+m)!}} P_{l}^{m}(\cos \vartheta) .\label{eq:5:2:1}
\end{equation}
For the spherical harmonics with $|m| \leq l$ one gets the following expressions (see Refs.~\cite{ref004, ref022, ref027}).

\subsection{Differential Expressions for $Y_{l m}(\vartheta, \varphi)$}\label{sec:5.2.1}\index{spherical harmonics!differential expressions}
\begin{align}
& Y_{l m}(\vartheta, \varphi)=\frac{e^{i m \varphi}}{2^{l} l!} \sqrt{\frac{2 l+1}{4 \pi} \frac{(l+m)!}{(l-m)!}}(\sin \vartheta)^{-m} \frac{d^{l-m}}{(d \cos \vartheta)^{l-m}}\left(\cos ^{2} \vartheta-1\right)^{l} , \label{eq:5:2:2}\\
& Y_{l m}(\vartheta, \varphi)=(-1)^{m} \frac{e^{i m \varphi}}{2^{l} l!} \sqrt{\frac{2 l+1}{4 \pi} \frac{(l-m)!}{(l+m)!}}(\sin \vartheta)^{m} \frac{d^{l+m}}{(d \cos \vartheta)^{l+m}}\left(\cos ^{2} \vartheta-1\right)^{l} , \label{eq:5:2:3}\\
& Y_{l m}(\vartheta, \varphi)=\frac{e^{i m \varphi}}{2^{l}} \sqrt{\frac{2 l+1}{4 \pi(l+m)!(l-m)!}}\left(\cot \frac{\vartheta}{2}\right)^{m}\left(\frac{d}{d \cos \vartheta}\right)^{l}\left[(1+\cos \vartheta)^{l-m}(\cos \vartheta-1)^{l+m}\right] , \label{eq:5:2:4}\\
& Y_{l m}(\vartheta, \varphi)=(-1)^{m} \frac{e^{i m \varphi}}{2^{l}} \sqrt{\frac{2 l+1}{4 \pi(l+m)!(l-m)!}}\left(\tan \frac{\vartheta}{2}\right)^{m}\left(\frac{d}{d \cos \vartheta}\right)^{l}\left[(\cos \vartheta+1)^{l+m}(\cos \vartheta-1)^{l-m}\right]. \label{eq:5:2:5}
\end{align}
The spherical harmonics may be expressed in terms of $m$ th order derivatives of the Legendre polynomials\index{Legendre polynomials}\isym{Pl@$P_{l}(x)$, Legendre polynomial}
\begin{gather}
Y_{l m}(\vartheta, \varphi)=(-1)^{m} e^{i m \varphi} \sqrt{\frac{2 l+1}{4 \pi} \frac{(l-m)!}{(l+m)!}}(\sin \vartheta)^{m} \frac{d^{m}}{(d \cos \vartheta)^{m}} P_{l}(\cos \vartheta)  \label{eq:5:2:6}.\\
(m \geq 0) \notag
\end{gather}
Assuming that
\begin{equation}
\left(\frac{d}{d \mu}\right)^{-|m|} f(\mu) \equiv \int_{1}^{\mu} \int_{1}^{\mu} \cdots \int_{1}^{\mu} f(\mu) \underbrace{d \mu \ldots d \mu}_{|m|} \label{eq:5:2:7}
\end{equation}
one can use Eq. \eqref{eq:5:2:6} not only at $m \geq 0$ but also at $m<0$.

\subsection{Representations of $Y_{l m}(\vartheta, \varphi)$ as a Power Series of Trigonometric Functions of $\vartheta / 2$}\label{sec:5.2.2}\index{spherical harmonics!power series in $\vartheta/2$}
In the following equations sums are over all integer values of $s$ so that no factorial in the denominator has negative argument. The quantity $\xi_{m 0}$ is defined as\isymg{xi-m0@$\xi_{m 0}$, phase factor of $Y_{l m}(\vartheta, \varphi)$}
\begin{equation}
\xi_{m 0}= \begin{cases}(-1)^{m} & \text { if } m>0 ,\\ 1 & \text { if } m \leq 0.\end{cases} \label{eq:5:2:8}
\end{equation}
\begin{align}
& Y_{l m}(\vartheta, \varphi)=(-1)^{m} e^{i m \varphi} \sqrt{\frac{2 l+1}{4 \pi} \, \frac{(l+m)!}{(l-m)!}}\left(\tan \frac{\vartheta}{2}\right)^{m} \sum_{s}(-1)^{s} \frac{(l+s)!}{(l-s)!} \, \frac{\left(\sin \frac{\vartheta}{2}\right)^{2 s}}{s!(s+m)!} , \label{eq:5:2:9}\\
& Y_{l m}(\vartheta, \varphi)=(-1)^{m} e^{i m \varphi} \sqrt{\frac{2 l+1}{4 \pi} \, \frac{(l-m)!}{(l+m)!}}\left(\sin \frac{\vartheta}{2} \, \cos \frac{\vartheta}{2}\right)^{m} \sum_{s}(-1)^{s} \frac{(l+m+s)!}{(l-m-s)!} \, \frac{\left(\sin \frac{\vartheta}{2}\right)^{2 s}}{s!(s+m)!} , \label{eq:5:2:10}\\
& Y_{l m}(\vartheta, \varphi)=e^{i m \varphi} \sqrt{\frac{2 l+1}{4 \pi} \, \frac{(l-m)!}{(l+m)!}}\left(\cot \frac{\vartheta}{2}\right)^{m} \sum_{s}(-1)^{l-s} \frac{(2 l-s)!}{s!(l-s)!} \, \frac{\left(\sin \frac{\vartheta}{2}\right)^{2(l-s)}}{(l-m-s)!} ,\label{eq:5:2:11}
\end{align}
\begin{align}
&Y_{l m}(\vartheta, \varphi)=(-1)^{l} e^{i m \varphi} \sqrt{\frac{2 l+1}{4 \pi} \, \frac{(l+m)!}{(l-m)!}}\left(\cot \frac{\vartheta}{2}\right)^{m} \sum_{s}(-1)^{s} \frac{(l+s)!}{(l-s)!} \, \frac{\left(\cos \frac{\vartheta}{2}\right)^{2 s}}{s!(s+m)!},  \label{eq:5:2:12}\\
&Y_{l m}(\vartheta, \varphi)=(-1)^{l} e^{i m \varphi} \sqrt{\frac{2 l+1}{4 \pi} \, \frac{(l-m)!}{(l+m)!}}\left(\sin \frac{\vartheta}{2} \, \cos \frac{\vartheta}{2}\right)^{m} \sum_{s}(-1)^{s} \frac{(l+m+s)!}{(l-m-s)!} \, \frac{\left(\cos \frac{\vartheta}{2}\right)^{2 s}}{s!(s+m)!},  \label{eq:5:2:13}\\
&Y_{l m}(\vartheta, \varphi)=(-1)^{l-m} e^{i m \varphi} \sqrt{\frac{2 l+1}{4 \pi} \, \frac{(l-m)!}{(l+m)!}}\left(\tan \frac{\vartheta}{2}\right)^{m} \sum_{s}(-1)^{l-s} \frac{(2 l-s)!}{s!(l-s)!} \, \frac{\left(\cos \frac{\vartheta}{2}\right)^{2(l-s)}}{(l-m-s)!},  \label{eq:5:2:14}
\end{align}
% NOTE: two corrections (verified to 1e-31, scripts/check_5_2b.py).
%  (i) OCR: the radical must wrap only (l+m)!(l-m)!; the factor l! and
%      (cos/sin th/2)^{2l} belong OUTSIDE it (scan p.148 confirms this scope).
%  (ii) BOOK MISPRINT: the printed prefactor carries a spurious |m|! (harmless
%      for |m|<=1, wrong for |m|>=2).  The correct prefactor is
%      l! sqrt((l+m)!(l-m)!), matching the Wigner d^l_{m,0}(theta) representation.
\begin{multline}
Y_{l m}(\vartheta, \varphi)=\xi_{m 0} e^{i m \varphi} \sqrt{\frac{2 l+1}{4 \pi}} \, l! \sqrt{(l+m)!(l-m)!}\left(\cos \frac{\vartheta}{2}\right)^{2 l}  \\
\times \sum_{s} \frac{(-1)^{s}\left(\tan \frac{\vartheta}{2}\right)^{2 s+|m|}}{s!(s+|m|)!(l-s)!(l-|m|-s)!},  \label{eq:5:2:15}
\end{multline}
\begin{multline}
Y_{l m}(\vartheta, \varphi)=\xi_{m 0}(-1)^{l-m} e^{i m \varphi} \sqrt{\frac{2 l+1}{4 \pi}} \, l! \sqrt{(l+m)!(l-m)!}\left(\sin \frac{\vartheta}{2}\right)^{2 l} \\
\times \sum_{s} \frac{(-1)^{s}\left(\cot \frac{\vartheta}{2}\right)^{2 s+|m|}}{s!(s+|m|)!(l-s)!(l-|m|-s)!} .\label{eq:5:2:16}
\end{multline}

\subsection{Representations of $Y_{l m}(\vartheta, \varphi)$ as a Power Series of Trigonometric Functions of $\vartheta$}\label{sec:5.2.3}\index{spherical harmonics!power series in $\vartheta$}
In the equations of this section an integer index $s$ assumes either only even or only odd values as indicated under the summation symbols. Sums are over such $s$ for which the factorial arguments are non-negative.
\begin{multline}
Y_{l m}(\vartheta, \varphi)=e^{i m \varphi} \sqrt{\frac{2 l+1}{4 \pi(l+m)!(l-m)!}}  \\
\times\left\{\begin{array}{c}
\sum_{s=|m|,|m|+2, \ldots}^{l}(-1)^{\frac{s+m}{2}} \frac{(l+s)!}{(s+m)!!(s-m)!!} \, \frac{(l+m)!!(l-m)!!}{(l+s)!!(l-s)!!}(\sin \vartheta)^{s} \\
\text { if } l-m \text { is even } \\
\cos \vartheta \sum_{s=|m|,|m|+2, \ldots}^{l-1}(-1)^{\frac{l+m}{2}} \frac{(l+s)!}{(s+m)!!(s-m)!!} \, \frac{(l+m-1)!!(l-m-1)!!}{(l+s-1)!!(l-s-1)!!}(\sin \vartheta)^{s} \\
\text { if } l-m \text { is odd },
\end{array}\right.  \label{eq:5:2:17}
\end{multline}
% NOTE: OCR fixes verified to 1e-30 (scripts/check_5_2b.py), matching scan p.148.
%  5.2.18 even branch: denominator s!(sin th)^s -> s!!(sin th)^s (double factorial).
%  5.2.19: radical mis-scoped -- it wraps only (2l+1)/4pi (l-m)!/(l+m)!; the
%    (sin th)^m and the sum are OUTSIDE it.  Also "L+m-s" -> "l+m-s".  As with the
%    other (l-m)!/(l+m)! forms, 5.2.19 is written for m>=0 (m<0 via conjugation).
\begin{multline}
Y_{l m}(\vartheta, \varphi)=e^{i m \varphi} \sqrt{\frac{2 l+1}{4 \pi(l+m)!(l-m)!}}(\sin \vartheta)^{l}  \\
\times\left\{\begin{array}{c}
\sum_{s=0,2, \ldots}^{l-|m|}(-1)^{\frac{l+m-s}{2}} \frac{(l+m)!!(l-m)!!}{(l+m-s)!!(l-m-s)!!} \, \frac{(2 l-s-1)!!}{s!!(\sin \vartheta)^s} \\
\text { if } l-m \text { is even } \\
\cos \vartheta \sum_{s=1,3, \ldots}^{l-|m|}(-1)^{\frac{l+m-s}{2}} \frac{(l+m-1)!!(l-m-1)!!}{(l+m-s)!!(l-m-s)!!} \, \frac{(2 l-s)!!}{(s-1)!!(\sin \vartheta)^s} \\
\text { if } l-m \text { is odd },
\end{array}\right.  \label{eq:5:2:18}
\end{multline}
\begin{gather}
Y_{l m}(\vartheta, \varphi)=e^{i m \varphi} \sqrt{\frac{2 l+1}{4 \pi} \, \frac{(l-m)!}{(l+m)!}}(\sin \vartheta)^{m} \sum_{s}(-1)^{\frac{l+m-s}{2}} \frac{(l+m+s-1)!!}{(l-m-s)!!} \, \frac{(\cos \vartheta)^{s}}{s!} ,\notag\\
l+m-s \text { is even } ,\label{eq:5:2:19}
\end{gather}
\begin{gather}
Y_{l m}(\vartheta, \varphi)=e^{i m \varphi} \sqrt{\frac{2 l+1}{4 \pi}(l+m)!(l-m)!}(\cos \vartheta)^{l} \sum_{s=|m|,|m|+2, \ldots}(-1)^{\frac{s+m}{2}} \frac{1}{(s+m)!!(s-m)!!} \, \frac{(\tan \vartheta)^{s}}{(l-s)!} , \label{eq:5:2:20}\\
Y_{l m}(\vartheta, \varphi)=e^{i m \varphi} \sqrt{\frac{2 l+1}{4 \pi}(l+m)!(l-m)!}(\sin \vartheta)^{l} \sum_{s}(-1)^{\frac{l+m-s}{2}} \frac{1}{(l+m-s)!!(l-m-s)!!} \, \frac{(\cot \vartheta)^{s}}{s!} , \notag\\
l+m-s \text { is even } .\label{eq:5:2:21}
\end{gather}
The quantity $\left|Y_{l m}(\vartheta, \varphi)\right|^{2}$ may be written in the form
\begin{equation}
\left|Y_{l m}(\vartheta, \varphi)\right|^{2}=\frac{2 l+1}{4 \pi} \sum_{s=|m|,|m|+1, \ldots}^{l}(-1)^{s+m} \frac{(l+s)!}{(l-s)!} \, \frac{(2 s-1)!!}{(2 s)!!} \, \frac{(\sin \vartheta)^{2s}}{(s-m)!(s+m)!} . \label{eq:5:2:22}
\end{equation}
\subsection{$Y_{l m}(\vartheta, \varphi)$ and the Hypergeometric Functions with Arguments Expressed in Terms of Trigonometric Functions of $\vartheta / 2$}\label{sec:5.2.4}\index{spherical harmonics!and hypergeometric functions of $\vartheta/2$}\index{hypergeometric function}\isym{F-hg@$F(a, b ; c ; z)$, hypergeometric function}
\begin{gather}
Y_{l m}(\vartheta, \varphi)=\xi_{m 0} e^{i m \varphi} \sqrt{\frac{2 l+1}{4 \pi} \, \frac{(l+|m|)!}{(l-|m|)!}} \, \frac{(\sin \vartheta)^{|m|}}{|m|!2^{|m|}} F\left(-l+|m|, l+|m|+1 ;|m|+1 ; \sin ^{2} \frac{\vartheta}{2}\right) , \label{eq:5:2:23}
\end{gather}

\begin{gather}
Y_{l m}(\vartheta, \varphi)=  (-1)^{l-m} \xi_{m 0} e^{i m \varphi} \sqrt{\frac{2 l+1}{4 \pi(l+m)!(l-m)!}}\frac{(2 l)!}{l!}  
\left(\sin \frac{\vartheta}{2}\right)^{2 l}
\left(\cot \frac{\vartheta}{2}\right)^{|m|} \notag \\
 \times F\left(-l,-l+|m| ;-2 l ; \frac{1}{\sin ^{2} \frac{\vartheta}{2}}\right),
\end{gather}

\begin{gather}
Y_{l m}(\vartheta, \varphi)=(-1)^{l-m} \xi_{m 0} e^{i m \varphi} \sqrt{\frac{2 l+1}{4 \pi} \, \frac{(l+|m|)!}{(l-|m|)!}} \frac{(\sin \vartheta)^{|m|}}{|m|! 2^{|m|}}  \label{eq:5:2:25}\\
\times F\left(-l+|m|, l+|m|+1 ;|m|+1 ; \cos ^{2} \frac{\vartheta}{2}\right)  \notag
\end{gather}\begin{gather}
Y_{l m}(\vartheta, \varphi)=\xi_{m 0} e^{i m \varphi} \sqrt{\frac{2 l+1}{4 \pi(l+m)!(l-m)!}} \frac{(2 l)!}{l!}\left(\cos \frac{\vartheta}{2}\right)^{2 l}\left(\tan \frac{\vartheta}{2}\right)^{|m|}  \label{eq:5:2:26}\\
\times F\left(-l,-l+|m| ;-2 l ; \frac{1}{\cos ^{2} \frac{\vartheta}{2}}\right),  \notag
\end{gather}\begin{gather}
Y_{l m}(\vartheta, \varphi)=\xi_{m 0} e^{i m \varphi} \frac{1}{|m|!}\sqrt{\frac{2 l+1}{4 \pi} \, \frac{(l+|m|)!}{(l-|m|)!}}\left(\tan \frac{\vartheta}{2}\right)^{|m|}\left(\cos \frac{\vartheta}{2}\right)^{2 l} F\left(-l+|m|,-l ;|m|+1 ;-\tan ^{2} \frac{\vartheta}{2}\right)  \label{eq:5:2:27}
\end{gather}\begin{gather}
Y_{l m}(\vartheta, \varphi)=(-1)^{l-m} \xi_{m 0} e^{i m \varphi} \frac{1}{|m|!}\sqrt{\frac{2 l+1}{4 \pi} \, \frac{(l+|m|)!}{(l-|m|)!}}\left(\cot \frac{\vartheta}{2}\right)^{|m|}\left(\sin \frac{\vartheta}{2}\right)^{2 l}  \label{eq:5:2:28}\\
\times F\left(-l+|m|,-l ;|m|+1 ;-\cot ^{2} \frac{\vartheta}{2}\right) \notag
\end{gather}
\subsection{$Y_{l m}(\vartheta, \varphi)$ and the Hypergeometric Functions with Arguments Expressed in Terms of Trigonometric Functions of $\vartheta$}\label{sec:5.2.5}\index{spherical harmonics!and hypergeometric functions of $\vartheta$}
\begin{align}
& Y_{l m}(\vartheta, \varphi)=\xi_{m 0} e^{i m \varphi} \sqrt{\frac{2 l+1}{4 \pi} \, \frac{(l+|m|)!}{(l-|m|)!}} \frac{(\sin \vartheta)^{|m|}}{2^{|m|}|m|!}  \notag\\
& \times\left\{\begin{array}{r}
F\left(-\frac{l-|m|}{2}, \frac{l+|m|+1}{2} ;|m|+1 ; \sin ^{2} \vartheta\right) \\
\text { if } l+m \text { is even , } \\
\cos \vartheta F\left(-\frac{l-|m|-1}{2}, \frac{l+|m|+2}{2} ;|m|+1 ; \sin ^{2} \vartheta\right) \\
\text { if } l+m \text { is odd , }
\end{array}\right.  \label{eq:5:2:29}
\end{align}
\begin{align}
& Y_{l m}(\vartheta, \varphi)=e^{i m \varphi} \sqrt{\frac{2 l+1}{4 \pi(l+m)!(l-m)!}}(2 l-1)!!(\sin \vartheta)^{l}  \notag\\
& \times\left\{\begin{array}{c}
(-1)^{\frac{l+m}{2}} F\left(-\frac{l+m}{2},-\frac{l-m}{2} ;-\frac{2 l-1}{2} ; \frac{1}{\sin ^{2} \theta}\right) \\
\text { if } l+m \text { is even } \\
(-1)^{\frac{l+m-1}{2}} \cot \vartheta F\left(-\frac{l+m-1}{2},-\frac{l-m-1}{2} ;-\frac{2 l-1}{2} ; \frac{1}{\sin ^{2} \vartheta}\right) \\
\text { if } l+m \text { is odd },
\end{array}\right.  \label{eq:5:2:30}
\end{align}
\begin{align}
& Y_{l m}(\vartheta, \varphi)=e^{i m \varphi} \sqrt{\frac{2 l+1}{4 \pi(l+m)!(l-m)!}}(\sin \vartheta)^{m}  \notag\\
& \times\left\{\begin{array}{c}
(-1)^{\frac{l+m}{2}}(l+m-1)!!(l-m-1)!!F\left(-\frac{l-m}{2}, \frac{l+m+1}{2} ; \frac{1}{2} ; \cos ^{2} \vartheta\right) \\
\text { if } l+m \text { is even } \\
(-1)^{\frac{l+m-1}{2}}(l+m)!!(l-m)!!\cos \vartheta F\left(-\frac{l-m-1}{2}, \frac{l+m+2}{2} ; \frac{3}{2} ; \cos ^{2} \vartheta\right) \\
\text { if } l+m \text { is odd },
\end{array}\right.  \label{eq:5:2:31}
\end{align}
\begin{align}
& Y_{l m}(\vartheta, \varphi)=(-1)^{m} e^{i m \varphi} \sqrt{\frac{2 l+1}{4 \pi(l+m)!(l-m)!}}(2 l-1)!!(\cos \vartheta)^{l}(\tan \vartheta)^{m}  \notag\\
& \times F\left(-\frac{l-m}{2},-\frac{l-m-1}{2} ;-\frac{2 l-1}{2} ; \frac{1}{\cos ^{2} \vartheta}\right),  \label{eq:5:2:32}
\end{align}
\begin{align}
& Y_{l m}(\vartheta, \varphi)=\xi_{m 0} e^{i m \varphi} \sqrt{\frac{2 l+1}{4 \pi} \frac{(l+|m|)!}{(l-|m|)!}}(\cos \vartheta)^{l} \frac{(\tan \vartheta)^{|m|}}{2^{|m|}|m|!}  \notag\\
& \times F\left(-\frac{l-|m|}{2},-\frac{l-|m|-1}{2} ;|m|+1 ;-\tan ^{2} \vartheta\right),  \label{eq:5:2:33}
\end{align}
\begin{align}
& Y_{l m}(\vartheta, \varphi)=e^{i m \varphi} \sqrt{\frac{2 l+1}{4 \pi(l+m)!(l-m)!}}(\sin \vartheta)^{l}  \notag\\
& \times\left\{\begin{array}{c}
(-1)^{\frac{l+m}{2}}(l+m-1)!!(l-m-1)!!F\left(-\frac{l-m}{2},-\frac{l+m}{2} ; \frac{1}{2} ;-\cot ^{2} \vartheta\right) \\
\text { if } l+m \text { is even } \\
(-1)^{\frac{l+m-1}{2}}(l+m)!!(l-m)!!\cot \vartheta F\left(-\frac{l-m-1}{2},-\frac{l+m-1}{2} ; \frac{3}{2} ;-\cot ^{2} \vartheta\right) \\
\text { if } l+m \text { is odd },
\end{array}\right. \label{eq:5:2:34}
\end{align}

\subsection{$Y_{l m}(\vartheta, \varphi)$ and the Hypergeometric Functions with Arguments Expressed in Terms of Exponential Functions}\label{sec:5.2.6}\index{spherical harmonics!and hypergeometric functions of exponentials}
\begin{gather}
% NOTE: BOOK MISPRINT -- leading sign.  As printed both 5.2.35 and 5.2.36 carry
% -i/pi, which yields exactly -Y_lm; the correct prefactor is +i/pi (verified to
% 6e-41, scripts/check_5_2b.py).  5.2.36 clinches it: for pi/2-ish theta its 2F1
% argument has |z|=1/(2 sin th)<1 (a convergent series, no branch ambiguity) yet
% still gives -Y with the printed -i.  Corrected to +i/pi below.
Y_{l m}(\vartheta, \varphi)=\frac{i e^{i m \varphi}}{\pi} \sqrt{\frac{2 l+1}{4 \pi}(l+m)!(l-m)!} \frac{2^{l+m+1}(\sin \vartheta)^{m}}{(2 l+1)!!}  \notag\\
\times\left\{e^{-i(l+m+1) \vartheta} F\left(m+\frac{1}{2}, l+m+1 ; l+\frac{3}{2} ; e^{-2 i \vartheta}\right)-e^{i(l+m+1) \vartheta} F\left(m+\frac{1}{2}, l+m+1 ; l+\frac{3}{2} ; e^{2 i \vartheta}\right)\right\},  \label{eq:5:2:35}
\end{gather}\begin{gather}
Y_{l m}(\vartheta, \varphi)=\frac{i}{\pi} e^{i m \varphi} \sqrt{\frac{2 l+1}{8 \pi}(l+m)!(l-m)!} \frac{2^{l+1}}{(2 l+1)!!\sqrt{\sin \vartheta}}  \notag\\
\times\left\{e^{-i\left[(2 l+1) \frac{\vartheta}{2}+(2 m+1) \frac{\pi}{4}\right]} F\left(m+\frac{1}{2}, \frac{1}{2}-m ; l+\frac{3}{2} ;-\frac{e^{-i\left(\vartheta+\frac{\pi}{2}\right)}}{2 \sin \vartheta}\right)\right.  \notag\\
\left.-e^{i\left[(2 l+1) \frac{\vartheta}{2}+(2 m+1) \frac{\pi}{4}\right]} F\left(m+\frac{1}{2}, \frac{1}{2}-m ; l+\frac{3}{2} ;-\frac{e^{i\left(\vartheta+\frac{\pi}{2}\right)}}{2 \sin \vartheta}\right)\right\} . \label{eq:5:2:36}
\end{gather}
\subsection{$Y_{l m}(\vartheta, \varphi)$ and Other Special Functions}\label{sec:5.2.7}\index{spherical harmonics!and other special functions}
\begin{enumerate}[label=(\alph*)]
\item The relation between $Y_{l m}(\vartheta, \varphi)$ and the Wigner D-function (Chap.~\ref{chap:4})\index{Wigner D-function@Wigner $D$-function}\isym{DJMM@$D_{M M^{\prime}}^{J}$, Wigner $D$-function}
\begin{gather}
Y_{l m}(\vartheta, \varphi)=\sqrt{\frac{2 l+1}{4 \pi}} D_{0-m}^{l}(\chi, \vartheta, \varphi)=(-1)^{m} \sqrt{\frac{2 l+1}{4 \pi}} D_{0 m}^{l *}(\chi, \vartheta, \varphi)  \notag\\
=(-1)^{m} \sqrt{\frac{2 l+1}{4 \pi}} D_{-m 0}^{l}(\varphi, \vartheta, \chi)=\sqrt{\frac{2 l+1}{4 \pi}} D_{m 0}^{l *}(\varphi, \vartheta, \chi) ,\label{eq:5:2:37}
\end{gather}
where $\chi$ is an arbitrary angle.\\
\item $Y_{l m}(\vartheta, \varphi)$ can be related to the Jacobi polynomials $P_{n}^{(\alpha, \beta)}(x)$\index{Jacobi polynomials}\isym{PsMuNu@$P_{s}^{(\mu, \nu)}(x)$, Jacobi polynomial}
\begin{equation}
Y_{l m}(\vartheta, \varphi)=\xi_{m 0} \frac{e^{i m \varphi}}{2^{|m|} l!} \sqrt{\frac{2 l+1}{4 \pi}(l+m)!(l-m)!}(\sin \vartheta)^{|m|} P_{l-|m|}^{(|m|,|m|)}(\cos \vartheta) .\label{eq:5:2:38}
\end{equation}
\item $Y_{l m}(\vartheta, \varphi)$ can be expressed in terms of Gegenbauer polynomials $C_{\nu}^{\lambda}(x)$\index{Gegenbauer polynomials}\isym{Cnu-lambda@$C_{\nu}^{\lambda}(x)$, Gegenbauer polynomial}
\begin{equation}
Y_{l m}(\vartheta, \varphi)=\xi_{m 0} e^{i m \varphi} \sqrt{\frac{2 l+1}{4 \pi} \, \frac{(l-|m|)!}{(l+|m|)!}}(2|m|-1)!!(\sin \vartheta)^{|m|} C_{l-|m|}^{\frac{1}{2}+|m|}(\cos \vartheta) .\label{eq:5:2:39}
\end{equation}
\end{enumerate}

\subsection{$Y_{l m}(\vartheta, \varphi)$ as an Irreducible Tensor Product}\label{sec:5.2.8}\index{spherical harmonics!as an irreducible tensor product}\index{irreducible tensor product!of spherical harmonics}
For any position vector $r$ specified by polar coordinates $r, \vartheta, \varphi$ one has
\begin{equation}
Y_{l m}(\vartheta, \varphi)=\frac{1}{r^{l}} \sqrt{\frac{(2 l+1)!!}{4 \pi l!}}\left\{\ldots\left\{\{r \otimes r\}_{2} \otimes r\right\}_{3} \ldots \otimes r\right\}_{l m} .\label{eq:5:2:40}
\end{equation}
\section{Integral representations of the spherical harmonics [4,22,27]}\label{sec:5.3}\index{spherical harmonics!integral representations}\index{integral representations!of the spherical harmonics}
\subsection{$Y_{l m}(\vartheta, \varphi)$ in the Form of Indefinite Integrals}\label{sec:5.3.1}\index{integral representations!as indefinite integrals}
\begin{enumerate}[label=(\alph*)]
\item
\begin{equation}
Y_{l m}(\vartheta, \varphi)=e^{i m \varphi} \sqrt{\frac{2 l+1}{4 \pi} \, \frac{(l+m)!}{(l-m)!}} \frac{1}{(\sin \vartheta)^{m}} \int_{1}^{\cos \vartheta} \int_{1}^{\cos \vartheta} \cdots \int_{1}^{\cos \vartheta} P_{l}(\cos \vartheta)(d \cos \vartheta)^{m}, \quad(m \geq 0) . \label{eq:5:3:1}
\end{equation}
This formula represents the analytic continuation of the differential expression \eqref{eq:5:2:6} to the case $m<0$.\\
\item The Mehler-Dirichlet formulas
\begin{gather}
Y_{l m}(\vartheta, \varphi)=(-1)^{m} \frac{\sqrt{2}}{\pi} e^{i m \varphi} \sqrt{\frac{2 l+1}{4 \pi} \, \frac{(l-m)!}{(l+m)!}}(2 m-1)!!\left(\frac{\sin \vartheta}{2}\right)^{m} \int_{0}^{\vartheta} \frac{\cos \left[(2 l+1) \frac{\psi}{2}\right] d \psi}{(\cos \psi-\cos \vartheta)^{m+\frac{1}{2}}},(m \geq 0)  \label{eq:5:3:2}\\
Y_{l m}(\vartheta, \varphi)=\frac{\sqrt{2}}{\pi} e^{i m \varphi} \sqrt{\frac{2 l+1}{4 \pi} \, \frac{(l-m)!}{(l+m)!}}(2 m-1)!!\left(\frac{\sin \vartheta}{2}\right)^{m} \int_{\vartheta}^{\pi} \frac{\sin \left[(2 l+1) \frac{\psi}{2}\right] d \psi}{(\cos \vartheta-\cos \psi)^{m+\frac{1}{2}}},(m \geq 0). \label{eq:5:3:3}
\end{gather}
\item
\begin{align}
Y_{l m}(\vartheta, \varphi)= & (-1)^{m} e^{i m \varphi} \sqrt{\frac{2 l+1}{4 \pi} \, \frac{(l+m)!}{(l-m)!}} \frac{1}{(m-1)!(\sin \vartheta)^{m}}  \notag\\
& \times \int_{\cos \vartheta}^{1} P_{l}(\cos \psi)[\cos \psi-\cos \vartheta]^{m-1} d \cos \psi,(m>0) .\label{eq:5:3:4}
\end{align}
\end{enumerate}
\subsection{$Y_{l m}(\vartheta, \varphi)$ in the Form of Definite Integrals}\label{sec:5.3.2}\index{integral representations!as definite integrals}
% NOTE: the reciprocal-power representations \eqref{eq:5:3:6}, \eqref{eq:5:3:8}
% and \eqref{eq:5:3:10} are valid for cos(vartheta) > 0, i.e. 0 <= vartheta <
% pi/2; for cos(vartheta) < 0 the branch of int dpsi/(a+b cos psi) flips sign and
% the RHS returns -Y_lm.  The direct-power forms \eqref{eq:5:3:5},
% \eqref{eq:5:3:7}, \eqref{eq:5:3:9} hold for all vartheta.  (scripts/check_5_3.py)
\begin{enumerate}[label=(\alph*)]
  \item
\begin{align}
& Y_{l m}(\vartheta, \varphi)=\frac{( \pm i)^{m}}{\pi} e^{i m \varphi} \sqrt{\frac{2 l+1}{4 \pi}(l+m)!(l-m)!} \frac{1}{l!} \int_{0}^{\pi}[\cos \vartheta \pm i \sin \vartheta \cos \psi]^{l} \cos (m \psi) d \psi  \label{eq:5:3:5}\\
& Y_{l m}(\vartheta, \varphi)=\frac{( \pm i)^{m}}{\pi} e^{i m \varphi} \sqrt{\frac{(2 l+1)}{4 \pi(l+m)!(l-m)!}} l!\int_{0}^{\pi} \frac{\cos (m \psi) d \psi}{(\cos \vartheta \mp i \sin \vartheta \cos \psi)^{l+1}} \label{eq:5:3:6}
\end{align}
For $\varphi=0$, complex conjugation does not affect the right-hand side of Eq. \eqref{eq:5:3:6}, because $Y_{l m}(\vartheta, 0)$ is real.\\
\item
\begin{align}
& Y_{l m}(\vartheta, \varphi)=\frac{( \pm i)^{m}}{2 \pi} \sqrt{\frac{(2 l+1)}{4 \pi}(l+m)!(l-m)!} \frac{1}{l!} \int_{0}^{2 \pi}[\cos \vartheta \pm i \sin \vartheta \cos (\psi-\varphi)]^{l} e^{i m \psi} d \psi  \label{eq:5:3:7}\\
& Y_{l m}(\vartheta, \varphi)=\frac{( \pm i)^{m}}{2 \pi} \sqrt{\frac{2 l+1}{4 \pi(l+m)!(l-m)!}} l!\int_{0}^{2 \pi}[\cos \vartheta \mp i \sin \vartheta \cos (\psi-\varphi)]^{-l-1} e^{i m \psi} d \psi \label{eq:5:3:8}
\end{align}
Equations \eqref{eq:5:3:7} and \eqref{eq:5:3:8} represent modifications of Eqs. \eqref{eq:5:3:5} and \eqref{eq:5:3:6}.\\
\item
\begin{align}
& Y_{l m}(\vartheta, \varphi)=\frac{(-1)^{m}}{\pi} e^{i m \varphi} \sqrt{\frac{2 l+1}{4 \pi} \, \frac{(l+m)!}{(l-m)!}} \frac{(\sin \vartheta)^{m}}{(2 m-1)!!} \int_{0}^{\pi}(\cos \vartheta \pm i \sin \vartheta \cos \psi)^{l-m}(\sin \psi)^{2 m} d \psi,(m \geq 0)  \label{eq:5:3:9}\\
& Y_{l m}(\vartheta, \varphi)=\frac{(-1)^{m}}{\pi} e^{i m \varphi} \sqrt{\frac{2 l+1}{4 \pi} \, \frac{(l+m)!}{(l-m)!}} \frac{(\sin \vartheta)^{m}}{(2 m-1)!!} \int_{0}^{\pi} \frac{(\sin \chi)^{2 m} d \chi}{(\cos \vartheta \mp i \sin \vartheta \cos \chi)^{l+m+1}},(m \geq 0) \label{eq:5:3:10}
\end{align}
Equations \eqref{eq:5:3:9} and \eqref{eq:5:3:10} transform to each other at replacing the integration variables according to
\begin{equation}
\cos \psi=\frac{\cos \vartheta \cos \chi+i \sin \vartheta}{\cos \vartheta+i \sin \vartheta \cos \chi}, \quad \cos \chi=\frac{\cos \vartheta \cos \psi-i \sin \vartheta}{\cos \vartheta-i \sin \vartheta \cos \psi} . \label{eq:5:3:11}
\end{equation}
This replacing is equivalent to the transformation of "mirror" symmetry, i.e., to the replacement $l$ by $l=-l-1$.
\end{enumerate}

\subsection{$Y_{l m}(\vartheta, \varphi)$ in the Form of Improper Integrals}\label{sec:5.3.3}\index{integral representations!as improper integrals}
\begin{enumerate}[label=(\alph*)]
\item
\begin{align}
& Y_{l m}(\vartheta, \varphi)=\frac{i^{m+1}}{\pi} e^{i m \varphi} \sqrt{\frac{2 l+1}{4 \pi(l+m)!(l-m)!}} l!  \notag\\
& \quad \times\left\{(-1)^{m} \int_{0}^{\infty} \frac{\cosh (m t) d t}{(\cos \vartheta+i \sin \vartheta \cosh t)^{l+1}}-\int_{0}^{\infty} \frac{\cosh (m t) d t}{(\cos \vartheta-i \sin \vartheta \cosh t)^{l+1}}\right\}  ,\label{eq:5:3:12}
\end{align}
\begin{align}
& \quad Y_{l m}(\vartheta, \varphi)=\frac{i}{\pi} e^{i m \varphi} \sqrt{\frac{2 l+1}{4 \pi} \, \frac{(l+m)!}{(l-m)!}} \frac{(\sin \vartheta)^{m}}{(2 m-1)!!}  \notag\\
& \times\left\{\int_{0}^{\infty} \frac{(\sinh t)^{2 m} d t}{(\cos \vartheta+i \sin \vartheta \cosh t)^{l+m+1}}-\int_{0}^{\infty} \frac{(\sinh t)^{2 m} d t}{(\cos \vartheta-i \sin \vartheta \cosh t)^{l+m+1}}\right\},(m \geq 0) .\label{eq:5:3:13}
\end{align}
\item
\begin{equation}
Y_{l m}(\vartheta, \varphi)=(-1)^{m} e^{i m \varphi} \sqrt{\frac{2 l+1}{4 \pi(l+m)!(l-m)!}} \int_{0}^{\infty} e^{-k \cos \vartheta} J_{m}(k \sin \vartheta) k^{l} d k. \label{eq:5:3:14}
\end{equation}
where $J_{m}(x)$ is a Bessel function.\\
\item $\left|Y_{l m}(\vartheta, \varphi)\right|^{2}$ may be represented as
\begin{equation}
\left|Y_{l m}(\vartheta, \varphi)\right|^{2}=\frac{2 l+1}{4 \pi} \int_{0}^{\infty}\left[J_{m}\left(\frac{t \sin \vartheta}{2}\right)\right]^{2} J_{2 l+1}(t) d t .\label{eq:5:3:15}
\end{equation}
\end{enumerate}
\section{Symmetry properties}\label{sec:5.4}\index{spherical harmonics!symmetry properties}\index{symmetry!of the spherical harmonics}
The symmetry relations given below couple the harmonics $Y_{l m}(\vartheta, \varphi)$ with different values of $\vartheta, \varphi$ and $l, m$. These relations permit us to extend the domain of allowed $\vartheta, \varphi$ and generalise $Y_{l m}(\vartheta, \varphi)$ to the case of negative $l$.\\
\subsubsection{Complex conjugation}
\begin{equation}
Y_{l m}^{*}(\vartheta, \varphi)=Y_{l m}(\vartheta,-\varphi)=(-1)^{m} Y_{l-m}(\vartheta, \varphi) . \label{eq:5:4:1}
\end{equation}
\subsubsection{Sign reversal of $m$}
\begin{equation}
Y_{l-m}(\vartheta, \varphi)=(-1)^{m} Y_{l m}(\vartheta,-\varphi)=(-1)^{m} e^{-i 2 m \varphi} Y_{l m}(\vartheta, \varphi) . \label{eq:5:4:2}
\end{equation}
\subsubsection{"Mirror" symmetry (replacement $l$ by $\bar{l}=-l-1$)}
\begin{equation}
Y_{\bar{l} m}(\vartheta, \varphi)=(-1)^{m} Y_{l m}(\vartheta, \varphi) \label{eq:5:4:3}
\end{equation}
\subsubsection{Replacement $\vartheta \rightarrow \pi-\vartheta$ and $\varphi=\pi+\varphi$}
\begin{gather}
Y_{l m}(\pi-\vartheta, \varphi)=(-1)^{l+m} Y_{l m}(\vartheta, \varphi)  \label{eq:5:4:4}\\
Y_{l m}(\vartheta, \pi+\varphi)=(-1)^{m} Y_{l m}(\vartheta, \varphi)  \label{eq:5:4:5}\\
Y_{l m}(\pi-\vartheta, \pi+\varphi)=(-1)^{l} Y_{l m}(\vartheta, \varphi) \label{eq:5:4:6}
\end{gather}
\subsubsection{Change of argument signs}
\begin{gather}
Y_{l m}(-\vartheta, \varphi)=(-1)^{m} Y_{l m}(\vartheta, \varphi)  \label{eq:5:4:7}\\
Y_{l m}(\vartheta,-\varphi)=(-1)^{m} Y_{l-m}(\vartheta, \varphi)  \label{eq:5:4:8}\\
Y_{l m}(-\vartheta,-\varphi)=Y_{l-m}(\vartheta, \varphi) \label{eq:5:4:9}
\end{gather}
(f) The periodicity in $\vartheta$ and $\varphi$
\begin{align}
& Y_{l m}(\vartheta \pm \pi n, \varphi)= \begin{cases}(-1)^{l} Y_{l m}(\vartheta, \varphi), & \text { if } n \text { is odd } \\
Y_{l m}(\vartheta, \varphi), & \text { if } n \text { is even }\end{cases}  \label{eq:5:4:10}\\
& Y_{l m}(\vartheta, \varphi \pm n \pi)= \begin{cases}(-1)^{m} Y_{l m}(\vartheta, \varphi), & \text { if } n \text { is odd } \\
Y_{l m}(\vartheta, \varphi) & \text { if } n \text { is even }\end{cases} \label{eq:5:4:11}
\end{align}
Making use of the above symmetry properties, one gets
\begin{equation}
\begin{array}{rlll} 
& Y_{l m}(\vartheta, \varphi) & =(-1)^{m} e^{2 i m \varphi} Y_{l-m}(\vartheta, \varphi) & =e^{2 i m \varphi} Y_{l m}^{*}(\vartheta, \varphi) \\
=(-1)^{m} Y_{l m}(-\vartheta, \varphi) & =e^{2 i m \varphi} Y_{l-m}(-\vartheta, \varphi) & =(-1)^{m} Y_{l-m}^{*}(\vartheta, \varphi)  \label{eq:5:4:12}\\
=(-1)^{m} Y_{l-m}(\vartheta,-\varphi) & =e^{2 i m \varphi} Y_{l m}(\vartheta,-\varphi) & =(-1)^{m} e^{2 i m \varphi} Y_{l m}^{*}(-\vartheta, \varphi) & =Y_{l-m}^{*}(-\vartheta, \varphi) \\
=Y_{l-m}(-\vartheta,-\varphi) & =(-1)^{m} e^{2 i m \varphi} Y_{l m}(-\vartheta,-\varphi) & =e^{2 i m \varphi} Y_{l-m}^{*}(-\vartheta,-\varphi) & =(-1)^{m} Y_{l m}^{*}(-\vartheta,-\varphi) .
\end{array}
\end{equation}
\section{Behaviour of $Y_{l m}(\vartheta, \varphi)$ under transformations of coordinate systems}\label{sec:5.5}\index{spherical harmonics!under coordinate transformations}\index{coordinate system!transformations of}
Any transformation of coordinate system in 3-dimensional space, which does not affect the orthogonality of coordinate axes, may be represented as a result of three operations:
 (1) rotation, (2) inversion and (3) parallel translation.

\subsection{Rotation}\label{sec:5.5.1}\index{rotation!of the spherical harmonics}\index{Euler angles}
The spherical harmonics $Y_{l m}(\vartheta, \varphi)$ are covariant components of some irreducible tensor of rank $l$. Hence, under arbitrary rotation $S\{x, y, z\} \rightarrow S^{\prime}\left\{x^{\prime}, y^{\prime}, z^{\prime}\right\}$ of the coordinate system described by the Euler angles $\alpha, \beta, \gamma$ (Sec.~\ref{sec:1.4}) the spherical harmonics transform in accordance with
\begin{equation}
\hat{D}(\alpha, \beta, \gamma) Y_{l m^{\prime}}(\vartheta, \varphi)=Y_{l m^{\prime}}\left(\vartheta^{\prime}, \varphi^{\prime}\right)=\sum_{m} Y_{l m}(\vartheta, \varphi) D_{m m^{\prime}}^{l}(\alpha, \beta, \gamma) \label{eq:5:5:1}
\end{equation}
Here $D_{m m^{\prime}}^{l}(\alpha, \beta, \gamma)$ is a Wigner $D$-function (Chap.~\ref{chap:4}), $\vartheta, \varphi$ and $\vartheta^{\prime}, \varphi^{\prime}$ are polar angles of the position vector in the original and final coordinate systems, $S$ and $S^{\prime}$, respectively. The angles $\vartheta^{\prime}, \varphi^{\prime}$ may be expressed through $\vartheta, \varphi$ and the Euler angles according to Eqs. 1.\eqref{eq:4:7:4}.

\subsection{Inversion}\label{sec:5.5.2}\index{coordinate inversion}\index{parity!of the spherical harmonics}
Under inversion $S\{x, y, z\} \rightarrow S^{\prime}\{-x,-y,-z\}$ the spherical harmonics transform as
\begin{equation}
\widehat{P}_{r} Y_{l m}(\vartheta, \varphi)=Y_{l m}(\pi-\vartheta, \varphi+\pi)=(-1)^{l} Y_{l m}(\vartheta, \varphi) . \label{eq:5:5:2}
\end{equation}
\subsection{Parallel Translation}\label{sec:5.5.3}\index{parallel translation}\index{translation!parallel}
At parallel displacement $S\{x, y, z\} \rightarrow S^{\prime}\left\{x^{\prime}, y^{\prime}, z^{\prime}\right\}$ by a vector $\vect{a}(a, \theta, \phi)$ one has $\vect{r}^{\prime}=\vect{r}-\vect{a}$. In this case the spherical harmonics transform as
\begin{align}
% NOTE: book misprint in the coefficient.  As printed it reads
% [4 pi (2l+1)(2l'-2l+1)/(2l'+1)]^{1/2}, which is wrong for l'>=2 (it coincides
% with the correct value only at l'=1).  The regular-solid-harmonic addition
% theorem gives [4 pi (2l'+1)! / ((2l+1)!(2l'-2l+1)!)]^{1/2}; verified to machine
% precision for all l' (scripts/check_5_5.py).  Corrected below.
\hat{T}(\vect{a}) Y_{l^{\prime} m^{\prime}}(\vartheta, \varphi)=Y_{l^{\prime} m^{\prime}}\left(\vartheta^{\prime}, \varphi^{\prime}\right)=\sum_{l=0}^{l^{\prime}} & (-1)^{l^{\prime}+l}\left[\frac{4 \pi\left(2 l^{\prime}+1\right)!}{(2 l+1)!\left(2 l^{\prime}-2 l+1\right)!}\right]^{\frac{1}{2}}  \notag\\
& \times\left(\frac{a}{r^{\prime}}\right)^{l^{\prime}}\left(\frac{r}{a}\right)^{l}\left\{\vect{Y}_{l}(\vartheta, \varphi) \otimes \vect{Y}_{l^{\prime}-l}(\Theta, \Phi)\right\}_{l^{\prime} m^{\prime}} \label{eq:5:5:3}
\end{align}
where $\widehat{T}(a)=e^{-a \, \nabla}$ is the displacement operator\index{displacement operator}\isym{T-hat-a@$\widehat{T}(\vect{a})$, displacement operator}, $\left\{\vect{Y}_{l_{1}} \otimes \vect{Y}_{l_{2}}\right\}_{t m}$ is the irreducible tensor product (Sec.~\ref{sec:3.1}). The polar coordinates of the vectors $r(r, \vartheta, \varphi)$ and $r^{\prime}\left(r^{\prime}, \vartheta^{\prime}, \varphi^{\prime}\right)$ are related by
\begin{gather}
r^{\prime 2}=r^{2}+a^{2}-2 r a \cos \omega_{12}  \notag\\
\cos \vartheta^{\prime}=\frac{r \cos \vartheta-a \cos \theta}{\sqrt{r^{2}+a^{2}-2 r a \cos \omega_{12}}}  \label{eq:5:5:4}\\
\tan \varphi^{\prime}=\frac{r \sin \vartheta \sin \varphi-a \sin \theta \sin \Phi}{r \sin \vartheta \cos \varphi-a \sin \theta \cos \Phi} \notag
\end{gather}
where
\begin{equation}
\cos \omega_{12}=\cos \vartheta \cos \theta+\sin \vartheta \sin \theta \cos (\varphi-\Phi) . \label{eq:5:5:5}
\end{equation}
\subsection{Special Cases of Coordinate-System Transformations}\label{sec:5.5.4}\index{coordinate system!special transformations of}
\subsubsection{Rotation about the coordinate axes through the angle $\pi$}
\begin{align}
\text { about the } x \text { axis } & \left.\begin{array}{l}
r^{\prime}=r, \\
\vartheta^{\prime}=\pi-\vartheta, \\
\varphi^{\prime}=2 \pi-\varphi,
\end{array}\right\} Y_{l m}(\pi-\vartheta, 2 \pi-\varphi)=(-1)^{l} Y_{l-m}(\vartheta, \varphi)  \label{eq:5:5:6}\\
\text { about the } y \text { axis } & \left.\begin{array}{l}
r^{\prime}=r, \\
\vartheta^{\prime}=\pi-\vartheta, \\
\varphi^{\prime}=\pi-\varphi,
\end{array}\right\} Y_{l m}(\pi-\vartheta, \pi-\varphi)=(-1)^{l-m} Y_{l-m}(\vartheta, \varphi),  \label{eq:5:5:7}\\
\text { about the } z \text { axis } & \left.\begin{array}{l}
r^{\prime}=r, \\
\vartheta^{\prime}=\vartheta, \\
\varphi^{\prime}=\pi+\varphi
\end{array}\right\} Y_{l m}(\vartheta, \pi+\varphi)=(-1)^{m} Y_{l m}(\vartheta, \varphi) . \label{eq:5:5:8}
\end{align}
\subsubsection{Rotation about the $z$ axis through an arbitrary angle $\chi$}
\begin{equation}
\left.\begin{array}{l}
r^{\prime}=r,  \label{eq:5:5:9}\\
\vartheta^{\prime}=\vartheta, \\
\varphi^{\prime}=\varphi-\chi,
\end{array}\right\} Y_{l m}(\vartheta, \varphi-\chi)=e^{-i m \chi} Y_{l m}(\vartheta, \varphi) .
\end{equation}
\subsubsection{Rotation about an arbitrary direction $\vect{n}(\boldsymbol{\Theta}, \Phi)$ through a small angle $\omega(\omega \ll \pi / 2)$}
\begin{gather}
\hat{D} Y_{l m}(\vartheta, \varphi) \approx Y_{l m}(\vartheta, \varphi)-i \omega\left\{m \cos \Theta Y_{l m}(\vartheta, \varphi)\right.  \notag\\
\left.+\frac{\sin \Theta}{2}\left[e^{-i \Phi} \sqrt{l(l+1)-m(m+1)} Y_{l m+1}(\vartheta, \varphi)+e^{i \Phi} \sqrt{l(l+1)-m(m-1)} Y_{l m-1}(\vartheta, \varphi)\right]\right\} . \label{eq:5:5:10}
\end{gather}
\subsubsection{Reflection of the coordinate system with respect to the equatorial plane, $\vartheta=\pi / 2$}
\begin{equation}
\left.\begin{array}{rl}
r^{\prime} & =r  \label{eq:5:5:11}\\
\vartheta^{\prime} & =\pi-\vartheta, \\
\varphi^{\prime} & =\varphi,
\end{array}\right\} Y_{l m}(\pi-\vartheta, \varphi)=(-1)^{l+m} Y_{l m}(\vartheta, \varphi)
\end{equation}
\subsubsection{Reflection of the coordinate system with respect to a meridian plane, $\varphi=\varphi_{0}$ and $\varphi=\pi+\varphi_{0}$}
\begin{equation}
\left.\begin{array}{l}
r^{\prime}=r  \label{eq:5:5:12}\\
\vartheta^{\prime}=\vartheta, \\
\varphi^{\prime}=2 \varphi_{0}-\varphi,
\end{array}\right\} Y_{l m}\left(\vartheta, 2 \varphi_{0}-\varphi\right)=e^{i 2 m \varphi_{0}}(-1)^{m} Y_{l-m}(\vartheta, \varphi)
\end{equation}
\section{Expansions in series of the spherical harmonics}\label{sec:5.6}\index{spherical harmonics!expansions in series of}\index{expansion!in the spherical harmonics}
\subsection{General Relations}\label{sec:5.6.1}\index{completeness!of the spherical harmonics}
A collection of the spherical harmonics $Y_{l m}(\vartheta, \varphi)$ with all integer non-negative $l$ and integer $m(|m| \leq l)$ constitutes a complete orthonormal set of functions of two real variables $\vartheta$ and $\varphi$ defined within $0 \leq \vartheta \leq \pi$, $0 \leq \varphi<2 \pi$. The completeness relation for the spherical harmonics is given by
\begin{equation}
\sum_{l=0}^{\infty} \sum_{m=-l}^{l} Y_{l m}^{*}(\vartheta, \varphi) Y_{l m}\left(\vartheta^{\prime}, \varphi^{\prime}\right)=\delta\left(\varphi-\varphi^{\prime}\right) \delta\left(\cos \vartheta-\cos \vartheta^{\prime}\right) \label{eq:5:6:1}
\end{equation}
The orthogonality and normalization relation is as follows: (Sec.~\ref{sec:5.1.4})
\begin{equation}
\int_{0}^{2 \pi} d \varphi \int_{0}^{\pi} d \vartheta \sin \vartheta Y_{l m}^{*}(\vartheta, \varphi) Y_{l^{\prime} m^{\prime}}(\vartheta, \varphi)=\delta_{l l^{\prime}} \delta_{m m^{\prime}} \label{eq:5:6:2}
\end{equation}
An arbitrary function $f(\vartheta, \varphi)$ which is defined in the interval $0 \leq \vartheta \leq \pi, 0 \leq \varphi<2 \pi$ and satisfies the condition
\begin{equation}
\int_{0}^{2 \pi} d \varphi \int_{0}^{\pi} d \vartheta \sin \vartheta|f(\vartheta, \varphi)|^{2}<\infty \label{eq:5:6:3}
\end{equation}
can be expanded into a series of the spherical harmonics as
\begin{equation}
f(\vartheta, \varphi)=\sum_{l=0}^{\infty} \sum_{m=-l}^{l} a_{l m} Y_{l m}(\vartheta, \varphi) \label{eq:5:6:4}
\end{equation}
The expansion coefficients $a_{l m}$ are given by the relation
\begin{equation}
a_{l m}=\int_{0}^{2 \pi} d \varphi \int_{0}^{\pi} d \vartheta \sin \vartheta Y_{l m}^{*}(\vartheta, \varphi) f(\vartheta, \varphi) \label{eq:5:6:5}
\end{equation}
This relation may be treated as an integral transformation of $f(\vartheta, \varphi)$ from the continuous variables $\vartheta, \varphi$ to the discrete variables $l, m$. In this case $Y_{l m}(\vartheta, \varphi) \equiv\braket{\vartheta, \varphi}{l m}$ plays the role of the transformation matrix
\begin{equation}
\braket{l m}{f}=\braket{l m}{\vartheta \varphi}\braket{\vartheta \varphi}{f} \label{eq:5:6:6}
\end{equation}
where
\[
\braket{l m}{f} \equiv a_{l m}, \quad\braket{l m}{\vartheta, \varphi} \equiv Y_{l m}^{*}(\vartheta, \varphi), \quad\braket{\vartheta, \varphi}{f} \equiv f(\vartheta, \varphi)
\]
As usual, in relations of this type, summation or integration is assumed over all variables which are repeated twice.

The transformation \eqref{eq:5:6:5} is unitary, i.e.,
\begin{equation}
\braket{f}{\operatorname{lm}}\braket{\operatorname{lm}}{f}=\braket{f}{\vartheta, \varphi}\braket{\vartheta, \varphi}{f} . \label{eq:5:6:7}
\end{equation}
The expansion coefficients $a_{l m}$ satisfy the Parseval condition\index{Parseval condition}
\begin{equation}
\sum_{l=0}^{\infty} \sum_{m=-l}^{l}\left|a_{l m}\right|^{2}=\int_{0}^{2 \pi} d \varphi \int_{0}^{\pi} d \vartheta \sin \vartheta|f(\vartheta, \varphi)|^{2} \label{eq:5:6:8}
\end{equation}
The expansion \eqref{eq:5:6:4} in terms of the spherical harmonics is widely used in different branches of physics. It is called the multipole expansion, and the $a_{l m}$ are called multipole moments.\index{multipole expansion}\index{multipole moments}\isym{alm@$a_{l m}$, multipole moment}

\subsection{Expansion of Products of the Spherical Harmonics}\label{sec:5.6.2}\index{spherical harmonics!expansion of products of}
\begin{enumerate}[label=(\alph*)]
\item A direct product of two spherical harmonics of the same arguments may be expanded in series as (the so-called Clebsch-Gordan series)\index{Clebsch-Gordan series}\index{Clebsch-Gordan coefficient}
\begin{equation}
Y_{l_{1} m_{1}}(\vartheta, \varphi) Y_{l_{2} m_{2}}(\vartheta, \varphi)=\sum_{L M} \sqrt{\frac{\left(2 l_{1}+1\right)\left(2 l_{2}+1\right)}{4 \pi(2 L+1)}} \clebsch{l_{1}}{0}{l_{2}}{0}{L}{0} \clebsch{l_{1}}{m_{1}}{l_{2}}{m_{2}}{L}{M} Y_{L M}(\vartheta, \varphi) . \label{eq:5:6:9}
\end{equation}
The inverse relation may be written as
\begin{equation}
\sqrt{\frac{4 \pi(2 L+1)}{\left(2 l_{1}+1\right)\left(2 l_{2}+1\right)}} \sum_{m_{1} m_{2}} \clebsch{l_{1}}{m_{1}}{l_{2}}{m_{2}}{L}{M} Y_{l_{1} m_{1}}(\vartheta, \varphi) Y_{l_{2} m_{2}}(\vartheta, \varphi)=\clebsch{l_{1}}{0}{l_{2}}{0}{L}{0} Y_{L M}(\vartheta, \varphi) . \label{eq:5:6:10}
\end{equation}
Products of three and more spherical harmonics are decomposed according to
\begin{gather}
Y_{l_{1} m_{1}}(\vartheta, \varphi) Y_{l_{2} m_{2}}(\vartheta, \varphi) Y_{l_{3} m_{3}}(\vartheta, \varphi)=\sum_{L M L^{\prime} M^{\prime}} \sqrt{\frac{\left(2 l_{1}+1\right)\left(2 l_{2}+1\right)\left(2 l_{3}+1\right)}{(4 \pi)^{2}(2 L+1)}}  \notag\\
\times \clebsch{l_{1}}{0}{l_{2}}{0}{L^{\prime}}{0} \clebsch{L^{\prime}}{0}{l_{3}}{0}{L}{0} \clebsch{l_{1}}{m_{1}}{l_{2}}{m_{2}}{L^{\prime}}{M^{\prime}} \clebsch{L^{\prime}}{M^{\prime}}{l_{3}}{m_{3}}{L}{M} Y_{L M}(\vartheta, \varphi)  \label{eq:5:6:11}\\
Y_{l_{1} m_{1}}(\vartheta, \varphi) Y_{l_{2} m_{2}}(\vartheta, \varphi) \times \ldots \times Y_{l_{n} m_{n}}(\vartheta, \varphi)=\sum_{L_{n}} B_{L_{n}} Y_{L_{n} M_{n}}(\vartheta, \varphi) \label{eq:5:6:12}
\end{gather}
where
\begin{equation}
B_{L_{n}}=\sqrt{\frac{4 \pi}{2 L_{n}+1}} \sum_{\substack{L_{1} L_{2} \ldots L_{n-1} \\ M_{1} M_{2}, \ldots M_{n-1}}} \prod_{i=1}^{n}\left(\sqrt{\frac{2 l_{i}+1}{4 \pi}} \clebsch{L_{i-1}}{0}{l_{i}}{0}{L_{i}}{0} \clebsch{L_{i-1}}{M_{i-1}}{l_{i}}{m_{i}}{L_{i}}{M_{i}}\right) \label{eq:5:6:13}
\end{equation}
with $L_{0} \equiv M_{0} \equiv 0$.\\
\item Irreducible tensor products (defined in Sec.~\ref{sec:3.1}) of the spherical harmonics may be expanded as\isym{Ybold@$\vect{Y}_{l}(\vartheta, \varphi)$, spherical harmonic as an irreducible tensor}
\begin{gather}
\left\{\vect{Y}_{l_{1}}(\vartheta, \varphi) \otimes \vect{Y}_{l_{2}}(\vartheta, \varphi)\right\}_{L M}=\sqrt{\frac{\left(2 l_{1}+1\right)\left(2 l_{2}+1\right)}{4 \pi(2 L+1)}} \clebsch{l_{1}}{0}{l_{2}}{0}{L}{0} Y_{L M}(\vartheta, \varphi)  \label{eq:5:6:14}\\
\left\{\left\{\vect{Y}_{l_{1}}(\vartheta, \varphi) \otimes \vect{Y}_{l_{2}}(\vartheta, \varphi)\right\}_{L^{\prime}} \otimes \vect{Y}_{l_{3}}(\vartheta, \varphi)\right\}_{L M}=\sqrt{\frac{\left(2 l_{1}+1\right)\left(2 l_{2}+1\right)\left(2 l_{3}+1\right)}{(4 \pi)^{2}(2 L+1)}} \clebsch{l_{1}}{0}{l_{2}}{0}{L^{\prime}}{0} \clebsch{L^{\prime}}{0}{l_{3}}{0}{L}{0} Y_{L M}(\vartheta, \varphi)  \label{eq:5:6:15}\\
\left\{\ldots\left\{\left\{\vect{Y}_{l_{1}}(\vartheta, \varphi) \otimes \vect{Y}_{l_{2}}(\vartheta, \varphi)\right\}_{L_{2}} \otimes \vect{Y}_{l_{3}}(\vartheta, \varphi)\right\}_{L_{2}} \ldots \otimes \vect{Y}_{l_{n}}(\vartheta, \varphi)\right\}_{L_{n} M_{n}}  \notag\\
=\sqrt{\frac{4 \pi}{2 L_{n}+1}} \prod_{i=1}^{n}\left(\sqrt{\frac{2 l_{i}+1}{4 \pi}} \clebsch{L_{i-1}}{0}{l_{i}}{0}{L_{i}}{0}\right) Y_{L_{n} M_{n}}(\vartheta, \varphi) \label{eq:5:6:16}
\end{gather}
In particular, when $l_{1}=l_{2}=\ldots=l_{n}=1$ and $L_{k}=k(k=1,2,3, \ldots, n)$ one has
\begin{equation}
\left\{\ldots\left\{\left\{\vect{Y}_{1}(\vartheta, \varphi) \otimes \vect{Y}_{1}(\vartheta, \varphi)\right\}_{2} \otimes \vect{Y}_{1}(\vartheta, \varphi)\right\}_{3} \ldots \otimes \vect{Y}_{1}(\vartheta, \varphi)\right\}_{n m}=\sqrt{\frac{3^{n}}{(4 \pi)^{n-1}} \frac{n!}{(2 n+1)!!}} Y_{n m}(\vartheta, \varphi) \label{eq:5:6:17}
\end{equation}
Note that when expanding the products of the spherical harmonics, the functions $C_{l m}(\vartheta, \varphi)$ which contain an additional factor $\sqrt{4 \pi /(2 l+1)}$ (Sec.~\ref{sec:5.1.4}), are more convenient.
\end{enumerate}

\section{Recursion relations}\label{sec:5.7}\index{spherical harmonics!recursion relations}\index{recursion relations!for the spherical harmonics}
Recursion relations for the spherical harmonics may be obtained, for example, from the Clebsch-Gordan series (see Eq. \eqref{eq:5:6:9}).
\begin{align}
-2 m \cot \vartheta Y_{l m}(\vartheta, \varphi) & =\sqrt{l(l+1)-m(m+1)} e^{-i \varphi} Y_{l m+1}(\vartheta, \varphi)+\sqrt{l(l+1)-m(m-1)} e^{i \varphi} Y_{l m-1}(\vartheta, \varphi)  \label{eq:5:7:1}\\
\cos \vartheta Y_{l m}(\vartheta, \varphi) & =\sqrt{\frac{(l-m+1)(l+m+1)}{(2 l+1)(2 l+3)}} Y_{l+1 m}(\vartheta, \varphi)+\sqrt{\frac{(l-m)(l+m)}{(2 l-1)(2 l+1)}} Y_{l-1 m}(\vartheta, \varphi)  \label{eq:5:7:2}\\
\sin \vartheta Y_{l m}(\vartheta, \varphi) e^{-i \varphi} & =\sqrt{\frac{(l-m+1)(l-m+2)}{(2 l+1)(2 l+3)}} Y_{l+1 m-1}(\vartheta, \varphi)-\sqrt{\frac{(l+m-1)(l+m)}{(2 l-1)(2 l+1)}} Y_{l-1 m-1}(\vartheta, \varphi)  \notag\\
& =\cos \vartheta \sqrt{\frac{l-m+1}{l+m}} Y_{l m-1}(\vartheta, \varphi)-\sqrt{\frac{2 l+1}{2 l-1} \, \frac{l+m-1}{l+m}} Y_{l-1 m-1}(\vartheta, \varphi)  \notag\\
& =-\cos \vartheta \sqrt{\frac{l+m}{l-m+1}} Y_{l m-1}(\vartheta, \varphi)+\sqrt{\frac{2 l+1}{2 l+3} \, \frac{l-m+2}{l-m+1}} Y_{l+1 m-1}(\vartheta, \varphi) \label{eq:5:7:3}
\end{align}
\begin{align}
\sin \vartheta Y_{l m}(\vartheta, \varphi) e^{i \varphi} & =-\sqrt{\frac{(l+m+1)(l+m+2)}{(2 l+1)(2 l+3)}} Y_{l+1 m+1}(\vartheta, \varphi)+\sqrt{\frac{(l-m-1)(l-m)}{(2 l-1)(2 l+1)}} Y_{l-1 m+1}(\vartheta, \varphi)  \notag\\
& =-\cos \vartheta \sqrt{\frac{l+m+1}{l-m}} Y_{l m+1}(\vartheta, \varphi)+\sqrt{\frac{2 l+1}{2 l-1} \, \frac{l-m-1}{l-m}} Y_{l-1 m+1}(\vartheta, \varphi)  \notag\\
& =\cos \vartheta \sqrt{\frac{l-m}{l+m+1}} Y_{l m+1}(\vartheta, \varphi)-\sqrt{\frac{2 l+1}{2 l+3} \, \frac{l+m+2}{l+m+1}} Y_{l+1 m+1}(\vartheta, \varphi) \label{eq:5:7:4}
\end{align}
\begin{align}
& (2 l-1)(2 l+3) \cos ^{2} \vartheta Y_{l m}(\vartheta, \varphi)=(2 l-1) \sqrt{\frac{(l+1)^{2}-m^{2}}{2 l+1} \, \frac{(l+2)^{2}-m^{2}}{2 l+5}} Y_{l+2 m}(\vartheta, \varphi)  \notag\\
& \quad+\left[2 l(l+1)-2 m^{2}-1\right] Y_{l m}(\vartheta, \varphi)+(2 l+3) \sqrt{\frac{l^{2}-m^{2}}{2 l+1} \, \frac{(l-1)^{2}-m^{2}}{2 l-3}} Y_{l-2 m}(\vartheta, \varphi) \label{eq:5:7:5}
\end{align}
\begin{align}
& (2 l-1)(2 l+3) \sin \vartheta \cos \vartheta e^{i \varphi} Y_{l m}(\vartheta, \varphi)=-(2 l-1) \sqrt{\frac{(l+1)^{2}-m^{2}}{2 l+1} \, \frac{(l+m+2)(l+m+3)}{2 l+5}} Y_{l+2 m+1}(\vartheta, \varphi)  \notag\\
& -(2 m+1) \sqrt{l(l+1)-m(m+1)} Y_{l m+1}(\vartheta, \varphi)+(2 l+3) \sqrt{\frac{l^{2}-m^{2}}{2 l+1} \, \frac{(l-m-1)(l-m-2)}{2 l-3}} Y_{l-2 m+1}(\vartheta, \varphi) \label{eq:5:7:6}
\end{align}
\begin{align}
& (2 l-1)(2 l+3) \sin \vartheta \cos \vartheta e^{-i \varphi} Y_{l m}(\vartheta, \varphi)=(2 l-1) \sqrt{\frac{(l+1)^{2}-m^{2}}{2 l+1} \, \frac{(l-m+2)(l-m+3)}{2 l+5}} Y_{l+2 m-1}(\vartheta, \varphi)  \notag\\
& -(2 m-1) \sqrt{l(l+1)-m(m-1)} Y_{l m-1}(\vartheta, \varphi)-(2 l+3) \sqrt{\frac{l^{2}-m^{2}}{2 l+1} \, \frac{(l+m-1)(l+m-2)}{(2 l-3)}} Y_{l-2 m-1}(\vartheta, \varphi) \label{eq:5:7:7}
\end{align}
\begin{align}
& (2 l-1)(2 l+3) \sin ^{2} \vartheta e^{i 2 \varphi} Y_{l m}(\vartheta, \varphi)=\frac{(2 l-1)}{\sqrt{(2 l+1)(2 l+5)}} \sqrt{\frac{(l+m+4)!}{(l+m)!}} Y_{l+2 m+2}(\vartheta, \varphi)  \notag\\
+ & \frac{(2 l+3)}{\sqrt{(2 l+1)(2 l-3)}} \sqrt{\frac{(l-m)!}{(l-m-4)!}} Y_{l-2 m+2}(\vartheta, \varphi)-2 \sqrt{\frac{(l+m+2)!(l-m)!}{(l-m-2)!(l+m)!}} Y_{l m+2}(\vartheta, \varphi) \label{eq:5:7:8}
\end{align}
\begin{align}
& (2 l-1)(2 l+3) \sin ^{2} \vartheta e^{-i 2 \varphi} Y_{l m}(\vartheta, \varphi)=\frac{(2 l-1)}{\sqrt{(2 l+1)(2 l+5)}} \sqrt{\frac{(l-m+4)!}{(l-m)!}} Y_{l+2 m-2}(\vartheta, \varphi)  \notag\\
& +\frac{(2 l+3)}{\sqrt{(2 l+1)(2 l-3)}} \sqrt{\frac{(l+m)!}{(l+m-4)!}} Y_{l-2 m-2}(\vartheta, \varphi)-2 \sqrt{\frac{(l-m+2)!(l+m)!}{(l+m-2)!(l-m)!}} Y_{l m-2}(\vartheta, \varphi) \label{eq:5:7:9}
\end{align}
\section{Differential relations}\label{sec:5.8}\index{spherical harmonics!differential relations}\index{differential relations!for the spherical harmonics}
\subsection{Action of the Operator of Orbital Angular Momentum on $Y_{l m}(\vartheta, \varphi)$}\label{sec:5.8.1}\index{orbital angular momentum operator}\index{angular momentum operator!orbital}
Spherical components of the operator $\overline{\vect{L}}$ (Eq. $\eqref{eq:2:2:18}$ ) act on $\vect{Y}_{l m}(\vartheta, \varphi)$ according to
\begin{align}
\widehat{L}_{ \pm 1} Y_{l m}(\vartheta, \varphi) & =\mp \sqrt{\frac{l(l+1)-m(m \pm 1)}{2}} Y_{l m \pm 1}(\vartheta, \varphi)  \label{eq:5:8:1}\\
\widehat{L}_{0} Y_{l m}(\vartheta, \varphi) & =m Y_{l m}(\vartheta, \varphi) \notag
\end{align}
or in a more compact form
\begin{equation}
\widehat{L}_{\mu} Y_{l m}(\vartheta, \varphi)=\sqrt{l(l+1)} \clebsch{l}{m}{1}{\mu}{l}{m+\mu} Y_{l m+\mu}(\vartheta, \varphi) \label{eq:5:8:2}
\end{equation}
The action of the operator $\hat{\vect{L}}^{2}$ on $Y_{l m}(\vartheta, \varphi)$ yields
\begin{equation}
\hat{\vect{L}}^{2} Y_{l m}(\vartheta, \varphi)=l(l+1) Y_{l m}(\vartheta, \varphi) \label{eq:5:8:3}
\end{equation}
\subsection{First and Second Order Derivatives of $Y_{l m}(\vartheta, \varphi)$}\label{sec:5.8.2}\index{spherical harmonics!derivatives of}\index{derivatives!of the spherical harmonics}
\begin{equation}
\frac{\partial}{\partial \varphi} Y_{l m}(\vartheta, \varphi)=i m Y_{l m}(\vartheta, \varphi) \label{eq:5:8:4}
\end{equation}
\begin{align}
\frac{\partial}{\partial \vartheta} Y_{l m}(\vartheta, \varphi) & =m \cot \vartheta Y_{l m}(\vartheta, \varphi)+\sqrt{l(l+1)-m(m+1)} Y_{l m+1}(\vartheta, \varphi) e^{-i \varphi}  \notag\\
& =-m \cot \vartheta Y_{l m}(\vartheta, \varphi)-\sqrt{l(l+1)-m(m-1)} Y_{l m-1}(\vartheta, \varphi) e^{i \varphi}  \notag\\
& =\frac{1}{2} \sqrt{l(l+1)-m(m+1)} Y_{l m+1}(\vartheta, \varphi) e^{-i \varphi}-\frac{1}{2} \sqrt{l(l+1)-m(m-1)} Y_{l m-1}(\vartheta, \varphi) e^{i \varphi} \label{eq:5:8:5}
\end{align}
\begin{gather}
\sin \vartheta \frac{\partial}{\partial \vartheta} Y_{l m}(\vartheta, \varphi)=-\sin ^{2} \vartheta \frac{\partial}{\partial \cos \vartheta} Y_{l m}(\vartheta, \varphi)  \notag\\
=l \cos \vartheta Y_{l m}(\vartheta, \varphi)-\sqrt{\frac{2 l+1}{2 l-1}\left(l^{2}-m^{2}\right)} Y_{l-1 m}(\vartheta, \varphi)  \notag\\
=-(l+1) \cos \vartheta Y_{l m}(\vartheta, \varphi)+\sqrt{\frac{2 l+1}{2 l+3}\left[(l+1)^{2}-m^{2}\right]} Y_{l+1 m}(\vartheta, \varphi)  \notag\\
=l \sqrt{\frac{(l+1)^{2}-m^{2}}{(2 l+1)(2 l+3)}} Y_{l+1 m}(\vartheta, \varphi)-(l+1) \sqrt{\frac{l^{2}-m^{2}}{(2 l+1)(2 l-1)}} Y_{l-1 m}(\vartheta, \varphi)  \label{eq:5:8:6}\\
\frac{\partial^{2}}{\partial \vartheta^{2}} Y_{l m}(\vartheta, \varphi)=-\left[l(l+1)-\frac{m^{2}}{\sin ^{2} \vartheta}\right] Y_{l m}(\vartheta, \varphi)-\cot \vartheta \frac{\partial}{\partial \vartheta} Y_{l m}(\vartheta, \varphi) .  \label{eq:5:8:7}\\
\sin ^{2} \vartheta \frac{\partial^{2}}{\partial(\cos \vartheta)^{2}} Y_{l m}(\vartheta, \varphi)=2 \cos \vartheta \frac{\partial}{\partial \cos \vartheta} Y_{l m}(\vartheta, \varphi)-\left[l(l+1)-\frac{m^{2}}{\sin ^{2} \vartheta}\right] Y_{l m}(\vartheta, \varphi) . \label{eq:5:8:8}
\end{gather}
\subsection{Vector Differentiation Operations}\label{sec:5.8.3}\index{differential operations!on the spherical harmonics}\index{gradient}
\begin{enumerate}[label=(\alph*)]
\item The gradient of a function $f(r) Y_{l m}(\vartheta, \varphi)$, where $f(r)$ is an arbitrary function of $r \equiv|\vect{r}|$ can be expanded in terms of vector spherical harmonics (Sec.~\ref{sec:7.3}) as follows:\index{vector spherical harmonics}
\begin{equation}
\boldsymbol{\nabla}\left[f(r) Y_{l m}(\vartheta, \varphi)\right]=-\sqrt{\frac{l+1}{2 l+1}}\left(\frac{d f}{d r}-\frac{l}{r} f\right) \vect{Y}_{l m}^{l+1}(\vartheta, \varphi)+\sqrt{\frac{l}{2 l+1}}\left(\frac{d f}{d r}+\frac{l+1}{r} f\right) \vect{Y}_{l m}^{l-1}(\vartheta, \varphi) \label{eq:5:8:9}
\end{equation}
or in component form
\begin{align}
\nabla_{0}\left[f(r) Y_{l m}(\vartheta, \varphi)\right]= & \sqrt{\frac{(l+1)^{2}-m^{2}}{(2 l+1)(2 l+3)}}\left(\frac{d f}{d r}-\frac{l}{r} f\right) Y_{l+1 m}(\vartheta, \varphi)  \notag\\
& +\sqrt{\frac{l^{2}-m^{2}}{(2 l-1)(2 l+1)}}\left(\frac{d f}{d r}+\frac{l+1}{r} f\right) Y_{l-1 m}(\vartheta, \varphi)  \label{eq:5:8:10}\\
\nabla_{ \pm 1}\left[f(r) Y_{l m}(\vartheta, \varphi)\right]= & \sqrt{\frac{(l \pm m+1)(l \pm m+2)}{2(2 l+1)(2 l+3)}}\left(\frac{d f}{d r}-\frac{l}{r} f\right) Y_{l+1 m \pm 1}(\vartheta, \varphi)  \notag\\
& -\sqrt{\frac{(l \mp m-1)(l \mp m)}{2(2 l-1)(2 l+1)}}\left(\frac{d f}{d r}+\frac{l+1}{r} f\right) Y_{l-1 m \pm 1}(\vartheta, \varphi) \label{eq:5:8:11}
\end{align}
Here $\nabla_{0, \pm 1}$ are spherical components of the operator $\boldsymbol{\nabla}$ (Sec.~\ref{sec:1.3}).\\
Let us consider some special cases of Eq. \eqref{eq:5:8:9}. If $f(r)=j(k r)$ is a spherical Bessel function, one has,
\begin{equation}
\frac{1}{k} \boldsymbol{\nabla}\left[j_{l}(k r) Y_{l m}(\vartheta, \varphi)\right]=\sqrt{\frac{l+1}{2 l+1}} j_{l+1}(k r) \vect{Y}_{l m}^{l+1}(\vartheta, \varphi)+\sqrt{\frac{l}{2 l+1}} j_{l-1}(k r) \vect{Y}_{l m}^{l-1}(\vartheta, \varphi) \label{eq:5:8:12}
\end{equation}
Putting $f(r)=r^{l}$, one gets
\begin{equation}
\boldsymbol{\nabla}\left[r^{l} Y_{l m}(\vartheta, \varphi)\right]=\sqrt{l(2 l+1)} r^{l-1} \vect{Y}_{l m}^{l-1}(\vartheta, \varphi) \label{eq:5:8:13}
\end{equation}
If $f(r)=r^{-l-1}$, Eq. \eqref{eq:5:8:9} is reduced to
\begin{equation}
\boldsymbol{\nabla}\left[r^{-l-1} Y_{l m}(\vartheta, \varphi)\right]=\sqrt{(l+1)(2 l+1)} r^{-l-2} \vect{Y}_{l m}^{l+1}(\vartheta, \varphi) \label{eq:5:8:14}
\end{equation}
\item The divergence of a vector function $f(r) \hat{\vect{L}} Y_{l m}(\vartheta, \varphi)$ vanishes for any function $f(r)$
\begin{equation}
\boldsymbol{\nabla}\left[f(r) \hat{\mathrm{L}} Y_{l m}(\vartheta, \varphi)\right]=0 . \label{eq:5:8:15}
\end{equation}
(c) The curl of a vector function $f(r) \hat{\vect{L}} Y_{ l m}(\vartheta, \varphi)$, where $f(r)$ is an arbitrary function of $r$, is expressible in terms of a gradient of the corresponding scalar function
\begin{equation}
\boldsymbol{\nabla} \times\left[f(\mathrm{r}) \hat{\mathrm{L}} Y_{l m}(\vartheta, \varphi)\right]=i \boldsymbol{\nabla}\left[\left(r \frac{d f}{d r}+f\right) Y_{l m}(\vartheta, \varphi)\right]-i \mathrm{r}\left[\frac{1}{r^{2}} \frac{d}{d r}\left(r^{2} \frac{d f}{d r}\right)-\frac{l(l+1)}{r^{2}} f\right] Y_{l m}(\vartheta, \varphi) . \label{eq:5:8:16}
\end{equation}
In particular, if $f(r)=j_{l}(k r)$, one gets
\begin{equation}
\operatorname{curl}\left[j_{l}(k r) \hat{L} Y_{l m}(\vartheta, \varphi)\right]=i \boldsymbol{\nabla}\left[\left(k r j_{l-1}(k r)-l j_{l}(k r)\right) Y_{l m}(\vartheta, \varphi)\right]+i r k^{2} j_{l}(k r) Y_{l m}(\vartheta, \varphi) . \label{eq:5:8:17}
\end{equation}
If $f(r)=r$,
\begin{equation}
\operatorname{curl}\left[r \hat{\vect{L}} Y_{l m}(\vartheta, \varphi)\right]=i(l+1) \sqrt{l(2 l+1)} r^{l-1} \vect{Y}_{l m}^{l-1}(\vartheta, \varphi) . \label{eq:5:8:18}
\end{equation}
and if $f(r)=r^{-l-1}$,
\begin{equation}
\operatorname{curl}\left|r^{-l-1} \hat{\vect{L}} Y_{l m}(\vartheta, \varphi)\right|=-i l \sqrt{(l+1)(2 l+1)} r^{-l-2} \vect{Y}_{l m}^{l+1}(\vartheta, \varphi) . \label{eq:5:8:19}
\end{equation}
Some additional formulas for the vector differentiation of functions which contain $Y_{l m}(\theta, \varphi)$ will be given in Sec.~\ref{sec:7.3.6}.
\end{enumerate}

\section{Some integrals involving spherical harmonics}\label{sec:5.9}\index{spherical harmonics!integrals involving}\index{integrals!of the spherical harmonics}
\subsection{Integrals over Total Solid Angle}\label{sec:5.9.1}\index{integrals!over the total solid angle}\isymg{Omega-solid@$\Omega$, solid angle}
\begin{gather}
\int_{0}^{2 \pi} d \varphi \int_{0}^{\pi} d \vartheta \sin \vartheta Y_{l m}(\vartheta, \varphi)=\sqrt{4 \pi} \delta_{l 0} \delta_{m 0}  \label{eq:5:9:1}\\
\int_{0}^{2 \pi} d \varphi \int_{0}^{\pi} d \vartheta \sin \vartheta Y_{l_{1} m_{1}}(\vartheta, \varphi) Y_{l_{2} m_{2}}^{*}(\vartheta, \varphi)=\delta_{l_{1} l_{2}} \delta_{m_{1} m_{2}}  \label{eq:5:9:2}\\
\int_{0}^{2 \pi} d \varphi \int_{0}^{\pi} d \vartheta \sin \vartheta Y_{l_{1} m_{1}}(\vartheta, \varphi) Y_{l_{2} m_{2}}(\vartheta, \varphi)=(-1)^{m_{2}} \delta_{l_{1} l_{2}} \delta_{-m_{1} m_{2}}  \label{eq:5:9:3}\\
\int_{0}^{2 \pi} d \varphi \int_{0}^{\pi} d \vartheta \sin \vartheta Y_{l_{1} m_{1}}(\vartheta, \varphi) Y_{l_{2} m_{2}}(\vartheta, \varphi) Y_{l_{3} m_{3}}^{*}(\vartheta, \varphi)=\sqrt{\frac{\left(2 l_{1}+1\right)\left(2 l_{2}+1\right)}{4 \pi\left(2 l_{3}+1\right)}} \clebsch{l_{1}}{0}{l_{2}}{0}{l_{3}}{0} \clebsch{l_{1}}{m_{1}}{l_{2}}{m_{2}}{l_{3}}{m_{3}}  \label{eq:5:9:4}\\
\int_{0}^{2 \pi} d \varphi \int_{0}^{\pi} d \vartheta \sin \vartheta Y_{l_{1} m_{1}}(\vartheta, \varphi) Y_{l_{2} m_{2}}(\vartheta, \varphi) Y_{l_{3} m_{3}}(\vartheta, \varphi)  \notag\\
=\sqrt{\frac{\left(2 l_{1}+1\right)\left(2 l_{2}+1\right)\left(2 l_{3}+1\right)}{4 \pi}}\left(\begin{array}{ccc}
l_{1} & l_{2} & l_{3} \\
0 & 0 & 0
\end{array}\right)\left(\begin{array}{ccc}
l_{1} & l_{2} & l_{3} \\
m_{1} & m_{2} & m_{3}
\end{array}\right) \label{eq:5:9:5}
\end{gather}
Integrals involving products of three and more spherical harmonics can be evaluated by reducing them to integrals which involve products of a smaller number of harmonics. For this purpose it is convenient to use the Clebsch-Gordan expansion (Eq. \eqref{eq:5:6:9}).\\
\subsection{Fourier Transformations for Some Functions Which Contain $Y_{l m}(\vartheta, \varphi)$}\label{sec:5.9.2}\index{Fourier transformation}
\begin{gather}
\int_{0}^{\infty} d r r^{2} \int_{0}^{2 \pi} d \varphi \int_{0}^{\pi} d \vartheta \sin \vartheta e^{i \vect{k} \vect{r}} j_{l}(q r) Y_{l m}(\vartheta, \varphi)=2 \pi^{2} i^{l} \frac{\delta(q-k)}{q^{2}} Y_{l m}\left(\vartheta_{k}, \varphi_{k}\right)  \label{eq:5:9:6}\\
\int_{0}^{\infty} d r r^{2} \int_{0}^{2 \pi} d \varphi \int_{0}^{\pi} d \vartheta \sin \vartheta e^{i \vect{k} \vect{r}} \hat{\vect{L}}\left [j_{l}(q r) Y_{l m}(\vartheta, \varphi)\right]=2 \pi^{2} i^{l} \frac{\delta(q-k)}{q^{2}} \hat{\vect{L}}_{\vect{k}} Y_{l m}\left(\vartheta_{k}, \varphi_{k}\right)  \label{eq:5:9:7}\\
\int_{0}^{\infty} d r r^{2} \int_{0}^{2 \pi} d \varphi \int_{0}^{\pi} d \vartheta \sin \vartheta e^{i \vect{k} \vect{r}}[\boldsymbol{\nabla} \times \hat{\vect{L}}]\left(j_{l}(q r) Y_{l m}(\vartheta, \varphi)\right)=2 \pi^{2} i^{l-1} \frac{\delta(q-k)}{q^{2}}\left[\vect{k} \times \hat{\vect{L}}_{\vect{k}}\right] Y_{l m}\left(\vartheta_{k}, \varphi_{k}\right) \label{eq:5:9:8}
\end{gather}
Here $j_{l}(x)$ is a spherical Bessel function, $\vartheta_{k}$ and $\varphi_{k}$ are polar angles of the vector $\vect{k} ; \hat{\vect{L}}$ and $\hat{\vect{L}}_{\vect{k}}$ are the operators of orbital angular momentum in the coordinate and momentum representations, respectively,
\begin{equation}
\hat{\vect{L}}=-i[\vect{r} \times \boldsymbol{\nabla}], \hat{\vect{L}}_{\vect{k}}=-i\left[\vect{k} \times \boldsymbol{\nabla}_{\vect{k}}\right] . \label{eq:5:9:9}
\end{equation}
\subsection{Integrals with Respect to $\vartheta$}\label{sec:5.9.3}\index{integrals!with respect to $\vartheta$}
\begin{gather}
\int_{0}^{\pi} Y_{l m}^{*}(\vartheta, \varphi) Y_{l^{\prime} m}(\vartheta, \varphi) \sin \vartheta d \vartheta=\frac{\delta_{l l^{\prime}}}{2 \pi} .  \label{eq:5:9:10}\\
\int_{0}^{\pi} Y_{l m}^{*}(\vartheta, \varphi) Y_{l m^{\prime}}(\vartheta, \varphi) \frac{d \vartheta}{\sin \vartheta}=\frac{2 l+1}{4 \pi m} \delta_{m m^{\prime}}\left(m, m^{\prime}>0\right) .  \label{eq:5:9:11}\\
\int_{0}^{\pi / 2}(\sin \vartheta)^{m+1}(\cos \vartheta)^{n} Y_{l m}(\vartheta, \varphi) d \vartheta  \notag\\
=\sqrt{\frac{2 l+1}{4 \pi} \cdot \frac{(l+m)!}{(l-m)!}}(-1)^{m} \frac{n!e^{i m \varphi}}{(n+l+m+1)!!(n-l+m)!!},(m, n \geq 0)  \label{eq:5:9:12}\\
\int_{a}^{b}\left[\left(l_{1}-l_{2}\right)\left(l_{1}+l_{2}+1\right)-\frac{m_{1}^{2}-m_{2}^{2}}{\sin ^{2} \vartheta}\right] Y_{l_{1} m_{1}}(\vartheta, \varphi) Y_{l_{2} m_{2}}(\vartheta, \varphi) d(\cos \vartheta)  \notag\\
=\left.\sin ^{2} \vartheta\left[Y_{l_{1} m_{1}}(\vartheta, \varphi) \frac{d}{d \cos \vartheta} Y_{l_{2} m_{2}}(\vartheta, \varphi)-Y_{l_{2} m_{2}}(\vartheta, \varphi) \frac{d}{d \cos \vartheta} Y_{l_{1} m_{1}}(\vartheta, \varphi)\right]\right|_{a} ^{b}  \notag\\
% NOTE: book misprint in the closed form.  As printed the coefficients are the
% bare (l_2+m_2),(l_1+m_1) -- correct for the *un-normalized* associated Legendre
% functions P_l^m, but the harmonics here are the normalized Y_lm, for which the
% Wronskian coefficient is sqrt((2l+1)(l^2-m^2)/(2l-1)).  Corrected below;
% verified numerically (scripts/check_5_9.py) for general m_1,m_2.
=\left[\cos \vartheta\left(l_{1}-l_{2}\right) Y_{l_{1} m_{1}}(\vartheta, \varphi) Y_{l_{2} m_{2}}(\vartheta, \varphi)+\sqrt{\frac{\left(2 l_{2}+1\right)\left(l_{2}^{2}-m_{2}^{2}\right)}{2 l_{2}-1}} Y_{l_{1} m_{1}}(\vartheta, \varphi) Y_{l_{2}-1 m_{2}}(\vartheta, \varphi)\right.  \notag\\
\left.-\sqrt{\frac{\left(2 l_{1}+1\right)\left(l_{1}^{2}-m_{1}^{2}\right)}{2 l_{1}-1}} Y_{l_{1}-1 m_{1}}(\vartheta, \varphi) Y_{l_{2} m_{2}}(\vartheta, \varphi)\right]\left.\right|_{a} ^{b} . \label{eq:5:9:13}
\end{gather}
\section{Sums involving spherical harmonics}\label{sec:5.10}\index{spherical harmonics!sums involving}\index{sums!of the spherical harmonics}
\subsection{Sums over $m$ (with fixed $l$ )}\label{sec:5.10.1}\index{sums!over $m$}
\begin{gather}
\sum_{m=-l}^{l}\left|Y_{l m}(\vartheta, \varphi)\right|^{2}=\frac{2 l+1}{4 \pi},  \label{eq:5:10:1}\\
\sum_{m=-l}^{l} m\left|Y_{l m}(\vartheta, \varphi)\right|^{2}=0,  \label{eq:5:10:2}\\
\sum_{m=-l}^{l} m^{2}\left|Y_{l m}(\vartheta, \varphi)\right|^{2}=\frac{l(l+1)(2 l+1)}{8 \pi} \sin ^{2} \vartheta,  \label{eq:5:10:3}\\
\sum_{m=-l}^{l} \sqrt{\frac{\left(l^{2}-m^{2}\right)\left[(l+1)^{2}-m^{2}\right]}{(2 l-1)(2 l+3)}} Y_{l-1 m}(\vartheta, \varphi) Y_{l+1 m}^{*}(\vartheta, \varphi)=\frac{l(l+1)}{8 \pi}\left(3 \cos ^{2} \vartheta-1\right),  \label{eq:5:10:4}\\
\sum_{m=-l}^{l} \frac{(\mp i)^{m}}{\sqrt{(l-m)!(l+m)!}} Y_{l m}(\vartheta, \varphi)=\sqrt{\frac{2 l+1}{4 \pi}} \, \frac{(\cos \vartheta \pm i \sin \vartheta \cos \varphi)^{l}}{l!} . \label{eq:5:10:5}
\end{gather}

\subsection{Sums over $l$ (with fixed $m \geq 0$)}\label{sec:5.10.2}\index{sums!over $l$}

In the equations given below \cite{ref027} we will use the notation $R=\sqrt{1-2 t \cos \vartheta+t^{2}}$.
\begin{gather}
\sum_{l=m}^{n+m}(-1)^{l} \frac{t^{n-l+m}}{(n-l+m)!\sqrt{(l-m)!(l+m)!}} \sqrt{\frac{4 \pi}{2 l+1}} Y_{l m}(\vartheta, \varphi)  \notag\\
=(-1)^{n} \frac{(2 m-1)!!}{(2 m+n)!}\left(\sin \vartheta e^{i \varphi}\right)^{m}\left(1-2 t \cos \vartheta+t^{2}\right)^{\frac{n}{2}} C_{n}^{m+\frac{1}{2}}\left(\frac{\cos \vartheta-t}{\sqrt{1-2 t \cos \vartheta+t^{2}}}\right),(|t|<1),  \label{eq:5:10:6}
\end{gather}

\begin{gather}
\sum_{l=m}^{\infty} \sqrt{\frac{4 \pi}{(2 l+1)(l-m)!(l+m)!}} \, \frac{1}{l!} t^{l-m} Y_{l m}(\vartheta, \varphi)  \notag\\
=\frac{\left(-\sin \vartheta e^{i \varphi}\right)^{m}}{2^{m}(m!)^{2}} \Fnm[l]{0}{1}{}{m+1}{t \cos ^{2} \frac{\vartheta}{2}}
\Fnm[l]{0}{1}{}{m+1}{-t \sin ^{2} \frac{\vartheta}{2}},(|t|<1),  \label{eq:5:10:7}
\end{gather}

\begin{gather}
\sum_{l=m}^{\infty} \frac{(n+l-m)!t^{l-m}}{\sqrt{(l+m)!(l-m)!}} \sqrt{\frac{4 \pi}{2 l+1}} Y_{l m}(\vartheta, \varphi)  \notag\\
=\frac{n!}{2^{m} m!} \, \frac{\left(-\sin \vartheta e^{i \varphi}\right)^{m}}{(1-t \cos \vartheta)^{n+1}} F\left\{\frac{n+1}{2}, \frac{n}{2}+1 ; m+1 ;-\left(\frac{t \sin \vartheta}{1-t \cos \vartheta}\right)^{2}\right\},(|t|<1), \label{eq:5:10:8}
\end{gather}

\begin{gather}
\sum_{l=m}^{\infty} \sqrt{\frac{4 \pi}{(2 l+1)(l-m)!(l+m)!}} \, \frac{(l+m-n)!(l-m+n-1)!}{l!} t^{l-m} Y_{l m}(\vartheta, \varphi)=\frac{(2 m-n)!(n-1)!}{2^{m}(m!)^{2}}  \notag\\
\times\left(-\sin \vartheta e^{i \varphi}\right)^{m} F\left(n, 2 m+1-n ; m+1 ; \frac{1-t-R}{2}\right) F\left(n, 2 m+1-n ; m+1 ; \frac{1+t-R}{2}\right),(|t|<1),  \label{eq:5:10:9}
\end{gather}

\begin{gather}
\sum_{l=m}^{\infty} \sqrt{\frac{4 \pi(l-m)!}{(2 l+1)(l+m)!}} L_{l-m}^{2 m}(y) t^{l-m} Y_{l m}(\vartheta, \varphi)  \notag\\
=\frac{1}{2^{m} m!} \, \frac{\left(-\sin \vartheta e^{i \varphi}\right)^{m}}{R^{2 m+1}} \exp \left[-\frac{y t(\cos \vartheta-t)}{R^{2}}\right] \Fnm[l]{0}{1}{}{m+1}{-\frac{y^{2} t^{2} \sin ^{2} \vartheta}{4 R^{4}}},(|t|<1),  \label{eq:5:10:10}
\end{gather}

\begin{gather}
\sum_{l=m}^{\infty} i^{l-m} \sqrt{\frac{4 \pi(2 l+1)(l+m)!}{(l-m)!}} j_{l}(t) Y_{l m}(\vartheta, \varphi)=\left(-t \sin \vartheta e^{i \varphi}\right)^{m} e^{i t \cos \vartheta},  \label{eq:5:10:11}\\
\sum_{l=m, m+2, \ldots}^{\infty} i^{l-m} \sqrt{\frac{4 \pi(2 l+1)(l+m)!}{(l-m)!}} j_{l}(t) Y_{l m}(\vartheta, \varphi)=\left(-t \sin \vartheta e^{i \varphi}\right)^{m} \cos (t \cos \vartheta),  \label{eq:5:10:12}\\
\sum_{l=m+1, m+3, \ldots}^{\infty} i^{l-m-1} \sqrt{\frac{4 \pi(2 l+1)(l+m)!}{(l-m)!}} j_{l}(t) Y_{l m}(\vartheta, \varphi)=\left(-t \sin \vartheta e^{i \varphi}\right)^{m} \sin (t \cos \vartheta),  \label{eq:5:10:13}\\
\sum_{l=m}^{\infty} \sqrt{\frac{4 \pi(2 l+1)(l+m)!}{(l-m)!}} j_{l}(t) z_{l}(t) Y_{l m}(\vartheta, \varphi)=\left(-t \cos \frac{\vartheta}{2} e^{i \varphi}\right)^{m} z_{m}\left(2 t \sin \frac{\vartheta}{2}\right),  \label{eq:5:10:14}\\
\sum_{l=m}^{\infty} \sqrt{\frac{4 \pi(2 l+1)(l+m)!}{(l-m)!}} j_{l}(x) z_{l}(y) Y_{l m}(\vartheta, \varphi)=\left(-\frac{x y \sin \vartheta e^{i \varphi}}{\sqrt{x^{2}+y^{2}-2 x y \cos \vartheta}}\right)^{m} z_{m}\left(\sqrt{x^{2}+y^{2}-2 x y \cos \vartheta}\right),  \label{eq:5:10:15}\\
\sum_{l=m}^{\infty} i^{l-m} j_{l}(t) Y_{l m}\left(\vartheta_{1}, \varphi_{1}\right) Y_{l m}^{*}\left(\vartheta_{2}, \varphi_{2}\right)=\frac{1}{4 \pi} J_{m}\left(t \sin \vartheta_{1} \sin \vartheta_{2}\right) e^{i t \cos \vartheta_{1} \cos \vartheta_{2}} e^{i m\left(\varphi_{1}-\varphi_{2}\right)} \label{eq:5:10:16}
\end{gather}

\section{Generating functions for $Y_{l m}(\vartheta, \varphi)$}\label{sec:5.11}\index{spherical harmonics!generating functions}\index{generating functions!for the spherical harmonics}
For fixed $m$ one has (see Refs.~\cite{ref004, ref022})
\begin{equation}
\frac{1}{R^{2 m+1}}=\left\{\begin{array}{l}
\frac{(-1)^{m}}{(2 m-1)!!(\sin \vartheta)^{m}} \sum_{l=m}^{\infty} t^{l-m} \sqrt{\frac{4 \pi}{2 l+1} \, \frac{(l+m)!}{(l-m)!}} Y_{l m}(\vartheta, 0),|t|<1  \label{eq:5:11:1}\\
\frac{(-1)^{m}}{(2 m-1)!!(\sin \vartheta)^{m}} \sum_{l=m}^{\infty} \frac{1}{t^{l+m+1}} \sqrt{\frac{4 \pi}{2 l+1} \, \frac{(l+m)!}{(l-m)!}} Y_{l m}(\vartheta, 0),|t|>1
\end{array}\right.
\end{equation}
where
\[
R=\sqrt{1-2 t \cos \vartheta+t^{2}}
\]
In particular, for $m=0$ one gets
\begin{gather}
\frac{1}{R}= \begin{cases}\sum_{l=0}^{\infty} t^{l} P_{l}(\cos \vartheta), & |t|<1, \\
\sum_{l=0}^{\infty} \frac{1}{t^{l+1}} P_{l}(\cos \vartheta), & |t|>1 .\end{cases}  \label{eq:5:11:2}\\
\frac{\left[(1+R)^{2}-t^{2}\right]^{-m}}{R}=\frac{(-1)^{m}}{2^{m}(\sin \vartheta)^{m}} \sum_{l=m}^{\infty} t^{l-m} \, l! \, \sqrt{\frac{4 \pi}{(2 l+1)(l+m)!(l-m)!}} Y_{l m}(\vartheta, 0),|t|<1 . \label{eq:5:11:3}
\end{gather}
In addition, we present the following expansions in terms of $Y_{l m}(\vartheta, \varphi)$ \cite{ref004}\index{step function}\isymg{Theta-step@$\Theta(x)$, step function}
\begin{align}
& \frac{(\sin \vartheta)^{m}}{(\cos \psi-\cos \vartheta)^{m+\frac{1}{2}}} \Theta(\cos \psi-\cos \vartheta)=\frac{(-1)^{m} 2^{m}}{(2 m-1)!!} \sum_{l=m}^{\infty} \sqrt{\frac{8 \pi}{2 l+1} \, \frac{(l+m)!}{(l-m)!}} \cos \left[(2 l+1) \frac{\psi}{2}\right] Y_{l m}(\vartheta, 0)  \label{eq:5:11:4}\\
& \frac{(\sin \vartheta)^{m}}{(\cos \vartheta-\cos \psi)^{m+\frac{1}{2}}} \Theta(\cos \vartheta-\cos \psi)=\frac{(-1)^{m} 2^{m}}{(2 m-1)!!} \sum_{l=m}^{\infty} \sqrt{\frac{8 \pi}{2 l+1} \, \frac{(l+m)!}{(l-m)!}} \sin \left[(2 l+1) \frac{\psi}{2}\right] Y_{l m}(\vartheta, 0) \label{eq:5:11:5}
\end{align}
where
\begin{equation}
\Theta(x)= \begin{cases}1 & \text { if } x \geq 0,  \label{eq:5:11:6}\\ 0 & \text { if } x<0 .\end{cases}
\end{equation}
In particular, when $m=0$ one has
\begin{equation}
\frac{\Theta(\cos \psi-\cos \vartheta)}{\sqrt{\cos \psi-\cos \vartheta}} \pm i \frac{\Theta(\cos \vartheta-\cos \psi)}{\sqrt{\cos \vartheta-\cos \psi}}=\sqrt{2} \sum_{l=0}^{\infty} e^{ \pm i(2 l+1) \frac{\psi}{2}} P_{l}(\cos \vartheta) . \label{eq:5:11:7}
\end{equation}
\section{Asymptotic expressions for $Y_{l m}(\vartheta, \varphi)$}\label{sec:5.12}\index{spherical harmonics!asymptotic expressions}\index{asymptotics!of the spherical harmonics}
\subsection{$Y_{l m}(\vartheta, \varphi)$ for Large $l$}\label{sec:5.12.1}\index{asymptotics!for large $l$}
If $l \gg 1$ and $l \gg m \geq 0$, a spherical harmonic $Y_{l m}(\vartheta, \varphi)$ for $\varepsilon \leq \vartheta \leq \pi-\varepsilon(0<\varepsilon \ll 1 / l)$ and $0 \leq \varphi<2 \pi$ can be approximated by (see Ref.~\cite{ref004})
\begin{equation}
Y_{l m}(\vartheta, \varphi) \approx \frac{e^{i m \varphi}}{\pi} \frac{\cos \left[(2 l+1) \frac{\vartheta}{2}+(2 m-1) \frac{\pi}{4}\right]}{\sqrt{\sin \vartheta}}+O\left(\frac{1}{l}\right) . \label{eq:5:12:1}
\end{equation}
A more exact formula \cite{ref022} is
% NOTE: book misprint -- the printed second term has a "+" sign, which makes this
% "more exact" formula WORSE than the leading \eqref{eq:5:12:1} for m>=1 (its
% error stays O(1/l)).  With a MINUS sign the error is genuinely O(1/l^2) for all
% m (2-27x smaller than the leading form, improvement growing with m).  Corrected
% to "-" below; verified numerically (scripts/check_5_12.py).
\begin{equation}
\begin{array}{r}
Y_{l m}(\vartheta, \varphi) \approx \frac{e^{i m \varphi}}{\pi \sqrt{\sin \vartheta}}\left\{\left(1+\frac{4 m^{2}-3}{8 l}\right) \cos \left[(2 l+1) \frac{\vartheta}{2}+(2 m-1) \frac{\pi}{4}\right]\right. \\
\left.-\frac{4 m^{2}-1}{8 l \sin \vartheta} \cos \left[(2 l+3) \frac{\vartheta}{2}+(2 m-3) \frac{\pi}{4}\right]\right\}+O\left(\frac{1}{l^{2}}\right) \label{eq:5:12:2}
\end{array}
\end{equation}
The terms $O(1 / l)$ and $O\left(1 / l^{2}\right)$ depend on $m$ and $\vartheta$.\\
If $l \gg 1$ and $l \gg m \geq 0$, one also obtains
\begin{equation}
\left|Y_{l \pm m}(\vartheta, \varphi)\right|<\frac{2}{\pi}(\sin \vartheta)^{-\left(m+\frac{1}{2}\right)} . \label{eq:5:12:3}
\end{equation}
\subsection{Behaviour of $Y_{l m}(\vartheta, \varphi)$ in Neighbourhoods of $\vartheta=0, \pi$ and $\pi / 2$}\label{sec:5.12.2}\index{asymptotics!near the poles and the equator}
For $0 \leq \vartheta \leq \varepsilon(\varepsilon \ll 1)$ one has
\begin{equation}
Y_{l \pm m}(\vartheta, \varphi) \approx(\mp 1)^{m} \frac{e^{ \pm i m \varphi}}{m!} \sqrt{\frac{2 l+1}{4 \pi} \, \frac{(l+m)!}{(l-m)!}}\left(\frac{\vartheta}{2}\right)^{m}\left[1-\frac{3 l(l+1)-m(m+1)}{3(m+1)}\left(\frac{\vartheta}{2}\right)^{2}\right] . \label{eq:5:12:4}
\end{equation}
For $\pi-\varepsilon \leq \vartheta \leq \pi$, one obtains
\begin{equation}
Y_{l \pm m}(\vartheta, \varphi) \approx(-1)^{l}( \pm 1)^{m} \frac{e^{ \pm i m \varphi}}{m!} \sqrt{\frac{2 l+1}{4 \pi} \, \frac{(l+m)!}{(l-m)!}}\left(\frac{\pi-\vartheta}{2}\right)^{m}\left[1-\frac{3 l(l+1)-m(m+1)}{3(m+1)}\left(\frac{\pi-\vartheta}{2}\right)^{2}\right] . \label{eq:5:12:5}
\end{equation}
In the case when $\pi / 2-\varepsilon \leq \vartheta \leq \pi / 2+\varepsilon$
\begin{equation}
Y_{l \pm m}(\vartheta, \varphi) \approx(-1)^{\frac{l \pm m}{2}} e^{ \pm i m \varphi} \sqrt{\frac{2 l+1}{4 \pi} \, \frac{(l-m-1)!!}{(l-m)!!} \, \frac{(l+m-1)!!}{(l+m)!!}}\left[1-\frac{l(l+1)-m^{2}}{2}\left(\frac{\pi}{2}-\vartheta\right)^{2}\right] \label{eq:5:12:6}
\end{equation}
if $l+m$ is even and
\begin{gather}
Y_{l \pm m}(\vartheta, \varphi) \approx(-1)^{\frac{l \pm m-1}{2}} e^{ \pm i m \varphi} \sqrt{\frac{2 l+1}{4 \pi} \, \frac{(l-m)!!}{(l-m-1)!!} \, \frac{(l+m)!!}{(l+m-1)!!}}\left(\frac{\pi}{2}-\vartheta\right)  \notag\\
\times\left[1-\frac{l(l+1)-\left(m^{2}+1\right)}{6}\left(\frac{\pi}{2}-\vartheta\right)^{2}\right] \label{eq:5:12:7}
\end{gather}
if $l+m$ is odd.\\
If $\vartheta$ is small, $Y_{l m}(\vartheta, \varphi)$ may be approximated by the Bessel function (McDonald formula \cite{ref004})
\begin{gather}
Y_{l-m}(\vartheta, \varphi)=\sqrt{\frac{2 l+1}{4 \pi} \frac{(l+m)!}{(l-m)!}} e^{-i m \varphi}\left[\left(l+\frac{1}{2}\right) \cos \frac{\vartheta}{2}\right]^{-m}  \notag\\
\times\left\{J_{m}(\alpha)+\left(\sin \frac{\vartheta}{2}\right)^{2}\left[\frac{\alpha}{6} J_{m+3}(\alpha)-J_{m+2}(\alpha)+\frac{1}{2 \alpha} J_{m+1}(\alpha)\right]+O\left[\left(\sin \frac{\vartheta}{2}\right)^{4}\right]\right\} \label{eq:5:12:8}
\end{gather}
Here $\alpha=(2 l+1) \sin (\vartheta / 2), J_{\lambda}(\alpha)$ is the Bessel function.

\subsection{Asymptotic Expression for Fixed $m, l \rightarrow \infty, \vartheta \rightarrow 0$ and Finite $l \vartheta$}\label{sec:5.12.3}\index{asymptotics!for fixed $m$ and large $l$}
In this case
\begin{equation}
Y_{l-m}(\vartheta, \varphi) \approx \sqrt{\frac{l}{2 \pi}} e^{-i m \varphi} J_{m}(l \vartheta) \label{eq:5:12:9}
\end{equation}
\begin{figure}[tbhp]
\begin{center}
  \includegraphics[alt={},max width=0.7\textwidth]{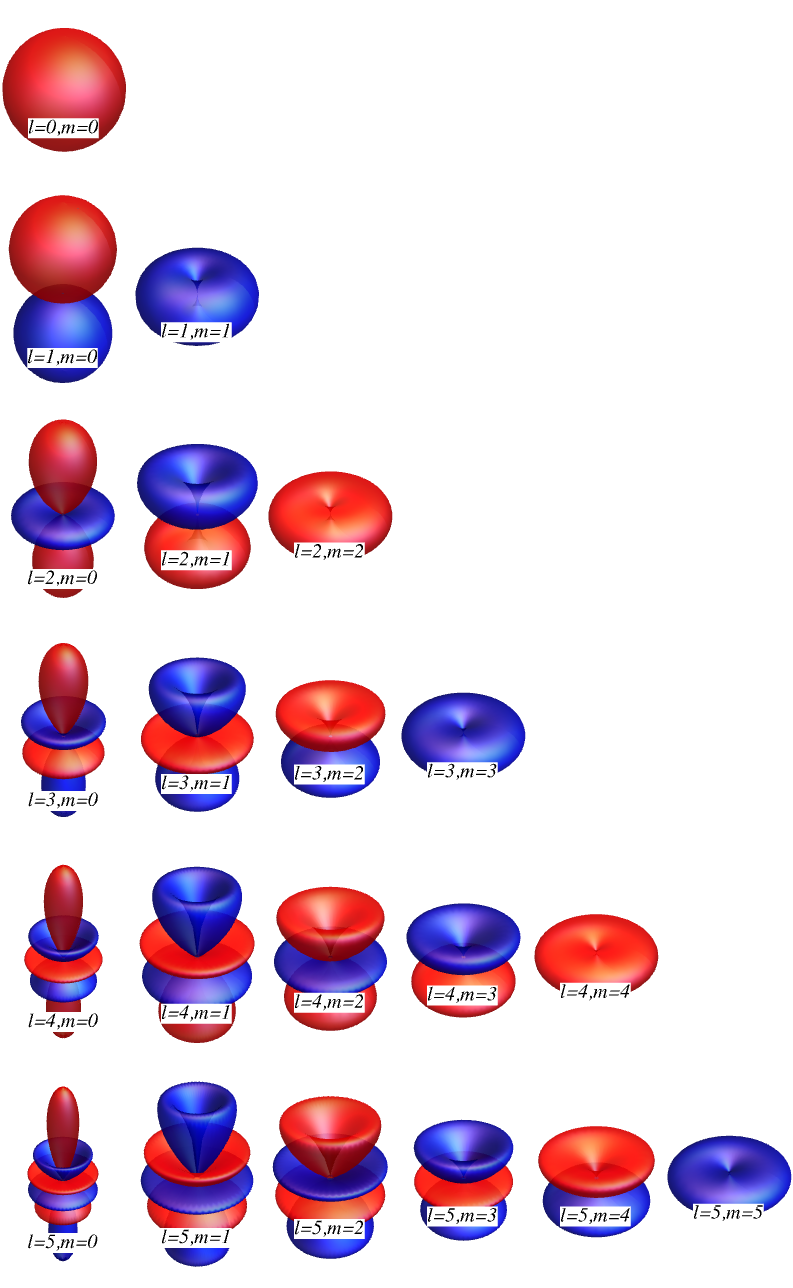}
\caption{Spherical harmonics $Y_{l m}(\vartheta, 0)$ are shown as functions of the polar angle $\vartheta$. Red  and Blue colours indicate the sign of the function (positive and negative respectively).}
\label{chap5:fig:1}
\end{center}
\end{figure}

\section{$Y_{l m}(\vartheta, \varphi)$ for special values of $l$ and $m$}\label{sec:5.13}\index{spherical harmonics!for special $l$ and $m$}
\subsection{$Y_{l m}(\vartheta, \varphi)$ for $l \leq 5$}\label{sec:5.13.1}\index{spherical harmonics!for $l \leq 5$}
Please use sympy or Mathematica to generate the explicit forms of the spherical harmonics for $l=0,1,2,3,4,5$ in terms of trigonometric functions.

\subsection{$Y_{l m}(\vartheta, \varphi)$ for $|m|=0,1,2,3,4$ and Any Integer $l$}\label{sec:5.13.2}\index{spherical harmonics!for small $m$ and any $l$}
In this case $Y_{l m}(\vartheta, \varphi)$ may be expressed in terms of the Legendre polynomials,
\begin{gather}
Y_{l 0}(\vartheta, \varphi)=\sqrt{\frac{2 l+1}{4 \pi}} P_{l}(\cos \vartheta),  \label{eq:5:13:1}\\
Y_{l \pm 1}(\vartheta, \varphi)=\mp \frac{e^{ \pm i \varphi}}{\sin \vartheta} \sqrt{\frac{l(l+1)}{4 \pi(2 l+1)}}\left[P_{l-1}(\cos \vartheta)-P_{l+1}(\cos \vartheta)\right],  \label{eq:5:13:2}\\
Y_{l \pm 2}(\vartheta, \varphi)=\frac{e^{ \pm i 2 \varphi}}{(\sin \vartheta)^{2}} \sqrt{\frac{(l-1) l(l+1)(l+2)}{4 \pi(2 l+1)}}\left[\frac{P_{l-2}(\cos \vartheta)}{2 l-1}\right.  \notag\\
\left.-\frac{2(2 l+1)}{(2 l-1)(2 l+3)} P_{l}(\cos \vartheta)+\frac{1}{2 l+3} P_{l+2}(\cos \vartheta)\right],  \label{eq:5:13:3}\\
Y_{l \pm 3}(\vartheta, \varphi)=\mp \frac{e^{ \pm i 3 \varphi}}{(\sin \vartheta)^{3}} \sqrt{\frac{1}{4 \pi(2 l+1)} \cdot \frac{(l+3)!}{(l-3)!}}\left[\frac{1}{(2 l-3)(2 l-1)} P_{l-3}(\cos \vartheta)-\frac{3}{(2 l-3)(2 l+3)} P_{l-1}(\cos \vartheta)\right.  \notag\\
\left.+\frac{3}{(2 l-1)(2 l+5)} P_{l+1}(\cos \vartheta)-\frac{1}{(2 l+3)(2 l+5)} P_{l+3}(\cos \vartheta)\right],  \label{eq:5:13:4}\\
Y_{l \pm 4}(\vartheta, \varphi)=\frac{e^{ \pm i 4 \varphi}}{(\sin \vartheta)^{4}} \sqrt{\frac{1}{4 \pi(2 l+1)} \cdot \frac{(l+4)!}{(l-4)!}}\left[\frac{1}{(2 l-5)(2 l-3)(2 l-1)} P_{l-4}(\cos \vartheta)\right.  \notag\\
-\frac{4}{(2 l-5)(2 l-1)(2 l+3)} P_{l-2}(\cos \vartheta)+\frac{6(2 l+1)}{(2 l-3)(2 l-1)(2 l+3)(2 l+5)} P_{l}(\cos \vartheta)  \notag\\
\left.-\frac{4}{(2 l-1)(2 l+3)(2 l+7)} P_{l+2}(\cos \vartheta)+\frac{1}{(2 l+3)(2 l+5)(2 l+7)} P_{l+4}(\cos \vartheta)\right] . \label{eq:5:13:5}
\end{gather}
\subsection{$Y_{l m}(\vartheta, \varphi)$ for $|m|=l, l-1, l-2, l-3, l-4, l-5$ and Any Integer $l$}\label{sec:5.13.3}\index{spherical harmonics!for $m$ near $l$}
In this case the spherical harmonics are conveniently expressed in terms of trigonometric functions.
\begin{align}
Y_{l \pm l}(\vartheta, \varphi) & =(\mp 1)^{l} e^{ \pm i l \varphi} \sqrt{\frac{(2 l+1)!!}{4 \pi(2 l)!!}}(\sin \vartheta)^{l}  \label{eq:5:13:6}\\
Y_{l \pm(l-1)}(\vartheta, \varphi) & =(\mp 1)^{l-1} e^{ \pm i(l-1) \varphi} \sqrt{\frac{(2 l+1)!!}{4 \pi(2 l-2)!!}} \cos \vartheta(\sin \vartheta)^{l-1}  \label{eq:5:13:7}\\
Y_{l \pm(l-2)}(\vartheta, \varphi) & =(\mp 1)^{l-2} e^{ \pm i(l-2) \varphi} \sqrt{\frac{2 l+1}{8 \pi} \, \frac{(2 l-3)!!}{(2 l-2)!!}}(\sin \vartheta)^{l-2}\left[(2 l-1) \cos ^{2} \vartheta-1\right]  \label{eq:5:13:8}\\
Y_{l \pm(l-3)}(\vartheta, \varphi) & =(\mp 1)^{l-3} e^{ \pm i(l-3) \varphi} \sqrt{\frac{2 l+1}{24 \pi} \, \frac{(2 l-3)!!}{(2 l-4)!!}}(\sin \vartheta)^{l-3} \cos \vartheta\left[(2 l-1) \cos ^{2} \vartheta-3\right] \label{eq:5:13:9}
\end{align}
\begin{align}
& Y_{l \pm(l-4)}(\vartheta, \varphi)=(\mp 1)^{l-4} e^{ \pm i(l-4) \varphi} \sqrt{\frac{2 l+1}{96 \pi} \, \frac{(2 l-5)!!}{(2 l-4)!!}}(\sin \vartheta)^{l-4}  \notag\\
& \times\left[(2 l-1)(2 l-3) \cos ^{4} \vartheta-6(2 l-3) \cos ^{2} \vartheta+3\right] \label{eq:5:13:10}
\end{align}
\begin{align}
& Y_{l \pm(l-5)}(\vartheta, \varphi)=(\mp 1)^{l-5} e^{ \pm i(l-5) \varphi} \sqrt{\frac{2 l+1}{480 \pi} \, \frac{(2 l-5)!!}{(2 l-6)!!}}(\sin \vartheta)^{l-5} \cos \vartheta  \notag\\
& \times\left[(2 l-1)(2 l-3) \cos ^{4} \vartheta-10(2 l-3) \cos ^{2} \vartheta+15\right] \label{eq:5:13:11}
\end{align}
\section{$Y_{l m}(\vartheta, \varphi)$ and $\frac{\partial}{\partial \vartheta} Y_{l m}(\vartheta, \varphi)$ for special $\vartheta$}\label{sec:5.14}\index{spherical harmonics!for special $\vartheta$}
\begin{enumerate}[label=(\alph*)]
\item
\begin{gather}
Y_{l m}(0, \varphi)=\delta_{m 0} \sqrt{\frac{2 l+1}{4 \pi}},  \label{eq:5:14:1}\\
Y_{l m}\left(\frac{\pi}{2}, \varphi\right)= \begin{cases}(-1)^{\frac{l+m}{2}} e^{i m \varphi} \sqrt{\frac{2 l+1}{4 \pi} \, \frac{(l+m-1)!!}{(l+m)!!} \, \frac{(l-m-1)!!}{(l-m)!!}} & \text { if } l+m \text { is even }, \\
0 & \text { if } l+m \text { is odd },\end{cases}  \label{eq:5:14:2}\\
Y_{l m}(\pi, \varphi)=\delta_{m 0}(-1)^{l} \sqrt{\frac{2 l+1}{4 \pi}},  \label{eq:5:14:3}\\
Y_{l m}( \pm \pi n, \varphi)=\delta_{m 0}(-1)^{n l} \sqrt{\frac{2 l+1}{4 \pi}} . \label{eq:5:14:4}
\end{gather}
\item
\begin{gather}
\left.\frac{\partial}{\partial \vartheta} Y_{l m}(\vartheta, \varphi)\right|_{\vartheta=0}=\left(\delta_{m-1} e^{-i \varphi}-\delta_{m 1} e^{i \varphi}\right) \sqrt{\frac{l(l+1)(2 l+1)}{16 \pi}}  \label{eq:5:14:5}\\
\left.\frac{\partial}{\partial \vartheta} Y_{l m}(\vartheta, \varphi)\right|_{\vartheta=\frac{\pi}{2}}= \begin{cases}(-1)^{\frac{l+m+1}{2}} e^{i m \varphi} \sqrt{\frac{2 l+1}{4 \pi} \, \frac{(l+m)!!(l-m)!!}{(l+m-1)!!(l-m-1)!!}} & \text { if } l+m \text { is odd } \\
0 & \text { if } l+m \text { is even }\end{cases}  \label{eq:5:14:6}\\
\left.\frac{\partial}{\partial \vartheta} Y_{l m}(\vartheta, \varphi)\right|_{\vartheta=\pi}=(-1)^{l}\left(\delta_{m-1} e^{-i \varphi}-\delta_{m 1} e^{i \varphi}\right) \sqrt{\frac{l(l+1)(2 l+1)}{16 \pi}}  \label{eq:5:14:7}\\
\left.\frac{\partial}{\partial \vartheta} Y_{l m}(\vartheta, \varphi)\right|_{\vartheta= \pm n \pi}=(-1)^{n l}\left(\delta_{m-1} e^{-i \varphi}-\delta_{m 1} e^{i \varphi}\right) \sqrt{\frac{l(l+1)(2 l+1)}{16 \pi}} \label{eq:5:14:8}
\end{gather}
\end{enumerate}
\section{Zeros of $Y_{l m}(\vartheta, \varphi)$ and $\frac{\partial}{\partial \vartheta} Y_{l m}(\vartheta, \varphi)$}\label{sec:5.15}\index{spherical harmonics!zeros of}\index{zeros!of the spherical harmonics}
In the interval $0 \leq \vartheta \leq \pi$ the spherical harmonics $Y_{l m}(\vartheta, \varphi)$ and their derivatives $\frac{\partial}{\partial \vartheta} Y_{l m}(\vartheta, \varphi)$ possess some zeros.

The number of zeros, i.e., the number of roots of the equations
\begin{equation}
Y_{l m}(\vartheta, \varphi)=0 \quad \text { and } \quad \frac{\partial}{\partial \vartheta} Y_{l m}(\vartheta, \varphi)=0 \label{eq:5:15:1}
\end{equation}
is finite in interval $0<\vartheta<\pi$. All these roots are non-degenerate. Below we denote the zeros of $Y_{l m}(\vartheta, \varphi)$ and $\frac{\partial}{\partial \vartheta} Y_{l m}(\vartheta, \varphi)$ by $\vartheta_{\alpha}$ and $\vartheta_{\beta}$, respectively.\isymg{theta-alpha-beta@$\vartheta_{\alpha}, \vartheta_{\beta}$, zeros of $Y_{l m}$ and of its derivative} Note that the functions in question have no roots with respect to $\varphi$ in the domain $0 \leq \varphi<2 \pi$.

\subsection{Zeros of $Y_{l m}(\vartheta, \varphi)$}\label{sec:5.15.1}\index{zeros!of $Y_{l m}$}
If $m=0, Y_{l m}(\vartheta, \varphi)$ has $l$ zeros in the interval $0<\vartheta<\pi$. If $m \neq 0, Y_{l m}(\vartheta, \varphi)$ has $(l-|m|)$ zeros inside the same interval of $\vartheta$ and also two zeros at $\vartheta_{\alpha}=0$ and $\vartheta_{\alpha}=\pi$. All these zeros lie symmetrically with respect to\\
$\vartheta=\pi / 2$. Hence, $Y_{l m}(\pi / 2, \varphi)=0$ provided $(l-m)$ is odd. If $m \neq 0$, the zeros of $Y_{l m}(\vartheta, \varphi)$ are determined by the equations

\begin{align}
\text{if }m & = \pm l, \quad \cos ^{2} \vartheta_{\alpha}=1 ;  \notag\\
m & = \pm(l-1) \quad \cos ^{2} \vartheta_{\alpha}=1,0 ;  \notag\\
m & = \pm(l-2) \quad \cos ^{2} \vartheta_{\alpha}=1, \frac{1}{2 l-1} ;  \notag\\
m & = \pm(l-3) \quad \cos ^{2} \vartheta_{\alpha}=1, \frac{3}{(2 l-1)}, 0 ;  \notag\\
m & = \pm(l-4) \quad \cos ^{2} \vartheta_{\alpha}=1, \frac{3 \pm 2 \sqrt{3(l-2) /(2 l-3)}}{2 l-1} ;  \notag\\
m & = \pm(l-5) \quad \cos ^{2} \vartheta_{\alpha}=1, \frac{5 \pm 2 \sqrt{5(l-3) /(2 l-3)}}{(2 l-1)}, 0 . \label{eq:5:15:2}
\end{align}
Numerical values of $\vartheta_{\alpha}$ for can be found using computer programs. 
If $l \gg 1$, approximate values of $\vartheta_{\alpha}$ may be found from the relation
\begin{equation}
\vartheta_{\alpha} \approx \frac{\pi}{2}\left(\frac{4 k+3-2 m}{2 l+1}\right), \label{eq:5:15:3}
\end{equation}
where $k$ is integer and $\frac{2|m|-3}{4} \leq k \leq \frac{4 l+2|m|-1}{4}$.

\subsection{Zeros of $\frac{\partial}{\partial \vartheta} Y_{l m}(\vartheta, \varphi)$}\label{sec:5.15.2}\index{zeros!of the derivative of $Y_{l m}$}
If $m=0, \frac{\partial}{\partial \vartheta} Y_{l m}(\vartheta, \varphi)$ has $l-1$ zeros in the interval $0<\vartheta<\pi$ and also two zeros at $\vartheta_{\beta}=0$ and $\vartheta_{\beta}=\pi$. If $m \neq 0, \frac{\partial}{\partial \vartheta} Y_{l m}(\vartheta, \varphi)$ has $(l-|m|+1)$ zeros in the same interval and two additional ones at $\vartheta_{\beta}=0$ and $\vartheta_{\beta}=\pi$ provided $m \geq 2$. All zeros of $\frac{\partial}{\partial \vartheta} Y_{l m}(\vartheta, \varphi)$ lie symmetrically with respect to $\vartheta=\pi / 2$. Hence $\frac{\partial}{\partial \vartheta} Y_{l m}(\vartheta, \varphi)=0$\\
for $\vartheta=\pi / 2$ provided $(l-m+1)$ is odd. When $m \neq 0$ the values of $\vartheta_{\beta}$ are determined by the equations

\begin{align}
\text{if }m & = \pm l \quad \cos ^{2} \vartheta_{\beta}=0 ; 1(l \geq 2) ;  \notag\\
m & = \pm(l-1) \quad \cos ^{2} \vartheta_{\beta}=\frac{1}{l} ; 1(l \geq 3) ;  \notag\\
m & = \pm(l-2) \quad \cos ^{2} \vartheta_{\beta}=\frac{5 l-4}{l(2 l-1)} ; 0 ; 1(l \geq 4) ;  \notag\\
m & = \pm(l-3) \quad \cos ^{2} \vartheta_{\beta}=\frac{9(l-1) \pm  \sqrt{3\left(19 l^{2}-50 l+27\right)}}{2 l(2 l-1)} ; 1(l \geq 5) \notag\\
m & = \pm(l-4) \quad \cos ^{2} \vartheta_{\beta}=\frac{7 l-8 \pm 2 \sqrt{\left(11 l^{3}-62 l^{2}+104 l-48\right) /(2 l-3)}}{ l(2 l-1)} ; 0 ; 1(l \geq 6) \label{eq:5:15:4}
\end{align}

%Numerical values for $\vartheta_{\beta}$ are given in Table~\ref{chap5:tab:1}.\\
It is evident that $Y_{l m}(\vartheta, \varphi)$ has an extremum at $\vartheta=\vartheta_{\beta}$. Zeros of $Y_{l m}(\vartheta, \varphi)$ and $\frac{\partial}{\partial \vartheta} Y_{l m}(\vartheta, \varphi)$ alternate in the domain $0 \leq \vartheta \leq \pi$.

All harmonics $Y_{lm}(\vartheta, 0)$ and $Y_{l+s, m+s}(\vartheta, 0)$ with $m>0$, and $m+s>0$ have the same number of maxima, minima and zeros, i.e., can be represented by topologically equivalent plots (Fig.~\ref{chap5:fig:1}). With increasing $l+s$ the positions of all maxima, minima and zeros of $Y_{l+s, m+s}(\vartheta, 0)$ shift to $\vartheta=\pi / 2$.

\section{Bipolar and tripolar spherical harmonics}\label{sec:5.16}\index{spherical harmonics!bipolar and tripolar}\index{bipolar spherical harmonics}\index{tripolar spherical harmonics}
In many applications one has to deal with functions which depend on two or three vector directions. Convenient bases for expansions of such functions are provided by the bipolar and tripolar harmonics.

\subsection{Bipolar Spherical Harmonics}\label{sec:5.16.1}\index{bipolar spherical harmonics!definition}
Bipolar spherical harmonics are given by irreducible tensor product of the spherical harmonics with different arguments\isym{Ybipolar@$\{\vect{Y}_{l_{1}}(\Omega_{1}) \otimes \vect{Y}_{l_{2}}(\Omega_{2})\}_{L M}$, bipolar spherical harmonic}
\begin{equation}
\left\{\vect{Y}_{l_{1}}\left(\vartheta_{1}, \varphi_{1}\right) \otimes \vect{Y}_{l_{2}}\left(\vartheta_{2}, \varphi_{2}\right)\right\}_{L M}=\sum_{m_{1} m_{2}} \clebsch{l_{1}}{m_{1}}{l_{2}}{m_{2}}{L}{M} Y_{l_{1} m_{1}}\left(\vartheta_{1}, \varphi_{1}\right) Y_{l_{2} m_{2}}\left(\vartheta_{2}, \varphi_{2}\right) \label{eq:5:16:1}
\end{equation}
The variety of bipolar spherical harmonics with different $l_{1}, l_{2}, L$ and $M$ forms a complete orthonormal set of functions which depend on two unit vectors, $\vect{n}_{1}$ and $\vect{n}_{2}$. These functions possess simple transformation properties under rotations and inversion of coordinate system.

The orthogonality and normalization relation for these harmonics is given by
\begin{equation}
\iint d \Omega_{1} d \Omega_{2}\left\{\vect{Y}_{l_{1}}\left(\Omega_{1}\right) \otimes \vect{Y}_{l_{2}}\left(\Omega_{2}\right)\right\}_{L M}\left\{\vect{Y}_{l_{1}^{\prime}}\left(\Omega_{1}\right) \otimes \vect{Y}_{l_{2}^{\prime}}\left(\Omega_{2}\right)\right\}_{L^{\prime} M^{\prime}}=\delta_{l_{1} l_{1}^{\prime}} \delta_{l_{2} l_{2}^{\prime}} \delta_{L L^{\prime}} \delta_{M M^{\prime}} \label{eq:5:16:2}
\end{equation}
where
\[
\Omega \equiv\{\vartheta, \varphi\}, \quad \int d \Omega \equiv \int_{0}^{2 \pi} d \varphi \int_{0}^{\pi} d \vartheta \sin \vartheta
\]
The completeness condition has the form
\begin{equation}
\sum_{l_{1} l_{2} L M}\left\{\vect{Y}_{l_{1}}\left(\Omega_{1}\right) \otimes \vect{Y}_{l_{2}}\left(\Omega_{2}\right)\right\}_{L M}\left\{\vect{Y}_{l_{1}}\left(\Omega_{1}^{\prime}\right) \otimes \vect{Y}_{l_{2}}\left(\Omega_{2}^{\prime}\right)\right\}_{L M}^{*}=\delta\left(\Omega_{1}-\Omega_{1}^{\prime}\right) \delta\left(\Omega_{2}-\Omega_{2}^{\prime}\right) \label{eq:5:16:3}
\end{equation}
where
\[
\delta\left(\Omega-\Omega^{\prime}\right) \equiv \delta\left(\cos \vartheta-\cos \vartheta^{\prime}\right) \delta\left(\varphi-\varphi^{\prime}\right) .
\]
In addition, let us note the following equation
\begin{equation}
\sum_{L M}\left|\left\{\vect{Y}_{l_{1}}\left(\Omega_{1}\right) \otimes \vect{Y}_{l_{2}}\left(\Omega_{2}\right)\right\}_{L M}\right|^{2}=\frac{\left(2 l_{1}+1\right)\left(2 l_{2}+1\right)}{(4 \pi)^{2}} \label{eq:5:16:4}
\end{equation}
Under a rotation of the coordinate system the bipolar harmonics transform according to
\begin{equation}
\hat{D}(\alpha, \beta, \gamma)\left\{\vect{Y}_{l_{1}}\left(\Omega_{1}\right) \otimes \vect{Y}_{l_{2}}\left(\Omega_{2}\right)\right\}_{L M^{\prime}}=\left\{\vect{Y}_{l_{1}}\left(\Omega_{1}^{\prime}\right) \otimes \vect{Y}_{l_{2}}\left(\Omega_{2}^{\prime}\right)\right\}_{L M^{\prime}}=\sum_{M} D_{M M^{\prime}}^{L}(\alpha, \beta, \gamma)\left\{\vect{Y}_{l_{1}}\left(\Omega_{1}\right) \otimes \vect{Y}_{l_{2}}\left(\Omega_{2}\right)\right\}_{L M} \label{eq:5:16:5}
\end{equation}
where the angles $\Omega_{i}^{\prime} \equiv\left\{\vartheta_{i}^{\prime}, \varphi_{i}^{\prime}\right\}$ and $\Omega_{i} \equiv\left\{\vartheta_{i}, \varphi_{i}\right\}$ are related by Eqs. \eqref{eq:1:4:2}, \eqref{eq:1:4:3}.\\
Under inversion of coordinate system the phases of the bipolar harmonics change according to
\begin{equation}
\widehat{P}_{r}\left\{\vect{Y}_{l_{1}}\left(\Omega_{1}\right) \otimes \vect{Y}_{l_{2}}\left(\Omega_{2}\right)\right\}_{L M}=(-1)^{l_{1}+l_{2}}\left\{\vect{Y}_{l_{1}}\left(\Omega_{1}\right) \otimes \vect{Y}_{l_{2}}\left(\Omega_{2}\right)\right\}_{L M} \label{eq:5:16:6}
\end{equation}
The Clebsch-Gordan series for product of the bipolar harmonics may be presented in the form
\begin{gather}
\left\{\vect{Y}_{l_{1}^{\prime}}\left(\Omega_{1}\right) \otimes \vect{Y}_{l_{2}^{\prime}}\left(\Omega_{2}\right)\right\}_{L^{\prime} M^{\prime}}\left\{\vect{Y}_{l_{1}^{\prime \prime}}\left(\Omega_{1}\right) \otimes \vect{Y}_{l_{2}^{\prime \prime}}\left(\Omega_{2}\right)\right\}_{L^{\prime \prime} M^{\prime \prime}}  \notag\\
=\sum_{L M} \clebsch{L^{\prime}}{M^{\prime}}{L^{\prime \prime}}{M^{\prime \prime}}{L}{M} \sum_{l_{1} l_{2}} B_{l_{1}^{\prime} l_{2}^{\prime} L^{\prime} l_{1}^{\prime \prime} l_{2}^{\prime \prime} L^{\prime \prime}}^{l_{1} l_{2} L}\left\{\vect{Y}_{l_{1}}\left(\Omega_{1}\right) \otimes \vect{Y}_{l_{2}}\left(\Omega_{2}\right)\right\}_{L M}, \label{eq:5:16:7}
\end{gather}
where
\begin{equation}
B_{l_{1}^{\prime} l_{2}^{\prime} L^{\prime} l_{1}^{\prime \prime} l_{2}^{\prime \prime} L^{\prime \prime}}^{l_{1} l_{2} L}=\sqrt{\frac{\left(2 l_{1}^{\prime}+1\right)\left(2 l_{2}^{\prime}+1\right)\left(2 l_{1}^{\prime \prime}+1\right)\left(2 l_{2}^{\prime \prime}+1\right)\left(2 L^{\prime}+1\right)\left(2 L^{\prime \prime}+1\right)}{(4 \pi)^{2}}} \clebsch{l_{1}^{\prime}}{0}{l_{1}^{\prime \prime}}{0}{l_{1}}{0} \clebsch{l_{2}^{\prime}}{0}{l_{2}^{\prime \prime}}{0}{l_{2}}{0}\left\{\begin{array}{lll}
l_{1}^{\prime} & l_{1}^{\prime \prime} & l_{1}  \label{eq:5:16:8}\\
l_{2}^{\prime} & l_{2}^{\prime \prime} & l_{2} \\
L^{\prime} & L^{\prime \prime} & L
\end{array}\right\} .
\end{equation}
For the important special case $L=0$ a bipolar harmonic is reduced to a scalar product of the spherical harmonics
\begin{equation}
\left\{\vect{Y}_{l_{1}}\left(\Omega_{1}\right) \otimes \vect{Y}_{l_{2}}\left(\Omega_{2}\right)\right\}_{00}=\frac{(-1)^{l_{1}}}{\sqrt{2 l_{1}+1}}\left(\vect{Y}_{l_{1}}\left(\Omega_{1}\right) \, \vect{Y}_{l_{1}}\left(\Omega_{2}\right)\right) \delta_{l_{1} l_{2}} \label{eq:5:16:9}
\end{equation}
In accordance with Eq. \eqref{eq:3:1:30} this scalar product is given by
\begin{equation}
\left(\vect{Y}_{l}\left(\Omega_{1}\right) \cdot \vect{Y}_{l}\left(\Omega_{2}\right)\right)=\sum_{m=-l}^{l} Y_{l m}^{*}\left(\Omega_{1}\right) Y_{l m}\left(\Omega_{2}\right) \label{eq:5:16:10}
\end{equation}
\subsection{Tripolar Spherical Harmonics}\label{sec:5.16.2}\index{tripolar spherical harmonics!definition}
A tripolar spherical harmonic is defined as an irreducible tensor product of three spherical harmonics with different arguments,
\begin{gather}
\left\{\vect{Y}_{l_{1}}\left(\vartheta_{1}, \varphi_{1}\right) \otimes\left\{\vect{Y}_{l_{2}}\left(\vartheta_{2}, \varphi_{2}\right) \otimes \vect{Y}_{l_{3}}\left(\vartheta_{3}, \varphi_{3}\right)\right\}_{l_{23}}\right\}_{L M}  \label{eq:5:16:11}\\
=\sum_{\substack{m_{1} m_{2} m_{3} \\
m_{23}}} \clebsch{l_{1}}{m_{1}}{l_{23}}{m_{23}}{L}{M} \clebsch{l_{2}}{m_{2}}{l_{3}}{m_{3}}{l_{23}}{m_{23}} Y_{l_{1} m_{1}}\left(\vartheta_{1}, \varphi_{1}\right) Y_{l_{2} m_{2}}\left(\vartheta_{2}, \varphi_{2}\right) Y_{l_{3} m_{3}}\left(\vartheta_{3}, \varphi_{3}\right) \notag
\end{gather}
In contrast to bipolar harmonics, the tripolar harmonics allow different coupling schemes of angular momenta, namely, $l_{1}+\left(l_{2}+l_{3}\right)_{l_{23}}=\vect{L},\left(l_{1}+l_{2}\right)_{l_{12}}+l_{3}=\vect{L}$ or $\left(l_{1}+l_{3}\right)_{l_{13}}+l_{2}=\vect{L}$. The relations between different coupling schemes are considered in Sec.~\ref{sec:3.3}.

The tripolar spherical harmonics form a complete orthonormal set of functions which depend on three unit vectors $n_{1}, n_{2}$ and $n_{3}$.

The orthogonality and normalization relations is as follows
\begin{gather}
\iiint d \Omega_{1} d \Omega_{2} d \Omega_{3}\left\{\vect{Y}_{l_{1}}\left(\Omega_{1}\right) \otimes\left\{\vect{Y}_{l_{2}}\left(\Omega_{2}\right) \otimes \vect{Y}_{l_{3}}\left(\Omega_{3}\right)\right\}_{\lambda}\right\}_{L M}\left\{\vect{Y}_{l_{1}^{\prime}}\left(\Omega_{1}\right) \otimes\left\{\vect{Y}_{l_{2}^{\prime}}\left(\Omega_{2}\right) \otimes \vect{Y}_{l_{3}^{\prime}}\left(\Omega_{3}\right)\right\}_{\lambda^{\prime}}\right\}_{L^{\prime} M^{\prime}}^{*}  \notag\\
=\delta_{l_{1} l_{1}^{\prime}} \delta_{l_{2} l_{2}^{\prime}} \delta_{l_{3} l_{3}^{\prime}} \delta_{\lambda \lambda^{\prime}} \delta_{L L^{\prime}} \delta_{M M^{\prime}} \label{eq:5:16:12}
\end{gather}
The completeness relation may be presented in the form
\begin{gather}
\sum_{\substack{l_{1} l_{2} l_{3} \\
\lambda L M}}\left\{\vect{Y}_{l_{1}}\left(\Omega_{1}\right) \otimes\left\{\vect{Y}_{l_{2}}\left(\Omega_{2}\right) \otimes \vect{Y}_{l_{3}}\left(\Omega_{3}\right)\right\}_{\lambda}\right\}_{L M}\left\{\vect{Y}_{l_{1}}\left(\Omega_{1}^{\prime}\right) \otimes\left\{\vect{Y}_{l_{2}}\left(\Omega_{2}^{\prime}\right) \otimes \vect{Y}_{l_{3}}\left(\Omega_{3}^{\prime}\right)\right\}_{\lambda}\right\}_{L M}^{*}  \notag\\
=\delta\left(\Omega_{1}-\Omega_{1}^{\prime}\right) \delta\left(\Omega_{2}-\Omega_{2}^{\prime}\right) \delta\left(\Omega_{3}-\Omega_{3}^{\prime}\right) \label{eq:5:16:13}
\end{gather}
In addition we also have the following equation
\begin{equation}
\sum_{\lambda L M}\left|\left\{\vect{Y}_{l_{1}}\left(\Omega_{1}\right) \otimes\left\{\vect{Y}_{l_{2}}\left(\Omega_{2}\right) \otimes \vect{Y}_{l_{3}}\left(\Omega_{3}\right)\right\}_{\lambda}\right\}_{L M}\right|^{2}=\frac{\left(2 l_{1}+1\right)\left(2 l_{2}+1\right)\left(2 l_{3}+1\right)}{(4 \pi)^{3}} \label{eq:5:16:14}
\end{equation}
Under a rotation of the coordinate system the tripolar harmonics transform according to
\begin{gather}
\hat{D}(\alpha, \beta, \gamma)\left\{\vect{Y}_{l_{1}}\left(\Omega_{1}\right) \otimes\left\{\vect{Y}_{l_{2}}\left(\Omega_{2}\right) \otimes \vect{Y}_{l_{3}}\left(\Omega_{3}\right)\right\}_{\lambda}\right\}_{L M^{\prime}}=\left\{\vect{Y}_{l_{1}}\left(\Omega_{1}^{\prime}\right) \otimes\left\{\vect{Y}_{l_{2}}\left(\Omega_{2}^{\prime}\right) \otimes \vect{Y}_{l_{3}}\left(\Omega_{3}^{\prime}\right)\right\}_{\lambda}\right\}_{L M^{\prime}}  \notag\\
=\sum_{M} D_{M M^{\prime}}^{L}(\alpha, \beta, \gamma)\left\{\vect{Y}_{l_{1}}\left(\Omega_{1}\right) \otimes\left\{\vect{Y}_{l_{2}}\left(\Omega_{2}\right) \otimes \vect{Y}_{l_{3}}\left(\Omega_{3}\right)\right\}_{\lambda}\right\}_{L M} \label{eq:5:16:15}
\end{gather}
where the angles $\Omega_{i}^{\prime} \equiv\left\{\vartheta_{i}^{\prime}, \varphi_{i}^{\prime}\right\}$ and $\Omega_{i} \equiv\left\{\vartheta_{i}, \varphi_{i}\right\}$ are related through Eqs. \eqref{eq:1:4:2}, \eqref{eq:1:4:3}.\\
Under inversion of the coordinate system the tripolar harmonics acquire the phase factor
\begin{equation}
\widehat{P}_{r}\left\{\vect{Y}_{l_{1}}\left(\Omega_{1}\right) \otimes\left\{\vect{Y}_{l_{2}}\left(\Omega_{2}\right) \otimes \vect{Y}_{l_{3}}\left(\Omega_{3}\right)\right\}_{\lambda}\right\}_{L M}=(-1)^{l_{1}+l_{2}+l_{3}}\left\{\vect{Y}_{l_{1}}\left(\Omega_{1}\right) \otimes\left\{\vect{Y}_{l_{2}}\left(\Omega_{2}\right) \otimes \vect{Y}_{l_{3}}\left(\Omega_{3}\right)\right\}_{\lambda}\right\}_{L M} \label{eq:5:16:16}
\end{equation}
The Clebsch-Gordan series for the tripolar harmonics is given by
\begin{align}
& \left\{\vect{Y}_{l_{1}^{\prime}}\left(\Omega_{1}\right) \otimes\left\{\vect{Y}_{l_{2}^{\prime}}\left(\Omega_{2}\right) \otimes \vect{Y}_{l_{3}^{\prime}}\left(\Omega_{3}\right)\right\}_{\lambda^{\prime}}\right\}_{L^{\prime} M^{\prime}}\left\{\vect{Y}_{l_{1}^{\prime \prime}}\left(\Omega_{1}\right) \otimes\left\{\vect{Y}_{l_{2}^{\prime \prime}}\left(\Omega_{2}\right) \otimes \vect{Y}_{l_{3}^{\prime \prime}}\left(\Omega_{3}\right)\right\}_{\lambda^{\prime \prime}}\right\}_{L^{\prime \prime} M^{\prime \prime}}  \notag\\
& =\sum_{L M} \clebsch{L^{\prime}}{M^{\prime}}{L^{\prime \prime}}{M^{\prime \prime}}{L}{M} \sum_{l_{1} l_{2} l_{3} \lambda} B_{l_{1}^{\prime} l_{2}^{\prime} l_{3}^{\prime} \lambda^{\prime} L^{\prime} l_{1}^{\prime \prime} l_{2}^{\prime \prime} l_{3}^{\prime \prime} \lambda^{\prime \prime} L^{\prime \prime}}^{l_{1} l_{2} l_{3} \lambda L}\left\{\vect{Y}_{l_{1}}\left(\Omega_{1}\right) \otimes\left\{\vect{Y}_{l_{2}}\left(\Omega_{2}\right) \otimes \vect{Y}_{l_{3}}\left(\Omega_{3}\right)\right\}_{\lambda}\right\}_{L M} \label{eq:5:16:17}
\end{align}
where
\begin{gather}
B_{l_{1}^{\prime} l_{2}^{\prime} l_{3}^{\prime} \lambda^{\prime} L^{\prime} l_{1}^{\prime \prime} l_{2}^{\prime \prime} l_{3}^{\prime \prime} \lambda^{\prime \prime} L^{\prime \prime}}^{l_{1} l_{2} l_{3} \lambda L}=\sqrt{\frac{\left(2 l_{1}^{\prime}+1\right)\left(2 l_{1}^{\prime \prime}+1\right)\left(2 l_{2}^{\prime}+1\right)\left(2 l_{2}^{\prime \prime}+1\right)\left(2 l_{3}^{\prime \prime}+1\right)\left(2 l_{3}^{\prime}+1\right)\left(2 L^{\prime}+1\right)\left(2 L^{\prime \prime}+1\right)}{(4 \pi)^{3}}}  \notag\\
\times \sqrt{\left(2 \lambda^{\prime}+1\right)\left(2 \lambda^{\prime \prime}+1\right)(2 \lambda+1)} \clebsch{l_{1}^{\prime}}{0}{l_{1}^{\prime \prime}}{0}{l_{1}}{0} \clebsch{l_{2}^{\prime}}{0}{l_{2}^{\prime \prime}}{0}{l_{2}}{0} \clebsch{l_{3}^{\prime}}{0}{l_{3}^{\prime \prime}}{0}{l_{3}}{0}\left\{\begin{array}{ccc}
l_{1}^{\prime} & l_{1}^{\prime \prime} & l_{1} \\
\lambda^{\prime} & \lambda^{\prime \prime} & \lambda \\
L^{\prime} & L^{\prime \prime} & L
\end{array}\right\}\left\{\begin{array}{ccc}
l_{2}^{\prime} & l_{2}^{\prime \prime} & l_{2} \\
l_{3}^{\prime} & l_{3}^{\prime \prime} & l_{3} \\
\lambda^{\prime} & \lambda^{\prime \prime} & \lambda
\end{array}\right\} \label{eq:5:16:18}
\end{gather}
Putting $L=0$ in Eq. \eqref{eq:5:16:11}, we get an important special case for the tripolar scalar harmonic:
\begin{gather}
\left\{\mathrm{Y}_{l_{1}}\left(\Omega_{1}\right) \otimes\left\{\mathrm{Y}_{l_{2}}\left(\Omega_{2}\right) \otimes \mathrm{Y}_{l_{3}}\left(\Omega_{3}\right)\right\}_{\lambda}\right\}_{00}  \notag\\
=(-1)^{l_{1}+l_{2}+l_{3}} \delta_{\lambda l_{1}} \sum_{m_{1} m_{2} m_{3}}\left(\begin{array}{ccc}
l_{1} & l_{2} & l_{3} \\
m_{1} & m_{2} & m_{3}
\end{array}\right) Y_{l_{1} m_{1}}\left(\Omega_{1}\right) Y_{l_{2} m_{2}}\left(\Omega_{2}\right) Y_{l_{3} m_{3}}\left(\Omega_{3}\right) \label{eq:5:16:19}
\end{gather}
\section{Expansions of functions which depend on two vectors}\label{sec:5.17}\index{expansion!of functions of two vectors}
\subsection{Preliminary Remarks}\label{sec:5.17.1}
\begin{figure}[tbh]
\begin{center}
  \includegraphics[alt={}]{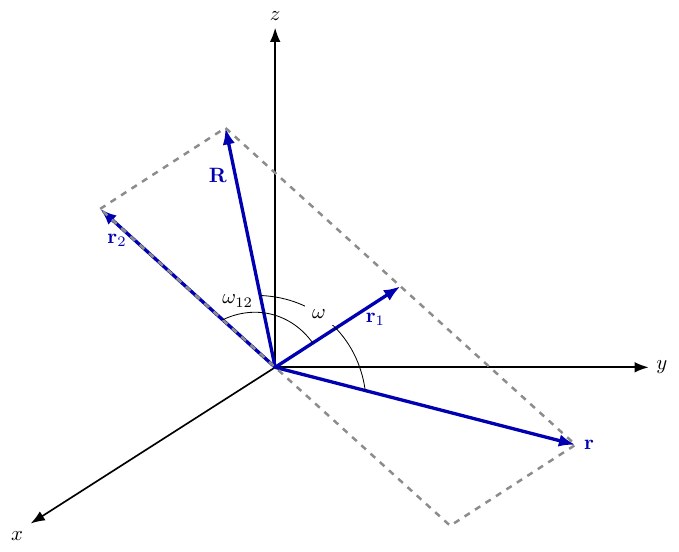}
\caption{Relations between the vectors $\vect{R}, \vect{r}$ and $\vect{r}_{1}, \vect{r}_{2}$.}
\label{chap5:fig:2}
\end{center}
\end{figure}

\index{expansion!in the bipolar harmonics}
Any function $f\left(\vect{r}_{1}, \vect{r}_{2}\right)$ which depends on two arbitrary vectors $\vect{r}_{1}\left(r_{1}, \vartheta_{1}, \varphi_{1}\right)$ and $\vect{r}_{2}\left(r_{2}, \vartheta_{2}, \varphi_{2}\right)$ may be expanded into a series of the bipolar harmonics (Sec.~\ref{sec:5.16}):
\begin{equation}
f\left(\vect{r}_{1}, \vect{r}_{2}\right)=\sum_{l_{1} l_{2} L M} C_{l_{1} l_{2}}^{L M}\left(r_{1}, r_{2}\right)\left\{\vect{Y}_{l_{1}}\left(\Omega_{1}\right) \otimes \vect{Y}_{l_{2}}\left(\Omega_{2}\right)\right\}_{L M} \label{eq:5:17:1}
\end{equation}
The function $f\left(\vect{r}_{1}, \vect{r}_{2}\right)$ may be presented in terms of $\vect{r}_{1}$ and $\vect{r}_{2}$ as well as in terms of $\vect{R}$ and $\vect{r}$, where
\begin{equation}
\vect{R}=\vect{r}_{1}+\vect{r}_{2}, \vect{r}=\vect{r}_{1}-\vect{r}_{2}, \vect{r}_{1}=\frac{1}{2}(\vect{R}+\vect{r}), \vect{r}_{2}=\frac{1}{2}(\vect{R}-\vect{r}) \label{eq:5:17:2}
\end{equation}
Polar coordinates of the vectors $\vect{R}(R, \Theta, \Phi)$ and $\vect{r}(r, \vartheta, \varphi)$ are related to polar coordinates of the vectors $\vect{r}_{1}\left(r_{1}, \vartheta_{1}, \varphi_{1}\right)$ and $\vect{r}_{2}\left(r_{2}, \vartheta_{2}, \varphi_{2}\right)$ by
\begin{align}
R^{2} & =r_{1}^{2}+r_{2}^{2}+2 r_{1} r_{2} \cos \omega_{12},  \notag\\
\cos \Theta & =\frac{r_{1} \cos \vartheta_{1}+r_{2} \cos \vartheta_{2}}{\sqrt{r_{1}^{2}+r_{2}^{2}+2 r_{1} r_{2} \cos \omega_{12}}},  \notag\\
\tan \Phi & =\frac{r_{1} \sin \vartheta_{1} \sin \varphi_{1}+r_{2} \sin \vartheta_{2} \sin \varphi_{2}}{r_{1} \sin \vartheta_{1} \cos \varphi_{1}+r_{2} \sin \vartheta_{2} \cos \varphi_{2}},  \notag\\
r^{2} & =r_{1}^{2}+r_{2}^{2}-2 r_{1} r_{2} \cos \omega_{12},  \label{eq:5:17:3}\\
\cos \vartheta & =\frac{r_{1} \cos \vartheta_{1}-r_{2} \cos \vartheta_{2}}{\sqrt{r_{1}^{2}+r_{2}^{2}-2 r_{1} r_{2} \cos \omega_{12}}},  \notag\\
\tan \varphi & =\frac{r_{1} \sin \vartheta_{1} \sin \varphi_{1}-r_{2} \sin \vartheta_{2} \sin \varphi_{2}}{r_{1} \sin \vartheta_{1} \cos \varphi_{1}-r_{2} \sin \vartheta_{2} \cos \varphi_{2}} . \notag
\end{align}
Here $\omega_{12}$ is the angle between $\vect{r}_{1}$ and $\vect{r}_{2}$ (Fig.~\ref{chap5:fig:2}),\isymg{omega12@$\omega_{12}$, angle between two vectors}
\begin{equation}
\cos \omega_{12}=\cos \vartheta_{1} \cos \vartheta_{2}+\sin \vartheta_{1} \sin \vartheta_{2} \cos \left(\varphi_{1}-\varphi_{2}\right) . \label{eq:5:17:4}
\end{equation}
The inverse relations to \eqref{eq:5:17:3} can be written in the form
\begin{align}
4 r_{1}^{2} & =R^{2}+r^{2}+2 R r \cos \omega, & 4 r_{2}^{2} & =R^{2}+r^{2}-2 R r \cos \omega, \notag\\
\cos \vartheta_{1} & =\frac{R \cos \Theta+r \cos \vartheta}{\sqrt{R^{2}+r^{2}+2 R r \cos \omega}}, & \cos \vartheta_{2} & =\frac{R \cos \Theta-r \cos \vartheta}{\sqrt{R^{2}+r^{2}-2 R r \cos \omega}}, \notag\\
\tan \varphi_{1} & =\frac{R \sin \Theta \sin \Phi+r \sin \vartheta \sin \varphi}{R \sin \Theta \cos \Phi+r \sin \vartheta \cos \varphi}, & \tan \varphi_{2} & =\frac{R \sin \Theta \sin \Phi-r \sin \vartheta \sin \varphi}{R \sin \Theta \cos \Phi-r \sin \vartheta \cos \varphi} .
\end{align}
$\omega$ being the angle between $\vect{R}$ and $\vect{r}$ (Fig.~\ref{chap5:fig:2})
\begin{equation}
\cos \omega=\cos \Theta \cos \vartheta+\sin \Theta \sin \vartheta \cos (\Phi-\varphi) \text {. } \label{eq:5:17:6}
\end{equation}
Note that the expressions for the polar coordinates of $\vect{R}$ and $\vect{r}$ are interchanged by the replacement $\vect{r}_{2} \rightarrow-\vect{r}_{2}$, i.e., $\vartheta_{2} \rightarrow \pi-\vartheta_{2}, \varphi_{2} \rightarrow \pi+\varphi_{2}$. Hence, the expansions of functions $F(\vect{R}, \vect{r})$ and $F(\vect{r}, \vect{R})$, which may be reduced to each other by the replacement $\vect{R} \rightleftarrows \vect{r}$ are mutually related. If
\begin{equation}
F(\vect{R}, \vect{r})=f\left(\vect{r}_{1}, \vect{r}_{2}\right)=\sum_{l_{1} l_{2} L M} C_{l_{1} l_{2}}^{L M}\left(r_{1}, r_{2}\right)\left\{\vect{Y}_{l_{1}}\left(\Omega_{1}\right) \otimes \vect{Y}_{l_{2}}\left(\Omega_{2}\right)\right\}_{L M} \label{eq:5:17:7}
\end{equation}

then
\begin{equation}
F(\vect{r}, \vect{R})=f\left(\vect{r}_{1},-\vect{r}_{2}\right)=\sum_{l_{1} l_{2} L M}(-1)^{l_{2}} C_{l_{1} l_{2}}^{L M}\left(r_{1}, r_{2}\right)\left\{\vect{Y}_{l_{1}}\left(\Omega_{1}\right) \otimes \vect{Y}_{l_{2}}\left(\Omega_{2}\right)\right\}_{L M} \label{eq:5:17:8}
\end{equation}
\subsection{Addition Theorem for the Spherical Harmonics}\label{sec:5.17.2}\index{addition theorem!for the spherical harmonics}\index{spherical harmonics!addition theorem for}
The functions $f\left(r_{1}, r_{2}\right)$ which are invariant under rotations of coordinate systems play an especially important role in applications. These functions depend only on three variables $r_{1}=\left|r_{1}\right|, r_{2}=\left|r_{2}\right|,\left(r_{1} \cdot r_{2}\right)= r_{1} r_{2} \cos \omega_{12}$ or, equivalently, on $R=\left|r_{1}+r_{2}\right|, r=\left|r_{1}-r_{2}\right|$ and $(\vect{R} \cdot \vect{r})=R r \cos \omega$. The expansions of these functions will evidently contain only the bipolar harmonics of zero rank ( $L=0$ ), i.e., only scalar products $\left(Y_{1}\left(\Omega_{1}\right) \cdot Y_{1}\left(\Omega_{2}\right)\right)$ (Eqs. \eqref{eq:5:16:10}). Such expansions are equivalent to the expansions in terms of the Legendre polynomials, which depend on $\cos \omega_{12}$ because
\begin{equation}
\left(Y_{l}\left(\Omega_{1}\right) \cdot Y_{l}\left(\Omega_{2}\right)\right)=\frac{2 l+1}{4 \pi} P_{l}\left(\cos \omega_{12}\right) . \label{eq:5:17:9}
\end{equation}
This relation is called the addition theorem for the spherical harmonics.

\subsection{Expansions of Some Functions Which Depend on $\left(\mathrm{r}_{1} \cdot \mathrm{r}_{2}\right)$}\label{sec:5.17.3}\index{expansion!of functions of a scalar product}
\begin{align}
\left(\vect{r}_{1} \cdot \vect{r}_{2}\right) & =\frac{4 \pi}{3} r_{1} r_{2}\left(\vect{Y}_{1}\left(\vartheta_{1}, \varphi_{1}\right) \cdot \vect{Y}_{1}\left(\vartheta_{2}, \varphi_{2}\right)\right)  \label{eq:5:17:10}\\
\left(\vect{r}_{1} \cdot \vect{r}_{2}\right)^{2} & =\frac{4 \pi}{3} r_{1}^{2} r_{2}^{2}\left[\left(\vect{Y}_{0}\left(\vartheta_{1}, \varphi_{1}\right) \cdot \vect{Y}_{0}\left(\vartheta_{2}, \varphi_{2}\right)\right)+\frac{2}{5}\left(\vect{Y}_{2}\left(\vartheta_{1}, \varphi_{1}\right) \cdot \vect{Y}_{2}\left(\vartheta_{2}, \varphi_{2}\right)\right)\right]  \label{eq:5:17:11}\\
\left(\vect{r}_{1} \cdot \vect{r}_{2}\right)^{3} & =\frac{4 \pi}{5} r_{1}^{3} r_{2}^{3}\left[\left(\vect{Y}_{1}\left(\vartheta_{1}, \varphi_{1}\right) \cdot \vect{Y}_{1}\left(\vartheta_{2}, \varphi_{2}\right)\right)+\frac{2}{7}\left(\vect{Y}_{3}\left(\vartheta_{1}, \varphi_{1}\right) \cdot \vect{Y}_{3}\left(\vartheta_{2}, \varphi_{2}\right)\right)\right] \label{eq:5:17:12}
\end{align}
In the general case \cite{ref069}, for integer $n$, one has
\begin{equation}
\left(\vect{r}_{1} \cdot \vect{r}_{2}\right)^{n}=4 \pi r_{1}^{n} r_{2}^{n} \sum_{l} \frac{n!}{(n-l)!!(n+l+1)!!}\left(\vect{Y}_{l}\left(\vartheta_{1}, \varphi_{1}\right) \cdot \vect{Y}_{l}\left(\vartheta_{2}, \varphi_{2}\right)\right) . \label{eq:5:17:13}
\end{equation}
In this formula the summation index $l$ assumes the values $l=0,2, \ldots, n-2, n$ if $n$ is even, and $l=1,3, \ldots$, $n-2, n$ if $n$ is odd.

The expansion of the exponential function is given by
\begin{equation}
e^{i\left(r_{1} \cdot r_{2}\right)}=4 \pi \sum_{l=0}^{\infty} i^{l} j_{l}\left(r_{1} r_{2}\right)\left(\vect{Y}_{l}\left(\Omega_{1}\right) \cdot \vect{Y}_{l}\left(\Omega_{2}\right)\right) \label{eq:5:17:14}
\end{equation}
where $j_l(x)$ is a spherical Bessel function.\\
An arbitrary function $f\left(\vect{r}_{1} \cdot \vect{r}_{2}\right)$ which can be written as a power series
\begin{equation}
f\left(r_{1} \cdot r_{2}\right)=\sum_{n=0}^{\infty} c_{n}\left(r_{1} \cdot r_{2}\right)^{n} \label{eq:5:17:15}
\end{equation}
may be expanded in terms of the spherical harmonics:
\begin{equation}
f\left(\vect{r}_{1} \cdot \vect{r}_{2}\right)=\sum_{l=0}^{\infty} f_{l}\left(r_{1} r_{2}\right)\left(\vect{Y}_{l}\left(\Omega_{1}\right) \cdot \vect{Y}_{l}\left(\Omega_{2}\right)\right) \label{eq:5:17:16}
\end{equation}
The expansion coefficients are determined by
\begin{equation}
f_{l}\left(r_{1} r_{2}\right)=4 \pi \sum_{n=l, l+2}^{\infty} c_{n} \frac{n!\left(r_{1} r_{2}\right)^{n}}{(n-1)!!(n+l+1)!!} \label{eq:5:17:17}
\end{equation}
\subsection{Expansions of Some Functions Which Depend on $r=\left|\vect{r}_{1}-\vect{r}_{2}\right|$}\label{sec:5.17.4}\index{expansion!of functions of the distance between two points}
An expansion of the delta function $\delta\left(\mathrm{r}_{1}-\mathrm{r}_{2}\right)$ is\index{delta function}
\begin{equation}
\delta\left(\vect{r}_{1}-\vect{r}_{2}\right)=\frac{\delta\left(r_{1}-r_{2}\right)}{r_{1}^{2}} \sum_{l=0}^{\infty}\left(\vect{Y}_{l}\left(\Omega_{1}\right) \cdot \vect{Y}_{l}\left(\Omega_{2}\right)\right) \label{eq:5:17:18}
\end{equation}
This equation results from the completeness relation for the spherical harmonics (Eqs. \eqref{eq:5:6:1}).\\
An expansion of the Green's function for the scalar Helmholtz equation may be written in the form\index{Green's function}\index{Helmholtz equation}
\begin{equation}
\frac{e^{i k r}}{r}=4 \pi i k \sum_{l=0}^{\infty} j_{l}\left(k r_{1}\right) h_{l}^{(1)}\left(k r_{2}\right)\left(\vect{Y}_{l}\left(\Omega_{1}\right) \cdot \vect{Y}_{l}\left(\Omega_{2}\right)\right), \quad\left(r_{1}<r_{2}\right) \label{eq:5:17:19}
\end{equation}
where $j_l(x)$ and $h_{l}^{(1)}(x)$ are the spherical Bessel function and the Hankel function of the first kind; respectively. Equation \eqref{eq:5:17:19} is valid at $r_{1}<r_{2}$; for $r_{1}>r_{2}$ one should make the replacement $r_{1} \rightleftarrows r_{2}$. Equation \eqref{eq:5:17:19} follows from a more general expansion which is given in Ref.~\cite{ref070}.
\begin{equation}
z_{0}(k r)=4 \pi \sum_{l=0}^{\infty} j_{l}\left(k r_{1}\right) z_{l}\left(k r_{2}\right)\left(\vect{Y}_{l}\left(\Omega_{1}\right) \cdot \vect{Y}_{l}\left(\Omega_{2}\right)\right), \quad\left(r_{1}<r_{2}\right) \label{eq:5:17:20}
\end{equation}
Here $z_{l}(x)=\sqrt{\pi / 2 x} Z_{l+\frac{1}{2}}(x)$, and $Z_{l+\frac{1}{2}}(x)$ is any of the cylinder functions.\\
When $k=0$ Eq. \eqref{eq:5:17:19} yields the important expansion of the Green's function for the scalar Laplace equation
\begin{equation}
\frac{1}{r}=\frac{4 \pi}{r_{2}} \sum_{l=0}^{\infty} \frac{1}{(2 l+1)}\left(\frac{r_{1}}{r_{2}}\right)^{l}\left(\vect{Y}_{l}\left(\Omega_{1}\right) \cdot \vect{Y}_{l}\left(\Omega_{2}\right)\right), \quad\left(r_{1}<r_{2}\right) \label{eq:5:17:21}
\end{equation}
This result as well as Eq. \eqref{eq:5:17:19} are valid when $r_{1}<r_{2}$, whereas if $r_{1}>r_{2}$ it becomes valid after the interchange $r_{1} \vec{\leftarrow} r_{2}$. We may also write Eq. \eqref{eq:5:17:21} in a symmetric form valid both at $r_{1}<r_{2}$ and at $r_{1}>r_{2}$ :
\begin{equation}
\frac{1}{r}=\frac{4 \pi}{\sqrt{r_{1}^{2}+r_{2}^{2}}} \sum_{l=0}^{\infty}\left\{\sum_{n=l, l+2}^{\infty} \frac{(2 n-1)!!}{(n-l)!!(l+n+1)!!}\left(\frac{r_{1} r_{2}}{r_{1}^{2}+r_{2}^{2}}\right)^{n}\right\}\left(\vect{Y}_{l}\left(\Omega_{1}\right) \cdot \vect{Y}_{l}\left(\Omega_{2}\right)\right) \label{eq:5:17:22}
\end{equation}
Below we give some examples of expansions of $r^{n}(n=1,-3,-5)$
\begin{align}
r & =4 \pi \sum_{l=0}^{\infty} \frac{1}{(2 l+1)}\left(\frac{r_{1}^{l}}{r_{2}^{l+1}}\right)\left(\frac{r_{1}^{2}}{2 l+3}-\frac{r_{2}^{2}}{2 l-1}\right)\left(\vect{Y}_{l}\left(\Omega_{1}\right) \cdot \vect{Y}_{l}\left(\Omega_{2}\right)\right)  \label{eq:5:17:23}\\
\frac{1}{r^{3}} & =\frac{4 \pi}{r_{2}^{2}-r_{1}^{2}} \sum_{l=0}^{\infty}\left(\frac{r_{1}^{l}}{r_{2}^{l+1}}\right)\left(\vect{Y}_{l}\left(\Omega_{1}\right) \cdot \vect{Y}_{l}\left(\Omega_{2}\right)\right)  \label{eq:5:17:24}\\
\frac{1}{r^{5}} & =\frac{4 \pi}{3\left(r_{2}^{2}-r_{1}^{2}\right)^{3}} \sum_{l=0}^{\infty}(2 l-1)(2 l+3)\left(\frac{r_{1}^{l}}{r_{2}^{l+1}}\right)\left(\frac{r_{2}^{2}}{2 l-1}-\frac{r_{1}^{2}}{2 l+3}\right)\left(\vect{Y}_{l}\left(\Omega_{1}\right) \cdot \vect{Y}_{l}\left(\Omega_{2}\right)\right) \label{eq:5:17:25}
\end{align}
These equations are valid when $r_{1}<r_{2}$, whereas for $r_{1}>r_{2}$ they are valid after interchanging $r_{1}$ and $r_{2}$. Equations \eqref{eq:5:17:23}-\eqref{eq:5:17:25} follow from the expansions which will be considered below.

\subsection{Expansions of $r^{n} \equiv\left|r_{1}-r_{2}\right|^{n}$}\label{sec:5.17.5}\index{expansion!of powers of the distance}
Let $n$ be an integer. Then the expansion under discussion is as follows
\begin{equation}
r^{n}=4 \pi \sum_{l=0}^{\infty} \frac{1}{2 l+1} a_{l}^{n}\left(r_{1}, r_{2}\right)\left(\vect{Y}_{l}\left(\Omega_{1}\right) \cdot \vect{Y}_{l}\left(\Omega_{2}\right)\right), \label{eq:5:17:26}
\end{equation}
where the expansion coefficients $a_{1}^{n}\left(r_{1}, r_{2}\right)$ are determined by (see Ref.~\cite{ref097})
\begin{align}
a_{l}^{n}\left(r_{1}, r_{2}\right)= & \frac{\left(-\frac{n}{2}\right)_{l}}{\left(\frac{1}{2}\right)_{l}} r_{2}^{n}\left(\frac{r_{1}}{r_{2}}\right)^{l} F\left(l-\frac{n}{2},-\frac{1}{2}-\frac{n}{2} ; l+\frac{3}{2} ; \frac{r_{1}^{2}}{r_{2}^{2}}\right),\left(r_{1}<r_{2}\right),  \label{eq:5:17:27}\\
a_{l}^{n}\left(r_{1}, r_{2}\right)= & \frac{\left(-\frac{n}{2}\right)_{l}}{\left(\frac{1}{2}\right)_{l}} \frac{r_{1}^{l}\left(r_{2}^{2}-r_{1}^{2}\right)^{n+2}}{r_{2}^{l+n+4}} F\left(l+2+\frac{n}{2}, \frac{3}{2}+\frac{n}{2} ; l+\frac{3}{2} ; \frac{r_{1}^{2}}{r_{2}^{2}}\right),\left(r_{1}<r_{2}\right),  \label{eq:5:17:28}\\
a_{l}^{n}\left(r_{1}, r_{2}\right)= & \frac{2^{n+1}\left(l+\frac{1}{2}\right)\left(-\frac{n}{2}\right)_{l}}{\left(1+\frac{n}{2}\right)_{l+1}^{l} r_{2}^{n-l} F\left(l-\frac{n}{2},-\frac{1}{2}-\frac{n}{2} ;-1-n ; \frac{r_{2}^{2}-r_{1}^{2}}{r_{2}^{2}}\right)}  \notag\\
& -\frac{2 l+1}{2^{n+3}(n+2)} \cdot \frac{r_{1}^{l}\left(r_{2}^{2}-r_{1}^{2}\right)^{n+2}}{r_{2}^{l+n+4}} F\left(l+\frac{n}{2}+2, \frac{3}{2}+\frac{n}{2} ; n+3 ; \frac{r_{2}^{2}-r_{1}^{2}}{r_{2}^{2}}\right),  \\
a_{l}^{n}\left(r_{1}, r_{2}\right)= & \frac{\left(-\frac{n}{2}\right)_{l}(-1)^{\frac{n}{2}} \cos \frac{n \pi}{2} r_{1}^{n-l} r_{2}^{l} F\left(l-\frac{n}{2},-\frac{1}{2}-\frac{n}{2} ; l+\frac{3}{2} ; \frac{r_{2}^{2}}{r_{1}^{2}}\right)}{\left(\frac{1}{2}\right)_{l}}  \notag\\
& +\frac{\Gamma\left(l+\frac{3}{2}\right) \sqrt{\pi}(-1)^{\frac{n+1}{2}}}{\Gamma\left(-\frac{n}{2}\right) \Gamma\left(2+l+\frac{n}{2}\right)} \frac{r_{1}^{n+l+1}}{r_{2}^{l+1}} F\left(-1-l-\frac{n}{2},-\frac{1}{2}-\frac{n}{2} ; \frac{1}{2}-l ; \frac{r_{2}^{2}}{r_{1}^{2}}\right), \label{eq:5:17:30}
\end{align}
If $r_{1}>r_{2}$, one should interchange $r_{1}$ and $r_{2}$ in Eqs.~\eqref{eq:5:17:27}-\eqref{eq:5:17:30}. Let us present also symmetric expressions for $a_{l}^{n}\left(r_{1}, r_{2}\right)$ which are valid when $r_{1}<r_{2}$ and $r_{1}>r_{2}$ :
\begin{align}
& a_{l}^{n}\left(r_{1}, r_{2}\right)=\frac{\left(-\frac{n}{2}\right)_{l}}{\left(\frac{1}{2}\right)_{l}} \cdot \frac{\left(r_{1} r_{2}\right)^{l}}{\left(r_{1}^{2}+r_{2}^{2}\right)^{l-\frac{n}{2}}} F\left(\frac{l}{2}-\frac{n}{4}, \frac{l}{2}-\frac{n}{4}+\frac{1}{2} ; l+\frac{3}{2} ;\left(\frac{2 r_{1} r_{2}}{r_{1}^{2}+r_{2}^{2}}\right)^{2}\right)  \label{eq:5:17:31}\\
& a_{l}^{n}\left(r_{1}, r_{2}\right)=\frac{\left(-\frac{n}{2}\right)_{l}}{\left(\frac{1}{2}\right)_{l}} \cdot \frac{\left(r_{1} r_{2}\right)^{l}}{\left(r_{1}+r_{2}\right)^{2 l-n}} F\left(l-\frac{n}{2}, 1+l ; 2 l+2 ; \frac{4 r_{1} r_{2}}{\left(r_{1}+r_{2}\right)^{2}}\right) \label{eq:5:17:32}
\end{align}
A comparison of Eqs. \eqref{eq:5:17:27} and \eqref{eq:5:17:28} shows that the expansion coefficient for $r^{n}$ and $r^{-n-4}$ are related by
\begin{equation}
a_{l}^{n}\left(r_{1}, r_{2}\right)=\frac{\left(-\frac{n}{2}\right)_{l}}{\left(\frac{n}{2}+2\right)_{l}}\left(r_{2}^{2}-r_{1}^{2}\right)^{n+2} a_{l}^{-n-4}\left(r_{1}, r_{2}\right) \label{eq:5:17:33}
\end{equation}
Note that all the expansions given in Secs. \ref{sec:5.17.4} and \ref{sec:5.17.5} remain valid after substituting $R=\left|\vect{r}_{1}+\vect{r}_{2}\right|$ for $r=\left|r_{1}-r_{2}\right|$ and multiplying the expansion coefficients by $(-1)^{l}$.

\subsection{Expansions of Spherical Waves}\label{sec:5.17.6}\index{spherical waves!expansions of}\index{expansion!of spherical waves}
The spherical waves have the form $z_{L}(k r) Y_{L M}(\vartheta, \varphi)$, where $\vect{r}(r, \vartheta, \varphi)=\vect{r}_{1}-\vect{r}_{2}, z_{L}(x)=\sqrt{\pi / 2 x} Z_{L+\frac{1}{2}}(x)$ and $Z_{L+\frac{1}{2}}(x)$ is any of the cylinder functions.\index{cylinder functions}\isym{ZL@$Z_{L+1/2}(x)$, cylinder function} The spherical waves are expressed in terms of the bipolar spherical harmonics of rank $L$ \cite{ref070} (Sec.~\ref{sec:5.16.1})
\begin{align}
z_{L}(k r) Y_{L M}(\vartheta, \varphi)= & \sqrt{\frac{4 \pi}{2 L+1}} \sum_{l_{1}, l_{2}=0}^{\infty} i^{l_{1}-l_{2}-L} \sqrt{\left(2 l_{1}+1\right)\left(2 l_{2}+1\right)}  \notag\\
& \times \clebsch{l_{1}}{0}{l_{2}}{0}{L}{0} j_{l_{1}}\left(k r_{1}\right) z_{l_{2}}\left(k r_{2}\right)\left\{\vect{Y}_{l_{1}}\left(\Omega_{1}\right) \otimes \vect{Y}_{l_{2}}\left(\Omega_{2}\right)\right\}_{L M},\left(r_{1}<r_{2}\right) \label{eq:5:17:34}
\end{align}
When $r_{1}>r_{2}$, one should interchange $r_{1}$ and $r_{2}$ in Eq. \eqref{eq:5:17:34}. If $z_{L}(k r)=j_{L}(k r), j_{L}(x)$ being a spherical Bessel function, one obtains as $k \rightarrow 0[22,96]$
\begin{equation}
r^{L} Y_{L M}(\vartheta, \varphi)=\sqrt{4 \pi(2 L+1)!} \sum_{\substack{l_{1}, l_{2}=0 \\ l_{1}+l_{2}=L}}^{L}(-1)^{l_{2}} \frac{r_{1}^{l_{1}} r_{2}^{l_{2}}}{\sqrt{\left(2 l_{1}+1\right)!\left(2 l_{2}+1\right)!}}\left\{\vect{Y}_{l_{1}}\left(\Omega_{1}\right) \otimes \vect{Y}_{l_{2}}\left(\Omega_{2}\right)\right\}_{L M} \label{eq:5:17:35}
\end{equation}
Similarly, if $z_{L}(k r)=n_{L}(k r), n_{L}(k r)$ being a spherical Neumann function, one obtains as $k \rightarrow 0[22,96]$\index{spherical Neumann function}
\begin{equation}
\frac{1}{r^{L+1}} Y_{L M}(\vartheta, \varphi)=\sqrt{\frac{4 \pi}{(2 L)!}} \sum_{\substack{l_{1}, l_{2}=0 \\ l_{2}-l_{1}=L}}^{\infty}(-1)^{l_{2}} \sqrt{\frac{\left(2 l_{2}\right)!}{\left(2 l_{1}+1\right)!}} \cdot \frac{r_{1}^{l_{1}}}{r_{2}^{l_{2}+1}}\left\{\vect{Y}_{l_{1}}\left(\Omega_{1}\right) \otimes \vect{Y}_{l_{2}}\left(\Omega_{2}\right)\right\}_{L M},\left(r_{1}<r_{2}\right) \label{eq:5:17:36}
\end{equation}
For example, spherical components of $r / r^{3}$ read
\begin{equation}
\frac{r_{\mu}}{r^{3}}=4 \pi \sum_{l=1}^{\infty}(-1)^{l} \sqrt{\frac{l}{3}} \cdot \frac{r_{1}^{l-1}}{r_{2}^{l+1}}\left\{\vect{Y}_{l-1}\left(\Omega_{1}\right) \otimes \vect{Y}_{l}\left(\Omega_{2}\right)\right\}_{1 \mu}, \quad\left(r_{1}<r_{2}\right) \label{eq:5:17:37}
\end{equation}
\subsection{Expansions of $r^{N} Y_{L M}(\vartheta, \varphi)$}\label{sec:5.17.7}\index{expansion!of $r^{N} Y_{L M}$}
The expansion of this function in terms of bipolar harmonics of rank $L$ may be written as
\begin{equation}
r^{N} Y_{L M}(\vartheta, \varphi)=4 \pi \sum_{l_{1}, l_{2}=0}^{\infty} a_{l_{1} l_{2}}^{N L}\left(r_{1}, r_{2}\right) \sqrt{\frac{\left(2 l_{1}+1\right)\left(2 l_{2}+1\right)}{4 \pi(2 L+1)}} \clebsch{l_{1}}{0}{l_{2}}{0}{L}{0}\left\{\vect{Y}_{l_{1}}\left(\Omega_{1}\right) \otimes \vect{Y}_{l_{2}}\left(\Omega_{2}\right)\right\}_{L M} \label{eq:5:17:38}
\end{equation}
The sum is over all positive integer $l_{1}$ and $l_{2}$ allowed by the momentum addition rules and the parity selection rule ( $l_{1}+l_{2}-L$ is even). The expansion coefficients $a_{l_{1} l_{2}}^{N L}\left(r_{1}, r_{2}\right)$ are given by (see Ref.~\cite{ref098})
\begin{align}
a_{l_{1} l_{2}}^{N L}\left(r_{1}, r_{2}\right)= & (-1)^{\frac{l_{1}-l_{2}-L}{2}} \frac{2^{l_{1}}}{\left(2 l_{1}+1\right)!!} \frac{\Gamma\left(\frac{l_{1}+l_{2}-N}{2}\right) \Gamma\left(\frac{L+N+3}{2}\right)}{\Gamma\left(\frac{L-N}{2}\right) \Gamma\left(\frac{-l_{1}+l_{2}+N+3}{2}\right)}  \notag\\
& \times r_{2}^{N}\left(\frac{r_{1}}{r_{2}}\right)^{l_{1}} F\left(\frac{l_{1}+l_{2}-N}{2}, \frac{l_{1}-l_{2}-N-1}{2} ; l_{1}+\frac{3}{2} ; \frac{r_{1}^{2}}{r_{2}^{2}}\right),\left(r_{1}<r_{2}\right) . \label{eq:5:17:39}
\end{align}
If $r_{1}>r_{2}$, the expression for the expansion coefficients may be obtained from Eq. \eqref{eq:5:17:39} by interchanging $r_{1}$ and $r_{2}, l_{1}$ and $l_{2}$ and introducing the additional factor $(-1)^{L}$. The expression \eqref{eq:5:17:39} was studied in detail in Ref.~\cite{ref078}. For different values of $N$ and $L$ the expansion coefficients are determined by the following relations. For $N=L, L+2, L+4, \ldots$ we have
\begin{equation}
a_{l_{1} l_{2}}^{N L}\left(r_{1}, r_{2}\right)= \begin{cases}(-1)^{l_{2}}(N+L+1)!!(N-L)!! & \sum_{e_{1}, e_{2}=0}^{\infty} \frac{r_{1}^{l_{1}+2 e_{1} r_{2}+2 e_{2}}}{\left(2 e_{1}\right) 11\left(2 e_{1}+2 l_{1}+1\right) 11\left(2 e_{2}\right) 11\left(2 e_{2}+2 l_{2}+1\right) 11}  \label{eq:5:17:40}\\ & \left(e_{1}+e_{2}=\frac{N-l_{1}-l_{2}}{5}\right) \\ & \text { if } l_{1}+l_{2} \leq N, \\ 0 & \text { if } l_{1}+l_{2}>N .\end{cases}
\end{equation}
For $N=L-2, L-4, L-6, \ldots$ one gets
\begin{gather}
a_{l_{1} l_{2}}^{N L}\left(r_{1}, r_{2}\right)=(-1)^{l_{2}+\frac{N-2}{j}} \frac{(N+L+1)!!}{2(L-N-2)!!\left(N+l_{1}+l_{2}+2\right)!} r_{1}^{l_{1}} r_{2}^{l_{2}}  \notag\\
\times\left(\frac{1}{r_{1}} \cdot \frac{\partial}{\partial r_{1}}\right)^{l_{1}}\left(\frac{1}{r_{2}} \cdot \frac{\partial}{\partial r_{2}}\right)^{l_{2}} \frac{1}{r_{1} r_{2}}\left[\left(r_{1}-r_{2}\right)^{N+l_{1}+l_{2}+2} \ln \left|r_{1}-r_{2}\right|-\left(r_{1}+r_{2}\right)^{N+l_{1}+l_{2}+2} \ln \left(r_{1}+r_{2}\right)\right] . \label{eq:5:17:41}
\end{gather}
If $N$ is integer, $N-L$ is odd, and $r_{1}<r_{2}$, the expansion coefficient is given by
\begin{equation}
a_{l_{1} l_{2}}^{N L}\left(r_{1}, r_{2}\right)= \begin{cases}(-1)^{N+l_{2}+1}+l_{1} & \frac{(N+L+1) \|}{(L-N-2) \|} r_{1}^{l_{1}} r_{2}^{N-l_{1}} \sum_{:=0}^{\frac{N-l_{1}+l_{2}+1}{2}} \frac{\left(\frac{r_{1}}{r_{2}}\right)^{2 \cdot}}{(2 \cdot)\left\|\left(N-l_{1}-l_{3}-2 \cdot\right)\right\|\left(N-l_{1}+l_{2}+1-2 \cdot\right)\left\|\left(2 l_{1}+2 \cdot+1\right)\right\|}  \label{eq:5:17:42}\\ 0 & \text { if } N-l_{1}+l_{2}+1 \geq 0, \\ & \text { if } N-l_{1}+l_{2}+1<0 .\end{cases}
\end{equation}
The expansion coefficient for $r_{1}>r_{2}$ may be obtained from \eqref{eq:5:17:42} by interchanging $r_{1}$ and $r_{2}, l_{1}$ and $l_{2}$ and introducing the additional factor $(-1)^{L}$. It should be emphasised that if $N \leq-3$ the expansion coefficients contain additional terms involving the delta function $\delta\left(r_{1}-r_{2}\right)$ and its derivatives \cite{ref078}
\begin{align}
a_{l_{1} l_{2}}^{N L}\left(r_{1}, r_{2}\right)_{\delta}= & (-1)^{N+l_{1}-1} \frac{(N+L+1)!!}{(L-N-2)!!} \sum_{e_{1}=0}^{l_{1}} \sum_{\substack{e_{2}=0 \\
e_{1}+e_{2}+N \leq-3}}^{l_{2}}(-1)^{e_{2}} \frac{\left(l_{1}+s_{1}\right)!\left(l_{2}+s_{2}\right)!}{\left(l_{1}-s_{1}\right)!\left(l_{2}-s_{2}\right)!}  \notag\\
& \times \frac{1}{\left(2 s_{1}\right)!!\left(2 s_{2}\right)!!} \cdot \frac{1}{r_{2}^{\beta_{2}+1}}\left(\frac{\partial}{\partial r_{2}}\right)^{-N-s_{1}-e_{2}-3}\left[\frac{\delta\left(r_{1}-r_{2}\right)}{r_{1}^{2} r_{2}^{\beta_{1}-1}}\right] . \label{eq:5:17:43}
\end{align}
These additional terms give finite contributions into the integrals involving the expansion \eqref{eq:5:17:38}.\\
Finally, if $N$ is not integer, the expansion coefficient is given by
\begin{gather}
a_{l_{1} l_{2}}^{N L}\left(r_{1}, r_{2}\right)=(-1)^{l_{1}+L+1} \frac{2^{L-1} \Gamma\left(\frac{N+L+3}{2}\right) \Gamma(N-L+2)}{\Gamma\left(\frac{N-L+3}{2}\right) \Gamma\left(N+l_{1}+l_{2}+3\right)} r_{1}^{l_{1}} r_{2}^{l_{2}}  \notag\\
\times\left(\frac{1}{r_{1}} \frac{\partial}{\partial r_{1}}\right)^{l_{1}}\left(\frac{1}{r_{2}} \frac{\partial}{\partial r_{2}}\right)^{l_{2}} \frac{1}{r_{1} r_{2}}\left\{\left|r_{1}+r_{2}\right|^{N+l_{1}+l_{2}+2}-\left|r_{1}-r_{2}\right|^{N+l_{1}+l_{2}+2}\right\} \label{eq:5:17:44}
\end{gather}
Note that all the expansions given in Secs.~\ref{sec:5.17.6} and \ref{sec:5.17.7} remain valid after substituting $\vect{R}(R, \Theta, \Phi)=\vect{r}_{1}+\vect{r}_{2}$ for $\vect{r}(r, \vartheta, \varphi)=\vect{r}_{1}-\vect{r}_{2}$ and introducing the additional factor $(-1)^{l_{2}}$ into the expansion coefficients.

\chapter{Spin functions}\label{chap:6}\index{spin function}
\section{Spin functions of particles with arbitrary spin}\label{sec:6.1}\index{spin function!for arbitrary spin}
\subsection{Definition}\label{sec:6.1.1}\index{spin function!definition}\index{spin variable}
Spin functions describe polarization states of particles ${ }^{1}$ of definite spin, i.e. definite intrinsic angular momentum.

The spin functions $\chi(\sigma)$ may be treated as functions of a discrete variable $\sigma$ which is the spin projection on the $z$-axis.\isymg{chi-sigma@$\chi(\sigma)$, spin function}\isymg{sigma-spin@$\sigma$, spin variable} The variable $\sigma$ takes $2 S+1$ values, $\sigma=-S,-S+1, \ldots, S-1, S$, where $S$ is the particle spin (integer or half integer non-negative number). The quantity $|\chi(\sigma)|^{2}$ gives the probability that the spin projection on the $z$-axis in a given state equals $\sigma$.

The spin functions are commonly written as column matrices which contain $2 S+1$ elements:
\begin{equation}
\chi=\begin{pmatrix}
\chi(S)  \\
\chi(S-1) \\
\vdots \\
\chi(-S+1) \\
\chi(-S)
\end{pmatrix}. \label{eq:6:1:1}
\end{equation}
The elements of this matrix give the values of the spin function $\chi(\sigma)$ for corresponding values of spin variable $\sigma$. The quantities $\chi(\sigma)$ are called the contravariant components of the spin function $\chi$ (see Sec.~\ref{sec:6.1.4} below).

In such a representation operators acting on the spin variables take the form of square $(2 S+1) \times(2 S+1)$ matrices, and summation over the spin variable is replaced by matrix multiplication. The Hermitian conjugate function $\chi^{\dagger}$ has the form of a row matrix\isymg{chi-dagger@$\chi^{\dagger}$, Hermitian adjoint spin function}
\begin{equation}
\chi^{\dagger}=\left(\chi^{*}(S), \chi^{*}(S-1), \ldots ,\chi^{*}(-S+1), \chi^{*}(-S)\right) . \label{eq:6:1:2}
\end{equation}
The interpretation of $|\chi(\sigma)|^{2}$ as the probability for the spin projection on the $z$-axis to be equal to $\sigma$ is possible only if $\chi(\sigma)$ satisfies the normalization condition
\begin{equation}
\sum_{\sigma=-S}^{S}|\chi(\sigma)|^{2}=1. \label{eq:6:1:3}
\end{equation}
\footnote{We use the name "particle" not only for an elementary particle (electron, nucleon, etc.) but also for any composite system (atom, molecule) which can be treated in phenomena under consideration as one object.}
This condition may be written also in a matrix form
\begin{equation}
\chi^{\dagger} \chi=1 . \label{eq:6:1:4}
\end{equation}
The above representation of spin functions is called the spherical basis representation.\index{spherical basis representation}\index{representation!spherical basis}\index{cartesian basis representation}\index{representation!cartesian basis} When the spin value is integer, one may also use the cartesian basis representation. The latter representation is considered below (Sec.~\ref{sec:6.3}) for spin-1 particles.

\subsection{Basis Spin Functions}\label{sec:6.1.2}\index{spin function!basis}\index{basis spin functions}
According to definition, the basis spin functions describe the states with definite spin and spin projection on the $z$-axis. The basis spin functions are eigenfunctions of the operators $\widehat{\vect{S}}^{2}$ and $\widehat{S}_{z}$, where $\widehat{\vect{S}}$ is the spin operator (see Sec.~\ref{sec:2.3})\index{spin operator}\isym{S-hat@$\widehat{\vect{S}}$, spin operator}
\begin{equation}
\widehat{\vect{S}}^{2} \chi_{S m}=S(S+1) \chi_{S m}, \widehat{S}_{z} \chi_{S m}=m \chi_{S m} . \label{eq:6:1:5}
\end{equation}
As follows from the definition, the dependence of the basis spin functions $\chi_{S m}(\sigma)$ on the spin variable $\sigma$ is given by
\begin{equation}
\chi_{S m}(\sigma)=\delta_{m \sigma} . \label{eq:6:1:6}
\end{equation}
In other words, contravariant components of the basis spin functions $\chi_{S m}$ have the form\isymg{chi-Sm@$\chi_{S m}$, spin function}
\begin{equation}
\left[\chi_{S m}\right]^{\sigma}=\delta_{m \sigma}. \label{eq:6:1:7}
\end{equation}
If the basis spin functions are written as column matrices, then
\begin{equation}
\chi_{S S}=\left(\begin{array}{c}
1 \\
0 \\
\vdots \\
0 \\
0
\end{array}\right), \quad \chi_{S\, S-1}=\left(\begin{array}{c}
0 \\
1 \\
\vdots \\
0 \\
0
\end{array}\right), \ldots, \quad \chi_{s-s}=\left(\begin{array}{c}
0 \\
0 \\
\vdots \\
0 \\
1
\end{array}\right) . \label{eq:6:1:8}
\end{equation}
The collection of $2 S+1$ basis functions $\chi_{S m}(m=S, S-1, \ldots,-S)$ constitutes a complete orthonormal set of functions. The orthonormality condition for the basis functions is
\begin{equation}
\chi_{S m}^{\dagger} \chi_{S m^{\prime}}=\delta_{m m^{\prime}} . \label{eq:6:1:9}
\end{equation}
The completeness condition for the basis spin functions may be written in matrix form
\begin{equation}
\sum_{m=-S}^{S} \chi_{S m} \chi_{S m}^{\dagger}=\widehat{I} . \label{eq:6:1:10}
\end{equation}
where $\widehat{I}$ is the unit $(2 S+1) \times(2 S+1)$ matrix.\\
Matrix products $\chi_{S m} \chi_{S m^{\prime}}^{\dagger}$ of the basis spin functions are the square $(2 S+1) \times(2 S+1)$ matrices. They may be expanded in terms of the polarization operators $\widehat{T}_{L M}(S)$ (see 2.4),
\begin{equation}
\chi_{S m} \chi_{S m^{\prime}}^{\dagger}=\sum_{L} \sqrt{\frac{2 L+1}{2 S+1}} \clebsch{S}{m^{\prime}}{L}{M}{S}{m} \hat{T}_{L M}(S) . \label{eq:6:1:11}
\end{equation}
Matrix elements of spherical components of the spin operator $\widehat{S}_{\mu}(\mu= \pm 1,0)$ between the basis spin functions are given by
\begin{equation}
\chi_{S m^{\prime}}^{\dagger} \widehat{S}_{\mu} \chi_{S m}=\sqrt{S(S+1)} \clebsch{S}{m}{1}{\mu}{S}{m^{\prime}} ; \label{eq:6:1:12}
\end{equation}
The only non-vanishing matrix elements are the following:
\begin{gather}
\chi_{S m+1}^{\dagger} \widehat{S}_{+1} \chi_{S m}=-\frac{1}{\sqrt{2}} \sqrt{(S-m)(S+m+1)} , \notag\\
\chi_{S m}^{\dagger} \widehat{S}_{0} \chi_{S m}=m , \label{eq:6:1:13}\\
\chi_{S m-1}^{\dagger} \widehat{S}_{-1} \chi_{S m}=\frac{1}{\sqrt{2}} \sqrt{(S+m)(S-m+1)}. \notag
\end{gather}
For cartesian components of the spin operator $\widehat{S}_{i}(i=x, y, z)$ the non-vanishing matrix elements are
\begin{align}
& \chi_{S m \pm 1}^{\dagger} \widehat{S}_{x} \chi_{S m}=\frac{1}{2} \sqrt{(S \mp m)(S \pm m+1)}. \notag\\
& \chi_{S m \pm 1}^{\dagger} \widehat{S}_{y} \chi_{S m}=\mp \frac{i}{2} \sqrt{(S \mp m)(S \pm m+1)}. \label{eq:6:1:14}\\
& \chi_{S m}^{\dagger} \widehat{S}_{z} \chi_{S m}=m. \notag
\end{align}
The matrix elements of the polarization operators $\hat{T}_{L M}(S)$ between the basis spin functions may be written as
\begin{equation}
\chi_{S m^{\prime}}^{\dagger} \hat{T}_{L M}(S) \chi_{S m}=\sqrt{\frac{2 L+1}{2 S+1}} \clebsch{S}{m}{L}{M}{S}{m^{\prime}}. \label{eq:6:1:15}
\end{equation}
Under rotations of the coordinate system the basis spin functions are transformed either by the rotation operator $\hat{D}^{S}(\alpha, \beta, \gamma)$ (Eq.~\eqref{eq:2:4:17}), if rotations are defined by the Euler angles $\alpha, \beta, \gamma$, or by the rotation operator $\hat{U}^{S}(\omega ; \theta, \Phi)$ (Eq.~\eqref{eq:2:4:18}) if rotations are described by the rotation axis $\vect{n}(\theta, \Phi)$ and the rotation angle $\omega$
\begin{align}
& \chi_{S m^{\prime}}^{\prime} \equiv \hat{D}^{S}(\alpha, \beta, \gamma) \chi_{S m^{\prime}}=\sum_{m} D_{m m^{\prime}}^{S}(\alpha, \beta, \gamma) \chi_{S m} , \notag\\
& \chi_{S m^{\prime}}^{\prime} \equiv \hat{U}^{S}(\omega ; \Theta, \Phi) \chi_{S m^{\prime}}=\sum_{m} U_{m m^{\prime}}^{S}(\omega ; \Theta, \Phi) \chi_{S m}, \label{eq:6:1:16}
\end{align}
where $D_{m m^{\prime}}^{S}$ are the Wigner $D$-functions (Chap.~\ref{chap:4}), and $U_{m m^{\prime}}^{S}$ are the functions defined in Sec.~\ref{sec:4.5}. The functions $\chi_{S m^{\prime}}^{\prime}$ describe quantum states with definite spin $S$ and spin projection $m^{\prime}$ on the new $z^{\prime}$-axis. They are eigenfunctions of the operators $\widehat{\vect{S}}^{\prime 2}$ and $\widehat{S}_{z}^{\prime}$,
\begin{gather}
\hat{\vect{S}}^{\prime 2} \chi_{S m^{\prime}}^{\prime}=S(S+1) \chi_{S m^{\prime}}^{\prime} , \label{eq:6:1:17}\\
\hat{S}_{z}^{\prime} \chi_{S m^{\prime}}^{\prime}=m^{\prime} \chi_{S m^{\prime}}^{\prime} . \notag
\end{gather}
Here $\widehat{\vect{S}}^{\prime}$ is the spin operator in the rotated coordinate system
\begin{equation}
\widehat{S}_{\mu}^{\prime}=\sum_{\nu} D_{\nu \mu}^{1}(\alpha, \beta, \gamma) \widehat{S}_{\nu}, \quad(\mu, \nu= \pm 1,0) . \label{eq:6:1:18}
\end{equation}
\subsection{Helicity Basis Functions}\label{sec:6.1.3}\index{helicity}\index{helicity basis functions}\index{spin function!helicity basis}
The spin projection on the linear momentum direction of a particle is called the helicity.\isymg{lambda-helicity@$\lambda$, helicity} For particles of spin $S$ the helicity $\lambda$ assumes $(2 S+1)$ values, $\lambda=S, S-1, \ldots,-S+1, S$.

The helicity basis functions $\chi_{S \lambda}(\vartheta, \varphi)$, by definition, are the eigenfunctions of the operators\isymg{chi-Slambda@$\chi_{S \lambda}(\vartheta, \varphi)$, helicity basis function} $\widehat{\vect{S}}^{2}$ and $\widehat{\vect{S}} \cdot \vect{n}$ where $\widehat{\vect{S}}$ is the spin operator, $\vect{n}=\vect{p} / \vect{p}$ is the momentum direction of the particles, $\vartheta$ and $\varphi$ are the polar angles of the vector $n$,
\begin{gather}
\widehat{\vect{S}}^{2} \chi_{S \lambda}(\vartheta, \varphi)=S(S+1) \chi_{S \lambda}(\vartheta, \varphi), \label{eq:6:1:19}\\
\widehat{\vect{S}} \cdot \vect{n} \chi_{S \lambda}(\vartheta, \varphi)=\lambda \chi_{S \lambda}(\vartheta, \varphi). \notag
\end{gather}
The function $\chi_{S \lambda}(\vartheta, \varphi)$ describes a quantum state with definite spin $S$ and helicity $\lambda$.\\
The helicity basis functions $\chi_{S \lambda}(\vartheta, \varphi)$ may be expressed in terms of the basis functions $\chi_{S m}$ by means of the rotation which turns the $z$-axis parallel to $\vect{n}(\vartheta, \varphi)$. Such a rotation is determined by the Euler angles $\alpha=\varphi$, $\beta=\vartheta$. The third Euler angle $\gamma$ (the angle of rotation about the new $z^{\prime}$-axis) is arbitrary. For simplicity, let us adopt $\gamma=0$. In this case the rotated cartesian basis vectors $\mathrm{e}_{x}^{\prime}, \mathrm{e}_{y}^{\prime}, \mathrm{e}_{z}^{\prime}$ will coincide with the polar basis vectors $\vect{e}_{\theta}, \vect{e}_{\varphi}, \vect{e}_{\rho}$, respectively (Sec.~\ref{sec:1.1.2}). Making use of the transformation properties of the basis spin functions $\chi_{S m}$ under the rotation \eqref{eq:6:1:16} we obtain
\begin{gather}
\chi_{S \lambda}(\vartheta, \varphi)=\sum_{m} D_{m \lambda}^{S}(\varphi, \vartheta, 0) \chi_{S m}, \label{eq:6:1:20}\\
\chi_{S m}=\sum_{\lambda} D_{-\lambda-m}^{S}(0, \vartheta, \varphi) \chi_{S \lambda}(\vartheta, \varphi). \notag
\end{gather}
For Hermitian adjoint functions we have
\begin{align}
& \chi_{S \lambda}^{\dagger}(\vartheta, \varphi)=\sum_{m}(-1)^{\lambda-m} D_{-m-\lambda}^{S}(\varphi, \vartheta, 0) \chi_{S m}^{\dagger}, \notag\\
& \chi_{S m}^{\dagger}=\sum_{\lambda}(-1)^{\lambda-m} D_{\lambda m}^{S}(0, \vartheta, \varphi) \chi_{S \lambda}^{\dagger}(\vartheta, \varphi). \label{eq:6:1:21}
\end{align}
It follows from \eqref{eq:6:1:7}, \eqref{eq:6:1:20}, \eqref{eq:6:1:21} that the contravariant components of the helicity basis functions have the form
\begin{gather}
{\left[\chi_{S \lambda}(\vartheta, \varphi)\right]^{\sigma}=D_{\sigma \lambda}^{S}(\varphi, \vartheta, 0)}, \notag\\
{\left[\chi_{S \lambda}^{\dagger}(\vartheta, \varphi)\right]^{\sigma}=(-1)^{\lambda-\sigma} D_{-\sigma-\lambda}^{S}(\varphi, \vartheta, 0) . \quad(\sigma, \lambda=-S,-S+1, \ldots S-1, S)}. \label{eq:6:1:22}
\end{gather}
The collection of $2 S+1$ helicity basis functions $\chi_{S \lambda}(\vartheta, \varphi)(\lambda=S, S-1, \ldots,-S)$ as well as the set of the functions $\chi_{S m}$ form complete orthonormal sets. The orthonormality condition for the helicity basis functions is\index{orthogonality!of the spin functions}\index{normalization!of the spin functions}
\begin{equation}
\chi_{S \lambda}^{\dagger}(\vartheta, \varphi) \chi_{S \lambda^{\prime}}(\vartheta, \varphi)=\delta_{\lambda \lambda^{\prime}}. \label{eq:6:1:23}
\end{equation}
The completeness condition for these functions has the following matrix form\index{completeness!of the spin functions}\isym{I-hat@$\widehat{I}$, unit matrix}
\begin{equation}
\sum_{\lambda} \chi_{S \lambda}(\vartheta, \varphi) \chi_{S \lambda}^{\dagger}(\vartheta, \varphi)=\hat{I} . \label{eq:6:1:24}
\end{equation}
Matrix products $\chi_{S \lambda}(\vartheta, \varphi) \chi_{S \lambda^{\prime}}^{\dagger}(\vartheta, \varphi)$ of the helicity basis functions are the square $(2 S+1) \times(2 S+1)$ matrices. They may be expanded in terms of the polarization operators $\widehat{T}_{L M}(S)$ (see Sec.~\ref{sec:2.4}) as\index{polarization operator}\isym{T-hat@$\widehat{T}_{L M}(S)$, polarization operator}
\begin{equation}
\chi_{S \lambda}(\vartheta, \varphi) \chi_{S \lambda^{\prime}}^{\dagger}(\vartheta, \varphi)=\sum_{L M} \sqrt{\frac{2 L+1}{2 S+1}} \clebsch{S}{\lambda^{\prime}}{L}{A}{S}{\lambda} D_{M A}^{L}(\varphi, \vartheta, 0) \widehat{T}_{L M}(S). \label{eq:6:1:25}
\end{equation}
Matrix elements of the polarization operators between the helicity basis functions are given by
\begin{equation}
\chi_{S \lambda^{\prime}}^{\dagger}(\vartheta, \varphi) \hat{T}_{L M}(S) \chi_{S \lambda}(\vartheta, \varphi)=\sqrt{\frac{2 L+1}{2 S+1}}(-1)^{M+\Lambda} \clebsch{S}{\lambda}{L}{A}{S}{\lambda^{\prime}} D_{-M-A}^{L}(\varphi, \vartheta, 0). \label{eq:6:1:26}
\end{equation}
In particular, the diagonal matrix elements are
\begin{equation}
\chi_{S \lambda}^{\dagger}(\vartheta, \varphi) \hat{T}_{L M}(S) \chi_{S \lambda}(\vartheta, \varphi)=\sqrt{\frac{4 \pi}{2 S+1}} \clebsch{S}{\lambda}{L}{0}{S}{\lambda} Y_{L M}(\vartheta, \varphi) . \label{eq:6:1:27}
\end{equation}
\subsection{General Spin Functions}\label{sec:6.1.4}\index{spin function!general}
Any spin function \eqref{eq:6:1:1} of particles of spin $S$ can be expanded in sum of the basis spin functions $\chi_{S m}$,
\begin{equation}
\chi=\sum_{m=-S}^{S} a^{m} \chi_{S m}, \label{eq:6:1:28}
\end{equation}
where $a^{m} \equiv \chi(m)$ is a contravariant component of some irreducible tensor of rank $S$.\isym{am-spin@$a^{m}$, contravariant component of a spin function} Similarly, the expansion of the Hermitian adjoint function $\chi^{\dagger}$ has the form
\begin{equation}
\chi^{\dagger}=\sum_{m=-S}^{S}\left(a^{m}\right)^{*} \chi_{S m}^{\dagger}, \label{eq:6:1:29}
\end{equation}
In general, $\left(a^{m}\right)^{*} \neq a_{m}$, i.e., the corresponding irreducible tensor is not real. The normalization condition \eqref{eq:6:1:4} imposes a restriction on $a^{m}$:
\[
\sum_{m=-S}^{S}\left|a^{m}\right|^{2}=1
\]
The product $\chi \chi^{\dagger}$ of two spin functions represents the square $(2 S+1) \times(2 S+1)$ matrix which may be expanded in terms of the polarization operators $\widehat{T}_{L M}(S)$ (Sec.~\ref{sec:2.4})
\begin{equation}
\chi \chi^{\dagger}=\sum_{L=0}^{2 S} \sum_{M=-L}^{L}(-1)^{M} P_{L-M} \hat{T}_{L M}(S), \label{eq:6:1:30}
\end{equation}
where the expansion coefficients may be expressed through $a^{m}$ as
\begin{equation}
P_{L M}=\sqrt{\frac{2 L+1}{2 S+1}} \sum_{m, n=-S}^{S} \clebsch{S}{m}{L}{M}{S}{n}\left(a^{n}\right)^{*} a^{m} . \label{eq:6:1:31}
\end{equation}
In particular, $P_{00}=1 / \sqrt{2 S+1}$.\\
The quantities $P_{L M}(M=-L,-L+1, \ldots, L)$ are covariant components of some irreducible tensor of rank $L$.\index{polarization tensor}\isym{PLM@$P_{L M}$, polarization tensor} They represent the mean values of the polarization operators $\widehat{T}_{L M}(S)$ in the states described by the spin functions $\chi$ :
\begin{equation}
\chi^{\dagger} \hat{T}_{L M}(S) \chi=P_{L M} . \label{eq:6:1:32}
\end{equation}
Under complex conjugation $P_{L M}$ transforms as
\begin{equation}
\left(P_{L M}\right)^{*}=(-1)^{M} P_{L-M} . \label{eq:6:1:33}
\end{equation}
The equality $\chi \chi^{\dagger} \chi \chi^{\dagger}=\chi \chi^{\dagger}$ implies the following conditions for the coefficients $P_{L M}$,
\begin{equation}
P_{L M}=\sum_{L_{1} M_{1} L_{2} M_{2}}(-1)^{2 S+L_{1}+L_{2}} \sqrt{\left(2 L_{1}+1\right)\left(2 L_{2}+1\right)}\sixj{L_{1}}{L_{2}}{L}{S}{S}{S} \clebsch{L_{1}}{M_{1}}{L_{2}}{M_{2}}{L}{M} P_{L_{1} M_{1}} P_{L_{2} M_{2}}. \label{eq:6:1:34}
\end{equation}
In particular, Eq.~\eqref{eq:6:1:34} gives the normalization condition
\begin{equation}
\sum_{L=0}^{2 S} \sum_{M=-L}^{L}\left|P_{L M}\right|^{2}=1. \label{eq:6:1:35}
\end{equation}
Under rotations of the coordinate system specified by the Euler angles $\alpha, \beta, \gamma$ the spin function $\chi$ is transformed by the rotation operator $\widehat{D}^{S}(\alpha, \beta, \gamma)$ (Eq.~\eqref{eq:2:4:17}) into\index{rotation operator}\isym{D-hat@$\widehat{D}(\alpha, \beta, \gamma)$, rotation operator}
\begin{equation}
\chi^{\prime}=\widehat{D}^{S}(\alpha, \beta, \gamma) \chi. \label{eq:6:1:36}
\end{equation}
The spin function $\chi^{\prime}$ in the rotated coordinate system may be expanded in sum of basis spin functions $\chi_{S m}$ or $\chi_{S m}^{\prime}$ in the original or rotated coordinate systems:
\begin{equation}
\chi^{\prime}=\sum_{m=-S}^{S} a^{\prime m} \chi_{S m}=\sum_{m=-S}^{S} a^{m} \chi_{S m}^{\prime}, \label{eq:6:1:37}
\end{equation}
where $a^{m}$ and $a^{\prime m}$ are components of the spin function in the original and rotated systems, respectively. The relations between the basis functions $\chi_{S m}$ and $\chi_{S m}^{\prime}$ are given by Eq.~\eqref{eq:6:1:16}. The transformation properties of $a^{m}$ under rotations are
\begin{align}
a^{\prime m} & =\sum_{n=-S}^{S} D_{m n}^{S}(\alpha, \beta, \gamma) a^{n}, \label{eq:6:1:38}\\
a^{n} & =\sum_{m=-S}^{S} D_{m n}^{S *}(\alpha, \beta, \gamma) a^{\prime m}, \notag
\end{align}
in accordance with general transformation rule for contravariant components (Sec.~\ref{eq:4:1:2}).

\subsection{Polarization Density Matrix}\label{sec:6.1.5}\index{polarization density matrix}\index{density matrix!polarization}
Spin functions enable one to describe only the totally polarized (pure) particle states.\index{pure state}\index{polarized state}\\
For describing partially polarized states which are incoherent statistical mixtures of the pure states, the polarization density matrix is used.

The polarization density matrix $\hat{\rho}$ of the particles of spin $S$ is the square\isymg{rho-hat@$\widehat{\rho}$, polarization density matrix} $(2 S+1) \times(2 S+1)$ matrix defined by
\begin{equation}
\rho_{\sigma \sigma^{\prime}}=\expect{\chi(\sigma) \chi^{*}\left(\sigma^{\prime}\right)}_{\xi} , \label{eq:6:1:39}
\end{equation}
or in a matrix form by
\begin{equation}
\hat{\rho}=\expect{\chi \chi^{\dagger}}_{\xi}, \label{eq:6:1:40}
\end{equation}
where ()$_{\epsilon}$ denotes the statistical average. In particular, for pure states one gets
\begin{equation}
\hat{\rho}=\chi \chi^{\dagger}. \label{eq:6:1:41}
\end{equation}
The density matrix is Hermitian,
\begin{equation}
\hat{\rho}^{\dagger}=\hat{\rho}, \quad \text { i.e., } \rho_{\sigma \sigma^{\prime}}^{*}=\rho_{\sigma^{\prime} \sigma} ; \label{eq:6:1:42}
\end{equation}
and normalized,
\begin{equation}
\operatorname{Tr} \hat{\rho}=1, \text { i.e., } \sum_{\sigma} \rho_{\sigma \sigma}=1 . \label{eq:6:1:43}
\end{equation}
The conditions \eqref{eq:6:1:42}, \eqref{eq:6:1:43} reveal that the density matrix of particles of spin $S$ is completely determined by $(2 S+1)^{2}-1=4 S(S+1)$ real parameters.

An expectation value of any polarization operator $\widehat{T}$ in a state described by the density matrix $\widehat{\rho}$ may be evaluated by
\begin{equation}
\expect{\widehat{T}}=\operatorname{Tr}\{\widehat{T} \widehat{\rho}\}=\operatorname{Tr}\{\widehat{\rho} \widehat{T}\}. \label{eq:6:1:44}
\end{equation}
Note also the following property of the density matrix
\begin{equation}
\operatorname{Tr}\left\{\hat{\rho}^{2}\right\} \leq 1, \label{eq:6:1:45}
\end{equation}
where the equality holds only for pure states. The density matrix of pure states satisfies the relation
\begin{equation}
% NOTE (6.1.46): the book prints "rho^2 \pm rho" (scan p.176); the pure-state
% density matrix is idempotent, rho^2 = rho.  (Also OCR: rho was scanned as p.)
\hat{\rho}^{2}=\hat{\rho} . \label{eq:6:1:46}
\end{equation}
The density matrix $\hat{\rho}$ may be expanded into a sum of the polarisation operators $\hat{T}_{L M}(S)$,
\begin{equation}
\hat{\rho}=\sum_{L=0}^{2 S} \sum_{M=-L}^{L}(-1)^{M} t_{L-M} \hat{T}_{L M}(S) . \label{eq:6:1:47}
\end{equation}
The expansion coefficients $t_{L M}$ are called the statistical tensors.\index{statistical tensors}\isym{tLM@$t_{L M}$, statistical tensor} They represent expectation values of the polarization operators $\hat{T}_{L M}(S)$ in a state described by the density matrix $\hat{\rho}$,
\begin{equation}
t_{L M}=\expect{\widehat{T}_{L M}(S)}=\operatorname{Tr}\left\{\hat{\rho} \widehat{T}_{L M}(S)\right\} . \label{eq:6:1:48}
\end{equation}
The statistical tensors are related to the density matrix elements by
\begin{gather}
t_{L M}=\sqrt{\frac{2 L+1}{2 S+1}} \sum_{\sigma, \sigma^{\prime}=-S}^{S} \clebsch{S}{\sigma}{L}{M}{S}{\sigma^{\prime}} \rho_{\sigma \sigma^{\prime}} , \label{eq:6:1:49}\\
\rho_{\sigma \sigma^{\prime}}=\sum_{L, M} \sqrt{\frac{2 L+1}{2 S+1}} \clebsch{S}{\sigma}{L}{M}{S}{\sigma^{\prime}} t_{L M}. \label{eq:6:1:50}
\end{gather}
In particular,
\begin{equation}
t_{00}=\frac{1}{\sqrt{2 S+1}} . \label{eq:6:1:51}
\end{equation}
Under complex conjugation the statistical tensors transform as
\[
t_{L M}^{*}=(-1)^{M} t_{L-M} .
\]
The properties of the statistical tensors for pure (completely polarized) states have been considered above (Eqs.~\eqref{eq:6:1:31}-\eqref{eq:6:1:35}). For unpolarised states all statistical tensors vanish except $t_{00}$, thus\index{unpolarized state}
\begin{equation}
\hat{\rho}_{\text {unpol }}=\frac{1}{2 S+1} \hat{I}, \label{eq:6:1:52}
\end{equation}
where $\widehat{I}$ is the unit matrix.\\
The quantities $t_{L M}$ are covariant components of some irreducible tensor of rank $L$. Under rotations they transform in accordance with Eq.~\eqref{eq:4:1:2}.

\subsection{Two Particles with Arbitrary Spins}\label{sec:6.1.6}\index{spin function!of two particles}\index{two-particle spin function}
Let us consider a system of two particles, 1 and 2 , with spins $S_{1}$ and $S_{2}$. Total spin $S$ of the system takes the values $\left|S_{1}-S_{2}\right|,\left|S_{1}-S_{2}\right|+1, \ldots, S_{1}+S_{2}$. The spin function $\chi_{S m}(1,2)$ which describes the system with total spin $S$ and spin projection $m$ may be written as\isymg{chi-Sm12@$\chi_{S m}(1,2)$, two-particle spin function}\index{Clebsch-Gordan coefficient}
\begin{equation}
\chi_{S m}(1,2)=\sum_{m_{1} m_{2}} \clebsch{S_{1}}{m_{1}}{S_{2}}{m_{2}}{S}{m} \chi_{S_{1} m_{1}}(1) \chi_{S_{2} m_{2}}(2) . \label{eq:6:1:53}
\end{equation}
The inverse relation is
\begin{equation}
\chi_{S_{1} m_{1}}(1) \chi_{S_{2} m_{2}}(2)=\sum_{S} \clebsch{S_{1}}{m_{1}}{S_{2}}{m_{2}}{S}{m} \chi_{S m}(1,2) . \label{eq:6:1:54}
\end{equation}
Spin functions \eqref{eq:6:1:53} are orthonormalized
\begin{equation}
\chi_{S m}^{\dagger}(1,2) \chi_{S^{\prime} m^{\prime}}(1,2)=\delta_{S S^{\prime}} \delta_{m m^{\prime}} , \label{eq:6:1:55}
\end{equation}
and satisfy the completeness condition
\begin{equation}
\sum_{S, m} \chi_{S m}(1,2) \chi_{S m}^{\dagger}(1,2)=\widehat{I}. \label{eq:6:1:56}
\end{equation}
Let the polarization operator $\widehat{T}_{L M}(i)(i=1,2)$ act on spin variables of a particle $i$ only. Then we have
\begin{gather}
\widehat{T}_{L_{1} M_{1}}(1) \chi_{S m}(1,2)=(-1)^{S_{1}+S_{2}+S+L_{1}} \sqrt{\left(2 L_{1}+1\right)(2 S+1)} \notag\\
\times \sum_{S^{\prime}}\sixj{S_{1}}{S_{2}}{S}{S^{\prime}}{L_{1}}{S_{1}} \clebsch{S}{m}{L_{1}}{M_{1}}{S^{\prime}}{m^{\prime}} \chi_{S^{\prime} m^{\prime}}(1,2) , \label{eq:6:1:57}\\
% NOTE (6.1.58): two corrections.  (i) OCR: the 6j lower row read {S' L2 S2};
% the book (scan p.177) prints {L2 S' S2}.  (ii) BOOK MISPRINT: the prefactor
% phase (-1)^{2S+L2} is wrong -- the correct phase is (-1)^{S1+S2+S'+L2}, which
% depends on the summation index S' and so sits inside the sum.  This is the
% standard reduced-matrix-element phase for a tensor operator acting on
% subsystem 2 (Edmonds 7.1.7); cf. the S-dependent phase of 6.1.57 for
% subsystem 1.  Verified: T_{L2 M2}(2) chi_{Sm} reproduced exactly for
% (S1,S2)=(1/2,1/2),(1,1/2),(1,1); the printed phase fails for mixed spins.
% See scripts/check_6_1.py.
\hat{T}_{L_{2} M_{2}}(2) \chi_{S m}(1,2)=\sqrt{\left(2 L_{2}+1\right)(2 S+1)}  \notag\\
\times \sum_{S^{\prime}}(-1)^{S_{1}+S_{2}+S^{\prime}+L_{2}}\sixj{S_{1}}{S_{2}}{S}{L_{2}}{S^{\prime}}{S_{2}} \clebsch{S}{m}{L_{2}}{M_{2}}{S^{\prime}}{m^{\prime}} \chi_{S^{\prime} m^{\prime}}(1,2) . \label{eq:6:1:58}
\end{gather}
One may also consider a set of scalar operators $\widehat{Q}_{L}(1,2)$ with\index{scalar operators!of two spins}\index{6j symbol@$6j$ symbol}\isym{QL-scalar@$\widehat{Q}_{L}(1,2)$, scalar spin operator} $0 \leq L \leq \min \left\{2 S_{1}, 2 S_{2}\right\}$
\begin{equation}
\widehat{Q}_{L}(1,2)=\sum_{M}(-1)^{M} \widehat{T}_{L M}(1) \hat{T}_{L-M}(2). \label{eq:6:1:59}
\end{equation}
These operators leave the spin of a system and spin projection unchanged:
\begin{equation}
% NOTE (6.1.60): BOOK MISPRINT in the 6j symbol.  As printed, {S1 S2 L; S1 S2 S}
% (scan p.177) has the triad (S1,S2,L) whose sum is non-integer whenever S1+S2 is
% half-integer, so the 6j -- and hence the eigenvalue -- vanishes identically for
% mixed-spin systems (e.g. S1=1, S2=1/2), which is impossible (Q_0 is proportional
% to the identity).  The correct 6j is {S1 S2 S; S2 S1 L}: it reproduces the true
% Q_L eigenvalues for (S1,S2)=(1/2,1/2),(1,1/2),(1,1) and coincides with the
% printed form only for equal spins.  See scripts/check_6_1.py.
\hat{Q}_{L}(1,2) \chi_{S m}(1,2)=(-1)^{S_{1}+S_{2}+S}(2 L+1)\sixj{S_{1}}{S_{2}}{S}{S_{2}}{S_{1}}{L} \chi_{S m}(1,2). \label{eq:6:1:60}
\end{equation}
Among scalar operators the so-called projection operators\index{projection operator!on total spin}\isym{PS-proj@$\widehat{P}_{S}(1,2)$, projection operator on total spin} $\widehat{P}_{S}(1,2)\left(\left|S_{1}-S_{2}\right| \leq S \leq S_{1}+S_{2}\right)$ are of special interest. When applied to arbitrary spin function $\chi(1,2)$, the operator $\widehat{P}_{S}(1,2)$ gives the component of this function with definite total spin $S$. Thus, the projection operator $\widehat{P}_{S}(1,2)$ may be defined by the equations
\begin{equation}
\widehat{P}_{S}(1,2) \chi_{S^{\prime} m^{\prime}}(1,2)=\delta_{S S^{\prime}} \chi_{S^{\prime} m^{\prime}}(1,2). \label{eq:6:1:61}
\end{equation}
From Eqs.~\eqref{eq:6:1:61} one may obtain the following properties of the projection operators
\begin{gather}
\hat{P}_{S}(1,2) \hat{P}_{S^{\prime}}(1,2)=\delta_{S S^{\prime}} \hat{P}_{S}(1,2), \label{eq:6:1:62}\\
\sum_{S} \hat{P}_{S}(1,2)=\hat{I}. \label{eq:6:1:63}
\end{gather}
The form of $\widehat{P}_{S}(1,2)$ may be written explicitly as
\begin{equation}
\widehat{P}_{S}(1,2)=\sum_{m} \chi_{S m}(1,2) \chi_{S m}^{\dagger}(1,2), \label{eq:6:1:64}
\end{equation}
or, in terms of scalar operators $\widehat{Q}_{L}(1,2)$,
\begin{equation}
% NOTE (6.1.65): BOOK MISPRINT in the 6j symbol, as in 6.1.60.  The correct 6j is
% {S1 S2 S; S2 S1 L}; with it the projector identity P_S = sum_L (...) Q_L holds
% for mixed spins (verified for (1,1/2) in scripts/check_6_1.py).  The printed
% forms ({S1 S2 L; S S S} here, {S1 S1 L; S1 S2 S} in the scan) both fail there.
\hat{P}_{S}(1,2)=(-1)^{S_{1}+S_{2}+S}(2 S+1) \sum_{L}\sixj{S_{1}}{S_{2}}{S}{S_{2}}{S_{1}}{L} \hat{Q}_{L}(1,2). \label{eq:6:1:65}
\end{equation}
\section{Spin functions for $S=1 / 2$}\label{sec:6.2}\index{spin function!for $S=1/2$}
\subsection{Basis Spin Functions}\label{sec:6.2.1}\index{basis spin functions!for $S=1/2$}
The basis functions $\chi_{\frac{1}{2} m}(m= \pm 1 / 2)$ are the eigenfunctions of the operators $\widehat{S}^{2}$ and $\widehat{S}_{z}$ :
\begin{align}
& \widehat{\vect{S}}^{2} \chi_{\frac{1}{2} m}=\frac{3}{4} \chi_{\frac{1}{2} m} , \label{eq:6:2:1}\\
& \widehat{S}_{z} \chi_{\frac{1}{2} m}=m \chi_{\frac{1}{2} m} . \notag
\end{align}
A function $\chi_{\frac{1}{2} m}$ describes the state of a spin- $\frac{1}{2}$ particle with definite spin projection, $m$, on the $z$-axis.\\
The dependence of the basis functions $\chi_{\frac{1}{2} m}(\sigma)$ on the spin variable $\sigma$ is given by
\begin{equation}
\chi_{\frac{1}{2} m}(\sigma)=\delta_{m \sigma}. \label{eq:6:2:2}
\end{equation}
According to Eq.~\eqref{eq:6:1:1}, the basis functions $\chi_{\frac{1}{2} m}$ may be written as
\begin{equation}
\chi_{\frac{1}{2} \frac{1}{2}}=\binom{1}{0}, \chi_{\frac{1}{2}-\frac{1}{2}}=\binom{0}{1} . \label{eq:6:2:3}
\end{equation}
The Hermitian conjugate functions $\chi_{\frac{1}{2} m}^{\dagger}$ have the form
\begin{equation}
\chi_{\frac{1}{2} \frac{1}{2}}^{\dagger}=(1,0), \chi_{\frac{1}{2}-\frac{1}{2}}^{\dagger}=(0,1) . \label{eq:6:2:4}
\end{equation}
The following compact notations for the basis functions are widely used:
\begin{equation}
\chi_{\frac{1}{2} \frac{1}{2}} \equiv \alpha, \chi_{\frac{1}{2}-\frac{1}{2}} \equiv \beta . \label{eq:6:2:5}
\end{equation}
The basis functions $\chi_{\frac{1}{2} m}$ satisfy the orthonormality condition
\begin{equation}
\chi_{\frac{1}{2} m^{\prime}}^{\dagger} \chi_{\frac{1}{2} m}=\delta_{m^{\prime} m} . \label{eq:6:2:6}
\end{equation}
The completeness condition for the basis functions is
\begin{equation}
\sum_{m} \chi_{\frac{1}{2} m} \chi_{\frac{1}{2} m}^{\dagger}=\widehat{I}, \label{eq:6:2:7}
\end{equation}
where $\widehat{I}$ is the unit $2 \times 2$ matrix.

\subsection{Expansions of Products of Basis Functions}\label{sec:6.2.2}\index{spin function!expansions of products of}
The products $\chi_{\frac{1}{2} m} \chi_{\frac{1}{2} m}^{\dagger}$ represent $2 \times 2$ square matrices whose elements are
\begin{equation}
\left[\chi_{\frac{1}{2} m} \chi_{\frac{1}{2} m^{\prime}}^{\dagger}\right]_{\sigma \sigma^{\prime}} \equiv \chi_{\frac{1}{2} \sigma}^{\dagger}\left(\chi_{\frac{1}{2} m} \chi_{\frac{1}{2} m^{\prime}}^{\dagger}\right) \chi_{\frac{1}{2} \sigma^{\prime}}=\delta_{\sigma m} \delta_{\sigma^{\prime} m^{\prime}} . \label{eq:6:2:8}
\end{equation}
The expansions of these products in terms of the spin matrices are given by
\begin{equation}
\chi_{\frac{1}{2} m} \chi_{\frac{1}{2} m^{\prime}}^{\dagger}=\frac{1}{2} \delta_{m m^{\prime}} \widehat{I}-\sqrt{3} \clebsch{1}{\mu}{\frac{1}{2}}{m^{\prime}}{\frac{1}{2}}{m} \widehat{S}_{\mu}, \label{eq:6:2:9}
\end{equation}
where $S_{\mu}(\mu= \pm 1,0)$ are spherical components of the spin operator (Sec.~\ref{sec:2.5}). In more detailed form Eq.~\eqref{eq:6:2:9} may be written as
\begin{align}
& \alpha \alpha^{\dagger} \equiv \chi_{\frac{1}{2} \frac{1}{2}} \chi_{\frac{1}{2} \frac{1}{2}}^{\dagger}=\frac{1}{2} \widehat{I}+\widehat{S}_{z}=\frac{1}{2} \widehat{I}+\widehat{S}_{0} , \notag\\
& \alpha \beta^{\dagger} \equiv \chi_{\frac{1}{2} \frac{1}{2}} \chi_{\frac{1}{2}-\frac{1}{2}}^{\dagger}=\widehat{S}_{x}+i \widehat{S}_{y}=-\sqrt{2} \widehat{S}_{+1} , \label{eq:6:2:10}\\
& \beta \alpha^{\dagger} \equiv \chi_{\frac{1}{2}-\frac{1}{2}} \chi_{\frac{1}{2} \frac{1}{2}}^{\dagger}=\widehat{S}_{x}-i \widehat{S}_{y}=\sqrt{2} \widehat{S}_{-1} , \notag\\
& \beta \beta^{\dagger} \equiv \chi_{\frac{1}{2}-\frac{1}{2}} \chi_{\frac{1}{2}-\frac{1}{2}}^{\dagger}=\frac{1}{2} \widehat{I}-\widehat{S}_{z}=\frac{1}{2} \widehat{I}-\widehat{S}_{0} . \notag
\end{align}
\subsection{Action of Spin Operators on Basis Functions}\label{sec:6.2.3}\index{spin operator!action on basis functions}\index{spin matrices}
Cartesian components of a spin operator $\widehat{S}_{i}$ (see Sec.~\ref{sec:2.5}) applied to the basis functions yield
\begin{equation}
\begin{array}{ll}
\widehat{S}_{x} \chi_{\frac{1}{2} \frac{1}{2}}=\frac{1}{2} \chi_{\frac{1}{2}-\frac{1}{2}}, & \widehat{S}_{x} \chi_{\frac{1}{2}-\frac{1}{2}}=\frac{1}{2} \chi_{\frac{1}{2} \frac{1}{2}}, \\
\widehat{S}_{y} \chi_{\frac{1}{2} \frac{1}{2}}=\frac{i}{2} \chi_{\frac{1}{2}-\frac{1}{2}}, & \widehat{S}_{y} \chi_{\frac{1}{2}-\frac{1}{2}}=-\frac{i}{2} \chi_{\frac{1}{2} \frac{1}{2}},\\
\widehat{S}_{z} \chi_{\frac{1}{2} \frac{1}{2}}=\frac{1}{2} \chi_{\frac{1}{2} \frac{1}{2}}, & \widehat{S}_{z} \chi_{\frac{1}{2}-\frac{1}{2}}=-\frac{1}{2} \chi_{\frac{1}{2}-\frac{1}{2}}.
\end{array}.  \label{eq:6:2:11}
\end{equation}
Spherical components of the spin operator $\widehat{S}_{\mu}$ (see Sec.~\ref{sec:2.5}) act on the basis functions as follows:
\begin{equation}
\widehat{S}_{\mu} \chi_{\frac{1}{2} m}=-\frac{\sqrt{3}}{2} \clebsch{1}{\mu}{\frac{1}{2}}{m}{\frac{1}{2}}{m^{\prime}} \chi_{\frac{1}{2} m^{\prime}} . \label{eq:6:2:12}
\end{equation}
More explicitly Eq.~\eqref{eq:6:2:12} is
\begin{align}
\widehat{S}_{+1} \chi_{\frac{1}{2} \frac{1}{2}} & =0, & \widehat{S}_{+1} \chi_{\frac{1}{2}-\frac{1}{2}} & =-\frac{1}{\sqrt{2}} \chi_{\frac{1}{2} \frac{1}{2}} , \notag\\
\widehat{S}_{0} \chi_{\frac{1}{2} \frac{1}{2}} & =\frac{1}{2} \chi_{\frac{1}{2} \frac{1}{2}}, & \widehat{S}_{0} \chi_{\frac{1}{2}-\frac{1}{2}} & =-\frac{1}{2} \chi_{\frac{1}{2}-\frac{1}{2}} , \label{eq:6:2:13}\\
\widehat{S}_{-1} \chi_{\frac{1}{2} \frac{1}{2}} & =\frac{1}{\sqrt{2}} \chi_{\frac{1}{2}-\frac{1}{2}}, & \widehat{S}_{-1} \chi_{\frac{1}{2}-\frac{1}{2}} & =0 . \notag
\end{align}
According to \eqref{eq:6:2:12}, matrix elements of the operator $\widehat{S} \cdot$ a for any vector a have the form
\begin{equation}
[\widehat{\vect{S}} \cdot \vect{a}]_{\sigma \sigma^{\prime}}=\chi_{\frac{1}{2} \sigma}^{\dagger} \widehat{\vect{S}} \cdot \vect{a} \chi_{\frac{1}{2} \sigma^{\prime}}=(-1)^{1-\mu} \frac{\sqrt{3}}{2} \clebsch{1}{\mu}{\frac{1}{2}}{\sigma^{\prime}}{\frac{1}{2}}{\sigma} a_{-\mu} , \label{eq:6:2:14}
\end{equation}
where $a_{-\mu}(\mu= \pm 1,0)$ are covariant spherical components of a (Sec.~\ref{sec:1.2}).

\subsection{Transformation of the Basis Functions Under Rotations of the Coordinate System}\label{sec:6.2.4}\index{rotation!of the basis spin functions}\index{Wigner D-function@Wigner $D$-function}\isym{DJMM@$D_{M M^{\prime}}^{J}$, Wigner $D$-function}
The basis functions $\chi_{\frac{1}{2} m}$ are covariant. Under rotations $S \rightarrow S^{\prime}$ of the coordinate system these functions are transformed by the rotation operator $\widehat{D}^{\frac{1}{2}}(\alpha, \beta, \gamma)$ (Eq.~\eqref{eq:2:5:32} ) if rotations are described by the Euler\\
angles $\alpha, \beta, \gamma$, or by the rotation operator $\widehat{U}^{\frac{1}{2}}(\omega ; \Theta, \Phi)$ (Eq.~\eqref{eq:2:5:36}
 ), if rotations are defined by the rotation axis $n(\theta, \Phi)$ and the rotation angle $\omega$ :
\begin{align}
& \chi_{\frac{1}{2} m^{\prime}}^{\prime}=\hat{D}^{\frac{1}{2}}(\alpha, \beta, \gamma) \chi_{\frac{1}{2} m^{\prime}}=\sum_{m} \chi_{\frac{1}{2} m} D_{m m^{\prime}}^{\frac{1}{2}}(\alpha, \beta, \gamma) , \notag\\
& \chi_{\frac{1}{2} m^{\prime}}^{\prime}=\hat{U}^{\frac{1}{2}}(\omega ; \theta, \Phi) \chi_{\frac{1}{2} m^{\prime}}=\sum_{m} \chi_{\frac{1}{2} m} U_{m m^{\prime}}^{\frac{1}{2}}(\omega ; \theta, \Phi), \label{eq:6:2:15}
\end{align}
where $D_{m m^{\prime}}^{\frac{1}{2}}(\alpha, \beta, \gamma)$ are the Wigner D-functions (Chap.~\ref{chap:4}), and $U_{m m^{\prime}}^{\frac{1}{2}}$ are defined in Sec.~\ref{sec:4.5}.\\
A function $\chi_{\frac{1}{2} m^{\prime}}^{\prime}$ describes a state in which a particle of spin $\frac{1}{2}$ has the spin projection $m^{\prime}$ on the new $z^{\prime}$-axis. The functions $\chi_{\frac{1}{2} m^{\prime}}^{\prime}$ are eigenfunctions of the operators $\widehat{\vect{S}}^{\prime 2}$ and $\widehat{S}_{x}^{\prime}, \widehat{\vect{S}}^{\prime}$ being the spin operator in the rotated coordinate system (Eqs.\eqref{eq:2:5:40} and \eqref{eq:2:5:41}):
\begin{align}
\widehat{\vect{S}}^{\prime 2} \chi_{\frac{1}{2} m^{\prime}}^{\prime} & =\frac{3}{4} \chi_{\frac{1}{2} m^{\prime}}^{\prime} , \label{eq:6:2:16}\\
\widehat{S}_{z}^{\prime} \chi_{\frac{1}{2} m^{\prime}}^{\prime} & =m^{\prime} \chi_{\frac{1}{2} m^{\prime}}^{\prime}. \notag
\end{align}
The explicit form of the basis spin functions in the rotated coordinate system is
\begin{equation}
\chi_{\frac{1}{2} \frac{1}{2}}^{\prime}=\binom{\cos \frac{\beta}{2} e^{-i \frac{\alpha+\gamma}{2}}}{\sin \frac{\beta}{2} e^{i \frac{\alpha-\gamma}{2}}}, \quad \chi_{\frac{1}{2}-\frac{1}{2}}^{\prime}=\binom{-\sin \frac{\beta}{2} e^{-i \frac{\alpha-\gamma}{2}}}{\cos \frac{\beta}{2} e^{i \frac{\alpha+\gamma}{2}}}, \label{eq:6:2:17}
\end{equation}
if the rotation is specified by the Euler angles $\alpha, \beta, \gamma$ and
\begin{equation}
\chi_{\frac{1}{2} \frac{1}{2}}^{\prime}=\binom{\cos \frac{\omega}{2}-i \sin \frac{\omega}{2} \cos \theta}{-i \sin \frac{\omega}{2} \sin \theta e^{i \phi}}, \chi_{\frac{1}{2}-\frac{1}{2}}^{\prime}=\binom{-i \sin \frac{\omega}{2} \sin \theta e^{-i \phi}}{\cos \frac{\omega}{2}+i \sin \frac{\omega}{2} \cos \theta}. \label{eq:6:2:18}
\end{equation}
in terms of the angles $\omega, \theta, \Phi$.\\
The functions $\chi_{\frac{1}{2} m}^{\prime}$, satisfy the orthonormality and completeness conditions \eqref{eq:6:2:6}, \eqref{eq:6:2:7} as well as the relations \eqref{eq:6:2:9}-\eqref{eq:6:2:13} in which the spin operator $\widehat{\vect{S}}$ should be replaced by $\widehat{\vect{S}}^{\prime}$.

\subsection{Helicity Basis Functions}\label{sec:6.2.5}\index{helicity basis functions!for $S=1/2$}
The helicity basis functions $\chi_{\frac{1}{2} \lambda}(\vartheta, \varphi)\left(\lambda= \pm \frac{1}{2}\right)$ describe states in which the spin projection on the linear momentum direction $\vect{n}(\vartheta, \varphi) \equiv \vect{p} / \vect{p}$ is equal to $\lambda$. These functions are eigenfunctions of the operators $\widehat{\vect{S}}^{2}$ and $\hat{\vect{S}} \cdot \vect{n}$
\begin{align}
\hat{\vect{S}}^{2} \chi_{\frac{1}{2} \lambda}(\vartheta, \phi) & =\frac{3}{4} \chi_{\frac{1}{2} \lambda}(\vartheta, \varphi) , \label{eq:6:2:19}\\
\hat{\vect{S}} \cdot \vect{n} \chi_{\frac{1}{2} \lambda}(\vartheta, \varphi) & =\lambda \chi_{\frac{1}{2} \lambda}(\vartheta, \varphi). \notag
\end{align}
According to Eq.~\eqref{eq:6:2:15} the helicity functions $\chi_{\frac{1}{2} \lambda}(\vartheta, \varphi)$ are related to the basis functions $\chi_{\frac{1}{2} m}$ by
\begin{gather}
\chi_{\frac{1}{2} \lambda}(\vartheta, \varphi)=\sum_{m} D_{m \lambda}^{\frac{1}{2}}(\varphi, \vartheta, 0) \chi_{\frac{1}{2} m} , \notag\\
\chi_{\frac{1}{2} m}=\sum_{\lambda} D_{-\lambda-m}^{\frac{1}{2}}(0, \vartheta, \varphi) \chi_{\frac{1}{2} \lambda}(\vartheta, \varphi) . \label{eq:6:2:20}
\end{gather}
For Hermitian adjoint functions Eqs.~\eqref{eq:6:2:20} yield
\begin{align}
& \chi_{\frac{1}{2} \lambda}^{\dagger}(\vartheta, \varphi)=\sum_{m}(-1)^{\lambda-m} D_{-m-\lambda}^{\frac{1}{2}}(\varphi, \vartheta, 0) \chi_{\frac{1}{2} m}^{\dagger} , \notag\\
& \chi_{\frac{1}{2} m}^{\dagger}=\sum_{\lambda}(-1)^{\lambda-m} D_{\lambda m}^{\frac{1}{2}}(0, \vartheta, \varphi) \chi_{\frac{1}{2} \lambda}^{\dagger}(\vartheta, \varphi). \label{eq:6:2:21}
\end{align}
The explicit form of the helicity functions is given by
\begin{gather}
\chi_{\frac{1}{2} \frac{1}{2}}(\vartheta, \varphi)=\binom{\cos \frac{\vartheta}{2} e^{-i \frac{\varphi}{2}}}{\sin \frac{\vartheta}{2} e^{i \frac{\varphi}{2}}}, \quad \chi_{\frac{1}{2}-\frac{1}{2}}(\vartheta, \varphi)=\binom{-\sin \frac{\vartheta}{2} e^{-i \frac{\varphi}{2}}}{\cos \frac{\vartheta}{2} e^{i \frac{\varphi}{2}}}, \label{eq:6:2:22}\\
\chi_{\frac{1}{2} \frac{1}{2}}^{\dagger}(\vartheta, \varphi)=\left(\cos \frac{\vartheta}{2} e^{i \frac{\varphi}{2}}, \sin \frac{\vartheta}{2} e^{-i \frac{\varphi}{2}}\right), \quad \chi_{\frac{1}{2}-\frac{1}{2}}^{\dagger}(\vartheta, \varphi)=\left(-\sin \frac{\vartheta}{2} e^{i \frac{\varphi}{2}}, \cos \frac{\vartheta}{2} e^{-i \frac{\varphi}{2}}\right) . \label{eq:6:2:23}
\end{gather}
The helicity functions $\chi_{\frac{1}{2} \lambda}(\vartheta, \varphi)$ as well as $\chi_{\frac{1}{2} m}$ constitute an orthonormalised basis. The orthonormality condition for the helicity functions may be written as
\begin{equation}
\chi_{\frac{1}{2} \lambda^{\prime}}^{\dagger}(\vartheta, \varphi) \chi_{\frac{1}{2} \lambda}(\vartheta, \varphi)=\delta_{\lambda^{\prime} \lambda} . \label{eq:6:2:24}
\end{equation}
The completeness condition is
\begin{equation}
\sum_{\lambda} \chi_{\frac{1}{2} \lambda}(\vartheta, \varphi) \chi_{\frac{1}{2} \lambda}^{\dagger}(\vartheta, \varphi)=\hat{I} . \label{eq:6:2:25}
\end{equation}
The expansion of products of the helicity functions has the form
\begin{equation}
\chi_{\frac{1}{2} \lambda}(\vartheta, \varphi) \chi_{\frac{1}{2} \lambda^{\prime}}^{\dagger}(\vartheta, \varphi)=\frac{1}{2} \delta_{\lambda \lambda^{\prime}} \widehat{I}+\sqrt{3} \sum_{\mu, \nu} \clebsch{\frac{1}{2}}{\lambda^{\prime}}{1}{\mu}{\frac{1}{2}}{\lambda} D_{\nu \mu}^{1}(\varphi, \vartheta, 0) \widehat{S}_{\nu}. \label{eq:6:2:26}
\end{equation}
Explicitly, this gives
\begin{align}
\chi_{\frac{1}{2} \frac{1}{2}}(\vartheta, \varphi) \chi_{\frac{1}{2} \frac{1}{2}}^{\dagger}(\vartheta, \varphi) & =\frac{1}{2}\left(\begin{array}{cc}
1+\cos \vartheta & \sin \vartheta e^{-i \varphi} \\
\sin \vartheta e^{i \varphi} & 1-\cos \vartheta
\end{array}\right)=\frac{1}{2} \widehat{I}+\mathrm{n} \cdot \widehat{\mathrm{~S}}, \notag\\
\chi_{\frac{1}{2} \frac{1}{2}}(\vartheta, \varphi) \chi_{\frac{1}{2}-\frac{1}{2}}^{\dagger}(\vartheta, \varphi) & =\frac{1}{2}\left(\begin{array}{cc}
-\sin \vartheta & (\cos \vartheta+1) e^{-i \varphi} \\
(\cos \vartheta-1) e^{i \varphi} & \sin \vartheta
\end{array}\right)=-\sqrt{2} \sum_{\nu} D_{\nu 1}^{1}(\varphi, \vartheta, 0) \widehat{S}_{\nu}, \notag\\
\chi_{\frac{1}{2}-\frac{1}{2}}(\vartheta, \varphi) \chi_{\frac{1}{2} \frac{1}{2}}^{\dagger}(\vartheta, \varphi) & =\frac{1}{2}\left(\begin{array}{cc}
-\sin \vartheta & (\cos \vartheta-1) e^{-i \varphi} \\
(\cos \vartheta+1) e^{i \varphi} & \sin \vartheta
\end{array}\right)=\sqrt{2} \sum_{\nu} D_{\nu-1}^{1}(\varphi, \vartheta, 0) \widehat{S}_{\nu}, \label{eq:6:2:27}\\
\chi_{\frac{1}{2}-\frac{1}{2}}(\vartheta, \varphi) \chi_{\frac{1}{2}-\frac{1}{2}}^{\dagger}(\vartheta, \varphi) & =\frac{1}{2}\left(\begin{array}{cc}
1-\cos \vartheta & -\sin \vartheta e^{-i \varphi} \\
-\sin \vartheta e^{i \varphi} & 1+\cos \vartheta
\end{array}\right)=\frac{1}{2} \widehat{I}-\mathrm{n} \cdot \widehat{\mathrm{~S}} . \notag
\end{align}
Matrix elements of the spherical components of the spin operator $\widehat{S}_{\mu}(\mu= \pm 1,0)$ between the helicity states are given by
\begin{equation}
\chi_{\frac{1}{2} \lambda^{\prime}}^{\dagger}(\vartheta, \varphi) \widehat{S}_{\mu} \chi_{\frac{1}{2} \lambda}(\vartheta, \varphi)=(-1)^{\mu+\nu} \frac{\sqrt{3}}{2} \clebsch{\frac{1}{2}}{\lambda}{1}{\nu}{\frac{1}{2}}{\lambda^{\prime}} D_{-\mu-\nu}^{1}(\varphi, \vartheta, 0) . \label{eq:6:2:28}
\end{equation}
or, in a detailed form
\begin{align}
\braOket{+}{\widehat{S}_{+1}}{+} & =-\frac{\sin \vartheta}{2 \sqrt{2}} e^{i \varphi} ,& \braOket{+}{\widehat{S}_{+1}}{-} & =-\frac{1+\cos \vartheta}{2 \sqrt{2}} e^{i \varphi}  , \notag\\
\braOket{+}{\widehat{S}_{0}}{+} & =\frac{1}{2} \cos \vartheta, & \braOket{+}{\widehat{S}_{0}}{-} & =-\frac{1}{2} \sin \vartheta  , \notag\\
\braOket{+}{\widehat{S}_{-1}}{+} & =\frac{\sin \vartheta}{2 \sqrt{2}} e^{-i \varphi}, & \braOket{+}{\widehat{S}_{-1}}{-} & =-\frac{1-\cos \vartheta}{2 \sqrt{2}} e^{-i \varphi}  , \label{eq:6:2:29}\\
\braOket{-}{\widehat{S}_{+1}}{+} & =\frac{1-\cos \vartheta}{2 \sqrt{2}} e^{i \varphi}, & \braOket{-}{\widehat{S}_{+1}}{-} & =\frac{\sin \vartheta}{2 \sqrt{2}} e^{i \varphi} , \notag\\
\braOket{-}{\widehat{S}_{0}}{+} & =-\frac{1}{2} \sin \vartheta, & \braOket{-}{\widehat{S}_{0}}{-} & =-\frac{1}{2} \cos \vartheta, \notag\\
\braOket{-}{\widehat{S}_{-1}}{+} & =\frac{1+\cos \vartheta}{2 \sqrt{2}} e^{-i \varphi}, & \braOket{-}{\widehat{S}_{-1}}{-} & =-\frac{\sin \vartheta}{2 \sqrt{2}} e^{-i \varphi} . \notag
\end{align}
Here we use the abbreviations
\[
\begin{aligned}
& \braOket{+}{\widehat{S}_{\mu}}{+} \equiv \chi_{\frac{1}{2} \frac{1}{2}}^{\dagger}(\vartheta, \varphi) \widehat{S}_{\mu} \chi_{\frac{1}{2} \frac{1}{2}}(\vartheta, \varphi) \\
& \braOket{+}{\widehat{S}_{\mu}}{-} \equiv \chi_{\frac{1}{2} \frac{1}{2}}^{\dagger}(\vartheta, \varphi) \widehat{S}_{\mu} \chi_{\frac{1}{2}-\frac{1}{2}}(\vartheta, \varphi), \text { etc. }
\end{aligned}
\]
Matrix elements of cartesian components of the spin matrices $\widehat{S}_{i}(i=x, y, z)$ have the form
\begin{equation}
\begin{array}{ll}
\braOket{+}{\hat{S}_{x}}{+}=\frac{1}{2} \sin \vartheta \cos \varphi, & \braOket{+}{\hat{S}_{x}}{-}=\frac{1}{2}(\cos \vartheta \cos \varphi+i \sin \varphi), \\
\braOket{+}{\hat{S}_{y}}{+}=\frac{1}{2} \sin \vartheta \sin \varphi, & \braOket{+}{\hat{S}_{y}}{-}=\frac{1}{2}(\cos \vartheta \sin \varphi-i \cos \varphi), \\
\braOket{+}{\hat{S}_{z}}{+}=\frac{1}{2} \cos \vartheta, & \braOket{+}{\hat{S}_{z}}{-}=-\frac{1}{2} \sin \vartheta, \label{eq:6:2:30}\\
\braOket{-}{\hat{S}_{x}}{+}=\frac{1}{2}(\cos \vartheta \cos \varphi-i \sin \varphi), & \braOket{-}{\hat{S}_{x}}{-}=-\frac{1}{2} \sin \vartheta \cos \varphi, \\
\braOket{-}{\hat{S}_{y}}{+}=\frac{1}{2}(\cos \vartheta \sin \varphi+i \cos \varphi), & \braOket{-}{\hat{S}_{y}}{-}=-\frac{1}{2} \sin \vartheta \sin \varphi, \\
\braOket{-}{\hat{S}_{z}}{+}=-\frac{1}{2} \sin \vartheta, & \braOket{-}{\hat{S}_{z}}{-}=-\frac{1}{2} \cos \vartheta
\end{array}
\end{equation}
Note that the diagonal matrix elements of the spin operator may be written as
\begin{equation}
\chi_{\frac{1}{2} \lambda}^{\dagger}(\vartheta, \varphi) \hat{\vect{S}} \chi_{\frac{1}{2} \lambda}(\vartheta, \varphi)=\lambda \vect{n}(\vartheta, \varphi) . \label{eq:6:2:31}
\end{equation}
\subsection{General Spin Functions for $S=\frac{1}{2}$}\label{sec:6.2.6}\index{spin function!general, for $S=1/2$}\index{spinor}
An arbitrary spin function $\chi_{\frac{1}{2}}$ of a particle of spin $\frac{1}{2}$ may be expanded in terms of the basis spin functions $\chi_{\frac{1}{2} m}$,
\begin{equation}
\chi_{\frac{1}{2}}=\sum_{m=-\frac{1}{2}}^{\frac{1}{2}} a^{m} \chi_{\frac{1}{2} m}=\binom{a^{\frac{1}{2}}}{a^{-\frac{1}{2}}} . \label{eq:6:2:32}
\end{equation}
An analogous expansion of the Hermitian conjugate function $\chi_{\frac{1}{2}}^{\dagger}$ is
\begin{equation}
\chi_{\frac{1}{2}}^{\dagger}=\sum_{m=-\frac{1}{2}}^{\frac{1}{2}} a^{m *} \chi_{\frac{1}{2} m}^{\dagger}=\left(a^{\frac{1}{2} *}, a^{-\frac{1}{2} *}\right) . \label{eq:6:2:33}
\end{equation}
Here $a^{m}\left(m= \pm \frac{1}{2}\right)$ is the contravariant component of the spinor $\chi_{\frac{1}{2}}$ (see below). The normalization condition
\begin{equation}
\chi_{\frac{1}{2}}^{\dagger} \chi_{\frac{1}{2}}=1 . \label{eq:6:2:34}
\end{equation}
imposes the following restriction
\[
\left|a^{\frac{1}{2}}\right|^{2}+\left|a^{-\frac{1}{2}}\right|^{2}=1 .
\]
The coefficient $a^{m}$ may be interpreted as the probability amplitude that the spin projection on the $z$-axis for the particle in a given state $\chi_{\frac{1}{2}}$ is equal to $m$.

\subsubsection{Special cases}
\subsubsection{Spin is directed along the $z$-axis}
\begin{equation}
a^{\frac{1}{2}}=1, a^{-\frac{1}{2}}=0, \quad \chi_{\frac{1}{2}}=\chi_{\frac{1}{2} \frac{1}{2}}=\binom{1}{0} . \label{eq:6:2:35}
\end{equation}
\subsubsection{Spin is directed along the negative $z$-axis}
\begin{equation}
a^{\frac{1}{2}}=0, a^{-\frac{1}{2}}=1, \quad \chi_{\frac{1}{2}}=\chi_{\frac{1}{2}-\frac{1}{2}}=\binom{0}{1} . \label{eq:6:2:36}
\end{equation}
\subsubsection{Spin is parallel to $\vect{n}(\vartheta, \varphi)$}
\begin{equation}
a^{\frac{1}{2}}=\cos \frac{\vartheta}{2} e^{-i \frac{\varphi}{2}}, a^{-\frac{1}{2}}=\sin \frac{\vartheta}{2} e^{i \frac{\varphi}{2}}, \chi_{\frac{1}{2}}=\chi_{\frac{1}{2} \frac{1}{2}}(\vartheta, \varphi)=\binom{\cos \frac{\vartheta}{2} e^{-i \frac{\varphi}{2}}}{\sin \frac{\vartheta}{2} e^{i \frac{\varphi}{2}}} . \label{eq:6:2:37}
\end{equation}
\subsubsection{Spin is antiparallel to $\vect{n}(\vartheta, \varphi)$}
\begin{equation}
a^{\frac{1}{2}}=-\sin \frac{\vartheta}{2} e^{-i \frac{\varphi}{2}}, a^{-\frac{1}{2}}=\cos \frac{\vartheta}{2} e^{i \frac{\varphi}{2}}, \chi_{\frac{1}{2}}=\chi_{\frac{1}{2}-\frac{1}{2}}(\vartheta, \varphi)=\binom{-\sin \frac{\vartheta}{2} e^{-i \frac{\varphi}{2}}}{\cos \frac{\vartheta}{2} e^{i \frac{\varphi}{2}}} . \label{eq:6:2:38}
\end{equation}
In general, the spin function \eqref{eq:6:2:32} describes a spin- $\frac{1}{2}$ particle whose spin is directed along some unit vector n with cartesian components
\begin{equation}
n_{x}=2 \operatorname{Re}\left(a^{\frac{1}{2} *} a^{-\frac{1}{2}}\right), n_{y}=2 \operatorname{Im}\left(a^{\frac{1}{2} *} a^{-\frac{1}{2}}\right), n_{z}=\left|a^{\frac{1}{2}}\right|^{2}-\left|a^{-\frac{1}{2}}\right|^{2} . \label{eq:6:2:39}
\end{equation}
The spherical components of this unit vector $n$ are given by
\begin{equation}
n_{\mu}=\sqrt{3} \sum_{m, m^{\prime}} \clebsch{\frac{1}{2}}{m}{1}{\mu}{\frac{1}{2}}{m^{\prime}} a^{m^{\prime} *} a^{m}, \label{eq:6:2:40}
\end{equation}
or more explicitly
\begin{equation}
n_{+1}=-\sqrt{2} a^{\frac{1}{2} *} a^{-\frac{1}{2}}, n_{0}=\left|a^{\frac{1}{2}}\right|^{2}-\left|a^{-\frac{1}{2}}\right|^{2}, n_{-1}=\sqrt{2} a^{-\frac{1}{2} *} a^{\frac{1}{2}} . \label{eq:6:2:41}
\end{equation}
The product of the spin functions $\chi_{\frac{1}{2}} \chi_{\frac{1}{2}}^{\dagger}$ may be expressed in terms of the spin matrices
\begin{equation}
\chi_{\frac{1}{2}} \chi_{\frac{1}{2}}^{\dagger}=\left(\begin{array}{cc}
\left|a^{\frac{1}{2}}\right|^{2} & a^{-\frac{1}{2} *} a^{\frac{1}{2}} \label{eq:6:2:42}\\
a^{\frac{1}{2} *} a^{-\frac{1}{2}} & \left|a^{-\frac{1}{2}}\right|^{2}
\end{array}\right)=\frac{1}{2} \widehat{I}+\mathrm{n} \cdot \widehat{\mathrm{~S}},
\end{equation}
where $n$ specifies the spin direction \eqref{eq:6:2:39}-\eqref{eq:6:2:41}.

Matrix elements of the spin operator $\widehat{\vect{S}}$ may be written in terms of the vector $\vect{n}$ as
\begin{equation}
\chi_{\frac{1}{2}}^{\dagger} \widehat{\vect{S}} \chi_{\frac{1}{2}}=\frac{1}{2} \mathrm{n} . \label{eq:6:2:43}
\end{equation}
The transformation of spin functions under rotations of coordinate system is effected by the rotation operators $\widehat{D}^{\frac{1}{2}}(\alpha, \beta, \gamma)$ (Eq.~\eqref{eq:2:5:32}) or $\widehat{U}^{\frac{1}{2}}(\omega ; \theta, \Phi)$ (Eq.~\eqref{eq:2:5:36}), depending on the choice of the parameters to describe rotations:
\begin{equation}
\chi_{\frac{1}{2}}^{\prime}=\widehat{D}^{\frac{1}{2}}(\alpha, \beta, \gamma) \chi_{\frac{1}{2}}=\widehat{U}^{\frac{1}{2}}(\omega ; \theta, \Phi) \chi_{\frac{1}{2}} . \label{eq:6:2:44}
\end{equation}
The spin functions $\chi_{\frac{1}{2}}^{\prime}$ in the rotated coordinate system may be expressed in terms of the basis spin functions, $\chi_{\frac{1}{2} m}$ or $\chi_{\frac{1}{2} m}^{\prime}$, referred to the original or rotated coordinate systems, respectively.
\begin{equation}
\chi_{\frac{1}{2}}^{\prime}=\binom{a^{\prime \frac{1}{2}}}{a^{\prime-\frac{1}{2}}}=\sum_{\dot{m}} a^{\prime m} \chi_{\frac{1}{2} m}=\sum_{m} a^{m} \chi_{\frac{1}{2} m}^{\prime} . \label{eq:6:2:45}
\end{equation}
In this case $a^{m}$ and $a^{m}$ are spinor components in the original and rotated coordinate systems. The relation between the basis functions $\chi_{\frac{1}{2} m}^{\prime}$ and $\chi_{\frac{1}{2} m}$ is given by Eq.~\eqref{eq:6:2:15}. The transformation properties of $a^{m}$ are as follows:
\begin{equation}
\begin{aligned}
a^{\prime m} & =\sum_{n} D_{m n}^{\frac{1}{2}}(\alpha, \beta, \gamma) a^{n}, \\
a^{n} & =\sum_{m} D_{m n}^{\frac{1}{2} *}(\alpha, \beta, \gamma) a^{\prime m} . 
\end{aligned}\quad\left(\dot{m}, n= \pm \frac{1}{2}\right) .
\end{equation}
Thus, $a^{m}$ are contravariant spinor components (see Eq.~\eqref{eq:4:1:2}).\index{spinor!contravariant components}

\subsection{Polarization Density Matrix}\label{sec:6.2.7}\index{polarization density matrix!for $S=1/2$}
The polarisation density matrix for particles of spin $\frac{1}{2}$ may be written in the form
\begin{equation}
\hat{\rho}=\frac{1}{2}\{\hat{I}+2 \vect{P} \hat{\vect{S}}\}, \label{eq:6:2:47}
\end{equation}
where the real vector $\vect{P}$ is called the polarisation vector.\index{polarization vector}\isym{P-vec@$\vect{P}$, polarization vector} This vector gives the expectation value of the spin operator multiplied by 2 ,
\begin{equation}
\vect{P}=2\expect{\vect{S}}=2 \operatorname{Tr}\{\hat{\rho} \widehat{\vect{S}}\} . \label{eq:6:2:48}
\end{equation}
The absolute value of the vector $\vect{P}$ is called the polarization degree;\index{polarization degree} it ranges from 0 (for unpolarised states) to 1 (for pure, i.e., totally polarized states):
\begin{equation}
0 \leq|P| \leq 1 . \label{eq:6:2:49}
\end{equation}
The spherical components of the polarization vector are related to the elements of the density matrix by
\begin{equation}
P_{\mu}=\sqrt{3} \sum_{\sigma \sigma^{\prime}} \clebsch{\frac{1}{2}}{\sigma}{1}{\mu}{\frac{1}{2}}{\sigma^{\prime}} \rho_{\sigma \sigma^{\prime}}, \label{eq:6:2:50}
\end{equation}
or in an expanded form
\begin{equation}
P_{+1}=-\sqrt{2} \rho_{-\frac{1}{2} \frac{1}{2}}, P_{0}=\rho_{\frac{1}{2} \frac{1}{2}}-\rho_{-\frac{1}{2}-\frac{1}{2}}, P_{-1}=\sqrt{2} \rho_{\frac{1}{2}-\frac{1}{2}} . \label{eq:6:2:51}
\end{equation}
Cartesian components of the polarization vector are given by\index{polarization vector!cartesian components}
\begin{equation}
P_{x}=\rho_{\frac{1}{2}-\frac{1}{2}}+\rho_{-\frac{1}{2} \frac{1}{2}}, P_{y}=i\left(\rho_{\frac{1}{2}-\frac{1}{2}}-\rho_{-\frac{1}{2} \frac{1}{2}}\right), P_{z}=\rho_{\frac{1}{2} \frac{1}{2}}-\rho_{-\frac{1}{2}-\frac{1}{2}} . \label{eq:6:2:52}
\end{equation}
For a pure state described by a spin function $\chi_{\frac{1}{2}}$ the density matrix $\hat{\rho}$ has the form
\begin{equation}
\hat{\rho}_{\text {pure }}=\chi_{\frac{1}{2}} \chi_{\frac{1}{2}}^{\dagger}, \label{eq:6:2:53}
\end{equation}
and the polarization vector $\vect{P}$ coincides with the unit vector $\vect{n}$ (see Eqs.~\eqref{eq:6:2:39}-\eqref{eq:6:2:41}). For an unpolarized state
\begin{equation}
\hat{\rho}_{\text {unpol }}=\frac{1}{2} \widehat{I} . \label{eq:6:2:54}
\end{equation}
\section{Spin functions for $S=1$}\label{sec:6.3}\index{spin function!for $S=1$}
\subsection{Basis Spin Functions}\label{sec:6.3.1}\index{basis spin functions!for $S=1$}
The basis functions $\chi_{1 m}(m= \pm 1,0)$ are eigenfunctions of the operators $\widehat{S}^{2}$ and $\widehat{S}_{z}$
\begin{align}
& \widehat{\vect{S}}^{2} \chi_{1 m}=2 \chi_{1 m}, \label{eq:6:3:1}\\
& \widehat{S}_{z} \chi_{1 m}=m \chi_{1 m} . \notag
\end{align}
The function $\chi_{1 m}$ describes the state of a spin- 1 particle with definite spin projection $m$ on the $z$-axis. The three basis functions $\chi_{1 m}(m= \pm 1,0)$ may be treated as spherical covariant basis vectors $\vect{e}_{m}$ (see Sec.~\ref{sec:1.1}) written in column form. Instead of $\chi_{1 m}$ one may use the basis functions $\chi_{i}(i=x, y, z)$ which are the cartesian basis vectors $\vect{e}_{i}$ :\index{basis vectors!Cartesian}\isymg{chi-i@$\chi_{i}$, cartesian basis spin function}
\begin{equation}
\begin{array}{ll}
\chi_{11}=-\frac{1}{\sqrt{2}}\left(\chi_{x}+i \chi_{y}\right), & \chi_{x}=\frac{1}{\sqrt{2}}\left(\chi_{1-1}-\chi_{11}\right) \\
\chi_{10}=\chi_{x}, & \chi_{y}=\frac{i}{\sqrt{2}}\left(\chi_{1-1}+\chi_{11}\right), \label{eq:6:3:2}\\
\chi_{1-1}=\frac{1}{\sqrt{2}}\left(\chi_{x}-i \chi_{y}\right), & \chi_{z}=\chi_{10} .
\end{array}
\end{equation}
The functions $\chi_{1 m}(m= \pm 1,0)$ as well as $\chi_{i}(i=x, y, z)$ constitute an orthonormal basis. The orthonormality conditions read
\begin{equation}
\chi_{1 m^{\prime}}^{\dagger} \chi_{1 m}=\delta_{m^{\prime} m}, \quad \chi_{i}^{\dagger} \chi_{k}=\delta_{i k} . \label{eq:6:3:3}
\end{equation}
The completeness condition for basis functions may be written as
\begin{equation}
\sum_{m= \pm 1,0} \chi_{1 m} \chi_{1 m}^{\dagger}=\widehat{I}, \quad \sum_{i=x, y, z} \chi_{i} \chi_{i}^{\dagger}=\widehat{I}, \label{eq:6:3:4}
\end{equation}
where $\widehat{I}$ is the unit $3 \times 3$ matrix.\\
To describe the states of particles with spin 1 one may use the spherical basis representation or the cartesian basis representation.

\subsubsection{Spherical basis representation}
In the spherical basis representation the dependence of the functions $\chi_{1 m}(\sigma)$ on the spin variable $\sigma$ is given by
\begin{equation}
\chi_{1 m}(\sigma)=\delta_{m \sigma} . \label{eq:6:3:5}
\end{equation}
According to Eq.~\eqref{eq:6:1:1}, the basis spin functions $\chi_{1 m}$ may be written as
\begin{equation}
\chi_{11}=\left(\begin{array}{l}
1 \\
0 \\
0
\end{array}\right), \quad \chi_{10}=\left(\begin{array}{l}
0 \\
1 \\
0
\end{array}\right), \quad \chi_{1-1}=\left(\begin{array}{l}
0 \\
0 \\
1
\end{array}\right), \label{eq:6:3:6}
\end{equation}
and the Hermitian conjugate functions $\chi_{1 m}^{\dagger}$ read
\begin{equation}
\chi_{11}^{\dagger}=(1,0,0), \quad \chi_{10}^{\dagger}=(0,1,0), \quad \chi_{1-1}^{\dagger}=(0,0,1) . \label{eq:6:3:7}
\end{equation}
In the spherical basis representation the functions $\chi_{i}(i=x, y, z)$ have the form
\begin{gather}
\chi_{x}=\frac{1}{\sqrt{2}}\left(\begin{array}{r}
-1 \\
0 \\
1
\end{array}\right), \quad \chi_{y}=\frac{i}{\sqrt{2}}\left(\begin{array}{l}
1 \\
0 \\
1
\end{array}\right), \quad \chi_{z}=\left(\begin{array}{l}
0 \\
1 \\
0
\end{array}\right), \label{eq:6:3:8}\\
\chi_{x}^{\dagger}=\frac{1}{\sqrt{2}}(-1,0,1), \quad \chi_{y}^{\dagger}=-\frac{i}{\sqrt{2}}(1,0,1), \quad \chi_{z}^{\dagger}=(0,1,0) . \label{eq:6:3:9}
\end{gather}
It follows from \eqref{eq:6:3:6}, \eqref{eq:6:3:8} that in the spherical basis representation the functions $\chi_{1 m}$ are real,
\begin{equation}
\chi_{1 m}^{*}=\chi_{1 m}, \quad(m= \pm 1,0), \label{eq:6:3:10}
\end{equation}
and the functions $\chi_{i}$ satisfy the relations
\begin{equation}
\chi_{x}^{*}=\chi_{x}, \quad \chi_{y}^{*}=-\chi_{y}, \quad \chi_{z}^{*}=\chi_{z} . \label{eq:6:3:11}
\end{equation}
Explicit forms of the spin matrices and the polarization operators in the spherical basis representation are given by Eqs.~\eqref{eq:2:6:9}-\eqref{eq:2:6:22}.

\subsubsection{Cartesian basis representation}
In the cartesian basis representation the spin variable $\sigma$ assumes three possible values, $\sigma=x, y, z$. The dependence of the functions $\chi_{i}(\sigma)$ on $\sigma$ is given by
\begin{equation}
\chi_{i}(\sigma)=\delta_{i \sigma} . \label{eq:6:3:12}
\end{equation}
Thus, the basis spin functions $\chi_{i}(i=x, y, z)$ in the cartesian basis representation may be written in the following form (see also Eq.~\eqref{eq:1:4:37})
\begin{equation}
\begin{array}{ccc}
\chi_{x}=\left(\begin{array}{l}
1 \\
0 \\
0
\end{array}\right), & \chi_{y}=\left(\begin{array}{l}
0 \\
1 \\
0
\end{array}\right), & \chi_{z}=\left(\begin{array}{l}
0 \\
0 \\
1
\end{array}\right) \end{array}\label{eq:6:3:13}
\end{equation}
\begin{equation}
    \begin{array}{ccc}
\chi_{x}^{\dagger}=(1,0,0), & \chi_{y}^{\dagger}=(0,1,0), & \chi_{z}^{\dagger}=(0,0,1) . \label{eq:6:3:14}
\end{array}
\end{equation}
The basis functions $\chi_{1 m}(m= \pm 1,0)$ read
\begin{align}
& \chi_{11}=-\frac{1}{\sqrt{2}}\left(\begin{array}{l}
1 \\
i \\
0
\end{array}\right), \quad \chi_{10}=\left(\begin{array}{l}
0 \\
0 \\
1
\end{array}\right), \quad \chi_{1-1}=\frac{1}{\sqrt{2}}\left(\begin{array}{r}
1 \\
-i \\
0
\end{array}\right), \label{eq:6:3:15}\\
& \chi_{11}^{\dagger}=-\frac{1}{\sqrt{2}}(1,-i, 0), \quad \chi_{10}^{\dagger}=(0,0,1), \quad \chi_{1-1}^{\dagger}=\frac{1}{\sqrt{2}}(1, i, 0) . \label{eq:6:3:16}
\end{align}
It follows from Eqs.~\eqref{eq:6:3:13}, \eqref{eq:6:3:15} that in the cartesian basis representation the functions $\chi_{i}$ are real,
\begin{equation}
\chi_{i}^{*}=\chi_{i} \quad(i=x, y, z), \label{eq:6:3:17}
\end{equation}
and the functions $\chi_{1 m}$ satisfy the relations
\begin{equation}
\chi_{1 m}^{*}=(-1)^{m} \chi_{1-m}(m= \pm 1,0) . \label{eq:6:3:18}
\end{equation}
The spin matrices and polarization operators in the cartesian basis representation are given by Eqs.~\eqref{eq:2:6:23} and \eqref{eq:2:6:35}.

The direct and reverse transformations from a spherical basis to a cartesian one can be performed by the use of the unitary $3 \times 3$ matrix $U$,
\begin{align}
\chi(\text { spherical basis }) & =U \chi(\text { cartesian basis })  , \notag\\
\chi(\text { cartesian basis }) & =U^{-1} \chi(\text { spherical basis }) . \label{eq:6:3:19}
\end{align}
The explicit form of $U$ is given by Eq.~\eqref{eq:2:6:37}. The expressions below are independent of the representation used unless the contrary is indicated.

\subsection{Expansions of Products of Spin Functions}\label{sec:6.3.2}\index{spin function!expansions of products of}
Products of the basis functions $\chi_{1 m} \chi_{1 m^{\prime}}^{\dagger}$ and $\chi_{i} \chi_{k_{1}}^{\dagger}$ are square $3 \times 3$ matrices which may be expanded in terms of the polarization matrices $\widehat{I}, \widehat{S}_{\mu}(\mu= \pm 1,0), \widehat{T}_{2 M}(M= \pm 2, \pm 1,0)$ or, equivalently, of the matrices $\widehat{I}, \widehat{S}_{i}, \widehat{Q}_{i k}(i, k=x, y, z)$ (Sec.~\ref{sec:2.6}). These expansions are given by
\begin{gather}
\chi_{1 m} \chi_{1 m^{\prime}}^{\dagger}=\frac{1}{3} \delta_{m m^{\prime}} \hat{I}+\frac{1}{\sqrt{2}} \clebsch{1}{m^{\prime}}{1}{\mu}{1}{m} \hat{S}_{\mu}+\sqrt{\frac{5}{3}} \clebsch{1}{m^{\prime}}{2}{M}{1}{m} \hat{T}_{2 M}, \label{eq:6:3:20}\\
\chi_{i} \chi_{k}^{\dagger}=\frac{1}{3} \delta_{i k} \hat{I}+\frac{i}{2} \varepsilon_{i k l} \widehat{S}_{l}-\widehat{Q}_{i k} . \label{eq:6:3:21}
\end{gather}
Written in component form these expansions yield
\begin{align}
\chi_{11} \chi_{11}^{\dagger} & =\frac{1}{3} \widehat{I}+\frac{1}{2} \widehat{S}_{0}+\frac{1}{\sqrt{6}} \widehat{T}_{20}=\frac{1}{3} \widehat{I}+\frac{1}{2} \widehat{S}_{z}+\frac{1}{2} \widehat{Q}_{z z}, \notag\\
\chi_{11} \chi_{10}^{\dagger} & =-\frac{1}{2} \widehat{S}_{+1}-\frac{1}{\sqrt{2}} \widehat{T}_{21}=\frac{1}{2 \sqrt{2}}\left(\widehat{S}_{x}+i \widehat{S}_{y}+2 \widehat{Q}_{x z}+2 i \widehat{Q}_{y z}\right), \notag\\
\chi_{11} \chi_{1-1}^{\dagger} & =\widehat{T}_{22}=\frac{1}{2}\left(\widehat{Q}_{x x}-\widehat{Q}_{y y}+2 i \widehat{Q}_{x y}\right), \notag\\
\chi_{10} \chi_{11}^{\dagger} & =\frac{1}{2} \widehat{S}_{-1}+\frac{1}{\sqrt{2}} \widehat{T}_{2-1}=\frac{1}{2 \sqrt{2}}\left(\widehat{S}_{x}-i \widehat{S}_{y}+2 \widehat{Q}_{x z}-2 i \widehat{Q}_{y z}\right), \notag\\
\chi_{10} \chi_{10}^{\dagger} & =\frac{1}{3} \widehat{I}-\sqrt{\frac{2}{3}} \widehat{T}_{20}=\frac{1}{3} \widehat{I}-\widehat{Q}_{z z}, \label{eq:6:3:22}\\
\chi_{10} \chi_{1-1}^{\dagger} & =-\frac{1}{2} \widehat{S}_{+1}+\frac{1}{\sqrt{2}} \widehat{T}_{21}=\frac{1}{2 \sqrt{2}}\left(\widehat{S}_{x}+i \widehat{S}_{y}-2 \widehat{Q}_{x z}-2 i \widehat{Q}_{y z}\right), \notag\\
\chi_{1-1} \chi_{11}^{\dagger} & =\widehat{T}_{2-2}=\frac{1}{2}\left(\widehat{Q}_{x x}-\widehat{Q}_{y y}-2 i \widehat{Q}_{x y}\right), \notag\\
\chi_{1-1} \chi_{10}^{\dagger} & =\frac{1}{2} \widehat{S}_{-1}-\frac{1}{\sqrt{2}} \widehat{T}_{2-1}=\frac{1}{2 \sqrt{2}}\left(\widehat{S}_{x}-i \widehat{S}_{y}-2 \widehat{Q}_{x z}+2 i \widehat{Q}_{y z}\right), \notag\\
\chi_{1-1} \chi_{1-1}^{\dagger} & =\frac{1}{3} \widehat{I}-\frac{1}{2} \widehat{S}_{0}+\frac{1}{\sqrt{6}} \widehat{T}_{20}=\frac{1}{3} \widehat{I}-\frac{1}{2} \widehat{S}_{z}+\frac{1}{2} \widehat{Q}_{z z} . \notag
\end{align}
\begin{align}
& \chi_{x} \chi_{x}^{\dagger}=\frac{1}{3} \widehat{I}-\frac{1}{2} \widehat{T}_{2-2}-\frac{1}{2} \widehat{T}_{22}+\frac{1}{\sqrt{6}} \widehat{T}_{20}=\frac{1}{3} \widehat{I}-\widehat{Q}_{x x}, \notag\\
& \chi_{x} \chi_{y}^{\dagger}=\frac{i}{2}\left(\widehat{S}_{0}-\widehat{T}_{2-2}+\widehat{T}_{22}\right)=\frac{i}{2} \widehat{S}_{z}-\widehat{Q}_{x y}, \notag\\
& \chi_{x} \chi_{z}^{\dagger}=\frac{1}{2 \sqrt{2}}\left(\widehat{S}_{-1}+\widehat{S}_{+1}\right)-\frac{1}{2}\left(\widehat{T}_{2-1}-\widehat{T}_{21}\right)=-\frac{i}{2} \widehat{S}_{y}-\widehat{Q}_{x z}, \notag\\
& \chi_{y} \chi_{x}^{\dagger}=-\frac{i}{2}\left(\widehat{S}_{0}+\widehat{T}_{2-2}-\widehat{T}_{22}\right)=-\frac{i}{2} \widehat{S}_{z}-\widehat{Q}_{y x}, \notag\\
& \chi_{y} \chi_{y}^{\dagger}=\frac{1}{3} \widehat{I}+\frac{1}{2} \widehat{T}_{2-2}+\frac{1}{2} \widehat{T}_{22}+\frac{1}{\sqrt{6}} \widehat{T}_{20}=\frac{1}{3} \widehat{I}-\widehat{Q}_{y y}, \label{eq:6:3:23}\\
& \chi_{y} \chi_{z}^{\dagger}=\frac{i}{2 \sqrt{2}}\left(\widehat{S}_{-1}-\widehat{S}_{+1}\right)-\frac{i}{2}\left(\widehat{T}_{2-1}+\widehat{T}_{21}\right)=\frac{i}{2} \widehat{S}_{x}-\widehat{Q}_{y z}, \notag\\
& \chi_{z} \chi_{x}^{\dagger}=-\frac{1}{2 \sqrt{2}}\left(\widehat{S}_{-1}+\widehat{S}_{+1}\right)-\frac{1}{2}\left(\widehat{T}_{2-1}-\widehat{T}_{21}\right)=\frac{i}{2} \widehat{S}_{y}-\widehat{Q}_{z x}, \notag\\
& \chi_{z} \chi_{y}^{\dagger}=-\frac{i}{2 \sqrt{2}}\left(\widehat{S}_{-1}-\widehat{S}_{+1}\right)-\frac{i}{2}\left(\widehat{T}_{2-1}+\widehat{T}_{21}\right)=-\frac{i}{2} \widehat{S}_{x}-\widehat{Q}_{z y}, \notag\\
& \chi_{z} \chi_{z}^{\dagger}=\frac{1}{3} \widehat{I}-\sqrt{\frac{2}{3}} \widehat{T}_{20}=\frac{1}{3} \widehat{I}-\widehat{Q}_{z z} . \notag
\end{align}
\subsection{Action of Spin Operators on Basis Functions}\label{sec:6.3.3}\index{spin operator!action on basis functions}
Spherical components of the spin operator $\widehat{S} \mu(\mu= \pm 1,0)$ and its cartesian components $\widehat{S}_{i}(i=x, y, z)$ act on the basis functions as follows
\begin{align}
\widehat{S}_{\mu} \chi_{1 m} & =\sqrt{2} \clebsch{1}{m}{1}{\mu}{1}{m^{\prime}} \chi_{1 m^{\prime}}, \label{eq:6:3:24}\\
\widehat{S}_{i} \chi_{k} & =i \varepsilon_{i k l} \chi_{l} . \notag
\end{align}
In a more detailed form
\begin{gather}
\widehat{S}_{+1} \chi_{11}=0 \quad \widehat{S}_{+1} \chi_{10}=-\chi_{11}, \quad \widehat{S}_{+1} \chi_{1-1}=-\chi_{10}, \notag\\
\widehat{S}_{0} \chi_{11}=\chi_{11}, \quad \widehat{S}_{0} \chi_{10}=0, \quad \widehat{S}_{0} \chi_{1-1}=-\chi_{1-1}, \label{eq:6:3:25}\\
\widehat{S}_{-1} \chi_{11}=\chi_{10} \quad \widehat{S}_{-1} \chi_{10}=\chi_{1-1}, \quad \widehat{S}_{-1} \chi_{1-1}=0 . \notag\\
\widehat{S}_{+1} \chi_{x}=-\frac{1}{\sqrt{2}} \chi_{z}, \quad \widehat{S}_{+1} \chi_{y}=-\frac{i}{\sqrt{2}} \chi_{z}, \quad \widehat{S}_{+1} \chi_{z}=\frac{1}{\sqrt{2}}\left(\chi_{x}+i \chi_{y}\right), \notag\\
\widehat{S}_{0} \chi_{x}=i \chi_{y}, \quad \widehat{S}_{0} \chi_{y}=-i \chi_{x}, \quad \widehat{S}_{0} \chi_{z}=0, \label{eq:6:3:26}\\
\widehat{S}_{-1} \chi_{x}=-\frac{1}{\sqrt{2}} \chi_{z}, \quad \widehat{S}_{-1} \chi_{y}=\frac{i}{\sqrt{2}} \chi_{z}, \quad \widehat{S}_{-1} \chi_{z}=\frac{1}{\sqrt{2}}\left(\chi_{x}-i \chi_{y}\right) . \notag\\
\widehat{S}_{x} \chi_{11}=\frac{1}{\sqrt{2}} \chi_{10}, \quad \widehat{S}_{x} \chi_{10}=\frac{1}{\sqrt{2}}\left(\chi_{1-1}+\chi_{11}\right), \quad \widehat{S}_{x} \chi_{1-1}=\frac{1}{\sqrt{2}} \chi_{10}, \notag\\
\widehat{S}_{y} \chi_{11}=\frac{i}{\sqrt{2}} \chi_{10}, \quad \widehat{S}_{y} \chi_{10}=\frac{i}{\sqrt{2}}\left(\chi_{1-1}-\chi_{11}\right), \quad \widehat{S}_{y} \chi_{1-1}=-\frac{i}{\sqrt{2}} \chi_{10}, \label{eq:6:3:27}\\
\widehat{S}_{z} \chi_{11}=\chi_{11}, \quad \widehat{S}_{z} \chi_{10}=0, \quad \widehat{S}_{z} \chi_{1-1}=-\chi_{1-1} . \notag\\
\widehat{S}_{x} \chi_{x}=0, \quad \widehat{S}_{x} \chi_{y}=i \chi_{z}, \quad \widehat{S}_{x} \chi_{z}=-i \chi_{y}, \notag\\
\widehat{S}_{y} \chi_{x}=-i \chi_{z}, \quad \widehat{S}_{y} \chi_{y}=0, \quad \widehat{S}_{y} \chi_{z}=i \chi_{x}, \label{eq:6:3:28}\\
\widehat{S}_{z} \chi_{x}=i \chi_{y}, \quad \widehat{S}_{z} \chi_{y}=-i \chi_{x}, \quad \widehat{S}_{z} \chi_{z}=0 . \notag
\end{gather}
\subsection{Action of Quadrupole Operators on Basis Functions}\label{sec:6.3.4}\index{quadrupole operator}\index{polarization operator!quadrupole}
The operators $\widehat{T}_{2 M}(M= \pm 2, \pm 1,0)$ and $\widehat{Q}_{i k}(i, k=x, y, z)$ act on basis functions in accordance with
\begin{gather}
\hat{T}_{2 M} \chi_{1 m}=\sqrt{\frac{5}{3}} \clebsch{1}{m}{2}{M}{1}{m^{\prime}} \chi_{1 m^{\prime}}, \label{eq:6:3:29}\\
\hat{Q}_{i k} \chi_{l}=\frac{1}{2}\left(\frac{2}{3} \delta_{i k} \chi_{l}-\delta_{i l} \chi_{k}-\delta_{k l} \chi_{i}\right) . \notag
\end{gather}
In detailed form we have
\begin{equation}
\begin{array}{lll}
\hat{T}_{22} \chi_{11}=0, & \hat{T}_{22} \chi_{10}=0, & \hat{T}_{22} \chi_{1-1}=\chi_{11}, \\
\hat{T}_{21} \chi_{11}=0, & \hat{T}_{21} \chi_{10}=-\frac{1}{\sqrt{2}} \chi_{11}, & \hat{T}_{21} \chi_{1-1}=\frac{1}{\sqrt{2}} \chi_{10}, \\
\hat{T}_{20} \chi_{11}=\frac{1}{\sqrt{6}} \chi_{11}, & \hat{T}_{20} \chi_{10}=-\sqrt{\frac{2}{3}} \chi_{10}, & \hat{T}_{20} \chi_{1-1}=\frac{1}{\sqrt{6}} \chi_{1-1}, \label{eq:6:3:30}\\
\hat{T}_{2-1} \chi_{11}=\frac{1}{\sqrt{2}} \chi_{10}, & \hat{T}_{2-1} \chi_{10}=-\frac{1}{\sqrt{2}} \chi_{1-1}, & \hat{T}_{2-1} \chi_{1-1}=0, \\
\hat{T}_{2-2} \chi_{11}=\chi_{1-1}, & \hat{T}_{2-2} \chi_{10}=0, & \hat{T}_{2-2} \chi_{1-1}=0 .
\end{array}
\end{equation}
\begin{equation}
\begin{array}{lll}
\hat{T}_{22} \chi_{x}=-\frac{1}{2}\left(\chi_{x}+i \chi_{y}\right), & \hat{T}_{22} \chi_{y}=\frac{1}{2}\left(\chi_{y}-i \chi_{x}\right), & \hat{T}_{22} \chi_{z}=0 \\
\hat{T}_{21} \chi_{x}=\frac{1}{2} \chi_{z}, & \hat{T}_{21} \chi_{y}=\frac{i}{2} \chi_{z}, & \hat{T}_{21} \chi_{z}=\frac{1}{2}\left(\chi_{x}+i \chi_{y}\right), \\
\hat{T}_{20} \chi_{x}=\frac{1}{\sqrt{6}} \chi_{x}, & \hat{T}_{20} \chi_{y}=\frac{1}{\sqrt{6}} \chi_{y}, & \hat{T}_{20} \chi_{z}=-\sqrt{\frac{2}{3}} \chi_{z}, \label{eq:6:3:31}\\
\hat{T}_{2-1} \chi_{x}=-\frac{1}{2} \chi_{z}, & \hat{T}_{2-1} \chi_{y}=\frac{i}{2} \chi_{z}, & \hat{T}_{2-1} \chi_{z}=-\frac{1}{2}\left(\chi_{x}-i \chi_{y}\right), \\
\hat{T}_{2-2} \chi_{x}=-\frac{1}{2}\left(\chi_{x}-i \chi_{y}\right), & \hat{T}_{2-2} \chi_{y}=\frac{1}{2}\left(\chi_{y}+i \chi_{x}\right) & \hat{T}_{2-2} \chi_{z}=0 .
\end{array}
\end{equation}
\begin{equation}
\begin{array}{lll}
\hat{Q}_{x x} \chi_{11}=\frac{1}{2} \chi_{1-1}-\frac{1}{6} \chi_{11}, & \hat{Q}_{x x} \chi_{10}=\frac{1}{3} \chi_{10}, & \hat{Q}_{x x} \chi_{1-1}=\frac{1}{2} \chi_{11}-\frac{1}{6} \chi_{1-1}, \\
\hat{Q}_{y y} \chi_{11}=-\frac{1}{2} \chi_{1-1}-\frac{1}{6} \chi_{11}, & \hat{Q}_{y y} \chi_{10}=\frac{1}{3} \chi_{10}, & \hat{Q}_{y y} \chi_{1-1}=-\frac{1}{2} \chi_{11}-\frac{1}{6} \chi_{1-1}, \\
\hat{Q}_{z z} \chi_{11}=\frac{1}{3} \chi_{11}, & \hat{Q}_{z z} \chi_{10}=-\frac{2}{3} \chi_{10}, & \hat{Q}_{z z} \chi_{1-1}=\frac{1}{3} \chi_{1-1}, \\
\hat{Q}_{x y} \chi_{11}=\frac{i}{2} \chi_{1-1}, & \hat{Q}_{x y} \chi_{10}=0, & \hat{Q}_{x y} \chi_{1-1}=-\frac{i}{2} \chi_{11}, \\
\hat{Q}_{x z} \chi_{11}=\frac{1}{2 \sqrt{2}} \chi_{10}, & \hat{Q}_{x z} \chi_{10}=\frac{1}{2 \sqrt{2}}\left(\chi_{11}-\chi_{1-1}\right), & \hat{Q}_{x z} \chi_{1-1}=-\frac{1}{2 \sqrt{2}} \chi_{10}, \\
\hat{Q}_{y z} \chi_{11}=\frac{i}{2 \sqrt{2}} \chi_{10}, & \hat{Q}_{y z} \chi_{10}=-\frac{i}{2 \sqrt{2}}\left(\chi_{11}+\chi_{1-1}\right), & \hat{Q}_{y z} \chi_{1-1}=\frac{i}{2 \sqrt{2}} \chi_{10} .
\end{array} \label{eq:6:3:32}
\end{equation}
\begin{equation}
\begin{array}{lll}
\hat{Q}_{x x} \chi_{x}=-\frac{2}{3} \chi_{x}, & \hat{Q}_{x x} \chi_{y}=\frac{1}{3} \chi_{y}, & \hat{Q}_{x x} \chi_{z}=\frac{1}{3} \chi_{z}, \\
\hat{Q}_{y y} \chi_{x}=\frac{1}{3} \chi_{x}, & \hat{Q}_{y y} \chi_{y}=-\frac{2}{3} \chi_{y}, & \hat{Q}_{y y} \chi_{z}=\frac{1}{3} \chi_{z} ,\\
\hat{Q}_{z z} \chi_{x}=\frac{1}{3} \chi_{x}, & \hat{Q}_{z z} \chi_{y}=\frac{1}{3} \chi_{y}, & \hat{Q}_{z z} \chi_{z}=-\frac{2}{3} \chi_{z}  ,\\
\hat{Q}_{x y} \chi_{x}=-\frac{1}{2} \chi_{y}, & \hat{Q}_{x y} \chi_{y}=-\frac{1}{2} \chi_{x}, & \hat{Q}_{x y} \chi_{z}=0 ,\\
\hat{Q}_{x z} \chi_{x}=-\frac{1}{2} \chi_{z}, & \hat{Q}_{x z} \chi_{y}=0, & \hat{Q}_{x z} \chi_{z}=-\frac{1}{2} \chi_{x} ,\\
\hat{Q}_{y z} \chi_{x}=0, & \hat{Q}_{y z} \chi_{y}=-\frac{1}{2} \chi_{z}, & \hat{Q}_{y z} \chi_{z}=-\frac{1}{2} \chi_{y}. 
\end{array}\label{eq:6:3:33}
\end{equation}
\subsection{Transformation of Basis Functions Under Rotations of the Coordinate Systems}\label{sec:6.3.5}\index{rotation!of the basis spin functions}
The spin functions $\chi_{1 m}(m= \pm 1,0)$ and $\chi_{i}(i=x, y, z)$ constitute the covariant spherical basis and cartesian basis, respectively. Under rotations they transform through the rotation operator $\widehat{D}^{1}(\alpha, \beta, \gamma)$ (Eqs.~\eqref{eq:2:6:75}, \eqref{eq:2:6:78}) if rotations are specified by the Euler angles $\alpha, \beta, \gamma$ or through the rotation operator $\widehat{U}^{1}(\omega ; \Theta, \Phi)$ (Eqs.~\eqref{eq:2:6:79}-\eqref{eq:2:6:83}) if rotations are described by the rotation angle $\omega$ and the rotation axis $n(\Theta, \Phi)$.
\begin{gather}
\chi_{1 m^{\prime}}^{\prime}=\hat{D}^{1}(\alpha, \beta, \gamma) \chi_{1 m^{\prime}}=\sum_{m= \pm 1,0} D_{m m^{\prime}}^{1}(\alpha, \beta, \gamma) \chi_{1 m}, \notag\\
\chi_{i}^{\prime}=\hat{D}^{1}(\alpha, \beta, \gamma) \chi_{i}=\sum_{k=x, y, z} a_{k i} \chi_{k} . \label{eq:6:3:34}
\end{gather}
Here $D_{m m^{\prime}}^{1}(\alpha, \beta, \gamma)$ are the Wigner $D$-functions (Chap.~\ref{chap:4}) and $a_{i k}$ are elements of the rotation matrix (Sec.~\ref{sec:1.4.6}).

The functions $\chi_{1 m^{\prime}}^{\prime}$ describe quantum states in which a particle of spin 1 has the spin projection $m^{\prime}$ on the new $z^{\prime}$-axis. These functions are eigenfunctions of the operators $\widehat{\vect{S}}^{\prime 2}$ and $\widehat{S}_{z}^{\prime}$ where $\widehat{\vect{S}}^{\prime}$ is the spin operator in the rotated coordinate system (Eqs.~\eqref{eq:2:6:85} and \eqref{eq:2:6:87})
\begin{align}
\widehat{\vect{S}}^{\prime 2} \chi_{1 m^{\prime}}^{\prime} & =2 \chi_{1 m^{\prime}}^{\prime}, \label{eq:6:3:35}\\
\widehat{S}_{z}^{\prime} \chi_{1 m^{\prime}}^{\prime} & =m^{\prime} \chi_{1 m^{\prime}}^{\prime} . \notag
\end{align}
In the spherical basis representation the functions $\chi_{1 m^{\prime}}^{\prime}$ have the form
\begin{equation}
\chi_{11}^{\prime}=\left(\begin{array}{c}
\frac{1+\cos \beta}{2} e^{-i(\alpha+\gamma)} \\
\frac{\sin \beta}{\sqrt{2}} e^{-i \gamma} \\
\frac{1-\cos \beta}{2} e^{i(\alpha-\gamma)}
\end{array}\right), \quad \chi_{10}^{\prime}=\left(\begin{array}{c}
-\frac{\sin \beta}{\sqrt{2}} e^{-i \alpha} \\
\cos \beta \\
\frac{\sin \beta}{\sqrt{2}} e^{i \alpha}
\end{array}\right), \quad \chi_{1-1}^{\prime}=\left(\begin{array}{c}
\frac{1-\cos \beta}{2} e^{-i(\alpha-\gamma)} \\
-\frac{\sin \beta}{\sqrt{2}} e^{i \gamma} \\
\frac{1+\cos \beta}{2} e^{i(\alpha+\gamma)}
\end{array}\right), \label{eq:6:3:36}
\end{equation}
and the functions $\chi_{i}^{\prime}$ are given by
\begin{equation}
\chi_{x}^{\prime}=\left(\begin{array}{c}
-\frac{\cos \beta \cos \gamma-i \sin \gamma}{\sqrt{2}} e^{-i \alpha} \\
-\sin \beta \cos \gamma \\
\frac{\cos \beta \cos \gamma+i \sin \gamma}{\sqrt{2}} e^{i \alpha}
\end{array}\right), \quad \chi_{y}^{\prime}=\left(\begin{array}{c}
\frac{\cos \beta \sin \gamma+i \cos \gamma}{\sqrt{2}} e^{-i \alpha} \\
\sin \beta \sin \gamma \\
-\frac{\cos \beta \sin \gamma-i \cos \gamma}{\sqrt{2}} e^{i \alpha}
\end{array}\right), \quad \chi_{z}^{\prime}=\left(\begin{array}{c}
-\frac{\sin \beta}{\sqrt{2}} e^{-i \alpha} \\
\cos \beta \\
\frac{\sin \beta}{\sqrt{2}} e^{i \alpha}
\end{array}\right) . \label{eq:6:3:37}
\end{equation}
In the cartesian basis representation these functions read
\begin{gather}
\chi_{11}^{\prime}=\left(\begin{array}{c}
-\frac{\cos \beta \cos \alpha-i \sin \alpha}{\sqrt{2}} e^{-i \gamma} \\
-\frac{\cos \beta \sin \alpha+i \cos \alpha}{\sqrt{2}} e^{-i \gamma} \\
\frac{\sin \beta}{\sqrt{2}} e^{-i \gamma}
\end{array}\right), \quad \chi_{10}^{\prime}=\left(\begin{array}{c}
\sin \beta \cos \alpha \\
\sin \beta \sin \alpha \\
\cos \beta
\end{array}\right), \quad \chi_{1-1}^{\prime}=\left(\begin{array}{c}
\frac{\cos \beta \cos \alpha+i \sin \alpha}{\sqrt{2}} e^{i \gamma} \\
\frac{\cos \beta \sin \alpha-i \cos \alpha}{\sqrt{2}} e^{i \gamma} \\
-\frac{\sin \beta}{\sqrt{2}} e^{i \gamma}
\end{array}\right), \label{eq:6:3:38}\\
\chi_{x}^{\prime}=\left(\begin{array}{c}
\cos \beta \cos \alpha \cos \gamma-\sin \alpha \sin \gamma \\
\cos \beta \sin \alpha \cos \gamma+\cos \alpha \sin \gamma \\
-\sin \beta \cos \gamma
\end{array}\right), \quad \chi_{y}^{\prime}=\left(\begin{array}{c}
-\cos \beta \cos \alpha \sin \gamma-\sin \alpha \cos \gamma \\
-\cos \beta \sin \alpha \sin \gamma+\cos \alpha \cos \gamma \\
\sin \beta \sin \gamma
\end{array}\right), \quad \chi_{z}^{\prime}=\left(\begin{array}{c}
\sin \beta \cos \alpha \\
\sin \beta \sin \alpha \\
\cos \beta
\end{array}\right) . \label{eq:6:3:39}
\end{gather}
The expansions for the basis functions in the rotated coordinate system in terms of the angles $\omega, \Theta, \Phi$ may be obtained by the use of the rotation operator $\hat{U}^{1}(\omega ; \Theta, \Phi)$ (Eqs.~\eqref{eq:2:6:80}-\eqref{eq:2:6:81}).

\subsection{Helicity Basis Functions for $S=1$}\label{sec:6.3.6}\index{helicity basis functions!for $S=1$}
The helicity basis functions $\chi_{1 \lambda}(\vartheta, \varphi)(\lambda= \pm 1,0)$ describe the states in which the spin projection on the linear momentum direction $\vect{n}(\vartheta, \varphi) \equiv \vect{p} /|\vect{p}|$ is equal to $\lambda$. The functions $\chi_{1 \lambda}(\vartheta, \varphi)$ are eigenfunctions of the operators $\widehat{\vect{S}}^{2}$ and $\widehat{\vect{S}} \cdot \vect{n}$
\begin{align}
\widehat{\vect{S}}^{2} \chi_{1 \lambda}(\vartheta, \varphi) & =2 \chi_{1 \lambda}(\vartheta, \varphi), \notag\\
\widehat{\vect{S}} \cdot \vect{n} \chi_{1 \lambda}(\vartheta, \varphi) & =\lambda \chi_{1 \lambda}(\vartheta, \varphi) . \label{eq:6:3:40}
\end{align}
According to Eq.~\eqref{eq:6:1:20}, the helicity basis functions may be derived from the functions $\chi_{1 m}$ by a coordinate rotation
\begin{gather}
\chi_{1 \lambda}(\vartheta, \varphi)=\sum_{m} D_{m \lambda}^{1}(\varphi, \vartheta, 0) \chi_{1 m}, \notag\\
\chi_{1 m}=\sum_{\lambda} D_{-\lambda-m}^{1}(0, \vartheta, \varphi) \chi_{1 \lambda}(\vartheta, \varphi) . \label{eq:6:3:41}
\end{gather}
For Hermitian adjoint functions Eqs.~\eqref{eq:6:3:41} assume the form
\begin{align}
& \chi_{1 \lambda}^{\dagger}(\vartheta, \varphi)=\sum_{m}(-1)^{\lambda-m} D_{-m-\lambda}^{1}(\varphi, \vartheta, 0) \chi_{1 m}^{\dagger}, \notag\\
& \chi_{1 m}^{\dagger}=\sum_{\lambda}(-1)^{\lambda-m} D_{\lambda m}^{1}(0, \vartheta, \varphi) \chi_{1 \lambda}^{\dagger}(\vartheta, \varphi) . \label{eq:6:3:42}
\end{align}
The helicity functions $\chi_{1 \lambda}(\vartheta, \varphi)$ may be treated as the covariant helicity basis vectors $\vect{e}_{\lambda}^{\prime}$ (see Sec.~\ref{sec:1.1.4}) written as column matrices
\begin{equation}
\chi_{1 \lambda}(\vartheta, \varphi)=\vect{e}_{\lambda}^{\prime} \quad(\lambda= \pm 1,0) . \label{eq:6:3:43}
\end{equation}
One can construct the following linear combinations $\chi_{i}(\vartheta, \varphi)(i=x, y, z)$ of the helicity functions $\chi_{1 \lambda}(\vartheta, \varphi)$ :
\begin{align}
& \chi_{x}(\vartheta, \varphi)=\frac{1}{\sqrt{2}}\left\{\chi_{1-1}(\vartheta, \varphi)-\chi_{11}(\vartheta, \varphi)\right\}, \notag\\
& \chi_{y}(\vartheta, \varphi)=\frac{i}{\sqrt{2}}\left\{\chi_{1-1}(\vartheta, \varphi)+\chi_{11}(\vartheta, \varphi)\right\}, \label{eq:6:3:44}\\
& \chi_{z}(\vartheta, \varphi)=\chi_{10}(\vartheta, \varphi) . \notag
\end{align}
The functions $\chi_{i}(\vartheta, \varphi)(i=x, y, z)$ coincide with the polar basis vectors (see Sec.~\ref{sec:1.1.3}):
\begin{equation}
\chi_{x}(\vartheta, \varphi) \equiv \vect{e}_{\vartheta}, \quad \chi_{y}(\vartheta, \varphi) \equiv \vect{e}_{\varphi}, \quad \chi_{z}(\vartheta, \varphi)=\vect{e}_{p}. \label{eq:6:3:45}
\end{equation}
It should be emphasized that the functions $\chi_{i}(\vartheta, \varphi)$ describe the states with no definite helicity. These functions are related to the basis functions $\chi_{k}(k=x, y, z)$ (Sec.~\ref{sec:6.3.1}) by
\begin{align}
& \chi_{i}(\vartheta, \varphi)=\sum_{k} a_{k i}(\varphi, \vartheta, 0) \chi_{k}, \notag\\
& \chi_{k}=\sum_{i} a_{i k}(\varphi, \vartheta, 0) \chi_{i}(\vartheta, \varphi), \label{eq:6:3:46}
\end{align}
where $a_{i k}(\alpha, \beta, \gamma)$ are elements of the rotation matrix given by Eqs.~\eqref{eq:1:4:54}. The relations between the Hermitian conjugate functions $\chi_{i}^{\dagger}(\vartheta, \varphi)$ and $\chi_{k}^{\dagger}$ are also given by Eq.~\eqref{eq:6:3:46} because the elements $a_{i k}$ are real.

Three helicity functions $\chi_{1 \lambda}(\vartheta, \varphi)$ with $\lambda= \pm 1,0$ or, equivalently, three functions $\chi_{i}(\vartheta, \varphi)$ with $i=x, y, z$ constitute an orthonormalized basis. The orthogonality and normalization conditions have the form
\begin{gather}
\chi_{1 \lambda}^{\dagger}(\vartheta, \varphi) \chi_{1 \lambda^{\prime}}(\vartheta, \varphi)=\delta_{\lambda \lambda^{\prime}}, \notag\\
\chi_{i}^{\dagger}(\vartheta, \varphi) \chi_{k}(\vartheta, \varphi)=\delta_{i k} . \label{eq:6:3:47}
\end{gather}
The completeness conditions are
\begin{align}
& \sum_{\lambda= \pm 1,0} \chi_{1 \lambda}(\vartheta, \varphi) \chi_{1 \lambda}^{\dagger}(\vartheta, \varphi)=\widehat{I}, \notag\\
& \sum_{i=x, y, z} \chi_{i}(\vartheta, \varphi) \chi_{i}^{\dagger}(\vartheta, \varphi)=\widehat{I} . \label{eq:6:3:48}
\end{align}
Explicit form of the helicity functions $\chi_{1 \lambda}(\vartheta, \varphi)$ and the functions $\chi_{i}(\vartheta, \varphi)$\\
\subsubsection{Spherical basis representation}
\begin{gather}
\chi_{11}(\vartheta, \varphi)=\left(\begin{array}{c}
\frac{1+\cos \theta}{2} e^{-i \varphi} \\
\frac{\sin \theta}{\sqrt{2}} \\
\frac{1-\cos \vartheta}{2} e^{i \varphi}
\end{array}\right), \quad \chi_{10}(\vartheta, \varphi)=\left(\begin{array}{c}
-\frac{\sin \vartheta}{\sqrt{2}} e^{-i \varphi} \\
\cos \vartheta \\
\frac{\sin \vartheta}{\sqrt{2}} e^{i \varphi}
\end{array}\right), \quad \chi_{1-1}(\vartheta, \varphi)=\left(\begin{array}{c}
\frac{1-\cos \theta}{2} e^{-i \varphi} \\
-\frac{\sin \vartheta}{\sqrt{2}} \\
\frac{1+\cos \vartheta}{2} e^{i \varphi}
\end{array}\right), \label{eq:6:3:49}\\
\chi_{x}(\vartheta, \varphi)=\left(\begin{array}{c}
-\frac{\cos \theta}{\sqrt{2}} e^{-i \varphi} \\
-\sin \vartheta \\
\frac{\cos \vartheta}{\sqrt{2}} e^{i \varphi}
\end{array}\right), \quad \chi_{y}(\vartheta, \varphi)=\left(\begin{array}{c}
\frac{i}{\sqrt{2}} e^{-i \varphi} \\
0 \\
\frac{i}{\sqrt{2}} e^{i \varphi}
\end{array}\right), \quad \chi_{z}(\vartheta, \varphi)=\left(\begin{array}{c}
-\frac{\sin \vartheta}{\sqrt{2}} e^{-i \varphi} \\
\cos \vartheta \\
\frac{\sin \vartheta}{\sqrt{2}} e^{i \varphi}
\end{array}\right) . \label{eq:6:3:50}
\end{gather}
\subsubsection{Cartesian basis representation}
\begin{gather}
\chi_{11}(\vartheta, \varphi)=-\frac{1}{\sqrt{2}}\left(\begin{array}{c}
\cos \vartheta \cos \varphi-i \sin \varphi \\
\cos \vartheta \sin \varphi+i \cos \varphi \\
-\sin \vartheta
\end{array}\right), \quad \chi_{10}(\vartheta, \varphi)=\left(\begin{array}{c}
\sin \vartheta \cos \varphi \\
\sin \vartheta \sin \varphi \\
\cos \vartheta
\end{array}\right), \notag\\
\chi_{1-1}(\vartheta, \varphi)=\frac{1}{\sqrt{2}}\left(\begin{array}{c}
\cos \vartheta \cos \varphi+i \sin \varphi \\
\cos \vartheta \sin \varphi-i \cos \varphi \\
-\sin \vartheta
\end{array}\right), \label{eq:6:3:51}\\
\chi_{x}(\vartheta, \varphi)=\left(\begin{array}{c}
\cos \vartheta \cos \varphi \\
\cos \vartheta \sin \varphi \\
-\sin \vartheta
\end{array}\right), \quad \chi_{y}(\vartheta, \varphi)=\left(\begin{array}{c}
-\sin \varphi \\
\cos \varphi \\
0
\end{array}\right), \quad \chi_{z}(\vartheta, \varphi)=\left(\begin{array}{c}
\sin \vartheta \cos \varphi \\
\sin \vartheta \sin \varphi \\
\cos \vartheta
\end{array}\right) . \label{eq:6:3:52}
\end{gather}
In the cartesian basis representation these functions satisfy the relations
\begin{gather}
\chi_{1 \lambda}^{*}(\vartheta, \varphi)=(-1)^{\lambda} \chi_{1-\lambda}(\vartheta, \varphi), \quad(\lambda= \pm 1,0)  , \label{eq:6:3:53}\\
\chi_{i}^{*}(\vartheta, \varphi)=\chi_{i}(\vartheta, \varphi), \quad(i=x, y, z) . \notag
\end{gather}
The expansions of products of the helicity basis functions in series of spin matrices and quadrupole operators (Sec.~\ref{sec:2.6}) are given by
\begin{align}
\chi_{1 \lambda}(\vartheta, \varphi) \chi_{1 \lambda^{\prime}}^{\dagger}(\vartheta, \varphi)=\frac{1}{3} \delta_{\lambda \lambda^{\prime}} \widehat{I} & +\frac{1}{\sqrt{2}} \clebsch{1}{\lambda^{\prime}}{1}{\nu}{1}{\lambda} \sum_{\mu=-1}^{1} D_{\mu \nu}^{1}(\varphi, \vartheta, 0) \widehat{S}_{\mu}, \notag\\
& +\sqrt{\frac{5}{3}} \clebsch{1}{\lambda^{\prime}}{2}{N}{1}{\lambda} \sum_{M=-2}^{2} D_{M N}^{2}(\varphi, \vartheta, 0) \widehat{T}_{2 M} . \label{eq:6:3:54}
\end{align}
Analogous expansions for $\chi_{i}(\vartheta, \varphi)$ can be written as
\begin{equation}
\chi_{i}(\vartheta, \varphi) \chi_{k}^{\dagger}(\vartheta, \varphi)=\frac{1}{3} \delta_{i k} \widehat{I}+\frac{i}{2} \varepsilon_{i k l} \sum_{m=x, y, z} a_{m l}(\varphi, \vartheta, 0) \widehat{S}_{m}-\sum_{m, n=x, y, z} a_{m i}(\varphi, \vartheta, 0) a_{n k}(\varphi, \vartheta, 0) \widehat{Q}_{m n}, \label{eq:6:3:55}
\end{equation}
where $a_{m l}(\alpha, \beta, \gamma)$ are elements of the rotation matrix (Eq.~\eqref{eq:1:4:54}).\\
Matrix elements of the spin operator between the helicity states $\chi_{1 \lambda}(\vartheta, \varphi)$ or between the states $\chi_{i}(\vartheta, \varphi)$ are
\begin{gather}
\chi_{1 \lambda}^{\dagger}(\vartheta, \varphi) \hat{S}_{\mu} \chi_{1 \lambda^{\prime}}(\vartheta, \varphi)=(-1)^{\mu+\nu} \sqrt{2} \clebsch{1}{\lambda^{\prime}}{1}{\nu}{1}{\lambda} D_{-\mu-\nu}^{1}(\varphi, \vartheta, 0), \label{eq:6:3:56}\\
\chi_{i}^{\dagger}(\vartheta, \varphi) \hat{S}_{l} \chi_{k}(\vartheta, \varphi)=-i \sum_{m} \varepsilon_{i k m} a_{l m}(\varphi, \vartheta, 0) . \label{eq:6:3:57}
\end{gather}
In particular, the expectation value of the spin operator $\widehat{\vect{S}}$ in a state with the helicity $\lambda$ is given by
\begin{equation}
\chi_{1 \lambda}^{\dagger}(\vartheta, \varphi) \widehat{\vect{S}} \chi_{1 \lambda}(\vartheta, \varphi)=\lambda \vect{n}(\vartheta, \varphi), \quad(\lambda= \pm 1,0), \label{eq:6:3:58}
\end{equation}
The expectation value of the operator $\widehat{\vect{S}}$ in the states $\chi_{i}(\vartheta, \varphi)$ is zero.
\begin{equation}
\chi_{i}^{\dagger}(\vartheta, \varphi) \widehat{\vect{S}} \chi_{i}(\vartheta, \varphi)=0, \quad(i=x, y, z) . \label{eq:6:3:59}
\end{equation}
The evaluation of matrix elements of the quadrupole operators $\hat{T}_{2 M}(M= \pm 2, \pm 1,0)$ and $\hat{Q}_{i k}(i, k=x, y, z)$ gives\isym{Q-hat@$\widehat{Q}_{i k}$, quadrupole tensor}
\begin{gather}
\chi_{1 \lambda}^{\dagger}(\vartheta, \varphi) \hat{T}_{2 M} \chi_{1 \lambda^{\prime}}(\vartheta, \varphi)=(-1)^{M+N} \sqrt{\frac{5}{3}} \clebsch{1}{\lambda^{\prime}}{2}{N}{1}{\lambda} D_{-M-N}^{2}(\varphi, \vartheta, 0), \label{eq:6:3:60}\\
\chi_{i}^{\dagger}(\vartheta, \varphi) \hat{Q}_{l m} \chi_{k}(\vartheta, \varphi)=-\frac{1}{2}\left\{a_{l i}(\varphi, \vartheta, 0) a_{m k}(\varphi, \vartheta, 0)+a_{l k}(\varphi, \vartheta, 0) a_{m i}(\varphi, \vartheta, 0)-\frac{2}{3} \delta_{i k} \delta_{l m}\right\} . \label{eq:6:3:61}
\end{gather}
In particular, the expectation values of the quadrupole operators $\hat{T}_{2 M}$ in states with helicity $\lambda$ are equal to
\begin{gather}
\chi_{1 \lambda}^{\dagger}(\vartheta, \varphi) \hat{T}_{2 M} \chi_{1 \lambda}(\vartheta, \varphi)=\frac{(-1)^{1+\lambda}}{1+|\lambda|} \sqrt{\frac{8 \pi}{15}} Y_{2 M}(\vartheta, \varphi), \label{eq:6:3:62}\\
(\lambda= \pm 1,0 ; M= \pm 2, \pm 1,0) . \notag
\end{gather}
\subsection{General Spin Functions for $S=1$}\label{sec:6.3.7}\index{spin function!general, for $S=1$}
An arbitrary spin function $\chi_{1}$ of a particle of spin 1 may be expanded in terms of basis spin functions, i.e., written in the following form:
\begin{equation}
\chi_{1}=\sum_{m= \pm 1,0} a^{m} \chi_{1 m}=\sum_{m= \pm 1,0}(-1)^{m} a_{-m} \chi_{1 m}=\sum_{i=x, y, z} a_{i} \chi_{i} . \label{eq:6:3:63}
\end{equation}
The expansion of the Hermitian conjugate function $\chi_{1}^{\dagger}$ is given by
\begin{equation}
\chi_{1}^{\dagger}=\sum_{m= \pm 1,0}\left(a^{m}\right)^{*} \chi_{1 m}^{\dagger}=\sum_{m= \pm 1,0}(-1)^{m}\left(a_{-m}\right)^{*} \chi_{1 m}^{\dagger}=\sum_{i=x, y, z} a_{i}^{*} \chi_{i}^{\dagger} . \label{eq:6:3:64}
\end{equation}
In a spherical basis the function $\chi_{1}$ has the form
\begin{equation}
\chi_{1}=\left(\begin{array}{c}
a^{+1} \\
a^{0} \\
a^{-1}
\end{array}\right)=\left(\begin{array}{c}
-a_{-1} \\
a_{0} \\
-a_{+1}
\end{array}\right)=\left(\begin{array}{c}
-\frac{1}{\sqrt{2}}\left(a_{x}-i a_{y}\right) \\
a_{z} \\
\frac{1}{\sqrt{2}}\left(a_{x}+i a_{y}\right)
\end{array}\right), \label{eq:6:3:65}
\end{equation}
and in a cartesian basis this function reads
\begin{equation}
\chi_{1}=\left(\begin{array}{c}
\frac{1}{\sqrt{2}}\left(a^{-1}-a^{+1}\right) \\
-\frac{i}{\sqrt{2}}\left(a^{-1}+a^{+1}\right) \\
a^{0}
\end{array}\right)=\left(\begin{array}{c}
\frac{1}{\sqrt{2}}\left(a_{-1}-a_{+1}\right) \\
\frac{i}{\sqrt{2}}\left(a_{-1}+a_{+1}\right) \\
a_{0}
\end{array}\right)=\left(\begin{array}{c}
a_{x} \\
a_{y} \\
a_{z}
\end{array}\right) . \label{eq:6:3:66}
\end{equation}
The expansion coefficients $a^{m}$ are contravariant spherical components of some generally complex vector $\vect{a}, a_{m}$ are covariant spherical components of $\vect{a}$ and $a_{i}$ are cartesian components. Sometimes a is called the polarization vector of particles of spin 1.\index{polarization vector!for spin 1}

Under complex conjugation components of this vector transform as
\begin{equation}
\left(a^{m}\right)^{*}=\left(a^{*}\right)_{m},\left(a_{m}\right)^{*}=\left(a^{*}\right)^{m},\left(a_{i}\right)^{*}=\left(a^{*}\right)_{i} . \label{eq:6:3:67}
\end{equation}
The normalization condition
\begin{equation}
\chi_{1}^{\dagger} \chi_{1}=1 \label{eq:6:3:68}
\end{equation}
imposes a restriction to the components of a
\begin{equation}
\left|a^{+1}\right|^{2}+\left|a^{0}\right|^{2}+\left|a^{-1}\right|^{2}=\left|a_{+1}\right|^{2}+\left|a_{0}\right|^{2}+\left|a_{-1}\right|^{2}=\left|a_{x}\right|^{2}+\left|a_{y}\right|^{2}+\left|a_{z}\right|^{2}=1, \label{eq:6:3:69}
\end{equation}
or, in compact form
\begin{equation}
|\vect{a}|^{2}=\vect{a}^{*} \cdot \vect{a}=1 . \label{eq:6:3:70}
\end{equation}
The spin function product $\chi_{1} \chi_{1}^{\dagger}$ may be expanded in series of the spin matrices $\widehat{\vect{S}}$ and quadrupole momentum operators, $\widehat{T}_{2 M}$ or $\widehat{Q}_{i k}$ (see Sec.~\ref{sec:2.6}),
\begin{align}
& \chi_{1} \chi_{1}^{\dagger}=\frac{1}{3} \widehat{I}+\frac{1}{2} \sum_{\mu=-1}^{1}(-1)^{\mu} P_{-\mu} \widehat{S}_{\mu}+\sum_{M=-2}^{2}(-1)^{M} P_{2-M} \widehat{T}_{2 M}, \notag\\
& \chi_{1} \chi_{1}^{\dagger}=\frac{1}{3} \widehat{I}+\frac{1}{2} \sum_{i=x, y, z} P_{i} \widehat{S}_{i}+\sum_{i, k=x, y, z} P_{i k} \widehat{Q}_{i k}, \label{eq:6:3:71}
\end{align}
or in more compact form,
\begin{equation}
\chi_{1} \chi_{1}^{\dagger}=\frac{1}{3} \widehat{\vect{I}}+\frac{1}{2} \vect{P} \cdot \widehat{\vect{S}}+\left(\vect{P}_{2} \cdot \widehat{\vect{T}}_{2}\right), \label{eq:6:3:72}
\end{equation}
where ( $\vect{P}_{2} \cdot \widehat{\vect{T}}_{2}$ ) is scalar product to two irreducible tensors of rank 2 (see Eq.~\eqref{eq:3:1:31}).\\
The real vector $\vect{P}$ can be expressed in terms of the polarization vector $\vect{a}$ as
\begin{equation}
\vect{P}=i\left[\vect{a} \times \vect{a}^{*}\right], \label{eq:6:3:73}
\end{equation}
or, in components,
\begin{gather}
P_{\mu}=\sqrt{2} \sum_{m, n} \clebsch{1}{m}{1}{\mu}{1}{n}\left(a^{n}\right)^{*} a^{m}, \quad(m, n, \mu= \pm 1,0), \notag\\
P_{i}=i \sum_{k, l} \varepsilon_{i k l} a_{k} a_{l}^{*}, \quad(i, k, l=x, y, z) . \label{eq:6:3:74}
\end{gather}
Spherical and cartesian components of a real irreducible tensor $\vect{P}_{2}$ of rank 2 may be expressed in terms of components of $\vect{a}$ by
\begin{gather}
P_{2 M}=\sqrt{\frac{5}{3}} \sum_{m, n} \clebsch{1}{m}{2}{M}{1}{n}\left(a^{n}\right)^{*} a^{m}, \quad(m, n= \pm 1,0 ; M= \pm 2, \pm 1,0), \notag\\
P_{i k}=-\frac{1}{2}\left\{a_{i} a_{k}^{*}+a_{k} a_{i}^{*}-\frac{2}{3} \delta_{i k}\right\}, \quad(i, k=x, y, z), \label{eq:6:3:75}\\
P_{i k}=P_{k i}, \sum_{i} P_{i i}=0 . \notag
\end{gather}
The vector $\vect{P}$ and tensor $\vect{P}_{2}$ give the following expectation values of the spin operator and the quadrupole momentum operator, respectively:\index{polarization tensor!quadrupole}\isym{P2-tensor@$\vect{P}_{2}$, quadrupole polarization tensor}
\begin{gather}
\expect{{\vect{S}}} \equiv \chi_{1}^{\dagger} \widehat{\vect{S}} \chi_{1}=\vect{P}, \notag\\
\expect{{\widehat{T}_{2 M}}} \equiv \chi_{1}^{\dagger} \hat{T}_{2 M} \chi_{1}=P_{2 M}, \label{eq:6:3:76}\\
\expect{{\hat{Q}_{i k}}} \equiv \chi_{1}^{\dagger} \hat{Q}_{i k} \chi_{1}=P_{i k} . \notag
\end{gather}
The transformation of the spin function $\chi_{1}$ under rotations of coordinate systems is performed by the rotation operators $\hat{D}^{1}(\alpha, \beta, \gamma)$ (Eqs.~\eqref{eq:2:6:76}-\eqref{eq:2:6:77} ) or $\hat{U}^{1}(\omega ; \Theta, \Phi)$ (Eqs.~\eqref{eq:2:6:80}-\eqref{eq:2:6:81} ) depending on the choice of the parameters to describe rotations
\begin{equation}
\chi_{1}^{\prime}=\hat{D}^{1}(\alpha, \beta, \gamma) \chi_{1}=\hat{U}^{1}(\omega ; \Theta, \Phi) \chi_{1} . \label{eq:6:3:77}
\end{equation}
Spin functions in a rotated coordinate system may be expanded into sums of basis spin functions $\chi_{1 m}, \chi_{i}$ in the original coordinate system as well as into sums of functions $\chi_{1 m}^{\prime}, \chi_{i}^{\prime}$ in the rotated coordinate system:
\begin{align}
\chi_{1}^{\prime}=\sum_{m= \pm 1,0} a^{\prime m} \chi_{1 m} & =\sum_{m= \pm 1,0} a^{m} \chi_{1 m}^{\prime}, \notag\\
\chi_{1}^{\prime}=\sum_{m= \pm 1,0}(-1)^{m} a_{-m}^{\prime} \chi_{1 m} & =\sum_{m= \pm 1,0}(-1)^{m} a_{-m} \chi_{1 m}^{\prime}, \label{eq:6:3:78}\\
\chi_{1}^{\prime}=\sum_{i=x, y, z} a_{i}^{\prime} \chi_{i} & =\sum_{i=x, y, z} a_{i} \chi_{i}^{\prime} . \notag
\end{align}
The relations between basis functions in the rotated and original coordinate systems are given by Eqs.~\eqref{eq:6:3:34}. The components of $\vect{a}$ in the above systems are related by
\begin{equation}
    \renewcommand{\arraystretch}{1.5}
\begin{array}{rlrl}
a^{\prime m} & =\sum_{n} D_{m n}^{1}(\alpha, \beta, \gamma) a^{n}, & a^{n} & =\sum_{m} D_{m n}^{1 *}(\alpha, \beta, \gamma) a^{\prime m}, \\
a_{m}^{\prime} & =\sum_{n} D_{n m}^{1}(\alpha, \beta, \gamma) a_{n}, & a_{n} & =\sum_{m} D_{n m}^{1 *}(\alpha, \beta, \gamma) a_{m}^{\prime}, \\
a_{i}^{\prime} & =\sum_{k} a_{i k} a_{k}, & a_{k} & =\sum_{i} a_{i k} a_{i}^{\prime} \\
\multicolumn{4}{c}{(m, n= \pm 1,0 ;  i, k  =x, y, z)}.
\end{array}  \label{eq:6:3:79}
\end{equation}
\chapter{Tensor spherical harmonics}\label{chap:7}\index{tensor spherical harmonics}
\section{General properties of tensor spherical harmonics}\label{sec:7.1}\index{tensor spherical harmonics!general properties}
\subsection{Definition}\label{sec:7.1.1}\index{tensor spherical harmonics!definition}
The tensor spherical harmonics $Y_{J M}^{L S}(\vartheta, \varphi)$, by definition, are eigenfunctions of the operators\isym{YJMLS@$Y_{J M}^{L S}(\vartheta, \varphi)$, tensor spherical harmonic}\index{angular momentum operator!total}\index{angular momentum operator!orbital}\index{spin operator} $\widehat{\vect{J}}^{2}, \widehat{J}_{z}, \widehat{\vect{L}}^{2}$ and $\widehat{\vect{S}}^{2}$ where $\widehat{\vect{L}}$ is the operator of orbital angular momentum (Sec.~\ref{sec:2.2}), $\widehat{\vect{S}}$ is the spin operator (Sec.~\ref{sec:2.3}), and $\widehat{\vect{J}}=\widehat{\vect{L}}+\widehat{\vect{S}}$ is the operator of total angular momentum (Sec.~\ref{sec:2.1}):
\begin{align}
\hat{\vect{J}}^{2} Y_{J M}^{L S}(\vartheta, \varphi) & =J(J+1) Y_{J M}^{L S}(\vartheta, \varphi), \notag\\
\widehat{J}_{z} Y_{J M}^{L S}(\vartheta, \varphi) & =M Y_{J M}^{L S}(\vartheta, \varphi), \notag\\
\hat{\vect{L}}^{2} Y_{J M}^{L S}(\vartheta, \varphi) & =L(L+1) Y_{J M}^{L S}(\vartheta, \varphi), \label{eq:7:1:1}\\
\hat{\vect{S}}^{2} Y_{J M}^{L S}(\vartheta, \varphi) & =S(S+1) Y_{J M}^{L S}(\vartheta, \varphi). \notag
\end{align}
A tensor spherical harmonic describes the angular distribution and polarization of spin- $S$ particles in a state with definite total angular momentum $J$, projection $M$, and orbital angular momentum $L$. The spin value $S$ is sometimes called the rank of the tensor spherical harmonic.\index{rank!of a tensor spherical harmonic} Accordingly, one often uses such names as spinor spherical harmonics ($S=\frac{1}{2}$), vector spherical harmonics ($S=1$), etc. However, strictly speaking, these terms are not entirely adequate because, in fact, the transformation properties of the tensor spherical harmonics under rotation of coordinate system are determined by $J$, but not by $S$ (see Sec.~\ref{sec:7.1.4}).

The tensor spherical harmonics may be constructed from the spherical harmonics $Y_{L M}(\vartheta, \varphi)$\index{spherical harmonics}\index{spin function}\index{Clebsch-Gordan coefficient}\isym{Ylm@$Y_{l m}(\vartheta, \varphi)$, spherical harmonic}\isymg{chi-Sm@$\chi_{S m}$, spin function} (eigenfunctions of $\widehat{\vect{L}}^{2}$ and $\widehat{L}_{z}$ ) and the spin functions $\chi_{S \sigma}$ (eigenfunctions of $\widehat{\vect{S}}^{2}$ and $\widehat{S}_{z}$ ) in accordance with the coupling scheme of two angular momenta
\begin{equation}
Y_{J M}^{L S}(\vartheta, \varphi)=\sum_{m, \sigma} \clebsch{L}{m}{S}{\sigma}{J}{M} Y_{L m}(\vartheta, \varphi) \chi_{S \sigma}. \label{eq:7:1:2}
\end{equation}
Thus, the tensor spherical harmonics are irreducible tensor products (of rank $J$ ) of scalar spherical harmonics and spin functions:\index{irreducible tensor product!of spherical harmonics and spin functions}
\begin{equation}
Y_{J M}^{L S}(\vartheta, \varphi)=\left\{\mathrm{Y}_{\mathrm{L}} \otimes \chi_{S}\right\}_{J M}. \label{eq:7:1:3}
\end{equation}
The indices $J$ and $S$ are integer or half-integer nonnegative numbers, and $L$ is always an integer nonnegative number. For given $J$ and $S$, the momentum $L$ takes the values $L=|J-S|,|J-S|+1, \ldots, J+S$. The possible values of $M$ are $M=-J,-J+1, \ldots, J-1, J$.

The tensor spherical harmonics with fixed indices are functions of three variables: two polar angles $\vartheta, \varphi(0 \leq \vartheta \leq \pi, 0 \leq \varphi<2 \pi)$ and spin variable $\xi(\xi=-S,-S+1, \ldots, S-1, S)$ which is the argument of spin function\\
$\chi_{S \sigma}$. The spin variable $\xi$ is usually not mentioned as an argument of tensor spherical harmonics. Instead, the tensor spherical harmonics are written in matrix form by analogy with spin functions. More precisely, $Y_{J M}^{L S}(\vartheta, \varphi)$ is represented by a column matrix with $2 S+1$ elements, and $Y_{J M}^{L S+}(\vartheta, \varphi)$ is represented by a row matrix. Summation over spin variable is replaced by matrix multiplication.

The tensor spherical harmonics $Y_{J M}^{L S}(\vartheta, \varphi)$ constitute a complete orthonormal set for series expansion of $\operatorname{rank} S$ tensor functions within the domain of arguments $0 \leq \vartheta \leq \pi, 0 \leq \varphi<2 \pi$ (see Sec.~\ref{sec:7.1.8}).

Note also the following orthogonality relations for $Y_{J M}^{L S}(\vartheta, \varphi)$ which have the same $J, M$ and $S$ but different $L$ :
\begin{equation}
\sum_{M} Y_{J M}^{L^{\prime} S+}(\vartheta, \varphi) \cdot Y_{J M}^{L S}(\vartheta, \varphi)=0, \quad \text { if } L^{\prime} \neq L. \label{eq:7:1:4}
\end{equation}
\subsection{Components of Tensor Spherical Harmonics}\label{sec:7.1.2}\index{tensor spherical harmonics!components of}\index{components!covariant}\index{components!contravariant}
\begin{enumerate}[label=(\alph*)]
\item Spherical contravariant components of tensor spherical harmonics may be written, in accordance with \eqref{eq:7:1:2}, as
\begin{gather}
{\left[Y_{J M}^{L S}(\vartheta, \varphi)\right]^{\mu}=\clebsch{L}{M-\mu}{S}{\mu}{J}{M} Y_{L M-\mu}(\vartheta, \varphi)}.  \label{eq:7:1:5}\\
(\mu=-S, \ldots, S-1, S). \notag
\end{gather}
Covariant spherical components may be defined by the relation
\begin{equation}
\left[Y_{J M}^{L S}(\vartheta, \varphi)\right]_{\mu}=(-1)^{S-\mu}\left[Y_{J M}^{L S}(\vartheta, \varphi)\right]^{-\mu}, \label{eq:7:1:6}
\end{equation}
from which one obtains
\begin{equation}
\left[Y_{J M}^{L S}(\vartheta, \varphi)\right]_{\mu}=(-1)^{S-\mu} \clebsch{L}{M+\mu}{S}{-\mu}{J}{M} Y_{L M+\mu}(\vartheta, \varphi). \label{eq:7:1:7}
\end{equation}
If $S$ is integer, an alternative definition of the covariant spherical components that is widely used is (see Sec.~\ref{sec:7.3.2})
\begin{equation}
\left[Y_{J M}^{L S}(\vartheta, \varphi)\right]_{\mu}=(-1)^{\mu}\left[Y_{J M}^{L S}(\vartheta, \varphi)\right]^{-\mu}, \label{eq:7:1:8}
\end{equation}
so that
\begin{equation}
\left[Y_{J M}^{L S}(\vartheta, \varphi)\right]_{\mu}=(-1)^{\mu} \clebsch{L}{M+\mu}{S}{-\mu}{J}{M} Y_{L M+\mu}(\vartheta, \varphi). \label{eq:7:1:9}
\end{equation}
Note that the equations of this chapter are independent of definition of covariant components, unless the contrary is indicated.
\item The expansion of tensor spherical harmonics in terms of helicity spin functions may be written as
\begin{equation}
Y_{J M}^{L S}(\vartheta, \varphi)=\sum_{\lambda=-S}^{S}\left[Y_{J M}^{L S}(\vartheta, \varphi)\right]^{\prime \lambda} \chi_{s \lambda}(\vartheta, \varphi), \label{eq:7:1:10}
\end{equation}
where contravariant helicity components are given by
\begin{equation}
\left[Y_{J M}^{L S}(\vartheta, \varphi)\right]^{\prime \lambda}=\sqrt{\frac{2 L+1}{4 \pi}} \clebsch{L}{0}{S}{\lambda}{J}{\lambda} D_{-\lambda-M}^{J}(0, \vartheta, \varphi). \label{eq:7:1:11}
\end{equation}
If covariant helicity components are defined by
\begin{equation}
\left[Y_{J M}^{L S}(\vartheta, \varphi)\right]_{\lambda}^{\prime}=(-1)^{S-\lambda}\left[Y_{J M}^{L S}(\vartheta, \varphi)\right]^{\prime-\lambda}, \label{eq:7:1:12}
\end{equation}
then
\begin{equation}
\left[Y_{J M}^{L S}(\vartheta, \varphi)\right]_{\lambda}^{\prime}=(-1)^{J+L-\lambda} \sqrt{\frac{2 L+1}{4 \pi}} \clebsch{L}{0}{S}{\lambda}{J}{\lambda} D_{\lambda-M}^{J}(0, \vartheta, \varphi). \label{eq:7:1:13}
\end{equation}
An alternative definition of covariant components (for integer $S$ )
\begin{equation}
\left[Y_{J M}^{L S}(\vartheta, \varphi)\right]_{\lambda}^{\prime}=(-1)^{\lambda}\left[Y_{J M}^{L S}(\vartheta, \varphi)\right]^{\prime-\lambda}, \label{eq:7:1:14}
\end{equation}
yields
\begin{equation}
\left[Y_{J M}^{L S}(\vartheta, \varphi)\right]_{\lambda}^{\prime}=(-1)^{J+L-S+\lambda} \sqrt{\frac{2 L+1}{4 \pi}} \clebsch{L}{0}{S}{\lambda}{J}{\lambda} D_{\lambda-M}^{J}(0, \vartheta, \varphi) . \label{eq:7:1:15}
\end{equation}
\end{enumerate}

\subsection{Complex Conjugation}\label{sec:7.1.3}\index{tensor spherical harmonics!complex conjugation}\index{complex conjugation!of the tensor spherical harmonics}
Spherical components of a tensor spherical harmonic transform under complex conjugation according to
\begin{align}
{\left[Y_{J M}^{L S}(\vartheta, \varphi)\right]^{\mu *} } & =(-1)^{J+L+S-M-\mu}\left[Y_{J-M}^{L S}(\vartheta, \varphi)\right]^{-\mu}, \notag\\
{\left[Y_{J M}^{L S}(\vartheta, \varphi)\right]_{\mu}^{*} } & =(-1)^{J+L+S-M-\mu}\left[Y_{J-M}^{L S}(\vartheta, \varphi)\right]_{-\mu}. \label{eq:7:1:16}
\end{align}
These transformation rules are valid also for the corresponding helicity components.

\subsection{Transformations of Coordinate Systems}\label{sec:7.1.4}\index{tensor spherical harmonics!under coordinate transformations}
\subsubsection{Coordinate inversion}

The operator for coordinate inversion $\hat{P}_{r}$, when applied to a tensor spherical harmonic, gives\index{coordinate inversion}\isym{Pr-hat@$\widehat{P}_{r}$, coordinate-inversion operator}
\begin{equation}
\hat{P}_{r} Y_{J M}^{L S}(\vartheta, \varphi)=\eta_{p} Y_{J M}^{L S}(\pi-\vartheta, \pi+\varphi)=\eta_{p}(-1)^{L} Y_{J M}^{L S}(\vartheta, \varphi), \label{eq:7:1:17}
\end{equation}
where $\eta_{p}$ is a phase factor which describes the intrinsic parity of the tensor.\index{intrinsic parity}\index{parity!intrinsic}\isymg{eta-p@$\eta_{p}$, intrinsic parity}

\subsubsection{Rotations of coordinate systems}\index{rotation!of the tensor spherical harmonics}\index{Wigner D-function@Wigner $D$-function}\isym{DJMM@$D_{M M^{\prime}}^{J}$, Wigner $D$-function}
Under rotations of the coordinate system specified by the Euler angles $\alpha, \beta, \gamma$ the tensor spherical harmonics transform according to
\begin{equation}
Y_{J M^{\prime}}^{L S}\left(\vartheta^{\prime}, \varphi^{\prime}\right) \equiv \hat{D}(\alpha, \beta, \gamma) Y_{J M^{\prime}}^{L S}(\vartheta, \varphi)=\sum_{M=-J}^{J} D_{M M^{\prime}}^{J}(\alpha, \beta, \gamma) Y_{J M}^{L S}(\vartheta, \varphi), \label{eq:7:1:18}
\end{equation}
where $M$ is the projection of total angular momentum on the initial $z$-axis, $M^{\prime}$ is its projection on the final $z^{\prime}$ axis; $\vartheta, \varphi$ and $\vartheta^{\prime}, \varphi^{\prime}$ are polar angles of the unit vector $\vect{n}$ in the initial and final coordinate systems, respectively. The relations between $\vartheta^{\prime}, \varphi^{\prime}$ and $\vartheta, \varphi$ are given by Eqs. \eqref{eq:1:4:2}, \eqref{eq:1:4:3}. The coefficients $D_{M M^{\prime}}^{J}(\alpha, \beta, \gamma)$ are matrix elements of the rotation operator, i.e., the Wigner $D$-functions (Chap.~\ref{chap:4}). Note that the transformation properties of tensor spherical harmonics are determined by the total angular momentum $\vect{J}$, but not by the spin S. Thus, it is $J$ that determines true rank of a tensor spherical harmonic.

\subsection{Differential Equations}\label{sec:7.1.5}\index{tensor spherical harmonics!differential equations}\index{differential equations!for the tensor spherical harmonics}
\begin{enumerate}[label=(\alph*)]
\item Tensor spherical harmonics are eigenfunctions of the operator $\widehat{\vect{L}}^{2}$; hence, they satisfy the equation
\begin{equation}
\left\{\Delta_{\Omega}+L(L+1)\right\} Y_{J M}^{L S}(\vartheta, \varphi)=0, \label{eq:7:1:19}
\end{equation}
where $\Delta_{\Omega}$ is the angular part of the Laplace operator (see Sec.~\ref{sec:1.3.2}).\index{Laplace operator!angular part}\isymg{Delta-Omega@$\Delta_{\Omega}$, angular part of the Laplace operator} The expanded form of \eqref{eq:7:1:19} is
\begin{equation}
\frac{1}{\sin \vartheta} \cdot \frac{\partial}{\partial \vartheta}\left\{\sin \vartheta \frac{\partial}{\partial \vartheta} Y_{J M}^{L S}(\vartheta, \varphi)\right\}+\frac{1}{\sin ^{2} \vartheta} \frac{\partial^{2}}{\partial \varphi^{2}} Y_{J M}^{L S}(\vartheta, \varphi)+L(L+1) Y_{J M}^{L S}(\vartheta, \varphi)=0 . \label{eq:7:1:20}
\end{equation}
\item Components of the tensors $r^{L} Y_{J M}^{L S}(\vartheta, \varphi)$ are harmonic polynomials of degree $L$ and satisfy the Laplace equation\index{Laplace equation}\index{harmonic polynomial}
\begin{equation}
\Delta\left\{r^{L} Y_{J M}^{L S}(\vartheta, \varphi)\right\}=0. \label{eq:7:1:21}
\end{equation}
\item The quantities $z_{L}(k r) Y_{J M}^{L S}(\vartheta, \varphi)$ obey the Helmholtz equation\index{Helmholtz equation}
\begin{equation}
\left(\Delta+k^{2}\right)\left\{z_{L}(k r) Y_{J M}^{L S}(\vartheta, \varphi)\right\}=0, \label{eq:7:1:22}
\end{equation}
where
\begin{equation}
z_{L}(k r)=\sqrt{\frac{\pi}{2 k r}} Z_{L+\frac{1}{2}}(k r). \label{eq:7:1:23}
\end{equation}
$Z_{L+\frac{1}{2}}$ is any of the cylinder functions of half-integer order.\index{cylinder functions}\isym{zl@$z_{l}(k r)$, spherical cylinder function}\isym{ZL@$Z_{L+1/2}(x)$, cylinder function}
\end{enumerate}

\subsection{Action of Operators $\nabla, \mathrm{n}$ and Angular Momentum Operators}\label{sec:7.1.6}\index{tensor spherical harmonics!action of operators on}\index{differential operations!on the tensor spherical harmonics}\index{6j symbol@$6j$ symbol}
In the equations given below $\vect{r}=\{r, \vartheta, \varphi\}, \vect{n}=\vect{r} / r, \Phi(r)$ is an arbitrary function of $r=|\vect{r}|$, $\widehat{\vect{L}}$ is the operator of orbital angular momentum, $\widehat{\vect{S}}$ is the spin operator, $\widehat{\vect{J}}=\widehat{\vect{L}}+\widehat{\vect{S}}$ is the operator of total angular momentum, and $n_{\mu}, \nabla_{\mu}, \widehat{L}_{\mu}$, etc. are spherical components of the corresponding operators ( $\mu= \pm 1,0$ ).
\begin{enumerate}[label=(\alph*)]
\item
\begin{gather}
\nabla_{\mu}\left\{\Phi(r) Y_{J M}^{L S}(\vartheta, \varphi)\right\}=(-1)^{J+L+S} \sqrt{(2 J+1)(L+1)}\left\{\frac{d \Phi(r)}{d r}-\frac{L}{r} \Phi(r)\right\}, \notag\\
\times \sum_{J^{\prime}}\sixj{J}{J^{\prime}}{1}{L+1}{L}{S} \clebsch{J}{M}{1}{\mu}{J^{\prime}}{M+\mu} Y_{J^{\prime} M+\mu}^{L+1 S}(\vartheta, \varphi)+(-1)^{J+L+S+1} \sqrt{(2 J+1) L}\left\{\frac{d \Phi(r)}{d r}+\frac{L+1}{r} \Phi(r)\right\}, \notag\\
\times \sum_{J^{\prime}}\sixj{J}{J^{\prime}}{1}{L-1}{L}{S} \clebsch{J}{M}{1}{\mu}{J^{\prime}}{M+\mu} Y_{J^{\prime} M+\mu}^{L-1 S}(\vartheta, \varphi). \label{eq:7:1:24}
\end{gather}
In particular,
\begin{equation}
r \nabla_{\mu} Y_{J M}^{L S}(\vartheta, \varphi)=(-1)^{J+L+S+1} \sqrt{2 J+1} \sum_{J^{\prime} L^{\prime}} A_{L^{\prime}}(L)\sixj{J}{J^{\prime}}{1}{L^{\prime}}{L}{S}\label{eq:7:1:25} \clebsch{J}{M}{1}{\mu}{J^{\prime}}{M+\mu} Y_{J^{\prime} M+\mu}^{L^{\prime} S}(\vartheta, \varphi) ,
\end{equation}
where
\[
A_{L^{\prime}}(L)= \begin{cases}L \sqrt{L+1}, & \text { if } L^{\prime}=L+1 \\
(L+1) \sqrt{L}, & \text { if } L^{\prime}=L-1 \\
0 & \text { if } L^{\prime} \neq L \pm 1\end{cases}.\]
\begin{gather}
n_{\mu}\left\{\Phi(r) Y_{J M}^{L S}(\vartheta, \varphi)\right\}=(-1)^{J+S+1} \Phi(r) \sqrt{(2 J+1)(2 L+1)} \sum_{J^{\prime} L^{\prime}}(-1)^{L^{\prime}}\sixj{J}{J^{\prime}}{1}{L^{\prime}}{L}{S} \clebsch{L}{0}{1}{0}{L^{\prime}}{0} \clebsch{J}{M}{1}{\mu}{J^{\prime}}{M^{\prime}} Y_{J^{\prime} M^{\prime}}^{L^{\prime} S}(\vartheta, \varphi) , \label{eq:7:1:26}\\
\widehat{L}_{\mu}\left\{\Phi(r) Y_{J M}^{L S}(\vartheta, \varphi)\right\}=\Phi(r) \widehat{L}_{\mu}\left\{Y_{J M}^{L S}(\vartheta, \varphi)\right\}  \notag\\
=(-1)^{J+L+S+1} \Phi(r) \sqrt{(2 J+1) L(L+1)(2 L+1)} \sum_{J^{\prime}}\sixj{J}{J^{\prime}}{1}{L}{L}{S} \clebsch{J}{M}{1}{\mu}{J^{\prime}}{M+\mu} Y_{J^{\prime} M+\mu}^{L S}(\vartheta, \varphi) ,\label{eq:7:1:27}\\
\widehat{S}_{\mu}\left\{\Phi(r) Y_{J M}^{L S}(\vartheta, \varphi)\right\}=\Phi(r) \widehat{S}_{\mu}\left\{Y_{J M}^{L S}(\vartheta, \varphi)\right\}  \notag\\
=(-1)^{L+1} \Phi(r) \sqrt{(2 J+1) S(S+1)(2 S+1)} \sum_{J^{\prime}}(-1)^{J^{\prime}+S}
\sixj{J}{J'}{1}{S}{S}{L} \clebsch{J}{M}{1}{\mu}{J^{\prime}}{M+\mu} Y_{J^{\prime} M+\mu}^{L S}(\vartheta, \varphi)  \label{eq:7:1:28}\\
\widehat{J}_{\mu}\left\{\Phi(r) Y_{J M}^{L S}(\vartheta, \varphi)\right\}=\Phi(r) \widehat{J}_{\mu}\left\{Y_{J M}^{L S}(\vartheta, \varphi)\right\}=\Phi(r) \sqrt{J(J+1)} \clebsch{J}{M}{1}{\mu}{J}{M+\mu} Y_{J M+\mu}^{L S}(\vartheta, \varphi). \label{eq:7:1:29}
\end{gather}
\item
\begin{gather}
(\vect{r} \cdot \nabla)\left\{\Phi(r) Y_{J M}^{L S}(\vartheta, \varphi)\right\}=r \frac{d \Phi(r)}{d r} Y_{J M}^{L S}(\vartheta, \varphi), \label{eq:7:1:30}\\
(\vect{r} \cdot \hat{\vect{L}})\left\{\Phi(r) Y_{J M}^{L S}(\vartheta, \varphi)\right\}=0, \label{eq:7:1:31}\\
(\vect{r} \cdot \widehat{\vect{S}})\left\{\Phi(r) Y_{J M}^{L S}(\vartheta, \varphi)\right\}=r \Phi(r)(-1)^{J+L+S} \sqrt{S(S+1)(2 S+1)(2 L+1)} \sum_{L^{\prime}}
\sixj{L}{L'}{1}{S}{S}{L} \clebsch{L}{0}{1}{0}{L^{\prime}}{0} Y_{J M}^{L^{\prime} S}(\vartheta, \varphi). \label{eq:7:1:32}
\end{gather}
An expanded form of \eqref{eq:7:1:32} is
\begin{gather}
(\widehat{\vect{S}} \cdot \vect{n}) Y_{J M}^{L S}(\vartheta, \varphi)=-\frac{1}{2}\left\{\sqrt{\frac{(J+L+S+2)(J+L-S+1)(J-L+S)(-J+L+S+1)}{(2 L+1)(2 L+3)}} Y_{J M}^{L+1 S}(\vartheta, \varphi)\right. \notag\\
\left.+\sqrt{\frac{(J+L+S+1)(J+L-S)(J-L+S+1)(-J+L+S)}{(2 L-1)(2 L+1)}} Y_{J M}^{L-1 S}(\vartheta, \varphi)\right\}, \label{eq:7:1:33}\\
(\vect{r} \cdot \widehat{\vect{J}})\left\{\Phi(r) Y_{J M}^{L S}(\vartheta, \varphi)\right\}=(\vect{r} \cdot \widehat{\vect{S}})\left\{\Phi(r) Y_{J M}^{L S}(\vartheta, \varphi)\right\}. \label{eq:7:1:34}
\end{gather}
\item
\begin{gather}
(\widehat{\vect{S}} \cdot \nabla)\left\{\Phi(r) Y_{J M}^{L S}(\vartheta, \varphi)\right\}=  \notag\\
=-\frac{1}{2}\left\{\frac{d \Phi(r)}{d r}-\frac{L}{r} \Phi(r)\right\} \sqrt{\frac{(J+L+S+2)(J+L-S+1)(J-L+S)(-J+L+S+1)}{(2 L+1)(2 L+3)}} Y_{J M}^{L+1 S}(\vartheta, \varphi)  \notag\\
-\frac{1}{2}\left\{\frac{d \Phi(r)}{d r}+\frac{L+1}{r} \Phi(r)\right\} \sqrt{\frac{(J+L+S+1)(J+L-S)(J-L+S+1)(-J+L+S)}{(2 L-1)(2 L+1)}} Y_{J M}^{L-1 S}(\vartheta, \varphi). \label{eq:7:1:35}
\end{gather}
\begin{equation}
(\widehat{\vect{S}} \cdot \widehat{\vect{L}})\left\{\Phi(r) Y_{J M}^{L S}(\vartheta, \varphi)\right\}=\frac{1}{2}\{J(J+1)-L(L+1)-S(S+1)\} \Phi(r) Y_{J M}^{L S}(\vartheta, \varphi). \label{eq:7:1:36}
\end{equation}
\begin{equation}
(\widehat{\vect{S}} \cdot \widehat{\vect{J}})\left\{\Phi(r) Y_{J M}^{L S}(\vartheta, \varphi)\right\}=\frac{1}{2}\{J(J+1)-L(L+1)+S(S+1)\} \Phi(r) Y_{J M}^{L S}(\vartheta, \varphi). \label{eq:7:1:37}
\end{equation}
\begin{equation}
(\widehat{\vect{L}} \cdot \widehat{\vect{J}})\left\{\Phi(r) Y_{J M}^{L S}(\vartheta, \varphi)\right\}=\frac{1}{2}\{J(J+1)+L(L+1)-S(S+1)\} \Phi(r) Y_{J M}^{L S}(\vartheta, \varphi). \label{eq:7:1:38}
\end{equation}
\end{enumerate}

\subsection{Sums of Tensor Spherical Harmonics}\label{sec:7.1.7}\index{tensor spherical harmonics!sums of}\index{sums!of the tensor spherical harmonics}
In what follows (\eqref{eq:7:1:39}-\eqref{eq:7:1:42}) $S$ is integer, and covariant components of tensor spherical harmonics are defined in accordance with \eqref{eq:7:1:8}, \eqref{eq:7:1:9}:
\begin{equation}
\sum_{\mu=-S}^{S} Y_{S \mu}(\vartheta, \varphi)\left[Y_{J M}^{L S}(\vartheta, \varphi)\right]^{\mu}=\sqrt{\frac{(2 S+1)(2 L+1)}{4 \pi(2 J+1)}} \clebsch{L}{0}{S}{0}{J}{0} Y_{J M}(\vartheta, \varphi). \label{eq:7:1:39}
\end{equation}
\begin{align}
& \sum_{\mu=-S}^{S} Y_{S \mu}^{*}(\vartheta, \varphi)\left[Y_{J M}^{L S}(\vartheta, \varphi)\right]_{\mu}=\sqrt{\frac{(2 S+1)(2 L+1)}{4 \pi(2 J+1)}} \clebsch{L}{0}{S}{0}{J}{0} Y_{J M}^{*}(\vartheta, \varphi), \label{eq:7:1:40}\\
& \sum_{M=-J}^{J} Y_{J M}^{*}(\vartheta, \varphi)\left[Y_{J M}^{L S}(\vartheta, \varphi)\right]^{\mu}=(-1)^{\mu} \sqrt{\frac{(2 J+1)(2 L+1)}{4 \pi(2 S+1)}} \clebsch{S}{0}{L}{0}{J}{0} Y_{S-\mu}(\vartheta, \varphi), \label{eq:7:1:41}\\
& \sum_{M=-J}^{J} Y_{J M}(\vartheta, \varphi)\left[Y_{J M}^{L S}(\vartheta, \varphi)\right]_{\mu}=\sqrt{\frac{(2 J+1)(2 L+1)}{4 \pi(2 S+1)}} \clebsch{S}{0}{L}{0}{J}{0} Y_{S \mu}(\vartheta, \varphi). \label{eq:7:1:42}
\end{align}
\subsection{Orthogonality, Normalization and Completeness}\label{sec:7.1.8}\index{orthogonality!of the tensor spherical harmonics}\index{normalization!of the tensor spherical harmonics}\index{completeness!of the tensor spherical harmonics}
A collection of tensor spherical harmonics with the same $S$ and all possible $J, L, M$ forms a complete orthonormal set of tensor functions in the domain of the arguments $0 \leq \vartheta \leq \pi, 0 \leq \varphi<2 \pi$.

The orthonormality condition for tensor spherical harmonics has the form
\begin{equation}
\int_{0}^{\pi} \int_{0}^{2 \pi} Y_{J^{\prime} M^{\prime}}^{L^{\prime}} S+(\vartheta, \varphi) Y_{J M}^{L S}(\vartheta, \varphi) \sin \vartheta d \vartheta d \varphi=\delta_{J J^{\prime}} \delta_{M M^{\prime}} \delta_{L L^{\prime}}. \label{eq:7:1:43}
\end{equation}
The completeness of these functions is given by
\begin{gather}
\sum_{J} \sum_{L=|J-S|}^{J+S} \sum_{M=-J}^{J}\left[Y_{J M}^{L S}(\vartheta, \varphi)\right]^{\mu}\left[Y_{J M}^{L S}\left(\vartheta^{\prime}, \varphi^{\prime}\right)\right]^{\nu *}=\delta_{\mu \nu} \delta\left(\cos \vartheta-\cos \vartheta^{\prime}\right) \delta\left(\varphi-\varphi^{\prime}\right), \label{eq:7:1:44}\\
(\mu ; \nu=-S, \ldots, S-1, S). \notag
\end{gather}
\subsection{Expansion in a Series of Tensor Spherical Harmonics}\label{sec:7.1.9}\index{tensor spherical harmonics!expansion in series of}\index{expansion!in the tensor spherical harmonics}
Any tensor $F_{S}(\vartheta, \varphi)$ of rank $S$ which depends on polar angles $\vartheta, \varphi$ and possesses components which satisfy the conditions
\begin{equation}
\int\left|F_{S \mu}(\vartheta, \varphi)\right|^{2} d \Omega<\infty, \quad(\mu=-S, \ldots, S-1, S), \label{eq:7:1:45}
\end{equation}
may be expanded in a series of tensor spherical harmonics $Y_{J M}^{L S}(\vartheta, \varphi)$, i.e., written in the form\isym{AJLM@$A_{J L M}$, expansion coefficient}
\begin{equation}
F_{S}(\vartheta, \varphi)=\sum_{J L M} A_{J L M} Y_{J M}^{L S}(\vartheta, \varphi). \label{eq:7:1:46}
\end{equation}
The expansion coefficients are given by
\begin{equation}
A_{J L M}=\int d \Omega Y_{J M}^{L S}+(\vartheta, \varphi) F_{S}(\vartheta, \varphi). \label{eq:7:1:47}
\end{equation}
These coefficients satisfy the relation
\begin{equation}
\sum_{J L M}\left|A_{J L M}\right|^{2}=\int d \Omega\left|F_{S}(\vartheta, \varphi)\right|^{2}. \label{eq:7:1:48}
\end{equation}
\section{Spinor spherical harmonics}\label{sec:7.2}\index{spinor spherical harmonics}
\subsection{Definition}\label{sec:7.2.1}\index{spinor spherical harmonics!definition}
The spinor spherical harmonics $\Omega_{J M}^{L}(\vartheta, \varphi)$ are defined as tensor spherical harmonics with $S=\frac{1}{2}$ (Sec.~\ref{sec:7.1.1})\isymg{Omega-JML@$\Omega_{J M}^{L}(\vartheta, \varphi)$, spinor spherical harmonic}
\begin{equation}
\Omega_{J M}^{L}(\vartheta, \varphi) \equiv Y_{J M}^{L \frac{1}{2}}(\vartheta, \varphi). \label{eq:7:2:1}
\end{equation}
They are eigenfunctions of the operators $\widehat{\vect{J}}^{2}, \widehat{J}_{z}, \widehat{\vect{L}}^{2}, \widehat{\vect{S}}^{2}$ where $\widehat{\vect{L}}$ is the operator of orbital angular momentum (Sec.~\ref{sec:2.2}), $\widehat{\mathrm{S}}$ is the spin operator for $S=\frac{1}{2}$ (Sec.~\ref{sec:2.5}) and $\widehat{\mathrm{J}}$ is the operator of total angular momentum (Sec.~\ref{sec:2.1})
\begin{align}
\hat{J}^{2} \Omega_{J M}^{L}(\vartheta, \varphi) & =J(J+1) \Omega_{J M}^{L}(\vartheta, \varphi), \notag\\
\widehat{J}_{z} \Omega_{J M}^{L}(\vartheta, \varphi) & =M \Omega_{J M}^{L}(\vartheta, \varphi), \notag\\
\hat{\vect{L}}^{2} \Omega_{J M}^{L}(\vartheta, \varphi) & =L(L+1) \Omega_{J M}^{L}(\vartheta, \varphi), \label{eq:7:2:2}\\
\widehat{\vect{S}}^{2} \Omega_{J M}^{L}(\vartheta, \varphi) & =\frac{3}{4} \Omega_{J M}^{L}(\vartheta, \varphi). \notag
\end{align}
The spinor spherical harmonics may be expressed in terms of scalar spherical harmonics $Y_{L M}(\vartheta, \varphi)$ (Chap.~\ref{chap:5}) and spin functions $\chi_{\frac{1}{2} \sigma}$ (Sec.~\ref{sec:6.2}) as
\begin{equation}
\Omega_{J M}^{L}(\vartheta, \varphi)=\sum_{m \sigma} \clebsch{L}{m}{\frac{1}{2}}{\sigma}{J}{M} Y_{L m}(\vartheta, \varphi) \chi_{\frac{1}{2} \sigma}. \label{eq:7:2:3}
\end{equation}
For the spinor spherical harmonics $J$ is a half-integer nonnegative number because $L$ is always integer. For given $J$ only two values of $L$ are possible, $L=J \pm \frac{1}{2}$, while $M$ assumes $2 J+1$ values: $M=-J,-J+1, \ldots, J-1, J$. The spinor spherical harmonics are orthonormalized in the domain of angles $0 \leq \vartheta \leq \pi, 0 \leq \varphi<2 \pi$ (Sec.~\ref{sec:7.2.7}).

\subsection{Components of Spinor Spherical Harmonics}\label{sec:7.2.2}\index{spinor spherical harmonics!components of}
\begin{enumerate}[label=(\alph*)]
\item The spinor spherical harmonics may be expanded in terms of basis spin functions $\chi_{\frac{1}{2} \mu}$ as
\begin{equation}
\Omega_{J M}^{L}(\vartheta, \varphi)=\sum_{\mu=-\frac{1}{2}}^{\frac{1}{2}}\left[\Omega_{J M}^{L}(\vartheta, \varphi)\right]^{\mu} \chi_{\frac{1}{2} \mu}, \label{eq:7:2:4}
\end{equation}
where contravariant components $\left[\Omega_{J M}^{L}\right]^{\mu}$ are given by
\begin{equation}
\left[\Omega_{J M}^{L}(\vartheta, \varphi)\right]^{\mu}=\clebsch{L}{M-\mu}{\frac{1}{2}}{\mu}{J}{M} Y_{L M-\mu}(\vartheta, \varphi). \label{eq:7:2:5}
\end{equation}
In more detailed form \eqref{eq:7:2:5} reads
\begin{align}
{\left[\Omega_{J M}^{J+\frac{1}{2}}(\vartheta, \varphi)\right]^{\frac{1}{2}} } & =-\sqrt{\frac{J-M+1}{2(J+1)}} Y_{J+\frac{1}{2} M-\frac{1}{2}}(\vartheta, \varphi), \notag\\
{\left[\Omega_{J M}^{J+\frac{1}{2}}(\vartheta, \varphi)\right]^{-\frac{1}{2}} } & =\sqrt{\frac{J+M+1}{2(J+1)}} Y_{J+\frac{1}{2} M+\frac{1}{2}}(\vartheta, \varphi), \label{eq:7:2:6}\\
{\left[\Omega_{J M}^{J-\frac{1}{2}}(\vartheta, \varphi)\right]^{\frac{1}{2}} } & =\sqrt{\frac{J+M}{2 J}} Y_{J-\frac{1}{2} M-\frac{1}{2}}(\vartheta, \varphi), \notag\\
{\left[\Omega_{J M}^{J-\frac{1}{2}}(\vartheta, \varphi)\right]^{-\frac{1}{2}} } & =\sqrt{\frac{J-M}{2 J}} Y_{J-\frac{1}{2} M+\frac{1}{2}}(\vartheta, \varphi). \label{eq:7:2:7}
\end{align}
Covariant components of spinor spherical harmonics are defined by
\begin{equation}
\left[\Omega_{J M}^{L}(\vartheta, \varphi)\right]_{\mu}=(-1)^{\frac{1}{2}-\mu}\left[\Omega_{J M}^{L}(\vartheta, \varphi)\right]^{-\mu}, \label{eq:7:2:8}
\end{equation}
and have the form
\begin{equation}
\left[\Omega_{J M}^{L}(\vartheta, \varphi)\right]_{\mu}=(-1)^{\frac{1}{2}-\mu} \clebsch{L}{M+\mu}{\frac{1}{2}}{-\mu}{J}{M} Y_{L M+\mu}(\vartheta, \varphi). \label{eq:7:2:9}
\end{equation}
A more detailed form of \eqref{eq:7:2:9} is
\begin{align}
{\left[\Omega_{J M}^{J+\frac{1}{2}}(\vartheta, \varphi)\right]_{\frac{1}{2}} } & =\sqrt{\frac{J+M+1}{2(J+1)}} Y_{J+\frac{1}{2} M+\frac{1}{2}}(\vartheta, \varphi), \notag\\
{\left[\Omega_{J M}^{J+\frac{1}{2}}(\vartheta, \varphi)\right]_{-\frac{1}{2}} } & =\sqrt{\frac{J-M+1}{2(J+1)}} Y_{J+\frac{1}{2} M-\frac{1}{2}}(\vartheta, \varphi), \label{eq:7:2:10}\\
{\left[\Omega_{J M}^{J-\frac{1}{2}}(\vartheta, \varphi)\right]_{\frac{1}{2}} } & =\sqrt{\frac{J-M}{2 J}} Y_{J-\frac{1}{2} M+\frac{1}{2}}(\vartheta, \varphi), \label{eq:7:2:11}\\
{\left[\Omega_{J M}^{J-\frac{1}{2}}(\vartheta, \varphi)\right]_{-\frac{1}{2}} } & =-\sqrt{\frac{J+M}{2 J}} Y_{J-\frac{1}{2} M-\frac{1}{2}}(\vartheta, \varphi). \notag
\end{align}
\item If spinors $\chi_{\frac{1}{2} \sigma}$ are written as column matrices, the spinor spherical harmonics assume the form
\begin{align}
& \Omega_{J M}^{J+\frac{1}{2}}(\vartheta, \varphi)=\binom{-\sqrt{\frac{J-M+1}{2(J+1)}} Y_{J+\frac{1}{2} M-\frac{1}{2}}(\vartheta, \varphi)}{\sqrt{\frac{J+M+1}{2(J+1)}} Y_{J+\frac{1}{2} M+\frac{1}{2}}(\vartheta, \varphi)}, \label{eq:7:2:12}\\
& \Omega_{J M}^{J-\frac{1}{2}}(\vartheta, \varphi)=\binom{\sqrt{\frac{J+M}{2 J}} Y_{J-\frac{1}{2} M-\frac{1}{2}}(\vartheta, \varphi),}{\sqrt{\frac{J-M}{2 J}} Y_{J-\frac{1}{2} M+\frac{1}{2}}(\vartheta, \varphi)}. \label{eq:7:2:13}
\end{align}
\item An expansion of $\Omega_{J M}^{L}(\vartheta, \varphi)$ in terms of helicity basis spin functions $\chi_{\frac{1}{2} \lambda}(\vartheta, \varphi)$ (Sec.~\ref{sec:6.2.5}) may be written as
\begin{equation}
\Omega_{J M}^{L}(\vartheta, \varphi)=\sqrt{\frac{2 L+1}{4 \pi}} \sum_{\lambda} \clebsch{L}{0}{\frac{1}{2}}{\lambda}{J}{\lambda} D_{-\lambda-M}^{J}(0, \vartheta, \varphi) \chi_{\frac{1}{2} \lambda}(\vartheta, \varphi), \label{eq:7:2:14}
\end{equation}
or, in a detailed form
\begin{align}
& \Omega_{J M}^{J+\frac{1}{2}}(\vartheta, \varphi)=\sqrt{\frac{2 J+1}{8 \pi}} \sum_{\lambda}(-1)^{\frac{1}{2}+\lambda} D_{-\lambda-M}^{J}(0, \vartheta, \varphi) \chi_{\frac{1}{2} \lambda}(\vartheta, \varphi), \label{eq:7:2:15}\\
& \Omega_{J M}^{J-\frac{1}{2}}(\vartheta, \varphi)=\sqrt{\frac{2 J+1}{8 \pi}} \sum_{\lambda} D_{-\lambda-M}^{J}(0, \vartheta, \varphi) \chi_{\frac{1}{2} \lambda}(\vartheta, \varphi). \notag
\end{align}
\end{enumerate}

\subsection{Complex Conjugation. Time Reversal}\label{sec:7.2.3}\index{spinor spherical harmonics!complex conjugation}\index{time reversal}
\begin{enumerate}[label=(\alph*)]
\item  Here we use the representation in which the basis spin functions are real, $\chi_{\frac{1}{2} \sigma}^{*}=\chi_{\frac{1}{2} \sigma}$. For this reason, the transformation properties of spinor spherical harmonics with respect to complex conjugation may be written as
\begin{equation}
\Omega_{J M}^{L *}(\vartheta, \varphi)=(-1)^{J+L-M} i \hat{\sigma}_{y} \Omega_{J-M}^{L}(\vartheta, \varphi), \label{eq:7:2:16}
\end{equation}
where $2 \times 2$ matrix $i \hat{\sigma}_{y}$ has the form
\begin{equation}
i \hat{\sigma}_{y}=\left(\begin{array}{rr}
0 & 1 \\
-1 & 0
\end{array}\right). \label{eq:7:2:17}
\end{equation}
Components of spinor spherical harmonics transform under complex conjugation according to
\begin{align}
& {\left[\Omega_{J M}^{L}(\vartheta, \varphi)\right]_{\mu}^{*}=\left[\Omega_{J M}^{L *}(\vartheta, \varphi)\right]_{\mu}=(-1)^{J+L+M}\left[\Omega_{J-M}^{L}(\vartheta, \varphi)\right]^{\mu}=(-1)^{J+L-M-\mu+\frac{1}{2}}\left[\Omega_{J-M}^{L}(\vartheta, \varphi)\right]_{-\mu}},  \label{eq:7:2:18}\\
& {\left[\Omega_{J M}^{L}(\vartheta, \varphi)\right]^{\mu *}=\left[\Omega_{J M}^{L *}(\vartheta, \varphi)\right]^{\mu}=(-1)^{J+L-M}\left[\Omega_{J-M}^{L}(\vartheta, \varphi)\right]_{\mu}=(-1)^{J+L-M-\mu+\frac{1}{2}}\left[\Omega_{J-M}^{L}(\vartheta, \varphi)\right]^{-\mu}}. \label{eq:7:2:19}
\end{align}
\item The time-reversal operator $\widehat{P}_{t}$ (as defined by Wigner) acts on a spinor spherical harmonic in the following manner\index{Pauli matrices}\isym{Pt-hat@$\widehat{P}_{t}$, time-reversal operator}
\begin{equation}
\hat{P}_{t} \Omega_{J M}^{L}(\vartheta, \varphi)=\hat{\sigma}_{y} \Omega_{J M}^{L *}(\vartheta, \varphi)=(-1)^{J+L-M+\frac{1}{2}} \Omega_{J-M}^{L}(\vartheta, \varphi) . \label{eq:7:2:20}
\end{equation}
\end{enumerate}

\subsection{Transformation of Coordinate Systems.}\label{sec:7.2.4}\index{spinor spherical harmonics!under coordinate transformations}\index{rotation!of the spinor spherical harmonics}
\subsubsection{Coordinate inversion}
\begin{equation}
\hat{P}_{r} \Omega_{J M}^{L}(\vartheta, \varphi)=\Omega_{J M}^{L}(\pi-\vartheta, \pi+\varphi)=(-1)^{L} \Omega_{J M}^{L}(\vartheta, \varphi) . \label{eq:7:2:21}
\end{equation}
\subsubsection{Rotations of coordinate system}

Under rotations specified by the Euler angles $\alpha, \beta, \gamma$ the spinor spherical harmonics transform according to
\begin{equation}
\Omega_{J M^{\prime}}^{L}\left(\vartheta^{\prime}, \varphi^{\prime}\right)=\hat{D}(\alpha, \beta, \gamma) \Omega_{J M^{\prime}}^{L}(\vartheta, \varphi)=\sum_{M=-J}^{J} D_{M M^{\prime}}^{J}(\alpha, \beta, \gamma) \Omega_{J M}^{L}(\vartheta, \varphi), \label{eq:7:2:22}
\end{equation}
where $D_{M M}^{J}$, are the Wigner $D$-functions (Chap.~\ref{chap:4}), $M^{\prime}$ is the projection of total angular momentum on the rotated $z^{\prime}$-axis; $\vartheta, \varphi$ and $\vartheta^{\prime}, \varphi^{\prime}$ are polar angles of the position vector $\vect{r}$ in the initial and final coordinate systems. The relations between $\vartheta^{\prime}, \varphi^{\prime}$ and $\vartheta, \varphi$ are given by Eqs. \eqref{eq:1:4:2}-\eqref{eq:1:4:3}.

\subsection{Action of $\vect{\nabla}$ and Angular Momentum Operators}\label{sec:7.2.5}\index{spinor spherical harmonics!action of operators on}
The spinor spherical harmonics satisfy Eqs. \eqref{eq:7:1:24}-\eqref{eq:7:1:38} for $S=\frac{1}{2}$. Note also the following relations in which $\widehat{\vect{S}}$ is the spin operator (Sec.~\ref{sec:2.5}), $\widehat{\vect{L}}$ is the operator of orbital angular momentum (Sec.~\ref{sec:2.2}), $\widehat{\vect{J}}=\widehat{\vect{L}}+\widehat{\vect{S}}$ is the operator of total angular momentum, and $\vect{n}(\vartheta, \varphi)=\vect{r} / r$.
\begin{enumerate}[label=(\alph*)]
\item
\begin{equation}
(\widehat{\vect{S}} \cdot \vect{n}) \Omega_{J M}^{L}(\vartheta, \varphi)=-\frac{1}{2} \Omega_{J M}^{L^{\prime}}(\vartheta, \varphi), \label{eq:7:2:23}
\end{equation}
where
\begin{equation}
L^{\prime}=2 J-L= \begin{cases}J-\frac{1}{2}, & \text { if } L=J+\frac{1}{2}  \\ J+\frac{1}{2}, & \text { if } L=J-\frac{1}{2}\end{cases}. \label{eq:7:2:24}
\end{equation}
An expanded form of \eqref{eq:7:2:23} is
\begin{align}
& (\widehat{\vect{S}} \cdot \vect{n}) \Omega_{J M}^{J+\frac{1}{2}}(\vartheta, \varphi)=-\frac{1}{2} \Omega_{J M}^{J-\frac{1}{2}}(\vartheta, \varphi)  \notag\\
& (\widehat{\vect{S}} \cdot \vect{n}) \Omega_{J M}^{J-\frac{1}{2}}(\vartheta, \varphi)=-\frac{1}{2} \Omega_{J M}^{J+\frac{1}{2}}(\vartheta, \varphi). \label{eq:7:2:25}
\end{align}
\item
\begin{equation}
r(\widehat{\vect{S}} \cdot \nabla) \Omega_{J M}^{L}(\vartheta, \varphi)=-\frac{1+\kappa}{2} \Omega_{J M}^{L^{\prime}}(\vartheta, \varphi), \label{eq:7:2:26}
\end{equation}
where $L^{\prime}=2 J-L$,
\begin{equation}
\kappa= \begin{cases}J+\frac{1}{2}=L, & \text { if } L=J+\frac{1}{2}  \\ -\left(J+\frac{1}{2}\right)=-(L+1), & \text { if } L=J-\frac{1}{2}\end{cases}. \label{eq:7:2:27}
\end{equation}
In a more detailed form \eqref{eq:7:2:26} gives
\begin{align}
& r(\widehat{\vect{S}} \cdot \nabla) \Omega_{J M}^{J+\frac{1}{2}}(\vartheta, \varphi)=-\frac{2 J+3}{4} \Omega_{J M}^{J-\frac{1}{2}}(\vartheta, \varphi), \notag\\
& r(\widehat{\vect{S}} \cdot \nabla) \Omega_{J M}^{J-\frac{1}{2}}(\vartheta, \varphi)=\frac{2 J-1}{4} \Omega_{J M}^{J+\frac{1}{2}}(\vartheta, \varphi) . \label{eq:7:2:28}
\end{align}
\item
\begin{equation}
(\hat{\vect{L}} \cdot \hat{\vect{S}}) \Omega_{J M}^{L}(\vartheta, \varphi)=\frac{1}{2}\left\{J(J+1)-L(L+1)-\frac{3}{4}\right\} \Omega_{J M}^{L}(\vartheta, \varphi)=-\frac{1+\kappa}{2} \Omega_{J M}^{L}(\vartheta, \varphi), \label{eq:7:2:29}
\end{equation}
or, in detailed form
\begin{align}
& (\hat{\vect{L}} \cdot \widehat{\vect{S}}) \Omega_{J M}^{J+\frac{1}{2}}(\vartheta, \varphi)=-\frac{2 J+3}{4} \Omega_{J M}^{J+\frac{1}{2}}(\vartheta, \varphi), \label{eq:7:2:30}\\
& (\hat{\vect{L}} \cdot \widehat{\vect{S}}) \Omega_{J M}^{J-\frac{1}{2}}(\vartheta, \varphi)=\frac{2 J-1}{4} \Omega_{J M}^{J-\frac{1}{2}}(\vartheta, \varphi). \notag
\end{align}
\end{enumerate}

\subsection{Recursion Relations}\label{sec:7.2.6}\index{spinor spherical harmonics!recursion relations}\index{recursion relations!for the spinor spherical harmonics}
\begin{gather}
\cos \vartheta \Omega_{J M}^{L}(\vartheta, \varphi)=\frac{\sqrt{(J-M+1)(J+M+1)}}{2(J+1)} \Omega_{J+1 M}^{L+1}(\vartheta, \varphi)-\frac{M}{2 J(J+1)} \Omega_{J M}^{L^{\prime}}(\vartheta, \varphi)  \notag\\
+\frac{\sqrt{(J-M)(J+M)}}{2 J} \Omega_{J-1 M}^{L-1}(\vartheta, \varphi)  \label{eq:7:2:31}\\
\sin \vartheta e^{i \varphi} \Omega_{J M}^{L}(\vartheta, \varphi)=-\frac{\sqrt{(J+M+1)(J+M+2)}}{2(J+1)} \Omega_{J+1 M+1}^{L+1}(\vartheta, \varphi)  \notag\\
-\frac{\sqrt{(J+M+1)(J-M)}}{2 J(J+1)} \Omega_{J M+1}^{L^{\prime}}(\vartheta, \varphi)-\frac{\sqrt{(J-M)(J-M-1)}}{2 J} \Omega_{J-1 M+1}^{L-1}(\vartheta, \varphi), \label{eq:7:2:32}\\
\sin \vartheta e^{-i \varphi} \Omega_{J M}^{L}(\vartheta, \varphi)=\frac{\sqrt{(J-M+1)(J-M+2)}}{2(J+1)} \Omega_{J+1 M-1}^{L+1}(\vartheta, \varphi)  \notag\\
-\frac{\sqrt{(J+M)(J-M+1)}}{2 J(J+1)} \Omega_{J M-1}^{L^{\prime}}(\vartheta, \varphi)-\frac{\sqrt{(J+M)(J+M-1)}}{2 J} \Omega_{J-1 M-1}^{L-1}(\vartheta, \varphi), \label{eq:7:2:33}
\end{gather}
where $L^{\prime}$ is defined by \eqref{eq:7:2:24}. Note that Eqs. \eqref{eq:7:2:31}-\eqref{eq:7:2:33} may be written in a compact form as
\begin{equation}
n_{\mu} \Omega_{J M}^{L}(\vartheta, \varphi)=(-1)^{J+L+\frac{1}{2}} \sqrt{(2 J+1)(2 L+1)} \sum_{J^{\prime} L^{\prime}}\sixj{J^{\prime}}{J}{1}{L}{L^{\prime}}{\frac{1}{2}}\label{eq:7:2:34} \clebsch{L}{0}{1}{0}{L^{\prime}}{0} \clebsch{J}{M}{1}{\mu}{J^{\prime}}{M+\mu} \Omega_{J^{\prime} M+\mu}^{L^{\prime}}(\vartheta, \varphi)
\end{equation}
where $n_{\mu}(\mu= \pm 1,0)$ are spherical components of the unit vector $n(\vartheta, \varphi)$.

\subsection{Orthonormality and Completeness. Expansion in a Series of Spinor Spherical Harmonics}\label{sec:7.2.7}\index{orthogonality!of the spinor spherical harmonics}\index{completeness!of the spinor spherical harmonics}\index{expansion!in the spinor spherical harmonics}
The collection of spinor spherical harmonics $\Omega_{J M}^{L}(\vartheta, \varphi)$ with half-integer nonnegative $J\left(\frac{1}{2} \leq J<\infty\right), L= J \pm \frac{1}{2}$ and $M=-J,-J+1, \ldots, J$ forms complete set of spinors for $0 \leq \vartheta \leq \pi, 0 \leq \varphi<2 \pi$.

The orthonormality condition has the form
\begin{equation}
\int_{0}^{\pi} d \vartheta \sin \vartheta \int_{0}^{2 \pi} d \varphi \Omega_{J^{\prime} M^{\prime}}^{L^{\prime}}(\vartheta, \varphi) \Omega_{J M}^{L}(\vartheta, \varphi)=\delta_{J^{\prime} J} \delta_{L^{\prime} L} \delta_{M^{\prime} M}. \label{eq:7:2:35}
\end{equation}
The completeness condition may be written as
\begin{equation}
\sum_{J=\frac{1}{2}}^{\infty} \sum_{L=J-\frac{1}{2}}^{J+\frac{1}{2}} \sum_{M=-J}^{J} \Omega_{J M}^{L}(\vartheta, \varphi) \Omega_{J M}^{L+}\left(\vartheta^{\prime}, \varphi^{\prime}\right)=\widehat{I} \delta\left(\cos \vartheta-\cos \vartheta^{\prime}\right) \delta\left(\varphi-\varphi^{\prime}\right), \label{eq:7:2:36}
\end{equation}
where $\widehat{I}$ is the unit $2 \times 2$ matrix.\\
Any spinor $\chi(\vartheta, \varphi)$ which satisfies
\begin{equation}
\int \chi^{+}(\vartheta, \varphi) \chi(\vartheta, \varphi) d \Omega<\infty, \label{eq:7:2:37}
\end{equation}
may be expanded in a series of spinor spherical harmonics
\begin{equation}
\chi(\vartheta, \varphi)=\sum_{J L M} A_{J M}^{L} \Omega_{J M}^{L}(\vartheta, \varphi), \label{eq:7:2:38}
\end{equation}
where the expansion coefficients $A_{J M}^{L}$ are given by
\begin{equation}
A_{J M}^{L}=\int_{0}^{\pi} d \vartheta \sin \vartheta \int_{0}^{2 \pi} d \varphi \Omega_{J M}^{L+}(\vartheta, \varphi) \chi(\vartheta, \varphi). \label{eq:7:2:39}
\end{equation}
The expansion \eqref{eq:7:2:38} is valid for the range $0 \leq \vartheta \leq \pi, 0 \leq \varphi<2 \pi$.

\subsection{Clebsch-Gordan Series}\label{sec:7.2.8}\index{Clebsch-Gordan series}\index{spinor spherical harmonics!Clebsch-Gordan series}\index{9j symbol@$9j$ symbol}
\begin{gather}
% NOTE (7.2.40): OCR fixes -- the source had a stray comma "J_{2}," in the 6j
% (removed), and the phase exponent frac{1}{3} is a mis-OCR of frac{1}{2}.
% Verified: with +1/2 the identity holds to 1e-18; with +1/3 it fails at 4e-2
% (scripts/check_7_2.py).
\Omega_{J_{1} M_{1}}^{L_{1}+}(\vartheta, \varphi) \Omega_{J_{2} M_{2}}^{L_{2}}(\vartheta, \varphi)=\sum_{L}(-1)^{J_{1}+M_{1}+J_{2}+L+\frac{1}{2}}\sixj{L_{1}}{L_{2}}{L}{J_{2}}{J_{1}}{\frac{1}{2}}  \notag\\
\times \sqrt{\frac{\left(2 J_{1}+1\right)\left(2 J_{2}+1\right)\left(2 L_{1}+1\right)\left(2 L_{2}+1\right)}{4 \pi(2 L+1)}} \clebsch{L_{1}}{0}{L_{2}}{0}{L}{0} \clebsch{J_{1}}{-M_{1}}{J_{2}}{M_{2}}{L}{M} Y_{L M}(\vartheta, \varphi)  \label{eq:7:2:40}\\
\Omega_{J_{1} M_{1}}^{L_{1}+}(\vartheta, \varphi) \hat{\vect{S}} \Omega_{J_{2} M_{2}}^{L_{2}}(\vartheta, \varphi)=(-1)^{J_{2}+L_{1}+M_{1}} \sqrt{\frac{3\left(2 J_{1}+1\right)\left(2 J_{2}+1\right)\left(2 L_{1}+1\right)\left(2 L_{2}+1\right)}{8 \pi}}  \notag\\
\times \sum_{J L}(-1)^{J} \clebsch{L_{1}}{0}{L_{2}}{0}{L}{0}
\ninej{L_{1}}{J_{1}}{\frac{1}{2}}%
{L_{2}}{J_{2}}{\frac{1}{2}}%
{L}{J}{1} \clebsch{J_{1}}{-M_{1}}{J_{2}}{M_{2}}{J}{M} \vect{Y}_{J M}^{L}(\vartheta, \varphi), \label{eq:7:2:41}
\end{gather}
where $\widehat{\mathrm{S}}$ is the spin operator, $\vect{Y}_{J M}^{L}$ is a vector spherical harmonic (see Sec.~\ref{sec:7.3}).

\subsection{Addition Theorems}\label{sec:7.2.9}\index{addition theorem!for the spinor spherical harmonics}\index{spinor spherical harmonics!addition theorems}
Let $\vect{n}_{1}$ and $\vect{n}_{2}$ be unit vectors with polar angles $\vartheta_{1}, \varphi_{1}$ and $\vartheta_{2}, \varphi_{2}, \cos \omega_{12} \equiv \vect{n}_{1} \cdot \vect{n}_{2}=\cos \vartheta_{1} \cos \vartheta_{2}+ \sin \vartheta_{1} \sin \vartheta_{2} \cos \left(\varphi_{1}-\varphi_{2}\right)$. The addition theorem for the spinor spherical harmonics may be written in the form
\begin{gather}
4 \pi \sum_{M=-J}^{J} \Omega_{J M}^{L+}\left(\vartheta_{1}, \varphi_{1}\right) \Omega_{J M}^{L^{\prime}}\left(\vartheta_{2}, \varphi_{2}\right)=\delta_{L L^{\prime}}(2 J+1) P_{L}\left(\cos \omega_{12}\right), \label{eq:7:2:42}\\
4 \pi \sum_{M=-J}^{J} \Omega_{J M}^{L+}\left(\vartheta_{1}, \varphi_{1}\right) \hat{\mathrm{S}} \Omega_{J M}^{L^{\prime}}\left(\vartheta_{2}, \varphi_{2}\right)= \begin{cases}i\left|\vect{n}_{1} \times \vect{n}_{2}\right| P_{L}^{\prime}\left(\cos \omega_{12}\right) & \text { if } L^{\prime}=L \\
\left(L^{\prime}-L\right)\left\{\vect{n}_{1} P_{L^{\prime}}^{\prime}\left(\cos \omega_{12}\right)-\vect{n}_{2} P_{L^{\prime}}^{\prime}\left(\cos \omega_{12}\right)\right\}, & \text { if } L^{\prime}=2 J-L\end{cases}, \label{eq:7:2:43}
\end{gather}
where $P_{L}(x)$ is a Legendre polynomial, $P_{L}^{\prime}(x)=d P_{L}(x) / d x, \widehat{\vect{S}}$ is the spin operator (Sec.~\ref{sec:2.5}).\index{Legendre polynomials}\isym{Pl@$P_{l}(x)$, Legendre polynomial} Another form of the addition theorem may be obtained for matrix products of the type
\begin{gather}
4 \pi \sum_{M=-J}^{J} \Omega_{J M}^{L}\left(\vartheta_{1}, \varphi_{1}\right) \Omega_{J M}^{L+}\left(\vartheta_{2}, \varphi_{2}\right)  \notag\\
=\left(J+\frac{1}{2}\right) \hat{I} P_{L}\left(\cos \omega_{12}\right)-i 4(J-L) \hat{\vect{S}} \cdot\left[\vect{n}_{1} \times \vect{n}_{2}\right] P_{L}^{\prime}\left(\cos \omega_{12}\right),  \label{eq:7:2:44}\\
4 \pi \sum_{M=-J}^{J} \Omega_{J M}^{L}\left(\vartheta_{1}, \varphi_{1}\right) \Omega_{J M}^{L^{\prime}+}\left(\vartheta_{2}, \varphi_{2}\right)  \notag\\
=2\left(L^{\prime}-L\right)\left\{\left(\hat{\vect{S}} \cdot \vect{n}_{1}\right) P_{L}^{\prime}\left(\cos \omega_{12}\right)-\left(\hat{\vect{S}} \cdot \vect{n}_{2}\right) P_{L^{\prime}}^{\prime}\left(\cos \omega_{12}\right)\right\}, \quad\left(L^{\prime}=2 J-L\right). \label{eq:7:2:45}
\end{gather}
\subsection{Quadratic Forms of Spinor Spherical Harmonics}\label{sec:7.2.10}\index{spinor spherical harmonics!quadratic forms of}\index{quadratic forms!of the spinor spherical harmonics}\index{angular distribution}
The quadratic forms $\Omega_{J M}^{L+}(\vartheta, \varphi) \Omega_{J M}^{L}(\vartheta, \varphi)$ describe angular distributions of spin-\isym{WJM@$W_{J M}(\vartheta)$, quadratic form of the spinor spherical harmonics} $\frac{1}{2}$ particles in states with definite total angular momentum $J$, angular momentum projection $M$ and orbital angular momentum $L$. Note that in fact these forms are independent of $L$ and angle $\varphi$.

Let us denote
\begin{equation}
W_{J M}(\vartheta) \equiv \Omega_{J M}^{L+}(\vartheta, \varphi) \Omega_{J M}^{L}(\vartheta, \varphi). \label{eq:7:2:46}
\end{equation}
Then one obtains
\begin{align}
W_{J M}(\vartheta) & =\frac{1}{2(J+1)}\left\{(J+M+1)\left|Y_{J+\frac{1}{2} M+\frac{1}{2}}(\vartheta, \varphi)\right|^{2}+(J-M+1)\left|Y_{J+\frac{1}{2} M-\frac{1}{2}}(\vartheta, \varphi)\right|^{2}\right\}  \notag\\
& =\frac{1}{2 J}\left\{(J+M)\left|Y_{J-\frac{1}{2} M-\frac{1}{2}}(\vartheta, \varphi)\right|^{2}+(J-M)\left|Y_{J-\frac{1}{2} M+\frac{1}{2}}(\vartheta, \varphi)\right|^{2}\right\}. \label{eq:7:2:47}
\end{align}
An expansion of $W_{J M}(\vartheta)$ in terms of the Legendre polynomials may be written as
\begin{equation}
W_{J M}(\vartheta)=\sum_{n=0}^{J-\frac{1}{2}} a_{n}(J, M) P_{2 n}(\cos \vartheta), \label{eq:7:2:48}
\end{equation}
where
\begin{align}
& a_{n}(J, M)=-\frac{\sqrt{2 J(2 J+1)}}{4 \pi}(4 n+1)\sixj{J}{J}{2 n}{J-\frac{1}{2}}{J-\frac{1}{2}}{\frac{1}{2}} \clebsch{J-\frac{1}{2}}{0}{2n}{0}{J-\frac{1}{2}}{0} \clebsch{J}{M}{2n}{0}{J}{M}  \notag\\
= & (-1)^{n} \frac{(4 n+1)(2 J+2 n+1)}{4 \pi \sqrt{2 J+1}} \cdot \frac{\left(J-\frac{1}{2}+n\right)!}{\left(J-\frac{1}{2}-n\right)!} \frac{(2 n)!}{(n!)^{2}} \sqrt{\frac{(2 J-2 n)!}{(2 J+2 n+1)!}} \clebsch{J}{M}{2n}{0}{J}{M}. \label{eq:7:2:49}
\end{align}
In particular,
\begin{align}
& a_{0}(J, M)=\frac{1}{4 \pi}, \quad a_{1}(J, M)=\frac{5}{16 \pi} \cdot \frac{J(J+1)-3 M^{2}}{J(J+1)}, \label{eq:7:2:50}\\
& a_{2}(J, M)=\frac{27}{256 \pi} \cdot \frac{3\left(J^{2}+2 J-5 M^{2}\right)\left(J^{2}-5 M^{2}-1\right)-10 M^{2}\left(4 M^{2}-1\right)}{(J-1) J(J+1)(J+2)}. \notag
\end{align}
The functions $W_{J M}(\vartheta)$ are normalized according to
\begin{gather}
\int W_{J M}(\vartheta) d \Omega=1, \label{eq:7:2:51}\\
\sum_{M=-J}^{J} W_{J M}(\vartheta)=\frac{2 J+1}{4 \pi}. \notag
\end{gather}
The symmetries of $W_{J M}(\vartheta)$ are
\begin{equation}
W_{J M}(\vartheta)=W_{J-M}(\vartheta)=W_{J M}(\pi-\vartheta)=W_{J-M}(\pi-\vartheta) . \label{eq:7:2:52}
\end{equation}
For $\vartheta=0$ and $\vartheta=\pi$ we have
\begin{equation}
W_{J M}(0)=W_{J M}(\pi)= \begin{cases}\frac{2 J+1}{8 \pi}, & \text { if } M= \pm \frac{1}{2} \\ 0, & \text { otherwise .}\end{cases} \label{eq:7:2:53}
\end{equation}
For special $M$ values one obtains
\begin{gather}
W_{J \pm \frac{1}{2}}(\vartheta)=\frac{1}{2 \pi(2 J+1)}\left\{\sin ^{2} \vartheta\left[P_{J-\frac{1}{2}}^{\prime}(\cos \vartheta)\right]^{2}+\left(J+\frac{1}{2}\right)^{2}\left[P_{J-\frac{1}{2}}(\cos \vartheta)\right]^{2}\right\}, \label{eq:7:2:54}\\
W_{J \pm(J-1)}(\vartheta)=\frac{(2 J-1)!(\sin \vartheta)^{2 J-3}}{\pi 2^{2 J+1}\left[\left(J-\frac{1}{2}\right)!\right]^{2}}\left\{1+4 J(J-1) \cos ^{2} \vartheta\right\}, \label{eq:7:2:55}\\
W_{J \pm J}(\vartheta)=\frac{(2 J)!}{\pi 2^{2 J+1}\left[\left(J-\frac{1}{2}\right)!\right]^{2}} \sin ^{2 J-1} \vartheta. \label{eq:7:2:56}
\end{gather}
The explicit forms of $W_{J M}(\vartheta)$ for $J=\frac{1}{2}, \frac{3}{2}, \frac{5}{2}, \frac{7}{2}, \frac{9}{2}, \frac{11}{2}\left(\frac{1}{2} \leq M \leq J\right)$ are given in Table~\ref{chap7:tab:1}. For negative $M$ one may use the relation $W_{J M}(\vartheta)=W_{J-M}(\vartheta)$.

\section{Vector spherical harmonics}\label{sec:7.3}\index{vector spherical harmonics}
\subsection{Definition}\label{sec:7.3.1}\index{vector spherical harmonics!definition}
The vector spherical harmonics $\vect{Y}_{J M}^{L}(\vartheta, \varphi)$ are defined as the tensor spherical harmonics at $S=1$ (Sec.~\ref{sec:7.1.1}).\isym{YJML-vec@$\vect{Y}_{J M}^{L}(\vartheta, \varphi)$, vector spherical harmonic}
\begin{equation}
\vect{Y}_{J M}^{L}(\vartheta, \varphi) \equiv Y_{J M}^{L 1}(\vartheta, \varphi). \label{eq:7:3:1}
\end{equation}
They are eigenfunctions of the operators $\widehat{\vect{J}}^{2}, \widehat{J}_{z}, \widehat{\vect{L}}^{2}, \widehat{\vect{S}}^{2}$ where $\widehat{\vect{L}}$ is the orbital angular momentum operator (Sec.~\ref{sec:2.2}), $\widehat{\vect{S}}$ is the spin operator for $S=1$ (Sec.~\ref{sec:2.6}), and $\widehat{\vect{J}}=\widehat{\vect{L}}+\widehat{\vect{S}}$ is the total angular momentum operator (Sec.~\ref{sec:2.1})
\begin{align}
\hat{\vect{J}}^{2} \vect{Y}_{J M}^{L}(\vartheta, \varphi) & =J(J+1) \vect{Y}_{J M}^{L}(\vartheta, \varphi), \notag\\
\widehat{J}_{z} \vect{Y}_{J M}^{L}(\vartheta, \varphi) & =M \vect{Y}_{J M}^{L}(\vartheta, \varphi), \notag\\
\hat{\vect{L}}^{2} \vect{Y}_{J M}^{L}(\vartheta, \varphi) & =L(L+1) \vect{Y}_{J M}^{L}(\vartheta, \varphi), \label{eq:7:3:2}\\
\widehat{\vect{S}}^{2} \vect{Y}_{J M}^{L}(\vartheta, \varphi) & =2 \vect{Y}_{J M}^{L}(\vartheta, \varphi) . \notag
\end{align}

\begin{table}[tbh]
\begin{center}
\caption{Explicit forms of $W_{J M}(\vartheta)$.}
\renewcommand{\arraystretch}{1.8}
\label{chap7:tab:1}
\begin{tabular}{l|l|c}
\hline
$J$ & M & $W_{J M}(\vartheta)=\Omega_{J M}^{L+}(\vartheta, \varphi) \Omega_{J M}^{L}(\vartheta, \varphi)$ \\
\hline
$\frac{1}{2}$ & $\frac{1}{2}$ & $\frac{1}{4 \pi}$ \\
$\frac{3}{2}$ & $\frac{1}{2}$ & $\frac{1}{8 \pi}\left\{3 \cos ^{2} \vartheta+1\right\}$ \\
$\frac{3}{2}$ & $\frac{3}{2}$ & $\frac{3}{8 \pi} \sin^2\vartheta$ \\
$\frac{5}{2}$ & $\frac{1}{2}$ & $\frac{3}{16 \pi}\left\{5 \cos ^{4} \vartheta-2 \cos ^{2} \vartheta+1\right\}$ \\
$\frac{5}{2}$ & $\frac{3}{2}$ & $\frac{3}{32 \pi} \sin ^{2} \vartheta\left(15 \cos ^{2} \vartheta+1\right)$ \\
$\frac{5}{2}$ & $\frac{5}{2}$ & $\frac{15}{32 \pi} \sin ^{4} \vartheta$ \\
$\frac{7}{2}$ & $\frac{1}{2}$ & $\frac{1}{64 \pi}\left\{175 \cos ^{6} \vartheta-165 \cos ^{4} \vartheta+45 \cos ^{2} \vartheta+9\right\}$ \\
$\frac{7}{2}$ & $\frac{3}{2}$ & $\frac{15}{64 \pi} \sin ^{2} \vartheta\left\{21 \cos ^{4} \vartheta-6 \cos ^{2} \vartheta+1\right\}$ \\
$\frac{7}{2}$ & $\frac{5}{2}$ & $\frac{5}{64 \pi} \sin ^{4} \vartheta\left\{35 \cos ^{2} \vartheta+1\right\}$ \\
$\frac{7}{2}$ & $\frac{7}{2}$ & $\frac{35}{64 \pi} \sin ^{6} \vartheta$ \\
$\frac{9}{2}$ & $\frac{1}{2}$ & $\frac{5}{256 \pi}\left\{441 \cos ^{8} \vartheta-644 \cos ^{6} \vartheta+294 \cos ^{4} \vartheta-36 \cos ^{2} \vartheta+9\right\}$ \\
$\frac{9}{2}$ & $\frac{3}{2}$ & $\frac{15}{128 \pi} \sin ^{2} \vartheta\left(147 \cos ^{6} \vartheta-105 \cos ^{4} \vartheta+21 \cos ^{2} \vartheta+1\right)$ \\
$\frac{9}{2}$ & $\frac{5}{2}$ & $\frac{35}{128 \pi} \sin ^{4} \vartheta\left(45 \cos ^{4} \vartheta-10 \cos ^{2} \vartheta+1\right)$ \\
$\frac{9}{2}$ & $\frac{7}{2}$ & $\frac{35}{512 \pi} \sin ^{6} \vartheta\left\{63 \cos ^{2} \vartheta+1\right\}$ \\
$\frac{9}{2}$ & $\frac{9}{2}$ & $\frac{315}{512 \pi} \sin ^{8} \vartheta$ \\
$\frac{11}{2}$ & $\frac{1}{2}$ & $\frac{3}{512 \pi}\left\{4851 \cos ^{10} \vartheta-9555 \cos ^{8} \vartheta+6510 \cos ^{6} \vartheta-1750 \cos ^{4} \vartheta+175 \cos ^{2} \vartheta+25\right\}$ \\
$\frac{11}{2}$ & $\frac{3}{2}$ & $\frac{105}{512 \pi} \sin ^{2} \vartheta\left\{297 \cos ^{8} \vartheta-348 \cos ^{6} \vartheta+126 \cos ^{4} \vartheta-12 \cos ^{2} \vartheta+1\right\}$ \\
$\frac{11}{2}$ & $\frac{5}{2}$ & $\frac{105}{1024 \pi} \sin ^{4} \vartheta\left\{495 \cos ^{6} \vartheta-285 \cos ^{4} \vartheta+45 \cos ^{2} \vartheta+1\right\}$ \\
$\frac{11}{2}$ & $\frac{7}{2}$ & $\frac{315}{1024 \pi} \sin ^{6} \vartheta\left\{77 \cos ^{4} \vartheta-14 \cos ^{2} \vartheta+1\right\}$ \\
$\frac{11}{2}$ & $\frac{9}{2}$ & $\frac{63}{1024 \pi} \sin ^{8} \vartheta\left(99 \cos ^{2} \vartheta+1\right)$ \\
$\frac{11}{2}$ & $\frac{11}{2}$ & $\frac{693}{1024 \pi} \sin ^{10} \vartheta$ 
\end{tabular}
\end{center}
\end{table}

According to \eqref{eq:7:1:2}, the vector spherical harmonics $\vect{Y}_{J M}^{L}$ may be written as
\begin{equation}
\vect{Y}_{J M}^{L}(\vartheta, \varphi)=\sum_{m, \sigma} \clebsch{L}{m}{1}{\sigma}{J}{M} Y_{L m}(\vartheta, \varphi) \vect{e}_{\sigma}, \label{eq:7:3:3}
\end{equation}
where $\clebsch{L}{m}{1}{\sigma}{J}{M}$ is the Clebsch-Gordan coefficient (Chap.~\ref{chap:8}), $Y_{L M}$ is a scalar spherical harmonic (Chap.~\ref{chap:5}), and $\vect{e}_{\sigma}$ is a covariant spherical basis vector (spin function for an $S=1$ particle, Sec.~\ref{sec:1.1.3})\index{basis vectors!spherical}; $J$ and $L$ are nonnegative integers. For a given $J$, three values of $L$ are possible: $L=J, J \pm 1$ (with the only exception that $L=1$ for $J=0$ ), and $M$ takes the values $M=-J,-J+1, \ldots, J-1, J$.

The vectors $\vect{Y}_{J M}^{L}(\vartheta, \varphi)$ which have the same $J, M$ but different $L$ satisfy the orthogonality relations
\begin{equation}
\sum_{M} \vect{Y}_{J M}^{L^{\prime} *}(\vartheta, \varphi) \cdot \vect{Y}_{J M}^{L}(\vartheta, \varphi)=0, \quad \text { if } L \neq L^{\prime}. \label{eq:7:3:4}
\end{equation}
Along with $\vect{Y}_{J M}^{L}$, other vector spherical harmonics $\vect{Y}_{J M}^{(\lambda)}(\lambda=0, \pm 1)$ are widely used.\index{vector spherical harmonics!transverse and longitudinal}\isym{YJMlambda-vec@$\vect{Y}_{J M}^{(\lambda)}(\vartheta, \varphi)$, transverse and longitudinal vector harmonics} The $\vect{Y}_{J M}^{(\lambda)}(\vartheta, \varphi)$ 's (unlike $\vect{Y}_{J M}^{L}$ ) are not eigenfunctions of the operator $\widehat{\vect{L}}^{2}$ but are suitably oriented with respect to the vector $\vect{n} \equiv \vect{r} / r$ specified by polar angles $\vartheta, \varphi$. The vector spherical harmonics $\vect{Y}_{J M}^{(1)}(\vartheta, \varphi)$ and $\vect{Y}_{J M}^{(0)}(\vartheta, \varphi)$ are transverse, and $\vect{Y}_{J M}^{(-1)}(\vartheta, \varphi)$ is longitudinal with respect to $\vect{n}(\vartheta, \varphi)$.
\begin{gather}
\vect{n} \cdot \vect{Y}_{J M}^{(1)}(\vartheta, \varphi)=\vect{n} \cdot \vect{Y}_{J M}^{(0)}(\vartheta, \varphi)=0, \notag\\
\vect{n} \times \vect{Y}_{J M}^{(-1)}(\vartheta, \varphi)=0. \label{eq:7:3:5}
\end{gather}
Sometimes $\vect{Y}_{J M}^{(1)}(\vartheta, \varphi)$ and $\vect{Y}_{J M}^{(0)}(\vartheta, \varphi)$ are called the electric and magnetic multipoles, respectively.\index{electric multipole}\index{magnetic multipole}\index{multipole!electric}\index{multipole!magnetic} They have the form
\begin{align}
& \vect{Y}_{J M}^{(1)}(\vartheta, \varphi)=\frac{1}{\sqrt{J(J+1)}} \nabla_{\Omega} Y_{J M}(\vartheta, \varphi), \notag\\
& \vect{Y}_{J M}^{(0)}(\vartheta, \varphi)=\frac{-i}{\sqrt{J(J+1)}}\left(\vect{n} \times \nabla_{\Omega}\right) Y_{J M}(\vartheta, \varphi)=\frac{\hat{\vect{L}}}{\sqrt{J(J+1)}} Y_{J M}(\vartheta, \varphi). \label{eq:7:3:6}
\end{align}
The longitudinal vector $\vect{Y}_{J M}^{(-1)}(\vartheta, \varphi)$ is given by
\begin{equation}
\vect{Y}_{J M}^{(-1)}(\vartheta, \varphi)=\mathrm{n} Y_{J M}(\vartheta, \varphi). \label{eq:7:3:7}
\end{equation}
In \eqref{eq:7:3:6} $\nabla_{\Omega}$ denotes the angular part of the $\nabla$ operator (see Eqs. \eqref{eq:1:3:9}-\eqref{eq:1:3:10}), and $\widehat{\vect{L}}$ is the orbital angular momentum operator.\index{operator!angular part of $\nabla$}\isymo{nabla-Omega@$\nabla_{\Omega}$, angular part of the $\nabla$ operator} Three vectors $\vect{Y}_{J M}^{(\lambda)}(\vartheta, \varphi)$ with different $\lambda$ are mutually orthogonal:
\begin{equation}
\vect{Y}_{J M}^{\left(\lambda^{\prime}\right)}(\vartheta, \varphi) \cdot \vect{Y}_{J M}^{(\lambda)}(\vartheta, \varphi)=0, \quad \text { if } \lambda^{\prime} \neq \lambda. \label{eq:7:3:8}
\end{equation}
The vector spherical harmonics $\vect{Y}_{J M}^{(\lambda)}(\vartheta, \varphi)$ are linear superpositions of $\vect{Y}_{J M}^{L}(\vartheta, \varphi)$ with different $L$. Thus
\begin{align}
\vect{Y}_{J M}^{(1)}(\vartheta, \varphi) & =\sqrt{\frac{J+1}{2 J+1}} \vect{Y}_{J M}^{J-1}(\vartheta, \varphi)+\sqrt{\frac{J}{2 J+1}} \vect{Y}_{J M}^{J+1}(\vartheta, \varphi), \notag\\
\vect{Y}_{J M}^{(0)}(\vartheta, \varphi) & =\vect{Y}_{J M}^{J}(\vartheta, \varphi), \label{eq:7:3:9}\\
\vect{Y}_{J M}^{(-1)}(\vartheta, \varphi) & =\sqrt{\frac{J}{2 J+1}} \vect{Y}_{J M}^{J-1}(\vartheta, \varphi)-\sqrt{\frac{J+1}{2 J+1}} \vect{Y}_{J M}^{J+1}(\vartheta, \varphi). \notag
\end{align}
The inverse relations are written as
\begin{align}
\vect{Y}_{J M}^{J+1}(\vartheta, \varphi) & =\sqrt{\frac{J}{2 J+1}} \vect{Y}_{J M}^{(1)}(\vartheta, \varphi)-\sqrt{\frac{J+1}{2 J+1}} \vect{Y}_{J M}^{(-1)}(\vartheta, \varphi), \notag\\
\vect{Y}_{J M}^{J}(\vartheta, \varphi) & =\vect{Y}_{J M}^{(0)}(\vartheta, \varphi), \label{eq:7:3:10}\\
\vect{Y}_{J M}^{J-1}(\vartheta, \varphi) & =\sqrt{\frac{J+1}{2 J+1}} \vect{Y}_{J M}^{(1)}(\vartheta, \varphi)+\sqrt{\frac{J}{2 J+1}} \vect{Y}_{J M}^{(-1)}(\vartheta, \varphi). \notag
\end{align}
The vector spherical harmonics $\vect{Y}_{J M}^{L}(\vartheta, \varphi)$, as well as $\vect{Y}_{J M}^{(\lambda)}(\vartheta, \varphi)$ constitute a complete orthonormal vector set for the range $0 \leq \vartheta \leq \pi, 0 \leq \varphi<2 \pi$ (see Sec.~\ref{sec:7.3.13} below).

\subsection{Components of Vector Spherical Harmonics}\label{sec:7.3.2}\index{vector spherical harmonics!components of}
\begin{enumerate}[label=(\alph*)]
\item The vector spherical harmonics $\vect{Y}_{J M}^{L}(\vartheta, \varphi)$ may be expanded in terms of the spherical covariant basis vectors $\vect{e}_{\mu}(\mu= \pm 1,0)$ as
\begin{equation}
\vect{Y}_{J M}^{L}(\vartheta, \varphi)=\sum_{\mu}\left[\vect{Y}_{J M}^{L}(\vartheta, \varphi)\right]^{\mu} \vect{e}_{\mu}=\sum_{\mu}(-1)^{\mu}\left[\vect{Y}_{J M}^{L}(\vartheta, \varphi)\right]_{-\mu} \vect{e}_{\mu}, \label{eq:7:3:11}
\end{equation}
where $\left[\vect{Y}_{J M}^{L}(\vartheta, \varphi)\right]^{\mu}$ and $\left[\vect{Y}_{J M}^{L}(\vartheta, \varphi)\right]_{\mu}$ are contravariant and covariant components, respectively.\\
The contravariant spherical components of $\vect{Y}_{J M}^{L}(\vartheta, \varphi)$ are given, according to \eqref{eq:7:1:5}, by
\begin{equation}
\left[\vect{Y}_{J M}^{L}(\vartheta, \varphi)\right]^{\mu}=\clebsch{L}{M-\mu}{1}{\mu}{J}{M} Y_{L M-\mu}(\vartheta, \varphi). \label{eq:7:3:12}
\end{equation}
The covariant spherical components of $\vect{Y}_{J M}^{L}(\vartheta, \varphi)$ are related to the contravariant ones by
\begin{equation}
\left[\vect{Y}_{J M}^{L}(\vartheta, \varphi)\right]_{\mu}=(-1)^{\mu}\left[\vect{Y}_{J M}^{L}(\vartheta, \varphi)\right]^{-\mu}, \label{eq:7:3:13}
\end{equation}
and are given by
\begin{equation}
\left[\vect{Y}_{J M}^{L}(\vartheta, \varphi)\right]_{\mu}=(-1)^{\mu} \clebsch{L}{M+\mu}{1}{-\mu}{J}{M} Y_{L M+\mu}(\vartheta, \varphi). \label{eq:7:3:14}
\end{equation}
In more detailed form Eqs. \eqref{eq:7:3:12}, \eqref{eq:7:3:14} may be written as
\begin{gather}
{\left[\vect{Y}_{J M}^{J+1}(\vartheta, \varphi)\right]^{+1}=-\left[\vect{Y}_{J M}^{J+1}(\vartheta, \varphi)\right]_{-1}=\left[\frac{(J-M+1)(J-M+2)}{2(J+1)(2 J+3)}\right]^{\frac{1}{2}} Y_{J+1 M-1}(\vartheta, \varphi)}, \notag\\
{\left[\vect{Y}_{J M}^{J+1}(\vartheta, \varphi)\right]^{0}=\left[\vect{Y}_{J M}^{J+1}(\vartheta, \varphi)\right]_{0}=-\left[\frac{(J-M+1)(J+M+1)}{(J+1)(2 J+3)}\right]^{\frac{1}{2}} Y_{J+1 M}(\vartheta, \varphi)}, \label{eq:7:3:15}\\
{\left[\vect{Y}_{J M}^{J+1}(\vartheta, \varphi)\right]^{-1}=-\left[\vect{Y}_{J M}^{J+1}(\vartheta, \varphi)\right]_{+1}=\left[\frac{(J+M+1)(J+M+2)}{2(J+1)(2 J+3)}\right]^{\frac{1}{2}} Y_{J+1 M+1}(\vartheta, \varphi)}, \notag\\
{\left[\vect{Y}_{J M}^{J}(\vartheta, \varphi)\right]^{+1}=-\left[\vect{Y}_{J M}^{J}(\vartheta, \varphi)\right]_{-1}=-\left[\frac{(J+M)(J-M+1)}{2 J(J+1)}\right]^{\frac{1}{2}} Y_{J M-1}(\vartheta, \varphi)}, \notag\\
{\left[\vect{Y}_{J M}^{J}(\vartheta, \varphi)\right]^{0}=\left[\vect{Y}_{J M}^{J}(\vartheta, \varphi)\right]_{0}=\frac{M}{\sqrt{J(J+1)}} Y_{J M}(\vartheta, \varphi)}, \label{eq:7:3:16}\\
{\left[\vect{Y}_{J M}^{J}(\vartheta, \varphi)\right]^{-1}=-\left[\vect{Y}_{J M}^{J}(\vartheta, \varphi)\right]_{+1}=\left[\frac{(J-M)(J+M+1)}{2 J(J+1)}\right]^{\frac{1}{2}} Y_{J M+1}(\vartheta, \varphi)}. \notag
\end{gather}
\begin{align}
{\left[\vect{Y}_{J M}^{J-1}(\vartheta, \varphi)\right]^{+1} } & =-\left[\vect{Y}_{J M}^{J-1}(\vartheta, \varphi)\right]_{-1}=\left[\frac{(J+M)(J+M-1)}{2 J(2 J-1)}\right]^{\frac{1}{2}} Y_{J-1 M-1}(\vartheta, \varphi), \notag\\
{\left[\vect{Y}_{J M}^{J-1}(\vartheta, \varphi)\right]^{0} } & =\left[\vect{Y}_{J M}^{J-1}(\vartheta, \varphi)\right]_{0}=\left[\frac{(J-M)(J+M)}{J(2 J-1)}\right]^{\frac{1}{2}} Y_{J-1 M}(\vartheta, \varphi), \label{eq:7:3:17}\\
{\left[\vect{Y}_{J M}^{J-1}(\vartheta, \varphi)\right]^{-1} } & =-\left[\vect{Y}_{J M}^{J-1}(\vartheta, \varphi)\right]_{+1}=\left[\frac{(J-M)(J-M-1)}{2 J(2 J-1)}\right]^{\frac{1}{2}} Y_{J-1 M+1}(\vartheta, \varphi). \notag
\end{align}
\item 
 An expansion of $\vect{Y}_{J M}^{(\lambda)}(\vartheta, \varphi)$ in terms of the spherical covariant basis vectors $\vect{e}_{\mu}(\mu= \pm 1,0)$ has the form
\begin{equation}
\vect{Y}_{J M}^{(\lambda)}(\vartheta, \varphi)=\sum_{\mu}\left[\vect{Y}_{J M}^{(\lambda)}(\vartheta, \varphi)\right]^{\mu} \vect{e}_{\mu}=\sum_{\mu}(-1)^{\mu}\left[\vect{Y}_{J M}^{(\lambda)}(\vartheta, \varphi)\right]_{-\mu} \vect{e}_{\mu}. \label{eq:7:3:18}
\end{equation}
The contravariant spherical components of $\vect{Y}_{J M}^{(\lambda)}(\vartheta, \varphi)$ may be written as
\begin{align}
{\left[\vect{Y}_{J M}^{(1)}(\vartheta, \varphi)\right]^{\mu} } & =\sqrt{\frac{J+1}{2 J+1}} \clebsch{J-1}{M-\mu}{1}{\mu}{J}{M} Y_{J-1 M-\mu}(\vartheta, \varphi)+\sqrt{\frac{J}{2 J+1}} \clebsch{J+1}{M-\mu}{1}{\mu}{J}{M} Y_{J+1 M-\mu}(\vartheta, \varphi), \notag\\
{\left[\vect{Y}_{J M}^{(0)}(\vartheta, \varphi)\right]^{\mu} } & =\clebsch{J}{M-\mu}{1}{\mu}{J}{M} Y_{J M-\mu}(\vartheta, \varphi), \label{eq:7:3:19}\\
{\left[\vect{Y}_{J M}^{(-1)}(\vartheta, \varphi)\right]^{\mu} } & =\sqrt{\frac{J}{2 J+1}} \clebsch{J-1}{M-\mu}{1}{\mu}{J}{M} Y_{J-1 M-\mu}(\vartheta, \varphi)-\sqrt{\frac{J+1}{2 J+1}} \clebsch{J+1}{M-\mu}{1}{\mu}{J}{M} Y_{J+1 M-\mu}(\vartheta, \varphi). \notag
\end{align}
The covariant spherical components of $\vect{Y}_{J M}^{(\lambda)}(\vartheta, \varphi)$ are related to contravariant ones by
\begin{equation}
\left[\vect{Y}_{J M}^{(\lambda)}(\vartheta, \varphi)\right]_{\mu}=(-1)^{\mu}\left[\vect{Y}_{J M}^{(\lambda)}(\vartheta, \varphi)\right]^{-\mu}, \label{eq:7:3:20}
\end{equation}
and may be written as
\begin{align}
{\left[\vect{Y}_{J M}^{(+1)}(\vartheta, \varphi)\right]^{+1}=-\left[\vect{Y}_{J M}^{(+1)}(\vartheta, \varphi)\right]_{-1} } & =\left[\frac{(J+M)(J+M-1)(J+1)}{2 J(2 J-1)(2 J+1)}\right]^{\frac{1}{2}} Y_{J-1 M-1}(\vartheta, \varphi), \notag\\
& +\left[\frac{(J-M+1)(J-M+2) J}{2(J+1)(2 J+1)(2 J+3)}\right]^{\frac{1}{2}} Y_{J+1 M-1}(\vartheta, \varphi), \notag\\
{\left[\vect{Y}_{J M}^{(+1)}(\vartheta, \varphi)\right]^{0}=\left[\vect{Y}_{J M}^{(+1)}(\vartheta, \varphi)\right]_{0} } & =\left[\frac{(J-M)(J+M)(J+1)}{J(2 J-1)(2 J+1)}\right]^{\frac{1}{2}} Y_{J-1 M}(\vartheta, \varphi), \label{eq:7:3:21}\\
& -\left[\frac{(J-M+1)(J+M+1) J}{(J+1)(2 J+1)(2 J+3)}\right]^{\frac{1}{2}} Y_{J+1 M}(\vartheta, \varphi), \notag\\
{\left[\vect{Y}_{J M}^{(+1)}(\vartheta, \varphi)\right]^{-1}=-\left[\vect{Y}_{J M}^{(+1)}(\vartheta, \varphi)\right]_{+1} } & =\left[\frac{(J-M)(J-M-1)(J+1)}{2 J(2 J-1)(2 J+1)}\right]^{\frac{1}{2}} Y_{J-1 M+1}(\vartheta, \varphi), \notag\\
& +\left[\frac{(J+M+1)(J+M+2) J}{2(J+1)(2 J+1)(2 J+3)}\right]^{\frac{1}{2}} Y_{J+1 M+1}(\vartheta, \varphi). \notag
\end{align}
\begin{align}
{\left[\vect{Y}_{J M}^{(0)}(\vartheta, \varphi)\right]^{+1}=-\left[\vect{Y}_{J M}^{(0)}(\vartheta, \varphi)\right]_{-1} } & =-\left[\frac{(J+M)(J-M+1)}{2 J(J+1)}\right]^{\frac{1}{2}} Y_{J M-1}(\vartheta, \varphi), \notag\\
{\left[\vect{Y}_{J M}^{(0)}(\vartheta, \varphi)\right]^{0}=\left[\vect{Y}_{J M}^{(0)}(\vartheta, \varphi)\right]_{0}=} & \frac{M}{\sqrt{J(J+1)}} Y_{J M}(\vartheta, \varphi), \label{eq:7:3:22}\\
{\left[\vect{Y}_{J M}^{(0)}(\vartheta, \varphi)\right]^{-1}=-\left[\vect{Y}_{J M}^{(0)}(\vartheta, \varphi)\right]_{+1} } & =\left[\frac{(J-M)(J+M+1)}{2 J(J+1)}\right]^{\frac{1}{2}} Y_{J M+1}(\vartheta, \varphi), \notag\\
{\left[\vect{Y}_{J M}^{(-1)}(\vartheta, \varphi)\right]^{+1}=-\left[\vect{Y}_{J M}^{(-1)}(\vartheta, \varphi)\right]_{-1} } & =\left[\frac{(J+M)(J+M-1)}{2(2 J-1)(2 J+1)}\right]^{\frac{1}{2}} Y_{J-1 M-1}(\vartheta, \varphi), \notag\\
& -\left[\frac{(J-M+1)(J-M+2)}{2(2 J+1)(2 J+3)}\right]^{\frac{1}{2}} Y_{J+1 M-1}(\vartheta, \varphi), \notag\\
{\left[\vect{Y}_{J M}^{(-1)}(\vartheta, \varphi)\right]^{0}=\left[\vect{Y}_{J M}^{(-1)}(\vartheta, \varphi)\right]_{0} } & =\left[\frac{(J-M)(J+M)}{(2 J-1)(2 J+1)}\right]^{\frac{1}{2}} Y_{J-1 M}(\vartheta, \varphi), \notag\\
& +\left[\frac{(J-M+1)(J+M+1)}{(2 J+1)(2 J+3)}\right]^{\frac{1}{2}} Y_{J+1 M}(\vartheta, \varphi), \label{eq:7:3:23}\\
{\left[\vect{Y}_{J M}^{(-1)}(\vartheta, \varphi)\right]^{-1}=-\left[\vect{Y}_{J M}^{(-1)}(\vartheta, \varphi)\right]_{+1} } & =\left[\frac{(J-M)(J-M-1)}{2(2 J-1)(2 J+1)}\right]^{\frac{1}{2}} Y_{J-1 M+1}(\vartheta, \varphi), \notag\\
& -\left[\frac{(J+M+1)(J+M+2)}{2(2 J+1)(2 J+3)}\right]^{\frac{1}{2}} Y_{J+1 M+1}(\vartheta, \varphi). \notag
\end{align}
\item An expansion of $\vect{Y}_{J M}^{(\lambda)}(\vartheta, \varphi)$ in terms of the polar basis vectors $\vect{e}_{r}$, $\vect{e}_{\vartheta}$ and $\vect{e}_{\varphi}$ (Sec.~\ref{sec:1.1.2}) is given by
\begin{equation}
\vect{Y}_{J M}^{(\lambda)}(\vartheta, \varphi)=\left[\vect{Y}_{J M}^{(\lambda)}(\vartheta, \varphi)\right]_{r} \vect{e}_{r}+\left[\vect{Y}_{J M}^{(\lambda)}(\vartheta, \varphi)\right]_{\vartheta} \vect{e}_{\vartheta}+\left[\vect{Y}_{J M}^{(\lambda)}(\vartheta, \varphi)\right]_{\varphi} \vect{e}_{\varphi}, \label{eq:7:3:24}
\end{equation}
where the polar components of $\vect{Y}_{J M}^{(\lambda)}(\vartheta, \varphi)$ have the form
\begin{align}
{\left[\vect{Y}_{J M}^{(1)}(\vartheta, \varphi)\right]_{\sigma} } & =0, \notag\\
{\left[\vect{Y}_{J M}^{(1)}(\vartheta, \varphi)\right]_{\vartheta} } & =\frac{1}{\sqrt{J(J+1)}} \cdot \frac{\partial}{\partial \vartheta} Y_{J M}(\vartheta, \varphi), \notag\\
& =\frac{1}{2} \sqrt{\frac{(J-M)(J+M+1)}{J(J+1)}} e^{-i \varphi} Y_{J M+1}(\vartheta, \varphi)-\frac{1}{2} \sqrt{\frac{(J+M)(J-M+1)}{J(J+1)}} e^{i \varphi} Y_{J M-1}(\vartheta, \varphi), \notag\\
{\left[\vect{Y}_{J M}^{(1)}(\vartheta, \varphi)\right]_{\varphi} } & =\frac{1}{\sqrt{(J(J+1)}} \cdot \frac{1}{\sin \vartheta} \cdot \frac{\partial}{\partial \varphi} Y_{J M}(\vartheta, \varphi)=\frac{i M}{\sqrt{J(J+1)}} \cdot \frac{1}{\sin \vartheta} Y_{J M}(\vartheta, \varphi). \label{eq:7:3:25}
\end{align}
\begin{align}
{\left[\vect{Y}_{J M}^{(0)}(\vartheta, \varphi)\right]_{r} } & =0, \notag\\
{\left[\vect{Y}_{J M}^{(0)}(\vartheta, \varphi)\right]_{\vartheta} } & =\frac{i}{\sqrt{J(J+1)}} \cdot \frac{1}{\sin \vartheta} \cdot \frac{\partial}{\partial \varphi} Y_{J M}(\vartheta, \varphi)=\frac{-M}{\sqrt{J(J+1)}} \cdot \frac{1}{\sin \vartheta} Y_{J M}(\vartheta, \varphi), \notag\\
{\left[\vect{Y}_{J M}^{(0)}(\vartheta, \varphi)\right]_{\varphi} } & =-\frac{i}{\sqrt{J(J+1)}} \cdot \frac{\partial}{\partial \vartheta} Y_{J M}(\vartheta, \varphi), \notag\\
& =-\frac{i}{2} \sqrt{\frac{(J-M)(J+M+1)}{J(J+1)}} e^{-i \varphi} Y_{J M+1}(\vartheta, \varphi)+\frac{i}{2} \sqrt{\frac{(J+M)(J-M+1)}{J(J+1)}} e^{i \varphi} Y_{J M-1}(\vartheta, \varphi). \label{eq:7:3:26}
\end{align}
\begin{align}
& {\left[\vect{Y}_{J M}^{(-1)}(\vartheta, \varphi)\right]_{r}=Y_{J M}(\vartheta, \varphi)}, \notag\\
& {\left[\vect{Y}_{J M}^{(-1)}(\vartheta, \varphi)\right]_{\vartheta}=0}, \label{eq:7:3:27}\\
& {\left[\vect{Y}_{J M}^{(-1)}(\vartheta, \varphi)\right]_{\varphi}=0}. \notag
\end{align}
\item An expansion of $\vect{Y}_{J M}^{L}(\vartheta, \varphi)$ in terms of helicity basis vectors $\vect{e}_{\nu}^{\prime}(\vartheta, \varphi)$ (Sec.~\ref{sec:1.1.4}) may be written as
\begin{equation}
\vect{Y}_{J M}^{L}(\vartheta, \varphi)=\sum_{\nu}\left[\vect{Y}_{J M}^{L}(\vartheta, \varphi)\right]^{\nu} \vect{e}_{\nu}^{\prime}(\vartheta, \varphi)=\sum_{\nu}(-1)^{\nu}\left[\vect{Y}_{J M}^{L}(\vartheta, \varphi)\right]_{-\nu}^{\prime} \vect{e}_{\nu}^{\prime}(\vartheta, \varphi), \label{eq:7:3:28}
\end{equation}
where $\left[\vect{Y}_{J M}^{L}(\vartheta, \varphi)\right]^{\nu}$ and $\left[\vect{Y}_{J M}^{L}(\vartheta, \varphi)\right]_{\nu}^{\prime}$ are contravariant and covariant helicity components, respectively.
\begin{equation}
\left|\vect{Y}_{J M}^{L}(\vartheta, \varphi)\right|^{\prime \nu}=\sqrt{\frac{2 L+1}{4 \pi}} \clebsch{L}{0}{1}{\nu}{J}{\nu} D_{-\nu-M}^{J}(0, \vartheta, \varphi). \label{eq:7:3:29}
\end{equation}
The covariant helicity components of $\vect{Y}_{J M}^{L}(\vartheta, \varphi)$ are related to contravariant ones by
\begin{equation}
\left[\vect{Y}_{J M}^{L}(\vartheta, \varphi)\right]_{\nu}^{\prime}=(-1)^{\nu}\left[\vect{Y}_{J M}^{L}(\vartheta, \varphi)\right]^{\prime-\nu}, \label{eq:7:3:30}
\end{equation}
and may be written as
\begin{equation}
\left[\vect{Y}_{J M}^{L}(\vartheta, \varphi)\right]_{\nu}^{\prime}=(-1)^{J+L+\nu+1} \sqrt{\frac{2 L+1}{4 \pi}} \clebsch{L}{0}{1}{\nu}{J}{\nu} D_{\nu-M}^{J}(0, \vartheta, \varphi) . \label{eq:7:3:31}
\end{equation}
Explicitly we may write Eqs. \eqref{eq:7:3:29}-\eqref{eq:7:3:31} as follows:
\begin{gather}
{\left[\vect{Y}_{J M}^{J+1}(\vartheta, \varphi)\right]^{\prime+1}=-\left[\vect{Y}_{J M}^{J+1}(\vartheta, \varphi)\right]_{-1}^{\prime}=\sqrt{\frac{J}{8 \pi}} D_{-1-M}^{J}(0, \vartheta, \varphi)}, \notag\\
{\left[\vect{Y}_{J M}^{J+1}(\vartheta, \varphi)\right]^{\prime 0}=\left[\vect{Y}_{J M}^{J+1}(\vartheta, \varphi)\right]_{0}^{\prime}=-\sqrt{\frac{J+1}{4 \pi}} D_{0-M}^{J}(0, \vartheta, \varphi)}, \label{eq:7:3:32}\\
{\left[\vect{Y}_{J M}^{J+1}(\vartheta, \varphi)\right]^{\prime-1}=-\left[\vect{Y}_{J M}^{J+1}(\vartheta, \varphi)\right]_{+1}^{\prime}=\sqrt{\frac{J}{8 \pi}} D_{1-M}^{J}(0, \vartheta, \varphi)}, \notag\\
{\left[\vect{Y}_{J M}^{J}(\vartheta, \varphi)\right]^{\prime+1}=-\left[\vect{Y}_{J M}^{J}(\vartheta, \varphi)\right]_{-1}^{\prime}=-\sqrt{\frac{2 J+1}{8 \pi}} D_{-1-M}^{J}(0, \vartheta, \varphi)}, \notag\\
{\left[\vect{Y}_{J M}^{J}(\vartheta, \varphi)\right]^{\prime 0}=\left[\vect{Y}_{J M}^{J}(\vartheta, \varphi)\right]_{0}^{\prime}=0}, \label{eq:7:3:33}\\
{\left[\vect{Y}_{J M}^{J}(\vartheta, \varphi)\right]^{\prime-1}=-\left[\vect{Y}_{J M}^{J}(\vartheta, \varphi)\right]_{+1}^{\prime}=\sqrt{\frac{2 J+1}{8 \pi}} D_{1-M}^{J}(0, \vartheta, \varphi)}. \notag
\end{gather}
\begin{align}
{\left[\vect{Y}_{J M}^{J-1}(\vartheta, \varphi)\right]^{\prime+1} } & =-\left[\vect{Y}_{J M}^{J-1}(\vartheta, \varphi)\right]_{-1}^{\prime}=\sqrt{\frac{J+1}{8 \pi}} D_{-1-M}^{J}(0, \vartheta, \varphi), \notag\\
{\left[\vect{Y}_{J M}^{J-1}(\vartheta, \varphi)\right]^{\prime 0} } & =\left[\vect{Y}_{J M}^{J-1}(\vartheta, \varphi)\right]_{0}^{\prime}=\sqrt{\frac{J}{4 \pi}} D_{0-M}^{J}(0, \vartheta, \varphi), \label{eq:7:3:34}\\
{\left[\vect{Y}_{J M}^{J-1}(\vartheta, \varphi)^{\prime-1}\right.} & =-\left[\vect{Y}_{J M}^{J-1}(\vartheta, \varphi)\right]_{+1}^{\prime}=\sqrt{\frac{J+1}{8 \pi}} D_{1-M}^{J}(0, \vartheta, \varphi). \notag
\end{align}
\item An expansion of $\vect{Y}_{J M}^{(\lambda)}(\vartheta, \varphi)$ in terms of the helicity basis vectors $\vect{e}_{\nu}^{\prime}(\vartheta, \varphi)(\nu= \pm 1,0)$ has the form
\begin{align}
\vect{Y}_{J M}^{(1)}(\vartheta, \varphi) & =\sqrt{\frac{2 J+1}{8 \pi}}\left\{D_{-1-M}^{J}(0, \vartheta, \varphi) \vect{e}_{1}^{\prime}(\vartheta, \varphi)+D_{1-M}^{J}(0, \vartheta, \varphi) \vect{e}_{-1}^{\prime}(\vartheta, \varphi)\right\}, \notag\\
\vect{Y}_{J M}^{(0)}(\vartheta, \varphi) & =\sqrt{\frac{2 J+1}{8 \pi}}\left\{-D_{-1-M}^{J}(0, \vartheta, \varphi) \vect{e}_{1}^{\prime}(\vartheta, \varphi)+D_{1-M}^{J}(0, \vartheta, \varphi) \vect{e}_{-1}^{\prime}(\vartheta, \varphi)\right\}, \label{eq:7:3:35}\\
\vect{Y}_{J M}^{(-1)}(\vartheta, \varphi) & =\sqrt{\frac{2 J+1}{4 \pi}} D_{0-M}^{J}(0, \vartheta, \varphi) \vect{e}_{0}^{\prime}(\vartheta, \varphi). \notag
\end{align}
The contravariant and covariant helicity components, $\left[\vect{Y}_{J M}^{(\lambda)}\right]^{\nu}$ and $\left[\vect{Y}_{J M}^{(\lambda)}\right]_{\nu}^{\prime}$, are given by
\begin{gather}
{\left[\vect{Y}_{J M}^{(1)}(\vartheta, \varphi)\right]^{\prime+1}=-\left[\vect{Y}_{J M}^{(1)}(\vartheta, \varphi)\right]_{-1}^{\prime}=\sqrt{\frac{2 J+1}{8 \pi}} D_{-1-M}^{J}(0, \vartheta, \varphi)}, \notag\\
{\left[\vect{Y}_{J M}^{(1)}(\vartheta, \varphi)\right]^{\prime 0}=\left[\vect{Y}_{J M}^{(1)}(\vartheta, \varphi)\right]_{0}^{\prime}=0}, \label{eq:7:3:36}\\
{\left[\vect{Y}_{J M}^{(1)}(\vartheta, \varphi)\right]^{\prime-1}=-\left[\vect{Y}_{J M}^{(1)}(\vartheta, \varphi)\right]_{+1}^{\prime}=\sqrt{\frac{2 J+1}{8 \pi}} D_{1-M}^{J}(0, \vartheta, \varphi)}, \notag\\
{\left[\vect{Y}_{J M}^{(0)}(\vartheta, \varphi)\right]^{\prime+1}=-\left[\vect{Y}_{J M}^{(0)}(\vartheta, \varphi)\right]_{-1}^{\prime}=-\sqrt{\frac{2 J+1}{8 \pi}} D_{-1-M}^{J}(0, \vartheta, \varphi)}, \notag\\
{\left[\vect{Y}_{J M}^{(0)}(\vartheta, \varphi)\right]^{\prime 0}=\left[\vect{Y}_{J M}^{(0)}(\vartheta, \varphi)\right]_{0}^{\prime}=0}, \label{eq:7:3:37}\\
{\left[\vect{Y}_{J M}^{(0)}(\vartheta, \varphi)\right]^{\prime-1}=-\left[\vect{Y}_{J M}^{(0)}(\vartheta, \varphi)\right]_{+1}^{\prime}=\sqrt{\frac{2 J+1}{8 \pi}} D_{1-M}^{J}(0, \vartheta, \varphi)}, \notag\\
{\left[\vect{Y}_{J M}^{(-1)}(\vartheta, \varphi)\right]^{\prime+1}=-\left[\vect{Y}_{J M}^{(-1)}(\vartheta, \varphi)\right]_{-1}^{\prime}=0}, \notag\\
{\left[\vect{Y}_{J M}^{(-1)}(\vartheta, \varphi)\right]^{\prime 0}=\left[\vect{Y}_{J M}^{(-1)}(\vartheta, \varphi)\right]_{0}^{\prime}=\sqrt{\frac{2 J+1}{4 \pi}} D_{0-M}^{J}(0, \vartheta, \varphi)}, \label{eq:7:3:38}\\
{\left[\vect{Y}_{J M}^{(-1)}(\vartheta, \varphi)\right]^{\prime-1}=-\left[\vect{Y}_{J M}^{(-1)}(\vartheta, \varphi)\right]_{+1}^{\prime}=0}. \notag
\end{gather}
\end{enumerate}

\subsection{Complex Conjugation}\label{sec:7.3.3}\index{vector spherical harmonics!complex conjugation}
The vector spherical harmonics $\vect{Y}_{J M}^{L}(\vartheta, \varphi)$ and $\vect{Y}_{J M}^{(\lambda)}(\vartheta, \varphi)$ are complex vectors. Under complex conjugation they transform as follows
\begin{align}
\vect{Y}_{J M}^{L *}(\vartheta, \varphi) & =(-1)^{J+L+M+1} \vect{Y}_{J-M}^{L}(\vartheta, \varphi), \notag\\
\vect{Y}_{J M}^{(\lambda) *}(\vartheta, \varphi) & =(-1)^{M+\lambda+1} \vect{Y}_{J-M}^{(\lambda)}(\vartheta, \varphi). \label{eq:7:3:39}
\end{align}
The complex conjugates of the spherical components of $\vect{Y}_{J M}^{L}$ and $\vect{Y}_{J M}^{(\lambda)}$ are given by
\begin{align}
{\left[\vect{Y}_{J M}^{L}(\vartheta, \varphi)\right]_{\mu}^{*} } & =\left[\vect{Y}_{J M}^{L *}(\vartheta, \varphi)\right]^{\mu}=(-1)^{J+L+M+\mu+1}\left[\vect{Y}_{J-M}^{L}(\vartheta, \varphi)\right]_{-\mu}=(-1)^{J+L+M+1}\left[\vect{Y}_{J-M}^{L}(\vartheta, \varphi)\right]^{\mu}, \label{eq:7:3:40}\\
{\left[\vect{Y}_{J M}^{L}(\vartheta, \varphi)\right]^{\mu *} } & =\left[\vect{Y}_{J M}^{L *}(\vartheta, \varphi)\right]_{\mu}=(-1)^{J+L+M+\mu+1}\left[\vect{Y}_{J-M}^{L}(\vartheta, \varphi)\right]^{-\mu}=(-1)^{J+L+M+1}\left[\vect{Y}_{J-M}^{L}(\vartheta, \varphi)\right]_{\mu}, \label{eq:7:3:41}\\
{\left[\vect{Y}_{J M}^{(\lambda)}(\vartheta, \varphi)\right]_{\mu}^{*} } & =\left[\vect{Y}_{J M}^{(\lambda) *}(\vartheta, \varphi)\right]^{\mu}=(-1)^{M+\mu+\lambda+1}\left[\vect{Y}_{J-M}^{(\lambda)}(\vartheta, \varphi)\right]_{-\mu}=(-1)^{M+\lambda+1}\left[\vect{Y}_{J-M}^{(\lambda)}(\vartheta, \varphi)\right]^{\mu}, \label{eq:7:3:42}\\
{\left[\vect{Y}_{J M}^{(\lambda)}(\vartheta, \varphi)\right]^{\mu *} } & =\left[\vect{Y}_{J M}^{(\lambda) *}(\vartheta, \varphi)\right]_{\mu}=(-1)^{M+\mu+\lambda+1}\left[\vect{Y}_{J-M}^{(\lambda)}(\vartheta, \varphi)\right]^{-\mu}=(-1)^{M+\lambda+1}\left[\vect{Y}_{J-M}^{(\lambda)}(\vartheta, \varphi)\right]_{\mu}. \label{eq:7:3:43}
\end{align}
\subsection{Transformations of Coordinate Systems}\label{sec:7.3.4}\index{vector spherical harmonics!under coordinate transformations}\index{rotation!of the vector spherical harmonics}
\subsubsection{Coordinate inversion}
\begin{align}
& \hat{P}_{r} \vect{Y}_{J M}^{L}(\vartheta, \varphi)=-\vect{Y}_{J M}^{L}(\pi-\vartheta, \pi+\varphi)=(-1)^{L+1} \vect{Y}_{J M}^{L}(\vartheta, \varphi)  \notag\\
& \hat{P}_{r} \vect{Y}_{J M}^{(\lambda)}(\vartheta, \varphi)=-\vect{Y}_{J M}^{(\lambda)}(\pi-\vartheta, \pi+\varphi)=(-1)^{J+\lambda+1} \vect{Y}_{J M}^{(\lambda)}(\vartheta, \varphi). \label{eq:7:3:44}
\end{align}
\subsubsection{Rotations of the coordinate system}

Under rotations specified by the Euler angles $\alpha, \beta, \gamma$ (Sec.~\ref{sec:1.4}) the vector spherical harmonics transform according to
\begin{align}
& \vect{Y}_{J M^{\prime}}^{L}\left(\vartheta^{\prime}, \varphi^{\prime}\right)=\hat{D}(\alpha, \beta, \gamma) \vect{Y}_{J M^{\prime}}^{L}(\vartheta, \varphi)=\sum_{M} D_{M M^{\prime}}^{J}(\alpha, \beta, \gamma) \vect{Y}_{J M}^{L}(\vartheta, \varphi)  \notag\\
& \vect{Y}_{J M^{\prime}}^{(\lambda)}\left(\vartheta^{\prime}, \varphi^{\prime}\right)=\hat{D}(\alpha, \beta, \gamma) \vect{Y}_{J M^{\prime}}^{(\lambda)}(\vartheta, \varphi)=\sum_{M}^{J} D_{M M^{\prime}}^{J}(\alpha, \beta, \gamma) \vect{Y}_{J M}^{(\lambda)}(\vartheta, \varphi). \label{eq:7:3:45}
\end{align}
where $D_{M M^{\prime}}^{J}(\alpha, \beta, \gamma)$ are the Wigner $D$-functions (Chap.~\ref{chap:4}), $M^{\prime}$ is the projection of total angular momentum on the new $z^{\prime}$-axis; $\vartheta, \varphi$ and $\vartheta^{\prime}, \varphi^{\prime}$ are polar angles of $r$ in the initial and final coordinate systems, respectively. The relations between $\vartheta^{\prime}, \varphi^{\prime}$ and $\vartheta, \varphi$ are given by eqs. \eqref{eq:1:4:2}, \eqref{eq:1:4:3}.

\subsection{Differential Equations}\label{sec:7.3.5}\index{vector spherical harmonics!differential equations}
\begin{enumerate}[label=(\alph*)]
\item The vector spherical harmonics $\vect{Y}_{J M}^{L}(\vartheta, \varphi)$ are eigenfunctions of the operator $\hat{\vect{L}}^{2}$; hence, they satisfy the equation
\begin{equation}
\frac{1}{\sin \vartheta} \cdot \frac{\partial}{\partial \vartheta}\left\{\sin \vartheta \frac{\partial}{\partial \vartheta} \vect{Y}_{J M}^{L}(\vartheta, \varphi)\right\}+\frac{1}{\sin ^{2} \vartheta} \frac{\partial^{2}}{\partial \varphi^{2}} \vect{Y}_{J M}^{L}(\vartheta, \varphi)+L(L+1) \vect{Y}_{J M}^{L}(\vartheta, \varphi)=0 . \label{eq:7:3:46}
\end{equation}
This equation may be rewritten in a compact form as
\begin{equation}
\left\{\Delta_{\Omega}+L(L+1)\right\} \vect{Y}_{J M}^{L}(\vartheta, \varphi)=0 \label{eq:7:3:47}
\end{equation}
where $\Delta_{\Omega} \equiv \nabla_{\Omega}^{2}$ is the angular part of the Laplace operator (see Eq. \eqref{eq:1:3:15}).
\item The vector functions $r^{L} \vect{Y}_{J M}^{L}(\vartheta, \varphi)$ are solutions of the Laplace equation
\begin{equation}
\Delta\left\{r^{L} \vect{Y}_{J M}^{J}(\vartheta, \varphi)\right\}=0. \label{eq:7:3:48}
\end{equation}
It follows from \eqref{eq:7:3:48} that the components of $r^{L} \vect{Y}_{J M}^{L}(\vartheta, \varphi)$ are harmonic polynomials of degree $L$.\\
\item The vectors $z_{L}(k r) \vect{Y}_{J M}^{L}(\vartheta, \varphi)$ satisfy the Helmholtz equation
\begin{equation}
\left(\Delta+k^{2}\right)\left\{z_{L}(k r) \vect{Y}_{J M}^{L}(\vartheta, \varphi)\right\}=0 \label{eq:7:3:49}
\end{equation}
where
\begin{equation}
z_{L}(k r)=\sqrt{\frac{\pi}{2 k r}} Z_{L+\frac{1}{2}}(k r), \label{eq:7:3:50}
\end{equation}
and $Z_{L+\frac{1}{2}}$ is any of the cylinder functions of order $L+\frac{1}{2}$.
\end{enumerate}

\subsection{Differential Operations}\label{sec:7.3.6}\index{vector spherical harmonics!differential operations}\index{gradient}\index{divergence}\index{curl}
Below we present the results of the action of the $\nabla$ operator on scalar and vector spherical harmonics; here $f(r)$ denotes an arbitrary function of $r \equiv|r|$.\\
\begin{enumerate}[label=(\alph*)]
\item 
\begin{align}
& \vect{\nabla}\left\{f(r) Y_{J M}(\vartheta, \varphi)\right\} \equiv \operatorname{grad}\left\{f(r) Y_{J M}(\vartheta, \varphi)\right\}  \notag\\
& =\sqrt{\frac{J}{2 J+1}}\left\{\frac{d}{d r}+\frac{J+1}{r}\right\} f(r) \vect{Y}_{J M}^{J-1}(\vartheta, \varphi)-\sqrt{\frac{J+1}{2 J+1}}\left\{\frac{d}{d r}-\frac{J}{r}\right\} f(r) \vect{Y}_{J M}^{J+1}(\vartheta, \varphi), \label{eq:7:3:51}\\
& \nabla\left\{f(r) Y_{J M}(\vartheta, \varphi)\right\} \equiv \operatorname{grad}\left\{f(r) Y_{J M}(\vartheta, \varphi)\right\}=\frac{d f(r)}{d r} \vect{Y}_{J M}^{(-1)}(\vartheta, \varphi)+\sqrt{J(J+1)} \frac{1}{r} f(r) \vect{Y}_{J M}^{(1)}(\vartheta, \varphi), \notag\\
& (\vect{r} \cdot \nabla)\left\{f(r) \vect{Y}_{J M}^{L}(\vartheta, \varphi)\right\}=r \frac{d}{d r} f(r) \vect{Y}_{J M}^{L}(\vartheta, \varphi), \label{eq:7:3:52}\\
& (\vect{r} \cdot \nabla)\left\{f(r) \vect{Y}_{J M}^{(\lambda)}(\vartheta, \varphi)\right\}=r \frac{d}{d r} f(r) \vect{Y}_{J M}^{(\lambda)}(\vartheta, \varphi). \notag
\end{align}
\item 
\begin{align}
\vect{\nabla} \cdot\left[f(r) \vect{Y}_{J M}^{J+1}(\vartheta, \varphi)\right] \equiv \operatorname{div}\left[f(r) \vect{Y}_{J M}^{J+1}(\vartheta, \varphi)\right]=-\sqrt{\frac{J+1}{2 J+1}}\left(\frac{d}{d r}+\frac{J+2}{r}\right) f(r) Y_{J M}(\vartheta, \varphi), \notag\\
\vect{\nabla} \cdot\left[f(r) \vect{Y}_{J M}^{J}(\vartheta, \varphi)\right] \equiv \operatorname{div}\left[f(r) \vect{Y}_{J M}^{J}(\vartheta, \varphi)\right]=0, \label{eq:7:3:53}\\
\vect{\nabla} \cdot\left[f(r) \vect{Y}_{J M}^{J-1}(\vartheta, \varphi)\right] \equiv \operatorname{div}\left[f(r) \vect{Y}_{J M}^{J-1}(\vartheta, \varphi)\right]=\sqrt{\frac{J}{2 J+1}}\left(\frac{d}{d r}-\frac{J-1}{r}\right) f(r) Y_{J M}(\vartheta, \varphi), \notag\\
\vect{\nabla} \cdot\left[f(r) \vect{Y}_{J M}^{(1)}(\vartheta, \varphi)\right] \equiv \operatorname{div}\left[f(r) \vect{Y}_{J M}^{(1)}(\vartheta, \varphi)\right]=-\sqrt{J(J+1)} \frac{1}{r} f(r) Y_{J M}(\vartheta, \varphi), \notag\\
\vect{\nabla} \cdot\left[f(r) \vect{Y}_{J M}^{(0)}(\vartheta, \varphi)\right] \equiv \operatorname{div}\left[f(r) \vect{Y}_{J M}^{(0)}(\vartheta, \varphi)\right]=0, \label{eq:7:3:54}\\
\vect{\nabla} \cdot\left[f(r) \vect{Y}_{J M}^{(-1)}(\vartheta, \varphi)\right] \equiv \operatorname{div}\left[f(r) \vect{Y}_{J M}^{(-1)}(\vartheta, \varphi)\right]=\left(\frac{d}{d r}+\frac{2}{r}\right) f(r) Y_{J M}(\vartheta, \varphi). \notag
\end{align}
\item 
\begin{align}
& \vect{\nabla} \times\left[f(r) \vect{Y}_{J M}^{J+1}(\vartheta, \varphi)\right] \equiv \operatorname{curl}\left[f(r) \vect{Y}_{J M}^{J+1}(\vartheta, \varphi)\right]=i \sqrt{\frac{J}{2 J+1}}\left(\frac{d}{d r}+\frac{J+2}{r}\right) f(r) \vect{Y}_{J M}^{J}(\vartheta, \varphi), \notag\\
& \vect{\nabla} \times\left[f(r) \vect{Y}_{J M}^{J}(\vartheta, \varphi)\right] \equiv \operatorname{curl}\left[f(r) \vect{Y}_{J M}^{J}(\vartheta, \varphi)\right], \notag\\
& \quad=i \sqrt{\frac{J}{2 J+1}}\left(\frac{d}{d r}-\frac{J}{r}\right) f(r) \vect{Y}_{J M}^{J+1}(\vartheta, \varphi)+i \sqrt{\frac{J+1}{2 J+1}}\left(\frac{d}{d r}+\frac{J+1}{r}\right) f(r) \vect{Y}_{J M}^{J-1}(\vartheta, \varphi), \label{eq:7:3:55}\\
& \vect{\nabla} \times\left[f(r) \vect{Y}_{J M}^{J-1}(\vartheta, \varphi)\right] \equiv \operatorname{curl}\left[f(r) \vect{Y}_{J M}^{J-1}(\vartheta, \varphi)\right]=i \sqrt{\frac{J+1}{2 J+1}}\left(\frac{d}{d r}-\frac{J-1}{r}\right) f(r) \vect{Y}_{J M}^{J}(\vartheta, \varphi), \notag\\
& \vect{\nabla} \times\left[f(r) \vect{Y}_{J M}^{(1)}(\vartheta, \varphi)\right] \equiv \operatorname{curl}\left[f(r) \vect{Y}_{J M}^{(1)}(\vartheta, \varphi)\right]=i\left(\frac{d}{d r}+\frac{1}{r}\right) f(r) \vect{Y}_{J M}^{(0)}(\vartheta, \varphi), \notag\\
& \vect{\nabla} \times\left[f(r) \vect{Y}_{J M}^{(0)}(\vartheta, \varphi)\right] \equiv \operatorname{curl}\left[f(r) \vect{Y}_{J M}^{(0)}(\vartheta, \varphi)\right], \notag\\
& \quad=i\left(\frac{d}{d r}+\frac{1}{r}\right) f(r) \vect{Y}_{J M}^{(1)}(\vartheta, \varphi)+i \sqrt{J(J+1)} \frac{1}{r} f(r) \vect{Y}_{J M}^{(-1)}(\vartheta, \varphi), \label{eq:7:3:56}\\
& \vect{\nabla} \times\left[f(r) \vect{Y}_{J M}^{(-1)}(\vartheta, \varphi)\right] \equiv \operatorname{curl}\left[f(r) \vect{Y}_{J M}^{(-1)}(\vartheta, \varphi)\right]=-i \sqrt{J(J+1)} \frac{1}{r} f(r) \vect{Y}_{J M}^{(0)}(\vartheta, \varphi). \notag
\end{align}
\item Some applications of spherical harmonics to physical problems involve the functions
\begin{align}
F_{J M}(\mathrm{r}) & \equiv z_{J}(k r) Y_{J M}(\vartheta, \varphi), \notag\\
\vect{F}_{J M}^{L}(\mathrm{r}) & \equiv z_{L}(k r) \vect{Y}_{J M}^{L}(\vartheta, \varphi). \label{eq:7:3:57}
\end{align}
where $z_{L}(k r) \equiv \sqrt{\pi /(2 k r)} Z_{L+\frac{1}{2}}(k r)$, and $Z_{L+\frac{1}{2}}$ is any cylinder functions; $k$ is an arbitrary parameter. For these special functions Eqs. \eqref{eq:7:3:51}-\eqref{eq:7:3:56} are reduced to the following
\begin{align}
& \frac{1}{k} \operatorname{grad}\left\{F_{J M}(\vect{r})\right\}=\sqrt{\frac{J}{2 J+1}} z_{J-1}(k r) \vect{Y}_{J M}^{J-1}(\vartheta, \varphi)+\sqrt{\frac{J+1}{2 J+1}} z_{J+1}(k r) \vect{Y}_{J M}^{J+1}(\vartheta, \varphi)  \notag\\
&=\sqrt{\frac{J}{2 J+1}} \vect{F}_{J M}^{J-1}(\vect{r})+\sqrt{\frac{J+1}{2 J+1}} \vect{F}_{J M}^{J+1}(\vect{r}), \label{eq:7:3:58}\\
& \frac{1}{k} \operatorname{div}\left\{\vect{F}_{J M}^{J+1}(\vect{r})\right\}=-\sqrt{\frac{J+1}{2 J+1}} z_{J}(k r) Y_{J M}(\vartheta, \varphi)=-\sqrt{\frac{J+1}{2 J+1}} F_{J M}(\vect{r}), \notag\\
& \frac{1}{k} \operatorname{div}\left\{\vect{F}_{J M}^{J}(\vect{r})\right\}=0, \label{eq:7:3:59}\\
& \frac{1}{k} \operatorname{div}\left\{\vect{F}_{J M}^{J-1}(\vect{r})\right\}=-\sqrt{\frac{J}{2 J+1}} z_{J}(k r) Y_{J M}(\vartheta, \varphi)=-\sqrt{\frac{J}{2 J+1}} F_{J M}(\vect{r}), \notag\\
&  \notag\\
& \begin{aligned}
\frac{1}{k} \operatorname{curl}\left\{\vect{F}_{J M}^{J+1}(\vect{r})\right\} & =i \sqrt{\frac{J}{2 J+1}} z_{J}(k r) \vect{Y}_{J M}^{J}(\vartheta, \varphi)=i \sqrt{\frac{J}{2 J+1}} \vect{F}_{J M}^{J}(\vect{r}) \\
\frac{1}{k} \operatorname{curl}\left\{\vect{F}_{J M}^{J}(\vect{r})\right\} & =-i \sqrt{\frac{J}{2 J+1}} z_{J+1}(k r) \vect{Y}_{J M}^{J+1}(\vartheta, \varphi)+i \sqrt{\frac{J+1}{2 J+1}} z_{J-1}(k r) \vect{Y}_{J M}^{J-1}(\vartheta, \varphi) \\
& =-i \sqrt{\frac{J}{2 J+1}} \vect{F}_{J M}^{J+1}(r)+i \sqrt{\frac{J+1}{2 J+1}} \vect{F}_{J M}^{J-1}(\vect{r}) \\
\frac{1}{k} \operatorname{curl}\left\{\vect{F}_{J M}^{J-1}(\vect{r})\right\} & =-i \sqrt{\frac{J+1}{2 J+1}} z_{J}(k r) \vect{Y}_{J M}^{J}(\vartheta, \varphi)=-i \sqrt{\frac{J+1}{2 J+1}} \vect{F}_{J M}^{J}(\vect{r}).
\end{aligned} \label{eq:7:3:60}
\end{align}
\end{enumerate}

\subsection{Action of Angular Momentum Operators}\label{sec:7.3.7}\index{vector spherical harmonics!action of angular momentum operators}
The action of the angular momentum operators on vector spherical harmonics is given by Eqs. \eqref{eq:7:1:27} and \eqref{eq:7:1:29} for $S=1$. In addition, we note the following relations in which $\widehat{\vect{S}}, \widehat{\vect{L}}$ and $\widehat{\vect{J}}=\widehat{\vect{L}}+\widehat{\vect{S}}$ are the spin, orbital and total angular momentum operators, respectively; $n$ is the unit vector defined by the polar angles $\vartheta, \varphi$.\\
\begin{enumerate}[label=(\alph*)]
\item
\begin{gather}
(\widehat{\vect{S}} \cdot \vect{n}) \vect{Y}_{J M}^{L}(\vartheta, \varphi)=-\frac{1}{2} \sqrt{\frac{(J+L+3)(J+L)(-J+L+2)(J-L+1)}{(2 L+1)(2 L+3)}} \vect{Y}_{J M}^{L+1}(\vartheta, \varphi), \notag\\
-\frac{1}{2} \sqrt{\frac{(J+L+2)(J+L-1)(-J+L+1)(J-L+2)}{(2 L-1)(2 L+1)}} \vect{Y}_{J M}^{L-1}(\vartheta, \varphi). \label{eq:7:3:61}
\end{gather}
In particular,
\begin{align}
&(\widehat{\vect{S}} \cdot \vect{n}) \vect{Y}_{J M}^{J+1}(\vartheta, \varphi)=-\sqrt{\frac{J}{2 J+1}} \vect{Y}_{J M}^{J}(\vartheta, \varphi), \notag\\
&(\widehat{\vect{S}} \cdot \vect{n}) \vect{Y}_{J M}^{J}(\vartheta, \varphi)=-\sqrt{\frac{J}{2 J+1}} \vect{Y}_{J M}^{J+1}(\vartheta, \varphi)-\sqrt{\frac{J+1}{2 J+1}} \vect{Y}_{J M}^{J-1}(\vartheta, \varphi), \label{eq:7:3:62}\\
&(\widehat{\vect{S}} \cdot \vect{n}) \vect{Y}_{J M}^{J-1}(\vartheta, \varphi)=-\sqrt{\frac{J+1}{2 J+1}} \vect{Y}_{J M}^{J}(\vartheta, \varphi), \notag\\
&(\widehat{\vect{S}} \cdot \vect{n}) \vect{Y}_{J M}^{(1)}(\vartheta, \varphi)=-\vect{Y}_{J M}^{(0)}(\vartheta, \varphi), \notag\\
&(\widehat{\vect{S}} \cdot \vect{n}) \vect{Y}_{J M}^{(0)}(\vartheta, \varphi)=-\vect{Y}_{J M}^{(1)}(\vartheta, \varphi), \label{eq:7:3:63}\\
&(\widehat{\vect{S}} \cdot \vect{n}) \vect{Y}_{J M}^{(-1)}(\vartheta, \varphi)=0. \notag
\end{align}
\item
\begin{align}
& r(\widehat{\vect{S}} \cdot \nabla) \vect{Y}_{J M}^{L}(\vartheta, \varphi)=\frac{L}{2} \sqrt{\frac{(J+L+3)(J+L)(-J+L+2)(J-L+1)}{(2 L+1)(2 L+3)}} \vect{Y}_{J M}^{L+1}(\vartheta, \varphi), \notag\\
& \quad-\frac{L+1}{2} \sqrt{\frac{(J+L+2)(J+L-1)(-J+L+1)(J-L+2)}{(2 L-1)(2 L+1)}} \vect{Y}_{J M}^{L-1}(\vartheta, \varphi). \label{eq:7:3:64}
\end{align}
In particular,
\begin{align}
r(\widehat{\vect{S}} \cdot \nabla) \vect{Y}_{J M}^{J+1}(\vartheta, \varphi) & =-(J+2) \sqrt{\frac{J}{2 J+1}} \vect{Y}_{J M}^{J}(\vartheta, \varphi), \notag\\
r(\widehat{\vect{S}} \cdot \nabla) \vect{Y}_{J M}^{J}(\vartheta, \varphi) & =J \sqrt{\frac{J}{2 J+1}} \vect{Y}_{J M}^{J+1}(\vartheta, \varphi)-(J+1) \sqrt{\frac{J+1}{2 J+1}} \vect{Y}_{J M}^{J-1}(\vartheta, \varphi), \label{eq:7:3:65}\\
r(\widehat{\vect{S}} \cdot \nabla) \vect{Y}_{J M}^{J-1}(\vartheta, \varphi) & =(J-1) \sqrt{\frac{J+1}{2 J+1}} \vect{Y}_{J M}^{J}(\vartheta, \varphi), \notag\\
r(\widehat{\vect{S}} \cdot \nabla) \vect{Y}_{J M}^{(1)}(\vartheta, \varphi) & =-\vect{Y}_{J M}^{(0)}(\vartheta, \varphi), \notag\\
r(\widehat{\vect{S}} \cdot \nabla) \vect{Y}_{J M}^{(0)}(\vartheta, \varphi) & =-\vect{Y}_{J M}^{(1)}(\vartheta, \varphi)-\sqrt{J(J+1)} \vect{Y}_{J M}^{(-1)}(\vartheta, \varphi), \label{eq:7:3:66}\\
r(\widehat{\vect{S}} \cdot \nabla) \vect{Y}_{J M}^{(-1)}(\vartheta, \varphi) & =\sqrt{J(J+1)} \vect{Y}_{J M}^{(0)}(\vartheta, \varphi). \notag
\end{align}
\item
\begin{equation}
(\widehat{\vect{S}} \widehat{\vect{L}}) \vect{Y}_{J M}^{L}(\vartheta, \varphi)=\frac{1}{2}\{J(J+1)-L(L+1)-2\} \vect{Y}_{J M}^{L}(\vartheta, \varphi). \label{eq:7:3:67}
\end{equation}
In particular,
\begin{align}
(\widehat{\vect{S}} \cdot \widehat{\vect{L}}) \vect{Y}_{J M}^{J+1}(\vartheta, \varphi) & =-(J+2) \vect{Y}_{J M}^{J+1}(\vartheta, \varphi), \notag\\
(\widehat{\vect{S}} \cdot \widehat{\vect{L}}) \vect{Y}_{J M}^{J}(\vartheta, \varphi) & =-\vect{Y}_{J M}^{J}(\vartheta, \varphi), \label{eq:7:3:68}\\
(\widehat{\vect{S}} \cdot \widehat{\vect{L}}) \vect{Y}_{J M}^{J-1}(\vartheta, \varphi) & =(J-1) \vect{Y}_{J M}^{J-1}(\vartheta, \varphi), \notag\\
(\widehat{\vect{S}} \cdot \widehat{\vect{L}}) \vect{Y}_{J M}^{(1)}(\vartheta, \varphi) & =-\vect{Y}_{J M}^{(1)}(\vartheta, \varphi)+\sqrt{J(J+1)} \vect{Y}_{J M}^{(-1)}(\vartheta, \varphi), \notag\\
(\widehat{\vect{S}} \cdot \widehat{\vect{L}}) \vect{Y}_{J M}^{(0)}(\vartheta, \varphi) & =-\vect{Y}_{J M}^{(0)}(\vartheta, \varphi), \label{eq:7:3:69}\\
(\widehat{\vect{S}} \cdot \widehat{\vect{L}}) \vect{Y}_{J M}^{(-1)}(\vartheta, \varphi) & =\sqrt{J(J+1)} \vect{Y}_{J M}^{(1)}(\vartheta, \varphi)-2 \vect{Y}_{J M}^{(-1)}(\vartheta, \varphi). \notag
\end{align}
\end{enumerate}

\subsection{Algebraic Relations}\label{sec:7.3.8}\index{vector spherical harmonics!algebraic relations}
In the equations given below $\vect{n}$ is the unit vector specified by the polar angles $\vartheta, \varphi$.\\
\begin{enumerate}[label=(\alph*)]
\item
\begin{align}
& \vect{n} Y_{J M}(\vartheta, \varphi)=\sqrt{\frac{J}{2 J+1}} \vect{Y}_{J M}^{J-1}(\vartheta, \varphi)-\sqrt{\frac{J+1}{2 J+1}} \vect{Y}_{J M}^{J+1}(\vartheta, \varphi), \notag\\
& \vect{n} Y_{J M}(\vartheta, \varphi)=\vect{Y}_{J M}^{(-1)}(\vartheta, \varphi). \label{eq:7:3:70}
\end{align}
\item
\begin{align}
\vect{n} \cdot \vect{Y}_{J M}^{J+1}(\vartheta, \varphi) & =-\sqrt{\frac{J+1}{2 J+1}} Y_{J M}(\vartheta, \varphi), \notag\\
\vect{n} \cdot \vect{Y}_{J M}^{J}(\vartheta, \varphi) & =0, \label{eq:7:3:71}\\
\vect{n} \cdot \vect{Y}_{J M}^{J-1}(\vartheta, \varphi) & =\sqrt{\frac{J}{2 J+1}} Y_{J M}(\vartheta, \varphi). \notag
\end{align}
\begin{align}
\vect{n} \cdot \vect{Y}_{J M}^{(1)}(\vartheta, \varphi) & =0, \notag\\
\vect{n} \cdot \vect{Y}_{J M}^{(0)}(\vartheta, \varphi) & =0, \label{eq:7:3:72}\\
\vect{n} \cdot \vect{Y}_{J M}^{(-1)}(\vartheta, \varphi) & =Y_{J M}(\vartheta, \varphi). \notag
\end{align}
\item
\begin{align}
\vect{n} \times \vect{Y}_{J M}^{J+1}(\vartheta, \varphi)= & i \sqrt{\frac{J}{2 J+1}} \vect{Y}_{J M}^{J}(\vartheta, \varphi), \notag\\
\vect{n} \times \vect{Y}_{J M}^{J}(\vartheta, \varphi)= & i \sqrt{\frac{J+1}{2 J+1}} \vect{Y}_{J M}^{J-1}(\vartheta, \varphi)+i \sqrt{\frac{J}{2 J+1}} \vect{Y}_{J M}^{J+1}(\vartheta, \varphi), \label{eq:7:3:73}\\
\vect{n} \times \vect{Y}_{J M}^{J-1}(\vartheta, \varphi)= & i \sqrt{\frac{J+1}{2 J+1}} \vect{Y}_{J M}^{J}(\vartheta, \varphi), \notag\\
& \vect{n} \times \vect{Y}_{J M}^{(1)}(\vartheta, \varphi)=i \vect{Y}_{J M}^{(0)}(\vartheta, \varphi), \notag\\
& \vect{n} \times \vect{Y}_{J M}^{(0)}(\vartheta, \varphi)=i \vect{Y}_{J M}^{(1)}(\vartheta, \varphi), \label{eq:7:3:74}\\
& \vect{n} \times \vect{Y}_{J M}^{(-1)}(\vartheta, \varphi)=0. \notag
\end{align}
\end{enumerate}
\subsection{Sums of Vector Spherical Harmonics}\label{sec:7.3.9}\index{vector spherical harmonics!sums of}\index{sums!of the vector spherical harmonics}
\begin{enumerate}[label=(\alph*)]
\item
\begin{align}
& \sum_{\mu=-1}^{1} Y_{1 \mu}^{*}(\vartheta, \varphi)\left[\vect{Y}_{J M}^{L}(\vartheta, \varphi)\right]_{\mu}=\sqrt{\frac{3}{4 \pi}} \vect{n} \cdot \vect{Y}_{J M}^{L}(\vartheta, \varphi), \notag\\
& \sum_{\mu=-1}^{1} Y_{1 \mu}^{*}(\vartheta, \varphi)\left[\vect{Y}_{J M}^{(\lambda)}(\vartheta, \varphi)\right]_{\mu}=\sqrt{\frac{3}{4 \pi}} \vect{n} \cdot \vect{Y}_{J M}^{(\lambda)}(\vartheta, \varphi). \label{eq:7:3:75}
\end{align}
From \eqref{eq:7:3:75} one obtains
\begin{align}
& \sum_{\mu=-1}^{1} Y_{1 \mu}^{*}(\vartheta, \varphi)\left[\vect{Y}_{J M}^{J+1}(\vartheta, \varphi)\right]_{\mu}=-\sqrt{\frac{3(J+1)}{4 \pi(2 J+1)}} Y_{J M}(\vartheta, \varphi), \notag\\
& \sum_{\mu=-1}^{1} Y_{1 \mu}^{*}(\vartheta, \varphi)\left[\vect{Y}_{J M}^{J}(\vartheta, \varphi)\right]_{\mu}=0, \label{eq:7:3:76}\\
& \sum_{\mu=-1}^{1} Y_{1 \mu}^{*}(\vartheta, \varphi)\left[\vect{Y}_{J M}^{J-1}(\vartheta, \varphi)\right]_{\mu}=\sqrt{\frac{3 J}{4 \pi(2 J+1)}} Y_{J M}(\vartheta, \varphi) . \notag\\
& \sum_{\mu=-1}^{1} Y_{1 \mu}^{*}(\vartheta, \varphi)\left[\vect{Y}_{J M}^{(1)}(\vartheta, \varphi)\right]_{\mu}=0, \notag\\
& \sum_{\mu=-1}^{1} Y_{1 \mu}^{*}(\vartheta, \varphi)\left[\vect{Y}_{J M}^{(0)}(\vartheta, \varphi)\right]_{\mu}=0, \label{eq:7:3:77}\\
& \sum_{\mu=-1}^{1} Y_{1 \mu}^{*}(\vartheta, \varphi)\left[\vect{Y}_{J M}^{(-1)}(\vartheta, \varphi)\right]_{\mu}=\sqrt{\frac{3}{4 \pi}} Y_{J M}(\vartheta, \varphi) . \notag
\end{align}
\item
\begin{equation}
\sum_{M=-J}^{J} Y_{J M}^{*}(\vartheta, \varphi) \vect{Y}_{J M}^{L}(\vartheta, \varphi)=\frac{\sqrt{(2 J+1)(2 L+1)}}{4 \pi} \clebsch{1}{0}{L}{0}{J}{0} \vect{n}. \label{eq:7:3:78}
\end{equation}
In more detailed form \eqref{eq:7:3:78} reads
\begin{align}
& \sum_{M=-J}^{J} Y_{J M}^{*}(\vartheta, \varphi) \vect{Y}_{J M}^{J+1}(\vartheta, \varphi)=-\frac{\sqrt{(J+1)(2 J+1)}}{4 \pi} \vect{n}, \notag\\
& \sum_{M=-J}^{J} Y_{J M}^{*}(\vartheta, \varphi) \vect{Y}_{J M}^{J}(\vartheta, \varphi)=0, \label{eq:7:3:79}\\
& \sum_{M=-J}^{J} Y_{J M}^{*}(\vartheta, \varphi) \vect{Y}_{J M}^{J-1}(\vartheta, \varphi)=\frac{\sqrt{J(2 J+1)}}{4 \pi} \vect{n}. \notag
\end{align}
Analogous relations for $\vect{Y}_{J M}^{(\lambda)}(\vartheta, \varphi)$ have the form
\begin{gather}
\sum_{M=-J}^{J} Y_{J M}^{*}(\vartheta, \varphi) \vect{Y}_{J M}^{(1)}(\vartheta, \varphi)=\sum_{M=-J}^{J} Y_{J M}^{*}(\vartheta, \varphi) \vect{Y}_{J M}^{(0)}(\vartheta, \varphi)=0, \label{eq:7:3:80}\\
\sum_{M=-J}^{J} Y_{J M}^{*}(\vartheta, \varphi) \vect{Y}_{J M}^{(-1)}(\vartheta, \varphi)=\frac{2 J+1}{4 \pi} \vect{n}. \notag
\end{gather}
\item
\begin{align}
& \sum_{M=-J}^{J} \vect{Y}_{J M}^{L^{\prime} *}(\vartheta, \varphi) \cdot \vect{Y}_{J M}^{L}(\vartheta, \varphi)=\frac{2 J+1}{4 \pi} \delta_{L L^{\prime}}, \label{eq:7:3:81}\\
& \sum_{M=-J}^{J} \vect{Y}_{J M}^{\left(\lambda^{\prime}\right) *}(\vartheta, \varphi) \cdot \vect{Y}_{J M}^{(\lambda)}(\vartheta, \varphi)=\frac{2 J+1}{4 \pi} \delta_{\lambda \lambda^{\prime}}. \notag
\end{align}
\item Below $\vect{a}$ is an arbitrary complex vector, $\vect{n}$ is the unit vector specified by the polar angles $\vartheta, \varphi$.
\begin{gather}
\sum_{M=-J}^{J}\left|\vect{a} \cdot \vect{Y}_{J M}^{J+1}(\vartheta, \varphi)\right|^{2}=\frac{1}{8 \pi}\left\{J|\vect{a}|^{2}+(J+2)|\vect{n} \cdot \vect{a}|^{2}\right\}, \label{eq:7:3:82}\\
\sum_{M=-J}^{J}\left|\vect{a} \cdot \vect{Y}_{J M}^{J}(\vartheta, \varphi)\right|^{2}=\frac{2 J+1}{8 \pi}\left\{|\vect{a}|^{2}-|\vect{n} \cdot \vect{a}|^{2}\right\}, \label{eq:7:3:83}\\
\sum_{M=-J}^{J}\left|\vect{a} \cdot \vect{Y}_{J M}^{J-1}(\vartheta, \varphi)\right|^{2}=\frac{1}{8 \pi}\left\{(J+1)|\vect{a}|^{2}+(J-1)|\vect{n} \cdot \vect{a}|^{2}\right\}, \label{eq:7:3:84}\\
\sum_{M=-J}^{J}\left(\vect{a} \cdot \vect{Y}_{J M}^{J+1}(\vartheta, \varphi)\right)^{*}\left(\vect{a} \cdot \vect{Y}_{J M}^{J}(\vartheta, \varphi)\right)=-i \frac{\sqrt{J(2 J+1)}}{8 \pi} \vect{n} \cdot\left[\vect{a}{ }^{*} \times \vect{a}\right], \label{eq:7:3:85}\\
\sum_{M=-J}^{J}\left(\vect{a} \cdot \vect{Y}_{J M}^{J+1}(\vartheta, \varphi)\right)^{*}\left(\vect{a} \cdot \vect{Y}_{J M}^{J-1}(\vartheta, \varphi)\right)=\frac{\sqrt{J(J+1)}}{8 \pi}\left\{|\vect{a}|^{2}-3|\vect{n} \cdot \vect{a}|^{2}\right\}, \label{eq:7:3:86}\\
\sum_{M=-J}^{J}\left(\vect{a} \cdot \vect{Y}_{J M}^{J}(\vartheta, \varphi)\right)^{*}\left(\vect{a} \cdot \vect{Y}_{J M}^{J+1}(\vartheta, \varphi)\right)=-i \frac{\sqrt{J(2 J+1)}}{8 \pi} \vect{n} \cdot\left[\vect{a}^{*} \times \vect{a}\right], \label{eq:7:3:87}\\
\sum_{M=-J}^{J}\left(\vect{a} \cdot \vect{Y}_{J M}^{J}(\vartheta, \varphi)\right)^{*}\left(\vect{a} \cdot \vect{Y}_{J M}^{J-1}(\vartheta, \varphi)\right)=-i \frac{\sqrt{(J+1)(2 J+1)}}{8 \pi} \vect{n} \cdot\left[\vect{a}^{*} \times \vect{a}\right]. \label{eq:7:3:88}
\end{gather}
\begin{align}
& \sum_{M=-J}^{J}\left(\vect{a} \cdot \vect{Y}_{J M}^{J-1}(\vartheta, \varphi)\right)^{*}\left(\vect{a} \cdot \vect{Y}_{J M}^{J+1}(\vartheta, \varphi)\right)=\frac{\sqrt{J(J+1)}}{8 \pi}\left\{|\vect{a}|^{2}-3|\vect{n} \cdot \vect{a}|^{2}\right\}, \label{eq:7:3:89}\\
& \sum_{M=-J}^{J}\left(\vect{a} \cdot \vect{Y}_{J M}^{J-1}(\vartheta, \varphi)\right)^{*}\left(\vect{a} \cdot \vect{Y}_{J M}^{J}(\vartheta, \varphi)\right)=-i \frac{\sqrt{(J+1)(2 J+1)}}{8 \pi} \vect{n} \cdot\left[\vect{a}^{*} \times \vect{a}\right]. \label{eq:7:3:90}
\end{align}
\item Analogous relations for $\vect{Y}_{J M}^{(\lambda)}(\vartheta, \varphi)$ have the form:
\begin{gather}
\sum_{M=-J}^{J}\left|\vect{a} \cdot \vect{Y}_{J M}^{(1)}(\vartheta, \varphi)\right|^{2}=\frac{2 J+1}{8 \pi}\left\{|\vect{a}|^{2}-|\vect{n} \cdot \vect{a}|^{2}\right\}, \label{eq:7:3:91}\\
\sum_{M=-J}^{J}\left|\vect{a} \cdot \vect{Y}_{J M}^{(0)}(\vartheta, \varphi)\right|^{2}=\frac{2 J+1}{8 \pi}\left\{|\vect{a}|^{2}-|\vect{n} \cdot \vect{a}|^{2}\right\}, \label{eq:7:3:92}\\
\sum_{M=-J}^{J}\left|\vect{a} \cdot \vect{Y}_{J M}^{(-1)}(\vartheta, \varphi)\right|^{2}=\frac{2 J+1}{4 \pi}|\vect{n} \cdot \vect{a}|^{2}, \label{eq:7:3:93}\\
\sum_{M=-J}^{J}\left(\vect{a} \cdot \vect{Y}_{J M}^{(1)}(\vartheta, \varphi)\right)^{*}\left(\vect{a} \cdot \vect{Y}_{J M}^{(0)}(\vartheta, \varphi)\right)=-i \frac{2 J+1}{8 \pi} \vect{n} \cdot\left[\vect{a}{ }^{*} \times \vect{a}\right], \label{eq:7:3:94}\\
\sum_{M=-J}^{J}\left(\vect{a} \cdot \vect{Y}_{J M}^{(1)}(\vartheta, \varphi)\right)^{*}\left(\vect{a} \cdot \vect{Y}_{J M}^{(-1)}(\vartheta, \varphi)\right)=0, \label{eq:7:3:95}\\
\sum_{M=-J}^{J}\left(\vect{a} \cdot \vect{Y}_{J M}^{(0)}(\vartheta, \varphi)\right)^{*}\left(\vect{a} \cdot \vect{Y}_{J M}^{(1)}(\vartheta, \varphi)\right)=-i \frac{2 J+1}{8 \pi} \vect{n}\left[\vect{a}^{*} \times \vect{a}\right], \label{eq:7:3:96}\\
\sum_{M=-J}^{J}\left(\vect{a} \cdot \vect{Y}_{J M}^{(0)}(\vartheta, \varphi)\right)^{*}\left(\vect{a} \cdot \vect{Y}_{J M}^{(-1)}(\vartheta, \varphi)\right)=0, \label{eq:7:3:97}\\
\sum_{M=-J}^{J}\left(\vect{a} \cdot \vect{Y}_{J M}^{(-1)}(\vartheta, \varphi)\right)^{*}\left(\vect{a} \cdot \vect{Y}_{J M}^{(1)}(\vartheta, \varphi)\right)=0, \label{eq:7:3:98}\\
\sum_{M=-J}^{J}\left(\vect{a} \cdot \vect{Y}_{J M}^{(-1)}(\vartheta, \varphi)\right)^{*}\left(\vect{a} \cdot \vect{Y}_{J M}^{(0)}(\vartheta, \varphi)\right)=0 . \label{eq:7:3:99}
\end{gather}
\end{enumerate}
\subsection{Clebsch-Gordan Series}\label{sec:7.3.10}\index{Clebsch-Gordan series}\index{vector spherical harmonics!Clebsch-Gordan series}
\begin{align}
& \vect{Y}_{J_{1} M_{1}}^{L_{1}}(\vartheta, \varphi) \cdot \vect{Y}_{J_{2} M_{2}}^{L_{2}}(\vartheta, \varphi)  \notag\\
& =\sum_{L}(-1)^{J_{2}+L_{1}+L} \sqrt{\frac{\left(2 J_{1}+1\right)\left(2 J_{2}+1\right)\left(2 L_{1}+1\right)\left(2 L_{2}+1\right)}{4 \pi(2 L+1)}}\sixj{L_{1}}{L_{2}}{L}{J_{2}}{J_{1}}{1} \clebsch{L_{1}}{0}{L_{2}}{0}{L}{0} \clebsch{J_{1}}{M_{1}}{J_{2}}{M_{2}}{L}{M} Y_{L M}(\vartheta, \varphi), \label{eq:7:3:100}\\
& \vect{Y}_{J_{1} M_{1}}^{L_{1}}(\vartheta, \varphi) \times \vect{Y}_{J_{2} M_{2}}^{L_{2}}(\vartheta, \varphi)=i \sqrt{\frac{3}{2 \pi}\left(2 J_{1}+1\right)\left(2 J_{2}+1\right)\left(2 L_{1}+1\right)\left(2 L_{2}+1\right)}  \notag\\
& \times \sum_{J L}\left\{\begin{array}{lll}
J_{1} & L_{1} & 1 \\
J_{2} & L_{2} & 1 \\
J & L & 1
\end{array}\right\} \clebsch{L_{1}}{0}{L_{2}}{0}{L}{0} \clebsch{J_{1}}{M_{1}}{J_{2}}{M_{2}}{J}{M} \vect{Y}_{J M}^{L}(\vartheta, \varphi) . \label{eq:7:3:101}
\end{align}
\subsection{Addition Theorems for Vector Spherical Harmonics}\label{sec:7.3.11}\index{addition theorem!for the vector spherical harmonics}\index{vector spherical harmonics!addition theorems}
Let $\vect{n}_{1}$ and $\vect{n}_{2}$ be the unit vectors determined by the polar angles $\vartheta_{1}, \varphi_{1}$ and $\vartheta_{2}, \varphi_{2}$, respectively; $\cos \omega_{12} \equiv \vect{n}_{1} \cdot \vect{n}_{2}=\cos \vartheta_{1} \cos \vartheta_{2}+\sin \vartheta_{1} \sin \vartheta_{2} \cos \left(\varphi_{1}-\varphi_{2}\right)$. Then we have the following addition theorems for the vector spherical harmonics $\vect{Y}_{J M}^{L}$.\\
\begin{enumerate}[label=(\alph*)]
\item
\begin{equation}
4 \pi \sum_{M=-J}^{J} \vect{Y}_{J M}^{L^{\prime} *}\left(\vartheta_{1}, \varphi_{1}\right) \cdot \vect{Y}_{J M}^{L}\left(\vartheta_{2}, \varphi_{2}\right)=\delta_{L^{\prime} L}(2 J+1) P_{L}\left(\cos \omega_{12}\right). \label{eq:7:3:102}
\end{equation}
\item
\begin{align}
& 4 \pi \sum_{M=-J}^{J} \vect{Y}_{J M}^{J+1 *}\left(\vartheta_{1}, \varphi_{1}\right) \times \vect{Y}_{J M}^{J+1}\left(\vartheta_{2}, \varphi_{2}\right)=\frac{2 J+1}{J+1}\left[\vect{n}_{1} \times \vect{n}_{2}\right] P_{J+1}^{\prime}\left(\cos \omega_{12}\right), \notag\\
& 4 \pi \sum_{M=-J}^{J} \vect{Y}_{J M}^{J+1 *}\left(\vartheta_{1}, \varphi_{1}\right) \times \vect{Y}_{J M}^{J}\left(\vartheta_{2}, \varphi_{2}\right)=-i \frac{\sqrt{J(2 J+1)}}{J+1}\left\{\vect{n}_{1} P_{J+1}^{\prime}\left(\cos \omega_{12}\right)-\vect{n}_{2} P_{J}^{\prime}\left(\cos \omega_{12}\right)\right\}, \label{eq:7:3:103}\\
& 4 \pi \sum_{M=-J}^{J} \vect{Y}_{J M}^{J+1 *}\left(\vartheta_{1}, \varphi_{1}\right) \times \vect{Y}_{J M}^{J-1}\left(\vartheta_{2}, \varphi_{2}\right)=0, \notag\\
& 4 \pi \sum_{M=-J}^{J} \vect{Y}_{J M}^{J *}\left(\vartheta_{1}, \varphi_{1}\right) \times \vect{Y}_{J M}^{J+1}\left(\vartheta_{2}, \varphi_{2}\right)=i \frac{\sqrt{J(2 J+1)}}{J+1}\left\{\vect{n}_{1} P_{J}^{\prime}\left(\cos \omega_{12}\right)-\vect{n}_{2} P_{J+1}^{\prime}\left(\cos \omega_{12}\right)\right\}, \notag\\
& 4 \pi \sum_{M=-J}^{J} \vect{Y}_{J M}^{J *}\left(\vartheta_{1}, \varphi_{1}\right) \times \vect{Y}_{J M}^{J}\left(\vartheta_{2}, \varphi_{2}\right)=\frac{(2 J+1)}{J(J+1)}\left[\vect{n}_{1} \times \vect{n}_{2}\right] P_{J}^{\prime}\left(\cos \omega_{12}\right), \label{eq:7:3:104}\\
& 4 \pi \sum_{M=-J}^{J} \vect{Y}_{J M}^{J *}\left(\vartheta_{1}, \varphi_{1}\right) \times \vect{Y}_{J M}^{J-1}\left(\vartheta_{2}, \varphi_{2}\right)=-i \frac{\sqrt{(J+1)(2 J+1)}}{J}\left\{\vect{n}_{1} P_{J}^{\prime}\left(\cos \omega_{12}\right)-\vect{n}_{2} P_{J-1}^{\prime}\left(\cos \omega_{12}\right)\right\}, \notag\\
& 4 \pi \sum_{M=-J}^{J} \vect{Y}_{J M}^{J-1 *}\left(\vartheta_{1}, \varphi_{1}\right) \times \vect{Y}_{J M}^{J+1}\left(\vartheta_{2}, \varphi_{2}\right)=0, \notag\\
& 4 \pi \sum_{M=-J}^{J} \vect{Y}_{J M}^{J-1 *}\left(\vartheta_{1}, \varphi_{1}\right) \times \vect{Y}_{J M}^{J}\left(\vartheta_{2}, \varphi_{2}\right)=i \frac{\sqrt{(J+1)(2 J+1)}}{J}\left\{\vect{n}_{1} P_{J-1}^{\prime}\left(\cos \omega_{12}\right)-\vect{n}_{2} P_{J}^{\prime}\left(\cos \omega_{12}\right)\right\}, \label{eq:7:3:105}\\
& 4 \pi \sum_{M=-J}^{J} \vect{Y}_{J M}^{J-1 *}\left(\vartheta_{1}, \varphi_{1}\right) \times \vect{Y}_{J M}^{J-1}\left(\vartheta_{2}, \varphi_{2}\right)=-\frac{2 J+1}{J}\left[\vect{n}_{1} \times \vect{n}_{2}\right] P_{J-1}^{\prime}\left(\cos \omega_{12}\right) . \notag
\end{align}
In these equations $P_{L}(x)$ is a Legendre polynomial, and $P_{L}^{\prime}(x) \equiv d P_{L}(x) / d x$. Analogous equations for $\vect{Y}_{J M}^{(\lambda)}$ may be obtained from Eqs. \eqref{eq:7:3:9}-\eqref{eq:7:3:10}.
\item The most general form of the addition theorems for $\vect{Y}_{J M}^{L}$ may be written by introducing arbitrary complex vectors $\vect{a}_{1}$ and $\vect{a}_{2}$. For brevity, we shall omit the arguments ( $\cos \omega_{12}$ ) of derivatives of the Legendre polynomials, writing $P_{L}^{\prime}$ instead of $P_{L}^{\prime}\left(\cos \omega_{12}\right)$, etc.
\begin{align}
& 4 \pi \sum_{M=-J}^{J}\left(\vect{a}_{1} \cdot \vect{Y}_{J M}^{J+1}\left(\vartheta_{1}, \varphi_{1}\right)\right)^{*}\left(\vect{a}_{2} \cdot \vect{Y}_{J M}^{J+1}\left(\vartheta_{2}, \varphi_{2}\right)\right)=\frac{1}{J+1}\left\{\left(\vect{a}_{1}^{*} \cdot \vect{n}_{1}\right)\left(\vect{a}_{2} \cdot \vect{n}_{2}\right)\left[P_{J}^{\prime \prime}+(2 J+1) P_{J+1}^{\prime}\right]\right. \notag\\
& \left.\quad+\left(\vect{a}_{1}^{*} \cdot \vect{n}_{2}\right)\left(\vect{a}_{2} \cdot \vect{n}_{1}\right) P_{J}^{\prime \prime}-\left[\left(\vect{a}_{1}^{*} \cdot \vect{n}_{1}\right)\left(\vect{a}_{2} \cdot \vect{n}_{1}\right)+\left(\vect{a}_{1}^{*} \cdot \vect{n}_{2}\right)\left(\vect{a}_{2} \cdot \vect{n}_{2}\right)\right] P_{J+1}^{\prime \prime}+\left(\vect{a}_{1}^{*} \cdot \vect{a}_{2}\right) P_{J}^{\prime}\right\}, \notag\\
& 4 \pi \sum_{M=-J}^{J}\left(\vect{a}_{1} \cdot \vect{Y}_{J M}^{J+1}\left(\vartheta_{1}, \varphi_{1}\right)\right)^{*}\left(\vect{a}_{2} \cdot \vect{Y}_{J M}^{J}\left(\vartheta_{2}, \varphi_{2}\right)\right)=\frac{i}{2(J+1)} \sqrt{\frac{2 J+1}{J}}, \notag\\
& \quad \times\left\{-\left[\vect{a}_{1}^{*} \times \vect{a}_{2}\right] \cdot \vect{n}_{1} J P_{J+1}^{\prime}+\left[\vect{a}_{1}^{*} \times \vect{a}_{2}\right] \cdot \vect{n}_{2} J P_{J}^{\prime}-\left[\left(\vect{a}_{1}^{*} \cdot \vect{n}_{1}\right)\left(\vect{a}_{2} \cdot\left[\vect{n}_{1} \times \vect{n}_{2}\right]\right)+\left(\vect{a}_{1}^{*} \cdot\left[\vect{n}_{1} \times \vect{n}_{2}\right]\right)\left(\vect{a}_{2} \cdot \vect{n}_{1}\right)\right] P_{J+1}^{\prime \prime}\right. \notag\\
& \left.\quad+\left[\left(\vect{a}_{1}^{*} \cdot \vect{n}_{2}\right)\left(\vect{a}_{2} \cdot\left[\vect{n}_{1} \times \vect{n}_{2}\right]\right)+\left(\vect{a}_{1}^{*} \cdot\left[\vect{n}_{1} \times \vect{n}_{2}\right]\right)\left(\vect{a}_{2} \cdot \vect{n}_{2}\right)\right] P_{J}^{\prime \prime}\right\}, \label{eq:7:3:106}\\
& 4 \pi \sum_{M=-J}^{J}\left(\vect{a}_{1} \cdot \vect{Y}_{J M}^{J+1}\left(\vartheta_{1}, \varphi_{1}\right)\right)^{*}\left(\vect{a}_{2} \cdot \vect{Y}_{J M}^{J-1}\left(\vartheta_{2}, \varphi_{2}\right)\right)=\frac{1}{\sqrt{J(J+1)}}, \notag\\
& \quad \times\left\{\left[\left(\vect{a}_{1}^{*} \cdot \vect{n}_{1}\right)\left(\vect{a}_{2} \cdot \vect{n}_{2}\right)+\left(\vect{a}_{1}^{*} \cdot \vect{n}_{2}\right)\left(\vect{a}_{2} \cdot \vect{n}_{1}\right)\right] P_{j}^{\prime \prime}-\left(\vect{a}_{1}^{*} \cdot \vect{n}_{1}\right)\left(\vect{a}_{2} \cdot \vect{n}_{1}\right) P_{J+1}^{\prime \prime}\right. \notag\\
& \left.\quad-\left(\vect{a}_{1}^{*} \cdot \vect{n}_{2}\right)\left(\vect{a}_{2} \cdot \vect{n}_{2}\right) P_{J-1}^{\prime \prime}+\left(\vect{a}_{1}^{*} \cdot \vect{a}_{2}\right) P_{J}^{\prime}\right\}. \notag
\end{align}
\[
\begin{aligned}
& 4 \pi \sum_{M=-J}^{J}\left(\vect{a}_{1} \cdot \vect{Y}_{J M}^{J}\left(\vartheta_{1}, \varphi_{1}\right)\right)^{*}\left(\vect{a}_{2} \cdot \vect{Y}_{J M}^{J+1}\left(\vartheta_{2}, \varphi_{2}\right)\right)=\frac{i}{2(J+1)} \sqrt{\frac{2 J+1}{J}} \\
& \quad \times\left\{\left[\vect{a}_{1}^{*} \times \vect{a}_{2}\right] \cdot \vect{n}_{1} J P_{J}^{\prime}-\left[\vect{a}_{1}^{*} \times \vect{a}_{2}\right] \cdot \vect{n}_{2} J P_{J+1}^{\prime}+\left[\left(\vect{a}_{1}^{*} \cdot \vect{n}_{1}\right)\left(\vect{a}_{2} \cdot\left[\vect{n}_{1} \times \vect{n}_{2}\right]\right)+\left(\vect{a}_{1}^{*} \cdot\left[\vect{n}_{1} \times \vect{n}_{2}\right]\right)\left(\vect{a}_{2} \cdot \vect{n}_{1}\right)\right] P_{J}^{\prime \prime}\right. \\
& \left.\quad-\left[\left(\vect{a}_{1}^{*} \cdot \vect{n}_{2}\right)\left(\vect{a}_{2} \cdot\left[\vect{n}_{1} \times \vect{n}_{2}\right]\right)+\left(\vect{a}_{1}^{*} \cdot\left[\vect{n}_{1} \times \vect{n}_{2}\right]\right)\left(\vect{a}_{2} \cdot \vect{n}_{2}\right)\right] P_{J+1}^{\prime \prime}\right\}
\end{aligned}
\]
\begin{align}
& 4 \pi \sum_{M=-J}^{J}\left(\vect{a}_{1} \cdot \vect{Y}_{J M}^{J}\left(\vartheta_{1}, \varphi_{1}\right)\right)^{*}\left(\vect{a}_{2} \cdot \vect{Y}_{J M}^{J}\left(\vartheta_{2}, \varphi_{2}\right)\right)=\frac{2 J+1}{J(J+1)}\left\{-\left(\vect{a}_{1}^{*} \cdot \vect{n}_{1}\right)\left(\vect{a}_{2} \cdot \vect{n}_{2}\right)\left[P_{J-1}^{\prime \prime}+(J-1) P_{J}^{\prime}\right]\right. \notag\\
& \left.-\left(\vect{a}_{1}^{*} \cdot \vect{n}_{2}\right)\left(\vect{a}_{2} \cdot \vect{n}_{1}\right)\left[P_{J-1}^{\prime \prime}+J P_{J}^{\prime}\right]+\left[\left(\vect{a}_{1}^{*} \cdot \vect{n}_{1}\right)\left(\vect{a}_{2} \cdot \vect{n}_{1}\right)+\left(\vect{a}_{1}^{*} \cdot \vect{n}_{2}\right)\left(\vect{a}_{2} \cdot \vect{n}_{2}\right)\right] P_{J}^{\prime \prime}+\left(\vect{a}_{1}^{*} \cdot \vect{a}_{2}\right)\left[J^{2} P_{J}-P_{J-1}^{\prime}\right]\right\}. \label{eq:7:3:107}
\end{align}
\[
\begin{aligned}
& 4 \pi \sum_{M=-J}^{J}\left(\vect{a}_{1} \cdot \vect{Y}_{J M}^{J}\left(\vartheta_{1}, \varphi_{1}\right)\right)^{*}\left(\vect{a}_{2} \cdot \vect{Y}_{J M}^{J-1}\left(\vartheta_{2}, \varphi_{2}\right)\right)=\frac{i}{2 J} \sqrt{\frac{2 J+1}{J+1}}\left\{-\left[\vect{a}_{1}^{*} \times \vect{a}_{2}\right] \cdot \vect{n}_{1}(J+1) P_{J}^{\prime}\right. \\
& \quad+\left[\vect{a}_{1}^{*} \times \vect{a}_{2}\right] \cdot \vect{n}_{2}(J+1) P_{J-1}^{\prime}+\left[\left(\vect{a}_{1}^{*} \cdot \vect{n}_{1}\right)\left(\vect{a}_{2} \cdot\left[\vect{n}_{1} \times \vect{n}_{2}\right]\right)+\left(\vect{a}_{1}^{*} \cdot\left[\vect{n}_{1} \times \vect{n}_{2}\right]\right)\left(\vect{a}_{2} \cdot \vect{n}_{1}\right)\right] P_{J}^{\prime \prime} \\
& \left.\quad-\left[\left(\vect{a}_{1}^{*} \cdot \vect{n}_{2}\right)\left(\vect{a}_{2} \cdot\left[\vect{n}_{1} \times \vect{n}_{2}\right]\right)+\left(\vect{a}_{1}^{*} \cdot\left[\vect{n}_{1} \times \vect{n}_{2}\right]\right)\left(\vect{a}_{2} \cdot \vect{n}_{2}\right)\right] P_{J-1}^{\prime \prime}\right\}
\end{aligned}
\]
\[
\begin{aligned}
& 4 \pi \sum_{M=-J}^{J}\left(\vect{a}_{1} \cdot \vect{Y}_{J M}^{J-1}\left(\vartheta_{1}, \varphi_{1}\right)\right)^{*}\left(\vect{a}_{2} \cdot \vect{Y}_{J M}^{J+1}\left(\vartheta_{2}, \varphi_{2}\right)\right)=\frac{1}{\sqrt{J(J+1)}}\left\{\left[\left(\vect{a}_{1}^{*} \cdot \vect{n}_{1}\right)\left(\vect{a}_{2} \cdot \vect{n}_{2}\right)+\left(\vect{a}_{1}^{*} \cdot \vect{n}_{2}\right)\left(\vect{a}_{2} \cdot \vect{n}_{1}\right)\right] P_{J}^{\prime \prime}\right. \\
& \left.\quad-\left(\vect{a}_{1}^{*} \cdot \vect{n}_{1}\right)\left(\vect{a}_{2} \cdot \vect{n}_{1}\right) P_{J-1}^{\prime \prime}-\left(\vect{a}_{1}^{*} \cdot \vect{n}_{2}\right)\left(\vect{a}_{2} \cdot \vect{n}_{2}\right) P_{J+1}^{\prime \prime}+\left(\vect{a}_{1}^{*} \cdot \vect{a}_{2}\right) P_{J}^{\prime}\right\}
\end{aligned}
\]
\begin{align}
& 4 \pi \sum_{M=-J}^{J}\left(\vect{a}_{1} \cdot \vect{Y}_{J M}^{J-1}\left(\vartheta_{1}, \varphi_{1}\right)\right)^{*}\left(\vect{a}_{2} \cdot \vect{Y}_{J M}^{J}\left(\vartheta_{2}, \varphi_{2}\right)\right)=\frac{i}{2 J} \sqrt{\frac{2 J+1}{J+1}}\left\{\left[\vect{a}_{1}^{*} \times \vect{a}_{2}\right] \cdot \vect{n}_{1}(J+1) P_{J-1}^{\prime}\right. \notag\\
& \quad-\left[\vect{a}_{1}^{*} \times \vect{a}_{2}\right] \cdot \vect{n}_{2}(J+1) P_{J}^{\prime}-\left[\left(\vect{a}_{1}^{*} \cdot \vect{n}_{1}\right)\left(\vect{a}_{2} \cdot\left[\vect{n}_{1} \times \vect{n}_{2}\right]\right)+\left(\vect{a}_{1}^{*} \cdot\left[\vect{n}_{1} \times \vect{n}_{2}\right]\right)\left(\vect{a}_{2} \cdot \vect{n}_{1}\right)\right] P_{J-1}^{\prime \prime}, \notag\\
& \left.\quad+\left[\left(\vect{a}_{1}^{*} \cdot \vect{n}_{2}\right)\left(\vect{a}_{2} \cdot\left[\vect{n}_{1} \times \vect{n}_{2}\right]\right)+\left(\vect{a}_{1}^{*} \cdot\left[\vect{n}_{1} \times \vect{n}_{2}\right]\right)\left(\vect{a}_{2} \cdot \vect{n}_{2}\right)\right] P_{J}^{\prime \prime}\right\}, \label{eq:7:3:108}\\
& 4 \pi \sum_{M=-J}^{J}\left(\vect{a}_{1} \cdot \vect{Y}_{J M}^{J-1}\left(\vartheta_{1}, \varphi_{1}\right)\right)^{*}\left(\vect{a}_{2} \cdot \vect{Y}_{J M}^{J-1}\left(\vartheta_{2}, \varphi_{2}\right)\right)=\frac{1}{J}\left\{\left(\vect{a}_{1}^{*} \cdot \vect{n}_{1}\right)\left(\vect{a}_{2} \cdot \vect{n}_{2}\right)\left[P_{J}^{\prime \prime}-(2 J-1) P_{J-1}^{\prime}\right]\right. \notag\\
& \left.\quad+\left(\vect{a}_{1}^{*} \cdot \vect{n}_{2}\right)\left(\vect{a}_{2} \cdot \vect{n}_{1}\right) P_{J}^{\prime \prime}-\left[\left(\vect{a}_{1}^{*} \cdot \vect{n}_{1}\right)\left(\vect{a}_{2} \cdot \vect{n}_{1}\right)+\left(\vect{a}_{1}^{*} \cdot \vect{n}_{2}\right)\left(\vect{a}_{2} \cdot \vect{n}_{2}\right)\right] P_{J-1}^{\prime \prime}+\left(\vect{a}_{1}^{*} \cdot \vect{a}_{2}\right) P_{J}^{\prime}\right\}. \notag
\end{align}
\item The case when the vectors $\vect{a}_{1}$ and $\vect{a}_{2}$ are transverse (i.e., $\vect{a}_{1} \cdot \vect{n}_{1}=\vect{a}_{2} \cdot \vect{n}_{2}=0$ ) is of special interest (e.g., to describe the multipole electromagnetic fields). In this case Eqs. \eqref{eq:7:3:106}-\eqref{eq:7:3:108} are considerably simplified to
\begin{gather}
4 \pi \sum_{M=-J}^{J}\left(\vect{a}_{1} \cdot \vect{Y}_{J M}^{J+1}\left(\vartheta_{1}, \varphi_{1}\right)\right)^{*}\left(\vect{a}_{2} \cdot \vect{Y}_{J M}^{J+1}\left(\vartheta_{2}, \varphi_{2}\right)\right)=\frac{1}{J+1}\left\{\left(\vect{a}_{1}^{*} \cdot \vect{n}_{2}\right)\left(\vect{a}_{2} \cdot \vect{n}_{1}\right) P_{J}^{\prime \prime}+\left(\vect{a}_{1}^{*} \cdot \vect{a}_{2}\right) P_{J}^{\prime}\right\}, \notag\\
4 \pi \sum_{M=-J}^{J}\left(\vect{a}_{1} \cdot \vect{Y}_{J M}^{J+1}\left(\vartheta_{1}, \varphi_{1}\right)\right)^{*}\left(\vect{a}_{2} \cdot \vect{Y}_{J M}^{J}\left(\vartheta_{2}, \varphi_{2}\right)\right), \label{eq:7:3:109}\\
=\frac{i}{J+1} \sqrt{\frac{2 J+1}{J}}\left\{\left[\vect{a}_{1}^{*} \times \vect{a}_{2}\right] \cdot \vect{n}_{1} P_{J}^{\prime \prime}+\left[\vect{a}_{1}^{*} \times \vect{a}_{2}\right] \cdot \vect{n}_{2}\left[(J+1) P_{J}^{\prime}-P_{J+1}^{\prime \prime}\right]\right\}, \notag\\
4 \pi \sum_{M=-J}^{J}\left(\vect{a}_{1} \cdot \vect{Y}_{J M}^{J+1}\left(\vartheta_{1}, \varphi_{1}\right)\right)^{*}\left(\vect{a}_{2} \cdot \vect{Y}_{J M}^{J-1}\left(\vartheta_{2}, \varphi_{2}\right)\right)=\frac{1}{\sqrt{J(J+1)}}\left\{\left(\vect{a}_{1}^{*} \cdot \vect{n}_{2}\right)\left(\vect{a}_{2} \cdot \vect{n}_{1}\right) P_{J}^{\prime \prime}+\left(\vect{a}_{1}^{*} \cdot \vect{a}_{2}\right) P_{J}^{\prime}\right\}, \notag\\
4 \pi \sum_{M=-J}^{J}\left(\vect{a}_{1} \cdot \vect{Y}_{J M}^{J}\left(\vartheta_{1}, \varphi_{1}\right)\right)^{*}\left(\vect{a}_{2} \cdot \vect{Y}_{J M}^{J+1}\left(\vartheta_{2}, \varphi_{2}\right)\right)  \notag\\
=\frac{i}{J+1} \sqrt{\frac{2 J+1}{J}}\left\{\left[\vect{a}_{1}^{*} \times \vect{a}_{2}\right] \cdot \vect{n}_{1}\left[(J+1) P_{J}^{\prime}-P_{J+1}^{\prime \prime}\right]+\left[\vect{a}_{1}^{*} \times \vect{a}_{2}\right] \cdot \vect{n}_{2} P_{J}^{\prime \prime}\right\}, \notag\\
4 \pi \sum_{M=-J}^{J}\left(\vect{a}_{1} \cdot \vect{Y}_{J M}^{J}\left(\vartheta_{1}, \varphi_{1}\right)\right)^{*}\left(\vect{a}_{2} \cdot \vect{Y}_{J M}^{J}\left(\vartheta_{2}, \varphi_{2}\right)\right)  \label{eq:7:3:110}\\
=\frac{2 J+1}{J(J+1)}\left\{-\left(\vect{a}_{1}^{*} \cdot \vect{n}_{2}\right)\left(\vect{a}_{2} \cdot \vect{n}_{1}\right)\left[P_{J-1}^{\prime \prime}+J P_{J}^{\prime}\right]+\left(\vect{a}_{1}^{*} \cdot \vect{a}_{2}\right)\left[J^{2} P_{J}-P_{J-1}^{\prime}\right]\right\}  \notag\\
4 \pi \sum_{M=-J}^{J}\left(\vect{a}_{1} \cdot \vect{Y}_{J M}^{J}\left(\vartheta_{1}, \varphi_{1}\right)\right)^{*}\left(\vect{a}_{2} \cdot \vect{Y}_{J M}^{J-1}\left(\vartheta_{2}, \varphi_{2}\right)\right)  \notag\\
=\frac{i}{J} \sqrt{\frac{2 J+1}{J+1}}\left\{-\left[\vect{a}_{1}^{*} \times \vect{a}_{2}\right] \cdot \vect{n}_{1}\left[P_{J-1}^{\prime \prime}+J P_{J}^{\prime}\right]+\left[\vect{a}_{1}^{*} \times \vect{a}_{2}\right] \cdot \vect{n}_{2} P_{J}^{\prime \prime}\right\}, \notag\\
4 \pi \sum_{M=-J}^{J}\left(\vect{a}_{1} \cdot \vect{Y}_{J M}^{J-1}\left(\vartheta_{1}, \varphi_{1}\right)\right)^{*}\left(\vect{a}_{2} \cdot \vect{Y}_{J M}^{J+1}\left(\vartheta_{2}, \varphi_{2}\right)\right)=\frac{1}{\sqrt{J(J+1)}}\left\{\left(\vect{a}_{1}^{*} \cdot \vect{n}_{2}\right)\left(\vect{a}_{2} \cdot \vect{n}_{1}\right) P_{J}^{\prime \prime}+\left(\vect{a}_{1}^{*} \cdot \vect{a}_{2}\right) P_{J}^{\prime}\right\}  \notag\\
4 \pi \sum_{M=-J}^{J}\left(\vect{a}_{1} \cdot \vect{Y}_{J M}^{J-1}\left(\vartheta_{1}, \varphi_{1}\right)\right)^{*}\left(\vect{a}_{2} \cdot \vect{Y}_{J M}^{J}\left(\vartheta_{2}, \varphi_{2}\right)\right)  \label{eq:7:3:111}\\
=\frac{i}{J} \sqrt{\frac{2 J+1}{J+1}}\left\{\left[\vect{a}_{1}^{*} \times \vect{a}_{2}\right] \cdot \vect{n}_{1} P_{J}^{\prime \prime}-\left[\vect{a}_{1}^{*} \times \vect{a}_{2}\right] \cdot \vect{n}_{2}\left[P_{J-1}^{\prime \prime}+J P_{J}^{\prime}\right]\right\}, \notag
\end{gather}
\[
4 \pi \sum_{M=-J}^{J}\left(\vect{a}_{1} \cdot \vect{Y}_{J M}^{J-1}\left(\vartheta_{1}, \varphi_{1}\right)\right)^{*}\left(\vect{a}_{2} \cdot \vect{Y}_{J M}^{J-1}\left(\vartheta_{2}, \varphi_{2}\right)\right)=\frac{1}{J}\left\{\left(\vect{a}_{1}^{*} \cdot \vect{n}_{2}\right)\left(\vect{a}_{2} \cdot \vect{n}_{1}\right) P_{J}^{\prime \prime}+\left(\vect{a}_{1}^{*} \cdot \vect{a}_{2}\right) P_{J}^{\prime}\right\}
\]
Note that $P_{J+1}^{\prime \prime}-(J+1) P_{J}^{\prime}=P_{J-1}^{\prime \prime}+J P_{J}^{\prime}$.
\item 
 Analogous relations for $\vect{Y}_{J M}^{(\lambda)}$, when $\vect{a}_{1} \cdot \vect{n}_{1}=\vect{a}_{2} \cdot \vect{n}_{2}=0$, acquire the form
\begin{gather}
4 \pi \sum_{M=-J}^{J}\left(\vect{a}_{1} \cdot \vect{Y}_{J M}^{(1)}\left(\vartheta_{1}, \varphi_{1}\right)\right)^{*}\left(\vect{a}_{2} \cdot \vect{Y}_{J M}^{(1)}\left(\vartheta_{2}, \varphi_{2}\right)\right)=\frac{2 J+1}{J(J+1)}\left\{\left(\vect{a}_{1}^{*} \cdot \vect{n}_{2}\right)\left(\vect{a}_{2} \cdot \vect{n}_{1}\right) P_{J}^{\prime \prime}+\left(\vect{a}_{1}^{*} \cdot \vect{a}_{2}\right) P_{J}^{\prime}\right\}  \notag\\
4 \pi \sum_{M=-J}^{J}\left(\vect{a}_{1} \cdot \vect{Y}_{J M}^{(1)}\left(\vartheta_{1}, \varphi_{1}\right)\right)^{*}\left(\vect{a}_{2} \cdot \vect{Y}_{J M}^{(0)}\left(\vartheta_{2}, \varphi_{2}\right)\right)  \label{eq:7:3:112}\\
=\frac{i}{J(J+1)}\left\{\left[\vect{a}_{1}^{*} \times \vect{a}_{2}\right] \cdot \vect{n}_{1}(2 J+1) P_{J}^{\prime \prime}-\left[\vect{a}_{1}^{*} \times \vect{a}_{2}\right] \cdot \vect{n}_{2}\left[(J+1) P_{J-1}^{\prime \prime}+J P_{J+1}^{\prime \prime}\right]\right\}  \notag\\
4 \pi \sum_{M=-J}^{J}\left(\vect{a}_{1} \cdot \vect{Y}_{J M}^{(0)}\left(\vartheta_{1}, \varphi_{1}\right)\right)^{*}\left(\vect{a}_{2} \cdot \vect{Y}_{J M}^{(1)}\left(\vartheta_{2}, \varphi_{2}\right)\right)  \notag\\
=\frac{i}{J(J+1)}\left\{-\left[\vect{a}_{1}^{*} \times \vect{a}_{2}\right] \cdot \vect{n}_{1}\left[(J+1) P_{J-1}^{\prime \prime}+J P_{J+1}^{\prime \prime}\right]+\left[\vect{a}_{1}^{*} \times \vect{a}_{2}\right] \cdot \vect{n}_{2}(2 J+1) P_{J}^{\prime \prime}\right\}  \notag\\
4 \pi \sum_{M=-J}^{J}\left(\vect{a}_{1} \cdot \vect{Y}_{J M}^{(0)}\left(\vartheta_{1}, \varphi_{1}\right)\right)^{*}\left(\vect{a}_{2} \cdot \vect{Y}_{J M}^{(0)}\left(\vartheta_{2}, \varphi_{2}\right)\right)  \label{eq:7:3:113}\\
=\frac{2 J+1}{J(J+1)}\left\{-\left(\vect{a}_{1}^{*} \cdot \vect{n}_{2}\right)\left(\vect{a}_{2} \cdot \vect{n}_{1}\right)\left[P_{J-1}^{\prime \prime}+J P_{J}^{\prime}\right]+\left(\vect{a}_{1}^{*} \cdot \vect{a}_{2}\right)\left[J^{2} P_{J}-P_{J-1}^{\prime}\right]\right\}. \notag
\end{gather}
The sums involving the longitudinal vectors $\vect{Y}_{J M}^{(-1)}$ are equal to zero, because
\[
\vect{a}_{1} \cdot \vect{Y}_{J M}^{(-1)}\left(\vartheta_{1}, \varphi_{1}\right)=\vect{a}_{2} \cdot \vect{Y}_{J M}^{(-1)}\left(\vartheta_{2}, \varphi_{2}\right)=0 .
\]
\end{enumerate}
\subsection{Integrals Involving Vector Spherical Harmonics}\label{sec:7.3.12}\index{vector spherical harmonics!integrals involving}\index{integrals!of the vector spherical harmonics}
Let us use the notation $\int f(\vartheta, \varphi) d \Omega \equiv \int_{0}^{\pi} d \vartheta \sin \vartheta \int_{0}^{2 \pi} d \varphi f(\vartheta, \varphi)$.
\begin{equation}
\int e^{i k r} \vect{Y}_{J M}^{L}(\vartheta, \varphi) d \Omega=i^{L} 4 \pi j_{L}(k r) \vect{Y}_{J M}^{L}\left(\vartheta_{k}, \varphi_{k}\right) \label{eq:7:3:114}
\end{equation}
where
\begin{gather}
\vect{k}=\left(k, \vartheta_{k}, \varphi_{k}\right), \vect{r}=(r, \vartheta, \varphi), j_{L}(x)=\sqrt{\frac{\pi}{2 x}} J_{L+\frac{1}{2}}(x)  \notag\\
\int e^{i k r} \vect{Y}_{J M}^{(1)}(\vartheta, \varphi) d \Omega=4 \pi i^{J+1}\left\{\sqrt{\frac{J}{2 J+1}} j_{J+1}(k r) \vect{Y}_{J M}^{J+1}\left(\vartheta_{k}, \varphi_{k}\right)-\sqrt{\frac{J+1}{2 J+1}} j_{J-1}(k r) \vect{Y}_{J M}^{J-1}\left(\vartheta_{k}, \varphi_{k}\right)\right\}  \notag\\
=\frac{4 \pi i^{J+1}}{2 J+1}\left\{\left[J \cdot j_{J+1}(k r)-(J+1) j_{J-1}(k r)\right] \vect{Y}_{J M}^{(1)}\left(\vartheta_{k}, \varphi_{k}\right)-\sqrt{J(J+1)}\left[j_{J+1}(k r)+j_{J-1}(k r)\right] \vect{Y}_{J M}^{(-1)}\left(\vartheta_{k}, \varphi_{k}\right)\right\}, \label{eq:7:3:115}\\
\int e^{i k r} \vect{Y}_{J M}^{(0)}(\vartheta, \varphi) d \Omega=4 \pi i^{J} j_{J}(k r) \vect{Y}_{J M}^{J}\left(\vartheta_{k}, \varphi_{k}\right)=4 \pi i^{J} j_{J}(k r) \vect{Y}_{J M}^{(0)}\left(\vartheta_{k}, \varphi_{k}\right), \label{eq:7:3:116}
\end{gather}
\begin{gather}
\int e^{i k r} \vect{Y}_{J M}^{(-1)}(\vartheta, \varphi) d \Omega=4 \pi i^{J-1}\left\{\sqrt{\frac{J+1}{2 J+1}} j_{J+1}(k r) \vect{Y}_{J M}^{J+1}\left(\vartheta_{k}, \varphi_{k}\right)+\sqrt{\frac{J}{2 J+1}} j_{J-1}(k r) \vect{Y}_{J M}^{J-1}\left(\vartheta_{k}, \varphi_{k}\right)\right\}  \notag\\
=\frac{4 \pi i^{J-1}}{2 J+1}\left\{\sqrt{J(J+1)}\left[j_{J+1}(k r)+j_{J-1}(k r)\right] \vect{Y}_{J M}^{(1)}\left(\vartheta_{k}, \varphi_{k}\right)-\left[(J+1) j_{J+1}(k r)-J j_{J-1}(k r)\right] \vect{Y}_{J M}^{(-1)}\left(\vartheta_{k}, \varphi_{k}\right)\right\}  \label{eq:7:3:117}\\
\int \vect{Y}_{J M}^{L}(\vartheta, \varphi) d \Omega=\sqrt{4 \pi} \delta_{J 1} \delta_{L 0} \vect{e}_{M}  \notag\\
\int \vect{Y}_{J M}^{(\lambda)}(\vartheta, \varphi) d \Omega=\sqrt{\frac{4 \pi}{3}} \delta_{J 1}\left(\sqrt{2} \delta_{\lambda 1}+\delta_{\lambda-1}\right) \vect{e}_{M} ;  \label{eq:7:3:118}\\
\int \vect{Y}_{J^{\prime} M^{\prime}}^{L^{\prime} *}(\vartheta, \varphi) \cdot \vect{Y}_{J M}^{L}(\vartheta, \varphi) d \Omega=\delta_{J^{\prime} J} \delta_{L^{\prime} L} \delta_{M^{\prime} M}  \label{eq:7:3:119}\\
\int \vect{Y}_{J^{\prime} M^{\prime}}^{\left(\lambda^{\prime}\right) *}(\vartheta, \varphi) \cdot \vect{Y}_{J M}^{(\lambda)}(\vartheta, \varphi) d \Omega=\delta_{J^{\prime} J} \delta_{\lambda^{\prime} \lambda} \delta_{M^{\prime} M} ;  \notag\\
\int \vect{Y}_{J^{\prime} M^{\prime}}^{L^{\prime} *}(\vartheta, \varphi) \times \vect{Y}_{J M}^{L}(\vartheta, \varphi) d \Omega=i(-1)^{J+L} \delta_{L L^{\prime}} \sqrt{6\left(2 J^{\prime}+1\right)}\sixj{1}{1}{1}{J^{\prime}}{J}{L} \clebsch{J^{\prime}}{M^{\prime}}{1}{m}{J}{M} \vect{e}_{m} . \label{eq:7:3:120}
\end{gather}
\subsection{Orthogonality, Normalization and Completeness}\label{sec:7.3.13}\index{orthogonality!of the vector spherical harmonics}\index{normalization!of the vector spherical harmonics}\index{completeness!of the vector spherical harmonics}
The collection of vector spherical harmonics $\vect{Y}_{J M}^{L}(\vartheta, \varphi)$ with integer nonnegative $J(0 \leq J<\infty), L= J, J \pm 1$ and integer $M(|M| \leq J)$ constitutes a complete orthonormal set of vector functions in the domain of arguments $0 \leq \vartheta \leq \pi, 0 \leq \varphi<2 \pi$. The orthonormality condition for $\vect{Y}_{J M}^{L}(\vartheta, \varphi)$ has the form
\begin{equation}
\int_{0}^{\pi} \int_{0}^{2 \pi} \vect{Y}_{J^{\prime} M^{\prime}}^{L^{\prime}}(\vartheta, \varphi) \vect{Y}_{J M}^{L}(\vartheta, \varphi) \sin \vartheta d \vartheta d \varphi=\delta_{J^{\prime} J} \delta_{L^{\prime} L} \delta_{M^{\prime} M}. \label{eq:7:3:121}
\end{equation}
The completeness condition for spherical components of $\vect{Y}_{J M}^{L}(\vartheta, \varphi)$ may be written as
\begin{equation}
\sum_{J=0}^{\infty} \sum_{L=J-1}^{J+1} \sum_{M=-J}^{J}\left[\vect{Y}_{J M}^{L}(\vartheta, \varphi)\right]_{\mu}\left[\vect{Y}_{J M}^{L}\left(\vartheta^{\prime}, \varphi^{\prime}\right)\right]_{\nu}^{*}=\delta_{\mu \nu} \delta\left(\cos \vartheta-\cos \vartheta^{\prime}\right) \delta\left(\varphi-\varphi^{\prime}\right). \label{eq:7:3:122}
\end{equation}
Similarly, the set of $\vect{Y}_{J M}^{(\lambda)}(\vartheta, \varphi)$ constitutes a complete orthonormal set. The orthonormality property for $\vect{Y}_{J M}^{(\lambda)}(\vartheta, \varphi)$ is the following
\begin{equation}
\int_{0}^{\pi} \int_{0}^{2 \pi} \vect{Y}_{J^{\prime} M^{\prime}}^{\left(\lambda^{\prime}\right) *}(\vartheta, \varphi) \vect{Y}_{J M}^{(\lambda)}(\vartheta, \varphi) \sin \vartheta d \vartheta d \varphi=\delta_{J^{\prime} J} \delta_{\lambda^{\prime} \lambda} \delta_{M^{\prime} M} \label{eq:7:3:123}
\end{equation}
and the completeness relation for spherical components of $\vect{Y}_{J M}^{(\lambda)}(\vartheta, \varphi)$ reads
\begin{equation}
\sum_{J=0}^{\infty} \sum_{\lambda=-1}^{1} \sum_{M=-J}^{J}\left[\vect{Y}_{J M}^{(\lambda)}(\vartheta, \varphi)\right]_{\mu}\left[\vect{Y}_{J M}^{(\lambda)}\left(\vartheta^{\prime}, \varphi^{\prime}\right)\right]_{\nu}^{*}=\delta_{\mu \nu} \delta\left(\cos \vartheta-\cos \vartheta^{\prime}\right) \delta\left(\varphi-\varphi^{\prime}\right). \label{eq:7:3:124}
\end{equation}
\subsection{Expansion in Series of Vector Spherical Harmonics}\label{sec:7.3.14}\index{vector spherical harmonics!expansion in series of}\index{expansion!in the vector spherical harmonics}
Any vector $\vect{F}(\vartheta, \varphi)$ which depends on polar angles $\vartheta, \varphi$ and satisfies the condition
\begin{equation}
\int|\vect{F}(\vartheta, \varphi)|^{2} d \Omega<\infty \label{eq:7:3:125}
\end{equation}
with $\int d \Omega \equiv \int_{0}^{\pi} d \vartheta \sin \vartheta \int_{0}^{2 \pi} d \varphi$, may be expanded in series of vector spherical harmonics $\vect{Y}_{J M}^{L}(\vartheta, \varphi)$ or $\vect{Y}_{J M}^{(\lambda)}(\vartheta, \varphi)$. In order words, for $0 \leq \vartheta \leq \pi, 0 \leq \varphi<2 \pi \vect{F}(\vartheta, \varphi)$ may be written as
\begin{equation}
\vect{F}(\vartheta, \varphi)=\sum_{J L M} A_{J M}^{L} \vect{Y}_{J M}^{L}(\vartheta, \varphi) \label{eq:7:3:126}
\end{equation}
or
\begin{equation}
\vect{F}(\vartheta, \varphi)=\sum_{J \lambda M} A_{J M}^{(\lambda)} \vect{Y}_{J M}^{(\lambda)}(\vartheta, \varphi) . \label{eq:7:3:127}
\end{equation}
The expansion coefficients are given by
\begin{align}
& A_{J M}^{L}=\int d \Omega \vect{F}(\vartheta, \varphi) \cdot \vect{Y}_{J M}^{L *}(\vartheta, \varphi), \notag\\
& A_{J M}^{(\lambda)}=\int d \Omega \vect{F}(\vartheta, \varphi) \cdot \vect{Y}_{J M}^{(\lambda) *}(\vartheta, \varphi) . \label{eq:7:3:128}
\end{align}
Note also the following useful rule. If a scalar function $\Phi\left(\vect{r}_{1}, \vect{r}_{2}\right)$ of vector arguments $\vect{r}_{1}\left(r_{1}, \vartheta_{1}, \varphi_{1}\right)$ and $r_{2}\left(r_{2}, \vartheta_{2}, \varphi_{2}\right)$ may be expanded in a series of scalar spherical harmonics as
\begin{equation}
\Phi\left(\vect{r}_{1}, \vect{r}_{2}\right)=\sum_{L m} A_{L}\left(r_{1}, r_{2}\right) Y_{L m}^{*}\left(\vartheta_{2}, \varphi_{2}\right) Y_{L m}\left(\vartheta_{1}, \varphi_{1}\right) \label{eq:7:3:129}
\end{equation}
then the expansion of the vector function $\vect{F}\left(\vect{r}_{2}\right) \Phi\left(\vect{r}_{1}, \vect{r}_{2}\right)$ in a series of $\vect{Y}_{J M}^{L}\left(\vartheta_{1}, \varphi_{1}\right)$ has the form
\begin{equation}
\vect{F}\left(\vect{r}_{2}\right) \Phi\left(\vect{r}_{1}, \vect{r}_{2}\right)=\sum_{J L M} A_{L}\left(r_{1}, r_{2}\right)\left[\vect{F}\left(\vect{r}_{2}\right) \cdot \vect{Y}_{J M}^{L *}\left(\vartheta_{2}, \varphi_{2}\right)\right] \vect{Y}_{J M}^{L}\left(\vartheta_{1}, \varphi_{1}\right). \label{eq:7:3:130}
\end{equation}
Some examples of such expansions are given below.\\
\subsubsection{Expansion of a plane wave}
\begin{equation}
\varepsilon(\vect{k}) e^{i \vect{k}\vect{r}}=4 \pi \sum_{J L M} i^{L} j_{L}(k r)\left\{\varepsilon(\vect{k}) \cdot \vect{Y}_{J M}^{L *}\left(\vartheta_{k}, \varphi_{k}\right)\right\} \vect{Y}_{J M}^{L}(\vartheta, \varphi) \label{eq:7:3:131}
\end{equation}
where $j_{L}(x)=\sqrt{\pi /(2 x)} J_{L+\frac{1}{2}}(x)$ is a spherical Bessel function, $\vect{k}=\left(k, \vartheta_{k}, \varphi_{k}\right), \vect{r}=(r, \vartheta, \varphi)$. The expansion in a series of $\vect{Y}_{J M}^{(\lambda)}(\vartheta, \varphi)$ has the form
\begin{equation}
\varepsilon(\vect{k}) e^{i \vect{k}\vect{r}}=\sum_{J \lambda M} A_{J M}^{(\lambda)} \vect{Y}_{J M}^{(\lambda)}(\vartheta, \varphi) \label{eq:7:3:132}
\end{equation}
where the expansion coefficients are given by
\begin{align}
A_{J M}^{(1)}=4 \pi i^{J+1} & \left\{\sqrt{\frac{J}{2 J+1}} j_{J+1}(k r) \varepsilon(\vect{k}) \cdot \vect{Y}_{J M}^{J+1 *}\left(\vartheta_{k}, \varphi_{k}\right)-\sqrt{\frac{J+1}{2 J+1}} j_{J-1}(k r) \varepsilon(\vect{k}) \cdot \vect{Y}_{J M}^{J-1 *}\left(\vartheta_{k}, \varphi_{k}\right)\right\}  \notag\\
= & \frac{4 \pi i^{J+1}}{2 J+1}\left\{\left[J \cdot j_{J+1}(k r)-(J+1) j_{J-1}(k r)\right] \varepsilon(\vect{k}) \cdot \vect{Y}_{J M}^{(1) *}\left(\vartheta_{k}, \varphi_{k}\right)\right. \label{eq:7:3:133}\\
& \left.-\sqrt{J(J+1)}\left[j_{J+1}(k r)+j_{J-1}(k r)\right] \varepsilon(\vect{k}) \cdot Y_{J M}^{(-1) *}\left(\vartheta_{k}, \varphi_{k}\right)\right\}. \notag
\end{align}
\begin{gather}
A_{J M}^{(0)}=4 \pi i^{J} j_{J}(k r) \vect{\epsilon}(\vect{k}) \cdot \vect{Y}_{J M}^{J *}\left(\vartheta_{k}, \varphi_{k}\right)=4 \pi i^{J} j_{J}(k r) \cdot \vect{\epsilon}(\vect{k}) \cdot \vect{Y}_{J M}^{(0) *}\left(\vartheta_{k}, \varphi_{k}\right)  \label{eq:7:3:134}\\
A_{J M}^{(-1)}=4 \pi i^{J-1}\left\{\sqrt{\frac{J+1}{2 J+1}} j_{J+1}(k r) \vect{\varepsilon}(\vect{k}) \cdot \vect{Y}_{J M}^{J+1 *}\left(\vartheta_{k}, \varphi_{k}\right)+\sqrt{\frac{J}{2 J+1}} j_{J-1}(k r) \vect{\varepsilon}(\vect{k}) \cdot \vect{Y}_{J M}^{J-1 *}\left(\vartheta_{k}, \varphi_{k}\right)\right\}  \notag\\
=\frac{4 \pi i^{J-1}}{2 J+1}\left\{\sqrt{J(J+1)}\left[j_{J+1}(k r)+j_{J-1}(k r)\right] \vect{\varepsilon}(\vect{k}) \cdot \vect{Y}_{J M}^{(1) *}\left(\vartheta_{k}, \varphi_{k}\right)\right. \notag\\
\left.-\left[(J+1) j_{J+1}(k r)-J \cdot j_{J-1}(k r)\right] \vect{\varepsilon}(\vect{k}) \cdot Y_{J M}^{(-1) *}\left(\vartheta_{k}, \varphi_{k}\right)\right\}. \label{eq:7:3:135}
\end{gather}
If the plane wave is transverse, i.e., $\vect{k} \cdot \vect{\varepsilon}(\vect{k})=0$, then $\vect{\varepsilon}(\vect{k}) \cdot \vect{Y}_{J M}^{(-1) *}\left(\vartheta_{k}, \varphi_{k}\right)=0$.
\subsubsection{Green's function for Laplace equation}
\begin{equation}
\vect{J}(\vect{R}) \frac{1}{|\vect{R}-\vect{r}|}=4 \pi \sum_{J L M} A_{L}(R, r) \frac{1}{(2 L+1)}\left\{\vect{J}(\vect{R}) \cdot \vect{Y}_{J M}^{L *}(\Theta, \Phi)\right\} \vect{Y}_{J M}^{L}(\vartheta, \varphi) \label{eq:7:3:136}
\end{equation}
where $\vect{r}=(r, \vartheta, \varphi), \vect{R}=(R, \Theta, \Phi)$,
\begin{equation}
A_{L}(R, r)= \begin{cases}\frac{r^{L}}{R^{L+1}}, & \text { if } r<R  ,
    \\ \frac{R^{L}}{r^{L+1}}, & \text { if } r>R.\end{cases} \label{eq:7:3:137}
\end{equation}
\subsubsection{Green's function for Helmholtz equation}
\begin{equation}
\vect{J}(\vect{R}) \frac{e^{i k|\vect{R}-\vect{r}|}}{|\vect{R}-\vect{r}|}=4 \pi i k \sum_{J L M} A_{L}(R, r)\left\{\vect{J}(\vect{R}) \vect{Y}_{J M}^{L *}(\Theta, \Phi)\right\} \vect{Y}_{J M}^{L}(\vartheta, \varphi) \label{eq:7:3:138}
\end{equation}
where
\begin{equation}
A_{L}(R, r)= \begin{cases}j_{L}(k r) h_{L}^{(1)}(k R), & \text { if } r<R,
    \\ h_{L}^{(1)}(k r) j_{L}(k R), & \text { if } r>R.\end{cases}
    \label{eq:7:3:139}
\end{equation}
$h_{L}^{(1)}(x)=\sqrt{\pi /(2 x)} H_{L+\frac{1}{2}}^{(1)}(x)$ is the spherical Hankel function.

\subsection{Vector Spherical Harmonics for $\vartheta=0$ or $\vartheta=\pi$}\label{sec:7.3.15}\index{vector spherical harmonics!at $\vartheta=0$ and $\vartheta=\pi$}
If $\vartheta=0$ or $\vartheta=\pi$, the vector spherical harmonics $\vect{Y}_{J M}^{L}(\vartheta, \varphi)$ and $\vect{Y}_{J M}^{(\lambda)}(\vartheta, \varphi)$ are equal to zero unless $M=0, \pm 1$. The expressions for these harmonics in terms of spherical basis vectors $\vect{e}_{\mu}(\mu=0, \pm 1)$ have the\\
form
\begin{gather}
\vect{Y}_{J M}^{J+1}(0, \varphi)=(-1)^{J-1} \vect{Y}_{J M}^{J+1}(\pi, \varphi)= \begin{cases}\sqrt{\frac{J}{8 \pi}} \vect{e}_{M}, & \text { if } M= \pm 1 ; \\
-\sqrt{\frac{J+1}{4 \pi}} \vect{e}_{0}, & \text { if } M=0\end{cases}  \label{eq:7:3:140}\\
\vect{Y}_{J M}^{J}(0, \varphi)=(-1)^{J} \vect{Y}_{J M}^{J}(\pi, \varphi)= \begin{cases}-M \sqrt{\frac{2 J+1}{8 \pi}} \vect{e}_{M}, & \text { if } M= \pm 1 ; \\
0, & \text { otherwise }\end{cases}  \label{eq:7:3:141}\\
\vect{Y}_{J M}^{J-1}(0, \varphi)=(-1)^{J-1} \vect{Y}_{J M}^{J-1}(\pi, \varphi)= \begin{cases}\sqrt{\frac{J+1}{8 \pi}} \vect{e}_{M}, & \text { if } M= \pm 1 ; \\
\sqrt{\frac{J}{4 \pi}} \vect{e}_{0}, & \text { if } M=0\end{cases}  \label{eq:7:3:142}\\
\vect{Y}_{J M}^{(1)}(0, \varphi)=(-1)^{J-1} \vect{Y}_{J M}^{(1)}(\pi, \varphi)= \begin{cases}\sqrt{\frac{2 J+1}{8 \pi}} \vect{e}_{M}, & \text { if } M= \pm 1 ; \\
0, & \text { otherwise }\end{cases}  \label{eq:7:3:143}\\
\vect{Y}_{J M}^{(0)}(0, \varphi)=(-1)^{J} \vect{Y}_{J M}^{(0)}(\pi, \varphi)= \begin{cases}-M \sqrt{\frac{2 J+1}{8 \pi}} \vect{e}_{M}, & \text { if } M= \pm 1 \\
0, & \text { otherwise }\end{cases}  \label{eq:7:3:144}\\
\vect{Y}_{J M}^{(-1)}(0, \varphi)=(-1)^{J-1} \vect{Y}_{J M}^{(-1)}(\pi, \varphi)= \begin{cases}\sqrt{\frac{2 J+1}{4 \pi}} \vect{e}_{0}, & \text { if } M=0 ; \\
0, & \text { otherwise }.\end{cases} \label{eq:7:3:145}
\end{gather}
\subsection{Vector Spherical Harmonics at $J=0,1$}\label{sec:7.3.16}\index{vector spherical harmonics!at $J=0,1$}
If $J=0$ or $J=1$, the vector spherical harmonics $\vect{Y}_{J M}^{L}(\vartheta, \varphi)$ may be expressed in terms of spherical basis vectors $\vect{e}_{\mu}(\mu=0, \pm 1)$ and the unit vector $\vect{n}(\vartheta, \varphi)$ as
\begin{align}
\vect{Y}_{00}^{1}(\vartheta, \varphi) & =-\frac{1}{\sqrt{4 \pi}} \vect{n}  \label{eq:7:3:146}\\
\vect{Y}_{1 M}^{0}(\vartheta, \varphi) & =\frac{1}{\sqrt{4 \pi}} \vect{e}_{M}  \label{eq:7:3:147}\\
\vect{Y}_{1 M}^{1}(\vartheta, \varphi) & =i \sqrt{\frac{3}{8 \pi}}\left[\vect{e}_{M} \times \vect{n}\right]  \label{eq:7:3:148}\\
\vect{Y}_{1 M}^{2}(\vartheta, \varphi) & =\frac{1}{\sqrt{8 \pi}}\left\{\vect{e}_{M}-3 \vect{n}\left(\vect{e}_{M} \cdot \vect{n}\right)\right\}. \label{eq:7:3:149}
\end{align}
Analogous expressions for $\vect{Y}_{J M}^{(\lambda)}(\vartheta, \varphi)$ are
\begin{align}
\vect{Y}_{00}^{(-1)}(\vartheta, \varphi) & =\frac{1}{\sqrt{4 \pi}} \vect{n}  \label{eq:7:3:150}\\
\vect{Y}_{1 M}^{(1)}(\vartheta, \varphi) & =\sqrt{\frac{3}{8 \pi}}\left\{\vect{e}_{M}-\vect{n}\left(\vect{e}_{M} \cdot \vect{n}\right)\right\}  \label{eq:7:3:151}\\
\vect{Y}_{1 M}^{(0)}(\vartheta, \varphi) & =i \sqrt{\frac{3}{8 \pi}}\left[\vect{e}_{M} \times \vect{n}\right]  \label{eq:7:3:152}\\
\vect{Y}_{1 M}^{(-1)}(\vartheta, \varphi) & =\sqrt{\frac{3}{4 \pi}} \vect{n}\left(\vect{e}_{M} \cdot \vect{n}\right). \label{eq:7:3:153}
\end{align}
\subsection{Quadratic Forms of the Vector Spherical Harmonics}\label{sec:7.3.17}\index{vector spherical harmonics!quadratic forms of}\index{quadratic forms!of the vector spherical harmonics}
The functions $\left|\vect{Y}_{J M}^{(\lambda)}(\vartheta, \varphi)\right|^{2}$ are important for physical applications.\index{angular distribution}\isym{WJM-perp@$W_{J M}^{\perp}(\vartheta)$, quadratic form of the vector spherical harmonics} In particular, they describe the angular distribution of spin-1 particles in quantum states with total angular momentum $J$. These functions are independent of $\varphi$ and are the same at $\lambda=1$ and $\lambda=0$. Let us use the notation
\begin{align}
& W_{J M}^{\perp}(\vartheta) \equiv\left|\vect{Y}_{J M}^{(1)}(\vartheta, \varphi)\right|^{2}=\left|\vect{Y}_{J M}^{(0)}(\vartheta, \varphi)\right|^{2}  \label{eq:7:3:154}\\
& W_{J M}^{\|}(\vartheta) \equiv\left|\vect{Y}_{J M}^{(-1)}(\vartheta, \varphi)\right|^{2}. \notag
\end{align}
Then one has
\begin{align}
W_{J M}^{\perp}(\vartheta)= & \frac{1}{2 J(J+1)}\left\{(J-M)(J+M+1)\left|Y_{J M+1}(\vartheta, \varphi)\right|^{2}+2 M^{2}\left|Y_{J M}(\vartheta, \varphi)\right|^{2}\right. \notag\\
& \left.\quad+(J+M)(J-M+1)\left|Y_{J M-1}(\vartheta, \varphi)\right|^{2}\right\}  \label{eq:7:3:155}\\
W_{J M}^{\|}(\vartheta)= & \left|Y_{J M}(\vartheta, \varphi)\right|^{2}. \notag
\end{align}
The expansions of $W_{J M}^{\perp, \|}(\vartheta)$ in terms of the Legendre polynomials have the form
\begin{align}
& W_{J M}^{\perp}(\vartheta)=\sum_{n=0}^{J} a_{n}(J, M) P_{2 n}(\cos \vartheta)  \notag\\
& W_{J M}^{\|}(\vartheta)=\sum_{n=0}^{J} b_{n}(J, M) P_{2 n}(\cos \vartheta) \label{eq:7:3:156}
\end{align}
where
\begin{align}
a_{n}(J, M) & =-\frac{(2 J+1)(4 n+1)}{4 \pi}\sixj{J}{J}{2 n}{J}{J}{1} \clebsch{J}{0}{2n}{0}{J}{0} \clebsch{J}{M}{2n}{0}{J}{M}  \notag\\
& =(-1)^{n} \frac{4 n+1}{4 \pi} \frac{J(J+1)-n(2 n+1)}{J(J+1)} \frac{(J+n)!(2 n)!}{(J-n)!(n!)^{2}} \sqrt{\frac{(2 J+1)(2 J-2 n)!}{(2 J+2 n+1)!}} \clebsch{J}{M}{2n}{0}{J}{M}, \label{eq:7:3:157}\\
b_{n}(J, M) & =\frac{4 n+1}{4 \pi} \clebsch{J}{0}{2n}{0}{J}{0} \clebsch{J}{M}{2n}{0}{J}{M}=(-1)^{n} \frac{4 n+1}{4 \pi} \frac{(J+n)!}{(J-n)!} \frac{(2 n)!}{(n!)^{2}} \sqrt{\frac{(2 J+1)(2 J-2 n)!}{(2 J+2 n+1)!}} \clebsch{J}{M}{2n}{0}{J}{M} . \label{eq:7:3:158}
\end{align}
In particular,
\begin{align}
& a_{0}(J, M)=b_{0}(J, M)=\frac{1}{4 \pi}  \label{eq:7:3:159}\\
& a_{1}(J, M)=\frac{5}{4 \pi} \frac{[J(J+1)-3]\left[J(J+1)-3 M^{2}\right]}{J(J+1)(2 J-1)(2 J+3)}  \label{eq:7:3:160}\\
& b_{1}(J, M)=\frac{5}{4 \pi} \frac{J(J+1)-3 M^{2}}{(2 J-1)(2 J+3)}  \label{eq:7:3:161}\\
& a_{2}(J, M)=\frac{27}{16 \pi} \frac{[J(J+1)-10]}{J(J+1)} \cdot \frac{\left[3\left(J^{2}+2 J-5 M^{2}\right)\left(J^{2}-5 M^{2}-1\right)-10 M^{2}\left(4 M^{2}-1\right)\right]}{(2 J-3)(2 J-1)(2 J+3)(2 J+5)}  \label{eq:7:3:162}\\
& b_{2}(J, M)=\frac{27}{16 \pi} \frac{\left[3\left(J^{2}+2 J-5 M^{2}\right)\left(J^{2}-5 M^{2}-1\right)-10 M^{2}\left(4 M^{2}-1\right)\right]}{(2 J-3)(2 J-1)(2 J+3)(2 J+5)}. \label{eq:7:3:163}
\end{align}
The functions $W_{J M}^{\perp \|}(\vartheta)$ satisfy the normalization conditions
\begin{gather}
\int W_{J M}^{\perp}(\vartheta) d \Omega=\int W_{J M}^{\|}(\vartheta) d \Omega=1  \notag\\
\sum_{M=-J}^{J} W_{J M}^{\perp}(\vartheta)=\sum_{M=-J}^{J} W_{J M}^{\|}(\vartheta)=\frac{2 J+1}{4 \pi}. \label{eq:7:3:164}
\end{gather}
Note the following symmetry properties of $W_{J M}^{\perp \|}(\vartheta)$
\begin{align}
& W_{J M}^{\perp}(\vartheta)=W_{J-M}^{\perp}(\vartheta)=W_{J M}^{\perp}(\pi-\vartheta)=W_{J-M}^{\perp}(\pi-\vartheta)  \notag\\
& W_{J M}^{\|}(\vartheta)=W_{J-M}^{\|}(\vartheta)=W_{J M}^{\|}(\pi-\vartheta)=W_{J-M}^{\|}(\pi-\vartheta). \label{eq:7:3:165}
\end{align}
For $\vartheta=0$ and $\vartheta=\pi$ we have
\begin{align}
& W_{J M}^{\perp}(0)=W_{J M}^{\perp}(\pi)= \begin{cases}\frac{2 J+1}{8 \pi}, & \text { if } M= \pm 1 \\
0, & \text { otherwise }\end{cases}  \label{eq:7:3:166}\\
& W_{J M}^{\|}(0)=W_{J M}^{\|}(\pi)= \begin{cases}\frac{2 J+1}{4 \pi}, & \text { if } M=0 \\
0, & \text { otherwise }\end{cases}. \label{eq:7:3:167}
\end{align}
The explicit forms of $W_{J M}^{\perp \|}(\vartheta)$ for $J=0,1,2,3,4,5(0 \leq M \leq J)$ are given in Table~\ref{chap7:tab:2}. For negative $M$ one may use the relation $W_{J-M}=W_{J M}$. Note also the explicit form of $W_{J M}^{\perp, \|}(\vartheta)$ for special values of $M$ :
\begin{gather}
W_{J 0}^{\perp}(\vartheta)=\frac{2 J+1}{4 \pi(J+1) J} \sin ^{2} \vartheta\left[P_{J}^{\prime}(\cos \vartheta)\right]^{2}  \label{eq:7:3:168}\\
W_{J 0}^{\|}(\vartheta)=\frac{2 J+1}{4 \pi}\left[P_{J}(\cos \vartheta)\right]^{2}  \notag\\
W_{J J}^{\perp}(\vartheta)=W_{J-J}^{\perp}(\vartheta)=\frac{1}{4 \pi} \frac{(2 J+1)!(\sin \vartheta)^{2 J-2}}{2^{2 J}(J+1)!(J-1)!}\left(1+\cos ^{2} \vartheta\right)  \notag\\
W_{J J}^{\|}(\vartheta)=W_{J-J}^{\|}(\vartheta)=\frac{1}{4 \pi} \frac{(2 J+1)!}{2^{2 J}(J!)^{2}}(\sin \vartheta)^{2 J}. \label{eq:7:3:169}
\end{gather}
\begin{table}[tbh!]
\begin{center}
\caption{Explicit forms of $W_{J M}^{\perp}(\vartheta)$ and $W_{J M}^{\|}(\vartheta)$.}
\label{chap7:tab:2}
\renewcommand{\arraystretch}{1.8}
\begin{tabular}{l|l|c|c}
\hline
J & M & $W_{J M}^{\perp}(\vartheta)$ & $W_{J M}^{\|}(\vartheta)$ \\
\hline
0 & 0 & $1 / 4 \pi$ & $1 / 4 \pi$ \\
1 & 0 & $\frac{3}{8 \pi} \sin ^{2} \vartheta$ & $\frac{3}{4 \pi} \cos ^{2} \vartheta$ \\
1 & 1 & $\frac{3}{16 \pi}\left(1+\cos ^{2} \vartheta\right)$ & $\frac{3}{8 \pi} \sin^2\vartheta$ \\
2 & 0 & $\frac{15}{8 \pi} \sin ^{2} \vartheta \cos ^{2} \vartheta$ & $\frac{5}{16 \pi}\left(1-3 \cos ^{2} \vartheta\right)^{2}$ \\
2 & 1 & $\frac{5}{16 \pi}\left(1-3 \cos ^{2} \vartheta+4 \cos ^{4} \vartheta\right)$ & $\frac{15}{8 \pi} \sin ^{2} \vartheta \cos ^{2} \vartheta$ \\
2 & 2 & $\frac{5}{16 \pi}\left(1-\cos ^{4} \vartheta\right)$ & $\frac{15}{32 \pi} \sin ^{4} \vartheta$ \\
3 & 0 & $\frac{21}{64 \pi} \sin ^{2} \vartheta\left(1-5 \cos ^{2} \vartheta\right)^{2}$ & $\frac{7}{16 \pi} \cos ^{2} \vartheta\left(3-5 \cos ^{2} \vartheta\right)^{2}$ \\
3 & 1 & $\frac{7}{256 \pi}\left(1+111 \cos ^{2} \vartheta-305 \cos ^{4} \vartheta+225 \cos ^{6} \vartheta\right)$ & $\frac{21}{64 \pi} \sin ^{2} \vartheta\left(1-5 \cos ^{2} \vartheta\right)^{2}$ \\
3 & 2 & $\frac{35}{128 \pi} \sin ^{2} \vartheta\left(1-2 \cos ^{2} \vartheta+9 \cos ^{4} \vartheta\right)$ & $\frac{105}{32 \pi} \sin ^{4} \vartheta \cos ^{2} \vartheta$ \\
3 & 3 & $\frac{105}{256 \pi} \sin ^{4} \vartheta\left(1+\cos ^{2} \vartheta\right)$ & $\frac{35}{64 \pi} \sin ^{6}\vartheta$ \\
4 & 0 & $\frac{45}{64 \pi} \sin ^{2} \vartheta \cos ^{2} \vartheta\left(3-7 \cos ^{2} \vartheta\right)^{2}$ & $\frac{9}{256 \pi}\left(3-30 \cos ^{2} \vartheta+35 \cos ^{4} \vartheta\right)^{2}$ \\
4 & 1 & $\frac{9}{256 \pi}\left(9-153 \cos ^{2} \vartheta+855 \cos ^{4} \vartheta-1463 \cos ^{6} \vartheta+784 \cos ^{8} \vartheta\right)$ & $\frac{45}{64 \pi} \sin ^{2} \vartheta \cos ^{2} \vartheta\left(3-7 \cos ^{2} \vartheta\right)^{2}$ \\
4 & 2 & $\frac{9}{128 \pi} \sin ^{2} \vartheta\left(1+50 \cos ^{2} \vartheta-175 \cos ^{4} \vartheta+196 \cos ^{6} \vartheta\right)$ & $\frac{45}{128 \pi} \sin ^{4} \vartheta\left(1-7 \cos ^{2} \vartheta\right)^{2}$ \\
4 & 3 & $\frac{63}{256 \pi} \sin ^{4} \vartheta\left(1+\cos ^{2} \vartheta+16 \cos ^{4} \vartheta\right)$ & $\frac{315}{64 \pi} \sin ^{6} \vartheta \cos ^{2} \vartheta$ \\
4 & 4 & $\frac{63}{128 \pi} \sin ^{6} \vartheta\left(1+\cos ^{2} \vartheta\right)$ & $\frac{315}{512 \pi} \sin ^{8} \vartheta$ \\
5 & 0 & $\frac{165}{512 \pi} \sin ^{2} \vartheta\left(1-14 \cos ^{2} \vartheta+21 \cos ^{4} \vartheta\right)^{2}$ & $\frac{11}{256 \pi} \cos ^{2} \vartheta\left(15-70 \cos ^{2} \vartheta+63 \cos ^{4} \vartheta\right)^{2}$ \\
5 & 1 & \( \begin{gathered} \frac{11}{1024 \pi}\left(1+813 \cos ^{2} \vartheta-7070 \cos ^{4} \vartheta+21378 \cos ^{6} \vartheta-\right. \\ \left.-26019 \cos ^{8} \vartheta+11025 \cos ^{10} \vartheta\right) \end{gathered} \) & $\frac{165}{512 \pi} \sin ^{2} \vartheta\left(1-14 \cos ^{2} \vartheta+21 \cos ^{4} \vartheta\right)^{2}$ \\
5 & 2 & $\frac{77}{256 \pi} \sin ^{2} \vartheta\left(1-20 \cos ^{2} \vartheta+150 \cos ^{4} \vartheta-324 \cos ^{6} \vartheta+225 \cos ^{8} \vartheta\right)$ & $\frac{1155}{128 \pi} \sin ^{4} \vartheta \cos ^{2} \vartheta\left(1-3 \cos ^{2} \vartheta\right)^{2}$ \\
5 & 3 & $\frac{231}{2048 \pi} \sin ^{4} \vartheta\left(1+31 \cos ^{2} \vartheta-129 \cos ^{4} \vartheta+225 \cos ^{6} \vartheta\right)$ & $\frac{385}{1024 \pi} \sin ^{6} \vartheta\left(1-9 \cos ^{2} \vartheta\right)^{2}$ \\
5 & 4 & $\frac{231}{1024 \pi} \sin ^{6} \vartheta\left(1+6 \cos ^{2} \vartheta+25 \cos ^{4} \vartheta\right)$ & $\frac{3465}{512 \pi} \sin ^{8} \vartheta \cos ^{2} \vartheta$ \\
5 & 5 & $\frac{1155}{2048 \pi} \sin ^{8} \vartheta\left(1+\cos ^{2} \vartheta\right)$ & $\frac{693}{1024 \pi} \sin ^{10} \vartheta$ \\
\end{tabular}
\end{center}
\end{table}

\section{Other notations for tensor spherical harmonics}\label{sec:7.4}\index{tensor spherical harmonics!notations of other authors}\index{notation!of other authors}
The spherical harmonics defined by other authors are presented on the left-hand sides of the equalities given below. On the right-hand sides of these equalities we exhibit the corresponding harmonics in our notation.

\subsubsection{Tensor spherical harmonics}
Berestetskii, Dolginov and Ter-Martirosian \cite{ref055}
\[
Y_{l m}^{L \lambda}=\eta Y_{l m}^{l+\lambda L}, \quad \text { where } \eta= \begin{cases}(-1)^{L} & \text { for integer } L \\ 1 & \text { for half-integer } L .\end{cases}
\]
Newton \cite{ref028}
\[
\mathcal{U}_{j l S}^{M}=i^{l+S} Y_{j M}^{l S}
\]
\subsubsection{Spinor spherical harmonics}
Akhiezer and Berestetskii \cite{ref002}: $\Omega_{j l m}=\Omega_{j m}^{l}$;\\[0pt]
Berestetskii, Dolginov and Ter-Martirosian \cite{ref055}: $Y_{j \mu}^{(\lambda)}=\Omega_{j \mu}^{j+\lambda}$;\\[0pt]
Berestetskii, Lifshitz and Pitaevsky \cite{ref006}: $\Omega_{j l m}=i^{l} \Omega_{j m}^{l}$.

\subsubsection{Vector spherical harmonics}
Akhiezer and Berestetskii \cite{ref002}: $\vect{Y}_{J M}^{(\lambda)}=\vect{Y}_{J M}^{(\lambda)}, \vect{Y}_{J L M}=\vect{Y}_{J M}^{L}$;\\[0pt]
Blatt and Weisskopf \cite{ref007}; Rose \cite{ref030}: $\mathrm{T}_{L \lambda}^{M}=(-1)^{\lambda+1-L} \vect{Y}_{L M}^{(\lambda)}$;\\[0pt]
Berestetskii, Dolginov and Ter-Martirosian \cite{ref055}: $\vect{Y}_{l m}^{(\lambda)}=-\vect{Y}_{l m}^{l+\lambda}$.\\[0pt]
Jackson \cite{ref023}: $\vect{X}_{l m}=\vect{Y}_{l m}^{(0)}$.\\[0pt]
Newton \cite{ref028}: $\vect{Y}_{J M}^{(e)}=i^{J} \vect{Y}_{J M}^{(1)}, \vect{Y}_{J M}^{(m)}=i^{J+1} \vect{Y}_{J M}^{(0)}, \vect{Y}_{J M}^{(0)}=i^{J} \vect{Y}_{J M}^{(-1)}, \vect{Y}_{J l}^{M}=i^{l+1} \vect{Y}_{J M}^{l}$.\\[0pt]
Berestetskii, Lifshitz and Pitaevskii \cite{ref006}: $\vect{Y}_{J M}^{(e)}=i^{J} \vect{Y}_{J M}^{(1)}, \vect{Y}_{J M}^{(M)}=i^{J+1} \vect{Y}_{J M}^{(0)}, \vect{Y}_{J M}^{(l)}=i^{J} \vect{Y}_{J M}^{(-1)}$.

\chapter{CLEBSCH-GORDAN COEFFICIENTS AND $3 j m$ SYMBOLS}\label{chap:8}\index{Clebsch-Gordan coefficient}\index{3jm symbol@$3jm$ symbol}
\section{DEFINITION}\label{sec:8.1}\index{vector-addition coefficients!definition}
The Clebsch-Gordan coefficients are vector addition coefficients.\index{vector-addition coefficients}\index{rotation group!irreducible representations of}\index{3nj symbols@$3nj$ symbols} They play an important role in the decomposition of reducible representations of rotation group into irreducible representations. All recoupling coefficients or $3 n j$ symbols can be determined as the sums of products of the Clebsch-Gordan coefficients. The aforesaid explains the extensive application of these coefficients in the quantum theory of angular momentum.

\subsection{The Clebsch-Gordan Coefficients}\label{sec:8.1.1}\index{Clebsch-Gordan coefficient!definition}
Let $\vect{j}_{1}$ and $\vect{j}_{2}$ be two angular momenta with projections $m_{1}$ and $m_{2}$ on the quantization axis. A ClebschGordan coefficients represents the probability amplitude that\index{probability amplitude}\isym{C-clebsch@$\clebsch{a}{\alpha}{b}{\beta}{c}{\gamma}$, Clebsch-Gordan coefficient} $\mathrm{j}_{1}$ and $\mathrm{j}_{2}$ are coupled into a resultant angular momentum $\vect{j}$ with projection $m$. In accordance with the vector addition rules $\vect{j}_{1}+\vect{j}_{2}=\vect{j}$, the Clebsch-Gordan coefficient vanishes unless the triangular conditions (triangular inequalities) are fulfilled\index{triangle condition}\index{triangular inequalities|see{triangle condition}}, i.e.,
\begin{equation}
\left|j_{1}-j_{2}\right| \leq j \leq j_{1}+j_{2}, \label{eq:8:1:1}
\end{equation}
and the requirement
\begin{equation}
m_{1}+m_{2}=m, \label{eq:8:1:2}
\end{equation}
is satisfied. It will also be assumed that arguments of the Clebsch-Gordan coefficients satisfy the following conditions:
\begin{enumerate}[label=(\alph*)]
\item $j_{1}, j_{2}, j$ are integer or half-integer non-negative numbers; \footnote{The extension of the Clebsch-Gordan coefficients to negative values of momenta will be considered in Sec.~\ref{sec:8.4.5}.
}
\item $m_{1}, m_{2}, m$ are integer or half-integer (positive or negative) numbers;
\item $\left|m_{1}\right| \leq j_{1},\left|m_{2}\right| \leq j_{2},|m| \leq j$;
\item $j_{1}+m_{1}, j_{2}+m_{2}, j+m, j_{1}+j_{2}+j$ are integer non-negative numbers.
\end{enumerate}
%no eq. 3 in book! conditions (a)-(d) above occupy equation number 8.1.3
\refstepcounter{equation}\label{eq:8:1:3}
The absolute value of a Clebsch-Gordan coefficient is given by
\begin{equation}
\left[\clebsch{j_{1}}{m_{1}}{j_{2}}{m_{2}}{j}{m}\right]^{2}=\frac{2 j+1}{8 \pi} \int_{0}^{2 \pi} d \alpha \int_{0}^{\pi} d \beta \sin \beta \int_{0}^{2 \pi} d \gamma D_{m_{1} m_{1}}^{j_{1}}(\alpha, \beta, \gamma) D_{m_{2} m_{2}}^{j_{2}}(\alpha, \beta, \gamma) D_{m m}^{j *}(\alpha, \beta, \gamma). \label{eq:8:1:4}
\end{equation}
The phase of the Clebsch-Gordan coefficients may be chosen in different ways. The phase convention proposed by Condon and Shortley \cite{ref010} is universally accepted.\index{Condon-Shortley phase convention}\index{phase!of the Clebsch-Gordan coefficients} In accordance with this convention the Clebsch-Gordan

coefficients are real and their phases are fixed by the additional relation
\begin{equation}
\clebsch{j_{1}}{m_{1}}{j_{2}}{m_{2}}{j}{m} \clebsch{j_{1}}{j_{1}}{j_{2}}{-j_{2}}{j}{j_{1}-j_{2}}=\frac{2 j+1}{8 \pi} \int_{0}^{2 \pi} d \alpha \int_{0}^{\pi} d \beta \sin \beta \int_{0}^{2 \pi} d \gamma D_{m_{1} j_{1}}^{j_{1}}(\alpha, \beta, \gamma) D_{m_{2}-j_{2}}^{j_{2}}(\alpha, \beta, \gamma) D_{m j_{1}-j_{2}}^{j *}(\alpha, \beta, \gamma), \label{eq:8:1:5}
\end{equation}
where
\[
\clebsch{j_{1}}{j_{1}}{j_{2}}{-j_{2}}{j}{j_{1}-j_{2}}>0 .
\]
Under such a phase definition the Clebsch-Gordan coefficients satisfy the relations
\begin{gather}
\clebsch{j}{m}{0}{0}{j}{m}=1, \quad \clebsch{j_{1}}{j_{1}}{j_{2}}{j_{2}}{j_{1}+j_{2}}{j_{1}+j_{2}}=1, \quad \clebsch{j_{1}}{m_{1}}{j_{2}}{m_{2}}{j_{1}+j_{2}}{m_{1}+m_{2}}>0, \quad \clebsch{j_{1}}{m_{1}}{j_{2}}{-j_{2}}{j}{m}>0, \quad(-1)^{j_{1}+j_{2}-j} \clebsch{j_{1}}{m_{1}}{j_{2}}{j_{2}}{j}{m}>0, \notag\\
(-1)^{j_{1}-m_{1}} \clebsch{j_{1}}{m_{1}}{j_{2}}{m_{2}}{j}{j}>0, \quad(-1)^{j_{2}+m_{2}} \clebsch{j_{1}}{m_{1}}{j_{2}}{m_{2}}{j_{1}-j_{2}}{m}>0 . \label{eq:8:1:6}
\end{gather}
The Clebsch-Gordan coefficients are elements of the unitary matrix which performs direct and inverse transformations between state vectors\index{Clebsch-Gordan coefficient!unitarity}\index{unitarity!of the Clebsch-Gordan coefficients} $\left|j_{1} m_{1} j_{2} m_{2}\right\rangle$ and $\left|j_{1} j_{2} j m\right\rangle$
\begin{equation}
\clebsch{j_{1}}{m_{1}}{j_{2}}{m_{2}}{j}{m}=\braket{j_{1} m_{1} j_{2} m_{2}}{j_{1} j_{2} j m}=\braket{j_{1} j_{2} j m}{j_{1} m_{1} j_{2} m_{2}} . \label{eq:8:1:7}
\end{equation}
The unitarity relation is
\begin{gather}
\sum_{m_{1} m_{2}} \clebsch{j_{1}}{m_{1}}{j_{2}}{m_{2}}{j}{m} \clebsch{j_{1}}{m_{1}}{j_{2}}{m_{2}}{j^{\prime}}{m^{\prime}}=\delta_{j j^{\prime}} \delta_{m m^{\prime}}. \notag\\
\sum_{j(m)} \clebsch{j_{1}}{m_{1}}{j_{2}}{m_{2}}{j}{m} \clebsch{j_{1}}{m_{1}^{\prime}}{j_{2}}{m_{2}^{\prime}}{j}{m}=\delta_{m_{1} m_{1}^{\prime}} \delta_{m_{2} m_{2}^{\prime}}. \label{eq:8:1:8}
\end{gather}
The direct product of two irreducible tensors $\mathfrak{M}_{j_{1} m_{1}}$ and $\mathfrak{N}_{j_{2} m_{2}}$ may be decomposed into irreducible tensors.\index{irreducible tensor product}\isym{Mfrak@$\mathfrak{M}_{J}$, irreducible tensor} The coefficients of this decomposition are just the Clebsch-Gordan coefficients:
\begin{equation}
\mathfrak{M}_{j_{1} m_{1}} \mathfrak{N}_{j_{2} m_{2}}=\sum_{j(m)} \clebsch{j_{1}}{m_{1}}{j_{2}}{m_{2}}{j}{m}\left\{\mathfrak{N}_{j_{1}} \otimes \mathfrak{N}_{j_{2}}\right\}_{j m} . \label{eq:8:1:9}
\end{equation}
The inverse relation is
\begin{equation}
\left\{\mathfrak{M}_{j_{1}} \otimes \mathfrak{N}_{j_{2}}\right\}_{j m}=\sum_{m_{1} m_{2}} \clebsch{j_{1}}{m_{1}}{j_{2}}{m_{2}}{j}{m_{1}} \mathfrak{M}_{j_{1} m_{1}} \mathfrak{R}_{j_{2} m_{2}} . \label{eq:8:1:10}
\end{equation}
\subsection{The Wigner 3jm Symbols}\label{sec:8.1.2}\index{3jm symbol@$3jm$ symbol!definition}
Not infrequently the Wigner $3 j \mathrm{~m}$ symbols \cite{ref110} are used instead of the Clebsch-Gordan coefficients.\isymo{3jm@$\bigl(\begin{smallmatrix} a & b & c\\ \alpha & \beta & \gamma\end{smallmatrix}\bigr)$, Wigner $3jm$ symbol} These symbols possess simpler symmetry properties. The $3 j \mathrm{~m}$ symbols are related to the Clebsch-Gordan coefficients by
\begin{equation}
\threej{j_{1}}{j_{2}}{j_{3}}{m_{1}}{m_{2}}{m_{3}}\label{eq:8:1:11}=(-1)^{j_{3}+m_{3}+2 j_{1}} \frac{1}{\sqrt{2 j_{3}+1}} \clebsch{j_{1}}{-m_{1}}{j_{2}}{-m_{3}}{j_{3}}{m_{3}} .
\end{equation}
The inverse relation is
\begin{equation}
\clebsch{j_{1}}{m_{1}}{j_{2}}{m_{2}}{j_{3}}{m_{3}}=(-1)^{j_{1}-j_{2}+m_{3}} \sqrt{2 j_{3}+1}\threej{j_{1}}{j_{2}}{j_{3}}{m_{1}}{m_{2}}{-m_{3}}\label{eq:8:1:12} .
\end{equation}
The $3 j \mathrm{~m}$ symbol represents the probability amplitude that three angular momenta $\mathrm{j}_{1}, \mathrm{j}_{2}$ and $\mathrm{j}_{3}$ with projections $m_{1}, m_{2}$ and $m_{3}$ are coupled to yield zero angular momentum:
\begin{equation}
\threej{j_{1}}{j_{2}}{j_{3}}{m_{1}}{m_{2}}{m_{3}}\label{eq:8:1:13}=\eta \sum_{j^{\prime} m^{\prime}} \clebsch{j_{1}}{m_{1}}{j_{2}}{m_{3}}{j^{\prime}}{m^{\prime}} \clebsch{j^{\prime}}{m^{\prime}}{j_{3}}{m_{3}}{0}{0}
\end{equation}
The phase factor $\eta=(-1)^{j_{1}-j_{2}+j_{3}}$ is chosen in such a way that any cyclic permutation of columns leaves the $3 j m$ symbol unchanged. Below we shall use, along with the letters $j, m$ other Latin letters ( $a, b, c$, etc.) to denote angular momenta, and Greek letters ( $\alpha, \beta, \gamma$, etc.) to denote momentum projections in the arguments of the $3 j m$ symbols and Clebsch-Gordan coefficients.

\subsection{Regge R-Symbols}\label{sec:8.1.3}\index{Regge R-symbol@Regge $R$-symbol}\index{R-symbol@$R$-symbol|see{Regge $R$-symbol}}
The Wigner $3 j m$ symbol may be represented by a square $3 \times 3$ array\isym{Rik@$R_{i k}$, Regge $R$-symbol}\isym{J-abc@$J=a+b+c$, sum of three momenta} $\left\|R_{i k}\right\|(i, k=1,2,3)$ which is called the Regge $R$-symbol \cite{ref094}
\begin{equation}
\threej{a}{b}{c}{\alpha}{\beta}{\gamma}\label{eq:8:1:14}=\left\|\begin{array}{lll}
R_{11} & R_{12} & R_{13} \\
R_{21} & R_{22} & R_{23} \\
R_{31} & R_{32} & R_{33}
\end{array}\right\|
\end{equation}
where
\begin{equation}
\begin{array}{lll}
R_{11}=-a+b+c, & R_{12}=a-b+c, & R_{13}=a+b-c, \\
R_{21}=a+\alpha, & R_{22}=b+\beta, & R_{23}=c+\gamma, \\
R_{31}=a-\alpha, & R_{32}=b-\beta, & R_{33}=c-\gamma. \label{eq:8:1:15}
\end{array}
\end{equation}
The inverse relations are
\begin{equation}
\begin{array}{ll}
2 a=R_{21}+R_{31}=R_{12}+R_{13}, & 2 \alpha=R_{21}-R_{31}=R_{32}-R_{22}+R_{33}-R_{23}, \\
2 b=R_{22}+R_{32}=R_{11}+R_{13}, & 2 \beta=R_{22}-R_{32}=R_{31}-R_{21}+R_{33}-R_{23}, \\
2 c=R_{23}+R_{33}=R_{11}+R_{12}, & 2 \gamma=R_{23}-R_{33}=R_{31}-R_{21}+R_{32}-R_{22} . \label{eq:8:1:16}
\end{array}
\end{equation}
All nine elements $R_{i k}$ are non-negative integers which satisfy the relations
\begin{equation}
\sum_{i} R_{i k}=J, \quad \sum_{k} R_{i k}=J, \label{eq:8:1:17}
\end{equation}
where
\begin{equation}
J=a+b+c. \label{eq:8:1:18}
\end{equation}
\section{EXPLICIT FORMS OF THE CLEBSCH-GORDAN COEFFICIENTS AND THEIR RELATIONS TO OTHER FUNCTIONS}\label{sec:8.2}\index{Clebsch-Gordan coefficient!explicit forms}
Below we shall assume that the arguments of the Clebsch-Gordan coefficients and $3 j m$ symbols satisfy Eqs.~\ref{eq:8:1:1}-\ref{eq:8:1:3}. In addition we introduce the $\Delta$-symbol defined by\index{Delta-symbol@$\Delta$-symbol}\index{triangle coefficient}\isymg{Delta-abc@$\Delta(a b c)$, triangle coefficient}
\begin{equation}
\Delta(a b c)=\left[\frac{(a+b-c)!(a-b+c)!(-a+b+c)!}{(a+b+c+1)!}\right]^{\frac{1}{2}}. \label{eq:8:2:1}
\end{equation}
The $\Delta$-symbol is invariant under permutations of $a, b, c$. Numerical values of the $\Delta$-symbol are given in Table 8.12 for $\frac{1}{2} \leq a, b, c \leq 5$. If one of the momenta $a, b, c$ equals zero, the $\Delta$-symbol is reduced to
\begin{equation}
\Delta(a a 0)=\frac{1}{\sqrt{2 a+1}}. \label{eq:8:2:2}
\end{equation}
\subsection{Representations of the Clebsch-Gordan Coefficients in the Form of Algebraic Sums}\label{sec:8.2.1}\index{Clebsch-Gordan coefficient!as algebraic sums}
\begin{align}
\clebsch{a}{\alpha}{b}{\beta}{c}{\gamma}= & \delta_{\gamma, \alpha+\beta} \Delta(a b c)[(a+\alpha)!(a-\alpha)!(b+\beta)!(b-\beta)!(c+\gamma)!(c-\gamma)!(2 c+1)]^{\frac{1}{2}}, \notag \\
& \times \sum_{z} \frac{(-1)^{z}}{z!(a+b-c-z)!(a-\alpha-z)!(b+\beta-z)!(c-b+\alpha+z)!(c-a-\beta+z)!}  \label{eq:8:2:3}
\end{align}
(Van der Waerden \cite{ref040}, Racah \cite{ref091}) 
\begin{align}
\clebsch{a}{\alpha}{b}{\beta}{c}{\gamma}= & \delta_{\gamma, \alpha+\beta} \frac{\Delta(a b c)}{(c+a-b)!(c-a+b)!}\left[\frac{(a-\alpha)!(b-\beta)!(c+\gamma)!(c-\gamma)!(2 c+1)}{(a+\alpha)!(b+\beta)!}\right]^{\frac{1}{2}} \notag \\
& \times \sum_{z} \frac{(-1)^{a-\alpha+z}(a+\alpha+z)!(c+b-\alpha-z)!}{z!(a-\alpha-z)!(c-\gamma-z)!(b-c+\alpha+z)!}  \label{eq:8:2:4}
\end{align}
(Racah \cite{ref091}, Fock \cite{ref068})
\begin{align}
\clebsch{a}{\alpha}{b}{\beta}{c}{\gamma}=&\delta_{\gamma, \alpha+\beta} \Delta(a b c)\left[\frac{(c+\gamma)!(c-\gamma)!(2 c+1)}{(a+\alpha)!(a-\alpha)!(b+\beta)!(b-\beta)!}\right]^{\frac{1}{2}}\notag \\ &
 \sum_{z} \frac{(-1)^{b+\beta+z}(c+b+\alpha-z)!(a-\alpha+z)!}{(c-a+b-z)!(c+\gamma-z)!(a-b-\gamma+z)!} \label{eq:8:2:5}
\end{align}
(Wigner \cite{ref043})
\begin{align}
\clebsch{a}{\alpha}{b}{\beta}{c}{\gamma}= & \delta_{\gamma, \alpha+\beta} \frac{\Delta(a b c)}{(a-b+c)!(a+b-c)!}\left[\frac{(a+\alpha)!(a-\alpha)!(b-\beta)!(c+\gamma)!(2 c+1)}{(b+\beta)!(c-\gamma)!}\right]^{\frac{1}{2}}  \notag\\
& \times \sum_{z} \frac{(-1)^{b+\beta+x}(2 c-z)!(a+b-c+z)!}{z!(c-a+b-z)!(c+\gamma-z)!(a-c-\beta+z)!}. \label{eq:8:2:6}
\end{align}
(Majumdar \cite{ref082})
\begin{align}
\clebsch{a}{\alpha}{b}{\beta}{c}{\gamma}= & \delta_{\gamma, \alpha+\beta} \frac{\Delta(a b c)(a+b+c+1)!}{(a-b+c)!}\left[\frac{(a-\alpha)!(c+\gamma)!(2 c+1)}{(a+\alpha)!(b+\beta)!(b-\beta)!(c-\gamma)!}\right]^{\frac{1}{2}}. \notag\\
& \times \sum_{z} \frac{(-1)^{b+\beta+z}(2 c-z)!(c+b+\alpha-z)!}{z!(c-a+b-z)!(c+\gamma-z)!(a+b+c+1-z)!}. \label{eq:8:2:7}
\end{align}
(Bandzaitis and Yutsis \cite{ref051})
\begin{align}
\clebsch{a}{\alpha}{b}{\beta}{c}{\gamma}= & \frac{\delta_{\gamma, \alpha+\beta}}{\Delta(a b c)}\left[\frac{(a+\alpha)!(a-\alpha)!(c+\gamma)!(c-\gamma)!(2 c+1)}{(b+\beta)!(b-\beta)!}\right]^{\frac{1}{2}}  \notag\\
& \times \sum_{z} \frac{(-1)^{a-\alpha+z}(a+b-\gamma-z)!(b+c-\alpha-z)!}{z!(a-\alpha-z)!(c-\gamma-z)!(a+b+c+1-z)!}. \label{eq:8:2:8}
\end{align}
(Bandzaitis and Yutsis \cite{ref051}).\\
In Eqs.~\eqref{eq:8:2:3}-\eqref{eq:8:2:8} the summation index $z$ assumes integer values for which all the factorial arguments are nonnegative.

\subsection{Quasi-Binomial Representation of the Clebsch-Gordan Coefficients}\label{sec:8.2.2}\index{Clebsch-Gordan coefficient!quasi-binomial representation}\index{quasi-binomial}
The Clebsch-Gordan coefficients can be presented in the form of quasi-binomials $(u \pm v)^{(k)}$ \cite{ref048}
\begin{equation}
\left.\left.\clebsch{a}{\alpha}{b}{\beta}{c}{\gamma}=\delta_{\gamma, \alpha+\beta} \frac{\Delta(a b c)}{(a+b-c)!}\left[\frac{(c-\gamma)!(c+\gamma)!(2 c+1)}{(a-\alpha)!(a+\alpha)!(b-\beta)!(b+\beta)!}\right]^{\frac{1}{2}} \right\rvert\,(a+\alpha)(b-\beta)-(a-\alpha)(b+\beta)\right]^{(a+b-c)} . \label{eq:8:2:9}
\end{equation}
For calculating the quasi-binomial one should use the binomial formula with all powers replaced by quasipowers.
\begin{equation}
(u \pm v)^{(k)}=\sum_{z}( \pm 1)^{z}\binom{k}{z} u^{(k-z)} v^{(z)}. \label{eq:8:2:10}
\end{equation}
Quasi-powers (or generalized powers) are defined by\index{quasi-power}\index{generalized power|see{quasi-power}}\isym{quasipower@$u^{(z)}$, quasi-power}
\begin{gather}
u^{(z)} \equiv u^{(1)(z)}=\frac{\Gamma(u+1)}{\Gamma(u-z+1)}, \notag\\
u^{(-1)(z)}=(u+z)^{(z)}=\frac{\Gamma(u+z+1)}{\Gamma(u+1)}. \label{eq:8:2:11}
\end{gather}
If $z$ is integer and positive, we get
\begin{align}
u^{(z)} & =u(u-1) \ldots(u-z+1), \notag\\
u^{(-1)(z)} & =(u+1)(u+2) \ldots(u+z). \label{eq:8:2:12}
\end{align}
\subsection{Clebsch-Gordan Coefficients and Finite Differences}\label{sec:8.2.3}\index{Clebsch-Gordan coefficient!and finite differences}\index{finite differences}
The Clebsch-Gordan coefficient are related to finite differences \cite{ref041}
\begin{gather}
\clebsch{a}{\alpha}{b}{\beta}{c}{\gamma}=\delta_{\gamma, \alpha+\beta}(-1)^{a-c+\beta}\left[\frac{(c+\gamma)!(a+b-c)!(2 c+1)}{(c-\gamma)!(c+a-b)!(c-a+b)!(a+b+c+1)!}\right]^{\frac{1}{2}}  \notag\\
\times\left[\frac{(a-\alpha)^{(a-b-\gamma)}}{(a+\alpha)^{(a-b+\gamma)}}\right]^{\frac{1}{2}} \Delta_{\alpha}^{c-\gamma}\left[\frac{(a+\alpha)^{(c+a-b)}}{(a-\alpha)^{(a-b-c)}}\right]. \label{eq:8:2:13}
\end{gather}
where the difference of order $k$ with respect to the argument $a$ is given by\isymg{Delta-fd@$\Delta_{a}^{k}$, finite-difference operator}
\begin{equation}
\Delta_{a}^{k} f(a)=\sum_{n=0}^{k}(-1)^{k-n} \frac{k!}{n!(k-n)!} f(a+n). \label{eq:8:2:14}
\end{equation}
In particular,
\begin{gather}
\Delta_{a} f(a)=f(a+1)-f(a), \notag\\
\Delta_{a} a^{(n)}=n a^{(n-1)} . \label{eq:8:2:15}
\end{gather}
Finite differences are analogous to differentials, while quasi-powers are analogous to ordinary powers. From this it can be concluded that the Clebsch-Gordan coefficients are analogous to the Wigner $D$-functions [20, 58].\index{Wigner D-function@Wigner $D$-function}\isym{DJMM@$D_{M M^{\prime}}^{J}$, Wigner $D$-function} To illustrate this similarity let us introduce the notations
\begin{equation}
j=b, \quad \mu=c-a, \quad \nu=\beta, \quad P=\frac{1}{2}(a+b+c), \quad Q=a+\alpha-\frac{1}{2}(a+b+c) . \label{eq:8:2:16}
\end{equation}
\footnotetext{${ }^{2}$ This follows from the fact that the Clebsch-Gordan coefficients may be expressed in terms of orthogonal polynomials of a discrete variable (Hahn polynomials); see Refs.~\cite{ref142, ref143}.
}Now, Eq.~\eqref{eq:8:2:3} may be rewritten in terms of quasi-powers
\begin{align}
\sqrt{\frac{a+b+c+1}{2 c+1}} \clebsch{a}{\alpha}{b}{\beta}{c}{\gamma}= & {\left[\frac{(j+\mu)!(j-\mu)!(j+\nu)!(j-\nu)!}{(2 P)^{(2 j)}(P-Q)^{(j+\mu)}(P-Q)^{(j+\nu)}(P+Q)^{(-\mu-\nu)}}\right]^{\frac{1}{2}} }  \notag\\
& \times \sum_{x}(-1)^{z} \frac{(P-Q)^{(j+\mu+z)}(P+Q)^{(j-\mu-z)}}{z!(j-\mu-z)!(j+\nu-z)!(\mu-\nu+z)!}. \label{eq:8:2:17}
\end{align}
Replacing quasi-powers in \eqref{eq:8:2:17} by ordinary powers and introducing the angle $\vartheta$ according to
\begin{equation}
\cos \theta=\frac{Q}{P}, \quad \cos \frac{\theta}{2}=\sqrt{\frac{P+Q}{2 P}}, \quad \sin \frac{\theta}{2}=\sqrt{\frac{P-Q}{2 P}}, \label{eq:8:2:18}
\end{equation}
we see that \eqref{eq:8:2:17} reduces to the following relation for the Wigner $D$-function
\begin{align}
& d_{\nu \mu}^{j}(\theta)=\left[\frac{(j+\mu)!(j-\mu)!(j+\nu)!(j-\nu)!}{(2 P)^{2 j}(P-Q)^{j+\mu}(P-Q)^{j+\nu}(P+Q)^{-\mu-\nu}}\right]^{\frac{1}{2}}  \notag\\
& \times \sum_{x}(-1)^{x} \frac{(P-Q)^{j+\mu+x}(P+Q)^{j-\mu-x}}{z!(j-\mu-z)!(j+\nu-z)!(\mu-\nu+z)!}. \label{eq:8:2:19}
\end{align}
\eqref{eq:8:2:19} is similar to Eq.~\eqref{eq:4:3:4}. The analogy between \eqref{eq:8:2:17} and \eqref{eq:8:2:19} permits us to connect the recursion relations, symmetry properties, etc. of the Wigner D-functions and the Clebsch-Gordan coefficients \cite{ref058}. Since for $u \gg z, u \gg 1$ quasi-powers became asymptotically equal to ordinary powers, a Clebsch-Gordan coefficient turns into a $D$-function in the large-momentum limit (\eqref{eq:8:9:1}).

\subsection{Expressions for the Clebsch-Gordan Coefficients in Terms of the Binomial Coefficients \cite{ref126}}\label{sec:8.2.4}\index{Clebsch-Gordan coefficient!in terms of binomial coefficients}\index{binomial coefficients}
\begin{equation}
\clebsch{a}{\alpha}{b}{\beta}{c}{\gamma}=\delta_{\gamma, \alpha+\beta}\left[\frac{\binom{2 a}{J-2 c}\binom{2 b}{J-2 c}}{\binom{J+1}{J-2 c}\binom{2 a}{a-\alpha}\binom{2 b}{b-\beta}\binom{2 c}{c-\gamma}}\right]^{\frac{1}{2}} \sum_{x}(-1)^{x}\binom{J-2 c}{z}\binom{J-2 b}{a-\alpha-z}\binom{J-2 a}{b+\beta-z}, \label{eq:8:2:20}
\end{equation}
where $J=a+b+c$.

\subsection{Representations of the Clebsch-Gordan Coefficients in Terms of of the Hypergeometric Functions}\label{sec:8.2.5}\index{Clebsch-Gordan coefficient!and hypergeometric functions}\index{hypergeometric function!generalized}
The Clebsch-Gordan coefficients may be expressed via the generalized hypergeometric functions $\Fhg{3}{2}$ of unit argument,\isym{F32@$\Fhg{3}{2}$, generalized hypergeometric function}
\begin{gather}
\clebsch{a}{\alpha}{b}{\beta}{c}{\gamma}=\delta_{\gamma, \alpha+\beta} \frac{\Delta(a b c)}{(a+b-c)!(-b+c+\alpha)!(-a+c-\beta)!}\left[\frac{(a+\alpha)!(b-\beta)!(c+\gamma)!(c-\gamma)!(2 c+1)}{(a-\alpha)!(b+\beta)!}\right]^{\frac{1}{2}}  \notag\\
\times\Fhg{3}{2}\left[\left.\begin{array}{c}
-a-b+c,-a+\alpha,-b-\beta \\
-a+c-\beta+1,-b+c+\alpha+1
\end{array} \right\rvert\, 1\right], \label{eq:8:2:21}
\end{gather}
\begin{align}
& \begin{aligned}
\clebsch{a}{\alpha}{b}{\beta}{c}{\gamma}= & \delta_{\gamma, \alpha+\beta}(-1)^{a-\alpha} \frac{\Delta(a b c)(b+c-\alpha)!}{(a-b+c)!(-a+b+c)!(b-c+\alpha)!}\left[\frac{(a+\alpha)!(b-\beta)!(c+\gamma)!(2 c+1)}{(a-\alpha)!(b+\beta)!(c-\gamma)!}\right]^{\frac{1}{2}} \\
& \times\Fhg{3}{2}\left[\left.\begin{array}{c}
a+\alpha+1,-a+\alpha,-c+\gamma \\
-b-c+\alpha, b-c+\alpha+1
\end{array} \right\rvert\, 1\right],
\end{aligned}, \\
& \begin{aligned}
\clebsch{a}{\alpha}{b}{\beta}{c}{\gamma}= & \delta_{\gamma, \alpha+\beta}(-1)^{b+\beta} \frac{\Delta(a b c)(b+c+\alpha)!}{(-a+b+c)!(a-b-\gamma)!}\left[\frac{(a-\alpha)!(c-\gamma)!(2 c+1)}{(a+\alpha)!(b+\beta)!(b-\beta)!(c+\gamma)!}\right]^{\frac{1}{2}} \\
& \quad \times\Fhg{3}{2}\left[\left.\begin{array}{c}
a-b-c, a-\alpha+1,-c-\gamma \\
a-b-\gamma+1,-b-c-\alpha \\
\end{array} \right\rvert\, 1\right]\\
&\qquad 
(\text { Rose \cite{ref030}})
\end{aligned}, \label{eq:8:2:23}\\
& \clebsch{a}{\alpha}{b}{\beta}{c}{\gamma}=\delta_{\gamma, \alpha+\beta}(-1)^{b+\beta} \frac{\Delta(a b c)(2 c)!}{(a-b+c)!(-a+b+c)!(a-c-\beta)!}\left[\frac{(a+\alpha)!(a-\alpha)!(b-\beta)!(2 c+1)}{(b+\beta)!(c+\gamma)!(c-\gamma)!}\right]^{\frac{1}{2}}  \notag\\
&  \label{eq:8:2:24}\\
& \quad \times\Fhg{3}{2}\left[\begin{array}{c}
a-b-c, a+b-c+1,-c-\gamma \mid 1], \\
-2 c, a-c-\beta+1
\end{array}\right. \notag\\
& \clebsch{a}{\alpha}{b}{\beta}{c}{\gamma}=\delta_{\gamma, \alpha+\beta}(-1)^{b+\beta} \frac{\Delta(a b c)(2 c)!(b+c+\alpha)!}{(a-b+c)!(-a+b+c)!}\left[\frac{(a+\alpha)!(b+\beta)!(b-\beta)!(c+\gamma)!(c-\gamma)!}{(a-\alpha)}\right]^{\frac{1}{2}}  \notag\\
& \quad \times\Fhg{3}{2}\left[\begin{array}{c}
a-b-c,-a-b-c-1,-c-\gamma \mid 1], \\
-2 c,-b-c-\alpha
\end{array}\right. \label{eq:8:2:25}\\
& \clebsch{a}{\alpha}{b}{\beta}{c}{\gamma}=\delta_{\gamma, \alpha+\beta}(-1)^{a-\alpha} \frac{(a+b-\gamma)!(b+c-\alpha)!}{\Delta(a b c)(a+b+c+1)!}\left[\frac{(a+\alpha)!(c+\gamma)!(2 c+1)}{(a-\alpha)!(b+\beta)!(b-\beta)!(c-\gamma)!}\right]^{\frac{1}{2}}  \notag\\
& \quad \times\Fhg{3}{2}\left[\begin{array}{c}
-a-b-c-1,-a+\alpha,-c+\gamma \mid 1 \\
-a-b+\gamma,-b-c+\alpha
\end{array}\right] . \label{eq:8:2:26}
\end{align}
Equations \eqref{eq:8:2:21}-\eqref{eq:8:2:26} may be associated with Eqs.~\eqref{eq:8:2:3}-\eqref{eq:8:2:8}, respectively.\\[0pt]
According to Ref.~\cite{ref036}, all $\Fhg{3}{2}$-functions of unit argument, which are given by finite sums, may be expressed in terms of the Clebsch-Gordan coefficients. Any Clebsch-Gordan coefficient may also be represented in the form of derivatives of the hypergeometric functions \cite{ref046}:
\begin{gather}
\clebsch{a}{\alpha}{b}{\beta}{c}{\gamma}=\delta_{\gamma, \alpha+\beta}(-1)^{a+b-c} \frac{\Delta(a b c)}{(a+b-c)!(-a+b+c)!(a-b+\gamma)!}\left[\frac{(a+\alpha)!(a-\alpha)!(b+\beta)!(c+\gamma)!(2 c+1)}{(b-\beta)!(c-\gamma)!}\right]^{\frac{1}{2}}  \notag\\
\times\left\{\frac{d^{b-\beta}}{(d t)^{b-\beta}}\left[(1-t)^{a+b-c} F(a-b-c,-c+\gamma ; a-b+\gamma+1 ; t)\right]\right\}_{t=0}. \label{eq:8:2:27}
\end{gather}
\subsection{Representations of the $3 j \mathrm{~m}$ Symbols in the Form of Algebraic Sums}\label{sec:8.2.6}\index{3jm symbol@$3jm$ symbol!as algebraic sums}
The explicit form of the Wigner $3 j m$ symbols may be obtained from Eqs.~\eqref{eq:8:2:3}-\eqref{eq:8:2:8} for the Clebsch-Gordan coefficients. Rewriting these equations in terms of the $R$-symbol elements, we get the following relations for\\
the $3 j m$ symbols \cite{ref045}
\begin{align}
 \threej{a}{b}{c}{\alpha}{\beta}{\gamma}&=(-1)^{R_{21}+R_{32}}\left[\frac{\prod_{i, j=1}^{3}\left(R_{i j}\right)!}{(J+1)!}\right]^{\frac{1}{2}} \notag \\
& \times \sum_{x}(-1)^{x} \frac{1}{z!\left(R_{13}-z\right)!\left(R_{22}-z\right)!\left(R_{31}-z\right)!\left(R_{33}-R_{22}+z\right)!\left(R_{11}-R_{22}+z\right)!}, \label{eq:8:2:28}\\
\threej{a}{b}{c}{\alpha}{\beta}{\gamma}&=(-1)^{R_{21}+R_{31}+R_{32}}\left[\frac{R_{13}!R_{23}!R_{31}!R_{32}!R_{33}!}{(J+1)!R_{11}!R_{12}!R_{21}!R_{22}!}\right]^{\frac{1}{2}}  \notag\\
 &\times \sum_{x}(-1)^{x} \frac{\left(R_{21}+z\right)!\left(R_{11}+R_{31}-z\right)!}{z!\left(R_{31}-z\right)!\left(R_{23}-z\right)!\left(R_{13}-R_{31}+z\right)!}, \label{eq:8:2:29}\\
 \threej{a}{b}{c}{\alpha}{\beta}{\gamma}&=(-1)^{R_{21}+R_{22}+R_{32}}\left[\frac{R_{11}!R_{12}!R_{13}!R_{23}!R_{33}!}{(J+1)!R_{21}!R_{22}!R_{31}!R_{32}!}\right]^{\frac{1}{2}} \sum_{z}(-1)^{z} \frac{\left(R_{31}+z\right)!\left(R_{32}+R_{33}-z\right)!}{z!\left(R_{11}-z\right)!\left(R_{33}-z\right)!\left(R_{12}-R_{33}+z\right)!}, \\
 \threej{a}{b}{c}{\alpha}{\beta}{\gamma}&=(-1)^{R_{21}+R_{22}+R_{32}}\left[\frac{R_{11}!R_{21}!R_{31}!R_{32}!R_{33}!}{(J+1)!R_{12}!R_{13}!R_{22}!R_{33}!}\right]^{\frac{1}{2}} \sum_{z}(-1)^{z} \frac{\left(R_{13}+z\right)!\left(R_{23}+R_{33}-z\right)!}{z!\left(R_{11}-z\right)!\left(R_{33}-z\right)!\left(R_{21}-R_{33}+z\right)!}, \label{eq:8:2:31}\\
 \threej{a}{b}{c}{\alpha}{\beta}{\gamma}&=(-1)^{R_{21}+R_{22}+R_{32}}\left[\frac{(J+1)!R_{11}!R_{13}!R_{31}!R_{39}!}{R_{12}!R_{21}!R_{22}!R_{23}!R_{32}!}\right]^{\frac{1}{2}} \sum_{z}(-1)^{z} \frac{\left(R_{11}+R_{21}-z\right)!\left(R_{23}+R_{33}-z\right)!}{z!\left(R_{11}-z\right)!\left(R_{33}-z\right)!(J+1-z)!}, \label{eq:8:2:32}\\
 \threej{a}{b}{c}{\alpha}{\beta}{\gamma}&=(-1)^{R_{21}+R_{31}+R_{32}}\left[\frac{(J+1)!R_{21}!R_{23}!R_{31}!R_{39}!}{R_{11}!R_{12}!R_{13}!R_{22}!R_{32}!}\right]^{\frac{1}{2}} \sum_{z}(-1)^{z} \frac{\left(R_{11}+R_{31}-z\right)!\left(R_{13}+R_{23}-z\right)!}{z!\left(R_{23}-z\right)!\left(R_{31}-z\right)!(J+1-z)!}, \label{eq:8:2:33}
\end{align}
where $J=a+b+c$.

\subsection{Quasi-binomial Representations of the $3jm$ Symbols}\label{sec:8.2.7}\index{3jm symbol@$3jm$ symbol!quasi-binomial representations}
\begin{equation}
\threej{a}{b}{c}{\alpha}{\beta}{\gamma}\label{eq:8:2:34}=(-1)^{R_{21}+R_{32}}\left[\frac{R_{21}^{(-1)\left(R_{31}-R_{13}\right)} R_{31}^{(-1)\left(R_{32}-R_{13}\right)} R_{32}^{(-1)\left(R_{22}-R_{13}\right)} R_{22}^{(-1)\left(R_{21}-R_{13}\right)}}{(J+1)!R_{13}!}\right]^{\frac{1}{4}}(u-v)^{\left(R_{13}\right)} .
\end{equation}
The quasi-binomial may be evaluated according to Eqs.~\eqref{eq:8:2:10} and \eqref{eq:8:2:11}. Variables $u$ and $v$ may be chosen in different ways \cite{ref045}:
\begin{align}
(a)& \qquad u=R_{21}^{(1)} R_{32}^{(1)}, && v=R_{22}^{(1)} R_{31}^{(1)},\notag\\
(b)& \qquad u=R_{21}^{(1)} R_{11}^{(-1)}, && v=R_{22}^{(1)} R_{12}^{(-1)},\notag\\
(c)& \qquad u=R_{11}^{(-1)} R_{23}^{(-1)}, && v=R_{32}^{(1)}(J+1)^{(1)}.
\end{align}

\section{INTEGRAL REPRESENTATIONS}\label{sec:8.3}\index{Clebsch-Gordan coefficient!integral representations}\index{integral representations!of the Clebsch-Gordan coefficients}
\subsection{Integrals Involving Algebraic Functions}\label{sec:8.3.1}\index{integral representations!with algebraic functions}
The Clebsch-Gordan coefficients may be represented by the following integrals \cite{ref041} ( $J=a+b+c$ )
\begin{gather}
\clebsch{a}{\alpha}{b}{\beta}{c}{\gamma}=\frac{(-1)^{a-c+\beta}}{2^{J+1}}\left[\frac{(c+\gamma)!(J-2 c)!(J+1)!(2 c+1)}{(a-\alpha)!(a+\alpha)!(b-\beta)!(b+\beta)!(c-\gamma)!(J-2 a)!(J-2 b)!}\right]^{\frac{1}{2}}  \notag\\
\times \int_{-1}^{1}(1-x)^{a-\alpha}(1+x)^{b-\beta} \frac{d^{c-\gamma}}{(d x)^{c-\gamma}}\left[(1-x)^{J-2 a}(1+x)^{J-2 b}\right] d x, \label{eq:8:3:1}\\
\clebsch{a}{\alpha}{b}{\beta}{c}{\gamma}=\frac{(-1)^{a-c+\beta}}{2^{J+1}}\left[\frac{(J-2 b)!(J-2 c)!(J+1)!(2 c+1)}{(a-\alpha)!(a+\alpha)!(b-\beta)!(b+\beta)!(c-\gamma)!(c+\gamma)!(J-2 a)!}\right]^{\frac{1}{2}}  \notag\\
\times \int_{-1}^{1}(1-x)^{b+\beta}(1+x)^{b-\beta} \frac{d^{J-2 a}}{(d x)^{J-2 a}}\left[(1-x)^{c-\gamma}(1+x)^{c+\gamma}\right] d x. \label{eq:8:3:2}
\end{gather}
In particular, if $\alpha=\beta=\gamma=0$, then
\begin{equation}
\clebsch{a}{0}{b}{0}{c}{0}=\frac{(-1)^{a-c}}{2^{J+1} a!b!c!}\left[\frac{(J-2 b)!(J-2 c)!(J+1)!(2 c+1)}{(J-2 a)!}\right]^{\frac{1}{2}} \int_{-1}^{1}\left(1-x^{2}\right)^{b} \frac{d^{J-2 a}}{(d x)^{J-2 a}}\left[\left(1-x^{2}\right)^{c}\right] d x. \label{eq:8:3:3}
\end{equation}
\subsection{Integrals Involving the Wigner D-Functions}\label{sec:8.3.2}\index{integral representations!with the Wigner $D$-functions}\isym{dR@$d R$, invariant volume element of the rotation group}
\begin{equation}
\clebsch{a}{\alpha}{b}{\beta}{c}{\gamma}=\frac{1}{8 \pi^{2}}\left[\frac{(a+b+c+1)!(a+b-c)!(2 c+1)}{(2 a)!(2 b)!}\right]^{\frac{1}{2}} \int d R D_{\alpha a}^{a}(R) D_{\beta-b}^{b}(R) D_{\gamma a-b}^{c *}(R), \label{eq:8:3:4}
\end{equation}
where $R$ means the set of the Euler angles (see Sec.~\ref{sec:4.12})
\begin{equation}
\begin{aligned}
\clebsch{a}{\alpha}{b}{\beta}{c}{\gamma}= & \frac{(-1)^{a+c-\alpha-\gamma}}{2^{a+b+1}}\left[\frac{(J-2 c)!(J+1)!(2 c+1)}{(a+\alpha)!(a-\alpha)!(b+\beta)!(b-\beta)!}\right]^{\frac{1}{2}} \\
& \times \int_{-1}^{1}(\sin \theta)^{a+b}\left(\frac{1-\cos \theta}{1+\cos \theta}\right)^{\frac{\alpha-\beta}{2}} d_{b-a \gamma}^{c}(\theta) d(\cos \theta)\\
&\qquad 
\text{(Akim and Levin, \cite{ref046})}
\end{aligned}
\end{equation}

\subsection{Integral Involving the Spherical Harmonics}\label{sec:8.3.3}\index{integral representations!with the spherical harmonics}\index{spherical harmonics}\isym{Ylm@$Y_{l m}(\vartheta, \varphi)$, spherical harmonic}
If $a, b, c$ are integers and $J=a+b+c=2 g$ is even (i.e., $g$ is integer), then
\begin{equation}
\clebsch{a}{\alpha}{b}{\beta}{c}{\gamma}=(-1)^{g-c}\left[\frac{4 \pi}{(2 a+1)(2 b+1)}\right]^{\frac{1}{2}} \frac{(g-a)!(g-b)!(g-c)!}{\Delta(a b c) g!} \int_{0}^{2 \pi} d \varphi \int_{0}^{\pi} d \vartheta \sin \vartheta Y_{a \alpha}(\vartheta, \varphi) Y_{b \beta}(\vartheta, \varphi) Y_{c \gamma}^{*}(\vartheta, \varphi). \label{eq:8:3:6}
\end{equation}
\subsection{Integral Representations for Products of the Clebsch-Gordan Coefficients}\label{sec:8.3.4}\index{integral representations!for products of Clebsch-Gordan coefficients}\index{Legendre polynomials}\isym{Pl@$P_{l}(x)$, Legendre polynomial}
\begin{equation}
\clebsch{a}{\alpha}{b}{\beta}{c}{\gamma} \clebsch{a}{\alpha^{\prime}}{b}{\beta^{\prime}}{c}{\gamma^{\prime}}=\frac{2 c+1}{8 \pi^{2}} \int d R D_{\alpha \alpha^{\prime}}^{a}(R) D_{\beta \beta^{\prime}}^{b}(R) D_{\gamma \gamma^{\prime}}^{c *}(R). \label{eq:8:3:7}
\end{equation}
In particular,
\begin{equation}
\left[\clebsch{a}{0}{b}{0}{c}{0}\right]^{2}=\frac{2 c+1}{2} \int_{0}^{\pi} d \theta \sin \vartheta P_{a}(\cos \vartheta) P_{b}(\cos \vartheta) P_{c}(\cos \vartheta). \label{eq:8:3:8}
\end{equation}
\section{SYMMETRY PROPERTIES}\label{sec:8.4}\index{vector-addition coefficients!symmetry properties}\index{symmetry!of the vector-addition coefficients}
The simplest way to formulate the symmetry properties of the Clebsch-Gordan coefficients and 3jm symbols is to use the representations of these coefficients in the form of the Regge $R$-symbols \cite{ref094} (see Sec.~\ref{sec:8.1.3}).

\subsection{Symmetry Properties of the $R$-Symbols}\label{sec:8.4.1}\index{Regge R-symbol@Regge $R$-symbol!symmetry properties}
The $R$-symbol possesses the following symmetry properties \cite{ref094}.\\
\subsubsection{Permutations of columns}
\begin{equation}
\left\|\begin{array}{lll}
R_{11} & R_{12} & R_{13}\\
R_{21} & R_{22} & R_{23} \\
R_{31} & R_{32} & R_{33}
\end{array}\right\|=\varepsilon\left\|\begin{array}{lll}
R_{1 i} & R_{1 k} & R_{1 i} \\
R_{2 i} & R_{2 k} & R_{2 l} \\
R_{3 i} & R_{3 k} & R_{3 l}
\end{array}\right\|,   \label{eq:8:4:1}
\end{equation}
where
\begin{equation}
c= \begin{cases}+1 & \text { for cyclic permutation (even number of permutations) }  \\ (-1)^{J} & \text { for non-cyclic permutation (odd number of permutations) }\end{cases}, \label{eq:8:4:2}
\end{equation}
and
\[
J \equiv \sum_{k} R_{i k}=\sum_{i} R_{i k} ;
\]
\subsubsection{Permutations of rows}
\begin{equation}
\left\|\begin{array}{lll}
R_{11} & R_{12} & R_{13}≈\\
R_{21} & R_{22} & R_{23} \\
R_{31} & R_{32} & R_{33}
\end{array}\right\|=\left\|\begin{array}{lll}
R_{i 1} & R_{i 2} & R_{i 3} \\
R_{k 1} & R_{k 2} & R_{k 3} \\
R_{i 1} & R_{i 2} & R_{i 3}
\end{array}\right\|, 
\end{equation}
\subsubsection{Transposition}
\begin{equation}
\left\|\begin{array}{lll}
R_{11} & R_{12} & R_{13} \\
R_{21} & R_{22} & R_{23} \\
R_{31} & R_{32} & R_{33}
\end{array}\right\|=\left\|\begin{array}{lll}
R_{11} & R_{21} & R_{31} \\
R_{12} & R_{22} & R_{32} \\
R_{13} & R_{23} & R_{33}
\end{array}\right\|.  \label{eq:8:4:4}
\end{equation}
These symmetry transformations relate $6 \times 6 \times 2=72$ generally different coefficients.

\subsection{Symmetry Properties of the $3 j m$ Symbols}\label{sec:8.4.2}\index{3jm symbol@$3jm$ symbol!symmetry properties}
The above relations are equivalent to the following symmetry properties of the $3 j m$ symbol.

\subsubsection{Classical symmetries \cite{ref110}}
\begin{enumerate}[label=(\roman*)]
\item Permutations of columns (corresponding to permutations of columns of the $R$-symbol):
\begin{align}
\threej{a}{b}{c}{\alpha}{\beta}{\gamma}=\threej{b}{c}{a}{\beta}{\gamma}{\alpha}=\threej{c}{a}{b}{\gamma}{\alpha}{\beta} & =(-1)^{a+b+c}\threej{a}{c}{b}{\alpha}{\gamma}{\beta}  \notag\\
& =(-1)^{\alpha+\beta+\gamma}\threej{b}{a}{c}{\beta}{\alpha}{\gamma}=(-1)^{\alpha+b+c}\threej{c}{b}{a}{\gamma}{\beta}{\alpha}. \label{eq:8:4:5}
\end{align}
\item Change of signs of momentum projections (corresponding to the permutation of the second and third rows of the $R$-symbol)
\begin{equation}
\threej{a}{b}{c}{\alpha}{\beta}{\gamma}\label{eq:8:4:6}=(-1)^{a+b+c}\threej{a}{b}{c}{-\alpha}{-\beta}{-\gamma}
\end{equation}
\subsubsection{Regge symmetries \cite{ref094}}
\item Replacement of arguments:
\begin{equation}
\threej{a}{b}{c}{\alpha}{\beta}{\gamma}\label{eq:8:4:7}=(-1)^{a+b+c}\threej{\frac{b+c+\alpha}{2}}{\frac{a+c+\beta}{2}}{\frac{a+b+\gamma}{2}}{a-\frac{b+c-\alpha}{2}}{b-\frac{a+c-\beta}{2}}{c-\frac{a+b-\gamma}{2}}
\end{equation}
(corresponding to the permutation of the first and third rows of the $R$-symbol) and\\
\item
\begin{equation}
\threej{a}{b}{c}{\alpha}{\beta}{\gamma}\label{eq:8:4:8}=\threej{a}{\frac{b+c-\alpha}{2}}{\frac{b+c+\alpha}{2}}{-b+c}{\frac{b-c-\alpha}{2}-\gamma}{\frac{b-c+\alpha}{2}+\gamma}
\end{equation}
This corresponds to the transposition of the $R$-symbol. According to the above symmetry properties, 72 formally different $3 j m$ symbols have the same absolute values. One can decompose these symbols into six groups ( 12 coefficients in each group). The $3 j m$ symbols which belong to different groups correspond to triangles of different forms but the same perimeter $J=a+b+c$. On the other hand, all $3 j m$ symbols which belong to one group are related by the classical symmetry properties \eqref{eq:8:4:5} and \eqref{eq:8:4:6}. In \eqref{eq:8:4:9} we list one $3 j m$ symbol from each group.
\begin{align}
& \threej{a}{b}{c}{\alpha}{\beta}{\gamma}=\threej{a}{\frac{b+c-\alpha}{2}}{\frac{b+c+\alpha}{2}}{c-b}{\frac{b+\beta-c-\gamma}{2}}{\frac{b-\beta-c+\gamma}{2}}=\threej{b}{\frac{a+c-\beta}{2}}{\frac{a+c+\beta}{2}}{a-c}{\frac{-a-\alpha+c+\gamma}{2}}{\frac{-a+\alpha+c-\gamma}{2}}  \notag\\
& =\threej{c}{\frac{a+b-\gamma}{2}}{\frac{a+b+\gamma}{2}}{b-a}{\frac{a+\alpha-b-\beta}{2}}{\frac{a-\alpha-b+\beta}{2}}=\threej{\frac{b+c-\alpha}{2}}{\frac{a+c-\beta}{2}}{\frac{a+b-\gamma}{2}}{a-\alpha-\frac{b+c-\alpha}{2}}{b-\beta-\frac{a+c-\beta}{2}}{c-\gamma-\frac{a+b-\gamma}{2}}  \notag\\
& =\threej{\frac{b+c+\alpha}{2}}{\frac{a+c+\beta}{2}}{\frac{a+b+\gamma}{2}}%
{\frac{b+c+\alpha}{2}-a-\alpha}{\frac{a+c+\beta}{2}-b-\beta}{\frac{a+b+\gamma}{2}-c-\gamma}
 . \label{eq:8:4:9}
\end{align}
\end{enumerate}
\subsection{Symmetry Properties of the Clebsch-Gordan Coefficients}\label{sec:8.4.3}\index{Clebsch-Gordan coefficient!symmetry properties}
\begin{enumerate}[label=(\roman*)]
\item
  \begin{align}
\clebsch{a}{\alpha}{b}{\beta}{c}{\gamma}=(-1)^{a+b-c} \clebsch{b}{\beta}{a}{\alpha}{c}{\gamma} & =(-1)^{a-\alpha} \sqrt{\frac{2 c+1}{2 b+1}} \clebsch{a}{\alpha}{c}{-\gamma}{b}{-\beta}=(-1)^{a-\alpha} \sqrt{\frac{2 c+1}{2 b+1}} \clebsch{c}{\gamma}{a}{-\alpha}{b}{\beta}  \notag\\
& =(-1)^{b+\beta} \sqrt{\frac{2 c+1}{2 a+1}} \clebsch{c}{-\gamma}{b}{\beta}{a}{-\alpha}=(-1)^{b+\beta} \sqrt{\frac{2 c+1}{2 a+1}} \clebsch{b}{-\beta}{c}{\gamma}{a}{\alpha}. \label{eq:8:4:10}
\end{align}
\item
\begin{equation}
\clebsch{a}{\alpha}{b}{\beta}{c}{\gamma}=(-1)^{a+b-c} \clebsch{a}{-\alpha}{b}{-\beta}{c}{-\gamma}. \label{eq:8:4:11}
\end{equation}
\item
\begin{equation}
\clebsch{a}{\alpha}{b}{\beta}{c}{\gamma}=\clebsch{a^{\prime}}{\alpha^{\prime}}{b^{\prime}}{\beta^{\prime}}{c^{\prime}}{\gamma^{\prime}}, \label{eq:8:4:12}
\end{equation}
where
\begin{align}
2 a^{\prime} & =(a+\alpha)+(b+\beta), & 2 a & =\left(a^{\prime}+a^{\prime}\right)+\left(b^{\prime}+\beta^{\prime}\right), \notag\\
2 \alpha^{\prime} & =(a+\alpha)-(b+\beta), & 2 \alpha & =\left(a^{\prime}+\alpha^{\prime}\right)-\left(b^{\prime}+\beta^{\prime}\right), \notag\\
2 b^{\prime} & =(a-\alpha)+(b-\beta), & 2 b & =\left(a^{\prime}-\alpha^{\prime}\right)+\left(b^{\prime}-\beta^{\prime}\right), \notag\\
2 \beta^{\prime} & =(a-\alpha)-(b-\beta), & 2 \beta & =\left(a^{\prime}-\alpha^{\prime}\right)-\left(b^{\prime}-\beta^{\prime}\right), \label{eq:8:4:13}
\end{align}
\[
\begin{aligned}
c^{\prime} & =c, & c & =c^{\prime} \\
\gamma^{\prime} & =a-b, & \gamma & =a^{\prime}-b^{\prime}
\end{aligned}
\]
\item
\begin{equation}
\clebsch{a}{\alpha}{b}{\beta}{c}{\gamma}=(-1)^{b+\beta} \sqrt{\frac{2 c+1}{2 c^{\prime}+1}} \clebsch{a^{\prime}}{\alpha^{\prime}}{b^{\prime}}{\beta^{\prime}}{c^{\prime}}{\gamma^{\prime}}, \label{eq:8:4:14}
\end{equation}
where
\begin{align}
2 a^{\prime} & =(b-\beta)+(c+\gamma), & 2 a & =\left(b^{\prime}-\beta^{\prime}\right)+\left(c^{\prime}+\gamma^{\prime}\right), \notag\\
2 \alpha^{\prime} & =2(a+\alpha)-(b-\beta)-(c+\gamma), & 2 \alpha & =2\left(a^{\prime}+\alpha^{\prime}\right)-\left(b^{\prime}-\beta^{\prime}\right)-\left(c^{\prime}+\gamma^{\prime}\right), \notag\\
2 b^{\prime} & =(a-\alpha)+(c+\gamma), & 2 b & =\left(a^{\prime}-\alpha^{\prime}\right)+\left(c^{\prime}+\gamma^{\prime}\right), \label{eq:8:4:15}\\
2 \beta^{\prime} & =2(b+\beta)-(a-\alpha)-(c+\gamma), & 2 \beta & =2\left(b^{\prime}+\beta^{\prime}\right)-\left(a^{\prime}-\alpha^{\prime}\right)-\left(c^{\prime}+\gamma^{\prime}\right), \notag\\
2 c^{\prime} & =(a-\alpha)+(b-\beta), & 2 c & =\left(a^{\prime}-\alpha^{\prime}\right)+\left(b^{\prime}-\beta^{\prime}\right), \notag\\
2 \gamma^{\prime} & =(a-\alpha)+(b-\beta)-2(c-\gamma), & 2 \gamma & =\left(a^{\prime}-\alpha^{\prime}\right)+\left(b^{\prime}-\beta^{\prime}\right)-2\left(c^{\prime}-\gamma^{\prime}\right) . \notag
\end{align}
\end{enumerate}
\subsection{"Mirror" Symmetry}\label{sec:8.4.4}\index{mirror symmetry}\index{symmetry!mirror}\index{negative momenta}
The original relations which define the $3 j m$ symbols and the Clebsch-Gordan coefficients may be extended to the domain of negative integer or half-integer arguments $a, b, c$. In this case the following symmetry relations are valid \cite{ref045} (here we denote $\bar{c} \equiv-c-1, \gamma \equiv-\gamma$, etc.).\\
\subsubsection{$3jm$ symbols}
The relations for the $3 j m$ symbols are
\begin{equation}
\threej{a}{b}{c}{\alpha}{\beta}{\gamma}\label{eq:8:4:16}=(-1)^{b-c-\alpha}\threej{a}{b}{c}{\alpha}{\beta}{\gamma}=i(-1)^{a+\alpha-\beta}\threej{a}{b}{c}{\alpha}{\beta}{\gamma}=i(-1)^{a+b+c}\threej{\bar{a}}{b}{\bar{c}}{\alpha}{\beta}{\gamma} .
\end{equation}
The $3 j \mathrm{~m}$ symbols with negative momenta obey the symmetry relations \eqref{eq:8:4:5} and \eqref{eq:8:4:6} provided $\bar{a}, \bar{b}, \bar{c}$ are replaced by $-a,-b,-c$, respectively, in the phase factors of these relations.\\
\subsubsection{Clebsch-Gordan symbols}
The relations for the Clebsch-Gordan coefficients are
\begin{align}
\clebsch{a}{\alpha}{b}{\beta}{c}{\gamma}=(-1)^{a+b-c} \clebsch{a}{\alpha}{\delta}{\beta}{c}{\gamma} & =(-1)^{b-c-\alpha} \clebsch{a}{\alpha}{b}{\beta}{c}{\gamma}=(-1)^{b+\beta} \clebsch{a}{\alpha}{b}{\beta}{c}{\gamma}=(-1)^{a-\alpha} \clebsch{a}{\alpha}{\delta}{\beta}{c}{\gamma}=\clebsch{a}{\alpha}{b}{\beta}{c}{\gamma}  \notag\\
& =\sqrt{\frac{2 c+1}{2 b+1}} \clebsch{a}{\alpha}{c}{\gamma}{b}{\beta}=\sqrt{\frac{2 c+1}{2 a+1}} \clebsch{c}{\gamma}{b}{\beta}{a}{\alpha}=\sqrt{\frac{2 c+1}{2 a+1}} \clebsch{c}{\gamma}{b}{\beta}{a}{\alpha}. \label{eq:8:4:17}
\end{align}
From \eqref{eq:8:4:17} one gets
\begin{align}
& \clebsch{a}{\bar{a}}{b}{\bar{\beta}}{c}{\bar{\gamma}}=\clebsch{\bar{a}}{\alpha}{\bar{b}}{\beta}{\bar{c}}{\gamma}, \label{eq:8:4:18}\\
& \clebsch{\bar{a}}{\bar{a}}{b}{\beta}{c}{\gamma}=\clebsch{a}{\alpha}{\bar{b}}{\bar{\beta}}{\bar{c}}{\bar{\gamma}}, \label{eq:8:4:19}\\
& \clebsch{a}{\alpha}{\bar{b}}{\bar{\beta}}{c}{\gamma}=\clebsch{\bar{a}}{\bar{a}}{b}{\beta}{\bar{c}}{\bar{\gamma}}. \label{eq:8:4:20}
\end{align}
When tabulating formulas for the Clebsch-Gordan coefficients, it is convenient to use the "mirror" symmetry properties in the form
\begin{equation}
\clebsch{a}{\alpha}{b}{\beta}{a+k}{\gamma}=(-1)^{b+\beta} \clebsch{a}{\alpha}{b}{\beta}{a-k}{\gamma}. \label{eq:8:4:21}
\end{equation}
\subsection{Properties of the Vector-Addition Coefficients under Transformations of the Coordinate System and Time Reversal}\label{sec:8.4.5}\index{vector-addition coefficients!under coordinate transformations}\index{time reversal}\index{coordinate inversion}\isym{Pr-hat@$\widehat{P}_{r}$, coordinate-inversion operator}\isym{Pt-hat@$\widehat{P}_{t}$, time-reversal operator}
The vector-addition coefficients are not invariant under transformations in question, although formally they are independent of coordinates and time.

\subsubsection{Rotation of the coordinate system.}

The projections of angular momenta vary under rotations of the coordinate system. The $3 j m$ symbols and the Clebsch-Gordan coefficients in a new coordinate system may be expressed as superpositions of the corresponding quantities in the initial coordinate system. Thus
\begin{align}
\hat{R}\threej{a}{b}{c}{\alpha}{\beta}{\gamma} & \equiv\threej{a}{b}{c}{\alpha^{\prime}}{\beta^{\prime}}{\gamma^{\prime}}=\sum_{\alpha \beta \gamma}\threej{a}{b}{c}{\alpha}{\beta}{\gamma} D_{\alpha \alpha^{\prime}}^{a}(R) D_{\beta \beta^{\prime}}^{b}(R) D_{\gamma \gamma^{\prime}}^{c}(R), \label{eq:8:4:22}\\
\widehat{R} \clebsch{a}{\alpha}{b}{\beta}{c}{\gamma} & \equiv \clebsch{a}{\alpha^{\prime}}{b}{\beta^{\prime}}{c}{\gamma^{\prime}}=\sum_{\alpha \beta \gamma} \clebsch{a}{\alpha}{b}{\beta}{c}{\gamma} D_{\alpha \alpha^{\prime}}^{a}(R) D_{\beta \beta^{\prime}}^{b}(R) D_{\gamma \gamma^{\prime}}^{c *}(R). \notag
\end{align}
\subsubsection{Inversion of the coordinate system.}

The $3 j m$ symbols and the Clebsch-Gordan coefficients are invariant under inversion of the coordinate system, because angular momentum is a pseudovector.
\begin{align}
\widehat{P}_{r}\threej{a}{b}{c}{\alpha}{\beta}{\gamma} & =\threej{a}{b}{c}{\alpha}{\beta}{\gamma}, \label{eq:8:4:23}\\
\widehat{P}_{r} \clebsch{a}{\alpha}{b}{\beta}{c}{\gamma} & =\clebsch{a}{\alpha}{b}{\beta}{c}{\gamma}. \notag
\end{align}
\subsubsection{(Time reversal.}
The time reversal changes signs of projections of all angular momenta (reversal of the rotation direction). Hence
\begin{align}
\hat{P}_{t}\threej{a}{b}{c}{\alpha}{\beta}{\gamma} & =\threej{a}{b}{c}{-\alpha}{-\beta}{-\gamma}=(-1)^{a+b+c}\threej{a}{b}{c}{\alpha}{\beta}{\gamma}, \label{eq:8:4:24}\\
\widehat{P}_{t} \clebsch{a}{\alpha}{b}{\beta}{c}{\gamma} & =\clebsch{a}{-\alpha}{b}{-\beta}{c}{-\gamma}=(-1)^{a+b-c} \clebsch{a}{\alpha}{b}{\beta}{c}{\gamma}. \notag
\end{align}
\section{EXPLICIT FORMS OF THE CLEBSCH-GORDAN COEFFICIENTS FOR SPECIAL VALUES OF THE ARGUMENTS}\label{sec:8.5}\index{Clebsch-Gordan coefficient!for special values of the arguments}
\subsection{Special Values of Momenta $a, b, c$}\label{sec:8.5.1}\index{Clebsch-Gordan coefficient!for special momenta}
\subsubsection{$c=0$ or $b=0$.}
\begin{align}
& \clebsch{a}{\alpha}{b}{\beta}{0}{0}=(-1)^{a-\alpha} \frac{\delta_{\alpha b} \delta_{\alpha,-\beta}}{\sqrt{2 a+1}}, \label{eq:8:5:1}\\
& \clebsch{a}{\alpha}{0}{0}{\alpha}{\gamma}=\delta_{a c} \delta_{\alpha \gamma}. \label{eq:8:5:2}
\end{align}
\subsubsection{$c=a+b$} 
\begin{equation}
\clebsch{a}{a}{b}{\beta}{a+b}{\alpha+\beta}=\left[\frac{(2 a)!(2 b)!(a+b+\alpha+\beta)!(a+b-\alpha-\beta)!}{(2 a+2 b)!(a+\alpha)!(a-\alpha)!(b+\beta)!(b-\beta)!}\right]^{\frac{1}{2}}. \label{eq:8:5:3}
\end{equation}
The following two formulas are equivalent to \eqref{eq:8:5:3} (i and $n$ are integer)
\begin{gather}
\clebsch{a}{a-n+i}{b}{b-i+1}{a+b}{a+b-n+1}=\left[\frac{\binom{2 a}{n-i}\binom{2 b}{i-1}}{\binom{2 a+2 b}{n-1}}\right]^{\frac{1}{2}}, \label{eq:8:5:4}\\
\clebsch{a}{a+n-i}{b}{-b+i-1}{a+b}{a-b+n-1}=\left[\frac{\binom{2 a}{i-n}\binom{2 b}{i-1}}{\binom{2 a+2 b}{2 a+n-1}}\right]^{\frac{1}{2}}. \label{eq:8:5:5}
\end{gather}
In particular,
\begin{gather}
\clebsch{a}{a}{b}{b}{a+b}{a+b}=1, \label{eq:8:5:6}\\
\clebsch{a}{a}{b}{-b}{a+b}{a-b}=\left[\frac{(2 a)!(2 b)!}{(2 a+2 b)!}\right]^{\frac{1}{2}}. \label{eq:8:5:7}
\end{gather}
\subsubsection{$c=a+b-1$}
\begin{equation}
\clebsch{a}{\alpha}{b}{\beta}{a+b-1}{\alpha+\beta}=2(b \alpha-a \beta)\left[\frac{(2 a+2 b-1)(2 a-1)!(2 b-1)!(a+b+\alpha+\beta-1)!(a+b-\alpha-\beta-1)!}{(a+\alpha)!(a-\alpha)!(b+\beta)!(b-\beta)!(2 a+2 b)!}\right]^{\frac{1}{2}} . \label{eq:8:5:8}
\end{equation}
In particular,
\begin{gather}
\clebsch{a}{\alpha}{b}{\beta}{a+b-1}{\alpha+\beta}=0, \quad \text { if } \quad \frac{\alpha}{\beta}=\frac{a}{b}. \label{eq:8:5:9}\\
\clebsch{a}{a-1}{b}{b}{a+b-1}{a+b-1}=-\sqrt{\frac{b}{a+b}}. \label{eq:8:5:10}\\
\clebsch{a}{a}{b}{b-1}{a+b-1}{a+b-1}=\sqrt{\frac{a}{a+b}}. \label{eq:8:5:11}\\
\clebsch{a}{a}{b}{-b}{a+b-1}{a-b}=\left[\frac{(2 a)!(2 b)!(2 a+2 b-1)}{(2 a+2 b)!}\right]^{\frac{1}{2}}, \label{eq:8:5:12}
\end{gather}
\subsubsection{$c=a-b(a \geq b)$}
\begin{equation}
\clebsch{a}{\alpha}{b}{\beta}{a-b}{\alpha+\beta}=(-1)^{b+\beta}\left[\frac{(a+\alpha)!(a-\alpha)!(2 b)!(2 a-2 b+1)!}{(2 a+1)!(b+\beta)!(b-\beta)!(a-b+\alpha+\beta)!(a-b-\alpha-\beta)!}\right]^{\frac{1}{2}}. \label{eq:8:5:13}
\end{equation}
The following two formulas are equivalent to \eqref{eq:8:5:13} ( $i$ and $n$ are integer)
\begin{gather}
\clebsch{a}{a-n+i}{b}{b-i+1}{a-b}{a+b-n+1}=(-1)^{2 b-i+1}\left[\frac{2 a-2 b+1}{2 a+1} \frac{\binom{2 b}{i-1}\binom{2 a-2 b}{2 a-n+1}}{\binom{2 a}{n-i}}\right]^{\frac{1}{2}}, \label{eq:8:5:14}\\
\clebsch{a}{a-n+i}{b}{-b-i+1}{a-b}{a-b-n+1}=(-1)^{i+1}\left[\frac{2 a-2 b+1}{2 a+1} \frac{\binom{2 b}{-i+1}\binom{2 a-2 b}{n-1}}{\binom{2 a}{n-i}}\right]^{\frac{1}{2}}. \label{eq:8:5:15}
\end{gather}
In particular
\begin{equation}
\clebsch{a}{a}{b}{-b}{a-b}{a-b}=\sqrt{\frac{2 a-2 b+1}{2 a+1}} . \label{eq:8:5:16}
\end{equation}
\subsubsection{$c=a-b+1(a+1 \geq b)$}
\begin{equation}
\clebsch{a}{\alpha}{b}{\beta}{a-b+1}{\alpha+\beta}=(-1)^{b+\beta+1} 2(a \beta+b \alpha+\beta)\left[\frac{(2 a-2 b+3)(2 b-1)!(2 a-2 b+1)!(a+\alpha)!(a-\alpha)!}{(2 a+2)!(b+\beta)!(b-\beta)!(a-b+\alpha+\beta+1)!(a-b-\alpha-\beta+1)!}\right]^{\frac{1}{2}} . \label{eq:8:5:17}
\end{equation}
In particular,
\begin{equation}
\clebsch{a}{a}{b}{-b}{a-b+1}{a-b}=\left[\frac{(2 a-2 b+3) 2 b}{(2 a+2)(2 a+1)}\right]^{\frac{1}{2}}. \label{eq:8:5:18}
\end{equation}
\subsubsection{$c=a+b-2$}
\begin{align}
\clebsch{a}{\alpha}{b}{\beta}{a+b-2}{\alpha+\beta} & =\left[\frac{2 a(2 a-1) 2 b(2 b-1)}{2(2 a+2 b-2)(2 a+2 b-1)}\right]^{\frac{1}{2}}\left[\binom{2 a}{a-\alpha}\binom{2 b}{b-\beta}\binom{2 a+2 b-4}{a+b-\alpha-\beta-2}\right]^{-\frac{1}{2}}  \notag\\
& \times\left\{\binom{2 a-2}{a-\alpha}\binom{2 b-2}{b+\beta}-2\binom{2 a-2}{a-\alpha-1}\binom{2 b-2}{b+\beta-1}+\binom{2 a-2}{a-\alpha-2}\binom{2 b-2}{b+\beta-2}\right\} . \label{eq:8:5:19}
\end{align}
In particular,
\begin{equation}
\clebsch{a}{a}{b}{-b}{a+b-2}{a-b}=\left[\frac{(2 a)!(2 b)!(2 a+2 b-3)}{2(2 a+2 b-1)!}\right]^{\frac{1}{2}}. \label{eq:8:5:20}
\end{equation}
\subsubsection{$c=a-b+2(a+2 \geq b)$}
\begin{gather}
\clebsch{a}{\alpha}{b}{\beta}{a-b+2}{\alpha+\beta}=\left[\frac{(2 b-1) 2 b(2 a-2 b+5)(2 a-2 b+4)(2 a-2 b+3)}{2(2 a+1)(2 a+2)(2 a+3)}\right]^{\frac{1}{2}}  \notag\\
\times\left[\binom{2 a}{a-\alpha}\binom{2 b}{b-\beta}\binom{2 a-2 b+4}{a-b-\alpha-\beta+2}\right]^{-\frac{1}{2}}(-1)^{b+\beta}\left\{\binom{2 b-2}{b+\beta}\binom{2 a-2 b+2}{a-b-\alpha-\beta}\right. \notag\\
\left.-2\binom{2 b-2}{b+\beta-1}\binom{2 a-2 b+2}{a-b-\alpha-\beta+1}+\binom{2 b-2}{b+\beta-2}\binom{2 a-2 b+2}{a-b-\alpha-\beta+2}\right\} . \label{eq:8:5:21}
\end{gather}
In particular,
\begin{equation}
\clebsch{a}{a}{b}{-b}{a-b+2}{a-b}=\left[\frac{(2 a-2 b+5) 2 b(2 b-1)}{(2 a+1)(2 a+2)(2 a+3)}\right]^{\frac{1}{2}}. \label{eq:8:5:22}
\end{equation}
\subsubsection{$a=b, \alpha=\beta$}
\begin{equation}
\clebsch{a}{\alpha}{a}{\alpha}{c}{\gamma}= \begin{cases}0, & \text { if } 2 a+c=2 g+1, \label{eq:8:5:23}\\ \delta_{\gamma, 2 \alpha} \frac{(-1)^{g-c} \sqrt{2 c+1} g!}{\left(\frac{c+\gamma}{2}\right)!\left(\frac{\varepsilon-\gamma}{2}\right)!(g-c)!}\left[\frac{(c+\gamma)!(c-\gamma)!(2 g-2 c)!}{(2 g+1)!}\right]^{\frac{1}{2}}, & \text { if } 2 a+c=2 g,\end{cases}
\end{equation}
$g$ is integer and positive.\\
\subsubsection{$b=a+\frac{1}{2}$, $\beta=\alpha \pm \frac{1}{2}$, $2 a+c=2 g$}
\begin{align}
& \clebsch{a}{\alpha}{a+\frac{1}{2}}{\alpha+\frac{1}{2}}{c+\frac{1}{2}}{2 \alpha+\frac{1}{2}}=\left[\frac{c+2 \alpha+1}{a+\alpha+1}\right]^{\frac{1}{2}} C, \label{eq:8:5:24}\\
& \clebsch{a}{a}{a+\frac{1}{2}}{\alpha-\frac{1}{2}}{c+\frac{1}{2}}{2 \alpha-\frac{1}{2}}=\left[\frac{c-2 \alpha+1}{a-\alpha+1}\right]^{\frac{1}{2}} C, \notag
\end{align}
where
\begin{equation}
C=\frac{(-1)^{c / 2-a}\left(a+\frac{c}{2}\right)!}{\left(a-\frac{c}{2}\right)!\left(\frac{c}{2}+\alpha\right)!\left(\frac{c}{2}-\alpha\right)!}\left[\frac{(2 a+c+2)(2 a-c)!(c+2 \alpha)!(c-2 \alpha)!}{2(2 a+c+1)!}\right]^{\frac{1}{2}}, \label{eq:8:5:25}
\end{equation}
\subsubsection{$b=a+\frac{1}{2}$, $\beta=\alpha \pm \frac{1}{2}$, $2 a+c=2 g+1$}
\begin{align}
& \clebsch{a}{\alpha}{a+\frac{1}{2}}{\alpha+\frac{1}{2}}{c+\frac{1}{2}}{2 \alpha+\frac{1}{2}}=\left[\frac{c-2 \alpha+1}{a+\alpha+1}\right]^{\frac{1}{2}} D, \notag\\
& \clebsch{a}{\alpha}{a+\frac{1}{2}}{\alpha-\frac{1}{2}}{c+\frac{1}{2}}{2 \alpha-\frac{1}{2}}=-\left[\frac{c+2 \alpha+1}{a-\alpha+1}\right]^{\frac{1}{2}} D, \label{eq:8:5:26}
\end{align}
where
\begin{equation}
D=\frac{(-1)^{\frac{a-1}{2}-a}\left(a+\frac{c+1}{2}\right)!}{\left(a-\frac{c+1}{2}\right)!\left(\frac{c+1}{2}+\alpha\right)!\left(\frac{c+1}{2}-\alpha\right)!}\left[\frac{(2 a-c)!(c+2 \alpha+1)!(c-2 \alpha+1)!}{2(2 a+c+2)!}\right]^{\frac{1}{2}}. \label{eq:8:5:27}
\end{equation}
\subsubsection{$b=a+1$; $\beta=\alpha$, $\alpha \pm 1$; $2 a+c=2 g$}
\begin{align}
\clebsch{a}{\alpha}{a+1}{a+1}{c}{2 a+1} & =-\left[\frac{(c+2 \alpha+1)(c-2 \alpha)}{(a+\alpha+2)(a+\alpha+1)}\right]^{\frac{1}{2}} E, \notag\\
\clebsch{a}{\alpha}{a+1}{\alpha}{c}{2 \alpha} & =\frac{2 \alpha}{[(a+\alpha+1)(a-\alpha+1)]^{\frac{1}{2}}} E, \label{eq:8:5:28}\\
\clebsch{a}{\alpha}{a+1}{\alpha-1}{c}{2 \alpha-1} & =\left[\frac{(c-2 \alpha+1)(c+2 \alpha)}{(a-\alpha+2)(a-\alpha+1)}\right]^{\frac{1}{2}} E, \notag
\end{align}
where
\begin{equation}
E=\frac{(-1)^{\frac{c}{2}-a}\left(a+\frac{c}{2}\right)!}{2\left(a-\frac{c}{2}\right)!\left(\frac{c}{2}+\alpha\right)!\left(\frac{c}{2}-\alpha\right)!}\left[\frac{(2 c+1)(2 a+c+2)(2 a-c+1)!(c+2 \alpha)!(c-2 \alpha)!}{c(c+1)(2 a+c+1)!}\right]^{\frac{1}{2}}. \label{eq:8:5:29}
\end{equation}
\subsubsection{$b=a+1$; $\beta=\alpha$, $\alpha \pm 1$; $2 a+c=2 g+1$}
\begin{align}
\clebsch{a}{\alpha}{a+1}{\alpha+1}{c}{2 \alpha+1} & =\frac{c(c+1)+(2 \alpha+1)(2 a+2)}{2\left(\frac{c+1}{2}+\alpha\right)!\left(\frac{c+1}{2}-\alpha\right)!}\left[\frac{(c+2 \alpha+1)!(c-2 \alpha-1)!}{(a+\alpha+2)(a+\alpha+1)}\right]^{\frac{1}{2}} F, \notag\\
\clebsch{a}{\alpha}{a+1}{\alpha}{c}{2 \alpha} & =\frac{2 a+2}{\left(\frac{c-1}{2}+\alpha\right)!\left(\frac{c-1}{2}-\alpha\right)!}\left[\frac{(c+2 \alpha)!(c-2 \alpha)!}{(a+\alpha+1)(a-\alpha+1)}\right]^{\frac{1}{2}} F, \label{eq:8:5:30}\\
\clebsch{a}{\alpha}{a+1}{\alpha-1}{c}{2 \alpha-1} & =\frac{c(c+1)-(2 \alpha-1)(2 a+2)}{2\left(\frac{c+1}{2}-\alpha\right)!\left(\frac{c-1}{2}+\alpha\right)!}\left[\frac{(c+2 \alpha-1)!(c-2 \alpha+1)!}{(a-\alpha+2)(a-\alpha+1)}\right]^{\frac{1}{2}} F, \notag
\end{align}
where
\begin{equation}
F=\frac{(-1)^{\frac{c-1}{2}-a}\left(a+\frac{c+1}{2}\right)!}{\left(a-\frac{c-1}{2}\right)!}\left[\frac{(2 c+1)(2 a-c+1)!}{c(c+1)(2 a+c+2)!}\right]^{\frac{1}{2}}. \label{eq:8:5:31}
\end{equation}
Equations \eqref{eq:8:5:24}-\eqref{eq:8:5:31} were obtained by Stone \cite{ref107}.

\subsection{Special Values of Momentum Projections}\label{sec:8.5.2}\index{Clebsch-Gordan coefficient!for special projections}
\subsubsection{$\alpha=\beta=\gamma=0$}
\begin{equation}
\clebsch{a}{0}{b}{0}{c}{0}= \begin{cases}0, & \text { if } a+b+c=2 g+1  \\ \frac{(-1)^{g-c} \sqrt{2 c+1} g!}{(g-a)!(g-b)!(g-c)!}\left[\frac{(2 g-2 a)!(2 g-2 b)!(2 g-2 c)!}{(2 g+1)!}\right]^{\frac{1}{2}}, & \text { if } a+b+c=2 g\end{cases}, \label{eq:8:5:32}
\end{equation}
$g$ is integer and positive. In particular,
\begin{gather}
\clebsch{a}{0}{b}{0}{a+b}{0}=\frac{(a+b)!}{a!b!}\left[\frac{(2 a)!(2 b)!}{(2 a+2 b)!}\right]^{\frac{1}{2}}, \label{eq:8:5:33}\\
\clebsch{a}{0}{b}{0}{a-b}{0}=(-1)^{b} \frac{a!}{b!(a-b)!}\left[\frac{(2 b)!(2 a-2 b+1)!}{(2 a+1)!}\right]^{\frac{1}{2}}. \label{eq:8:5:34}
\end{gather}
\subsubsection{ $\gamma=c$ or $\alpha=a$ :}
\begin{gather}
\clebsch{a}{\alpha}{b}{\beta}{c}{c}=\delta_{\alpha+\beta, c}(-1)^{a-\alpha}\left[\frac{(2 c+1)!(a+b-c)!(a+\alpha)!(b+\beta)!}{(a+b+c+1)!(a-b+c)!(-a+b+c)!(a-\alpha)!(b-\beta)!}\right]^{\frac{1}{2}}, \label{eq:8:5:35}\\
\clebsch{a}{a}{b}{\beta}{c}{\gamma}=\delta_{\gamma-\beta, \alpha}\left[\frac{(2 c+1)(2 a)!(-a+b+c)!(b-\beta)!(c+\gamma)!}{(a+b+c+1)!(a-b+c)!(a+b-c)!(b+\beta)!(c-\gamma)!}\right]^{\frac{1}{2}}. \label{eq:8:5:36}
\end{gather}
In particular,
\begin{gather}
\clebsch{a}{a}{b}{b}{c}{c}=\delta_{a+b, c}, \label{eq:8:5:37}\\
\clebsch{a}{a}{b}{-b}{c}{c}=\delta_{a-b, c}\left[\frac{2 c+1}{2 a+1}\right]^{\frac{1}{2}}, \label{eq:8:5:38}\\
\clebsch{a}{a}{b}{c-a}{c}{c}=\left[\frac{(2 a)!(2 c+1)!}{(a+b+c+1)!(a-b+c)!}\right]^{\frac{1}{2}}, \label{eq:8:5:39}\\
\clebsch{a}{a-1}{b}{c-a+1}{c}{c}=-\left[\frac{(2 a-1)!(2 c+1)!(a+b-c)(-a+b+c+1)}{(a+b+c+1)!(a-b+c)!}\right]^{\frac{1}{2}}, \label{eq:8:5:40}\\
\clebsch{a}{a}{b}{c-a-1}{c}{c-1}=\left[\frac{(2 a)!(2 c+1)!(a+b-c+1)(-a+b+c)}{(a+b+c+1)!(a-b+c)!2 c}\right]^{\frac{1}{2}}, \label{eq:8:5:41}\\
\clebsch{c}{c}{b}{0}{c}{c}=(2 c)!\left[\frac{2 c+1}{(2 c-b)!(2 c+b+1)!}\right]^{\frac{1}{2}}, \label{eq:8:5:42}\\
\clebsch{c}{c-b}{b}{b}{c}{c}=(-1)^{b}\left[\frac{(2 c+1)!(2 b)!}{(2 c+b+1)!b!}\right]^{\frac{1}{2}}, \label{eq:8:5:43}
\end{gather}
\subsubsection{ $\gamma=c-1$ or $\alpha=a-1$ :}
\begin{align}
\clebsch{a}{\alpha}{b}{\beta}{c}{c-1}= & \delta_{\alpha+\beta, c-1}(-1)^{a-\alpha}\{(b-\beta)(b+\beta+1)-(a-\alpha)(a+\alpha+1)\}  \notag\\
& \times\left[\frac{(2 c+1)(2 c-1)!(a+b-c)!(a+\alpha)!(b+\beta)!}{(a+b+c+1)!(a-b+c)!(-a+b+c)!(a-\alpha)!(b-\beta)!}\right]^{\frac{1}{2}}, \label{eq:8:5:44}\\
\clebsch{a}{a-1}{b}{\beta}{c}{\gamma}= & \delta_{\gamma-\beta, a-1}\{(c-\gamma)(c+\gamma+1)-(b+\beta)(b-\beta+1)\}  \notag\\
& \times\left[\frac{(2 c+1)(2 a-1)!(-a+b+c)!(b-\beta)!(c+\gamma)!}{(a+b+c+1)!(a-b+c)!(a+b-c)!(b+\beta)!(c-\gamma)!}\right]^{\frac{1}{2}}, \label{eq:8:5:45}
\end{align}
In particular,
\begin{gather}
\clebsch{a}{\alpha}{a}{\alpha}{c}{c-1}=0, \label{eq:8:5:46}\\
\clebsch{a}{a-1}{c}{-\gamma}{c}{\gamma}=0, \label{eq:8:5:47}\\
\clebsch{a}{a-1}{b}{\beta}{c}{c-1}=\delta_{\beta, c-a}\{a(a+1)-b(b+1)+c(c+1)-2 a c\}\left[\frac{(2 c+1)(2 c-1)!(2 a-1)!}{(a+b+c+1)!(a-b+c)!}\right]^{\frac{1}{2}} . \label{eq:8:5:48}
\end{gather}
\section{RECURSION RELATIONS FOR THE CLEBSCH-GORDAN COEFFICIENTS}\label{sec:8.6}\index{Clebsch-Gordan coefficient!recursion relations}\index{recursion relations!for the Clebsch-Gordan coefficients}
\subsection{General Recursion Relations}\label{sec:8.6.1}\index{recursion relations!general}
\begin{align}
\clebsch{a}{\alpha}{b}{\beta}{c}{\gamma}= & {\left[\frac{(b+\beta-2 k)!(c+\gamma)!(a+b-c)!(-a+b+c)!(a+b+c+1)!(2 c+1)}{(b+\beta)!(c-\gamma)!(a-b+c)!}\right]^{\frac{1}{2}} }  \notag\\
& \times \sum_{c^{\prime}=c-k}^{c+k}(-1)^{c^{\prime}+k-c} \clebsch{a}{\alpha}{b-k}{\beta-k}{c^{\prime}}{\gamma-k} \frac{\left(c^{\prime}-k+c\right)!(2 k)!}{\left(c+k-c^{\prime}\right)!\left(c+c^{\prime}+k+1\right)!\left(c^{\prime}+k-c\right)!}  \notag\\
& \times\left[\frac{\left(c^{\prime}+k-\gamma\right)!\left(a-b+c^{\prime}+k\right)!\left(2 c^{\prime}+1\right)}{\left(c^{\prime}-k+\gamma\right)!\left(-a+b+c^{\prime}-k\right)!\left(a+b-c^{\prime}-k\right)!\left(a+b+c^{\prime}-k+1\right)!}\right]^{\frac{1}{2}}. \label{eq:8:6:1}
\end{align}
(Yutsis and Bandzaitis \cite{ref045}) where $k$ is integer or half-integer, $0 \leq k \leq(b+\beta) / 2$. In accordance with the choice of $k$ \eqref{eq:8:6:1} yields different recursion relations.

Note also the relation obtained by Stone \cite{ref107}
\begin{align}
\clebsch{a}{\alpha}{b}{\beta}{c}{\gamma}= & {\left[\frac{(a+b+c+1)!(a-b+c)!(a+b-c)!(b+\beta)!(b-\beta)!(a-b+\alpha-\beta)!(a-b-\alpha+\beta)!}{(-a+b+c)!(a+\alpha)!(a-\alpha)!}\right]^{\frac{1}{2}} }  \notag\\
& \times \sum_{b^{\prime}}(-1)^{b-b^{\prime} / 2} 
\clebsch{a-b}{\alpha-\beta}{b^{\prime}}{2 \beta}{c}{\gamma} \frac{\left(2 b^{\prime}+1\right)\left(b+\frac{b^{\prime}}{2}\right)!}{\left(2 b+b^{\prime}+1\right)!\left(b-\frac{b^{\prime}}{2}\right)!\left(\frac{b^{\prime}}{2}+\beta\right)!\left(\frac{b^{\prime}}{2}-\beta\right)!}  \notag\\
& \times\left[\frac{\left(-a+b+c+b^{\prime}\right)!\left(b^{\prime}+2 \beta\right)!\left(b^{\prime}-2 \beta\right)!}{\left(a-b+c+b^{\prime}+1\right)!\left(a-b+c-b^{\prime}\right)!\left(a-b-c+b^{\prime}\right)!}\right]^{\frac{1}{2}}. \label{eq:8:6:2}
\end{align}
Here $a-b \geq|\alpha-\beta| \geq 0, b^{\prime}$ is integer, $b^{\prime}+2 b$ is even, and $a-b+c \geq b^{\prime} \geq|-a+b+c|, 2 b \geq b^{\prime} \geq|2 \beta|$.\\[0pt]
Equation \eqref{eq:8:6:1} may be rewritten by the use of quasi-powers in the form \cite{ref045}
\begin{align}
\clebsch{a}{\alpha}{b}{\beta}{c}{\gamma}= & { \left.\left[\frac{2 c+1}{(b+\beta)^{(2 k)}}\right]^{\frac{1}{2}} \sum_{k^{\prime}=-k}^{k}(-1)^{k+k^{\prime}} \clebsch{a}{\alpha}{b-k}{\beta-k}{c+k^{\prime}}{\gamma-k} \right\rvert\,(c+\gamma)^{\left(k-k^{\prime}\right)}\left(c-\gamma+k+k^{\prime}\right)^{\left(k+k^{\prime}\right)}(a+b-c)^{\left(k+k^{\prime}\right)} }  \notag\\
& \left.\times(-a+b+c)^{\left(k-k^{\prime}\right)}(a+b+c+1)^{\left(k-k^{\prime}\right)}\left(a-b+c+k+k^{\prime}\right)^{\left(k+k^{\prime}\right)}\left(2 c+2 k^{\prime}+1\right)\right]^{\frac{1}{2}}  \notag\\
& \times \frac{(2 k)^{\left(k+k^{\prime}\right)}}{\left(2 c+k+k^{\prime}+1\right)^{(2 k+1)}\left(k+k^{\prime}\right)^{\left(k+k^{\prime}\right)}}, \label{eq:8:6:3}
\end{align}
where
\[
a^{(n)}=\frac{a!}{(a-n)!}, \quad \frac{1}{2} \leq k \leq \frac{b-\kappa}{2}, \quad \kappa= \begin{cases}0 & \text { if } b \text { is integer }, \\ \frac{1}{2} & \text { if } b \text { is half-integer } .\end{cases}
\]
\subsection{Arguments $\alpha, \beta, \gamma$ Change by 1}\label{sec:8.6.2}\index{recursion relations!in the projections}
\begin{equation}
[(c \pm \gamma)(c \mp \gamma+1)]^{\frac{1}{2}} \clebsch{a}{\alpha}{b}{\beta}{c}{\gamma \mp 1}=[(a \mp \alpha)(a \pm \alpha+1)]^{\frac{1}{2}} \clebsch{a}{\alpha \pm 1}{b}{\beta}{c}{\gamma}+[(b \mp \beta)(b \pm \beta+1)]^{\frac{1}{2}} \clebsch{a}{\alpha}{b}{\beta \pm 1}{c}{\gamma}. \label{eq:8:6:4}
\end{equation}
In particular, for $|\gamma|=c$ one has
\begin{equation}
\clebsch{a}{\alpha \mp 1}{ b}{\beta}{c}{ \pm c}=-
\clebsch{a}{\alpha}{ b}{\beta\mp 1}{c}{\pm c}\left[\frac{(b \pm \beta)(b \mp \beta+1)}{(a \pm \alpha)(a \mp \alpha+1)}\right]^{\frac{1}{2}}, \label{eq:8:6:5}
\end{equation}
when $|\alpha|=|\beta|=\frac{1}{2}$
\begin{equation}
\clebsch{a}{\frac{1}{2}}{b}{\frac{1}{2}}{c}{1}=\clebsch{a}{\frac{1}{2}}{b}{-\frac{1}{2}}{c}{0} \frac{(2 b+1)+(-1)^{a+b-c}(2 a+1)}{2[c(c+1)]^{\frac{1}{2}}}, \label{eq:8:6:6}
\end{equation}
and for $|\alpha|=|\beta|=1, a+b+c$ even, one gets
\begin{gather}
\clebsch{a}{1}{b}{-1}{c}{0}=\clebsch{a}{0}{b}{0}{c}{0} \frac{c(c+1)-a(a+1)-b(b+1)}{2[a(a+1) b(b+1)]^{\frac{1}{2}}}, \label{eq:8:6:7}\\
\clebsch{a}{1}{b}{1}{c}{2}=\clebsch{a}{0}{b}{0}{c}{0} \frac{a(a+1)[c(c+1)-a(a+1)+b(b+1)]+b(b+1)[c(c+1)+a(a+1)-b(b+1)]}{2[a(a+1) b(b+1)(c-1) c(c+1)(c+2)]^{\frac{1}{2}}}. \label{eq:8:6:8}
\end{gather}
\subsection{Arguments Change by $1/2$}\label{sec:8.6.3}\index{recursion relations!in half-integer steps}
\begin{gather}
(2 a+1)[b \pm \beta]^{\frac{1}{2}} \clebsch{a}{\alpha}{b}{\beta}{c}{\gamma}=\mp[(a \mp \alpha)(a+b-c)(a+b+c+1)]^{\frac{1}{2}} \clebsch{a-\frac{1}{2}}{\alpha \pm \frac{1}{2}}{b-\frac{1}{2}}{\beta \mp \frac{1}{2}}{c}{\gamma}  \notag\\
+[(a \pm \alpha+1)(-a+b+c)(a-b+c+1)]^{\frac{1}{2}} \clebsch{a+\frac{1}{2}}{\alpha \pm \frac{1}{2}}{b-\frac{1}{2}}{\beta \mp \frac{1}{2}}{c}{\gamma}  \label{eq:8:6:9}
\end{gather}
\begin{gather}
(2 a+1)[b \mp \beta+1]^{\frac{1}{2}} \clebsch{a}{\alpha}{b}{\beta}{c}{\gamma}=[(a \mp \alpha)(a-b+c)(-a+b+c+1)]^{\frac{1}{2}} \clebsch{a-\frac{1}{2}}{\alpha \pm \frac{1}{2}}{b+\frac{1}{2}}{\beta \mp \frac{1}{2}}{c}{\gamma}  \notag\\
\pm[(a \pm \alpha+1)(a+b-c+1)(a+b+c+2)]^{\frac{1}{2}} \clebsch{a+\frac{1}{2}}{\alpha \pm \frac{1}{2}}{b+\frac{1}{2}}{\beta \mp \frac{1}{2}}{c}{\gamma}  \label{eq:8:6:10}
\end{gather}
\begin{multline}
{[(-a+b+c)(a-b+c+1)]^{\frac{1}{2}} \clebsch{a}{\alpha}{b}{\beta}{c}{\gamma}=[(a-\alpha+1)(b-\beta)]^{\frac{1}{2}} \clebsch{a+\frac{1}{2}}{\alpha-\frac{1}{2}}{b-\frac{1}{2}}{\beta+\frac{1}{2}}{c}{\gamma}}  \\
+[(a+\alpha+1)(b+\beta)]^{\frac{1}{2}} \clebsch{a+\frac{1}{2}}{\alpha+\frac{1}{2}}{b-\frac{1}{2}}{\beta-\frac{1}{2}}{c}{\gamma}  \label{eq:8:6:11}
\end{multline}
\begin{multline}
{[(a-b+c)(-a+b+c+1)]^{\frac{1}{2}} \clebsch{a}{\alpha}{b}{\beta}{c}{\gamma}=[(a-\alpha)(b-\beta+1)]^{\frac{1}{2}} \clebsch{a-\frac{1}{2}}{\alpha+\frac{1}{2}}{b+\frac{1}{2}}{\beta-\frac{1}{2}}{c}{\gamma}}  \\
+[(a+\alpha)(b+\beta+1)]^{\frac{1}{2}} \clebsch{a-\frac{1}{2}}{\alpha-\frac{1}{2}}{b+\frac{1}{2}}{\beta+\frac{1}{2}}{c}{\gamma}  \label{eq:8:6:12}
\end{multline}
\begin{multline}
\left[\frac{2 c(-a+b+c)(a+b+c+1)}{2 c+1}\right]^{\frac{1}{2}}
\clebsch{a}{\alpha}{b}{\beta}{c}{\gamma}=
[(b-\beta)(c-\gamma)]^{\frac{1}{2}} 
\clebsch{a}{\alpha}{b-\frac{1}{2}}{\beta+\frac{1}{2}}{c-\frac{1}{2}}{\gamma+\frac{1}{2}}
+[(b+\beta)(c+\gamma)]^\frac{1}{2} \clebsch{a}{\alpha}{b-\frac{1}{2}}{\beta-\frac{1}{2}}{c-\frac{1}{2}}{\gamma-\frac{1}{2}} , \label{eq:8:6:13}
\end{multline}
\begin{multline}
{\left[\frac{2 c(c \mp \gamma)(a+b+c+1)}{2 c+1}\right]}^{\frac{1}{2}}
\clebsch{a}{\alpha}{b}{\beta}{c}{\gamma}
=[(a \mp \alpha)(a-b+c)]^{\frac{1}{2}} 
\clebsch{a-\frac{1}{2}}{\alpha\pm\frac{1}{2}}{b}{\beta}{c-\frac{1}{2}}{\gamma\pm\frac{1}{2}}+
[(b\mp\beta)(-a+b+c)]^{\frac{1}{2}} 
\clebsch{a}{\alpha}{b-\frac{1}{2}}{\beta\pm\frac{1}{2}} {c-\frac{1}{2}}{\gamma\pm\frac{1}{2}}, \label{eq:8:6:14}
\end{multline}
\begin{multline}
  {\left[\frac{2 c(c \mp \gamma)(a+b-c+1)}{2 c+1}\right]}^{\frac{1}{2}}
\clebsch{a}{\alpha}{b}{\beta}{c}{\gamma}
=\pm[(a \pm \alpha+1)(-a+b+c)]^{\frac{1}{2}} 
\clebsch{a+\frac{1}{2}}{\alpha\pm\frac{1}{2}}{b}{\beta}{c-\frac{1}{2}}{\gamma\pm\frac{1}{2}}\\
\mp[(b \pm \beta+1)(a-b+c)]^{\frac{1}{2}} \clebsch{a}{\alpha}{b+\frac{1}{2}}{\beta \pm \frac{1}{2}}{c-\frac{1}{2}}{\gamma \pm \frac{1}{2}} , \label{eq:8:6:15}
\end{multline}
\begin{multline}
{\left[\frac{2(c+1)(c \pm \gamma+1)(a+b-c)}{2 c+1}\right]^{\frac{1}{2}}\clebsch{a}{\alpha}{b}{\beta}{c}{\gamma}
=\mp[(a \mp \alpha)(-a+b+c+1)]^{\frac{1}{2}} \clebsch{a-\frac{1}{2}}{\alpha \pm \frac{1}{2}}{b}{\beta}{c+\frac{1}{2}}{\gamma \pm \frac{1}{2}}} \\
\pm[(b \mp \beta)(a-b+c+1)]^{\frac{1}{2}} \clebsch{a}{\alpha}{b-\frac{1}{2}}{\beta \pm \frac{1}{2}}{c+\frac{1}{2}}{\gamma \pm \frac{1}{2}}, \label{eq:8:6:16}
\end{multline}
\begin{multline}
{\left[(2c+1)(b\mp\beta)\right]^{\frac{1}{2}}\clebsch{a}{\alpha}{b}{\beta}{c}{\gamma}=\left[\frac{(c\mp\gamma)(-a+b+c)(a+b+c+1)}{2c}\right]^{\frac{1}{2}} \clebsch{a}{\alpha}{b-\frac{1}{2}}{\beta \pm \frac{1}{2}}{c-\frac{1}{2}}{\gamma \pm \frac{1}{2}}} \\
\pm\left[\frac{(c \pm \gamma+1)(a-b+c+1)(a+b-c)}{2(c+1)}\right]^{\frac{1}{2}}
\clebsch{a}{\alpha}{b-\frac{1}{2}}{\beta \pm \frac{1}{2}}{c+\frac{1}{2}}{\gamma\pm\frac{1}{2}}. 
\label{eq:8:6:17}
\end{multline}
In particular, Eqs.~\eqref{eq:8:6:9} and \eqref{eq:8:6:10} yield
\begin{equation}
\clebsch{a}{\pm \frac{1}{2}}{b}{\mp \frac{1}{2}}{c}{0}= \pm \clebsch{a-\frac{1}{2}}{0}{b-\frac{1}{2}}{0}{c}{0}\left[\frac{(a+b-c)(a+b+c+1)}{(2 a+1)(2 b+1)}\right]^{\frac{1}{2}}=\mp \clebsch{a+\frac{1}{2}}{0}{b+\frac{1}{2}}{0}{c}{0}\left[\frac{(a+b-c+1)(a+b+c+2)}{(2 a+1)(2 b+1)}\right]^{\frac{1}{2}}, \label{eq:8:6:18}
\end{equation}
if $a+b+c$ is odd, and
\begin{equation}
\clebsch{a}{\pm \frac{1}{2}}{b}{\mp \frac{1}{2}}{c}{0}=\clebsch{a+\frac{1}{2}}{0}{b-\frac{1}{2}}{0}{c}{0}\left[\frac{(-a+b+c)(a-b+c+1)}{(2 a+1)(2 b+1)}\right]^{\frac{1}{2}}=\clebsch{a-\frac{1}{2}}{0}{b+\frac{1}{2}}{0}{c}{0}\left[\frac{(a-b+c)(-a+b+c+1)}{(2 a+1)(2 b+1)}\right]^{\frac{1}{2}}, \label{eq:8:6:19}
\end{equation}
if $a+b+c$ is even.

\subsection{The Case When $\alpha=\beta=\gamma=0$}\label{sec:8.6.4}\index{recursion relations!for zero projections}
Let $p$ be integer and $2 g=a+b+c$. Then
\begin{equation}
\clebsch{a+p}{0}{b-p}{0}{c}{0}=\clebsch{a}{0}{b}{0}{c}{0} \frac{(g-a)!(g-b)!}{(g-a-p)!(g-b+p)!}\left[\frac{(2 g-2 a-2 p)!(2 g-2 b+2 p)!}{(2 g-2 a)!(2 g-2 b)!}\right]^{\frac{1}{2}}. \label{eq:8:6:20}
\end{equation}
In particular, \eqref{eq:8:6:20} yields
\begin{gather}
\clebsch{a+1}{0}{b-1}{0}{c}{0}=\clebsch{a}{0}{b}{0}{c}{0}\left[\frac{(-a+b+c)(a-b+c+1)}{(-a+b+c-1)(a-b+c+2)}\right]^{\frac{1}{2}}, \label{eq:8:6:21}\\
\clebsch{a}{0}{b+2p}{0}{c}{0}=\clebsch{a}{0}{b}{0}{c}{0}(-1)^{p} \frac{(g+p)!(g-a)!(g-b)!(g-c)!}{g!(g-a+p)!(g-b-p)!(g-c+p)!}, \notag\\
\times\left[\frac{(a+b-c+2 p)!(a-b+c-2 p)!(-a+b+c+2 p)!(a+b+c+1)!}{(a+b-c)!(a-b+c)!(-a+b+c)!(a+b+c+2 p+1)!}\right]^{\frac{1}{2}}. \label{eq:8:6:22}
\end{gather}
Moreover, from \eqref{eq:8:6:22} one gets
\begin{equation}
\clebsch{a}{0}{b+2}{0}{c}{0}=-\clebsch{a}{0}{b}{0}{c}{0}\left[\frac{(a+b+c+2)(a+b-c+1)(a-b+c)(-a+b+c+1)}{(a+b+c+3)(a+b-c+2)(a-b+c-1)(-a+b+c+2)}\right]^{\frac{1}{2}}. \label{eq:8:6:23}
\end{equation}
\subsection{Arguments $a, b, c$ Change by 1}\label{sec:8.6.5}\index{recursion relations!in the momenta}
\begin{align}
& 2\left[b^{2}-\beta^{2}\right]^{\frac{1}{2}} \clebsch{a}{\alpha}{b-1}{\beta}{c}{\gamma}=\frac{1}{a(2 a+1)}\left[\left(a^{2}-\alpha^{2}\right)(-a+b+c)(-a+b+c+1)(a-b+c)(a-b+c+1)\right]^{\frac{1}{2}} \notag\\
& \quad \times \clebsch{a-1}{\alpha}{b}{\beta}{c}{\gamma}+\frac{\alpha}{a(a+1)}[(-a+b+c)(a-b+c+1)(a+b-c)(a+b+c+1)]^{\frac{1}{2}} \clebsch{a}{\alpha}{b}{\beta}{c}{\gamma}  \notag\\
& -\frac{1}{(a+1)(2 a+1)}\left\{\left[(a+1)^{2}-\alpha^{2}\right](a+b-c)(a+b-c+1)(a+b+c+1)(a+b+c+2)\right\}^{\frac{1}{2}} \clebsch{a+1}{\alpha}{b}{\beta}{c}{\gamma}, \label{eq:8:6:24}\\
& 2 \beta \clebsch{a}{\alpha}{b}{\beta}{c}{\gamma}=-\frac{1}{a(2 a+1)}\left[\left(a^{2}-\alpha^{2}\right)(-a+b+c+1)(a-b+c)(a+b-c)(a+b+c+1)\right]^{\frac{1}{2}} \clebsch{a-1}{\alpha}{b}{\beta}{c}{\gamma}  \notag\\
& \quad-\frac{\alpha}{a(a+1)}\{a(a+1)+b(b+1)-c(c+1)\} \clebsch{a}{\alpha}{b}{\beta}{c}{\gamma}  \notag\\
& -\frac{1}{(a+1)(2 a+1)}\left\{\left[(a+1)^{2}-\alpha^{2}\right](-a+b+c)(a-b+c+1)(a+b-c+1)(a+b+c+2)\right\}^{\frac{1}{2}} \clebsch{a+1}{\alpha}{b}{\beta}{c}{\gamma}, \label{eq:8:6:25}
\end{align}
\begin{gather}
2\left[(b+1)^{2}-\beta^{2}\right]^{\frac{1}{2}} \clebsch{a}{\alpha}{b+1}{\beta}{c}{\gamma}=-\frac{1}{a(2 a+1)}\left[\left(a^{2}-\alpha^{2}\right)(a+b-c)(a+b-c+1)(a+b+c+1)(a+b+c+2)\right]^{\frac{1}{2}} \notag\\
\times \clebsch{a-1}{\alpha}{b}{\beta}{c}{\gamma}+\frac{\alpha}{a(a+1)}[(-a+b+c+1)(a-b+c)(a+b-c+1)(a+b+c+2)]^{\frac{1}{2}} \clebsch{a}{\alpha}{b}{\beta}{c}{\gamma}  \notag\\
+\frac{1}{(a+1)(2 a+1)}\left\{\left[(a+1)^{2}-\alpha^{2}\right](-a+b+c)(-a+b+c+1)(a-b+c)(a-b+c+1)\right\}^{\frac{1}{2}} \clebsch{a+1}{\alpha}{b}{\beta}{c}{\gamma}, \label{eq:8:6:26}\\
\clebsch{a}{\alpha}{b}{\beta}{c}{\gamma}=\left[\frac{4 c^{2}(2 c+1)(2 c-1)}{(c+\gamma)(c-\gamma)(-a+b+c)(a-b+c)(a+b-c+1)(a+b+c+1)}\right]^{\frac{1}{2}}  \notag\\
\times\left\{\frac{(\alpha-\beta) c(c-1)-\gamma a(a+1)+\gamma b(b+1)}{2 c(c-1)} \clebsch{a}{\alpha}{b}{\beta}{c-1}{\gamma}\right. \notag\\
\left.-\left[\frac{(c-\gamma-1)(c+\gamma-1)(-a+b+c-1)(a-b+c-1)(a+b-c+2)(a+b+c)}{4(c-1)^{2}(2 c-3)(2 c-1)}\right]^{\frac{1}{2}} \clebsch{a}{\alpha}{b}{\beta}{c-2}{\gamma}\right\} . \label{eq:8:6:27}
\end{gather}
\subsection{Arguments $a, b, \alpha, \beta$ Change by 1}\label{sec:8.6.6}\index{recursion relations!in $a, b, \alpha, \beta$}
\begin{gather}
{[(b \pm \beta)(b \pm \beta+1)]^{\frac{1}{2}} \clebsch{a}{\alpha}{b-1}{\beta}{c}{\gamma}}  \notag\\
=\frac{1}{2 a(2 a+1)}[(a \pm \alpha)(a \pm \alpha-1)(-a+b+c)(-a+b+c+1)(a-b+c)(a-b+c+1)]^{\frac{1}{2}} \clebsch{a-1}{\alpha \mp 1}{b}{\beta \pm 1}{c}{\gamma}  \notag\\
\mp \frac{1}{2 a(a+1)}[(a \pm \alpha)(a \mp \alpha+1)(-a+b+c)(a-b+c+1)(a+b-c)(a+b+c+1)]^{\frac{1}{2}} \clebsch{a}{\alpha \mp 1}{b}{\beta \pm 1}{c}{\gamma}  \notag\\
+\frac{1}{2(a+1)(2 a+1)}[(a \mp \alpha+1)(a \mp \alpha+2)(a+b-c)(a+b-c+1)(a+b+c+1)(a+b+c+2)]^{\frac{1}{2}}  \notag\\
\times \clebsch{a+1}{\alpha \mp 1}{b}{\beta \pm 1}{c}{\gamma} . \label{eq:8:6:28}\end{gather}
\begin{gather}
[(b\mp\beta)(b\pm\beta+1)]^{\frac{1}{2}}\clebsch{a}{\alpha}{b}{\beta}{c}{\gamma}\notag\\
= \pm \frac{1}{2 a(2 a+1)}[(a \pm \alpha)(a \pm \alpha-1)(-a+b+c+1)(a-b+c)(a+b-c)(a+b+c+1)]^{\frac{1}{2}} \clebsch{a-1}{\alpha \mp 1}{b}{\beta \pm 1}{c}{\gamma}  \notag\\
\quad-\frac{1}{2 a(a+1)}\{a(a+1)+b(b+1)-c(c+1)\}\left[(a \pm \alpha)(a \mp \alpha+1)\right]^{\frac{1}{2}} \clebsch{a}{\alpha \mp 1}{b}{\beta \pm 1}{c}{\gamma}  \notag\\
\mp \frac{1}{2(a+1)(2 a+1)}[(a \mp \alpha+1)(a \mp \alpha+2)(-a+b+c)(a-b+c+1)(a+b-c+1)(a+b+c+2)]^{\frac{1}{2}}  \notag\\
\quad \times \clebsch{a+1}{\alpha \mp 1}{b}{\beta \pm 1}{c}{\gamma} . \label{eq:8:6:29}
\end{gather}
\begin{gather}
{[(b \mp \beta)(b \mp \beta+1)]^{\frac{1}{2}} \clebsch{a}{\alpha}{b+1}{\beta}{c}{\gamma}}  \notag\\
=\frac{1}{2 a(2 a+1)}[(a \pm \alpha)(a \pm \alpha-1)(a+b-c)(a+b-c+1)(a+b+c+1)(a+b+c+2)]^{\frac{1}{2}} \clebsch{a-1}{\alpha \mp 1}{b}{\beta \pm 1}{c}{\gamma}  \notag\\
\pm \frac{1}{2 a(a+1)}[(a \pm \alpha)(a \mp \alpha+1)(-a+b+c+1)(a-b+c)(a+b-c+1)(a+b+c+2)]^{\frac{1}{2}} \clebsch{a}{\alpha \mp 1}{b}{\beta \pm 1}{c}{\gamma}  \notag\\
+\frac{1}{2(a+1)(2 a+1)}[(a \mp \alpha+1)(a \mp \alpha+2)(-a+b+c)(-a+b+c+1)(a-b+c)(a-b+c+1)]^{\frac{1}{2}}  \notag\\
\times \clebsch{a+1}{\alpha \mp 1}{b}{\beta \pm 1}{c}{\gamma} \label{eq:8:6:30}
\end{gather}
\subsection{Arguments $c, b, \gamma, \beta$ Change by 1}\label{sec:8.6.7}\index{recursion relations!in $c, b, \gamma, \beta$}
\begin{gather}
{\left[\frac{(-a+b+c)(a-b+c)(a+b-c+1)(a+b+c+1)(2 c-1)}{(2 c+1)}\right]^{\frac{1}{2}} \clebsch{a}{\alpha}{b}{\beta}{c}{\gamma}}  \notag\\
=[(b+\beta)(b-\beta+1)(c+\gamma)(c+\gamma-1)]^{\frac{1}{2}} \clebsch{a}{\alpha}{b}{\beta-1}{c-1}{\gamma-1}-2 \beta\left[c^{2}-\gamma^{2}\right]^{\frac{1}{2}} \clebsch{a}{\alpha}{b}{\beta}{c-1}{\gamma}  \notag\\
-[(b-\beta)(b+\beta+1)(c-\gamma)(c-\gamma-1)]^{\frac{1}{2}} \clebsch{a}{\alpha}{b}{\beta+1}{c-1}{\gamma+1} . \label{eq:8:6:31}
\end{gather}
\begin{gather}
{[(b \mp \beta)(b \mp \beta+1)]^{\frac{1}{2}} \clebsch{a}{\alpha}{b-1}{\beta}{c}{\gamma}}  \notag\\
=\frac{1}{2 c}\left[\frac{(c \pm \gamma)(c \pm \gamma-1)(a-b+c)(a-b+c+1)(a+b-c)(a+b-c+1)}{(2 c-1)(2 c+1)}\right]^{\frac{1}{2}} \clebsch{a}{\alpha}{b}{\beta \mp 1}{c-1}{\gamma \mp 1}  \notag\\
\pm \frac{1}{2 c(c+1)}[(c \pm \gamma)(c \mp \gamma+1)(-a+b+c)(a-b+c+1)(a+b-c)(a+b+c+1)]^{\frac{1}{2}} \clebsch{a}{\alpha}{b}{\beta \mp 1}{c}{\gamma \mp 1}  \notag\\
+\frac{1}{2(c+1)}\left[\frac{(c \mp \gamma+1)(c \mp \gamma+2)(-a+b+c)(-a+b+c+1)(a+b+c+1)(a+b+c+2)}{(2 c+1)(2 c+3)}\right]^{\frac{1}{2}} \clebsch{a}{\alpha}{b}{\beta \mp 1}{c+1}{\gamma \mp 1}  \label{eq:8:6:32}
\end{gather}
\begin{gather}
  [(b\pm\beta)(b\mp\beta+1)]^{\frac{1}{2}}\clebsch{a}{\alpha}{b}{\beta}{c}{\gamma}\notag\\
= \pm \frac{1}{2 c}\left[\frac{(c \pm \gamma)(c \pm \gamma-1)(-a+b+c)(a-b+c)(a+b-c+1)(a+b+c+1)}{(2 c-1)(2 c+1)}\right]^{\frac{1}{2}} \clebsch{a}{\alpha}{b}{\beta \mp 1}{c-1}{\gamma \mp 1}  \notag\\
+\frac{1}{2 c(c+1)}\{-a(a+1)+b(b+1)+c(c+1)\}[(c \pm \gamma)(c \mp \gamma+1)]^{\frac{1}{2}} \clebsch{a}{\alpha}{b}{\beta \mp 1}{c}{\gamma \mp 1}  \notag\\
\mp \frac{1}{2(c+1)}\left[\frac{(c \mp \gamma+1)(c \mp \gamma+2)(-a+b+c+1)(a-b+c+1)(a+b-c)(a+b+c+2)}{(2 c+1)(2 c+3)}\right]^{\frac{1}{2}} \clebsch{a}{\alpha}{b}{\beta \mp 1}{c+1}{\gamma \mp 1} \label{eq:8:6:33}
\end{gather}
\begin{gather}
  [(b\pm\beta)(b\pm\beta+1)]^{\frac{1}{2}}\clebsch{a}{\alpha}{b+1}{\beta}{c}{\gamma}\notag\\
=\frac{1}{2 c}\left[\frac{(c \pm \gamma)(c \pm \gamma-1)(-a+b+c)(-a+b+c+1)(a+b+c+1)(a+b+c+2)}{(2 c-1)(2 c+1)}\right]^{\frac{1}{2}} \clebsch{a}{\alpha}{b}{\beta \mp 1}{c-1}{\gamma \mp 1} \notag\\
\mp \frac{1}{2 c(c+1)}[(c \pm \gamma)(c \mp \gamma+1)(-a+b+c+1)(a-b+c)(a+b-c+1)(a+b+c+2)]^{\frac{1}{2}} \clebsch{a}{\alpha}{b}{\beta \mp 1}{c}{\gamma \mp 1}  \notag\\
+\frac{1}{2(c+1)}\left[\frac{(c \mp \gamma+1)(c \mp \gamma+2)(a-b+c)(a-b+c+1)(a+b-c)(a+b-c+1)}{(2 c+1)(2 c+3)}\right]^{\frac{1}{2}} \clebsch{a}{\alpha}{b}{\beta \mp 1}{c+1}{\gamma \mp 1} .  \label{eq:8:6:34}
\end{gather}
\subsection{Recursion Relations for the Regge Symbols}\label{sec:8.6.8}\index{Regge R-symbol@Regge $R$-symbol!recursion relations}
Recursion relations for the Regge $R$-symbols are very convenient because each relation is equivalent to a variety of formally different expressions for the Clebsch-Gordan coefficients. The formulas given below represent some recursion relations $[86,103]$. In these formulas we use the notation $J=\sum_{k=1}^{3} R_{i k}=\sum_{i=1}^{3} R_{i k}=a+b+c$.
\begin{align}
& {\left[R_{11}(J+1)\right]^{\frac{1}{2}}\left\|\begin{array}{lll}
R_{11} & R_{12} & R_{13} \\
R_{21} & R_{22} & R_{23} \\
R_{31} & R_{32} & R_{33}
\end{array}\right\|-\left[R_{22} R_{33}\right]^{\frac{1}{2}}\left\|\begin{array}{lll}
R_{11}-1 & R_{12} & R_{13} \\
R_{21} & R_{22}-1 & R_{23} \\
R_{31} & R_{32} & R_{33}-1
\end{array}\right\|}  \notag\\
& +\left[R_{23} R_{32}\right]^{\frac{1}{2}}\left\|\begin{array}{lll}
R_{11}-1 & R_{12} & R_{13} \\
R_{21} & R_{22} & R_{23}-1 \\
R_{31} & R_{32}-1 & R_{33}
\end{array}\right\|=0 . \label{eq:8:6:35}\\
& {\left[\left(R_{11}+1\right) R_{21}\right]^{\frac{1}{2}}\left\|\begin{array}{lll}
R_{11}+1 & R_{12} & R_{13} \\
R_{21}-1 & R_{22} & R_{23} \\
R_{31} & R_{32} & R_{33}
\end{array}\right\|+\left[\left(R_{12}+1\right) R_{22}\right]^{\frac{1}{2}}\left\|\begin{array}{lll}
R_{11} & R_{12}+1 & R_{13} \\
R_{21} & R_{22}-1 & R_{23} \\
R_{31} & R_{32} & R_{33}
\end{array}\right\|}  \notag\\
& +\left[\left(R_{13}+1\right) R_{23}\right]^{\frac{1}{2}}\left\|\begin{array}{lll}
R_{11} & R_{12} & R_{13}+1 \\
R_{21} & R_{22} & R_{23}-1 \\
R_{31} & R_{32} & R_{33}
\end{array}\right\|=0 .
\footnote{Column recursion — it shifts R₁ₖ+1 and R₂ₖ−1, so each array has equal column sums but row sums J+1, J−1, J. A Regge symbol requires all six line-sums equal, so every term vanishes as a standard 3jm and the relation is only vacuously satisfied.  It's not an OCR issue; the transcription matches the scan.}
\label{eq:8:6:36}\\
& {\left[R_{11} R_{22}(J+1)\right]^{\frac{1}{2}}\left\|\begin{array}{lll}
R_{11} & R_{12} & R_{13} \\
R_{21} & R_{22} & R_{23} \\
R_{31} & R_{32} & R_{33}
\end{array}\right\|-\left(R_{22}+R_{23}\right)\left[R_{33}\right]^{\frac{1}{2}}\left\|\begin{array}{lll}
R_{11}-1 & R_{12} & R_{13} \\
R_{21} & R_{22}-1 & R_{23} \\
R_{31} & R_{32} & R_{33}-1
\end{array}\right\|}  \notag\\
& -\left[R_{23} R_{31}\left(R_{21}+1\right)\right]^{\frac{1}{2}}\left\|\begin{array}{lll}
R_{11}-1 & R_{12} & R_{13} \\
R_{21}+1 & R_{22}-1 & R_{23}-1 \\
R_{31}-1 & R_{32} & R_{33}
\end{array}\right\|=0 . \label{eq:8:6:37}\\
& \left(R_{11}+R_{12}+1\right)\left[R_{22}\right]^{\frac{1}{2}}\left\|\begin{array}{lll}
R_{11} & R_{12} & R_{13} \\
R_{21} & R_{22} & R_{23} \\
R_{31} & R_{32} & R_{33}
\end{array}\right\|+\left[R_{13}\left(R_{23}+1\right)\left(R_{12}+1\right)\right]^{\frac{1}{2}}\left\|\begin{array}{lll}
R_{11} & R_{12}+1 & R_{13}-1 \\
R_{21} & R_{22}-1 & R_{23}+1 \\
R_{31} & R_{32} & R_{33}
\end{array}\right\|  \notag\\
& -\left[R_{11} R_{33}(J+1)\right]^{\frac{1}{2}}\left\|\begin{array}{lll}
R_{11}-1 & R_{12} & R_{13} \\
R_{21} & R_{22}-1 & R_{23} \\
R_{31} & R_{32} & R_{33}-1
\end{array}\right\|=0 . \label{eq:8:6:38}\\
& \left(R_{22}-R_{33}\right)\left[R_{11} R_{23} R_{32}\right]^{\frac{1}{2}}\left\|\begin{array}{lll}
R_{11}-1 & R_{12} & R_{13} \\
R_{21} & R_{22} & R_{23}-1 \\
R_{31} & R_{32}-1 & R_{33}
\end{array}\right\|  \notag\\
& +\left(R_{33}-R_{11}\right)\left[R_{13} R_{22} R_{31}\right]^{\frac{1}{2}}\left\|\begin{array}{lll}
R_{11} & R_{12} & R_{13}-1 \\
R_{21} & R_{22}-1 & R_{23} \\
R_{31}-1 & R_{32} & R_{33}
\end{array}\right\|  \notag\\
& +\left(R_{11}-R_{22}\right)\left[R_{12} R_{21} R_{33}\right]^{\frac{1}{2}}\left\|\begin{array}{lll}
R_{11} & R_{12}-1 & R_{13} \\
R_{21}-1 & R_{22} & R_{23} \\
R_{31} & R_{32} & R_{33}-1
\end{array}\right\|=0 . \label{eq:8:6:39}
\end{align}
\section{SUMS OF PRODUCTS OF THE CLEBSCH-GORDAN COEFFICIENTS}\label{sec:8.7}\index{Clebsch-Gordan coefficient!sums of products of}\index{sums!of the Clebsch-Gordan coefficients}
Sums of products of the vector-addition coefficients are more conveniently represented in terms of the $3 j m$ symbols than in terms of the Clebsch-Gordan coefficients, because the $3 j m$ symbols have simpler symmetry relations. The sums expressed in terms of the $3 j m$ symbols will be given in Sec.~\ref{sec:12.2}. Nevertheless the ClebschGordan coefficients are also widely used. That is why we present the formulas which seem to be most useful for sums of products containing the Clebsch-Gordan coefficients. For practical convenience we consider various versions of the formulas which differ by permutations of upper and lower arguments. In these formulas we use the notation\index{dimension factor}\isymg{Pi-dim@$\Pi_{a b c \ldots}$, dimension factor $[(2a+1)(2b+1)\cdots]^{1/2}$}
\[
\Fact{a b \ldots c}=[(2 a+1)(2 b+1) \ldots(2 c+1)]^{\frac{1}{2}}
\]
\subsection{Sums Involving One Clebsch-Gordan Coefficient}\label{sec:8.7.1}\index{sums!with one Clebsch-Gordan coefficient}
\begin{gather}
\sum_{\alpha} \clebsch{a}{\alpha}{b}{0}{a}{\alpha}=\Fact{a}^{2} \delta_{b 0}, \label{eq:8:7:1}\\
\sum_{\alpha}(-1)^{a-\alpha} \clebsch{a}{\alpha}{a}{-\alpha}{c}{0}=\Fact{a} \delta_{c 0}, \label{eq:8:7:2}\\
\sum_{\alpha \beta \gamma}(-1)^{c+\gamma} \clebsch{a}{\alpha}{b}{\beta}{c}{\gamma}[(a+\alpha)!(a-\alpha)!(b+\beta)!(b-\beta)!(c+\gamma)!(c-\gamma)!]^{-\frac{1}{2}}=0. \label{eq:8:7:3}
\end{gather}
\subsection{Sums Involving Products of Two Clebsch-Gordan Coefficients}\label{sec:8.7.2}\index{sums!with two Clebsch-Gordan coefficients}
\begin{gather}
\sum_{\alpha \beta} \clebsch{a}{\alpha}{b}{\beta}{c}{\gamma} \clebsch{a}{\alpha}{b}{\beta}{c^{\prime}}{\gamma^{\prime}}=\delta_{c c^{\prime}} \delta_{\gamma \gamma^{\prime}}, \label{eq:8:7:4}\\
\sum_{\alpha \gamma} \clebsch{a}{\alpha}{b}{\beta}{c}{\gamma} \clebsch{a}{\alpha}{b^{\prime}}{\beta^{\prime}}{c}{\gamma}=\frac{\Fact{c}^{2}}{\Fact{b}^{2}} \delta_{b b^{\prime}} \delta_{\beta \beta^{\prime}}, \label{eq:8:7:5}\\
\sum_{\alpha \beta}(-1)^{b+\beta} \clebsch{a}{\alpha}{b}{\beta}{c}{\gamma} \clebsch{c^{\prime}}{-\gamma^{\prime}}{a}{\alpha}{b}{-\beta}=(-1)^{c+\gamma} \frac{\Fact{b}}{\Fact{c}} \delta_{c c^{\prime}} \delta_{\gamma \gamma^{\prime}}, \label{eq:8:7:6}\\
\sum_{\alpha \beta}(-1)^{a+\alpha} \clebsch{b}{\beta}{a}{\alpha}{c}{\gamma} \clebsch{a}{-\alpha}{c^{\prime}}{\gamma^{\prime}}{b}{\beta}=\frac{\Fact{b}}{\Fact{c}} \delta_{c c^{\prime}} \delta_{\gamma \gamma^{\prime}}, \label{eq:8:7:7}\\
\sum_{\alpha \beta} \clebsch{c}{\gamma}{b}{\beta}{a}{\alpha} \clebsch{c^{\prime}}{-\gamma^{\prime}}{a}{\alpha}{b}{\beta}=(-1)^{b-a-\gamma} \frac{\Fact{a b}}{\Fact{c}^{2}} \delta_{c c^{\prime}} \delta_{\gamma \gamma^{\prime}}, \label{eq:8:7:8}\\
\sum_{c \gamma} \clebsch{a}{\alpha}{b}{\beta}{c}{\gamma} \clebsch{a}{\alpha^{\prime}}{b}{\beta^{\prime}}{c}{\gamma}=\delta_{\alpha \alpha^{\prime}} \delta_{\beta \beta^{\prime}}, \label{eq:8:7:9}\\
\sum_{c \gamma} \Fact{c}^{2} \clebsch{a}{\alpha}{c}{\gamma}{b}{\beta} \clebsch{a}{\alpha^{\prime}}{c}{\gamma}{b}{\beta^{\prime}}=\Fact{b}^{2} \delta_{\alpha \alpha^{\prime}} \delta_{\beta \beta^{\prime}}, \label{eq:8:7:10}\\
\sum_{a \alpha}(-1)^{a-\alpha} \Fact{a}^{2} \clebsch{a}{\alpha}{b}{\beta}{c}{\gamma} \clebsch{a}{\alpha}{c}{\gamma^{\prime}}{b}{\beta^{\prime}}=\Fact{b c} \delta_{\gamma,-\gamma^{\prime}} \delta_{\beta,-\beta^{\prime}}. \label{eq:8:7:11}
\end{gather}
\subsection{Sums Involving Products of Three Clebsch-Gordan Coefficients}\label{sec:8.7.3}\index{sums!with three Clebsch-Gordan coefficients}
\begin{gather}
\sum_{\alpha \beta \delta} \clebsch{a}{\alpha}{b}{\beta}{c}{\gamma} \clebsch{d}{\delta}{b}{\beta}{e}{\varepsilon} \clebsch{a}{\alpha}{f}{\varphi}{d}{\delta}=\kappa_{1} \Fact{c d} \clebsch{c}{\gamma}{f}{\varphi}{e}{\varepsilon}\sixj{a}{b}{c}{e}{f}{d}, \label{eq:8:7:12}\\
\sum_{\alpha \beta \delta} \clebsch{b}{\beta}{c}{\gamma}{a}{\alpha} \clebsch{b}{\beta}{e}{\varepsilon}{d}{\delta} \clebsch{a}{\alpha}{f}{\varphi}{d}{\delta}=\kappa_{1} \frac{\Fact{a d d}}{\Fact{e}} \clebsch{c}{\gamma}{f}{\varphi}{e}{\varepsilon}\sixj{a}{b}{c}{e}{f}{d}, \label{eq:8:7:13}\\
\sum_{\alpha \beta \delta} 
\clebsch{b}{\beta}{a}{\alpha}{c}{\gamma} 
\clebsch{b}{\beta}{d}{\delta}{e}{\varepsilon} 
\clebsch{a}{\alpha}{f}{\varphi}{d}{\delta}
=\kappa_{2} \Fact{c d} 
\clebsch{c}{\gamma}{f}{\varphi}{e}{\varepsilon}
\sixj{a}{b}{c}{e}{f}{d}, \label{eq:8:7:14}\\
\sum_{\alpha \beta \delta}(-1)^{a-\alpha} \clebsch{a}{\alpha}{b}{\beta}{c}{\gamma} \clebsch{d}{\delta}{b}{\beta}{e}{\varepsilon} \clebsch{d}{\delta}{a}{-\alpha}{f}{\varphi}=\kappa_{1} \Fact{c f} \clebsch{c}{\gamma}{f}{\varphi}{e}{\varepsilon}\sixj{a}{b}{c}{e}{f}{d}, \label{eq:8:7:15}\\
\sum_{\alpha \beta \delta}(-1)^{b+\beta} \clebsch{a}{\alpha}{b}{\beta}{c}{\gamma} \clebsch{b}{-\beta}{e}{\varepsilon}{d}{\delta} \clebsch{a}{\alpha}{f}{\varphi}{d}{\delta}=\kappa_{1} \frac{\Fact{c d d}}{\Fact{e}} \clebsch{c}{\gamma}{f}{\varphi}{e}{\varepsilon}\sixj{a}{b}{c}{e}{f}{d}, \label{eq:8:7:16}\\
\sum_{\alpha \beta \delta}(-1)^{a-\alpha} \clebsch{b}{\beta}{a}{\alpha}{c}{\gamma} \clebsch{b}{\beta}{d}{\delta}{e}{\varepsilon} \clebsch{d}{\delta}{a}{-\alpha}{f}{\varphi}=\kappa_{2} \Fact{c f} \clebsch{c}{\gamma}{f}{\varphi}{e}{\varepsilon}
\sixj{a}{b}{c}{e}{f}{d}, \label{eq:8:7:17}\\
\sum_{\alpha \beta \delta}(-1)^{b+\beta} \clebsch{b}{-\beta}{c}{\gamma}{a}{\alpha} \clebsch{d}{\delta}{b}{\beta}{e}{\varepsilon} \clebsch{a}{\alpha}{f}{\varphi}{d}{\delta}=\kappa_{1} \Fact{a d} \clebsch{c}{\gamma}{f}{\varphi}{e}{\varepsilon}
\sixj{a}{b}{c}{e}{f}{d}, \label{eq:8:7:18}\\
\sum_{\alpha \beta \delta}(-1)^{a-\alpha} \clebsch{a}{\alpha}{c}{-\gamma}{b}{\beta} \clebsch{d}{\delta}{b}{-\beta}{e}{\varepsilon} \clebsch{a}{\alpha}{f}{\varphi}{d}{\delta}=\kappa_{1} \Fact{b d} \clebsch{c}{\gamma}{f}{\varphi}{e}{\varepsilon}\sixj{a}{b}{c}{e}{f}{d}, \label{eq:8:7:19}
\end{gather}
where $\kappa_{1}=(-1)^{b+c+d+f}, \kappa_{2}=(-1)^{a+b+e+f}$.\\
\subsection{Sums Involving Products of Four Clebsch-Gordan Coefficients}\label{sec:8.7.4}\index{sums!with four Clebsch-Gordan coefficients}
\begin{align}
& \sum_{\beta \gamma \varepsilon \varphi} \clebsch{b}{\beta}{c}{\gamma}{a}{\alpha} \clebsch{e}{\varepsilon}{f}{\varphi}{d}{\delta} \clebsch{e}{\varepsilon}{b}{\beta}{g}{\eta} \clebsch{f}{\varphi}{c}{\gamma}{j}{\mu}  \notag\\
& =(-1)^{a-b+c+d+e-f} \sum_{s \sigma} \Fact{s s a g} \clebsch{a}{\alpha}{s}{\sigma}{j}{\mu} \clebsch{g}{\eta}{s}{\sigma}{d}{\delta}\sixj{b}{c}{a}{j}{s}{f}\sixj{b}{e}{g}{d}{s}{f}=\Fact{a d g j} \sum_{k \kappa} \clebsch{g}{\eta}{j}{\mu}{k}{\kappa} \clebsch{d}{\delta}{a}{\alpha}{k}{\kappa}\ninej{c}{b}{a}{f}{e}{d}{j}{g}{k}, \label{eq:8:7:20}\\
& \sum_{\beta \gamma \varepsilon \varphi} \clebsch{b}{\beta}{c}{-\gamma}{a}{\alpha} \clebsch{e}{\varepsilon}{f}{-\varphi}{d}{\delta} \clebsch{b}{\beta}{e}{\varepsilon}{g}{\eta} \clebsch{c}{\gamma}{f}{\varphi}{j}{\mu}=(-1)^{b+e-g} \Fact{a d g j} \sum_{k \kappa} \clebsch{g}{\eta}{j}{-\mu}{k}{\kappa} \clebsch{d}{\delta}{a}{\alpha}{k}{\kappa}\ninej{c}{b}{a}{f}{e}{d}{j}{g}{k}, \label{eq:8:7:21}\\
& \sum_{\beta \gamma \varepsilon \varphi} \clebsch{b}{\beta}{a}{\alpha}{c}{\gamma} \clebsch{f}{\varphi}{j}{\mu}{c}{\gamma} \clebsch{b}{\beta}{g}{\eta}{e}{\varepsilon} \clebsch{f}{\varphi}{d}{\delta}{e}{\varepsilon}=(-1)^{a-b+f-j} \Fact{c e}^{2} \sum_{k \kappa} \clebsch{g}{\eta}{j}{\mu}{k}{\kappa} \clebsch{d}{\delta}{a}{\alpha}{k}{\kappa}\ninej{c}{b}{a}{f}{e}{d}{j}{g}{k}, \label{eq:8:7:22}\\
& \sum_{\beta \gamma \varepsilon \varphi} \clebsch{a}{\alpha}{b}{\beta}{c}{\gamma} \clebsch{g}{\eta}{e}{\varepsilon}{b}{\beta} \clebsch{d}{\delta}{f}{\varphi}{e}{\varepsilon} \clebsch{j}{\mu}{c}{\gamma}{f}{\varphi}=(-1)^{d+e-c-j} \Fact{b c e f} \sum_{k \kappa}(-1)^{k-\kappa} \clebsch{g}{\eta}{j}{\mu}{k}{-\kappa} \clebsch{d}{\delta}{a}{\alpha}{k}{\kappa}\ninej{c}{b}{a}{f}{e}{d}{j}{g}{k}, \label{eq:8:7:23}\\
& \sum_{\beta \gamma \varepsilon \varphi} \clebsch{b}{\beta}{a}{-\alpha}{c}{\gamma} \clebsch{e}{\varepsilon}{d}{-\delta}{f}{\varphi} \clebsch{g}{\eta}{b}{-\beta}{e}{\varepsilon} \clebsch{j}{\mu}{c}{-\gamma}{f}{\varphi}=(-1)^{b-c-g-\alpha+\eta} \Fact{c e f f} \sum_{k \kappa} \clebsch{g}{\eta}{j}{-\mu}{k}{\kappa} \clebsch{d}{\delta}{a}{\alpha}{k}{\kappa}\ninej{c}{b}{a}{f}{e}{d}{j}{g}{k}, \label{eq:8:7:24}
\end{align}
\begin{align}
& \sum_{\beta \gamma \varepsilon \varphi} \clebsch{b}{\beta}{c}{-\gamma}{a}{\alpha} \clebsch{e}{\varepsilon}{f}{\varphi}{d}{\delta} \clebsch{e}{-\varepsilon}{g}{\eta}{b}{\beta} \clebsch{f}{\varphi}{j}{\mu}{c}{\gamma}=(-1)^{b+f-g-\delta} \Fact{a b c d} \sum_{k \kappa} \clebsch{g}{\eta}{j}{-\mu}{k}{\kappa} \clebsch{d}{\delta}{a}{\alpha}{k}{\kappa}\ninej{c}{b}{a}{f}{e}{d}{j}{g}{k}, \label{eq:8:7:25}\\
& \sum_{\beta \gamma \varepsilon \varphi} \clebsch{b}{\beta}{c}{\gamma}{a}{\alpha} \clebsch{e}{\varepsilon}{f}{\varphi}{d}{\delta} \clebsch{e}{\varepsilon}{g}{\eta}{b}{\beta} \clebsch{f}{\varphi}{j}{\mu}{c}{\gamma}=\sum_{k \kappa} \Fact{b c d k} \clebsch{g}{\eta}{j}{\mu}{k}{\kappa} \clebsch{d}{ \delta}{k}{\kappa}{a}{\alpha}\ninej{a}{b}{c}{d}{e}{f}{k}{g}{j}, \label{eq:8:7:26}\\
& \sum_{\beta \gamma \varepsilon \varphi} \clebsch{b}{\beta}{c}{-\gamma}{a}{\alpha} \clebsch{e}{\varepsilon}{f}{-\varphi}{d}{\delta} \clebsch{g}{\eta}{b}{-\beta}{e}{\varepsilon} \clebsch{j}{\mu}{f}{-\varphi}{c}{\gamma}=(-1)^{c+e-g+j+\alpha-\mu} \Fact{a d e c} \sum_{k \kappa} \clebsch{g}{\eta}{j}{-\mu}{k}{\kappa} \clebsch{d}{\delta}{a}{\alpha}{k}{\kappa}\ninej{c}{b}{a}{f}{e}{d}{j}{g}{k}, \label{eq:8:7:27}\\
& \sum_{\beta \gamma \varepsilon \varphi} \clebsch{b}{\beta}{c}{\gamma}{a}{\alpha} \clebsch{b}{\beta}{g}{\eta}{e}{\varepsilon} \clebsch{f}{\varphi}{d}{\delta}{e}{\varepsilon} \clebsch{f}{\varphi}{c}{\gamma}{j}{\mu}=(-1)^{j-a+\delta-\eta} \Fact{a e e j} \sum_{k \kappa} \clebsch{g}{\eta}{j}{-\mu}{k}{\kappa} \clebsch{d}{\delta}{a}{-\alpha}{k}{\kappa}\ninej{c}{b}{a}{f}{e}{d}{j}{g}{k}, \label{eq:8:7:28}\\
& \sum_{\beta \gamma \varepsilon \varphi} \clebsch{b}{\beta}{c}{\gamma}{a}{\alpha} \clebsch{g}{\eta}{e}{\varepsilon}{b}{\beta} \clebsch{f}{\varphi}{d}{\delta}{e}{\varepsilon} \clebsch{f}{\varphi}{c}{\gamma}{j}{\mu}=(-1)^{j-a+g+\delta} \Fact{a b e j} \sum_{k \kappa} \clebsch{g}{-\eta}{j}{-\mu}{k}{\kappa} \clebsch{d}{\delta}{a}{-\alpha}{k}{\kappa}\ninej{c}{b}{a}{f}{e}{d}{j}{g}{k}, \label{eq:8:7:29}\\
& \sum_{\beta \gamma \varepsilon\varphi}(-1)^{c-\gamma+e-\varepsilon} \clebsch{a}{\alpha}{b}{\beta}{c}{\gamma} \clebsch{d}{\delta}{f}{\varphi}{e}{\varepsilon} \clebsch{e}{\varepsilon}{b}{\beta}{g}{\eta} \clebsch{c}{\gamma}{f}{\varphi}{j}{\mu}=(-1)^{a+d-\alpha-\delta} \Fact{c e g j} \sum_{k \kappa} \clebsch{g}{\eta}{j}{-\mu}{k}{\kappa} \clebsch{d}{\delta}{a}{-\alpha}{k}{\kappa}\ninej{c}{b}{a}{f}{e}{d}{j}{g}{k} . \label{eq:8:7:30}
\end{align}
\subsection{Sums Involving Products of the Clebsch-Gordan Coefficients and One 6j Symbol}\label{sec:8.7.5}\index{sums!with a 6j symbol@with a $6j$ symbol}\index{6j symbol@$6j$ symbol}
\begin{align}
& \sum_{e \varepsilon}(-1)^{2 e} \Fact{c d} \clebsch{b}{\beta}{d}{\delta}{e}{\varepsilon} \clebsch{f}{\varphi}{c}{\gamma}{e}{\varepsilon}
\sixj{a}{b}{c}{e}{f}{d}
=\clebsch{a}{\alpha}{b}{\beta}{c}{\gamma} 
\clebsch{a}{\alpha}{f}{\varphi}{d}{\delta}, \label{eq:8:7:31}\\
& \sum_{f \varphi}(-1)^{c+d+f} \Fact{c e} \clebsch{e}{\varepsilon}{a}{\alpha}{f}{\varphi} \clebsch{d}{\delta}{c}{\gamma}{f}{\varphi}\sixj{b}{a}{c}{f}{d}{e}=\clebsch{a}{\alpha}{b}{\beta}{c}{\gamma} \clebsch{d}{\delta}{b}{\beta}{e}{\varepsilon}, \label{eq:8:7:32}\\
& \sum_{c \gamma}(-1)^{2 e-d+\alpha+\varphi} \Fact{a e} \clebsch{f}{-\varphi}{b}{\beta}{c}{\gamma} \clebsch{e}{\varepsilon}{a}{-\alpha}{c}{\gamma}\sixj{c}{f}{b}{d}{e}{a}=\clebsch{b}{\beta}{d}{\delta}{e}{\varepsilon} \clebsch{f}{\varphi}{d}{\delta}{a}{\alpha}, \label{eq:8:7:33}\\
& \sum_{c \gamma}(-1)^{c+d-\beta-\varphi} \Fact{d}^{2} \clebsch{a}{\alpha}{b}{\beta}{c}{\gamma} \clebsch{f}{-\varphi}{e}{\varepsilon}{c}{\gamma}\sixj{a}{b}{c}{e}{f}{d}=\clebsch{a}{\alpha}{f}{\varphi}{d}{\delta} \clebsch{b}{-\beta}{e}{\varepsilon}{d}{\delta}, \label{eq:8:7:34}\\
& \sum_{c \gamma}(-1)^{2 e} \Fact{c d} \clebsch{a}{\alpha}{b}{\beta}{c}{\gamma} \clebsch{f}{\varphi}{c}{\gamma}{e}{\varepsilon}\sixj{a}{b}{c}{e}{f}{d}=\clebsch{b}{\beta}{d}{\delta}{e}{\varepsilon} \clebsch{a}{\alpha}{f}{\varphi}{d}{\delta}, \label{eq:8:7:35}\\
& \sum_{f \varphi}(-1)^{2 c} \Fact{e f} \clebsch{b}{\beta}{d}{\delta}{f}{\varphi} \clebsch{a}{\alpha}{f}{\varphi}{c}{\gamma}\sixj{a}{b}{e}{d}{c}{f}=\clebsch{b}{\beta}{a}{\alpha}{e}{\varepsilon} \clebsch{d}{\delta}{e}{\varepsilon}{c}{\gamma}, \label{eq:8:7:36}\\
& \sum_{c \gamma}(-1)^{e+d-\beta} \frac{\Fact{c d d}}{\Fact{e}} \clebsch{a}{\alpha}{b}{\beta}{c}{\gamma} \clebsch{f}{\varphi}{c}{\gamma}{e}{\varepsilon}\sixj{a}{b}{c}{e}{f}{d}=\clebsch{b}{-\beta}{e}{\varepsilon}{d}{\delta} \clebsch{a}{\alpha}{f}{\varphi}{d}{\delta}, \label{eq:8:7:37}\\
& \sum_{c \gamma}(-1)^{2 e} \frac{\Fact{a c c}}{\Fact{b}} \clebsch{f}{\varphi}{c}{\gamma}{b}{\beta} \clebsch{c}{\gamma}{a}{\alpha}{e}{\varepsilon}\sixj{c}{f}{b}{d}{e}{a}=\clebsch{d}{\delta}{b}{\beta}{e}{\varepsilon} \clebsch{f}{\varphi}{d}{\delta}{a}{\alpha} . \label{eq:8:7:38}
\end{align}
\subsection{Sums Involving Products of the Clebsch-Gordan Coefficients and One 9j Symbol}\label{sec:8.7.6}\index{sums!with a 9j symbol@with a $9j$ symbol}\index{9j symbol@$9j$ symbol}
\begin{align}
& \sum_{a k} \Fact{a d g j} \clebsch{b}{\beta}{c}{\gamma}{a}{\alpha} \clebsch{g}{\eta}{j}{\mu}{k}{\kappa} \clebsch{d}{\delta}{a}{\alpha}{k}{\kappa}\ninej{c}{b}{a}{f}{e}{d}{j}{g}{k}=\clebsch{e}{\varepsilon}{f}{\varphi}{d}{\delta} \clebsch{e}{\varepsilon}{b}{\beta}{g}{\eta} \clebsch{f}{\varphi}{c}{\gamma}{j}{\mu}, \label{eq:8:7:39}\\
& \sum_{g j} \Fact{a d g j} \clebsch{g}{\eta}{j}{\mu}{k}{\kappa} \clebsch{e}{\varepsilon}{b}{\beta}{g}{\eta} \clebsch{f}{\varphi}{c}{\gamma}{j}{\mu}\ninej{c}{b}{a}{f}{e}{d}{j}{g}{k}=\clebsch{d}{\delta}{a}{\alpha}{k}{\kappa} \clebsch{e}{\varepsilon}{f}{\varphi}{d}{\delta} \clebsch{b}{\beta}{c}{\gamma}{a}{\alpha}, \label{eq:8:7:40}\\
& \sum_{g j}(-1)^{b+e-g} \Fact{a d g j} \clebsch{g}{\eta}{j}{-\mu}{k}{\kappa} \clebsch{b}{\beta}{e}{\varepsilon}{g}{\eta} \clebsch{c}{\gamma}{f}{\varphi}{j}{\mu}\ninej{c}{b}{a}{f}{e}{d}{j}{g}{k}=\clebsch{d}{\delta}{a}{\alpha}{k}{\kappa} \clebsch{e}{\varepsilon}{f}{-\varphi}{d}{\delta} \clebsch{b}{\beta}{c}{-\gamma}{a}{\alpha}, \label{eq:8:7:41}\\
& \sum_{a d}(-1)^{b+f-g-\delta} \Fact{a b c d} \clebsch{b}{\beta}{c}{-\gamma}{a}{\alpha} \clebsch{e}{\varepsilon}{f}{\varphi}{d}{\delta} \clebsch{d}{\delta}{a}{\alpha}{k}{\kappa}\ninej{c}{b}{a}{f}{e}{d}{j}{g}{k}=\clebsch{g}{\eta}{j}{-\mu}{k}{\kappa} \clebsch{e}{-\varepsilon}{g}{\eta}{b}{\beta} \clebsch{f}{\varphi}{j}{\mu}{c}{\gamma}, \label{eq:8:7:42}\\
& \sum_{g j}(-1)^{g-b-f+\varphi+\varepsilon} \frac{\Fact{a d j j g g}}{\Fact{b c}} \clebsch{g}{\eta}{j}{-\mu}{k}{\kappa} \clebsch{e}{-\varepsilon}{g}{\eta}{b}{\beta} \clebsch{f}{\varphi}{j}{\mu}{c}{\gamma}\ninej{c}{b}{a}{f}{e}{d}{j}{g}{k}=\clebsch{b}{\beta}{c}{-\gamma}{a}{\alpha} \clebsch{e}{\varepsilon}{f}{\varphi}{d}{\delta} \clebsch{d}{\delta}{a}{\alpha}{k}{\kappa} . \label{eq:8:7:43}
\end{align}
Sums of larger numbers of the Clebsch-Gordan coefficients may be obtained from the above equations using the orthogonality relation (\eqref{eq:8:1:8}).

\subsection{Some Additional Sums of Products of Two Clebsch-Gordan Coefficients}\label{sec:8.7.7}\index{sums!more, with two Clebsch-Gordan coefficients}
We note also some other sums of products of the Clebsch-Gordan coefficients. The first two sums, obtained by Morgan \cite{ref136}, are written as
\begin{equation}
\sum_{l^{\prime}=0}^{l} \frac{1}{\left(2 l^{\prime}-1\right)\left(2 l-2 l^{\prime}+2 J+1\right)}\left(\clebsch{l}{0}{l^{\prime}+J}{0}{l-l^{\prime}+J}{0}\right)^{2}=-\frac{\delta_{l0}}{2 J+1}, \label{eq:8:7:44}
\end{equation}
and
\begin{equation}
\sum_{l^{\prime}=0}^{l}\left(\frac{1}{2 l^{\prime}+3}-\frac{l+1}{2 l+3} \frac{1}{2 l^{\prime}+1}\right) \frac{1}{2 l-2 l^\prime+2 J+1}
\left(
  \clebsch{l}{0}{l^\prime+J}{0}{l-l^\prime+J}{0}
\right)^{2}=0 . \label{eq:8:7:45}
\end{equation}
In these equations $l, l^{\prime}, J$ are integers.\\[0pt]
Din proved the following identity \cite{ref134}
\begin{equation}
\sum_{i=|c-b|}^{c+b} \frac{2 i+1}{i(i+1)-a(a+1)}\left(\clebsch{i}{0}{b}{0}{c}{0}\right)^{2}=0, \label{eq:8:7:46}
\end{equation}
where $a, b, c$ are non-negative integers, with $a+b+c$ odd, which satisfy the triangular condition
\[
|c-b| \leq a \leq c+b .
\]
The sum runs only over values of $i$ for which the Clebsch-Gordan coefficient does not vanish. Dunlop and Judd \cite{ref135} give the following relation
\begin{equation}
\sum_{m} \clebsch{a}{m}{k}{0}{a}{m} \clebsch{c}{M-m}{k}{0}{c}{M-m}=(-1)^{k} \frac{\Fact{a c}}{2 k+1} \sqrt{\frac{(2 a-k)!(2 c+k+1)!}{(2 c-k)!(2 a+k+1)!}}, \label{eq:8:7:47}
\end{equation}
which is valid for $a-c \geq|M|$.

\section{GENERATING FUNCTIONS}\label{sec:8.8}\index{Clebsch-Gordan coefficient!generating functions}\index{generating functions!for the Clebsch-Gordan coefficients}
The Clebsch-Gordan coefficients may be generated by expansions of some functions in power series or finite sums. Below we present several generating functions. In these relations $J \equiv a+b+c$.

\subsection{Regge Determinant to Power $J$}\label{sec:8.8.1}\index{Regge determinant}\index{generating functions!Regge determinant}
\begin{equation}
\left|\begin{array}{ccc}
u_{1} & u_{2} & u_{3}  \\
v_{1} & v_{2} & v_{3} \\
w_{1} & w_{2} & w_{3}
\end{array}\right|^{J}=J!\sqrt{(J+1)!} \sum_{R_{i k}}\left\|\begin{array}{lll}
R_{11} & R_{12} & R_{13} \\
R_{21} & R_{22} & R_{23} \\
R_{31} & R_{32} & R_{33}
\end{array}\right\| \frac{u_{1}^{R_{11}} u_{2}^{R_{12}} u_{3}^{R_{13}} v_{1}^{R_{21}} v_{2}^{R_{22}} v_{3}^{R_{23}} w_{1}^{R_{31}} w_{2}^{R_{32}} w_{3}^{R_{33}}}{\left[R_{11}!R_{12}!R_{13}!R_{21}!R_{22}!R_{23}!R_{31}!R_{32}!R_{33}!\right]^{\frac{1}{2}}} . \label{eq:8:8:1}
\end{equation}
(Regge \cite{ref094}, Shelepin \cite{ref106}). Here $R_{i k}$ are the elements of the $R$-symbol (see Sec.~\ref{sec:8.1.3}). The $R$-symbols and the Clebsch-Gordan coefficients are related by Eqs.~\eqref{eq:8:1:14} and \eqref{eq:8:1:11}.

\subsection{Products of Binomials}\label{sec:8.8.2}\index{generating functions!products of binomials}
\begin{align}
\MoveEqLeft{(t_1-t_2)^{J-2c} (t_2-t_3)^{J-2a}(t_3-t_1)^{J-2b}}\notag\\
= & \left[(J+1)!\frac{(J-2 a)!(J-2 b)!(J-2 c)!}{2 c+1}\right]^{\frac{1}{2}} 
\sum_{\substack{\alpha, \beta, \gamma \\ \alpha+\beta+\gamma=0}}(-1)^{a-b-\gamma}   %\notag\\\times & 
\frac{t_{1}^{a+\alpha} t_{2}^{b+\beta} t_{3}^{c+\gamma}}{[(a+\alpha)!(a-\alpha)!(b+\beta)!(b-\beta)!(c+\gamma)!(c-\gamma)!]^{\frac{1}{2}}} \clebsch{a}{\alpha}{b}{\beta}{c}{-\gamma}, \label{eq:8:8:2}\\
\MoveEqLeft{\left(v_{1} u_{2}-u_{1} v_{2}\right)^{J-2 c}\left(w_{1} v_{2}-w_{2} v_{1}\right)^{J-2 a}\left(u_{1} w_{2}-w_{1} u_{2}\right)^{J-2 b}} \notag \\
= & \left[\frac{(J+1)!(J-2 a)!(J-2 b)!(J-2 c)!}{2 c+1}\right]^{\frac{1}{2}}\notag\\& \times 
\sum_{\substack{\alpha, \beta, \gamma \\ \alpha+\beta+\gamma=0}}
(-1)^{a-b-\gamma} 
\frac{u_{1}^{a-\alpha} u_{2}^{a+\alpha} v_{1}^{b-\beta} v_{2}^{b+\beta} w_{1}^{c-\gamma} w_{2}^{c+\gamma}}{[(a+\alpha)!(a-\alpha)!(b+\beta)!(b-\beta)!(c+\gamma)!(c-\gamma)!]^{\frac{1}{2}}} \clebsch{a}{\alpha}{b}{\beta}{c}{-\gamma} . \label{eq:8:8:3}
\end{align}

\subsection{Exponential Function}\label{sec:8.8.3}\index{generating functions!exponential function}
\begin{align}
\MoveEqLeft{\exp\{x_1(t_2-t_3)+x_2(t_3-t_1)+x_3(t_1-t_2)\}}\notag\\
  &=\sum_{\substack{a, b, c \\ \alpha, \beta, \gamma}} \frac{(-1)^{a-b-\gamma}}{\Delta(a b c)} \frac{x_1^{-a+b+c}x_2^{a-b+c}x_3^{a+b-c}t_1^{a+\alpha}t_2^{b+\beta}t_3^{c+\gamma}}{[(2 c+1)(a+\alpha)!(a-\alpha)!(b+\beta)!(b-\beta)!(c+\gamma)!(c-\gamma)!]^{\frac{1}{2}}} \clebsch{a}{\alpha}{b}{\beta}{c}{-\gamma}, \label{eq:8:8:4}
\end{align}
where $\Delta(a b c)$ is given by \eqref{eq:8:2:1}.

\subsection{Hypergeometric Function}\label{sec:8.8.4}\index{generating functions!hypergeometric function}
\begin{gather}
{\left[\frac{(J-2 b)!(c+\gamma)!(2 c+1)}{(J-2 a)!(J-2 c)!(c-\gamma)!(J+1)!}\right]^{\frac{1}{2}} \frac{(t-1)^{J-2 c}}{(a-b+\gamma)!} F(\gamma-c, a-b-c ; a-b+\gamma+1 ; t)}  \notag\\
=\sum_{\substack{\alpha, \beta \\
\alpha+\beta=\gamma}} \clebsch{a}{\alpha}{b}{\beta}{c}{\gamma} \frac{t^{b-\beta}}{[(a+\alpha)!(a-\alpha)!(b+\beta)!(b-\beta)!]^{\frac{1}{2}}}. \label{eq:8:8:5}
\end{gather}
(Akim and Levin \cite{ref046})

\subsection{Wigner D-Function}\label{sec:8.8.5}\index{generating functions!Wigner $D$-function}\isym{dJMM@$d_{M M^{\prime}}^{J}(\beta)$, Wigner (small) $d$-function}
\begin{equation}
\left[\frac{(2 c+1)}{(J+1)!(J-2 c)!}\right]^{\frac{1}{2}} d_{\gamma \gamma^{\prime}}^{c}(\theta)=\sum_{\substack{\alpha, \beta \\ \alpha+\beta=\gamma}}(-1)^{b+\beta} \clebsch{a}{\alpha}{b}{\beta}{c}{\gamma} \frac{\left(\cos \frac{\theta}{2}\right)^{a+b+\alpha-\beta}\left(\sin \frac{\theta}{2}\right)^{a+b-\alpha+\beta}}{[(a+\alpha)!(a-\alpha)!(b+\beta)!(b-\beta)!]^{\frac{1}{2}}}, \label{eq:8:8:6}
\end{equation}
where $a-b=\gamma^{\prime},|a-b| \leq c \leq a+b$.

\subsection{Generating Function for the Coefficients $\clebsch{a}{0}{b}{0}{c}{0}$}\label{sec:8.8.6}\index{generating functions!for coefficients with zero projections}
\begin{equation}
\frac{1}{1+u^{2}+v^{2}+w^{2}}=\sum_{a b c}(-1)^{a-b}\left[\frac{(J+1)!}{2 c+1}\right]^{\frac{1}{2}} \clebsch{a}{0}{b}{0}{c}{0} \frac{u^{J-2 a} v^{J-2 b} w^{J-2 c}}{[(J-2 a)!(J-2 b)!(J-2 c)!]^{\frac{1}{2}}} . \label{eq:8:8:7}
\end{equation}
(Schwinger \cite{ref101})

\section{CLASSICAL LIMIT AND ASYMPTOTIC EXPRESSIONS FOR THE CLEBSCH-GORDAN COEFFICIENTS}\label{sec:8.9}\index{Clebsch-Gordan coefficient!classical limit}\index{Clebsch-Gordan coefficient!asymptotic expressions}\index{asymptotics!of the Clebsch-Gordan coefficients}\index{classical limit}
Asymptotic expressions permit us to simplify calculations of the Clebsch-Gordan coefficients. Note that these expressions appear to be accurate even if the arguments are not very large.\\
\subsection{Asymptotic Expressions for $a, c \gg b$}\label{sec:8.9.1}\index{asymptotics!for two large momenta}
\begin{multline}
\clebsch{a}{\alpha}{b}{\beta}{c}{\gamma} \approx[(b+\beta)!(b-\beta)!(b+\delta)!(b-\delta)!]^{\frac{1}{2}}  \\
\times \sum_{s} \frac{(-1)^{b-\delta-s}\left(\cos \frac{\vartheta}{2}\right)^{2s+\beta+\delta}\left(\sin \frac{\vartheta}{2}\right)^{2 b-\beta-\delta-2 s}}{s!(s+\beta+\delta)!(b-\beta-s)!(b-\delta-s)!}=\delta_{\gamma-\alpha, \beta} D_{\beta \delta}^{b}(0, \vartheta, 0), \label{eq:8:9:1}
\end{multline}
where $\delta=c-a$, and $\vartheta$ is the angle between the momentum $c$ and the $z$ axis,
\begin{equation}
\cos \frac{\vartheta}{2}=\sqrt{\frac{c+\gamma+\frac{1}{2}}{2 c+1}}=\clebsch{c}{\gamma}{\frac{1}{2}}{\frac{1}{2}}{c+\frac{1}{2}}{\gamma+\frac{1}{2}}, \quad \sin \frac{\vartheta}{2}=\sqrt{\frac{c-\gamma+\frac{1}{2}}{2 c+1}}=\clebsch{c}{\gamma}{\frac{1}{2}}{-\frac{1}{2}}{c+\frac{1}{2}}{\gamma-\frac{1}{2}}, \quad \cos \vartheta=\frac{\gamma}{c+\frac{1}{2}} . \label{eq:8:9:2}
\end{equation}
This formula was obtained by Edmonds \cite{ref064}, Brussaard and Tolhoek \cite{ref060}. In particular,
\begin{align}
\clebsch{a}{\alpha}{b}{b}{c}{\gamma} &\approx \delta_{b, \gamma-\alpha}(-1)^{b-\delta}\left[\frac{(2 b)!}{(b+\delta) !(b-\delta)!}\right]^{\frac{1}{2}}\left(\cos \frac{\vartheta}{2}\right)^{b+\delta}\left(\sin \frac{\vartheta}{2}\right)^{b-\delta}, \label{eq:8:9:3}\\
\clebsch{a}{\alpha}{b}{\beta}{c}{c} &\approx \delta_{\beta, c-\alpha} D_{\beta c-a}^{b}(0,0,0)=\delta_{\beta, c-\alpha} \delta_{\beta, c-a}, \label{eq:8:9:4}\\
\clebsch{a}{-\beta}{b}{\beta}{c}{0} &\approx D_{\beta c-a}^{b}\left(0, \frac{\pi}{2}, 0\right), \label{eq:8:9:5}\\
\clebsch{c}{\gamma}{b}{0}{c}{\gamma} &\approx P_{b}(\cos \vartheta), \label{eq:8:9:6}\\
\clebsch{c}{\alpha}{b}{\beta}{c}{\gamma} &\approx \delta_{\beta, \gamma-\alpha} \sqrt{\frac{4 \pi}{2 b+1}} Y_{b \beta}(\vartheta, 0), \label{eq:8:9:7}\\
\clebsch{c}{\pm c}{b}{0}{c}{\pm c} &\approx( \pm 1)^{b}. \label{eq:8:9:8}
\end{align}
If $a, c \gg b \gg 1$ one can use in \eqref{eq:8:9:1} the asymptotic expressions for the Wigner $D$-function (Eqs.~\eqref{eq:4:18:1}\eqref{eq:4:18:4}). Then the Clebsch-Gordan coefficients may be represented in the form
\begin{gather}
\clebsch{a}{\alpha}{b}{\beta}{c}{\gamma} \approx \frac{(-1)^{\beta-\delta}}{(\beta-\delta)!}\left[\frac{2}{\pi} \frac{(b+\beta)!(b-\beta+1)!}{(b+\delta)!(b-\delta)!}\right]^{\frac{1}{2}} \frac{\cos \left[(2 b+1) \frac{\vartheta}{2}-\frac{\pi}{2}\left(\beta-\delta+\frac{1}{2}\right)\right]}{\sqrt{\sin \vartheta}}, \label{eq:8:9:9}\\
\text { if } \beta \geq \delta, \beta \geq 0, \notag\\
\clebsch{a}{\alpha}{b}{\beta}{c}{\gamma} \approx \frac{1}{(\delta-\beta)!}\left[\frac{2}{\pi} \frac{(b+\delta)!(b-\delta+1)!}{(b+\beta)!(b-\beta)!}\right]^{\frac{1}{2}} \frac{\cos \left[(2 b+1) \frac{\vartheta}{2}-\frac{\pi}{2}\left(\delta-\beta+\frac{1}{2}\right)\right]}{\sqrt{\sin \vartheta}}, \label{eq:8:9:10}\\
\text { if } \delta \geq \beta, \delta \geq 0 . \notag
\end{gather}
\subsection{Asymptotic Expressions for $a, b, c, \gamma \gg a+b-\gamma$}\label{sec:8.9.2}\index{asymptotics!for large momenta and projections}
\begin{equation}
\clebsch{a}{\alpha}{b}{\beta}{c}{\gamma} \approx(-1)^{a+b-c} \frac{(c+\gamma+1)^{b-\beta}}{(a-b-\alpha)!}\left[\frac{(2 c+1)(a-b+c)!(a-\alpha)!(b+\beta)!(c-\gamma)!(c+\gamma)!}{(a+b-c)!(-a+b+c)!(a+b+c+1)!(a+\alpha)!(b-\beta)!}\right]^{\frac{1}{2}}. \label{eq:8:9:11}
\end{equation}
(Akim and Levin \cite{ref046}).

\subsection{Semiclassical Formulas for $a, b, c \gg 1$}\label{sec:8.9.3}
\begin{figure}[tbh]
\begin{center}
  \includegraphics{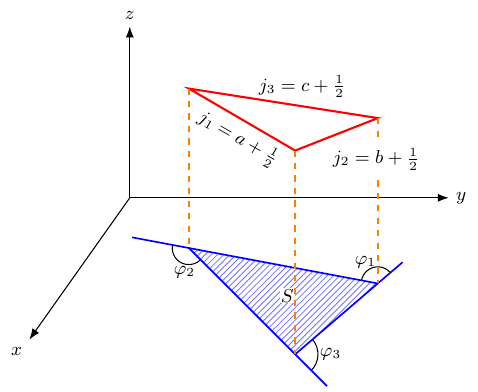}
\caption{Geometric interpretation of the parameters in the asymptotic relation \eqref{eq:8:9:12}.}
\label{chap8:fig:1}
\end{center}
\end{figure}

\index{semiclassical formulas}
\begin{equation}
\clebsch{a}{\alpha}{b}{\beta}{c}{\gamma} \approx(-1)^{2 b+c+\gamma+1} \sqrt{\frac{2 c+1}{2 \pi S}} \cos \left[j_{1} \theta_{1}+j_{2} \theta_{2}+j_{3} \theta_{3}-m_{2} \varphi_{1}+m_{1} \varphi_{2}+\frac{\pi}{4}\right], \label{eq:8:9:12}
\end{equation}
(Ponzano and Regge \cite{ref089}), where
\begin{gather}
j_{1} \equiv a+\frac{1}{2}, j_{2} \equiv b+\frac{1}{2}, j_{3} \equiv c+\frac{1}{2}, m_{1} \equiv \alpha, \quad m_{2} \equiv \beta, \quad m_{3} \equiv-\gamma, \label{eq:8:9:13}\\
S^{2}=\frac{1}{16}\left\{-j_{1}^{4}-j_{2}^{4}-j_{3}^{4}+2\left(j_{1}^{2} j_{2}^{2}+j_{1}^{2} j_{3}^{2}+j_{2}^{2} j_{3}^{2}\right)+4\left(j_{1}^{2} m_{2} m_{3}+j_{2}^{2} m_{1} m_{3}+j_{3}^{2} m_{1} m_{2}\right)\right\}  \notag\\
=-\frac{1}{16}\left|\begin{array}{cccc}
0 & j_{1}^{2}-m_{1}^{2} & j_{2}^{2}-m_{2}^{2} & 1 \\
j_{1}^{2}-m_{1}^{2} & 0 & j_{3}^{2}-m_{3}^{2} & 1 \\
j_{2}^{2}-m_{2}^{2} & j_{3}^{2}-m_{3}^{2} & 0 & 1 \\
1 & 1 & 1 & 0
\end{array}\right| . \label{eq:8:9:14}
\end{gather}
If $m_{1}=m_{2}=m_{3}=0$, \eqref{eq:8:9:14} reduces to
\begin{gather}
S^{2}=\frac{1}{16}\left(a+b+c+\frac{3}{2}\right)\left(-a+b+c+\frac{1}{2}\right)\left(a-b+c+\frac{1}{2}\right)\left(a+b-c+\frac{1}{2}\right), \label{eq:8:9:15}
\end{gather}
\begin{equation}\begin{gathered}
\cos \theta_{i}=\frac{2 j_{i}^{2} m_{l}+m_{i}\left(j_{i}^{2}-j_{k}^{2}+j_{l}^{2}\right)}{\sqrt{\left(j_{i}^{2}-m_{i}^{2}\right)\left(4 j_{i}^{2} j_{l}^{2}-\left(j_{i}^{2}-j_{k}^{2}+j_{l}^{2}\right)^{2}\right)}}, \\
\cos \varphi_{i}=\frac{1}{2} \frac{j_{i}^{2}-j_{k}^{2}-j_{l}^{2}-2 m_{k} m_{l}}{\sqrt{\left(j_{k}^{2}-m_{k}^{2}\right)\left(j_{l}^{2}-m_{l}^{2}\right)}}. 
\end{gathered} \label{eq:8:9:16}
\end{equation}
Indices $i, k, l$ in \eqref{eq:8:9:16} are those which may be obtained by cyclic permutation of $1,2,3$. The quantities which enter Eqs.~\eqref{eq:8:9:14}-\eqref{eq:8:9:16} have a simple geometrical interpretation. Three momenta $j_{1}=a+\frac{1}{2}, j_{2}=b+\frac{1}{2}$ and $j_{3}=c+\frac{1}{2}$ form a triangle oriented in space in such a manner that the projections of its sides on the $z$ axis are\\
equal to $m_{1}, m_{2}$ and $m_{3}$, respectively (Fig.~\ref{chap8:fig:1}). $S$ is an area of the triangle projection onto the ( $x, y$ )-plane. Let us construct a trihedral prism whose sidelong edges are parallel to the $z$ axis and one of the cross sections coincides with the oriented triangle. Then $\theta_{i}$ is the angle between the normals to the planes adjacent to the edge $j_{i}$. The angles $\varphi_{i}$ are shown in Fig.~\ref{chap8:fig:1}.

In particular, if $\alpha=\beta=\gamma=0$, we get $\theta_{1}=\theta_{2}=\theta_{3}=\pi / 2$.
\begin{equation}
\clebsch{a}{0}{b}{0}{c}{0} \approx(-1)^{(a+b-c)/2}\left[\frac{1+(-1)^{a+b-c}}{2}\right] \sqrt{\frac{2 c+1}{2 \pi S}} . \label{eq:8:9:17}
\end{equation}
Equations \eqref{eq:8:9:12}-\eqref{eq:8:9:17} are valid, if $S^{2}>0$.

More accurate expressions valid not only for $S^{2}>0$ but also if $S^{2}<0$ or $S^{2} \approx 0$ were obtained by Schulten and Gordon \cite{ref140}. In the case of $S^{2} \geq 0$ they derived
\begin{subequations}
\begin{gather}
\clebsch{a}{\alpha}{b}{\beta}{c}{\gamma}=(-1)^{2 b+c+\gamma+1} \sqrt{\frac{2 c+1}{2 S}} Z^{\frac{1}{4}} \notag\\
\times \begin{cases}\cos \Omega_{0} A i(-Z)-\sin \Omega_{0} B i(-Z), & \text { if } \Omega-\Omega_{0}<0, \\
\cos \Omega_{0} B i(-Z)-\sin \Omega_{0} A i(-Z), & \text { if } \Omega-\Omega_{0}>0,\end{cases}, \label{eq:8:9:18a}
\end{gather}
while for $S^{2} \leq 0$
\begin{gather}
\clebsch{a}{\alpha}{b}{\beta}{c}{\gamma}=(-1)^{2 b+c+\gamma+1} \sqrt{\frac{2 c+1}{2|S|}} Z^{\frac{1}{4}} \notag\\
\times \begin{cases}\cos \Omega_{0} A i(Z)-\sin \Omega_{0} B i(Z), & \text { if } \Omega-\Omega_{0}<0, \\
\cos \Omega_{0} B i(Z)-\sin \Omega_{0} A i(Z), & \text { if } \Omega-\Omega_{0}>0 .\end{cases}. \label{eq:8:9:18b}
\end{gather}
\end{subequations}
In these equations $A i(Z)$ and $B i(Z)$ are the Airy functions
\[
\begin{aligned}
Z & =\left(\frac{3}{2}\left|\Omega-\Omega_{0}\right|\right)^{\frac{1}{3}}, \\
\Omega_{0} & =j_{1} \theta_{1}^{(0)}+j_{2} \theta_{2}^{(0)}+j_{3} \theta_{3}^{(0)}+m_{1} \varphi_{2}^{(0)}-m_{2} \varphi_{1}^{(0)}, \\
\Omega & =j_{1} \theta_{1}+j_{2} \theta_{2}+j_{3} \theta_{3}-m_{2} \varphi_{1}+m_{1} \varphi_{2},
\end{aligned}
\]
$\theta_{i}^{(0)}$ and $\varphi_{i}^{(0)}$ being determined by the relations
\[
\begin{aligned}
& \theta_{i}^{(0)}= \begin{cases}0 & \text { if } 0 \leq \operatorname{Re} \theta_{i} \leq \pi / 2 \\
\pi & \text { if } \pi / 2<\operatorname{Re} \theta_{i} \leq \pi\end{cases} \\
& \varphi_{i}^{(0)}= \begin{cases}0 & \text { if } 0 \leq \operatorname{Re} \varphi_{i} \leq \pi / 2 \\
\pi & \text { if } \pi / 2<\operatorname{Re} \varphi_{i} \leq \pi\end{cases}
\end{aligned}
\]
\subsection{Squares of the Clebsch-Gordan Coefficients in the Classical Limit}\label{sec:8.9.4}\index{Clebsch-Gordan coefficient!squares in the classical limit}\index{probability amplitude}
According to Sec.~\ref{sec:8.1.1} the square of a Clebsch-Gordan coefficient represents the probability of the coupling of two angular momenta $\mathrm{j}_{1}$ and $\mathrm{j}_{2}$ with projections $m_{1}$ and $m_{2}$ into the resultant angular momentum j with , projection $m$. On the other hand, the same square is equal to the probability that the momenta $\vect{j}_{1}$ and $\vect{j}_{2}$ coupled into the momentum $j$ with the projection $m$ have the projections $m_{1}$ and $m_{2}$, respectively. Let us consider the square of the Clebsch-Gordan coefficients in the classical limit \cite{ref060}.

\begin{figure}[tbh!]
\begin{center}
\includegraphics{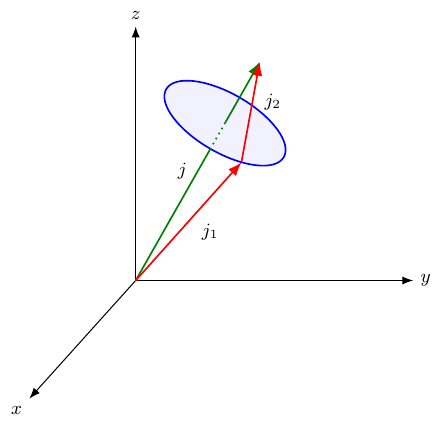}
\caption{Addition of angular momenta in the classical limit.}
\label{chap8:fig:2}
\end{center}
\end{figure}

If the vector $j$ as well as the magnitudes of $j_{1}$ and $j_{2}$ are given, the possible positions of the end of the vector $\mathrm{j}_{1}$ form a circle (Fig.~\ref{chap8:fig:2}). The probability in question is proportional to the length of the circle arc confined between the planes $z=m_{1}$ and $z=m_{2}$. In the classical limit (for $\left.j_{1}, j_{2}, j \gg 1\right)$ this probability may be expressed by
\begin{equation}
\left(\clebsch{j_{1}}{m_{1}}{j_{2}}{m_{2}}{j}{m}\right)^{2} \approx \frac{2 j+1}{\pi}\left[-\left(j_{1}^{4}+j_{2}^{4}+j^{4}\right)+2\left(j_{1}^{2} j_{2}^{2}+j_{1}^{2} j^{2}+j_{2}^{2} j^{2}\right)-4\left(j_{1}^{2} m_{2} m+j_{2}^{2} m_{1} m-j^{2} m_{1} m_{2}\right)\right]^{-\frac{1}{2}}. \label{eq:8:9:19}
\end{equation}
\eqref{eq:8:9:19} is valid only on the average, because if $j_{1}, j_{2}$ and $j$ are large, even fairly minor relative variations of these quantities lead to rapid oscillations of the Clebsch-Gordan coefficients (See \eqref{eq:8:9:12}). When $m_{1}=m_{2}=m=0$ and $j_{1}+j_{2}-j$ is even, one has
\begin{equation}
\clebsch{j_{1}}{0}{j_{2}}{0}{j}{0} \approx(-1)^{(j_1+j_{2}-j)/2} \sqrt{\frac{2 j+1}{\pi}}\left[-\left(j_{1}^{4}+j_{2}^{4}+j^{4}\right)+2\left(j_{1}^{2} j_{2}^{2}+j_{1}^{2} j^{2}+j_{2}^{2} j^{2}\right)\right]^{-\frac{1}{4}}. \label{eq:8:9:20}
\end{equation}

\section{ZEROS OF THE VECTOR-ADDITION COEFFICIENTS}\label{sec:8.10}\index{Clebsch-Gordan coefficient!zeros of}\index{zeros!of the vector-addition coefficients}\index{selection rules}
A Clebsch-Gordan coefficient $\clebsch{a}{\alpha}{b}{\beta}{c}{\gamma}$ and associated $3 j m$ symbol $\threej{a}{b}{c}{\alpha}{\beta}{-\gamma}$ vanish unless triangular conditions are satisfied (Eqs.~\eqref{eq:8:1:1} and \eqref{eq:8:1:2}). However, there exist non-trivial cases when the Clebsch-Gordan coefficients are equal to zero although the triangular conditions are fulfilled. This leads to some additional selection rules which forbid the quantum transitions whose amplitudes are proportional to the appropriate Clebsch-Gordan coefficients. Here we point out the most important relations between the arguments for which the Clebsch-Gordan coefficients vanish.
\begin{enumerate}[label=(\alph*)]
\item $\clebsch{a}{0}{b}{0}{c}{0}=0$, if $J \equiv a+b+c$ is odd;
\item $\clebsch{a}{\alpha}{a}{\alpha}{c}{\gamma}=0$, $\clebsch{a}{-\alpha}{c}{\gamma}{a}{\alpha}=0$, $\clebsch{c}{\gamma}{a}{-\alpha}{a}{\alpha}=0$, if $J=2 a+c$ is odd;
\item $\clebsch{a}{J-3 a}{b}{J-3 b}{c}{-(J-3 c)}=0$, if $J$ is odd;
\item $\clebsch{a}{\alpha}{b}{\beta}{a+b-1}{\gamma}=0$, if $\frac{\alpha}{\beta}=\frac{a}{b}$;
\item $\clebsch{a}{\alpha}{b}{\beta}{a-b+1}{\gamma}=0$ if $\frac{\alpha}{\beta}=-\frac{a+1}{b}$;
\item $\clebsch{a}{\alpha}{b}{\beta}{c}{c-1}=0$ if $a(a+1)-b(b+1)=(\alpha-\beta) c$, in particular
\[
\clebsch{a}{\alpha}{a}{\alpha}{c}{c-1}=0 .
\]
\item $\clebsch{a}{\alpha}{b}{b-1}{c}{\gamma}=0$ if $a(a+1)-c(c+1)=-(\alpha+\gamma) b$ in particular,
\[
\clebsch{a}{\alpha}{b}{b-1}{a}{-\alpha}=0 .
\]
\end{enumerate}
Below we list particular 3jm symbols with $J=a+b+c \leq 17$ which vanish in nontrivial cases. They are arranged in the order of increasing $J$ and $a \leq b \leq c$. Roots of other 3jm symbols may be obtained by the use of the simplest symmetry properties of the $3 j \mathrm{~m}$ symbols, concerned with the permutations of momenta $a, b, c$ and sign changes of the projections $\alpha, \beta, \gamma$ (see Eqs.~\eqref{eq:8:4:5} and \eqref{eq:8:4:6}), Hence, any $3 j m$ symbols from the table given below generally represents 12 formally different versions of the $3 j m$ symbols or the Clebsch-Gordan coefficients which are equal to zero. One should bear in mind that the $3 j \mathrm{~m}$ symbols and the Clebsch-Gordan coefficients are related by (see Sec.~\ref{sec:8.1.2})
\[
\threej{a}{b}{c}{\alpha}{\beta}{-\gamma}=(-1)^{a+b+\gamma} \frac{1}{\sqrt{2 c+1}} \clebsch{a}{\alpha}{b}{\beta}{c}{\gamma} .
\]
\section{CONNECTION OF THE CLEBSCH-GORDAN COEFFICIENTS AND THE 3 jm SYMBOLS WITH ANALOGOUS FUNCTIONS OF OTHER AUTHORS}\label{sec:8.11}\index{Clebsch-Gordan coefficient!notations of other authors}\index{notation!of other authors}
\[
\clebsch{j_{1}}{m_{1}}{j_{2}}{m_{2}}{j}{m}= \begin{cases}S_{j m_{1} m_{2}}^{j j_{2}} & \text { - Wigner \cite{ref043}, } \\
A_{m m_{1} m_{2}}^{j j_{1} j_{2}} & \text { - Eckart \cite{ref015}, } \\
C_{m_{1} m_{2}}^{j} & \text { - Vander Waerden \cite{ref040}, } \\
\left(j_{1} j_{2} m_{1} m_{2} \mid j_{1} j_{2} j m\right) & \text { - Condon and Shortley \cite{ref010}, } \\
C_{j m}^{j_{1} j_{2}}\left(m_{1} m_{2}\right) & \text { - Fock \cite{ref068}, } \\
X\left(j, m_{1}, j_{1}, j_{2}, m_{1}\right) & \text { - Boys \cite{ref059}, } \\
C\left(j m_{1} m_{1} m_{2}\right) & \text { - Blatt and Weisskopf \cite{ref007}, } \\
C_{m_{1} m_{2} m}^{j_{1} j_{2} j} & \text { - Biedenharn \cite{ref115}, } \\
C\left(j_{1} j_{2} j, m_{1} m_{2}\right) & \text { - Rose \cite{ref030}, } \\
{\left[\begin{array}{lll}
j_{1} & j_{2} & j \\
m_{1} & m_{2} & m
\end{array}\right]} & \text { - Yutsis and Bandzaitis \cite{ref045}, } \\
\braket{j_{1} m_{1} j_{2} m_{2}}{\left(j_{1} j_{2}\right) j m} & \text { - Fano \cite{ref066}. }\end{cases}
\]
\begin{table}[tbhp!]
  \caption{Accidental zeroes of the $3j$ coefficients for $J=j_1+j_2+j_3\leq 17$}
$\small\setlength{\arraycolsep}{2pt}
\begin{aligned}
 J=3& \threej{1}{1}{1}{0}{0}{0} . \\
J=5 &\threej{1}{2}{2}{0}{0}{0},\threej{3 / 2}{3 / 2}{2}{1 / 2}{1 / 2}{-1} . \\
J=7 &\threej{1}{3}{3}{0}{0}{0},\threej{2}{2}{3}{0}{0}{0},\threej{2}{2}{3}{1}{1}{-2},\threej{2}{5 / 2}{5 / 2}{-1}{1 / 2}{1 / 2} . \\
J=8 &\threej{2}{3}{3}{0}{2}{-2},\threej{3 / 2}{3}{7 / 2}{1 / 2}{1}{-3 / 2} \text {. } \\
J=9 &\threej{1}{4}{4}{0}{0}{0},\threej{2}{3}{4}{0}{0}{0},\threej{3}{3}{3}{0}{0}{0},\threej{3}{3}{3}{1}{1}{-2}, \\
& \threej{2}{7 / 2}{7 / 2}{-1}{1 / 2}{1 / 2},\threej{5 / 2}{3}{7 / 2}{3 / 2}{0}{-3 / 2},\threej{5 / 2}{5 / 2}{4}{1 / 2}{1 / 2}{-1},\threej{5 / 2}{5 / 2}{4}{3 / 2}{3 / 2}{-3} . \\
J=11 &\threej{1}{5}{5}{0}{0}{0},\threej{2}{4}{5}{0}{0}{0},\threej{2}{4}{5}{1}{2}{-3},\threej{3}{3}{5}{0}{0}{0},\threej{3}{3}{5}{1}{1}{-2},\threej{3}{3}{5}{2}{2}{-4}, \\
& \threej{3}{4}{4}{0}{0}{0},\threej{3}{4}{4}{-2}{1}{1},\threej{2}{9 / 2}{9 / 2}{-1}{1 / 2}{1 / 2},\threej{3}{7 / 2}{9 / 2}{2}{1 / 2}{-5 / 2},\threej{7 / 2}{7 / 2}{4}{1 / 2}{1 / 2}{-1}, \\
& \threej{7 / 2}{7 / 2}{4}{3 / 2}{3 / 2}{-3},\threej{3 / 2}{9 / 2}{5}{1 / 2}{3 / 2}{-2},\threej{5 / 2}{4}{9 / 2}{1 / 2}{3}{-7 / 2} \text {. } \\
J=13 &\threej{1}{6}{6}{0}{0}{0},\threej{2}{5}{6}{0}{0}{0},\threej{3}{4}{6}{0}{0}{0},\threej{3}{5}{5}{0}{0}{0},\threej{3}{5}{5}{-2}{1}{1},\threej{4}{4}{5}{0}{0}{0}, \\
& \threej{4}{4}{5}{1}{1}{-2},\threej{4}{4}{5}{2}{2}{-4},\threej{2}{11 / 2}{11 / 2}{-1}{1 / 2}{1 / 2},\threej{7 / 2}{9 / 2}{5}{5 / 2}{-1 / 2}{-2},\threej{7 / 2}{4}{11 / 2}{5 / 2}{1}{-7 / 2}, \\
& \threej{4}{9 / 2}{9 / 2}{-1}{1 / 2}{1 / 2},\threej{4}{9 / 2}{9 / 2}{-3}{3 / 2}{3 / 2},\threej{7 / 2}{7 / 2}{6}{1 / 2}{1 / 2}{-1},\threej{7 / 2}{7 / 2}{6}{3 / 2}{3 / 2}{-3},\threej{7 / 2}{7 / 2}{6}{5 / 2}{5 / 2}{-5} . \\
J=14 &\threej{3}{5}{6}{1}{-1}{0},\threej{3}{5}{6}{2}{1}{-3},\threej{3}{5}{6}{1}{4}{-5},\threej{4}{4}{6}{0}{2}{-2},\threej{4}{5}{5}{1}{3}{-4},\threej{4}{5}{5}{2}{1}{-3}, \\
& \threej{3}{9 / 2}{13 / 2}{1}{3 / 2}{-5 / 2},\threej{5 / 2}{5}{13 / 2}{1 / 2}{1}{-3 / 2},\threej{5 / 2}{5}{13 / 2}{3 / 2}{3}{-9 / 2},\threej{4}{9 / 2}{11 / 2}{0}{7 / 2}{-7 / 2},\threej{3 / 2}{6}{13 / 2}{1 / 2}{2}{-5 / 2},\threej{5 / 2}{11 / 2}{6}{3 / 2}{1 / 2}{-2} . \\
J=15 &\threej{1}{7}{7}{0}{0}{0},\threej{2}{6}{7}{0}{0}{0},\threej{2}{6}{7}{1}{3}{-4},\threej{3}{5}{7}{0}{0}{0},\threej{3}{6}{6}{0}{0}{0},\threej{3}{6}{6}{0}{5}{-5},\threej{3}{6}{6}{-2}{1}{1}, \\
& \threej{4}{5}{6}{0}{0}{0},\threej{4}{5}{6}{3}{0}{-3},\threej{4}{4}{7}{0}{0}{0},\threej{4}{4}{7}{1}{1}{-2},\threej{4}{4}{7}{2}{2}{-4},\threej{4}{4}{7}{3}{3}{-6},\threej{5}{5}{5}{0}{0}{0}, \\
& \threej{5}{5}{5}{1}{1}{-2},\threej{5}{5}{5}{2}{2}{-4},\threej{2}{13 / 2}{13 / 2}{-1}{1 / 2}{1 / 2},\threej{4}{9 / 2}{13 / 2}{3}{3 / 2}{-9 / 2},\threej{4}{11 / 2}{11 / 2}{-1}{1 / 2}{1 / 2},\threej{4}{11 / 2}{11 / 2}{-3}{3 / 2}{3 / 2}, \\
& \threej{9 / 2}{5}{11 / 2}{3 / 2}{0}{-3 / 2},\threej{9 / 2}{9 / 2}{6}{1 / 2}{1 / 2}{-1},\threej{9 / 2}{9 / 2}{6}{3 / 2}{3 / 2}{-3},\threej{9 / 2}{9 / 2}{6}{5 / 2}{5 / 2}{-5} . \\
J=17 &\threej{1}{8}{8}{0}{0}{0},\threej{2}{7}{8}{0}{0}{0},\threej{3}{6}{8}{0}{0}{0},\threej{3}{6}{8}{2}{4}{-6},\threej{3}{7}{7}{0}{0}{0},\threej{3}{7}{7}{-2}{1}{1},\threej{4}{5}{8}{0}{0}{0},\threej{4}{6}{7}{0}{0}{0}, \\
& \threej{5}{5}{7}{0}{0}{0},\threej{5}{5}{7}{1}{1}{-2},\threej{5}{5}{7}{2}{2}{-4},\threej{5}{5}{7}{3}{3}{-6},\threej{5}{6}{6}{0}{0}{0},\threej{5}{6}{6}{-2}{1}{1},\threej{5}{6}{6}{-4}{2}{2}, \\
& \threej{2}{15 / 2}{15 / 2}{-1}{1 / 2}{1 / 2},\threej{3}{13 / 2}{15 / 2}{2}{3 / 2}{-7 / 2},\threej{4}{13 / 2}{13 / 2}{-1}{1 / 2}{1 / 2},\threej{4}{13 / 2}{13 / 2}{-3}{3 / 2}{3 / 2},\threej{5}{11 / 2}{13 / 2}{2}{1 / 2}{-5 / 2},\threej{7 / 2}{6}{15 / 2}{3 / 2}{5}{-13 / 2}, \\
& \threej{9 / 2}{6}{13 / 2}{1 / 2}{-5}{9 / 2},\threej{9 / 2}{6}{13 / 2}{-7 / 2}{1}{5 / 2},\threej{11 / 2}{11 / 2}{6}{1 / 2}{1 / 2}{-1},\threej{11 / 2}{11 / 2}{6}{3 / 2}{3 / 2}{-3},\threej{11 / 2}{11 / 2}{6}{5 / 2}{5 / 2}{-5},\threej{9 / 2}{11 / 2}{7}{7 / 2}{1 / 2}{-4}, \\
& \threej{3 / 2}{15 / 2}{8}{1 / 2}{5 / 2}{-3},\threej{9 / 2}{9 / 2}{8}{1 / 2}{1 / 2}{-1},\threej{9 / 2}{9 / 2}{8}{3 / 2}{3 / 2}{-3},\threej{9 / 2}{9 / 2}{8}{5 / 2}{5 / 2}{-5},\threej{9 / 2}{9 / 2}{8}{7 / 2}{7 / 2}{-7},\threej{9 / 2}{5}{15 / 2}{7 / 2}{2}{-11 / 2} .
\end{aligned}
$
\end{table}
\[
\threej{j_{1}}{j_{2}}{j_{3}}{m_{1}}{m_{2}}{m_{3}}= \begin{cases}(-1)^{-j_{1}+j_{2}+j_{3}} V\left(j_{1} j_{2} j_{3}, m_{1} m_{2} m_{3}\right) & - \text { Racah \cite{ref091}} \\
(-1)^{-j_{1}+j_{2}+j_{3}} \bar{V}\left(j_{1} j_{2} j_{3}, m_{1} m_{2} m_{3}\right) & - \text { Fano and Racah \cite{ref018} } \\
(-1)^{j_{1}-j_{2}+j_{3}} S_{j_{1} m_{1} j_{2} m_{2} j_{3} m_{3}} & - \text { Landau and Lifshitz \cite{ref025} } \\
(-1)^{j_{1}-j_{2}+j_{3}}\left(j_{1} m_{1}, j_{2} m_{2}, j_{3} m_{3}|0\rangle\right. & - \text { Fano \cite{ref066} } \\
X\left(j_{1} j_{2} j_{3}, m_{1} m_{2} m_{3}\right) & - \text { Schwinger \cite{ref101} } \\
(-1)^{j_{1}-j_{2}+j_{3}} U\left(j_{1} j_{2} m_{1} m_{2} \mid j_{3}-m_{3}\right) & - \text { Lubarskii \cite{ref026} } \\
\left(j_{1} j_{2} j_{3}\right)_{m_{1} m_{2} m_{3}} & - \text { Sharp \cite{ref102} }\end{cases}
\]
\section{ALGEBRAIC TABLES OF THE CLEBSCH-GORDAN COEFFICIENTS}\label{sec:8.12}\index{Clebsch-Gordan coefficient!algebraic tables}\index{tables!of the Clebsch-Gordan coefficients}
Below we present tables of algebraic formulas for the Clebsch-Gordan coefficients $\clebsch{a}{\alpha}{b}{\beta}{c}{\gamma}$ with $b=1 / 2,1,3 / 2,2$, $5 / 2,3,7 / 2,4,9 / 2,5$. If $b=3,7 / 2,4,9 / 2,5$, we give $\clebsch{a}{\alpha}{b}{\beta}{c}{\gamma}$ with $\beta \geq 0$, while $\clebsch{a}{\alpha}{b}{\beta}{c}{\gamma}$ with $\beta<0$ may be obtained from the symmetry property
\[
\clebsch{a}{\alpha}{b}{\beta}{c}{\gamma}=(-1)^{a+b-c} \clebsch{a}{-\alpha}{b}{-\beta}{c}{-\gamma} .
\]
The algebraic tables of the Clebsch-Gordan coefficients are also available in Ref.~\cite{ref010} for $b=1 / 2,1,3 / 2,2$; in Ref.~$[121,125]$ for $b=5 / 2$; in Refs.~\cite{ref117,ref130} with $b=3$; and in Ref.~\cite{ref126} with $b=7 / 2,4,9 / 2,5$.

\section{NUMERICAL TABLES OF THE CLEBSCH-GORDAN COEFFICIENTS}\label{sec:8.13}\index{Clebsch-Gordan coefficient!numerical tables}\index{tables!numerical}
In Table 8.11 we present numerical values of the Clebsch-Gordan coefficients for $a, b, c \leq 3$. These values are given in the form of square roots of rational fractions and in decimals. The Clebsch-Gordan coefficients are separated into groups with respect to the values of $c$. The arguments $a, \alpha, b, \beta$ for the Clebsch-Gordan coefficients given in Table 8.11 satisfy the inequalities (a) $a \geq b, \alpha \geq 0$; (b) $\alpha \geq \beta$ for $a=b$. Other ClebschGordan coefficients with $a, b, c \leq 3$ may be reduced to those in Table 8.11 by the use of the symmetry properties
\[
\clebsch{a}{\alpha}{b}{\beta}{c}{\alpha+\beta}=(-1)^{a+b-c} \clebsch{b}{\beta}{a}{\alpha}{c}{\alpha+\beta}=(-1)^{a+b-c} \clebsch{a}{-\alpha}{b}{-\beta}{c}{-\alpha-\beta}=\clebsch{b}{-\beta}{a}{-\alpha}{c}{-\alpha-\beta}
\]
Numerical values of the Clebsch-Gordan coefficients are also given in Refs.~\cite{ref126, ref127}. Reference \cite{ref127} presents the Clebsch-Gordan coefficients for $\frac{1}{2} \leq a \leq 4, \frac{1}{2} \leq b, c \leq 9 / 2$ in decimals. Reference \cite{ref126} gives the Clebsch-Gordan coefficients for $\frac{1}{2} \leq b \leq 6, a=5,11 / 2,6$ in the form of rational fractions. Numerical values of the $3 j m$ symbols are also available in Ref.~\cite{ref113}.

Tables~\ref{chap8:tab:1}--\ref{chap8:tab:2}. Algebraic Formulas for the Clebsch-Gordan Coefficients.
Tables not reproduced here can be found in the original book.
\begin{table}[tbh!]
\begin{center}
\caption{$\clebsch{a}{\alpha}{\frac{1}{2}}{\beta}{c}{\gamma}$}
\label{chap8:tab:1}
\begin{tabular}{c|c|c}
\hline
$c$ & $\beta=1 / 2$ & $\beta=-1 / 2$ \\
\hline
$a+1 / 2$ & $\left[\frac{c+\gamma}{2 c}\right]^{1 / 2}$ & $\left[\frac{c-\gamma}{2 c}\right]^{1 / 2}$ \\
$a-1 / 2$ & $-\left[\frac{c-\gamma+1}{2 c+2}\right]^{1 / 2}$ & $\left[\frac{c+\gamma+1}{2 c+2}\right]^{1 / 2}$ \\
\end{tabular}
\end{center}
\end{table}

\begin{table}[tbh!]
\begin{center}
\caption{$\clebsch{a}{\alpha}{1}{\beta}{c}{\gamma}$}
\label{chap8:tab:2}
\begin{tabular}{c|c|c|c}
\hline
c & $\beta=1$ & $\beta=0$ & $\beta=-1$ \\
\hline
$a+1$ & $\left[\frac{(c+\gamma-1)(c+\gamma)}{(2 c-1) 2 c}\right]^{1 / 2}$ & 
\(\left[\frac{(c+\gamma)(c-\gamma)}{(2 c-1) c}\right]^{1 / 2}\)& 
\(\left[\frac{(c-\gamma-1)(c-\gamma)}{(2 c-1) 2c}\right]^{1 / 2}\)
\\
$a$ & \(-\left[\frac{(c+\gamma)(c-\gamma+1)}{2 c(c+1)}\right]^{1 / 2 }\) &\(\frac{\gamma}{[c(c+1)]^{1 / 2}}\)&
$\left[\frac{(c+\gamma+1)(c-\gamma)}{2 c(c+1)}\right]^{1 / 2}$\\
$a-1$ & \(  \left[\frac{(c-\gamma+1)(c-\gamma+2)}{(2 c+2)(2 c+3)}\right]^{1 / 2} \) 
 & \(-\left[\frac{(c+\gamma+1)(c-\gamma+1)}{( c+1)(2 c+3)}\right]^{1 / 2} \)
 & \( \left[\frac{(c+\gamma+2)(c+\gamma+1)}{(2 c+2)(2 c+3)}\right]^{1 / 2} \) 
\end{tabular}
\end{center}
\end{table}

\chapter{6j Symbols and the Racah Coefficients}\label{chap:9}\index{6j symbol@$6j$ symbol}\index{Racah coefficient}
% styles for the Sec.~\ref{sec:9.9} tetrahedron figures (Figs. 9.1-9.3)

\section{Definition}\label{sec:9.1}\index{6j symbol@$6j$ symbol!definition}
\subsection{6j Symbols}\label{sec:9.1.1}\index{recoupling!of three angular momenta}\index{coupling schemes}
The Wigner $6 j$ symbols \cite{ref110} are related to the coefficients of transformations between different coupling schemes of three angular momenta.\isymo{6j@$\bigl\{\begin{smallmatrix} a & b & c\\ d & e & f\end{smallmatrix}\bigr\}$, Wigner $6j$ symbol}\index{addition of angular momenta} The angular momenta $\mathrm{j}_{1}, \mathrm{j}_{2}, \mathrm{j}_{3}$ may be coupled to give a resultant angular momentum $j$ and its projection $m$ in three different ways:
\begin{equation}
\begin{array}{rlrl}
\text { I) } \vect{j}_{1}+\vect{j}_{2}  =\vect{j}_{12}, & \vect{j}_{12}+\vect{j}_{3}=\vect{j}, \\
\text { II) } \vect{j}_{2}+\vect{j}_{3}=\vect{j}_{23}, & \vect{j}_{1}+\vect{j}_{23}=\vect{j}, \\
\text { III) } \vect{j}_{1}+\vect{j}_{3}=\vect{j}_{13}, & \vect{j}_{13}+\vect{j}_{2}=\vect{j} . \label{eq:9:1:1}
\end{array}
\end{equation}
Let $\ket{j_{1} j_{2}\left(j_{12}\right) j_{3} j m}$ denote the state vectors corresponding to the coupling scheme I. These vectors are eigenvectors of the operators $\widehat{j}_{1}^{2}, \widehat{j}_{2}^{2}, \hat{j}_{3}^{2}, \hat{j}_{12}^{2}, \hat{j}^{2}, \hat{j}_{x}$ and may be written as
\begin{equation}
\ket{j_{1} j_{2}\left(j_{12}\right) j_{3} j m}=\sum_{m_{1} m_{2} m_{3}} \clebsch{j_{12}}{m_{12}}{j_{3}}{m_{3}}{j}{m} \clebsch{j_{1}}{m_{1}}{j_{2}}{m_{2}}{j_{12}}{m_{12}}\ket{j_{1} m_{1}, j_{2} m_{2}, j_{3} m_{3}} . \label{eq:9:1:2}
\end{equation}
The state vectors corresponding to the coupling scheme II are eigenvectors of the operators $\widehat{j}_{1}^{2}, \widehat{j}_{2}^{2}, \widehat{j}_{3}^{2}, \widehat{j}_{23}^{2}, \widehat{j}^{2}, \widehat{j}_{x}$,
\begin{equation}
\ket{j_{1}, j_{2} j_{3}\left(j_{23}\right) j m}=\sum_{m_{1} m_{2} m_{3}} \clebsch{j_{1}}{m_{1}}{j_{23}}{m_{23}}{j}{m} \clebsch{j_{2}}{m_{2}}{j_{3}}{m_{3}}{j_{23}}{m_{23}}\ket{j_{1} m_{1}, j_{2} m_{2}, j_{3} m_{3}} . \label{eq:9:1:3}
\end{equation}
Similarly, the state vectors corresponding to the coupling scheme III are eigenvectors of the operators $\widehat{\vect{j}}_{1}^{2}, \widehat{\vect{j}}_{2}^{2}, \widehat{\vect{j}}_{3}^{2}$, $\widehat{j}_{13}^{2}, \widehat{j}^{2}, \widehat{j}_{x}$,
\begin{equation}
\ket{j_{1} j_{3}\left(j_{13}\right) j_{2} j m}=\sum_{m_{1} m_{2} m_{3}} \clebsch{j_{13}}{m_{13}}{j_{2}}{m_{2}}{j}{m} \clebsch{j_{1}}{m_{1}}{j_{3}}{m_{3}}{j_{13}}{m_{13}}\ket{j_{1} m_{1}, j_{2} m_{2}, j_{3} m_{3}} . \label{eq:9:1:4}
\end{equation}
States belonging to each coupling scheme form a complete set of states. A transition from one coupling scheme to another is performed by some unitary transformation which relates the states with the same total angular momentum $j$ and projection $m$. The coefficients $U$ of this transformation differ from the $6 j$ symbols only by normalization and phase factors. These factors are chosen in such a way to make the $6 j$ symbols more symmetric (Sec.~\ref{sec:9.4}).

One defines the Wigner $6 j$ symbols $\sixj{j_{1}}{j_{2}}{j_{12}}{j_{3}}{j}{j_{23}}$ by the relation
\begin{align}
 \braket{j_{1} j_{2}\left(j_{12}\right) j_{3} j m}{j_{1}, j_{2} j_{3}\left(j_{23}\right) j^{\prime} m^{\prime}}=&\delta_{j j^{\prime}} \delta_{m m^{\prime}} U\left(j_{1} j_{2} j j_{3} ; j_{12} j_{23}\right)  \notag\\
= & \delta_{j j^{\prime}} \delta_{m m^{\prime}}(-1)^{j_{1}+j_{2}+j_{3}+j} \sqrt{\left(2 j_{12}+1\right)\left(2 j_{23}+1\right)}\sixj{j_{1}}{j_{2}}{j_{12}}{j_{3}}{j}{j_{23}} . \label{eq:9:1:5}
\end{align}
From Eq.~\eqref{eq:9:1:5} one may obtain [92,64]
\begin{align}
\braket{j_{1} j_{2}\left(j_{12}\right) j_{3} j m}{j_{1} j_{3}\left(j_{13}\right) j_{2} j^{\prime} m^{\prime}}=&\delta_{j j^{\prime}} \delta_{m m^{\prime}}(-1)^{j+j_{1}-j_{12}-j_{13}} U\left(j_{2} j_{1} j j_{3} ; j_{12} j_{13}\right)  \notag\\
=&\delta_{j j^{\prime}} \delta_{m m^{\prime}}(-1)^{j_{2}+j_{3}+j_{12}+j_{13}} \sqrt{\left(2 j_{12}+1\right)\left(2 j_{13}+1\right)}\sixj{j_{2}}{j_{1}}{j_{12}}{j_{3}}{j}{j_{13}}  \label{eq:9:1:6}\\
\braket{j_{1}, j_{2} j_{3}\left(j_{23}\right) j m}{j_{1} j_{3}\left(j_{13}\right) j_{2} j^{\prime} m^{\prime}}=&\delta_{j j^{\prime}} \delta_{m m^{\prime}}(-1)^{j_{2}+j_{3}-j_{23}} U\left(j_{1} j_{3} j j_{2} ; j_{13} j_{23}\right)  \notag\\
=&\delta_{j j^{\prime}} \delta_{m m^{\prime}}(-1)^{j_{1}+j+j_{23}} \sqrt{\left(2 j_{13}+1\right)\left(2 j_{23}+1\right)}\sixj{j_{1}}{j_{3}}{j_{13}}{j_{2}}{j}{j_{23}} . \label{eq:9:1:7}
\end{align}
According to the definition \eqref{eq:9:1:5} the $6 j$ symbols may be given in terms of the Clebsch-Gordan coefficients\index{Clebsch-Gordan coefficient}\isym{C-clebsch@$\clebsch{a}{\alpha}{b}{\beta}{c}{\gamma}$, Clebsch-Gordan coefficient}
\begin{multline}
\sum \clebsch{j_{12}}{m_{12}}{j_{3}}{m_{3}}{j}{m} \clebsch{j_{1}}{m_{1}}{j_{2}}{m_{2}}{j_{12}}{m_{12}} \clebsch{j_{1}}{m_{1}}{j_{23}}{m_{23}}{j^{\prime}}{m^{\prime}} \clebsch{j_{2}}{m_{2}}{j_{3}}{m_{3}}{j_{23}}{m_{23}}  \\
=\delta_{j j^{\prime}} \delta_{m m^{\prime}}(-1)^{j_{1}+j_{2}+j_{3}+j} \sqrt{\left(2 j_{12}+1\right)\left(2 j_{23}+1\right)}\sixj{j_{1}}{j_{2}}{j_{12}}{j_{3}}{j}{j_{23}} . \label{eq:9:1:8}
\end{multline}
Here the sum is over $m_{1}, m_{2}, m_{3}, m_{12}, m_{23}$ while $m$ and $m^{\prime}$ are fixed. This relation completely determines absolute values and phases of the $6 j$ symbols. The $6 j$ symbols turn out to be real just as the Clebsch-Gordan coefficients are.

The quantum-mechanical rules of vector addition impose some restrictions on possible values of momenta which are arguments of the $6 j$ symbol $\sixj{j_{1}}{j_{2}}{j_{12}}{j_{3}}{j}{j_{23}}$.
\begin{enumerate}[label=(\alph*)]
\item All momenta are integer or half-integer nonnegative numbers (with one exception considered in Sec.~\ref{sec:9.4}).
\item Each triad $\left(j_{1} j_{2} j_{12}\right),\left(j_{12} j_{3} j\right),\left(j_{2} j_{3} j_{23}\right)$ and $\left(j_{23} j_{1} j\right)$ should satisfy the triangular condition (Eq.~\eqref{eq:8:1:1}).\index{triangle condition}\index{triad} 
\end{enumerate}
The unitarity of the recoupling transformations implies the orthogonality and normalization conditions of the $6 j$ symbols,\index{orthogonality!of the $6j$ symbols}\index{normalization!of the $6j$ symbols}\index{unitarity!of the recoupling transformations}
\begin{align}
& \sum_{j_{12}}\left(2 j_{12}+1\right)\left(2 j_{23}+1\right)\sixj{j_{1}}{j_{2}}{j_{12}}{j_{3}}{j}{j_{23}}\sixj{j_{1}}{j_{2}}{j_{12}}{j_{3}}{j}{j_{23}^{\prime}}=\delta_{j_{23} j_{23}^{\prime}} , \label{eq:9:1:9}\\
& \sum_{j_{23}}\left(2 j_{12}+1\right)\left(2 j_{23}+1\right)\sixj{j_{1}}{j_{2}}{j_{12}}{j_{3}}{j}{j_{23}}\sixj{j_{1}}{j_{2}}{j_{12}^{\prime}}{j_{3}}{j}{j_{23}}=\delta_{j_{12} j_{12}^{\prime}} . \label{eq:9:1:10}
\end{align}
Below we shall use Latin letters $a, b, c$, etc., to denote arguments of the $6 j$ symbols.

\subsection{Racah Coefficients}\label{sec:9.1.2}\index{Racah coefficient!definition}
Instead of the Wigner $6 j$ symbols the Racah coefficients \cite{ref091} are often used, especially in spectroscopy theory.\isym{W-racah@$W(a b e d ; c f)$, Racah coefficient} These coefficients differ from the $6 j$ symbols only by a phase factor:
\begin{equation}
\sixj{a}{b}{c}{d}{e}{f}  \label{eq:9:1:11} \equiv(-1)^{a+b+d+e} W(a b e d ; c f) .
\end{equation}
The Racah coefficients were introduced independently of the $6 j$ symbols. The phase of the Racah coefficients coincides with the phase of the coefficients which describe the transformation between I and II coupling schemes (Eq.~\eqref{eq:9:1:5}).

\subsection{$R$ Symbols}\label{sec:9.1.3}\index{R-symbol@$R$-symbol!of the $6j$ symbol}\index{Shelepin R-symbol@Shelepin $R$-symbol}
The $6 j$ symbols and the Racah coefficients may be written in the form of a $3 \times 4$ array\isym{R-ialpha@$R_{i \alpha}$, $R$-symbol of the $6j$ symbol} $\left\|R_{i \alpha}\right\|(i=1,2,3 ; \alpha= 1,2,3,4)$ which is called the $R$-symbol (Shelepin \cite{ref105})
\begin{equation}
\left\|\begin{array}{llll}
R_{11} & R_{12} & R_{13} & R_{14}  \label{eq:9:1:12}\\
R_{21} & R_{22} & R_{23} & R_{24} \\
R_{31} & R_{32} & R_{33} & R_{34}
\end{array}\right\| \equiv\sixj{a}{b}{c}{d}{e}{f} \equiv(-1)^{a+b+d+e} W(a b e d; c f) ,
\end{equation}
where
\begin{equation}
\begin{array}{llll}
R_{11}=-c+d+e, & R_{12}=b+d-f, & R_{13}=a+e-f, & R_{14}=a+b-c, \\
R_{21}=-b+d+f, & R_{22}=c+d-e . & R_{23}=a-b+c, & R_{24}=a-e+f, \\
R_{31}=-a+e+f, & R_{32}=-a+b+c, & R_{33}=c-d+e, & R_{34}=b-d+f . \label{eq:9:1:13}
\end{array}
\end{equation}
The inverse relations are
\begin{equation}
\begin{array}{ll}
2 a=R_{13}+R_{24}=R_{14}+R_{23}, & 2 d=R_{11}+R_{22}=R_{12}+R_{21}, \\
2 b=R_{12}+R_{34}=R_{14}+R_{32}, & 2 e=R_{11}+R_{33}=R_{13}+R_{31}, \\
2 c=R_{22}+R_{33}=R_{23}+R_{32}, & 2 f=R_{21}+R_{34}=R_{24}+R_{31} . \label{eq:9:1:14}
\end{array}
\end{equation}
All 12 elements $R_{i \alpha}$ are integer nonnegative numbers. The differences between corresponding elements of rows and columns are constant:
\begin{equation}
\begin{aligned}
R_{i \alpha}-R_{k \alpha} & =R_{i \beta}-R_{k \beta}, \\
R_{i \alpha}-R_{i \beta} & =R_{k \alpha}-R_{k \beta}, 
\end{aligned}\quad(i, k=1,2,3 ; \alpha, \beta=1,2,3,4) .
\end{equation}
Note the following relations:
\begin{gather}
\sum_{i=1}^{3} R_{i 1}=2(d+e+f)-a-b-c, \quad \sum_{i=1}^{3} R_{i 3}=2(a+c+e)-b-d-f,  \notag\\
\sum_{i=1}^{3} R_{i 2}=2(b+c+d)-a-e-f, \quad \sum_{i=1}^{3} R_{i 4}=2(a+b+f)-c-d-e,  \label{eq:9:1:16}\\
\sum_{i, \alpha} R_{i \alpha}=2(a+b+c+d+e+f) . \notag
\end{gather}
One may also use the following parametrisation of the elements $R_{i \alpha}$ \cite{ref045}:
\begin{equation}
R_{i \alpha}=A_{i}-B_{\alpha} . \label{eq:9:1:17}
\end{equation}
Here $A_{i}, B_{\alpha}$ are integer nonnegative numbers
\begin{equation}
\begin{array}{ll}
A_{1}=a+b+d+e, & B_{1}=a+b+c,  \\
A_{2}=a+c+d+f, & B_{2}=a+e+f, \\
A_{3}=b+c+e+f, & B_{3}=b+d+f, \\
& B_{4}=c+d+e, \label{eq:9:1:18}
\end{array}
\end{equation}
with
\begin{equation}
\sum_{i=1}^{3} A_{i}=\sum_{\alpha=1}^{4} B_{\alpha}=2(a+b+c+d+e+f). \label{eq:9:1:19}
\end{equation}
The inverse relations are
\begin{align}
2 a & =A_{1}+A_{2}-B_{3}-B_{4}, & 2 d & =A_{1}+A_{2}-B_{1}-B_{2},  \notag\\
2 b & =A_{1}+A_{3}-B_{2}-B_{4}, & 2 e & =A_{1}+A_{3}-B_{1}-B_{3},  \notag\\
2 c & =A_{2}+A_{3}-B_{2}-B_{3}, & 2 f & =A_{2}+A_{3}-B_{1}-B_{4} . \label{eq:9:1:20}
\end{align}
The $R$ symbols provide the simplest formulation of the symmetry properties of the $6 j$ symbols and Racah coefficients.

\section{General Expressions for the $6 j$ Symbols. Relations Between the $6 j$ Symbols and Other Functions}\label{sec:9.2}\index{6j symbol@$6j$ symbol!general expressions}\index{6j symbol@$6j$ symbol!relations to other functions}
The $6 j$ symbols $\sixj{a}{b}{c}{d}{e}{f}$ vanish if at least one of the triads $(a b c),(c d e),(a e f)$ and $(b d f)$ does not obey the triangular conditions \eqref{eq:8:1:1}. The expressions for the $6 j$ symbols given below are valid if all these conditions are satisfied. Corresponding expressions for the Racah coefficients may be obtained by the use of the relations between these coefficients and the $6 j$ symbols \eqref{eq:9:1:11}.

\subsection{Expressions for the 6j Symbols in Terms of Finite Sums}\label{sec:9.2.1}\index{6j symbol@$6j$ symbol!as finite sums}
In the expressions presented below the sums are over all integer nonnegative values of $n$ so that no factorial in denominators has a negative argument. The quantities $\Delta(a b c)$ are defined in Eq.~$\eqref{eq:8:2:1}$.\index{triangle coefficient}\index{Delta-symbol@$\Delta$-symbol}\isymg{Delta-abc@$\Delta(a b c)$, triangle coefficient} %Numerical values of $\Delta(a b c)$ are given in Table 8.12.
\begin{gather}
\sixj{a}{b}{c}{d}{e}{f}=\Delta(a b c) \Delta(c d e) \Delta(a e f) \Delta(b d f)  \notag\\
\times \sum_{n} \frac{(-1)^{n}(n+1)!}{(n-a-b-c)!(n-c-d-e)!(n-a-e-f)!(n-b-d-f)!(a+b+d+e-n)!}  \notag\\
\times \frac{1}{(a+c+d+f-n)!(b+c+e+f-n)!}  \notag\\
\text { (Racah \cite{ref091}) } \label{eq:9:2:1}
\end{gather}
By the replacement $n \rightarrow a+b+d+e-n$ one can rewrite Eq.~\eqref{eq:9:2:1} in the form
\begin{gather}
\sixj{a}{b}{c}{d}{e}{f}=(-1)^{a+b+d+e} \Delta(a b c) \Delta(c d e) \Delta(a e f) \Delta(b d f)  \notag\\
\times \sum_{n} \frac{(-1)^{n}(a+b+d+e+1-n)!}{n!(a+b-c-n)!(-c+d+e-n)!(a+e-f-n)!(b+d-f-n)!}  \notag\\
\times \frac{1}{(-a+c-d+f+n)!(-b+c-e+f+n)!}  \notag\\
\text { (Racah \cite{ref091}) } \label{eq:9:2:2}
\end{gather}
Some other expressions for the $6 j$ symbols which cannot be easily reduced to \eqref{eq:9:2:1} and \eqref{eq:9:2:2} are [45,50].
\begin{align}
& \sixj{a}{b}{c}{d}{e}{f}=(-1)^{a+c+d+f} \frac{\Delta(a e f) \Delta(b d f)}{\Delta(a b c) \Delta(c d e)}  \notag\\
& \times \sum_{n}(-1)^{n} \frac{(-a+b+c+n)!(c-d+e+n)!(a-c+d+f-n)!}{n!(a-e+f-n)!(-b+d+f-n)!(-a+b-d+e+n)!(b+c+e-f+1+n)!},  \label{eq:9:2:3}\\
& \sixj{a}{b}{c}{d}{e}{f}=(-1)^{a+c+d+f} \frac{\Delta(a b c) \Delta(b d f)}{\Delta(a e f) \Delta(c d e)}  \notag\\
& \times \sum_{n}(-1)^{n} \frac{(a-b+d+e-n)!(-b+c+e+f-n)!(a+c+d+f+1-n)!}{n!(a-b+c-n)!(-b+d+f-n)!(a+e+f+1-n)!(c+d+e+1-n)!},  \label{eq:9:2:4}\\
& \sixj{a}{b}{c}{d}{e}{f}=(-1)^{a+b+d+e} \frac{\Delta(a b c) \Delta(c d e) \Delta(a e f) \Delta(b d f)(a+e+f+1)!(b+d+f+1)!}{(a+b-c)!(a-b+c)!(-c+d+e)!(c+d-e)!(-a+e+f)!(b-d+f)!}  \notag\\
& \times \sum_{n}(-1)^{n} \frac{(-a+e+f+n)!(b-d+f+n)!(a+c+d-f-n)!}{n!(a+e-f-n)!(b+d-f-n)!(-a+c-d+f+n)!(2 f+1+n)!},  \label{eq:9:2:5}\\
& \sixj{a}{b}{c}{d}{e}{f}=(-1)^{b+c+e+f} \frac{\Delta(a b c) \Delta(c d e) \Delta(a e f) \Delta(b d f)(a+b+c+1)!(b+d+f+1)!}{(a+b-c)!(c-d+e)!(c+d-e)!(a-e+f)!(-a+e+f)!(b+d-f)!}  \notag\\
& \times \sum_{n}(-1)^{n} \frac{(2 b-n)!(b+c-e+f-n)!(b+c+e+f+1-n)!}{n!(-a+b+c-n)!(b-d+f-n)!(a+b+c+1-n)!(b+d+f+1-n)!} . \label{eq:9:2:6}
\end{align}
\subsection{Bargmann Formula \cite{ref058}}\label{sec:9.2.2}\index{Bargmann formula}\index{6j symbol@$6j$ symbol!Bargmann formula}
\begin{equation}
\sixj{a}{b}{c}{d}{e}{f}  \label{eq:9:2:7} \equiv\left\|\begin{array}{llll}
R_{11} & R_{12} & R_{13} & R_{14} \\
R_{21} & R_{22} & R_{23} & R_{24} \\
R_{31} & R_{32} & R_{33} & R_{34}
\end{array}\right\|=\left[\frac{\prod_{i=1}^{3} \prod_{\alpha=1}^{4}\left(R_{i \alpha}\right)!}{\prod_{\alpha=1}^{4}\left(B_{\alpha}+1\right)!}\right]^{\frac{1}{2}} \sum_{x_{i}, y_{\alpha}} \frac{(-1)^{n}(n+1)!}{\prod_{i=1}^{3}\left(x_{i}\right)!\prod_{\alpha=1}^{4}\left(y_{\alpha}\right)!}
\end{equation}
Here $R_{i \alpha}$ are elements of the $R$ symbol (Sec.~\ref{sec:9.1.3}), $B_{\alpha}$ are given by Eq.~\eqref{eq:9:1:18}, $x_{i}, y_{\alpha}$ are summation indices, $n \equiv \sum_{i=1}^{3} x_{i}+\sum_{\alpha=1}^{4} y_{\alpha}$. The sums are over all integer nonnegative values of $x_{i}, y_{\alpha}$ which satisfy the conditions $x_{i}+y_{\alpha}=R_{i \alpha}$. These conditions show that only one of the summation indices is independent. The sum in \eqref{eq:9:2:7} contains $r+1$ terms where $r=\min \left\{R_{i \alpha}\right\}$. If we take the quantity $n$ (integer nonnegative) as an independent summation index, then $x_{i}=A_{i}-n, y_{\alpha}=n-B_{\alpha}, A_{i}$ and $B_{\alpha}$ being given by Eq.~\eqref{eq:9:1:18}. In this case the Bargmann formula \eqref{eq:9:2:7} reduces to the Racah formula \eqref{eq:9:2:1}.

\subsection{Relations Between the $6 j$ Symbols and the Generalized Hypergeometric Functions}\label{sec:9.2.3}\index{6j symbol@$6j$ symbol!and hypergeometric functions}\index{hypergeometric function!generalized}
The $6 j$ symbols may be written in terms of the hypergeometric functions ${ }\Fhg{4}{3}$ with unit argument:\isym{F43@$\Fhg{4}{3}$, generalized hypergeometric function}
\begin{align}
\sixj{a}{b}{c}{d}{e}{f}&=(-1)^{a+b+d+e} \frac{\Delta(a b c) \Delta(c d e) \Delta(a e f) \Delta(b d f)(a+b+d+e+1)!}{(a+b-c)!(-c+d+e)!(a+e-f)!(b+d-f)!(-a+c-d+f)!(-b+c-e+f)!}  \notag\\
& \times\Fhg{4}{3}\left[\left.\begin{array}{c}
-a-b+c, c-d-e,-a-e+f,-b-d+f \\
-a-b-d-e-1,-a+c-d+f+1,-b+c-e+f+1
\end{array} \right\rvert\, 1\right], \quad(\text { Rose \cite{ref030}) }  \label{eq:9:2:8}\\
\sixj{a}{b}{c}{d}{e}{f}&=(-1)^{a+c+d+f} \frac{\Delta(a e f) \Delta(b d f)(-a+b+c)!(c-d+e)!(a-c+d+f)!}{\Delta(a b c) \Delta(c d e)(a-e+f)!(-b+d+f)!(-a+b-d+e)!(b+c+e-f+1)!}  \notag\\
& \times \Fhg{4}{3}\left[\left.\begin{array}{c}
-a+b+c+1, c-d+e+1,-a+e-f, b-d-f \\
-a+c-d-f,-a+b-d+e+1, b+c+e-f+2
\end{array} \right\rvert\, 1\right],  \label{eq:9:2:9}\\
 \sixj{a}{b}{c}{d}{e}{f}&=(-1)^{a+c+d+f} \frac{\Delta(a b c) \Delta(b d f)(a-b+d+e)!(-b+c+e+f)!(a+c+d+f+1)!}{\Delta(a e f) \Delta(c d e)(a-b+c)!(-b+d+f)!(a+e+f+1)!(c+d+e+1)!}  \notag\\
& \times \Fhg{4}{3}\left[\left.\begin{array}{c}
-a+b-c, b-d-f,-a-e-f-1,-c-d-e-1 \\
-a+b-d-e, b-c-e-f,-a-c-d-f-1
\end{array} \right\rvert\, 1\right],  \label{eq:9:2:10}\\
 \sixj{a}{b}{c}{d}{e}{f}&=(-1)^{a+b+d+e}  \notag\\
& \times \frac{\Delta(a b c) \Delta(c d e) \Delta(a e f) \Delta(b d f)(a+e+f+1)!(b+d+f+1)!(a+c+d-f)!}{(a+b-c)!(a-b+c)!(-c+d+e)!(c+d-e)!(a+e-f)!(b+d-f)!(-a+c-d+f)!(2 f+1)!}  \notag\\
& \times\Fhg{4}{3}\left[\left.\begin{array}{c}
-a-e+f,-b-d+f,-a+e+f+1, b-d+f+1 \\
-a-c-d+f,-a+c-d+f+1,2 f+2
\end{array} \right\rvert\, 1\right],  \label{eq:9:2:11}\\
 \sixj{a}{b}{c}{d}{e}{f}&=(-1)^{b+c+e+f}  \notag\\
& \times \frac{\Delta(a b c) \Delta(c d e) \Delta(a e f) \Delta(b d f)(2 b)!(b+c-e+f)!(b+c+e+f+1)!}{(-a+b+c)!(a+b-c)!(c-d+e)!(c+d-e)!(a-e+f)!(-a+e+f)!(b+d-f)!(b-d+f)!}  \notag\\
& \times\Fhg{4}{3}\left[\left.\begin{array}{c}
a-b-c,-b+d-f,-a-b-c-1,-b-d-f-1 \\
-2 b,-b-c+e-f,-b-c-e-f-1
\end{array} \right\rvert\, 1\right] . \label{eq:9:2:12}
\end{align}
Equations \eqref{eq:9:2:8}-\eqref{eq:9:2:12} present Eqs.~\eqref{eq:9:2:2}-\eqref{eq:9:2:6} in terms of the hypergeometric functions.

\subsection{Relations Between the $6 j$ Symbols and the $3jm$ Symbols}\label{sec:9.2.4}\index{6j symbol@$6j$ symbol!and the $3jm$ symbols}\index{3jm symbol@$3jm$ symbol}
The $6 j$ symbols may be written as sums of products of the Clebsch-Gordan coefficients (Eq.~\eqref{eq:9:1:8}) or $3jm$ symbols. The relations between the $6 j$ symbols and the $3 j m$ symbols are
\begin{equation}
\sixj{a}{b}{c}{d}{e}{f}  \label{eq:9:2:13}=\sum(-1)^{d+e+f+\delta+\varepsilon+\varphi}\threej{a }{ b }{ c }{\alpha }{ \beta }{ \gamma}\threej{a }{ e }{ f }{\alpha }{ \varepsilon }{ -\varphi}\threej{d }{ b }{ f }{-\delta }{ \beta }{ \varphi}\threej{d }{ e }{ c }{\delta }{ -\varepsilon }{ \gamma} .
\end{equation}
In Eq.~\eqref{eq:9:2:13} the sum is over all possible values of $\alpha, \beta, \gamma, \delta, \varepsilon, \varphi$ with only three summation indices being independent. Some other sums of the $3 j m$ symbols which yield the $6 j$ symbols will be considered in Chap.~\ref{chap:12} .

\subsection{Quasi-Binomial Representation of the 6j Symbols}\label{sec:9.2.5}\index{6j symbol@$6j$ symbol!quasi-binomial representation}\index{quasi-binomial}
The $6 j$ symbols may be written in terms of quasi-binomials [45, 99] which are defined in Sec.~\ref{sec:8.2.2}. These representations are widely used in tabulating the formulas for the $6 j$ symbols.

Let us introduce the following definitions:
\begin{equation}
\begin{array}{lll}
k_{1} \equiv e-d, & B \equiv R_{21}=-b+d+f, & F \equiv R_{12}=b+d-f, \\
k_{2} \equiv a-b, & D \equiv R_{34}=b-d+f, & E \equiv R_{12}+R_{21}+R_{34}+1=b+d+f+1 \label{eq:9:2:14}
\end{array}
\end{equation}
Then the dependence of the $6 j$ symbols on $k_{1}$ and $k_{2}$ is given by
\begin{gather}
\sixj{a}{b}{c}{d}{e}{f} \equiv\sixj{b+k_{2}}{b}{c}{d}{d+k_{1}}{f}  \notag\\
=(-1)^{E+k_{1}+k_{2}+1}\left[\frac{\left(c+k_{2}\right)^{\left(2 k_{2}\right)} B^{\left(k_{2}-k_{1}\right)} D^{(-1)\left(k_{2}-k_{1}\right)} E^{(-1)\left(k_{2}+k_{1}\right)} F^{(-1)\left(k_{2}+k_{1}\right)}}{\left(c+k_{1}\right)!\left(c-k_{1}\right)!\left(2 b+c+k_{2}+1\right)^{(2 c+1)}\left(2 d+c+k_{1}+1\right)^{(2 c+1)}}\right]^{\frac{1}{2}}(u-v)^{\left(c-k_{2}\right)} . \label{eq:9:2:15}
\end{gather}
The quantities $u$ and $v$ may be chosen in different ways. This depends on which equations in Sec.~\ref{sec:9.2.1}. are supposed to be written in a quasi-binomial form \cite{ref045}.

Equations \eqref{eq:9:2:1} and \eqref{eq:9:2:2} are obtained by putting
\begin{align}
& u=\left(c+k_{1}\right)^{(1)}\left(B-k_{2}+k_{1}\right)^{(1)} D^{(1)}  \notag\\
& v=\left(c-k_{1}\right)^{(1)} F^{(1)}\left(E+k_{1}+k_{2}\right)^{(-1)} \label{eq:9:2:16}
\end{align}
Equation \eqref{eq:9:2:3} is obtained, if
\begin{align}
& u=\left(c+k_{1}\right)^{(1)} B^{(-1)}\left(D+k_{2}-k_{1}\right)^{(-1)}  \notag\\
& v=\left(c-k_{1}\right)^{(1)}\left(F+k_{2}+k_{1}\right)^{(-1)} E^{(1)} \label{eq:9:2:17}
\end{align}
or
\begin{align}
& u=\left(2 d-c+k_{1}\right)^{(-1)} D^{(1)}\left(F+k_{2}+k_{1}\right)^{(-1)}  \notag\\
& v=\left(2 d+c+k_{1}+1\right)^{(1)}\left(D+k_{2}-k_{1}\right)^{(-1)} F^{(1)} \label{eq:9:2:18}
\end{align}
Equation \eqref{eq:9:2:4} corresponds to
\begin{align}
& u=\left(2 d+c+k_{1}+1\right)^{(1)}\left(B-k_{2}+k_{1}\right)^{(1)} E^{(1)}  \notag\\
& v=\left(2 d-c+k_{1}\right)^{(-1)} B^{(-1)}\left(E+k_{2}+k_{1}\right)^{(-1)} \label{eq:9:2:19}
\end{align}
Equation \eqref{eq:9:2:5} is obtained provided
\begin{align}
& u=\left(c+k_{2}\right)^{(-1)}\left(B-k_{2}+k_{1}\right)^{(1)}\left(D+k_{2}-k_{1}\right)^{(-1)}  \notag\\
& v=\left(c-k_{1}\right)^{(1)}\left(2 b+c+k_{2}+1\right)^{(1)}\left(2 d-c+k_{1}\right)^{(-1)} \label{eq:9:2:20}
\end{align}
or
\begin{align}
& u=\left(c+k_{1}\right)^{(1)}\left(2 b-c+k_{2}\right)^{(-1)}\left(2 d-c+k_{1}\right)^{(-1)}  \label{eq:9:2:21}\\
& v=\left(c+k_{2}\right)^{(-1)} F^{(1)} E^{(1)} \notag
\end{align}
Equation \eqref{eq:9:2:6} corresponds to
\begin{align}
& u=D^{(1)} E^{(1)}\left(2 b+c+k_{2}+1\right)^{(1)}  \notag\\
& v=\left(D+k_{2}-k_{1}\right)^{(-1)}\left(E+k_{2}+k_{1}\right)^{(-1)}\left(2 b-c+k_{2}\right)^{(-1)} \label{eq:9:2:22}
\end{align}
or
\begin{align}
& u=\left(c+k_{1}\right)^{(1)}\left(2 b+c+k_{2}+1\right)^{(1)}\left(2 d+c+k_{1}+1\right)^{(1)}  \label{eq:9:2:23}\\
& v=\left(c+k_{2}\right)^{(-1)}\left(F+k_{2}+k_{1}\right)^{(-1)}\left(E+k_{2}+k_{1}\right)^{(-1)} \notag
\end{align}
Equation \eqref{eq:9:2:15} for the $6 j$ symbols is valid, if all the exponents $2 k_{2}, k_{2}-k_{1}$ and $k_{2}+k_{1}$ are integer nonnegative numbers, i.e., if $k_{2} \geq\left|k_{1}\right| \geq 0$. If some of the exponents are negative, the corresponding quasi-power should be replaced in accordance with
\begin{equation}
p^{(\sigma)} \rightarrow \frac{1}{p^{(-1)(|\sigma|)}}, \quad p^{(-1)(\sigma)} \rightarrow \frac{1}{p^{(|\sigma|)}} \quad \text { for } \sigma<0 \label{eq:9:2:24}
\end{equation}
\section{Integral Representations of the $6 j$ Symbols}\label{sec:9.3}\index{6j symbol@$6j$ symbol!integral representations}\index{integral representations!of the $6j$ symbols}
Squares of the $6 j$ symbols may be expressed by integrals involving the characters of the representations of the rotation group \cite{ref110}\index{character!of an irreducible representation}\index{rotation group}\isymg{chi-J@$\chi^{J}(\omega)$, character of an irreducible representation}\isym{dR@$d R$, invariant volume element of the rotation group}
\begin{equation}
\sixj{a}{b}{c}{d}{e}{f}^{2}  \label{eq:9:3:1}=\frac{1}{\left(8 \pi^{2}\right)^{3}} \int d R_{1} d R_{2} d R_{3} \chi^{a}\left(R_{1}\right) \chi^{b}\left(R_{2}\right) \chi^{c}\left(R_{3}\right) \chi^{d}\left(R_{2} R_{3}^{-1}\right) \chi^{e}\left(R_{3} R_{1}^{-1}\right) \chi^{f}\left(R_{1} R_{2}^{-1}\right)
\end{equation}
Here
\[
\chi^{j}(R) \equiv \sum_{m} D_{m m}^{j}(R) \equiv \sum_{m} D_{m m}^{j}(\alpha, \beta, \gamma)
\]
is the character of the representation of rank $j$ (Sec.~\ref{sec:4.14})
\[
\int f(R) d R \equiv \int_{0}^{2 \pi} d \alpha \int_{0}^{\pi} \sin \beta d \beta \int_{0}^{2 \pi} d \gamma f(\alpha, \beta, \gamma)
\]
Note also the following integral representations for some special $6 j$ symbols
\begin{align}
\MoveEqLeft{\sixj{a}{b}{c}{a}{b}{f}=\frac{(-1)^{2 c}}{\left(8 \pi^{2}\right)^{2}} \int d R_{1} d R_{2} \chi^{c}\left(R_{1}\right) \chi^{f}\left(R_{2}\right) \chi^{a}\left(R_{2} R_{1}\right) \chi^{b}\left(R_{2}^{-1} R_{1}\right),}  \label{eq:9:3:2}\\
\MoveEqLeft{\sixj{a}{b}{g}{d}{b}{c}\sixj{a}{b}{g}{d}{b}{f}\sixj{a}{b}{c}{d}{b}{f} } \notag\\
=&\frac{(-1)^{2 a}}{\left(8 \pi^{2}\right)^{4}} \int d R_{1} d R_{2} d R_{3} d R_{4} \chi^{c}\left(R_{1}\right) \chi^{g}\left(R_{2}\right) \chi^{f}\left(R_{3}\right) \chi^{a}\left(R_{4} R_{2}\right) \chi^{b}\left(R_{4} R_{3} R_{2} R_{1} R_{4}\right) \chi^{d}\left(R_{1} R_{4} R_{3}\right),  \label{eq:9:3:3}\\
\MoveEqLeft{\sixj{a}{b}{c}{d}{e}{f}\sixj{a}{e}{f}{a}{b}{c}\sixj{d}{b}{f}{a}{b}{c}\sixj{d}{e}{c}{a}{b}{c}}  \notag\\
=&\frac{(-1)^{2 d}}{\left(8 \pi^{2}\right)^{5}} \int d R_{1} d R_{2} d R_{3} d R_{4} d R_{5} \chi^{d}\left(R_{1}\right) \chi^{e}\left(R_{2}\right) \chi^{f}\left(R_{3}\right) \chi^{a}\left(R_{4}\right) 
\notag\\&\qquad\qquad \qquad\times
\chi^{a}\left(R_{5}^{-1} R_{3} R_{2}\right) \chi^{b}\left(R_{5} R_{4} R_{2} R_{1}\right) \chi^{c}\left(R_{4}^{-1} R_{5} R_{1} R_{3}\right). \label{eq:9:3:4}
\end{align}
\section{Symmetries of the $6 j$ Symbols and the Racah Coefficients}\label{sec:9.4}\index{6j symbol@$6j$ symbol!symmetry properties}\index{Racah coefficient!symmetry properties}\index{symmetry!of the $6j$ symbols}
\subsection{R-Symbols}\label{sec:9.4.1}\index{R-symbol@$R$-symbol!symmetry properties}
The symmetry properties of the $6 j$ symbols and the $W$-coefficients may be formulated in a fairly simple way if these coefficients are written in terms of the $R$ symbols (see Sec.~\ref{sec:9.1.3}).

The value of the $R$ symbol is invariant under any permutation of its rows or columns \cite{ref105}
\begin{equation}
\left\|\begin{array}{llll}
R_{11} & R_{12} & R_{13} & R_{14}  \label{eq:9:4:1}\\
R_{21} & R_{22} & R_{23} & R_{24} \\
R_{31} & R_{32} & R_{33} & R_{34}
\end{array}\right\|=\left\|\begin{array}{llll}
R_{i 1} & R_{i 2} & R_{i 3} & R_{i 4} \\
R_{k 1} & R_{k 2} & R_{k 3} & R_{k 4} \\
R_{l 1} & R_{l 2} & R_{l 3} & R_{l 4}
\end{array}\right\|=\left\|\begin{array}{llll}
R_{1 \alpha} & R_{1 \beta} & R_{1 \gamma} & R_{1 \delta} \\
R_{2 \alpha} & R_{2 \beta} & R_{2 \gamma} & R_{2 \delta} \\
R_{3 \alpha} & R_{3 \beta} & R_{3 \gamma} & R_{3 \delta}
\end{array}\right\| .
\end{equation}
In other words, any permutation of parameters $A_{i}$ or $B_{\alpha}$ (see Sec.~\ref{sec:9.1.3}) leaves the value of the $R$ symbol unchanged. These symmetry relations involve $3!\times 4!=144$ generally different Racah coefficients.

\subsection{6j Symbols}\label{sec:9.4.2}\index{6j symbol@$6j$ symbol!permutation symmetries}
The above-mentioned symmetries of the $R$ symbol are equivalent to the following symmetries of the $6 j$ symbols.\\[0pt]
\subsubsection{Classical Symmetries \cite{ref110}}
The $6 j$ symbol is invariant under any permutation of its columns or under interchange of the upper and lower arguments in each of any two columns:
\begin{equation}
\begin{aligned}
& \sixj{a}{b}{c}{d}{e}{f}=\sixj{a}{c}{b}{d}{f}{e}=\sixj{b}{a}{c}{e}{d}{f}=\sixj{b}{c}{a}{e}{f}{d}=\sixj{c}{a}{b}{f}{d}{e}=\sixj{c}{b}{a}{f}{e}{d}  \\
 =&\sixj{a}{e}{f}{d}{b}{c}=\sixj{a}{f}{e}{d}{c}{b}=\sixj{e}{a}{f}{b}{d}{c}=\sixj{e}{f}{a}{b}{c}{d}=\sixj{f}{a}{e}{c}{d}{b}=\sixj{f}{e}{a}{c}{b}{d}  \\
=&\sixj{d}{e}{c}{a}{b}{f}=\sixj{d}{c}{e}{a}{f}{b}=\sixj{e}{d}{c}{b}{a}{f}=\sixj{e}{c}{d}{b}{f}{a}=\sixj{c}{d}{e}{f}{a}{b}=\sixj{c}{e}{d}{f}{b}{a}  \\
 =&\sixj{d}{b}{f}{a}{e}{c}=\sixj{d}{f}{b}{a}{c}{e}=\sixj{b}{d}{f}{e}{a}{c}=\sixj{b}{f}{d}{e}{c}{a}=\sixj{f}{d}{b}{c}{a}{e}=\sixj{f}{b}{d}{c}{e}{a} . 
\end{aligned}
\label{eq:9:4:2}
\end{equation}

These relations involve $3!\times 4=24$ different $6 j$ symbols.\\[0pt]
\subsubsection{Regge Symmetries \cite{ref095}}
The relations below are functional ones, i.e. in general they cannot be obtained by interchanging the $6 j$ symbol arguments.
\begin{gather}
\sixj{a}{b}{c}{d}{e}{f}
=\sixj{a}{s_{1}-b}{s_{1}-c}{d}{s_{1}-e}{s_{1}-f}
=\sixj{s_{2}-a}{b}{s_{2}-c}{s_{2}-d}{e}{s_{2}-f}  \notag\\
=\sixj{s_{3}-a}{s_{3}-b}{c}{s_{3}-d}{s_{3}-e}{f}
=\sixj{s_{2}-d}{s_{3}-e}{s_{1}-f}{s_{2}-a}{s_3-b}{s_1-c}
=\sixj{s_{3}-d}{s_{1}-e}{s_{2}-f}{s_{3}-a}{s_1-b}{s_2-c} , \label{eq:9:4:3}
\end{gather}
where
\begin{equation}
s_{1}=\frac{1}{2}(b+c+e+f), \quad s_{2}=\frac{1}{2}(a+c+d+f), \quad s_{3}=\frac{1}{2}(a+b+d+e) . \label{eq:9:4:4}
\end{equation}
These relations are especially useful when $s_{i}$ equals one of the $6 j$ symbol arguments. Combining the Regge symmetries and the classical symmetries, one gets all 144 symmetry relations.

\subsection{Racah Coefficients}\label{sec:9.4.3}\index{Racah coefficient!permutation symmetries}
For the Racah coefficients the symmetry relations are the following.\\[0pt]
\subsubsection{Classical Symmetries \cite{ref091}}
\begin{alignat}{4}
& W(a b e d ; c f)&=&W(d e b a ; c f)&=&W(e d a b ; c f)&=&W(b a d e ; c f)  \notag\\
=&W(a e b d ; f c)&=&W(d b e a ; f c)&=&W(b d a e ; f c)&=&W(e a d b ; f c)  \notag\\
=\varepsilon_{1} &W(a c f d ; b e)&=\varepsilon_{1} &W(d f c a ; b e)&=\varepsilon_{1} &W(f d a c ; b e)&=\varepsilon_{1} &W(c a d f ; b e)  \notag\\
=\varepsilon_{1}& W(a f c d ; e b)&=\varepsilon_{1} &W(d c f a ; e b)&=\varepsilon_{1} &W(c d a f ; e b)&=\varepsilon_{1} &W(f a d c ; e b)  \label{eq:9:4:5}\\
=\varepsilon_{2}& W(c b e f ; a d)&=\varepsilon_{2} &W(f e b c ; a d)&=\varepsilon_{2} &W(e f c b ; a d)&=\varepsilon_{2} &W(b c f e ; a d)  \notag\\
=\varepsilon_{2}& W(c e b f ; d a)&=\varepsilon_{2} &W(f b e c ; d a)&=\varepsilon_{2} &W(b f c e ; d a)&=\varepsilon_{2} &W(e c f b ; d a), \notag
\end{alignat}
where
\begin{equation}
\varepsilon_{1}=(-1)^{b+e-c-f}, \quad \varepsilon_{2}=(-1)^{a+d-c-f} \label{eq:9:4:6}
\end{equation}
\subsubsection{Regge Symmetries}
\begin{align}
W(a b e d ; c f)&=W\left(s_{3}-a, s_{3}-b, s_{3}-e, s_{3}-d ; c f\right)=\varepsilon_{1} W\left(a, s_{1}-b, s_{1}-e, d ; s_{1}-c, s_{1}-f\right)  \notag\\
&=\varepsilon_{1} W\left(s_{2}-d, s_{3}-e, s_{3}-b, s_{2}-a ; s_{1}-f, s_{1}-c\right)=\varepsilon_{2} W\left(s_{2}-a, b, e, s_{2}-d ; s_{2}-c, s_{2}-f\right)  \notag\\
&=\varepsilon_{2} W\left(s_{3}-d, s_{1}-e, s_{1}-b, s_{3}-a ; s_{2}-f, s_{2}-c\right) \label{eq:9:4:7}
\end{align}
Here $s_{1}, s_{2}, s_{3}$ are given by Eq.~\eqref{eq:9:4:4} and $\varepsilon_{1}, \varepsilon_{2}$ by Eq.~\eqref{eq:9:4:6}.

\subsection{"Mirror" Symmetry}\label{sec:9.4.4}\index{mirror symmetry}\index{symmetry!mirror}\index{negative momenta}
The formulas for the $6 j$ symbols may be extended to include negative integer or half-integer values of arguments. In this case one has the following symmetry properties \cite{ref045} corresponding to the replacement $j \rightarrow-j-1$,
\begin{align}
& \sixj{a}{b}{c}{d}{e}{f}
=-\sixj{\bar{a}}{\bar{b}}{\bar{c}}{\bar{d}}{\bar{e}}{\bar{f}}
=(-1)^{\varphi_{1}}\left\{\begin{array}{lll}
\bar{a} & b & c\\
d & e & f
\end{array}\right\}
=(-1)^{\varphi_{1}+1}\left\{\begin{array}{lll}
a & \bar{b}&\bar{c}\\
\bar{d}& \bar{e}&\bar{f}
\end{array}\right\}\notag\\
& =(-1)^{\varphi_{2}}\left\{\begin{array}{lll}
\bar{a} & b & c\\
\bar{d} & e & f
\end{array}\right\}
=(-1)^{\varphi_{2}+1}\left\{\begin{array}{lll}
a & \bar{b} & \bar{c}\\
\bar{d} & \bar{e} & \bar{f}
\end{array}\right\}
=i(-1)^{\varphi_{3}}\left\{\begin{array}{lll}
\bar{a}& \bar{b} &c\\
d & e & f
\end{array}\right\}
=i(-1)^{\varphi_{3}}\left\{\begin{array}{lll}
a & b& \bar{c} \\
\bar{d} & \bar{e} &\bar{f}
\end{array}\right\}  \notag\\
& =i(-1)^{\varphi_{4}}\sixj{\bar{a}}{\bar{b}}{\bar{c}}{d}{e}{f}
=i(-1)^{\varphi_{4}}\sixj{a}{b}{c}{\bar{d}}{\bar{e}}{\bar{f}}
=(-1)^{\varphi_{5}}\sixj{\bar{a}}{\bar{b}}{c}{\bar{d}}{e}{f}
=(-1)^{\varphi_{5}+1}\sixj{a}{b}{\bar{c}}{d}{\bar{e}}{\bar{f}} . \label{eq:9:4:8}
\end{align}
Here
\begin{gather}
\bar{a} \equiv-a-1, \quad \bar{b} \equiv-b-1, \text { etc. }  \notag\\
\varphi_{1}=b-c-e+f, \varphi_{2}=2(a+d), \varphi_{3}=c+d+e+2 f, \varphi_{4}=a+b+c, \varphi_{5}=2(c+f)+1 \label{eq:9:4:9}
\end{gather}
Similarly, for the Racah coefficients one gets
\begin{gather}
W(a b e d ; c f)=-W(\bar{a} \bar{b} \bar{e} \bar{d} ; \bar{c} \bar{f})=W(\bar{a} b e \bar{d} ; c f)=-W(a \bar{b} \bar{e} d ; \bar{c} \bar{f})  \notag\\
=(-1)^{\psi_{1}+1} W(\bar{a} b e d ; c f)=(-1)^{\psi_{1}} W(a \bar{b} \bar{e} \bar{d} ; \bar{c} \bar{f})=i(-1)^{\psi_{2}} W(\bar{a} \bar{b} e d ; c f)=i(-1)^{\psi_{2}} W(a b \bar{e} d ; \bar{c} \bar{f})  \notag\\
=i(-1)^{\psi_{3}} W(\bar{a} \bar{b} e d ; \bar{c} f)=i(-1)^{\psi_{3}} W(a b \bar{e} \bar{d} ; c \bar{f})=(-1)^{\psi_{4}} W(\bar{a} \bar{b} e \bar{d} ; c f)=(-1)^{\psi_{4}+1} W(a b \bar{e} d ; \bar{c} f) . \label{eq:9:4:10}
\end{gather}
Here
\begin{equation}
\psi_{1}=-b+c-e+f, \quad \psi_{2}=-c+d+e+2 f, \quad \psi_{3}=a+b-c, \quad \psi_{4}=2(d+f) \label{eq:9:4:11}
\end{equation}
\section{Explicit Forms of the $6 j$ Symbols for Certain Arguments}\label{sec:9.5}\index{6j symbol@$6j$ symbol!explicit forms}
\subsection{One of Arguments is Equal to Zero}\label{sec:9.5.1}\index{6j symbol@$6j$ symbol!with a zero argument}
For the $6 j$ symbols one obtains
\begin{alignat}{2}
\sixj{0}{b}{c}{d}{e}{f}&=(-1)^{b+e+d} \frac{\delta_{b c} \delta_{e f}}{\sqrt{(2 b+1)(2 e+1)}}, \quad& \sixj{a}{b}{c}{0}{e}{f}&=(-1)^{a+b+e} \frac{\delta_{b f} \delta_{c e}}{\sqrt{(2 b+1)(2 c+1)}},\notag\\
\sixj{a}{0}{c}{d}{e}{f}&=(-1)^{a+d+e} \frac{\delta_{a c} \delta_{d f}}{\sqrt{(2 a+1)(2 d+1)}}, & \sixj{a}{b}{c}{d}{0}{f}&=(-1)^{a+b+d} \frac{\delta_{a f} \delta_{c d}}{\sqrt{(2 a+1)(2 c+1)}}, \notag\\
\sixj{a}{b}{0}{d}{e}{f}&=(-1)^{a+e+f} \frac{\delta_{a b} \delta_{d e}}{\sqrt{(2 a+1)(2 d+1)}}, & \sixj{a}{b}{c}{d}{e}{0}&=(-1)^{a+b+c} \frac{\delta_{a e} \delta_{b d}}{\sqrt{(2 a+1)(2 b+1)}} . \label{eq:9:5:1}
\end{alignat}
Analogous relations for the Racah coefficients are
\begin{alignat}{2}
W(0 b e d ; c f)&=\frac{\delta_{b c} \delta_{e f}}{\sqrt{(2 b+1)(2 e+1)}}, \quad& W(a 0 e d ; c f)&=\frac{\delta_{a c} \delta_{d f}}{\sqrt{(2 a+1)(2 d+1)}}, \notag\\
W(a b 0 d ; c f)&=\frac{\delta_{a f} \delta_{c d}}{\sqrt{(2 a+1)(2 c+1)}}, & W(a b e 0 ; c f)&=\frac{\delta_{b f} \delta_{c e}}{\sqrt{(2 b+1)(2 c+1)}}, \\
W(a b e d ; 0 f)&=(-1)^{a+e-f} \frac{\delta_{a b} \delta_{d e}}{\sqrt{(2 a+1)(2 d+1)}}, & W(a b e d ; c 0)&=(-1)^{a+b-c} \frac{\delta_{a e} \delta_{b d}}{\sqrt{(2 a+1)(2 b+1)}}\notag
\end{alignat}

In this case all other arguments are supposed to satisfy the triangular condition.

\subsection{One of Arguments is Equal to the Sum of Two Others}\label{sec:9.5.2}\index{6j symbol@$6j$ symbol!with an argument equal to a sum of two others}
If one of the $6 j$ symbol arguments is equal to sum of two others from the same triad (abc), (cde), (aef), (bdf), one may use the classical symmetries of the $6 j$ symbol (Eqs.~\eqref{eq:9:4:2}) to express it in the form
\begin{gather}
\sixj{a}{b}{a+b}{d}{e}{f}=(-1)^{a+b+d+e} W(a b e d ; a+b f)=(-1)^{a+b+d+e}  \notag\\
\times\left[\frac{(2 a)!(2 b)!(a+b+d+e+1)!(a+b-d+e)!(a+b+d-e)!(-a+e+f)!(-b+d+f)!}{(2 a+2 b+1)!(-a-b+d+e)!(a+e-f)!(a-e+f)!(a+e+f+1)!(b+d-f)!(b-d+f)!(b+d+f+1)!}\right]^{\frac{1}{2}} \label{eq:9:5:3}
\end{gather}
In particular,
\begin{align}
&\sixj{a}{b}{a+b}{d}{e}{a+e}
=(-1)^{a+b+d+e}\left[\frac{(2 b)!(2 e)!(a+b+d-e)!(a-b+d+e)!}{(2 a+2 b+1)!(2 a+2 e+1)!(-a-b+d+e)!(-a+b+d-e)!}\right]^{\frac{1}{2}}  \label{eq:9:5:4}\\
&\sixj{a}{b}{a+b}{d}{e}{a+e}=(-1)^{a+b+d+e} \notag\\
& \times\left[2 a(a+b+d+e+1)(a+b-d+e) \frac{(2 b)!(2 e-1)!(a+b+d-e)!(a-b+d+e-1)!}{(2 a+2 b+1)!(2 a+2 e)!(-a-b+d+e)!(-a+b+d-e+1)!}\right]^{\frac{1}{2}}, \\
&\sixj{a}{b}{a+b}{a}{e}{a+b}=(-1)^{2a+b+e} 
\frac{(2 a)!(b+e)!}{(2 a+2 b+1)!(-b+e)!},\\
&\sixj{a}{b}{a+b}{a}{e}{a+b-1}=(-1)^{2 a+b+e} \frac{(2 a-1)!(b+e-1)!}{(2 a+2 b)!(-b+e)!}\left[\frac{2 a \cdot 2 b(b+e)(2 a+b-e)(2 a+b+e+1)}{(2 a+2 b+1)(-b+e+1)}\right]^{\frac{1}{2}}  \label{eq:9:5:7}\\
&\sixj{a}{b}{a+b}{a}{b}{f}=(-1)^{2 a+2 b} \frac{(2 a)!(2 b)!}{(a+b-f)!(a+b+f+1)!},  \label{eq:9:5:8}\\
&\sixj{a}{b}{a+b}{b}{a}{f}=(-1)^{2 a+2 b} \frac{(2 a)!(2 b)!}{[(2 a-f)!(2 a+f+1)!(2 b-f)!(2 b+f+1)!]^{\frac{1}{2}}},  \label{eq:9:5:9}\\
&\sixj{a}{b}{a+b}{a}{b}{a+b}=(-1)^{2 a+2 b} \frac{(2 a)!(2 b)!}{(2 a+2 b+1)!}, \label{eq:9:5:10}\\
&\sixj{a}{b}{a+b}{b}{b}{a+b-1}=(-1)^{2 a+2 b} \frac{(2 a)!(2 b)!}{(2 a+2 b)!}, \label{eq:9:5:11} \\
&\sixj{a}{b}{a+b}{a}{b}{a-b}=(-1)^{2 a+2 b} \frac{1}{2 a+1}, \quad(a \geq b), \label{eq:9:5:12}\\
&\sixj{a}{b}{a+b}{a}{b}{a-b+1}=(-1)^{2 a+2 b} \frac{2 b}{(2 a+1)(2 a+2)}, \quad(a \geq b-1) , \label{eq:9:5:13}\\
&\sixj{a}{b}{a+b}{a+b+e}{e}{a+e}=(-1)^{2(a+b+e)} \frac{1}{[(2 a+2 b+1)(2 a+2 e+1)]^{\frac{1}{2}}}. \label{eq:9:5:14}
\end{align}

\subsection{One of Arguments is Smaller by Unity than the Sum of Two Others}\label{sec:9.5.3}\index{6j symbol@$6j$ symbol!with an argument one less than a sum of two others}
If the $6 j$ symbol has one argument which is one less than the sum of two others from the same triad $(a b c),(c d e),(a e f),(b d f)$, one may use the classical symmetries of the $6 j$ symbol (Eqs.~\eqref{eq:9:4:2}) to bring it into the form
\begin{gather}
\sixj{a}{b}{a+b-1}{d}{e}{f}=(-1)^{a+b+d+e} W(a b e d ; a+b-1 f)  \notag\\
=(-1)^{a+b+d+e} 2\{a b(a+b)+(a+b) f(f+1)-a d(d+1)-b e(e+1)\}  \notag\\
\times\left[\frac{(2 a-1)!(2 b-1)!(a+b+d+e)!(a+b-d+e-1)!(a+b+d-e-1)!(-a+e+f)!(-b+d+f)!}{(2 a+2 b)!(-a-b+d+e+1)!(a+e-f)!(a-e+f)!(a+e+f+1)!(b+d-f)!(b-d+f)!(b+d+f+1)!}\right]^{\frac{1}{2}} \label{eq:9:5:15}
\end{gather}
In particular
\begin{align}
& \sixj{a}{b}{a+b-1}{d}{e}{a+e-1}=(-1)^{a+b+d+e} 2\{a(a+b+e-1)(a+b+e)-a d(d+1)-2 b e\}  \notag\\
& \times\left[\frac{(2 b-1)!(2 e-1)!(a+b+d-e-1)!(a-b+d+e-1)!}{(2 a+2 b)!(2 a+2 e)!(-a-b+d+e+1)!(-a+b+d-e+1)!}\right]^{\frac{1}{2}},  \label{eq:9:5:16}\\
& \sixj{a}{b}{a+b-1}{a}{e}{a+b-1}=(-1)^{2 a+b+e} 2\left\{b(2 a+b-1)(2 a+b)-b e(e+1)-2 a^{2}\right\} \frac{(2 a-1)!(b+e-1)!}{(2 a+2 b)!(-b+e+1)!},  \label{eq:9:5:17}\\
& \sixj{a}{b}{a+b-1}{a}{b}{f}=(-1)^{2 a+2 b} 2\left\{a b(a+b)+(a+b) f(f+1)-a^{2}(a+1)\right.  \notag\\
& \left.-b^{2}(b+1)\right\} \frac{(2 a-1)!(2 b-1)!}{(a+b-f)!(a+b+f+1)!},  \label{eq:9:5:18}\\
& \sixj{a}{b}{a+b-1}{b}{a}{f}=(-1)^{2 a+2 b} 2\{(a+b) f(f+1)-2 a b\} \frac{(2 a-1)!(2 b-1)!}{[(2 a-f)!(2 a+f+1)!(2 b-f)!(2 b+f+1)!]^{\frac{1}{2}}} ;  \label{eq:9:5:19}\\
& \sixj{a}{b}{a+b-1}{a+b+e-1}{e}{a+e-1}=(-1)^{2(a+b+e)}\left[\frac{2 b \cdot 2 e}{(2 a+2 b)(2 a+2 b-1)(2 a+2 e)(2 a+2 e-1)}\right]^{\frac{1}{2}} . \label{eq:9:5:20}
\end{align}
Some other cases are given by Eqs.~\eqref{eq:9:5:7}, \eqref{eq:9:5:9}, \eqref{eq:9:5:11}.

\subsection{Arguments $a, b, d, e$ are Equal in Pairs}\label{sec:9.5.4}\index{6j symbol@$6j$ symbol!with arguments equal in pairs}
If $a=b$ and $d=e$ or $a=e$ and $b=d$, the Wigner $6 j$ symbol may be rewritten as \cite{ref056}
\begin{align}
\sixj{a}{a}{c}{b}{b}{f} & =\sixj{a}{b}{f}{b}{a}{c}=(-1)^{2 a+2 b} W(a a b b ; c f)=(-1)^{2 a+2 b} W(a b a b ; f c)  \notag\\
& =(-1)^{a+b+c+f}\left[\frac{(2 a-c)!(2 b-c)!}{(2 a+c+1)!(2 b+c+1)!}\right]^{\frac{1}{2}} V_{c}(a, f, b) \label{eq:9:5:21}
\end{align}
where $c$ is integer, and $V_{c}(a, f, b)=V_{c}(b, f, a)$. According to Eq.~\eqref{eq:9:6:6}, the quantities $V_{c}$ satisfy the recursion relation
\begin{equation}
V_{c+1}=\frac{2 c+1}{c+1} V_{1} V_{c}-c(2 c+1) V_{c}-\frac{c}{c+1}\left[4 a(a+1)+1-c^{2}\right]\left[4 b(b+1)+1-c^{2}\right] V_{c-1} \label{eq:9:5:22}
\end{equation}
Let us denote
\begin{equation}
\tilde{a} \equiv a(a+1), \quad \tilde{b} \equiv b(b+1), \quad x \equiv f(f+1)-a(a+1)-b(b+1)=\tilde{f}-\tilde{a}-\tilde{b} \label{eq:9:5:23}
\end{equation}
Then for some special values of $c$ the functions $V_{c}$ are given by
\begin{align}
V_{0}(a, f, b)&=1, \\
V_{1}(a, f, b)&=-2 x, \\
V_{2}(a, f, b)&=6 x^{2}+6 x-8 \tilde{a} \tilde{b}, \\
V_{3}(a, f, b)&=-20 x^{3}-80 x^{2}-16 x[3+\tilde{a}+\tilde{b}-3 \tilde{a} \tilde{b}]+80 \tilde{a} \tilde{b}, \\
V_{4}(a, f, b)&=70 x^{4}+700 x^{3}+40 x^{2}[39+5 \tilde{a}+5 \tilde{b}-6 \tilde{a} \tilde{b}] \notag\\
&\quad+80 x[9+6 \tilde{a}+6 \tilde{b}-17 \tilde{a} \tilde{b}]-48 \tilde{a} \tilde{b}[27+4 \tilde{a}+4 \tilde{b}-2 \tilde{a} \tilde{b}],
  \label{eq:9:5:28}
\end{align}
\begin{align}
  & \begin{aligned}
& V_{2 a-1}(a, f, b)=(-1)^{1+a-b+f} 2\left\{a^{2}+\tilde{f}-\tilde{b}\right\} \frac{(2 a)!(2 a-1)!(2 a+2 b)!(-a+b+f)!}{(2 b-2 a+1)!(a+b-f)!(a-b+f)!(a+b+f+1)!} \\
& \quad\left(a \leq b+\frac{1}{2}\right),
\end{aligned}  \label{eq:9:5:29}\\
& \begin{aligned}
& V_{2 b-1}(a, f, b)=(-1)^{1-a+b+f} 2\left\{b^{2}+\tilde{f}-\tilde{a}\right\} \frac{(2 b)!(2 b-1)!(2 a+2 b)!(a-b+f)!}{(2 a-2 b+1)!(a+b-f)!(-a+b+f)!(a+b+f+1)!} \\
& \quad\left(b \leq a+\frac{1}{2}\right),
\end{aligned}  \label{eq:9:5:30}\\
& V_{2 a}(a, f, b)=(-1)^{a-b+f} \frac{(2 a)!(2 a)!(2 a+2 b+1)!(-a+b+f)!}{(2 b-2 a)!(a+b-f)!(a-b+f)!(a+b+f+1)!}  \label{eq:9:5:31}\\
& \quad(a \leq b),  \notag\\
& V_{2 b}(a, f, b)=(-1)^{-a+b+f} \frac{(2 b)!(2 b)!(2 a+2 b+1)!(a-b+f)!}{(2 a-2 b)!(a+b-f)!(-a+b+f)!(a+b+f+1)!}  \label{eq:9:5:32}\\
& (b \leq a) . \notag
\end{align}
For special values of $f$ one has
\begin{align}
& V_{c}(a, a-b, b)=\frac{(2 b)!(2 a+c+1)!}{(2 a+1)!(2 b-c)!}, \quad(a \geq b)  \label{eq:9:5:33}\\
& V_{c}(a, b-a, b)=\frac{(2 a)!(2 b+c+1)!}{(2 b+1)!(2 a-c)!}, \quad(a \leq b)  \label{eq:9:5:34}\\
& V_{c}(a, a-b+1, b)=2\{2 b(a+1)-(a-b+1) c(c+1)\} \frac{(2 b-1)!(2 a+c+1)!}{(2 a+2)!(2 b-c)!},  \label{eq:9:5:35}\\
&(a \geq b-1)  \notag\\
& V_{c}(a, b-a+1, b)=2\{2 a(b+1)+(a-b-1) c(c+1)\} \frac{(2 a-1)!(2 b+c+1)!}{(2 b+2)!(2 a-c)!},  \label{eq:9:5:36}\\
&(b \geq a-1),  \notag\\
& V_{c}(a, a+b-1, b)=(-1)^{c+1} 2\{(a+b) c(c+1)-2 a b\} \frac{(2 a-1)!(2 b-1)!}{(2 a-c)!(2 b-c)!},  \label{eq:9:5:37}\\
& V_{c}(a, a+b, b)=(-1)^{c} \frac{(2 a)!(2 b)!}{(2 a-c)!(2 b-c)!} \label{eq:9:5:38}
\end{align}
See also Eqs.~\eqref{eq:9:5:9} and \eqref{eq:9:5:19}.

\section{Recursion Relations}\label{sec:9.6}\index{6j symbol@$6j$ symbol!recursion relations}\index{recursion relations!for the $6j$ symbols}
\subsection{Relations in Which Arguments are Changed by $1 / 2$}\label{sec:9.6.1}\index{recursion relations!in half-integer steps}
\begin{gather}
{[(a+b+c+1)(-a+b+c)(c+d+e+1)(c+d-e)]^{\frac{1}{2}}\sixj{a}{b}{c}{d}{e}{f}}  \notag\\
=-2 c[(b+d+f+1)(b+d-f)]^{\frac{1}{2}}\sixj{a}{b-\frac{1}{2}}{c-\frac{1}{2}}{d-\frac{1}{2}}{e}{f}  \notag\\
+[(a+b-c+1)(a-b+c)(-c+d+e+1)(c-d+e)]^{\frac{1}{2}}\sixj{a}{b}{c-1}{d}{e}{f},  \label{eq:9:6:1}\\
(a-b-d+e)[(a+b+c+1)(c+d+e+1)]^{\frac{1}{2}}\sixj{a}{b}{c}{d}{e}{f}  \notag\\
=-[(a-b+c)(c-d+e)(a+e-f)(a+e+f+1)]^{\frac{1}{2}}\sixj{a-\frac{1}{2}}{b}{c-\frac{1}{2}}{d}{e-\frac{1}{2}}{f}  \notag\\
+[(-a+b+c)(c+d-e)(b+d-f)(b+d+f+1)]^{\frac{1}{2}}\sixj{a}{b-\frac{1}{2}}{c-\frac{1}{2}}{d-\frac{1}{2}}{e}{f},  \label{eq:9:6:2}\\
{[(-a+b+c)(a-b+c+1)(a+e-f+1)(b+d+f+1)]^{\frac{1}{2}}\sixj{a}{b}{c}{d}{e}{f},}  \notag\\
=[(c+d-e)(c-d+e+1)(a+e+f+2)(b+d-f)]^{\frac{1}{2}}\sixj{a+\frac{1}{2}}{b-\frac{1}{2}}{c}{d-\frac{1}{2}}{e+\frac{1}{2}}{f}  \notag\\
+(a-b-d+e+1)[(-a+b+c)(b-d+f)]^{\frac{1}{2}}\sixj{a+\frac{1}{2}}{b-\frac{1}{2}}{c}{d}{e}{f-\frac{1}{2}}, \label{eq:9:6:3}
\end{gather}
\emph{Note.} Relation~\eqref{eq:9:6:3}, as given in the original source, does not hold identically: a direct evaluation of both sides shows that they differ for generic physical values of the arguments. For example, at $a=b=\frac{1}{2},\ c=1,\ d=\frac{1}{2},\ e=\frac{3}{2},\ f=1$ the left-hand side equals $-\frac{2\sqrt{3}}{3}$ whereas the right-hand side equals $-\frac{\sqrt{6}}{3}$. The equation therefore appears to contain a misprint whose correct form we have been unable to reconstruct; the neighbouring recursions \eqref{eq:9:6:1}, \eqref{eq:9:6:2}, \eqref{eq:9:6:4} and \eqref{eq:9:6:5} have all been verified numerically.
\begin{align}
& (2 d+1)(2 f+1)[(a+b+c+1)(a-b+c)]^{\frac{1}{2}}\sixj{a}{b}{c}{d}{e}{f}  \notag\\
& =-[(a+e+f+1)(a-e+f)(b+d+f+1)(-b+d+f)(c+d+e+1)(c+d-e)]^{\frac{1}{2}}\sixj{a-\frac{1}{2}}{b}{c-\frac{1}{2}}{d-\frac{1}{2}}{e}{f-\frac{1}{2}}  \notag\\
& -[(-a+e+f+1)(a+e-f)(b-d+f+1)(b+d-f)(c+d+e+1)(c+d-e)]^{\frac{1}{2}}\sixj{a-\frac{1}{2}}{b}{c-\frac{1}{2}}{d-\frac{1}{2}}{e}{f+\frac{1}{2}}  \notag\\
& -\left[(a+e+f+1)(a-e+f)(b+d-f+1)(b-d+f)(-c+d+e+1)(c-d+e)\right]^{\frac{1}{2}}\sixj{a-\frac{1}{2}}{b}{c-\frac{1}{2}}{d+\frac{1}{2}}{e}{f-\frac{1}{2}}  \notag\\
& +[(-a+e+f+1)(a+e-f)(b+d+f+2)(-b+d+f+1)(-c+d+e+1)(c-d+e)]^{\frac{1}{2}}  \notag\\
& \times\sixj{a-\frac{1}{2}}{b}{c-\frac{1}{2}}{d+\frac{1}{2}}{e}{f+\frac{1}{2}} . \label{eq:9:6:4}
\end{align}
\subsection{Relations in Which Arguments are Changed by 1}\label{sec:9.6.2}\index{recursion relations!in unit steps}
\begin{gather}
(2 c+1)\{2[a(a+1) d(d+1)+b(b+1) e(e+1)-c(c+1) f(f+1)]  \notag\\
-[a(a+1)+b(b+1)-c(c+1)][d(d+1)+e(e+1)-c(c+1)]\}\sixj{a}{b}{c}{d}{e}{f}  \notag\\
=-c[(a+b+c+2)(-a+b+c+1)(a-b+c+1)(a+b-c)  \notag\\
\times(d+e+c+2)(-d+e+c+1)(d-e+c+1)(d+e-c)]^{\frac{1}{2}}\sixj{a}{b}{c+1}{d}{e}{f}  \notag\\
-(c+1)[(a+b+c+1)(-a+b+c)(a-b+c)(a+b-c+1)  \notag\\
\times(d+e+c+1)(-d+e+c)(d-e+c)(d+e-c+1)]^{\frac{1}{2}}\sixj{a}{b}{c-1}{d}{e}{f} \label{eq:9:6:5}
\end{gather}
In particular
\begin{align}
& (2 c+1)\{-2 a(a+1)-2 b(b+1)+2 f(f+1)+c(c+1)\}\sixj{a}{a}{c}{b}{b}{f}  \notag\\
& =(c+1)[(2 a+c+2)(2 a-c)(2 b+c+2)(2 b-c)]^{\frac{1}{2}}\sixj{a}{a}{c+1}{b}{b}{f}  \notag\\
& +c[(2 a+c+1)(2 a-c+1)(2 b+c+1)(2 b-c+1)]^{\frac{1}{2}}\sixj{a}{a}{c-1}{b}{b}{f} \label{eq:9:6:6}
\end{align}
\begin{gather}
(2 c+1)\left\{[a(a+1)+b(b+1)-c(c+1)]^{2}-2\left[a^{2}(a+1)^{2}+b^{2}(b+1)^{2}-c(c+1) f(f+1)\right]\right\}\sixj{a}{b}{c}{a}{b}{f}  \notag\\
=c(a+b+c+2)(-a+b+c+1)(a-b+c+1)(a+b-c)\sixj{a}{b}{c+1}{a}{b}{f}  \notag\\
+(c+1)(a+b+c+1)(-a+b+c)(a-b+c)(a+b-c+1)\sixj{a}{b}{c-1}{a}{b}{f} \label{eq:9:6:7}
\end{gather}
\section{Generating Function}\label{sec:9.7}\index{6j symbol@$6j$ symbol!generating function}\index{generating functions!for the $6j$ symbols}
The $6 j$ symbols turn out to be the coefficients of a power series expansion of the generating function $f\left(\tau_{i \alpha}\right)$ \cite{ref053}, which depends on 12 variables $\tau_{i \alpha}(i=1,2,3 ; \alpha=1,2,3,4)$,
\begin{equation}
f\left(r_{i \alpha}\right) \equiv\left[1+\sum_{i=1}^{3} \prod_{\alpha=1}^{4} r_{i \alpha}+\sum_{\alpha=1}^{4} \prod_{i=1}^{3} r_{i \alpha}\right]^{-2}=\sum_{R_{i \alpha}} N\left(R_{i \alpha}\right)\sixj{a}{b}{c}{d}{e}{f}  \label{eq:9:7:1} \prod_{i=1}^{3} \prod_{\alpha=1}^{4}\left(r_{i \alpha}\right)^{R_{i \alpha}}
\end{equation}
The exponents $R_{i \alpha}$ are elements of the $R$ symbol. The relation between $R_{i \alpha}$ and arguments $a, b, c$, etc. is given by Eqs.~\eqref{eq:9:1:13}-\eqref{eq:9:1:14}. The normalization factors $N$ are
\begin{equation}
N\left(R_{i \alpha}\right)=\left[\frac{\prod_{\alpha=1}^{4}\left(B_{\alpha}+1\right)!}{\prod_{i=1}^{3} \prod_{\alpha=1}^{4}\left(R_{i \alpha}\right)!}\right]^{\frac{1}{2}}, \label{eq:9:7:2}
\end{equation}
where $B_{\alpha}$ are given by Eqs.~\eqref{eq:9:1:18}.

\section{Sums Involving the $6 j$ Symbols}\label{sec:9.8}\index{6j symbol@$6j$ symbol!sums involving}\index{sums!of the $6j$ symbols}\index{9j symbol@$9j$ symbol}\index{12j symbol@$12j$ symbol}
In this section only the most important sums involving products of the $6 j$ symbols are presented. We shall use the notation $\{a b c\}$ for the symbol which is equal to 1 if $a, b, c$ satisfy the triangular conditions $\eqref{eq:8:1:1}$ and is zero otherwise.\index{triangle symbol}\isymo{braces-jjj@$\{j_{1} j_{2} j_{3}\}$, triangle symbol} In the equations below the sum is over all possible values (integer or half-integer) of $X$ which obey all triangular conditions.
\begin{align}
& \sum_{X}(2 X+1)\sixj{a}{b}{X}{a}{b}{c}=(-1)^{2 c}\{a b c\},  \label{eq:9:8:1}\\
& \sum_{X}(-1)^{a+b+X}(2 X+1)\sixj{a}{b}{X}{b}{a}{c}=\delta_{c 0} \sqrt{(2 a+1)(2 b+1)},  \label{eq:9:8:2}\\
& \sum_{X}(2 X+1)\sixj{a}{b}{X}{c}{d}{p}\sixj{a}{b}{X}{c}{d}{q}=\delta_{p q} \frac{\{a d p\}\{b c p\}}{2 p+1},  \label{eq:9:8:3}\\
& \sum_{X}(-1)^{p+q+X}(2 X+1)\sixj{a}{b}{X}{c}{d}{p}\sixj{a}{b}{X}{d}{c}{q}=\sixj{a}{c}{q}{b}{d}{p},  \label{eq:9:8:4}\\
& \sum_{X}(-1)^{2 X}(2 X+1)\sixj{a}{b}{X}{c}{d}{p}\sixj{c}{d}{X}{e}{f}{q}\sixj{e}{f}{X}{a}{b}{r}=\ninej{a}{f}{r}{d}{q}{e}{p}{c}{b},  \label{eq:9:8:5}\\
& \sum_{X}(-1)^{R+X}(2 X+1)\sixj{a}{b}{X}{c}{d}{p}\sixj{c}{d}{X}{e}{f}{q}\sixj{e}{f}{X}{b}{a}{r}=\sixj{p}{q}{r}{e}{a}{d}\sixj{p}{q}{r}{f}{b}{c},  \label{eq:9:8:6}\\
& (R=a+b+c+d+e+f+p+q+r)  \notag\\
& \sum_{X}(2 X+1)\sixj{a}{b}{X}{c}{d}{p}\sixj{c}{d}{X}{e}{f}{q}\sixj{e}{f}{X}{g}{h}{r}\sixj{g}{h}{X}{a}{b}{s}=(-1)^{p-s-q+r}\twelvejII{h,a,s,c,p,b,f,r,g,q,e,d},  \label{eq:9:8:7}\\
& \sum_{X}(-1)^{T-X}(2 X+1)\sixj{a}{b}{X}{c}{d}{p}\sixj{c}{d}{X}{e}{f}{q}\sixj{e}{f}{X}{g}{h}{r}\sixj{g}{h}{X}{b}{a}{s}=\twelvejI{a,d,e,h,p,q,r,s,b,c,f,g},  \notag\\
& (T=a+b+c+d+e+f+g+h+p+q+r+s) . \label{eq:9:8:8}
\end{align}
Many other sums, involving the $6 j$ symbols as well as the $3 j m$ symbols and the $9 j$ symbols, will be given in Chap.~\ref{chap:12}.

\section{Asymptotics of the $6 j$ Symbols for Large Angular Momenta}\label{sec:9.9}\index{6j symbol@$6j$ symbol!asymptotics of}\index{asymptotics!of the $6j$ symbols}
\subsection{Asymptotic Relations Between the 6j Symbols and the Clebsch-Gordan Coefficients}\label{sec:9.9.1}\index{asymptotics!relation to the Clebsch-Gordan coefficients}
If $R \gg 1$ and $a, b, c$ etc. are arbitrary, one gets the following asymptotic relation
\begin{equation}
\sixj{a}{b}{c}{d+R}{e+R}{f+R}  \label{eq:9:9:1} \approx \frac{(-1)^{a+b+d+e}}{\sqrt{2 R(2 c+1)}} \clebsch{a}{\alpha}{b}{\beta}{c}{\gamma}
\end{equation}
where $\alpha=f-e, \beta=d-f, \gamma=d-e$.\\
For the $6 j$ symbols and the $3 j m$ symbols this relation assumes the form \cite{ref047}
\begin{equation}
\sixj{a}{b}{c}{d+R}{e+R}{f+R}  \label{eq:9:9:2} \approx \frac{(-1)^{a+b+c+2(d+e+f)}}{\sqrt{2 R}}\threej{a }{ b }{ c }{e-f }{ f-d }{ d-e} .
\end{equation}
The asymptotic relation between the $R$ symbols which correspond to the $6 j$ symbols and the Clebsch-Gordan coefficients is written as
\begin{align}
& \left\|\begin{array}{llll}
-c+d+e+2 R & b+d-f & a+e-f & a+b-c \\
-b+d+f+2 R & c+d-e & a-b+c & a-e+f \\
-a+e+f+2 R & -a+b+c & c-d+e & b-d+f
\end{array}\right\|  \notag\\
& \approx \frac{(-1)^{a+b+c+2(d+e+f)}}{\sqrt{2 R}}\left\|\begin{array}{ccc}
-a+b+c & a-b+c & a+b-c \\
a+e-f & b-d+f & c+d-e \\
a-e+f & b+d-f & c-d+e
\end{array}\right\| . \label{eq:9:9:3}
\end{align}
In particular, when $d=e=f=0$ one obtains \cite{ref060}
\begin{equation}
\sixj{a}{b}{c}{R}{R}{R}  \label{eq:9:9:4} \approx \frac{(-1)^{c}}{\sqrt{2 R(2 c+1)}} \clebsch{a}{0}{b}{0}{c}{0}
\end{equation}
\subsection{Asymptotic Expressions for the $6 j$ Symbols}\label{sec:9.9.2}
\begin{figure}[tbh!]
\begin{center}
\includegraphics{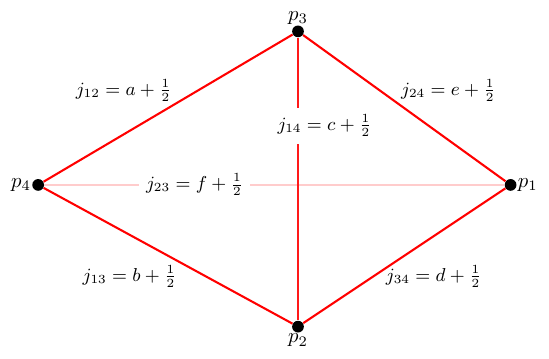}
\caption{The tetrahedron associated with asymptotic behaviour of the $6 j$ symbols.}
\label{chap9:fig:1}
\end{center}
\end{figure}

\index{asymptotics!for large angular momenta}
The asymptotic behaviour of the $6 j$ symbols $\sixj{a}{b}{c}{d}{e}{f}$ for large angular momenta is closely associated with geometric properties of the tetrahedron whose edges are\index{tetrahedron}\index{6j symbol@$6j$ symbol!and the tetrahedron} $a+\frac{1}{2}, b+\frac{1}{2}$, etc. (Fig.~\ref{chap9:fig:1}).

\subsubsection{The Ponzano-Regge Formula \cite{ref089} (semiclassical approximation to the $6 j$ symbols)}\index{Ponzano-Regge formula}\index{semiclassical formulas}\isym{V-tetra@$V$, volume of the tetrahedron}
If $a, b, c, d, e, f \gg 1$, then
\begin{equation}
\sixj{a}{b}{c}{d}{e}{f}  \label{eq:9:9:5} \approx \frac{1}{\sqrt{12 \pi V}} \cos \left(\sum_{i, k=1}^{4} j_{i k} \Theta_{i k}+\frac{\pi}{4}\right), \quad\left(V^{2}>0\right)
\end{equation}
Here
\begin{gather}
j_{12}=a+\frac{1}{2}, \quad j_{13}=b+\frac{1}{2}, \quad j_{14}=c+\frac{1}{2},  \notag\\
j_{23}=f+\frac{1}{2}, \quad j_{24}=e+\frac{1}{2}, \quad j_{34}=d+\frac{1}{2},  \label{eq:9:9:6}\\
j_{i k}=j_{k i}, \quad j_{i i}=0 . \notag
\end{gather}
$V$ is the volume of the tetrahedron, $\Theta_{i k}$ is the angle between two external normals to the planes adjacent to the edge $j_{i k}$.

The tetrahedron volume is equal to
\begin{equation}
V^{2}=\frac{1}{2^{3}(3!)^{2}}\left|\begin{array}{ccccc}
0 & j_{34}^{2} & j_{24}^{2} & j_{23}^{2} & 1  \label{eq:9:9:7}\\
j_{34}^{2} & 0 & j_{14}^{2} & j_{13}^{2} & 1 \\
j_{24}^{2} & j_{14}^{2} & 0 & j_{12}^{2} & 1 \\
j_{23}^{2} & j_{13}^{2} & j_{12}^{2} & 0 & 1 \\
1 & 1 & 1 & 1 & 0
\end{array}\right|
\end{equation}
The angles $\Theta_{i k}$ are given by
\begin{equation}
S_{i} S_{k} \sin \Theta_{i k}=\frac{3}{2} V j_{i k}, \quad(i \neq k) \label{eq:9:9:8}
\end{equation}
Here $S_{i}$ is the area of the triangle opposite to the vertex $p_{i}$ (Fig.~\ref{chap9:fig:1}). One can evaluate $S_{i}$, using the standard formulas. For example,
\begin{equation}
S_{1}^{2}=\frac{1}{16}\left(j_{12}+j_{13}+j_{14}\right)\left(j_{12}+j_{13}-j_{14}\right)\left(j_{12}-j_{13}+j_{14}\right)\left(-j_{12}+j_{13}+j_{14}\right)=-\frac{1}{16}\left|\begin{array}{cccc}
0 & j_{12}^{2} & j_{13}^{2} & 1  \label{eq:9:9:9}\\
j_{12}^{2} & 0 & j_{14}^{2} & 1 \\
j_{13}^{2} & j_{14}^{2} & 0 & 1 \\
1 & 1 & 1 & 0
\end{array}\right| .
\end{equation}
The asymptotic expressions \eqref{eq:9:9:5} are valid only if $V^{2}>0$ (classically allowed domain). If $V^{2}<0$ (classically forbidden domain, when an associated tetrahedron does not exist), the asymptotic expression becomes
\begin{equation}
\sixj{a}{b}{c}{d}{e}{f}  \label{eq:9:9:10} \approx \frac{1}{2 \sqrt{12 \pi|V|}} \cos \Phi \exp \left\{-\left|\sum_{i, k=1}^{4} j_{i k} \operatorname{Im} \Theta_{i k}\right|\right\}, \quad\left(V^{2}<0\right)
\end{equation}
where
\begin{equation}
\Phi=\sum_{i, k=1}^{4}\left(j_{i k}-\frac{1}{2}\right) \operatorname{Re} \Theta_{i k} \label{eq:9:9:11}
\end{equation}
In this case the $6 j$ symbols are exponentially small even if the triangular condition is satisfied.\\
Near the classical domain boundary, where $V^{2} \approx 0$, Eqs.~\eqref{eq:9:9:5}, \eqref{eq:9:9:10} are of little use, although one may use the improved expressions \cite{ref089}
\begin{equation}
\sixj{a}{b}{c}{d}{e}{f}  \label{eq:9:9:12} \approx 2^{-\frac{4}{3}}\left(\prod_{i=1}^{4} S_{i}\right)^{-\frac{1}{6}}\{\cos \Phi \mathrm{Ai}(z)+\sin \Phi \mathrm{Bi}(z)\}
\end{equation}
$\mathrm{Ai}(z)$ and $\mathrm{Bi}(z)$ being the Airy functions \cite{ref027},
\begin{equation}
z= \begin{cases}-(3 V)^{2}\left(4 \prod_{i=1}^{4} S_{i}\right)^{-\frac{2}{3}} & \text { if } V^{2}>0,  \label{eq:9:9:13}\\ (3|V|)^{2}\left(4 \prod_{i=1}^{4} S_{i}\right)^{-\frac{2}{3}} & \text { if } V^{2}<0 .\end{cases}
\end{equation}
Note that the Ponzano-Regge approximation is sufficiently accurate even at comparatively small angular momenta $a, b, c$ etc. Asymptotic expressions similar to \eqref{eq:9:9:12} but extended over the entire domain of angular momenta are obtained in Ref.~\cite{ref140}.\\[0pt]
\subsubsection{The Edmonds' Formula \cite{ref016}}
\begin{figure}[tbh!]
\begin{center}
\includegraphics{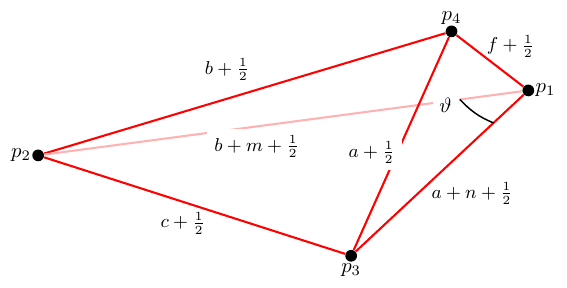}
\caption{The angle $\Theta$ which enters Eq.~\eqref{eq:9:9:14}.}
\label{chap9:fig:2}
\end{center}
\end{figure}

If $f, m, n$ are arbitrary integers or half-integers and $a, b, c \gg f, m, n$, then
\begin{equation}
\sixj{a}{b}{c}{b+m}{a+n}{f}  \label{eq:9:9:14} \approx \frac{(-1)^{a+b+c+f+m}}{\sqrt{(2 a+1)(2 b+1)}} d_{m n}^{f}(\theta),
\end{equation}
where $d_{m n}^{f}(\Theta)$ is the rotation matrix (Chap.~\ref{chap:4}), $\Theta$ is an angle between the tetrahedron edges $a+n+\frac{1}{2}$ and $b+m+\frac{1}{2}$ (Fig.~\ref{chap9:fig:2})
\begin{equation}
\cos \theta=\frac{a(a+1)+b(b+1)-c(c+1)}{2 \sqrt{a(a+1) b(b+1)}} . \label{eq:9:9:15}
\end{equation}

In particular, if $m=n=0, a, b, c \gg f$ and $f$ is an arbitrary integer, then \eqref{eq:9:9:14} turns into the Racah formula \cite{ref093},
\begin{equation}
\sixj{a}{b}{c}{b}{a}{f}  \label{eq:9:9:16} \approx \frac{(-1)^{a+b+c+f}}{\sqrt{(2 a+1)(2 b+1)}} P_{f}(\cos \theta),
\end{equation}
where $P_{f}$ is the Legendre polynomial. If, in addition, $f$ is large ( $a, b, c \gg f \gg 1$ ), one can substitute into \eqref{eq:9:9:16} asymptotic expressions for the Legendre polynomials to obtain
\begin{equation}
\sixj{a}{b}{c}{b}{a}{f}  \label{eq:9:9:17} \approx(-1)^{a+b+c+f}\left[\frac{4}{\pi(2 a+1)(2 b+1)(2 f+1) \sin \theta}\right]^{\frac{1}{3}} \cos \left[\left(f+\frac{1}{2}\right) \theta-\frac{\pi}{4}\right] .
\end{equation}
This expression may also be written in the form (cf. Eq.~\eqref{eq:9:9:5})
\begin{equation}
\sixj{a}{b}{c}{b}{a}{f}  \label{eq:9:9:18} \approx \frac{(-1)^{a+b+c+f}}{\sqrt{12 \pi V}} \cos \left[\left(f+\frac{1}{2}\right) \Theta-\frac{\pi}{4}\right],
\end{equation}
where the tetrahedron volume is given by
\begin{equation}
V=\frac{1}{6}\left(a+\frac{1}{2}\right)\left(b+\frac{1}{2}\right)\left(f+\frac{1}{2}\right) \sin \Theta . \label{eq:9:9:19}
\end{equation}
\subsubsection{$a, b, c, d, e, f \gg 1, m, n, p$}
\begin{figure}[tbh!]
\begin{center}
\includegraphics{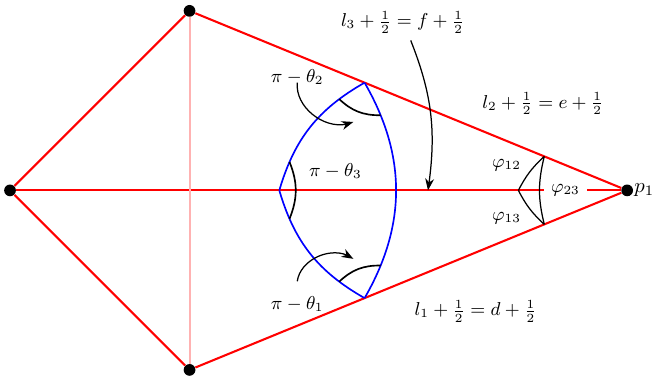}
\caption{Geometrical interpretation of the angles in Eqs.~\eqref{eq:9:9:20}-\eqref{eq:9:9:23}.}
\label{chap9:fig:3}
\end{center}
\end{figure}

In this case,  one has
\begin{equation}
\sixj{a}{b}{c}{d}{e}{f}  \label{eq:9:9:20}\sixj{a}{b}{c}{d+m}{e+n}{f+p} \approx \frac{\Theta\left(V^{2}\right)}{24 \pi V} \cos \left(m \Theta_{1}+n \Theta_{2}+p \Theta_{3}\right) .
\end{equation}
where $V$ is the tetrahedron volume \eqref{eq:9:9:7}
\begin{equation}
\Theta\left(V^{2}\right)= \begin{cases}1, & \text { if } V^{2}>0,  \label{eq:9:9:21}\\ 0, & \text { if } V^{2}<0,\end{cases}
\end{equation}
$\Theta_{i}$ is the angle between two external normals to the planes adjacent to the edge $l_{i}+\frac{1}{2}$ (Fig.~\ref{chap9:fig:3}), with $l_{1} \equiv d, l_{2} \equiv e, l_{3} \equiv f$. The angles $\Theta_{i}$ can be evaluated from
\begin{equation}
\cos \Theta_{i}=\frac{\cos \varphi_{i k} \cos \varphi_{i l}-\cos \varphi_{k l}}{\sin \varphi_{i k} \sin \varphi_{i l}} \label{eq:9:9:22}
\end{equation}
where combinations $i, k, l$ are obtained by cyclic permutations of $1,2,3$.

In Eq.~\eqref{eq:9:9:22} $\varphi_{i k} \equiv \varphi_{k i}$ is the angle between the edges $l_{i}+\frac{1}{2}$ and $l_{k}+\frac{1}{2}$ of the tetrahedron (Fig.~\ref{chap9:fig:3})
\begin{equation}
\cos \varphi_{i k}=\frac{l_{i}\left(l_{i}+1\right)+l_{k}\left(l_{k}+1\right)-j_{l}\left(j_{l}+1\right)}{2 \sqrt{l_{i}\left(l_{i}+1\right) l_{k}\left(l_{k}+1\right)}} \label{eq:9:9:23}
\end{equation}
with $i \neq k \neq l$ and $j_{1} \equiv a, j_{2} \equiv b, j_{3} \equiv c$.\\
\subsubsection{$m=n=p=0$}In particular, for $m=n=p=0$ Eq.~\protect\eqref{eq:9:9:20} yields the Wigner formula \cite{ref043}
\begin{equation}
\sixj{a}{b}{c}{d}{e}{f}^{2}  \label{eq:9:9:24} \approx \frac{\Theta\left(V^{2}\right)}{24 \pi V}, \quad(a, b, c, d, e, f \gg 1)
\end{equation}
This formula is valid only on the average because the $6 j$ symbols oscillate rapidly with momentum variations in the region of large angular momenta.\\
\subsubsection{$R \rightarrow \infty$}
If $a, b, c$, etc. are fixed and $R \rightarrow \infty$, one has \cite{ref089}
\begin{align}
\sixj{a}{b+R}{c+R}{d}{e+R}{f+R} \approx & (-1)^{\varphi}\left[\frac{(a-b+c)!(a-e+f)!(c+d-e)!(-b+d+f)!}{(a+b-c)!(a+e-f)!(-c+d+e)!(b+d-f)!}\right]^{\frac{1}{2} \operatorname{sign}(c+f-b-e)}  \notag\\
& \times \frac{(2 R)^{-1-|b+e-c-f|}}{|b+e-c-f|!}\left[1+O\left(\frac{1}{R^{2}}\right)\right] . \label{eq:9:9:25}
\end{align}
In this case
\begin{gather}
\varphi=a+d+\min \{b+e, c+f\},  \notag\\
\operatorname{sign} x= \begin{cases}1, & \text { if } x \geq 0 \\
-1, & \text { if } x<0\end{cases} \label{eq:9:9:26}
\end{gather}
\section{Relations Between the Wigner $6 j$ Symbols and Analogous Functions of Other Authors}\label{sec:9.10}\index{6j symbol@$6j$ symbol!notations of other authors}\index{notation!of other authors}\isym{U-jahn@$U(a b e d ; c f)$, normalized Racah coefficient}
Racah \cite{ref091}:
\[
W(a b e d ; c f)=(-1)^{a+b+d+e}\sixj{a}{b}{c}{d}{e}{f}
\]
Jahn \cite{ref073}:
\[
U(a b e d ; c f)=[(2 c+1)(2 f+1)]^{\frac{1}{2}} W(a b e d ; c f)
\]
Biedenharn, Blatt and Rose \cite{ref056}:
\[
Z(a b e d ; c f)=i^{-a+e+f} \left[(2 a+1)(2 b+1)(2 d+1)(2 e+1)\right]^{\frac{1}{2}} \clebsch{a}{0}{e}{0}{f}{0} W(a b e d ; c f)
\]
\section{Tables of Algebraic Expressions for the $6 j$ Symbols}\label{sec:9.11}\index{6j symbol@$6j$ symbol!algebraic tables}\index{Racah coefficient!algebraic tables}\index{tables!of the $6j$ symbols}
Algebraic expressions for the $6 j$ symbols $\sixj{a}{b}{c}{d}{e}{f}$ with $d=\frac{1}{2}, 1, \frac{3}{2}, 2, \frac{5}{2}, 3, \frac{7}{2}, 4$ are presented in Tables 9.1-9.8. We use the following notations
\[
\begin{gathered}
s=a+b+c \\
X=-a(a+1)+b(b+1)+c(c+1)
\end{gathered}
\]
The tables of algebraic formulas for the $6 j$ symbols and Racah coefficients are also given in Refs.~\cite{ref003,ref045, ref056}.

\chapter{9j and 12j Symbols}\label{chap:10}\index{9j symbol@$9j$ symbol}\index{12j symbol@$12j$ symbol}
\section{Definition of the $9 j$ Symbols}\label{sec:10.1}\index{9j symbol@$9j$ symbol!definition}
\subsection{9j Symbols as Recoupling Coefficients}\label{sec:10.1.1}\index{recoupling!of four angular momenta}\index{coupling schemes}
The Wigner $9 j$ symbols \cite{ref110} (or Fano coefficients \cite{ref066}) are associated with the coefficients of unitary transformations which connect state vectors corresponding to different coupling schemes of four angular momenta.\isymo{9j@$\bigl\{\begin{smallmatrix} a & b & c\\ d & e & f\\ g & h & j\end{smallmatrix}\bigr\}$, Wigner $9j$ symbol}\index{Fano coefficients|see{$9j$ symbol}}\index{addition of angular momenta}

Let us consider an addition of four angular momenta $\vect{j}_{1}, \vect{j}_{2}, \vect{j}_{3}$ and $\vect{j}_{4}$ to form a resultant angular momentum $j$ with projection $m$. There exist different coupling schemes of these angular momenta. We pay attention to the following schemes\footnote{Along with the coupling schemes \eqref{eq:10:1:1} there exist some other ones, e.g., $\vect{j}_{1}+\vect{j}_{2}=\vect{j}_{12}, \vect{j}_{12}+\vect{j}_{3}=\vect{j}_{123}, \vect{j}_{123}+\vect{j}_{4}=\vect{j}$. However, the coefficients of transformation which connects a state vector in this coupling scheme with vectors in coupling schemes, I, II or III are related to the products of two $6 j$ symbols rather than to the $9 j$ symbols.
}
\begin{equation}
\begin{array}{rll}
\text { I) } \vect{j}_{1}+\vect{j}_{2}=\vect{j}_{12}, & \vect{j}_{3}+\vect{j}_{4}=\vect{j}_{34}, & \vect{j}_{12}+\vect{j}_{34}=\vect{j} ; \\
\text { II) } \vect{j}_{1}+\vect{j}_{3}=\vect{j}_{13}, & \vect{j}_{2}+\vect{j}_{4}=\vect{j}_{24}, & \vect{j}_{13}+\vect{j}_{24}=\vect{j} ; \\
\text { III) } \vect{j}_{1}+\vect{j}_{4}=\vect{j}_{14}, & \vect{j}_{2}+\vect{j}_{3}=\vect{j}_{23}, & \vect{j}_{14}+\vect{j}_{23}=\vect{j} ; \label{eq:10:1:1}
\end{array}
\end{equation}
Let $\ket{j_{1} j_{2}\left(j_{12}\right) j_{3} j_{4}\left(j_{34}\right) j m}$ be a state vector which corresponds to scheme I. This vector is an eigenvector of the operators $\widehat{\mathrm{j}}_{1}^{2}, \widehat{\mathrm{j}}_{2}^{2}, \widehat{\mathrm{j}}_{3}^{2}, \widehat{\mathrm{j}}_{4}^{2}, \widehat{\mathrm{j}}_{12}^{2}, \widehat{\mathrm{j}}_{34}^{2}, \widehat{\mathrm{j}}^{2}$ and $\widehat{j}_{x}$ and may be written as
\begin{equation}
\ket{j_{1} j_{2}\left(j_{12}\right) j_{3} j_{4}\left(j_{34}\right) j m}=\sum_{\substack{m_{1} m_{2} m_{3} m_{4} \\ m_{12} m_{34}}} \clebsch{j_{12}}{m_{12}}{j_{34}}{m_{34}}{j}{m} \clebsch{j_{1}}{m_{1}}{j_{2}}{m_{2}}{j_{12}}{m_{12}} \clebsch{j_{3}}{m_{3}}{j_{4}}{m_{4}}{j_{34}}{m_{34}}\ket{j_{1} m_{1}, j_{2} m_{2}, j_{3} m_{3}, j_{4} m_{4}} . \label{eq:10:1:2}
\end{equation}
Here $\ket{j_{1} m_{1}, j_{2} m_{2}, j_{3} m_{3}, j_{4} m_{4}}$ is eigenvector of the operators $\hat{j}_{i}^{2}$ and $\hat{j}_{i x}(i=1,2,3,4)$. A state vector which corresponds to scheme II is the eigenvector of the operators $\widehat{\mathrm{j}}_{1}^{2}, \widehat{\mathrm{j}}_{2}^{2}, \widehat{\mathrm{j}}_{3}^{2}, \widehat{\mathrm{j}}_{4}^{2}, \widehat{\mathrm{j}}_{13}^{2}, \widehat{\mathrm{j}}_{24}^{2}, \widehat{\mathrm{j}}^{2}$ and $\widehat{\mathrm{j}}_{z}$. It can be represented in the form
\begin{equation}
\ket{j_{1} j_{3}\left(j_{13}\right) j_{2} j_{4}\left(j_{24}\right) j m}=\sum_{\substack{m_{1} m_{2} m_{3} m_{4} \\ m_{13} m_{24}}} \clebsch{j_{13}}{m_{13}}{j_{24}}{m_{24}}{j}{m} \clebsch{j_{1}}{m_{1}}{j_{3}}{m_{3}}{j_{13}}{m_{13}} \clebsch{j_{2}}{m_{2}}{j_{4}}{m_{4}}{j_{24}}{m_{24}}\ket{j_{1} m_{1}, j_{2} m_{2}, j_{3} m_{3}, j_{4} m_{4}} \label{eq:10:1:3}
\end{equation}
Similarly, a state vector, which corresponds to scheme III, is the eigenvector of the operators $\widehat{\vect{j}}_{1}^{2}, \widehat{\vect{j}}_{2}^{2}, \widehat{\vect{j}}_{3}^{2}, \widehat{\vect{j}}_{4}^{2}, \widehat{\vect{j}}_{14}^{2}, \widehat{\vect{j}}_{23}^{2}$, $\widehat{\mathrm{j}}^{2}$ and $\widehat{j}_{x}$. This state vector is given by
\begin{equation}
\ket{j_{1} j_{4}\left(j_{14}\right) j_{2} j_{3}\left(j_{23}\right) j m}=\sum_{\substack{m_{1} m_{2} m_{3} m_{4} \\ m_{14} m_{23}}} \clebsch{j_{14}}{m_{14}}{j_{23}}{m_{23}}{j}{m} \clebsch{j_{1}}{m_{1}}{j_{4}}{m_{4}}{j_{14}}{m_{14}} \clebsch{j_{2}}{m_{2}}{j_{3}}{m_{3}}{j_{23}}{m_{23}}\ket{j_{1} m_{1}, j_{2} m_{2}, j_{3} m_{3}, j_{4} m_{4}} \label{eq:10:1:4}
\end{equation}
The state vectors associated with each of these coupling schemes form a complete set of functions. Any transformation which connects the state vectors in different coupling schemes, is determined by a unitary matrix. Such a matrix relates the state vectors with the same resultant angular momentum $j$ and its projection $m$. It may be shown that the transformation matrix is independent of $m$. The Wigner $9 j$ symbols are proportional to the matrix elements (transformation coefficient). Thus, the Wigner $9 j$ symbols (or, equivalently, the Fano coefficients)
\[
\ninej{j_{1}}{j_{2}}{j_{12}}{j_{3}}{j_{4}}{j_{34}}{j_{13}}{j_{24}}{j}
\]
are related to the coefficients of the transformation between state vectors in schemes I and II by
\begin{gather}
\braket{ j_{1} j_{2}\left(j_{12}\right) j_{3} j_{4}\left(j_{34}\right) j m }{ j_{1} j_{3}\left(j_{13}\right) j_{2} j_{4}\left(j_{24}\right) j^{\prime} m^{\prime}}  \notag\\
=\delta_{j j^{\prime}} \delta_{m m^{\prime}}\left[\left(2 j_{12}+1\right)\left(2 j_{13}+1\right)\left(2 j_{24}+1\right)\left(2 j_{34}+1\right)\right]^{\frac{1}{2}}\ninej{j_{1}}{j_{2}}{j_{12}}{j_{3}}{j_{4}}{j_{34}}{j_{13}}{j_{24}}{j} \label{eq:10:1:5}
\end{gather}
Using this definition, we may also obtain
\begin{gather}
\braket{ j_{1} j_{2}\left(j_{12}\right) j_{3} j_{4}\left(j_{34}\right) j m }{ j_{1} j_{4}\left(j_{14}\right) j_{2} j_{3}\left(j_{23}\right) j^{\prime} m^{\prime}}  \notag\\
=\delta_{j j^{\prime}} \delta_{m m^{\prime}}(-1)^{j_{3}+j_{4}-j_{34}}\left[\left(2 j_{12}+1\right)\left(2 j_{14}+1\right)\left(2 j_{23}+1\right)\left(2 j_{34}+1\right)\right]^{\frac{1}{2}}\ninej{j_{1}}{j_{2}}{j_{12}}{j_{4}}{j_{3}}{j_{34}}{j_{14}}{j_{23}}{j},  \label{eq:10:1:6}\\
\braket{ j_{1} j_{3}\left(j_{13}\right) j_{2} j_{4}\left(j_{24}\right) j m }{ j_{1} j_{4}\left(j_{14}\right) j_{2} j_{3}\left(j_{23}\right) j^{\prime} m^{\prime}}  \notag\\
=\delta_{j j^{\prime}} \delta_{m m^{\prime}}(-1)^{j_{3}-j_{4}-j_{23}+j_{24}}\left[\left(2 j_{13}+1\right)\left(2 j_{14}+1\right)\left(2 j_{24}+1\right)\left(2 j_{23}+1\right)\right]^{\frac{1}{2}}\ninej{j_{1}}{j_{3}}{j_{13}}{j_{4}}{j_{2}}{j_{24}}{j_{14}}{j_{23}}{j} . \label{eq:10:1:7}
\end{gather}
As follows from definition \eqref{eq:10:1:5} a $9 j$ symbol can be represented as a sum of products of the Clebsch-Gordan coefficients:\index{Clebsch-Gordan coefficient}\isym{C-clebsch@$\clebsch{a}{\alpha}{b}{\beta}{c}{\gamma}$, Clebsch-Gordan coefficient}
\begin{align}
& \sum_{\substack{m_{1} m_{2} m_{3} m_{4} \\ m_{12} m_{34} m_{13} m_{24}}} \clebsch{j_{1}}{m_{1}}{j_{2}}{m_{2}}{j_{12}}{m_{12}} \clebsch{j_{3}}{m_{3}}{j_{4}}{m_{4}}{j_{34}}{m_{34}} \clebsch{j_{12}}{m_{12}}{j_{34}}{m_{34}}{j}{m} \clebsch{j_{1}}{m_{1}}{j_{3}}{m_{3}}{j_{13}}{m_{13}} \clebsch{j_{2}}{m_{2}}{j_{4}}{m_{4}}{j_{24}}{m_{24}} \clebsch{j_{13}}{m_{13}}{j_{24}}{m_{24}}{j^{\prime}}{m^{\prime}}  \notag\\
& \quad=\delta_{j j^{\prime}} \delta_{m m^{\prime}}\left[\left(2 j_{12}+1\right)\left(2 j_{13}+1\right)\left(2 j_{24}+1\right)\left(2 j_{34}+1\right)\right]^{\frac{1}{2}}\ninej{j_{1}}{j_{2}}{j_{12}}{j_{3}}{j_{4}}{j_{34}}{j_{13}}{j_{24}}{j} \label{eq:10:1:8}
\end{align}
This equation unambiguously fixes the absolute value and phase of the $9 j$ symbol. The $9 j$ symbols are real.\\
Below we shall also use Latin letters ( $a, b, c$, etc.) to denote arguments of the $9 j$ symbols.\\
In accordance with the quantum mechanical rules of angular momentum addition, arguments of the $9 j$ symbol
\[
\ninej{a}{b}{c}{d}{e}{f}{g}{h}{j}
\]
satisfy the following conditions.\\
(a) Arguments $a, b, c \ldots, j$ are integer or half-integer non-negative numbers (generalization of the $9 j$ symbols to the case of negative arguments will be considered in Sec.~\ref{sec:10.4.3}).\\
(b) A $9 j$ symbol vanishes unless the triangular conditions (see Sec.~\ref{sec:8.1.1}) are fulfilled for triads\index{triangle condition}\index{triad} (abc), (def), $(g h j),(a d g),(b e h)$ and (cfj).

The $9 j$ symbols satisfy the orthogonality and normalization relations\index{orthogonality!of the $9j$ symbols}\index{normalization!of the $9j$ symbols}\index{triangle symbol}\isymo{braces-jjj@$\{j_{1} j_{2} j_{3}\}$, triangle symbol}
\begin{align}
& \sum_{g h}(2 g+1)(2 h+1)\ninej{a}{b}{c}{d}{e}{f}{g}{h}{j}\ninej{a}{b}{c^{\prime}}{d}{e}{f^{\prime}}{g}{h}{j}=\delta_{c c^{\prime}} \delta_{f f^{\prime}}\{a b c\}\{d e f\}\{c f j\} \frac{1}{(2 c+1)(2 f+1)},  \label{eq:10:1:9}\\
& \sum_{c f}(2 c+1)(2 f+1)\ninej{a}{b}{c}{d}{e}{f}{g}{h}{j}\ninej{a}{b}{c}{d}{e}{f}{g^{\prime}}{h^{\prime}}{j}=\delta_{g g^{\prime}} \delta_{h h^{\prime}}\{a d g\}\{b e h\}\{g h j\} \frac{1}{(2 g+1)(2 h+1)} . \label{eq:10:1:10}
\end{align}
These relations follow from the unitarity of transformations which relate state vectors in different coupling schemes.

\subsection{9j Symbol and $r$ Symbol}\label{sec:10.1.2}\index{r-symbol@$r$-symbol}\index{9j symbol@$9j$ symbol!and the $r$-symbol}
A $9 j$ symbol may be represented as a $3 \times 6$ array\isym{r-ik@$r_{i k}$, $r$-symbol of the $9j$ symbol} $\left\|r_{i k}, r_{i k}^{\prime}\right\|$ (Shelepin \cite{ref105}, Wu \cite{ref111}) which is called the $r$-symbol: \footnote{The array $\left\|r_{i k}, r_{i k}^{\prime}\right\|$ under discussion (Wu \cite{ref111}) differs from the $r$-symbol introduced by Shelepin \cite{ref105}.
}
\begin{equation}
\ninej{a}{b}{c}{d}{e}{f}{g}{h}{j} \label{eq:10:1:11}\equiv\left\|\begin{array}{llllll}
r_{11} & r_{12} & r_{13} & r_{11}^{\prime} & r_{12}^{\prime} & r_{13}^{\prime} \\
r_{21} & r_{22} & r_{23} & r_{21}^{\prime} & r_{22}^{\prime} & r_{23}^{\prime} \\
r_{31} & r_{32} & r_{33} & r_{31}^{\prime} & r_{32}^{\prime} & r_{33}^{\prime}
\end{array}\right\|,
\end{equation}
where
\begin{equation}
\begin{array}{lll}
r_{11}=-a+b+c, & r_{12}=a-b+c, & r_{13}=a+b-c, \\
r_{21}=-d+e+f, & r_{22}=d-e+f, & r_{23}=d+e-f, \\
r_{31}=-g+h+j, & r_{32}=g-h+j, & r_{33}=g+h-j, \\
r_{11}^{\prime}=-a+d+g, & r_{12}^{\prime}=-b+e+h, & r_{13}^{\prime}=-c+f+j, \\
r_{21}^{\prime}=a-d+g, & r_{22}^{\prime}=b-e+h, & r_{23}^{\prime}=c-f+j, \\
r_{31}^{\prime}=a+d-g, & r_{32}^{\prime}=b+e-h, & r_{33}^{\prime}=c+f-j . \label{eq:10:1:12}
\end{array}
\end{equation}
The inverse relations are as follows
\begin{equation}
\begin{array}{lll}
2 a=r_{12}+r_{13}=r_{21}^{\prime}+r_{31}^{\prime}, & 2 d=r_{22}+r_{23}=r_{11}^{\prime}+r_{31}^{\prime}, & 2 g=r_{32}+r_{33}=r_{11}^{\prime}+r_{21}^{\prime}, \\
2 b=r_{11}+r_{13}=r_{22}^{\prime}+r_{32}^{\prime}, & 2 e=r_{21}+r_{23}=r_{12}^{\prime}+r_{32}^{\prime}, & 2 h=r_{31}+r_{33}=r_{12}^{\prime}+r_{22}^{\prime}, \\
2 c=r_{11}+r_{12}=r_{23}^{\prime}+r_{33}^{\prime}, & 2 f=r_{21}+r_{22}=r_{13}^{\prime}+r_{33}^{\prime}, & 2 j=r_{31}+r_{32}=r_{13}^{\prime}+r_{23}^{\prime} . \label{eq:10:1:13}
\end{array}
\end{equation}
All eighteen elements $r_{i k}, r_{i k}^{\prime}$ are integer and non-negative. Only nine of them are linearly independent. The

elements $r_{i k}$ and $r_{i k}^{\prime}$ satisfy the following relations
\begin{gather}
\sum_{i=1}^{3} r_{1 i}=a+b+c, \quad \sum_{i=1}^{3} r_{2 i}=d+e+f, \quad \sum_{i=1}^{3} r_{3 i}=g+h+j,  \notag\\
\sum_{i=1}^{3} r_{i 1}=R-2(a+d+g), \quad \sum_{i=1}^{3} r_{i 2}=R-2(b+e+h), \quad \sum_{i=1}^{3} r_{i 3}=R-2(c+f+j),  \notag\\
\sum_{i=1}^{3} r_{1 i}^{\prime}=R-2(a+b+c), \quad \sum_{i=1}^{3} r_{2 i}^{\prime}=R-2(d+e+f), \quad \sum_{i=1}^{3} r_{3 i}^{\prime}=R-2(g+h+j),  \label{eq:10:1:14}\\
\sum_{i=1}^{3} r_{i 1}^{\prime}=a+d+g, \quad \sum_{i=1}^{3} r_{i 2}^{\prime}=b+e+h, \quad \sum_{i=1}^{3} r_{i 3}^{\prime}=c+f+j,  \notag\\
\sum_{i, k=1}^{3} r_{i k}=R, \quad \sum_{i, k=1}^{3} r_{i k}^{\prime}=R, \notag
\end{gather}
where
\begin{equation}
R=a+b+c+d+e+f+g+h+j . \label{eq:10:1:15}
\end{equation}
\section{Explicit Forms of the $9 j$ Symbols and Their Relations to Other Functions}\label{sec:10.2}\index{9j symbol@$9j$ symbol!explicit forms}\index{9j symbol@$9j$ symbol!relations to other functions}
The expressions given below represent the $9 j$ symbols in the forms of algebraic sums and sums involving products of the Clebsch-Gordan coefficients, $3 j m$ and $6 j$ symbols. Unlike the $3 j m$ and $6 j$ symbols the expressions for the $9 j$ symbols in terms of the generalized hypergeometric functions are still unknown. Convenient expressions for the $9 j$ symbols in terms of quasi-binomials can be obtained only in some special cases.\\
\subsection{Expressions for the $9 j$ Symbols in the Forms of Algebraic Sums}\label{sec:10.2.1}\index{9j symbol@$9j$ symbol!as algebraic sums}
\begin{gather}
\ninej{a}{b}{c}{d}{e}{f}{g}{h}{j}=\frac{\Delta(a b c) \Delta(d e f) \Delta(b e h) \Delta(g h j)}{\Delta(a d g) \Delta(c f j)}  \notag\\
\times \frac{(a+d-g)!(c+f-j)!(g+h+j+1)!}{(a+d+g+1)!(a-b+c)!(-a+b+c)!(d-e+f)!(-d+e+f)!(b-e+h)!(-b+e+h)!}  \notag\\
\times \sum_{x y z t}(-1)^{a-c+e-g+j+x+y+z+t} \frac{(2 a-x)!(2 b-y)!(2 d-z)!(2 e-t)!}{x!y!z!t!}  \notag\\
\times \frac{(-a+b+c+x)!(-b+e+h+y)!(-d+e+f+z)!(b-e+g-j+t)!(-a-e+f+g+x+t)!}{(a+b-c-x)!(b+e-h-y)!(d+e-f-z)!(b+e-g+j-t)!(a+d-g-x-z)!(e-b+h+y-t)!}  \notag\\
\times \frac{(c-d+e+j+z-t)!}{(-d+e+f+z-t)!(b-e+g-j-y+t)!(-a+c-e+g-j+x+t)!(-a+c-d+f+g+j+1+x+z)!} . \label{eq:10:2:1}
\end{gather}
In Eq. \eqref{eq:10:2:1} the sums are over all integer values of $x, y, z, t$ for which the factorial arguments are greater than or equal to zero. An expression for the $9 j$ symbols in the form of triple sum was proposed by Alisauskas and Yutsis \cite{ref131}.

\subsection{Wu Formulas \cite{ref111}}\label{sec:10.2.2}\index{Wu formulas}\index{9j symbol@$9j$ symbol!Wu formulas}\index{triangle coefficient}\isymg{Delta-abc@$\Delta(a b c)$, triangle coefficient}
\begin{align}
& \ninej{j_{11}}{j_{12}}{j_{13}}{j_{21}}{j_{22}}{j_{23}}{j_{31}}{j_{32}}{j_{33}}=\Delta\left(j_{11} j_{12} j_{13}\right) \Delta\left(j_{21} j_{22} j_{23}\right) \Delta\left(j_{31} j_{32} j_{33}\right) \Delta\left(j_{11} j_{21} j_{31}\right)  \notag\\
& \times \Delta\left(j_{12} j_{22} j_{32}\right) \Delta\left(j_{13} j_{23} j_{33}\right) \sum_{\nu, \omega}(-1)^{\omega_{4}+\omega_{8}+\omega_{6}} \frac{(n+1)!}{\prod_{p, q=1}^{3} \nu_{p q}!\prod_{\alpha=1}^{6} \omega_{\alpha}!} \label{eq:10:2:2}
\end{align}
where
\begin{equation}
n=\sum_{p, q=1}^{3} \nu_{p q}+\sum_{\alpha=1}^{6} \omega_{\alpha} \label{eq:10:2:3}
\end{equation}
In Eq. \eqref{eq:10:2:3} the summation is over 15 integer and non-negative variables $\nu_{p q}(p, q=1,2,3), \omega_{\alpha}(\alpha=1,2, \ldots, 6)$. These variables satisfy the following conditions owing to which only six variables are independent:
\begin{equation}
\begin{array}{ll}
\omega_{1}+\omega_{4}+\nu_{21}+\nu_{31}=r_{11}, & \omega_{1}+\omega_{4}+\nu_{12}+\nu_{13}=r_{11}^{\prime} \\
\omega_{3}+\omega_{6}+\nu_{22}+\nu_{32}=r_{12}, & \omega_{3}+\omega_{6}+\nu_{11}+\nu_{13}=r_{12}^{\prime} \\
\omega_{2}+\omega_{5}+\nu_{23}+\nu_{33}=r_{13}, & \omega_{2}+\omega_{5}+\nu_{11}+\nu_{12}=r_{13}^{\prime} \\
\omega_{3}+\omega_{5}+\nu_{11}+\nu_{31}=r_{21}, & \omega_{3}+\omega_{5}+\nu_{22}+\nu_{23}=r_{21}^{\prime} \\
\omega_{2}+\omega_{4}+\nu_{12}+\nu_{32}=r_{22}, & \omega_{2}+\omega_{4}+\nu_{21}+\nu_{23}=r_{22}^{\prime}  \label{eq:10:2:4}\\
\omega_{1}+\omega_{6}+\nu_{13}+\nu_{33}=r_{23}, & \omega_{1}+\omega_{6}+\nu_{21}+\nu_{22}=r_{23}^{\prime} \\
\omega_{2}+\omega_{6}+\nu_{11}+\nu_{21}=r_{31}, & \omega_{2}+\omega_{6}+\nu_{32}+\nu_{33}=r_{31}^{\prime} \\
\omega_{1}+\omega_{5}+\nu_{12}+\nu_{22}=r_{32}, & \omega_{1}+\omega_{5}+\nu_{31}+\nu_{33}=r_{32}^{\prime} \\
\omega_{3}+\omega_{4}+\nu_{13}+\nu_{23}=r_{33}, & \omega_{3}+\omega_{4}+\nu_{31}+\nu_{32}=r_{33}^{\prime}
\end{array}
\end{equation}
$\nu_{p q}$ and $\omega_{\alpha}$ are related to arguments of the $9 j$ symbols by
\begin{align}
\sum_{q=1}^{3} \nu_{p q}=n-\sum_{q=1}^{3} r_{p q}, & \sum_{q=1}^{3} \nu_{q p}=n-\sum_{q=1}^{3} r_{q p}^{\prime} . \\
n+\omega_{1}+\omega_{4}-\nu_{11}=j_{12}+j_{13}+j_{21}+j_{31}, & n+\omega_{3}+\omega_{5}-\nu_{21}=j_{22}+j_{23}+j_{11}+j_{31},\notag \\
n+\omega_{3}+\omega_{6}-\nu_{12}=j_{11}+j_{13}+j_{22}+j_{32}, & n+\omega_{2}+\omega_{4}-\nu_{22}=j_{21}+j_{23}+j_{12}+j_{32}, \notag\\
n+\omega_{2}+\omega_{5}-\nu_{13}=j_{11}+j_{12}+j_{23}+j_{33}, & n+\omega_{1}+\omega_{6}-\nu_{23}=j_{21}+j_{22}+j_{13}+j_{33},  \label{eq:10:2:6}\\
n+\omega_{2}+\omega_{6}-\nu_{31}=j_{32}+j_{33}+j_{11}+j_{21}, \notag\\
n+\omega_{1}+\omega_{5}-\nu_{32}=j_{31}+j_{33}+j_{12}+j_{22}, \notag\\
n+\omega_{3}+\omega_{4}-\nu_{33}=j_{31}+j_{32}+j_{13}+j_{23} . \notag
\end{align}
A choice of independent summation variables can be made in many ways, leading to formally different representations of the $9 j$ symbols. Some of them are given below.\\
(a)
\begin{align}
\ninej{j_{11}}{j_{12}}{j_{13}}{j_{21}}{j_{22}}{j_{23}}{j_{31}}{j_{32}}{j_{33}}= & \Delta\left(j_{11} j_{12} j_{13}\right) \Delta\left(j_{21} j_{22} j_{23}\right) \Delta\left(j_{31} j_{32} j_{33}\right) \Delta\left(j_{11} j_{21} j_{31}\right) \Delta\left(j_{12} j_{22} j_{32}\right) \Delta\left(j_{13} j_{23} j_{33}\right)  \notag\\
& \times \sum_{x_{\alpha}} \frac{(-1)^{\frac{1}{2} R+\frac{1}{2} \sum_{\beta=1}^{3} x_{\beta}+x_{6}+x_{5}+x_{6}}\left(1+\sum_{\beta=1}^{6} z_{\beta}\right)!}{\prod_{p, q=1}^{3}\left(z_{p q}-j_{p q}\right)!\prod_{\alpha=1}^{6}\left\{\frac{1}{2}\left(R-\sum_{\beta=1}^{6} z_{\beta}\right)+z_{\alpha}-h_{\alpha}\right\}!} \label{eq:10:2:7}
\end{align}
where
\begin{gather}
\left(z_{p q}\right) \equiv\left(\begin{array}{lll}
z_{11} & z_{12} & z_{13} \\
z_{21} & z_{22} & z_{23} \\
z_{31} & z_{32} & z_{33}
\end{array}\right)=\left(\begin{array}{lll}
z_{1}+z_{4} & z_{3}+z_{6} & z_{2}+z_{5} \\
z_{3}+z_{5} & z_{2}+z_{4} & z_{1}+z_{6} \\
z_{2}+z_{6} & z_{1}+z_{5} & z_{3}+z_{4}
\end{array}\right)  \label{eq:10:2:8}\\
h_{1}=j_{11}+j_{23}+j_{32}, \quad h_{4}=j_{11}+j_{22}+j_{33}  \notag\\
h_{2}=j_{31}+j_{22}+j_{13}, \quad h_{5}=j_{21}+j_{32}+j_{13}  \label{eq:10:2:9}\\
h_{3}=j_{21}+j_{12}+j_{33}, \quad h_{6}=j_{31}+j_{12}+j_{23}  \notag\\
R=\sum_{p, q=1}^{3} j_{p q} \label{eq:10:2:10}
\end{gather}
Summation in Eq. \eqref{eq:10:2:7} is over all integer and half-integer $z_{\alpha} \geq 0$.\\
(b)
\begin{gather}
\ninej{j_{11}}{j_{12}}{j_{13}}{j_{21}}{j_{22}}{j_{23}}{j_{31}}{j_{32}}{j_{33}}=\Delta\left(j_{11} j_{12} j_{13}\right) \Delta\left(j_{21} j_{22} j_{23}\right) \Delta\left(j_{31} j_{32} j_{33}\right) \Delta\left(j_{11} j_{21} j_{31}\right) \Delta\left(j_{12} j_{22} j_{32}\right) \Delta\left(j_{13} j_{23} j_{33}\right)  \notag\\
\quad \times(-1)^{R} \sum_{z_{\alpha}} \frac{(-1)^{z_{4}+z_{s}+z_{6}}\left(1+\sum_{\beta=1}^{6} z_{\beta}\right)!}{\prod_{p, q=1}^{3}\left(z_{p q}-j_{p q}\right)!\prod_{p=1}^{3}\left[\left(R-\sum_{\beta=1}^{6} z_{\beta}+z_{p}-h_{p}\right)!\left(z_{p+3}-h_{p+3}\right)!\right]} \label{eq:10:2:11}
\end{gather}
Summation in \eqref{eq:10:2:11} is over all integer and half-integer $z_{\alpha} \geq 0 ; z_{p q}$ and $h_{p}$ are determined by Eqs. \eqref{eq:10:2:8} and \eqref{eq:10:2:9}.\\
(c)
\begin{align}
& \ninej{j_{11}}{j_{12}}{j_{13}}{j_{21}}{j_{22}}{j_{23}}{j_{31}}{j_{32}}{j_{33}}=\Delta\left(j_{11} j_{12} j_{13}\right) \Delta\left(j_{21} j_{22} j_{23}\right) \Delta\left(j_{31} j_{32} j_{33}\right) \Delta\left(j_{11} j_{21} j_{31}\right) \Delta\left(j_{12} j_{22} j_{32}\right) \Delta\left(j_{13} j_{23} j_{33}\right)  \notag\\
& \times \sum_{x_{\alpha}} \frac{(-1)^{x_{4}+x_{5}+x_{6}}\left(1+R-2 t+\sum_{\beta=1}^{6} x_{\beta}-x_{4}\right)!}{\left(\prod_{p=1}^{3} x_{p}!\right) x_{4}!\left(x_{5}+x_{4}\right)!\left(x_{6}+x_{4}\right)!\prod_{p \neq q=1}^{3}\left(x_{p q}+t_{p q}\right)!\prod_{p=1}^{3}\left(t-j_{q r}-j_{r q}+x_{p}-\sum_{\beta=1}^{6} x_{\beta}\right)!} \label{eq:10:2:12}
\end{align}
where
\begin{equation}
\left(x_{p q}\right) \equiv\left(\begin{array}{lll}
x_{11} & x_{12} & x_{13}  \label{eq:10:2:13}\\
x_{21} & x_{22} & x_{23} \\
x_{31} & x_{32} & x_{33}
\end{array}\right)=\left(\begin{array}{ccc}
x_{1} & x_{3}+x_{6} & x_{2}+x_{5} \\
x_{3}+x_{5} & x_{2} & x_{1}+x_{6} \\
x_{2}+x_{6} & x_{1}+x_{5} & x_{3}
\end{array}\right)
\end{equation}
Summation in Eq. \eqref{eq:10:2:12} is over all integer $x_{\alpha} \geq 0$.
\begin{align}
t_{p q} & \equiv j_{q r}+j_{r p}-j_{p p}-j_{q q}, \quad(p \neq q \neq r=1,2,3)  \label{eq:10:2:14}\\
t & =j_{11}+j_{22}+j_{33} \notag
\end{align}
\subsection{9j Symbols as Sums of Products of the Clebsch-Gordan Coefficients or the 3jm Symbols \cite{ref110}}\label{sec:10.2.3}\index{9j symbol@$9j$ symbol!as sums of Clebsch-Gordan coefficients}\index{3jm symbol@$3jm$ symbol}
\begin{gather}
\ninej{a}{b}{c}{d}{e}{f}{g}{h}{j}=\left[(2 c+1)(2 f+1)(2 g+1)(2 h+1)\right]^{-\frac{1}{2}}(2 j+1)^{-1} \sum_{\substack{\alpha \beta \gamma \delta \varepsilon \varphi \\
\eta \mu \nu}} \clebsch{a}{\alpha}{b}{\beta}{c}{\gamma} \clebsch{d}{\delta}{e}{\varepsilon}{f}{\varphi} \clebsch{c}{\gamma}{f}{\varphi}{j}{\nu} \clebsch{a}{\alpha}{d}{\delta}{g}{\eta} \clebsch{b}{\beta}{e}{\varepsilon}{h}{\mu} \clebsch{g}{\eta}{h}{\mu}{j}{\nu}  \label{eq:10:2:15}\\
\ninej{a}{b}{c}{d}{e}{f}{g}{h}{j}=\frac{(-1)^{2(c+g)}}{(2 a+1)(2 e+1)(2 j+1)} \sum_{\substack{\alpha \beta \gamma \\
\delta \varepsilon \varphi \\
\eta \mu \nu}} \clebsch{c}{\gamma}{b}{\beta}{a}{\alpha} \clebsch{g}{\eta}{d}{\delta}{a}{\alpha} \clebsch{b}{\beta}{h}{\mu}{e}{\varepsilon} \clebsch{d}{\delta}{f}{\varphi}{e}{\varepsilon} \clebsch{h}{\mu}{g}{\eta}{j}{\nu} \clebsch{f}{\varphi}{c}{\gamma}{j}{\nu}  \label{eq:10:2:16}\\
\ninej{a}{b}{c}{d}{e}{f}{g}{h}{j}=\sum_{\substack{\alpha \beta \gamma \delta \varepsilon \varphi \\
\eta \mu \nu}}\threej{a }{ b }{ c }{\alpha }{ \beta }{ \gamma}\threej{d }{ e }{ f }{\delta }{ \varepsilon }{ \varphi}\threej{g }{ h }{ j }{\eta }{ \mu }{ \nu}\threej{a }{ d }{ g }{\alpha }{ \delta }{ \eta}\threej{b }{ e }{ h }{\beta }{ \varepsilon }{ \mu}\threej{c }{ f }{ j }{\gamma }{ \varphi }{ \nu} \label{eq:10:2:17}
\end{gather}
The following two relations were obtained in Ref. \cite{ref045}:
\begin{gather}
\ninej{a}{b}{c}{d}{e}{f}{g}{h}{j}=\frac{\Delta(a b c) \Delta(d e f)}{\Delta(a d g) \Delta(c f j)}\left[\frac{(g+h+j+1)!(g-h+j)!}{(2 h+1)}\right]^{\frac{1}{2}}  \notag\\
\times \frac{(c+f-j)!(a+d-g)!}{(a+d+g+1)!(a-b+c)!(-a+b+c)!(d-e+f)!(-d+e+f)!}  \notag\\
\times \sum_{x y \beta}(-1)^{a-c+\beta+x+y} \frac{(2 a-x)!(2 d-y)!(-a+b+c+x)!(-d+e+f+y)!}{x!y!(a+b-c-x)!(d+e-f-y)!(-a+c-\beta+x)!(-d+f+g-j+\beta+y)!}  \notag\\
\times \frac{(-a+f+j-\beta+x)!(-d+c+g+\beta+y)!}{(a+d-g-x-y)!(-a+c-d+f+g+j+1+x+y)!}\left[\frac{(b-\beta)!(e+g-j+\beta)!}{(b+\beta)!(e-g+j-\beta)!}\right]^{\frac{1}{2}} \clebsch{b}{\beta}{e}{j-g-\beta}{h}{j-g}  \label{eq:10:2:18}\\
\ninej{a}{b}{c}{d}{e}{f}{g}{h}{j}=\left[\frac{(a+d-g)!(c+f-j)!(g-h+j)!(g+h+j+1)!}{(-a+d+g)!(a-d+g)!(a+d+g+1)!(c+f+j+1)!(c-f+j)!(-c+f+j)!}\right]^{\frac{1}{2}}  \notag\\
\times \frac{1}{[(2 c+1)(2 f+1)(2 h+1)]^{\frac{1}{2}}} \sum_{\alpha \beta}(-1)^{a-c+\beta}\left[\frac{(a+\alpha)!(d+g-\alpha)!(c+\alpha+\beta)!(f+j-\alpha-\beta)!}{(a-\alpha)!(d-g+\alpha)!(c-\alpha-\beta)!(f-j+\alpha+\beta)!}\right]^{\frac{1}{2}}  \notag\\
\times \clebsch{b}{\beta}{e}{j-g-\beta}{h}{j-g} \clebsch{a}{\alpha}{b}{\beta}{c}{\alpha+\beta} \clebsch{d}{g-\alpha}{e}{j-g-\beta}{f}{j-\alpha-\beta} \label{eq:10:2:19}
\end{gather}
10.2.4. 9j Symbols as Sums of Products of the 6j Symbols or the Racah Coefficients [110,74]
\begin{align}
& \ninej{a}{b}{c}{d}{e}{f}{g}{h}{j}=\sum_{x}(-1)^{2 x}(2 x+1)\sixj{a}{b}{c}{f}{j}{x}\sixj{d}{e}{f}{b}{x}{h}\sixj{g}{h}{j}{x}{a}{d},  \label{eq:10:2:20}\\
& \ninej{a}{b}{c}{d}{e}{f}{g}{h}{j}=\sum_{x}(2 x+1) W( aech;  x b) W( aegf  ; x d) W(ch f g ; x j) . \label{eq:10:2:21}
\end{align}
\section{Integral Representations of the 9j Symbols}\label{sec:10.3}\index{9j symbol@$9j$ symbol!integral representations}\index{integral representations!of the $9j$ symbols}
\subsection{Representation Involving the Generalised Hypergeometric Function}\label{sec:10.3.1}\index{integral representations!with the generalized hypergeometric function}\index{hypergeometric function!generalized}
\begin{gather}
\ninej{a}{b}{c}{d}{e}{f}{g}{h}{j}=\frac{(-1)^{a+b-c-g-h+j}}{2^{b+e+h+1}} \cdot \frac{\Delta(a b c) \Delta(d e f) \Delta(b e h)}{\Delta(a d g) \Delta(c f j) \Delta(g h j)}  \notag\\
\times \frac{(a+d-g)!(c+f-j)!(g-h+j)!(b+e+h+1)!}{(a-b+c)!(-a+b+c)!(d-e+f)!(-d+e+f)!(a+d+g+1)!(-b+e+h)!(b+e+j-g)!}  \notag\\
\times \sum_{x y} \frac{(-1)^{x+y}(2 a-x)!(-a+b+c+x)!(-a-e+f+g+x)!(2 d-y)!(c-d+e+j+y)!}{x!y!(a+b-c-x)!(-a+c-e+g-j+x)!(d+e-f-y)!}  \notag\\
\times \frac{1}{(a+d-g-x-y)!(-a-d+g+c+f+j+1+x+y)!}  \notag\\
\times \int_{-1}^{1}(1+z)^{2 e}{ }_{3} F_{2}\left[\begin{array}{c}
-b-e+g-j,-a-e+f+g+1+x, d-e-f-y ; z-1 \\
-a+c-e+g-j+1+x,-c+d-e-j-y ; \\
z+1
\end{array}\right]  \notag\\
\times \frac{d^{-b+e+h}}{(d z)^{-b+e+h}}\left[(1-z)^{g+h-j}(1+z)^{-g+h+j}\right] d z . \label{eq:10:3:1}
\end{gather}
\subsection{Representation Involving the Wigner $D$ Functions}\label{sec:10.3.2}\index{integral representations!with the Wigner $D$-functions}\index{Wigner D-function@Wigner $D$-function}\isym{DJMM@$D_{M M^{\prime}}^{J}$, Wigner $D$-function}\isym{dR@$d R$, invariant volume element of the rotation group}
\begin{align}
& \ninej{a}{b}{c}{d}{e}{f}{g}{h}{j} \clebsch{c}{\gamma}{\rho}{\varphi}{j}{\mu}=(-1)^{c-f-g+h} \frac{[(2 j+1)(a+b-c)!(a+b+c+1)!(d+e-f)!(d+e+f+1)!]^{\frac{1}{2}}}{(a+d+g+1)!(b+e+h+1)!\Delta(a d g) \Delta(b e h)}  \notag\\
& \times[(g+h-j)!(g+h+j+1)!]^{\frac{1}{2}} \int d R_{1} d R_{2} d R_{3} B_{b c h}^{a d g}\left(R_{1}, R_{2}, R_{3}\right) D_{b-a, \gamma}^{c}\left(R_{1}\right) D_{c-d, \varphi}^{j}\left(R_{2}\right) D_{g-h, \mu}^{j *}\left(R_{3}\right) \label{eq:10:3:2}
\end{align}
In this equation $\gamma, \varphi, \mu$ are arbitrary integer or half-integer numbers and
\begin{equation}
d R_{i}=\frac{1}{8 \pi^{2}} \sin \beta_{i} d \alpha_{i} d \beta_{i} d \gamma_{i}, \quad(i=1,2,3) \label{eq:10:3:3}
\end{equation}
The integration limits are standard (see Eq. \eqref{eq:4:10:4}). $B_{b e h}^{a d g}\left(R_{1}, R_{2}, R_{3}\right)$ may be presented in the form
\begin{gather}
B_{b c h}^{a d g}\left(R_{1}, R_{2}, R_{3}\right)=B_{12}^{a+d-g} B_{12}^{* b+e-h} B_{31}^{a-d+g} B_{31}^{* b-e+h} B_{32}^{-a+d+g} B_{32}^{*-b+e+h}  \label{eq:10:3:4}\\
B_{i k} \equiv e^{-\frac{i}{2}\left(\alpha_{i}+\alpha_{k}\right)}\left\{\sin \frac{\beta_{i}}{2} \cos \frac{\beta_{k}}{2} e^{\frac{i}{2}\left(\gamma_{i}-\gamma_{k}\right)}-\cos \frac{\beta_{i}}{2} \sin \frac{\beta_{k}}{2} e^{-\frac{i}{2}\left(\gamma_{i}-\gamma_{k}\right)}\right\}  \notag\\
=\sum_{M}(-1)^{\frac{1}{2}-M} D_{\frac{1}{2} M}^{\frac{1}{2}}\left(R_{i}\right) D_{\frac{1}{2}-M}^{\frac{1}{2}}\left(R_{k}\right)=-D_{\frac{1}{2}-\frac{1}{2}}^{\frac{1}{2}}\left(R_{i} R_{k}^{-1}\right) \label{eq:10:3:5}
\end{gather}
The coefficients $B_{i k}$ possess the following properties:
\begin{gather}
\left(B_{i k}\right)^{2 j}=\sum_{m}(-1)^{j-m} D_{j m}^{j}\left(R_{i}\right) D_{j-m}^{j}\left(R_{k}\right)=(-1)^{2 j} D_{j-j}^{j}\left(R_{i} R_{k}^{-1}\right)  \label{eq:10:3:6}\\
\left(B_{i k}^{*}\right)^{2 j}=\sum_{m}(-1)^{j-m} D_{-j m}^{j}\left(R_{i}\right) D_{-j-m}^{j}\left(R_{k}\right)=D_{-j j}^{j}\left(R_{i} R_{k}^{-1}\right)  \label{eq:10:3:7}\\
\left|B_{i k}\right|=\sin \frac{\Omega_{i k}}{2} \label{eq:10:3:8}
\end{gather}
where
\begin{equation}
\cos \Omega_{i k}=\cos \beta_{i} \cos \beta_{k}+\sin \beta_{i} \sin \beta_{k} \cos \left(\gamma_{i}-\gamma_{k}\right) \label{eq:10:3:9}
\end{equation}
In particular, when $a=b, d=e, g=h$, from Eq. \eqref{eq:10:3:2}, one has \cite{ref063}
\begin{gather}
\ninej{a}{a}{c}{d}{d}{f}{g}{g}{j} \clebsch{c}{0}{\rho}{0}{j}{0}=(-1)^{c+f} \frac{1}{[(a+d-g+1)!]^{2} \Delta^{2}(a d g)}  \notag\\
\times[(2 j+1)(2 a-c)!(2 a+c+1)!(2 d-f)!(2 d+f+1)!(2 g-j)!(2 g+j+1)!]^{\frac{1}{2}}  \notag\\
\times \int d \Omega_{1} d \Omega_{2} d \Omega_{3} B_{a d g}\left(\Omega_{1}, \Omega_{2}, \Omega_{3}\right) P_{c}\left(\cos \beta_{1}\right) P_{f}\left(\cos \beta_{2}\right) P_{j}\left(\cos \beta_{3}\right) \label{eq:10:3:10}
\end{gather}
Here
\begin{gather}
d \Omega_{i}=\frac{1}{4 \pi} \sin \beta_{i} d \beta_{i} d \gamma_{i}  \label{eq:10:3:11}\\
B_{a d g}\left(\Omega_{1}, \Omega_{2}, \Omega_{3}\right)=\left(\sin \frac{\Omega_{12}}{2}\right)^{2(a+d-g)}\left(\sin \frac{\Omega_{31}}{2}\right)^{2(a-d+g)}\left(\sin \frac{\Omega_{32}}{2}\right)^{2(-a+d+g)} . \label{eq:10:3:12}
\end{gather}
\subsection{Integrals Involving Characters of Irreducible Representations of Rotation Group}\label{sec:10.3.3}\index{integral representations!with the characters}\index{character!of an irreducible representation}\index{rotation group}\isymg{chi-J@$\chi^{J}(\omega)$, character of an irreducible representation}
\begin{gather}
\ninej{a}{b}{c}{d}{e}{f}{g}{h}{j}^{2}=\frac{1}{\left(8 \pi^{2}\right)^{5}} \int d R_{1} d R_{2} d R_{3} d R_{4} d R_{5} \chi^{c}\left(R_{1}\right) \chi^{f}\left(R_{2}\right) \chi^{j}\left(R_{3}\right) \chi^{a}\left(R_{1} R_{4}\right) \chi^{b}\left(R_{1} R_{5}\right) \chi^{d}\left(R_{2} R_{4}\right)  \notag\\
\times \chi^{e}\left(R_{2} R_{5}\right) \chi^{g}\left(R_{3} R_{4}\right) \chi^{h}\left(R_{3} R_{5}\right)  \label{eq:10:3:13}\\
\ninej{a}{a}{b}{a}{a}{c}{b}{c}{d}=\frac{1}{\left(8 \pi^{2}\right)^{3}} \int d R_{1} d R_{2} d R_{3} \chi^{a}\left(R_{1}\right) \chi^{a}\left(R_{2}\right) \chi^{d}\left(R_{3}\right) \chi^{a}\left(R_{2} R_{1}\right) \chi^{b}\left(R_{3} R_{1}^{-1}\right) \chi^{c}\left(R_{3} R_{2}^{-1}\right) \label{eq:10:3:14}
\end{gather}
\section{Symmetry Properties of the 9j Symbols}\label{sec:10.4}\index{9j symbol@$9j$ symbol!symmetry properties}\index{symmetry!of the $9j$ symbols}
\subsection{Permutation Symmetry}\label{sec:10.4.1}\index{9j symbol@$9j$ symbol!permutation symmetry}
Any $9 j$ symbol possesses the following symmetry properties with respect to permutations \cite{ref074}: even permutations of columns or rows as well as transposition leave the $9 j$ symbol unchanged; odd permutations of columns or rows introduce the phase factor $(-1)^{R}$, where $R$ is sum of all arguments of the $9 j$ symbol. These symmetry properties may be written as follows\\
\subsubsection{Column permutations}
\begin{equation}
\ninej{j_{11}}{j_{12}}{j_{13}}{j_{21}}{j_{22}}{j_{23}}{j_{31}}{j_{32}}{j_{33}} \label{eq:10:4:1}=\varepsilon\ninej{j_{1 i}}{j_{1 k}}{j_{1 l}}{j_{2 i}}{j_{2 k}}{j_{2 l}}{j_{3 i}}{j_{3 k}}{j_{3 l}}
\end{equation}
\subsubsection{Row permutations}
\begin{equation}
\ninej{j_{11}}{j_{12}}{j_{13}}{j_{21}}{j_{22}}{j_{23}}{j_{31}}{j_{32}}{j_{33}} \label{eq:10:4:2}=\varepsilon\ninej{j_{i 1}}{j_{i 2}}{j_{i 3}}{j_{k 1}}{j_{k 2}}{j_{k 3}}{j_{l 1}}{j_{l 2}}{j_{l 3}}
\end{equation}
\subsubsection{Transposition}
\begin{equation}
\ninej{j_{11}}{j_{12}}{j_{13}}{j_{21}}{j_{22}}{j_{23}}{j_{31}}{j_{32}}{j_{33}} \label{eq:10:4:3}=\ninej{j_{11}}{j_{21}}{j_{31}}{j_{12}}{j_{22}}{j_{32}}{j_{13}}{j_{23}}{j_{33}}
\end{equation}
Here
\begin{gather}
\varepsilon= \begin{cases}1 & \text { for even permutations } \\
(-1)^{R} & \text { (cyclic permutations) } \\
& \text { for odd permutations } \\
& \text { (non-cyclic permutations). }\end{cases}  \label{eq:10:4:4}\\
R=\sum_{i, k=1}^{3} j_{i k} \label{eq:10:4:5}
\end{gather}
All these symmetry properties are independent and relate $3!\times 3!\times 2=729 j$ symbols given below (Eq. \eqref{eq:10:4:6}). The $9 j$ symbols obtained by column permutations are arranged horizontally. The $9 j$ symbols obtained by row permutations are arranged vertically. The six lower rows represent the transposed $9 j$ symbols.
\[
\begin{aligned}
& \ninej{a}{b}{c}{d}{e}{f}{g}{h}{j}=\ninej{c}{a}{b}{f}{d}{e}{j}{g}{h}=\ninej{b}{c}{a}{e}{f}{d}{h}{j}{g}=\varepsilon\ninej{b}{a}{c}{e}{d}{f}{h}{g}{j}=\varepsilon\ninej{c}{b}{a}{f}{e}{d}{j}{h}{g}=\varepsilon\ninej{a}{c}{b}{d}{f}{e}{g}{j}{h} \\
& =\ninej{g}{h}{j}{a}{b}{c}{d}{e}{f}=\ninej{j}{g}{h}{c}{a}{b}{f}{d}{e}=\ninej{h}{j}{g}{b}{c}{a}{e}{f}{d}=\varepsilon\ninej{h}{g}{j}{b}{a}{c}{e}{d}{f}=\varepsilon\ninej{j}{h}{g}{c}{b}{a}{f}{e}{d}=\varepsilon\ninej{g}{j}{h}{a}{c}{b}{d}{f}{e} \\
& =\ninej{d}{e}{f}{g}{h}{j}{a}{b}{c}=\ninej{f}{d}{e}{j}{g}{h}{c}{a}{b}=\ninej{e}{f}{d}{h}{j}{g}{b}{c}{a}=\varepsilon\ninej{e}{d}{f}{h}{g}{j}{b}{a}{c}=\varepsilon\ninej{f}{e}{d}{j}{h}{g}{c}{b}{a}=\varepsilon\ninej{d}{f}{e}{g}{j}{h}{a}{c}{b}
\end{aligned}
\]
\begin{align}
& =\varepsilon\ninej{d}{e}{f}{a}{b}{c}{g}{h}{j}=\varepsilon\ninej{f}{d}{e}{c}{a}{b}{j}{g}{h}=\varepsilon\ninej{e}{f}{d}{b}{c}{a}{h}{j}{g}=\ninej{e}{d}{f}{b}{a}{c}{h}{g}{j}=\ninej{f}{e}{d}{c}{b}{a}{j}{h}{g}=\ninej{d}{f}{e}{a}{c}{b}{g}{j}{h}  \notag\\
& =\varepsilon\ninej{g}{h}{j}{d}{e}{f}{a}{b}{c}=\varepsilon\ninej{j}{g}{h}{f}{d}{e}{c}{a}{b}=\varepsilon\ninej{h}{j}{g}{e}{f}{d}{b}{c}{a}=\ninej{h}{g}{j}{e}{d}{f}{b}{a}{c}=\ninej{j}{h}{g}{f}{e}{d}{c}{b}{a}=\ninej{g}{j}{h}{d}{f}{e}{a}{c}{b}  \notag\\
& =\varepsilon\ninej{a}{b}{c}{g}{h}{j}{d}{e}{f}=\varepsilon\ninej{c}{a}{b}{j}{g}{h}{f}{d}{e}=\varepsilon\ninej{b}{c}{a}{h}{j}{g}{e}{f}{d}=\ninej{b}{a}{c}{h}{g}{j}{e}{d}{f}=\ninej{c}{b}{a}{j}{h}{g}{f}{e}{d}=\ninej{a}{c}{b}{g}{j}{h}{d}{f}{e}  \notag\\
& =\ninej{a}{d}{g}{b}{e}{h}{c}{f}{j}=\ninej{g}{a}{d}{h}{b}{e}{j}{c}{f}=\ninej{d}{g}{a}{e}{h}{b}{f}{j}{c}=\varepsilon\ninej{d}{a}{g}{e}{b}{h}{f}{c}{j}=\varepsilon\ninej{g}{d}{a}{h}{e}{b}{j}{f}{c}=\varepsilon\ninej{a}{g}{d}{b}{h}{e}{c}{j}{f}  \notag\\
& =\ninej{c}{f}{j}{a}{d}{g}{b}{e}{h}=\ninej{j}{c}{f}{g}{a}{d}{h}{b}{e}=\ninej{f}{j}{c}{d}{g}{a}{e}{h}{b}=\varepsilon\ninej{f}{c}{j}{d}{a}{g}{e}{b}{h}=\varepsilon\ninej{j}{f}{c}{g}{d}{a}{h}{e}{b}=\varepsilon\ninej{c}{j}{f}{a}{g}{d}{b}{h}{e}  \notag\\
& =\ninej{b}{e}{h}{c}{f}{j}{a}{d}{g}=\ninej{h}{b}{e}{j}{c}{f}{g}{a}{d}=\ninej{e}{h}{b}{f}{j}{c}{d}{g}{a}=\varepsilon\ninej{e}{b}{h}{f}{c}{j}{d}{a}{g}=\varepsilon\ninej{h}{e}{b}{j}{f}{c}{g}{d}{a}=\varepsilon\ninej{b}{h}{e}{c}{j}{f}{a}{g}{d}  \notag\\
& =\varepsilon\ninej{b}{e}{h}{a}{d}{g}{c}{f}{j}=\varepsilon\ninej{h}{b}{e}{g}{a}{d}{j}{c}{f}=\varepsilon\ninej{e}{h}{b}{d}{g}{a}{f}{j}{c}=\ninej{e}{b}{h}{d}{a}{g}{f}{c}{j}=\ninej{h}{e}{b}{g}{d}{a}{j}{f}{c}=\ninej{b}{h}{e}{a}{g}{d}{c}{j}{f}  \notag\\
& =\varepsilon\ninej{c}{f}{j}{b}{e}{h}{a}{d}{g}=\varepsilon\ninej{j}{c}{f}{h}{b}{e}{g}{a}{d}=\varepsilon\ninej{f}{j}{c}{e}{h}{b}{d}{g}{a}=\ninej{f}{c}{j}{e}{b}{h}{d}{a}{g}=\ninej{j}{f}{c}{h}{e}{b}{g}{d}{a}=\ninej{c}{j}{f}{b}{h}{e}{a}{g}{d} .  \label{eq:10:4:6}\\
& =\varepsilon\ninej{a}{d}{g}{c}{f}{j}{b}{e}{h}=\varepsilon\ninej{g}{a}{d}{j}{c}{f}{h}{b}{e}=\varepsilon\ninej{d}{g}{a}{f}{j}{c}{e}{h}{b}=\ninej{d}{a}{g}{f}{c}{j}{e}{b}{h}=\ninej{g}{d}{a}{j}{f}{c}{h}{e}{b}=\ninej{a}{g}{d}{c}{j}{f}{b}{h}{e} \notag
\end{align}
\subsection{Symmetries of the $r$ Symbol}\label{sec:10.4.2}\index{r-symbol@$r$-symbol!symmetry properties}
Any $r$ symbol possesses the following symmetry properties:\\
\subsubsection{Permutations of rows introduce the phase factor $\varepsilon$ (see Eq. \protect\eqref{eq:10:4:4})}
\begin{equation}
\left\|\begin{array}{llllll}
r_{11} & r_{12} & r_{13} & r_{11}^{\prime} & r_{12}^{\prime} & r_{13}^{\prime}  \label{eq:10:4:7}\\
r_{21} & r_{22} & r_{23} & r_{21}^{\prime} & r_{22}^{\prime} & r_{23}^{\prime} \\
r_{31} & r_{32} & r_{33} & r_{31}^{\prime} & r_{32}^{\prime} & r_{33}^{\prime}
\end{array}\right\|=\varepsilon\left\|\begin{array}{llllll}
r_{i 1} & r_{i 2} & r_{i 3} & r_{i 1}^{\prime} & r_{i 2}^{\prime} & r_{i 3}^{\prime} \\
r_{k 1} & r_{k 2} & r_{k 3} & r_{k 1}^{\prime} & r_{k 2}^{\prime} & r_{k 3}^{\prime} \\
r_{l 1} & r_{l 2} & r_{l 3} & r_{l 1}^{\prime} & r_{l 2}^{\prime} & r_{l 3}^{\prime}
\end{array}\right\|,
\end{equation}
These permutations correspond to permutations of rows for the $9 j$ symbols.\\
\subsubsection{Simultaneous permutations of appropriate columns in both parts of the $r$ symbol (i.e., $r_{i k}$ and $r_{i k}^{\prime}$ ) introduce the same phase factor $\varepsilon$}
\begin{equation}
\left\|\begin{array}{llllll}
r_{11} & r_{12} & r_{13} & r_{11}^{\prime} & r_{12}^{\prime} & r_{13}^{\prime}  \label{eq:10:4:8}\\
r_{21} & r_{22} & r_{23} & r_{21}^{\prime} & r_{22}^{\prime} & r_{23}^{\prime} \\
r_{31} & r_{32} & r_{33} & r_{31}^{\prime} & r_{32}^{\prime} & r_{33}^{\prime}
\end{array}\right\|=\varepsilon\left\|\begin{array}{llllll}
r_{1 i} & r_{1 k} & r_{1 l} & r_{1 i}^{\prime} & r_{1 k}^{\prime} & r_{1 l}^{\prime} \\
r_{2 i} & r_{2 k} & r_{2 l} & r_{2 i}^{\prime} & r_{2 k}^{\prime} & r_{2 l}^{\prime} \\
r_{3 i} & r_{3 k} & r_{3 l} & r_{3 i}^{\prime} & r_{3 k}^{\prime} & r_{3 l}^{\prime}
\end{array}\right\| .
\end{equation}
This symmetry corresponds to the symmetry of the $9 j$ symbols with respect to column permutations.\\
\subsubsection{Transposition of both parts of the $r$ symbol with their simultaneous permutation is equivalent to transposition of the $9 j$ symbols}
\begin{equation}
\left\|\begin{array}{llllll}
r_{11} & r_{12} & r_{13} & r_{11}^{\prime} & r_{12}^{\prime} & r_{13}^{\prime}  \label{eq:10:4:9}\\
r_{21} & r_{22} & r_{23} & r_{21}^{\prime} & r_{22}^{\prime} & r_{23}^{\prime} \\
r_{31} & r_{32} & r_{33} & r_{31}^{\prime} & r_{32}^{\prime} & r_{33}^{\prime}
\end{array}\right\|=\left\|\begin{array}{llllll}
r_{11}^{\prime} & r_{21}^{\prime} & r_{31}^{\prime} & r_{11} & r_{21} & r_{31} \\
r_{12}^{\prime} & r_{22}^{\prime} & r_{32}^{\prime} & r_{12} & r_{22} & r_{32} \\
r_{13}^{\prime} & r_{23}^{\prime} & r_{33}^{\prime} & r_{13} & r_{23} & r_{33}
\end{array}\right\| .
\end{equation}
\subsection{"Mirror" Symmetry}\label{sec:10.4.3}\index{mirror symmetry}\index{symmetry!mirror}\index{negative momenta}
The original relations which define the $9 j$ symbols may be extended to the domain of negative integer or half-integer arguments. Introducing the notations $a=-a-1, b=-b-1, \ldots$, etc., we obtain the following symmetry properties \cite{ref045}
\begin{align}
&\left\{\begin{array}{lll}
a & b & c \\
d & e & f \\
g & h & j
\end{array}\right\}=\left\{\begin{array}{lll}
a & b & c \\
d & e & f \\
g & h & j
\end{array}\right\}=\eta_{1}\left\{\begin{array}{lll}
a & b & c \\
d & e & f \\
g & h & j
\end{array}\right\}=\eta_{1}\left\{\begin{array}{lll}
a & b & c \\
d & e & f \\
g & h & j
\end{array}\right\}  \notag\\
&=i \eta_{2}\left\{\begin{array}{ll}
a & b \\
d & c \\
g & e
\end{array}\right\}=-i \eta_{2}\left\{\begin{array}{lll}
a & b & c \\
d & e & f \\
g & h & j
\end{array}\right\}=\eta_{3}\left\{\begin{array}{lll}
a & b & c \\
d & e & f \\
g & h & j
\end{array}\right\}=\eta_{3}\left\{\begin{array}{lll}
a & b & c \\
d & e & f \\
g & h & j
\end{array}\right\}  \notag\\
&=i \eta_{4}\left\{\begin{array}{lll}
a & b & c \\
d & e & f \\
g & h & j
\end{array}\right\}=-i \eta_{4}\left\{\begin{array}{lll}
a & b & c \\
d & e & f \\
g & h & j
\end{array}\right\}=\eta_{5}\left\{\begin{array}{lll}
a & b & c \\
d & e & f \\
g & h & j
\end{array}\right\}=\eta_{5}\left\{\begin{array}{lll}
a & b & c \\
d & e & f \\
g & h & j
\end{array}\right\}  \notag\\
&=i \eta_{6}\left\{\begin{array}{lll}
a & b & c \\
d & e & f \\
g & h & j
\end{array}\right\}=-i \eta_{6}\left\{\begin{array}{lll}
a & b & c \\
d & e & f \\
g & h & j
\end{array}\right\}=\eta_{7}\left\{\begin{array}{lll}
a & b & c \\
d & e & f \\
g & h & j
\end{array}\right\}=\eta_{7}\left\{\begin{array}{lll}
a & b & c \\
d & e & f \\
g & h & j
\end{array}\right\}  \notag\\
&=\eta_{5}\left\{\begin{array}{lll}
a & b & c \\
d & c & f \\
g & h & j
\end{array}\right\}=\eta_{5}\left\{\begin{array}{lll}
a & b & c \\
d & e & f \\
g & h & j
\end{array}\right\}=\eta_{8}\left\{\begin{array}{lll}
a & b & c \\
d & e & f \\
g & h & j
\end{array}\right\}=\eta_{8}\left\{\begin{array}{lll}
a & b & c \\
d & e & f \\
g & h & j
\end{array}\right\}  \notag\\
&=-i \eta_{6}\left\{\begin{array}{lll}
a & b & c \\
d & e & f \\
g & h & j
\end{array}\right\}=-i \eta_{6}\left\{\begin{array}{lll}
a & b & c \\
d & e & f \\
g & h & j
\end{array}\right\}=-\left\{\begin{array}{lll}
a & b & c \\
d & e & f \\
g & h & j
\end{array}\right\}=\left\{\begin{array}{lll}
a & b & c \\
d & e & f \\
g & h & j
\end{array}\right\}  \notag\\
&=-\left\{\begin{array}{lll}
a & b & c \\
d & e & f \\
g & h & j
\end{array}\right\}=-\left\{\begin{array}{lll}
a & b & c \\
d & e & f \\
g & h & j
\end{array}\right\} . \label{eq:10:4:10}
\end{align}
In these equations
\begin{equation}
\begin{array}{ll}
\eta_{1}=(-1)^{b-c-d+g}, & \eta_{5}=(-1)^{c-f-g+h+1}, \\
\eta_{2}=(-1)^{c+d-e-g+h}, & \eta_{6}=(-1)^{g-h-j}, \\
\eta_{3}=(-1)^{c+f-g-h}, & \eta_{7}=(-1)^{2(a+e+j)}, \\
\eta_{4}=(-1)^{R}, & \eta_{8}=(-1)^{R-d+e-f+1} . \label{eq:10:4:11}
\end{array}
\end{equation}
When tabulating formulas for the $9 j$ symbols, it is convenient to use the "mirror" symmetry properties in\\[0pt]
the form \cite{ref045}
\begin{align}
& \ninej{d-\delta}{e+\varepsilon}{f+\varphi}{d}{e}{f}{g}{h}{j}=i(-1)^{g+\epsilon-\varphi}\ninej{\bar{d}+\delta}{e+\varepsilon}{f+\varphi}{\bar{d}}{e}{f}{g}{h}{j},  \label{eq:10:4:12}\\
& \ninej{d+\delta}{e-\varepsilon}{f+\varphi}{d}{e}{f}{g}{h}{j}=i(-1)^{h+\varphi-\delta}\ninej{d+\delta}{\bar{e}+\varepsilon}{f+\varphi}{d}{\bar{e}}{f}{g}{h}{j},  \label{eq:10:4:13}\\
& \ninej{d+\delta}{e+\varepsilon}{f-\varphi}{d}{e}{f}{g}{h}{j}=i(-1)^{j+\delta-e}\ninej{d+\delta}{e+\varepsilon}{\bar{f}+\varphi}{d}{e}{\bar{f}}{g}{h}{j},  \label{eq:10:4:14}\\
& \ninej{d-\delta}{e-\varepsilon}{f+\varphi}{d}{e}{f}{g}{h}{j}=(-1)^{h+\varphi-\theta+1}\ninej{d+\delta}{\bar{e}+\varepsilon}{f+\varphi}{\bar{d}}{\bar{e}}{f}{g}{h}{j},  \label{eq:10:4:15}\\
& \ninej{d-\delta}{e+\varepsilon}{f-\varphi}{d}{e}{f}{g}{h}{j}=(-1)^{g+\epsilon-j+1}\ninej{d+\delta}{e+\varepsilon}{\bar{f}+\varphi}{\bar{d}}{e}{\bar{f}}{g}{h}{j},  \label{eq:10:4:16}\\
& \ninej{d+\delta}{e-\varepsilon}{f-\varphi}{d}{e}{f}{g}{h}{j}=(-1)^{j+\delta-h+1}\ninej{d+\delta}{\bar{e}+\varepsilon}{\bar{f}+\varphi}{d}{\bar{e}}{\bar{f}}{g}{h}{j},  \label{eq:10:4:17}\\
& \ninej{d-\delta}{e-\varepsilon}{f-\varphi}{d}{e}{f}{g}{h}{j}=i(-1)^{g+h+j+\delta+\varepsilon+\varphi+1}\ninej{\bar{d}+\delta}{\bar{e}+\varepsilon}{\bar{f}+\varphi}{\bar{d}}{\bar{e}}{\bar{f}}{g}{h}{j}  \label{eq:10:4:18}\\
& l \notag
\end{align}
\section{Recursion Relations for the $9 j$ Symbols}\label{sec:10.5}\index{9j symbol@$9j$ symbol!recursion relations}\index{recursion relations!for the $9j$ symbols}
\subsection{General Form of the Recursion Relation}\label{sec:10.5.1}\index{recursion relations!general}
Recursion relations for the $9 j$ symbols may be obtained for instance, using the following equation \cite{ref077}
\begin{align}
& (-1)^{a+f+b+j} \sum_{x}(-1)^{2 x}(2 x+1)\ninej{a}{b}{x}{d}{e}{f}{g}{h}{j}\sixj{a}{b}{x}{c}{\lambda}{a^{\prime}}\sixj{j}{f}{x}{\lambda}{c}{f^{\prime}}  \notag\\
= & (-1)^{a^{\prime}+f^{\prime}+g+e} \sum_{y}(-1)^{2 y}(2 y+1)\ninej{a^{\prime}}{b}{c}{y}{e}{f^{\prime}}{g}{h}{j}\sixj{f^{\prime}}{e}{y}{d}{\lambda}{f}\sixj{g}{a^{\prime}}{y}{\lambda}{d}{a} . \label{eq:10:5:1}
\end{align}
Specifying $\lambda=\frac{1}{2}, 1, \frac{3}{2}, 2$, etc. and using explicit forms of the $6 j$ symbols, we may obtain various recursion relations. Examples of such relations are given below (Eqs. \eqref{eq:10:5:2}-\eqref{eq:10:5:5}).

\subsection{Relations Involving Four $9 j$ Symbols}\label{sec:10.5.2}\index{recursion relations!with four $9j$ symbols}
Equations \eqref{eq:10:5:2}-\eqref{eq:10:5:4} are derived by putting $\lambda=\frac{1}{2}$ in Eq. \eqref{eq:10:5:1} and specifying the values of $a^{\prime}$ and $f^{\prime}$. In particular\\
(a) for $a^{\prime}=a-\frac{1}{2}, f^{\prime}=f-\frac{1}{2}$, one has
\begin{gather}
\frac{1}{2 c+1}\left[\left(a-b+c+\frac{1}{2}\right)\left(a+b+c+\frac{3}{2}\right)\left(c+f-j+\frac{1}{2}\right)\left(c+f+j+\frac{3}{2}\right)\right]^{\frac{1}{2}}\ninej{a}{b}{c+\frac{1}{2}}{d}{e}{f}{g}{h}{j}  \notag\\
+\frac{1}{2 c+1}\left[\left(-a+b+c+\frac{1}{2}\right)\left(a+b-c+\frac{1}{2}\right)\left(-c+f+j+\frac{1}{2}\right)\left(c-f+j+\frac{1}{2}\right)\right]^{\frac{1}{2}}\ninej{a}{b}{c-\frac{1}{2}}{d}{e}{f}{g}{h}{j}  \notag\\
=\frac{1}{2 d+1}[(d+e-f+1)(-d+e+f)(-a+g+d+1)(a-d+g)]^{\frac{1}{2}}\ninej{a-\frac{1}{2}}{b}{c}{d+\frac{1}{2}}{e}{f-\frac{1}{2}}{g}{h}{j}  \notag\\
+\frac{1}{2 d+1}[(d-e+f)(d+e+f+1)(a+d-g)(a+d+g+1)]^{\frac{1}{2}}\ninej{a-\frac{1}{2}}{b}{c}{d-\frac{1}{2}}{e}{f-\frac{1}{2}}{g}{h}{j} ; \label{eq:10:5:2}
\end{gather}
(b) for $a^{\prime}=a+\frac{1}{2}, f^{\prime}=f+\frac{1}{2}$, one obtains
\begin{align}
& \frac{1}{2 c+1}\left[\left(-a+b+c+\frac{1}{2}\right)\left(a+b-c+\frac{1}{2}\right)\left(-c+f+j+\frac{1}{2}\right)\left(c-f+j+\frac{1}{2}\right)\right]^{\frac{1}{2}}\ninej{a}{b}{c+\frac{1}{2}}{d}{e}{f}{g}{h}{j}  \notag\\
& +\frac{1}{2 c+1}\left[\left(a-b+c+\frac{1}{2}\right)\left(a+b+c+\frac{3}{2}\right)\left(c+f-j+\frac{1}{2}\right)\left(c+f+j+\frac{3}{2}\right)\right]^{\frac{1}{2}}\ninej{a}{b}{c-\frac{1}{2}}{d}{e}{f}{g}{h}{j}  \notag\\
& \quad=\frac{1}{2 d+1}[(d-e+f+1)(d+e+f+2)(a+d-g+1)(a+d+g+2)]^{\frac{1}{2}}\ninej{a+\frac{1}{2}}{b}{c}{d+\frac{1}{2}}{e}{f+\frac{1}{2}}{g}{h}{j}  \notag\\
& \quad+\frac{1}{2 d+1}[(-d+e+f+1)(d+e-f)(-a+d+g)(a-d+g+1)]^{\frac{1}{2}}\ninej{a+\frac{1}{2}}{b}{c}{d-\frac{1}{2}}{e}{f+\frac{1}{2}}{g}{h}{j} \label{eq:10:5:3}
\end{align}
(c) for $a^{\prime}=a-\frac{1}{2}, f^{\prime}=f+\frac{1}{2}$, one gets
\begin{gather}
\frac{1}{2 c+1}\left[\left(a-b+c+\frac{1}{2}\right)\left(a+b+c+\frac{3}{2}\right)\left(-c+f+j+\frac{1}{2}\right)\left(c-f+j+\frac{1}{2}\right)\right]^{\frac{1}{2}}\ninej{a}{b}{c+\frac{1}{2}}{d}{e}{f}{g}{h}{j}  \notag\\
-\frac{1}{2 c+1}\left[\left(-a+b+c+\frac{1}{2}\right)\left(a+b-c+\frac{1}{2}\right)\left(c+f-j+\frac{1}{2}\right)\left(c+f+j+\frac{3}{2}\right)\right]^{\frac{1}{2}}\ninej{a}{b}{c-\frac{1}{2}}{d}{e}{f}{g}{h}{j}  \notag\\
=\frac{1}{2 d+1}[(d-e+f+1)(d+e+f+2)(-a+d+g+1)(a-d+g)]^{\frac{1}{2}}\ninej{a-\frac{1}{2}}{b}{c}{d+\frac{1}{2}}{e}{f+\frac{1}{2}}{g}{h}{j}  \notag\\
-\frac{1}{2 d+1}[(-d+e+f+1)(d+e-f)(a+d-g)(a+d+g+1)]^{\frac{1}{2}}\ninej{a-\frac{1}{2}}{b}{c}{d-\frac{1}{2}}{e}{f+\frac{1}{2}}{g}{h}{j} . \label{eq:10:5:4}
\end{gather}
The relation for $a^{\prime}=a+\frac{1}{2}$ and $f^{\prime}=f-\frac{1}{2}$ is equivalent to Eq. \eqref{eq:10:5:4}.

\subsection{Relations Involving Five $9 j$ Symbols \cite{ref045}}\label{sec:10.5.3}\index{recursion relations!with five $9j$ symbols}
Putting $\lambda=1, a^{\prime}=a$ and $f^{\prime}=f$ in Eq. \eqref{eq:10:5:1}, we get
\begin{gather}
\frac{A_{c+1}(a b, f j)}{(c+1)(2 c+1)}\ninej{a}{b}{c+1}{d}{e}{f}{g}{h}{j}+\frac{A_{c}(a b, f j)}{c(2 c+1)}\ninej{a}{b}{c-1}{d}{e}{f}{g}{h}{j}-\frac{A_{d+1}(e f, a g)}{(d+1)(2 d+1)}\ninej{a}{b}{c}{d+1}{e}{f}{g}{h}{j}  \notag\\
-\frac{A_{d}(e f, a g)}{d(2 d+1)}\ninej{a}{b}{c}{d-1}{e}{f}{g}{h}{j}  \notag\\
=\left\{\frac{[a(a+1)+d(d+1)-g(g+1)][d(d+1)-e(e+1)+f(f+1)]}{d(d+1)}\right.  \notag\\
\left.-\frac{[a(a+1)-b(b+1)+c(c+1)][c(c+1)+f(f+1)-j(j+1)]}{c(c+1)}\right\}\ninej{a}{b}{c}{d}{e}{f}{g}{h}{j}, \label{eq:10:5:5}
\end{gather}
where
\begin{align}
A_{q}(p r, s t)= & {[(-p+r+q)(p-r+q)(p+r-q+1)(p+r+q+1)}  \notag\\
& \times(-s+t+q)(s-t+q)(s+t-q+1)(s+t+q+1)]^{\frac{1}{2}} \label{eq:10:5:6}
\end{align}
We also present the relations of another type between five $9 j$ symbols \cite{ref075}:
\begin{gather}
{\left[\frac{(g+h+j+1)(g+h-j)(-b+e+h)(b+e-h+1)}{(a+d+g+2)(a-d+g+1)}\right]^{\frac{1}{2}}\ninej{a+\frac{1}{2}}{b+\frac{1}{2}}{c}{d}{e}{f}{g-\frac{1}{2}}{h-\frac{1}{2}}{j}}  \notag\\
+\left[\frac{(g-h+j)(-g+h+j+1)(b+e+h+2)(b-e+h+1)}{(a+d+g+2)(a-d+g+1)}\right]^{\frac{1}{2}}\ninej{a+\frac{1}{2}}{b+\frac{1}{2}}{c}{d}{e}{f}{g-\frac{1}{2}}{h+\frac{1}{2}}{j}  \notag\\
+\left[\frac{(g+h+j+2)(g+h-j+1)(b+e+h+2)(b-e+h+1)}{(a+d-g+1)(-a+d+g)}\right]^{\frac{1}{2}}\ninej{a+\frac{1}{2}}{b+\frac{1}{2}}{c}{d}{e}{f}{g+\frac{1}{2}}{h+\frac{1}{2}}{j}  \notag\\
-\left[\frac{(g-h+j+1)(-g+h+j)(b+e-h+1)(-b+e+h)}{(a+d-g+1)(-a+d+g)}\right]^{\frac{1}{2}}\ninej{a+\frac{1}{2}}{b+\frac{1}{2}}{c}{d}{e}{f}{g+\frac{1}{2}}{h-\frac{1}{2}}{j}  \notag\\
=(2 g+1)(2 h+1)\left[\frac{(a+b+c+2)(a+b-c+1)}{(a+d+g+2)(a-d+g+1)(a+d-g+1)(-a+d+g)}\right]^{\frac{1}{2}}\ninej{a}{b}{c}{d}{e}{f}{g}{h}{j} . \label{eq:10:5:7}
\end{gather}
\begin{align}
& {\left[\frac{(g+h+j+1)(g+h-j)(b+e+h+1)(b+h-e)}{(a+d-g)(-a+d+g+1)}\right]^{\frac{1}{2}}\ninej{a-\frac{1}{2}}{b-\frac{1}{2}}{c}{d}{e}{f}{g-\frac{1}{2}}{h-\frac{1}{2}}{j}}  \notag\\
& -\left[\frac{(-g+h+j+1)(g-h+j)(-b+e+h+1)(b+e-h)}{(a+d-g)(-a+d+g+1)}\right]^{\frac{1}{2}}\ninej{a-\frac{1}{2}}{b-\frac{1}{2}}{c}{d}{e}{f}{g-\frac{1}{2}}{h+\frac{1}{2}}{j}  \notag\\
& +\left[\frac{(g-h+j+1)(-g+h+j)(b+e+h+1)(b-e+h)}{(a+d+g+1)(a-d+g)}\right]^{\frac{1}{2}}\ninej{a-\frac{1}{2}}{b-\frac{1}{2}}{c}{d}{e}{f}{g+\frac{1}{2}}{h-\frac{1}{2}}{j}  \notag\\
& +\left[\frac{(g+h+j+2)(g+h-j+1)(-b+e+h+1)(b+e-h)}{(a+d+g+1)(a-d+g)}\right]^{\frac{1}{2}}\ninej{a-\frac{1}{2}}{b-\frac{1}{2}}{c}{d}{e}{f}{g+\frac{1}{2}}{h+\frac{1}{2}}{j}  \notag\\
& =(2 g+1)(2 h+1)\left[\frac{(a+b+c+1)(a+b-c)}{(a+d+g+1)(a-d+g)(a+d-g)(-a+d+g+1)}\right]^{\frac{1}{2}}\ninej{a}{b}{c}{d}{e}{f}{g}{h}{j} . \label{eq:10:5:8}
\end{align}
\begin{align}
& {\left[\frac{(-g+h+j+1)(g-h+j)(-b+e+h+1)(b+e-h)}{(a+d+g+2)(a-d+g+1)}\right]^{\frac{1}{2}}\ninej{a+\frac{1}{2}}{b-\frac{1}{2}}{c}{d}{e}{f}{g-\frac{1}{2}}{h+\frac{1}{2}}{j}}  \notag\\
& -\left[\frac{(g+h+j+1)(g+h-j)(b+e+h+1)(b-e+h)}{(a+d+g+2)(-a+d+g+1)}\right]^{\frac{1}{2}}\ninej{a+\frac{1}{2}}{b-\frac{1}{2}}{c}{d}{e}{f}{g-\frac{1}{2}}{h-\frac{1}{2}}{j}  \notag\\
& +\left[\frac{(g-h+j+1)(-g+h+j)(b+e+h+1)(b-e+h)}{(a+d-g+1)(-a+d+g)}\right]^{\frac{1}{2}}\ninej{a+\frac{1}{2}}{b-\frac{1}{2}}{c}{d}{e}{f}{g+\frac{1}{2}}{h-\frac{1}{2}}{j}  \notag\\
& +\left[\frac{(g+h+j+2)(g+h-j+1)(-b+e+h+1)(b+e-h)}{(a+d-g+1)(-a+d+g)}\right]^{\frac{1}{2}}\ninej{a+\frac{1}{2}}{b-\frac{1}{2}}{c}{d}{e}{f}{g+\frac{1}{2}}{h+\frac{1}{2}}{j}  \notag\\
& =(2 g+1)(2 h+1)\left[\frac{(a-b+c+1)(-a+b+c)}{(a+d+g+2)(a-d+g+1)(a+d-g+1)(-a+d+g)}\right]^{\frac{1}{2}}\ninej{a}{b}{c}{d}{e}{f}{g}{h}{j} . \label{eq:10:5:9}
\end{align}

\subsection{Relation Involving Six $9 j$ Symbols}\label{sec:10.5.4}\index{recursion relations!with six $9j$ symbols}
\begin{gather}
\frac{\left[(a+b+1)^{2}-(c+1)^{2}\right]^{\frac{1}{2}}\left[(c+1)^{2}-(a-b)^{2}\right]^{\frac{1}{2}}}{(2 c+1)}\left[\frac{f+j-c+1}{f+j-c}\right]^{\frac{1}{2}}\ninej{a}{b}{c+1}{d}{e}{f}{g}{h}{j}  \notag\\
+\frac{\left[(d+e+1)^{2}-(f+1)^{2}\right]^{\frac{1}{2}}\left[(f+1)^{2}-(d-e)^{2}\right]^{\frac{1}{2}}}{(2 f+1)}\left[\frac{c-f+j+1}{c-f+j}\right]^{\frac{1}{2}}\ninej{a}{b}{c}{d}{e}{f+1}{g}{h}{j}  \notag\\
+\frac{\left[(g+h+1)^{2}-(j+1)^{2}\right]^{\frac{1}{2}}\left[(j+1)^{2}-(g-h)^{2}\right]^{\frac{1}{2}}}{(2 j+1)}\left[\frac{c+f-j+1}{c+f-j}\right]^{\frac{1}{2}}\ninej{a}{b}{c}{d}{e}{f}{g}{h}{j+1}  \notag\\
=\left[\frac{(-c+f+j+1)(c-f+j+1)(c+f-j+1)(c+f+j+2)}{(-c+f+j)(c-f+j)(c+f-j)(c+f+j+1)}\right]^{\frac{1}{2}}  \notag\\
\times\left\{\frac{\left[(a+b+1)^{2}-c^{2}\right]^{\frac{1}{2}}\left[c^{2}-(a-b)^{2}\right]^{\frac{1}{2}}}{2 c+1}\left[\frac{-c+f+j}{-c+f+j+1}\right]^{\frac{1}{2}}\ninej{a}{b}{c-1}{d}{e}{f}{g}{h}{j}\right.  \notag\\
+\frac{\left[(d+e+1)^{2}-f^{2}\right]^{\frac{1}{2}}\left[f^{2}-(d-e)^{2}\right]^{\frac{1}{2}}}{2 f+1}\left[\frac{c-f+j}{c-f+j+1}\right]^{\frac{1}{2}}\ninej{a}{b}{c}{d}{e}{f-1}{g}{h}{j}  \notag\\
\left.+\frac{\left[(g+h+1)^{2}-j^{2}\right]^{\frac{1}{2}}\left[j^{2}-(g-h)^{2}\right]^{\frac{1}{2}}}{2 j+1}\left[\frac{c+f-j}{c+f-j+1}\right]^{\frac{1}{2}}\ninej{a}{b}{c}{d}{e}{f}{g}{h}{j-1}\right\} \label{eq:10:5:10}
\end{gather}
\subsection{Recursion Relations for Some Special Cases}\label{sec:10.5.5}\index{recursion relations!for special cases}
Somewhat simpler recursion relations may be obtained, if arguments of the $9 j$ symbols satisfy some special requirements.\\
(a) If two columns (or two rows) are the same, then
\begin{align}
& \frac{1}{2 c+1}[(2 a+c+2)(2 a-c)(-c+f+j)(c-f+j+1)(c+f-j+1)(c+f+j+2)]^{\frac{1}{2}}\ninej{a}{a}{c+1}{d}{d}{f}{g}{g}{j}  \notag\\
+ & \frac{1}{2 c+1}[(2 a+c+1)(2 a-c+1)(-c+f+j+1)(c-f+j)(c+f-j)(c+f+j+1)]^{\frac{1}{2}}\ninej{a}{a}{c-1}{d}{d}{f}{g}{g}{j}  \notag\\
= & \frac{1}{2 f+1}[(2 d+f+2)(2 d-f)(-c+f+j+1)(c-f+j)(c+f-j+1)(c+f+j+2)]^{\frac{1}{2}}\ninej{a}{a}{c}{d}{d}{f+1}{g}{g}{j}  \notag\\
+ & \frac{1}{2 f+1}[(2 d+f+1)(2 d-f+1)(-c+f+j)(c-f+j+1)(c+f-j)(c+f+j+1)]^{\frac{1}{2}}\ninej{a}{a}{c}{d}{d}{f-1}{g}{g}{j} \label{eq:10:5:11}
\end{align}
(b) If one of arguments in any triad is equal to the sum of two other arguments, one has \cite{ref045}
\begin{align}
& {\left[\frac{(-c+g+h-f)(c-g+h+f+1)(c+g-h+f+1)(c+g+h+f+2)}{(2 c+2 f+2)(2 c+2 f+3)}\right]^{\frac{1}{2}}\ninej{a}{b}{c}{d}{e}{f}{g}{h}{c+f}}  \notag\\
& \quad=\left[\frac{(-a+b+c+1)(a-b+c+1)(a+b-c)(a+b+c+2)}{(2 c+1)(2 c+2)}\right]^{\frac{1}{2}}\ninej{a}{b}{c+1}{d}{e}{f}{g}{h}{c+f+1}  \notag\\
& \quad+\left[\frac{(-d+e+f+1)(d-e+f+1)(d+e-f)(d+e+f+2)}{(2 f+1)(2 f+2)}\right]^{\frac{1}{2}}\ninej{a}{b}{c}{d}{e}{f+1}{g}{h}{c+f+1} \label{eq:10:5:12}
\end{align}
(c) If one of arguments in any triad is equal to 0 or 1 , one gets [32, 75]:
\begin{gather}
\ninej{a}{b}{c}{d}{e}{c}{g}{g}{1}=\frac{a(a+1)+e(e+1)-d(d+1)-b(b+1)}{2[c(c+1) g(g+1)]^{\frac{1}{2}}}\ninej{a}{b}{c}{d}{e}{c}{g}{g}{0},  \label{eq:10:5:13}\\
{[g(g+1)(c+1)(2 c+3)(2 d+c+2)(2 d-c)]^{\frac{1}{2}}\ninej{a}{a}{c}{d}{d}{c+1}{g}{g}{1}}  \notag\\
=[g(g+1) c(2 c-1)(2 d+c+1)(2 d-c+1)]^{\frac{1}{2}}\ninej{a}{a}{c}{d}{d}{c}{g}{g}{1}  \notag\\
+(2 c+1)[d(d+1)+g(g+1)-a(a+1)]\ninej{a}{a}{c}{d}{d}{c}{g}{g}{0},  \label{eq:10:5:14}\\
{[c(2 c+3)(d+e+c+2)(-d+e+c+1)(d-e+c+1)(d+e-c)]^{\frac{1}{2}}\ninej{a}{b}{c}{d}{e}{c+1}{g}{g}{1}}  \notag\\
-[(c+1)(2 c-1)(d+e+c+1)(-d+e+c)(d-e+c)(d+e-c+1)]^{\frac{1}{2}}\ninej{a}{b}{c}{d}{e}{c-1}{g}{g}{1}  \notag\\
=(2 c+1)\left\{d(d+1)-e(e+1)-c(c+1)+\frac{2 c(c+1)[e(e+1)+g(g+1)-b(b+1)]}{a(a+1)+e(e+1)-d(d+1)-b(b+1)}\right\}\ninej{a}{b}{c}{d}{e}{c}{g}{g}{1}, \label{eq:10:5:15}
\end{gather}
where either $a \neq b$ and $d \neq e$, or $a \neq d$ and $b \neq e$. In particular,
\begin{gather}
\ninej{a}{b}{c}{d}{e}{c+1}{\frac{1}{2}}{\frac{1}{2}}{1}=\left[\frac{2 c+1}{c+(a-d)(2 a+1)+(b-e)(2 b+1)}-1\right]  \notag\\
\times\left[\frac{c(2 c-1)(d+e-c)(-d+e+c+1)(d-e+c+1)(d+e+c+2)}{(c+1)(2 c+3)(d+e-c+1)(-d+e+c)(d-e+c)(d+e+c+1)}\right]^{\frac{1}{2}}\ninej{a}{b}{c}{d}{e}{c-1}{\frac{1}{2}}{\frac{1}{2}}{1} . \label{eq:10:5:16}
\end{gather}
\section{Generating Function of the $9 j$ Symbols}\label{sec:10.6}\index{9j symbol@$9j$ symbol!generating function}\index{generating functions!for the $9j$ symbols}
The following function of eighteen variables $\tau_{i k}, \tau_{i k}^{\prime}(i, k=1,2,3)$ \cite{ref111}
\begin{equation}
\left[1-\sum_{p, q=1}^{3} a_{p q}-\sum_{\alpha=1}^{6} b_{\alpha}\right]^{-1} \label{eq:10:6:1}
\end{equation}
may be regarded as a generating function for the $9 j$ symbols. Here $a_{p q}$ and $b_{\alpha}$ are given by the arrays
\begin{align}
& \left(a_{p q}\right)=\left(\begin{array}{lll}
a_{11} & a_{12} & a_{13} \\
a_{21} & a_{22} & a_{23} \\
a_{31} & a_{32} & a_{33}
\end{array}\right) \equiv\left(\begin{array}{llllllllllll}
\tau_{21} & \tau_{31} & \tau_{12}^{\prime} & \tau_{13}^{\prime} & \tau_{22} & \tau_{32} & \tau_{11}^{\prime} & \tau_{13}^{\prime} & \tau_{23} & \tau_{33} & \tau_{11}^{\prime} & \tau_{12}^{\prime} \\
\tau_{11} & \tau_{31} & \tau_{22}^{\prime} & \tau_{23}^{\prime} & \tau_{12} & \tau_{32} & \tau_{21}^{\prime} & \tau_{23}^{\prime} & \tau_{13} & \tau_{33} & \tau_{21}^{\prime} & \tau_{22}^{\prime} \\
\tau_{11} & \tau_{21} & \tau_{32}^{\prime} & \tau_{33}^{\prime} & \tau_{12} & \tau_{22} & \tau_{31}^{\prime} & \tau_{33}^{\prime} & \tau_{13} & \tau_{23} & \tau_{31}^{\prime} & \tau_{32}^{\prime}
\end{array}\right),  \label{eq:10:6:2}\\
& \left(b_{\alpha}\right)=\left(\begin{array}{l}
b_{1} \\
b_{2} \\
b_{3} \\
b_{4} \\
b_{5} \\
b_{6}
\end{array}\right)=\left(\begin{array}{rrrrrr}
\tau_{11} & \tau_{11}^{\prime} & \tau_{32} & \tau_{32}^{\prime} & \tau_{23} & \tau_{23}^{\prime} \\
\tau_{31} & \tau_{31}^{\prime} & \tau_{22} & \tau_{22}^{\prime} & \tau_{13} & \tau_{13}^{\prime} \\
\tau_{21} & \tau_{21}^{\prime} & \tau_{12} & \tau_{12}^{\prime} & \tau_{33} & \tau_{33}^{\prime} \\
-\tau_{11} & \tau_{11}^{\prime} & \tau_{22} & \tau_{22}^{\prime} & \tau_{33} & \tau_{33}^{\prime} \\
-\tau_{21} & \tau_{21}^{\prime} & \tau_{32} & \tau_{32}^{\prime} & \tau_{13} & \tau_{13}^{\prime} \\
-\tau_{31} & \tau_{31}^{\prime} & \tau_{12} & \tau_{12}^{\prime} & \tau_{23} & \tau_{23}^{\prime}
\end{array}\right) . \label{eq:10:6:3}
\end{align}
The elements $b_{\alpha}$ satisfy the equation
\begin{equation}
\sum_{\alpha=1}^{6} b_{\alpha}=-\left|\begin{array}{llllll}
\tau_{11} & \tau_{11}^{\prime} & \tau_{12} & \tau_{12}^{\prime} & \tau_{13} & \tau_{13}^{\prime}  \label{eq:10:6:4}\\
\tau_{21} & \tau_{21}^{\prime} & \tau_{22} & \tau_{22}^{\prime} & \tau_{23} & \tau_{23}^{\prime} \\
\tau_{31} & \tau_{31}^{\prime} & \tau_{32} & \tau_{32}^{\prime} & \tau_{33} & \tau_{33}^{\prime}
\end{array}\right|
\end{equation}
The function \eqref{eq:10:6:1} can be expanded into a power series whose expansion coefficients are proportional to the $9 j$ symbols:
\begin{equation}
\frac{1}{1-\sum_{p, q=1}^{3} a_{p q}-\sum_{\alpha=1}^{6} b_{\alpha}}=\sum_{r, r^{\prime}}\left[\frac{\prod_{p=1}^{3}\left(\sum_{q=1}^{3} r_{p q}+1\right)!\left(\sum_{q=1}^{3} r_{q p}^{\prime}+1\right)!}{\prod_{p, q=1}^{3} r_{p q}!r_{p q}^{\prime}!}\right]^{\frac{1}{2}}\ninej{a}{b}{c}{d}{e}{f}{g}{h}{j} \label{eq:10:6:5}\prod_{p, q=1}^{3}\left(\tau_{p q}\right)^{r_{p q}}\left(\tau_{p q}^{\prime}\right)^{r_{p q}^{\prime}} .
\end{equation}
The exponents $r_{p q}$ and $r_{p q}^{\prime}$ are the elements of some $r$ symbol.

\section{Asymptotic Expression for a $9 j$ Symbol}\label{sec:10.7}\index{9j symbol@$9j$ symbol!asymptotic expression}\index{asymptotics!of the $9j$ symbols}
If $a, b, c, d, e, f, g, h, j \gg 1$, we obtain\index{step function}\isymg{Theta-step@$\Theta(x)$, step function}\isym{B-gram@$B$, Gram determinant of the direction cosines}
\begin{equation}
\ninej{a}{b}{c}{d}{e}{f}{g}{h}{j} \label{eq:10:7:1} \approx \frac{2 \Theta(B)(-1)^{a+c-e+g+j} \cos \left[(d-h) \delta_{1}+(a-j) \delta_{2}+(b-f) \delta_{3}\right]}{\pi \sqrt{B(2 a+1)(2 b+1)(2 d+1)(2 f+1)(2 h+1)(2 j+1)}}
\end{equation}
In this equation
\begin{gather}
B=\left|\begin{array}{ccc}
1 & \cos \vartheta_{3} & \cos \vartheta_{1} \\
\cos \vartheta_{3} & 1 & \cos \vartheta_{2} \\
\cos \vartheta_{1} & \cos \vartheta_{2} & 1
\end{array}\right|,  \label{eq:10:7:2}\\
\Theta(B)=\left\{\begin{array}{cc}
1 & \text { if } B>0 \\
0 & \text { if } B<0
\end{array}\right. \label{eq:10:7:3}
\end{gather}
The angles $\delta_{1}, \delta_{2}$ and $\delta_{3}$ are determined by the relations
\begin{equation}
\begin{array}{ll}
\cos \delta_{1}=\frac{\cos \vartheta_{1} \cos \vartheta_{3}-\cos \vartheta_{2}}{\sin \vartheta_{1} \sin \vartheta_{3}}, & \frac{\sin \delta_{2}}{\sin \delta_{3}}=\frac{\sin \vartheta_{1}}{\sin \vartheta_{3}} \\
\cos \delta_{2}=\frac{\cos \vartheta_{2} \cos \vartheta_{3}-\cos \vartheta_{1}}{\sin \vartheta_{2} \sin \vartheta_{3}}, & \frac{\sin \delta_{1}}{\sin \delta_{2}}=\frac{\sin \vartheta_{2}}{\sin \vartheta_{1}}  \label{eq:10:7:4}\\
\cos \delta_{3}=\frac{\cos \vartheta_{1} \cos \vartheta_{2}-\cos \vartheta_{3}}{\sin \vartheta_{1} \sin \vartheta_{2}}, & \frac{\sin \delta_{3}}{\sin \delta_{1}}=\frac{\sin \vartheta_{3}}{\sin \vartheta_{2}}
\end{array}
\end{equation}
The angles $\vartheta_{1}, \vartheta_{2}$ and $\vartheta_{3}$ may in turn be evaluated using the equations
\begin{equation}
\begin{array}{ll}
\cos \vartheta_{1}=\frac{a(a+1)+b(b+1)-c(c+1)}{2 \sqrt{a(a+1) b(b+1)}}, & 0 \leq \vartheta_{1} \leq \pi, \\
\cos \vartheta_{2}=-\frac{d(d+1)-e(e+1)+f(f+1)}{2 \sqrt{d(d+1) f(f+1)}}, & 0 \leq \vartheta_{2} \leq \pi,  \label{eq:10:7:5}\\
\cos \vartheta_{3}=\frac{-g(g+1)+h(h+1)+j(j+1)}{2 \sqrt{h(h+1) j(j+1)}}, & 0 \leq \vartheta_{3} \leq \pi .
\end{array}
\end{equation}
\section{Explicit Forms of the 9j Symbols at Some Relations Between Arguments}\label{sec:10.8}\index{9j symbol@$9j$ symbol!for special relations between arguments}
\subsection{Two Rows (or Columns) are Identical}\label{sec:10.8.1}\index{9j symbol@$9j$ symbol!with two identical rows}
In this case a $9 j$ symbol vanishes unless the sum of all arguments is even \cite{ref049}, or
\begin{gather}
\ninej{a}{b}{c}{a}{b}{c}{g}{h}{j}=0, \text { if } g+h+j=2 k+1,  \notag\\
\ninej{a}{a}{c}{d}{d}{f}{g}{g}{j}=0, \text { if } c+f+j=2 k+1,  \label{eq:10:8:1}\\
k=0,1, \ldots \notag
\end{gather}
\subsection{One Degenerate Triad}\label{sec:10.8.2}\index{degenerate triad}\index{9j symbol@$9j$ symbol!with one degenerate triad}
Let one argument of a triad be equal to the sum of the other two. For brevity, we shall refer to such a triad as a "degenerate" triad.\index{triad!degenerate} If some $9 j$ symbol has one degenerate triad, it may be expressed as
\begin{gather}
\ninej{a}{b}{c}{d}{e}{f}{a+d}{h}{j}=\frac{\Delta(h, j, a+d)}{\Delta(a b c) \Delta(b e h) \Delta(d e f) \Delta(c f j)}\left[\frac{(2 a)!(2 d)!}{(2 a+2 d+1)!}\right]^{\frac{1}{2}}  \notag\\
\times \frac{(b+c-a)!(b+e-h)!(e+f-d)!(c+f-j)!(a+d+h+j+1)!}{(a+b+c+1)!(b+e+h+1)!(d+e+f+1)!(c+f+j+1)!}  \notag\\
\times \sum_{x y} \frac{(-1)^{x+y}(2 c-x)!(2 e-y)!(j+f-c+x)!(h+b-e+y)!}{x!y!(c+f-j-x)!(b+e-h-y)!(c-e-a+h-x+y)!(e-c-d+j+x-y)!} \label{eq:10:8:2}
\end{gather}
This equation may be rewritten in the form of a sum involving the Clebsch-Gordan coefficients \cite{ref104}
\begin{align}
&\ninej{a}{b}{c}{d}{e}{f}{a+d}{h}{j}= \frac{[(a+d+j-h)!(h+j+a+d+1)!]^{\frac{1}{2}}}{\Delta(a b c) \Delta(d e f) \Delta(c f j)}\left[\frac{(2 a)!(2 d)!}{(2 a+2 d+1)!(2 h+1)}\right]^{\frac{1}{2}}  \notag\\
& \times \frac{(c+b-a)!(f+e-d)!(f+c-j)!}{(a+b+c+1)!(f+e+d+1)!(f+c+j+1)!}  \notag\\
& \times \sum_{x}(-1)^{x} \frac{(f+j-c+x)!(2 c-x)!}{x!(f+c-j-x)!}\left[\frac{(b+a-c+x)!(e+d+c-j-x)!}{(b+c-a-x)!(e+j-c-d+x)!}\right]^{\frac{1}{2}} \clebsch{b}{c-a-x}{e}{j-d-c+x}{h}{j-a-d} \label{eq:10:8:3}
\end{align}
Equation \eqref{eq:10:8:2} may also be rewritten in the quasi-trinomial form
\begin{equation}
\left(A^{(1)}-B^{(1)}-C^{(1)}\right)^{(k)}=\sum_{x, y}\binom{k}{x}\binom{k-x}{y}(-1)^{x+y} C^{(x)} B^{(y)} A^{(k-x-y)}, \label{eq:10:8:4}
\end{equation}
where $A^{(x)}, B^{(y)}$ and $C^{(k-x-y)}$ are quasi-powers (Eq. \eqref{eq:8:2:11}). Using Eq. \eqref{eq:10:8:4}, one gets \cite{ref045}
\begin{align}
& \ninej{a}{b}{c}{d}{e}{f}{a+d}{h}{j}=(-1)^{b+e-h}\nonumber\\
& \times
\left[\frac{(2d)!}{(e-d+f)!(d-e+f)!(d+e-f)!(e+d+f+1)!(f-j+c)!(e+h-b)!(e-h+b)!(f+j-c)!}\right]^\frac12 \notag\\
&\times\left[\frac{(a+b+c+1)^{(-1)(h+j-b-c+d)}(a-b+c)^{(-1)(b+j-h-c+d)}(b-c+a)^{(-1)(h+c-b-j+d)}(b+c-a)^{(1)(b+c-h-j+d)} }{(b+e+h+1)^{(2 e+1)} (c+f+j+1)^{(2 f+1)}(2 a)^{(-1)(2 d+1)}}\right]^{\frac{1}{2}}  \notag\\
& \times(2 e)!(2 f)!\left[(h+j-a-d)^{(1)}-\frac{(e-h+b)^{(1)}(b-e+h)^{(-1)}}{(2 e)^{(1)}}-\frac{(f-j+c)^{(1)}(c-f+j)^{(-1)}}{(2 f)^{(1)}}\right]^{(e+f-d)} . \label{eq:10:8:5}
\end{align}
If one argument is equal to the difference of the other two from the same triad, the expression for the $9 j$ symbol may be obtained from the above equations either by using the "mirror" symmetry properties (Sec.~\ref{sec:10.4.3}),
\begin{equation}
\ninej{a}{b}{c  \label{eq:10:8:6}}{d}{e}{f}{a-d}{h}{j}=i(-1)^{d+h-b-j+c}\ninej{\bar{a}}{b}{c}{d}{e}{f}{\bar{a}+d}{h}{j},
\end{equation}
or by renaming and permuting the arguments:
\begin{equation}
\ninej{a}{b}{c  \label{eq:10:8:7}}{d}{e}{f}{a-d}{h}{j} \underset{a=d+g}{\longrightarrow}\ninej{d+g}{b}{c}{d}{e}{f}{g}{h}{j}=\ninej{d}{e}{f}{g}{h}{j}{d+g}{b}{c} .
\end{equation}
The equations of Sections 10.8.3-10.8.6 may be derived from Eqs. \eqref{eq:10:8:2}-\eqref{eq:10:8:5}.

\subsection{Two Degenerate Triads}\label{sec:10.8.3}\index{9j symbol@$9j$ symbol!with two degenerate triads}
In this case there exist five types of relations which cannot be transformed to one another by renaming and permuting the arguments [45, 104]. All these relations are given below
\begin{gather}
\ninej{a}{b}{c}{d}{e}{f}{a+d}{b+e}{j}=\frac{\Delta(a+d b+e j)}{\Delta(a b c) \Delta(d e f) \Delta(c f j)}\left[\frac{(2 a)!(2 b)!(2 d)!(2 e)!}{(2 a+2 b+1)!(2 d+2 e+1)!}\right]^{\frac{1}{2}}  \notag\\
\times \frac{(a+b+d+e+j+1)!(a-b+c)!(d-e+f)!(c+f-j)!}{(a+b+c+1)!(d+e+f+1)!(c+f+j+1)!}  \notag\\
\times \sum_{z} \frac{(-1)^{z}(2 f-z)!(j+c-f+z)!}{z!(c+f-j-z)!(d+f-e-z)!(a-b-f+j+z)!} . \label{eq:10:8:8}
\end{gather}
Equation \eqref{eq:10:8:8} may be rewritten in terms of the Clebsch-Gordan coefficients:
\begin{align}
& \ninej{a}{b}{c}{d}{e}{f}{a+d}{b+e}{j}  \notag\\
& =\left[\frac{(2 a)!(2 b)!(2 d)!(2 e)!(a+b+d+e+j+1)!(a+d+e+b-j)!}{(2 a+2 d+1)!(2 b+2 e+1)!(a+b+c+1)!(a+b-c)!(d+e+f+1)!(d+e-f)!(2 j+1)}\right]^{\frac{1}{2}}  \notag\\
& \times \clebsch{c}{a-b}{f}{d-e}{j}{a-b+d-e},  \label{eq:10:8:9}\\
& \ninej{d+g}{b}{c}{d}{e}{f}{g}{h}{c+f}=(-1)^{d+h-b-f}\left[\frac{(2 c)!(2 f)!(2 d)!(2 g)!}{(2 c+2 f+1)!(2 d+2 g+1)!}\right]^{\frac{1}{2}} \frac{(d+e-f)!}{\Delta(d+g b c) \Delta(c+f g h)}  \notag\\
& \times \frac{(b+e-h)!(d+g+b-c)!(c+f+h-g)!}{\Delta(b e h) \Delta(d e f)(d+e+f+1)!(b+e+h+1)!(d+g+b+c+1)!}  \notag\\
& \times \sum_{x} \frac{(2 e-x)!(h+b-e+x)!(d+g+e+c-h-x)!}{x!(b+e-h-x)!(d+e-f-x)!} .  \label{eq:10:8:10}\\
& \ninej{a}{b}{a+b}{d}{e}{f}{a+d}{h}{j}=\frac{\Delta(a+b f j) \Delta(a+d h j)}{\Delta(b e h) \Delta(d e f)}\left[\frac{(2 b)!(2 d)!}{(2 a+2 b+1)!(2 a+2 d+1)!}\right]^{\frac{1}{2}}  \notag\\
& \times \frac{(a+b+f+j+1)!(a+d+h+j+1)!(h-b+e)!(e-d+f)!}{(f+j-a-b)!(h+j-a-d)!(b+e+h+1)!(d+e+f+1)!}  \notag\\
& \times \sum_{x} \frac{(-1)^{x}(a+b+e+d-j-x)!(j+f-a-b+x)!(j+h-a-d+x)!}{x!(a+b+f-j-x)!(a+d+h-j-x)!(j+e-a-b-d+x)!(2 j+1+x)!},  \label{eq:10:8:11}\\
& \ninej{a}{b}{a-b}{d}{e}{f}{a-d}{h}{j}=(-1)^{b+f-d-h} \frac{[(2 a-2 b)!(2 a-2 d)!(2 b)!(2 d)!]^{\frac{1}{2}}}{\Delta(a-b f j) \Delta(a-d h j) \Delta(b e h) \Delta(d e f)}  \notag\\
& \times \frac{(b+f+j-a)!(d+h+j-a)!(h-b+e)!(e-d+f)!}{(a-b+f+j+1)!(a-d+h+j+1)!(b+e+h+1)!(d+e+f+1)!}  \notag\\
& \times \sum_{x} \frac{(x-1)!(b+e+d-a-j+x-1)!(a-d+h+j-x+1)!(a-b+f+j-x+1)!}{(2 j+1-x)!(b+f-a-j+x-1)!(d+h-a-j+x-1)!(a+e+j-b-d-x+1)!},  \label{eq:10:8:12}\\
& \ninej{a}{b}{a+b}{d}{e}{f}{g}{h}{a+b+f}=(-1)^{a+d-g} \frac{\Delta(a+b+f g h)}{\Delta(a d g) \Delta(b e h) \Delta(d e f)}\left[\frac{(2 a)!(2 b)!(2 f)!}{(2 a+2 b+1)(2 a+2 b+2 f+1)!}\right]^{\frac{1}{2}}  \notag\\
& \times \frac{(a+b+g+h+f+1)!(g-a+d)!(e-b+h)!(d+e-f)!}{(g+h-a-b-f)!(a+g+d+1)!(b+e+h+1)!(d+e+f+1)!} . \label{eq:10:8:13}
\end{align}
\subsection{Three Degenerate Triads}\label{sec:10.8.4}\index{9j symbol@$9j$ symbol!with three degenerate triads}
Below we present the simplest expressions of $9 j$ symbols
\begin{align}
& \ninej{a}{b}{a+b}{d}{e}{d+e}{g}{h}{g+h}=\frac{\Delta(a+b d+e g+h)}{\Delta(a d g) \Delta(b e h)} \cdot \frac{(a+b+e+d+g+h+1)!}{(a+d+g+1)!(b+e+h+1)!}  \notag\\
& \times\left[\frac{(2 a)!(2 b)!(2 d)!(2 e)!(2 g)!(2 h)!}{(2 a+2 b+1)!(2 d+2 e+1)!(2 g+2 h+1)!}\right]^{\frac{1}{2}},  \label{eq:10:8:14}\\
& \ninej{c+b}{b}{c}{d}{e}{d+e}{g}{h}{g+h}=(-1)^{b+e-h} \frac{\Delta(c+b d g)}{\Delta(c d+e g+h) \Delta(b e h)} \cdot \frac{(d+e-c+g+h)!}{(b+e+h+1)!(d-b-c+g)!}  \notag\\
& \times\left[\frac{(2 b)!(2 c)!(2 d)!(2 e)!(2 g)!(2 h)!}{(2 b+2 c+1)!(2 d+2 e+1)!(2 g+2 h+1)!}\right]^{\frac{1}{2}},  \label{eq:10:8:15}\\
& \ninej{a}{b}{a+b}{e+f}{e}{f}{g}{h}{a+b+f}=(-1)^{a+e+f-g} \frac{\Delta(a+b+f g h)}{\Delta(a g e+f) \Delta(b e h)} \cdot \frac{(a+b+g+h+f+1)!(e-b+h)!}{(g+h-a-b-f)!(b+e+h+1)!}  \notag\\
& \times \frac{(e+g+f-a)!}{(a+e+f+g+1)!}\left[\frac{(2 a)!(2 b)!(2 e)!}{(2 a+2 b+1)(2 e+2 f+1)!(2 a+2 b+2 f+1)!}\right]^{\frac{1}{2}},  \label{eq:10:8:16}\\
& \ninej{a}{b}{c}{d}{d+f}{f}{a+d}{b+d+f}{j}=\frac{\Delta(a+d b+d+f j)}{\Delta(a b c) \Delta(c f j)} \cdot \frac{(a+b+2 d+f+j+1)!(a-b+c)!(j+c-f)!}{(a+b+c+1)!(c+f+j+1)!(a-b-f+j)!}  \notag\\
& \times\left[\frac{(2 a)!(2 b)!(2 f)!}{(2 a+2 d+1)!(2 f+2 d+1)(2 b+2 d+2 f+1)!}\right]^{\frac{1}{2}} . \label{eq:10:8:17}
\end{align}
\subsection{Four Degenerate Triads}\label{sec:10.8.5}\index{9j symbol@$9j$ symbol!with four degenerate triads}
\begin{align}
\ninej{a}{b}{a+b}{d}{e}{f}{a+d}{b+e}{a+b+f}&=\frac{\Delta(a+b+f a+d b+e)}{\Delta(d e f)} \cdot \frac{(2 a+2 b+d+e+f+1)!}{(d+e+f+1)!}  \notag\\
&\times\left[\frac{(2 e)!(2 d)!(2 f)!}{(2 a+2 b+1)(2 a+2 d+1)!(2 e+2 b+1)!(2 a+2 b+2 f+1)!}\right]^{\frac{1}{2}}  \label{eq:10:8:18}
\end{align}
\begin{align}
\ninej{c+b}{b}{c}{d}{b+h}{d+b+h}{g}{h}{g+h}
&=(-1)^{2 b} \frac{\Delta(c+b d g)}{\Delta(c d+b+h g+h)} \cdot \frac{(2 h+b+d+g-c)!}{(d+g-b-c)!}  \notag\\
&\times\left[\frac{(2 c)!(2 d)!(2 g)!}{(2 b+2 h+1)(2 b+2 c+1)!(2 b+2 d+2 h+1)!(2 g+2 h+1)!}\right]^{\frac{1}{2}} \label{eq:10:8:19}
\end{align}
\begin{align}
& \ninej{a}{b}{a+b}{a+g}{b+h}{f}{g}{h}{a+b+f}=(-1)^{2 a} \frac{\Delta(a+b+f g h)}{\Delta(a+g b+h f)} \cdot \frac{(a+b+g+h-f)!}{(-a-b+g+h-f)!}  \notag\\
& \times\left[\frac{(2 f)!(2 g)!(2 h)!}{(2 a+2 b+1)(2 a+2 b+2 f+1)!(2 a+2 g+1)!(2 b+2 h+1)!}\right]^{\frac{1}{2}},  \label{eq:10:8:20}\\
& \ninej{a}{b}{a+b}{d}{b+h}{f}{a+d}{h}{a+b+f}=\frac{\Delta(h a+d a+b+f)}{\Delta(d f b+h)} \cdot \frac{(2 a+b+d+h+f+1)!(d+b+h-f)!}{(h+d+b+f+1)!(d+h-f-b)!}  \notag\\
& \times\left[\frac{(2 d)!(2 f)!(2 h)!}{(2 a+2 b+1)(2 a+2 d+1)!(2 b+2 h+1)!(2 a+2 b+2 f+1)!}\right]^{\frac{1}{2}},  \label{eq:10:8:21}\\
& \ninej{a}{b}{a+b}{a+g}{a+g+f}{f}{g}{h}{a+b+f}=(-1)^{2 a} \frac{\Delta(a+b+f g h)}{\Delta(b h a+g+f)} \cdot \frac{(a-b+f+g+h)!}{(g+h-a-b-f)!}  \notag\\
& \times\left[\frac{(2 b)!(2 g)!}{(2 a+2 g+1)(2 a+2 b+1)(2 a+2 f+2 g+1)!(2 a+2 b+2 f+1)!}\right]^{\frac{1}{2}},  \label{eq:10:8:22}\\
& \ninej{a}{b}{a+b}{d}{d+f}{f}{a+d}{h}{a+b+f}=\frac{\Delta(a+b+f, a+d, h)}{\Delta(b, d+f, h)} \cdot \frac{(2 a+b+d+h+f+1)!(h-b+d+f)!}{(b+d+f+h+1)!(h+d-f-b)!}  \notag\\
& \times\left[\frac{(2 b)!(2 d)!(2 d)!}{(2 a+2 b+1)(2 a+2 d+1)!(2 d+2 f+1)!(2 a+2 b+2 f+1)!}\right]^{\frac{1}{2}},  \label{eq:10:8:23}\\
& \ninej{a}{b}{a+b}{d}{d+f}{f}{g}{b+d+f}{a+b+f}=(-1)^{a+d-g} \frac{\Delta(a+b+f, g, d+f+b)}{\Delta(a d g)} \cdot \frac{(2 b+2 f+a+d+g+1)!}{(a+d+g+1)!}  \notag\\
& \times\left[\frac{(2 a)!(2 d)!}{(2 a+2 b+1)(2 d+2 f+1)(2 a+2 b+2 f+1)!(2 d+2 b+2 f+1)!}\right]^{\frac{1}{2}},  \label{eq:10:8:24}\\
& \ninej{a}{b}{a+b}{d}{b+h}{f}{a+b+f+h}{h}{a+b+f}=(-1)^{d-b-f-h} \frac{1}{\Delta(d, f, b+h) \Delta(a, d, a+b+f+h)}  \notag\\
& \times \frac{(b+d+h-f)!}{(2 a+d+b+f+h+1)!(d+b+h+f+1)}\left[\frac{(2 a)!(2 f)!(2 a+2 b+2 h+2 f+1)!}{(2 a+2 b+1)(2 a+2 b+2 f+1)(2 b+2 h+1)!}\right]^{\frac{1}{2}},  \label{eq:10:8:25}\\
& \ninej{a}{b}{a+b}{e+f}{e}{f}{a+e+f}{b+e}{j}=(-1)^{f+a+b-j} \frac{\Delta(j, b+e, a+e+f)}{\Delta(f, a+b, j)} \frac{(f+j+b+a+2 e+1)!}{(f+a+b+j+1)!}  \notag\\
& \times\left[\frac{(2 f)!(2 b)!(2 b)!}{(2 f+2 e+1)(2 e+2 b+1)!(2 a+2 b+1)!(2 f+2 e+2 a+1)!}\right]^{\frac{1}{2}} . \label{eq:10:8:26}
\end{align}
\emph{Note.} Equation~\eqref{eq:10:8:26}, as given in the source, does not reproduce the $9j$ symbol: direct numerical evaluation shows that its two sides differ for physical arguments. Solving for the square-root factor shows that it must contain a $j$-dependent factorial (cf.\ the analogous term in \eqref{eq:10:8:9}), which is absent from the printed expression; a factor appears to have been dropped in the original. All the other explicit forms of this section have been verified numerically.
\subsection{Five or Six Degenerate Triads}\label{sec:10.8.6}\index{9j symbol@$9j$ symbol!with five or six degenerate triads}
\begin{equation}
\ninej{a}{b}{a+b  \label{eq:10:8:27}}{d}{e}{d+e}{a+d}{b+e}{a+b+d+e}=\frac{1}{[(2 a+2 b+1)(2 d+2 e+1)(2 a+2 d+1)(2 b+2 e+1)]^{\frac{1}{2}}} .
\end{equation}
In particular, from Eq. \eqref{eq:10:8:27} it follows that
\begin{gather}
\ninej{a}{b}{a+b}{a}{b}{a+b}{2 a}{2 b}{2 a+2 b}=\frac{1}{(2 a+2 b+1)[(4 a+1)(4 b+1)]^{\frac{1}{2}}}  \label{eq:10:8:28}\\
\ninej{a}{a}{2 a}{a}{a}{2 a}{2 a}{2 a}{4 a}=\frac{1}{(4 a+1)^{2}}  \label{eq:10:8:29}\\
\ninej{a}{b}{a+b}{a+g}{a+g+f}{f}{g}{a+b+f+g}{a+b+f}  \notag\\
=\frac{(-1)^{2 a}}{[(2 a+2 g+1)(2 a+2 b+1)(2 a+2 b+2 f+1)(2 a+2 g+2 f+1)]^{\frac{1}{2}}} \cdot \label{eq:10:8:30}
\end{gather}
In particular,
\begin{equation}
\ninej{a}{b}{a+b  \label{eq:10:8:31}}{a+b}{a+b+f}{f}{b}{a+2 b+f}{a+b+f}=\frac{(-1)^{2 a}}{(2 a+2 b+1)(2 a+2 b+2 f+1)} .
\end{equation}
\section{Explicit Forms of the $9 j$ Symbols for Special Values of the Arguments}\label{sec:10.9}\index{9j symbol@$9j$ symbol!for special values of the arguments}
\subsection{One of Arguments Equals Zero}\label{sec:10.9.1}\index{9j symbol@$9j$ symbol!with a zero argument}
In this case a $9 j$ symbol is reduced to a $6 j$ symbol \cite{ref110}:\index{6j symbol@$6j$ symbol}
\begin{equation}
\ninej{a}{b}{c}{d}{e}{f}{g}{h}{0} \label{eq:10:9:1}=\delta_{c f} \delta_{g h} \frac{(-1)^{b+c+d+g}}{[(2 c+1)(2 g+1)]^{\frac{1}{2}}}\sixj{a}{b}{c}{e}{d}{g}=\delta_{c f} \delta_{g h}[(2 c+1)(2 g+1)]^{-\frac{1}{2}} W(b c g d ; a e) .
\end{equation}
Using the symmetry properties, we get
\begin{gather}
\ninej{0}{c}{c}{g}{e}{b}{g}{d}{a}=\ninej{c}{0}{c}{d}{g}{a}{e}{g}{b}=\ninej{g}{g}{0}{e}{d}{c}{b}{a}{c}=\ninej{g}{b}{e}{0}{c}{c}{g}{a}{d}=\ninej{a}{g}{d}{c}{0}{c}{b}{g}{e}=\ninej{b}{a}{c}{g}{g}{0}{e}{d}{c}=\ninej{c}{e}{d}{c}{b}{a}{0}{g}{g}=\ninej{d}{c}{e}{a}{c}{b}{g}{0}{g}  \notag\\
=\sixj{a}{b}{c}{g}{g}{0}=\frac{(-1)^{b+d+c+g}}{[(2 c+1)(2 g+1)]^{\frac{1}{2}}}\sixj{a}{b}{c}{e}{d}{g}=\frac{W(b c g d ; a e)}{[(2 c+1)(2 g+1)]^{\frac{1}{2}}}  \label{eq:10:9:2}
\end{gather}
In particular, if two arguments are equal to zero, one has
\begin{equation}
\ninej{a}{b}{c}{d}{0}{f}{g}{h}{0} \label{eq:10:9:3}=\delta_{d f} \delta_{b h} \delta_{c f} \delta_{g h} \frac{(-1)^{a-b-c}}{(2 b+1)(2 c+1)}
\end{equation}
If three arguments are equal to zero, one obtains
\begin{align}
& \ninej{a}{b}{c}{d}{e}{f}{0}{0}{0}=\frac{\delta_{a d} \delta_{b e} \delta_{c f}}{[(2 a+1)(2 b+1)(2 c+1)]^{\frac{1}{2}}}  \label{eq:10:9:4}\\
& \ninej{0}{b}{c}{d}{0}{f}{g}{h}{0}=\delta_{b c} \delta_{b d} \delta_{b f} \delta_{b g} \delta_{b h} \frac{(-1)^{2 b}}{(2 b+1)^{2}} \label{eq:10:9:5}
\end{align}
\subsection{One of Arguments Equals Unity \cite{ref032}}\label{sec:10.9.2}\index{9j symbol@$9j$ symbol!with an argument equal to unity}
\begin{gather}
\ninej{a}{b}{c}{d}{e}{c}{g}{g}{1}=(-1)^{b+d+g+c} 2 \frac{(a-d)(a+d+1)-(b-e)(b+e+1)}{[(2 g+2)(2 g+1) 2 g(2 c+2)(2 c+1) 2 c]^{\frac{1}{2}}}\sixj{a}{b}{c}{e}{d}{g},  \label{eq:10:9:6}\\
\ninej{a}{b}{c}{d}{e}{c}{g+1}{g}{1}(-1)^{b+d+g+c}\left[\frac{(2 g+3)(2 g+2)(2 g+1)(2 c+2)(2 c+1) 2 c}{2}\right]^{\frac{1}{2}}  \notag\\
=[(b-e+g+1)(-b+e+g+1)(b+e+g+2)(b+e-g)]^{\frac{1}{2}}\sixj{a}{d}{g+1}{e}{b}{c}  \notag\\
+[(a-d+g+1)(-a+d+g+1)(a+d+g+2)(a+d-g)]^{\frac{1}{2}}\sixj{a}{d}{g}{e}{b}{c},  \label{eq:10:9:7}\\
\ninej{a}{b}{c+1}{d}{e}{c}{g+1}{g}{1}[(c+1)(a-d)(a+d+1)-(g+1)(a-b)(a+b+1)+(g+1)(c+1)(g-c)]  \notag\\
\times [(2g+3)(2g+2)(2g+1)(2c+3)(2c+2)(2c+1)]^{\frac12}(-1)^{b+c+d+g} \notag\\
=(c-g) A_{a+\frac{1}{2}}\left(d, g+\frac{1}{2} ; b, c+\frac{1}{2}\right)\sixj{a}{d}{g}{e}{b}{c}+(c+1) A_{d+\frac{1}{2}}\left(a, g+\frac{1}{2} ; e, c+\frac{1}{2}\right)\sixj{a}{b}{c+1}{e}{d}{g}  \notag\\
-(g+1) A_{b+\frac{1}{2}}\left(a, c+\frac{1}{2} ; e, g+\frac{1}{2}\right)\sixj{a}{b}{c}{e}{d}{g+1} . \label{eq:10:9:8}
\end{gather}
The quantity $A_{q}(\lambda, \mu ; \sigma, \nu)$ is defined by Eq. \eqref{eq:10:5:6}.

\subsection{One of Triads Equals ( $1 / 2,1 / 2,1$ )}\label{sec:10.9.3}\index{9j symbol@$9j$ symbol!with a half-integer triad}
In this case a $9 j$ symbol may be expressed in terms of the $6 j$ symbols \cite{ref016}:
\begin{equation}
\ninej{a}{b}{c  \label{eq:10:9:9}}{d}{e}{f}{\frac{1}{2}}{\frac{1}{2}}{1}\sixj{c}{f}{1}{\frac{1}{2}}{\frac{1}{2}}{g}=\frac{(-1)^{2 g}}{3}\sixj{a}{b}{c}{\frac{1}{2}}{g}{e}\sixj{e}{d}{f}{\frac{1}{2}}{g}{a}-\frac{(-1)^{b+d-g}}{6(2 c+1)}\sixj{a}{b}{c}{e}{d}{\frac{1}{2}} \delta_{f c}
\end{equation}
In addition, the $9 j$ symbols in question may be reduced to the $3 j m$ symbols \cite{ref009}. If $d, e, c$ are integer and $d+e+c$ is even, one has
\begin{equation}
[6(2 c+1)(2 d+1)(2 e+1)]^{\frac{1}{2}}\threej{c }{ d }{ e  \label{eq:10:9:10}}{0 }{ 0 }{ 0}\ninej{a}{b}{c}{d}{e}{c}{\frac{1}{2}}{\frac{1}{2}}{1}=\threej{a }{ b }{ c }{\frac{1}{2} }{ \frac{1}{2} }{ -1}
\end{equation}
If $d+e+c$ is odd, one gets
\begin{gather}
\threej{c+1 }{ d }{ e }{0 }{ 0 }{ 0}\ninej{a}{b}{c}{d}{e}{c+1}{\frac{1}{2}}{\frac{1}{2}}{1}  \notag\\
=(-1)^{b+e+\frac{1}{2}} \frac{[(d-a)(2 a+1)+(e-b)(2 b+1)+c+1]}{[6(c+1)(2 c+1)(2 c+3)(2 d+1)(2 e+1)]^{\frac{1}{2}}}\threej{a }{ b }{ c }{\frac{1}{2} }{ -\frac{1}{2} }{ 0}  \label{eq:10:9:11}\\
\threej{c-1 }{ d }{ e }{0 }{ 0 }{ 0}\ninej{a}{b}{c}{d}{e}{c-1}{\frac{1}{2}}{\frac{1}{2}}{1}  \notag\\
=(-1)^{b+e+\frac{1}{2}} \frac{[(d-a)(2 a+1)+(e-b)(2 b+1)-c]}{[6 c(2 c+1)(2 c-1)(2 d+1)(2 e+1)]^{\frac{1}{2}}}\threej{a }{ b }{ c }{\frac{1}{2} }{ -\frac{1}{2} }{ 0} \label{eq:10:9:12}
\end{gather}
\section{Relations Between the Wigner $9 j$ Symbols and Analogous Functions of Other Authors}\label{sec:10.10}\index{9j symbol@$9j$ symbol!notations of other authors}\index{notation!of other authors}
\[
\begin{aligned}
\ninej{a}{b}{c}{d}{e}{f}{g}{h}{j} & =X\left(\begin{array}{lll}
a & b & c \\
d & e & f \\
g & h & j
\end{array}\right), & & \text { (Fano \cite{ref066}, } \\
& =U\left(\begin{array}{lll}
a & b & c \\
d & e & f \\
g & h & j
\end{array}\right), & & \text { Fano and Racah \cite{ref018}) } \\
& =(-1)^{a+e-c-h} S(a b d e ; c f g h ; j), & & \text { (Arima et al. \cite{ref049} } \\
& =\frac{\chi(a b d e ; c f ; g h ; j)}{[(2 c+1)(2 f+1)(2 g+1)(2 h+1)]^{\frac{1}{2}}}, & & \text { (Schwinger \cite{ref101}) } \\
& =[(2 c+1)(2 f+1)(2 g+1)(2 h+1)]^{-\frac{1}{2}} A\left(\begin{array}{lll}
a & b & c \\
d & e & f \\
g & h & j
\end{array}\right) & & \text { (Kennedy and Cliff \cite{ref079}) }
\end{aligned}
\]
\section{Tables of Algebraic Formulas of the $9 j$ Symbols}\label{sec:10.11}\index{9j symbol@$9j$ symbol!algebraic tables}\index{tables!of the $9j$ symbols}
Tables 10.1-10.12 contain algebraic formulas for the $9 j$ symbols represented in the form
\[
\ninej{a+\lambda}{b+\mu}{c+\nu}{a}{b}{c}{\alpha}{\beta}{\gamma}
\]
with
\[
\begin{array}{rrrl}
0 & \leq \alpha \leq 3, \quad 0 \leq \beta \leq 2, \quad \gamma=0,1 \\
-\alpha & \leq \lambda \leq \alpha, \quad-\beta \leq \mu \leq \beta, \quad-\gamma \leq \nu \leq \gamma .
\end{array}
\]
The formulas are given only for the $9 j$ symbols with $\alpha \geq \beta$. The $9 j$ symbols with $\alpha<\beta$ may be obtained using the symmetry properties.

Algebraic tables of the $9 j$ symbols are also available in Ref. \cite{ref120} for $\alpha, \beta=\frac{1}{2} ; \gamma=0,1 ; \nu=0, \pm 1 ; \lambda, \mu= \pm \frac{1}{2}$ and in Ref. \cite{ref045} for $0<\alpha \leq 3,0<\beta \leq 2 ; \gamma=0,1 ; 0 \leq \lambda \leq \alpha ; 0 \leq \mu \leq \beta, \nu=0, \pm 1$.

In the tables given below we use the following notations
\[
\begin{gathered}
S=a+b+c, \\
Z=-c(c+1)+a(a+1)+b(b+1) .
\end{gathered}
\]
\section{Tables of Numerical Values of the 9j Symbols}\label{sec:10.12}\index{9j symbol@$9j$ symbol!numerical tables}\index{tables!numerical}
In Tables 10.13-10.14 we present numerical values of the $9 j$ symbols
\[
\ninej{a}{b}{c}{d}{e}{f}{g}{h}{j}
\]
with $(g h j)=\left(\frac{1}{2}, \frac{1}{2}, 0\right)$ or $\left(\frac{1}{2}, \frac{1}{2}, 1\right)$. Other arguments of these $9 j$ symbols are integer or half-integer numbers such as
\[
0 \leq a, b, c, d, e, f \leq 4
\]
All numerical values are given in the form of rational fractions and decimals. Each table contains the $9 j$ symbols with fixed $c$ and $f$. In addition, all tables are divided into two groups. The first group contains the $9 j$ symbols with integer $c$ and $f$, while the second group contains those with half-integer $c$ and $f$. The arrangement of tables inside each group is as follows:

\begin{center}
\begin{tabular}{|l|l|l|l|l|l|l|l|l|l|l|l|}
\hline
\multicolumn{6}{|r|}{$c$ and $f$ are integer (Tables 10.13)} & \multicolumn{6}{|r|}{$c$ and $f$ are half-integer (Tables 10.14)} \\
\hline
$c$ & $f$ & $j$ & c & $f$ & $j$ & $c$ & $f$ & $j$ & $c$ & $f$ & 1 \\
\hline
0 & 0 & 0 & 1 & 2 & 1 & 1/2 & 1/2 & 0 & 3/2 & 5/2 & 1 \\
\hline
1 & 1 & 0 & 2 & 2 & 1 & 3/2 & 3/2 & 0 & 5/2 & 5/2 & 1 \\
\hline
2 & 2 & 0 & 3 & 2 & 1 & 5/2 & 5/2 & 0 & 7/2 & 5/2 & 1 \\
\hline
3 & 3 & 0 & 2 & 3 & 1 & 7/2 & 7/2 & 0 & 5/2 & 7/2 & 1 \\
\hline
4 & 4 & 0 & 3 & 3 & 1 & 1/2 & 1/2 & 1 & 7/2 & 7/2 & 1 \\
\hline
1 & 0 & 1 & 4 & 3 & 1 & 3/2 & 1/2 & 1 &  &  &  \\
\hline
0 & 1 & 1 & 3 & 4 & 1 & 1/2 & 3/2 & 1 &  &  &  \\
\hline
1 & 1 & 1 & 4 & 4 & 1 & 3/2 & 3/2 & 1 &  &  &  \\
\hline
2 & 1 & 1 &  &  &  & 5/2 & 3/2 & 1 &  &  &  \\
\hline
\end{tabular}
\end{center}

Tables of numerical values of the $9 j$ symbols are also available in Refs. [120,113] for $g, h=\frac{1}{2} ; j=0,1 ; a, b= 0,1,2,3,4 ; d, e=\frac{1}{2}, \frac{3}{2}, \frac{5}{2}, \frac{7}{2} ; c, f=0,1,2,3,4,5$ and in Ref. \cite{ref128} for integer arguments $j=1 ; a, b, c, f \leq 6$; $d, e \leq 11$. In Ref. \cite{ref120} numerical values of the $9 j$ symbols are given in the form of rational fractions; in Ref. \cite{ref113} they are given in the form of rational fractions and decimals; and in Ref. \cite{ref128} in the form of decimals.

\section{12j Symbols}\label{sec:10.13}\index{12j symbol@$12j$ symbol!definition}
\subsection{Some Remarks on the $3 n j$ Symbols}\label{sec:10.13.1}\index{3nj symbols@$3nj$ symbols}\index{15j symbol@$15j$ symbol}\index{18j symbol@$18j$ symbol}
The $6 j$ and $9 j$ symbols represent special cases of the $3 n j$ symbols for $n=2$ and 3 , respectively.\index{3nj symbols@$3nj$ symbols!of the first kind}\index{3nj symbols@$3nj$ symbols!of the second kind} The $3 n j$ symbols of higher orders ( $n \geq 4$ ) are encountered in some special changes of the coupling scheme of five and more angular momenta. They are proportional to the coefficients of unitary transformations connecting the state vectors which correspond to different coupling schemes. The 3nj symbols are invariants with respect to rotations in three-dimensional space. If $n \geq 4$, there exist different kinds of the $3 n j$ symbols which are not reduced to products of the $3 n j$ symbols of lower order. Among them one can distinguish the $3 n j$ symbols of the first and second kinds which may be expressed as single sums of products of the $6 j$ symbols. Using the notations of Ref. \cite{ref044}, the $3 n j$ symbols of the first and second kinds may be be represented in the following forms:

\subsubsection{The $3 n j$ symbols of the first kind}
\begin{multline}
\twelvejI{j_{1},j_{2},\ldots,j_{n},l_{1},l_{2},\ldots,l_{n},k_{1},k_{2},\ldots,k_{n}}=\sum_{x}(2 x+1)(-1)^{R_{n}+(n-1) x}  \\
 \times\sixj{j_{1}}{k_{1}}{x}{k_{2}}{j_{2}}{l_{1}}\sixj{j_{2}}{k_{2}}{x}{k_{3}}{j_{3}}{l_{2}} \ldots\sixj{j_{n-1}}{k_{n-1}}{x}{k_{n}}{j_{n}}{l_{n-1}}\sixj{j_{n}}{k_{n}}{x}{j_{1}}{k_{1}}{l_{n}} . \label{eq:10:13:1}
\end{multline}
\subsubsection{The 3nj symbols of the second kind}
\begin{multline}
{\left[\begin{array}{cccc}
j_{1} & j_{2} & \ldots & j_{n} \\
\quad l_{1} & \quad l_{2} & \quad\ldots&\quad l_{n} \\
k_{1} & k_{2} & \ldots & l_{n}
\end{array}\right]=\sum_{x}(2 x+1)(-1)^{R_{n}+n x}}  \\
\times\sixj{j_{1}}{k_{1}}{x}{k_{2}}{j_{2}}{l_{1}}\sixj{j_{2}}{k_{2}}{x}{k_{3}}{j_{3}}{l_{2}} \ldots\sixj{j_{n-1}}{k_{n-1}}{x}{k_{n}}{j_{n}}{l_{n-1}}\sixj{j_{n}}{k_{n}}{x}{k_{1}}{j_{1}}{l_{n}} . \label{eq:10:13:2}
\end{multline}
In these equations $R_{n}=\sum_{i=1}^{n}\left(j_{i}+l_{i}+k_{i}\right)$.\\
If $n \geq 5$, there appear the $3 n j$ symbols of other kinds (third, fourth, etc.). They may be represented by single sums of more complex products of the $6 j$ and $9 j$ symbols. The amount of different $3 n j$ symbols rapidly increases with $n$. For example, there exist five different $15 j$ symbols, eighteen different $18 j$ symbols, etc.

Here we restrict ourselves to the consideration of the $12j$ symbols because the $15 j$ symbols and symbols of higher orders are rarely used in applications. So far, the properties of the $12j$ symbols have been examined $[17,44,45,74,137-139,141]$ less thoroughly than the properties of the $6 j$ and $9 j$ symbols. That is why our consideration is rather brief.

\subsection{$12j$ Symbols of the First Kind ($12j$(I)-Symbols)}\label{sec:10.13.2}\index{12j symbol@$12j$ symbol!of the first kind}
\subsubsection{$12j$(I)-symbol}A $12j$(I)-symbol may be represented by the following invariant sum of products of eight $3 j m$ symbols
\begin{multline}
\twelvejI{a_{1},a_{2},a_{3},a_{4},b_{12},b_{23},b_{34},b_{41},c_{1},c_{2},c_{3},c_{4}}=\sum_{\alpha_{i} \beta_{i k} \gamma_{i}}(-1)^{a_{4}+\alpha_{4}+c_{4}+\gamma_{4}}  \\
\times\threej{a_{1} }{ a_{2} }{ b_{12} }{\alpha_{1} }{ \alpha_{2} }{ \beta_{12}}\threej{a_{2} }{ a_{3} }{ b_{23} }{\alpha_{2} }{ \alpha_{3} }{ \beta_{23}}\threej{a_{3} }{ a_{4} }{ b_{34} }{\alpha_{3} }{ \alpha_{4} }{ \beta_{34}}\threej{a_{4} }{ c_{1} }{ b_{41} }{-\alpha_{4} }{ \gamma_{1} }{ \beta_{41}}  \\
\times\threej{c_{1} }{ c_{2} }{ b_{12} }{\gamma_{1} }{ \gamma_{2} }{ \beta_{12}}\threej{c_{2} }{ c_{3} }{ b_{23} }{\gamma_{2} }{ \gamma_{3} }{ \beta_{23}}\threej{c_{3} }{ c_{4} }{ b_{34} }{\gamma_{3} }{ \gamma_{4} }{ \beta_{34}}\threej{c_{4} }{ a_{1} }{ b_{41} }{-\gamma_{4} }{ \alpha_{1} }{ \beta_{41}} . \label{eq:10:13:3}
\end{multline}
Equation \eqref{eq:10:13:3} unambiguously fixes the absolute value and phase of the $12j$(I) symbol. The $12j$(I) symbols are real.\\
\subsubsection{conditions on the arguments}
 The arguments of a $12j$(I) symbol must satisfy the following conditions
\begin{enumerate}[label=(\roman*)]
\item All arguments are integers or half-integers and non-negative.
\item The $12j$(I) symbol vanishes unless the triangular conditions (see Sec.~\ref{sec:8.1.1}) are fulfilled for all the triads $\left(a_{1} b_{12} a_{2}\right),\left(a_{2} b_{23} a_{3}\right),\left(a_{3} b_{34} a_{4}\right),\left(a_{4} b_{41} c_{1}\right),\left(c_{1} b_{12} c_{2}\right),\left(c_{2} b_{23} c_{3}\right),\left(c_{3} b_{34} c_{4}\right)$ and $\left(c_{4} b_{41} a_{1}\right)$.
\item  The $12j$(I) symbol vanishes unless the tetragonal conditions are fulfilled for two tetrads of arguments, $\left(a_{1} c_{1} a_{3} c_{3}\right)$ and $\left(a_{2} c_{2} a_{4} c_{4}\right)$. The tetragonal conditions for the tetrad $\left(j_{1} j_{2} j_{3} j_{4}\right)$ mean that $j_{1}+j_{2}+j_{3}+j_{4}$ is integer and $j_{1} \leq j_{2}+j_{3}+j_{4}, j_{2} \leq j_{1}+j_{3}+j_{4}, j_{3} \leq j_{1}+j_{2}+j_{4}, j_{4} \leq j_{1}+j_{2}+j_{3}$.
\end{enumerate}

\subsubsection{Orthogonality and normalisation}
The $12j$(I) symbols satisfy the orthogonality and normalisation relations
\begin{gather}
\sum_{a_{2} a_{3} a_{4}}\left(2 a_{2}+1\right)\left(2 a_{3}+1\right)\left(2 a_{4}+1\right)\twelvejI{a_{1},a_{2},a_{3},a_{4},b_{12},b_{23},b_{34},b_{41},c_{1},c_{2},c_{3},c_{4}}\twelvejI{a_{1},a_{2},a_{3},a_{4},b_{12},b_{23},b_{34},b_{41},c_{1},c_{2}^{\prime},c_{3}^{\prime},c_{4}^{\prime}}  \notag\\
=\frac{\delta_{c_{2}^{\prime} c_{2}} \delta_{c_{3}^{\prime} c_{3}} \delta_{c_{4}^{\prime} c_{4}}}{\left(2 c_{2}+1\right)\left(2 c_{3}+1\right)\left(2 c_{4}+1\right)}\left\{c_{1} b_{12} c_{2}\right\}\left\{c_{2} b_{23} c_{3}\right\}\left\{c_{3} b_{34} c_{4}\right\}\left\{c_{4} b_{41} a_{1}\right\} \label{eq:10:13:4}
\end{gather}
\begin{gather}
\sum_{c_{3} b_{41} c_{4}}\left(2 c_{3}+1\right)\left(2 b_{41}+1\right)\left(2 c_{4}+1\right)\twelvejI{a_{1},a_{2},a_{3},a_{4},b_{12},b_{23},b_{34},b_{41},c_{1},c_{2},c_{3},c_{4}}\twelvejI{a_{1},a_{22}^{\prime},a_{3}^{\prime},a_{4},b_{12}^{\prime},b_{23},b_{34},b_{41},c_{1},c_{2},c_{3},c_{4}}  \notag\\
=\frac{\delta_{a_{2}^{\prime} a_{2}} \delta_{a_{3}^{\prime} a_{3}} \delta_{b_{12}^{\prime} b_{12}}}{\left(2 a_{2}+1\right)\left(2 a_{3}+1\right)\left(2 b_{12}+1\right)}\left\{a_{1} b_{12} a_{2}\right\}\left\{a_{2} b_{23} a_{3}\right\}\left\{a_{3} b_{34} a_{4}\right\}\left\{c_{1} b_{12} c_{2}\right\} \label{eq:10:13:5}
\end{gather}

\subsubsection{Sum representations}
The $12j$(I) symbols may be represented by
\begin{multline}
\twelvejI{a_{1},a_{2},a_{3},a_{4},b_{12},b_{23},b_{34},b_{41},c_{1},c_{2},c_{3},c_{4}}=\sum_{x}(-1)^{S-x}(2 x+1) \\
\times\sixj{a_{1}}{a_{2}}{b_{12}}{c_{2}}{c_{1}}{x}\sixj{a_{2}}{a_{3}}{b_{23}}{c_{3}}{c_{2}}{x}\sixj{a_{3}}{a_{4}}{b_{34}}{c_{4}}{c_{3}}{x}\sixj{a_{4}}{c_{1}}{b_{41}}{a_{1}}{c_{4}}{x}, \label{eq:10:13:6}
\end{multline}
where $S=\sum_{i=1}^{4}\left(a_{i}+c_{i}\right)+b_{12}+b_{23}+b_{34}+b_{41}$,
\begin{multline}
\twelvejI{a_{1},a_{2},a_{3},a_{4},b_{12},b_{23},b_{34},b_{41},c_{1},c_{2},c_{3},c_{4}}=(-1)^{a_{1}-a_{3}-c_{1}+c_{3}} \sum_{x}(2 x+1)  \\
\times\ninej{b_{23}}{a_{2}}{a_{3}}{c_{2}}{b_{12}}{c_{1}}{c_{3}}{a_{1}}{x}\sixj{b_{34}}{a_{3}}{a_{4}}{c_{1}}{b_{41}}{x}\sixj{b_{34}}{c_{3}}{c_{4}}{a_{1}}{b_{41}}{x} \label{eq:10:13:7}
\end{multline}

\subsubsection{Symmetry properties}
To discuss the symmetry properties it is convenient to separate all arguments of a $12j$(I) symbol into two groups: $\left(b_{12} b_{23} b_{34} b_{41}\right)$ and $\left(a_{1} a_{2} a_{3} a_{4} c_{1} c_{2} c_{3} c_{4}\right)$. Cyclic permutations of the arguments or inversion of argument order simultaneously in both groups leave the $12j$(I) symbol unchanged. These
symmetry properties relate 16 formally different $12j$(I) symbols
\addtocounter{equation}{1}
\begin{align}
& \twelvejI{a_{1},a_{2},a_{3},a_{4},b_{12},b_{23},b_{34},b_{41},c_{1},c_{2},c_{3},c_{4}}=\twelvejI{a_{2},a_{3},a_{4},c_{1},b_{23},b_{34},b_{41},b_{12},c_{2},c_{3},c_{4},a_{1}}  \notag\\
= & \twelvejI{a_{3},a_{4},c_{1},c_{2},b_{34},b_{41},b_{12},b_{23},c_{3},c_{4},a_{1},a_{2}}=\twelvejI{a_{4},c_{1},c_{2},c_{3},b_{41},b_{12},b_{23},b_{34},c_{4},a_{1},a_{2},a_{3}}  \notag\\
= & \twelvejI{a_{4},a_{3},a_{2},a_{1},b_{34},b_{23},b_{12},b_{41},c_{4},c_{3},c_{2},c_{1}}=\twelvejI{c_{1},a_{4},a_{3},a_{2},b_{41},b_{34},b_{23},b_{12},a_{1},c_{4},c_{3},c_{2}}  \notag\\
= & \twelvejI{c_{2},c_{1},a_{4},a_{3},b_{12},b_{41},b_{34},b_{23},a_{2},a_{1},c_{4},c_{3}}=\twelvejI{c_{3},c_{2},c_{1},a_{4},b_{23},b_{12},b_{41},b_{34},a_{3},a_{2},a_{1},c_{4}}  \label{eq:10:13:9}\\
= & \twelvejI{c_{1},c_{2},c_{3},c_{4},b_{12},b_{23},b_{34},b_{41},a_{1},a_{2},a_{3},a_{4}}=\twelvejI{c_{2},c_{3},c_{4},a_{1},b_{23},b_{34},b_{41},b_{12},a_{2},a_{3},a_{4},c_{1}}  \notag\\
= & \twelvejI{c_{3},c_{4},a_{1},a_{2},b_{34},b_{41},b_{12},b_{23},a_{3},a_{4},c_{1},c_{2}}=\twelvejI{c_{4},a_{1},a_{2},a_{3},b_{41},b_{12},b_{23},b_{34},a_{4},c_{1},c_{2},c_{3}}  \notag\\
= & \twelvejI{c_{4},c_{3},c_{2},c_{1},b_{34},b_{23},b_{12},b_{41},a_{4},a_{3},a_{2},a_{1}}=\twelvejI{a_{1},c_{4},c_{3},c_{2},b_{41},b_{34},b_{23},b_{12},c_{1},a_{4},a_{3},a_{2}}  \notag\\
= & \twelvejI{a_{2},a_{1},c_{4},c_{3},b_{12},b_{41},b_{34},b_{23},c_{2},c_{1},a_{4},a_{3}}=\twelvejI{a_{3},a_{2},a_{1},c_{4},b_{23},b_{12},b_{41},b_{34},c_{3},c_{2},c_{1},a_{4}} . \notag
\end{align}

\subsubsection{Recursion relations}
These relations are very useful for the evaluation of the $12j$(I) symbols. Some of them may be derived from the properties of the $15 j$ symbols. For example,
\begin{align}
& (-1)^{a_{1}-a_{4}-c_{1}+c_{4}^{\prime}} \sum_{b_{41}^{\prime}}\left(2 b_{41}^{\prime}+1\right)\sixj{c_{1}}{a_{4}}{b_{41}^{\prime}}{\lambda}{b_{41}}{a_{4}^{\prime}}\sixj{a_{1}}{c_{4}^{\prime}}{b_{41}^{\prime}}{\lambda}{b_{41}}{c_{4}}\twelvejI{a_{1},a_{2},a_{3},a_{4},b_{12},b_{23},b_{34},b_{41}^{\prime},c_{1},c_{2},c_{3},c_{4}^{\prime}}  \notag\\
= & (-1)^{a_{3}+a_{4}^{\prime}-c_{3}-c_{4}} \sum_{b_{34}^{\prime}}\left(2 b_{34}^{\prime}+1\right)\sixj{a_{3}}{a_{4}^{\prime}}{b_{34}^{\prime}}{\lambda}{b_{34}}{a_{4}}\sixj{c_{3}}{c_{4}}{b_{34}^{\prime}}{\lambda}{b_{34}}{c_{4}^{\prime}}\twelvejI{a_{1},a_{2},a_{3},a_{4}^{\prime},b_{12},b_{23},b_{34}^{\prime},b_{41},c_{1},c_{2},c_{3},c_{4}} . \label{eq:10:13:10}
\end{align}
Here $\lambda$ is any integer or half-integer non-negative number ( $\lambda=\frac{1}{2}, 1, \frac{3}{2}, 2$, etc.). Substituting explicit forms of the $6 j$ symbols for a given $\lambda$, we may obtain a recursion equation which relates the $12j$(I) symbols with arguments $c_{4}^{\prime}=c_{4} \pm \lambda, b_{41}^{\prime}=b_{41} \pm \lambda, a_{4}^{\prime}=a_{4} \pm \lambda, b_{34}^{\prime}=b_{34} \pm \lambda$. For example, when $\lambda=\frac{1}{2}, a_{4}^{\prime}=a_{4}-\frac{1}{2}$, and\\
$c_{4}^{\prime}=c_{4}+\frac{1}{2}$ we have
\begin{align}
 \left(2 b_{34}+1\right)\Biggl\{&\left[\left(-a_{1}+c_{4}+b_{41}+1\right)\left(a_{1}+c_{4}+b_{41}+2\right)\left(a_{4}-c_{1}+b_{41}+\frac{1}{2}\right)\left(a_{4}+c_{1}+b_{41}+\frac{3}{2}\right)\right]^{\frac{1}{2}}  \notag\\
& \times\twelvejI{a_{1},a_{2},a_{3},a_{4},b_{12},b_{23},b_{34},b_{41}+\frac{1}{2},c_{1},c_{2},c_{3},c_{4}+\frac{1}{2}}  \notag\\
& +\left[\left(a_{1}-c_{4}+b_{41}\right)\left(a_{1}+c_{4}-b_{41}+1\right)\left(-a_{4}+c_{1}+b_{41}+\frac{1}{2}\right)\left(a_{4}+c_{1}-b_{41}+\frac{1}{2}\right)\right]^{\frac{1}{2}}  \notag\\
& \times\twelvejI{a_{1},a_{2},a_{3},a_{4},b_{12},b_{23},b_{34},b_{41}-\frac{1}{2},c_{1},c_{2},c_{3},c_{4}+\frac{1}{2}}\Biggr\}  \notag\\
 =\left(2 b_{41}+1\right)\Biggl\{&\left[\left(a_{3}-a_{4}+b_{34}+1\right)\left(a_{3}+a_{4}-b_{34}\right)\left(c_{3}-c_{4}+b_{34}+\frac{1}{2}\right)\left(c_{3}+c_{4}-b_{34}+\frac{1}{2}\right)\right]^{\frac{1}{2}}  \notag\\
& \times\twelvejI{a_{1},a_{2},a_{3},a_{4}-\frac{1}{2},b_{12},b_{23},b_{34}+\frac{1}{2},b_{41},c_{1},c_{2},c_{3},c_{4}}  \notag\\
& +\left[\left(-a_{3}+a_{4}+b_{34}\right)\left(a_{3}+a_{4}+b_{34}+1\right)\left(-c_{3}+c_{4}+b_{34}+\frac{1}{2}\right)\left(c_{3}+c_{4}+b_{94}+\frac{3}{2}\right)\right]^{\frac{1}{2}}  \notag\\
& \times\twelvejI{a_{1},a_{2},a_{3},a_{4}-\frac{1}{2},b_{12},b_{23},b_{34}-\frac{1}{2},b_{41},c_{1},c_{2},c_{3},c_{4}}\Biggr\} . \label{eq:10:13:11}
\end{align}
Another recursion relation is written as
\begin{align}
& (-1)^{a_{1}+b_{41}^{\prime}-c_{3}-b_{34}} \sum_{c_{4}^{\prime}}\left(2 c_{4}^{\prime}+1\right)\sixj{a_{1}}{c_{4}^{\prime}}{b_{41}^{\prime}}{\lambda}{b_{41}}{c_{4}}\sixj{c_{3}}{b_{34}}{c_{4}^{\prime}}{\lambda}{c_{4}}{b_{34}^{\prime}}\twelvejI{a_{1},a_{2},a_{3},a_{4},b_{12},b_{23},b_{34},b_{41}^{\prime},c_{1},c_{2},c_{3},c_{4}^{\prime}}  \notag\\
= & (-1)^{c_{1}+b_{41}-a_{3}-b_{34}^{\prime}} \sum_{a_{4}^{\prime}}\left(2 a_{4}^{\prime}+1\right)\sixj{c_{1}}{a_{4}}{b_{41}^{\prime}}{\lambda}{b_{41}}{a_{4}^{\prime}}\sixj{a_{3}}{b_{34}}{a_{4}}{\lambda}{a_{4}^{\prime}}{b_{34}^{\prime}}\twelvejI{a_{1},a_{2},a_{3},a_{4}^{\prime},b_{12},b_{23},b_{34}^{\prime},b_{41},c_{1},c_{2},c_{3},c_{4}} . \label{eq:10:13:12}
\end{align}

\subsubsection{Explicit forms of the $12j$(I) symbols at some special relations between arguments}
may be obtained from Eqs. \eqref{eq:10:13:6} and \eqref{eq:10:13:7}. Here we give some of such forms.

If some argument in any tetrad is equal to the sum of three other arguments from the same tetrad, the sum in Eq. \eqref{eq:10:13:7} contains only one term. For example,
\begin{align}
& \twelvejI{a_{1},a_{2},a_{3},a_{4},b_{12},b_{23},b_{34},b_{41},c_{1},c_{2},a_{1}+a_{3}+c_{1},c_{4}}=(-1)^{a_{1}+b_{41}-c_{4}+c_{1}+b_{12}-c_{2}} \frac{\left(2 a_{1}\right)!\left(2 a_{3}\right)!\left(2 c_{1}\right)!}{\left(2 a_{1}+2 a_{3}+2 c_{1}+1\right)!}  \notag\\
& \times \frac{D\left(a_{1} a_{2} b_{12}\right) D\left(c_{1} c_{2} b_{12}\right) D\left(a_{3} a_{2} b_{23}\right) D\left(a_{3} a_{4} b_{34}\right) D\left(a_{1} c_{4} b_{41}\right) D\left(c_{1} a_{4} b_{41}\right)}{D\left(a_{1}+a_{3}+c_{1} c_{2} b_{23}\right) D\left(a_{1}+a_{3}+c_{1} c_{4} b_{34}\right)} . \label{eq:10:13:13}
\end{align}
The quantity $D(a b c)$ is defined by
\begin{equation}
D(a b c)=\left[\frac{(-a+b+c)!}{(a+b+c+1)!(a-b+c)!(a+b-c)!}\right]^{\frac{1}{2}} \label{eq:10:13:14}
\end{equation}
Equation \eqref{eq:10:13:7} for the $12j$(I) symbol may also be reduced to one term. For example,
\begin{gather}
\twelvejI{a_{1},a_{2},a_{3},a_{4},b_{12},b_{23},b_{34},b_{41},c_{1},c_{2},a_{1}+b_{34}+b_{41},c_{4}}=\left[\frac{\left(2 b_{34}\right)!\left(2 b_{41}\right)!}{\left(2 b_{34}+2 b_{41}\right)!\left(2 a_{1}+2 b_{41}+1\right)}\right]^{\frac{1}{2}}  \notag\\
\times \frac{D\left(b_{34} a_{3} a_{4}\right) D\left(b_{41} c_{1} a_{4}\right)}{D\left(b_{34}+b_{41} c_{1} a_{3}\right)}\ninej{b_{23}}{a_{2}}{a_{3}}{c_{2}}{b_{12}}{c_{1}}{a_{1}+b_{34}+b_{41}}{a_{1}}{b_{34}+b_{41}} \delta_{c_{4} a_{1}+b_{41}} \label{eq:10:13:15}
\end{gather}
Similar expressions may be obtained from Eq. \eqref{eq:10:13:15} using the symmetries of the $12j$(I) symbols (see Eqs. \eqref{eq:10:13:9}).\\
If arguments $a_{i}$ and $c_{i}$ satisfy the equalities $a_{i}=b_{i k} \pm a_{k}, c_{i}=b_{i k} \pm c_{k},(k=i \pm 1)$ one obtains
\begin{gather}
\twelvejI{b_{12}+a_{2},a_{2},a_{3},a_{4},b_{12},b_{23},b_{34},b_{41},b_{12}-c_{2},c_{2},c_{3},c_{4}}=\frac{(-1)^{2 c_{3}+b_{23}-b_{34}-a_{2}-a_{4}}}{2 a_{2}+2 c_{2}+1}\sixj{a_{3}}{c_{3}}{a_{2}+c_{2}}{c_{4}}{a_{4}}{b_{34}}  \notag\\
\times\left[\frac{\left(2 a_{2}\right)!\left(2 c_{2}\right)!\left(2 b_{12}-2 c_{2}\right)!}{\left(2 b_{12}+1\right)\left(2 b_{12}+2 a_{2}+1\right)!}\right]^{\frac{1}{2}} \frac{D\left(a_{2} a_{3} b_{23}\right) D\left(c_{2} c_{3} b_{23}\right) D\left(b_{12}-c_{2} b_{41} a_{4}\right) D\left(a_{2}+c_{2} c_{4} a_{4}\right)}{D\left(a_{2}+c_{2} c_{3} a_{3}\right) D\left(b_{12}+a_{2} c_{4} b_{41}\right)} \label{eq:10:13:16}
\end{gather}

\subsubsection{Explicit forms of the $12j$(I) symbols for special values of arguments}
If one argument is equal to zero, the $12j$(I) symbol is reduced to a $9 j$ symbol or to a product of two $6 j$ symbols:
\begin{gather}
\twelvejI{a_{1},a_{2},a_{3},a_{4},b_{12},b_{23},b_{34},0,c_{1},c_{2},c_{3},c_{4}}=\frac{\delta_{a_{1} c_{4}} \delta_{a_{4} c_{1}}}{\sqrt{\left(2 a_{1}+1\right)\left(2 c_{1}+1\right)}}\ninej{a_{1}}{b_{12}}{a_{2}}{c_{3}}{c_{2}}{b_{23}}{b_{34}}{c_{1}}{a_{3}},  \label{eq:10:13:17}
\end{gather}
\begin{gather}
\twelvejI{a_{1},a_{2},a_{3},a_{4},b_{12},b_{23},b_{34},b_{41},c_{1},c_{2},c_{3},0}=(-1)^{b_{12}+b_{23}+b_{34}+b_{41}-a_{2}-c_{2}} \frac{1}{\sqrt{\left(2 b_{41}+1\right)\left(2 b_{34}+1\right)}}  \notag\\
\times\sixj{b_{41}}{a_{2}}{b_{12}}{c_{2}}{c_{1}}{a_{4}}\sixj{a_{2}}{a_{3}}{b_{23}}{b_{34}}{c_{2}}{a_{4}} . \label{eq:10:13:18}
\end{gather}
Simple expressions for the $12j$(I) symbols may also be obtained, if two of four arguments from one tetrad are equal to $\frac{1}{2}$, i.e., in the cases: (1) $a_{i}=c_{i}=\frac{1}{2}(i=1,2,3,4)$; (2) $a_{i}=a_{i \pm 2}=\frac{1}{2}$; (3) $c_{i}=c_{i \pm 2}=\frac{1}{2}$; (4) $c_{i}=a_{i \pm 2}=\frac{1}{2}(i \pm 2 \leq 4, i-2 \geq 1)$. For example, when $a_{4}=c_{4}=\frac{1}{2}$, one has
\begin{gather}
\twelvejI{a_{1},a_{2},a_{3},\frac{1}{2},b_{12},b_{23},b_{34},b_{41},c_{1},c_{2},c_{3},\frac{1}{2}}=\frac{\delta_{a_{1} c_{1}} \delta_{a_{2} c_{2}} \delta_{a_{3} c_{3}}}{2\left(2 a_{1}+1\right)\left(2 a_{2}+1\right)\left(2 a_{3}+1\right)}  \notag\\
+(-1)^{c} \frac{D\left(\frac{1}{2} a_{1} b_{41}\right) D\left(\frac{1}{2} c_{1} b_{41}\right) D\left(\frac{1}{2} a_{3} b_{34}\right) D\left(\frac{1}{2} c_{3} b_{34}\right)}{2 D\left(1 a_{1} c_{1}\right) D\left(1 a_{3} c_{3}\right)}\sixj{a_{1}}{a_{2}}{b_{12}}{c_{2}}{c_{1}}{1}\sixj{a_{2}}{a_{3}}{b_{23}}{c_{3}}{c_{2}}{1}, \label{eq:10:13:19}
\end{gather}
where $\varepsilon=a_{2}+c_{2}+b_{12}+b_{23}+b_{34}+b_{41}$. If $a_{1}=c_{3}=\frac{1}{2}$, then
\begin{gather}
\twelvejI{\frac{1}{2},a_{2},a_{3},a_{4},b_{12},b_{23},b_{34},b_{41},c_{1},c_{2},\frac{1}{2},c_{4}}=(-1)^{a_{2}+c_{2}+1} \frac{\delta_{a_{3} c_{1}}}{2\left(2 a_{3}+1\right)}\sixj{a_{2}}{a_{3}}{b_{23}}{c_{2}}{\frac{1}{2}}{b_{12}}  \notag\\
\times\left\{\frac{(-1)^{a_{4}-c_{4}}}{2 b_{34}+1} \delta_{b_{34} b_{41}}+\frac{(-1)^{b_{34}+b_{41}-g+1}}{2 D^{2}\left(\frac{1}{2} a_{3} g\right)}\left[\frac{\left(2 a_{3}-1\right)!}{\left(2 a_{3}+2\right)!}\right]^{\frac{1}{2}} \frac{D\left(\frac{1}{2} b_{41} c_{4}\right) D\left(\frac{1}{2} b_{34} c_{4}\right)}{D\left(1 b_{34} b_{41}\right)}\sixj{b_{34}}{b_{41}}{1}{a_{3}}{a_{3}}{a_{4}}\right\}  \notag\\
+(-1)^{b_{34}+b_{41}+2 g+1} \frac{D\left(\frac{1}{2} b_{41} c_{4}\right) D\left(\frac{1}{2} b_{34} c_{4}\right) D\left(1 a_{3} c_{1}\right)}{D\left(\frac{1}{2} c_{1} g\right) D\left(\frac{1}{2} a_{3} g\right) D\left(1 b_{34} b_{41}\right)}  \notag\\
\times\sixj{b_{23}}{a_{2}}{a_{3}}{\frac{1}{2}}{g}{b_{12}}\sixj{b_{12}}{c_{2}}{c_{1}}{\frac{1}{2}}{g}{b_{23}}\sixj{b_{34}}{b_{41}}{1}{c_{1}}{a_{3}}{a_{4}}, \label{eq:10:13:20}
\end{gather}
where $g$ is an arbitrary parameter which satisfies the triangular conditions for appropriate triads of the $6 j$ symbols in Eq. \eqref{eq:10:13:20}. The expressions for the $12j$(I) symbols are more simplified in the cases: (i) $a_{i}=a_{i \pm 2}=\frac{1}{2}$ and $c_{i}=c_{i \pm 2}$; (ii) $c_{i}=c_{i \pm 2}=\frac{1}{2}$ and $a_{i}=a_{i \pm 2}$; (iii) $a_{i}=c_{i \pm 2}=\frac{1}{2}$ and $c_{i}=a_{i \pm 2}$. In addition, if $a_{3}+b_{12}+c_{2}$ is even, then
\begin{align}
& \twelvejI{\frac{1}{2},a_{2},a_{3},a_{4},b_{12},b_{23},b_{34},b_{41},a_{3},c_{2},\frac{1}{2},c_{4}}=(-1)^{c_{4}-a_{4}+a_{2}+c_{2}+1} \frac{\delta_{b_{34} b_{41}}}{2\left(2 a_{3}+1\right)\left(2 b_{34}+1\right)}\sixj{b_{23}}{a_{2}}{a_{3}}{b_{12}}{c_{2}}{\frac{1}{2}}  \notag\\
& +\frac{(-1)^{b_{34}+b_{41}+2 a_{3}}}{2 \sqrt{\left(2 a_{3}+1\right)\left(2 c_{2}+1\right)\left(2 b_{12}+1\right)}} \frac{D\left(\frac{1}{2} b_{34} c_{4}\right) D\left(\frac{1}{2} b_{41} c_{4}\right)}{D\left(1 b_{34} b_{41}\right)}\sixj{b_{34}}{b_{41}}{1}{a_{3}}{a_{3}}{a_{4}} \frac{\threej{a_{2} }{ a_{3} }{ b_{23} }{\frac{1}{2} }{ -1 }{ \frac{1}{2}}}{\threej{a_{3} }{ c_{2} }{ b_{12} }{0 }{ 0 }{ 0}} \label{eq:10:13:21}
\end{align}
and if $a_{3}+b_{12}+c_{2}$ is odd, one has
\begin{gather}
\twelvejI{\frac{1}{2},a_{2},a_{3},a_{4},b_{12},b_{23},b_{34},b_{41},a_{3}+1,c_{2},\frac{1}{2},c_{4}}=(-1)^{a_{2}+a_{3}-a_{4}+b_{12}+b_{41}+\frac{2}{2}} \frac{D\left(\frac{1}{2} b_{34} c_{4}\right) D\left(\frac{1}{2} b_{41} c_{4}\right) D\left(a_{3} a_{4} b_{34}\right)}{D\left(a_{3}+1 a_{4} b_{41}\right)}  \notag\\
\times \frac{\left(c_{2}-b_{23}\right)\left(2 b_{23}+1\right)+\left(b_{12}-a_{2}\right)\left(2 a_{2}+1\right)+a_{3}+1}{\left(2 a_{3}+1\right)\left(2 a_{3}+2\right)\left(2 a_{3}+3\right) \sqrt{2\left(2 c_{2}+1\right)\left(2 b_{12}+1\right)}} \frac{\threej{a_{2} }{ a_{3} }{ b_{23} }{-\frac{1}{2} }{ 0 }{ \frac{1}{2}}}{\threej{a_{3}+1 }{ c_{2} }{ b_{12} }{0 }{ 0 }{ 0}}  \label{eq:10:13:22}\\
\twelvejI{\frac{1}{2},a_{2},a_{3},a_{4},b_{12},b_{23},b_{34},b_{41},a_{3}-1,c_{2},\frac{1}{2},c_{4}}=(-1)^{a_{2}+a_{3}-a_{4}+b_{12}+b_{41}+\frac{3}{2}} \frac{D\left(\frac{1}{2} b_{34} c_{4}\right) D\left(\frac{1}{2} b_{41} c_{4}\right) D\left(a_{3}-1 a_{4} b_{34}\right)}{D\left(a_{3} a_{4} b_{34}\right)}  \notag\\
\times \frac{\left(c_{2}-b_{23}\right)\left(2 b_{23}+1\right)+\left(b_{12}-a_{2}\right)\left(2 a_{2}+1\right)-a_{3}}{\left(2 a_{3}-1\right)\left(2 a_{3}\right)\left(2 a_{3}+1\right) \sqrt{2\left(2 c_{2}+1\right)\left(2 b_{12}+1\right)}} \frac{\threej{a_{2} }{ a_{3} }{ b_{23} }{-\frac{1}{2} }{ 0 }{ \frac{1}{2}}}{\threej{a_{3}-1 }{ b_{12} }{ c_{2} }{0 }{ 0 }{ 0}} \label{eq:10:13:23}
\end{gather}
\subsubsection{Relations between the $12j$(I) symbols introduced in this book and analogous functions of other authors}
are given by
\[
\begin{aligned}
\twelvejI{a_{1},a_{2},a_{3},a_{4},b_{12},b_{23},b_{34},b_{41},c_{1},c_{2},c_{3},c_{4}} & =\twelvejI{b_{12},a_{1},a_{2},a_{3},c_{1},b_{41},a_{4},c_{3},c_{2},c_{4},b_{23},b_{34}} \\
& \text { (Jahn and Hope \cite{ref074})} \\
& \left.=(-1)^{2 c_{1}}\twelvejI{a_{1},a_{2},a_{3},a_{4},b_{12},b_{23},b_{34},b_{41},c_{1},c_{2},c_{3},c_{4}} 1\right\} \\
& \text { (Elbaz and Castel \cite{ref017}). }
\end{aligned}
\]

\subsubsection{Tables of numerical values of the $12j$(I) symbols}
Some tables are presented in Ref.~\cite{ref139}. These tables contain the $12j$(I) symbols whose arguments are less than or equal 2 . Numerical values are given in decimals. However, they can easily be generated by using the recursion relations (see Sec.~\ref{sec:10.13.2}) and the explicit forms of the $12j$(I) symbols for special values of arguments (see Sec.~\ref{sec:10.13.2}).

\subsection{$12j$ Symbols of the Second Kind ($12j$(II) Symbols)}\label{sec:10.13.3}\index{12j symbol@$12j$ symbol!of the second kind}

\subsubsection{Alternative form considered here}
Instead of the $12j$(II) symbols determined by Eq.~\eqref{eq:10:13:2}, we shall consider the somewhat more symmetric $12j$(II) symbols introduced in Ref. \cite{ref141}.
The latter $12j$(II)-symbols may be represented by invariant sums of products of eight $3 j m$ symbols:
\begin{equation}
\begin{aligned}
\twelvejII{a_{2},a_{3},a_{4},b_{1},b_{3},b_{4},c_{1},c_{2},c_{4},d_{1},d_{2},d_{3}}
=(-1)^{a_{2}+c_{1}-c_{4}} \sum_{\alpha_{i} \beta_{i} \gamma_{i} \delta_{i}}
&\threej{a_{2} }{ a_{3} }{ a_{4} }{\alpha_{2} }{ \alpha_{3} }{ \alpha_{4}}\threej{a_{4} }{ b_{4} }{ c_{4} }{\alpha_{4} }{ \beta_{4} }{ \gamma_{4}}\threej{c_{4} }{ c_{1} }{ c_{2} }{\gamma_{4} }{ \gamma_{1} }{ \gamma_{2}}\threej{c_{2} }{ d_{2} }{ a_{2} }{\gamma_{2} }{ \delta_{2} }{ \alpha_{2}} \\
\times&\threej{b_{4} }{ b_{3} }{ b_{1} }{\beta_{4} }{ \beta_{3} }{ \beta_{1}}\threej{b_{1} }{ c_{1} }{ d_{1} }{\beta_{1} }{ \gamma_{1} }{ \delta_{1}}\threej{d_{1} }{ d_{2} }{ d_{3} }{\delta_{1} }{ \delta_{2} }{ \delta_{3}}\threej{d_{3} }{ b_{3} }{ a_{3} }{\delta_{3} }{ \beta_{3} }{ \alpha_{3}} .
\end{aligned}
\label{eq:10:13:24}
\end{equation}
Equation \eqref{eq:10:13:24} unambiguously fixes the absolute values and phases of the $12j$(II) symbols. These symbols are real.\\
\subsubsection{COnditions on the arguments} 
Arguments of the $12j$(II) symbols satisfy the following conditions:
\begin{enumerate}[label=(\roman*)]
\item All arguments are non-negative integers or half-integers.
\item A $12j$(II) symbol vanishes unless the triangular conditions (see Sec.~\ref{sec:8.1.1}) are fulfilled for the triads $\left(a_{2} a_{3} a_{4}\right),\left(b_{1} b_{3} b_{4}\right),\left(c_{1} c_{2} c_{4}\right),\left(d_{1} d_{2} d_{3}\right),\left(b_{1} c_{1} d_{1}\right),\left(a_{2} c_{2} d_{2}\right),\left(a_{3} b_{3} d_{3}\right)$ and $\left(a_{4} b_{4} c_{4}\right)$.\\
\item A $12j$(II) symbol vanishes unless the tetragonal conditions (see Sec.~\ref{sec:10.13.2}) are fulfilled for three tetrads $\left(a_{2} c_{4} d_{3} b_{1}\right),\left(a_{3} b_{4} d_{2} c_{1}\right)$ and $\left(a_{4} b_{3} c_{2} d_{1}\right)$.
\end{enumerate}

\subsubsection{Orthogonality conditions}
The $12j$(II) symbols satisfy the orthogonality and normalization condition
\begin{align}
& \sum_{x_{1} x_{2} x_{3}}\left(2 x_{1}+1\right)\left(2 x_{2}+1\right)\left(2 x_{3}+1\right)\twelvejII{a_{2},a_{3},a_{4},b_{1},b_{3},b_{4},c_{1},x_{1},c_{4},d_{1},x_{2},x_{3}}\twelvejII{a_{2},a_{3},a_{4}^{\prime},b_{1}^{\prime},b_{3},b_{4}^{\prime},c_{1},x_{1},c_{4},d_{1},x_{2},x_{3}}  \notag\\
& =\frac{\delta_{a_{4} a_{4}^{\prime}} \delta_{b_{1} b_{1}^{\prime}} \delta_{b_{4} b_{4}^{\prime}}}{\left(2 a_{4}+1\right)\left(2 b_{1}+1\right)\left(2 b_{4}+1\right)}\left\{a_{2} a_{3} a_{4}\right\}\left\{b_{1} b_{3} b_{4}\right\}\left\{b_{1} c_{1} d_{1}\right\}\left\{a_{4} b_{4} c_{4}\right\} \label{eq:10:13:25}
\end{align}

\subsubsection{Sum representation}
The $12j$(II)-symbols may be represented as the following sums
\begin{gather}
\twelvejII{a_{2},a_{3},a_{4},b_{1},b_{3},b_{4},c_{1},c_{2},c_{4},d_{1},d_{2},d_{3}}=(-1)^{b_{3}-a_{4}-d_{1}+c_{2}} \sum_{x}(2 x+1)  \notag\\
\times\sixj{a_{3}}{b_{4}}{x}{b_{1}}{d_{3}}{b_{3}}\sixj{a_{3}}{b_{4}}{x}{c_{4}}{a_{2}}{a_{4}}\sixj{b_{1}}{d_{3}}{x}{d_{2}}{c_{1}}{d_{1}}\sixj{c_{4}}{a_{2}}{x}{d_{2}}{c_{1}}{c_{2}}  \label{eq:10:13:26}\\
\twelvejII{a_{2},a_{3},a_{4},b_{1},b_{3},b_{4},c_{1},c_{2},c_{4},d_{1},d_{2},d_{3}}=(-1)^{b_{3}-a_{4}-d_{1}+c_{2}} \sum_{x}(2 x+1)\ninej{a_{3}}{b_{3}}{d_{3}}{a_{4}}{b_{4}}{c_{4}}{a_{2}}{b_{1}}{x}\ninej{d_{2}}{d_{1}}{d_{3}}{c_{2}}{c_{1}}{c_{4}}{a_{2}}{b_{1}}{x} . \notag
\end{gather}

\subsubsection{Symmetry properties}
Any $12j$(II) symbol is symmetric with respect to a permutation of any two columns and simultaneous permutation of the rows whose numbers coincide with the numbers of permuted columns. In addition, the $12j$(II) symbol does not change under transposition. These symmetry properties relate 48 formally different $12j$(II) symbols:
\begin{align}
& \twelvejII{a_{2},a_{3},a_{4},b_{1},b_{3},b_{4},c_{1},c_{2},c_{4},d_{1},d_{2},d_{3}}=\twelvejII{b_{1},b_{3},b_{4},a_{2},a_{3},a_{4},c_{2},c_{1},c_{4},d_{2},d_{1},d_{3}}=\twelvejII{c_{1},c_{2},c_{4},a_{3},a_{2},a_{4},b_{3},b_{1},b_{4},d_{3},d_{1},d_{2}}=\twelvejII{d_{1},d_{2},d_{3},a_{4},a_{2},a_{3},b_{4},b_{1},b_{3},c_{4},c_{1},c_{2}}  \notag\\
& =\twelvejII{a_{4},a_{2},a_{3},d_{1},d_{2},d_{3},b_{1},b_{4},b_{3},c_{1},c_{4},c_{2}}=\twelvejII{b_{4},b_{1},b_{3},d_{2},d_{1},d_{3},a_{2},a_{4},a_{3},c_{2},c_{4},c_{1}}=\twelvejII{c_{4},c_{1},c_{2},d_{3},d_{1},d_{2},a_{3},a_{4},a_{2},b_{3},b_{4},b_{1}}=\twelvejII{d_{3},d_{1},d_{2},c_{4},c_{1},c_{2},a_{4},a_{3},a_{2},b_{4},b_{3},b_{1}}  \notag\\
& =\twelvejII{a_{3},a_{4},a_{2},c_{1},c_{4},c_{2},d_{1},d_{3},d_{2},b_{1},b_{3},b_{4}}=\twelvejII{b_{3},b_{4},b_{1},c_{2},c_{4},c_{1},d_{2},d_{3},d_{1},a_{2},a_{3},a_{4}}=\twelvejII{c_{2},c_{4},c_{1},b_{3},b_{4},b_{1},d_{3},d_{2},d_{1},a_{3},a_{2},a_{4}}=\twelvejII{d_{2},d_{3},d_{1},b_{4},b_{3},b_{1},c_{4},c_{2},c_{1},a_{4},a_{2},a_{3}}  \notag\\
& =\twelvejII{a_{2},a_{4},a_{3},b_{1},b_{4},b_{3},d_{1},d_{2},d_{3},c_{1},c_{2},c_{4}}=\twelvejII{b_{1},b_{4},b_{3},a_{2},a_{4},a_{3},d_{2},d_{1},d_{3},c_{2},c_{1},c_{4}}=\twelvejII{c_{1},c_{4},c_{2},a_{3},a_{4},a_{2},d_{3},d_{1},d_{2},b_{3},b_{1},b_{4}}=\twelvejII{d_{1},d_{3},d_{2},a_{4},a_{3},a_{2},c_{4},c_{1},c_{2},b_{4},b_{1},b_{3}}  \notag\\
& =\twelvejII{a_{3},a_{2},a_{4},c_{1},c_{2},c_{4},b_{1},b_{3},b_{4},d_{1},d_{3},d_{2}}=\twelvejII{b_{3},b_{1},b_{4},c_{2},c_{1},c_{4},a_{2},a_{3},a_{4},d_{2},d_{3},d_{1}}=\twelvejII{c_{2},c_{1},c_{4},b_{3},b_{1},b_{4},a_{3},a_{2},a_{4},d_{3},d_{2},d_{1}}=\twelvejII{d_{2},d_{1},d_{3},b_{4},b_{1},b_{3},a_{4},a_{2},a_{3},c_{4},c_{2},c_{1}}  \notag\\
& =\twelvejII{a_{4},a_{3},a_{2},d_{1},d_{3},d_{2},c_{1},c_{4},c_{2},b_{1},b_{4},b_{3}}=\twelvejII{b_{4},b_{3},b_{1},d_{2},d_{3},d_{1},c_{2},c_{4},c_{1},a_{2},a_{4},a_{3}}=\twelvejII{c_{4},c_{2},c_{1},d_{3},d_{2},d_{1},b_{3},b_{4},b_{1},a_{3},a_{4},a_{2}}=\twelvejII{d_{3},d_{2},d_{1},c_{4},c_{2},c_{1},b_{4},b_{3},b_{1},a_{4},a_{3},a_{2}}  \notag\\
& =\twelvejII{b_{1},c_{1},d_{1},a_{2},c_{2},d_{2},a_{3},b_{3},d_{3},a_{4},b_{4},c_{4}}=\twelvejII{a_{2},c_{2},d_{2},b_{1},c_{1},d_{1},b_{3},a_{3},d_{3},b_{4},a_{4},c_{4}}=\twelvejII{a_{3},b_{3},d_{3},c_{1},b_{1},d_{1},c_{2},a_{2},d_{2},c_{4},a_{4},b_{4}}=\twelvejII{a_{4},b_{4},c_{4},d_{1},b_{1},c_{1},d_{2},a_{2},c_{2},d_{3},a_{3},b_{3}}  \notag\\
& =\twelvejII{d_{1},b_{1},c_{1},a_{4},b_{4},c_{4},a_{2},d_{2},c_{2},a_{3},d_{3},b_{3}}=\twelvejII{d_{2},a_{2},c_{2},b_{4},a_{4},c_{4},b_{1},d_{1},c_{1},b_{3},d_{3},a_{3}}=\twelvejII{d_{3},a_{3},b_{3},c_{4},a_{4},b_{4},c_{1},d_{1},b_{1},c_{2},d_{2},a_{2}}=\twelvejII{c_{4},a_{4},b_{4},d_{3},a_{3},b_{3},d_{1},c_{1},b_{1},d_{2},c_{2},a_{2}}  \notag\\
& =\twelvejII{c_{1},d_{1},b_{1},a_{3},d_{3},b_{3},a_{4},c_{4},b_{4},a_{2},c_{2},d_{2}}=\twelvejII{c_{2},d_{2},a_{2},b_{3},d_{3},a_{3},b_{4},c_{4},a_{4},b_{1},c_{1},d_{1}}=\twelvejII{b_{3},d_{3},a_{3},c_{2},d_{2},a_{2},c_{4},b_{4},a_{4},c_{1},b_{1},d_{1}}=\twelvejII{b_{4},c_{4},a_{4},d_{2},c_{2},a_{2},d_{3},b_{3},a_{3},d_{1},b_{1},c_{1}}  \notag\\
& =\twelvejII{b_{1},d_{1},c_{1},a_{2},d_{2},c_{2},a_{4},b_{4},c_{4},a_{3},b_{3},d_{3}}=\twelvejII{a_{2},d_{2},c_{2},b_{1},d_{1},c_{1},b_{4},a_{4},c_{4},b_{3},a_{3},d_{3}}=\twelvejII{a_{3},d_{3},b_{3},c_{1},d_{1},b_{1},c_{4},a_{4},b_{4},c_{2},a_{2},d_{2}}=\twelvejII{a_{4},c_{4},b_{4},d_{1},c_{1},b_{1},d_{3},a_{3},b_{3},d_{2},a_{2},c_{2}}  \notag\\
& =\twelvejII{c_{1},b_{1},d_{1},a_{3},b_{3},d_{3},a_{2},c_{2},d_{2},a_{4},c_{4},b_{4}}=\twelvejII{c_{2},a_{2},d_{2},b_{3},a_{3},d_{3},b_{1},c_{1},d_{1},b_{4},c_{4},a_{4}}=\twelvejII{b_{3},a_{3},d_{3},c_{2},a_{2},d_{2},c_{1},b_{1},d_{1},c_{4},b_{4},a_{4}}=\twelvejII{b_{4},a_{4},c_{4},d_{2},a_{2},c_{2},d_{1},b_{1},c_{1},d_{3},b_{3},a_{3}}  \notag\\
& =\twelvejII{d_{1},c_{1},b_{1},a_{4},c_{4},b_{4},a_{3},d_{3},b_{3},a_{2},d_{2},c_{2}}=\twelvejII{d_{2},c_{2},a_{2},b_{4},c_{4},a_{4},b_{3},d_{3},a_{3},b_{1},d_{1},c_{1}}=\twelvejII{d_{3},b_{3},a_{3},c_{4},b_{4},a_{4},c_{2},d_{2},a_{2},c_{1},d_{1},b_{1}}=\twelvejII{c_{4},b_{4},a_{4},d_{3},b_{3},a_{3},d_{2},c_{2},a_{2},d_{1},c_{1},b_{1}} . \label{eq:10:13:27}
\end{align}
\subsubsection{Recursion relations}
These relations are useful for evaluation of the $12j$(II) symbols. We present only one relation:
\begin{gather}
\sum_{a_{3}^{\prime}}\left(2 a_{3}^{\prime}+1\right)\sixj{\lambda}{a_{3}}{a_{3}^{\prime}}{d_{3}}{b_{3}^{\prime}}{b_{3}}\sixj{\lambda}{a_{3}}{a_{3}^{\prime}}{a_{2}}{a_{4}}{a_{4}^{\prime}}\twelvejII{a_{2},a_{3}^{\prime},a_{4},b_{1},b_{3}^{\prime},b_{4},c_{1},c_{2},c_{4},d_{1},d_{2},d_{3}}  \notag\\
=(-1)^{b_{1}+a_{2}-d_{3}-c_{4}} \sum_{b_{4}^{\prime}}\left(2 b_{4}^{\prime}+1\right)\sixj{\lambda}{b_{4}}{b_{4}^{\prime}}{c_{4}}{a_{4}^{\prime}}{a_{4}}\sixj{\lambda}{b_{4}}{b_{4}^{\prime}}{b_{1}}{b_{3}}{b_{3}^{\prime}}\twelvejII{a_{2},a_{3},a_{4}^{\prime},b_{1},b_{3},b_{4}^{\prime},c_{1},c_{2},c_{4},d_{1},d_{2},d_{3}} . \label{eq:10:13:28}
\end{gather}
In particular, putting $\lambda=\frac{1}{2}, a_{4}^{\prime}=a_{4}-\frac{1}{2}$ and $b_{3}^{\prime}=b_{3}+\frac{1}{2}$ in Eq. \eqref{eq:10:13:28} we obtain
\begin{align}
& \left(2 b_{4}+1\right)\left\{\left[\left(a_{3}+b_{3}-d_{3}+1\right)\left(a_{3}+b_{3}+d_{3}+2\right)\left(-a_{2}+a_{3}+a_{4}+\frac{1}{2}\right)\left(a_{2}+a_{3}+a_{4}+\frac{3}{2}\right)\right]^{\frac{1}{2}}\twelvejII{a_{2},a_{3}+\frac{1}{2},a_{4},b_{1},b_{3}+\frac{1}{2},b_{4},c_{1},c_{2},c_{4},d_{1},d_{2},d_{3}}\right. \notag \\
& \left.+\left[\left(-a_{3}+b_{3}+d_{3}+1\right)\left(a_{3}-b_{3}+d_{3}\right)\left(a_{2}-a_{3}+a_{4}+\frac{1}{2}\right)\left(a_{2}+a_{3}-a_{4}+\frac{1}{2}\right)\right]^{\frac{1}{2}}\twelvejII{a_{2},a_{3}-\frac{1}{2},a_{4},b_{1},b_{3}+\frac{1}{2},b_{4},c_{1},c_{2},c_{4},d_{1},d_{2},d_{3}}\right\} \notag \\
& =\left(2 a_{3}+1\right)\left\{\left[\left(-a_{4}+b_{4}+c_{4}+1\right)\left(a_{4}-b_{4}+c_{4}\right)\left(b_{1}-b_{3}+b_{4}+\frac{1}{2}\right)\left(b_{1}+b_{3}-b_{4}+\frac{1}{2}\right)\right]{ }^{\frac{1}{2}}\twelvejII{a_{2},a_{3},a_{4},b_{1},b_{3},b_{4}-\frac{1}{2},c_{1},c_{2},c_{4},d_{1},d_{2},d_{3}}\right\} \notag \\
& \left.+\left[\left(a_{4}+b_{4}-c_{4}\right)\left(a_{4}+b_{4}+c_{4}+1\right)\left(-b_{1}+b_{3}+b_{4}+\frac{1}{2}\right)\left(b_{1}+b_{3}+b_{4}+\frac{3}{2}\right)\right]^{\frac{1}{2}}\twelvejII{a_{2},a_{3},a_{4}-\frac{1}{2},b_{1},b_{3},b_{4}-\frac{1}{2},c_{1},c_{2},c_{4},d_{1},d_{2},d_{3}}\right\} .
\end{align}

\subsubsection{Explicit forms of the $12j$(II)-symbols for some relations between arguments}
may be obtained, using Eqs. \eqref{eq:10:13:26} and \eqref{eq:10:13:27}.

If one argument in any tetrad is equal to the sum of three other arguments from the same tetrad, one obtains
\begin{align}
& \twelvejII{b_{1}+c_{4}+d_{3},a_{3},a_{4},b_{1},b_{3},b_{4},c_{1},c_{2},c_{4},d_{1},d_{2},d_{3}}=(-1)^{b_{3}-b_{4}+c_{1}-d_{1}} \frac{\left(2 b_{1}\right)!\left(2 c_{4}\right)!\left(2 d_{3}\right)!}{\left(2 b_{1}+2 c_{4}+2 d_{3}+1\right)!}  \notag\\
& \quad \times \frac{D\left(b_{1} b_{3} b_{4}\right) D\left(c_{4} c_{1} c_{2}\right) D\left(d_{3} d_{1} d_{2}\right) D\left(b_{1} d_{1} c_{1}\right) D\left(c_{4} b_{4} a_{4}\right) D\left(d_{3} b_{3} a_{3}\right)}{D\left(b_{1}+d_{3}+c_{4} a_{4} a_{3}\right) D\left(b_{1}+d_{3}+c_{4} c_{2} d_{2}\right)} \label{eq:10:13:30}
\end{align}
Let each argument in any tetrad be equal to the sum of two arguments which form a triad with this argument. Then, the $12j$(II) symbol may be expressed as a sum involving products of two Clebsch-Gordan\\
coefficients.
\begin{gather}
\twelvejII{a_{3}+a_{4},a_{3},a_{4},b_{3}+b_{4},b_{3},b_{4},c_{1},c_{2},c_{1}+c_{2},d_{1},d_{2},d_{1}+d_{2}}=(-1)^{b_{3}-a_{4}-d_{1}+c_{2}} f\left(a_{3} b_{3} d_{1}+d_{2}\right) f\left(a_{4} b_{4} c_{1}+c_{2}\right)  \notag\\
\times f\left(d_{1} c_{1} b_{3}+b_{4}\right) f\left(d_{2} c_{2} a_{3}+a_{4}\right)\left[\frac{\left(2 a_{3}\right)!\left(2 a_{4}\right)!\left(2 b_{3}\right)!\left(2 b_{4}\right)!\left(2 c_{1}\right)!\left(2 c_{2}\right)!\left(2 d_{1}\right)!\left(2 d_{2}\right)!}{\left(2 a_{3}+2 a_{4}+1\right)!\left(2 b_{3}+2 b_{4}+1\right)!\left(2 c_{1}+2 c_{2}+1\right)!\left(2 d_{1}+2 d_{2}+1\right)!}\right]^{\frac{1}{2}}  \notag\\
\times \sum_{x}\left[f\left(a_{3}+a_{4} b_{3}+b_{4} x\right) f\left(d_{1}+d_{2} c_{1}+c_{2} x\right)\right]^{-1} \clebsch{d_{1}+d_{2}}{a_{3}-b_{3}}{c_{1}+c_{2}}{a_{4}-b_{4}}{x}{a_{3}+a_{4}-b_{3}-b_{4}} \clebsch{a_{3}+a_{4}}{d_{2}-c_{2}}{b_{3}+b_{4}}{d_{1}-c_{1}}{x}{d_{1}+d_{2}-c_{1}-c_{2}} \label{eq:10:13:31}
\end{gather}
where
\begin{equation}
f(a b c)=[(a+b+c+1)!(a+b-c)!]^{-\frac{1}{2}} \label{eq:10:13:32}
\end{equation}
\subsubsection{Explicit forms of the $12j$(II) symbols for special values of the arguments}

If one argument is equal to zero, a $12j$(II) symbol is reduced to product of two $6 j$ symbols:
\begin{align}
\twelvejII{0,a_{3},a_{4},b_{1},b_{3},b_{4},c_{1},c_{2},c_{4},d_{1},d_{2},d_{3}}= & (-1)^{b_{3}+b_{4}-d_{1}-c_{1}} \frac{\delta_{a_{3} a_{4}} \delta_{c_{2} d_{2}}}{\sqrt{\left(2 a_{3}+1\right)\left(2 c_{2}+1\right)}}  \notag\\
& \times\sixj{b_{1}}{b_{3}}{b_{4}}{a_{3}}{c_{4}}{d_{3}}\sixj{c_{1}}{c_{2}}{c_{4}}{d_{3}}{b_{1}}{d_{1}} \label{eq:10:13:33}
\end{align}
If two arguments from any tetrad are equal to $\frac{1}{2}$ we obtain
\begin{gather}
% Source misprints corrected for numerical consistency (verified 2445/2445,
% scripts/check_10_13.py): third Kronecker delta was \delta_{c_1 d_3} -> \delta_{c_1 d_2};
% denominator D-symbols were D(1 b_1 b_3) D(1 a_2 c_4) -> D(1 b_1 d_3) D(1 c_4 a_2)
% (D is the asymmetric symbol of Eq. (10.13.14), so argument order matters).
\twelvejII{a_{2},\frac{1}{2},a_{4},b_{1},b_{3},\frac{1}{2},c_{1},c_{2},c_{4},d_{1},d_{2},d_{3}}=\frac{\delta_{b_{1} d_{3}} \delta_{c_{4} a_{2}} \delta_{c_{1} d_{2}}}{2\left(2 b_{1}+1\right)\left(2 c_{1}+1\right)\left(2 c_{4}+1\right)}  \notag\\
+(-1)^{\varepsilon}\sixj{b_{1}}{d_{3}}{1}{d_{2}}{c_{1}}{d_{1}}\sixj{c_{4}}{a_{2}}{1}{d_{2}}{c_{1}}{c_{2}} \frac{D\left(\frac{1}{2} a_{2} a_{4}\right) D\left(\frac{1}{2} b_{1} b_{3}\right) D\left(\frac{1}{2} c_{4} a_{4}\right) D\left(\frac{1}{2} b_{3} d_{3}\right)}{2 D\left(1 b_{1} d_{3}\right) D\left(1 c_{4} a_{2}\right)}  \label{eq:10:13:34}\\
\varepsilon=b_{1}+b_{3}+c_{2}+c_{4}+a_{2}-a_{4}-d_{1}+d_{3}  \notag\\
\twelvejII{\frac{1}{2},a_{3},a_{4},\frac{1}{2},b_{3},b_{4},c_{1},c_{2},c_{4},d_{1},d_{2},d_{3}}=(-1)^{2 b_{3}+2 d_{1}} \frac{\delta_{d_{3} c_{4}}}{2\left(2 d_{3}+1\right)}\sixj{a_{3}}{a_{4}}{\frac{1}{2}}{b_{4}}{b_{3}}{d_{3}}\sixj{d_{2}}{c_{2}}{\frac{1}{2}}{c_{1}}{d_{1}}{d_{3}}  \notag\\
+3(-1)^{b_{3}-a_{4}-d_{1}+c_{2}}\ninej{a_{3}}{b_{3}}{d_{3}}{a_{4}}{b_{4}}{c_{4}}{\frac{1}{2}}{\frac{1}{2}}{1}\ninej{d_{2}}{d_{1}}{d_{3}}{c_{2}}{c_{1}}{c_{4}}{\frac{1}{2}}{\frac{1}{2}}{1} . \label{eq:10:13:35}
\end{gather}
At $c_{4}=d_{3}, d_{3} \pm 1$, a $12j$(II) symbol may be expressed in terms of the $3 j m$ and $6 j$ symbols:
\begin{align}
& \twelvejII{\frac{1}{2},a_{3},a_{4},\frac{1}{2},b_{3},b_{4},c_{1},c_{2},d_{3},d_{1},d_{2},d_{3}}=(-1)^{2 a_{3}+2 d_{1}} \frac{1}{2\left(2 d_{3}+1\right)}\sixj{a_{3}}{a_{4}}{\frac{1}{2}}{b_{4}}{b_{3}}{d_{3}}\sixj{d_{2}}{c_{2}}{\frac{1}{2}}{c_{1}}{d_{1}}{d_{3}}  \notag\\
& +\frac{(-1)^{b_{3}-a_{4}-d_{1}+c_{2}}}{2\left(2 d_{3}+1\right) \sqrt{\left(2 a_{4}+1\right)\left(2 b_{4}+1\right)\left(2 c_{1}+1\right)\left(2 c_{2}+1\right)}} \frac{\threej{a_{3} }{ b_{3} }{ d_{3} }{\frac{1}{2} }{ \frac{1}{2} }{ -1}\threej{d_{2} }{ d_{1} }{ d_{3} }{\frac{1}{2} }{ \frac{1}{2} }{ -1}}{\threej{a_{4} }{ b_{4} }{ d_{3} }{0 }{ 0 }{ 0}\threej{c_{1} }{ c_{2} }{ d_{3} }{0 }{ 0 }{ 0}} \label{eq:10:13:36}
\end{align}
This equation is valid, if $a_{4}+b_{4}+d_{3}$ and $c_{1}+c_{2}+d_{3}$ are even. If these numbers are odd, one has
\begin{gather}
\twelvejII{\frac{1}{2},a_{3},a_{4},\frac{1}{2},b_{3},b_{4},c_{1},c_{2},d_{3}+1,d_{1},d_{2},d_{3}}=\frac{\left(a_{4}-a_{3}\right)\left(2 a_{3}+1\right)+\left(b_{4}-b_{3}\right)\left(2 b_{3}+1\right)+d_{3}+1}{\left(2 d_{3}+1\right)\left(2 d_{3}+2\right)\left(2 d_{3}+3\right)}  \notag\\
\times \frac{\left(c_{2}-d_{2}\right)\left(2 d_{2}+1\right)+\left(c_{1}-d_{1}\right)\left(2 d_{1}+1\right)+d_{3}+1}{\sqrt{\left(2 a_{4}+1\right)\left(2 b_{4}+1\right)\left(2 c_{1}+1\right)\left(2 c_{2}+1\right)}} \frac{\threej{a_{3} }{ b_{3} }{ d_{3} }{\frac{1}{2} }{ -\frac{1}{2} }{ 0}\threej{d_{2} }{ d_{1} }{ d_{3} }{\frac{1}{2} }{ -\frac{1}{2} }{ 0}}{\threej{a_{4} }{ b_{4} }{ d_{3}+1 }{0 }{ 0 }{ 0}\threej{c_{1} }{ c_{2} }{ d_{3}+1 }{0 }{ 0 }{ 0}}  \label{eq:10:13:37}\\
\twelvejII{\frac{1}{2},a_{3},a_{4},\frac{1}{2},b_{3},b_{4},c_{1},c_{2},d_{3}-1,d_{1},d_{2},d_{3}}=\frac{\left(a_{4}-a_{3}\right)\left(2 a_{3}+1\right)+\left(b_{4}-b_{3}\right)\left(2 b_{3}+1\right)-d_{3}}{\left(2 d_{3}-1\right)\left(2 d_{3}\right)\left(2 d_{3}+1\right)}  \notag\\
\times \frac{\left(c_{2}-d_{2}\right)\left(2 d_{2}+1\right)+\left(c_{1}-d_{1}\right)\left(2 d_{1}+1\right)-d_{3}}{\sqrt{\left(2 a_{4}+1\right)\left(2 b_{4}+1\right)\left(2 c_{1}+1\right)\left(2 c_{2}+1\right)}} \frac{\threej{a_{3} }{ b_{3} }{ d_{3} }{\frac{1}{2} }{ -\frac{1}{2} }{ 0}\threej{d_{2} }{ d_{1} }{ d_{3} }{\frac{1}{2} }{ -\frac{1}{2} }{ 0}}{\threej{a_{4} }{ b_{4} }{ d_{3}-1 }{0 }{ 0 }{ 0}\threej{c_{1} }{ c_{2} }{ d_{3}-1 }{0 }{ 0 }{ 0}} \label{eq:10:13:38}
\end{gather}
\subsubsection{Relations between the $12j$(II) symbols considered above and analogous functions of other authors}
\[
\begin{gathered}
\twelvejII{a_{2},a_{3},a_{4},b_{1},b_{3},b_{4},c_{1},c_{2},c_{4},d_{1},d_{2},d_{3}}=\left(\begin{array}{cccc}
\cdot & a_{2} & a_{3} & a_{4} \\
b_{1} & \cdot & b_{3} & b_{4} \\
c_{1} & c_{2} & \cdot & c_{4} \\
d_{1} & d_{2} & d_{3} & \cdot
\end{array}\right), \\
\quad\text{ (Sharp \cite{ref141})} \\
1)^{a_{2}+a_{3}+d_{2}+d_{3}-b_{1}-b_{4}-c_{1}-c_{4}}\left[\begin{array}{cccc}
a_{3} & d_{3} & d_{2} & a_{2} \\
b_{3} & d_{1} & c_{2} & a_{4} \\
b_{4} & b_{1} & c_{1} & c_{4}
\end{array}\right],
\end{gathered}
\]
(Yutsis, Levinson and Vanagas \cite{ref044}),
\[
=(-1)^{a_{2}+a_{3}+d_{2}+d_{3}-b_{1}-b_{4}-c_{1}-c_{4}}\left\{\left.\begin{array}{cccc}
a_{3} & d_{3} & d_{2} & a_{2} \\
b_{3} & d_{1} & c_{2} & a_{4} \\
b_{4} & b_{1} & c_{1} & c_{4}
\end{array} \right\rvert\, 2\right\}
\]
(Elbaz and Castel \cite{ref017}).

\chapter{The graphical method in angular momentum theory}\label{chap:11}\index{graphical method}\index{diagram technique|see{graphical method}}
To solve many quantum mechanical problems involving angular and spin dependences one has to integrate a product of the spherical harmonics or the Wigner D-functions and to sum products of the Clebsch-Gordan coefficients and \(3 n j\) symbols. Expressions of this kind are, as a rule, rather complicated. They contain so many arguments and parameters that it is difficult to understand interrelations between them, to find out the symmetry of the expression, and to see its invariance with respect to a particular transformation.

The representation of such expressions in diagrammatic form facilitates their general analysis. A graphical representation is more compact and clear. All arguments and parameter interrelations are obvious. The structure and symmetries of the expression as a whole and each part of the expression are evident. Moreover, the graphical technique substantially reduces the calculation job.

There are several versions of the graphical method.\index{graphical method!versions of} A handy technique for summing the \(3 j m\) symbol products was suggested by Levinson \cite{ref080}, Yutsis, Levinson and Vanagas \cite{ref044} and Yutsis and Bandzaitis \cite{ref045}. A slightly modified version was proposed by Brink and Satchler \cite{ref009}. A certain development of this diagram technique was presented by El-Baz and Castel \cite{ref017} who extended it to expressions involving continuous variables. This version is now commonly used.\footnote{A somewhat different graphical method was suggested by Shelepin \cite{ref105}, but it is not very widespread now.}
 However, various books and papers concerning such diagrams are somewhat different in definitions, phase conventions, and rules of the graphical technique, although the notations, sometimes, are similar. Therefore, one should be careful in using formulas from various sources.

In this chapter we represent the unified graphical method in detail. The basic elements of diagrams are defined in Sections \ref{sec:11.1} and \ref{sec:11.2} where the graphical representation of the standard functions and main operations of the angular momentum theory are given. Section \ref{sec:11.3} exhibits the rules of the diagram technique, i.e., the graphical method of calculation. For practical convenience, all the definitions and rules are arranged in tables. The general scheme of application of the diagram technique is formulated in Sec.~\ref{sec:11.4}.

\section{Graphical representation of functions}\label{sec:11.1}\index{graphical method!representation of functions}\index{diagrams}
A principle of any graphical method is in establishing a one-to-one correspondence between elements of a diagram and constituents of the analytic expression. Any expression inherent in the theory may be represented by a definite diagram, and vice versa, any such diagram represents a definite analytic expression.

Moreover, a transformation of the expression under some analytic operation corresponds to a certain transformation of the associated diagram. That is why all calculations of spin angular quantities may be performed by the graphical method instead of the analytic one.
\subsection{Basic Elements of Diagrams}\label{sec:11.1.1}\index{diagrams!basic elements of}
The Dirac representation of quantum mechanical quantities seems to be most adequate to the graphical method. This representation operates with state vectors: the ``kets'', \(\ket{\Psi}\) and the Hermitian conjugate ``bras'', \(\bra{\psi}\). Graphically, these vectors are represented by directed lines connected with nodes (dot or cross bar).\index{node}\index{node!dot}\index{node!bar}\index{state vector}\index{ket}\index{bra}

A line with an arrow coming out of a node represents the ``ket'' \(\ket{\Psi}\), i.e., a covariant (standard) state vector, whereas, a line with double arrow coming into the node represents the ``bra'' \(\bra{\Psi}\), i.e. a contravariant (Hermitian conjugate) state vector.

A thin solid line corresponds to a state with some definite angular momentum \(j\) and projection \(m\) : 
\begin{equation}
\begin{aligned}
& \includegraphics[valign=t]{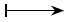}  \qquad \ket{jm}, \\
& \includegraphics[valign=t]{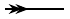}  \qquad \bra{j m}, 
\end{aligned}
\end{equation}
Such lines are referred to as the \(j\)-lines.\index{j-line@$j$-line}\index{lines!$j$-line}\\
A dashed line corresponds to a state with some definite direction of the position vector (or linear momentum) \(\Omega\{\vartheta, \varphi\}\)
\begin{equation}
\begin{aligned}
& \includegraphics[valign=t]{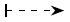}  \qquad \ket{\Omega}, \\
& \includegraphics[valign=t]{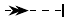}  \qquad \bra{\Omega}, 
\end{aligned}
\end{equation}
Such lines are referred to as the \(\Omega\)-lines.\index{Omega-line@$\Omega$-line}\index{lines!$\Omega$-line}\isymg{Omega-angles@$\Omega\{\vartheta, \varphi\}$, direction angles}\\
A double dashed line corresponds to a set of the Euler angles \(\alpha, \beta, \gamma\) (or the angles \(\omega, \Theta, \Phi\) ) which specify a rotation of the coordinate system, \(R\). Such lines are called the \(R\)-lines\index{R-line@$R$-line}\index{lines!$R$-line}\\
\centerline{\includegraphics[valign=t]{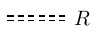}}

The symbols \(j m, \Omega\) and \(R\) are usually indicated near the corresponding lines. Therefore, length, curvature, orientation, and position of the line are inessential. For example, 
\begin{center}
\includegraphics[valign=t]{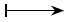} 
is equivalent to 
\includegraphics[valign=t]{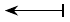} and \includegraphics[valign=t]{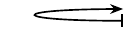} .
\end{center}

However, the arrow direction with respect to the node is of importance.\\
The normalization of state vectors may be different. In diagrams it is indicated by the type of node.\index{normalization!indicated by node type} A state vector \(\ket{j m}\) normalized according to
\begin{equation}
\braket{j m}{j^{\prime} m^{\prime}}=\delta_{j j^{\prime}} \delta_{m m^{\prime}}
\end{equation}
is represented by the \(j\)-line with a bar node, whereas, a state vector \(\ket{j m}\), which obeys the normalization condition
\begin{equation}
\rbraket{j m}{j^{\prime} m^{\prime}}=(2 j+1)^{-1} \delta_{j j^{\prime}} \delta_{m m^{\prime}}
\end{equation}
is represented by the \(j\)-line with a dot node. The completeness condition for wave functions \(\ket{j m}\) or \(\ket{j m}\)  reads
\begin{equation}
\sum_{j m}\ket{j m}\bra{j m}=\sum_{j m}\ket{j m}(2 j+1)\rbra{j m}=1
\end{equation}
Similarly, a state vector \(\ket{\Omega}\) normalized by the Dirac \(\delta\)-function
\begin{equation}
\braket{\Omega}{\Omega^{\prime}}=\delta\left(\Omega-\Omega^{\prime}\right)
\end{equation}
corresponds to the \(\Omega\)-line with the bar node, whereas, a state vector \(\ket{\Omega}\) normalized by
\begin{equation}
\rbraket{\Omega}{\Omega^{\prime}}=4 \pi \delta\left(\Omega-\Omega^{\prime}\right)
\end{equation}
corresponds to the \(\Omega\)-line with a dot node. The completeness condition for the functions \(\ket{\Omega}\) or \(\ket{\Omega}\) reads
\begin{equation}
\int\ket{\Omega} d \Omega\bra{\Omega}=1 \quad \text { and } \quad \int \ket{\Omega} \frac{d \Omega}{4 \pi}\rbra{\Omega}=1\,,
\end{equation}
respectively.

\subsection{Diagrams of the Basic Functions of the Theory}\label{sec:11.1.2}\index{diagrams!of the basic functions}
Quantum mechanical quantities, i.e., state vectors and operators, may depend on many spin and angular variables (such as \(j m\) or \(\Omega)\) and also on coordinate system rotation \(R\). Accordingly, these state vectors and matrix elements of the operators may be represented graphically as blocks (i.e., as subdiagrams) with external \(j m\)-, \(\Omega\) - and/or \(R\)-lines associated with the corresponding spin angular variables. A diagram corresponding to a composite expression will be a combination of such blocks connected by some of jm-, \(\Omega\) - and \(R\)-lines.

Diagrams of some spin and angular functions are listed in Tables \ref{tab:11.1}-\ref{tab:11.3}; graphical displays of their properties and relations is given in Tables \ref{tab:11.4}-\ref{tab:11.14}.

A spherical harmonic \(Y_{l m}(\vartheta, \varphi) \equiv\braket{\Omega}{l m}\) is represented by a junction of a thin solid and dashed lines\index{spherical harmonics!diagram of}\isym{Ylm@$Y_{l m}(\vartheta, \varphi)$, spherical harmonic} which correspond to \(\ket{l m}\) and \(\bra{\Omega}\).

A Wigner \(3 j m\) symbol \(\threej{a}{b}{c}{\alpha}{\beta}{\gamma} \equiv\rbraket{00}{a \alpha b \beta c \gamma}\) is represented by a dot-node junction of three thin solid lines\index{3jm symbol@$3jm$ symbol!diagram of} which correspond to \(\ket{a \alpha}, \ket{b \beta}\) and \(\ket{c \gamma}\). The sign of the node determines the cyclic order of arguments, i.e., the sequence of coupling of momenta: plus corresponds to the anticlockwise one.

The \(3 j\), \(6 j\), and \(9 j\) symbols, or, more generally speaking, the \(3 n j\) symbols, may be represented graphically by closed diagrams without external lines because they are invariant under coordinate rotation.\index{closed diagram}\index{3nj symbols@$3nj$ symbols!diagrams of}\index{6j symbol@$6j$ symbol!diagram of}\index{9j symbol@$9j$ symbol!diagram of} Such closed diagrams consist of \(3 n j\)-lines connected in \(2 n\) dot nodes by three's, because the \(3 n j\) symbols are actually the sums of \(2 n\) various \(3 j m\) symbols (Sec.~\ref{sec:10.13}). Any \(3 n j\) symbol may be represented by several equivalent diagrams which differ in the arrow directions of internal the \(j\) lines and in the node signs since each \(3 n j\) symbol may be expressed through different sums of products of the \(3 jm\) symbols; such sums turn into one another by replacing summation indices.

Matrix elements of the rotation operator in the \(\Omega\)-representation, \(\braOket{\Omega}{\hat{R}}{\Omega^{\prime}}=\delta\left(\Omega-\hat{R} \Omega^{\prime}\right)\), and in the \(j m\) representation, \(\braOket{j m}{\widehat{R}}{j^{\prime} m^{\prime}}=\delta_{j j^{\prime}} D_{m m^{\prime}}^{j}(R)\), are displayed as junctions of three lines. One of them is a double dashed \(R\)-line\index{rotation operator!diagram of}\index{Wigner D-function@Wigner $D$-function!diagram of}\isym{DJMM@$D_{M M^{\prime}}^{J}$, Wigner $D$-function} and the two others are either dashed lines corresponding to \(\ket{\Omega^{\prime}}\) and \(\bra{\Omega}\) states, or thin solid lines corresponding to \(\ket{j^{\prime} m^{\prime}}\) and \(\bra{ j m}\) states. The sign of the junction node determines the order of arguments; strictly speaking, it shows which of two states is initial and which is final. Plus corresponds to the anticlockwise reading of arguments, and minus to the clockwise one. Therefore, a sign reversal means conversion from the rotation \(R\) to the inverse rotation \(R^{-1}\).

Matrix elements of any irreducible tensor operator\index{irreducible tensor!diagram of}\index{block}\index{subdiagram} \(\braOket{\Psi}{T_{\lambda \mu}}{\Psi^{\prime}}\) are represented by a block with three external lines corresponding to \(\bra{\Psi}\), \(ket{\Psi^{\prime}}\) and \(\ket{\lambda \mu}\); the operator symbol (and other possible quantum numbers) are indicated inside the block.

\begin{table}[ptbh!]
\begin{center}
\captionsetup{justification=centering}%,labelformat=empty}
\caption{Basic Elements of Diagrams.}\label{tab:11.1}
\renewcommand{\arraystretch}{1.5}
\begin{tabular}{@{}p{0.65\textwidth}c@{}}
\hline
{Analytical Expression} & Graphical Representation \\
\hline
\multicolumn{2}{c}{\(jm\)-State Vector}\\
``ket''  \( \begin{aligned} & \ket{j m} \\ & \rket{j m} \end{aligned} \) & \includegraphics[valign=c]{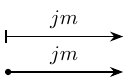}
\\
``bra''  \( \begin{aligned} & \bra{j m} \\ & \rbra{j m} \end{aligned} \) & 
\includegraphics[valign=c]{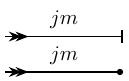}  \\
\multicolumn{2}{c}{Kronecker \(\delta\)-Symbol}\\
\( \braket{j m}{j^{\prime} m^{\prime}}=\delta_{j j^{\prime}} \delta_{m m^{\prime}}\) &
\includegraphics[valign=c]{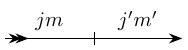}
 \\
\( \rbraket{j m}{j^{\prime} m^{\prime}}=\frac{\delta_{j j^{\prime}} \delta_{m m^{\prime}}}{2 j+1}
\) &
\includegraphics[valign=c]{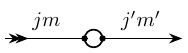}
 \\
\multicolumn{2}{c}{\(\Omega\)-State Vector}\\
``ket''  \( \begin{aligned} & \ket{\Omega} \\ & \rket{\Omega} \end{aligned} \) & \includegraphics[valign=c]{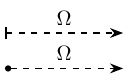}
\\
``bra''  \( \begin{aligned} & \bra{\Omega} \\ & \rbra{\Omega} \end{aligned} \) & 
\includegraphics[valign=c]{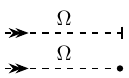}  \\
\multicolumn{2}{c}{Dirac \(\delta\)-function}\\
\( \braket{\Omega}{\Omega^{\prime}}=\delta(\Omega-\Omega')\) &
\includegraphics[valign=c]{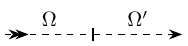}
 \\
\( \rbraket{\Omega}{\Omega^{\prime}}={\delta(\Omega-\Omega')}{4 \pi}
\) &
\includegraphics[valign=c]{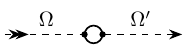}
 \\\( \braket{\Omega _1\Omega_2}{\Omega}=\delta(\Omega_1-\Omega)\delta(\Omega_2-\Omega)\) &
\includegraphics[valign=c]{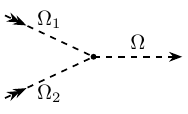}
 \\
 \multicolumn{2}{c}{Spherical Harmonics}\\
\( \braket{\Omega}{lm}=Y_{lm}(\Omega)\) &
\includegraphics[valign=c]{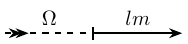}
 \\
 \( \braket{lm}{\Omega}=Y^*_{lm}(\Omega)\) &
\includegraphics[valign=c]{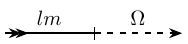}
 \\
\( \rbraket{\Omega}{lm}=C_{lm}(\Omega)\) &
\includegraphics[valign=c]{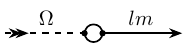}
 \\
\( \rbraket{lm}{\Omega}=C_{lm}^*(\Omega)\) & 
\includegraphics[valign=c]{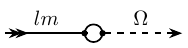}
 \\
\multicolumn{2}{c}{Clebsch-Gordan Coefficients}\\
\( \clebsch{j_1}{m_1}{j_2}{m_2}{J}{M}=\braket{j_1 m_1 j_2 m_2}{j_1 j_2 J M}=\braket{j_1j_2JM}{j_1 m_1 j_2 m_2}\) &
\includegraphics[valign=c]{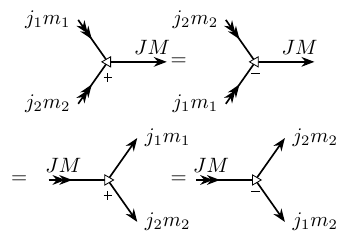}
\end{tabular}
\end{center}
\end{table}

\begin{table}[ptbh!]
\begin{center}
\captionsetup{justification=centering,labelformat=empty}
\caption*{Table \ref{tab:11.1}. (Cont'd).}
\renewcommand{\arraystretch}{1.5}
\begin{tabular}{@{}p{0.65\textwidth}c@{}}
\hline
{Analytical Expression} & Graphical Representation \\
\hline
\multicolumn{2}{c}{Wigner \(3jm\) Symbols}\\
\( \threej{j_1}{n_1}{j_2}{m_2}{j_3}{m_3}=\rbraket{j_1 m_1 j_2 m_2 j_3 m_3}{00}=\rbraket{00}{j_1 m_1 j_2 m_2 j_3 m_3}\) &
\includegraphics[valign=c]{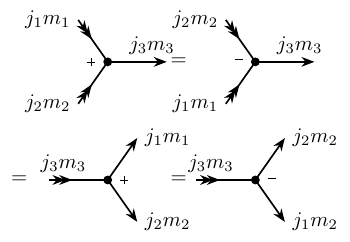}\\
\multicolumn{2}{c}{Metric tensor}\\
\(\displaystyle\binom{j}{m m^{\prime}}=(-1)^{j+m} \delta_{m,-m^{\prime}}\)&
%NOTE These next two diagrams need reviewing!
\includegraphics[valign=c]{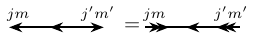} \\
\(\displaystyle\binom{j}{m^{\prime} m}=(-1)^{j-m} \delta_{-m, m^{\prime}}\)&
\includegraphics[valign=c]{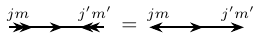} \\
\hline
\end{tabular}
\end{center}
\end{table}

\begin{table}[tbph!]
\begin{center}
\captionsetup{justification=centering}%,labelformat=empty}
\caption{Invariant Functions.}\label{tab:11.2}
\renewcommand{\arraystretch}{1.5}
\begin{tabular}{@{}p{0.65\textwidth}c@{}}
%\begin{tabular}{lc}
\hline
{Analytical Expression} & Graphical Representation \\
\hline
\((2 a+1)\) & \includegraphics[valign=t]{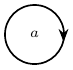}\\[1em]
\(\sqrt{2 a+1}\)&
\includegraphics[valign=t]{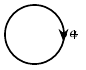}\\[1em]
\(\delta_{a b} \)&
\includegraphics[valign=t]{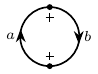}\\[1em]
\(\displaystyle\frac{\delta_{a b} \delta_{bc}}{\sqrt{2 a+1}}\)&
\includegraphics[valign=t]{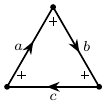}\\[1em]
\(\displaystyle\frac{\delta_{a} \delta_{b a} \delta_{c d}}{(2 a+1)}\)&
\includegraphics[valign=t]{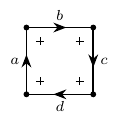}\\
\multicolumn{2}{c}{\(3 j\) Symbol}\\
\(\{a b c\}\)&
\includegraphics[valign=t]{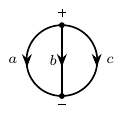}\\
\multicolumn{2}{c}{\(6 j\) Symbol} \\
\(\sixj{a}{b}{c}{A}{B}{C}\) & \includegraphics[valign=t]{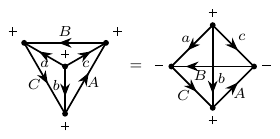} \\
\multicolumn{2}{c}{\(9 j\) Symbol} \\
\(\ninej{a_1}{a_2}{a_3}{b_1}{b_2}{b_3}{c_1}{c_2}{c_3}\) & \includegraphics[valign=t]{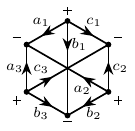}
 \\
 \multicolumn{2}{c}{\(12 j\)(I) Symbol} \\
 \(\twelvejI{a_1,a_2,a_3,a_4,b_1,b_2,b_3,b_4,c_1,c_2,c_3,c_4}\)& \includegraphics[valign=c]{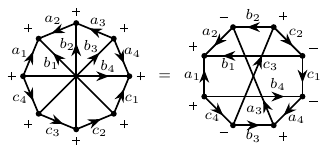}
 \\
 \multicolumn{2}{c}{\(12 j\)(II) Symbol} \\
\(\twelvejII{c2,b_3,a_4,c_3,a_2,b_4,b_2,a_3,c_4,a_1,b_1,c_1}\)& \includegraphics[valign=c]{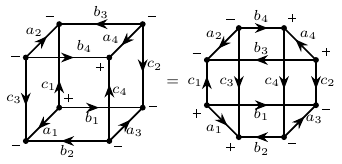}
 \\
 \hline
\end{tabular}
\end{center}
\end{table}
\begin{table}[ptbh!]
\begin{center}
\captionsetup{justification=centering}%,labelformat=empty}
\caption{Matrix Elements.}\label{tab:11.3}
\renewcommand{\arraystretch}{1.3}
\begin{tabular}{@{}m{0.7\textwidth}m{0.3\textwidth}@{}}
\hline
{Analytical Expression} & Graphical Representation \\
\hline
\multicolumn{2}{c}{Irreducible Tensor Operator}\\
\(\Omega\)-Representation 
\[\braOket{\gamma \Omega}{\mathfrak{M}_{\lambda \mu}}{\gamma^{\prime} \Omega^{\prime}}\] 
\(\gamma\) denotes all quantum numbers except \(\Omega\)
& 
\includegraphics[valign=t]{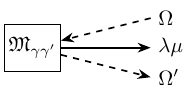}\\
Adjoint Matrix \[\braOket{\gamma \Omega}{\mathfrak{M}_{\lambda \mu}^{\dagger}}{\gamma^{\prime} \Omega^{\prime}}=(-1)^{\lambda-\mu}\braOket{\gamma'\Omega'}{\mathfrak{M}_{\lambda-\mu}}{\gamma\Omega}^*\] & 
\includegraphics[valign=t]{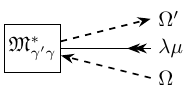}\\
\(jm\)-Representation
 \[\braOket{\gamma j m}{\mathfrak{M}_{\lambda \mu}}{\gamma^{\prime} j^{\prime} m^{\prime}}\] 
\(\gamma\) denotes all quantum numbers except \(jm\)
& 
\includegraphics[valign=t]{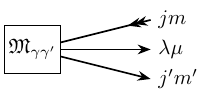}\\
Adjoint Matrix
 \[\braOket{\gamma j m}{\mathfrak{M}_{\lambda \mu}^{\dagger}}{\gamma^{\prime} j^{\prime} m^{\prime}}=(-1)^{\lambda-\mu}\braOket{\gamma'j'm'}{\mathfrak{M}_{\lambda-\mu}}{\gamma jm}^*\] & 
\includegraphics[valign=t]{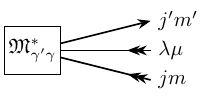}\\ 
Reduced Matrix Element
\textcolor{red}{Here there are some subtleties--the book uses both sharp and round bras and kets. The mathematical relation holds for both--it is equivalent to the one from Edmunds. However, the graphical method works most simply with round brackets--the vertical lines for the sharp brackets are hard to see, and the diagram shown is the one for that case. This has been corrected from the book. }
\[
\begin{aligned}
% PROOFREAD: reduced m.e. bracket convention -- book prints round parens (a||O||b); \braOketred renders angle. Kept angle here per instruction; revisit vs the round \rbraOketred used in Tables 11.14/11.15.
\rbraOketred{\gamma j}{\mathfrak{M}_{\lambda}}{\gamma^{\prime} j^{\prime}}= & \sum_{m \mu m^{\prime}}\frac{\braOket{\gamma j m}{\mathfrak{M}_{\lambda \mu}}{\gamma^{\prime} j^{\prime} m^{\prime}}}{\sqrt{(2j+1)(2j'+1)}}(-1)^{j-m} \\
& \times\threej{j}{\lambda}{j^{\prime}}{-m}{\mu}{m^{\prime}}
\end{aligned}
\] &
\includegraphics[valign=t]{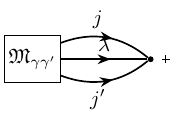}\\
\multicolumn{2}{c}{Rotation Operator} \\
\(\Omega\) representation
\[\braOket{\gamma'\Omega'}{D(R)}{\gamma\Omega}=\delta_{\gamma\gamma'}(\Omega'-R\Omega)\]
&
\includegraphics[valign=t]{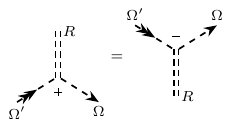}
\\
\(jm\)-Representation
\[\braOket{\gamma j m}{D(R)}{\gamma^{\prime} j^{\prime} m^{\prime}}=\delta_{\gamma \gamma^{\prime}} \delta_{j j^{\prime}} D_{m m^{\prime}}^{j}(R)\] 
& \includegraphics[valign=t]{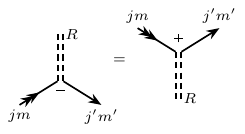}
 \\
Inverse Rotation Matrix   
\[ \begin{gathered} \braOket{\gamma j m}{D^{\dagger}(R)}{\gamma^{\prime} j^{\prime} m^{\prime}}=\delta_{\gamma \gamma^{\prime}} \delta_{j j^{\prime}} D_{m^{\prime} m}^{j *}(R) \\ =\braOket{\gamma j m}{D^{-1}(R)}{\gamma^{\prime} j^{\prime} m^{\prime}}=\delta_{\gamma \gamma^{\prime}} \delta_{j j^{\prime}} D_{m m^{\prime}}^{j}\left(R^{-1}\right)\end{gathered} \] & 
\includegraphics[valign=t]{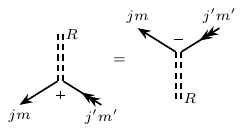}
 \\
\hline
\end{tabular}
\end{center}
\end{table}

\section{Graphical representation of the main operations of the theory}\label{sec:11.2}\index{graphical method!representation of operations}
Expressions which occur in calculations based on the quantum theory of angular momentum have a specific character. They contain spin angular variables of two kinds, i.e., external and internal ones.
\begin{enumerate}
\item External variables are actual arguments of the expression under consideration.\index{variables!external}\index{variables!internal} The evaluation of an expression does not involve any integration or summation over external variables, i.e., they appear as fixed (but, in principle, changeable) parameters.
\item Internal variables are those over which integration or summation is assumed. It is essential that each internal variable which enters an expression inherent in the theory corresponds to some scalar product \(\ket{\Psi}\bra{\Psi}\). This is why one does not need to take into account any transformation of internal variables under rotation of the coordinate system. This specific feature is important for practical calculations.
\end{enumerate}
In other words, quantum mechanical problems are mainly concerned with invariant integration and/or summation of bilinear forms over total ranges of all internal variables. Accordingly, the graphical method under consideration may be applied only to the expressions of such a type.

This section describes the graphical representation of those analytic operations which are specific to angular momentum theory, while the rules of calculations by the graphical method will be formulated in the next section.

\subsection{Multiplication}\label{sec:11.2.1}\index{multiplication!graphical}\index{diagrams!multiplication of}
The product (direct product) of factors \normalsize\(\mathfrak{M}\) and \normalsize\(\mathfrak{N}\) is represented graphically by unlinked subdiagrams (blocks) which correspond to factors \normalsize\(\mathfrak{M}\) and \normalsize\(\mathfrak{N}\). The mutual disposition and orientation of these subdiagrams are inessential. For example, the product
\begin{equation}
\threej{A}{B}{C}{\alpha}{\beta}{\gamma}\sixj{A}{B}{C}{a}{b}{c}
\end{equation}
may be represented by\\
\begin{center}
  \includegraphics[valign=t]{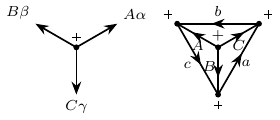}
\end{center}

\subsection{Invariant Summation over Projections}\label{sec:11.2.2}\index{invariant summation!over projections}\index{summation!over projections}
The scalar product of two irreducible tensors of rank \(j\), \normalsize\(\mathfrak{M}_{j m}\) and \normalsize\(\mathfrak{N}_{j m}\), i.e., the sum of bilinear combinations over all possible values of the projection \(m\)
\begin{equation}
\sum_{m=-j}^{j}\braOket{}{\mathfrak{M}}{j m}\braOket{j m}{\mathfrak{N}}{},
\end{equation}
is represented by the conjunction (closure) of \(\ket{j m}\) - and \(\rbra{j m}\)-lines of subdiagrams corresponding to \(\braOket{}{\mathfrak{M}}{j m}\) and \(\braket{j m}{\mathfrak{N}}\)\\
\begin{center}
\includegraphics[valign=t]{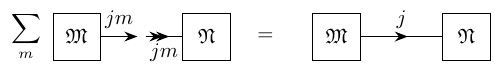}
\end{center}

The resultant linking \(j\)-line has the same direction as the original \(\ket{j m}\) - and \(\bra{j m}\)-lines.\\
For example, the sum of product of two \(3jm\) symbols over two projections, \(\alpha\) and \(\beta\)
\begin{equation}
\sum_{\alpha, \beta}\threej{a}{b}{c}{\alpha}{\beta}{\gamma}\threej{a}{b}{c^{\prime}}{\alpha}{\beta}{\gamma^{\prime}}=\sum_{\alpha, \beta}(-1)^{a-\alpha+b-\beta+c-\gamma}\threej{a}{b}{c}{-\alpha}{-\beta}{-\gamma}\threej{a}{b}{c^{\prime}}{\alpha}{\beta}{\gamma^{\prime}}
\end{equation}
may be displayed as\\
\centerline{
      \includegraphics[valign=t]{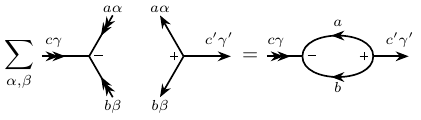}
    }

Thus, each internal \(j\)-line of a diagram represents a scalar product (contraction over \(m\) ) of two irreducible tensors of rank \(j\). To construct such a scalar product one of these tensors should be written in a covariant form \(\braOket{}{\mathfrak{M}}{j m} \equiv \mathfrak{M}_{j m}\) while the other should be written in the form \(\braOket{j m}{\mathfrak{N}}{}=(-1)^{j-m} \mathfrak{N}_{j-m}\) (Sec.~\ref{sec:3.1.8}). Graphically, this means that one of the corresponding subdiagrams has the external \(\ket{j m}\)-line (with single outward arrow) and the other has the  \(\bra{j m}\)-line (\emph{with}  double inward arrow).

Two diagrams which differ only by the direction of an internal \(j\)-line, correspond to analytic expressions which differ only by the phase factor \((-1)^{2 j}\), because
\begin{equation}
\sum_{m}(-1)^{j-m} \mathfrak{M}_{j m} \mathfrak{N}_{j-m}=(-1)^{2 j} \sum_{m}(-1)^{j-m} \mathfrak{M}_{j-m} \mathfrak{N}_{j m} .
\end{equation}
One should especially mention the entirely invariant sums of products of two, four, six, or more \(3 j m\) symbols over all projections. Corresponding diagrams are closed with respect to all \(j\)-lines. These sums represent the Wigner \(3 n j\) symbols (Table~\ref{tab:11.4}).

\begin{table}[tbhp!]
\begin{center}
\captionsetup{justification=centering}%,labelformat=empty}
\renewcommand{\arraystretch}{1.5}
\caption{Sums of \(3 j m\) Symbol Products.}\label{tab:11.4}
\small 
\begin{tabular}{@{}p{0.8\textwidth}>{\centering\arraybackslash}m{0.2\textwidth}@{}}
\hline
{Analytical Expression} & Graphical Representation \\
\hline
\multicolumn{2}{c}{\(3j\) Symbol}\\
\[
\sum_{\mu_{i}}\threej{j_{1}}{j_{2}}{j_{3}}{\mu_{1}}{\mu_{2}}{\mu_{3}}\threej{j_{1}}{j_{2}}{j_{3}}{\mu_{1}}{\mu_{2}}{\mu_{3}}=\sum_{\mu_{i}}(-1)^{j_{1}-\mu_{1}+j_{2}-\mu_{2}+j_{3}-\mu_{3}}\threej{j_{1}}{j_{2}}{j_{3}}{-\mu_{1}}{-\mu_{2}}{-\mu_{3}}\threej{j_{1}}{j_{2}}{j_{3}}{\mu_{1}}{\mu_{2}}{\mu_{3}}=\left\{j_{1} j_{2} j_{3}\right\}
\] &
\vspace*{2em}\includegraphics[valign=t]{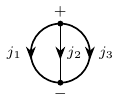}
\\
\multicolumn{2}{c}{6j Symbol}\\
\[
\begin{gathered}
\sum_{\mu_{i} x_{i}}(-1)^{k_{1}-x_{1}+k_{2}-x_{2}+k_{3}-x_{3}}\threej{j_{1}}{j_{2}}{j_{3}}{\mu_{1}}{\mu_{2}}{\mu_{3}}\threej{j_{1}}{k_{2}}{k_{3}}{\mu_{1}}{-x_{2}}{x_{3}}\threej{k_{1}}{j_{2}}{k_{3}}{x_{1}}{\mu_{2}}{-x_{3}}\threej{k_{1}}{k_{2}}{j_{3}}{-x_{1}}{x_{2}}{\mu_{3}} \\
=\sum_{\mu_{i} x_{i}}(-1)^{j_{1}-\mu_{1}+j_{2}-\mu_{2}+j_{3}-\mu_{3}+k_{1}-x_{1}+k_{2}-x_{2}+k_{3}-x_{3}}\threej{j_{1}}{j_{2}}{j_{3}}{\mu_{1}}{\mu_{2}}{\mu_{3}}\threej{j_{1}}{k_{2}}{k_{3}}{-\mu_{1}}{-x_{2}}{x_{3}}\threej{k_{1}}{j_{2}}{k_{3}}{x_{1}}{-\mu_{2}}{-x_{3}} \\
\times\threej{k_{1}}{k_{2}}{j_{3}}{-x_{1}}{x_{2}}{-\mu_{3}}=\sixj{j_{1}}{j_{2}}{j_{3}}{k_{1}}{k_{2}}{k_{3}}
\end{gathered}
\]&
\includegraphics[valign=t]{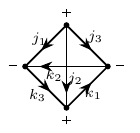}\par
\includegraphics[valign=t]{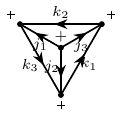}\\
\multicolumn{2}{c}{9j Symbol}\\
\[
\begin{gathered}
\sum_{\mu_{i} x_{i} \lambda_{i}}\threej{l_{1}}{j_{2}}{l_{3}}{\mu_{1}}{\mu_{2}}{\mu_{3}}\threej{k_{1}}{k_{2}}{k_{3}}{x_{1}}{x_{2}}{x_{3}}\threej{l_{1}}{l_{2}}{l_{3}}{\lambda_{1}}{\lambda_{2}}{\lambda_{3}}\threej{j_{1}}{k_{1}}{l_{1}}{\mu_{1}}{x_{1}}{\lambda_{1}}\threej{j_{2}}{k_{2}}{l_{2}}{\mu_{2}}{x_{2}}{\lambda_{2}}\threej{j_{3}}{k_{3}}{l_{3}}{\mu_{3}}{x_{3}}{\lambda_{3}} \\
=\sum_{\mu_{3} x_{i} \lambda_{i}}(-1)^{j_{1}-\mu_{1}+j_{2}-\mu_{2}+j_{3}-\mu_{2}+h_{1}-x_{1}+k_{2}-x_{3}+\mu_{3}-x_{3}+l_{1}-\lambda_{1}+l_{2}-\lambda_{2}+l_{3}-\lambda_{3}}\threej{j_{1}}{j_{2}}{j_{3}}{\mu_{1}}{\mu_{2}}{\mu_{3}}\threej{k_{1}}{k_{2}}{k_{3}}{x_{1}}{x_{2}}{x_{3}} \\
\times\threej{l_{1}}{l_{2}}{l_{3}}{\lambda_{1}}{\lambda_{2}}{\lambda_{3}}\threej{j_{1}}{k_{1}}{l_{1}}{-\mu_{1}}{-x_{1}}{-\lambda_{1}}\threej{j_{2}}{k_{2}}{l_{2}}{-\mu_{2}}{-x_{2}}{-\lambda_{2}}\threej{j_{3}}{k_{3}}{l_{3}}{-\mu_{3}}{-x_{3}}{-\lambda_{3}}=\ninej{j_{1}}{j_{2}}{j_{3}}{k_{1}}{k_{2}}{k_{3}}{l_{1}}{l_{2}}{l_{3}}
\end{gathered}
\]
&
\includegraphics[valign=t]{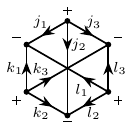}
\\
\hline
\end{tabular}
\end{center}
\end{table}

\subsection{Summation over Angular Momentum}\label{sec:11.2.3}\index{summation!over angular momentum}\index{thick j-line@thick $j$-line}
A thick \(j\)-line linking two subdiagrams, as in
\begin{center}
\includegraphics[valign=t]{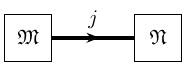}
\end{center}
represents the complete sum over \(j\) and \(m\) of scalar products of irreducible tensors of rank \(j\)
\[
\sum_{j=0}^{\infty} N_{j} \sum_{m=-j}^{j}\braOket{}{\mathfrak{M}}{j m}\braOket{j m}{\mathfrak{N}}{},
\]
where \(N_{j}\) is the weight factor; \(N_{j}=2 j+1\) for a summation involving the \(3 j m\) symbols, \(3 n j\) symbols and the\\
functions of the type \(D_{m m^{\prime}}^{j}(R), C_{j m}(\Omega) \equiv[4 \pi /(2 j+1)]^{1 / 2} Y_{j m}(\Omega)\), or \(P_{j}(\cos \omega)\). On the other hand, \(N_{j}=1\) for summations involving \(\clebsch{j_{1}}{m_{1}}{j_{2}}{m_{2}}{j}{m}, U\left(j_{1} j_{2} j_{3} j_{4} ; j j^{\prime}\right)\), or \(Y_{j m}(\Omega)\). For example,
\[
  \arraycolsep=10pt
\begin{array}{cc}
\sum_{e \varepsilon}(2 e+1)(-1)^{e-\varepsilon}\threej{a}{b}{e}{\alpha}{\beta}{-\varepsilon}\threej{a^{\prime}}{b^{\prime}}{e}{\alpha^{\prime}}{\beta^{\prime}}{\varepsilon} 
& \vcenter{\hbox{
    \includegraphics[valign=t]{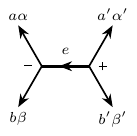}
    }
}
\\
\sum_{l m} Y_{l m}(\Omega) Y_{l m}^{*}(\Omega)=\braket{\Omega}{l m}\braket{l m}{\Omega^{\prime}} 
& 
\vcenter{\hbox{\includegraphics[valign=t]{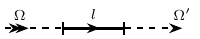}}}
\\
\sum_{C}(2 C+1)\sixj{A}{B}{C}{a}{b}{c}\sixj{A}{B}{C}{a}{b}{c^{\prime}} 
& 
\vcenter{\hbox{\includegraphics[valign=t]{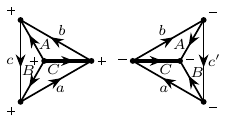}}}
\end{array}
\]
A Clebsch-Gordan coefficient may be represented as a sum over \(c^{\prime} \gamma^{\prime}\) of products of two \(3 j m\) symbols (Table~\ref{tab:11.10a})
\[
\begin{aligned}
\clebsch{a}{\alpha}{b}{\beta}{c}{\gamma} & \equiv\braket{a \alpha b \beta}{c \gamma}=(-1)^{a-b-c} \sum_{c^{\prime} \gamma^{\prime}}\left(2 c^{\prime}+1\right)\threej{a}{b}{c^{\prime}}{\alpha}{\beta}{\gamma^{\prime}}\threej{c^{\prime}}{c}{0}{\gamma^{\prime}}{\gamma}{0} \\
& =(-1)^{2 a} \sum_{c^{\prime} \gamma^{\prime}}\left(2 c^{\prime}+1\right)(-1)^{a-\alpha+b-\beta+c^{\prime}-\gamma^{\prime}}\threej{b}{a}{c^{\prime}}{-\beta}{-\alpha}{-\gamma^{\prime}}\threej{c^{\prime}}{c}{0}{\gamma^{\prime}}{\gamma}{0},
\end{aligned}
\]
\centerline{\includegraphics{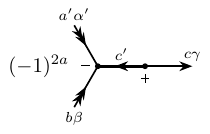}}

    \subsection{Invariant Integration over Directions}\label{sec:11.2.4}\index{invariant integration!over directions}\index{integration!over directions}
A double integral of a bilinear form
\begin{equation}
\int\braOket{}{\mathfrak{N}}{\Omega} \frac{d \Omega}{\Omega_{0}}\braket{\Omega}{\mathfrak{N}},
\end{equation}
over the total solid angle \(4 \pi(0 \leq \vartheta \leq \pi, 0 \leq \varphi<2 \pi)\) is represented by a dashed \(\Omega\) line which links \(\ket{\Omega}\) and \(\bra{\Omega}\) lines of subdiagrams corresponding to the integrand factors\footnote{Note this is not denoted in boldface!}\\
\centerline{\includegraphics[valign=t]{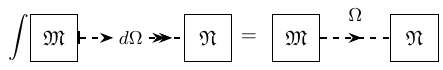}}
The resultant \(\Omega\)-line has the same direction as the original \(\ket{\Omega}\) - and \(\rbra{\Omega}\)-lines. In this case \(\Omega_{0}\) is the normalization constant; for integration of the standard spherical harmonics we have \(\Omega_{0}=1\),
\begin{equation}
\int Y_{l m}^{*}(\Omega) Y_{l^{\prime} m^{\prime}}(\Omega) d \Omega=\braket{l m}{\Omega}\braket{\Omega}{l^{\prime} m^{\prime}} ,
%\vcenter{\hbox{
  \qquad\includegraphics[valign=b]{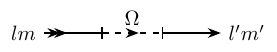}
%  }}
  \end{equation}
whereas \(\Omega_{0}=4 \pi\) for integration of the functions \(C_{l m}(\Omega)=[4 \pi /(2 l+1)]^{1 / 2} Y_{l m}(\Omega)\).

\subsection{Integration over Rotation Parameters}\label{sec:11.2.5}\index{integration!over rotation parameters}\index{Euler angles}

\begin{table}[tbhp!]
\begin{center}
\captionsetup{justification=centering}
\renewcommand{\arraystretch}{1.2}
\caption{Basic Operations of the Theory.}\label{tab:11.5}
\small 
\begin{tabular}{@{}>{\centering\arraybackslash}m{0.5\textwidth}>{\centering\arraybackslash}m{0.5\textwidth}@{}}
\hline
{Analytical Expression} & Graphical Representation \\
\hline
\multicolumn{2}{c}{Multiplication}\\
\(\displaystyle \braket{}{\mathfrak{M}}\braOket{}{\mathfrak{N}}{}\) &
  \includegraphics[valign=t]{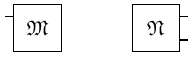}
\\
\multicolumn{2}{c}{Invariant Summation over Momentum Projection}\\
\multicolumn{2}{c}{(Scalar Product of irreducible tensors)}\\
\[
\sum_{m}\braOket{}{\mathfrak{N}}{j m}\braOket{j m}{\mathfrak{N}}{} \equiv\left(\mathfrak{N}_{j} \cdot \mathfrak{N}_{j}\right)\]
&\includegraphics[valign=c]{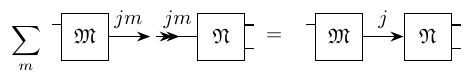}
\\
\multicolumn{2}{c}{Complete Summation over Angular Momentum and Projection}\\
\(
\begin{gathered}
\sum_{j m}\braOket{}{\mathfrak{M}}{j m} N_{j}\braOket{j m}{\mathfrak{N}}{} \\
\text { Completeness Condition } \\
\sum_{j m}\ket{j m} N_{j}\bra{ j m}=1
\end{gathered}
\)
&\includegraphics[valign=c]{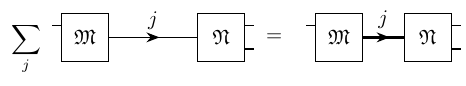}
\\
\multicolumn{2}{c}{Irreducible Tensor Product of Irreducible Tensors}\\
\(
\begin{aligned}
\sum_{m_{1} m_{2}}\braOket{}{\mathfrak{M}}{i_{1} m_{1}} & \braOket{}{\mathfrak{N}}{j_{2} m_{2}} \clebsch{j_1}{m_1}{j_2}{m_2}{J}{M} \equiv \\
& \equiv\left\{\mathfrak{M}_{j_{1}} \otimes \mathfrak{N}_{j_{2}}\right\}_{J M}
\end{aligned}
\)&
\includegraphics[valign=c]{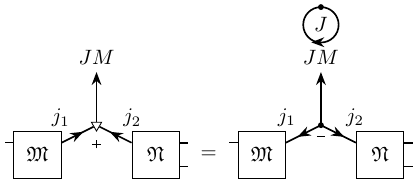}
\\
\multicolumn{2}{c}{Complete Invariant Integration over Direction Angles}\\
\[
\int\braOket{}{\mathfrak{M}}{\Omega} d \Omega\braOket{\Omega}{\mathfrak{N}}{} 
\]
Completeness Condition  
\[\int\ket{\Omega} d \Omega\bra{\Omega}=1\]
&\includegraphics[valign=c]{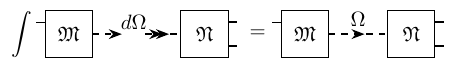}\\
\multicolumn{2}{c}{Complete Integration over Rotation Parameters}\\
\[
\int\braOket{}{\mathfrak{M}(R^{-1})}{} \frac{d R}{R_{0}}
\braOket{}{\mathfrak{N}(R)}{}
\]
Completeness Condition 
\[
\int \hat{R}^{-1} \hat{R} \frac{d R}{R_{0}}=1
\]
&\includegraphics[valign=c]{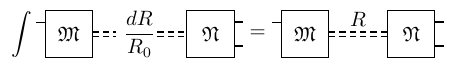}\\
\hline
\end{tabular}
\end{center}
\end{table}

The integration of a bilinear form with respect to parameters which specify the rotation \(R\) (the Euler angles \(\{\alpha, \beta, \gamma\}\) or the angles \(\{\omega, \theta, \Phi\})\) over the entire domain of their definition \(\left(R_{0} \equiv \int d R\right.\), see Sec.~\ref{sec:4.10})
\begin{equation}
\int\braOket{}{\mathfrak{M}(R)}{}\braOket{}{\mathfrak{N}\left(R^{-1}\right)}{} \frac{d R}{R_{0}}
\end{equation}
may be represented by a conjunction of corresponding \(R\) - and \(R^{-1}\)-lines of subdiagrams which display the integrand factors.\\
\centerline{\includegraphics{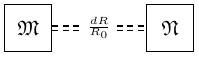}}

Note that integrals of three or more functions of \(R\) may be reduced to standard bilinear forms, using the Clebsch-Gordan expansion.

\section{Rules of the graphical technique}\label{sec:11.3}\index{graphical method!rules of}\index{diagram technique!rules of}
In this section the rules of calculation by the graphical method will be formulated, specifically, the rules for the reduction of diagrams to standard diagrams which represent the basis functions of the angular momentum theory.

\subsection{Deformation of Diagrams}\label{sec:11.3.1}\index{diagrams!deformation of}\index{diagrams!identical}\index{lines!external}\index{lines!internal}
Expressions which appear in calculations based on the theory of angular momentum can be rather complicated. Corresponding diagrams consist of many lines, thin and thick, solid and dashed, single and double. These lines are partly linked to some nodes. One can clearly distinguish two types of lines: internal and external. Internal lines have both ends linked to nodes, whereas external lines have one end linked to a node but the other free. Any identical transformation of an analytic expression corresponds to some transformation of the internal lines of the diagram with external lines remaining unchanged.

If the structure of the inner part of a diagram is unimportant, it may be replaced by a block with the same external lines, because these are just the external lines which determine the transformation properties of corresponding diagram (or sub-diagram) under rotation of the coordinate system. For instance, any closed diagram without external lines is invariant under such rotation and may be reduced to some of the \(3 n j\) symbols or their combinations.

As pointed out in Sec.~\ref{sec:11.1.1}, the length of the lines, their curvature and orientation are not important. Consequently, any diagram may be arbitrarily rotated and deformed, although the node signs have to be reversed for those nodes where the order of momentum coupling becomes changed.

Thus, any deformation of a diagram performed under the proper check for the node signs does not change the meaning of the diagram, i.e., the deformed and original diagrams represent the same expression. Such diagrams will be called identical. An example of identical diagrams is given below:\\
\[ 
\vcenter{\hbox{
    \includegraphics[valign=t]{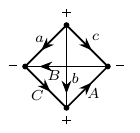}
    }
}
=
\vcenter{\hbox{
    \includegraphics[valign=t]{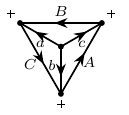}
    }
}
=
\vcenter{\hbox{
    \includegraphics[valign=t]{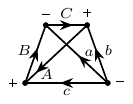}
    }
}
=
\vcenter{\hbox{
    \includegraphics[valign=t]{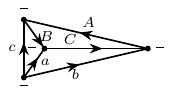}
    }
}
\]
Two diagrams will be called topologically similar, if it is possible to make them coincident by means of rotations, deformations and/or reflections. Topologically similar diagrams have the same numbers of nodes and of lines, solid and dashed, but they may differ by directions of lines and by node signs.

The expressions corresponding to two topologically similar diagrams are equal in absolute value, but may differ by a phase factor. One may find this factor by means of successive reductions of one diagram to the other, i.e., by reversals of node signs and line directions (where necessary) in accordance with the rules formulated in Secs.~\ref{sec:11.3.2} and \ref{sec:11.3.3}. In this way one may reduce any diagram to the topologically similar diagram of some basis function of the theory, and determine the additional phase factor. Several examples of such transformation for the \(3 j\), \(6 j\), and \(9 j\) symbols are displayed in Table~\ref{tab:11.6}.

\begin{table}[tbhp!]
\begin{center}
\captionsetup{justification=centering}
\caption{Symmetry of \(3nj\) Symbol Diagrams.}\label{tab:11.6}
\label{chap11:tab:2}
\renewcommand{\arraystretch}{1.5}
\begin{tabular}{c}
  \hline
\(3j\) Symbol \(\{a b c\}\)\\[2pt]
Reversal of Node Sign and Line Direction, and Argument Permutation\\[4pt]
\begin{tabular}{ccccc}
\includegraphics[valign=c]{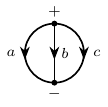} & 
\(=(-1)^{a+b+c}\) & \includegraphics[valign=c]{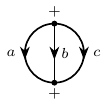} 
& \(=(-1)^{a+b+c}\) & \includegraphics[valign=c]{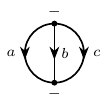}\;\(=\) \\[1.5em]
\includegraphics[valign=c]{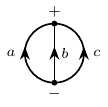} & \(=\) 
& \includegraphics[valign=c]{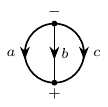} & \(=(-1)^{2a}\) & \includegraphics[valign=c]{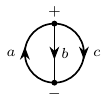}\end{tabular}\\
\(6j\) Symbol \(\sixj{a}{b}{c}{A}{B}{C}\)\\[6pt]
\multicolumn{1}{l}{Reversal of Node Sign}\\[4pt]
\includegraphics[valign=c]{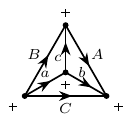} \(=(-1)^{2(A+B+C)}\)
\includegraphics[valign=c]{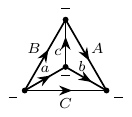} \(=(-1)^{a+b+c}\)
\includegraphics[valign=c]{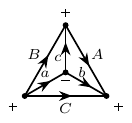}\\[10pt]
\multicolumn{1}{l}{Reversal of Line Direction}\\[4pt]
\includegraphics[valign=c]{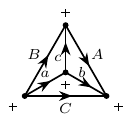} \(=(-1)^{2(A+B+C)}\)
\includegraphics[valign=c]{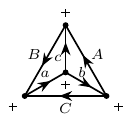} \(=\)
\includegraphics[valign=c]{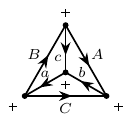} \(=(-1)^{2(A+B+C)}\)
\includegraphics[valign=c]{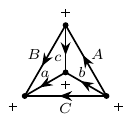}\\
\begin{tabular}{cccc}
\multicolumn{4}{c}{Argument Permutation}\\[4pt]
\includegraphics[valign=c]{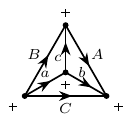} &\(=\)
\includegraphics[valign=c]{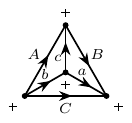} &\(=\)
\includegraphics[valign=c]{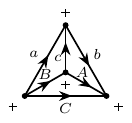} &\(=\)
\includegraphics[valign=c]{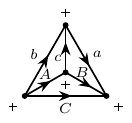} \(=\)\\[1.5em]
\includegraphics[valign=c]{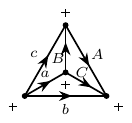} &\(=\)
\includegraphics[valign=c]{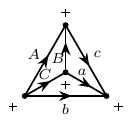} &\(=\)
\includegraphics[valign=c]{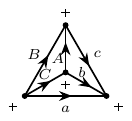} &\(=\)
\includegraphics[valign=c]{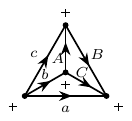}
\end{tabular}
\end{tabular}
\end{center}
\end{table}

\begin{table}[tbhp!]
\begin{center}
\captionsetup{justification=centering}
\caption*{Table \ref{tab:11.6}. (Cont'd)}
\renewcommand{\arraystretch}{1.5}
\begin{tabular}{cccc}
  \hline
  \multicolumn{4}{c}{
\(9j\) Symbol \(\ninej{a_{1}}{a_{2}}{a_{3}}{c_{1}}{c_{2}}{c_{3}}{b_{1}}{b_{2}}{b_{3}}\)}\\[6pt]
\multicolumn{4}{l}{Reversal of Node Sign}\\[4pt]
\includegraphics[valign=c]{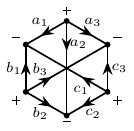} &\(=\)
\includegraphics[valign=c]{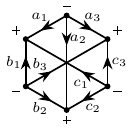} &\(=(-1)^{J_0}\)
\includegraphics[valign=c]{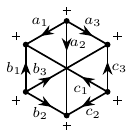} &\(=(-1)^{J_0}\)
\includegraphics[valign=c]{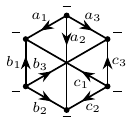}\\[10pt]
\multicolumn{4}{l}{Reversal of Line Direction}\\[4pt]
\includegraphics[valign=c]{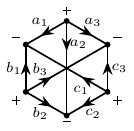} &\(=\)
\includegraphics[valign=c]{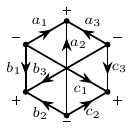} &\(=(-1)^{2(a_2+b_3+c_1)}\)
\includegraphics[valign=c]{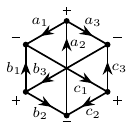} &\(=(-1)^{2(a_2+b_3+c_1)}\)
\includegraphics[valign=c]{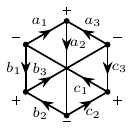}\\
  \multicolumn{4}{l}{Argument Permutation}\\[4pt]
\includegraphics[valign=c]{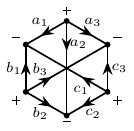} &\(=\)
\includegraphics[valign=c]{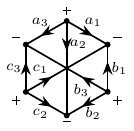} &\(=(-1)^{J_0}\)
\includegraphics[valign=c]{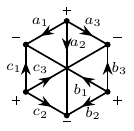} &\(=(-1)^{J_0}\)
\includegraphics[valign=c]{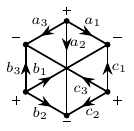}\\[1.5em]
\includegraphics[valign=c]{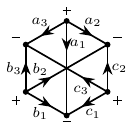} &\(=\)
\includegraphics[valign=c]{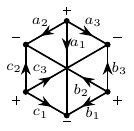} &\(=(-1)^{J_0}\)
\includegraphics[valign=c]{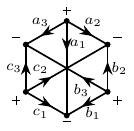} &\(=(-1)^{J_0}\)
\includegraphics[valign=c]{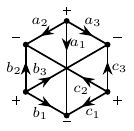}\\[1.5em]
\includegraphics[valign=c]{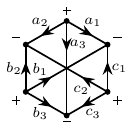} &\(=\)
\includegraphics[valign=c]{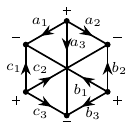} &\(=(-1)^{J_0}\)
\includegraphics[valign=c]{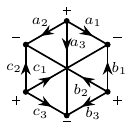} &\(=(-1)^{J_0}\)
\includegraphics[valign=c]{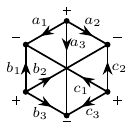}\\[6pt]
\multicolumn{4}{l}{where \(J_{0} \equiv a_{1}+a_{2}+a_{3}+b_{1}+b_{2}+b_{3}+c_{1}+c_{2}+c_{3}\)}\\
\hline
\end{tabular}
\end{center}
\end{table}

\subsection{Change of Node Sign}\label{sec:11.3.2}\index{node!change of sign}\index{phase factor!from node-sign reversal}
The reversal of the sign in a node joining \(\left(j_{1} j_{2} j_{3}\right)\)-lines corresponds to the change of the momentum-coupling order. This transformation is equivalent to introducing the additional phase factor \((-1)^{j_{1}+j_{2}+j_{3}}\). For example,\\
\centerline{\includegraphics[valign=c]{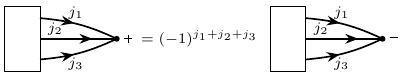}}
The reversal of signs of all the nodes in a diagram produces an additional phase factor of \((-1)^{2 J_{0}+J_{1}}\), where \(J_{0}\) and \(J_{1}\) are the algebraic sums of all the momenta for the internal and external lines, respectively. For example,\\
\centerline{\includegraphics{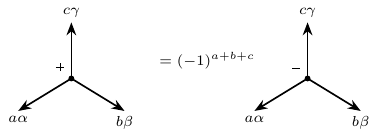}}\\
\centerline{\includegraphics{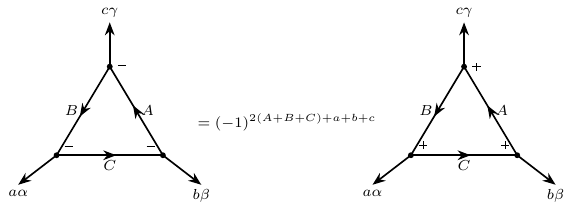}}

\subsection{Change of Direction of External Lines}\label{sec:11.3.3}\index{lines!change of direction of external}\index{metric tensor}\isym{jmm-metric@$\binom{j}{m m^{\prime}}$, metric tensor of the $(2j+1)$-dimensional space}
The direction of an external \(j\)-line specifies the form (covariant or contravariant, i.e. \(\ket{j m}\) or \(\bra{ j m})\) of an irreducible tensor of rank \(j\) (Sec.~\ref{sec:11.1.1}). The conversion of covariant components of a tensor into contravariant ones, as well as the inverse transformation, is carried out by contraction (over \(m\) ) of the tensor in question with the tensor \(\binom{j}{m m^{\prime}}=(-1)^{j+m} \delta_{m-m^{\prime}}\), which may be considered as the metric tensor for the \((2 j+1)\) dimensional space of the functions \(\ket{j m}\). Such a transformation reads
\begin{equation}
\begin{gathered}
\sum_{m^{\prime}}\braOket{}{\mathfrak{M}}{j m^{\prime}}\binom{ j}{m^{\prime} m}=\braOket{j m}{\mathfrak{M}}{}
\end{gathered}
\end{equation}
\centerline{\includegraphics[valign=t]{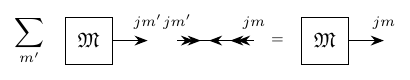}}
and the inverse transformation is
\begin{equation}
\begin{gathered}
\sum_{m^{\prime}}\binom{j}{m m^{\prime}}\braOket{j m^{\prime}}{\mathfrak{M}}{}=\braOket{}{\mathfrak{M}}{j m} 
\end{gathered}
\end{equation}
\centerline{\includegraphics[valign=t]{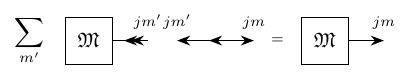}}
Thus, a reversal of the direction of any external \(j\)-line ( \(j m\) being fixed) corresponds to the transformation
\begin{equation}
\braOket{}{\mathfrak{M}}{j m} \equiv \mathfrak{M}_{j m} \rightleftarrows\braOket{j m}{\mathfrak{M}}{} \equiv(-1)^{j-m} \mathfrak{M}_{j-m},
\end{equation}
in the associated analytic expression. In other words, the replacement of a single outward arrow by a double inward one corresponds to the introduction of the phase factor \((-1)^{j-m}\), and substitution of \(-m\) for \(m\), and vice versa.

For example,
\begin{equation}
\sum_{\alpha^{\prime}}\binom{a}{\alpha^{\prime} \alpha}\threej{a}{b}{c}{\alpha^{\prime}}{\beta}{\gamma}=(-1)^{a-\alpha}\threej{a}{b}{c}{-\alpha}{\beta}{\gamma},
\end{equation}
and, conversely,
\begin{equation}
\sum_{\alpha^{\prime}}\binom{a}{\alpha \alpha^{\prime}}(-1)^{a-\alpha^{\prime}}\threej{a}{b}{c}{-\alpha^{\prime}}{\beta}{\gamma}=\threej{a}{b}{c}{\alpha}{\beta}{\gamma} .
\end{equation}
\subsection{Change of Direction of Integral Lines}\label{sec:11.3.4}\index{lines!change of direction of internal}\index{phase factor!from line reversal}
The reversal of the direction of any internal \(j\)-line produces the phase factor \((-1)^{2 j}\) in the associated analytic expression, because
\begin{equation}
\sum_{m}(-1)^{j-m} \mathfrak{M}_{j m} \mathfrak{N}_{j-m}=(-1)^{2 j} \sum_{m}(-1)^{j-m} \mathfrak{M}_{j-m} \mathfrak{N}_{j m} .
\end{equation}
For a closed diagram which represents a \(3 n j\) symbol, reversal of the directions of the \(j\)-lines produces the factor \((-1)^{2 J}\) in the associated expression, where \(J\) is the algebraic sum of all the momenta of the \(3 n j\) symbol (Table~\ref{tab:11.6}).

The reversal of the directions of three internal \(j\)-lines ( \(j_{1} j_{2} j_{3}\) ) coupled together in a node does not produce any additional phase factor, since \((-1)^{2\left(j_{1}+j_{2}+j_{3}\right)}=1\). For example,

\centerline{\includegraphics[valign=t]{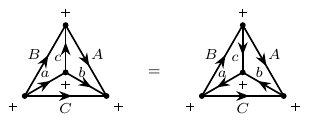}}

For a closed diagram associated with a \(3 n j\) symbol the reversal of directions of all the \(j\)-lines and simultaneous reversal of signs of all the nodes do not change the meaning of the expression.

\subsection{Linking Subdiagrams}\label{sec:11.3.5}\index{subdiagram!linking of}\index{diagrams!linking of}
In this Section the graphical rules for the multiplication of subdiagrams are formulated. These are the rules for linking the subdiagrams (blocks) into one united diagram.
\begin{enumerate}[label=(\alph*)]
  \item \label{schemea}
Two subdiagrams representing factors \normalsize\(\mathfrak{M}\) and \normalsize\(\mathfrak{N}\), which have at least one identical (for both subdiagrams) \(j\)-line, may be linked together into the combined diagram which displays the product of these factors:

\centerline{\includegraphics[valign=t]{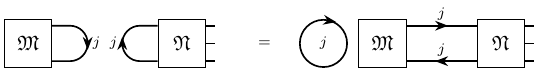}}

The necessary condition for such a linking is that at least one of these subdiagrams should have no external line. The graphical rule is equivalent to the relation
\begin{equation}
\left(\sum_{m}\braOket{j m}{\mathfrak{M}}{j m}\right) \times\left(\sum_{m^{\prime}}\braOket{j m^{\prime}}{\mathfrak{M}}{j m^{\prime}}\right)=(2 j+1) \sum_{m m^{\prime}}\braOket{j m}{\mathfrak{M}}{j m^{\prime}}\braOket{j m^{\prime}}{\mathfrak{M}}{j m} .
\end{equation}
\item \label{schemeb}
Two subdiagrams representing the factors \normalsize\(\mathfrak{N}\) and \normalsize\(\mathfrak{N}\) which have at least one identical (for both subdiagrams) node, i.e., three coupled \(j\)-lines ( \(j_{1} j_{2} j_{3}\) ) may be linked together into the combined diagram which corresponds to the product of these factors:

\centerline{\includegraphics[valign=t]{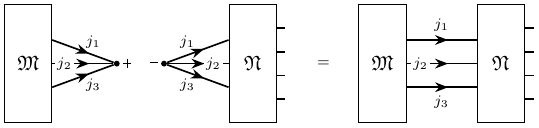}}

This is equivalent to the relation
\begin{equation}
\begin{gathered}
{\left[\sum_{m_{1} m_{2} m_{3}}\braOket{00}{\mathfrak{M}}{j_{1} m_{1} j_{2} m_{2} j_{3} m_{3}}\threej{j_{1}}{j_{2}}{j_{3}}{m_{1}}{m_{2}}{m_{3}}\right] \times\left[\sum_{m_{1}^{\prime} m_{3}^{\prime} m_{3}^{\prime}}\threej{j_{1}}{j_{2}}{j_{3}}{m_{1}^{\prime}}{m_{2}^{\prime}}{m_{3}^{\prime}}\braOket{j_{1} m_{1}^{\prime} j_{2} m_{2}^{\prime} j_{3} m_{3}^{\prime}}{\mathfrak{M}}{}\right]} \\
=\left\{j_{1} j_{2} j_{3}\right\} \sum_{m_{1} m_{3} m_{3}}\braOket{00}{\mathfrak{M}}{j_{1} m_{1} j_{2} m_{2} j_{3} m_{3}}\braOket{j_{1} m_{1} j_{2} m_{2} j_{3} m_{3}}{\mathfrak{M}}{}
\end{gathered}
\end{equation}
For example,\\
\centerline{\includegraphics[valign=t]{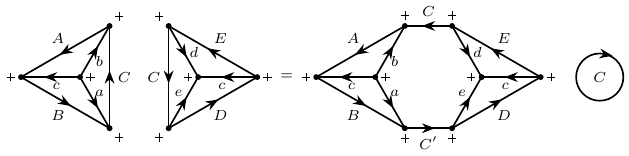}}
\centerline{\includegraphics[valign=t]{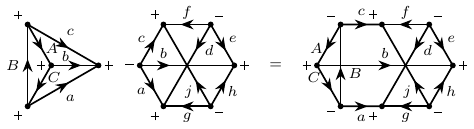}}

As in Scheme \ref{schemea}, at least one of these subdiagrams should have no external lines. However, in the other subdiagrams the joined \(j\)-lines may be external. For instance,\\
\centerline{\includegraphics[valign=t]{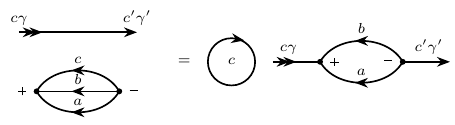}}
\centerline{\includegraphics[valign=t]{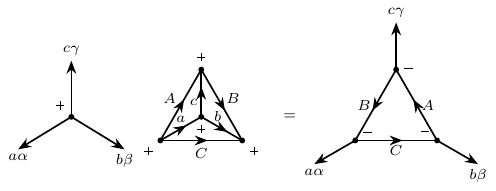}}
\end{enumerate}
Note that Scheme \ref{schemea} is a special case of Scheme \ref{schemeb} when one of the momenta \(j_{1}, j_{2}, j_{3}\) equals zero.

\subsection{Cutting Diagram into Subdiagrams}\label{sec:11.3.6}\index{diagrams!cutting of}\index{subdiagram!cutting into}
The scalar product of two \(n\)-fold irreducible tensors of rank \(j_{1}, j_{2}, \ldots, j_{n}\), i.e., the \(n\)-fold invariant sum over projections
\begin{equation}
\left(\mathfrak{M}_{j_{1} j_{2} \ldots j_{n}} \cdot \mathfrak{N}_{j_{1} j_{2} \ldots j_{n}}\right)=\sum_{m_{1} m_{2} \ldots m_{n}}\braOket{00}{\mathfrak{M}}{j_{1} m_{1} j_{2} m_{2} \ldots j_{n} m_{n}}\braOket{j_{1} m_{1} j_{2} m_{2} \ldots j_{n} m_{n}}{\mathfrak{N}}{}
\end{equation}
may be transformed into a \(k\)-fold complete sum (over momenta \(X_{1}, X_{2}, \ldots, X_{n}\) with \(k=n-3\) and \(n \geq 4\) ) of the product of factors:
\begin{equation}
\left(\mathfrak{M}_{j_{1} j_{2} \ldots j_{n}} \cdot \mathfrak{N}_{j_{1} j_{2} \ldots j_{n}}\right)=\sum_{X_{1} X_{2} \ldots X_{k}}^{k=n-3}\left(2 X_{1}+1\right)\left(2 X_{2}+1\right) \ldots\left(2 X_{k}+1\right) \mathfrak{M}_{j_{1} j_{2} \ldots j_{n}}\left(X_{1} X_{2} \ldots X_{k}\right) \mathfrak{N}_{j_{1} j_{2} \ldots j_{n}}\left(X_{1} X_{2} \ldots X_{k}\right) .
\end{equation}
These factors represent contractions of the original tensors and \(3 j m\) symbols:
\begin{equation}
\begin{gathered}
\mathfrak{M}_{j_{1} j_{2} \ldots j_{n}}\left(X_{1} X_{2} \ldots X_{k}\right) \\
\equiv \sum_{m_{i} M_{i}} \mathfrak{M}_{j_{1} m_{1} j_{2} m_{2} \ldots j_{n} m_{n}}\threej{j_{1}}{j_{2}}{X_{1}}{m_{1}}{m_{2}}{-M_{1}}(-1)^{X_{1}-M_{1}}\threej{X_{1}}{j_{3}}{X_{2}}{M_{1}}{m_{3}}{-M_{2}}(-1)^{X_{2}-M_{2}} \ldots\threej{X_{k}}{j_{n-1}}{j_{n}}{M_{k}}{m_{n-1}}{m_{n}} \\
\left.\equiv \sum_{m_{i} M_{i}} \mathfrak{N}_{j_{1} j_{2} \ldots j_{n}}^{m_{1} m_{2} \ldots m_{n}}\threej{j_{1}}{j_{2}}{X_{1}}{m_{1}}{m_{2}}{-M_{1}}(-1)^{\mathfrak{N}_{j_{1} j_{2} \ldots j_{n}}\left(X_{1} X_{2} \ldots\right.} X_{k}\right) \\
\threej{X_{1}}{j_{3}}{X_{2}}{M_{1}}{m_{3}}{-M_{2}}(-1)^{X_{2}-M_{2}} \ldots\threej{X_{k}}{j_{n-1}}{j_{n}}{M_{k}}{m_{n-1}}{m_{n}}
\end{gathered}
\end{equation}
Again, at least one of these contracted quantities must be a true invariant, i.e., should not depend on any additional momentum projection or direction.

For \(n \leq 3\) the sum over \(X_{i}\) may be reduced to the direct product of the contracted tensors. For example,
\begin{equation}
\left(\mathfrak{M}_{j_{1} j_{2} \ldots j_{n}} \cdot \mathfrak{N}_{j_{1} j_{2} \ldots j_{n}}\right)=\mathfrak{M}_{j_{1} j_{2} \ldots j_{n}} \cdot \mathfrak{N}_{j_{1} j_{2} \ldots j_{n}} .
\end{equation}
This means that the original sum becomes factorized. Therefore, the cases \(n \leq 3\) are the most important ones for practical applications.

The analytic transformations under study correspond to cutting the associated diagram into two subdiagrams which represent two factors. The necessary condition is that at least one of these subdiagrams should have no more external lines (apart from those \(j\)-lines which link the subdiagram together)\\
\centerline{\includegraphics[valign=t]{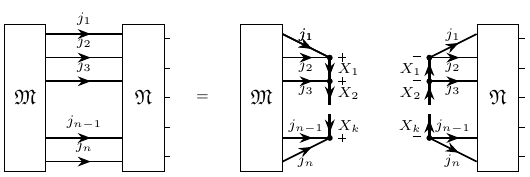}}

The most important cases of cutting of one, two, three, as well as four, five, and six \(j\)-lines are presented in Table~\ref{tab:11.9}.

In many cases, the successive use of these rules allows one to divide a diagram into several (three or more) subdiagrams representing invariant factors. For example,\\
\centerline{\includegraphics[valign=t]{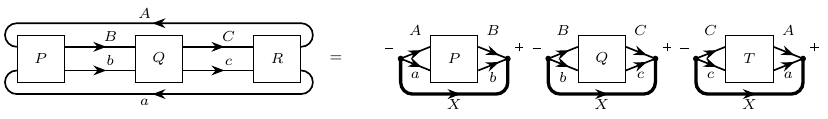}}

The most important cases for the application of the cutting rules are in the separation of external lines of a diagram, i.e., in extracting the dependence on all the external variables in the form of the standard functions of the theory ( \(\clebsch{j_{1}}{m_{1}}{j_{2}}{m_{2}}{j}{m}, D_{m m^{\prime}}^{j}(R), C_{j m}(\Omega)\), etc.). Such a separation is the essence of the Wigner-Eckart theorem which will be discussed in Sec.~\ref{sec:11.4.2}.

\subsection{Graphical Method of Summation}\label{sec:11.3.7}\index{summation!graphical method of}\index{summation rule}
According to Sec.~\ref{sec:11.2.3}, a complete invariant sum over angular momentum \(j\) and projection \(m\) is graphically represented by a thick \(j\)-line. Such a summation may be easily performed by the diagram technique, if this thick \(j\)-line connects two identical nodes.

Rule \(\sum\) : Any thick \(j\)-line which connects identical pairs of thin \(j\)-lines, may be removed provided the ends of the thin lines with equal momenta are linked together. This is displayed below:\\
\centerline{\includegraphics[valign=t]{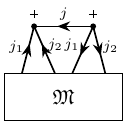}}
Such a graphical operation does not change the meaning of a diagram and corresponds to the equation
\begin{equation}
\sum_{j m}(2 j+1)\braOket{j_{1} j_{2} j m}{\mathfrak{M}}{j_{1} j_{2} j m}=\sum_{m_{1} m_{2}}\braOket{j_{1} m_{1} j_{2} m_{2}}{\mathfrak{M}}{j_{1} m_{1} j_{2} m_{2}}
\end{equation}
where
\begin{equation}
\braOket{j_{1} m_{1} j_{2} m_{2}}{\mathfrak{M}}{j_{1}^{\prime} m_{1}^{\prime} j_{2}^{\prime} m_{2}^{\prime}}=\sum_{j m}(2 j+1)\threej{j_{1}}{j_{2}}{j}{m_{1}}{m_{2}}{m}\braOket{j_{1} j_{2} j m}{\mathfrak{M}}{j_{1}^{\prime} j_{2}^{\prime} j m}\threej{j_{1}^{\prime}}{j_{2}^{\prime}}{j}{m_{1}^{\prime}}{m_{2}^{\prime}}{m} .
\end{equation}
For example,
\begin{equation}
\sum_{c \gamma}(2 c+1)(-1)^{a-\alpha+b-\beta+c-\gamma}\threej{a}{b}{c}{-\alpha}{-\beta}{-\gamma}\threej{a}{b}{c}{\alpha^{\prime}}{\beta^{\prime}}{\gamma^{\prime}}=\sum_{c \gamma}(2 c+1)\threej{a}{b}{c}{\alpha}{\beta}{\gamma}\threej{a}{b}{c}{\alpha^{\prime}}{\beta^{\prime}}{\gamma^{\prime}}=\delta_{\alpha \alpha^{\prime}} \delta_{\beta \beta^{\prime}},
\end{equation}
\centerline{\includegraphics[valign=t]{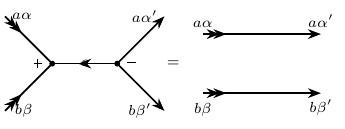}}

\begin{equation}
\sum_{X}(2 X+1)\ninej{a}{b}{c}{b^{\prime}}{d}{e}{c}{e}{X}=\{a b c\}\{b d e\} \frac{\delta_{b b^{\prime}}}{(2 b+1)}
\end{equation}
\begin{center}
\includegraphics[valign=t]{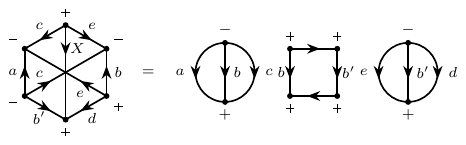}
\end{center}

When a common thick \(j\)-line is included into two or more unlinked subdiagrams, one should preliminarily link these subdiagrams with the aid of the rules of subdiagram multiplication (Sec.~\ref{sec:11.3.5}) in order to arrive at one thick \(j\)-line. Only after this operation can one perform graphical summation over \(j\) and \(m\) according to Rule \(\sum\). These combined graphical operations are displayed in Table~\ref{tab:11.9}.

The complete invariant summation over \(j\) and \(m\) may also be performed graphically in the cases when the lines adjacent to the thick \(j\)-line are dashed \(\Omega\)-lines. For instance,
\begin{equation}
\sum_{l m} Y_{l m}(\Omega) Y_{l m}^{*}\left(\Omega^{\prime}\right)=\delta\left(\Omega-\Omega^{\prime}\right)
\end{equation}
\begin{center}
\includegraphics[valign=t]{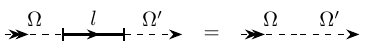}
\end{center}
\begin{equation}
\sum_{J M M^{\prime}}(2 J+1) D_{M M^{\prime}}^{J}\left(R_{1}\right) D_{M^{\prime} M}^{J}\left(R_{2}\right)=16 \pi^{2} \delta\left(R_{1}-R_{2}\right),
\end{equation}
\begin{center}
\includegraphics[valign=t]{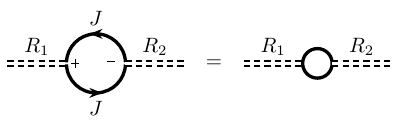}
\end{center}

In fact, all the rules for complete invariant summation formulated in this section follow from the completeness of \(\ket{j m}\)-functions.
\begin{equation}
\sum_{j m}\ket{j m}\bra{j m}=\sum_{j m}\ket{j m}(2 j+1)\rbra{j m}=1,
\end{equation}
\subsection{Replacing An Internal \(\Omega\)-Line by a Thick \(j\)-Line}\label{sec:11.3.8}\index{Omega-line@$\Omega$-line!replacement by a $j$-line}\index{integration!graphical}
Each internal \(\Omega\)-line which represents invariant integration over direction angles \(\Omega\{\vartheta, \varphi\}\) may be replaced by a thick \(j\)-line representing the complete invariant sum over \(j\) and \(m\), according to
\begin{equation}
\int d \Omega\braOket{\Omega}{\mathfrak{M}}{\Omega}=\sum_{j m}\braOket{jm}{\mathfrak{M}}{ j m},
\end{equation}
where
\begin{equation}
\braOket{j m}{\mathfrak{M}}{j^{\prime} m^{\prime}} \equiv \iint d \Omega d \Omega^{\prime}\braket{j m}{\Omega^{\prime}}\braOket{\Omega^{\prime}}{\mathfrak{M}}{\Omega}\braket{\Omega}{j^{\prime} m^{\prime}} .
\end{equation}
Thus, all the \(\Omega\)-lines in the inner part of a diagram may be converted into the \(j\)-lines. Hence, it is sufficient to formulate the rules of graphical operations only for diagrams in which all the internal lines are \(j\)-lines.

\subsection{Elimination of \(j=0\) Line}\label{sec:11.3.9}\index{j-line@$j$-line!elimination of a $j=0$ line}\index{elimination!of a $j=0$ line}
In the particular case of \(j=0\) the corresponding expression is independent of the angular variables \(\Omega\{\vartheta, \varphi\}\). Hence, the corresponding \(j\)-line may be removed from a diagram without changing its meaning. However, there remain the nodes which were initially connected with this \(j=0\) line. These nodes may also be extracted from the diagram in the form of a separated subdiagram corresponding to the factor\isym{K-factor@$K$, factor from eliminating a $j=0$ line}
\begin{equation}
K=( \pm 1)^{2 j_{1}} \frac{\delta_{j_{1} j_{2}}}{\sqrt{2 j_{1}+1}} .
\end{equation}
The phase of \(K\) depends on the node sign and line direction. \(K\) is positive if the node sign corresponds to the cyclic order of the coupling momenta \(\left(j_{1}, j_{1}, j_{2}\right)\), the \(j_{1}\)-line is directed toward the node and the \(j_{2}\)-line is directed outward from the node. In this case the resultant \(j_{1}=j_{2}\)-line has the same direction as that before the node's removal. In other cases the phase of \(K\) may be determined by means of the reduction of the diagram under study to the above case, using the rules of Sec.~\ref{sec:11.3.2}.

Thus elimination of the \(j=0\) line may be illustrated by\\
\begin{center}
\includegraphics[valign=t]{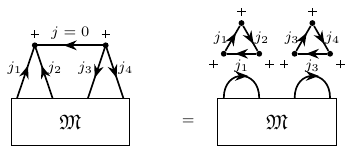}\\
\includegraphics[valign=t]{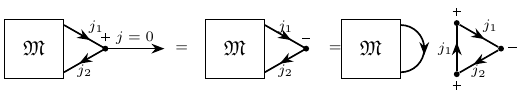}
\end{center}

Some special cases of the \(3 j m\) and \(3 n j\) symbols with \(j=0\) are presented in Table~\ref{tab:11.7}.

\subsection{Elimination of the \(R=1\) Line}\label{sec:11.3.10}\index{R-line@$R$-line!elimination of an $R=1$ line}\index{elimination!of an $R=1$ line}
In the particular case \(R=1\), a double dashed \(R\)-line corresponds to no rotation, i.e., it represents a fictitious dependence. Hence, an \(R=1\) line may be removed and the lines adjacent to the \(R\)-line have to be closed. This operation does not change the meaning of the diagram.\\
\begin{center}
\includegraphics[valign=t]{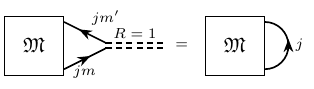}\\
\includegraphics[valign=t]{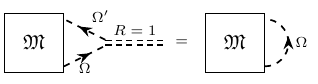}
\end{center}
Some examples of the expressions and diagrams containing \(R=1\) are given in Table~\ref{tab:11.8}.\\
A summary of the rules of graphical operations is presented in Table~\ref{tab:11.9}.
\begin{table}[ptbh!]
\begin{center}
\captionsetup{justification=centering}
\caption{Zero Angular Momentum (\(J=0\)).}\label{tab:11.7}
\small
\setlength{\tabcolsep}{4pt}
\renewcommand{\arraystretch}{1.4}
\begin{tabular}{@{}p{0.42\textwidth}c@{}}
\hline
{Analytical Expression} & Graphical Representation \\
\hline
\multicolumn{2}{c}{\(jm\) Symbols}\\
\(\threej{a}{0}{b}{\alpha}{0}{\beta}=\dfrac{\delta_{ab}}{\sqrt{2a+1}}\dbinom{a}{\alpha\beta}=(-1)^{a+\alpha}\dfrac{\delta_{ab}\delta_{\alpha,-\beta}}{\sqrt{2a+1}}\)
& \includegraphics[valign=t]{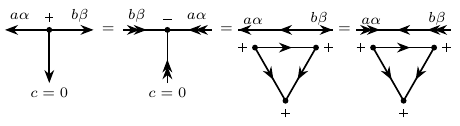}\\
\(\threej{a}{b}{0}{\alpha}{\beta}{0}=\dfrac{\delta_{ab}}{\sqrt{2a+1}}\dbinom{a}{\beta\alpha}=(-1)^{a-\alpha}\dfrac{\delta_{ab}\delta_{\alpha,-\beta}}{\sqrt{2a+1}}\)
& \includegraphics[valign=t]{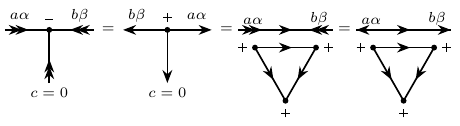} \\
\((-1)^{a-\alpha}\threej{a}{0}{b}{-\alpha}{0}{\beta}=\dfrac{\delta_{ab}\delta_{\alpha\beta}}{\sqrt{2a+1}}\)
& \includegraphics[valign=t]{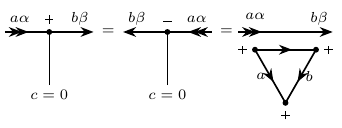}\\
\((-1)^{a-\alpha}\threej{a}{b}{0}{-\alpha}{\beta}{0}=(-1)^{2a}\dfrac{\delta_{ab}\delta_{\alpha\beta}}{\sqrt{2a+1}}\)
& \includegraphics[valign=t]{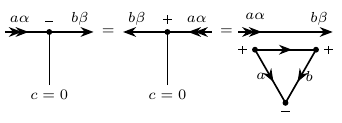}\\
\multicolumn{2}{c}{\(nj\) Symbols}\\
\(\{a\,b\,0\}=\delta_{ab}\)
& \includegraphics[valign=t]{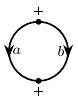}\(\;=\;\)\includegraphics[valign=t]{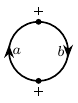} \\
\(\sixj{0}{0}{0}{a}{b}{c}=(-1)^{2a}\dfrac{\delta_{ab}\delta_{bc}}{\sqrt{2a+1}}\)
& \includegraphics[valign=t]{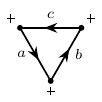}\(\;=\;\)\includegraphics[valign=t]{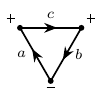} \\
\(\sixj{a}{a^{\prime}}{0}{b}{b^{\prime}}{0}=(-1)^{2a}\dfrac{\delta_{aa^{\prime}}\delta_{bb^{\prime}}\delta_{ab}}{2a+1}\)
& \includegraphics[valign=t]{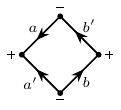}\(\;=\;\)\includegraphics[valign=t]{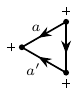}\,\includegraphics[valign=t]{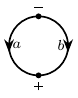}\,\includegraphics[valign=t]{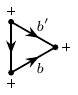} \\
\(\sixj{a}{a^{\prime}}{0}{b}{b^{\prime}}{c}=(-1)^{a+b+c}\dfrac{\delta_{aa^{\prime}}\delta_{bb^{\prime}}}{\sqrt{(2a+1)(2b+1)}}\{a\,b\,c\}\)
& \includegraphics[valign=t]{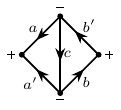}\(\;=\;\)\includegraphics[valign=t]{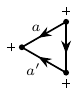}\,\includegraphics[valign=t]{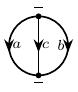},\includegraphics[valign=t]{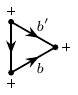} \\
\(\ninej{a}{b}{c}{e}{0}{e^{\prime}}{f}{b^{\prime}}{d}=(-1)^{b+c+e+f}\dfrac{\delta_{bb^{\prime}}\delta_{ee^{\prime}}}{\sqrt{(2b+1)(2e+1)}}\sixj{a}{b}{c}{d}{e}{f}\)
& \includegraphics[valign=t]{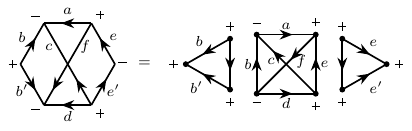} \\
\hline
\end{tabular}
\end{center}
\end{table}

\begin{table}[tbhp!]
\begin{center}
\captionsetup{justification=centering}
\caption{Zero Rotation (\(R=0\)).}\label{tab:11.8}
\small
\setlength{\tabcolsep}{4pt}
\renewcommand{\arraystretch}{1.5}
\begin{tabular}{@{}p{0.46\textwidth}c@{}}
\hline
\multicolumn{1}{c|}{Analytical Expression} & Graphical Representation \\
\hline
(a) For \(R=1\): \[\delta\left(\Omega^{\prime}-\hat{R}\Omega\right)=\delta\left(\Omega^{\prime}-\Omega\right),\] i.e.,
\[\braOket{\Omega^{\prime}}{R}{\Omega}=\braket{\Omega^{\prime}}{\Omega}.\]
& \includegraphics[valign=t]{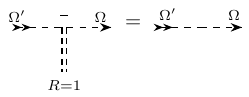} \\
(b) For \(R=1\): \[\delta_{jj^{\prime}}D_{mm^{\prime}}^{j}(000)=\delta_{jj^{\prime}}\delta_{mm^{\prime}},\] i.e.,
\[\braOket{jm}{D(1)}{j^{\prime}m^{\prime}}=\braket{jm}{j^{\prime}m^{\prime}}.\]
& \includegraphics[valign=t]{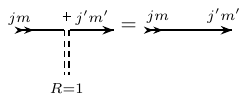}  \\
(c) For \(R^{-1}R=1\): \[\int\delta\!\left(\Omega^{\prime\prime}-R^{-1}\Omega^{\prime}\right)d\Omega^{\prime}\,\delta\!\left(\Omega^{\prime}-R\Omega\right)=\delta\!\left(\Omega^{\prime\prime}-\Omega\right),\]
i.e., \[\braOket{\Omega^{\prime\prime}}{R^{-1}}{\Omega^{\prime}}\braOket{\Omega^{\prime}}{R}{\Omega}=\braket{\Omega^{\prime\prime}}{\Omega}.\].
& \includegraphics[valign=t]{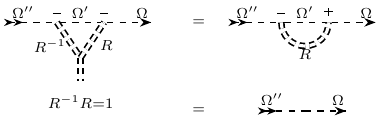}  \\
(d) For \(R^{-1}R=1\): \[\sum_{m^{\prime}}D_{m^{\prime\prime}m^{\prime}}^{j}(R^{-1})D_{m^{\prime}m}^{j}(R)=\delta_{m^{\prime\prime}m},\]
i.e., \[\braOket{jm^{\prime\prime}}{D(R^{-1})}{j^{\prime}m^{\prime}}\braOket{j^{\prime}m^{\prime}}{D(R)}{jm}=\braket{jm^{\prime\prime}}{jm}.\]
& \includegraphics[valign=t]{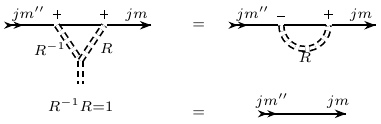}  \\
\hline
\end{tabular}
\end{center}
\end{table}

\begin{table}[tbhp!]
\begin{center}
\captionsetup{justification=centering}
\caption{
Rules for the Transformation of Diagrams}\label{tab:11.9}
\small
\setlength{\tabcolsep}{4pt}
\renewcommand{\arraystretch}{1.5}
\setlength{\parskip}{16pt}
\renewcommand{\arraystretch}{1.5}
\begin{tabular}{@{}p{0.5\textwidth}p{0.5\textwidth}@{}}
\hline
\multicolumn{1}{c|}{Analytical Expression} & \multicolumn{1}{c}{Graphical Representation} \\
\hline
Invariant integration over directions
\[
\int d\Omega\braOket{\Omega}{\mathfrak{M}}{\Omega}=\sum_{JM}\braOket{JM}{\mathfrak{M}}{JM}
\]
where
\[\braOket{JM}{\mathfrak{M}}{J^{\prime}M^{\prime}}\equiv\iint\braket{JM}{\Omega}\,d\Omega\braOket{\Omega}{\mathfrak{M}}{\Omega^{\prime}}\,d\Omega^{\prime}\braket{\Omega^{\prime}}{J^{\prime}M^{\prime}}.\]
Completeness Condition
\[
\int\ket{\Omega}\,d\Omega\,\bra{\Omega}=\sum_{JM}\ket{JM}\bra{JM}=1
\]
&
\centering Replacing dashed \(\Omega\)-line by thick solid \(j\)-line\par\medskip
\includegraphics[valign=c]{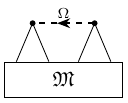}\;\(=\)\;%
\includegraphics[valign=c]{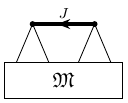}
\tabularnewline
\hline
\end{tabular}
\end{center}
\end{table}

\begin{table}[tbhp!]
\begin{center}
\captionsetup{justification=centering,labelformat=empty}
\caption*{Table \ref{tab:11.9}. Cont'd}
\small
\setlength{\tabcolsep}{4pt}
\renewcommand{\arraystretch}{1.5}
\begin{tabular}{@{}p{0.5\textwidth}|p{0.5\textwidth}}
\hline
\multicolumn{1}{c|}{Analytical Expression} & Graphical Representation \\
\hline
Complete summation over \(J\) and \(M\)
\[
\begin{aligned}
& \sum_{J M}(2 J+1)\braOket{j_{1} j_{2} ; J M}{\mathfrak{M}}{j_{1} j_{2} ; J M} \\
& =\sum_{m_{1} m_{2}}\braOket{j_{1} m_{1} j_{2} m_{2}}{\mathfrak{M}}{j_{1} m_{1} j_{2} m_{2}},
\end{aligned} \]&
Elimination of thick solid \(j\)-line\par
\includegraphics[valign=c]{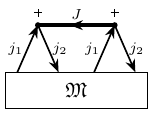} \(=\) \includegraphics[valign=c]{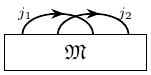}\\
\[\begin{aligned}
 \sum_{J M M^{\prime}}(2 J+1)&\braOket{j_{1} j_{2} ; J M}{\mathfrak{M}}{j_{3} j_{4} ; J M} \\
& \times\braOket{j_{3} j_{4} ; J M^{\prime}}{\mathfrak{N}}{j_{1} j_{2} ; J M^{\prime}} \\
 =\sum_{\substack{m_{1} m_{2} \\m_{3} m_{4}}}&\braOket{j_{1} m_{1} j_{2} m_{2}}{\mathfrak{M}}{j_{3} m_{3} j_{4} m_{4}}\\&\times\braOket{j_{3} m_{3} j_{4} m_{4}}{\mathfrak{N}}{j_{1} m_{1} j_{2} m_{2}} .
\end{aligned} \]&
\includegraphics[valign=t]{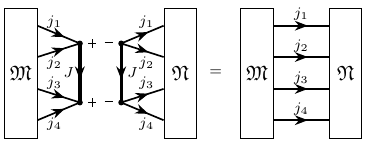}\\
\[\begin{aligned}
& \sum_{J M M^{\prime} M^{\prime\prime}}(2 J+1)\braOket{j_{1} j_{2} ; J M}{\mathfrak{M}}{j_{3} j_{4} ; J M} \\
& \times\braOket{j_{3} j_{4} ; J M^{\prime}}{\mathfrak{N}}{j_{5} j_{6} ; J M^{\prime}}\braOket{j_{5} j_{6} ; J M^{\prime \prime}}{\mathfrak{L}}{j_{1} j_{2} ; J M^{\prime \prime}} \\
& =\sum_{\substack{m_{1} m_{2} \\
m_{3} m_{4} \\
m_{5} m_{6}}}\braOket{j_{1} m_{1} j_{2} m_{2}}{\mathfrak{M}}{j_{3} m_{3} j_{4} m_{4}} \times \\
& \times\braOket{j_{3} m_{3} j_{4} m_{4}}{\mathfrak{N}}{j_{5} m_{5} j_{6} m_{6}}\braOket{j_{5} m_{5} j_{6} m_{6}}{\mathfrak{L}}{j_{1} m_{1} j_{2} m_{2}} .
\end{aligned}
\]where
\[
\begin{gathered}
\braOket{j_{1} j_{2} ; J M}{\mathfrak{M}}{j_{3} j_{4} ; J M} \\
\equiv \sum_{m_{i}}\threej{j_{1}}{j_{2}}{J}{m_{1}}{m_{2}}{M}\braOket{j_{1} m_{1} j_{2} m_{2}}{\mathfrak{M}}{j_{3} m_{3} j_{4} m_{4}}\threej{j_{3}}{j_{4}}{J}{m_{3}}{m_{4}}{M} .
\end{gathered}
\]&
\includegraphics[valign=t]{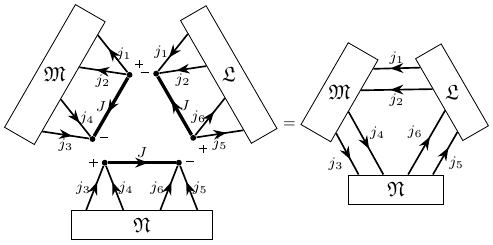}\\

Transformation of product of invariant sums into invariant sum of products (at least one of two factors must be a perfect invariant).&Linking subdiagrams containing identical \(j\)-line (at least one of subdiagrams should have no other external lines)\\
\(
\begin{aligned}
&\left(\sum_{m}\braOket{j m}{\mathfrak{M}}{j m}\right)\left(\sum_{m^{\prime}}\braOket{j m^{\prime}}{\mathfrak{N}}{j m^{\prime}}\right) \\
&=(2 j+1) \sum_{m m^{\prime}}\braOket{j m}{\mathfrak{M}}{j m^{\prime}}\braOket{j m^{\prime}}{\mathfrak{N}}{j m} \end{aligned} \)&
\includegraphics[valign=t]{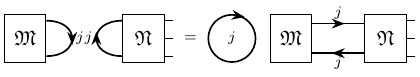}\\
\(\begin{aligned}
& {\left[\sum_{m_{1} m_{2} m_{3}}\braOket{}{\mathfrak{M}}{j_{1} m_{1} j_{2} m_{2} j_{3} m_{3}}\threej{j_{1}}{j_{2}}{j_{3}}{m_{1}}{m_{2}}{m_{3}}\right] } \\
& \times {\left[\sum_{m_{1} m_{2} m_{3}}\threej{j_{1}}{j_{2}}{j_{3}}{m_{1}}{m_{2}}{m_{3}}\braOket{j_{1} m_{1} j_{2} m_{2} j_{3} m_{3}}{\mathfrak{N}}{}\right] } \\
&= \sum_{m_{1} m_{2} m_{3}}\braOket{}{\mathfrak{M}}{j_{1} m_{1} j_{2} m_{2} j_{3} m_{3}}\braOket{j_{1} m_{1} j_{2} m_{2} j_{3} m_{3}}{\mathfrak{N}}{} .
\end{aligned}
\)&
\includegraphics[valign=t]{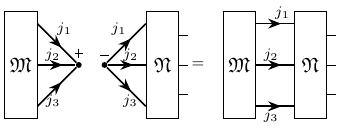}\\
Expression reduction for \(J=0\)&Elimination of a \(J=0\) line in a diagram\\
\(
\begin{gathered}
\sum_{m^{\prime}}\braOket{j m}{\mathfrak{M}}{j^{\prime} m^{\prime}}(-1)^{j^{\prime}-m^{\prime}}\threej{j^{\prime}}{0}{j}{-m^{\prime}}{0}{m} \\
=\frac{\delta_{j j^{\prime}}}{\sqrt{2 j+1}} \sum_{m}\braOket{j m}{\mathfrak{M}}{j m}
\end{gathered}
\)&
\includegraphics[valign=t]{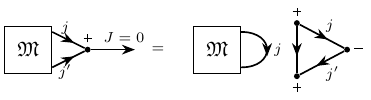}
\end{tabular}
\end{center}
\end{table}

\begin{table}[tbhp!]
\begin{center}
\captionsetup{justification=centering,labelformat=empty}
\caption*{Table \ref{tab:11.9}. Cont'd}
\small
\setlength{\tabcolsep}{4pt}
\renewcommand{\arraystretch}{1.5}
\begin{tabular}{@{}p{0.5\textwidth}p{0.5\textwidth}}
\hline
\multicolumn{1}{c|}{Analytical Expression} & Graphical Representation \\
\hline
Expression reduction for \(R=1\)
\[
\begin{gathered}
\sum_{m m^{\prime}}\braOket{j m}{\mathfrak{M}}{j^{\prime} m^{\prime}}\braOket{j^{\prime} m^{\prime}}{D(000)}{j m} \\
=\delta_{j j^{\prime}} \sum_{m}\braOket{j m}{\mathfrak{M}}{j m}
\end{gathered}
\] &
 Elimination of the \(R=1\) line\par
\centerline{\includegraphics[valign=t]{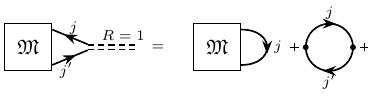}}
\\
Transformation of single, double, or triple invariant sum into direct product of two factors (at least one of the factors must be perfect invariant)&Cutting of a diagram into subdiagrams linked by one, two, or three \(j\)-lines (at least one of subdiagrams should have no additional external lines)\\
\[
\begin{gathered}
\sum_{m}\braOket{00}{\mathfrak{M}}{j m}\braOket{j m q}{\mathfrak{N}}{p}=\delta_{j 0} \delta_{m 0}\braOket{00}{\mathfrak{M}}{00} \\
\times\braOket{00 q}{\mathfrak{N}}{p} .
\end{gathered}
\] &
\centerline{\includegraphics[valign=t]{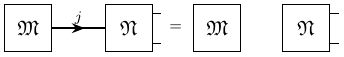}}
\\
\[
\begin{gathered}
\sum_{m_{1} m_{2}}\braOket{j_{1} m_{1}}{\mathfrak{M}}{j_{2} m_{2}}\braOket{j_{2} m_{2} q}{\mathfrak{N}}{j_{1} m_{1} p} \\
=\frac{\delta_{j_{1} j_{2}}}{\left(2 j_{1}+1\right)}\left(\sum_{m_{1}}\braOket{j_{1} m_{1}}{\mathfrak{M}}{j_{1} m_{1}}\right) \\
\times\left(\sum_{m_{2}}\braOket{j_{2} m_{2} q}{\mathfrak{N}}{j_{2} m_{2} p}\right)
\end{gathered}
\] &
\centerline{\includegraphics[valign=t]{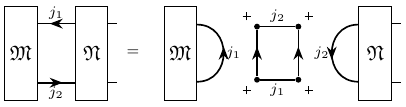}}
\\
\[
\begin{gathered}
\sum_{m_{1} m_{2} m_{3}}\braOket{00}{\mathfrak{M}}{j_{1} m_{1} j_{2} m_{2} j_{3} m_{3}}\braOket{j_{1} m_{1} j_{2} m_{2} j_{3} m_{3} q}{\mathfrak{N}}{p} \\
=\braOketred{0}{\mathfrak{M}}{j_{1} j_{2} j_{3}}\braOketred{j_{1} j_{2} j_{3} q}{\mathfrak{N}}{p}
\end{gathered}
\]
Here the reduced matrix elements are
\[
\begin{gathered}
\braOketred{0}{\mathfrak{M}}{j_{1} j_{2} j_{3}} \equiv \sum_{m_{1} m_{2} m_{3}}\braOket{00}{\mathfrak{M}}{j_{1} m_{1} j_{2} m_{2} j_{3} m_{3}} \\
\times\threej{j_{1}}{j_{2}}{j_{3}}{m_{1}}{m_{2}}{m_{3}},\quad\braOketred{j_{1} j_{2} j_{3} q}{\mathfrak{N}}{p} \equiv \sum_{m_{1} m_{2} m_{3}}\threej{j_{1}}{j_{2}}{j_{3}}{m_{1}}{m_{2}}{m_{3}} \\
\times\braOket{j_{1} m_{1} j_{2} m_{2} j_{3} m_{3} q}{\mathfrak{N}}{p} .
\end{gathered}
\] &
\centerline{\includegraphics[valign=t]{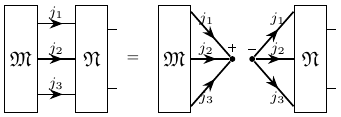}}
\\
Transformation of \(n\)-fold invariant sum ( \(n>4\) ) over projections into a \(k\)-fold sum over momenta ( \(k=n-3\) ) (at least one of factors under summation must be perfect invariant)&Cutting of a diagram into two subdiagrams linked by \(n j\)-lines \((n>4)\) (at least one of subdiagrams should have no additional external lines)\\
\[
\begin{aligned}
& \sum_{\substack{m_{1} m_{2} \\
m_{3} m_{4}}}\braOket{j_{1} m_{1} j_{2} m_{2}}{\mathfrak{M}}{j_{3} m_{3} j_{4} m_{4}}\braOket{j_{3} m_{3} j_{4} m_{4} q}{\mathfrak{N}}{j_{1} m_{1} j_{2} m_{2} p} \\
& \quad=\sum_{J}(2 J+1)\braOketred{j_{1} j_{2} J}{\mathfrak{M}}{j_{3} j_{4} J}\braOketred{j_{3} j_{4} J q}{\mathfrak{N}}{j_{1} j_{2} J p}
\end{aligned}
\]
where
\[
\begin{gathered}
\braOketred{j_{1} j_{2} J}{\mathfrak{M}}{j_{3} j_{4} J} \equiv \sum_{\substack{m_{1} m_{2} M \\
m_{3} m_{4}}}\threej{j_{1}}{j_{2}}{J}{m_{1}}{m_{2}}{M}\\
\times\braOket{j_{1} m_{1} j_{2} m_{2}}{\mathfrak{M}}{j_{3} m_{3} j_{4} m_{4}}\threej{j_{3}}{j_{4}}{J}{m_{3}}{m_{4}}{M}
\end{gathered}
\]
and similarly for \(\braOketred{j_{3} j_{4} J q}{\mathfrak{N}}{j_{1} j_{2} J p}\).&
\centerline{\includegraphics[valign=t]{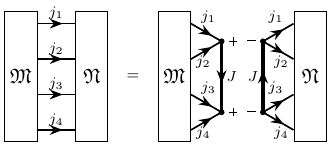}}
\end{tabular}
\end{center}
\end{table}

\begin{table}[tbhp!]
\begin{center}
\captionsetup{justification=centering,labelformat=empty}
\caption*{Table \ref{tab:11.9}. Cont'd}
\small
\setlength{\tabcolsep}{4pt}
\renewcommand{\arraystretch}{1.5}
\begin{tabular}{@{}p{0.5\textwidth}p{0.5\textwidth}}
\hline
\multicolumn{1}{c|}{Analytical Expression} & Graphical Representation \\
\hline
\[
\begin{gathered}
\sum_{m_{1} \ldots m_{5}}\braOket{00}{\mathfrak{M}}{j_{1} m_{1} \ldots j_{5} m_{5}}\braOket{j_{1} m_{1} \ldots j_{5} m_{5} q}{\mathfrak{N}}{p} \\
=\sum_{J_{12} J_{45}}\left(2 J_{12}+1\right)\left(2 J_{45}+1\right)\braOketred{0}{\mathfrak{M}}{J_{12} j_{3} J_{45}} \\
\times\braOketred{J_{12} j_{3} J_{45} q}{\mathfrak{N}}{p}
\end{gathered}
\]
where
\[
\begin{gathered}
\braOketred{0}{\mathfrak{M}}{J_{12} j_{3} J_{45}}=\sum_{m_{i} M_{i k}}\braOket{00}{\mathfrak{M}}{j_{1} m_{1} \ldots j_{5} m_{5}} \\
\times\threej{j_{1}}{j_{2}}{J_{12}}{m_{1}}{m_{2}}{M_{12}}(-1)^{j_{3}-m_{3}}\threej{J_{12}}{j_{3}}{J_{45}}{M_{12}}{-m_{3}}{M_{45}}\threej{j_{4}}{j_{5}}{J_{45}}{m_{4}}{m_{5}}{M_{45}},
\end{gathered}
\]
and similarly for \(\braOketred{J_{12} j_{3} J_{45} q}{\mathfrak{N}}{p}\).
&
\centerline{\includegraphics[valign=t]{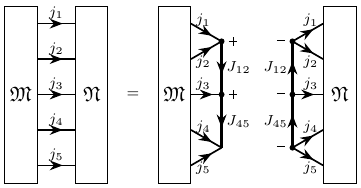}}\\
\[\sum_{m_{1} \ldots m_{6}}\braOket{00}{\mathfrak{M}}{j_{1} m_{1} \ldots j_{6} m_{6}}\braOket{j_{1} m_{1} \ldots j_{6} m_{6} q}{\mathfrak{N}}{p}\]
\[
\begin{gathered}
=\sum_{J_{i k}}\left(2 J_{12}+1\right)\left(2 J_{34}+1\right)\left(2 J_{56}+1\right)\braOketred{0}{\mathfrak{M}}{J_{12} J_{34} J_{56}} \\
\times\braOketred{J_{12} J_{34} J_{56} q}{\mathfrak{N}}{p}
\end{gathered}
\]
where
\[
\begin{gathered}
\braOketred{0}{\mathfrak{M}}{J_{12} J_{34} J_{56}}=\sum_{m_{i} M_{i k}}\braOket{00}{\mathfrak{M}}{j_{1} m_{1} \ldots j_{6} m_{6}} \\
\times\threej{j_{1}}{j_{2}}{J_{12}}{m_{1}}{m_{2}}{M_{12}}\threej{j_{3}}{j_{4}}{J_{34}}{m_{3}}{m_{4}}{M_{34}}\threej{j_{5}}{j_{6}}{J_{56}}{m_{5}}{m_{6}}{M_{56}}\threej{J_{12}}{J_{34}}{J_{56}}{M_{12}}{M_{34}}{M_{56}},
\end{gathered}
\]
and similarly for \(\braOketred{J_{12} J_{34} J_{56} q}{\mathfrak{N}}{p}\). &
\centerline{\includegraphics[valign=t]{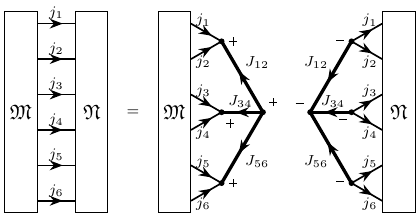}}\\
\hline
\end{tabular}\end{center}
\end{table}

\begin{table}[tbhp!]
\begin{center}
\captionsetup{justification=centering}
\caption{\(3jm \) symbols}\label{tab:11.10}
\small
\setlength{\tabcolsep}{4pt}
\renewcommand{\arraystretch}{1.5}
\begin{tabular}{@{}p{0.7\textwidth}p{0.3\textwidth}}
\hline
\multicolumn{1}{c|}{Analytical Expression} & Graphical Representation \\
\hline
\multicolumn{2}{c}{Reversal of Cyclic Order of Momentum Coupling}\\
\[
\threej{j_{1}}{j_{2}}{j_{3}}{m_{1}}{m_{2}}{m_{3}}=(-1)^{j_{1}+j_{2}+j_{3}}\threej{j_{3}}{j_{2}}{j_{1}}{m_{3}}{m_{2}}{m_{1}}
\]
&
\includegraphics[valign=t]{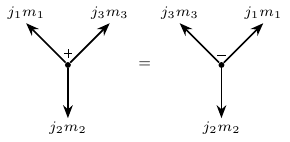}\\
\multicolumn{2}{c}{Hermitian Conjugation}\\
\[
\begin{gathered}
\sum_{m_{1}^{\prime}}\binom{j_{1}}{m_{1}^{\prime} m_{1}}\threej{j_{1}}{j_{2}}{j_{3}}{m_{1}^{\prime}}{m_{2}}{m_{3}}=(-1)^{j_{1}-m_{1}}\threej{j_{1}}{j_{2}}{j_{3}}{-m_{1}}{m_{2}}{m_{3}} \\
=\sum_{m_{2}^{\prime} m_{3}^{\prime}}\threej{j_{1}}{j_{2}}{j_{3}}{m_{1}}{m_{2}^{\prime}}{m_{3}^{\prime}}\binom{j_{2}}{m_{2} m_{2}^{\prime}}\binom{j_{3}}{m_{3} m_{3}^{\prime}}=(-1)^{j_{2}+m_{2}+j_{3}+m_{3}}\threej{j_{1}}{j_{2}}{j_{3}}{m_{1}}{-m_{2}}{-m_{3}}
\end{gathered}
\] &
\includegraphics[valign=t]{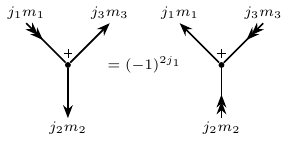}
\\
\[
\begin{gathered}
\sum_{m_{1}^{\prime} m_{2}^{\prime} m_{3}^{\prime}}\binom{j_{1}}{m_{1}^{\prime} m_{1}}\binom{j_{2}}{m_{2}^{\prime} m_{2}}\binom{j_{3}}{m_{3}^{\prime} m_{3}}\threej{j_{1}}{j_{2}}{j_{3}}{m_{1}^{\prime}}{m_{2}^{\prime}}{m_{3}^{\prime}} \\
=(-1)^{j_{1}-m_{1}+j_{2}-m_{2}+j_{3}-m_{3}}\threej{j_{1}}{j_{2}}{j_{3}}{-m_{1}}{-m_{2}}{-m_{3}}=\threej{j_{1}}{j_{2}}{j_{3}}{m_{1}}{m_{2}}{m_{3}}
\end{gathered}
\] &
\includegraphics[valign=t]{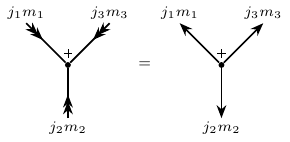}
\end{tabular}\end{center}\end{table}

\begin{table}[tbhp!]
\begin{center}
\captionsetup{justification=centering,labelformat=empty}
\caption*{Table \ref{tab:11.10}. Cont'd}
\small
\setlength{\tabcolsep}{4pt}
\renewcommand{\arraystretch}{1.5}
\begin{tabular}{@{}p{0.65\textwidth}c@{}}
\hline
\multicolumn{1}{c}{Analytical Expression} & Graphical Representation \\
\hline
\multicolumn{2}{c}{Orthonormality}\\
\[\begin{gathered}
\sum_{m_{1} m_{2}}\threej{j_{1}}{j_{2}}{j_{3}}{m_{1}}{m_{2}}{m_{3}}\threej{j_{1}}{j_{2}}{j_{3}^{\prime}}{m_{1}}{m_{2}}{m_{3}^{\prime}} \\
=\sum_{m_{1} m_{2}}(-1)^{j_{1}-m_{1}+j_{2}-m_{2}+j_{3}-m_{3}}\threej{j_{1}}{j_{2}}{j_{3}}{-m_{1}}{-m_{2}}{-m_{3}}\threej{j_{1}}{j_{2}}{j_{3}^{\prime}}{m_{1}}{m_{2}}{m_{3}^{\prime}} \\
=\frac{\delta_{j_{3} j_{3}^{\prime}} \delta_{m_{3} m_{3}^{\prime}}}{\left(2 j_{3}+1\right)}\left\{j_{1} j_{2} j_{3}\right\}
\end{gathered}
\]
&
\includegraphics[valign=t]{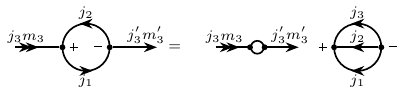}
\\
\multicolumn{2}{c}{Completeness}\\
\[
\begin{gathered}
\sum_{j_{3} m_{3}}\left(2 j_{3}+1\right)\threej{j_{1}}{j_{2}}{j_{3}}{m_{1}}{m_{2}}{m_{3}}\threej{j_{1}}{j_{2}}{j_{3}}{m_{1}^{\prime}}{m_{2}^{\prime}}{m_{3}} \\
=\sum_{j_{3} m_{3}}\left(2 j_{3}+1\right)(-1)^{j_{1}-m_{1}+j_{2}-m_{2}+j_{3}-m_{3}}\threej{j_{1}}{j_{2}}{j_{3}}{-m_{1}}{-m_{2}}{-m_{3}}\threej{j_{1}}{j_{2}}{j_{3}}{m_{1}^{\prime}}{m_{2}^{\prime}}{m_{3}} \\
=\delta_{m_{1} m_{1}^{\prime}} \delta_{m_{2} m_{2}^{\prime}}
\end{gathered}
\]
&
\includegraphics[valign=t]{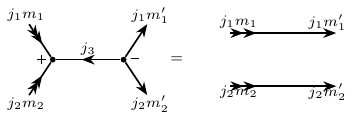}
\end{tabular}
\end{center}
\end{table}

\begin{table}[tbhp!]
\begin{center}
\captionsetup{justification=centering,labelformat=empty}
\caption{ Table 11.10a.\\ Relation of Clebsch-Gordan Coefficients to 3jm Symbols.}\label{tab:11.10a}
\small
\setlength{\tabcolsep}{5pt}
\renewcommand{\arraystretch}{1.3}
\begin{tabular}{@{}p{0.56\textwidth}p{0.44\textwidth}@{}}
\hline
\multicolumn{1}{c|}{Analytical Expression} & Graphical Representation \\
\hline
\[\begin{aligned}[t]
\clebsch{j_1}{m_1}{j_2}{m_2}{J}{M}&=\braket{j_1m_1j_2m_2}{j_1j_2JM}=\braket{j_1j_2JM}{j_1m_1j_2m_2}\\
&=(-1)^{2j_1}\sum_{j_3m_3}(2j_3+1)\threej{J}{0}{j_3}{M}{0}{m_3}\threej{j_3}{j_2}{j_1}{m_3}{m_2}{m_1}\\
&=(-1)^{2j_1}\sqrt{2J+1}\,(-1)^{J+M}\threej{J}{j_2}{j_1}{-M}{m_2}{m_1}\\
&=(-1)^{2j_2}\sum_{j_3m_3}(2j_3+1)\threej{0}{J}{j_3}{0}{M}{m_3}\threej{j_3}{j_2}{j_1}{m_3}{m_2}{m_1}\\
&=(-1)^{2j_2}\sqrt{2J+1}\,(-1)^{J-M}\threej{J}{j_2}{j_1}{-M}{m_2}{m_1}
\end{aligned}\]
& 
\begin{center}
\includegraphics[valign=t]{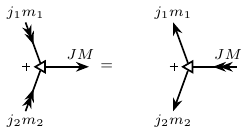}\\
\includegraphics[valign=t]{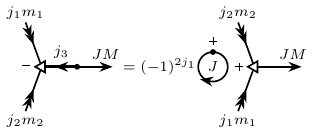}\\
\includegraphics[valign=t]{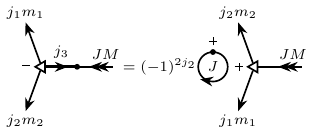}\\
\vspace*{5mm}
\end{center}
\\
\hline
\end{tabular}
\end{center}
\end{table}

\begin{table}[tbhp!]
\begin{center}
\captionsetup{justification=centering}
\caption{Metric Tensor for \((2j+1)\)-Dimensional Space of Functions \(\ket{jm}\).}\label{tab:11.11}
\small
\setlength{\tabcolsep}{5pt}
\renewcommand{\arraystretch}{1}
\begin{tabular}{@{}m{0.44\textwidth}m{0.56\textwidth}@{}}
\hline
\multicolumn{1}{c|}{Analytical Expression} & Graphical Representation \\
\hline
\multicolumn{2}{c}{Definition and Properties}\\
\[\ket{jm\,jm^{\prime}}=\bra{ jm\,jm^{\prime}}=\dbinom{j}{m\ m^{\prime}}=(-1)^{j+m}\delta_{m,-m^{\prime}}\]
& \centerline{\includegraphics[valign=t]{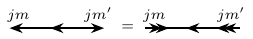}} \\
\[\bra{ jm^{\prime}\,jm}=\ket{jm^{\prime}\,jm}=\dbinom{j}{m^{\prime}\ m}=(-1)^{j-m}\delta_{m^{\prime},-m}\]
& \centerline{\includegraphics[valign=t]{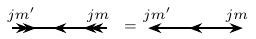}} \\
\begin{gather*}\ket{jm^{\prime}\,jm}=(-1)^{2j}\ket{jm\,jm^{\prime}}\\
=\bra{ jm^{\prime}\,jm}=(-1)^{2j}\bra{ jm\,jm^{\prime}}\end{gather*}
& \centerline{\includegraphics[valign=t]{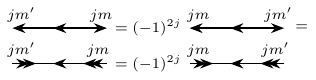}} \\
\[\displaystyle\sum_{m^{\prime}}\ket{jm\,jm^{\prime}}\bra{ jm^{\prime\prime}\,jm^{\prime}}=\braket{ jm^{\prime\prime}}{jm}\]
& \centerline{\includegraphics[valign=t]{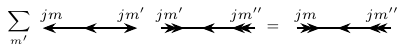}} \\
\multicolumn{2}{c}{Covariant \(\rightleftarrows\) Contravariant.}\\
\multicolumn{2}{c}{Transformation of the Irreducible Tensor}\\
\[\displaystyle\sum_{m^{\prime}}\ket{jm\,jm^{\prime}}\braOket{jm^{\prime}}{\mathfrak{M}}{}=\braOket{}{\mathfrak{M}}{jm}\]
& \centerline{\includegraphics[valign=t]{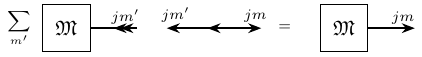}} \\
\[\displaystyle\sum_{m^{\prime}}\braOket{}{\mathfrak{M}}{jm^{\prime}}\bra{ jm^{\prime}\,jm}=\braOket{jm}{\mathfrak{M}}{}\]
& \centerline{\includegraphics[valign=t]{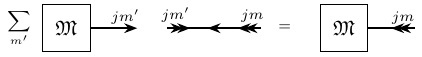}} \\
\multicolumn{2}{c}{Scalar Product of Irreducible Tensors}\\
\[\begin{aligned}
&\sum_{m}\braOket{}{\mathfrak{M}}{jm}\braOket{jm}{\mathfrak{N}}{}\\
&=\sum_{mm^{\prime}}\braOket{}{\mathfrak{M}}{jm}\braOket{}{\mathfrak{N}}{jm^{\prime}}\bra{ jm^{\prime}\,jm}\\
&=\sum_{mm^{\prime}}\ket{jm\,jm^{\prime}}\braOket{jm^{\prime}}{\mathfrak{M}}{}\braOket{jm}{\mathfrak{N}}{}.
\end{aligned}\]
& \centerline{\includegraphics[valign=t]{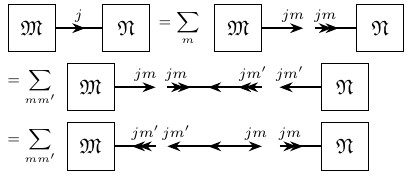}} \\
\[\begin{aligned}
&\sum_{m}\braOket{jm}{\mathfrak{M}}{}\braOket{}{\mathfrak{N}}{jm}\\
&=\sum_{mm^{\prime}}\braOket{}{\mathfrak{M}}{jm^{\prime}}\braOket{}{\mathfrak{N}}{jm}\bra{ jm^{\prime}\,jm}\\
&=\sum_{mm^{\prime}}\ket{jm\,jm^{\prime}}\braOket{jm}{\mathfrak{M}}{}\braOket{jm^{\prime}}{\mathfrak{N}}{}.
\end{aligned}\]
& \centerline{\includegraphics[valign=t]{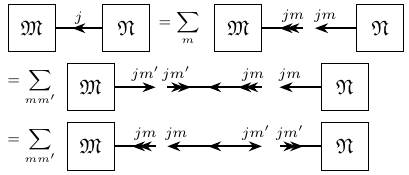}} \\
\[\displaystyle\sum_{m}\braOket{}{\mathfrak{M}}{jm}\braOket{jm}{\mathfrak{N}}{}=(-1)^{2j}\sum_{m}\braOket{jm}{\mathfrak{M}}{}\braOket{}{\mathfrak{N}}{jm}\]
& \centerline{\includegraphics[valign=t]{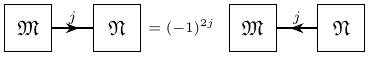}} \\
\hline
\end{tabular}
\end{center}
\end{table}

\begin{table}[tbhp!]
\begin{center}
\captionsetup{justification=centering}
\caption{Spherical Harmonics.}\label{tab:11.12}
\small
\setlength{\tabcolsep}{5pt}
\renewcommand{\arraystretch}{1.3}
\begin{tabular}{@{}m{0.5\textwidth}m{0.5\textwidth}@{}}
\hline
\multicolumn{1}{c|}{Analytical Expression} & Graphical Representation \\
\hline
\multicolumn{2}{c}{Definition}\\
\[C_{lm}(\Omega)\equiv\sqrt{\dfrac{4\pi}{2l+1}}\,Y_{lm}(\Omega)=\rbraket{\Omega}{lm},\] &
 \centerline{\includegraphics[valign=t]{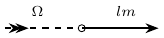}}
\\
\[C_{lm}^{*}(\Omega)=(-1)^{m}C_{l-m}(\Omega)=\rbraket{lm}{\Omega}.\] &
\centerline{\includegraphics[valign=t]{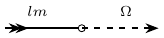}} \\
\multicolumn{2}{c}{Addition Theorem}\\
\[\displaystyle\sum_{m}C_{lm}(\Omega_1)C_{lm}^{*}(\Omega_2)=P_l(\cos\omega_{12}),\]
where \(\omega_{12}\) is the angle between \(\Omega_1\) and \(\Omega_2\).
& \centerline{\includegraphics[valign=t]{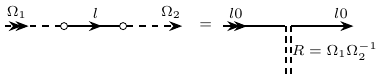}} \\
\multicolumn{2}{c}{Completeness}\\
\[\displaystyle\sum_{lm}(2l+1)C_{lm}(\Omega_1)C_{lm}^{*}(\Omega_2)=4\pi\,\delta(\Omega_1-\Omega_2).\]
& \centerline{\includegraphics[valign=t]{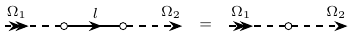}} \\
\multicolumn{2}{c}{Orthonormality}\\
\[\displaystyle\int C_{l_1m_1}^{*}(\Omega)C_{l_2m_2}(\Omega)\,\dfrac{d\Omega}{4\pi}=\dfrac{\delta_{l_1l_2}\delta_{m_1m_2}}{(2l_1+1)}.\]
& \centerline{\includegraphics[valign=t]{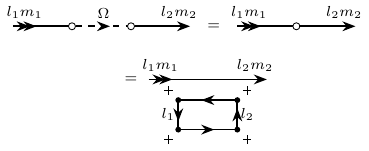}} \\
\multicolumn{2}{c}{Integral of a Product of Three Harmonics\footnote{This defines the "dot in a circle" notation.}}\\
\[\displaystyle\int C_{l_1m_1}(\Omega)C_{l_2m_2}(\Omega)C_{l_3m_3}(\Omega)\,\dfrac{d\Omega}{4\pi}=\threej{l_1}{l_2}{l_3}{0}{0}{0}\threej{l_1}{l_2}{l_3}{m_1}{m_2}{m_3}.\]
& \centerline{\includegraphics[valign=t]{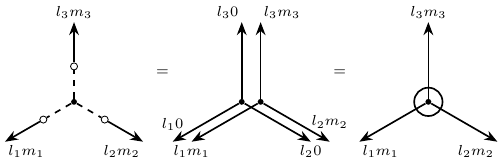}} \\
\multicolumn{2}{c}{Clebsch-Gordan Expansion}\\
\[\displaystyle C_{l_1m_1}(\Omega)C_{l_2m_2}(\Omega)=\sum_{l_3m_3}\threej{l_1}{l_2}{l_3}{0}{0}{0}\threej{l_1}{l_2}{l_3}{m_1}{m_2}{m_3}C_{l_3m_3}^{*}(\Omega)\].
& \centerline{\includegraphics[valign=t]{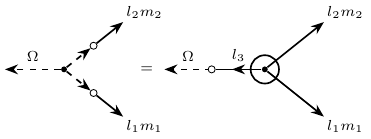}} \\
\multicolumn{2}{c}{Sum of Scalar Products of Three Harmonics}\\
\[\begin{aligned}
&\sum_{\substack{m_1m_2m_3\\ l_1}}(2l_1+1)\threej{l_1}{l_2}{l_3}{0}{0}{0}\threej{l_1}{l_2}{l_3}{m_1}{m_2}{m_3}\\
&\quad\times C_{l_1m_1}(\Omega_1)C_{l_2m_2}(\Omega_2)C_{l_3m_3}(\Omega_3)\\
&=P_{l_2}(\cos\omega_{21})\,P_{l_3}(\cos\omega_{31}).
\end{aligned}\]
& \centerline{\includegraphics[valign=t]{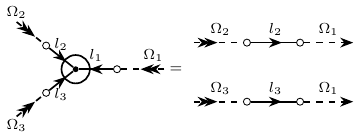}} \\
\hline
\end{tabular}
\end{center}
\end{table}

\begin{table}[tbhp!]
\begin{center}
\captionsetup{justification=centering}
\caption{Wigner $D$-Functions.}\label{tab:11.13}
\label{chap11:tab:9}
\small
\setlength{\tabcolsep}{5pt}
\renewcommand{\arraystretch}{1.3}
\begin{tabular}{@{}m{0.46\textwidth}m{0.54\textwidth}@{}}
\hline
\multicolumn{1}{c}{Analytical Expression} & Graphical Representation \\
\hline
\multicolumn{2}{c}{Adjoint Matrix}\\
\[\begin{aligned}
\left(D^{j}_{mm'}(R)\right)^{*}&=\sum_{\mu\mu'}\binom{j}{m\,\mu}D^{j}_{\mu\mu'}(R)\binom{j}{\mu'\ m'}\\
&=(-1)^{m-m'}D^{j}_{-m,-m'}(R).
\end{aligned}\]
&\centerline{\includegraphics[valign=t]{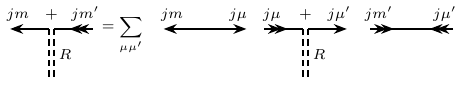}}\\
\multicolumn{2}{c}{Inverse Matrix}\\
\[(D^{j}_{mm'}(R^{-1})=\left(D^{j}_{m'm}(R)\right)^{*}.\]
&\centerline{\includegraphics[valign=t]{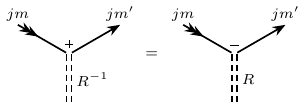}}\\
\multicolumn{2}{c}{Unitarity}\\
\[\displaystyle\sum_{m}\left(D^{j}_{mm'}(R)\right)^{*}D^{j}_{m\bar m'}(R)=\sum_{m}D^{j}_{m'm}(R^{-1})D^{j}_{m\bar m'}(R)=\delta_{m'\bar m'}\]
&\centerline{\includegraphics[valign=t]{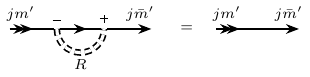}}\\
\[\displaystyle\sum_{m'}D^{j}_{mm'}(R)\left(D^{j}_{\bar m m'}(R)\right)^{*}=\sum_{m'}D^{j}_{mm'}(R)D^{j}_{m'\bar m}(R^{-1})=\delta_{m\bar m}\]
&\centerline{\includegraphics[valign=t]{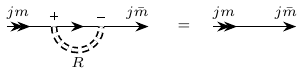}}\\
\multicolumn{2}{c}{Orthonormality}\\
\[\displaystyle\int D^{j_1}_{m_1m'_1}(R)\,D^{j_2\,*}_{m_2m'_2}(R)\,\frac{dR}{16\pi^2}=\frac{\delta_{j_1j_2}\delta_{m_1m_2}\delta_{m'_1m'_2}}{(2j_1+1)}\]
&\centerline{\includegraphics[valign=t]{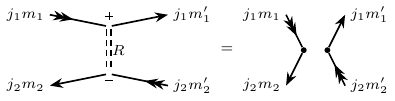}}\\
\multicolumn{2}{c}{Completeness}\\
\[\displaystyle\sum_{j=0,1/2,1}^{\infty}(2j+1)\sum_{mm'}D^{j}_{mm'}(R_1)\,D^{j\,*}_{mm'}(R_2)=16\pi^2\,\delta(R_1-R_2)\]
&\centerline{\includegraphics[valign=t]{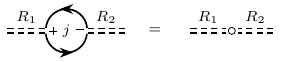}}\\
\multicolumn{2}{c}{Rotation Matrix in \(\Omega\)-Representation}\\
\[\begin{aligned}
\delta_{ll'}D^{l}_{mm'}(R)&=\frac{2l+1}{4\pi}\iint d\Omega\,d\Omega'\,C^{*}_{lm}(\Omega)\,\delta(\Omega-R^{-1}\Omega')\,C_{l'm'}(\Omega'),\\
\braOket{ lm}{R}{l'm'}&=\braket{ lm}{\Omega}\braOket{\Omega}{R^{-1}}{\Omega'}
\braket{\Omega'}{l'm'}.
\end{aligned}\]&\centerline{\includegraphics[valign=t]{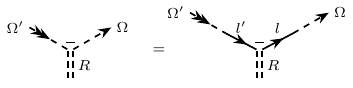}}\\
\[\begin{aligned}
\delta(\Omega'-R\Omega)&=\sum_{lmm'}\frac{2l+1}{4\pi}\,C_{lm'}(\Omega')\,D^{l}_{m'm}(R^{-1})\,C^{*}_{lm}(\Omega),\\
\braOket{\Omega'}{R}{\Omega}&=\braket{\Omega'}{l'm'}\braOket{ l'm'}{R^{-1}}{lm}\braket{ lm}{\Omega}.
\end{aligned}\]
&\centerline{\includegraphics[valign=t]{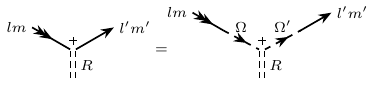}}\\
\hline
\end{tabular}
\end{center}
\end{table}

\begin{table}[tbhp!]
\begin{center}
\captionsetup{justification=centering,labelformat=empty}
\caption*{Table \ref{tab:11.13}\\Cnt'ed.}
\small
\setlength{\tabcolsep}{5pt}
\renewcommand{\arraystretch}{1.3}
\begin{tabular}{@{}m{0.5\textwidth}|m{0.5\textwidth}@{}}
\hline
\multicolumn{1}{c}{Analytical Expression} & Graphical Representation \\
\hline
\multicolumn{2}{c}{Clebsch-Gordan Expansion and Associated Relations}\\
% ---- CG row 1 ----
\begin{gather}
D^{j_1}_{m_1m'_1}(R)\,D^{j_2}_{m_2m'_2}(R)\notag\\
=\sum_{j_3m_3m'_3}(2j_3+1)\threej{j_1}{j_2}{j_3}{m_1}{m_2}{m_3}D^{j_3*}_{m_3m'_3}(R)\threej{j_1}{j_2}{j_3}{m'_1}{m'_2}{m'_3}\notag\end{gather}
&\centerline{\includegraphics[valign=t]{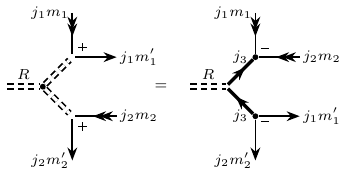}}\\
% ---- CG row 2 (bubble) ----
\begin{gather}\sum_{\substack{m_1m_2\\ m'_1m'_2}}\threej{j_1}{j_2}{j_3}{m_1}{m_2}{m_3}D^{j_1}_{m_1m'_1}(R)D^{j_2}_{m_2m'_2}(R)\threej{j_1}{j_2}{j'_3}{m'_1}{m'_2}{m'_3}\notag\\=\frac{\delta_{j_3j'_3}}{(2j_3+1)}\{j_1j_2j_3\}D^{j_3*}_{m_3m'_3}(R)\notag\end{gather}
&\centerline{\includegraphics[valign=t]{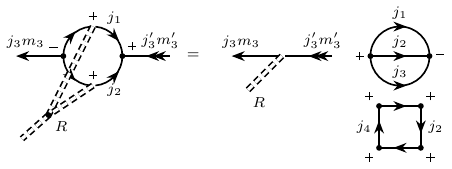}}\\
% ---- CG row 3 ----
\begin{gather}\sum_{m_1m_2m_3}\threej{j_1}{j_2}{j_3}{m_1}{m_2}{m_3}D^{j_1}_{m_1m'_1}(R)D^{j_2}_{m_2m'_2}(R)D^{j_3}_{m_3m'_3}(R)\notag\\
  =\threej{j_1}{j_2}{j_3}{m'_1}{m'_2}{m'_3}\notag\end{gather}
&\centerline{\includegraphics[valign=t]{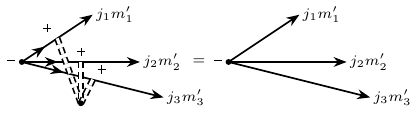}}\\
% ---- CG row 4 ----
\begin{gather}\sum_{\substack{m_1m_2\\ m'_1m'_2\\ m''_3}}\threej{j_1}{j_2}{j'_3}{m_1}{m_2}{m'_3}D^{j_1}_{m_1m'_1}(R)D^{j_2}_{m_2m'_2}(R)D^{j_3}_{m_3m''_3}(R)\threej{j_1}{j_2}{j_3}{m'_1}{m'_2}{m''_3}\notag\\
  =\frac{\delta_{j_3j'_3}\delta_{m_3m'_3}}{(2j_3+1)}\{j_1j_2j_3\}\notag\end{gather}
&\centerline{\includegraphics[valign=t]{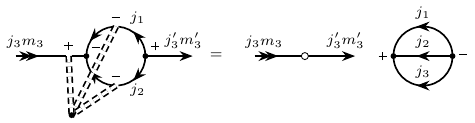}}\\
% ---- CG row 5 ----
\begin{gather} D^{j_3*}_{m_3m'_3}(R)\threej{j_1}{j_2}{j_3}{m'_1}{m'_2}{m'_3}\notag\\
  =\sum_{m_1m_2}\threej{j_1}{j_2}{j_3}{m_1}{m_2}{m_3}D^{j_1}_{m_1m'_1}(R)D^{j_2}_{m_2m'_2}(R)\notag\end{gather}
&\centerline{\includegraphics[valign=t]{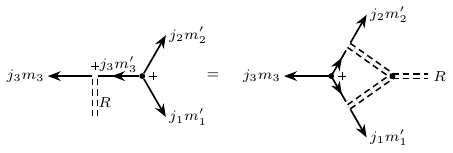}}\\
\multicolumn{2}{c}{Integral of Three D-Functions}\\
\begin{gather*}\int D^{j_1}_{m_1m'_1}(R)\,D^{j_2}_{m_2m'_2}(R)\,D^{j_3}_{m_3m'_3}(R)\,\frac{dR}{16\pi^2}\notag\\
  =\threej{j_1}{j_2}{j_3}{m_1}{m_2}{m_3}\threej{j_1}{j_2}{j_3}{m'_1}{m'_2}{m'_3}\end{gather*}
&\centerline{\includegraphics[valign=t]{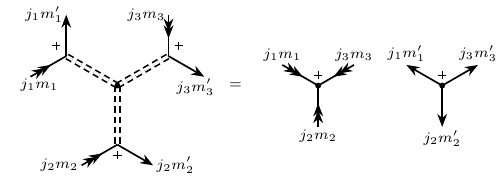}}\\
\end{tabular}
\end{center}
\end{table}

\begin{table}[tbhp!]
\begin{center}
\captionsetup{justification=centering,labelformat=empty}
\caption*{Table \ref{tab:11.13}\\Cnt'ed.}
\small
\setlength{\tabcolsep}{5pt}
\renewcommand{\arraystretch}{1.3}
\begin{tabular}{@{}m{0.5\textwidth}m{0.5\textwidth}@{}}
\hline
\multicolumn{1}{c}{Analytical Expression} & Graphical Representation \\
\hline
\multicolumn{2}{c}{Addition Theorems}\\
% ---- row 1: composition of rotations ----
\[\sum_{m'}D^{j}_{mm'}(R_2)\,D^{j}_{m'm''}(R_1)=D^{j}_{mm''}(R_2R_1)\]
&\centerline{\includegraphics[valign=t]{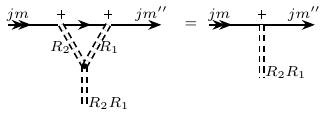}}\\
% ---- row 2: 6j H-tree = 2x2 D-grid ----
\[\begin{aligned}
&\sum_{\substack{j_3m_3\\ m'_1m'_2m'_3}}(2j_3+1)\threej{j_1}{j_2}{j_3}{\tilde m_1}{\tilde m_2}{m_3}\threej{j_1}{j_2}{j_3}{m'_1}{m'_2}{m'_3}\\
&\quad\times D^{j_1}_{m_1m'_1}(R_1)D^{j_2}_{m_2m'_2}(R_2)D^{j_3}_{m_3m'_3}(R_3)\\
&=D^{j_1}_{m_1\tilde m_1}(R_1R_3^{-1})D^{j_2}_{m_2\tilde m_2}(R_2R_3^{-1})
\end{aligned}\]
&\centerline{\includegraphics[valign=t]{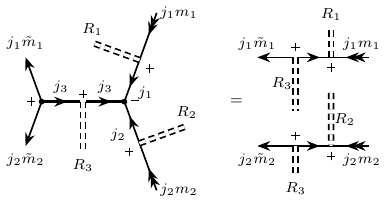}}\\
% ---- row 3: three-D star, R3 redistribution ----
\begin{gather*}\sum_{m'_1m'_2m'_3}D^{j_1}_{m_1m'_1}(R_1)D^{j_2}_{m_2m'_2}(R_2)D^{j_3}_{m_3m'_3}(R_3)\threej{j_1}{j_2}{j_3}{m'_1}{m'_2}{m'_3}\\
  =\sum_{m'_1m'_2}D^{j_1}_{m_1m'_1}(R_1R_3^{-1})D^{j_2}_{m_2m'_2}(R_2R_3^{-1})\threej{j_1}{j_2}{j_3}{m'_1}{m'_2}{m_3}\end{gather*}
&\centerline{\includegraphics[valign=t]{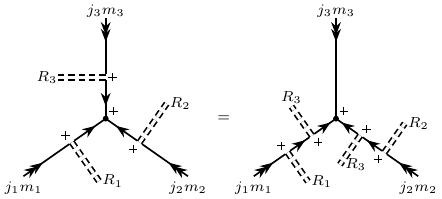}}\\
% ---- row 4: 6j-triangle recoupling ----
\[\begin{aligned}
&\sixj{j_1}{j_2}{j_3}{J_{23}}{J_{31}}{J_{12}}\sum_{m_1m_2m_3}\threej{j_1}{j_2}{j_3}{m_1}{m_2}{m_3}\\
&\quad\times D^{j_1}_{m_1m'_1}(R_1)D^{j_2}_{m_2m'_2}(R_2)D^{j_3}_{m_3m'_3}(R_3)\\
&=\sum_{\substack{M_{12}M_{23}M_{31}\\ M'_{12}M'_{23}M'_{31}}}(-1)^{J_{12}-M_{12}+J_{23}-M_{23}+J_{31}-M_{31}}\\
&\quad\times\threej{J_{12}}{j_1}{J_{31}}{M_{12}}{m'_1}{-M'_{31}}\threej{J_{23}}{j_2}{J_{12}}{M_{23}}{m'_2}{-M'_{12}}
%\\&\qquad\times
\threej{J_{31}}{j_3}{J_{23}}{M_{31}}{m'_3}{-M'_{23}}\\
&\quad\times D^{J_{31}}_{M_{31}M'_{31}}(R_3^{-1}R_1)D^{J_{12}}_{M_{12}M'_{12}}(R_1^{-1}R_2)
%\\&\qquad\times 
D^{J_{23}}_{M_{23}M'_{23}}(R_2^{-1}R_3)
\end{aligned}\]
&\centerline{\includegraphics[valign=t]{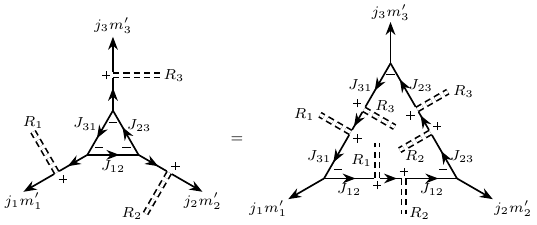}}\\
% ---- row 5: full double-triangle recoupling ----
\[\begin{aligned}
&\sum_{j_1j_2j_3}(2j_1+1)(2j_2+1)(2j_3+1)\sixj{j_1}{j_2}{j_3}{J_{23}}{J_{31}}{J_{12}}\\
&\times\sum_{\substack{m_1m_2m_3\\ m'_1m'_2m'_3}}(-1)^{J_{12}-M_{12}+J_{23}-M_{23}+J_{31}-M_{31}}\\
&\times\threej{J_{31}}{j_1}{J_{12}}{M'_{31}}{m_1}{-M_{12}}\threej{J_{12}}{j_2}{J_{23}}{M'_{12}}{m_2}{-M_{23}}
%\\&\qquad\times
\threej{J_{23}}{j_3}{J_{31}}{M'_{23}}{m_3}{-M_{31}}\\
&\times D^{j_1}_{m_1m'_1}(R_1)D^{j_2}_{m_2m'_2}(R_2)D^{j_3}_{m_3m'_3}(R_3)\threej{j_1}{j_2}{j_3}{m'_1}{m'_2}{m'_3}\\
&=D^{J_{12}}_{M_{12}M'_{12}}(R_1R_2^{-1})D^{J_{23}}_{M_{23}M'_{23}}(R_2R_3^{-1})
%\\&\qquad\times 
D^{J_{31}}_{M_{31}M'_{31}}(R_3R_1^{-1})
\end{aligned}\]
&\centerline{\includegraphics[valign=t]{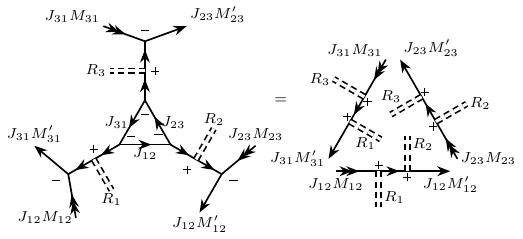}}\\
\hline
\end{tabular}
\end{center}
\end{table}

\begin{table}[tbhp!]
\begin{center}
\captionsetup{justification=centering}
\caption{Matrix Elements of Irreducible Tensor Operators.}\label{tab:11.14}
\small
\setlength{\tabcolsep}{5pt}
\renewcommand{\arraystretch}{1.3}
\begin{tabular}{@{}m{0.4\textwidth}|m{0.6\textwidth}@{}}
\hline
\multicolumn{1}{c}{Analytical Expression} & Graphical Representation \\
\hline
\multicolumn{2}{c}{\parbox{0.93\textwidth}{\centering Matrix element of an irreducible tensor operator of rank \(k\) in the JM-representation
(Wigner-Eckart theorem)}}\\
\[
\braOket{\gamma J M}{\mathfrak{M}_{kq}}{\gamma^{\prime} J^{\prime} M^{\prime}}=(-1)^{J-M}\threej{J}{k}{J^{\prime}}{-M}{q}{M^{\prime}}\rbraOketred{\gamma J}{\mathfrak{M}_{k}}{\gamma^{\prime} J^{\prime}} .
\]
where \(\rbraOketred{\gamma J}{\mathfrak{M}_{k}}{\gamma^{\prime} J^{\prime}}\) is the reduced matrix element with the renormalised states
&\centerline{\includegraphics[valign=t]{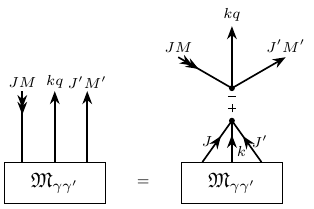}}\\
\multicolumn{2}{c}{\parbox{0.93\textwidth}{\centering Reduced Matrix elements of irreducible tensor product of irreducible tensor operators.}}\\
\[
\begin{gathered}
\rbraOketred{\gamma J}{\left\{\mathfrak{M}_{k_{1}} \otimes \mathfrak{N}_{k_{2}}\right\}_{k}}{\gamma^{\prime} J^{\prime}} \\
=(-1)^{J+k+J^{\prime}} \sqrt{2 k+1} \sum_{\gamma^{\prime\prime} J^{\prime \prime}}\sixj{J}{k}{J^{\prime}}{k_{2}}{J^{\prime \prime}}{k_{1}} \\
\times\rbraOketred{\gamma J}{\mathfrak{M}_{k_{1}}}{\gamma^{\prime \prime} J^{\prime \prime}}\rbraOketred{\gamma^{\prime\prime} J^{\prime \prime}}{\mathfrak{N}_{k_{2}}}{\gamma^{\prime} J^{\prime}} .
\end{gathered}
\] &\centerline{\includegraphics[valign=t]{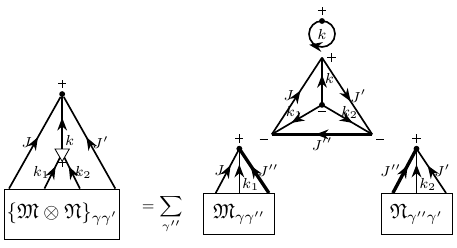}}\\
\multicolumn{2}{c}{\parbox{0.93\textwidth}{\centering Two coupled subsystems in the \(j_{1} j_{2};JM\)-representation. Reduced matrix element of the irreducible tensor product of two irreducible tensor operators, one of which depends only on variables of the first subsystem and the other depends only on variables of the second subsystem.}}\\
\[
\begin{aligned}
& \rbraOketred{\gamma j_{1} j_{2} J}{\left\{\mathfrak{M}_{k_{1}}(1) \otimes \mathfrak{N}_{k_{2}}(2)\right\}_{k}}{\gamma^{\prime} j_{1}^{\prime} j_{2}^{\prime} J^{\prime}} \\
= & \sqrt{(2 J+1)(2 k+1)\left(2 J^{\prime}+1\right)}\ninej{j_{1}}{j_{1}^{\prime}}{k_{1}}{j_{2}}{j_{2}^{\prime}}{k_{2}}{J}{J^{\prime}}{k} \\
& \times \sum_{\gamma^{\prime \prime}}\rbraOketred{\gamma j_{1}}{\mathfrak{M}_{k_{1}}}{\gamma^{\prime \prime} j_{1}^{\prime}}\rbraOketred{\gamma^{\prime \prime} j_{2}}{\mathfrak{N}_{k_{2}}}{\gamma^{\prime} j_{2}^{\prime}}
\end{aligned}
\]&\centerline{\includegraphics[valign=t]{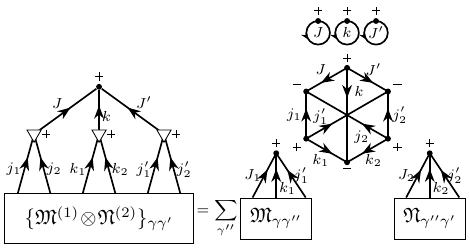}}\\
\end{tabular}
\end{center}
\end{table}

\begin{table}[ptbh!]
\begin{center}
\captionsetup{justification=centering,labelformat=empty}
\caption*{Table \ref{tab:11.14}\\Cnt'ed.}
\small
\setlength{\tabcolsep}{5pt}
\renewcommand{\arraystretch}{1.3}
\begin{tabular}{@{}m{0.4\textwidth}|m{0.6\textwidth}@{}}
\hline
\multicolumn{1}{c|}{Analytical Expression} & Graphical Representation \\
\hline
\multicolumn{2}{c}{\parbox{0.93\textwidth}{\centering Matrix element of the scalar product of two irreducible tensor operators which
depend on variables of the first and the second subsystems, respectively.}}\\
\[
\begin{gathered}
\braOket{\gamma j_{1} j_{2} J M}{\left(\mathfrak{M}_{k}(1) \cdot \mathfrak{N}_{k}(2)\right)}{\gamma^{\prime} j_{1}^{\prime} j_{2}^{\prime} J^{\prime} M^{\prime}} \\
=(-1)^{j_{1}^{\prime}+j_{2}+J}\delta_{J J^{\prime}} \delta_{M M^{\prime}}\sixj{j_{1}}{j_{2}}{J}{j_{2}^{\prime}}{j_{1}^{\prime}}{k} \\
\times \sum_{\gamma^{\prime \prime}}\rbraOketred{\gamma j_{1}}{\mathfrak{M}_{k}}{\gamma^{\prime \prime} j_{1}^{\prime}}\rbraOketred{\gamma^{\prime \prime} j_{2}}{\mathfrak{N}_{k}}{\gamma^{\prime} j_{2}^{\prime}}
\end{gathered}
\] &\centerline{\includegraphics[valign=t]{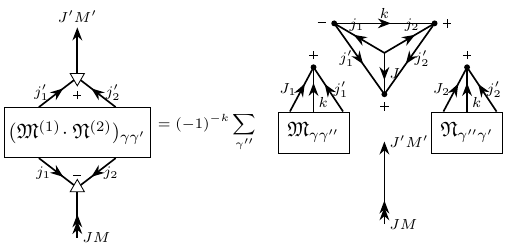}}\\
\multicolumn{2}{c}{\parbox{0.93\textwidth}{\centering Reduced matrix element of an irreducible tensor operator which depends only on variables of the first subsystem.}}\\
\[
\begin{gathered}
\rbraOketred{\gamma j_{1} j_{2} J}{\mathfrak{M}_{k}(1)}{\gamma^{\prime} j_{1}^{\prime} j_{2}^{\prime} J^{\prime}} \\
=(-1)^{j_{1}+j_{2}+J^{\prime}+k} \delta_{j_{2} j_{2}^{\prime}} \sqrt{(2 J+1)\left(2 J^{\prime}+1\right)} \\
\times\sixj{J}{k}{J^{\prime}}{j_{1}^{\prime}}{j_{2}}{j_{1}}\rbraOketred{\gamma j_{1}}{\mathfrak{M}_{k}}{\gamma^{\prime} j_{1}^{\prime}} .
\end{gathered}
\] &\centerline{\includegraphics[valign=t]{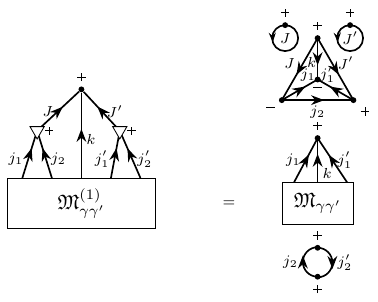}}\\
\[
\begin{gathered}
\rbraOketred{\gamma j_{1} j_{2} J}{\mathfrak{N}_{k}(2)}{\gamma^{\prime} j_{1}^{\prime} j_{2}^{\prime} J^{\prime}} \\
=(-1)^{j_{1}^{\prime}+j_{2}^{\prime}+J+k} \delta_{j_{1} j_{1}^{\prime}} \sqrt{(2 J+1)\left(2 J^{\prime}+1\right)} \\
\times\sixj{J}{k}{J^{\prime}}{j_{2}^{\prime}}{j_{1}}{j_{2}}\rbraOketred{\gamma j_{2}}{\mathfrak{N}_{k}}{\gamma^{\prime} j_{2}^{\prime}} .
\end{gathered}
\] &\centerline{\includegraphics[valign=t]{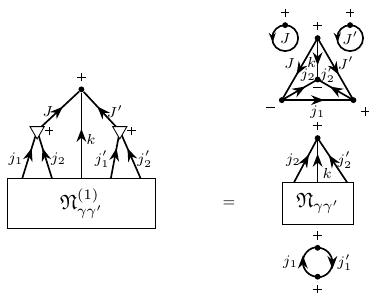}}\\
\hline
\end{tabular}
\end{center}
\end{table}

\section{Summary of the graphical technique}\label{sec:11.4}\index{graphical method!summary of}
The preceding sections present the principles of the graphical methods of angular momentum theory, i.e., graphical representations of the basis functions and operations (Secs. \ref{sec:11.1} and \ref{sec:11.2}) as well as the rules of handling them (Sec.~\ref{sec:11.3}). This section resumes the discussion of the general properties of diagrams, the Wigner-Eckart theorem in diagrammatic form and the scheme for the application of the graphical technique.

\subsection{General Properties of Diagrams}\label{sec:11.4.1}\index{diagrams!general properties of}
Let us summarize the main properties of diagrams which are consequences of the graphical relationships discussed above.
\begin{enumerate}
\item Any diagram may be arbitrarily turned and deformed, bent or stretched. Along with these continuous transformations, discrete transformations such as reflection and inversion are also allowed.

A transformed diagram corresponds to the same analytic expression provided node signs are reversed for those nodes in which the cyclic order of the coupled momenta becomes opposite to the initial one.
\item Each \(j\)-line holds a fixed value of angular momentum \(j\) in spite of any deformation of the line. For an external \(j\)-line, the projection \(m\) is fixed as well as the \(j\)-value. Moreover, \(j\) and \(m\) remain unchanged along the line even if any block (subdiagram) is inserted into the line, provided this block has no additional external \(j\)-lines.
\item The total angular momentum of any diagram or subdiagram, i.e., the vector sum of angular momenta of all external lines, is equal to zero.
\begin{equation}
\sum_{i} \mathrm{j}_{i}=0, \quad \sum_{i} m_{i}=0 .
\end{equation}
Hence, for a diagram with one external \(j\)-line \(j\) has to be zero; for a diagram which has two external \(j\)-lines associated with \(j_{1}\) and \(j_{2}\) one has \(j_{1}=j_{2}\).
\item Any diagram may be cut into two subdiagrams in many different ways. The total angular momenta of both these subdiagrams are identical. Moreover, the angular momenta will be the same for all the diagram cross-sections with fixed sets of external lines for each subdiagram.
\end{enumerate}

\subsection{Generalized Wigner-Eckart Theorem in Diagrammatic Form}\label{sec:11.4.2}\index{Wigner-Eckart theorem}\index{Wigner-Eckart theorem!in diagrammatic form}
Standard functions of angular momentum theory may be used as a basis for the expansion of any expression. However, sometimes all the dependence of the expression on external variables may be factorized in the form of some standard function. Graphically such a function may be separated according to the rules of Sec.~\ref{sec:11.3.6}. The most important special cases are the following.
\begin{enumerate}[label=(\alph*)]
\item A diagram without external lines (i.e. a closed diagram) is invariant with respect to coordinate-system rotations.
\item A diagram with one external line also is invariant. This is a consequence of the fact that, in principle, one cannot distinguish any preferred direction in isotropic space. Therefore, any dependence on one direction is fictitious, so that such an external \(j\)-line has to correspond to \(j=0\) and may be eliminated in accordance with the rule of Sec.~\ref{sec:11.3.9}.
\begin{equation}
\braOket{00}{\mathfrak{M}}{j m}=\delta_{j 0} \delta_{m 0}\braOket{00}{\mathfrak{N}}{00}
\end{equation}
\centerline{\includegraphics[valign=t]{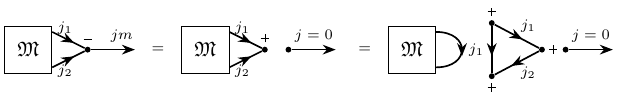}}
\item A diagram with two external lines corresponding to \(\bra{j_{1} m_{1}}\) and \(\ket{j_{2} m_{2}}\) is non-zero only if \(j_{1}=j_{2}\) and \(m_{1}=m_{2}\)
\begin{equation}
\braOket{j_{1} m_{1}}{\mathfrak{M}}{j_{2} m_{2}}=\frac{\delta_{j_{1} j_{2}} \delta_{m_{1} m_{2}}}{2 j_{1}+1} \sum_{m}\braOket{j_{1} m}{\mathfrak{M}}{j_{1} m}
\end{equation}
\begin{center}
\includegraphics[valign=t]{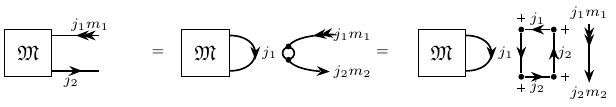}
\end{center}

This follows from the fact that any dependence on two directions is, actually, the dependence on the angle between these directions, i.e. it may be represented by the scalar product of two functions \(\ket{j m}\) and \(\bra{ j m}\).
\item A diagram with three external \(j\)-lines which represent \(\ket{j_{1} m_{1}}, \ket{j_{2} m_{2}}\) and \(\ket{j_{3} m_{3}}\) is proportional to the \(3 j m\) symbol \(\threej{j_{1}}{j_{2}}{j_{3}}{m_{1}}{m_{2}}{m_{3}}\)
\begin{equation}
\braOket{00}{\mathfrak{M}}{j_{1} m_{1} j_{2} m_{2} j_{3} m_{3}}=\braOketred{0}{\mathfrak{M}}{j_{1} j_{2} j_{3}}\threej{j_{1}}{j_{2}}{j_{3}}{m_{1}}{m_{2}}{m_{3}},
\end{equation}
where \(\braOketred{0}{\mathfrak{M}}{j_{1} j_{2} j_{3}}\) is the reduced matrix element defined by
\begin{equation}
\braOketred{0}{\mathfrak{M}}{j_{1} j_{2} j_{3}}=
\sum_{m_{1} m_{2} m_{3}}
\braOket{00}{\mathfrak{M}}{j_1 m_{1} j_{2} m_{2} j_{3} m_{3}}
\threej{j_{1}}{j_{2}}{j_{3}}{m_{1}}{m_{2}}{m_{3}},
\end{equation}
\centerline{\includegraphics[valign=t]{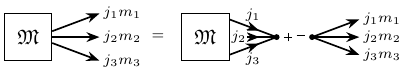}}
\item A diagram with four external \(j\)-lines corresponding \(j_{1} m_{1}, j_{2} m_{2}, j_{3} m_{3}\) and \(j_{4} m_{4}\) may be presented as a sum (over \(j\) ) of the product of two \(3 j m\) symbols
\begin{equation}
\braOket{j_{1} m_{1} j_{2} m_{2}}{\mathfrak{M}}{ j_{3} m_{3} j_{4} m_{4}}=\sum_{j m}(2 j+1)\threej{j_{1}}{j_{2}}{j}{m_{1}}{m_{2}}{m}\braOket{j_{1} j_{2} j m}{\mathfrak{M}}{j_{3} j_{4} j m}\threej{j_{3}}{j_{4}}{j}{m_{3}}{m_{4}}{m},
\end{equation}
where
\begin{equation}
\braOket{j_{1} j_{2} j m}{\mathfrak{M}}{j_{3} j_{4} j m} \equiv \sum_{m_{1} m_{2} m_{3} m_{4}}\threej{j_{1}}{j_{2}}{j}{m_{1}}{m_{2}}{m}\braOket{j_{1} m_{1} j_{2} m_{2}}{\mathfrak{M}}{j_{3} m_{3} j_{4} m_{4}}\threej{j_{3}}{j_{4}}{j^{\prime}}{m_{3}}{m_{4}}{m^{\prime}}
\end{equation}
\begin{center}
\includegraphics[valign=t]{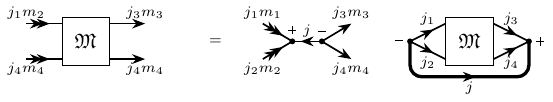}
\end{center}
\end{enumerate}

\begin{table}[tbhp!]
\begin{center}
\captionsetup{justification=centering}
\caption{Generalized Wigner-Eckart Theorem.}\label{tab:11.15}
\small
\setlength{\tabcolsep}{5pt}
\renewcommand{\arraystretch}{1.3}
\begin{tabular}{@{}m{0.42\textwidth}m{0.48\textwidth}@{}}
\hline
\multicolumn{1}{c|}{Analytical Expression} & Graphical Representation \\
\hline
% ---- row 1 ----
\[\braOket{00}{\mathfrak{M}}{lm}=\braOket{00}{\mathfrak{M}}{00}\,\delta_{l0}\delta_{m0}\]
&
\centerline{\includegraphics[valign=t]{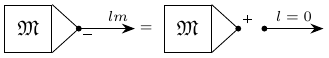}}
\\
% ---- row 2 ----
\[\braOket{jm}{\mathfrak{M}}{j'm'}=\dfrac{\delta_{jj'}\delta_{mm'}}{(2j+1)}\braOketred{j}{\mathfrak{M}}{j},\]
where \[\braOketred{j}{\mathfrak{M}}{j}\equiv\sum_{m}\braOket{jm}{\mathfrak{M}}{jm}\]
&
\centerline{\includegraphics[valign=t]{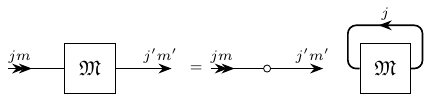}}
\\
% ---- row 3 ----
\[\braOket{\Omega}{\mathfrak{M}}{jm}=\braket{\Omega}{jm}\dfrac{\braOketred{j}{\mathfrak{M}}{j}}{(2j+1)}\]
&
\centerline{\includegraphics[valign=t]{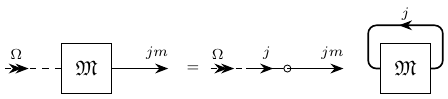}}
\\

% ---- row 4 ----
\[\braOket{\Omega_1}{\mathfrak{M}}{\Omega_2}=\dfrac{1}{4\pi}\sum_l P_l(\cos\omega_{12})\braOketred{l}{\mathfrak{M}}{l},\]
where \(\omega_{12}\) is the angle \(\widehat{\Omega_1\Omega_2}\) and
\[P_l(\cos\omega_{12})=\sum_m\braket{\Omega_1}{lm}\braket{lm}{\Omega_2}\]
&
\centerline{\includegraphics[valign=t]{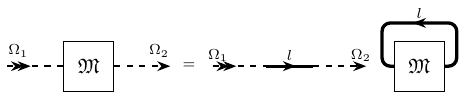}}
\\

% ---- row 5 ----
\[\begin{aligned}[t]
&\braOket{j_1m_1}{\mathfrak{M}}{j_2m_2 j_3m_3}\\
&=(-1)^{j_1-m_1}\threej{j_1}{j_2}{j_3}{-m_1}{m_2}{m_3}\braOketred{j_1}{\mathfrak{M}_{j_2}}{j_3},
\end{aligned}\]
where
\[\begin{aligned}[t]
&\braOketred{j_1}{\mathfrak{M}_{j_2}}{j_3}\equiv\sum_{m_1m_2m_3}(-1)^{j_1-m_1}\\
&\quad\times\threej{j_1}{j_2}{j_3}{-m_1}{m_2}{m_3}\braOket{j_1m_1}{\mathfrak{M}}{j_2m_2 j_3m_3}
\end{aligned}\]
&
\centerline{\includegraphics[valign=t]{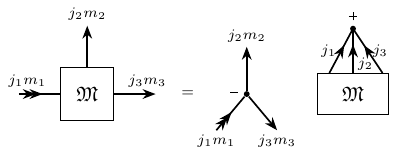}}
\\

% ---- row 6 ----
\[\begin{aligned}[t]
&\braOket{\Omega}{\mathfrak{M}}{j_2m_2 j_3m_3}=\sum_{j_1m_1}\braket{\Omega}{j_1m_1}\\
&\quad\times(-1)^{j_1-m_1}\threej{j_1}{j_2}{j_3}{-m_1}{m_2}{m_3}\braOketred{j_1}{\mathfrak{M}_{j_2}}{j_3}
\end{aligned}\]
&
\centerline{\includegraphics[valign=t]{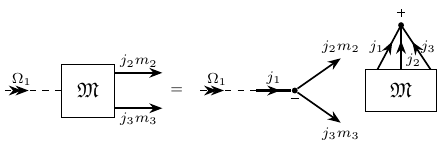}}
\\

% ---- row 7 ----
\[\begin{aligned}[t]
&\braOket{\Omega_1}{\mathfrak{M}}{j_2m_2 \Omega_3}=\sum_{j_1m_1 j_3m_3}\braket{\Omega_1}{j_1m_1}(-1)^{j_1-m_1}\\
&\quad\times\threej{j_1}{j_2}{j_3}{-m_1}{m_2}{m_3}\braket{j_3m_3}{\Omega_3}\braOketred{j_1}{\mathfrak{M}_{j_2}}{j_3}
\end{aligned}\]
&
\centerline{\includegraphics[valign=t]{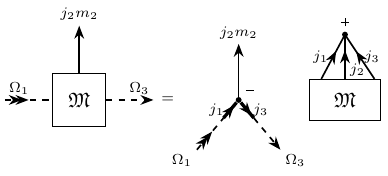}}
\\
\end{tabular}
\end{center}
\end{table}

\begin{table}[tbhp!]
\begin{center}
\captionsetup{justification=centering,labelformat=empty}
\caption*{Table \ref{tab:11.15}. (Cont'd)}
\small
\setlength{\tabcolsep}{5pt}
\renewcommand{\arraystretch}{1.3}
\begin{tabular}{@{}m{0.42\textwidth}m{0.48\textwidth}@{}}
\hline
\multicolumn{1}{c|}{Analytical Expression} & Graphical Representation \\
\hline
\[
\begin{gathered}
\braOket{j_{1} m_{1} j_{2} m_{2}}{\mathfrak{M}}{j_{3} m_{3} j_{4} m_{4}} \\
=\sum_{J M}(2 J+1)\threej{j_{1}}{j_{2}}{J}{m_{1}}{m_{2}}{M}\braOket{j_{1} j_{2} ; J M}{\mathfrak{M}}{j_{3} j_{4} ; J M} \\
\times\threej{j_{3}}{j_{4}}{J}{m_{3}}{m_{4}}{M}
\end{gathered}
\] &
\centerline{\includegraphics[valign=t]{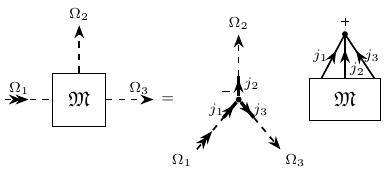}}
\\
\[
\begin{gathered}
\braOket{\Omega_{1}}{\mathfrak{M}}{\Omega_{2} \Omega_{3}}=\sum_{j_{1} j_{2} j_{3}}\braOketred{j_{1}}{\mathfrak{M}_{j_{2}}}{j_{3}} \\
\times \sum_{m_{1} m_{2} m_{3}}\braket{\Omega_{1}}{j_{1} m_{1}}(-1)^{j_{1}-m_{1}}\threej{j_{1}}{j_{2}}{j_{3}}{-m_{1}}{m_{2}}{m_{3}} \\
\times\braket{j_{2} m_{2}}{\Omega_{2}}\braket{j_{3} m_{3}}{\Omega_{3}}
\end{gathered}
\] &
\centerline{\includegraphics[valign=t]{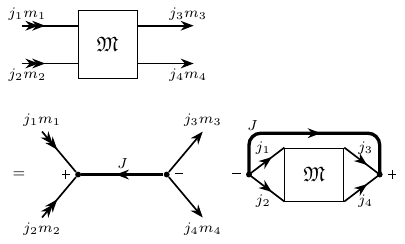}}
\\
hline
\end{tabular}
\end{center}
\end{table}

\subsection{Scheme for the Application of the Graphical Technique}\label{sec:11.4.3}\index{graphical method!scheme of application}\index{Clebsch-Gordan coefficient}\isym{C-clebsch@$\clebsch{a}{\alpha}{b}{\beta}{c}{\gamma}$, Clebsch-Gordan coefficient}
The aim of calculations inherent in the angular momentum theory is to transform an original formula in such a way as to express all the dependence on the orientation of the coordinate system in the form of a simple standard function \(\left(\clebsch{j_{1}}{m_{1}}{j_{2}}{m_{2}}{j}{m}, Y_{l m}(\Omega), D_{M M^{\prime}}^{J}(R)\right.\), or some related combination). The remaining factor (or the expansion coefficient) is invariant and can be represented as a product (or sum of products) of the standard \(3 j, 6 j, 9 j\) symbols.

To carry out these calculations by the graphical method one has to perform the following operations:
\begin{enumerate}[label=(\alph*)]
  \item
To represent the original analytic expression in diagrammatic form using the graphical representation of functions and operations in accordance with the rules formulated in Sections \ref{sec:11.1} and \ref{sec:11.2}.
\item To eliminate all fictitious \(j=0\) lines and \(R=1\) lines according to the rules of Sections \ref{sec:11.3.9} and \ref{sec:11.3.10}.\\
\item To link step-by-step, all the subdiagrams which have a common thick \(j\)-line, using the rules for graphical
multiplication of subdiagrams (Sec.~\ref{sec:11.3.5}).
\item To replace each internal \(\Omega\)-line by the corresponding thick \(j\)-line, using the rule of graphical integration (Sec.~\ref{sec:11.3.8}).
\item To remove thick \(j\)-lines with identical nodes \(\left(j_{1} j_{2} j_{3}\right)\) and reconnect the remaining thin \(j_{1}\) - and \(j_{2}\)-lines in accordance with the rule of graphical summation (Sec.~\ref{sec:11.3.7}).
\item To cut the resultant diagram (according to the rules of Sec.~\ref{sec:11.3.6}) into subdiagrams which are topologically similar to diagrams of the standard functions. In particular, one should separate all the external lines in the form of the simplest possible subdiagram.
\item To determine the phase factor of a diagram by reducing each subdiagram to some standard diagram (Tables 11.1 and 11.2) by means of reversal (when necessary) of node signs and line directions according to the rules from Section \ref{sec:11.3.2} and \ref{sec:11.3.3}.
\item To represent the final diagram in analytic form.
\end{enumerate}
One can use a simplified version of the graphical method without taking the signs of the nodes and directions of the lines into account, if one is interested only in the absolute value of an expression but not in the phase.
Such an  approach is taken in the next chapter.
\chapter{Sums involving vector addition and recoupling coefficients}\label{chap:12}\index{sums!of vector-addition and recoupling coefficients}\index{recoupling}\index{vector-addition coefficients}
Many quantum-mechanical problems are concerned with calculations of sums containing products of the Clebsch-Gordan coefficients and \(3nj\) symbols\index{Clebsch-Gordan coefficient}\index{3nj symbols@$3nj$ symbols}\isym{C-clebsch@$\clebsch{a}{\alpha}{b}{\beta}{c}{\gamma}$, Clebsch-Gordan coefficient} whose summation indices represent quantum numbers of angular momenta and momentum projections. Summation over different indices may be performed independently and successively. It is more convenient to start with the summation of 3jm symbol products over projections. This summation may be carried out with the aid of the equations presented in Sec.~\ref{sec:12.1}. Summation over projections yields invariant \(3nj\) symbols. The next step is to sum the products of \(3nj\) symbols over momenta; this may be done with the aid of the equations of Sec.~\ref{sec:12.2}.

\section{Summation of products of $3 j m$ symbols}\label{sec:12.1}\index{3jm symbol@$3jm$ symbol!sums of products of}\index{sums!of the $3jm$ symbols}\isymo{3jm@$\bigl(\begin{smallmatrix} a & b & c\\ \alpha & \beta & \gamma\end{smallmatrix}\bigr)$, Wigner $3jm$ symbol}
This section contains a collection of sums involving the 3jm symbols. Each sum is illustrated by a diagram which shows the coupling scheme of angular momenta involved in the sum.\index{graphical method}\index{diagrams}\index{coupling schemes} The equations are grouped in accordance with the number of $3 j m$ symbols in the products. Within each group the equations are arranged according to the number of fixed indices, i.e., the number of external lines of the corresponding diagram. The summations are of two kinds:\\
\begin{enumerate}[label=(\alph*)]
\item Summation index $j$ represents an angular momentum and runs over all possible values consistent with the triangle condition.\index{triangle condition}\index{summation!over angular momentum}\index{thick j-line@thick $j$-line} Summations of this kind are shown by thick internal $j$-lines in the diagrams.
\item Summation index $m_{j}$ is a projection of $j$ and runs over all allowed values\index{summation!over projections}\index{j-line@$j$-line}, $j, j-1, j-2, \ldots,-j$. Summations of such kind are shown by thin internal $j$-lines in the diagrams.
\end{enumerate}

The sums are written in such a form that each summation index $m_{j}$ is included in two 3jm symbols, although with the opposite signs. In addition, the sums contain the phase factors $(-1)^{j-m_{j}}$ which ensure that the results of summation are invariant under coordinate rotations (Sec.~\ref{sec:11.2}). Any sum over projections, which is invariant with respect to rotations, may be converted into such a form (Sec.~\ref{sec:8.4}).

Hereafter, we use the notation\index{dimension factor}\isymg{Pi-dim@$\Pi_{a b c \ldots}$, dimension factor $[(2a+1)(2b+1)\cdots]^{1/2}$}

\begin{equation}
\Fact{a b \ldots f}=\sqrt{(2 a+1)(2 b+1) \ldots(2 f+1)} . \label{chap12:eq:1}
\end{equation}

\subsection{Sum Involving One $3 j m$ Symbol}\label{sec:12.1.1}\index{sums!with one $3jm$ symbol}
\begin{equation}
\sum_{\alpha}(-1)^{a-\alpha}\threej{a}{a}{c}{\alpha}{-\alpha}{\gamma} \label{chap12:eq:2}=\Fact{a} \delta_{c 0} \delta_{\gamma 0}.
\end{equation}
\subsection{Sums Involving Products of Two $3 j m$ Symbols}\label{sec:12.1.2}\index{sums!with two $3jm$ symbols}\index{6j symbol@$6j$ symbol}\isymo{6j@$\bigl\{\begin{smallmatrix} a & b & c\\ d & e & f\end{smallmatrix}\bigr\}$, Wigner $6j$ symbol}
\begin{equation}
\sum_{\psi \kappa}(-1)^{p-\psi+q-\kappa}\threej{a}{p}{q}{-\alpha}{\psi}{\kappa} \label{chap12:eq:3}\threej{p}{q}{a^{\prime}}{-\psi}{-\kappa}{\alpha^{\prime}}=\frac{(-1)^{a+\alpha}}{\Fact{a}^{2}}\{a p q\} \delta_{a a^{\prime}} \delta_{\alpha \alpha^{\prime}},
\end{equation}

\centerline{\includegraphics[alt={}]{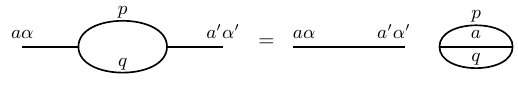}}
\begin{equation}
\sum_{q \kappa}(-1)^{q-\kappa} \Fact{q}^{2}\threej{a}{b}{q}{-\alpha}{-\beta}{\kappa} \label{chap12:eq:4}\threej{q}{a}{b}{-\kappa}{\alpha^{\prime}}{\beta^{\prime}}=(-1)^{a+\alpha+b+\beta} \delta_{\alpha \alpha^{\prime}} \delta_{\beta \beta^{\prime}},
\end{equation}

\centerline{\includegraphics[alt={}]{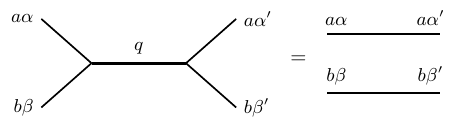}}
\begin{equation}
\sum_{\kappa}(-1)^{q-\kappa}\threej{a}{b}{q}{\alpha}{\beta}{-\kappa} \label{chap12:eq:5}\threej{q}{d}{c}{\kappa}{\delta}{\gamma}=(-1)^{2 a} \sum_{x \xi}(-1)^{x-\xi} \Fact{x}^{2}\threej{a}{c}{x}{\alpha}{\gamma}{-\xi}\threej{x}{d}{b}{\xi}{\delta}{\beta}\sixj{b}{d}{x}{c}{a}{q},
\end{equation}

\centerline{\includegraphics[alt={}]{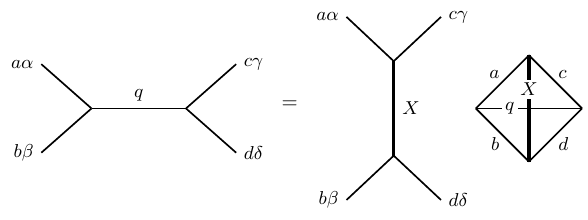}}
\subsection{Sums Involving Products of Three $3 j m$ Symbols}\label{sec:12.1.3}\index{sums!with three $3jm$ symbols}\index{9j symbol@$9j$ symbol}\isymo{9j@$\bigl\{\begin{smallmatrix} a & b & c\\ d & e & f\\ g & h & j\end{smallmatrix}\bigr\}$, Wigner $9j$ symbol}
\begin{equation}
\sum_{\kappa \psi \rho}(-1)^{p-\psi+q-\kappa+r-\rho}\threej{p}{a}{q}{\psi}{\alpha}{-\kappa} \label{chap12:eq:6}\threej{q}{b}{r}{\kappa}{\beta}{-\rho}\threej{r}{c}{p}{\rho}{\gamma}{-\psi}=\threej{a}{b}{c}{-\alpha}{-\beta}{-\gamma}\sixj{a}{b}{c}{r}{p}{q},
\end{equation}

\centerline{\includegraphics[alt={}]{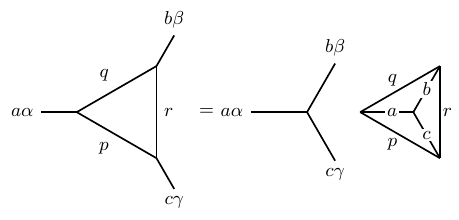}}

\begin{multline}
\sum_{\psi \kappa}\threej{a}{b}{p}{\alpha}{\beta}{\psi}\threej{p}{c}{q}{-\psi}{-\gamma}{-\kappa}\threej{q}{d}{e}{\kappa}{\delta}{\varepsilon}  \\
=\sum_{\substack{x y \\
\varepsilon \eta}} \Fact{x y}^{2}\threej{a}{d}{x}{\alpha}{\delta}{\xi}\threej{x}{c}{y}{-\xi}{-\gamma}{-\eta}\threej{y}{b}{e}{\eta}{\beta}{\varepsilon}\ninej{a}{b}{p}{d}{e}{q}{x}{y}{c}, \label{chap12:eq:7}
\end{multline}

\centerline{\includegraphics[alt={}]{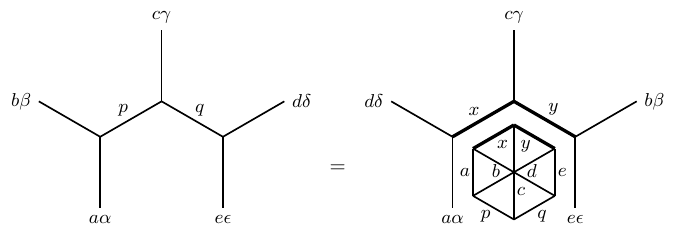}}

\subsection{Sums Involving Products of Four $3 j m$ Symbols}\label{sec:12.1.4}\index{sums!with four $3jm$ symbols}\index{12j symbol@$12j$ symbol}
% PROOFREAD: contains OCR-garbled or 12j Wigner array(s) not auto-converted -- verify/reconstruct against orig.
\begin{gather}
\sum_{\kappa \psi \rho \sigma \tau}(-1)^{p-\psi+q-\kappa+r-\rho+s-\sigma+t-\tau}
\threej{p}{a}{q}{\psi}{-\alpha}{\kappa}
\threej{q}{r}{t}{-\kappa}{\rho}{\tau}
\threej{r}{a^{\prime}}{s}{-\rho}{\alpha^{\prime}}{\sigma}
\threej{s}{p}{t}{-\sigma}{-\psi}{-\tau}  \notag\\
=\frac{(-1)^{a-\alpha}}{\Fact{a}^{2}}\sixj{q}{p}{a}{s}{r}{t} \delta_{a a^{\prime}} \delta_{\alpha \alpha^{\prime}}, \label{chap12:eq:8}
\end{gather}

\centerline{\includegraphics[alt={}]{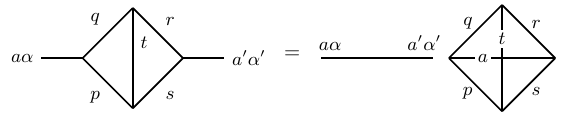}}

\begin{gather}
\sum_{\rho \psi \kappa \sigma}(-1)^{p-\psi+q-\kappa+r-\rho+s-\sigma}\threej{p}{a}{q}{\psi}{\alpha}{-\kappa}\threej{q}{b}{r}{\kappa}{\beta}{-\rho}\threej{r}{s}{p}{\rho}{\sigma}{-\psi}\threej{s}{c}{d}{-\sigma}{\gamma}{\delta}  \notag\\
=\sixj{a}{b}{s}{r}{p}{q} \sum_{\sigma}(-1)^{s-\sigma}\threej{a}{s}{b}{\alpha}{\sigma}{\beta}\threej{d}{s}{c}{\delta}{-\sigma}{\gamma}, \label{chap12:eq:9}
\end{gather}

\centerline{\includegraphics[alt={}]{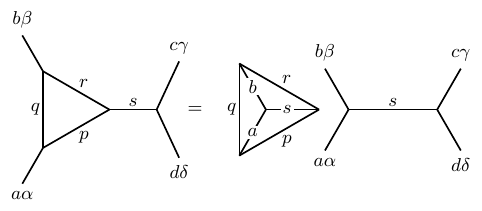}}

\begin{align}
& \sum_{\psi \kappa \rho \sigma}(-1)^{p-\psi+q-\kappa+r-\rho+s-\sigma}\threej{p}{a}{q}{\psi}{\alpha}{-\kappa}\threej{q}{b}{r}{\kappa}{\beta}{-\rho}\threej{r}{c}{s}{\rho}{\gamma}{-\sigma}\threej{s}{d}{p}{\sigma}{\delta}{-\psi}  \notag\\
& =(-1)^{s-a-d-q} \sum_{x \xi}(-1)^{x-\xi} \Fact{x}^{2}\threej{a}{x}{d}{\alpha}{-\xi}{\delta}\threej{b}{x}{c}{\beta}{\xi}{\gamma}\sixj{a}{x}{d}{s}{p}{q}\sixj{b}{x}{c}{s}{r}{q} \label{chap12:eq:10}
\end{align}

\centerline{\includegraphics[alt={}]{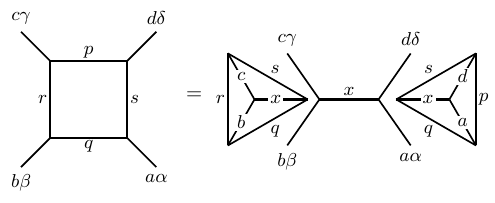}}
\begin{equation}
=(-1)^{2 r-s-d-a-q} \sum_{x \xi}(-1)^{x-\xi} \Fact{x}^{2}\threej{a}{x}{c}{\alpha}{-\xi}{\gamma} \label{chap12:eq:11}\threej{b}{x}{d}{\beta}{\xi}{\delta}\ninej{a}{p}{q}{x}{d}{b}{c}{s}{r},
\end{equation}

\centerline{\includegraphics[alt={}]{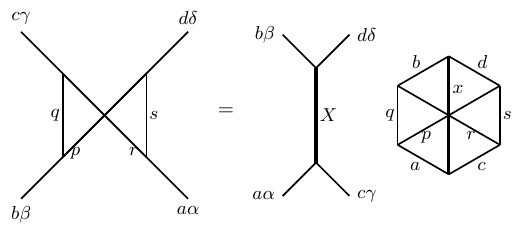}}

\subsection{Sums Involving Products of Five 3jm Symbols}\label{sec:12.1.5}\index{sums!with five $3jm$ symbols}
\begin{align}
& \sum_{\substack{\psi \kappa \rho \sigma \\
\tau \nu \mu}}(-1)^{p-\psi+q-\kappa+r-\rho+s-\sigma+t-\tau+u-\nu+v-\mu}\threej{r}{t}{q}{\rho}{\tau}{-\kappa}\threej{q}{p}{u}{\kappa}{\psi}{-\nu}\threej{u}{a}{v}{\nu}{\alpha}{-\mu}\threej{v}{s}{r}{\mu}{\sigma}{-\rho}\threej{s}{p}{t}{-\sigma}{-\psi}{-\tau}  \notag\\
&=(-1)^{2 u} \frac{1}{\Fact{u}} \delta_{u v} \delta_{a 0} \delta_{\alpha 0}\sixj{q}{p}{u}{s}{r}{t}, \label{chap12:eq:12}
\end{align}

\centerline{\includegraphics[alt={}]{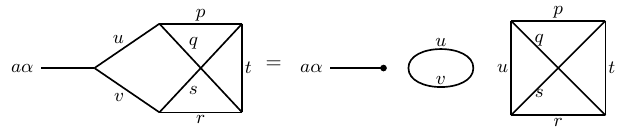}}

\begin{align}
\sum_{\substack{\psi \kappa p \sigma \\
\tau \nu}}(-1)^{p-\psi+q-\kappa+r-\rho+s-\sigma+t-\tau+u-\nu} & \threej{p}{a}{q}{\psi}{\alpha}{-\chi}\threej{q}{r}{t}{\kappa}{\rho}{-\tau}\threej{t}{s}{p}{\tau}{-\sigma}{-\psi}\threej{s}{c}{u}{\sigma}{\gamma}{-\nu}\threej{u}{b}{r}{\nu}{\beta}{-\rho}  \notag\\
= & \threej{a}{b}{c}{\alpha}{\beta}{\gamma}\sixj{a}{b}{c}{u}{s}{r}\sixj{a}{p}{q}{t}{r}{s}, \label{chap12:eq:13}
\end{align}

\centerline{\includegraphics[alt={}]{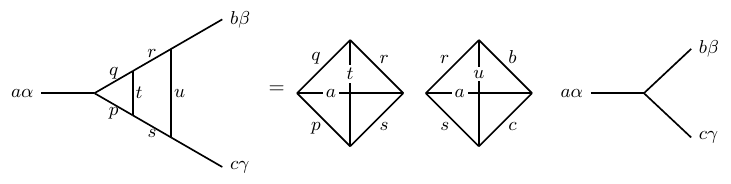}}

\begin{gather}
\sum_{\psi \kappa \rho r}(-1)^{p-\psi+q-\kappa+r-\rho+\rho-\sigma+t-r+u-\nu}\threej{p}{a}{q}{\psi}{\alpha}{\kappa}\threej{q}{t}{r}{-\kappa}{-\tau}{-\rho}\threej{r}{b}{s}{\rho}{\beta}{\sigma}\threej{s}{p}{u}{-\sigma}{-\psi}{-\nu}\threej{u}{c}{t}{\nu}{\gamma}{r}  \notag\\
=(-1)^{r+b+\theta}\threej{a}{b}{c}{-\alpha}{-\beta}{-\gamma}\ninej{a}{b}{c}{p}{s}{u}{q}{r}{t}, \label{chap12:eq:14}
\end{gather}

\centerline{\includegraphics[alt={}]{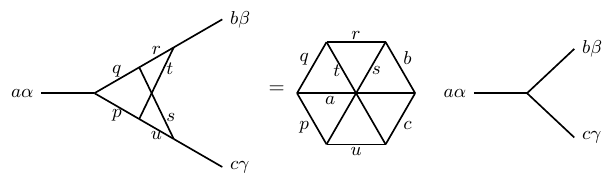}}

\begin{align}
& \sum_{\psi \kappa \rho}(-1)^{p-\psi+q-\kappa+r-\rho+s-\sigma+t-\tau}\threej{p}{a}{q}{\psi}{\alpha}{-\kappa}\threej{q}{b}{r}{\kappa}{\beta}{-\rho}\threej{r}{c}{s}{\rho}{\gamma}{-\sigma}\threej{s}{d}{t}{\sigma}{\delta}{-\tau}\threej{t}{e}{p}{\tau}{\varepsilon}{-\psi}  \notag\\
&=(-1)^{t-p-b-a-d-c} \sum_{x \xi y \eta}(-1)^{x-\xi+y-\eta} \Fact{x y}^{2}\threej{a}{b}{x}{\alpha}{\beta}{\xi}\threej{x}{e}{y}{-\xi}{\varepsilon}{-\eta}\threej{y}{c}{d}{\eta}{\gamma}{\delta}\sixj{a}{b}{x}{r}{p}{q}\sixj{x}{e}{y}{t}{r}{p}\sixj{y}{c}{d}{s}{t}{r} \label{chap12:eq:15}
\end{align}

\centerline{\includegraphics[alt={}]{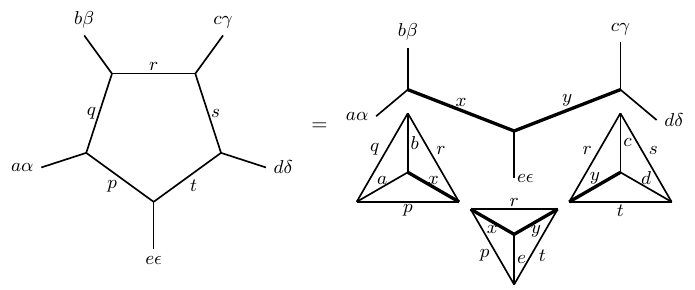}}
\begin{equation}
=(-1)^{p-t-b-a-d-c} \sum_{x \xi y \eta}(-1)^{x-\xi+y-\eta} \Fact{x y}^{2}\threej{a}{b}{x}{\alpha}{\beta}{\xi} \label{chap12:eq:16}\threej{x}{d}{y}{-\xi}{\delta}{-\eta}\threej{y}{c}{e}{\eta}{\gamma}{\varepsilon}\sixj{a}{b}{x}{r}{p}{q}\ninej{c}{e}{y}{r}{p}{x}{s}{t}{d},
\end{equation}

\centerline{\includegraphics[alt={}]{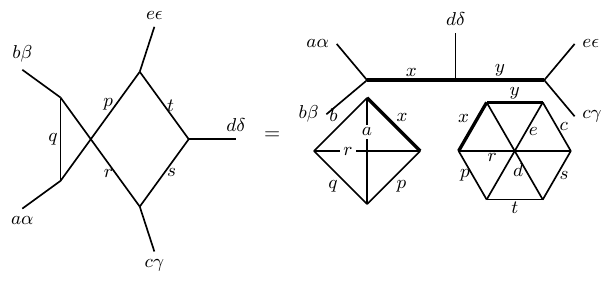}}

\begin{gather}
=(-1)^{r-t+p+a+d} \sum_{x \xi y \eta z}(-1)^{x-\xi+y-\eta+x} \Fact{x y z}^{2}\threej{a}{c}{x}{\alpha}{\gamma}{\xi}\threej{x}{e}{y}{-\xi}{\varepsilon}{-\eta}\threej{y}{b}{d}{\eta}{\beta}{\delta}  \notag\\
\sixj{c}{q}{z}{b}{s}{r}\sixj{a}{c}{x}{z}{p}{q}\sixj{x}{e}{y}{t}{z}{p}\sixj{y}{b}{d}{s}{t}{z}, \label{chap12:eq:17}
\end{gather}

\centerline{\includegraphics[alt={}]{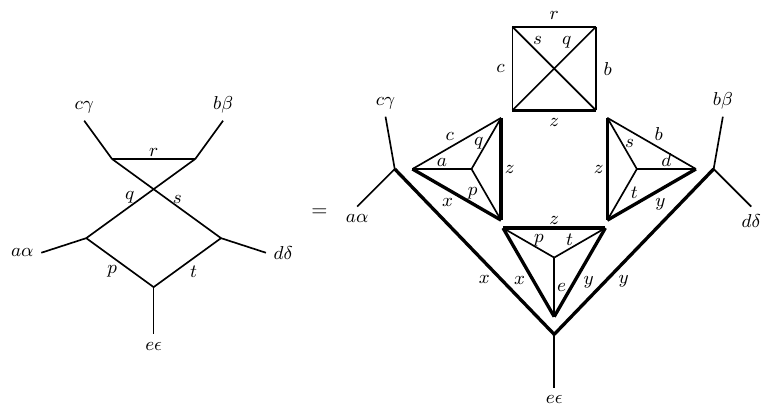}}

\subsection{Sums Involving Products of Six 3jm Symbols}\label{sec:12.1.6}\index{sums!with six $3jm$ symbols}
\begin{gather}
\sum_{\substack{\psi \kappa \rho \sigma \\ \tau \nu \mu \lambda}}
(-1)^{p-\psi+q-\kappa+r-\rho+\vartheta-\sigma+t-\tau+u-\nu+v-\mu+t^{\prime}-\lambda}
\threej{p}{a}{q}{\psi}{-\alpha}{\kappa}  
\threej{q}{r}{t'}{-\kappa}{\rho}{\lambda}
\threej{t^{\prime}}{u}{v}{-\lambda}{\nu}{\mu}
\threej{u}{v}{t}{-\nu}{-\mu}{\tau}\notag\\
\times
\threej{r}{a'}{s}{-\rho}{\alpha'}{\sigma}
\threej{s}{p}{t}{-\sigma}{-\psi}{-r}
= \frac{(-1)^{2t+a-\alpha}}{\Fact{at}^2}
\sixj{a}{p}{q}{t}{r}{s} {tuv} \delta+_{t't}\delta_{aa'}\delta_{\alpha\alpha'}
\label{chap12:eq:18}
\end{gather}

\centerline{\includegraphics[alt={}]{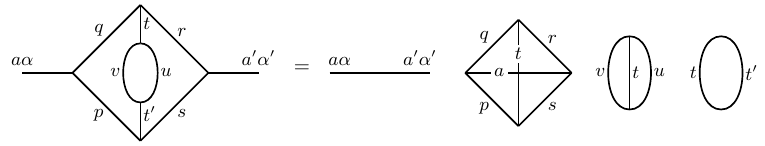}}

\begin{gather}
\sum_{\substack{\psi \kappa p \sigma \\
\nu \mu \tau \lambda}}(-1)^{p-\psi+q-\kappa+q^{\prime}-\rho+p^{\prime}-\sigma+u-\nu+v-\mu+t-\tau+\rho-\lambda}
\threej{u}{v}{q}{\nu}{\mu}{\kappa}
\threej{u}{v}{q^{\prime}}{-\nu}{-\mu}{-\rho}\threej{t}{s}{p}{\tau}{\lambda}{\psi}\threej{t}{s}{p^{\prime}}{-\tau}{-\lambda}{-\sigma}  \notag\\
\times\threej{q}{p}{a}{\kappa}{\psi}{\alpha}\threej{q^{\prime}}{p^{\prime}}{a^{\prime}}{-\rho}{-\sigma}{-\alpha^{\prime}}=\frac{(-1)^{\alpha-a}}{\Fact{a p q}^{2}} \delta_{q q^{\prime}} \delta_{p p^{\prime}} \delta_{a a^{\prime}} \delta_{\alpha \alpha^{\prime}}\{u v q\}\{p t s\}\{a p q\}, \label{chap12:eq:19}
\end{gather}

\centerline{\includegraphics[alt={}]{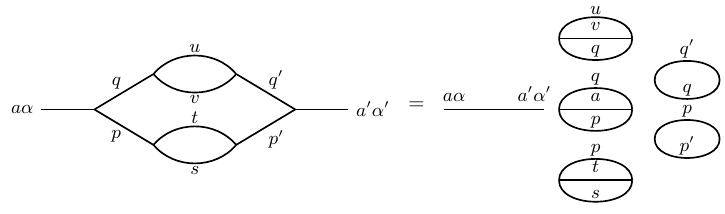}}

\begin{gather}
\sum_{\psi \kappa \rho \sigma}(-1)^{p-\psi+q-\kappa+r-\rho+s-\sigma+t-\tau+v-\mu+u-\nu+g-\lambda}\threej{a}{q}{p}{-\alpha}{\kappa}{\psi}\left(\begin{array}{rrr}
p & t &g\\
-\psi & r &-\lambda
\end{array}\right)\left(\begin{array}{rrr}
g & u  & v\\
\lambda&-\nu & \mu
\end{array}\right)\left(\begin{array}{rrr}
v & r &q\\
-\mu & \rho&-\kappa
\end{array}\right)  \notag\\
\times\threej{r}{u}{s}{-\rho}{\nu}{-\sigma}\left(\begin{array}{rrr}
s & t &a^{\prime} \\
\sigma & -\tau &\alpha^{\prime}
\end{array}\right)=\frac{(-1)^{s-p-\alpha-g}}{\Fact{a}^{2}}\sixj{q}{g}{s}{u}{r}{v}\sixj{a}{p}{q}{g}{s}{t} \delta_{a a^{\prime}} \delta_{\alpha \alpha^{\prime}} \label{chap12:eq:20}
\end{gather}

\centerline{\includegraphics[alt={}]{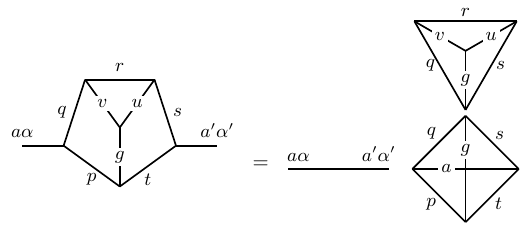}}
\begin{gather}
\sum_{\substack{\psi \kappa \rho \sigma \\
\nu \tau \mu \lambda}}
(-1)^{p-\psi+q-\kappa+r-\rho+s-\sigma+u-\nu+t-r+v-\mu+g-\lambda}\threej{p}{a}{q}{\psi}{\alpha}{-\kappa}\threej{q}{r}{v}{\kappa}{\rho}{-\mu}\threej{v}{s}{p}{\mu}{-\sigma}{-\psi}\threej{u}{r}{g}{\nu}{-\rho}{-\lambda} \notag\\
\times\threej{t}{a'}{u}{-\tau}{-\alpha'}{-\nu}
\threej{s}{t}{a^\prime}{\sigma}{-\tau}{-\alpha^{\prime}}
=
\frac{(-1)^{2 g+a-\alpha}}{\Fact{a}^{2}}
\sixj{a}{r}{s}{g}{t}{u}
\sixj{a}{r}{s}{v}{p}{q} 
\delta_{a a^{\prime}} \delta_{\alpha \alpha^{\prime}}, .\label{chap12:eq:21}
\end{gather}

\centerline{\includegraphics[alt={}]{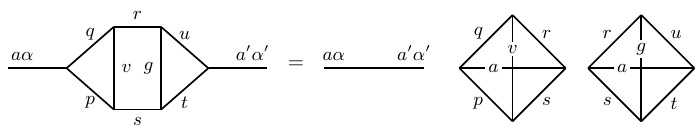}}
\begin{gather}
\sum_{\substack{\psi \kappa \rho \sigma \\
\nu \tau \mu \lambda}}
(-1)^{p-\psi+q-\kappa+r-\rho+s-\sigma+u-\nu+t-r+v-\mu+g-\lambda}
\threej{a}{p}{q}{\alpha}{\psi}{\kappa}
\threej{q}{g}{s}{-\kappa}{-\lambda}{-\sigma}
\threej{s}{t}{v}{\sigma}{\tau}{\mu}
\threej{v}{p}{r}{-\mu}{-\psi}{-\rho} \notag\\
\times
\threej{r}{g}{u}{\rho}{\lambda}{\nu}
\threej{u}{t}{a'}{-\nu}{-\tau}{-\alpha'}=
\frac{(-1)^{a-\alpha}}{\Fact{a}^{2}}
\ninej{a}{p}{q}{u}{r}{g}{t}{v}{s}
 \delta_{a a^{\prime}} \delta_{\alpha \alpha^{\prime}}, .\label{chap12:eq:21a}
\end{gather}

\centerline{\includegraphics[alt={}]{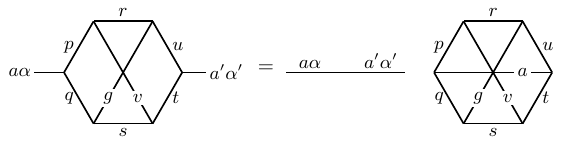}}
\begin{gather}
\sum_{\substack{\psi \kappa \nu \rho  \\
\mu \sigma \lambda}}
(-1)^{p-\psi+q-\kappa+u-\nu+r-\rho+v-\mu+s-\sigma+g-\lambda}
\threej{p}{a}{q}{\psi}{\alpha}{-\kappa}
\threej{q}{u}{g}{-\kappa}{\nu}{-\lambda}
\threej{g}{v}{s}{\lambda}{\mu}{-\sigma}
\threej{s}{d}{p}{\sigma}{\delta}{-\psi} \notag\\
\times
\threej{u}{b}{r}{-\nu}{\beta}{\rho}
\threej{r}{c}{v}{-\rho}{\gamma}{-\mu}\notag\\
=
(-1)^{s+b+u+v-q-a-d-c}\sum_{x\xi}{\Fact{x}^{2}}
\threej{a}{x}{d}{\alpha }{-\xi}{\beta}
\threej{c}{x}{b}{\gamma}{\xi}{\beta}
\sixj{a}{x}{d}{s}{p}{q}
\sixj{c}{x}{b}{u}{r}{v}
\sixj{u}{x}{v}{s}{g}{q}
.\label{chap12:eq:22a}
\end{gather}

\centerline{\includegraphics[alt={}]{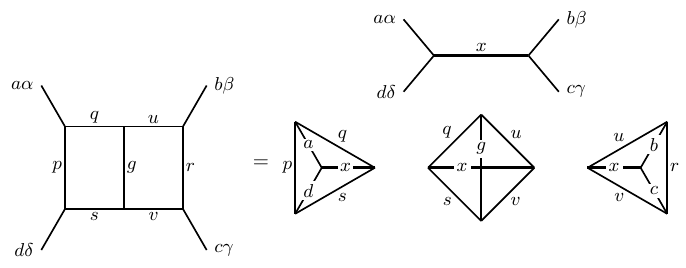}}

\begin{equation}
=(-1)^{v+u+r+p-g-a-c} \sum_{x y \eta}(-1)^{y-\eta} \Fact{x y}^{2}
\threej{a}{y}{b}{\alpha}{-\eta}{\beta} 
\threej{d}{y}{c}{\delta}{\eta}{\gamma}
\sixj{a}{y}{b}{r}{u}{x}
\sixj{d}{y}{c}{r}{v}{x}
\sixj{a}{p}{q}{g}{u}{x}
\sixj{v}{g}{s}{p}{d}{x},
\label{chap12:eq:22}
\end{equation}

\centerline{\includegraphics[alt={}]{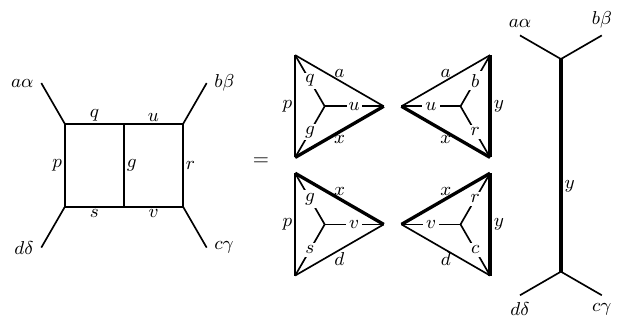}}
% PROOFREAD: contains OCR-garbled or 12j Wigner array(s) not auto-converted -- verify/reconstruct against orig.
\begin{equation}
=(-1)^{r+s+d-q-c} \sum_{x y \eta}(-1)^{y-\eta+x} \Fact{x y}^{2}
\threej{d}{y}{b}{\delta}{-\eta}{\beta}
\threej{a}{y}{c}{\alpha}{\eta}{\gamma}
\sixj{d}{y}{b}{r}{u}{x}
\sixj{a}{y}{c}{r}{v}{x}
\ninej{a}{p}{q}{x}{d}{u}{v}{s}{g},
\label{chap12:eq:23}
\end{equation}

\centerline{\includegraphics[alt={}]{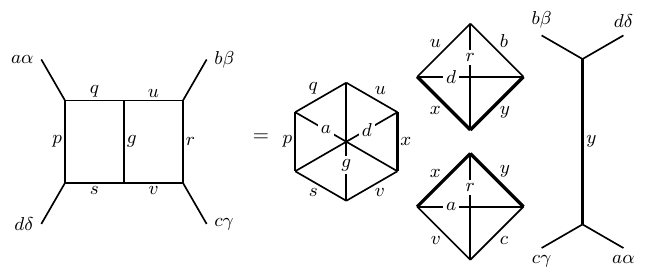}}
% PROOFREAD: contains OCR-garbled or 12j Wigner array(s) not auto-converted -- verify/reconstruct against orig.
\[
\begin{aligned}
\sum_{\psi \kappa \mu \rho}(-1)^{p-\psi+q-\kappa+v-\mu+r-\rho+s-\sigma+u-\nu+g-\lambda} & \threej{p}{a}{q}{\psi}{\alpha}{-\kappa}\threej{q}{v}{g}{\kappa}{-\mu}{-\lambda}\left(\begin{array}{rr}
g & u \\
\lambda & \nu \\
\nu & -\psi
\end{array}\right)\left(\begin{array}{rr}
v & b \\
\mu & \beta
\end{array}-\rho\right) \\
& \times\left(\begin{array}{rrr}
r & c & s \\
\rho & \gamma & -\sigma
\end{array}\right)\threej{s}{d}{u}{\sigma}{\delta}{-\nu}
\end{aligned}
\]
\begin{equation}
=(-1)^{a+b+c+u+s}\sixj{a}{p}{q}{g}{v}{u} \label{chap12:eq:24} \sum_{x \xi}(-1)^{x-\xi} \Fact{x}^{2}\threej{a}{d}{x}{\alpha}{\delta}{\xi}\threej{x}{b}{c}{-\xi}{\beta}{\gamma}\sixj{a}{d}{x}{s}{v}{u}\sixj{x}{b}{c}{r}{s}{v},
\end{equation}

\centerline{\includegraphics[alt={}]{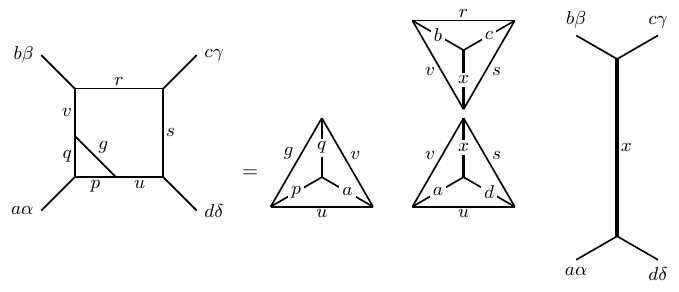}}
\[
\begin{gathered}
\sum_{\substack{\psi \kappa \rho \sigma \\
r \nu \mu}}(-1)^{p-\psi+q-\kappa+r-\rho+s-\sigma+t-r+u-\nu+v-\mu}\threej{p}{a}{q}{\psi}{\alpha}{-\kappa}\threej{q}{u}{r}{\kappa}{\nu}{-\rho}\threej{r}{c}{s}{\rho}{\gamma}{-\sigma}\threej{s}{v}{p}{\sigma}{\mu}{-\psi} \\
\times\threej{u}{b}{t}{-\nu}{\beta}{-\tau}\threej{t}{d}{v}{\tau}{\delta}{-\mu}
\end{gathered}
\]
\begin{equation}
=(-1)^{q+v+t-o-d+2 u} \sum_{x y \eta}(-1)^{x+y-\eta} \Fact{x y}^{2}\threej{a}{b}{y}{\alpha}{\beta}{-\eta} \label{chap12:eq:25}\threej{y}{c}{d}{\eta}{\gamma}{\delta}\sixj{p}{r}{x}{c}{v}{s}\sixj{p}{r}{x}{u}{a}{q}\sixj{a}{b}{y}{t}{x}{u}\sixj{y}{c}{d}{v}{t}{x}=
\end{equation}

\centerline{\includegraphics[alt={}]{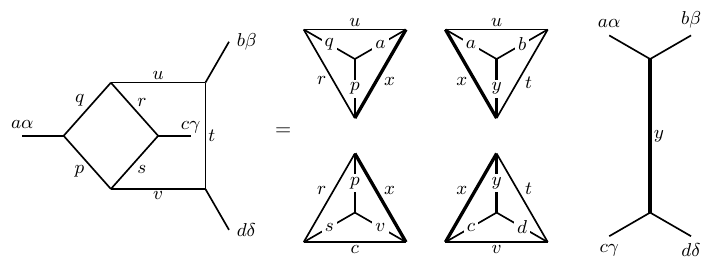}}
\begin{equation}
=(-1)^{q+u-a-s-b-d} \sum_{x \xi}(-1)^{x-\xi} \Fact{x}^{2}\threej{a}{x}{c}{\alpha}{-\xi}{\gamma} \label{chap12:eq:26}\threej{d}{x}{b}{\delta}{\xi}{\beta}\sixj{d}{x}{b}{u}{t}{v}\ninej{a}{x}{c}{p}{v}{s}{q}{u}{r},
\end{equation}

\centerline{\includegraphics[alt={}]{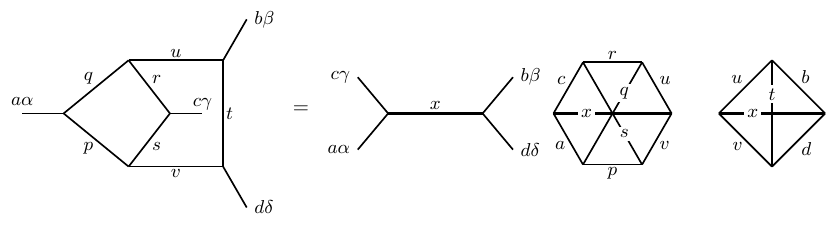}}

\begin{align}
& \sum_{\substack{\psi \kappa \rho \\
\sigma \tau \nu}}(-1)^{p-\psi+q-\kappa+r-\rho+\rho-\sigma+t-\tau+u-\nu}\threej{p}{a}{q}{\psi}{\alpha}{-\kappa}\threej{q}{b}{r}{\kappa}{\beta}{-\rho}\threej{r}{c}{s}{\rho}{\gamma}{-\sigma}\threej{s}{d}{t}{\sigma}{\delta}{-\tau}\threej{t}{e}{u}{\tau}{\varepsilon}{-\nu}\threej{u}{f}{p}{\nu}{\varphi}{-\psi}  \notag\\
& \quad=\sum_{\substack{x y s \\
\xi \eta s}}(-1)^{x-\xi+y-\eta+x-s} \Fact{x y z}^{2}\threej{a}{x}{b}{\alpha}{\xi}{\beta}\threej{c}{y}{d}{\gamma}{\eta}{\delta}\threej{e}{z}{f}{\varepsilon}{\zeta}{\varphi}\threej{x}{y}{z}{-\xi}{-\eta}{-s}\sixj{a}{x}{b}{r}{q}{p}\sixj{c}{y}{d}{t}{s}{r} \label{chap12:eq:27}
\end{align}

\centerline{\includegraphics[alt={}]{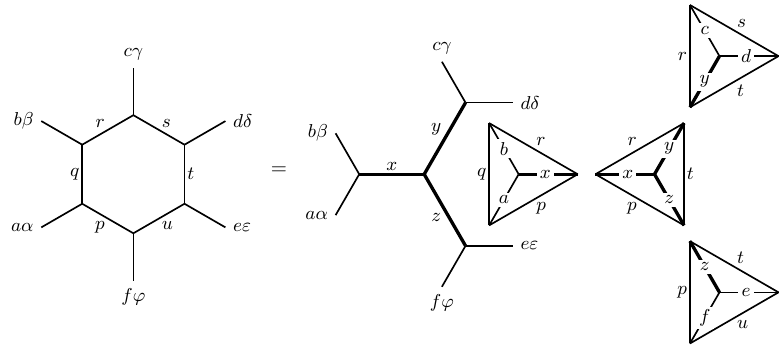}}
\[
=(-1)^{2 p-u-r+a+b+d+e-c} \sum_{\substack{x y z \\ \xi \eta \xi}}(-1)^{x} \Fact{x y z}^{2}(-1)^{x-\xi+y-\eta+x-\xi}
\]
\begin{equation}
\times\threej{a}{b}{x}{\alpha}{\beta}{-\xi} \label{chap12:eq:28}\threej{x}{c}{y}{\xi}{\gamma}{-\eta}\threej{y}{f}{z}{\eta}{\varphi}{-\zeta}\threej{z}{d}{e}{\zeta}{\delta}{e}\sixj{a}{b}{x}{r}{p}{q}\sixj{x}{c}{y}{s}{p}{r}\sixj{y}{f}{z}{u}{s}{p}\sixj{z}{d}{e}{t}{u}{s}=
\end{equation}

\centerline{\includegraphics[alt={}]{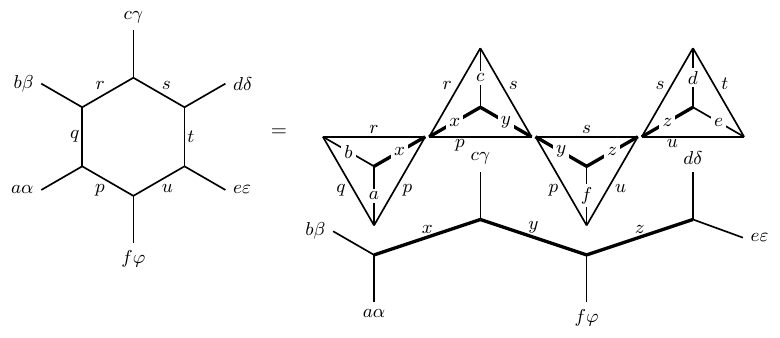}}

\begin{gather}
=(-1)^{u-d-c-c-r} \sum_{\substack{x y z \\
\xi \eta s}}(-1)^{x-\xi+y-\eta+x-\varsigma} \Fact{x y s}^{2}:\threej{a}{x}{b}{\alpha}{\xi}{\beta}\threej{e}{y}{d}{e}{\eta}{s}\threej{c}{z}{f}{\gamma}{\varsigma}{\varphi}\threej{x}{y}{z}{-\xi}{-\eta}{-\varsigma}  \notag\\
\times\sixj{a}{x}{b}{r}{q}{p}\sixj{e}{y}{d}{s}{t}{u}\ninej{x}{y}{z}{p}{u}{f}{r}{s}{c} \label{chap12:eq:29}
\end{gather}

\centerline{\includegraphics[alt={}]{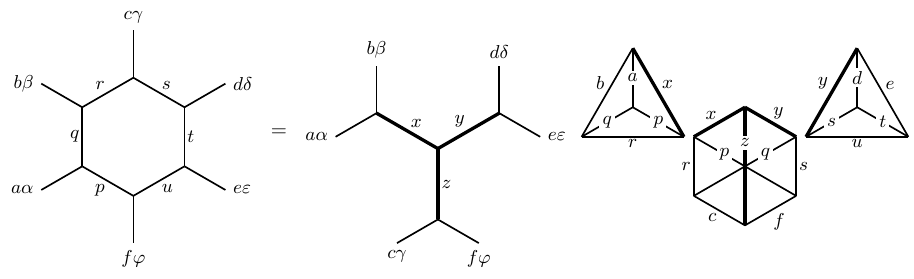}}

\section{Summation of products of $6 j$ and $9 j$ symbols}\label{sec:12.2}\index{sums!of the $6j$ and $9j$ symbols}\index{6j symbol@$6j$ symbol!sums of products of}\index{9j symbol@$9j$ symbol!sums of products of}
This section presents a collection of sums involving $6 j$ and, $9 j$ symbols. All the sums are illustrated by diagrams which display the coupling scheme for angular moments involved in the equation.\index{diagrams}\index{coupling schemes} The sums are grouped in accordance with the number of $3 n j$ symbols in the products. Within each group the sums are arranged according to the number of summation indices. The summation indices are printed by capital letters in equations and marked by thick $j$-lines. In Sec.~\ref{sec:12.2}, just as in Sec.~\ref{sec:12.1}, we use the notation
\[
\Fact{a b \ldots f}=[(2 a+1)(2 b+1) \ldots(2 f+1)]^{\frac{1}{2}} .
\]
Some of the sums presented are taken from Refs. \cite{ref024} and \cite{ref044}.\\
\subsection{Sums Involving One $3 n j$ Symbol}\label{sec:12.2.1}\index{sums!with one $3nj$ symbol}\index{triangle symbol}\isymo{braces-jjj@$\{j_{1} j_{2} j_{3}\}$, triangle symbol}

\begin{equation}
\sum_{X} \Fact{X}^{2}\{a b X\}=\Fact{a b}^{2} \label{chap12:eq:30}
\end{equation}

\centerline{\includegraphics[alt={}]{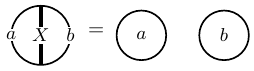}}

\begin{equation}
\sum_{x}(-1)^{2 a+\dot{x}} \Fact{x}^{2}\{a a X\}=\Fact{a}^{2} \label{chap12:eq:31}
\end{equation}

\centerline{\includegraphics[alt={}]{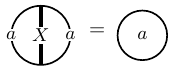}}
\begin{equation}
\sum_{x} \Fact{x}^{2}\sixj{a}{b}{X}{a}{b}{c} \label{chap12:eq:32}=(-1)^{2 c}\{a b c\},
\end{equation}

\centerline{\includegraphics[alt={}]{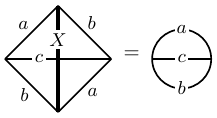}}
\begin{equation}
\sum_{X}(-1)^{a+b+X} \Fact{X}^{2}\sixj{a}{b}{x}{b}{a}{c} \label{chap12:eq:33}=\Fact{a b} \delta_{c 0}
\end{equation}

\centerline{\includegraphics[alt={}]{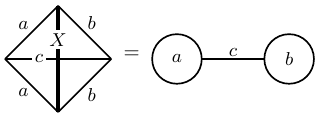}}
\begin{equation}
\sum_{X} \Fact{X}^{2}\ninej{a}{b}{e}{c}{d}{f}{e}{f}{X} \label{chap12:eq:34}=\frac{\delta_{b c}}{\Fact{b}^{2}}\{a b e\}\{b d f\}
\end{equation}

\centerline{\includegraphics[alt={}]{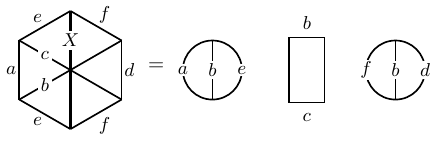}}
\begin{equation}
\sum_{X}(-1)^{a+b+c+d-X} \Fact{X}^{2}\ninej{a}{b}{e}{c}{d}{f}{f}{e}{X} \label{chap12:eq:35}=\frac{\delta_{a d}}{\Fact{a}^{2}}\{d b e\}\{a c f\}
\end{equation}

\centerline{\includegraphics[alt={}]{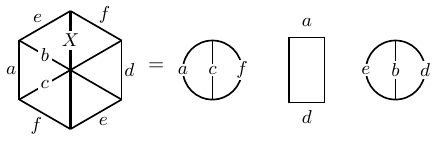}}
\subsection{Sums Involving Products of Two $3 n j$ Symbols}\label{sec:12.2.2}\index{sums!with two $3nj$ symbols}
\subsubsection{(i) Single summation}
\begin{equation}
\sum_{X} \Fact{X}^{2}\sixj{a}{b}{X}{c}{d}{p} \sixj{c}{d}{X}{a}{b}{q}=\frac{\delta_{p q}}{\Fact{p}^{2}}\{a d p\}\{b c p\}
\label{chap12:eq:36}
\end{equation}

\centerline{\includegraphics[alt={}]{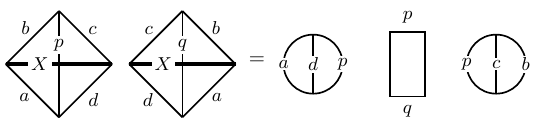}}
\begin{equation}
\sum_{X}(-1)^{p+q+X} \Fact{X}^{2}\sixj{a}{b}{X}{c}{d}{p} \sixj{c}{d}{X}{b}{a}{q}=\sixj{c}{a}{q}{d}{b}{p}
\label{chap12:eq:37}
\end{equation}

\centerline{\includegraphics[alt={}]{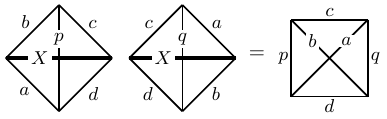}}
\begin{equation}
\sum_{X} \Fact{X}^{2}
\ninej{a}{f}{X}{d}{q}{e}{p}{c}{b}
\sixj{a}{f}{X}{e}{b}{s}=(-1)^{2 s}\sixj{a}{b}{s}{c}{d}{p}\sixj{c}{d}{s}{e}{f}{q},
 \label{chap12:eq:38}
\end{equation}

\centerline{\includegraphics[alt={}]{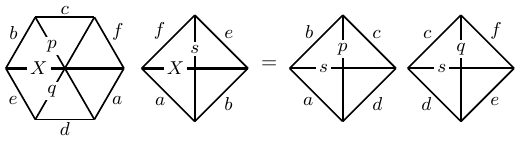}}
\begin{equation}
\sum_X (-1)^{R+X}\Fact{X}^2 
\ninej{a}{f}{X}{d}{q}{e}{p}{c}{b}
\sixj{a}{f}{X}{b}{e}{s}
=
(-1)^{2s} 
\sixj{p}{q}{s}{e}{a}{d}
\sixj{p}{q}{s}{f}{b}{c}\,,
\end{equation}
where $R=a+b+c+d+e+f+p+q$\\

\centerline{\includegraphics[alt={}]{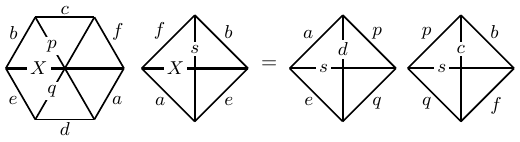}}
\begin{equation}
\sum_X\Fact{X}^2\ninej{a}{b}{p}{c}{d}{q}{r}{s}{X}
\ninej{e}{f}{p}{g}{h}{q}{r}{s}{X}
=
(-1)^{b-c-f+g}\twelvejII{r,a,c,s,b,d,h,g,q,f,e,p}
\end{equation}

\centerline{\includegraphics[alt={}]{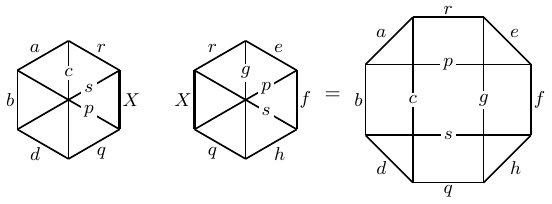}}
\begin{equation}
(-1)^{p+q-r-s} \sum_{X} \Fact{X}^{2}
\ninej{a}{b}{p}{c}{d}{q}{r}{s}{X} 
\ninej{e}{f}{p}{g}{h}{q}{r}{s}{X}=(-1)^{b+g-c-f} \sum_{Y} \Fact{Y}^{2}\ninej{a}{p}{b}{r}{e}{g}{c}{f}{Y}\ninej{d}{s}{b}{q}{h}{g}{c}{f}{Y},
\label{chap12:eq:39}
\end{equation}

\centerline{\includegraphics[alt={}]{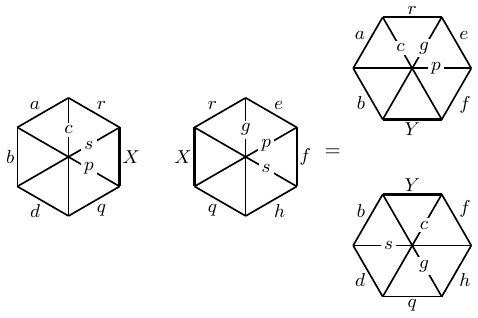}}
\subsubsection{(ii) Double summation}
\begin{equation}
\sum_{X Y} \Fact{X Y}^{2}\ninej{a}{b}{X}{c}{d}{Y}{e}{f}{j} \label{chap12:eq:40}\ninej{a}{b}{X}{c}{d}{Y}{g}{h}{j}=\frac{\delta_{e g} \delta_{f h}}{\Fact{e f}^{2}}\{a c e\}\{b d h\}\{g f j\},
\end{equation}

\centerline{\includegraphics[alt={}]{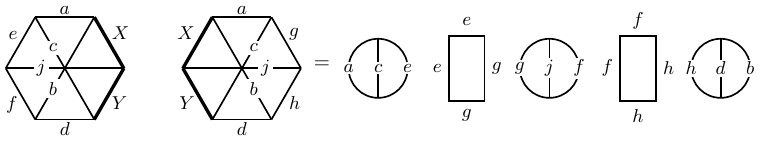}}

\begin{equation}
 \sum_{XY} (-1)^{Y} \Fact{XY}^2 \ninej{a}{b}{X}{c}{d}{Y}{e}{f}{j}
 \ninej{a}{b}{X}{d}{c}{Y}{g}{h}{j}
 =
 (-1)^{2b+f+h} \ninej{a}{d}{g}{c}{b}{h}{e}{f}{j}\,,
\end{equation}

\centerline{\includegraphics[alt={}]{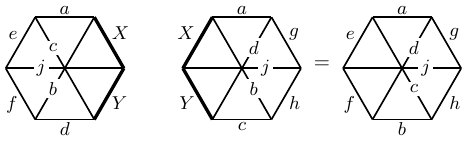}}

\subsubsection{(iii) Triple summation}
\begin{equation}
  \sum_{XYZ} \Fact{XYZ}^2 \sixj{X}{Y}{Z}{a}{b}{c}^2=\Fact{abc}^2
\end{equation}

\centerline{\includegraphics[alt={}]{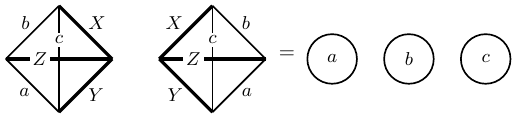}}
\begin{equation}
  \sum_{XYZ} \Fact{XYZ}^2 \ninej{X}{Y}{Z}{a}{b}{c}{b}{c}{a}
  \sixj{X}{Y}{Z}{c}{a}{b}= \{abc\}
\end{equation}

\centerline{\includegraphics[alt={}]{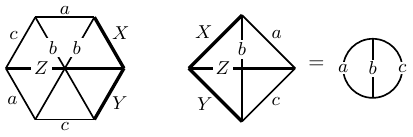}}
\begin{equation}
\sum_{X Y Z} \Fact{X Y Z}^{2}\ninej{X}{Y}{Z}{a}{b}{c}{d}{e}{f}^2=\{a b c\}\{d e f\}
\label{chap12:eq:41}
\end{equation}

\centerline{\includegraphics[alt={}]{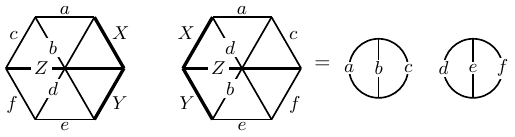}}

\subsection{Sums Involving Products of Three $3 n j$ Symbols}\label{sec:12.2.3}\index{sums!with three $3nj$ symbols}
\subsubsection{Single summation}
\begin{equation}\label{chap12:eq:42}\\
\sum_{X}(-1)^{R+X} \Fact{X}^{2}\sixj{a}{b}{X}{c}{d}{p}
\sixj{c}{d}{X}{e}{f}{q}
\sixj{e}{f}{X}{b}{a}{r}
=
\sixj{p}{q}{r}{e}{a}{d}
\sixj{p}{q}{r}{f}{b}{c}
,
\end{equation}
where $R=a+b+c+d+e+f+p+q+r$,\\

\centerline{\includegraphics[alt={}]{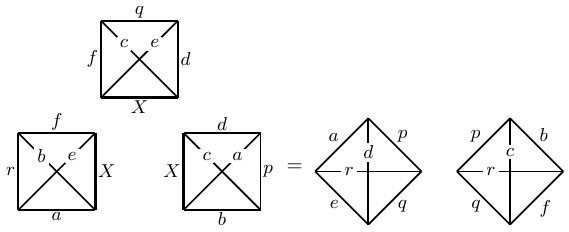}}
\begin{equation}
\sum_{X}(-1)^{2 X} \Fact{X}^{2}\sixj{a}{b}{X}{c}{d}{p} \label{chß12:eq:43}\sixj{c}{d}{X}{e}{f}{q}\sixj{e}{f}{X}{a}{b}{r}=\ninej{a}{f}{r}{d}{q}{e}{p}{c}{b},
\end{equation}

\centerline{\includegraphics[alt={}]{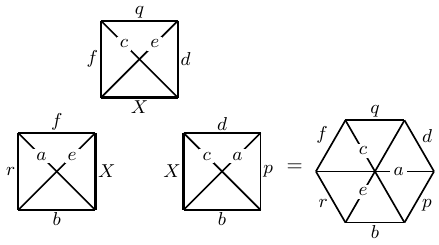}}
\begin{equation}
(-1)^{p+q-e-f} \sum_{X} \Fact{X}^{2}\ninej{a}{b}{p}{c}{d}{q}{e}{f}{X} \sixj{p}{q}{X}{k}{l}{g}\sixj{e}{f}{X}{k}{l}{h}=
\twelvejI{f,b,p,g,d,a,l,k,q,c,e,h}
\label{chap12:eq:44}
\end{equation}

\centerline{\includegraphics[alt={}]{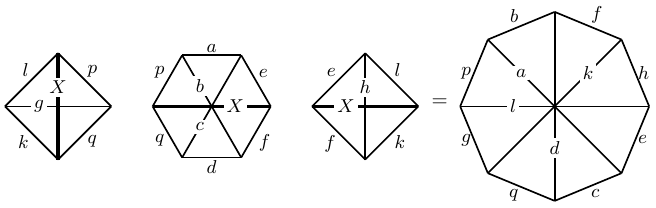}}
\begin{align}
&(-1)^{p+q-e-f} \sum_{X} \Fact{X}^{2}\ninej{a}{b}{p}{c}{d}{q}{e}{f}{X}
\sixj{p}{q}{X}{k}{l}{g}
\sixj{e}{f}{X}{k}{l}{h}\notag\\
&=(-1)^{b+g-c-h} \sum_{Y} \Fact{Y}^{2}\ninej{d}{f}{b}{q}{k}{g}{c}{h}{Y}
\sixj{b}{g}{Y}{l}{a}{p}
\sixj{c}{h}{Y}{l}{a}{e}
.
 \label{chap12:eq:45}
\end{align}

\centerline{\includegraphics[alt={}]{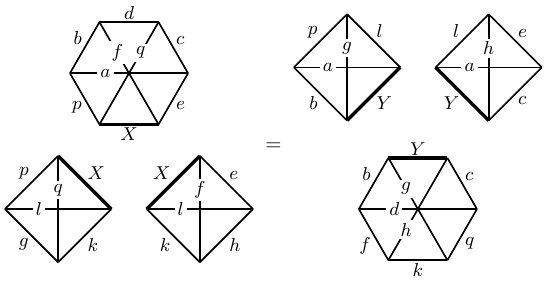}}

\subsubsection{Double summation}
\begin{equation}
\sum_{X Y} \Fact{X Y}^{2}\sixj{a}{b}{X}{c}{d}{p} \label{chap12:eq:46}\sixj{c}{d}{X}{a}{b}{Y}\sixj{a}{b}{q}{c}{d}{Y}=\sixj{a}{b}{q}{c}{d}{p},
\end{equation}

\centerline{\includegraphics[alt={}]{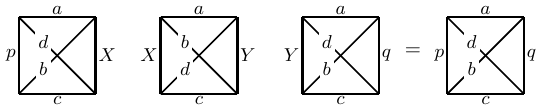}}
\begin{equation}
\sum_{X Y}(-1)^{X+Y+p} \Fact{X Y}^{2}\sixj{a}{b}{X}{c}{d}{p} \label{chap12:eq:47}\sixj{a}{c}{Y}{d}{b}{X}\sixj{a}{d}{q}{b}{c}{Y}=\frac{\delta_{p q}}{\Fact{p}^{2}}\{a d p\}\{b c p\}
\end{equation}

\centerline{\includegraphics[alt={}]{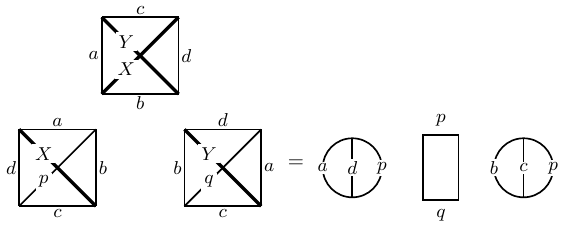}}
\begin{equation}
\sum_{X Y}(-1)^{X+Y+2 a-s-t} \Fact{X Y}^{2}\ninej{a}{b}{p}{c}{d}{X}{q}{Y}{g} \label{chap12:eq:48}\sixj{c}{d}{X}{p}{g}{s}\sixj{b}{d}{Y}{q}{g}{t}=\ninej{a}{b}{p}{c}{g}{s}{q}{t}{d},
\end{equation}

\centerline{\includegraphics[alt={}]{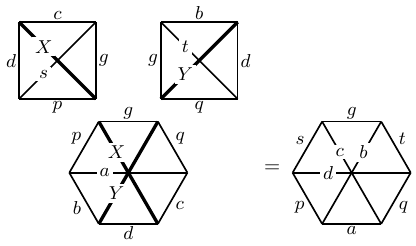}}
\begin{equation}
\sum_{X Y} \Fact{X Y}^{2}\ninej{a}{b}{p}{c}{d}{X}{q}{Y}{g} \sixj{c}{d}{X}{g}{p}{s}
\sixj{b}{d}{Y}{g}{q}{t}
=(-1)^{2 s} \frac{\delta_{s t}}{\Fact{s}^{2}}\{d g s\}\sixj{a}{b}{p}{s}{c}{q},
\label{chap12:eq:49}
\end{equation}

\centerline{\includegraphics[alt={}]{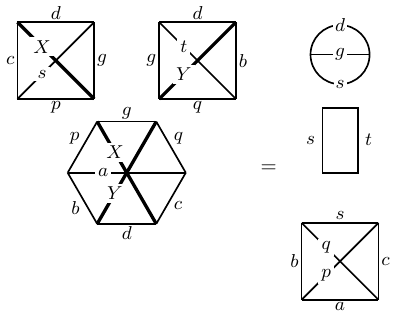}}
\begin{equation}
\sum_{X Y} \Fact{X Y}^{2}\ninej{a}{b}{X}{c}{d}{Y}{e}{f}{g} 
\sixj{a}{b}{X}{Y}{g}{h}
\sixj{c}{d}{Y}{b}{h}{j}
=(-1)^{2 h} \frac{\delta_{fj}}{\Fact{f}^{2}}\{b d f\}\sixj{e}{j}{g}{h}{a}{c},
\label{chap12:eq:49a}
\end{equation}

\centerline{\includegraphics[alt={}]{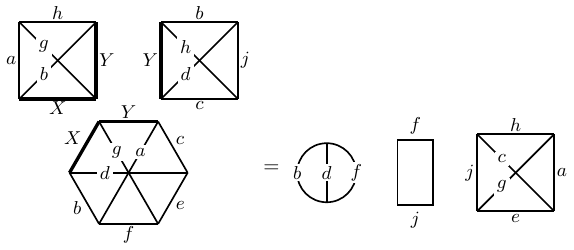}}
\begin{equation}
\sum_{X Y} \Fact{X Y}^{2}
\ninej{a}{f}{X}{d}{q}{e}{p}{c}{b}
\ninej{a}{f}{X}{h}{r}{e}{Y}{g}{b}
\sixj{a}{h}{Y}{g}{b}{s}
=
\sixj{a}{b}{s}{c}{d}{p}
\sixj{c}{d}{s}{e}{f}{q}
\sixj{e}{f}{s}{g}{h}{r}
,
\label{chap12:eq:49b}
\end{equation}

\centerline{\includegraphics[alt={}]{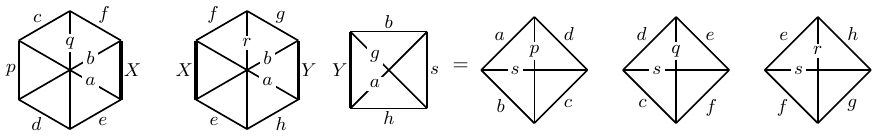}}

\subsubsection{(iii) Triple summation}
\begin{equation}
\sum_{X Y Z} \Fact{X Y Z}^{2}
\ninej{a}{b}{p}{c}{d}{X}{q}{Y}{Z} 
\sixj{p}{X}{Z}{d}{e}{c}
\sixj{q}{Y}{Z}{d}{e}{b}
=
(-1)^{2e}{\Fact{d}^{2}}
\sixj{a}{b}{p}{e}{c}{q},
\end{equation}

\centerline{\includegraphics[alt={}]{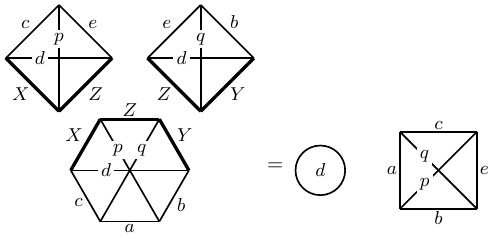}}
\begin{equation}
\sum_{X Y Z}(-1)^{X+Y-e-h} \Fact{X Y Z}^{2}
\ninej{a}{b}{X}{c}{d}{Z}{e}{f}{g} 
\ninej{g}{b}{Y}{c}{d}{Z}{h}{j}{a} 
\sixj{a}{b}{X}{g}{Z}{Y}
=
\frac{\delta_{fj}}{\Fact{d}^{2}}
\{bdf\}
\sixj{e}{f}{g}{h}{c}{a},
\end{equation}

\centerline{\includegraphics[alt={}]{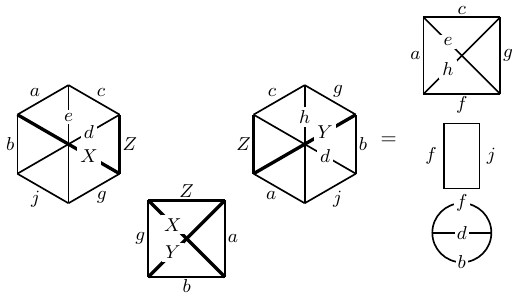}}
\begin{equation}
\sum_{X Y Z}(-1)^{X+f-a-l} \Fact{X Y Z}^{2}\ninej{a}{b}{p}{c}{d}{X}{q}{Y}{Z} \label{chap12:eq:50}\ninej{b}{d}{Y}{c}{f}{Z}{l}{k}{q}\sixj{p}{X}{Z}{c}{f}{d}=\frac{\delta_{p k}}{\Fact{k}^{2}}\{d f p\}\sixj{a}{b}{p}{l}{q}{c},
\end{equation}

\centerline{\includegraphics[alt={}]{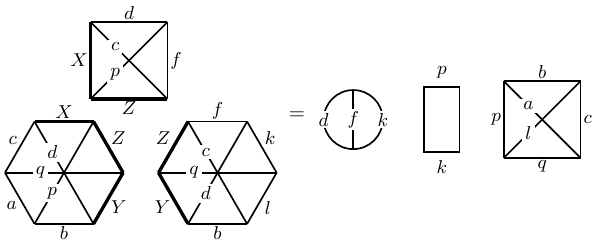}}
\begin{equation}
\sum_{X Y Z} \Fact{X Y Z}^{2}\ninej{a}{b}{p}{c}{d}{X}{q}{Y}{Z} \label{chap12:eq:51}\ninej{c}{d}{X}{e}{f}{Z}{g}{h}{p}\ninej{b}{d}{Y}{e}{f}{Z}{j}{k}{q}=\frac{\delta_{h k}}{\Fact{h}^{2}}\{d f h\}\ninej{a}{b}{p}{c}{e}{g}{q}{j}{h},
\end{equation}
% eq 51: three flat 9j-hex (X,Y,Z bold) = theta(d f h) . rect(h,k) . 9j-hex
% LHS 9j's: hex1 {b d Y;e f Z;j k q}, hex2 {a b p;c d X;q Y Z}, hex3 {c d X;e f Z;g h p}

\centerline{\includegraphics[alt={}]{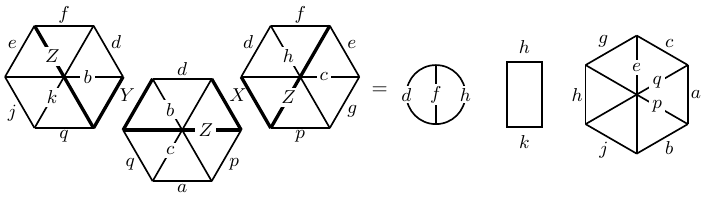}}
%checked figures from here onward: lessons learned: mainly issue with swaps of lines, which can be recovered from triangle relations
% of the 3nj coefficients.

\subsection{Sums Involving Products of Four $3 n j$ Symbols}\label{sec:12.2.4}\index{sums!with four $3nj$ symbols}\index{12j symbol@$12j$ symbol!sums involving}
\subsubsection{(i) Single summation}
\begin{gather}
\sum_{X}(-1)^{R-X} \Fact{X}^{2}\sixj{a}{b}{X}{c}{d}{p}\sixj{c}{d}{X}{e}{f}{q}\sixj{e}{f}{X}{g}{h}{r}\sixj{g}{h}{X}{b}{a}{s}\notag\\
=\sum_{Y}(-1)^{2 Y+a+b+e+\rho} \Fact{Y}^{2}\ninej{s}{h}{b}{g}{r}{f}{a}{e}{Y} \sixj{b}{f}{Y}{q}{p}{c}  \sixj{a}{e}{Y}{q}{p}{d} =
\twelvejI{e,h,b,c,r,s,p,q,f,g,a,d},
\label{chap12:eq:52}
\end{gather}
where $R=a+b+c+d+e+f+g+h+p+q+r+s$.\\

\centerline{\includegraphics[alt={},max width=\textwidth]{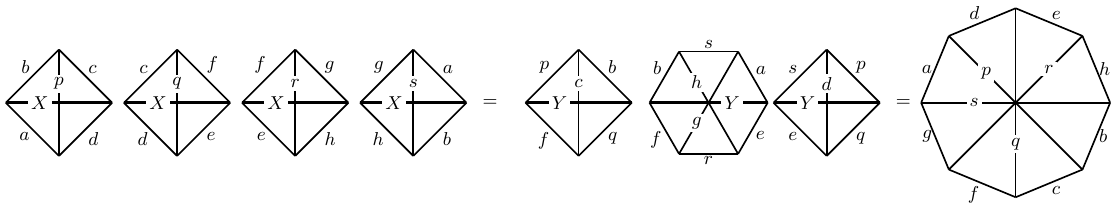}}

% PROOFREAD: contains OCR-garbled or 12j Wigner array(s) not auto-converted -- verify/reconstruct against orig.
\begin{align}
& \sum_{X} \Fact{X}^{2}
\sixj{a}{b}{X}{c}{d}{p}
\sixj{c}{d}{X}{e}{f}{q}
\sixj{e}{f}{X}{g}{h}{r}
\sixj{g}{h}{X}{a}{b}{s}\notag\\
& =\sum_{X} \Fact{X}^{2}
\ninej{a}{f}{X}{d}{q}{e}{p}{c}{b}
\ninej{a}{f}{X}{h}{r}{e}{s}{g}{b}=(-1)^{-p+q-r+}
\twelvejII{a,d,p,f,q,a,g,s,b,r,h,e}, \label{chap12:eq:53}
\end{align}

\centerline{\includegraphics[alt={}]{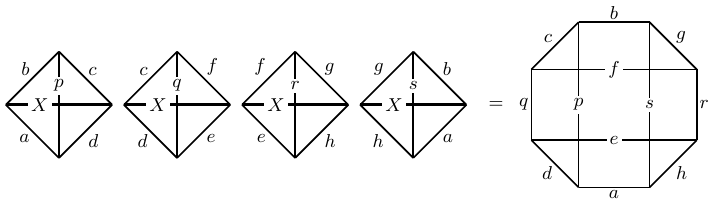}}

\subsubsection{(ii) Double summation}
\begin{equation}
\sum_{XY}\Fact{XY}^2
\sixj{a}{b}{c}{d}{X}{Y}
\sixj{a}{e}{f}{g}{X}{Y}
\sixj{c}{g}{p}{f}{d}{X}
\sixj{b}{g}{q}{e}{d}{Y}
=
(-1){2(c+e)} \sixj{a}{b}{c}{g}{p}{q}
\sixj{a}{e}{f}{d}{p}{q},
\end{equation}

\centerline{\includegraphics[alt={}]{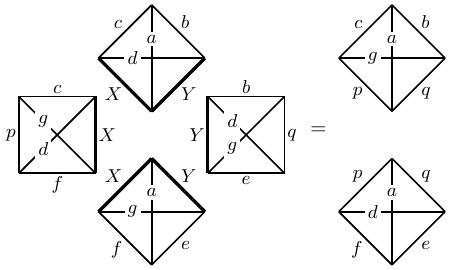}}

\begin{align}
& \sum_{X Y}(-1)^{X+Y} \Fact{X Y}^{2}\sixj{a}{b}{c}{d}{X}{Y}\sixj{a}{e}{f}{g}{X}{Y}\sixj{c}{f}{p}{g}{d}{X}\sixj{b}{e}{q}{g}{d}{Y}  \notag\\
& =(-1)^{-a+b+y c-d+e+f-e-p} \frac{\delta_{p q}}{\Fact{p}^{2}}\{d g p\}\sixj{a}{b}{c}{p}{f}{e} \label{chap12:eq:54}
\end{align}

\begin{center}

\centerline{\includegraphics[alt={}]{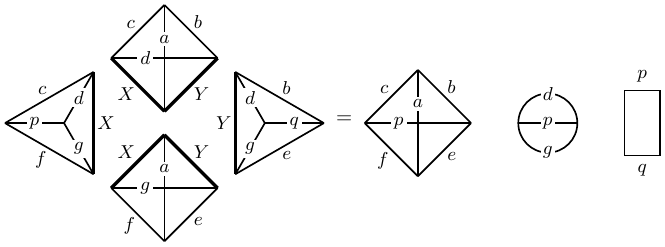}}
\end{center}
% PROOFREAD: contains OCR-garbled or 12j Wigner array(s) not auto-converted -- verify/reconstruct against orig.
\begin{equation}
\sum_{X Y}(-1)^{Y} \Fact{X Y}^{2}\ninej{a}{b}{X}{c}{d}{Y}{p}{q}{r} \label{chap12:eq:55}\sixj{X}{Y}{r}{j}{h}{g}\sixj{a}{b}{X}{h}{g}{e}
\sixj{c}{d}{Y}{j}{g}{f}=(-1)^{a-b+d+e-j-p+r}\ninej{e}{b}{h}{f}{d}{j}{p}{q}{r}\sixj{a}{c}{p}{f}{e}{g},
\end{equation}

\centerline{\includegraphics[alt={}]{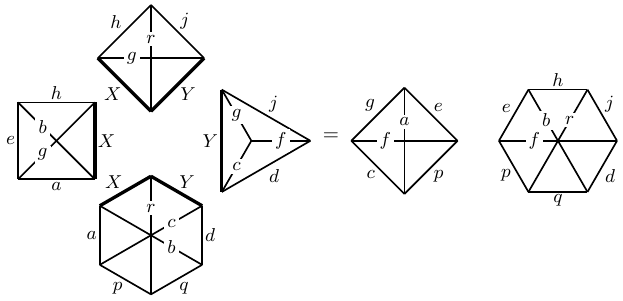}}

% PROOFREAD: contains OCR-garbled or 12j Wigner array(s) not auto-converted -- verify/reconstruct against orig.
\begin{gather}
\sum_{X Y}(-1)^{X+Y} \Fact{X Y}^{2}\ninej{a}{b}{X}{c}{d}{Y}{p}{q}{s}\ninej{e}{f}{X}{g}{h}{Y}{r}{t}{s}\sixj{a}{b}{X}{f}{e}{k}\sixj{c}{d}{Y}{h}{g}{l}  \notag\\
=(-1)^{k+l+a+c+p-s-h-t-f} \sum_{Z}(-1)^{Z} \Fact{Z}^{2}\ninej{c}{g}{l}{a}{e}{k}{p}{r}{Z}\ninej{f}{b}{k}{h}{d}{l}{t}{q}{Z}
\sixj{p}{r}{Z}{t}{q}{s}, \label{chap12:eq:56}
\end{gather}

\centerline{\includegraphics[alt={}]{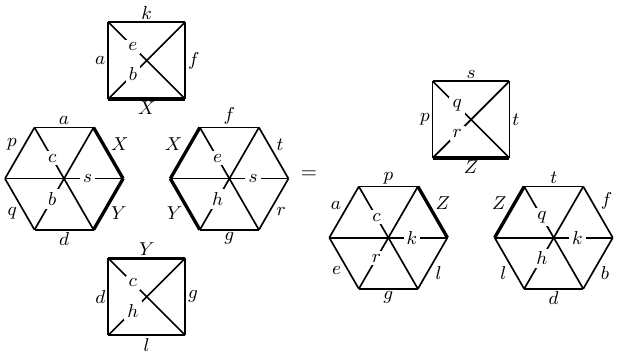}}
\subsubsection{(iii) Triple summation}

\begin{gather}
\sum_{X_{1} X_{2} X_{3}}(-1)^{X_{1}+X_{2}+X_{3}+a_{1}+a_{2}+a_{3}+b_{1}+b_{3}+b_{3}+c} \Fact{X_{1} X_{2} X_{3}}^{2}\sixj{a_{1}}{a_{2}}{a_{3}}{X_{1}}{X_{2}}{X_{3}}\sixj{a_{1}}{b_{2}}{b_{3}}{c}{X_{3}}{X_{2}}  \notag\\
\times\sixj{a_{2}}{b_{3}}{b_{1}}{c}{X_{1}}{X_{3}}\sixj{a_{3}}{b_{1}}{b_{2}}{c}{X_{2}}{X_{1}}=\Fact{c}^{2}\sixj{a_{1}}{a_{2}}{a_{3}}{b_{1}}{b_{2}}{b_{3}}, \label{chap12:eq:57}
\end{gather}

\centerline{\includegraphics[alt={}]{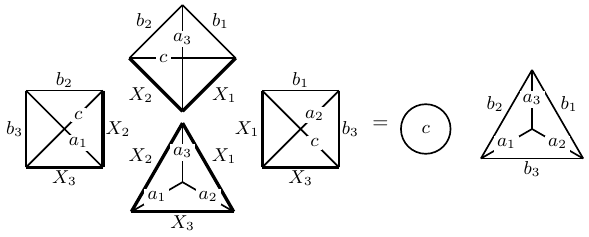}}
\begin{equation}
\sum_{X Y Z} \Fact{X Y Z}^{2}\ninej{X}{Y}{Z}{d}{e}{f}{a}{b}{c} \label{chap12:eq:58}\sixj{X}{Y}{Z}{f}{c}{g}\sixj{X}{a}{d}{b^{\prime}}{g}{c}\sixj{Y}{b}{e}{d^{\prime}}{f}{g}=(-1)^{2(b+d)} \frac{\delta_{b b^{\prime}} \delta_{d d^{\prime}}}{\Fact{b d}^{2}}\{a b c\}\{d e f\}\{b d g\},
\end{equation}

\centerline{\includegraphics[alt={}]{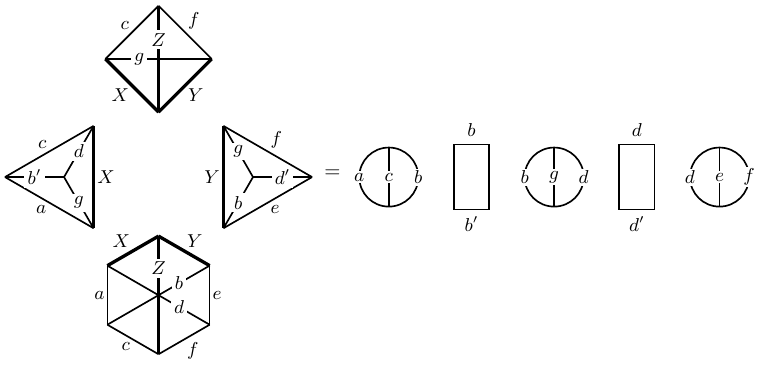}}
% PROOFREAD: contains OCR-garbled or 12j Wigner array(s) not auto-converted -- verify/reconstruct against orig.
\begin{equation}
\sum_{X Y Z}(-1)^{2 Z} \Fact{X Y Z}^{2}\ninej{a}{b}{X}{c}{d}{Y}{e^{\prime}}{f^{\prime}}{g} \sixj{a}{b}{X}{f}{Z}{d}
\sixj{c}{d}{Y}{Z}{e}{a}
\sixj{e}{f}{g}{X}{Y}{Z}=\frac{\delta_{e e^{\prime}} \delta_{f f^{\prime}}}{\Fact{e f}^{2}}\{a c e\}\{b d f\}\{e g f\},
\label{chap12:eq:59}
\end{equation}
\begin{center}

\centerline{\includegraphics[alt={}]{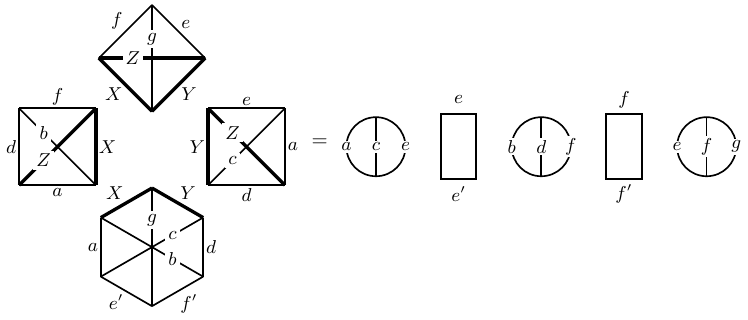}}
\end{center}
\begin{gather}
\sum_{X Y Z}(-1)^{Z} \Fact{X Y Z}^{2}\ninej{a}{b}{c^{\prime}}{d}{e}{f}{X}{Y}{g}\ninej{l}{e}{j}{a}{b}{c}{Z}{Y}{h}\sixj{a}{d}{X}{k}{Z}{l}\sixj{X}{Y}{g}{h}{k}{Z}  \notag\\
=(-1)^{2 c+b-d+g+j} \frac{\delta_{c c^{\prime}}}{\Fact{c}^{2}}\{a b c\}\sixj{k}{j}{f}{c}{g}{h}\sixj{k}{j}{f}{e}{d}{l}, \label{chap12:eq:60}
\end{gather}

\begin{center}

\centerline{\includegraphics[alt={}]{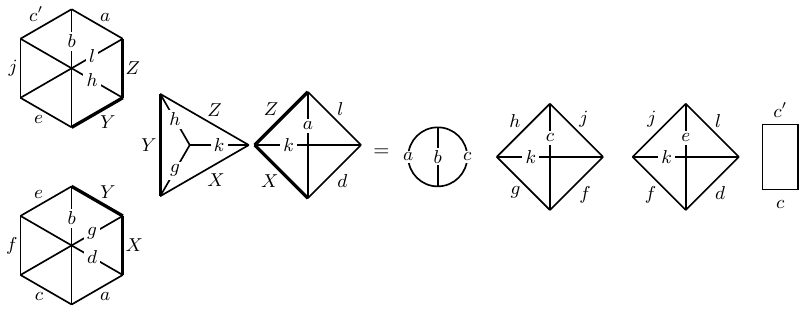}}
\end{center}

% PROOFREAD: contains OCR-garbled or 12j Wigner array(s) not auto-converted -- verify/reconstruct against orig.
\begin{gather}
\sum_{X Y Z}(-1)^{2 Y-Z} \Fact{X Y Z}^{2}\ninej{a}{b}{X}{c}{d}{Y}{e}{f}{Z}\ninej{g}{h}{X}{k}{l}{Y}{f}{e}{Z}\sixj{a}{b}{X}{g}{h}{j}
\sixj{c}{d}{Y}{k}{l}{j'}
=(-1)^{-b+c-h+k} \frac{\delta_{j j^{\prime}}}{\Fact{j}^{2}}\sixj{a}{c}{e}{l}{h}{j}\sixj{b}{d}{f}{k}{g}{j}, \label{chap12:eq:61}
\end{gather}

\centerline{\includegraphics[alt={}]{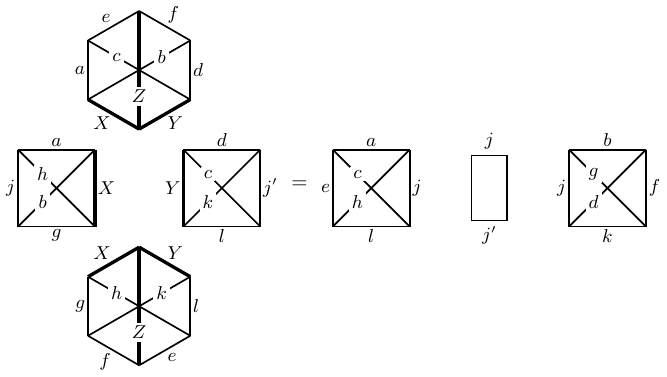}}

\begin{gather}
\sum_{X Y Z}(-1)^{Y-a-b-c-f-h-p+q} \Fact{X Y Z}^{2}\ninej{a}{b}{X}{g}{c}{q}{p}{Z}{Y}\ninej{b}{d}{f^{\prime}}{a}{h}{j}{Z}{Y}{p}\sixj{a}{b}{X}{d}{e}{f}\sixj{d}{h}{Y}{q}{X}{e}  \notag\\
=(-1)^{2 j} \frac{\delta_{f f^{\prime}}}{\Fact{f}^{2}}\{b d f\}\sixj{e}{g}{j}{p}{f}{a}\sixj{e}{g}{j}{c}{h}{q}, \label{chap12:eq:62}
\end{gather}

\begin{center}
  \includegraphics[alt={}]{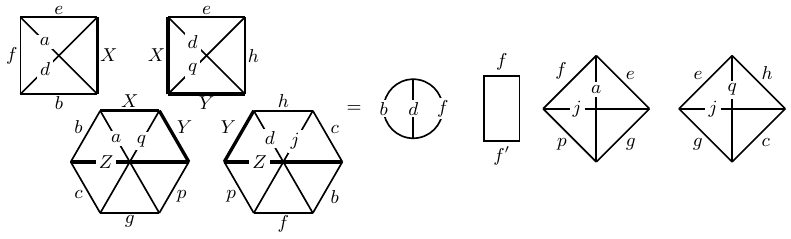}
\end{center}

\begin{gather}
  \sum_{XYZ}\Fact{XYZ}^2 
  \ninej{a}{b}{X}{c}{d}{Y}{t}{s}{r}
  \ninej{a}{b}{X}{h}{j}{q}{s}{f}{Z}
  \ninej{k}{l}{p}{c}{d}{Y}{e}{f}{Z}
  \sixj{p}{q}{r}{X}{Y}{Z}
  =
  \ninej{k}{l}{p}{h}{j}{q}{t}{s}{r}
  \sixj{k}{h}{t}{a}{c}{e}
  \sixj{l}{j}{s}{b}{d}{f},
\end{gather}

\centerline{\includegraphics[alt={}]{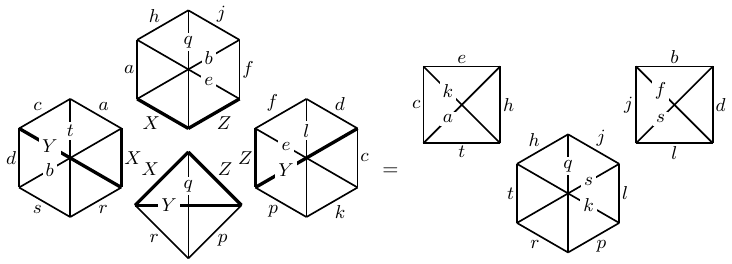}}

\chapter{Matrix elements of irreducible tensor operators}\label{chap:13}\index{matrix elements}\index{irreducible tensor operator}
\textcolor{red}{We have added explicit expressions for reduced matrix elements each time the book gave an expression for a matrix element of an irreducible tensor operator. These equations are as yet unnumbered.}
\section{The Wigner-Eckart theorem and the evaluation of matrix elements}\label{sec:13.1}\index{Wigner-Eckart theorem}\index{matrix elements!evaluation of}
In this section we consider the evaluation of matrix elements of irreducible tensor operators in a general form.\\
Let a quantum mechanical system be characterized by a set of the quantum numbers ( \(n j m\) ), where \(j\) and \(m\) are the angular momentum and its projection on the quantization axis, respectively. The symbol \(n\) stands for all other quantum numbers required to define the state of the system. Let \(\Psi_{n j m}\) be the wave function of this state. The matrix element of any irreducible tensor operator \(\widehat{\mathfrak{M}}_{k}\) of rank \(k\) between the states \(\ket{n^{\prime} j^{\prime} m^{\prime}}\) and \(\ket{n j m}\) is defined as
\begin{equation}
\braOket{ n^{\prime} j^{\prime} m^{\prime}}{ \widehat{\mathfrak{M}}_{k \kappa}}{n j m}=\int \Psi_{n^{\prime} j^{\prime} m^{\prime}}^{*} \widehat{\mathfrak{M}}_{k \kappa} \Psi_{n j m} d \tau ,\label{eq:13:1:1}
\end{equation}
where \(\widehat{\mathfrak{M}}_{k \kappa}\) is a tensor component.\isym{Mfrak-hat@$\widehat{\mathfrak{M}}_{k \kappa}$, irreducible tensor operator}

\subsection{Wigner-Eckart Theorem}\label{sec:13.1.1}\index{Wigner-Eckart theorem!statement of}
The Wigner-Eckart theorem is widely used for evaluation of matrix elements. According to this theorem, the dependence of any matrix element on the orientation of the coordinate system, i.e., on the projections\index{Clebsch-Gordan coefficient}\index{3jm symbol@$3jm$ symbol}\isym{C-clebsch@$\clebsch{a}{\alpha}{b}{\beta}{c}{\gamma}$, Clebsch-Gordan coefficient}\isymo{3jm@$\bigl(\begin{smallmatrix} a & b & c\\ \alpha & \beta & \gamma\end{smallmatrix}\bigr)$, Wigner $3jm$ symbol} \(m\), \(m^{\prime}\) and \(\kappa\), is entirely included in the \(3 j m\) symbol or the Clebsch-Gordan coefficient,
\begin{equation}
\braOket{ n^{\prime} j^{\prime} m^{\prime}}{ \widehat{\mathfrak{M}}_{k \kappa}}{n j m}=(-1)^{j^{\prime}-m^{\prime}}\threej{j^{\prime}}{k}{j}{-m^{\prime}}{\kappa}{m}
\braOketred{ n^{\prime} j^{\prime}}{\widehat{\mathfrak{M}}_{k}}{ n j}=(-1)^{2 k} 
\clebsch{j}{m}{k}{\kappa}{j^{\prime}}{m^{\prime}} \frac{
\braOketred{ n^{\prime} j^{\prime}}{\widehat{\mathfrak{M}}_{k}}{ n j}}{\Fact{j^{\prime}}}. \label{eq:13:1:2}
\end{equation}

The invariant factor \(
\braOketred{ n^{\prime} j^{\prime}}{\widehat{\mathfrak{M}}_{k}}{ n j}\) is called the reduced matrix element of the set of tensor operators\index{reduced matrix element}\index{matrix elements!reduced}\isym{reduced-me@$\braOketred{n^{\prime} j^{\prime}}{\widehat{\mathfrak{M}}_{k}}{n j}$, reduced matrix element} \(\widehat{\mathfrak{M}}_{k \kappa}\).\\
As follows from the definition \eqref{eq:13:1:2}, the reduced matrix element of the unit operator is given by\index{unit operator}\isym{I-vec-hat@$\widehat{\vect{I}}$, unit operator}
\begin{equation}
\braOketred{ n^{\prime} j^{\prime}}{\widehat{\vect{I}}}{ n j}=\sqrt{2 j+1} \delta_{n^{\prime} n} \delta_{j^{\prime} j}.  \label{eq:13:1:3}
\end{equation}
In this section, as well as in Chap.~\ref{chap:12} we shall again use the notation\index{dimension factor}\isymg{Pi-dim@$\Pi_{a b c \ldots}$, dimension factor $[(2a+1)(2b+1)\cdots]^{1/2}$}
\begin{equation*}
\Fact{a b \ldots c} \equiv[(2 a+1)(2 b+1) \ldots(2 c+1)]^{\frac{1}{2}}.
\end{equation*}

\subsection{Sum Rules}\label{sec:13.1.2}\index{sum rules!for matrix elements}\index{matrix elements!sum rules}
Making use of the definition \eqref{eq:13:1:2}, one can obtain the following rules for sums of matrix elements of irreducible tensor operators:
\begin{align}
\sum_{m \kappa}\left|
\braOket{n^{\prime} j^{\prime}m^{\prime}}{\widehat{\mathfrak{M}}_{k \kappa}}{ n j m}\right|^{2} & =\frac{\left|
\braOketred{n^{\prime} j^{\prime}}{\widehat{\mathfrak{M}}_{k}}{ n j}\right|^{2}}{2 j^{\prime}+1},  \label{eq:13:1:4}\\
\sum_{m^{\prime} \kappa}\left|
\braOket{n^{\prime} j^{\prime}m^{\prime}}{\widehat{\mathfrak{M}}_{k \kappa}}{ n j m}\right|^{2} & =\frac{\left|
\braOketred{n^{\prime} j^{\prime}}{\widehat{\mathfrak{M}}_{k}}{ n j}\right|^{2}}{2 j+1},  \label{eq:13:1:5}\\
\sum_{m m^{\prime}}\left|
\braOket{n^{\prime} j^{\prime}m^{\prime}}{\widehat{\mathfrak{M}}_{k \kappa}}{ n j m}\right|^{2} & =\frac{\left|
\braOketred{n^{\prime} j^{\prime}}{\widehat{\mathfrak{M}}_{k}}{ n j}\right|^{2}}{2 k+1},  \label{eq:13:1:6}\\
\sum_{m m^{\prime} \kappa}\left|
\braOket{n^{\prime} j^{\prime}m^{\prime}}{\widehat{\mathfrak{M}}_{k \kappa}}{ n j m}\right|^{2} & =\left|
\braOketred{n^{\prime} j^{\prime}}{\widehat{\mathfrak{M}}_{k}}{ n j}\right|^{2}. \label{eq:13:1:7}
\end{align}
\subsection{Matrix Elements of Products of Irreducible Tensor Operators}\label{sec:13.1.3}\index{matrix elements!of products of tensor operators}\index{irreducible tensor product!of operators}
From two irreducible tensor operators \(\widehat{P}_{a \alpha}\) and \(\widehat{Q}_{b \beta}\) one can compose the direct or reducible product \(\widehat{P}_{a \alpha}\, \widehat{Q}_{b \beta}\) and the irreducible product \(\bigl\{\widehat{\vect{P}}_{a} \otimes \widehat{\vect{Q}}_{b}\bigr\}_{c \gamma}\) of rank \(c\) (see Chap.~\ref{chap:3}). A matrix element of the direct tensor product may be written in the form
\begin{equation}
\braOket{n^{\prime} j^{\prime} m^{\prime}}{\widehat{\mathbf{P}}_{a \alpha} \, \widehat{\mathbf{Q}}_{b \beta}}{n j m}=\frac{(-1)^{2 a+2 b}}{\Fact{j^{\prime}}} \sum_{n_{1} j_{1} m_{1}} 
\clebsch{j_{1}}{m_{1}}{a}{\alpha}{j^{\prime}}{m^{\prime}} 
\clebsch{j}{m}{b}{\beta}{j_{1}}{m_{1}} \frac{1}{\Fact{j_1}}
\braOketred{n^{\prime} j^{\prime}}{\widehat{\mathbf{P}}_{a}}{n_{1} j_{1}}
\braOketred{n_{1} j_{1}}{\widehat{\mathbf{Q}}_{b}}{n j} . \label{eq:13:1:8}
\end{equation}
Another frequently used relationship for the same matrix element is as follows:
\begin{equation}
\braOket{n^{\prime} j^{\prime} m^{\prime}}{\widehat{\mathbf{P}}_{a \alpha} \, \widehat{\mathbf{Q}}_{b \beta}}{n j m}=\frac{(-1)^{j^{\prime}+j}}{\Fact{j^{\prime}}} \sum_{c \gamma}(-1)^{-c} \Fact{c} 
\clebsch{j}{m}{c}{\gamma}{j^{\prime}}{m^{\prime}} 
\clebsch{a}{\alpha}{b}{\beta}{c}{\gamma} 
\sum_{n_{1} j_{1}}
\sixj{a}{b}{c}{j}{j^\prime}{j_1}
\braOketred{n^\prime j^{\prime}}{\widehat{\mathbf{P}}_{a}}{n_{1} j_{1}}
\braOketred{n_{1} j_{1}}{\widehat{\mathbf{Q}}_{b}}{n j} .\label{eq:13:1:9}
\end{equation}

Matrix elements of irreducible tensor products may be evaluated from the equation
\begin{align}
\braOket{n^{\prime} j^{\prime} m^{\prime}}{\bigl\{\widehat{\mathbf{P}}_{a} \otimes \widehat{\mathbf{Q}}_{b}\bigr\}_{c\gamma}}{n j m}&
=(-1)^{j^{\prime}+j-c} \frac{\Fact{c}}{\Fact{j^{\prime}}}
\clebsch{j}{m}{c}{\gamma}{j^{\prime}}{m^{\prime}}
\sum_{n_{1} j_{1}}
\sixj{a}{b}{c}{j}{j^\prime}{j_1}
\braOketred{n^\prime j^{\prime}}{\widehat{\mathbf{P}}_{a}}{n_{1} j_{1}}
\braOketred{n_{1} j_{1}}{\widehat{\mathbf{Q}}_{b}}{n j} , \label{eq:13:1:10}\\
\braOketred{n^{\prime} j^{\prime}}{\bigl\{\widehat{\mathbf{P}}_{a} \otimes \widehat{\mathbf{Q}}_{b}\bigr\}_{c}}{n j }&
=(-1)^{j^{\prime}+j+c} \Fact{c} 
\sum_{n_{1} j_{1}}
\sixj{a}{b}{c}{j}{j^\prime}{j_1}
\braOketred{n^\prime j^{\prime}}{\widehat{\mathbf{P}}_{a}}{n_{1} j_{1}}
\braOketred{n_{1} j_{1}}{\widehat{\mathbf{Q}}_{b}}{n j} . \notag
\end{align}
Putting \(c=0\) and \(b=a\) in Eq. \eqref{eq:13:1:10} and multiplying the resulting expression by \((-1)^{-a} \Fact{a}\) (in accordance with Eq.~\eqref{eq:3:1:34}) we get a matrix element of scalar product of two tensors of rank a
\begin{align}
\braOket{n^{\prime} j^{\prime} m^{\prime}}{\widehat{\mathbf{P}}_{a} \cdot \widehat{\mathbf{Q}}_{a}}{n j m}
&=\delta_{j j^{\prime}} \delta_{m m^{\prime}} \frac{1}{\Fact{j}^{2}} \sum_{n_{1} j_{1}}(-1)^{-j+j_{1}}
\braOketred{n^\prime j}{\widehat{\mathbf{P}}_{a}}{n_{1} j_{1}}
\braOketred{n_{1} j_{1}}{\widehat{\mathbf{Q}}_{a}}{n j} , \label{eq:13:1:11}\\
\braOketred{n^{\prime} j^{\prime}}{\widehat{\mathbf{P}}_{a} \cdot \widehat{\mathbf{Q}}_{a}}{n j}
&=\delta_{j j^{\prime}} \frac{1}{\Fact{j}} \sum_{n_{1} j_{1}}(-1)^{-j+j_{1}}
\braOketred{n^\prime j}{\widehat{\mathbf{P}}_{a}}{n_{1} j_{1}}
\braOketred{n_{1} j_{1}}{\widehat{\mathbf{Q}}_{a}}{n j} ,\notag
\end{align}
From the three irreducible tensor operators \(\widehat{P}_{a \alpha}\), \(\widehat{Q}_{b \beta}\) and \(\widehat{R}_{d \delta}\), one can compose the products
\begin{equation*}
\begin{array}{ccc}
\widehat{P}_{a \alpha} \, \widehat{Q}_{b \beta} \, \widehat{R}_{d \delta}, & \bigl\{\widehat{\vect{P}}_{a} \otimes \widehat{\mathbf{Q}}_{b}\bigr\}_{c \gamma} \, \widehat{R}_{d \delta}, & \widehat{P}_{a \alpha} \,\bigl\{\widehat{\mathbf{Q}}_{b} \otimes \widehat{\vect{R}}_{d}\bigr\}_{f \varphi}, \\
\bigl\{\bigl\{\widehat{\vect{P}}_{a} \otimes \widehat{\mathbf{Q}}_{b}\bigr\}_{c} \otimes \widehat{\vect{R}}_{d}\bigr\}_{f\varphi}, & \bigl\{\widehat{\vect{P}}_{a} \otimes\bigl\{\widehat{\mathbf{Q}}_{b} \otimes \widehat{\vect{R}}_{d}\bigr\}_{c}\bigr\}_{f \varphi} .
\end{array}
\end{equation*}

Making permutations of operators one can construct other products which differ from the given above only by commutators of operators. The matrix element of the direct product of three tensor operators is given by
\begin{align}
\braOket{n^{\prime} j^{\prime} m^{\prime}}{\widehat{P}_{a \alpha} \, \widehat{Q}_{b \beta} \, \widehat{R}_{d \delta}}{n j m}
=&\frac{(-1)^{2 a+2 b+2 d}}{\Fact{j^{\prime}}} \sum_{n_{1} j_{1} n_{2} j_{2}} \frac{1}{\Fact{j_{1} j_{2}}}
\braOketred{n^\prime j^{\prime}}{\widehat{\mathbf{P}}_{a}}{n_{1} j_{1}}
\braOketred{n_{1} j_{1}}{\widehat{\mathbf{Q}}_{b}}{n_2 j_2}\notag\\
&\times
\braOketred{n_{2} j_{2}}{\widehat{\mathbf{R}}_{d}}{n j}
\sum_{m_{1} m_{2}} 
\clebsch{j_{1}}{m_{1}}{a}{\alpha}{j^{\prime}}{m^{\prime}} 
\clebsch{j_{2}}{m_{2}}{b}{\beta}{j_{1}}{m_{1}} 
\clebsch{j}{m}{d}{\delta}{j_{2}}{m_{2}}. \label{eq:13:1:12}
\end{align}
The evaluation of matrix elements of the products \(\bigl\{\widehat{\vect{P}}_{a} \otimes \widehat{\vect{Q}}_{b}\bigr\}_{c \gamma} \, \widehat{R}_{d \delta}\) and \(\widehat{P}_{a \alpha} \,\bigl\{\widehat{\vect{Q}}_{b} \otimes \widehat{\vect{R}}_{d}\bigr\}_{c \gamma}\) is reduced to the successive use of Eqs. \eqref{eq:13:1:8}-\eqref{eq:13:1:10}.

The matrix element of the irreducible tensor product of three tensors may be represented in the form
\begin{multline}
\braOket{n^{\prime} j^{\prime} m^{\prime}}{\bigl\{\bigl\{\widehat{\vect{P}}_{a} \otimes \widehat{\vect{Q}}_{b}\bigr\}_{c} \otimes \widehat{\vect{R}}_{d}\bigr\}_{f \varphi}}{n j m}
=(-1)^{f+c-j} \frac{\Fact{f c}}{\Fact{j^{\prime}}} 
\clebsch{j}{m}{f}{\varphi}{j^{\prime}}{m^{\prime}} \\
\times \sum_{n_{1} j_{1} n_{2} j_{2}}(-1)^{j_{2}}
\sixj{a}{b}{c}{j_{2}}{j^{\prime}}{j_{1}}
\sixj{d}{c}{f}{j^{\prime}}{j}{j_{2}}
\braOketred{ n^{\prime} j^{\prime}}{\widehat{\vect{P}}_{a}}{ n_{1} j_{1}}
\braOketred{ n_{1} j_{1}}{\widehat{\vect{Q}}_{b}}{ n_{2} j_{2}}
\braOketred{ n_{2} j_{2}}{\widehat{\vect{R}}_{d}}{ n j}. \label{eq:13:1:13}
\end{multline}
\begin{multline*}
\braOketred{n^{\prime} j^{\prime}}{\bigl\{\bigl\{\widehat{\vect{P}}_{a} \otimes \widehat{\vect{Q}}_{b}\bigr\}_{c} \otimes \widehat{\vect{R}}_{d}\bigr\}_{f}}{n j}
=(-1)^{f+c-j} \Fact{f c} \sum_{n_{1} j_{1} n_{2} j_{2}}(-1)^{j_{2}}
\sixj{a}{b}{c}{j_{2}}{j^{\prime}}{j_{1}}
\sixj{d}{c}{f}{j^{\prime}}{j}{j_{2}} \\
\times
\braOketred{ n^{\prime} j^{\prime}}{\widehat{\vect{P}}_{a}}{ n_{1} j_{1}}
\braOketred{ n_{1} j_{1}}{\widehat{\vect{Q}}_{b}}{ n_{2} j_{2}}
\braOketred{ n_{2} j_{2}}{\widehat{\vect{R}}_{d}}{ n j}.
\end{multline*}
By the use of Eq.~\eqref{eq:3:3:1} the evaluation of the matrix elements of \(\bigl\{\widehat{\vect{P}}_{a} \otimes\bigl\{\widehat{\vect{Q}}_{b} \otimes \widehat{\vect{R}}_{d}\bigr\}_{c}\bigr\}_{f \varphi}\) is reduced to applying Eq. \eqref{eq:13:1:13}.

The matrix element of a scalar product may be obtained from Eq. \eqref{eq:13:1:13}. Taking into account the factor \((-1)^{-c} \Fact{c}\), one gets
\begin{multline}
\braOket{n^{\prime} j^{\prime} m^{\prime}}{\bigl(\bigl\{\widehat{\vect{P}}_{a} \otimes \widehat{\vect{Q}}_{b}\bigr\}_{c} \cdot \widehat{\vect{R}}_{c}\bigr)}{n j m}=\delta_{j j^{\prime}} \delta_{m m^{\prime}}(-1)^{2 j-c} \frac{\Fact{c}}{\Fact{j}^{2}} \\
\times \sum_{n_{1} j_{1} n_{2} j_{2}}
\sixj{a}{b}{c}{j_{2}}{j}{j_{1}}
\braOketred{ n^{\prime} j}{\widehat{\vect{P}}_{a}}{ n_{1} j_{1}}
\braOketred{ n_{1} j_{1}}{\widehat{\vect{Q}}_{b}}{ n_{2} j_{2}}
\braOketred{ n_{2} j_{2}}{\widehat{\vect{R}}_{c}}{ n j} . \label{eq:13:1:14}
\end{multline}
\begin{multline*}
\braOketred{n^{\prime} j^{\prime}}{\bigl(\bigl\{\widehat{\vect{P}}_{a} \otimes \widehat{\vect{Q}}_{b}\bigr\}_{c} \cdot \widehat{\vect{R}}_{c}\bigr)}{n j}=\delta_{j j^{\prime}}(-1)^{2 j-c} \frac{\Fact{c}}{\Fact{j}}
\sum_{n_{1} j_{1} n_{2} j_{2}}
\sixj{a}{b}{c}{j_{2}}{j}{j_{1}} \\
\times
\braOketred{ n^{\prime} j}{\widehat{\vect{P}}_{a}}{ n_{1} j_{1}}
\braOketred{ n_{1} j_{1}}{\widehat{\vect{Q}}_{b}}{ n_{2} j_{2}}
\braOketred{ n_{2} j_{2}}{\widehat{\vect{R}}_{c}}{ n j} .
\end{multline*}
Of the four irreducible tensors \(\widehat{P}_{a \alpha}, \widehat{Q}_{b \beta}, \widehat{R}_{d \delta}\) and \(\widehat{S}_{e \varepsilon}\) one can compose the following types of tensor products:
\begin{enumerate}[label=(\alph*)]
\item \(\widehat{P}_{a \alpha} \, \widehat{Q}_{b \beta} \, \widehat{R}_{d \delta} \, \widehat{S}_{e \varepsilon}\);
\item \(\bigl\{\widehat{\vect{P}}_{a} \otimes \widehat{\vect{Q}}_{b}\bigr\}_{c \gamma} \, \widehat{R}_{d \delta} \, \widehat{S}_{e \varepsilon}\), \(\widehat{P}_{a \alpha} \,\bigl\{\widehat{\vect{Q}}_{b} \otimes \widehat{\vect{R}}_{d}\bigr\}_{k \kappa} \, \widehat{S}_{e \varepsilon}\), \(\widehat{P}_{a \alpha} \, \widehat{Q}_{b \beta} \,\bigl\{\widehat{\vect{R}}_{d} \otimes \widehat{\vect{S}}_{e}\bigr\}_{h \sigma}\);
\item \(\bigl\{\bigl\{\widehat{\vect{P}}_{a} \otimes \widehat{\vect{Q}}_{b}\bigr\}_{c} \otimes \widehat{\vect{R}}_{d}\bigr\}_{f \varphi} \, \widehat{S}_{e \varepsilon}\), \(\bigl\{\widehat{\vect{P}}_{a} \otimes\bigl\{\widehat{\vect{Q}}_{b} \otimes \widehat{\vect{R}}_{d}\bigr\}_{h}\bigr\}_{f \varphi} \, \widehat{S}_{e \varepsilon}\),  \(\widehat{P}_{a \alpha} \,\bigl\{\bigl\{\widehat{\vect{Q}}_{b} \otimes \widehat{\vect{R}}_{d}\bigr\}_{h} \otimes \widehat{\vect{S}}_{e}\bigr\}_{f \varphi}\), \par \(\widehat{P}_{a \alpha} \,\bigl\{\widehat{\vect{Q}}_{b} \otimes\bigl\{\widehat{\vect{R}}_{d} \otimes \widehat{\vect{S}}_{e}\bigr\}_{h}\bigr\}_{f \varphi}\), \(\bigl\{\widehat{\vect{P}}_{a} \otimes \widehat{\vect{Q}}_{b}\bigr\}_{c \gamma} \,\bigl\{\widehat{\vect{R}}_{d} \otimes \widehat{\vect{S}}_{e}\bigr\}_{h \sigma}\);
\item \(\bigl\{\bigl\{\bigl\{\widehat{\vect{P}}_{a} \otimes \widehat{\vect{Q}}_{b}\bigr\}_{c} \otimes \widehat{\vect{R}}_{d}\bigr\}_{f} \otimes \widehat{\vect{S}}_{e}\bigr\}_{k \kappa}\), \(\bigl\{\widehat{\vect{P}}_{a} \otimes\bigl\{\widehat{\vect{Q}}_{b} \otimes\bigl\{\widehat{\vect{R}}_{d} \otimes \widehat{\vect{S}}_{e}\bigr\}_{h}\bigr\}_{f}\bigr\}_{k}\), \(\bigl\{\bigl\{\widehat{\vect{P}}_{a} \otimes \widehat{\vect{Q}}_{b}\bigr\}_{c} \otimes\bigl\{\widehat{\vect{R}}_{d} \otimes \widehat{\vect{S}}_{e}\bigr\}_{f}\bigr\}_{k \kappa}\).
\end{enumerate}
Products with other orders of the operators can be reduced to the products listed here using the commutation rules.

A matrix element of a direct product may be evaluated as follows:
\begin{multline}
\braOket{ n^{\prime} j^{\prime} m^{\prime}}{ \widehat{P}_{a \alpha} \widehat{Q}_{b \beta} \widehat{R}_{d \delta} \widehat{S}_{e \varepsilon}}{n j m}=(-1)^{2 a+2 b+2 d+2 e} \frac{1}{\Fact{j^{\prime}}} \sum_{\substack{n_{1} n_{2} n_{3} \\
j_{1} j_{2} j_{3}}}
\braOketred{ n^{\prime} j^{\prime}}{\widehat{\vect{P}}_{a}}{ n_{1} j_{1}}
\braOketred{ n_{1} j_{1}}{\widehat{\vect{Q}}_{b}}{ n_{2} j_{2}}
\braOketred{ n_{2} j_{2}}{\widehat{\vect{R}}_{d}}{ n_{3} j_{3}} \\
\times
\braOketred{ n_{3} j_{3}}{\widehat{\vect{S}}_{e}}{ n j} \frac{1}{\Fact{j_{1} j_{2} j_{3}}} \sum_{m_{1} m_{2} m_{3}} 
\clebsch{j_{1}}{m_{1}}{a}{\alpha}{j^{\prime}}{m^{\prime}} 
\clebsch{j_{2}}{m_{2}}{b}{\beta}{j_{1}}{m_{1}} 
\clebsch{j_{3}}{m_{3}}{d}{\delta}{j_{2}}{m_{2}}
\clebsch{j}{m}{e}{\varepsilon}{j_{3}}{m_{3}}. \label{eq:13:1:15}
\end{multline}
Matrix elements of products of (b)- and (c)-type products can be evaluated by using Eqs. \eqref{eq:13:1:8}-\eqref{eq:13:1:14}.

The evaluation of matrix elements of irreducible tensor products yields
\begin{gather}
\braOket{ n^{\prime} j^{\prime} m^{\prime}}{\bigl\{\bigl\{\bigl\{\widehat{\vect{P}}_{a} \otimes \widehat{\vect{Q}}_{b}\bigr\}_{c} \otimes \widehat{\vect{R}}_{d}\bigr\}_{f} \otimes \widehat{\vect{S}}_{e}\bigr\}_{k \kappa}}{n j m}=(-1)^{j-j^{\prime}+f+c-k} \frac{\Fact{f c k}}{\Fact{j^{\prime}}} 
\clebsch{j}{m}{k}{\kappa}{j^{\prime}}{m^{\prime}} \sum_{\substack{n_{1} n_{2} n_{3} \\
j_{1} j_{2} j_{3}}}(-1)^{j_{2}+j_{3}}
\sixj{a}{b}{c}{j_{2}}{j^{\prime}}{j_{1}} \notag\\
\times
\sixj{d}{c}{f}{j^{\prime}}{j_{3}}{j_{2}}
\sixj{e}{f}{k}{j^{\prime}}{j}{j_{3}}
\braOketred{ n^{\prime} j^{\prime}}{\widehat{\vect{P}}_{a}}{ n_{1} j_{1}}
\braOketred{ n_{1} j_{1}}{\widehat{\vect{Q}}_{b}}{ n_{2} j_{2}}
\braOketred{ n_{2} j_{2}}{\widehat{\vect{R}}_{d}}{ n_{3} j_{3}}
\braOketred{ n_{3} j_{3}}{\widehat{\vect{S}}_{e}}{ n j},  \label{eq:13:1:16}\\
\braOket{ n^{\prime} j^{\prime} m^{\prime}}{\bigl(\bigl\{\bigl\{\widehat{\vect{P}}_{a} \otimes \widehat{\vect{Q}}_{b}\bigr\}_{c} \otimes \widehat{\vect{R}}_{d}\bigr\}_{f}  \widehat{\vect{S}}_{f}\bigr)}{n j m}=(-1)^{c-f-j} \frac{\Fact{f c}}{\Fact{j}^{2}} \delta_{j^{\prime} j} \delta_{m^{\prime} m} \sum_{\substack{n_{1} n_{2} n_{3} \\
j_{1} j_{2} j_{3}}}(-1)^{j_{2}} 
\sixj{a}{b}{c}{j_{2}}{j}{j_{1}}
\sixj{d}{c}{f}{j}{j_{3}}{j_{2}} \notag\\
\times
\braOketred{ n^{\prime} j^{\prime}}{\widehat{\vect{P}}_{a}}{ n_{1} j_{1}}
\braOketred{ n_{1} j_{1}}{\widehat{\vect{Q}}_{b}}{ n_{2} j_{2}}
\braOketred{ n_{2} j_{2}}{\widehat{\vect{R}}_{d}}{ n_{3} j_{3}}
\braOketred{ n_{3} j_{3}}{\widehat{\vect{S}}_{f}}{ n j}. \label{eq:13:1:17}
\end{gather}
\begin{multline*}
\braOketred{ n^{\prime} j^{\prime}}{\bigl\{\bigl\{\bigl\{\widehat{\vect{P}}_{a} \otimes \widehat{\vect{Q}}_{b}\bigr\}_{c} \otimes \widehat{\vect{R}}_{d}\bigr\}_{f} \otimes \widehat{\vect{S}}_{e}\bigr\}_{k}}{n j}=(-1)^{j-j^{\prime}+f+c-k} \Fact{f c k}
\sum_{\substack{n_{1} n_{2} n_{3} \\ j_{1} j_{2} j_{3}}}(-1)^{j_{2}+j_{3}}
\sixj{a}{b}{c}{j_{2}}{j^{\prime}}{j_{1}}
\sixj{d}{c}{f}{j^{\prime}}{j_{3}}{j_{2}}
\sixj{e}{f}{k}{j^{\prime}}{j}{j_{3}} \\
\times
\braOketred{ n^{\prime} j^{\prime}}{\widehat{\vect{P}}_{a}}{ n_{1} j_{1}}
\braOketred{ n_{1} j_{1}}{\widehat{\vect{Q}}_{b}}{ n_{2} j_{2}}
\braOketred{ n_{2} j_{2}}{\widehat{\vect{R}}_{d}}{ n_{3} j_{3}}
\braOketred{ n_{3} j_{3}}{\widehat{\vect{S}}_{e}}{ n j},
\end{multline*}
\begin{multline*}
\braOketred{ n^{\prime} j^{\prime}}{\bigl(\bigl\{\bigl\{\widehat{\vect{P}}_{a} \otimes \widehat{\vect{Q}}_{b}\bigr\}_{c} \otimes \widehat{\vect{R}}_{d}\bigr\}_{f} \cdot \widehat{\vect{S}}_{f}\bigr)}{n j}=(-1)^{c-f-j} \frac{\Fact{f c}}{\Fact{j}} \delta_{j^{\prime} j}
\sum_{\substack{n_{1} n_{2} n_{3} \\ j_{1} j_{2} j_{3}}}(-1)^{j_{2}}
\sixj{a}{b}{c}{j_{2}}{j}{j_{1}}
\sixj{d}{c}{f}{j}{j_{3}}{j_{2}} \\
\times
\braOketred{ n^{\prime} j^{\prime}}{\widehat{\vect{P}}_{a}}{ n_{1} j_{1}}
\braOketred{ n_{1} j_{1}}{\widehat{\vect{Q}}_{b}}{ n_{2} j_{2}}
\braOketred{ n_{2} j_{2}}{\widehat{\vect{R}}_{d}}{ n_{3} j_{3}}
\braOketred{ n_{3} j_{3}}{\widehat{\vect{S}}_{f}}{ n j}.
\end{multline*}
Finally Eq.~\eqref{eq:3:3:5} allows one to simplify the calculation of matrix elements of the tensor product \(\bigl\{\widehat{\vect{P}}_{a} \otimes\bigl\{\widehat{\vect{Q}}_{b} \otimes\) \(\bigl\{\widehat{\vect{R}}_{d} \otimes \widehat{\vect{S}}_{e}\bigr\}_{h}\bigr\}_{f}\bigr\}_{k \kappa}\) to Eq. \eqref{eq:13:1:16}. Thus
\begin{gather}
\braOket{ n^{\prime} j^{\prime} m^{\prime}}{\bigl\{\bigl\{\widehat{\vect{P}}_{a} \otimes \widehat{\vect{Q}}_{b}\bigr\}_{c} \otimes\bigl\{\widehat{\vect{R}}_{d} \otimes \widehat{\vect{S}}_{e}\bigr\}_{f}\bigr\}_{k \kappa}}{n j m}=(-1)^{2 j^{\prime}+c-k-f} \frac{\Fact{c f k}}{\Fact{j^{\prime}}} 
\clebsch{j}{m}{k}{\kappa}{j^{\prime}}{m^{\prime}} \sum_{\substack{n_{1} n_{2} n_{3} \notag\\
j_{1} j_{2} j_{3}}}
\sixj{a}{b}{c}{j_{2}}{j^{\prime}}{j_{1}}
\sixj{d}{e}{f}{j}{j_{2}}{j_{3}} \\
\times
\sixj{f}{c}{k}{j^{\prime}}{j}{j_{2}}
\braOketred{ n^{\prime} j^{\prime}}{\widehat{\vect{P}}_{a}}{ n_{1} j_{1}}
\braOketred{ n_{1} j_{1}}{\widehat{\vect{Q}}_{b}}{ n_{2} j_{2}}
\braOketred{ n_{2} j_{2}}{\widehat{\vect{R}}_{d}}{ n_{3} j_{3}}
\braOketred{ n_{3} j_{3}}{\widehat{\vect{S}}_{e}}{ n j},  \label{eq:13:1:18}
\end{gather}
\begin{multline*}
\braOketred{ n^{\prime} j^{\prime}}{\bigl\{\bigl\{\widehat{\vect{P}}_{a} \otimes \widehat{\vect{Q}}_{b}\bigr\}_{c} \otimes\bigl\{\widehat{\vect{R}}_{d} \otimes \widehat{\vect{S}}_{e}\bigr\}_{f}\bigr\}_{k}}{n j}=(-1)^{2 j^{\prime}+c-k-f} \Fact{c f k}
\sum_{\substack{n_{1} n_{2} n_{3} \\ j_{1} j_{2} j_{3}}}
\sixj{a}{b}{c}{j_{2}}{j^{\prime}}{j_{1}}
\sixj{d}{e}{f}{j}{j_{2}}{j_{3}}
\sixj{f}{c}{k}{j^{\prime}}{j}{j_{2}} \\
\times
\braOketred{ n^{\prime} j^{\prime}}{\widehat{\vect{P}}_{a}}{ n_{1} j_{1}}
\braOketred{ n_{1} j_{1}}{\widehat{\vect{Q}}_{b}}{ n_{2} j_{2}}
\braOketred{ n_{2} j_{2}}{\widehat{\vect{R}}_{d}}{ n_{3} j_{3}}
\braOketred{ n_{3} j_{3}}{\widehat{\vect{S}}_{e}}{ n j},
\end{multline*}
\begin{gather}
%\left.
\braOket{n^{\prime} j^{\prime} m^{\prime}}{\bigl(\bigl\{\widehat{\vect{P}}_{a} \otimes \widehat{\vect{Q}}_{b}\bigr\}_{c} \cdot\bigl\{\widehat{\vect{R}}_{d} \otimes \widehat{\vect{S}}_{e}\bigr\}_{c}\bigr)}{n j m}=\delta_{j^{\prime} j} \delta_{m^{\prime} m} \frac{\Fact{c}^{2}}{\Fact{j}^{2}} \sum_{\substack{n_{1} n_{2} n_{3} \\
j_{1} j_{2} j_{3}}}(-1)^{-j+j_{2}}
\sixj{a}{b}{c}{j_{2}}{j}{j_{1}}
\sixj{d}{e}{c}{j}{j_{2}}{j_{3}} \notag\\
\times
\braOketred{ n^{\prime} j}{\widehat{\vect{P}}_{a}}{ n_{1} j_{1}}
\braOketred{ n_{1} j_{1}}{\widehat{\vect{Q}}_{b}}{ n_{2} j_{2}}
\braOketred{ n_{2} j_{2}}{\widehat{\vect{R}}_{d}}{ n_{3} j_{3}}
\braOketred{ n_{3} j_{3}}{\widehat{\vect{S}}_{e}}{ n j} . \label{eq:13:1:19}
\end{gather}
\begin{multline*}
\braOketred{n^{\prime} j^{\prime}}{\bigl(\bigl\{\widehat{\vect{P}}_{a} \otimes \widehat{\vect{Q}}_{b}\bigr\}_{c} \cdot\bigl\{\widehat{\vect{R}}_{d} \otimes \widehat{\vect{S}}_{e}\bigr\}_{c}\bigr)}{n j}=\delta_{j^{\prime} j} \frac{\Fact{c}^{2}}{\Fact{j}} \sum_{\substack{n_{1} n_{2} n_{3} \\ j_{1} j_{2} j_{3}}}(-1)^{-j+j_{2}}
\sixj{a}{b}{c}{j_{2}}{j}{j_{1}}
\sixj{d}{e}{c}{j}{j_{2}}{j_{3}} \\
\times
\braOketred{ n^{\prime} j}{\widehat{\vect{P}}_{a}}{ n_{1} j_{1}}
\braOketred{ n_{1} j_{1}}{\widehat{\vect{Q}}_{b}}{ n_{2} j_{2}}
\braOketred{ n_{2} j_{2}}{\widehat{\vect{R}}_{d}}{ n_{3} j_{3}}
\braOketred{ n_{3} j_{3}}{\widehat{\vect{S}}_{e}}{ n j} .
\end{multline*}
\subsection{Matrix Elements of Operators Which Depend on Variables of Two Subsystems}\label{sec:13.1.4}\index{matrix elements!for two subsystems}\index{subsystems}
If a quantum-mechanical system consists of two subsystems, 1 and 2 , and eigenstates of these subsystems are characterized by the sets of quantum numbers \(\bigl(n_{1} j_{1} m_{1}\bigr)\) and \(\bigl(n_{2} j_{2} m_{2}\bigr)\), it is often convenient to introduce the two following representations of matrix elements.\\
(a) The angular momenta of both subsystems are not coupled. Then\index{representation!uncoupled}\index{representation!coupled}
\begin{equation}
\braOket{n_{1}^{\prime} j_{1}^{\prime} m_{1}^{\prime} ; n_{2}^{\prime} j_{2}^{\prime} m_{2}^{\prime}}{\widehat{\vect{P}}(1,2)}{n_{1} j_{1} m_{1} ; n_{2} j_{2} m_{2}}
=\int \Psi_{n_{1}^{\prime} j_{1}^{\prime} m_{1}^{\prime}}^{*}(1) \Psi_{n_{2}^{\prime} j_{2}^{\prime} m_{2}^{\prime}}^{*}(2) \widehat{\vect{P}}(1,2) \Psi_{n_{1} j_{1} m_{1}}(1) \Psi_{n_{2} j_{2} m_{2}}(2) d r_{1} d \tau_{2}, \label{eq:13:1:20}
\end{equation}
where \(\Psi_{n_{1} j_{1} m_{1}}\) and \(\Psi_{n_{2} j_{2} m_{2}}\) are wave functions that describe the eigenstates of each subsystem. Such a representation of matrix elements is called the \(\bigl(n_{1} j_{1} m_{1} ; n_{2} j_{2} m_{2}\bigr)\)-representation.\\
(b) If angular momenta \(\vect{j}_{1}\) and \(\vect{j}_{2}\) are coupled to the resultant angular momentum j with projection \(m= m_{1}+m_{2}\), then
\begin{equation}
\braOket{ n_{1}^{\prime} j_{1}^{\prime} n_{2}^{\prime} j_{2}^{\prime} j^{\prime} m^{\prime}}{\widehat{\vect{P}}(1,2)}{n_{1} j_{1} n_{2} j_{2} j m}=\int \Psi_{n_{1}^{\prime} j_{1}^{\prime} n_{2}^{\prime} j_{2}^{\prime} j^{\prime} m^{\prime}}^{*}(1,2) \widehat{\vect{P}}(1,2) \Psi_{n_{1} j_{1} n_{2} j_{2} j m}(1,2) d \tau_{1} d \tau_{2}. \label{eq:13:1:21}
\end{equation}
Here
\begin{equation}
\Psi_{n_{1} j_{1} n_{2} j_{2} j m}(1,2)=\sum_{m_{1} m_{2}} 
\clebsch{j_{1}}{m_{1}}{j_{2}}{m_{2}}{j}{m} \Psi_{n_{1} j_{1} m_{1}}(1) \Psi_{n_{2} j_{2} m_{2}}(2). \label{eq:13:1:22}
\end{equation}
This representation of matrix elements is called the \(\bigl(n_{1} j_{1} n_{2} j_{2} j m\bigr)\)-representation.

In general case, we have
\begin{multline}
\braOket{ n_{1}^{\prime} j_{1}^{\prime} m_{1}^{\prime} ; n_{2}^{\prime} j_{2}^{\prime} m_{2}^{\prime}}{ \widehat{\vect{P}}_{a \alpha}(1,2)}{n_{1} j_{1} m_{1} ; n_{2} j_{2} m_{2}} \\
=(-1)^{2 a} \sum_{j m j^{\prime} m^{\prime}} \frac{1}{\Fact{j^{\prime}}} 
\clebsch{j_{1}}{m_{1}}{j_{2}}{m_{2}}{j}{m} 
\clebsch{j_{1}^{\prime}}{m_{1}^{\prime}}{j_{2}^{\prime}}{m_{2}^{\prime}}{j^{\prime}}{m^{\prime}} 
\clebsch{j}{m}{a}{\alpha}{j^{\prime}}{m^{\prime}}
\braOketred{ n_{1}^{\prime} j_{1}^{\prime} n_{2}^{\prime} j_{2}^{\prime} j^{\prime}}{\widehat{\vect{P}}_{a}(1,2)}{ n_{1} j_{1} n_{2} j_{2} j}. \label{eq:13:1:23}
\end{multline}
Any operator \(\widehat{A}(1,2)\) which depends on variables of two subsystems can frequently be expressed as a direct or irreducible product of two operators \(\widehat{P}_{a \alpha}(1)\) and \(\widehat{Q}_{b \beta}(2)\) which depend only on variables of the first and second subsystem, respectively:
\begin{equation*}
\widehat{A}(1,2)=\widehat{P}_{a \alpha}(1) \widehat{Q}_{b \beta}(2) \quad \text { or } \quad \widehat{A}(1,2)=\bigl\{\widehat{\vect{P}}_{a}(1) \otimes \widehat{\vect{Q}}_{b}(2)\bigr\}_{c \gamma}.
\end{equation*}
The matrix element for the direct-product case is given by
\begin{multline}
\braOket{ n_{1}^{\prime} j_{1}^{\prime} m_{1}^{\prime} ; n_{2}^{\prime} j_{2}^{\prime} m_{2}^{\prime}}{ \widehat{P}_{a \alpha}(1)  \widehat{Q}_{b \beta}(2)}{n_{1} j_{1} m_{1} ; n_{2} j_{2} m_{2}} \\
=\frac{(-1)^{2 a+2 b}}{\Fact{j_{1}^{\prime} j_{2}^{\prime}}} 
\clebsch{j_{1}}{m_{1}}{a}{\alpha}{j_{1}^{\prime}}{m_{1}^{\prime}} 
\clebsch{j_{2}}{m_{2}}{b}{\beta}{j_{2}^\prime}{m_{2}^{\prime}}
\braOketred{ n_{1}^{\prime} j_{1}^{\prime}}{\widehat{\vect{P}}_{a}(1)}{ n_{1} j_{1}}
\braOketred{n_{2}^{\prime} j_{2}^{\prime}}{\widehat{\vect{Q}}_{b}(2)}{n_{2} j_{2}}, \label{eq:13:1:24}
\end{multline}
whereas for the irreducible tensor product case it may be expressed in the form
\begin{multline}
\braOket{ n_{1}^{\prime} j_{1}^{\prime} m_{1}^{\prime} ; n_{2}^{\prime} j_{2}^{\prime} m_{2}^{\prime}}{\bigl\{\widehat{\vect{P}}_{a}(1) \otimes \widehat{\vect{Q}}_{b}(2)\bigr\}_{c \gamma}}{n_{1} j_{1} m_{1} ; n_{2} j_{2} m_{2}} \\
=\frac{(-1)^{2 c}}{\Fact{j_{1}^{\prime} j_{2}^{\prime}}} 
\sum_{\alpha \beta} 
\clebsch{j_{1}}{m_{1}}{a}{\alpha}{j_{1}^{\prime}}{m_{1}^{\prime}} 
\clebsch{a}{\alpha}{b}{\beta}{c}{\gamma} 
\clebsch{j_{2}}{m_{2}}{b}{\beta}{j_{2}^{\prime}}{m_{2}^{\prime}}
\braOketred{ n_{1}^{\prime} j_{1}^{\prime}}{\widehat{\vect{P}}_{a}(1)}{ n_{1} j_{1}}
\braOketred{ n_{2}^{\prime} j_{2}^{\prime}}{\widehat{\vect{Q}}_{b}(2)}{ n_{2} j_{2}}. \label{eq:13:1:25}
\end{multline}
The matrix element of a scalar product is given as follows:
\begin{multline}
\braOket{ n_{1}^{\prime} j_{1}^{\prime} m_{1}^{\prime} ; n_{2}^{\prime} j_{2}^{\prime} m_{2}^{\prime}}{\bigl(\widehat{\vect{P}}_{a}(1) \cdot \widehat{\vect{Q}}_{a}(2)\bigr)}{n_{1} j_{1} m_{1} ; n_{2} j_{2} m_{2}} \\
=\frac{1}{\Fact{j_{1}^{\prime} j_{2}^{\prime}}} \sum_{\alpha}(-1)^{-\alpha} 
\clebsch{j_{1}}{m_{1}}{a}{\alpha}{j_{1}^{\prime}}{m_{1}^{\prime}} 
\clebsch{j_{2}}{m_{2}}{a}{-\alpha}{j_{2}^{\prime}}{m_{2}^{\prime}}
\braOketred{ n_{1}^{\prime} j_{1}^{\prime}}{\widehat{\vect{P}}_{a}(1)}{ n_{1} j_{1}}
\braOketred{ n_{2}^{\prime} j_{2}^{\prime}}{\widehat{\vect{Q}}_{b}(2)}{ n_{2} j_{2}}. \label{eq:13:1:26}
\end{multline}

Matrix elements of the same operators can be evaluated in the \(n_{1} j_{1} n_{2} j_{2} j m\) representation. This yields
\begin{multline}
\braOket{ n_{1}^{\prime} j_{1}^{\prime} n_{2}^{\prime} j_{2}^{\prime} j^{\prime} m^{\prime}}{ \widehat{P}_{a \alpha}(1)  \widehat{Q}_{b \beta}(2)}{n_{1} j_{1} n_{2} j_{2} j m} \\
=(-1)^{2 a+2 b} \Fact{j}
\braOketred{ n_{1}^{\prime} j_{1}^{\prime}}{\widehat{\vect{P}}_{a}(1)}{ n_{1} j_{1}}
\braOketred{ n_{2}^{\prime} j_{2}^{\prime}}{\widehat{\vect{Q}}_{b}(2)}{ n_{2} j_{2}} \sum_{c \gamma} \Fact{c} 
\clebsch{j}{m}{c}{\gamma}{j^{\prime}}{m^{\prime}} 
\clebsch{a}{\alpha}{b}{\beta}{c}{\gamma}
\ninej{a}{b}{c}{j_{1}^{\prime}}{j_{2}^{\prime}}{j^{\prime}}{j_{1}}{j_{2}}{j},  \label{eq:13:1:27}
\end{multline}
\begin{multline}
\braOket{ n_{1}^{\prime} j_{1}^{\prime} n_{2}^{\prime} j_{2}^{\prime} j^{\prime} m^{\prime}}{\bigl\{\widehat{\vect{P}}_{a}(1) \otimes \widehat{\vect{Q}}_{b}(2)\bigr\}_{c \gamma}}{n_{1} j_{1} n_{2} j_{2} j m} \\
=(-1)^{2 c} \Fact{c j} 
\clebsch{j}{m}{c}{\gamma}{j^{\prime}}{m^{\prime}}
\ninej{a}{b}{c}{j_{1}^{\prime}}{j_{2}^{\prime}}{j^{\prime}}{j_{1}}{j_{2}}{j}
\braOketred{ n_{1}^{\prime} j_{1}^{\prime}}{\widehat{\vect{P}}_{a}(1)}{ n_{1} j_{1}}
\braOketred{ n_{2}^{\prime} j_{2}^{\prime}}{\widehat{\vect{Q}}_{b}(2)}{ n_{2} j_{2}},  \label{eq:13:1:28}
\end{multline}
\begin{multline*}
\braOketred{ n_{1}^{\prime} j_{1}^{\prime} n_{2}^{\prime} j_{2}^{\prime} j^{\prime}}{\bigl\{\widehat{\vect{P}}_{a}(1) \otimes \widehat{\vect{Q}}_{b}(2)\bigr\}_{c}}{n_{1} j_{1} n_{2} j_{2} j}
=\Fact{c j j^{\prime}}
\ninej{a}{b}{c}{j_{1}^{\prime}}{j_{2}^{\prime}}{j^{\prime}}{j_{1}}{j_{2}}{j} \\
\times
\braOketred{ n_{1}^{\prime} j_{1}^{\prime}}{\widehat{\vect{P}}_{a}(1)}{ n_{1} j_{1}}
\braOketred{ n_{2}^{\prime} j_{2}^{\prime}}{\widehat{\vect{Q}}_{b}(2)}{ n_{2} j_{2}},
\end{multline*}
\begin{multline}
\braOket{ n_{1}^{\prime} j_{1}^{\prime} n_{2}^{\prime} j_{2}^{\prime} j^{\prime} m^{\prime}}{\bigl(\widehat{\vect{P}}_{a}(1) \cdot \widehat{\vect{Q}}_{a}(2)\bigr)}{n_{1} j_{1} n_{2} j_{2} j m} \\
=\delta_{j^{\prime} j} \delta_{m^{\prime} m}(-1)^{j+j_{1}+j_{2}^{\prime}}
\sixj{j_{1}^{\prime}}{j_{1}}{a}{j_{2}}{j_{2}^{\prime}}{j}
\braOketred{ n_{1}^{\prime} j_{1}^{\prime}}{\widehat{\vect{P}}_{a}(1)}{ n_{1} j_{1}}
\braOketred{ n_{2}^{\prime} j_{2}^{\prime}}{\widehat{\vect{Q}}_{a}(2)}{ n_{2} j_{2}}. \label{eq:13:1:29}
\end{multline}
\begin{multline*}
\braOketred{ n_{1}^{\prime} j_{1}^{\prime} n_{2}^{\prime} j_{2}^{\prime} j^{\prime}}{\bigl(\widehat{\vect{P}}_{a}(1) \cdot \widehat{\vect{Q}}_{a}(2)\bigr)}{n_{1} j_{1} n_{2} j_{2} j}
=\delta_{j^{\prime} j}(-1)^{j+j_{1}+j_{2}^{\prime}} \Fact{j}
\sixj{j_{1}^{\prime}}{j_{1}}{a}{j_{2}}{j_{2}^{\prime}}{j} \\
\times
\braOketred{ n_{1}^{\prime} j_{1}^{\prime}}{\widehat{\vect{P}}_{a}(1)}{ n_{1} j_{1}}
\braOketred{ n_{2}^{\prime} j_{2}^{\prime}}{\widehat{\vect{Q}}_{a}(2)}{ n_{2} j_{2}}.
\end{multline*}

Let us consider the products of three operators which depend only on variables of one of the subsystems. In this case the matrix elements of direct products of these operators can be evaluated by the successive use of Eqs. \eqref{eq:13:1:24}, \eqref{eq:13:1:27} and \eqref{eq:13:1:8}. If two of three operators form an irreducible tensor product, Eqs. \eqref{eq:13:1:25}, \eqref{eq:13:1:28}, \eqref{eq:13:1:9} and \eqref{eq:13:1:10} must additionally be applied.

Matrix elements of irreducible tensor products of three tensor operators may be represented in the forms
\begin{multline}
\braOket{ n_{1}^{\prime} j_{1}^{\prime} m_{1}^{\prime} ; n_{2}^{\prime} j_{2}^{\prime} m_{2}^{\prime}}{\bigl\{\bigl\{\widehat{\vect{P}}_{a}(1) \otimes \widehat{\vect{Q}}_{b}(2)\bigr\}_{c} \otimes \widehat{\vect{R}}_{d}(1)\bigr\}_{e \varepsilon}}{n_{1} j_{1} m_{1} ; n_{2} j_{2} m_{2}} \\
 =(-1)^{b-c-d-j_{1}-j_{1}^{\prime}} \frac{\Fact{c}}{\Fact{j_{1}^{\prime} j_{2}^{\prime}}} \sum_{f N J} \Fact{f}^{2}
\clebsch{f}{\varphi}{b}{\beta}{e}{\varepsilon} 
\clebsch{j_{1}}{m_{1}}{f}{\varphi}{j_{1}^{\prime}}{m_{1}^{\prime}} 
\clebsch{j_{2}}{m_{2}}{b}{\beta}{j_{2}^{\prime}}{m_{2}^{\prime}}
\sixj{a}{b}{c}{e}{d}{f}
\sixj{d}{a}{f}{j_{1}^{\prime}}{j_{1}}{J} \\
 \times
\braOketred{ n_{1}^{\prime} j_{1}^{\prime}}{\widehat{\vect{P}}_{a}(1)}{ N J}
\braOketred{ N J}{\widehat{\vect{R}}_{d}(1)}{ n_{1} j_{1}}
\braOketred{ n_{2}^{\prime} j_{2}^{\prime}}{\widehat{\vect{Q}}_{b}(2)}{ n_{2} j_{2}},  \label{eq:13:1:30}
\end{multline}
\begin{multline}
\braOket{ n_{1}^{\prime} j_{1}^{\prime} n_{2}^{\prime} j_{2}^{\prime} j^{\prime} m^{\prime}}{\bigl\{\bigl\{\widehat{\vect{P}}_{a}(1) \otimes \widehat{\vect{Q}}_{b}(2)\bigr\}_{c} \otimes \widehat{\vect{R}}_{d}(1)\bigr\}_{e \varepsilon}}{n_{1} j_{1} n_{2} j_{2} j m}=(-1)^{b-c-d-j_{1}-j_{1}^{\prime}} \Fact{c j e} 
\clebsch{j}{m}{e}{\varepsilon}{j^{\prime}}{m^{\prime}} 
\sum_{f N J} \Fact{f}^{2} \\
\sixj{a}{b}{c}{e}{d}{f}
\sixj{d}{a}{f}{j_{1}^{\prime}}{j_{1}}{J}
\ninej{j_{1}}{f}{j_{1}^{\prime}}{j_{2}}{b}{j_{2}^{\prime}}{j}{e}{j^{\prime}}
\braOketred{ n_{1}^{\prime} j_{1}^{\prime}}{\widehat{P}_{a}(1)}{ N J}
\braOketred{ N J}{\widehat{R}_{d}(1)}{ n_{1} j_{1}}
\braOketred{n_{2}^{\prime} j_{2}^{\prime}}{\widehat{Q}_{b}(2)}{ n_{2} j_{2}},  \label{eq:13:1:31}
\end{multline}
\begin{multline*}
\braOketred{ n_{1}^{\prime} j_{1}^{\prime} n_{2}^{\prime} j_{2}^{\prime} j^{\prime}}{\bigl\{\bigl\{\widehat{\vect{P}}_{a}(1) \otimes \widehat{\vect{Q}}_{b}(2)\bigr\}_{c} \otimes \widehat{\vect{R}}_{d}(1)\bigr\}_{e}}{n_{1} j_{1} n_{2} j_{2} j}=(-1)^{b-c-d-j_{1}-j_{1}^{\prime}} \Fact{c j e j^{\prime}}
\sum_{f N J} \Fact{f}^{2} \\
\sixj{a}{b}{c}{e}{d}{f}
\sixj{d}{a}{f}{j_{1}^{\prime}}{j_{1}}{J}
\ninej{j_{1}}{f}{j_{1}^{\prime}}{j_{2}}{b}{j_{2}^{\prime}}{j}{e}{j^{\prime}}
\braOketred{ n_{1}^{\prime} j_{1}^{\prime}}{\widehat{P}_{a}(1)}{ N J}
\braOketred{ N J}{\widehat{R}_{d}(1)}{ n_{1} j_{1}}
\braOketred{n_{2}^{\prime} j_{2}^{\prime}}{\widehat{Q}_{b}(2)}{ n_{2} j_{2}},
\end{multline*}
\begin{multline}
\braOket{ n_{1}^{\prime} j_{1}^{\prime} n_{2}^{\prime} j_{2}^{\prime} j^{\prime} m^{\prime}}{\bigl(\bigl\{\widehat{\vect{P}}_{a}(1) \otimes \widehat{\vect{Q}}_{b}(2)\bigr\}_{c} \cdot \widehat{\vect{R}}_{c}(1)\bigr)}{n_{1} j_{1} n_{2} j_{2} j m}
\\=\delta_{j^{\prime} j} \delta_{m^{\prime} m}(-1)^{-a+b+j+j_{2}^{\prime}-j_{1}^{\prime}} \Fact{c} \sixj{j_{1}^{\prime}}{j_{1}}{b}{j_{2}}{j_{2}^{\prime}}{j} \sum_{N J}
\sixj{a}{b}{c}{j_{1}}{J}{j_{1}^{\prime}}
\braOketred{ n_{1}^{\prime} j_{1}^{\prime}}{\widehat{\vect{P}}_{a}(1)}{ N J}
\braOketred{ N J}{\widehat{\vect{R}}_{c}(1)}{ n_{1} j_{1}}
\braOketred{ n_{2}^{\prime} j_{2}^{\prime}}{\widehat{\vect{Q}}_{b}(2)}{ n_{2} j_{2}} . \label{eq:13:1:32}
\end{multline}
\begin{multline*}
\braOketred{ n_{1}^{\prime} j_{1}^{\prime} n_{2}^{\prime} j_{2}^{\prime} j^{\prime}}{\bigl(\bigl\{\widehat{\vect{P}}_{a}(1) \otimes \widehat{\vect{Q}}_{b}(2)\bigr\}_{c} \cdot \widehat{\vect{R}}_{c}(1)\bigr)}{n_{1} j_{1} n_{2} j_{2} j}
=\delta_{j^{\prime} j}(-1)^{-a+b+j+j_{2}^{\prime}-j_{1}^{\prime}} \Fact{c j} \sixj{j_{1}^{\prime}}{j_{1}}{b}{j_{2}}{j_{2}^{\prime}}{j} \sum_{N J}
\sixj{a}{b}{c}{j_{1}}{J}{j_{1}^{\prime}} \\
\times
\braOketred{ n_{1}^{\prime} j_{1}^{\prime}}{\widehat{\vect{P}}_{a}(1)}{ N J}
\braOketred{ N J}{\widehat{\vect{R}}_{c}(1)}{ n_{1} j_{1}}
\braOketred{ n_{2}^{\prime} j_{2}^{\prime}}{\widehat{\vect{Q}}_{b}(2)}{ n_{2} j_{2}} .
\end{multline*}
The evaluation of matrix elements of the product \(\bigl\{\widehat{\vect{P}}_{a}(1) \otimes\bigl\{\widehat{\vect{Q}}_{b}(2) \otimes \widehat{\vect{R}}_{d}(1)\bigr\}_{f}\bigr\}_{k}\) can be performed using Eq.~\eqref{eq:3:3:1}, while, for the product \(\bigl\{\bigl\{\widehat{\vect{P}}_{a}(1) \otimes \widehat{\vect{R}}_{d}(1)\bigr\}_{f} \otimes \widehat{\vect{Q}}_{b}(2)\bigr\}_{k}\), one can use Eqs. \eqref{eq:13:1:25}, \eqref{eq:13:1:28}, \eqref{eq:13:1:10}.

Matrix elements of direct products of four tensor operators can be obtained by the successive use of the foregoing relations.

We give below expressions for matrix elements of irreducible tensor products of four tensor operators:
\begin{multline}
\braOket{ n_{1}^{\prime} j_{1}^{\prime} m_{1}^{\prime} ; n_{2}^{\prime} j_{2}^{\prime} m_{2}^{\prime}}{\bigl\{\bigl\{\widehat{\vect{P}}_{a}(1) \otimes \widehat{\vect{Q}}_{b}(2)\bigr\}_{c} \otimes\bigl\{\widehat{\vect{R}}_{d}(1) \otimes \widehat{\vect{S}}_{e}(2)\bigr\}_{f}\bigr\}_{k \kappa}}{n_{1} j_{1} m_{1} ; n_{2} j_{2} m_{2}}
\\
=(-1)^{2 j_{1}^{\prime}+2 j_{2}^{\prime}+2 k} \frac{\Fact{c j}}{\Fact{j_{1}^{\prime} j_{2}^{\prime}}} \sum_{g h \sigma \eta} \Fact{g h}^{2} 
\clebsch{h}{\sigma}{g}{\eta}{k}{\kappa} 
\clebsch{h}{\sigma}{j_{1}}{m_{1}}{j_{1}^{\prime}}{m_{1}^{\prime}} 
\clebsch{g}{\eta}{j_{2}}{m_{2}}{j_{2}^{\prime}}{m_{2}^{\prime}}
\ninej{a}{b}{c}{d}{e}{f}{h}{g}{k} \sum_{N_{1} J_{1} N_{2} J_{2}}
\sixj{a}{d}{h}{j_{1}}{j_{1}^{\prime}}{J_{1}}
\sixj{b}{e}{g}{j_{2}}{j_{2}^{\prime}}{J_{2}} \\
 \times
\braOketred{ n_{1}^{\prime} j_{1}^{\prime}}{\widehat{\vect{P}}_{a}(1)}{ N_{1} J_{1}}
\braOketred{ N_{1} J_{1}}{\widehat{\vect{R}}_{d}(1)}{ n_{1} j_{1}}
\braOketred{ n_{2}^{\prime} j_{2}^{\prime}}{\widehat{\vect{Q}}_{b}(2)}{ N_{2} J_{2}}
\braOketred{ N_{2} J_{2}}{\widehat{\vect{S}}_{e}(2)}{ n_{2} j_{2}}, \label{eq:13:1:33}
\end{multline}
\begin{multline} 
\braOket{n_{1}^{\prime} j_{1}^{\prime} n_{2}^{\prime} j_{2}^{\prime} j^{\prime} m^{\prime}}{\bigl\{\bigl\{\widehat{\vect{P}}_{a}(1) \otimes \widehat{\vect{Q}}_{b}(2)\bigr\}_{c} \otimes\bigl\{\widehat{\vect{R}}_{d}(1) \otimes \widehat{\vect{S}}_{e}(2)\bigr\}_{f}\bigr\}_{k \kappa}}{n_{1} j_{1} n_{2} j_{2} j m} \\
 =(-1)^{j+j^{\prime}-k} \Fact{c f j k} 
\clebsch{j}{m}{k}{\kappa}{j^{\prime}}{m^{\prime}}
\sum_{g h} \Fact{g h}^{2}
\ninej{a}{b}{c}{d}{e}{f}{h}{g}{k}
\ninej{j_{1}}{j_{1}^{\prime}}{h}{j_{2}}{j_{2}^{\prime}}{g}{j}{j^{\prime}}{k}\sum_{N_{1} J_{1} N_{2} J_{2}}
\sixj{d}{a}{h}{j_{1}^{\prime}}{j_{1}}{J_{1}}
\sixj{e}{b}{g}{j_{2}^{\prime}}{j_{2}}{J_{2}} \\
 \times
\braOketred{ n_{1}^{\prime} j_{1}^{\prime}}{\widehat{\vect{P}}_{a}(1)}{ N_{1} J_{1}}
\braOketred{ N_{1} J_{1}}{\widehat{\vect{R}}_{d}(1)}{ n_{1} j_{1}}
\braOketred{ n_{2}^{\prime} j_{2}^{\prime}}{\widehat{\vect{Q}}_{b}(2)}{ N_{2} J_{2}}
\braOketred{ N_{2} J_{2}}{\widehat{\vect{S}}_{e}(2)}{ n_{2} j_{2}},  \label{eq:13:1:34}
\end{multline}
\begin{multline*}
\braOketred{n_{1}^{\prime} j_{1}^{\prime} n_{2}^{\prime} j_{2}^{\prime} j^{\prime}}{\bigl\{\bigl\{\widehat{\vect{P}}_{a}(1) \otimes \widehat{\vect{Q}}_{b}(2)\bigr\}_{c} \otimes\bigl\{\widehat{\vect{R}}_{d}(1) \otimes \widehat{\vect{S}}_{e}(2)\bigr\}_{f}\bigr\}_{k}}{n_{1} j_{1} n_{2} j_{2} j}\\
=(-1)^{j+j^{\prime}-k} \Fact{c f j k j^{\prime}}
\sum_{g h} \Fact{g h}^{2}
\ninej{a}{b}{c}{d}{e}{f}{h}{g}{k}
\ninej{j_{1}}{j_{1}^{\prime}}{h}{j_{2}}{j_{2}^{\prime}}{g}{j}{j^{\prime}}{k} 
 \sum_{N_{1} J_{1} N_{2} J_{2}}
\sixj{d}{a}{h}{j_{1}^{\prime}}{j_{1}}{J_{1}}
\sixj{e}{b}{g}{j_{2}^{\prime}}{j_{2}}{J_{2}}\\
\times\braOketred{ n_{1}^{\prime} j_{1}^{\prime}}{\widehat{\vect{P}}_{a}(1)}{ N_{1} J_{1}}
\braOketred{ N_{1} J_{1}}{\widehat{\vect{R}}_{d}(1)}{ n_{1} j_{1}}
\braOketred{ n_{2}^{\prime} j_{2}^{\prime}}{\widehat{\vect{Q}}_{b}(2)}{ N_{2} J_{2}}
\braOketred{ N_{2} J_{2}}{\widehat{\vect{S}}_{e}(2)}{ n_{2} j_{2}},
\end{multline*}
\begin{multline} 
\braOket{ n_{1}^{\prime} j_{1}^{\prime} n_{2}^{\prime} j_{2}^{\prime} j^{\prime} m^{\prime}}{\bigl(\bigl\{\widehat{\vect{P}}_{a}(1) \otimes \widehat{\vect{Q}}_{b}(2)\bigr\}_{c} \cdot\bigl\{\widehat{\vect{R}}_{d}(1) \otimes \widehat{\vect{S}}_{e}(2)\bigr\}_{c}\bigr)}{n_{1} j_{1} n_{2} j_{2} j m}=\delta_{j^{\prime} j} \delta_{m^{\prime} m}(-1)^{2 c+j_{1}^{\prime}+j_{2}-j-b-d} \Fact{c}^{2} \\
 \times \sum_{g} \Fact{g}^{2}
\sixj{a}{b}{c}{e}{d}{g}
\sixj{j_{1}^{\prime}}{j_{1}}{g}{j_{2}}{j_{2}^{\prime}}{j} \sum_{N_{1} J_{1} N_{2} J_{2}}
\sixj{d}{a}{g}{j_{1}^{\prime}}{j_{1}}{J_{1}}
\sixj{e}{b}{g}{j_{2}^{\prime}}{j_{2}}{J_{2}} \\
 \times
\braOketred{ n_{1}^{\prime} j_{1}^{\prime}}{\widehat{\vect{P}}_{a}(1)}{ N_{1} J_{1}}
\braOketred{ N_{1} J_{1}}{\widehat{\vect{R}}_{d}(1)}{ n_{1} j_{1}}
\braOketred{ n_{2}^{\prime} j_{2}^{\prime}}{\widehat{\vect{Q}}_{b}(2)}{ N_{2} J_{2}}
\braOketred{N_{2} J_{2}}{\widehat{\vect{S}}_{e}(2)}{ n_{2} j_{2}}, \label{eq:13:1:35}
\end{multline}
\begin{multline*}
\braOketred{ n_{1}^{\prime} j_{1}^{\prime} n_{2}^{\prime} j_{2}^{\prime} j^{\prime}}{\bigl(\bigl\{\widehat{\vect{P}}_{a}(1) \otimes \widehat{\vect{Q}}_{b}(2)\bigr\}_{c} \cdot\bigl\{\widehat{\vect{R}}_{d}(1) \otimes \widehat{\vect{S}}_{e}(2)\bigr\}_{c}\bigr)}{n_{1} j_{1} n_{2} j_{2} j}=\delta_{j^{\prime} j}(-1)^{2 c+j_{1}^{\prime}+j_{2}-j-b-d} \Fact{j} \Fact{c}^{2} \\
\times \sum_{g} \Fact{g}^{2}
\sixj{a}{b}{c}{e}{d}{g}
\sixj{j_{1}^{\prime}}{j_{1}}{g}{j_{2}}{j_{2}^{\prime}}{j} \sum_{N_{1} J_{1} N_{2} J_{2}}
\sixj{d}{a}{g}{j_{1}^{\prime}}{j_{1}}{J_{1}}
\sixj{e}{b}{g}{j_{2}^{\prime}}{j_{2}}{J_{2}} \\
\times
\braOketred{ n_{1}^{\prime} j_{1}^{\prime}}{\widehat{\vect{P}}_{a}(1)}{ N_{1} J_{1}}
\braOketred{ N_{1} J_{1}}{\widehat{\vect{R}}_{d}(1)}{ n_{1} j_{1}}
\braOketred{ n_{2}^{\prime} j_{2}^{\prime}}{\widehat{\vect{Q}}_{b}(2)}{ N_{2} J_{2}}
\braOketred{N_{2} J_{2}}{\widehat{\vect{S}}_{e}(2)}{ n_{2} j_{2}},
\end{multline*}
\begin{multline} 
\braOket{ n_{1}^{\prime} j_{1}^{\prime} m_{1}^{\prime} ; n_{2}^{\prime} j_{2}^{\prime} m_{2}^{\prime}}{\bigl\{\bigl\{\bigl\{\widehat{\vect{P}}_{a}(1) \otimes \widehat{\vect{Q}}_{b}(2)\bigr\}_{c} \otimes \widehat{\vect{R}}_{d}(1)\bigr\}_{f} \otimes \widehat{\vect{S}}_{e}(2)\bigr\}_{k \kappa}}{n_{1} j_{1} m_{1} ; n_{2} j_{2} m_{2}} \\
 =(-1)^{c+d+k-e+2 j_{1}^{\prime}+2 j_{2}^{\prime}} \frac{\Fact{c f}}{\Fact{j_{1}^{\prime} j_{2}^{\prime}}} \sum_{g h \sigma \eta} \Fact{g h}^{2} 
\clebsch{h}{\sigma}{g}{\eta}{k}{\kappa} 
\clebsch{h}{\sigma}{j_{1}}{m_{1}}{j_{1}^{\prime}}{m_{1}^{\prime}} 
\clebsch{g}{\eta}{j_{2}}{m_{2}}{j_{2}^{\prime}}{m_{2}^{\prime}}
\sixj{a}{d}{h}{f}{b}{c}
\sixj{b}{e}{g}{k}{h}{f} \\
 \times \sum_{N_{1} J_{1} N_{2} J_{2}}
\sixj{d}{a}{h}{j_{1}^{\prime}}{j_{1}}{J_{1}}
\sixj{e}{b}{g}{j_{2}^{\prime}}{j_{2}}{J_{2}}
\braOketred{ n_{1}^{\prime} j_{1}^{\prime}}{\widehat{\vect{P}}_{a}(1)}{ N_{1} J_{1}}
\braOketred{ N_{1} J_{1}}{\widehat{\vect{R}}_{d}(1)}{ n_{1} j_{1}} \\
 \times
\braOketred{ n_{2}^{\prime} j_{2}^{\prime}}{\widehat{\vect{Q}}_{b}(2)}{ N_{2} J_{2}}
\braOketred{ N_{2} J_{2}}{\widehat{\vect{S}}_{e}(2)}{ n_{2} j_{2}},  \label{eq:13:1:36}
\end{multline}
\begin{multline} 
\braOket{ n_{1}^{\prime} j_{1}^{\prime} n_{2}^{\prime} j_{2}^{\prime} j^{\prime} m^{\prime}}{\bigl\{\bigl\{\bigl\{\widehat{\vect{P}}_{a}(1) \otimes \widehat{\vect{Q}}_{b}(2)\bigr\}_{c} \otimes \widehat{\vect{R}}_{d}(1)\bigr\}_{f} \otimes \widehat{\vect{S}}_{e}(2)\bigr\}_{k \kappa}}{n_{1} j_{1} n_{2} j_{2} j m} \\
 =(-1)^{c+d-e-j-j^{\prime}} \Fact{c f j k} 
\clebsch{j}{m}{k}{\kappa}{j^{\prime}}{m^{\prime}} \sum_{g h} \Fact{g h}^{2}
\sixj{a}{d}{h}{f}{b}{c}
\sixj{b}{e}{g}{k}{h}{f}
\ninej{j_{1}}{j_{1}^{\prime}}{h}{j_{2}}{j_{2}^{\prime}}{g}{j}{j^{\prime}}{k} \sum_{N_{1} J_{1} N_{2} J_{2}}
\sixj{d}{a}{h}{j_{1}^{\prime}}{j_{1}}{J_{1}}
\sixj{e}{b}{g}{j_{2}^{\prime}}{j_{2}}{J_{2}} \\
\braOketred{ n_{1}^{\prime} j_{1}^{\prime}}{\widehat{\vect{P}}_{a}(1)}{ N_{1} J_{1}}
\braOketred{ N_{1} J_{1}}{\widehat{\vect{R}}_{d}(1)}{ n_{1} j_{1}}
\braOketred{ n_{2}^{\prime} j_{2}^{\prime}}{\widehat{\vect{Q}}_{b}(2)}{ N_{2} J_{2}}
\braOketred{ N_{2} J_{2}}{\widehat{\vect{S}}_{e}(2)}{ n_{2} j_{2}},  \label{eq:13:1:37}
\end{multline}
\begin{multline*}
\braOketred{ n_{1}^{\prime} j_{1}^{\prime} n_{2}^{\prime} j_{2}^{\prime} j^{\prime}}{\bigl\{\bigl\{\bigl\{\widehat{\vect{P}}_{a}(1) \otimes \widehat{\vect{Q}}_{b}(2)\bigr\}_{c} \otimes \widehat{\vect{R}}_{d}(1)\bigr\}_{f} \otimes \widehat{\vect{S}}_{e}(2)\bigr\}_{k}}{n_{1} j_{1} n_{2} j_{2} j}
\\=(-1)^{c+d-e-j-j^{\prime}} \Fact{c f j k j^{\prime}} \sum_{g h} \Fact{g h}^{2}
\sixj{a}{d}{h}{f}{b}{c}
\sixj{b}{e}{g}{k}{h}{f}
\ninej{j_{1}}{j_{1}^{\prime}}{h}{j_{2}}{j_{2}^{\prime}}{g}{j}{j^{\prime}}{k} \\
 \sum_{N_{1} J_{1} N_{2} J_{2}}
\sixj{d}{a}{h}{j_{1}^{\prime}}{j_{1}}{J_{1}}
\sixj{e}{b}{g}{j_{2}^{\prime}}{j_{2}}{J_{2}}
\\
\times\braOketred{ n_{1}^{\prime} j_{1}^{\prime}}{\widehat{\vect{P}}_{a}(1)}{ N_{1} J_{1}}
\braOketred{ N_{1} J_{1}}{\widehat{\vect{R}}_{d}(1)}{ n_{1} j_{1}}
\braOketred{ n_{2}^{\prime} j_{2}^{\prime}}{\widehat{\vect{Q}}_{b}(2)}{ N_{2} J_{2}}
\braOketred{ N_{2} J_{2}}{\widehat{\vect{S}}_{e}(2)}{ n_{2} j_{2}},
\end{multline*}
\begin{multline} 
\braOket{ n_{1}^{\prime} j_{1}^{\prime} n_{2}^{\prime} j_{2}^{\prime} j^{\prime} m^{\prime}}{\bigl(\bigl\{\bigl\{\widehat{\vect{P}}_{a}(1) \otimes \widehat{\vect{Q}}_{b}(2)\bigr\}_{c} \otimes \widehat{\vect{R}}_{d}(1)\bigr\}_{f} \cdot \widehat{\vect{S}}_{f}(2)\bigr)}{n_{1} j_{1} n_{2} j_{2} j m} \\
 =\delta_{j^{\prime} j} \delta_{m^{\prime} m}(-1)^{c+d-b+f-j+j_{1}^{\prime}+j_{2}} \Fact{c f} \sum_{g} \Fact{g}^{2}
\sixj{a}{b}{c}{f}{d}{g}
\sixj{j_{1}^{\prime}}{j_{1}}{g}{j_{2}}{j_{2}^{\prime}}{j} \sum_{N_{1} J_{1} N_{2} J_{2}}
\sixj{a}{d}{g}{j_{1}}{j_{1}^{\prime}}{J_{1}}
\sixj{b}{f}{g}{j_{2}}{j_{2}^{\prime}}{J_{2}} \\
 \times
\braOketred{ n_{1}^{\prime} j_{1}^{\prime}}{\widehat{\vect{P}}_{a}(1)}{ N_{1} J_{1}}
\braOketred{ N_{1} J_{1}}{\widehat{\vect{R}}_{d}(1)}{ n_{1} j_{1}}
\braOketred{ n_{2}^{\prime} j_{2}^{\prime}}{\widehat{\vect{Q}}_{b}(2)}{ N_{2} J_{2}}
\braOketred{ N_{2} J_{2}}{\widehat{\vect{S}}_{f}(2)}{ n_{2} j_{2}} . \label{eq:13:1:38}
\end{multline}
\begin{multline*}
\braOketred{ n_{1}^{\prime} j_{1}^{\prime} n_{2}^{\prime} j_{2}^{\prime} j^{\prime}}{\bigl(\bigl\{\bigl\{\widehat{\vect{P}}_{a}(1) \otimes \widehat{\vect{Q}}_{b}(2)\bigr\}_{c} \otimes \widehat{\vect{R}}_{d}(1)\bigr\}_{f} \cdot \widehat{\vect{S}}_{f}(2)\bigr)}{n_{1} j_{1} n_{2} j_{2} j}\\
=\delta_{j^{\prime} j}(-1)^{c+d-b+f-j+j_{1}^{\prime}+j_{2}} \Fact{c f j} \sum_{g} \Fact{g}^{2}
\sixj{a}{b}{c}{f}{d}{g}
\sixj{j_{1}^{\prime}}{j_{1}}{g}{j_{2}}{j_{2}^{\prime}}{j} 
 \sum_{N_{1} J_{1} N_{2} J_{2}}
\sixj{a}{d}{g}{j_{1}}{j_{1}^{\prime}}{J_{1}}
\sixj{b}{f}{g}{j_{2}}{j_{2}^{\prime}}{J_{2}}
\\ \times
\braOketred{ n_{1}^{\prime} j_{1}^{\prime}}{\widehat{\vect{P}}_{a}(1)}{ N_{1} J_{1}}
\braOketred{ N_{1} J_{1}}{\widehat{\vect{R}}_{d}(1)}{ n_{1} j_{1}}
\braOketred{ n_{2}^{\prime} j_{2}^{\prime}}{\widehat{\vect{Q}}_{b}(2)}{ N_{2} J_{2}}
\braOketred{ N_{2} J_{2}}{\widehat{\vect{S}}_{f}(2)}{ n_{2} j_{2}} .
\end{multline*}

Matrix elements of irreducible tensor products with other coupling schemes of operators may be evaluated using the formulas for the recoupling of four irreducible tensors (see Eq.~\eqref{eq:3:3:3}-\eqref{eq:3:3:5}).

If the wave functions and operators depend on variables of three or more subsystems, the evaluation of matrix elements is reduced to the successive use of the above equations.

\subsection{Matrix Elements of Operators Which Depend on  Variables of One of the Subsystems}\label{sec:13.1.5}\index{matrix elements!for one subsystem}
Let \(\widehat{P}_{a \alpha}(1)\) be an operator which depends on variables of subsystem 1. Matrix elements of this operator are diagonal with respect to quantum numbers which characterize eigenstates of subsystem 2 .
\begin{equation}
\braOket{ n_{1}^{\prime} j_{1}^{\prime} m_{1}^{\prime} ; n_{2}^{\prime} j_{2}^{\prime} m_{2}^{\prime}}{ \widehat{P}_{a \alpha}(1)}{n_{1} j_{1} m_{1} ; n_{2} j_{2} m_{2}}=\delta_{n_{2}^{\prime} n_{2}} \delta_{j_{2}^{\prime} j_{2}} \delta_{m_{2}^{\prime} m_{2}}
\braOket{ n_{1}^{\prime} j_{1}^{\prime} m_{1}^{\prime}}{ \widehat{P}_{a \alpha}(1)}{n_{1} j_{1} m_{1}}. \label{eq:13:1:39}
\end{equation}
Further evaluation is reduced to the use of the foregoing relations.\\
If an eigenstate corresponds to some fixed resultant angular momentum \(\vect{j}=\vect{j}_{1}+\vect{j}_{2}\) and its projection \(m=m_{1}+m_{2}\), one has
\begin{equation}
\braOket{ n_{1}^{\prime} j_{1}^{\prime} n_{2}^{\prime} j_{2}^{\prime} j^{\prime} m^{\prime}}{ \widehat{P}_{a \alpha}(1)}{n_{1} j_{1} n_{2} j_{2} j m}=\delta_{j_{2}^{\prime} j_{2}} \delta_{n_{2}^{\prime} n_{2}}(-1)^{j+j_{1}^{\prime}+j_{2}-a} \Fact{j} 
\clebsch{j}{m}{a}{\alpha}{j^{\prime}}{m^{\prime}}
\sixj{j_{1}}{j_{2}}{j}{j^{\prime}}{a}{j_{1}^{\prime}}
\braOketred{ n_{1}^{\prime} j_{1}^{\prime}}{\widehat{\vect{P}}_{a}(1)}{ n_{1} j_{1}}.
\label{eq:13:1:40}
\end{equation}
\begin{equation*}
\braOketred{ n_{1}^{\prime} j_{1}^{\prime} n_{2}^{\prime} j_{2}^{\prime} j^{\prime}}{ \widehat{\vect{P}}_{a}(1)}{n_{1} j_{1} n_{2} j_{2} j}=\delta_{j_{2}^{\prime} j_{2}} \delta_{n_{2}^{\prime} n_{2}}(-1)^{j+j_{1}^{\prime}+j_{2}-a} \Fact{j j^{\prime}}
\sixj{j_{1}}{j_{2}}{j}{j^{\prime}}{a}{j_{1}^{\prime}}
\braOketred{ n_{1}^{\prime} j_{1}^{\prime}}{\widehat{\vect{P}}_{a}(1)}{ n_{1} j_{1}}.
\end{equation*}
A matrix element of the direct tensor product of two operators may be represented in the form
\begin{multline}
\braOket{ n_{1}^{\prime} j_{1}^{\prime} n_{2}^{\prime} j_{2}^{\prime} j^{\prime} m^{\prime}}{ \widehat{P}_{a \alpha}(1) \widehat{Q}_{b \beta}(1)}{n_{1} j_{1} n_{2} j_{2} j m}=\delta_{n_{2}^{\prime} n_{2}} \delta_{j_{2}^{\prime} j_{2}}(-1)^{j_{2}+j-j_{1}+2 a+2 b} \Fact{j} \sum_{c \gamma} \Fact{c} 
\clebsch{j}{m}{c}{\gamma}{j^{\prime}}{m^{\prime}} 
\clebsch{a}{\alpha}{b}{\beta}{c}{\gamma}
\sixj{j_{1}}{j_{2}}{j}{j^{\prime}}{c}{j_{1}^{\prime}} \\
\times \sum_{N J}
\sixj{a}{b}{c}{j_{1}}{j_{1}^{\prime}}{J}
\braOketred{ n_{1}^{\prime} j_{1}^{\prime}}{\widehat{\vect{P}}_{a}(1)}{ N J}
\braOketred{ N J}{\widehat{Q}_{b}(1)}{ n_{1} j_{1}}. \label{eq:13:1:41}
\end{multline}
For an irreducible tensor product one gets the expressions
\begin{multline}
\braOket{ n_{1}^{\prime} j_{1}^{\prime} n_{2}^{\prime} j_{2}^{\prime} j^{\prime} m^{\prime}}{\bigl\{\widehat{\vect{P}}_{a}(1) \otimes \widehat{\vect{Q}}_{b}(1)\bigr\}_{c \gamma}}{n_{1} j_{1} n_{2} j_{2} j m}=\delta_{n_{2}^{\prime} n_{2}} \delta_{j_{2}^{\prime} j_{2}}(-1)^{j_{2}+j-j_{1}+2 c} \Fact{j c} 
\clebsch{j}{m}{c}{\gamma}{j^{\prime}}{m^{\prime}}
\sixj{j_{1}}{j_{2}}{j}{j^{\prime}}{c}{j_{1}^{\prime}} \\
\times \sum_{N J}
\sixj{a}{b}{c}{j_{1}}{j_{1}^{\prime}}{J}
\braOketred{ n_{1}^{\prime} j_{1}^{\prime}}{\widehat{\vect{P}}_{a}(1)}{ N J}
\braOketred{ N J}{\widehat{\vect{Q}}_{b}(1)}{ n_{1} j_{1}},  \label{eq:13:1:42}
\end{multline}
\begin{multline*}
\braOketred{ n_{1}^{\prime} j_{1}^{\prime} n_{2}^{\prime} j_{2}^{\prime} j^{\prime}}{\bigl\{\widehat{\vect{P}}_{a}(1) \otimes \widehat{\vect{Q}}_{b}(1)\bigr\}_{c}}{n_{1} j_{1} n_{2} j_{2} j}=\delta_{n_{2}^{\prime} n_{2}} \delta_{j_{2}^{\prime} j_{2}}(-1)^{j_{2}+j-j_{1}} \Fact{j c j^{\prime}}
\sixj{j_{1}}{j_{2}}{j}{j^{\prime}}{c}{j_{1}^{\prime}} \\
\times \sum_{N J}
\sixj{a}{b}{c}{j_{1}}{j_{1}^{\prime}}{J}
\braOketred{ n_{1}^{\prime} j_{1}^{\prime}}{\widehat{\vect{P}}_{a}(1)}{ N J}
\braOketred{ N J}{\widehat{\vect{Q}}_{b}(1)}{ n_{1} j_{1}},
\end{multline*}
\begin{multline}
\braOket{ n_{1}^{\prime} j_{1}^{\prime} n_{2}^{\prime} j_{2}^{\prime} j^{\prime} m^{\prime}}{\bigl(\widehat{\vect{P}}_{a}(1) \cdot \widehat{\vect{Q}}_{a}(1)\bigr)}{n_{1} j_{1} n_{2} j_{2} j m}=\delta_{n_{2}^{\prime} n_{2}} \delta_{j_{2}^{\prime} j_{2}} \delta_{j_{1}^{\prime} j_{1}} \delta_{j^{\prime} j} \delta_{m^{\prime} m} \\
\times \frac{1}{\Fact{j_{1}}^{2}}(-1)^{-j_{1}} \sum_{J N}(-1)^{J}
\braOketred{n_{1}^{\prime} j_{1}}{\widehat{\vect{P}}_{a}(1)}{ N J}
\braOketred{ N J}{\widehat{\vect{Q}}_{a}(1)}{ n_{1} j_{1}}. \label{eq:13:1:43}
\end{multline}
\begin{multline*}
\braOketred{ n_{1}^{\prime} j_{1}^{\prime} n_{2}^{\prime} j_{2}^{\prime} j^{\prime}}{\bigl(\widehat{\vect{P}}_{a}(1) \cdot \widehat{\vect{Q}}_{a}(1)\bigr)}{n_{1} j_{1} n_{2} j_{2} j}=\delta_{n_{2}^{\prime} n_{2}} \delta_{j_{2}^{\prime} j_{2}} \delta_{j_{1}^{\prime} j_{1}} \delta_{j^{\prime} j} \frac{\Fact{j}}{\Fact{j_{1}}^{2}}(-1)^{-j_{1}} \\
\times \sum_{J N}(-1)^{J}
\braOketred{n_{1}^{\prime} j_{1}}{\widehat{\vect{P}}_{a}(1)}{ N J}
\braOketred{ N J}{\widehat{\vect{Q}}_{a}(1)}{ n_{1} j_{1}}.
\end{multline*}
In the case of the irreducible tensor product of three operators we obtain
\begin{multline}
\braOket{ n_{1}^{\prime} j_{1}^{\prime} n_{2}^{\prime} j_{2}^{\prime} j^{\prime} m^{\prime}}{\bigl\{\bigl\{\widehat{\vect{P}}_{a}(1) \otimes \widehat{\vect{Q}}_{b}(1)\bigr\}_{c} \otimes \widehat{\vect{R}}_{d}(1)\bigr\}_{f \varphi}}{n_{1} j_{1} n_{2} j_{2} j m} \\
=\delta_{n_{2}^{\prime} n_{2}} \delta_{j_{2}^{\prime} j_{2}}(-1)^{j+j_{1}-j_{1}^{\prime}+j_{2}+c} \Fact{j f c} 
\clebsch{j}{m}{f}{\varphi}{j^{\prime}}{m^{\prime}}
\sixj{j_{1}}{j_{2}}{j}{j^{\prime}}{f}{j_{1}^{\prime}} \times \sum_{N_{1} J_{1} N_{2} J_{2}}(-1)^{J_{2}}
\sixj{a}{b}{c}{J_{2}}{j_{1}^{\prime}}{J_{1}}
\sixj{d}{c}{f}{j_{1}^{\prime}}{j_{1}}{J_{2}} \\
\times
\braOketred{ n_{1}^{\prime} j_{1}^{\prime}}{\widehat{\vect{P}}_{a}(1)}{ N_{1} J_{1}}
\braOketred{ N_{1} J_{1}}{\widehat{\vect{Q}}_{b}(1)}{ N_{2} J_{2}}
\braOketred{ N_{2} J_{2}}{\widehat{\vect{R}}_{d}(1)}{ n_{1} j_{1}} , \label{eq:13:1:44}
\end{multline}
\begin{multline*}
\braOketred{ n_{1}^{\prime} j_{1}^{\prime} n_{2}^{\prime} j_{2}^{\prime} j^{\prime}}{\bigl\{\bigl\{\widehat{\vect{P}}_{a}(1) \otimes \widehat{\vect{Q}}_{b}(1)\bigr\}_{c} \otimes \widehat{\vect{R}}_{d}(1)\bigr\}_{f}}{n_{1} j_{1} n_{2} j_{2} j}
=\delta_{n_{2}^{\prime} n_{2}} \delta_{j_{2}^{\prime} j_{2}}(-1)^{j+j_{1}-j_{1}^{\prime}+j_{2}+c} \Fact{j f c j^{\prime}}
\sixj{j_{1}}{j_{2}}{j}{j^{\prime}}{f}{j_{1}^{\prime}} \\
\times \sum_{N_{1} J_{1} N_{2} J_{2}}(-1)^{J_{2}}
\sixj{a}{b}{c}{J_{2}}{j_{1}^{\prime}}{J_{1}}
\sixj{d}{c}{f}{j_{1}^{\prime}}{j_{1}}{J_{2}}
\braOketred{ n_{1}^{\prime} j_{1}^{\prime}}{\widehat{\vect{P}}_{a}(1)}{ N_{1} J_{1}}
\braOketred{ N_{1} J_{1}}{\widehat{\vect{Q}}_{b}(1)}{ N_{2} J_{2}}
\braOketred{ N_{2} J_{2}}{\widehat{\vect{R}}_{d}(1)}{ n_{1} j_{1}} ,
\end{multline*}
\begin{multline}
\braOket{ n_{1}^{\prime} j_{1}^{\prime} n_{2}^{\prime} j_{2}^{\prime} j^{\prime} m^{\prime}}{\bigl(\bigl\{\widehat{\vect{P}}_{a}(1) \otimes \widehat{\vect{Q}}_{b}(1)\bigr\}_{c} \cdot \widehat{\vect{R}}_{c}(1)\bigr)}{n_{1} j_{1} n_{2} j_{2} j m}=\delta_{n_{2}^{\prime} n_{2}} \delta_{j_{1}^{\prime} j_{1}} \delta_{j_{2}^{\prime} j_{2}} \delta_{j^{\prime} j} \delta_{m^{\prime} m}(-1)^{-c+2 j_{1}}\\
\times \frac{\Fact{c}}{\Fact{j_{1}}^{2}} \sum_{\substack{N_{1} N_{2} \\ J_{1} J_{2}}}
\sixj{a}{b}{c}{J_{2}}{j_{1}}{J_{1}}
\braOketred{ n_{1}^{\prime} j_{1}}{\widehat{\vect{P}}_{a}(1)}{ N_{1} J_{1}}
\braOketred{ N_{1} J_{1}}{\widehat{\vect{Q}}_{b}(1)}{ N_{2} J_{2}}
\braOketred{ N_{2} J_{2}}{\widehat{\vect{R}}_{c}(1)}{ n_{1} j_{1}}. \label{eq:13:1:45}
\end{multline}
\begin{multline*}
\braOketred{ n_{1}^{\prime} j_{1}^{\prime} n_{2}^{\prime} j_{2}^{\prime} j^{\prime}}{\bigl(\bigl\{\widehat{\vect{P}}_{a}(1) \otimes \widehat{\vect{Q}}_{b}(1)\bigr\}_{c} \cdot \widehat{\vect{R}}_{c}(1)\bigr)}{n_{1} j_{1} n_{2} j_{2} j}=\delta_{n_{2}^{\prime} n_{2}} \delta_{j_{1}^{\prime} j_{1}} \delta_{j_{2}^{\prime} j_{2}} \delta_{j^{\prime} j}(-1)^{-c+2 j_{1}}
\frac{\Fact{c j}}{\Fact{j_{1}}^{2}} \\
\times \sum_{\substack{N_{1} N_{2} \\ J_{1} J_{2}}}
\sixj{a}{b}{c}{J_{2}}{j_{1}}{J_{1}}
\braOketred{ n_{1}^{\prime} j_{1}}{\widehat{\vect{P}}_{a}(1)}{ N_{1} J_{1}}
\braOketred{ N_{1} J_{1}}{\widehat{\vect{Q}}_{b}(1)}{ N_{2} J_{2}}
\braOketred{ N_{2} J_{2}}{\widehat{\vect{R}}_{c}(1)}{ n_{1} j_{1}}.
\end{multline*}
The matrix elements of irreducible tensor products of four operators may be evaluated using the expressions given below:
\begin{multline}
\braOket{ n_{1}^{\prime} j_{1}^{\prime} n_{2}^{\prime} j_{2}^{\prime} j^{\prime} m^{\prime}}{\bigl\{\bigl\{\widehat{\vect{P}}_{a}(1) \otimes \widehat{\vect{Q}}_{b}(1)\bigr\}_{c} \otimes\bigl\{\widehat{\vect{R}}_{d}(1) \otimes \widehat{\vect{S}}_{e}(1)\bigr\}_{f}\bigr\}_{k \kappa}}{n_{1} j_{1} n_{2} j_{2} j m} \\
=\delta_{n_{2}^{\prime} n_{2}} \delta_{j_{2}^{\prime} j_{2}}(-1)^{j+j_{1}^{\prime}+j_{2}-c-f} \Fact{c f k j} 
\clebsch{j}{m}{k}{\kappa}{j^{\prime}}{m^{\prime}}
\sixj{j_{1}}{j_{2}}{j}{j^{\prime}}{k}{j_{1}^{\prime}} \sum_{\substack{N_{1} N_{2} N_{3} \\ J_{1} J_{2} J_{3}}}(-1)^{2 J_{2}}
\sixj{a}{b}{c}{J_{2}}{j_{1}^{\prime}}{J_{1}}
\sixj{e}{d}{f}{J_{2}}{j_{1}}{J_{3}}
\sixj{f}{c}{k}{j_{1}^{\prime}}{j_{1}}{J_{2}} \\
\times
\braOketred{n_{1}^{\prime} j_{1}^{\prime}}{\widehat{\vect{P}}_{a}(1)}{ N_{1} J_{1}}
\braOketred{ N_{1} J_{1}}{\widehat{\vect{Q}}_{b}(1)}{ N_{2} J_{2}}
\braOketred{ N_{2} J_{2}}{\widehat{\vect{R}}_{d}(1)}{ N_{3} J_{3}}
\braOketred{ N_{3} J_{3}}{\widehat{\vect{S}}_{e}(1)}{ n_{1} j_{1}},  \label{eq:13:1:46}
\end{multline}
\begin{multline*}
\braOketred{ n_{1}^{\prime} j_{1}^{\prime} n_{2}^{\prime} j_{2}^{\prime} j^{\prime}}{\bigl\{\bigl\{\widehat{\vect{P}}_{a}(1) \otimes \widehat{\vect{Q}}_{b}(1)\bigr\}_{c} \otimes\bigl\{\widehat{\vect{R}}_{d}(1) \otimes \widehat{\vect{S}}_{e}(1)\bigr\}_{f}\bigr\}_{k}}{n_{1} j_{1} n_{2} j_{2} j}\\
=\delta_{n_{2}^{\prime} n_{2}} \delta_{j_{2}^{\prime} j_{2}}(-1)^{j+j_{1}^{\prime}+j_{2}-c-f} \Fact{c f k j j^{\prime}}
\sixj{j_{1}}{j_{2}}{j}{j^{\prime}}{k}{j_{1}^{\prime}}  \sum_{\substack{N_{1} N_{2} N_{3} \\ J_{1} J_{2} J_{3}}}(-1)^{2 J_{2}}
\sixj{a}{b}{c}{J_{2}}{j_{1}^{\prime}}{J_{1}}
\sixj{e}{d}{f}{J_{2}}{j_{1}}{J_{3}}
\sixj{f}{c}{k}{j_{1}^{\prime}}{j_{1}}{J_{2}}\\
\times
\braOketred{n_{1}^{\prime} j_{1}^{\prime}}{\widehat{\vect{P}}_{a}(1)}{ N_{1} J_{1}}
\braOketred{ N_{1} J_{1}}{\widehat{\vect{Q}}_{b}(1)}{ N_{2} J_{2}}
\braOketred{ N_{2} J_{2}}{\widehat{\vect{R}}_{d}(1)}{ N_{3} J_{3}}
\braOketred{ N_{3} J_{3}}{\widehat{\vect{S}}_{e}(1)}{ n_{1} j_{1}},
\end{multline*}
\begin{multline} 
\braOket{ n_{1}^{\prime} j_{1}^{\prime} n_{2}^{\prime} j_{2}^{\prime} j^{\prime} m^{\prime}}{\bigl(\bigl\{\widehat{\vect{P}}_{a}(1) \otimes \widehat{\vect{Q}}_{b}(1)\bigr\}_{c} \cdot\bigl\{\widehat{\vect{R}}_{d}(1) \otimes \widehat{\vect{S}}_{e}(1)\bigr\}_{c}\bigr)}{n_{1} j_{1} n_{2} j_{2} j m} \\
=\delta_{n_{2}^{\prime} n_{2}} \delta_{j_{1}^{\prime} j_{1}} \delta_{j_{2}^{\prime} j_{2}} \delta_{j^{\prime} j} \delta_{m^{\prime} m} \frac{(-1)^{-j_{1}}}{\Fact{j_{1}}^{2}} \sum_{\substack{N_{1} N_{2} N_{3} \\ J_{1} J_{2} J_{3}}}(-1)^{J_{2}}
\sixj{a}{b}{c}{J_{2}}{j_{1}}{J_{1}}
\sixj{e}{d}{c}{J_{2}}{j_{1}}{J_{3}} \\
\times \Fact{c}^{2}
\braOketred{ n_{1}^{\prime} j_{1}}{\widehat{\vect{P}}_{a}(1)}{ N_{1} J_{1}}
\braOketred{ N_{1} J_{1}}{\widehat{\vect{Q}}_{b}(1)}{ N_{2} J_{2}}
\braOketred{ N_{2} J_{2}}{\widehat{\vect{R}}_{d}(1)}{ N_{3} J_{3}}
\braOketred{ N_{3} J_{3}}{\widehat{\vect{S}}_{e}(1)}{ n_{1} j_{1}},  \label{eq:13:1:47}
\end{multline}
\begin{multline*}
\braOketred{ n_{1}^{\prime} j_{1}^{\prime} n_{2}^{\prime} j_{2}^{\prime} j^{\prime}}{\bigl(\bigl\{\widehat{\vect{P}}_{a}(1) \otimes \widehat{\vect{Q}}_{b}(1)\bigr\}_{c} \cdot\bigl\{\widehat{\vect{R}}_{d}(1) \otimes \widehat{\vect{S}}_{e}(1)\bigr\}_{c}\bigr)}{n_{1} j_{1} n_{2} j_{2} j} \\
=\delta_{n_{2}^{\prime} n_{2}} \delta_{j_{1}^{\prime} j_{1}} \delta_{j_{2}^{\prime} j_{2}} \delta_{j^{\prime} j} \frac{(-1)^{-j_{1}} \Fact{j}}{\Fact{j_{1}}^{2}} \sum_{\substack{N_{1} N_{2} N_{3} \\ J_{1} J_{2} J_{3}}}(-1)^{J_{2}}
\sixj{a}{b}{c}{J_{2}}{j_{1}}{J_{1}}
\sixj{e}{d}{c}{J_{2}}{j_{1}}{J_{3}} \\
\times \Fact{c}^{2}
\braOketred{ n_{1}^{\prime} j_{1}}{\widehat{\vect{P}}_{a}(1)}{ N_{1} J_{1}}
\braOketred{ N_{1} J_{1}}{\widehat{\vect{Q}}_{b}(1)}{ N_{2} J_{2}}
\braOketred{ N_{2} J_{2}}{\widehat{\vect{R}}_{d}(1)}{ N_{3} J_{3}}
\braOketred{ N_{3} J_{3}}{\widehat{\vect{S}}_{e}(1)}{ n_{1} j_{1}},
\end{multline*}
\begin{multline} 
\braOket{ n_{1}^{\prime} j_{1}^{\prime} n_{2}^{\prime} j_{2}^{\prime} j^{\prime} m^{\prime}}{\bigl\{\bigl\{\bigl\{\widehat{\vect{P}}_{a}(1) \otimes \widehat{\vect{Q}}_{b}(1)\bigr\}_{c} \otimes \widehat{\vect{R}}_{d}(1)\bigr\}_{f} \otimes \widehat{\vect{S}}_{e}(1)\bigr\}_{k \kappa}}{n_{1} j_{1} n_{2} j_{2} j m} \\
=\delta_{n_{2}^{\prime} n_{2}} \delta_{j_{2}^{\prime} j_{2}}(-1)^{j_{1}+j_{2}+j+j^{\prime}+c} \Fact{f c k j} 
\clebsch{j}{m}{k}{\kappa}{j^{\prime}}{m^{\prime}}
\sixj{j_{1}}{j_{2}}{j}{j^{\prime}}{k}{j_{1}^{\prime}}\sum_{\substack{N_{1} N_{2} N_{3} \\ J_{1} J_{2} J_{3}}}(-1)^{J_{2}+J_{3}}
\sixj{a}{b}{c}{J_{2}}{j_{1}^{\prime}}{J_{1}}
\sixj{d}{c}{f}{j_{1}^{\prime}}{J_{3}}{J_{2}}
\sixj{e}{f}{k}{j_{1}^{\prime}}{j_{1}}{J_{3}} \\
\times
\braOketred{ n_{1}^{\prime} j_{1}^{\prime}}{\widehat{\vect{P}}_{a}(1)}{ N_{1} J_{1}}
\braOketred{ N_{1} J_{1}}{\widehat{\vect{Q}}_{b}(1)}{ N_{2} J_{2}}
\braOketred{ N_{2} J_{2}}{\widehat{\vect{R}}_{d}(1)}{ N_{3} J_{3}}
\braOketred{N_{3} J_{3}}{\widehat{\vect{S}}_{e}(1)}{ n_{1} j_{1}}, \label{eq:13:1:48}
\end{multline}
\begin{multline*}
\braOketred{ n_{1}^{\prime} j_{1}^{\prime} n_{2}^{\prime} j_{2}^{\prime} j^{\prime}}{\bigl\{\bigl\{\bigl\{\widehat{\vect{P}}_{a}(1) \otimes \widehat{\vect{Q}}_{b}(1)\bigr\}_{c} \otimes \widehat{\vect{R}}_{d}(1)\bigr\}_{f} \otimes \widehat{\vect{S}}_{e}(1)\bigr\}_{k}}{n_{1} j_{1} n_{2} j_{2} j} \\
=\delta_{n_{2}^{\prime} n_{2}} \delta_{j_{2}^{\prime} j_{2}}(-1)^{j_{1}+j_{2}+j+j^{\prime}+c} \Fact{f c k j j^{\prime}}
\sixj{j_{1}}{j_{2}}{j}{j^{\prime}}{k}{j_{1}^{\prime}} \sum_{\substack{N_{1} N_{2} N_{3} \\ J_{1} J_{2} J_{3}}}(-1)^{J_{2}+J_{3}}
\sixj{a}{b}{c}{J_{2}}{j_{1}^{\prime}}{J_{1}}
\sixj{d}{c}{f}{j_{1}^{\prime}}{J_{3}}{J_{2}}
\sixj{e}{f}{k}{j_{1}^{\prime}}{j_{1}}{J_{3}} \\
\times
\braOketred{ n_{1}^{\prime} j_{1}^{\prime}}{\widehat{\vect{P}}_{a}(1)}{ N_{1} J_{1}}
\braOketred{ N_{1} J_{1}}{\widehat{\vect{Q}}_{b}(1)}{ N_{2} J_{2}}
\braOketred{ N_{2} J_{2}}{\widehat{\vect{R}}_{d}(1)}{ N_{3} J_{3}}
\braOketred{N_{3} J_{3}}{\widehat{\vect{S}}_{e}(1)}{ n_{1} j_{1}},
\end{multline*}
\begin{multline}
\braOket{ n_{1}^{\prime} j_{1}^{\prime} n_{2}^{\prime} j_{2}^{\prime} j^{\prime} m^{\prime}}{\bigl(\bigl\{\bigl\{\widehat{\vect{P}}_{a}(1) \otimes \widehat{\vect{Q}}_{b}(1)\bigr\}_{c} \otimes \widehat{\vect{R}}_{d}(1)\bigr\}_{f} \cdot \widehat{\vect{S}}_{f}(1)\bigr)}{n_{1} j_{1} n_{2} j_{2} j m} \\
=\delta_{n_{2}^{\prime} n_{2}} \delta_{j_{1}^{\prime} j_{1}} \delta_{j_{2}^{\prime} j_{2}} \delta_{j^{\prime} j} \delta_{m^{\prime} m}(-1)^{c-f-j_{1}} \frac{\Fact{f c}}{\Fact{j_{1}}^{2}} \sum_{\substack{N_{1} N_{2} N_{3} \\
J_{1} J_{2} J_{3}}}(-1)^{J_{2}}
\sixj{a}{b}{c}{J_{2}}{j_{1}}{J_{1}}
\sixj{d}{c}{f}{j_{1}}{J_{3}}{J_{2}} \\
\times
\braOketred{ n_{1}^{\prime} j_{1}}{\widehat{\vect{P}}_{a}(1)}{ N_{1} J_{1}}
\braOketred{ N_{1} J_{1}}{\widehat{\vect{Q}}_{b}(1)}{ N_{2} J_{2}}
\braOketred{ N_{2} J_{2}}{\widehat{\vect{R}}_{d}(1)}{ N_{3} J_{3}}
\braOketred{ N_{3} J_{3}}{\widehat{\vect{S}}_{f}(1)}{ n_{1} j_{1}}. \label{eq:13:1:49}
\end{multline}
\begin{multline*}
\braOketred{ n_{1}^{\prime} j_{1}^{\prime} n_{2}^{\prime} j_{2}^{\prime} j^{\prime}}{\bigl(\bigl\{\bigl\{\widehat{\vect{P}}_{a}(1) \otimes \widehat{\vect{Q}}_{b}(1)\bigr\}_{c} \otimes \widehat{\vect{R}}_{d}(1)\bigr\}_{f} \cdot \widehat{\vect{S}}_{f}(1)\bigr)}{n_{1} j_{1} n_{2} j_{2} j}\\
=\delta_{n_{2}^{\prime} n_{2}} \delta_{j_{1}^{\prime} j_{1}} \delta_{j_{2}^{\prime} j_{2}} \delta_{j^{\prime} j}(-1)^{c-f-j_{1}} \frac{\Fact{f c j}}{\Fact{j_{1}}^{2}} \sum_{\substack{N_{1} N_{2} N_{3} \\
J_{1} J_{2} J_{3}}}(-1)^{J_{2}}
\sixj{a}{b}{c}{J_{2}}{j_{1}}{J_{1}}
\sixj{d}{c}{f}{j_{1}}{J_{3}}{J_{2}} \\
\times
\braOketred{ n_{1}^{\prime} j_{1}}{\widehat{\vect{P}}_{a}(1)}{ N_{1} J_{1}}
\braOketred{ N_{1} J_{1}}{\widehat{\vect{Q}}_{b}(1)}{ N_{2} J_{2}}
\braOketred{ N_{2} J_{2}}{\widehat{\vect{R}}_{d}(1)}{ N_{3} J_{3}}
\braOketred{ N_{3} J_{3}}{\widehat{\vect{S}}_{f}(1)}{ n_{1} j_{1}}.
\end{multline*}
Matrix elements for tensor products with other coupling schemes of operators may be obtained from the foregoing equations by the use of the recoupling rules and the commutation relations for irreducible tensors (Chap.~\ref{chap:3}).

\section{Matrix elements of basic tensor operators}\label{sec:13.2}\index{matrix elements!of the basic tensor operators}
\subsection{Some Introductory Remarks}\label{sec:13.2.1}\index{matrix elements!introductory remarks}\index{wave function}\index{spin function}\isymg{chi-Sm@$\chi_{S m}$, spin function}\isym{Ylm@$Y_{l m}(\vartheta, \varphi)$, spherical harmonic}
In this section we shall consider matrix elements of the following operators and their tensor products:
\begin{enumerate}[label=(\alph*)]
\item the unit operator \(\widehat{\vect{I}}\);
\item the unit vector operator \(\widehat{\vect{n}} \equiv \widehat{\vect{n}}_{1}\);
\item the gradient operators \(\delop \equiv \delop_{1}\) and \(\delop_{\Omega} \equiv\bigl(\delop_{\Omega}\bigr)_{1}\);
\item the total angular momentum operator \(\widehat{\vect{J}} \equiv \widehat{\vect{J}}_{1}\);
\item the orbital angular momentum operator \(\widehat{\vect{L}} \equiv \widehat{\vect{L}}_{1}\);
\item the spin angular momentum operator \(\widehat{\vect{S}} \equiv \widehat{\vect{S}}_{1}\);
\item the spherical harmonic operator \(Y_{l m}(\vartheta, \varphi)\).
\end{enumerate}
All matrix elements may be evaluated from the equations of Sec.~\ref{sec:13.1}.\\
A wave function which describes an eigenstate of some quantum system will generally depend on the position \((r, \vartheta, \varphi)\) as well as on the spin variables (\(\xi\)). It may be defined by
\begin{equation}
\Phi(r, \vartheta, \varphi ; \xi)=\Psi_{n}(r) Y_{l m}(\vartheta, \varphi) \chi_{s m_{s}}(\xi)=
\braket{r, \vartheta, \varphi ; \xi}{n, l, m ; s m_{s}}. \label{eq:13:2:1}
\end{equation}
where \(l\) and \(m\) are the orbital momentum and its projection; \(s\) and \(m_{s}\) are the spin momentum and its projection. The symbol \(n\) has been defined previously in Sec.~\ref{sec:13.1.1}.

The wave functions \eqref{eq:13:2:1} determine the $nlmsm_s$-representation of operators.\\
If the angular momenta \(\widehat{\vect{L}}\) and \(\widehat{\vect{S}}\) are coupled into the total angular momentum \(\widehat{\vect{J}}\) with its projection \(M\), the wave function which describes an eigenstate of the quantum system is defined as
\begin{equation}
\Phi(r, \vartheta, \varphi ; \xi)=\Psi_{n}(r) \sum_{m m_{s}}
\clebsch{l}{m}{s}{m_{s}}{J}{M} Y_{l m}(\vartheta, \varphi) \chi_{s m_{s}}(\xi)=\Psi_{n}(r) Y_{J M}^{l s}(\vartheta, \varphi)=
\braket{r, \vartheta, \varphi ; \xi}{ n, l, s, J, M}. \label{eq:13:2:2}
\end{equation}
These wave functions determine the \(n l s J M\)-representation of operators.

If an operator acts only on position variables, the evaluation of matrix elements in the \(nlmsm_{s}\)-representation is reduced to calculating matrix elements in the \(nlm\)-representation.
\begin{equation}
\braOket{ n^{\prime} l^{\prime} m^{\prime} ; s^{\prime} m_{s}^{\prime}}{ \widehat{\mathfrak{M}}_{k \kappa}(r, \vartheta, \varphi)}{n l m ; s m_{s}}=\delta_{s s^{\prime}} \delta_{m_{s} m_{s}^{\prime}}
\braOket{ n^{\prime} l^{\prime} m^{\prime}}{ \widehat{\mathfrak{M}}_{k \kappa}(r, \vartheta, \varphi)}{n l m} . \label{eq:13:2:3}
\end{equation}
If an operator depends only on spin variables, we get a similar relationship
\begin{equation}
\braOket{ n^{\prime} l^{\prime} m^{\prime} ; s^{\prime} m_{s}^{\prime}}{ \widehat{\mathfrak{N}}_{k \kappa}(\xi)}{n l m ; s m_{s}}=\delta_{l l^{\prime}} \delta_{m m^{\prime}}
\braOket{ n^{\prime} s^{\prime} m_{s}^{\prime}}{ \widehat{\mathfrak{N}}_{k \kappa}(\xi)}{n s m_{s}}. \label{eq:13:2:4}
\end{equation}
Hereafter matrix elements of operators which depend only on position variables or only on spin variables will be evaluated in the \(n l m\)  or the \(n s m_{s}\) representation, respectively.

To determine matrix elements of operators \(\widehat{\mathfrak{M}}_{k \kappa}(r, \vartheta, \varphi)\) and \(\widehat{\mathfrak{N}}_{k \kappa}(\xi)\) in the \(nlsJM\) representation one should use the equations
\begin{align}
& 
\braOketred{ n^{\prime} l^{\prime} s^{\prime} J^{\prime}}{\widehat{\mathfrak{M}}_{k}}{ n l s J}=\delta_{s s^{\prime}}(-1)^{J+l^{\prime}+s+k} \Fact{J J^{\prime}}
\sixj{l}{s}{J}{J^{\prime}}{k}{l^{\prime}}
\braOketred{ n^{\prime} l^{\prime}}{\mathfrak{M}_{k}}{ n l},  \label{eq:13:2:5}\\
& 
\braOketred{ n^{\prime} l^{\prime} s^{\prime} J^{\prime}}{\widehat{\mathfrak{N}}_{k}}{ n l s J}=\delta_{l l^{\prime}}(-1)^{J^{\prime}+l+s+k} \Fact{J J^{\prime}}
\sixj{s}{l}{J}{J^{\prime}}{k}{s^{\prime}}
\braOketred{ n^{\prime} s^{\prime}}{\widehat{\mathfrak{N}}_{k}}{ n s}. \label{eq:13:2:6}
\end{align}
If some operator \(\widehat{R}_{a \alpha}(r, \vartheta, \varphi ; \xi)\) acts on position as well as on spin variables, its matrix elements will be given in the $nlsJM$-representation. These matrix elements may be evaluated in the \(nlmsm_{s}\) representation with aid of
\begin{equation}
\braOket{ n^{\prime} l^{\prime} m^{\prime} ; s^{\prime} m_{s}^{\prime}}{ \widehat{R}_{a \alpha}(r, \vartheta, \varphi ; \xi)}{n l m ; s m_{s}}=\sum_{\substack{J M \\ J^{\prime} M^{\prime}}}
\clebsch{l}{m}{s}{m_{s}}{J}{M}
\clebsch{l^{\prime}}{m^{\prime}}{s^{\prime}}{m_{s}^{\prime}}{J^{\prime}}{M^{\prime}}
\braOket{n^{\prime} l^{\prime} s^{\prime} J^{\prime} M^{\prime}}{\widehat{R}_{a \alpha}(r, \vartheta, \varphi ; \xi)}{n l s J M}. \label{eq:13:2:7}
\end{equation}
or from the corresponding equations of the section above.\\
From now on, we will omit the arguments of operators in the expressions for matrix elements to simplify the equations. In addition, the index \(n\) will also be dropped in all the cases for which it is not essential.

\subsection{Matrix Elements of the Unit Operator \(\widehat{\vect{I}}\)}\label{sec:13.2.2}\index{matrix elements!of the unit operator}\index{unit operator}
The unit operator \(\widehat{\vect{I}}\) is an irreducible tensor operator of rank zero. It does not act on the variables \(r, \vartheta, \varphi, \xi\). Hence, in any representation its diagonal matrix elements are determined by
\begin{equation}
\braOket{\lambda^{\prime}}{ \widehat{\vect{I}}}{\lambda}=
\braket{\lambda^{\prime}}{\lambda}=\delta_{\lambda \lambda^{\prime}}. \label{eq:13:2:8}
\end{equation}
All non-diagonal matrix elements vanish. A reduced matrix element of the operator \(\widehat{\vect{I}}\) may be represented in the form
\begin{equation}
\braOketred{l}{\widehat{\vect{I}}}{l}=\Fact{l}, \quad\braOketred{s}{\widehat{\vect{I}}}{s}=\Fact{s}, \quad
\braOketred{ l^{\prime} s^{\prime} J}{\widehat{\vect{I}}}{ l s J}=\Fact{J} \delta_{l l^{\prime}} \delta_{s s^{\prime}}. \label{eq:13:2:9}
\end{equation}
\subsection{Matrix Elements of the Unit Vector \(\widehat{\vect{n}}(\vartheta, \varphi) \equiv \widehat{\vect{n}}_{1}(\vartheta, \varphi)\)}\label{sec:13.2.3}\index{matrix elements!of the unit vector operator}\index{unit vector operator}\isym{n-hat@$\widehat{\vect{n}}(\vartheta, \varphi)$, unit vector operator}
The unit vector operator \(\widehat{\vect{n}}_{1}(\vartheta, \varphi)\) acts only on variables \(\vartheta, \varphi\). In the (lm)-representation matrix elements of spherical components of this operator are given by
\begin{equation}
\braOket{ l^{\prime} m^{\prime}}{ \widehat{n}_{1 \mu}}{l m}=
\frac{
\braOketred{ l^{\prime}}{\widehat{n}_{1}}{ l}}{\sqrt{2 l^{\prime}+1}} 
\clebsch{l}{m}{1}{\mu}{l^{\prime}}{m^{\prime}}, \label{eq:13:2:10}
\end{equation}
where
\begin{equation}
\braOketred{ l^{\prime}}{\widehat{\vect{n}}_{1}}{ l}=\Fact{l} 
\clebsch{l}{0}{1}{0}{l^{\prime}}{0}. \label{eq:13:2:11}
\end{equation}
Taking into account Eq. \eqref{eq:13:2:5}, we obtain a reduced matrix element in the \(lsJ\)-representation
\begin{equation}
\braOketred{ l^{\prime} s^{\prime} J^{\prime}}{\widehat{\vect{n}}_{1}}{ l s J}=\delta_{s s^{\prime}}(-1)^{s+J+l^{\prime}+1} \Fact{J l}\sqrt{(2 J^{\prime}+1)}
\clebsch{l}{0}{1}{0}{l^{\prime}}{0}
\sixj{l}{s}{J}{J^{\prime}}{1}{l^{\prime}}. \label{eq:13:2:12}
\end{equation}
The matrix element \eqref{eq:13:2:10} vanishes unless \(l^{\prime}=l \pm 1\). Using Eq. \eqref{eq:13:2:11} one can rewrite Eq. \eqref{eq:13:2:10} in component form as
\begin{align}
\braOket{ l+1 m^{\prime}}{ \widehat{n}_{1 \pm 1}}{l m} & =\sqrt{\frac{(l \pm m+1)(l \pm m+2)}{2(2 l+1)(2 l+3)}} \delta_{m^{\prime} m \pm 1}, \notag\\
\braOket{ l+1 m^{\prime}}{ \widehat{n}_{10}}{l m} & =\sqrt{\frac{(l-m+1)(l+m+1)}{(2 l+1)(2 l+3)}} \delta_{m^{\prime} m},  \label{eq:13:2:13}\\
\braOket{ l-1 m^{\prime}}{ \widehat{n}_{1 \pm 1}}{l m} & =-\sqrt{\frac{(l \mp m-1)(l \mp m)}{2(2 l+1)(2 l-1)}} \delta_{m^{\prime} m \pm 1}, \notag \\
\braOket{ l-1 m^{\prime}}{ \widehat{n}_{10}}{l m} & =\sqrt{\frac{(l-m)(l+m)}{(2 l+1)(2 l-1)}} \delta_{m^{\prime} m}. \notag
\end{align}
All other matrix elements of the spherical components of the operator in question are equal to zero.\\
The matrix elements of the cartesian components of the operator \(\widehat{\vect{n}}_{1}(\vartheta, \varphi)\) are given by
\begin{align}
\braOket{ l^{\prime} m \pm 1}{ \widehat{n}_{x}}{l m} & =\mp \frac{1}{2} \sqrt{\frac{(l \pm m+1)(l \pm m+2)}{(2 l+1)(2 l+3)}} \delta_{l^{\prime} l+1} \pm \frac{1}{2} \sqrt{\frac{(l \mp m-1)(l \mp m)}{(2 l+1)(2 l-1)}} \delta_{l^{\prime} l-1},  \label{eq:13:2:14}\\
\braOket{ l^{\prime} m \pm 1}{ \widehat{n}_{y}}{l m} & =\frac{i}{2} \sqrt{\frac{(l \pm m+1)(l \pm m+2)}{(2 l+1)(2 l+3)}} \delta_{l^{\prime} l+1}-\frac{i}{2} \sqrt{\frac{(l \mp m-1)(l \mp m)}{(2 l+1)(2 l-1)}} \delta_{l^{\prime} l-1},  \label{eq:13:2:15}\\
\braOket{ l^{\prime} m}{ \widehat{n}_{z}}{l m} & =\sqrt{\frac{(l-m+1)(l+m+1)}{(2 l+1)(2 l+3)}} \delta_{l^{\prime} l+1}+\sqrt{\frac{(l-m)(l+m)}{(2 l+1)(2 l-1)}} \delta_{l^{\prime} l-1}. \label{eq:13:2:16}
\end{align}
All other matrix elements of the cartesian components of \(\widehat{\vect{n}}_{1}(\vartheta, \varphi)\) vanish.\\
It is easy to obtain the matrix elements of direct products of two operators involving \(\widehat{\vect{n}}_{1}(\vartheta, \varphi)\) :
\begin{equation}
\braOket{ l^{\prime} m^{\prime}}{ \widehat{n}_{1 \mu} \widehat{n}_{1 \nu}}{l m}=\frac{(-1)^{\mu}}{3} \delta_{l l^{\prime}} \delta_{m m^{\prime}} \delta_{\mu,-\nu}+\sqrt{\frac{2}{3}} \frac{\Fact{l}}{\Fact{l^{\prime}}} 
\clebsch{l}{0}{2}{0}{l^{\prime}}{0} 
\clebsch{1}{\mu}{1}{\nu}{2}{\kappa} 
\clebsch{l}{m}{2}{\kappa}{l^{\prime}}{m^{\prime}}. \label{eq:13:2:17}
\end{equation}
These matrix elements vanish unless \(l^{\prime}=l, l \pm 2\). To derive the matrix elements of direct products of cartesian components of \(\widehat{\vect{n}}_{1}(\vartheta, \varphi)\) one may express these products in terms of the products of spherical components. In this way we get
\begin{align}
\braOket{ l^{\prime} m \pm 2}{ \widehat{n}_{x} \widehat{n}_{x}}{l m} & =\frac{1}{2}
\braOket{ l^{\prime} m \pm 2}{ \widehat{n}_{1 \pm 1} \widehat{n}_{1 \pm 1}}{l m}, \notag\\
\braOket{ l^{\prime} m}{ \widehat{n}_{x} \widehat{n}_{x}}{l m} & =-
\braOket{ l^{\prime} m}{ \widehat{n}_{1 \pm 1} \widehat{n}_{1 \mp 1}}{l m}, \notag\\
\braOket{ l^{\prime} m \pm 2}{ \widehat{n}_{y} \widehat{n}_{y}}{l m} & =-
\braOket{ l^{\prime} m \pm 2}{ \widehat{n}_{1 \pm 1} \widehat{n}_{1 \pm 1}}{l m}, \notag\\
\braOket{ l^{\prime} m}{ \widehat{n}_{y} \widehat{n}_{y}}{l m} & =-
\braOket{ l^{\prime} m}{ \widehat{n}_{1 \pm 1} \widehat{n}_{1 \mp 1}}{l m}, \notag \\
\braOket{ l^{\prime} m}{ \widehat{n}_{x} \widehat{n}_{z}}{l m} & =
\braOket{ l^{\prime} m}{ \widehat{n}_{10} \widehat{n}_{10}}{l m},  \label{eq:13:2:18}\\
\braOket{ l^{\prime} m \pm 2}{ \widehat{n}_{x} \widehat{n}_{y}}{l m} & =\mp \frac{i}{2}
\braOket{ l^{\prime} m \pm 2}{ \widehat{n}_{1 \pm 1} \widehat{n}_{1 \pm 1}}{l m}, \notag\\
\braOket{ l^{\prime} m \pm 1}{ \widehat{n}_{x} \widehat{n}_{z}}{l m} & =\mp \frac{1}{\sqrt{2}}
\braOket{ l^{\prime} m \pm 1}{ \widehat{n}_{1 \pm 1} \widehat{n}_{10}}{l m}, \notag\\
\braOket{ l^{\prime} m \pm 1}{ \widehat{n}_{y} \widehat{n}_{z}}{l m} & =\frac{i}{\sqrt{2}}
\braOket{ l^{\prime} m \pm 1}{ \widehat{n}_{1 \pm 1} \widehat{n}_{10}}{l m}. \notag
\end{align}
The matrix elements of the irreducible tensor product of \(k\) unit vector operators \(\widehat{\vect{n}}_{1}(\vartheta, \varphi)\) may be represented in the form
\begin{equation}
\braOket{ l^{\prime} m^{\prime}}{\bigl\{\ldots\bigl\{\bigl\{\widehat{\vect{n}}_{1} \otimes \widehat{\vect{n}}_{1}\bigr\}_{q_{2}} \otimes \widehat{\vect{n}}_{1}\bigr\}_{q_{3}} \otimes \ldots \otimes \widehat{\vect{n}}_{1}\bigr\}_{q_{k} \kappa_{k}}}{l m}=\frac{\Fact{l}}{\Fact{l^{\prime}}} 
\clebsch{l}{0}{q_{k}}{0}{l^{\prime}}{0} \prod_{i=2}^{k} 
\clebsch{q_{i-1}}{0}{1}{0}{q_{i}}{0} 
\clebsch{l}{m}{q_{k}}{\kappa_{k}}{l^{\prime}}{m^{\prime}}, \label{eq:13:2:19}
\end{equation}
\begin{equation*}
\braOketred{ l^{\prime}}{\bigl\{\ldots\bigl\{\bigl\{\widehat{\vect{n}}_{1} \otimes \widehat{\vect{n}}_{1}\bigr\}_{q_{2}} \otimes \widehat{\vect{n}}_{1}\bigr\}_{q_{3}} \otimes \ldots \otimes \widehat{\vect{n}}_{1}\bigr\}_{q_{k}}}{ l}=\Fact{l}
\clebsch{l}{0}{q_{k}}{0}{l^{\prime}}{0} \prod_{i=2}^{k}
\clebsch{q_{i-1}}{0}{1}{0}{q_{i}}{0},
\end{equation*}
where \(q_{1}=1\). In particular, if \(q_{2}=2, q_{3}=3, q_{4}=4, \ldots\) one has
\begin{equation}
\braOket{ l^{\prime} m^{\prime}}{\bigl\{\ldots\bigl\{\bigl\{\widehat{\vect{n}}_{1} \otimes \widehat{\vect{n}}_{1}\bigr\}_{2} \otimes \widehat{\vect{n}}_{1}\bigr\}_{3} \otimes \ldots \otimes \widehat{\vect{n}}_{1}\bigr\}_{k \kappa}}{l m}=\sqrt{\frac{k!}{(2 k-1)!!}} \frac{\Fact{l}}{\Fact{l^{\prime}}}
\clebsch{l}{0}{k}{0}{l^{\prime}}{0} 
\clebsch{l}{m}{k}{\kappa}{l^{\prime}}{m^{\prime}}. \label{eq:13:2:20}
\end{equation}
\begin{equation*}
\braOketred{ l^{\prime}}{\bigl\{\ldots\bigl\{\bigl\{\widehat{\vect{n}}_{1} \otimes \widehat{\vect{n}}_{1}\bigr\}_{2} \otimes \widehat{\vect{n}}_{1}\bigr\}_{3} \otimes \ldots \otimes \widehat{\vect{n}}_{1}\bigr\}_{k}}{ l}=\sqrt{\frac{k!}{(2 k-1)!!}}
/\Fact{l}
\clebsch{l}{0}{k}{0}{l^{\prime}}{0}.
\end{equation*}
From Eq. \eqref{eq:13:2:19} one obtains two obvious equalities,
\begin{gather*}
\braOket{ l^{\prime} m^{\prime}}{(\widehat{\vect{n}} \cdot \widehat{\vect{n}})}{l m}=\delta_{l l^{\prime}} \delta_{m m^{\prime}} ,\\
\braOket{ l^{\prime} m^{\prime}}{[\widehat{\vect{n}} \times \widehat{\vect{n}}]}{l m}=0.
\end{gather*}

\subsection{Matrix Elements of the Operator \(\delop(r, \vartheta, \varphi) \equiv \delop_{1}(r, \vartheta, \varphi)\)}\label{sec:13.2.4}\index{matrix elements!of the gradient operator}\index{gradient}\isymo{delop@$\delop$, gradient (del) operator}
The operator \(\delop_{1}(r, \vartheta, \varphi)\) acts only on variables \(r, \vartheta, \varphi\). The matrix elements of its spherical components are determined by
\begin{equation}
\braOket{ n^{\prime} l^{\prime} m^{\prime}}{ \delop_{1 \mu}}{n l m}=
\frac{
\braOketred{ n^{\prime} l^{\prime}}{\delop_{1}}{ n l}}{\sqrt{2 l^{\prime}+1}} 
\clebsch{l}{m}{1}{\mu}{l^{\prime}}{m^{\prime}}, \label{eq:13:2:21}
\end{equation}
where
\begin{equation}
\braOketred{ n^{\prime} l^{\prime}}{\delop_{1}}{ n l}=\sqrt{l+1} A_{n^{\prime} l^{\prime} n l} \delta_{l^{\prime} l+1}-\sqrt{l} B_{n^{\prime} l^{\prime} n l} \delta_{l^{\prime} l-1}. \label{eq:13:2:22}
\end{equation}
The quantities \(A_{n^{\prime} l^{\prime} n l}\) and \(B_{n^{\prime} l^{\prime} n l}\) in Eq. \eqref{eq:13:2:22} may be represented as the integrals
\begin{align}
& A_{n^{\prime} l^{\prime} n l}=\int_{0}^{\infty} \Psi_{n^{\prime} l^{\prime}}^{*}(r)\bigl(\frac{\partial}{\partial r}-\frac{l}{r}\bigr) \Psi_{n l}(r) r^{2} d r ,\notag\\
& B_{n^{\prime} l^{\prime} n l}=\int_{0}^{\infty} \Psi_{n^{\prime} l^{\prime}}^{*}(r)\bigl(\frac{\partial}{\partial r}+\frac{l+1}{r}\bigr) \Psi_{n l}(r) r^{2} d r. \label{eq:13:2:23}
\end{align}
Here \(\Psi_{n l}(r)\) is the radial part of the corresponding wave function.\\
Reduced matrix elements of the operator \(\delop_{1}(r, \vartheta, \varphi)\) in the $nlsJM$ representation may be obtained from Eq. \eqref{eq:13:2:5} by replacing \(\widehat{\mathfrak{M}}_{k} \rightarrow \delop_{1}\).

The matrix elements \eqref{eq:13:2:21} vanish unless \(l^{\prime}=l \pm 1\). In detailed form Eq. \eqref{eq:13:2:21} reads
\begin{align}
\braOket{ n^{\prime} l+1 m^{\prime}}{ \delop_{1 \pm 1}}{n l m} & =\sqrt{\frac{(l \pm m+1)(l \pm m+2)}{2(2 l+1)(2 l+3)}} A_{n^{\prime} l+1, n l} \delta_{m^{\prime} m \pm 1}, \notag\\
\braOket{ n^{\prime} l+1 m^{\prime}}{ \delop_{10}}{n l m} & =\sqrt{\frac{(l-m+1)(l+m+1)}{(2 l+1)(2 l+3)}} A_{n^{\prime} l+1, n l} \delta_{m^{\prime} m}, \notag\\
\braOket{ n^{\prime} l-1 m^{\prime}}{ \delop_{1 \pm 1}}{n l m} & =-\sqrt{\frac{(l \mp m-1)(l \mp m)}{2(2 l+1)(2 l-1)}} B_{n^{\prime} l-1, n l} \delta_{m^{\prime} m \pm 1},  \label{eq:13:2:24}\\
\braOket{ n^{\prime} l-1 m^{\prime}}{ \delop_{10}}{n l m} & =\sqrt{\frac{(l-m)(l+m)}{(2 l+1)(2 l-1)}} B_{n^{\prime} l-1, n l} \delta_{m^{\prime} m}. \notag
\end{align}
All other matrix elements equal zero. The matrix elements of the cartesian components of \(\delop_{1}(r, \vartheta, \varphi)\) are as\\
follows:
\begin{align}
\braOket{ n^{\prime} l^{\prime} m \pm 1}{ \delop_{x}}{n l m}= & \mp \frac{1}{2} \sqrt{\frac{(l \pm m+1)(l \pm m+2)}{(2 l+1)(2 l+3)}} A_{n^{\prime} l+1, n l} \delta_{l^{\prime} l+1},\notag \\
& \pm \frac{1}{2} \sqrt{\frac{(l \mp m-1)(l \mp m)}{(2 l+1)(2 l-1)}} B_{n^{\prime} l-1, n l} \delta_{l^{\prime} l-1}, \notag \\
\braOket{ n^{\prime} l^{\prime} m \pm 1}{ \delop_{y}}{n l m}= & \frac{i}{2} \sqrt{\frac{(l \pm m+1)(l \pm m+2)}{(2 l+1)(2 l+3)}} A_{n^{\prime} l+1, n l} \delta_{l^{\prime} l+1}, \notag\\
& -\frac{i}{2} \sqrt{\frac{(l \mp m-1)(l \mp m)}{(2 l+1)(2 l-1)}} B_{n^{\prime} l-1, n l} \delta_{l^{\prime} l-1}, \label{eq:13:2:25}\\
\braOket{ n^{\prime} l^{\prime} m}{ \delop_{z}}{n l m}= & \sqrt{\frac{(l-m+1)(l+m+1)}{(2 l+1)(2 l+3)}} A_{n^{\prime} l+1, n l} \delta_{l^{\prime} l+1}, \notag\\
& +\sqrt{\frac{(l-m)(l+m)}{(2 l+1)(2 l-1)}} B_{n^{\prime} l-1, n l} \delta_{l^{\prime} l-1}. \notag
\end{align}
All other matrix elements are equal zero.\\
The matrix elements of the angular part of \(\delop(r, \vartheta, \varphi)\) may be presented in a simpler form. According to Eq.~\eqref{eq:1:3:10},
\begin{equation*}
\delop_{\Omega}=-i[\widehat{\vect{n}} \times \widehat{\vect{L}}].
\end{equation*}
This operator is independent of \(r\). The use of the Wigner-Eckart theorem yields the following expression for the matrix elements of the spherical components of \(\bigl(\delop_{\Omega}\bigr)_{1}\),
\begin{equation}
\braOket{ l^{\prime} m^{\prime}}{\bigl(\delop_{\Omega}\bigr)_{1 \mu}}{l m}=\frac{
\braOketred{ l^{\prime}}{\bigl(\delop_{\Omega}\bigr)_{1}}{ l}}{\sqrt{2 l^{\prime}+1}} 
\clebsch{l}{m}{1}{\mu}{l^{\prime}}{m^{\prime}} \label{eq:13:2:26}
\end{equation}
where
\begin{equation}
\braOketred{ l^{\prime}}{\bigl(\delop_{\Omega}\bigr)_{1}}{ l}=-\bigl\{l \sqrt{l+1} \delta_{l^{\prime} l+1}+(l+1) \sqrt{l} \delta_{l^{\prime} l-1}\bigr\}. \label{eq:13:2:27}
\end{equation}
Equation \eqref{eq:13:2:27} may be obtained from \eqref{eq:13:2:22} by the formal substitution
\begin{equation}
A_{n^{\prime} l^{\prime}, n l} \rightarrow-l, \quad B_{n^{\prime} l^{\prime}, n l} \rightarrow l+1 . \label{eq:13:2:28}
\end{equation}
Similarly, Eqs. \eqref{eq:13:2:24} and \eqref{eq:13:2:25} become valid for \(\bigl(\delop_{\Omega}\bigr)_{1}\) after the same substitution \eqref{eq:13:2:28}.\\
Let us consider the matrix elements of products of two operators \(\bigl(\delop_{\Omega}\bigr)_{1}\). In the case of the direct product we obtain
\begin{align}
& 
\braOket{ l^{\prime} m^{\prime}}{\bigl(\delop_{\Omega}\bigr)_{1 \mu}\bigl(\delop_{\Omega}\bigr)_{1 \nu}}{l m}=\frac{(-1)^{1-\mu}}{3} l(l+1) \delta_{l l^{\prime}} \delta_{m m^{\prime}} \delta_{\mu,-\nu}+
\clebsch{1}{\mu}{1}{\nu}{2}{\kappa}\bigl\{\frac{l(l+1) \sqrt{(l+1)(l+2)}}{\sqrt{(2 l+3)(2 l+5)}} 
\clebsch{l}{m}{2}{\kappa}{l+2}{m+\kappa} \delta_{l^{\prime} l+2} \notag\\
& \quad+\frac{1}{2 l+1} \sqrt{\frac{l(l+1)}{6(2 l-1)(2 l+3)}}\left[4 l^{3}+6 l^{2}-4 l-3\right] 
\clebsch{l}{m}{2}{\kappa}{l}{m+\kappa} \delta_{l l^{\prime}}+\frac{l(l+1) \sqrt{l(l-1)}}{\sqrt{(2 l-1)(2 l-3)}} 
\clebsch{l}{m}{2}{\kappa}{l-2}{m+\kappa} \delta_{l^{\prime} l-2}\bigr\}. \label{eq:13:2:29}
\end{align}
The matrix elements of the product of cartesian components \(\bigl(\delop_{\Omega}\bigr)_{i}\bigl(\delop_{\Omega}\bigr)_{k}(i, k=x, y, z)\) are related to the matrix elements \eqref{eq:13:2:29} by means of Eqs. \eqref{eq:13:2:18} where one should make the replacement \(\widehat{n}_{i} \widehat{n}_{k} \rightarrow\bigl(\delop_{\Omega}\bigr)_{i}\bigl(\delop_{\Omega}\bigr)_{k}(i, k=x, y, z)\). In the case of the irreducible tensor product of two operators \(\bigl(\delop_{\Omega}\bigr)_{1}\) one has
\begin{gather}
\braOket{ l^{\prime} m^{\prime}}{\bigl\{\bigl(\delop_{\Omega}\bigr)_{1} \otimes\bigl(\delop_{\Omega}\bigr)_{1}\bigr\}_{k \kappa}}{l m} \notag\\
=(-1)^{k+l+l^{\prime}} \Fact{k}/\Fact{l^\prime}
\clebsch{l}{m}{k}{\kappa}{l^{\prime}}{m^{\prime}} \sum_{l^{\prime \prime}}
\sixj{1}{1}{k}{l^{\prime}}{l}{l^{\prime \prime}}
\braOketred{ l^{\prime}}{\bigl(\delop_{\Omega}\bigr)_{1}}{ l^{\prime \prime}}
\braOketred{ l^{\prime \prime}}{\bigl(\delop_{\Omega}\bigr)_{1}}{ l} . \label{eq:13:2:30}
\end{gather}
\begin{equation*}
\braOketred{ l^{\prime}}{\bigl\{\bigl(\delop_{\Omega}\bigr)_{1} \otimes\bigl(\delop_{\Omega}\bigr)_{1}\bigr\}_{k}}{ l}=(-1)^{k+l+l^{\prime}} \Fact{k} \sum_{l^{\prime \prime}}
\sixj{1}{1}{k}{l^{\prime}}{l}{l^{\prime \prime}}
\braOketred{ l^{\prime}}{\bigl(\delop_{\Omega}\bigr)_{1}}{ l^{\prime \prime}}
\braOketred{ l^{\prime \prime}}{\bigl(\delop_{\Omega}\bigr)_{1}}{ l} .
\end{equation*}
In particular, Eq. \eqref{eq:13:2:30} yields the well known expression for the matrix elements of scalar product of two operators \(\bigl(\delop_{\Omega}\bigr)_{1}\)
\begin{equation*}
\braOket{ l^{\prime} m^{\prime}}{\bigl(\bigl(\delop_{\Omega}\bigr)_{1} \cdot\bigl(\delop_{\Omega}\bigr)_{1}\bigr)}{l m}=-\delta_{l^{\prime} l} \delta_{m^{\prime} m} l(l+1)
\end{equation*}

Matrix elements of products of the operators \(\widehat{\vect{n}}_{1}\) and \(\bigl(\delop_{\Omega}\bigr)_{1}\) are of interest for application. For the matrix elements of the direct product of these operators one obtains
\begin{align}
& 
\braOket{ l^{\prime} m^{\prime}}{ \widehat{n}_{1 \mu}\bigl(\delop_{\Omega}\bigr)_{1 \nu}}{l m}=-\delta_{l^{\prime} l} \sqrt{\frac{l(l+1)}{2}}\bigl\{
\clebsch{1}{\mu}{1}{\nu}{1}{\kappa} 
\clebsch{l}{m}{1}{\kappa}{l}{m^{\prime}}+\sqrt{\frac{3}{(2 l-1)(2 l+3)}} 
\clebsch{1}{\mu}{1}{\nu}{2}{\kappa} 
\clebsch{l}{m}{2}{\kappa}{l}{m^{\prime}}\bigr\} \notag\\
& -\delta_{l^{\prime} l+2} l \sqrt{\frac{(l+1)(l+2)}{(2 l+3)(2 l+5)}} 
\clebsch{1}{\mu}{1}{\nu}{2}{\kappa} 
\clebsch{l}{m}{2}{\kappa}{l^{\prime}}{m^{\prime}}+\delta_{l^{\prime} l-2}(l+1) \sqrt{\frac{l(l-1)}{(2 l-1)(2 l-3)}}
\clebsch{1}{\mu}{1}{\nu}{2}{\kappa} 
\clebsch{l}{m}{2}{\kappa}{l^{\prime}}{m^{\prime}}. \label{eq:13:2:31}
\end{align}
The matrix element of the direct product \(\bigl(\delop_{\Omega}\bigr)_{1 \mu} \widehat{n}_{1 \nu}\) differs from the foregoing one by the factor
\begin{equation*}
\braOket{ l^{\prime} m^{\prime}}{ \mathfrak{R}_{1 \nu 1 \mu}\bigl(\widehat{n}_{1 \nu,}\bigl(\nabla_{\Omega}\bigr)_{1 \mu}\bigr)}{l m},
\end{equation*}
where \(\mathfrak{R}_{1 \nu 1 \mu}\) is the commutator of the operator components in question (see Sec.~\ref{sec:3.1.7})
\begin{equation}
\mathfrak{R}_{1 \nu 1 \mu}\bigl(\widehat{n}_{1 \nu},\bigl(\delop_{\Omega}\bigr)_{1 \mu}\bigr)=-\bigl(\delop_{\Omega}\bigr)_{1 \mu} \widehat{n}_{1 \nu}=(-1)^{1+\nu} \frac{2}{3} \delta_{\nu,-\mu}+2 \sqrt{\frac{2 \pi}{15}} 
\clebsch{1}{\mu}{1}{\nu}{2}{\mu+\nu} Y_{2 \mu+\nu}(\vartheta, \varphi) \label{eq:13:2:32}
\end{equation}
and
\begin{equation}
\braOket{l^{\prime} m^{\prime}}{\mathfrak{R}_{1 \mu 1 \nu}\bigl(\widehat{n}_{1 \nu},\bigl(\delop_{\Omega}\bigr)_{1 \mu}\bigr)}{l m}=(-1)^{1+\nu} \frac{2}{3} \delta_{\nu,-\mu} \delta_{l l^{\prime}} \delta_{m m^{\prime}}+\sqrt{\frac{2}{3}} \Fact{l}/\Fact{l'}
\clebsch{l}{0}{2}{0}{l^{\prime}}{0} 
\clebsch{1}{\mu}{1}{\nu}{2}{\mu+\nu} 
\clebsch{l}{m}{2}{\mu+\nu}{l^{\prime}}{m^{\prime}} \tag{13.2.32a}
\end{equation}
To determine the matrix elements of products of cartesian components of the operators \(\widehat{\vect{n}}_{1}\) and \(\bigl(\delop_{\Omega}\bigr)_{1}\) one can transform these products to the products of spherical components (see Eq.~\eqref{eq:1:2:11}) and use Eq. \eqref{eq:13:2:31}.

For the irreducible tensor product of \(\widehat{\vect{n}}_{1}(\vartheta, \varphi)\) and \(\bigl(\delop_{\Omega}\bigr)_{1}\) one obtains

\begin{equation}
\braOket{ l^{\prime} m^{\prime}}{\bigl\{\widehat{\vect{n}}_{1} \otimes\bigl(\delop_{\Omega}\bigr)_{1}\bigr\}_{k \kappa}}{l m}=(-1)^{l^{\prime}+l+k} 
\clebsch{l}{m}{k}{\kappa}{l^{\prime}}{m^{\prime}} \Fact{k}/\Fact{l^\prime} \sum_{l^{\prime \prime}} \Fact{l^{\prime \prime}}
\sixj{1}{l}{l^{\prime \prime}}{l^\prime}{1}{k} \quad
\clebsch{l^{\prime\prime}}{0}{1}{0}{l^{\prime}}{0}
\braOketred{ l^{\prime \prime}}{\bigl(\delop_{\Omega}\bigr)_{1}}{ l} .\label{eq:13:2:33}
\end{equation}
\begin{equation*}
\braOketred{ l^{\prime}}{\bigl\{\widehat{\vect{n}}_{1} \otimes\bigl(\delop_{\Omega}\bigr)_{1}\bigr\}_{k}}{ l}=(-1)^{l^{\prime}+l+k} \Fact{k} \sum_{l^{\prime \prime}} \Fact{l^{\prime \prime}}
\sixj{1}{l}{l^{\prime \prime}}{l^\prime}{1}{k}
\clebsch{l^{\prime\prime}}{0}{1}{0}{l^{\prime}}{0}
\braOketred{ l^{\prime \prime}}{\bigl(\delop_{\Omega}\bigr)_{1}}{ l} .
\end{equation*}
The reduced matrix element of \(\bigl(\delop_{\Omega}\bigr)_{1}\) in \eqref{eq:13:2:33} is given by Eq. \eqref{eq:13:2:27}.
\begin{equation}
\braOket{ l^{\prime} m^{\prime}}{\bigl\{\bigl(\delop_{\Omega}\bigr)_{1} \otimes \widehat{\vect{n}}_{1}\bigr\}_{k \kappa}}{l m}=(-1)^{k}
\braOket{ l^{\prime} m^{\prime}}{\bigl\{\widehat{\vect{n}}_{1} \otimes\bigl(\nabla_{\Omega}\bigr)_{1}\bigr\}_{k \kappa}}{l m}-(-1)^{k}
\braOket{ l^{\prime} m^{\prime}}{ \mathfrak{R}_{k \kappa}^{11}\bigl(\widehat{\vect{n}}_{1},\bigl(\delop_{\Omega}\bigr)_{1}\bigr)}{l m}. \label{eq:13:2:34}
\end{equation}
The second term in Eq. \eqref{eq:13:2:34} is the matrix element of the commutator of the irreducible tensor product in question,
\begin{equation}
\mathfrak{R}_{k \kappa}^{11}\bigl(\widehat{\vect{n}}_{1},\bigl(\delop_{\Omega}\bigr)_{1}\bigr)=\frac{2}{\sqrt{3}} \delta_{k 0}+2 \sqrt{\frac{2 \pi}{15}} Y_{2 \kappa}(\vartheta, \varphi) \delta_{k 2}. \label{eq:13:2:35}
\end{equation}
Using Eqs. \eqref{eq:13:2:104} and \eqref{eq:13:2:105}, we find
\begin{equation}
\braOket{ l^{\prime} m^{\prime}}{ \mathfrak{R}_{k \kappa}^{11}\bigl(\widehat{\vect{n}}_{1},\bigl(\delop_{\Omega}\bigr)_{1}\bigr)}{l m}=\frac{2}{\sqrt{3}} \delta_{k 0} \delta_{l l^{\prime}} \delta_{m m^{\prime}}+\sqrt{\frac{2}{3}} \Fact{l}/\Fact{l^\prime}
\clebsch{l}{0}{2}{0}{l^{\prime}}{0}
\clebsch{l}{m}{2}{\kappa}{l^{\prime}}{m^{\prime}} \delta_{k 2}. \label{eq:13:2:36}
\end{equation}
\subsection{Matrix Elements of the Total Angular Momentum Operator \(\widehat{\vect{J}} \equiv \widehat{\vect{J}}_{1}\)}\label{sec:13.2.5}\index{matrix elements!of the total angular momentum operator}\index{angular momentum operator!total}\isym{J-hat@$\widehat{\vect{J}}$, total angular momentum operator}
The operator of the total angular momentum \(\widehat{\vect{J}}=\widehat{\vect{L}}+\widehat{\vect{S}}\) acts on position as well as spin variables. Making use of the Wigner-Eckart theorem, we get
\begin{equation}
\braOket{ l^{\prime} s^{\prime} J^{\prime} M^{\prime}}{ \widehat{\vect{J}}_{1 \mu}}{l s J M}=\frac{
\braOketred{ l^{\prime} s^{\prime} J^{\prime}}{\widehat{\vect{J}}_{1}}{ l s J}}{\sqrt{2 J^{\prime}+1}} 
\clebsch{J}{M}{1}{\mu}{J^{\prime}}{M^{\prime}} \label{eq:13:2:37}
\end{equation}
where
\begin{equation}
\braOketred{ l^{\prime} s^{\prime} J^{\prime}}{\widehat{\vect{J}}_{1}}{ l s J}=\delta_{l l^{\prime}} \delta_{s s^{\prime}} \delta_{J J^{\prime}} \Fact{J}\sqrt{J(J+1)}. \label{eq:13:2:38}
\end{equation}
For simplicity in equations \eqref{eq:13:2:41}-\eqref{eq:13:2:47} we shall omit the quantum numbers \(l\) and \(s\) in notation for matrix elements. Thus,
\begin{equation}
\braOket{ J^{\prime} M^{\prime}}{ \widehat{J}_{1 \mu}}{J M}=\frac{
\braOketred{ J^{\prime}}{\widehat{\vect{J}}_{1}}{ J}}{\sqrt{2 J^{\prime}+1}} 
\clebsch{J}{M}{1}{\mu}{J^{\prime}}{M^{\prime}} \label{eq:13:2:39}
\end{equation}
where
\begin{equation}
\braOketred{ J^{\prime}}{\widehat{\vect{J}}_{1}}{ J}=\delta_{J J^{\prime}}\Fact{J} \sqrt{J(J+1)}. \label{eq:13:2:40}
\end{equation}
In expanded form
\begin{align}
\braOket{ J M \pm 1}{ \widehat{J}_{1 \pm 1}}{J M} & =\mp \sqrt{\frac{(J \pm M+1)(J \mp M)}{2}} , \label{eq:13:2:41}\\
\braOket{ J M}{ \widehat{J}_{10}}{J M} & =M. \notag
\end{align}
All other matrix elements are equal to zero. Let us present the non-vanishing matrix elements of the cartesian components of \(\widehat{\vect{J}}\) :
\begin{align}
\braOket{ J M \pm 1}{ \widehat{J}_{x}}{J M} & =\frac{1}{2} \sqrt{(J \pm M+1)(J \mp M)}, \notag \\
\braOket{ J M \pm 1}{ \widehat{J}_{y}}{J M} & =\mp \frac{i}{2} \sqrt{(J \pm M+1)(J \mp M)},  \label{eq:13:2:42}\\
\braOket{ J M}{ \widehat{J}_{z}}{J M} & =M. \notag
\end{align}
The matrix element of the operator \(\widehat{J}^{2}\) is as follows:
\begin{equation}
\braOket{ J^{\prime} M^{\prime}}{ \widehat{\vect{J}}^{2}}{J M}=\delta_{J J^{\prime}} \delta_{M M^{\prime}} J(J+1). \label{eq:13:2:43}
\end{equation}
Let us consider matrix elements of products of the \(\widehat{\vect{J}}_{1}\) components. Matrix elements of direct product of spherical components may be represented in the form
\begin{equation}
\braOket{ J^{\prime} M^{\prime}}{ \widehat{J}_{1 \mu} \widehat{J}_{1 \nu}}{J M}=J(J+1) 
\clebsch{J}{M^{\prime}-\mu}{1}{\mu}{J}{M^{\prime}} 
\clebsch{J}{M}{1}{\nu}{J}{M^{\prime}-\mu} \delta_{J J^{\prime}}. \label{eq:13:2:44}
\end{equation}
In component form
\begin{align}
\braOket{ J M \pm 2}{ \widehat{J}_{1 \pm 1} \widehat{J}_{1 \pm 1}}{J M} & =\frac{1}{2} \sqrt{(J \pm M+2)(J \pm M+1)(J \mp M-1)(J \mp M)}, \notag\\
\braOket{ J M \pm 1}{ \widehat{J}_{1 \pm 1} \widehat{J}_{10}}{J M} & =\mp M \sqrt{\frac{(J \pm M+1)(J \mp M)}{2}}, \notag\\
\braOket{ J M \pm 1}{ \widehat{J}_{10} \widehat{J}_{1 \pm 1}}{J M} & =\mp(M \pm 1) \sqrt{\frac{(J \pm M+1)(J \mp M)}{2}},  \label{eq:13:2:45}\\
\braOket{ J M}{ \widehat{J}_{1 \pm 1} \widehat{J}_{1 \mp 1}}{J M} & =-\frac{1}{2}(J \pm M)(J \mp M+1), \notag\\
\braOket{ J M}{ \widehat{J}_{10} \widehat{J}_{10}}{J M} & =M^{2}. \notag
\end{align}
The non-vanishing matrix elements of product of the cartesian components are as follows:
\begin{align}
\braOket{ J M \pm 2}{ \widehat{J}_{x} \widehat{J}_{x}}{J M} & =\frac{1}{4} \sqrt{(J \pm M+2)(J \pm M+1)(J \mp M-1)(J \mp M)}, \notag\\
\braOket{ J M}{ \widehat{J}_{x} \widehat{J}_{x}}{J M} & =\frac{1}{2}\left[J(J+1)-M^{2}\right], \notag\\
\braOket{ J M \pm 2}{ \widehat{J}_{y} \widehat{J}_{y}}{J M} & =-\frac{1}{4} \sqrt{(J \pm M+2)(J \pm M+1)(J \mp M-1)(J \mp M)}, \notag\\
\braOket{ J M}{ \widehat{J}_{y} \widehat{J}_{y}}{J M} & =\frac{1}{2}\left[J(J+1)-M^{2}\right], \notag\\
\braOket{ J M}{ \widehat{J}_{z} \widehat{J}_{z}}{J M} & =M^{2}, \notag\\
\braOket{ J M \pm 2}{ \widehat{J}_{x} \widehat{J}_{y}}{J M} & =\mp \frac{i}{4} \sqrt{(J \pm M+2)(J \pm M+1)(J \mp M-1)(J \mp M)}, \notag\\
\braOket{ J M}{ \widehat{J}_{x} \widehat{J}_{y}}{J M} & =\frac{i}{2} M  \label{eq:13:2:46},\\
\braOket{ J M \pm 2}{ \widehat{J}_{y} \widehat{J}_{x}}{J M} & =\mp \frac{i}{4} \sqrt{(J \pm M+2)(J \pm M+1)(J \mp M-1)(J \mp M)}, \notag\\
\braOket{ J M}{ \widehat{J}_{y} \widehat{J}_{x}}{J M} & =-\frac{i}{2} M, \notag\\
\braOket{ J M \pm 1}{ \widehat{J}_{x} \widehat{J}_{z}}{J M} & =\frac{M}{2} \sqrt{(J \pm M+1)(J \mp M)}, \notag\\
\braOket{ J M \pm 1}{ \widehat{J}_{z} \widehat{J}_{x}}{J M} & =\frac{M \pm 1}{2} \sqrt{(J \pm M+1)(J \mp M)}, \notag\\
\braOket{ J M \pm 1}{ \widehat{J}_{y} \widehat{J}_{z}}{J M} & =\mp \frac{i}{2} M \sqrt{(J \pm M+1)(J \mp M)}, \notag\\
\braOket{ J M \pm 1}{ \widehat{J}_{z} \widehat{J}_{y}}{J M} & =\mp \frac{i}{2}(M \pm 1) \sqrt{(J \pm M+1)(J \mp M)}. \notag
\end{align}
In the case of the irreducible tensor product of two operators \(\widehat{\vect{J}}\) one obtains
\begin{equation}
\braOket{ J^{\prime} M^{\prime}}{\bigl\{\widehat{\vect{J}}_{1} \otimes \widehat{\vect{J}}_{1}\bigr\}_{k \kappa}}{J M}=\delta_{J J^{\prime}}(-1)^{2 J+k} J(J+1) \Fact{k J}
\sixj{1}{1}{k}{J}{J}{J}
\clebsch{J}{M}{k}{\kappa}{J}{M^{\prime}}. \label{eq:13:2:47}
\end{equation}
\begin{equation*}
\braOketred{ J^{\prime}}{\bigl\{\widehat{\vect{J}}_{1} \otimes \widehat{\vect{J}}_{1}\bigr\}_{k}}{ J}=\delta_{J J^{\prime}}(-1)^{2 J+k} J(J+1) \Fact{k J J^{\prime}}
\sixj{1}{1}{k}{J}{J}{J}.
\end{equation*}
In particular, from this equation one has two obvious relations
\begin{align*}
\braOket{ J^{\prime} M^{\prime}}{ (\widehat{\vect{J}} \cdot \widehat{\vect{J}})}{J M} & =\delta_{J J^{\prime}} \delta_{M M^{\prime}} J(J+1), \\
\braOket{ J^{\prime} M^{\prime}}{[\widehat{\vect{J}} \times \widehat{\vect{J}}]_{\kappa}}{J M} & =\delta_{J J^{\prime}} i \sqrt{J(J+1)} \clebsch{J}{M}{1}{\kappa}{J}{M^{\prime}}.
\end{align*}

In applications, the expressions for matrix elements of products of the operator \(\widehat{\vect{J}}\) and the operators \(\widehat{\vect{n}}_{1}\) or \(\bigl(\delop_{\Omega}\bigr)_{1}\) are of particular interest. Some relations for the matrix elements in questions will be considered below. The matrix elements of the direct tensor product of operators \(\widehat{\vect{n}}_{1}\) and \(\widehat{\vect{J}}_{1}\) are given by
\begin{gather}
\braOket{ l^{\prime} s^{\prime} J^{\prime} M^{\prime}}{ \widehat{n}_{1 \mu} \widehat{J}_{1 \nu}}{l s J M}=\delta_{s s^{\prime}}(-1)^{s+J+l^{\prime}+1} \sqrt{J(J+1)}\Fact{l J}
\clebsch{l}{0}{1}{0}{l^{\prime}}{0}
\sixj{l}{s}{J}{J^{\prime}}{1}{l^{\prime}} 
\clebsch{J}{M+\nu}{1}{\mu}{J^{\prime}}{M^{\prime}} 
\clebsch{J}{M}{1}{\nu}{J}{M+\nu},  \label{eq:13:2:48}\\
\braOket{ l^{\prime} s^{\prime} J^{\prime} M^{\prime}}{ \widehat{J}_{1 \nu} \widehat{n}_{1 \mu}}{l s J M}=
\braOket{ l^{\prime} s^{\prime} J^{\prime} M^{\prime}}{ \widehat{n}_{1 \mu} \widehat{J}_{1 \nu}}{l s J M}-
\braOket{ l^{\prime} s^{\prime} J^{\prime} M^{\prime}}{ \mathfrak{R}_{1 \mu 1 \nu}\bigl(n_{1 \mu}, J_{1 \nu}\bigr)}{l s J M}. \label{eq:13:2:49}
\end{gather}
The commutator of components of these operators is as follows:
\begin{equation}
\mathfrak{R}_{1 \mu 1 \nu}\bigl(\widehat{n}_{1 \mu}, \widehat{J}_{1 \nu}\bigr)=-\sqrt{\frac{8 \pi}{3}} 
\clebsch{1}{\mu}{1}{\nu}{1}{\mu+\nu} Y_{1 \mu+\nu}(\vartheta, \varphi). \label{eq:13:2:50}
\end{equation}
Using Eqs. \eqref{eq:13:2:104} and \eqref{eq:13:2:105}, one finds

\begin{equation}
\braOket{ l^{\prime} s^{\prime} J^{\prime} M^{\prime}}{ \mathfrak{R}_{1 \mu 1 \nu}\bigl(\widehat{n}_{1 \mu}, \widehat{J}_{1 \nu}\bigr)}{l s J M}=(-1)^{J+l^{\prime}+s} \delta_{s s^{\prime}} \sqrt{2}\Fact{l J}
\clebsch{l}{0}{1}{0}{l^{\prime}}{0}
\sixj{l}{s}{J}{J^{\prime}}{1}{l^{\prime}} 
\clebsch{1}{\mu}{1}{\nu}{1}{\mu+\nu} 
\clebsch{J}{M}{1}{\mu+\nu}{J^{\prime}}{M^{\prime}}. \label{eq:13:2:51} 
\end{equation}

In the case of the irreducible tensor product of the operators \(\widehat{\vect{n}}_{1}\) and \(\widehat{\vect{J}}_{1}\) one obtains
\begin{multline}
\braOket{ l^{\prime} s^{\prime} J^{\prime} M^{\prime}}{\bigl\{\widehat{\vect{n}}_{1} \otimes \widehat{\vect{J}}_{1}\bigr\}_{k \kappa}}{l s J M}=  (-1)^{s+l^{\prime}-J^{\prime}+k+1} \delta_{s s^{\prime}} \sqrt{J(J+1)}\Fact{k l} \\
 \times \Fact{J J}
\clebsch{l}{0}{1}{0}{l^{\prime}}{0}
\sixj{1}{1}{k}{J^{\prime}}{J}{J}
\sixj{l}{s}{J}{J^{\prime}}{1}{l^{\prime}}
\clebsch{J}{M}{k}{\kappa}{J^{\prime}}{M^{\prime}}, \label{eq:13:2:52}
\end{multline}
\begin{multline*}
\braOketred{ l^{\prime} s^{\prime} J^{\prime}}{\bigl\{\widehat{\vect{n}}_{1} \otimes \widehat{\vect{J}}_{1}\bigr\}_{k}}{l s J}=(-1)^{s+l^{\prime}-J^{\prime}+k+1} \delta_{s s^{\prime}} \sqrt{J(J+1)}\Fact{k l J^{\prime}}\Fact{J J} 
\clebsch{l}{0}{1}{0}{l^{\prime}}{0}
\sixj{1}{1}{k}{J^{\prime}}{J}{J}
\sixj{l}{s}{J}{J^{\prime}}{1}{l^{\prime}} .
\end{multline*}
The commutator of the irreducible tensor product may be expressed as
\begin{equation}
\mathfrak{R}_{k \kappa}^{11}\bigl(\widehat{\vect{n}}_{1}, \widehat{\vect{J}}_{1}\bigr)=-\sqrt{\frac{8 \pi}{3}} Y_{1 \kappa}(\vartheta, \varphi) \delta_{k 1}. \label{eq:13:2:53}
\end{equation}
Taking Eqs. \eqref{eq:13:2:104} and \eqref{eq:13:2:105} into account, we get
\begin{equation}
\braOket{ l^{\prime} s^{\prime} J^{\prime} M^{\prime}}{ \mathfrak{R}_{k \kappa}^{11}\bigl(\widehat{\vect{n}}_{1}, \widehat{\vect{J}}_{1}\bigr)}{l s J M}=(-1)^{J+l^{\prime}+s} \delta_{s s^{\prime}} \sqrt{2}\Fact{J l}
\sixj{l}{s}{J}{J^{\prime}}{1}{l^{\prime}} 
\clebsch{l}{0}{1}{0}{l^{\prime}}{0} 
\clebsch{J}{M}{1}{\kappa}{J^{\prime}}{M^{\prime}} \delta_{k 1}. \label{eq:13:2:54}
\end{equation}
\begin{equation*}
\braOketred{ l^{\prime} s^{\prime} J^{\prime}}{ \mathfrak{R}_{k}^{11}\bigl(\widehat{\vect{n}}_{1}, \widehat{\vect{J}}_{1}\bigr)}{l s J}=(-1)^{J+l^{\prime}+s} \delta_{s s^{\prime}} \sqrt{2}\Fact{J l l^{\prime}}
\sixj{l}{s}{J}{J^{\prime}}{1}{l^{\prime}} 
\clebsch{l}{0}{1}{0}{l^{\prime}}{0} 
\delta_{k 1}.
\end{equation*}

The matrix elements of \(\bigl\{\widehat{\vect{J}}_{1} \otimes \widehat{\vect{n}}_{1}\bigr\}_{k \kappa}\) are determined by
\begin{multline}
\braOket{ l^{\prime} s^{\prime} J^{\prime} M^{\prime}}{\bigl\{\widehat{\vect{J}}_{1} \otimes \widehat{\vect{n}}_{1}\bigr\}_{k \kappa}}{l s J M}\\
=(-1)^{k}
\braOket{ l^{\prime} s^{\prime} J^{\prime} M^{\prime}}{\bigl\{\widehat{\vect{n}}_{1} \otimes \widehat{\vect{J}}_{1}\bigr\}_{k \kappa}}{l s J M}+(-1)^{k+1}
\braOket{ l^{\prime} s^{\prime} J^{\prime} M^{\prime}}{ \mathfrak{R}_{k \kappa}^{11}\bigl(\widehat{\vect{n}}_{1}, \widehat{\vect{J}}_{1}\bigr)}{l s J M} . \label{eq:13:2:55}
\end{multline}
This expression may also be rewritten in the form
\begin{multline}
\braOket{ l^{\prime} s^{\prime} J^{\prime} M^{\prime}}{\bigl\{\widehat{\vect{J}}_{1} \otimes \widehat{\vect{n}}_{1}\bigr\}_{k \kappa}}{l s J M}= (-1)^{s+l^{\prime}-J^{\prime}+k+1} \delta_{s s^{\prime}} \Fact{l k J^{\prime} J}\\
\times \sqrt{J^{\prime}\bigl(J^{\prime}+1\bigr)} 
\clebsch{l}{0}{1}{0}{l^{\prime}}{0}
\sixj{1}{1}{k}{J^{\prime}}{J}{J^{\prime}}
\sixj{l}{s}{J}{J^{\prime}}{1}{l^{\prime}}
\clebsch{J}{M}{k}{\kappa}{J^{\prime}}{M^{\prime}}, \label{eq:13:2:56}
\end{multline}
\begin{multline*}
\braOketred{ l^{\prime} s^{\prime} J^{\prime}}{\bigl\{\widehat{\vect{J}}_{1} \otimes \widehat{\vect{n}}_{1}\bigr\}_{k}}{l s J}=(-1)^{s+l^{\prime}-J^{\prime}+k+1} \delta_{s s^{\prime}} \Fact{l k J}(2J^{\prime}+1) \sqrt{J^{\prime}\bigl(J^{\prime}+1\bigr)} 
\clebsch{l}{0}{1}{0}{l^{\prime}}{0}
\sixj{1}{1}{k}{J^{\prime}}{J}{J^{\prime}}
\sixj{l}{s}{J}{J^{\prime}}{1}{l^{\prime}} .
\end{multline*}
Let us consider the matrix elements of products of the operators \(\bigl(\delop_{\Omega}\bigr)_{1}\) and \(\widehat{\vect{J}}_{1}\). In the case of the direct product of these operators we have
\begin{equation}
\braOket{ l^{\prime} s^{\prime} J^{\prime} M^{\prime}}{\bigl(\delop_{\Omega}\bigr)_{1 \mu} \widehat{J}_{1 \nu}}{l s J M}=(-1)^{J+l^{\prime}+s+1} \sqrt{J(J+1)}\Fact{J}
\sixj{J}{J^{\prime}}{1}{l^{\prime}}{l}{s}
\braOketred{ l^{\prime}}{\bigl(\delop_{\Omega}\bigr)_{1}}{ l} 
\clebsch{J}{M+\nu}{1}{\mu}{J^{\prime}}{M^{\prime}} 
\clebsch{J}{M}{1}{\nu}{J}{M+\nu}. \label{eq:13:2:57}
\end{equation}
As in the foregoing case,
\begin{equation}
\braOket{ l^{\prime} s^{\prime} J^{\prime} M^{\prime}}{ \widehat{J}_{1 \nu}\bigl(\delop_{\Omega}\bigr)_{1 \mu}}{l s J M}=
\braOket{ l^{\prime} s^{\prime} J^{\prime} M^{\prime}}{\bigl(\delop_{\Omega}\bigr)_{1 \mu} \widehat{J}_{1 \nu}}{l s J M}-
\braOket{ l^{\prime} s^{\prime} J^{\prime} M^{\prime}}{ \mathfrak{R}_{1 \mu 1 \nu}\bigl(\bigl(\delop_{\Omega}\bigr)_{1 \mu}, J_{1 \nu}\bigr)}{l s J M}. \label{eq:13:2:58}
\end{equation}
The commutator of components of the operators \(\bigl(\delop_{\Omega}\bigr)_{1}\) and \(\widehat{J}_{1}\) is
\begin{equation}
\mathfrak{R}_{1 \mu 1 \nu}\bigl(\bigl(\delop_{\Omega}\bigr)_{1 \mu}, \widehat{J}_{1 \nu}\bigr)=-\sqrt{2} 
\clebsch{1}{\mu}{1}{\nu}{1}{\lambda}\bigl(\delop_{\Omega}\bigr)_{\lambda}. \label{eq:13:2:59}
\end{equation}
Using Eq. \eqref{eq:13:2:26}, we find
\begin{equation}
\braOket{ l^{\prime} s^{\prime} J^{\prime} M^{\prime}}{ \mathfrak{R}_{1 \mu 1 \nu}\bigl(\bigl(\delop_{\Omega}\bigr)_{1 \mu}, J_{1 \nu}\bigr)}{l s J M}=(-1)^{J+l^{\prime}-s} \delta_{s s^{\prime}} \sqrt{2}\Fact{J}
\sixj{l}{s}{J}{J^{\prime}}{1}{l^{\prime}} 
\clebsch{1}{\mu}{1}{\nu}{1}{\mu+\nu} 
\clebsch{J}{M}{1}{\mu+\nu}{J^{\prime}}{M^{\prime}}
\braOketred{ l^{\prime}}{\bigl(\nabla_{\Omega}\bigr)_{1}}{ l}. \label{eq:13:2:60}
\end{equation}

In the case of the irreducible tensor product of these operators, \(\bigl\{\bigl(\nabla_{\Omega}\bigr)_{1} \otimes \widehat{J}_{1}\bigr\}_{k \kappa}\) one has
\begin{multline}
\braOket{ l^{\prime} s^{\prime} J^{\prime} M^{\prime}}{\bigl\{\bigl(\delop_{\Omega}\bigr)_{1} \otimes \widehat{\vect{J}}_{1}\bigr\}_{k \kappa}}{l s J M}=(-1)^{s+l^{\prime}-J^{\prime}+k+1} \delta_{s s^{\prime}} \sqrt{J(J+1)} \Fact{k} \\
 \times\Fact{J J}
\sixj{J}{J^{\prime}}{1}{l^{\prime}}{l}{s}
\sixj{1}{1}{k}{J^{\prime}}{J}{J}
\braOketred{ l^{\prime}}{\bigl(\nabla_{\Omega}\bigr)_{1}}{ l}
\clebsch{J}{M}{k}{\kappa}{J^{\prime}}{M^{\prime}}, \label{eq:13:2:61}
\end{multline}
\begin{multline*}
\braOketred{ l^{\prime} s^{\prime} J^{\prime}}{\bigl\{\bigl(\delop_{\Omega}\bigr)_{1} \otimes \widehat{\vect{J}}_{1}\bigr\}_{k}}{l s J}=(-1)^{s+l^{\prime}-J^{\prime}+k+1} \delta_{s s^{\prime}} \sqrt{J(J+1)} \Fact{k J^{\prime}}\Fact{J J} 
%\\\times
\sixj{J}{J^{\prime}}{1}{l^{\prime}}{l}{s}
\sixj{1}{1}{k}{J^{\prime}}{J}{J}
\braOketred{ l^{\prime}}{\bigl(\nabla_{\Omega}\bigr)_{1}}{ l} ,
\end{multline*}
while for the irreducible tensor product \(\bigl\{\widehat{\vect{J}}_{1} \otimes\bigl(\delop_{\Omega}\bigr)_{1}\bigr\}_{k \kappa}\) we have
\begin{multline}
\braOket{ l^{\prime} s^{\prime} J^{\prime} M^{\prime}}{\bigl\{\widehat{\vect{J}}_{1} \otimes\bigl(\delop_{\Omega}\bigr)_{1}\bigr\}_{k \kappa}}{l s J M}=  (-1)^{k}
\braOket{l^{\prime} s^{\prime} J^{\prime} M^{\prime}}{\bigl\{\bigl(\delop_{\Omega}\bigr)_{1} \otimes \widehat{\vect{J}}_{1}\bigr\}_{k \kappa}}{l s J M} \\
 +(-1)^{k+1}
\braOket{ l^{\prime} s^{\prime} J^{\prime} M^{\prime}}{ \mathfrak{R}_{k \kappa}^{11}\bigl(\bigl(\delop_{\Omega}\bigr)_{1}, \widehat{\vect{J}}_{1}\bigr)}{l s J M}. \label{eq:13:2:62}
\end{multline}
The commutator of \(\bigl\{\bigl(\delop_{\Omega}\bigr)_{1} \otimes \widehat{\vect{J}}_{1}\bigr\}_{k \kappa}\) is determined by
\begin{equation}
\mathfrak{R}_{k \kappa}^{11}\bigl(\bigl(\delop_{\Omega}\bigr)_{1}, \widehat{\vect{J}}_{1}\bigr)=-\sqrt{2}\bigl(\delop_{\Omega}\bigr)_{1 \kappa} \delta_{k 1}, \label{eq:13:2:63}
\end{equation}
i.e.,
\begin{equation}
\braOket{ l^{\prime} s^{\prime} J^{\prime} M^{\prime}}{ \mathfrak{R}_{k \kappa}^{11}\bigl(\bigl(\delop_{\Omega}\bigr)_{1}, \widehat{J}_{1}\bigr)}{l s J M}=(-1)^{J+l^{\prime}+s} \delta_{s s^{\prime}} \delta_{k 1} \sqrt{2}\Fact{J}
\sixj{l}{s}{J}{J^{\prime}}{1}{l^{\prime}}
\braOketred{ l^{\prime}}{\bigl(\delop_{\Omega}\bigr)_{1}}{ l} 
\clebsch{J}{M}{1}{\kappa}{J^{\prime}}{M^{\prime}}, \label{eq:13:2:64}
\end{equation}
\begin{equation*}
\braOketred{ l^{\prime} s^{\prime} J^{\prime}}{ \mathfrak{R}_{k }^{11}\bigl(\bigl(\delop_{\Omega}\bigr)_{1}, \widehat{J}_{1}\bigr)}{l s J}=(-1)^{J+l^{\prime}+s} \delta_{s s^{\prime}} \delta_{k 1} \sqrt{2}\Fact{J J^{\prime}}
\sixj{l}{s}{J}{J^{\prime}}{1}{l^{\prime}}
\braOketred{ l^{\prime}}{\bigl(\delop_{\Omega}\bigr)_{1}}{ l} 
\clebsch{J}{M}{1}{\kappa}{J^{\prime}}{M^{\prime}}. 
\end{equation*}

Equation \eqref{eq:13:2:62} may also be represented by
\begin{multline}
\braOket{ l^{\prime} s^{\prime} J^{\prime} M^{\prime}}{\bigl\{\widehat{\vect{J}}_{1} \otimes\bigl(\delop_{\Omega}\bigr)_{1}\bigr\}_{k \kappa}}{l s J M}=(-1)^{s+l^{\prime}-J^{\prime}+k+1} \delta_{s s^{\prime}} \Fact{k l J^{\prime}} \\
 \times \sqrt{J^{\prime}\bigl(J^{\prime}+1\bigr)}
\sixj{J}{J^{\prime}}{1}{l^{\prime}}{l}{s}
\sixj{1}{1}{k}{J^{\prime}}{J}{J^{\prime}}
\braOketred{ l^{\prime}}{\bigl(\nabla_{\Omega}\bigr)_{1}}{ l}
\clebsch{J}{M}{k}{\kappa}{J^{\prime}}{M^{\prime}}, \label{eq:13:2:65}
\end{multline}
\begin{multline*}
\braOketred{ l^{\prime} s^{\prime} J^{\prime}}{\bigl\{\widehat{\vect{J}}_{1} \otimes\bigl(\delop_{\Omega}\bigr)_{1}\bigr\}_{k}}{l s J}=(-1)^{s+l^{\prime}-J^{\prime}+k+1} \delta_{s s^{\prime}} \Fact{k l J^{\prime} J^{\prime}} \sqrt{J^{\prime}\bigl(J^{\prime}+1\bigr)}
% \\\times
\sixj{J}{J^{\prime}}{1}{l^{\prime}}{l}{s}
\sixj{1}{1}{k}{J^{\prime}}{J}{J^{\prime}}
\braOketred{ l^{\prime}}{\bigl(\nabla_{\Omega}\bigr)_{1}}{ l} .
\end{multline*}

\subsection{Matrix Elements of the Orbital Angular Momentum Operator \(\widehat{\vect{L}} \equiv \widehat{\vect{L}}_{1}\)}\label{sec:13.2.6}\index{matrix elements!of the orbital angular momentum operator}\index{angular momentum operator!orbital}\isym{L-hat@$\widehat{\vect{L}}$, orbital angular momentum operator}
This operator acts only on the position variables \(\theta, \varphi\). In accordance with the Wigner-Eckart theorem, we get
\begin{equation}
\braOket{ l^{\prime} m^{\prime}}{ \widehat{L}_{1 \mu}}{l m}=\frac{
\braOketred{ l^{\prime}}{\widehat{\vect{L}}_{1}}{ l}}{\Fact{l^{\prime}}} 
\clebsch{l}{m}{1}{\mu}{l^{\prime}}{m^{\prime}} \label{eq:13:2:66}
\end{equation}
where
\begin{equation}
\braOketred{ l^{\prime}}{\widehat{\vect{L}}_{1}}{ l}=\delta_{l l^\prime} \sqrt{l(l+1)}\Fact{l}. \label{eq:13:2:67}
\end{equation}
Equations \eqref{eq:13:2:39}-\eqref{eq:13:2:47} are obtained for the matrix elements of the total angular momentum operator \(\widehat{\vect{J}}\) and remain valid for the operator \(\widehat{\vect{L}}\) after the replacement,
\begin{equation*}
J \rightarrow l, \quad M \rightarrow m, \quad \widehat{J}_{1 \mu} \rightarrow \widehat{L}_{1 \mu}
\end{equation*}
A reduced matrix element in the \(lsJ\)-representation is given by
\begin{equation}
\braOketred{ l^{\prime} s^{\prime} J^{\prime}}{\widehat{\vect{L}}_{1}}{ l s J}=(-1)^{s+J+l+1} \delta_{s s^{\prime}} \delta_{l l^{\prime}} \Fact{J J^{\prime} l} \sqrt{l(l+1)}
\sixj{l}{s}{J}{J^{\prime}}{1}{l}.\label{eq:13:2:68}
\end{equation}

We present matrix elements of products of the operators \(\widehat{\vect{L}}_{1}\) and \(\vect{n}_{1}\). For the direct product of these operators we have
\begin{equation}
\braOket{ l^{\prime} m^{\prime}}{ \widehat{n}_{1 \mu} \widehat{L}_{1 \nu}}{l m}=\sqrt{l(l+1)}\Fact{l}/\Fact{l^{\prime}}
\clebsch{l}{0}{1}{0}{l^{\prime}}{0}
\clebsch{l}{m+\nu}{1}{\mu}{l^{\prime}}{m^{\prime}} 
\clebsch{l}{m}{1}{\nu}{l}{m+\nu}. \label{eq:13:2:69}
\end{equation}
The commutator of the components \(\widehat{n}_{1 \mu}\) and \(\widehat{L}_{1 \nu}\) may be written in the form
\begin{equation}
\mathfrak{R}_{1 \mu 1 \nu}\bigl(\widehat{n}_{1 \mu}, \widehat{L}_{1 \nu}\bigr)=-\sqrt{\frac{8 \pi}{3}} 
\clebsch{1}{\mu}{1}{\nu}{1}{\mu+\nu} Y_{1 \mu+\nu}(\vartheta, \varphi). \label{eq:13:2:70}
\end{equation}
Taking Eqs. \eqref{eq:13:2:104} and \eqref{eq:13:2:105} in account one finds
\begin{equation}
\braOket{ l^{\prime} m^{\prime}}{ \mathfrak{R}_{1 \mu 1 \nu}\bigl(\widehat{n}_{1 \mu}, \widehat{L}_{1 \nu}\bigr)}{l m}=-\sqrt{2}\Fact{l}/\Fact{l^{\prime}}
\clebsch{l}{0}{1}{0}{l^{\prime}}{0}
\clebsch{1}{\mu}{1}{\nu}{1}{\mu+\nu} 
\clebsch{l}{m}{1}{\mu+\nu}{l^{\prime}}{m^{\prime}}. \label{eq:13:2:71}
\end{equation}
The matrix element of the direct product \(\widehat{L}_{1 \nu} \widehat{n}_{1 \mu}\) is
\begin{equation}
\braOket{ l^{\prime} m^{\prime}}{ \widehat{L}_{1 \nu} \widehat{n}_{1 \mu}}{l m}=\Fact{l}/\Fact{l^\prime}
\sqrt{ l^{\prime}\bigl(l^{\prime}+1\bigr)} 
\clebsch{l}{0}{1}{0}{l^{\prime}}{0} 
\clebsch{l^{\prime}}{m+\mu}{1}{\nu}{l^{\prime}}{m^{\prime}} 
\clebsch{l}{m}{1}{\mu}{l^{\prime}}{m+\mu}. \label{eq:13:2:72}
\end{equation}
In the case of irreducible tensor products of the operators in question one gets
\begin{align}
& 
\braOket{ l^{\prime} m^{\prime}}{\bigl\{\widehat{\vect{n}}_{1} \otimes \widehat{\vect{L}}_{1}\bigr\}_{k \kappa}}{l m}=(-1)^{l^{\prime}+l+k} 
\clebsch{l}{m}{k}{\kappa}{l^{\prime}}{m^{\prime}} \Fact{k l l}/\Fact{l^\prime}\sqrt{ l(l+1)}
\clebsch{l}{0}{1}{0}{l^{\prime}}{0}
\sixj{1}{1}{k}{l^{\prime}}{l}{l},  \label{eq:13:2:73}\\
& 
\braOket{ l^{\prime} m^{\prime}}{\bigl\{\widehat{\vect{L}}_{1} \otimes \widehat{\vect{n}}_{1}\bigr\}_{k \kappa}}{l m}=(-1)^{l^{\prime}+l+k} 
\clebsch{l}{m}{k}{\kappa}{l^{\prime}}{m^{\prime}} \Fact{k l }\sqrt{ l^{\prime}\bigl(l^{\prime}+1\bigr)} 
\clebsch{l}{0}{1}{0}{l^{\prime}}{0}
\sixj{1}{1}{k}{l^{\prime}}{l}{l^{\prime}}, \label{eq:13:2:74}
\end{align}
\begin{equation*}
\braOketred{ l^{\prime}}{\bigl\{\widehat{\vect{n}}_{1} \otimes \widehat{\vect{L}}_{1}\bigr\}_{k}}{ l}=(-1)^{l^{\prime}+l+k} \Fact{k l l}\sqrt{ l(l+1)}
\clebsch{l}{0}{1}{0}{l^{\prime}}{0}
\sixj{1}{1}{k}{l^{\prime}}{l}{l},
\end{equation*}
\begin{equation*}
\braOketred{ l^{\prime}}{\bigl\{\widehat{\vect{L}}_{1} \otimes \widehat{\vect{n}}_{1}\bigr\}_{k}}{ l}=(-1)^{l^{\prime}+l+k} \Fact{k l l^{\prime}}\sqrt{ l^{\prime}\bigl(l^{\prime}+1\bigr)}
\clebsch{l}{0}{1}{0}{l^{\prime}}{0}
\sixj{1}{1}{k}{l^{\prime}}{l}{l^{\prime}} .
\end{equation*}
Both equations are related by the expression
\begin{equation}
\braOket{ l^{\prime} m^{\prime}}{\bigl\{\widehat{\vect{n}}_{1} \otimes \widehat{\vect{L}}_{1}\bigr\}_{k \kappa}}{l m}=(-1)^{k}
\braOket{ l^{\prime} m^{\prime}}{\bigl\{\widehat{\vect{L}}_{1} \otimes \widehat{\vect{n}}_{1}\bigr\}_{k \kappa}}{l m}+
\braOket{ l^{\prime} m^{\prime}}{ \mathfrak{R}_{k \kappa}^{11}\bigl(\widehat{\vect{n}}_{1}, \widehat{\vect{L}}_{1}\bigr)}{l m}, \label{eq:13:2:75}
\end{equation}
where
\begin{align}
\mathfrak{R}_{k \kappa}^{11}\bigl(\widehat{\vect{n}}_{1}, \widehat{\vect{L}}_{1}\bigr) & =-\sqrt{\frac{8 \pi}{3}} Y_{1 \kappa} \delta_{k 1},  \label{eq:13:2:76}\\
\braOket{ l^{\prime} m^{\prime}}{ \mathfrak{R}_{k \kappa}^{11}\bigl(\widehat{\vect{n}}_{1}, \widehat{\vect{L}}_{1}\bigr)}{l m} & =-\frac{\Fact{l}}{\Fact{l^{\prime}}}
\clebsch{l}{0}{1}{0}{l^{\prime}}{0} 
\clebsch{l}{m}{1}{\kappa}{l^{\prime}}{m^{\prime}}, \label{eq:13:2:77}
\end{align}
\begin{equation*}
\braOketred{ l^{\prime}}{ \mathfrak{R}_{k}^{11}\bigl(\widehat{\vect{n}}_{1}, \widehat{\vect{L}}_{1}\bigr)}{l}=-\Fact{l}
\clebsch{l}{0}{1}{0}{l^{\prime}}{0} \delta_{k 1}.
\end{equation*}

Let us now consider matrix elements of products of the operators \(\bigl(\delop_{\Omega}\bigr)_{1}\) and \(\widehat{\vect{L}}_{1}\). The matrix elements of the direct product of these operators are
\begin{align}
& 
\braOket{ l^{\prime} m^{\prime}}{\bigl(\delop_{\Omega}\bigr)_{1 \mu} \widehat{L}_{1 \nu}}{l m}=\frac{\sqrt{l(l+1)}}{\Fact{l^{\prime}}}
\braOketred{ l^{\prime}}{\bigl(\delop_{\Omega}\bigr)_{1}}{ l} 
\clebsch{l}{m+\nu}{1}{\mu}{l^{\prime}}{m^{\prime}} 
\clebsch{l}{m}{1}{\nu}{l}{m+\nu},  \label{eq:13:2:78}\\
& 
\braOket{ l^{\prime} m^{\prime}}{ \widehat{L}_{1 \nu}\bigl(\delop_{\Omega}\bigr)_{1 \mu}}{l m}=\frac{\sqrt{l(l+1)}}{\Fact{l^{\prime}}}
\braOketred{ l^{\prime}}{\bigl(\delop_{\Omega}\bigr)_{1}}{ l} 
\clebsch{l^{\prime}}{m+\mu}{1}{\nu}{l^{\prime}}{m^{\prime}} 
\clebsch{l}{m}{1}{\mu}{l^{\prime}}{m+\mu}. \label{eq:13:2:79}
\end{align}
Equations \eqref{eq:13:2:78} and \eqref{eq:13:2:79} differ by a quantity which is the matrix element of the commutator of \(\widehat{L}_{1 \nu}\) and \(\bigl(\delop_{\Omega}\bigr)_{1 \mu}\),
\begin{gather}
\mathfrak{R}_{1 \mu 1 \nu}\bigl(\bigl(\delop_{\Omega}\bigr)_{1 \mu}, \widehat{L}_{1 \nu}\bigr)=-\sqrt{2} 
\clebsch{1}{\mu}{1}{\nu}{1}{\mu+\nu}\bigl(\delop_{\Omega}\bigr)_{1 \mu+\nu},  \label{eq:13:2:80}\\
\braOket{ l^{\prime} m^{\prime}}{ \mathfrak{R}_{1 \mu 1 \nu}\bigl(\bigl(\delop_{\Omega}\bigr)_{1 \mu}, \widehat{L}_{1 \nu}\bigr)}{l m}=-\frac{\sqrt{2}}{\Fact{l^{\prime}}}
\braOketred{ l^{\prime}}{\bigl(\delop_{\Omega}\bigr)_{1}}{ l} 
\clebsch{1}{\mu}{1}{\nu}{1}{\mu+\nu} 
\clebsch{l}{m}{1}{\mu+\nu}{l^{\prime}}{m^{\prime}}. \label{eq:13:2:81}
\end{gather}
In the case of the irreducible tensor product of the operators \(\bigl(\delop_{\Omega}\bigr)_{1}\) and \(\widehat{\vect{L}}_{1}\) we find
\begin{equation}
\braOket{ l^{\prime} m^{\prime}}{\bigl\{\bigl(\delop_{\Omega}\bigr)_{1} \otimes \widehat{\vect{L}}_{1}\bigr\}_{k \kappa}}{l m}=(-1)^{l^{\prime}+l+k} \Fact{k l}/\Fact{l^\prime}\sqrt{l(l+1)}
\braOketred{ l^{\prime}}{\bigl(\delop_{\Omega}\bigr)_{1}}{ l}
\sixj{1}{1}{k}{l^{\prime}}{l}{l} 
\clebsch{l}{m}{k}{\kappa}{l^{\prime}}{m^{\prime}},  \label{eq:13:2:82}
\end{equation}
\begin{equation*}
\braOketred{ l^{\prime}}{\bigl\{\bigl(\delop_{\Omega}\bigr)_{1} \otimes \widehat{\vect{L}}_{1}\bigr\}_{k}}{ l}=(-1)^{l^{\prime}+l+k} \Fact{k l}\sqrt{l(l+1)}
\braOketred{ l^{\prime}}{\bigl(\delop_{\Omega}\bigr)_{1}}{ l}
\sixj{1}{1}{k}{l^{\prime}}{l}{l},
\end{equation*}
\begin{equation} 
\braOket{ l^{\prime} m^{\prime}}{\bigl\{\widehat{\vect{L}}_{1} \otimes\bigl(\delop_{\Omega}\bigr)_{1}\bigr\}_{k \kappa}}{l m}=(-1)^{l^{\prime}+l+k} \Fact{k}\sqrt{ l^{\prime}\bigl(l^{\prime}+1\bigr)}
\braOketred{ l^{\prime}}{\bigl(\delop_{\Omega}\bigr)_{1}}{ l}
\sixj{1}{1}{k}{l^{\prime}}{l}{l^{\prime}}
\clebsch{l}{m}{k}{\kappa}{l^{\prime}}{m^{\prime}}, \label{eq:13:2:83}
\end{equation}
\begin{equation*}
\braOketred{ l^{\prime}}{\bigl\{\widehat{\vect{L}}_{1} \otimes\bigl(\delop_{\Omega}\bigr)_{1}\bigr\}_{k}}{ l}=(-1)^{l^{\prime}+l+k} \Fact{k l^{\prime}}\sqrt{ l^{\prime}\bigl(l^{\prime}+1\bigr)}
\braOketred{ l^{\prime}}{\bigl(\delop_{\Omega}\bigr)_{1}}{ l}
\sixj{1}{1}{k}{l^{\prime}}{l}{l^{\prime}} .
\end{equation*}
The commutator of the product \(\bigl\{\bigl(\delop_{\Omega}\bigr)_{1} \otimes \widehat{\vect{L}}_{1}\bigr\}_{k \kappa}\) may be represented in the form
\begin{equation}
\mathfrak{R}_{k \kappa}^{11}\bigl(\bigl(\delop_{\Omega}\bigr)_{1}, \widehat{\vect{L}}_{1}\bigr)=-\sqrt{2}\bigl(\delop_{\Omega}\bigr)_{1 \kappa} \delta_{k 1}, \label{eq:13:2:84}
\end{equation}
Using Eq. \eqref{eq:13:2:26}, one obtains
\begin{equation}
\braOket{ l^{\prime} m^{\prime}}{ \mathfrak{R}_{k \kappa}^{11}\bigl(\bigl(\delop_{\Omega}\bigr)_{1}, \widehat{\vect{L}}_{1}\bigr)}{l m}=-\frac{\sqrt{2}}{\Fact{l^{\prime}}}
\braOketred{ l^{\prime}}{\bigl(\delop_{\Omega}\bigr)_{1}}{ l}
\clebsch{l}{m}{1}{\kappa}{l^{\prime}}{m^{\prime}} , \label{eq:13:2:85}
\end{equation}
\begin{equation*}
\braOketred{ l^{\prime}}{ \mathfrak{R}_{k \kappa}^{11}\bigl(\bigl(\delop_{\Omega}\bigr)_{1}, \widehat{\vect{L}}_{1}\bigr)}{l}=-\sqrt{2}
\braOketred{ l^{\prime}}{\bigl(\delop_{\Omega}\bigr)_{1}}{ l}
 . 
\end{equation*}
The matrix elements of the irreducible tensor product of the operators \(\widehat{\vect{L}}_{1}\) and \(\widehat{\vect{J}}_{1}\) are
\begin{multline}
\braOket{ l^{\prime} s^{\prime} J^{\prime} M^{\prime}}{\bigl\{\widehat{\vect{L}}_{1} \otimes \widehat{\vect{J}}_{1}\bigr\}_{k \kappa}}{l s J M}=(-1)^{s+l-J^{\prime}+k+1} \delta_{l l^{\prime}} \delta_{s s^{\prime}}
\clebsch{J}{M}{k}{\kappa}{J^{\prime}}{M^{\prime}}\Fact{J J} \\
\times \Fact{k l}\sqrt{ l(l+1) J(J+1)}
\sixj{1}{1}{k}{J^{\prime}}{J}{J}
\sixj{l}{s}{J}{J^{\prime}}{1}{l},  \label{eq:13:2:86}
\end{multline}
\begin{multline*}
\braOketred{ l^{\prime} s^{\prime} J^{\prime}}{\bigl\{\widehat{\vect{L}}_{1} \otimes \widehat{\vect{J}}_{1}\bigr\}_{k}}{l s J}=(-1)^{s+l-J^{\prime}+k+1} \delta_{l l^{\prime}} \delta_{s s^{\prime}}
\Fact{J J}\Fact{J^{\prime}} \Fact{k l}\sqrt{ l(l+1) J(J+1)} 
\sixj{1}{1}{k}{J^{\prime}}{J}{J}
\sixj{l}{s}{J}{J^{\prime}}{1}{l},
\end{multline*}
\begin{multline}
\braOket{ l^{\prime} s^{\prime} J^{\prime} M^{\prime}}{\bigl\{\widehat{\vect{J}}_{1} \otimes \widehat{\vect{L}}_{1}\bigr\}_{k \kappa}}{l s J M}=(-1)^{s+l-J^{\prime}+k+1} \delta_{l^{\prime} l} \delta_{s s^{\prime}}
\clebsch{J}{M}{k}{\kappa}{J^{\prime}}{M^{\prime}}\\
\times \Fact{k l J J^\prime}\sqrt{ l(l+1) J^{\prime}\bigl(J^{\prime}+1\bigr)}
\sixj{1}{1}{k}{J^{\prime}}{J}{J^{\prime}}
\sixj{l}{s}{J}{J^{\prime}}{1}{l} . \label{eq:13:2:87}
\end{multline}
\begin{multline*}
\braOketred{ l^{\prime} s^{\prime} J^{\prime}}{\bigl\{\widehat{\vect{J}}_{1} \otimes \widehat{\vect{L}}_{1}\bigr\}_{k}}{l s J}=(-1)^{s+l-J^{\prime}+k+1} \delta_{l^{\prime} l} \delta_{s s^{\prime}}
\Fact{k l J J^\prime}\Fact{J^{\prime}}\sqrt{ l(l+1) J^{\prime}\bigl(J^{\prime}+1\bigr)} 
\sixj{1}{1}{k}{J^{\prime}}{J}{J^{\prime}}
\sixj{l}{s}{J}{J^{\prime}}{1}{l} .
\end{multline*}
The commutator of the irreducible tensor product \(\bigl\{\widehat{\vect{L}}_{1} \otimes \widehat{\vect{J}}_{1}\bigr\}_{k \kappa}\) is defined as
\begin{equation}
\mathfrak{R}_{k \kappa}^{11}\bigl(\widehat{\vect{L}}_{1}, \widehat{\vect{J}}_{1}\bigr)=-\sqrt{2} \widehat{L}_{1 \kappa} \delta_{k 1}, \label{eq:13:2:88}
\end{equation}
To evaluate its matrix elements one can use Eqs. \eqref{eq:13:2:66} and \eqref{eq:13:2:67}. Then

\begin{equation}
\braOket{ l^{\prime} s^{\prime} J^{\prime} M^{\prime}}{ \mathfrak{R}_{k \kappa}^{11}\bigl(\widehat{\vect{L}}_{1}, \widehat{\vect{J}}_{1}\bigr)}{l s J M}=(-1)^{s+J+l} \delta_{l l^{\prime}} \delta_{s s^{\prime}} \delta_{k 1} \Fact{J l}\sqrt{2 l(l+1)}
\sixj{l}{s}{J}{J^{\prime}}{1}{l} 
\clebsch{J}{M}{1}{\kappa}{J^{\prime}}{M^{\prime}}. \label{eq:13:2:89}
\end{equation}

To determine the matrix elements of the direct tensor product of the operators \(\widehat{\vect{L}}_{1}\) and \(\widehat{\vect{J}}_{1}\) one can decompose this direct product into irreducible products. In this way one obtains
\begin{align}
& 
\braOket{ l^{\prime} s^{\prime} J^{\prime} M^{\prime}}{ \widehat{L}_{1 \nu} \widehat{J}_{1 \mu}}{l s J M}=\sum_{k \kappa} 
\clebsch{1}{\mu}{1}{\nu}{k}{\kappa}
\braOket{ l^{\prime} s^{\prime} J^{\prime} M^{\prime}}{\bigl\{\widehat{\vect{L}}_{1} \otimes \widehat{\vect{J}}_{1}\bigr\}_{k \kappa}}{l s J M} .  \label{eq:13:2:90}\\
& 
\braOket{ l^{\prime} s^{\prime} J^{\prime} M^{\prime}}{ \widehat{J}_{1 \mu} \widehat{L}_{1 \nu}}{l s J M}=\sum_{k \kappa} 
\clebsch{1}{\mu}{1}{\nu}{k}{\kappa}
\braOket{ l^{\prime} s^{\prime} J^{\prime} M^{\prime}}{\bigl\{\widehat{\vect{J}}_{1} \otimes \widehat{\vect{L}}_{1}\bigr\}_{k \kappa}}{l s J M} . \label{eq:13:2:91}
\end{align}
The commutator of the operator components \(\widehat{L}_{1 \mu}\) and \(\widehat{J}_{1 \nu}\) may be presented in the form
\begin{equation}
\mathfrak{R}_{1 \mu 1 \nu}\bigl(\widehat{L}_{1 \mu}, \widehat{J}_{1 \nu}\bigr)=-\sqrt{2} 
\clebsch{1}{\mu}{1}{\nu}{1}{\mu+\nu} \widehat{L}_{1 \mu+\nu}. \label{eq:13:2:92}
\end{equation}
From Eqs. \eqref{eq:13:2:92}, \eqref{eq:13:2:66} and \eqref{eq:13:2:67} we find

\begin{equation}
\braOket{ l^{\prime} s^{\prime} J^{\prime} M^{\prime}}{ \mathfrak{R}_{1 \mu 1 \nu}\bigl(\widehat{L}_{1 \mu}, \widehat{J}_{1 \nu}\bigr)}{l s J M}=(-1)^{s+J+l} \delta_{l l^{\prime}} \delta_{s s^{\prime}} \Fact{J l}\sqrt{2 l(l+1)}
\sixj{l}{s}{J}{J^{\prime}}{1}{l}
\clebsch{1}{\mu}{1}{\nu}{1}{\mu+\nu} 
\clebsch{J}{M}{1}{\mu+\nu}{J^{\prime}}{M^{\prime}}. \label{eq:13:2:93} 
\end{equation}

\subsection{Matrix Elements of the Spin Angular Momentum Operator \(\widehat{\vect{S}} \equiv \widehat{\vect{S}}_{1}\)}\label{sec:13.2.7}\index{matrix elements!of the spin operator}\index{spin operator}\isym{S-hat@$\widehat{\vect{S}}$, spin operator}
This operator acts only on spin variables. Using the Wigner-Eckart theorem, we get
\begin{gather}
\braOket{ s^{\prime} m_{s}^{\prime}}{ \widehat{S}_{1 \mu}}{s m_{s}}=\frac{
\braOketred{ s^{\prime}}{\widehat{\vect{S}}_{1}}{ s}}{\Fact{s^{\prime}}} 
\clebsch{s}{m_{s}}{1}{\mu}{s^{\prime}}{m_{s}^{\prime}},  \label{eq:13:2:94}\\
\braOketred{ s^{\prime}}{\widehat{\vect{S}}_{1}}{ s}=\delta_{s s^{\prime}} \Fact{s}\sqrt{s(s+1)}. \label{eq:13:2:95}
\end{gather}
Equations \eqref{eq:13:2:39}-\eqref{eq:13:2:47} were obtained for the matrix elements of the operator \(\widehat{\vect{J}}_{1}\) and they remain valid for the spin operator \(\widehat{\vect{S}}_{1}\) after the replacement

\begin{equation*}
J \rightarrow s, \quad M \rightarrow m_{s}, \quad \widehat{\vect{J}}_{1} \rightarrow \widehat{\vect{S}}_{1}
\end{equation*}

The use of Eqs. \eqref{eq:13:2:6} and \eqref{eq:13:2:95} yields an expression for the reduced matrix element of the spin operator in the ( \(l s J\) )-representation
\begin{equation}
\braOketred{ l^{\prime} s^{\prime} J^{\prime}}{\widehat{\vect{S}}_{1}}{ l s J}=(-1)^{l+s+J^{\prime}+1} \delta_{l l^{\prime}} \delta_{s s^{\prime}} \Fact{s J J^\prime}\sqrt{s(s+1)}
\sixj{s}{l}{J}{J^{\prime}}{1}{s}. \label{eq:13:2:96}
\end{equation}
The operators \(\widehat{\vect{n}}_{1},\bigl(\delop_{\Omega}\bigr)_{1}, \widehat{\vect{L}}_{1}\) commute with the spin operator \(\widehat{\vect{S}}_{1}\). Matrix elements of the irreducible tensor products of these operators and the spin operator are given by
\begin{align}
& 
\braOket{ l^{\prime} s^{\prime} J^{\prime} M^{\prime}}{\bigl\{\widehat{\vect{n}}_{1} \otimes \widehat{\vect{S}}_{1}\bigr\}_{k \kappa}}{l s J M}=\delta_{s s^{\prime}} \Fact{k J l s}\sqrt{ s(s+1)} 
\clebsch{l}{0}{1}{0}{l^{\prime}}{0} 
\clebsch{J}{M}{k}{\kappa}{J^{\prime}}{M^{\prime}}
\ninej{l}{1}{l^{\prime}}{s}{1}{s}{J}{k}{J^{\prime}},  \label{eq:13:2:97}\\
& 
\braOket{ l^{\prime} s^{\prime} J^{\prime} M^{\prime}}{\bigl\{\bigl(\delop_{\Omega}\bigr)_{1} \otimes \widehat{\vect{S}}_{1}\bigr\}_{k \kappa}}{l s J M}=\delta_{s s^{\prime}}\Fact{k J s} \sqrt{ s(s+1)}\braOketred{ l^{\prime}}{\bigl(\delop_{\Omega}\bigr)_{1}}{ l}
\clebsch{J}{M}{k}{\kappa}{J^{\prime}}{M^{\prime}}
\ninej{l}{1}{l^{\prime}}{s}{1}{s}{J}{k}{J^{\prime}},  \label{eq:13:2:98}\\
& 
\braOket{ l^{\prime} s^{\prime} J^{\prime} M^{\prime}}{\bigl\{\widehat{\vect{L}}_{1} \otimes \widehat{\vect{S}}_{1}\bigr\}_{k \kappa}}{l s J M}=\delta_{l l^{\prime}} \delta_{s s^{\prime}} \Fact{k J l s}\sqrt{ l(l+1) s(s+1)}
\clebsch{J}{M}{k}{\kappa}{J^{\prime}}{M^{\prime}}
\ninej{l}{1}{l}{s}{1}{s}{J}{k}{J^{\prime}}, \label{eq:13:2:99}
\end{align}
\begin{equation*}
\braOketred{ l^{\prime} s^{\prime} J^{\prime}}{\bigl\{\widehat{\vect{n}}_{1} \otimes \widehat{\vect{S}}_{1}\bigr\}_{k}}{l s J}=\delta_{s s^{\prime}} \Fact{k J l s J^{\prime}}\sqrt{ s(s+1)}
\clebsch{l}{0}{1}{0}{l^{\prime}}{0}
\ninej{l}{1}{l^{\prime}}{s}{1}{s}{J}{k}{J^{\prime}},
\end{equation*}
\begin{equation*}
\braOketred{ l^{\prime} s^{\prime} J^{\prime}}{\bigl\{\bigl(\delop_{\Omega}\bigr)_{1} \otimes \widehat{\vect{S}}_{1}\bigr\}_{k}}{l s J}=\delta_{s s^{\prime}}\Fact{k J s J^{\prime}} \sqrt{ s(s+1)}\braOketred{ l^{\prime}}{\bigl(\delop_{\Omega}\bigr)_{1}}{ l}
\ninej{l}{1}{l^{\prime}}{s}{1}{s}{J}{k}{J^{\prime}},
\end{equation*}
\begin{equation*}
\braOketred{ l^{\prime} s^{\prime} J^{\prime}}{\bigl\{\widehat{\vect{L}}_{1} \otimes \widehat{\vect{S}}_{1}\bigr\}_{k}}{l s J}=\delta_{l l^{\prime}} \delta_{s s^{\prime}} \Fact{k J l s J^{\prime}}\sqrt{ l(l+1) s(s+1)}
\ninej{l}{1}{l}{s}{1}{s}{J}{k}{J^{\prime}}.
\end{equation*}
The operators \(\widehat{\vect{S}}_{1}\) and \(\widehat{\vect{J}}_{1}\) do not commute. Matrix elements of irreducible tensor products of these operators depend on the order of the operators.
\begin{align}
\braOket{ l^{\prime} s^{\prime} J^{\prime} M^{\prime}}{\bigl\{\widehat{\vect{S}}_{1} \otimes \widehat{\vect{J}}_{1}\bigr\}_{k \kappa}}{l s J M}= & (-1)^{J+k-l-s+1} \delta_{s^{\prime} s} \delta_{l^{\prime} l}\Fact{J J k s} \sqrt{ s(s+1) J(J+1)}\notag \\
& \times
\sixj{1}{1}{k}{J^{\prime}}{J}{J}
\sixj{s}{l}{J}{J^{\prime}}{1}{s} 
\clebsch{J}{M}{k}{\kappa}{J^{\prime}}{M^{\prime}},  \label{eq:13:2:100}\\
\braOket{ l^{\prime} s^{\prime} J^{\prime} M^{\prime}}{\bigl\{\widehat{\vect{J}}_{1} \otimes \widehat{\vect{S}}_{1}\bigr\}_{k \kappa}}{l s J M}= & (-1)^{J+k-l-s+1} \delta_{s^{\prime} s} \delta_{l^{\prime} l} \Fact{k J J^\prime s}\sqrt{s(s+1) J^{\prime}\bigl(J^{\prime}+1\bigr)}\notag \\
& \times 
\sixj{1}{1}{k}{J^{\prime}}{J}{J^{\prime}}
\sixj{s}{l}{J}{J^{\prime}}{1}{s}
\clebsch{J}{M}{k}{\kappa}{J^{\prime}}{M^{\prime}}, \label{eq:13:2:101}
\end{align}
\begin{multline*}
\braOketred{ l^{\prime} s^{\prime} J^{\prime}}{\bigl\{\widehat{\vect{S}}_{1} \otimes \widehat{\vect{J}}_{1}\bigr\}_{k}}{l s J}=(-1)^{J+k-l-s+1} \delta_{s^{\prime} s} \delta_{l^{\prime} l}\Fact{J J k s J^{\prime}} \sqrt{ s(s+1) J(J+1)} 
%\\\times
\sixj{1}{1}{k}{J^{\prime}}{J}{J}
\sixj{s}{l}{J}{J^{\prime}}{1}{s},
\end{multline*}
\begin{multline*}
\braOketred{ l^{\prime} s^{\prime} J^{\prime}}{\bigl\{\widehat{\vect{J}}_{1} \otimes \widehat{\vect{S}}_{1}\bigr\}_{k}}{l s J}=(-1)^{J+k-l-s+1} \delta_{s^{\prime} s} \delta_{l^{\prime} l} \Fact{J k s J^\prime J^{\prime}}\sqrt{ s(s+1)J^{\prime}\bigl(J^{\prime}+1\bigr)} % \\\times
\sixj{1}{1}{k}{J^{\prime}}{J}{J^{\prime}}
\sixj{s}{l}{J}{J^{\prime}}{1}{s} .
\end{multline*}
The commutator of the product \(\bigl\{\widehat{\vect{S}}_{1} \otimes \widehat{\vect{J}}_{1}\bigr\}_{k \kappa}\) is as follows:
\begin{equation}
\mathfrak{R}_{k \kappa}^{11}\bigl(\widehat{\vect{S}}_{1}, \widehat{\vect{J}}_{1}\bigr)=-\sqrt{2} \widehat{S}_{1 \kappa} \delta_{k 1}, \label{eq:13:2:102}
\end{equation}
Using Eqs. \eqref{eq:13:2:94} and \eqref{eq:13:2:95}, we find
\begin{equation}
\braOket{ l^{\prime} s^{\prime} J^{\prime} M^{\prime}}{ \mathfrak{R}_{k \kappa}^{11}\bigl(\widehat{\vect{S}}_{1}, \widehat{\vect{J}}_{1}\bigr)}{l s J M}=(-1)^{l+s+J^{\prime}} \delta_{l l^{\prime}} \delta_{s s^{\prime}} \delta_{k 1} \Fact{s J}\sqrt{2 s(s+1)}
\sixj{s}{l}{J}{J^{\prime}}{1}{s}
\clebsch{J}{M}{k}{\kappa}{J^{\prime}}{M^{\prime}}, \label{eq:13:2:103} 
\end{equation}
\begin{equation*}
\braOketred{ l^{\prime} s^{\prime} J^{\prime}}{ \mathfrak{R}_{k}^{11}\bigl(\widehat{\vect{S}}_{1}, \widehat{\vect{J}}_{1}\bigr)}{l s J}=(-1)^{l+s+J^{\prime}} \delta_{l l^{\prime}} \delta_{s s^{\prime}} \delta_{k 1} \Fact{s J J^{\prime}}\sqrt{2 s(s+1)}
\sixj{s}{l}{J}{J^{\prime}}{1}{s}
\clebsch{J}{M}{k}{\kappa}{J^{\prime}}{M^{\prime}}. 
\end{equation*}
To obtain matrix elements of direct products of the operator \(\widehat{\vect{S}}_{1}\) and the operators \(\widehat{\vect{n}}_{1},\bigl(\delop_{\Omega}\bigr)_{1}, \widehat{\vect{L}}_{1}\) or \(\widehat{\vect{J}}_{1}\) one should decompose the direct products into irreducible products in accordance with Eq.~\eqref{eq:3:1:22} and use Eqs. \eqref{eq:13:2:97}-\eqref{eq:13:2:101}.

\subsection{Matrix Elements of the Spherical Harmonic Operator \(\widehat{Y}_{L \nu} \equiv Y_{L \nu}(\vartheta, \varphi)\)}\label{sec:13.2:8}\index{matrix elements!of the spherical harmonic operator}\index{spherical harmonics}\isym{Y-hat@$\widehat{Y}_{L \nu}$, spherical harmonic operator}

A spherical harmonic operator, \(Y_{L \nu}\), depends only on position variables \(\vartheta, \varphi\). From the Wigner-Eckart theorem it follows that
\begin{gather}
\braOket{ l^{\prime} m^{\prime}}{ \widehat{Y}_{L \nu}}{l m}=\frac{
\braOketred{ l^{\prime}}{\widehat{\vect{Y}}_{L}}{ l}}{\Fact{l^{\prime}}}
\clebsch{l}{m}{L}{\nu}{l^{\prime}}{m^{\prime}},  \label{eq:13:2:104}\\
\braOketred{ l^{\prime}}{\widehat{\vect{Y}}_{L}}{ l}=\Fact{L l}\sqrt{\frac{1}{4 \pi}} 
\clebsch{l}{0}{L}{0}{l^{\prime}}{0}. \label{eq:13:2:105}
\end{gather}
Similar relationships for an operator \(\widehat{C}_{L \nu}(\vartheta, \varphi)\) (see Eq.~\eqref{eq:5:1:7}) may be represented in the form\index{spherical harmonics!normalized form $C_{l m}$}\isym{Clm@$C_{l m}(\vartheta, \varphi)$, normalized spherical harmonic}
\begin{align}
\braOket{ l^{\prime} m^{\prime}}{ \widehat{C}_{L \nu}}{l m} & =\frac{
\braOketred{ l^{\prime}}{\widehat{\vect{C}}_{L}}{ l}}{\Fact{l^{\prime}}} 
\clebsch{l}{m}{L}{\nu}{l^{\prime}}{m^{\prime}},  \label{eq:13:2:106}\\
\braOketred{ l^{\prime}}{\widehat{\vect{C}}_{L}}{ l} & =\Fact{l}
\clebsch{l}{0}{L}{0}{l^{\prime}}{0}. \label{eq:13:2:107}
\end{align}

Let us consider the matrix elements of some tensor products involving \(\widehat{Y}_{L \nu}(\vartheta, \varphi)\).\\
For products of the commuting operators \(\widehat{\vect{n}}_{1}\) and \(\widehat{\vect{Y}}_{L}\), we get\index{matrix elements!of products with the unit vector}\index{operators!commuting}
\begin{multline}
\braOket{l^{\prime} m^{\prime}}{\widehat{n}_{1 \mu} \widehat{Y}_{L \nu}}{ l m}=-\frac{\Fact{L l}}{\Fact{l^\prime}}\sqrt{\frac{1}{4 \pi}}\bigl\{\sqrt{\frac{l^{\prime}+1}{2 l^{\prime}+3}}
\clebsch{l}{0}{L}{0}{l^{\prime}+1}{0}
\clebsch{l}{m}{L}{\nu}{l^{\prime}+1}{m+\nu} \clebsch{l^{\prime}+1}{m+\nu}{1}{\mu}{l^{\prime}}{m^{\prime}}\\
-\sqrt{\frac{l^{\prime}}{2 l^{\prime}-1}}
\clebsch{l}{0}{L}{0}{l^{\prime}-1}{0}
\clebsch{l}{m}{L}{\nu}{l^{\prime}-1}{m+\nu}
\clebsch{l^{\prime}-1}{m+\nu}{1}{\mu}{l^{\prime}}{m^{\prime}}\bigr\},  \label{eq:13:2:108}
\end{multline}
\begin{equation}
\braOket{ l^{\prime} m^{\prime}}{\bigl\{\widehat{\vect{n}}_{1} \otimes \widehat{\vect{Y}}_{L}\bigr\}_{L^{\prime} \nu^{\prime}}}{l m}=(-1)^{L^{\prime}+l^{\prime}+l} \frac{\Fact{ L L^\prime l}}{\Fact{l^\prime}}\sqrt{\frac{1}{4 \pi}}
\clebsch{l}{m}{L^{\prime}}{\nu^{\prime}}{l^{\prime}}{m^{\prime}}
\times \sum_{k} \Fact{k}
\clebsch{k}{0}{1}{0}{l^{\prime}}{0} 
\clebsch{l}{0}{L}{0}{k}{0}
\sixj{L}{1}{L^{\prime}}{l^{\prime}}{l}{k}, \label{eq:13:2:109}
\end{equation}
\begin{equation*}
\braOketred{ l^{\prime}}{\bigl\{\widehat{\vect{n}}_{1} \otimes \widehat{\vect{Y}}_{L}\bigr\}_{L^{\prime}}}{ l}=(-1)^{L^{\prime}+l^{\prime}+l} \Fact{ L L^\prime l}\sqrt{\frac{1}{4 \pi}} 
 \sum_{k} \Fact{k}
\clebsch{k}{0}{1}{0}{l^{\prime}}{0}
\clebsch{l}{0}{L}{0}{k}{0}
\sixj{L}{1}{L^{\prime}}{l^{\prime}}{l}{k}.
\end{equation*}
The matrix elements of products of the non-commuting operators \(\bigl(\delop_{\Omega}\bigr)_{1}\) and \(\widehat{\Psi}_{L}\) are determined by\index{operators!non-commuting}\isymo{delop-Omega@$\delop_{\Omega}$, angular part of the gradient operator}
\begin{multline}
\braOket{ l^{\prime} m^{\prime}}{\bigl(\delop_{\Omega}\bigr)_{1 \mu} \widehat{Y}_{L \nu}}{l m}=-\Fact{L l}/\Fact{l^\prime}\sqrt{\frac{1}{4 \pi}}\biggl\{\bigl(l^{\prime}-1\bigr) \sqrt{\frac{l^{\prime}}{2 l^{\prime}-1}} 
\clebsch{l}{0}{L}{0}{l^{\prime}-1}{0} 
\clebsch{l}{m}{L}{\nu}{l^{\prime}-1}{m+\nu} 
\clebsch{l^{\prime}-1}{m+\nu}{1}{\mu}{l^{\prime}}{m^{\prime}}
\\ +
\bigl(l^{\prime}+2\bigr) \sqrt{\frac{l^{\prime}+1}{2 l^{\prime}+3}}
\clebsch{l}{0}{L}{0}{l^{\prime}+1}{0}
\clebsch{l}{m}{L}{\nu}{l^{\prime}+1}{m+\nu}
\clebsch{l^{\prime}+1}{m+\nu}{1}{\mu}{l^{\prime}}{m^{\prime}}\biggr\},  \label{eq:13:2:110}
\end{multline}
\begin{equation}
\braOket{ l^{\prime} m^{\prime}}{\bigl\{\bigl(\delop_{\Omega}\bigr)_{1} \otimes \widehat{\vect{Y}}_{L}\bigr\}_{L^{\prime} \nu^{\prime}}}{l m}=(-1)^{L^{\prime}+l^{\prime}+1} 
\Fact{L l L^\prime}/\Fact{l^\prime}\sqrt{\frac{1}{4 \pi}} \sum_{k}
\braOketred{ l^{\prime}}{\bigl(\delop_{\Omega}\bigr)_{1}}{ k} 
\clebsch{l}{0}{L}{0}{k}{0}
\sixj{L}{1}{L^{\prime}}{l^{\prime}}{l}{k} 
\clebsch{l}{m}{L^{\prime}}{\nu^{\prime}}{l^{\prime}}{m^{\prime}}, \label{eq:13:2:111}
\end{equation}
\begin{equation*}
\braOketred{ l^{\prime}}{\bigl\{\bigl(\delop_{\Omega}\bigr)_{1} \otimes \widehat{\vect{Y}}_{L}\bigr\}_{L^{\prime}}}{ l}=(-1)^{L^{\prime}+l^{\prime}+1}
\Fact{L l L^\prime}\sqrt{\frac{1}{4 \pi}} 
 \sum_{k}
\braOketred{ l^{\prime}}{\bigl(\delop_{\Omega}\bigr)_{1}}{ k}
\clebsch{l}{0}{L}{0}{k}{0}
\sixj{L}{1}{L^{\prime}}{l^{\prime}}{l}{k}.
\end{equation*}
The commutator of components of these operators is as follows
\begin{equation}
\mathfrak{R}_{1 \mu L \nu}\bigl(\bigl(\delop_{\Omega}\bigr)_{1 \mu} \widehat{Y}_{L \nu}\bigr)=-\sqrt{6 L(L+1)}\Fact{L L} \sum_{L^{\prime}} \frac{1}{\sqrt{2 L^{\prime}+1}} 
\clebsch{L}{0}{1}{0}{L^{\prime}}{0} 
\clebsch{1}{\mu}{L}{\nu}{L^{\prime}}{\mu+\nu}
\sixj{1}{1}{1}{L^{\prime}}{L}{L}
 \widehat{Y}_{L^{\prime} \mu+\nu}(\vartheta, \varphi). \label{eq:13:2:112}
\end{equation}

The commutator of the irreducible product of the same operators may be expressed as
\begin{equation}
\mathfrak{R}_{L^{\prime} \nu^{\prime}}^{1 L}\bigl(\bigl(\delop_{\Omega}\bigr)_{1}, \widehat{\vect{Y}}_{L}\bigr)=-\sqrt{6 L(L+1)}\Fact{L L}/\Fact{L^\prime} 
\clebsch{L}{0}{1}{0}{L^{\prime}}{0}
\sixj{1}{1}{1}{L^{\prime}}{L}{L} 
\widehat{Y}_{L^{\prime} \nu^{\prime}}(\vartheta, \varphi). \label{eq:13:2:113}
\end{equation}
One can easily obtain the matrix elements of these commutators:
\begin{multline}
\braOket{ l^{\prime} m^{\prime}}{ \mathfrak{R}_{1 \mu L \nu}\bigl(\bigl(\delop_{\Omega}\bigr)_{1 \mu}, \widehat{Y}_{L \nu}\bigr)}{l m}=-\Fact{l L L}/\Fact{l^\prime}\sqrt{\frac{6 L(L+1)}{4 \pi}} \\
\clebsch{L}{0}{1}{0}{L^{\prime}}{0} 
\clebsch{1}{0}{L^{\prime}}{0}{l^{\prime}}{0} 
\clebsch{1}{\mu}{L}{\nu}{L^{\prime}}{\mu+\nu} 
\clebsch{l}{m}{L^{\prime}}{\mu+\nu}{l^{\prime}}{m^{\prime}}
\sixj{1}{1}{1}{L^{\prime}}{L}{L},  \label{eq:13:2:114}
\end{multline}
\begin{equation}
\braOket{ l^{\prime} m^{\prime}}{ \mathfrak{R}_{L^{\prime} \nu^{\prime}}^{1 L}\bigl(\bigl(\delop_{\Omega}\bigr)_{1}, \widehat{\vect{Y}}_{L^{\prime}}\bigr)}{l m}=-\Fact{l L L}/\Fact{l^\prime}\sqrt{\frac{6 L(L+1)}{4 \pi}} 
\clebsch{l}{m}{L^{\prime}}{\nu^{\prime}}{l^{\prime}}{m^{\prime}} 
\clebsch{L}{0}{1}{0}{L^{\prime}}{0} 
\clebsch{l}{0}{L^{\prime}}{0}{l^{\prime}}{0}
\sixj{1}{1}{1}{L^{\prime}}{L}{L} , \label{eq:13:2:115}
\end{equation}
\begin{equation*}
\braOketred{ l^{\prime}}{ \mathfrak{R}_{L^{\prime}}^{1 L}\bigl(\bigl(\delop_{\Omega}\bigr)_{1}, \widehat{\vect{Y}}_{L^{\prime}}\bigr)}{l}=-\Fact{l L L}\sqrt{\frac{6 L(L+1)}{4 \pi}} 
\clebsch{L}{0}{1}{0}{L^{\prime}}{0} 
\clebsch{l}{0}{L^{\prime}}{0}{l^{\prime}}{0}
\sixj{1}{1}{1}{L^{\prime}}{L}{L} .
\end{equation*}
Of particular interest are the matrix elements of direct and irreducible products of the angular momentum operators and the spherical harmonic operator.\index{matrix elements!of products with the angular momentum operators}

The matrix elements of products of the operators \(\widehat{\vect{L}}_{1}\) and \(\widehat{\vect{Y}}_{L}\) are given by\index{matrix elements!of products with $\widehat{\vect{L}}$}
\begin{gather}
\braOket{l^{\prime} m^{\prime}}{\widehat{L}_{1 \mu} \widehat{Y}_{L \nu}}{ l m}=\Fact{L l}/\Fact{l^\prime}\sqrt{\frac{ l^{\prime}\bigl(l^{\prime}+1\bigr)}{4 \pi}} 
\clebsch{l}{0}{L}{0}{l^{\prime}}{0} 
\clebsch{l}{m}{L}{\nu}{l^{\prime}}{m+\nu} 
\clebsch{l^{\prime}}{m+\nu}{1}{\mu}{l^{\prime}}{m^{\prime}},  \label{eq:13:2:116}\\
\braOket{ l^{\prime} m^{\prime}}{\bigl\{\widehat{\vect{L}}_{1} \otimes \widehat{\vect{Y}}_{L}\bigr\}_{L^{\prime} \nu^{\prime}}}{l m}=(-1)^{L^{\prime}+l^{\prime}+l} \Fact{L^\prime L l}\sqrt{\frac{\bigl(l^{\prime}+1\bigr) l^{\prime}}{4 \pi}} 
\clebsch{l}{m}{L^{\prime}}{\nu^{\prime}}{l^{\prime}}{m^{\prime}} 
\clebsch{l}{0}{L}{0}{l^{\prime}}{0}
\sixj{L}{1}{L^{\prime}}{l^{\prime}}{l}{l^{\prime}}, \label{eq:13:2:117}
\end{gather}
\begin{equation*}
\braOketred{ l^{\prime}}{\bigl\{\widehat{\vect{L}}_{1} \otimes \widehat{\vect{Y}}_{L}\bigr\}_{L^{\prime}}}{ l}=(-1)^{L^{\prime}+l^{\prime}+l} \Fact{L^\prime L l l^{\prime}}\sqrt{\frac{\bigl(l^{\prime}+1\bigr) l^{\prime}}{4 \pi}}
\clebsch{l}{0}{L}{0}{l^{\prime}}{0}
\sixj{L}{1}{L^{\prime}}{l^{\prime}}{l}{l^{\prime}}.
\end{equation*}
The commutators of these operators are
\begin{gather}
\mathfrak{R}_{1 \mu L \nu}\bigl(\widehat{L}_{1 \mu}, \widehat{Y}_{L \nu}\bigr)=\sqrt{L(L+1)} \clebsch{L}{\nu}{1}{\mu}{L}{\mu+\nu} \widehat{Y}_{L \mu+\nu},  \label{eq:13:2:118}\\
\mathfrak{R}_{L^{\prime} \nu^{\prime}}^{1 L}\bigl(\widehat{\vect{L}}_{1}, \widehat{\vect{Y}}_{L}\bigr)=-\sqrt{L(L+1)} \widehat{Y}_{L \mu+\nu} \delta_{L^{\prime} L} \delta_{\nu^{\prime} \mu+\nu}. \label{eq:13:2:119}
\end{gather}
Using Eqs. \eqref{eq:13:2:104} and \eqref{eq:13:2:105}, we find the expressions for the matrix elements of the commutators,\index{commutator!matrix elements of}\index{matrix elements!of commutators}
\begin{align}
& 
\braOket{ l^{\prime} m^{\prime}}{ \mathfrak{R}_{1 \mu L \nu}\bigl(\widehat{L}_{1 \mu}, \widehat{\vect{Y}}_{L \nu}\bigr)}{l m}=\Fact{L l}/\Fact{l^\prime}\sqrt{\frac{L(L+1)}{4 \pi}} 
\clebsch{l}{m}{L}{\mu+\nu}{l^{\prime}}{m^{\prime}} 
\clebsch{L}{\nu}{1}{\mu}{L}{\nu+\mu} 
\clebsch{l}{0}{L}{0}{l^{\prime}}{0},  \label{eq:13:2:120}\\
& 
\braOket{ l^{\prime} m^{\prime}}{ \mathfrak{R}_{L^{\prime} \nu^{\prime}}^{1 L}\bigl(\widehat{\vect{L}}_{1}, \widehat{\vect{Y}}_{L}\bigr)}{l m}=-\Fact{L l}/\Fact{l^\prime}\sqrt{\frac{L(L+1)}{4 \pi}}
\clebsch{l}{0}{L}{0}{l^{\prime}}{0} 
\clebsch{l}{m}{L}{\mu+\nu}{l^{\prime}}{m^{\prime}} \delta_{L^{\prime} L}. \label{eq:13:2:121}
\end{align}
In the case of products of the operators \(\widehat{\vect{S}}_{1}\) and \(\widehat{\vect{Y}}_{L}\) the matrix elements may be evaluated in the $lsJM$ representation:\index{representation!$lsJM$}\index{matrix elements!of products with $\widehat{\vect{S}}$}
\begin{align}
\braOket{ l^{\prime} s^{\prime} J^{\prime} M^{\prime}}{ \widehat{S}_{1 \mu} \widehat{Y}_{L \nu}}{l s J M}= & \delta_{s s^{\prime}} \Fact{J l L s}\sqrt{\frac{ s(s+1)}{4 \pi}}
\clebsch{l}{0}{L}{0}{l^{\prime}}{0} \notag\\
& \times \sum_{k}(-1)^{L+k+1} \Fact{k}
\clebsch{J}{M}{k}{\mu+\nu}{J^{\prime}}{M^{\prime}} 
\clebsch{1}{\mu}{L}{\nu}{k}{\mu+\nu}
\ninej{l}{L}{l^{\prime}}{s}{1}{s}{J}{k}{J^{\prime}}; \label{eq:13:2:122}
\end{align}

\begin{equation}
\braOket{ l^{\prime} s^{\prime} J^{\prime} M^{\prime}}{\bigl\{\widehat{\vect{Y}}_{L} \otimes \widehat{\vect{S}}_{1}\bigr\}_{L^{\prime} \nu^{\prime}}}{l s J M}=  \delta_{s s^{\prime}} \Fact{J L^\prime L l s}\sqrt{\frac{ s(s+1)}{4 \pi}} \clebsch{l}{0}{L}{0}{l^{\prime}}{0}
\clebsch{J}{M}{L^{\prime}}{\nu^{\prime}}{J^{\prime}}{M^{\prime}}
\ninej{l}{L}{l^{\prime}}{s}{1}{s}{J}{L^{\prime}}{J^{\prime}}, \label{eq:13:2:123}
\end{equation}
\begin{equation*}
\braOketred{ l^{\prime} s^{\prime} J^{\prime}}{\bigl\{\widehat{\vect{Y}}_{L} \otimes \widehat{\vect{S}}_{1}\bigr\}_{L^{\prime}}}{l s J}=\delta_{s s^{\prime}} \Fact{J L^\prime L l s J^{\prime}}\sqrt{\frac{ s(s+1)}{4 \pi}} 
\clebsch{l}{0}{L}{0}{l^{\prime}}{0}
\ninej{l}{L}{l^{\prime}}{s}{1}{s}{J}{L^{\prime}}{J^{\prime}}.
\end{equation*}
The matrix elements of products of the operators \(\widehat{\vect{J}}_{1}\) and \(\widehat{\vect{Y}}_{L}\) in the same representation may be obtained by using the relations\index{matrix elements!of products with $\widehat{\vect{J}}$}
\begin{multline}
\braOket{ l^{\prime} s^{\prime} J^{\prime} M^{\prime}}{ \widehat{J}_{1 \mu} \widehat{Y}_{L \nu}}{l s J M}=(-1)^{l^{\prime}+s+J+L} \delta_{s s^{\prime}} \Fact{L l J}\sqrt{\frac{ J^{\prime}\bigl(J^{\prime}+1\bigr)}{4 \pi}}\\
\times
\clebsch{J}{M}{L}{\nu}{J^{\prime}}{M+\nu}
\clebsch{J^{\prime}}{M+\nu}{1}{\mu}{J^{\prime}}{M^{\prime}}
\clebsch{l}{0}{L}{0}{l^{\prime}}{0}
\sixj{l}{s}{J}{J^{\prime}}{L}{l^{\prime}};  \label{eq:13:2:124}
\end{multline}

\begin{multline}
\braOket{ l^{\prime} s^{\prime} J^{\prime} M^{\prime}}{\bigl\{\widehat{\vect{J}}_{1} \otimes \widehat{\vect{Y}}_{L}\bigr\}_{L^{\prime} \nu^{\prime}}}{l s J M}=(-1)^{l+s+L^{\prime}-J^{\prime}} \delta_{s s^{\prime}} \Fact{L L^\prime l J J^\prime}\sqrt{\frac{ J^{\prime}\bigl(J^{\prime}+1\bigr)}{4 \pi}} \\
\times
\clebsch{J}{M}{L^{\prime}}{\nu^{\prime}}{J^{\prime}}{M^{\prime}}
\sixj{L}{1}{L^{\prime}}{J^{\prime}}{J}{J^{\prime}}
\clebsch{l}{0}{L}{0}{l^{\prime}}{0}
\sixj{l}{s}{J}{J^{\prime}}{L}{l^{\prime}}, \label{eq:13:2:125}
\end{multline}
\begin{equation*}
\braOketred{ l^{\prime} s^{\prime} J^{\prime}}{\bigl\{\widehat{\vect{J}}_{1} \otimes \widehat{\vect{Y}}_{L}\bigr\}_{L^{\prime}}}{l s J}=(-1)^{l+s+L^{\prime}-J^{\prime}} \delta_{s s^{\prime}} \Fact{L L^\prime l J J^\prime J^{\prime}}\sqrt{\frac{ J^{\prime}\bigl(J^{\prime}+1\bigr)}{4 \pi}}
\sixj{L}{1}{L^{\prime}}{J^{\prime}}{J}{J^{\prime}}
\clebsch{l}{0}{L}{0}{l^{\prime}}{0}
\sixj{l}{s}{J}{J^{\prime}}{L}{l^{\prime}}.
\end{equation*}

The commutators of these operators are reduced to the commutators of \(\widehat{\vect{L}}_{1}\) and \(\widehat{\vect{Y}}_{L}\). Namely,
\begin{align*}
\mathfrak{R}_{1 \mu L \nu}\bigl(\widehat{J}_{1 \mu}, \widehat{Y}_{L \nu}\bigr) & =\mathfrak{R}_{1 \mu L \nu}\bigl(\widehat{L}_{1 \mu}, \widehat{Y}_{L \nu}\bigr) \\
\mathfrak{R}_{L^{\prime} \nu^{\prime}}^{L}\bigl(\widehat{J}_{1}, \widehat{\vect{Y}}_{L}\bigr) & =\mathfrak{R}_{L^{\prime} \nu^{\prime}}^{L}\bigl(\widehat{\vect{L}}_{1}, \widehat{\vect{Y}}_{L}\bigr).
\end{align*}
The matrix elements of these commutators in the $lsJM$-representation are given by
\begin{align}
\braOket{ l^{\prime} s^{\prime} J^{\prime} M^{\prime}}{ \mathfrak{R}_{1 \mu L \nu}\bigl(\widehat{J}_{1 \mu}, \widehat{Y}_{L \nu}\bigr)}{l s J M}=&(-1)^{J+l^{\prime}+s+L} \delta_{s s^{\prime}} \Fact{L l J}\sqrt{\frac{L(L+1)}{4 \pi}}\notag \\
&\times
\clebsch{l}{0}{L}{0}{l^{\prime}}{0}
\sixj{l}{s}{J}{J^{\prime}}{L}{l^{\prime}} 
\clebsch{L}{\nu}{1}{\mu}{L}{\mu+\nu} 
\clebsch{J}{M}{L}{\mu+\nu}{J^{\prime}}{M^{\prime}} , \label{eq:13:2:126}\\
\braOket{ l^{\prime} s^{\prime} J^{\prime} M^{\prime}}{ \mathfrak{R}_{L^{\prime} \nu^{\prime}}^{1 L}\bigl(\widehat{\vect{J}}_{1}, \widehat{\vect{Y}}_{L}\bigr)}{l s J M}=&(-1)^{J+l^{\prime}+s+L+1} \delta_{s s^{\prime}} \delta_{L L^{\prime}} \Fact{L l J}\sqrt{\frac{L(L+1)}{4 \pi}}\notag \\
& \times
\clebsch{l}{0}{L}{0}{l^{\prime}}{0}
\sixj{l}{s}{J}{J^{\prime}}{L}{l^{\prime}} 
\clebsch{J}{M}{L}{\nu^{\prime}}{J^{\prime}}{M^{\prime}}. \label{eq:13:2:127}
\end{align}
\begin{equation*} 
    \braOketred{ l^{\prime} s^{\prime} J^{\prime}}{ \mathfrak{R}_{L^{\prime}}^{1 L}\bigl(\widehat{\vect{J}}_{1}, \widehat{\vect{Y}}_{L}\bigr)}{l s J}=(-1)^{J+l^{\prime}+s+L+1} \delta_{s s^{\prime}} \delta_{L L^{\prime}} \Fact{l^\prime L l J}\sqrt{\frac{L(L+1)}{4 \pi}}
\clebsch{l}{0}{L}{0}{l^{\prime}}{0}
\sixj{l}{s}{J}{J^{\prime}}{L}{l^{\prime}} .
\end{equation*}

Below we also present expressions for matrix elements of some more complex tensor products which include the spherical harmonic operator:\index{matrix elements!of complex tensor products}
\begin{multline}
\braOket{ l^{\prime} m^{\prime}}{\bigl\{\widehat{\vect{n}}_{1} \otimes\bigl(\delop_{\Omega}\bigr)_{1}\bigr\}_{k \kappa} \widehat{Y}_{L \nu}}{l m}=\sum_{L^{\prime} \nu^{\prime} l^{\prime \prime}}(-1)^{l^{\prime}+L^{\prime}+k} \Fact{L l k l^{\prime \prime}}/\Fact{l^\prime L^\prime}\sqrt{\frac{1}{4 \pi}} \\
\times 
\clebsch{l}{m}{L}{\nu}{L^{\prime}}{\nu^{\prime}} 
\clebsch{L^{\prime}}{\nu^{\prime}}{k}{\kappa}{l^{\prime}}{m^{\prime}} 
\clebsch{l}{0}{L}{0}{L^{\prime}}{0} 
\clebsch{l^{\prime\prime}}{0}{1}{0}{l^{\prime}}{0} 
\sixj{1}{L^{\prime}}{l^{\prime \prime}}{l^{\prime}}{1}{k}
\braOketred{l^{\prime \prime}}{\bigl(\delop_{\Omega}\bigr)_{1}}{ L^{\prime}},  \label{eq:13:2:128}
\end{multline}

\begin{multline}
\braOket{ l^{\prime} m^{\prime}}{\bigl\{\bigl\{\widehat{\vect{n}}_{1} \otimes\bigl(\delop_{\Omega}\bigr)_{1}\bigr\}_{k} \otimes \widehat{\vect{Y}}_{L}\bigr\}_{F \varphi}}{l m}=(-1)^{F+l+k} \Fact{L l k F}/\Fact{l^\prime}\sqrt{\frac{1}{4 \pi}} \\
 \times 
\clebsch{l}{m}{F}{\varphi}{l^{\prime}}{m^{\prime}} \sum_{L^{\prime} l^{\prime \prime}}(-1)^{L^{\prime}} \Fact{l^{\prime \prime}} 
\clebsch{l}{0}{L}{0}{L^{\prime}}{0} 
\clebsch{l^{\prime\prime}}{0}{1}{0}{l^{\prime}}{0} 
\sixj{1}{L^{\prime}}{l^{\prime \prime}}{l^{\prime}}{1}{k}
\sixj{L}{k}{F}{l^{\prime}}{1}{L^{\prime}}
\braOketred{l^{\prime \prime}}{\bigl(\delop_{\Omega}\bigr)_{1}}{ L^{\prime}}, \label{eq:13:2:129}
\end{multline}
\begin{multline*}
\braOketred{ l^{\prime}}{\bigl\{\bigl\{\widehat{\vect{n}}_{1} \otimes\bigl(\delop_{\Omega}\bigr)_{1}\bigr\}_{k} \otimes \widehat{\vect{Y}}_{L}\bigr\}_{F}}{ l}=(-1)^{F+l+k} \Fact{L l k F}\sqrt{\frac{1}{4 \pi}} \\
\times \sum_{L^{\prime} l^{\prime \prime}}(-1)^{L^{\prime}} \Fact{l^{\prime \prime}}
\clebsch{l}{0}{L}{0}{L^{\prime}}{0}
\clebsch{l^{\prime\prime}}{0}{1}{0}{l^{\prime}}{0}
\sixj{1}{L^{\prime}}{l^{\prime \prime}}{l^{\prime}}{1}{k}
\sixj{L}{k}{F}{l^{\prime}}{1}{L^{\prime}}
\braOketred{l^{\prime \prime}}{\bigl(\delop_{\Omega}\bigr)_{1}}{ L^{\prime}}.
\end{multline*}

\begin{multline}
\braOket{ l^{\prime} m^{\prime}}{\bigl\{\widehat{\vect{n}}_{1} \otimes \widehat{\vect{L}}_{1}\bigr\}_{k \kappa} \widehat{Y}_{L \nu}}{l m}=\sum_{L^{\prime} \nu^{\prime}}(-1)^{l^{\prime}+L^{\prime}+k} \Fact{L l k L^{\prime}}/\Fact{l^\prime}\sqrt{\frac{L^{\prime}\bigl(L^{\prime}+1\bigr)}{4 \pi}} \\
 \times 
\clebsch{l}{m}{L}{\nu}{L^{\prime}}{\nu^{\prime}} 
\clebsch{L^{\prime}}{\nu^{\prime}}{k}{\kappa}{l^{\prime}}{m^{\prime}} 
\clebsch{l}{0}{L}{0}{L^{\prime}}{0} 
\clebsch{L^{\prime}}{0}{1}{0}{l^{\prime}}{0}
\sixj{1}{1}{k}{l^{\prime}}{L^{\prime}}{L^{\prime}},  \label{eq:13:2:130}
\end{multline}

\begin{multline} 
\braOket{ l^{\prime} m^{\prime}}{\bigl\{\bigl\{\widehat{\vect{n}}_{1} \otimes \widehat{\vect{L}}_{1}\bigr\}_{k} \otimes \vect{Y}_{L}\bigr\}_{F \varphi}}{l m}
=(-1)^{F+l+k} \Fact{L l k F}/\Fact{l^\prime}\sqrt{\frac{1}{4 \pi}} \\
 \times 
\clebsch{l}{m}{F}{\varphi}{l^{\prime}}{m^{\prime}} 
\sum_{L^{\prime}}(-1)^{L^{\prime}} \Fact{L^\prime}\sqrt{L^{\prime}(L^{\prime}+1)}
\clebsch{l}{0}{L}{0}{L^{\prime}}{0} 
\clebsch{L^{\prime}}{0}{1}{0}{l^{\prime}}{0}
\sixj{1}{1}{k}{l^{\prime}}{L^{\prime}}{L^{\prime}}
\sixj{L}{k}{F}{l^{\prime}}{l}{L^{\prime}},  \label{eq:13:2:131}
\end{multline}
\begin{multline*}
\braOketred{ l^{\prime}}{\bigl\{\bigl\{\widehat{\vect{n}}_{1} \otimes \widehat{\vect{L}}_{1}\bigr\}_{k} \otimes \vect{Y}_{L}\bigr\}_{F}}{ l}
=(-1)^{F+l+k} \Fact{L l k F}\sqrt{\frac{1}{4 \pi}} \\
\times
\sum_{L^{\prime}}(-1)^{L^{\prime}} \Fact{L^\prime}\sqrt{L^{\prime}(L^{\prime}+1)}
\clebsch{l}{0}{L}{0}{L^{\prime}}{0}
\clebsch{L^{\prime}}{0}{1}{0}{l^{\prime}}{0}
\sixj{1}{1}{k}{l^{\prime}}{L^{\prime}}{L^{\prime}}
\sixj{L}{k}{F}{l^{\prime}}{l}{L^{\prime}}.
\end{multline*}

\begin{multline}
\braOket{ l^{\prime} m^{\prime}}{\bigl\{\bigl(\delop_{\Omega}\bigr)_{1} \otimes \widehat{\vect{L}}_{1}\bigr\}_{k \kappa} \widehat{Y}_{L \nu}}{l m}=\sum_{L^{\prime} \nu^{\prime}}(-1)^{l^{\prime}+L^{\prime}+k} \Fact{k L l}/\Fact{l^\prime}\sqrt{\frac{L^{\prime}\bigl(L^{\prime}+1\bigr)}{4 \pi}} \\
 \times 
\clebsch{l}{m}{L}{\nu}{L^{\prime}}{\nu^{\prime}} 
\clebsch{L^{\prime}}{\nu^{\prime}}{k}{\kappa}{l^{\prime}}{m^{\prime}} 
\clebsch{l}{0}{L}{0}{l^{\prime}}{0}
\sixj{1}{1}{k}{l^{\prime}}{L^{\prime}}{L^{\prime}}
\braOketred{ l^{\prime}}{\bigl(\delop_{\Omega}\bigr)_{1}}{ L^{\prime}},  \label{eq:13:2:132}
\end{multline}

\begin{multline}
\braOket{ l^{\prime} m^{\prime}}{\bigl\{\bigl\{\bigl(\delop_{\Omega}\bigr)_{1} \otimes \widehat{\vect{L}}_{1}\bigr\}_{k} \otimes \widehat{\vect{Y}}_{L}\bigr\}_{F_{\varphi}}}{l m}=(-1)^{F+l+k} \Fact{k L l F}/\Fact{l^\prime}\sqrt{\frac{1}{4 \pi}} \\
 \times 
\clebsch{l}{m}{F}{\varphi}{l^{\prime}}{m^{\prime}} \sum_{L^{\prime}}(-1)^{L^{\prime}} \Fact{L^{\prime}}\sqrt{L^{\prime}\bigl(L^{\prime}+1\bigr)}
\clebsch{l}{0}{L}{0}{l^{\prime}}{0}
\sixj{1}{1}{k}{l^{\prime}}{L^{\prime}}{L^{\prime}}
\sixj{L}{k}{F}{l^{\prime}}{l}{L^{\prime}}
\braOketred{ l^{\prime}}{\bigl(\delop_{\Omega}\bigr)_{1}}{ L^{\prime}},  \label{eq:13:2:133}
\end{multline}
\begin{multline*}
\braOketred{ l^{\prime}}{\bigl\{\bigl\{\bigl(\delop_{\Omega}\bigr)_{1} \otimes \widehat{\vect{L}}_{1}\bigr\}_{k} \otimes \widehat{\vect{Y}}_{L}\bigr\}_{F}}{ l}=(-1)^{F+l+k} \Fact{k L l F}\sqrt{\frac{1}{4 \pi}} \\
\times \sum_{L^{\prime}}(-1)^{L^{\prime}} \Fact{L^{\prime}}\sqrt{L^{\prime}\bigl(L^{\prime}+1\bigr)}
\clebsch{l}{0}{L}{0}{l^{\prime}}{0}
\sixj{1}{1}{k}{l^{\prime}}{L^{\prime}}{L^{\prime}}
\sixj{L}{k}{F}{l^{\prime}}{l}{L^{\prime}}
\braOketred{ l^{\prime}}{\bigl(\delop_{\Omega}\bigr)_{1}}{ L^{\prime}}.
\end{multline*}

\begin{multline}
\braOket{ l^{\prime} s^{\prime} J^{\prime} M^{\prime}}{\bigl\{\widehat{\vect{n}}_{1} \otimes \widehat{\vect{S}}_{1}\bigr\}_{k \kappa} \widehat{Y}_{L \nu}}{l s J M}=(-1)^{l+s+J} \delta_{s s^{\prime}} \Fact{L l J k s}\sqrt{\frac{s(s+1)}{4 \pi}} \\
 \times \sum_{J_{1} L_{1} M_{1}} \Fact{J_1 L_1} 
\clebsch{J}{M}{L}{\nu}{J_{1}}{M_{1}} 
\clebsch{J_{1}}{M_{1}}{k}{\kappa}{J^{\prime}}{M^{\prime}} 
\clebsch{L}{0}{l}{0}{L_{1}}{0} 
\clebsch{L_{1}}{0}{1}{0}{l^{\prime}}{0}
\sixj{l}{s}{J}{J_{1}}{L}{L_{1}}
\ninej{L_{1}}{1}{l^{\prime}}{s}{1}{s}{J_{1}}{k}{J^{\prime}},  \label{eq:13:2:134}
\end{multline}

\begin{multline}
\braOket{ l^{\prime} s^{\prime} J^{\prime} M^{\prime}}{\bigl\{\bigl\{\widehat{\vect{n}}_{1} \otimes \widehat{\vect{S}}_{1}\bigr\}_{k} \otimes \vect{Y}_{L}\bigr\}_{F \varphi}}{l s J M}=(-1)^{l+s-J^{\prime}-F} \delta_{s s^{\prime}} 
\clebsch{J}{M}{F}{\varphi}{J^{\prime}}{M^{\prime}} \\
 \times \Fact{L J l k F s}\sqrt{\frac{ s(s+1)}{4 \pi}} \sum_{J_{1} L_{1}}
\Fact{L_1 J_1 J_1} \\
 \times
\clebsch{L}{0}{l}{0}{L_{1}}{0}
\clebsch{L_{1}}{0}{1}{0}{l^{\prime}}{0}
\sixj{l}{s}{J}{J_{1}}{L}{L_{1}}
\sixj{L}{k}{F}{J^{\prime}}{J}{J_{1}}
\ninej{L_{1}}{1}{l^{\prime}}{s}{1}{s}{J_{1}}{k}{J^{\prime}},  \label{eq:13:2:135}
\end{multline}
\begin{multline*}
\braOketred{ l^{\prime} s^{\prime} J^{\prime}}{\bigl\{\bigl\{\widehat{\vect{n}}_{1} \otimes \widehat{\vect{S}}_{1}\bigr\}_{k} \otimes \vect{Y}_{L}\bigr\}_{F}}{l s J}=(-1)^{l+s-J^{\prime}-F} \delta_{s s^{\prime}} \Fact{L J l k F s J^{\prime}}\sqrt{\frac{ s(s+1)}{4 \pi}} \\
\times \sum_{J_{1} L_{1}}
\Fact{L_1 J_1 J_1}
\clebsch{L}{0}{l}{0}{L_{1}}{0}
\clebsch{L_{1}}{0}{1}{0}{l^{\prime}}{0}
\sixj{l}{s}{J}{J_{1}}{L}{L_{1}}
\sixj{L}{k}{F}{J^{\prime}}{J}{J_{1}}
\ninej{L_{1}}{1}{l^{\prime}}{s}{1}{s}{J_{1}}{k}{J^{\prime}}.
\end{multline*}

\begin{multline}
    \braOket{l's'J'M'}{\bigl\{\bigl(\delop_{\Omega}\bigr)_{1}\otimes\widehat{\vect{S}}_1\bigr\}_{k\kappa}\widehat{Y}_{L\nu}}{lsJM}=(-1)^{l+s+J}\delta_{ss'}
    \Fact{L l J k s}\sqrt{\frac{s(s+1)}{4\pi}}\\
 \times \sum_{J_{1} L_{1} M_{1}} \Fact{J_{1}} 
\clebsch{J}{M}{L}{\nu}{J_{1}}{M_{1}} 
\clebsch{J_{1}}{M_{1}}{k}{\kappa}{J^{\prime}}{M^{\prime}} 
\clebsch{L}{0}{l}{0}{L_{1}}{0}
\sixj{l}{s}{J}{J_{1}}{L}{L_{1}}
\ninej{L_{1}}{1}{l^{\prime}}{s}{1}{s}{J_{1}}{k}{J^{\prime}}
\braOketred{ l^{\prime}}{\bigl(\delop_{\Omega}\bigr)_{1}}{L_{1}}, \label{eq:13:2:136}
\end{multline}

\begin{multline}
\braOket{ l^{\prime} s^{\prime} J^{\prime} M^{\prime}}{\bigl\{\bigl\{\bigl(\delop_{\Omega}\bigr)_{1} \otimes \widehat{\vect{S}}_{1}\bigr\}_{k} \otimes \widehat{\vect{Y}}_{L}\bigr\}_{F_{\varphi}}}{l s J M}=(-1)^{l+s-J^{\prime}-F} \delta_{s s^{\prime}}
\clebsch{J}{M}{F}{\varphi}{J^{\prime}}{M^{\prime}} \\
 \times \Fact{L l J k F s}\sqrt{\frac{s(s+1)}{4 \pi}} \sum_{J_{1} L_{1}}(2 J_1+1)
\clebsch{L}{0}{l}{0}{L_{1}}{0} \\
 \times
\sixj{l}{s}{J}{J_{1}}{L}{L_{1}}
\sixj{L}{k}{F}{J^{\prime}}{J}{J_{1}}
\ninej{L_{1}}{1}{l^{\prime}}{s}{1}{s}{J_{1}}{k}{J^{\prime}}
\braOketred{ l^{\prime}}{\bigl(\delop_{\Omega}\bigr)_{1}}{ L_{1}},  \label{eq:13:2:137}
\end{multline}
\begin{multline*}
\braOketred{ l^{\prime} s^{\prime} J^{\prime}}{\bigl\{\bigl\{\bigl(\delop_{\Omega}\bigr)_{1} \otimes \widehat{\vect{S}}_{1}\bigr\}_{k} \otimes \widehat{\vect{Y}}_{L}\bigr\}_{F}}{l s J}=(-1)^{l+s-J^{\prime}-F} \delta_{s s^{\prime}} \Fact{L l J k F s J^{\prime}}\sqrt{\frac{s(s+1)}{4 \pi}} \\
\times \sum_{J_{1} L_{1}}(2 J_1+1)
\clebsch{L}{0}{l}{0}{L_{1}}{0}
\sixj{l}{s}{J}{J_{1}}{L}{L_{1}}
\sixj{L}{k}{F}{J^{\prime}}{J}{J_{1}}
\ninej{L_{1}}{1}{l^{\prime}}{s}{1}{s}{J_{1}}{k}{J^{\prime}}
\braOketred{ l^{\prime}}{\bigl(\delop_{\Omega}\bigr)_{1}}{ L_{1}}.
\end{multline*}

\begin{multline}
\braOket{ l^{\prime} s^{\prime} J^{\prime} M^{\prime}}{\bigl\{\widehat{\vect{n}}_{1} \otimes \widehat{\vect{J}}_{1}\bigr\}_{k \kappa} \widehat{Y}_{L \nu}}{l s J M}=(-1)^{2 s+J-J^{\prime}+k+L} \delta_{s s^{\prime}} \\
 \times \Fact{L l J k}\sqrt{\frac{1}{4 \pi}} \sum_{L_{1} J_{1} M_{1}} \sqrt{J_{1}\bigl(J_{1}+1\bigr)}\Fact{L_1 J_1 J_1} \\
 \times 
\clebsch{J}{M}{L}{\nu}{J_{1}}{M_{1}} 
\clebsch{J_{1}}{M_{1}}{k}{\kappa}{J^{\prime}}{M^{\prime}} 
\clebsch{L}{0}{l}{0}{L_{1}}{0}
\clebsch{L_{1}}{0}{1}{0}{l^{\prime}}{0}
\sixj{l}{s}{J}{J_{1}}{L}{L_{1}}
\sixj{1}{1}{k}{J^{\prime}}{J_{1}}{J_{1}}
\sixj{L_{1}}{s}{J_{1}}{J^{\prime}}{1}{l^{\prime}},  \label{eq:13:2:138}
\end{multline}

\begin{multline}
\braOket{ l^{\prime} s^{\prime} J^{\prime} M^{\prime}}{\bigl\{\bigl\{\widehat{\vect{n}}_{1} \otimes \widehat{\vect{J}}_{1}\bigr\}_{k} \otimes \widehat{\vect{Y}}_{L}\bigr\}_{F \varphi}}{l s J M}=(-1)^{F+k+L} \delta_{s s^{\prime}}
\clebsch{J}{M}{F}{\varphi}{J^{\prime}}{M^{\prime}} \\
 \times \Fact{F L l J k}\sqrt{\frac{1}{4 \pi}} \sum_{L_{1} J_{1}} \sqrt{J_{1}\bigl(J_{1}+1\bigr)}\Fact{L_{1}} \\
 \times(2 J_{1}+1)^{\frac{3}{2}}
\clebsch{L}{0}{l}{0}{L_{1}}{0}
\clebsch{L_{1}}{0}{1}{0}{l^{\prime}}{0}
\sixj{l}{s}{J}{J_{1}}{L}{L_{1}}
\sixj{1}{1}{k}{J^{\prime}}{J_{1}}{J_{1}}
\sixj{L_{1}}{s}{J_{1}}{J^{\prime}}{1}{l^{\prime}}
\sixj{L}{k}{F}{J^{\prime}}{J}{J_{1}}, \label{eq:13:2:139}
\end{multline}
\begin{multline*}
\braOketred{ l^{\prime} s^{\prime} J^{\prime}}{\bigl\{\bigl\{\widehat{\vect{n}}_{1} \otimes \widehat{\vect{J}}_{1}\bigr\}_{k} \otimes \widehat{\vect{Y}}_{L}\bigr\}_{F}}{l s J}=(-1)^{F+k+L} \delta_{s s^{\prime}}
\Fact{F L l J k J^{\prime}}\sqrt{\frac{1}{4 \pi}} \\
\times \sum_{L_{1} J_{1}} \sqrt{J_{1}\bigl(J_{1}+1\bigr)}\Fact{L_{1}}(2 J_{1}+1)^{\frac{3}{2}}
\clebsch{L}{0}{l}{0}{L_{1}}{0}
\clebsch{L_{1}}{0}{1}{0}{l^{\prime}}{0}
\sixj{l}{s}{J}{J_{1}}{L}{L_{1}}
\sixj{1}{1}{k}{J^{\prime}}{J_{1}}{J_{1}}
\sixj{L_{1}}{s}{J_{1}}{J^{\prime}}{1}{l^{\prime}}
\sixj{L}{k}{F}{J^{\prime}}{J}{J_{1}}.
\end{multline*}

\begin{multline}
\braOket{ l^{\prime} s^{\prime} J^{\prime} M^{\prime}}{\bigl\{\bigl(\delop_{\Omega}\bigr)_{1} \otimes \widehat{\vect{J}}_{1}\bigr\}_{k\kappa} \widehat{Y}_{L\nu}}{l s J M}=(-1)^{l+l^{\prime}-J-J^{\prime}+k+1}\delta_{ss'} \Fact{L l J k}\frac{1}{\sqrt{4\pi}}
\\
\times
 \sum_{L_{1} J_{1}} \sqrt{J_{1}(J_{1}+1)}\Fact{J_{1}}^2 
 \clebsch{J}{M}{L}{\nu}{J_1}{M_1}
 \clebsch{J_1}{M_1}{k}{\kappa}{J'}{M'}
\clebsch{L}{0}{l}{0}{L_{1}}{0}
\braOketred{l^{\prime}}{\bigl(\delop_{\Omega}\bigr)_{1}}{ L_{1}}
\sixj{l}{s}{J}{J_{1}}{L}{L_{1}}
\sixj{J_{1}}{J^{\prime}}{1}{l^{\prime}}{L_{1}}{s}
\sixj{1}{1}{k}{J^{\prime}}{J_{1}}{J_{1}},  \label{eq:13:2:140}
\end{multline}

\begin{multline}
\braOket{ l^{\prime} s^{\prime} J^{\prime} M^{\prime}}{\bigl\{\bigl\{\bigl(\delop_{\Omega}\bigr)_{1} \otimes \widehat{\vect{J}}_{1}\bigr\}_{k} \otimes \widehat{\vect{Y}}_{L}\bigr\}_{F \varphi}}{l s J M}=(-1)^{l+l^{\prime}+F+k+1}
\clebsch{J}{M}{F}{\varphi}{J^{\prime}}{M^{\prime}} \\
\times \Fact{L l J k F}\sqrt{\frac{1}{4 \pi}} \sum_{L_{1} J_{1}} \sqrt{J_{1}\bigl(J_{1}+1\bigr)}(2 J_{1}+1)^{\frac{3}{2}} 
\clebsch{L}{0}{l}{0}{L_{1}}{0}  \\
\times
\braOketred{l^{\prime}}{\bigl(\delop_{\Omega}\bigr)_{1}}{ L_{1}}
\sixj{l}{s}{J}{J_{1}}{L}{L_{1}}
\sixj{J_{1}}{J^{\prime}}{1}{l^{\prime}}{L_{1}}{s}
\sixj{1}{1}{k}{J^{\prime}}{J_{1}}{J_{1}}
\sixj{L}{k}{F}{J^{\prime}}{J}{J_{1}},  \label{eq:13:2:141}
\end{multline}
\begin{multline*}
\braOketred{ l^{\prime} s^{\prime} J^{\prime}}{\bigl\{\bigl\{\bigl(\delop_{\Omega}\bigr)_{1} \otimes \widehat{\vect{J}}_{1}\bigr\}_{k} \otimes \widehat{\vect{Y}}_{L}\bigr\}_{F}}{l s J}=(-1)^{l+l^{\prime}+F+k+1}
\Fact{L l J k F J^{\prime}}\sqrt{\frac{1}{4 \pi}} \\
\times \sum_{L_{1} J_{1}} \sqrt{J_{1}\bigl(J_{1}+1\bigr)}(2 J_{1}+1)^{\frac{3}{2}}
\clebsch{L}{0}{l}{0}{L_{1}}{0}
\braOketred{l^{\prime}}{\bigl(\delop_{\Omega}\bigr)_{1}}{ L_{1}}
\sixj{l}{s}{J}{J_{1}}{L}{L_{1}}
\sixj{J_{1}}{J^{\prime}}{1}{l^{\prime}}{L_{1}}{s}
\sixj{1}{1}{k}{J^{\prime}}{J_{1}}{J_{1}}
\sixj{L}{k}{F}{J^{\prime}}{J}{J_{1}}.
\end{multline*}

\begin{multline}
\braOket{l^{\prime} s^{\prime} J^{\prime} M^{\prime}}{\bigl\{\widehat{\vect{L}}_{1} \otimes \widehat{\vect{S}}_{1}\bigr\}_{k \kappa} \widehat{Y}_{L \nu}}{ l s J M}=
(-1)^{l+s+J^{\prime}} \delta_{ss^{\prime}}
\clebsch{L}{0}{l}{0}{l^{\prime}}{0}
\Fact{L l k  s l^\prime J}
\sqrt{\frac{ s(s+1) l^{\prime}(l^{\prime}+1)}{4 \pi}} \\
\times \sum_{J_{1} M_{1}} \Fact{J_{1}}
\clebsch{J}{M}{L}{\nu}{J_{1}}{M_{1}} 
\clebsch{J_{1}}{M_{1}}{k}{\kappa}{J^{\prime}}{M^{\prime}}
\sixj{l}{s}{J}{J_{1}}{L}{l^{\prime}}
\ninej{l^{\prime}}{1}{l^{\prime}}{s}{1}{s}{J_{1}}{k}{J^{\prime}}, \label{eq:13:2:142}
\end{multline}

\begin{multline}
\braOket{ l^{\prime} s^{\prime} J^{\prime} M^{\prime}}{\bigl\{\bigl\{\widehat{\vect{L}}_{1} \otimes \widehat{\vect{S}}_{1}\bigr\}_{k} \otimes \widehat{\vect{Y}}_{L}\bigr\}_{F \varphi}}{l s J M}=(-1)^{l+s-J^{\prime}-F} \delta_{s s^{\prime}} 
\clebsch{J}{M}{F}{\varphi}{J^{\prime}}{M^{\prime}} 
\clebsch{L}{0}{l}{0}{l^{\prime}}{0} \\
 \times \Fact{L l k F s l^{\prime} J}\sqrt{\frac{s(s+1) l^{\prime}\bigl(l^{\prime}+1\bigr)}{4 \pi}} \\
 \times \sum_{J_{1}}(2 J_{1}+1)
\sixj{l}{s}{J}{J_{1}}{L}{l^{\prime}}
\sixj{L}{k}{F}{J^{\prime}}{J}{J_{1}}
\ninej{l^{\prime}}{1}{l^{\prime}}{s}{1}{s}{J_{1}}{k}{J^{\prime}},  \label{eq:13:2:143}
\end{multline}
\begin{multline*}
\braOketred{ l^{\prime} s^{\prime} J^{\prime}}{\bigl\{\bigl\{\widehat{\vect{L}}_{1} \otimes \widehat{\vect{S}}_{1}\bigr\}_{k} \otimes \widehat{\vect{Y}}_{L}\bigr\}_{F}}{l s J}=(-1)^{l+s-J^{\prime}-F} \delta_{s s^{\prime}}
\clebsch{L}{0}{l}{0}{l^{\prime}}{0}
\Fact{L l k F s l^{\prime} J J^{\prime}}\sqrt{\frac{s(s+1) l^{\prime}\bigl(l^{\prime}+1\bigr)}{4 \pi}} \\
\times \sum_{J_{1}}(2 J_{1}+1)
\sixj{l}{s}{J}{J_{1}}{L}{l^{\prime}}
\sixj{L}{k}{F}{J^{\prime}}{J}{J_{1}}
\ninej{l^{\prime}}{1}{l^{\prime}}{s}{1}{s}{J_{1}}{k}{J^{\prime}}.
\end{multline*}

\begin{multline}
 \braOket{ l^{\prime} s^{\prime} J^{\prime} M^{\prime}}{\bigl\{\widehat{\vect{L}}_{1} \otimes \widehat{\vect{J}}_{1}\bigr\}_{k \kappa} \widehat{Y}_{L \nu}}{l_{s} J M}=(-1)^{l+l^{\prime}+J+J^{\prime}+k+1} \delta_{s s^{\prime}} 
\clebsch{L}{0}{l}{0}{l^{\prime}}{0} \\
\times \Fact{L l J k l^{\prime}}\sqrt{\frac{l^{\prime}\bigl(l^{\prime}+1\bigr)}{4 \pi}} \sum_{J_{1} M_{1}}(2 J_{1}+1) \sqrt{J_{1}\bigl(J_{1}+1\bigr)} 
\clebsch{J}{M}{L}{\nu}{J_{1}}{M_{1}} 
\clebsch{J_{1}}{M_{1}}{k}{\kappa}{J^{\prime}}{M^{\prime}} \\
 \times
\sixj{l}{s}{J}{J_{1}}{L}{l^{\prime}}
\sixj{1}{1}{k}{J^{\prime}}{J_{1}}{J_{1}}
\sixj{l^{\prime}}{s}{J_{1}}{J^{\prime}}{1}{l^{\prime}},  \label{eq:13:2:144}
\end{multline}

\begin{multline}
\braOket{ l^{\prime} s^{\prime} J^{\prime} M^{\prime}}{\bigl\{\bigl\{\widehat{\vect{L}}_{1} \otimes \widehat{\vect{J}}_{1}\bigr\}_{k} \otimes \widehat{\vect{Y}}_{L}\bigr\}_{F \varphi}}{l s J M}=(-1)^{l+l^{\prime}-F+k+1} \delta_{s s^{\prime}} 
\clebsch{L}{0}{l}{0}{l^{\prime}}{0} 
\clebsch{J}{M}{F}{\varphi}{J^{\prime}}{M^{\prime}} \\
 \times \Fact{L l J k F l^{\prime}}\sqrt{\frac{ l^{\prime}\bigl(l^{\prime}+1\bigr)}{4 \pi}} \sum_{J_{1}} \sqrt{J_{1}\bigl(J_{1}+1\bigr)}(2 J_{1}+1)^{\frac{3}{2}} \\
 \times
\sixj{l}{s}{J}{J_{1}}{L}{l^{\prime}}
\sixj{1}{1}{k}{J^{\prime}}{J_{1}}{J_{1}}
\sixj{l^{\prime}}{s}{J_{1}}{J^{\prime}}{1}{l^{\prime}}
\sixj{L}{k}{F}{J^{\prime}}{J}{J_{1}},  \label{eq:13:2:145}
\end{multline}
\begin{multline*}
\braOketred{ l^{\prime} s^{\prime} J^{\prime}}{\bigl\{\bigl\{\widehat{\vect{L}}_{1} \otimes \widehat{\vect{J}}_{1}\bigr\}_{k} \otimes \widehat{\vect{Y}}_{L}\bigr\}_{F}}{l s J}=(-1)^{l+l^{\prime}-F+k+1} \delta_{s s^{\prime}}
\clebsch{L}{0}{l}{0}{l^{\prime}}{0}
\Fact{L l J k F l^{\prime} J^{\prime}}\sqrt{\frac{ l^{\prime}\bigl(l^{\prime}+1\bigr)}{4 \pi}} \\
\times \sum_{J_{1}} \sqrt{J_{1}\bigl(J_{1}+1\bigr)}(2 J_{1}+1)^{\frac{3}{2}}
\sixj{l}{s}{J}{J_{1}}{L}{l^{\prime}}
\sixj{1}{1}{k}{J^{\prime}}{J_{1}}{J_{1}}
\sixj{l^{\prime}}{s}{J_{1}}{J^{\prime}}{1}{l^{\prime}}
\sixj{L}{k}{F}{J^{\prime}}{J}{J_{1}}.
\end{multline*}

\begin{multline}
\braOket{ l^{\prime} s^{\prime} J^{\prime} M^{\prime}}{\bigl\{\widehat{\vect{S}}_{1} \otimes \widehat{\vect{J}}_{1}\bigr\}_{k \kappa} \widehat{Y}_{L \nu}}{l s J M}=(-1)^{J+L+k+1} \delta_{s s^{\prime}} 
\clebsch{L}{0}{l}{0}{l^{\prime}}{0} \\
 \times \Fact{L l J k s}\sqrt{\frac{s(s+1)}{4 \pi}} \sum_{J_{1} M_{1}} 
\clebsch{J}{M}{L}{\nu}{J_{1}}{M_{1}} 
\clebsch{J_{1}}{M_{1}}{k}{\kappa}{J^{\prime}}{M^{\prime}} \sqrt{J_{1}\bigl(J_{1}+1\bigr)}(2 J_{1}+1) \\
 \times(-1)^{J_{1}}
\sixj{l}{s}{J}{J_{1}}{L}{l^{\prime}}
\sixj{1}{1}{k}{J^{\prime}}{J_{1}}{J_{1}}
\sixj{s}{l^{\prime}}{J_{1}}{J^{\prime}}{1}{s},  \label{eq:13:2:146}
\end{multline}
\begin{multline}
\braOket{ l^{\prime} s^{\prime} J^{\prime} M^{\prime}}{\bigl\{\bigl\{\widehat{\vect{S}}_{1} \otimes \widehat{\vect{J}}_{1}\bigr\}_{k} \otimes \widehat{\vect{Y}}_{L}\bigr\}_{F \varphi}}{l s J M}=(-1)^{L+F-J^{\prime}+k+1} \delta_{s s^{\prime}} 
\clebsch{L}{0}{l}{0}{l^{\prime}}{0} 
\clebsch{J}{M}{F}{\varphi}{J^{\prime}}{M^{\prime}} \\
 \times \Fact{L l J k F s}\sqrt{\frac{s(s+1)}{4 \pi}} \sum_{J_{1}} \sqrt{J_{1}\bigl(J_{1}+1\bigr)}(2 J_{1}+1)^{\frac{3}{2}} \\
\times(-1)^{J_{1}}
\sixj{l}{s}{J}{J_{1}}{L}{l^{\prime}}
\sixj{1}{1}{k}{J^{\prime}}{J_{1}}{J_{1}}
\sixj{s}{l^{\prime}}{J_{1}}{J^{\prime}}{1}{s}
\sixj{L}{k}{F}{J^{\prime}}{J}{J_{1}} , \label{eq:13:2:147}
\end{multline}
\begin{multline*}
\braOketred{ l^{\prime} s^{\prime} J^{\prime}}{\bigl\{\bigl\{\widehat{\vect{S}}_{1} \otimes \widehat{\vect{J}}_{1}\bigr\}_{k} \otimes \widehat{\vect{Y}}_{L}\bigr\}_{F}}{l s J}=(-1)^{L+F-J^{\prime}+k+1} \delta_{s s^{\prime}}
\clebsch{L}{0}{l}{0}{l^{\prime}}{0}
\Fact{L l J k F s J^{\prime}}\sqrt{\frac{s(s+1)}{4 \pi}} \\
\times \sum_{J_{1}} \sqrt{J_{1}\bigl(J_{1}+1\bigr)}(2 J_{1}+1)^{\frac{3}{2}}(-1)^{J_{1}}
\sixj{l}{s}{J}{J_{1}}{L}{l^{\prime}}
\sixj{1}{1}{k}{J^{\prime}}{J_{1}}{J_{1}}
\sixj{s}{l^{\prime}}{J_{1}}{J^{\prime}}{1}{s}
\sixj{L}{k}{F}{J^{\prime}}{J}{J_{1}}.
\end{multline*}

The commutators of the operators in Eqs. \eqref{eq:13:2:128}-\eqref{eq:13:2:147} may be represented in a more general form as
\begin{gather}
\mathfrak{R}_{k \kappa, L \nu}\bigl(\bigl\{\widehat{\vect{P}}_{a} \otimes \widehat{\vect{Q}}_{b}\bigr\}_{k \kappa}, \widehat{Y}_{L \nu}\bigr)=\sum_{\alpha \beta} 
\clebsch{a}{\alpha}{b}{\beta}{k}{\kappa}\bigl\{\mathfrak{R}_{a \alpha, L \nu}\bigl(\widehat{P}_{a \alpha}, \widehat{Y}_{L \nu}\bigr) \widehat{Q}_{b \beta}+\widehat{P}_{a \alpha} \mathfrak{R}_{b \beta, L \nu}\bigl(\widehat{Q}_{b \beta}, \widehat{Y}_{L \nu}\bigr)\bigr\},  \label{eq:13:2:148}
\end{gather}
\begin{gather}
\mathfrak{R}_{F \varphi}^{k L}\bigl(\bigl\{\widehat{\vect{P}}_{a} \otimes \widehat{\vect{Q}}_{b}\bigr\}_{k}, \widehat{\vect{Y}}_{L}\bigr)=\sum_{\alpha \beta \kappa \nu} 
\clebsch{a}{\alpha}{b}{\beta}{k}{\kappa} 
\clebsch{k}{\kappa}{L}{\nu}{F}{\varphi}\bigl\{\mathfrak{R}_{a \alpha, L \nu}\bigl(\widehat{P}_{a \alpha}, \widehat{Y}_{L \nu}\bigr) \widehat{Q}_{b \beta}+\widehat{P}_{a \alpha} \mathfrak{R}_{b \beta L \nu}\bigl(\widehat{Q}_{b \beta}, \widehat{Y}_{L \nu}\bigr)\bigr\}. \label{eq:13:2:149}
\end{gather}
The evaluation of matrix elements of these commutators may be performed using the foregoing equations.

\subsection{Matrix Elements of Some Scalar and Vector Products}\label{sec:13.2.9}\index{matrix elements!of scalar and vector products}\index{scalar product}\index{vector product}
Using Eq. \eqref{eq:13:2:33}, one can easily obtain matrix elements of scalar and vector products of the operators \(\widehat{\vect{n}}\) and \(\delop_{\Omega}\)
\begin{gather}
\braOket{l^{\prime} m^{\prime}}{\bigl(\widehat{\vect{n}} \cdot \delop_{\Omega}\bigr)}{l m}=0,  \label{eq:13:2:150}\\
\braOket{ l^{\prime} m^{\prime}}{\bigl(\widehat{\vect{n}} \times\delop_{\Omega}\bigr)_{\mu}}{l m}=i \sqrt{l(l+1)} 
\clebsch{l}{m}{1}{\mu}{l^{\prime}}{m^{\prime}} \delta_{l l^{\prime}}. \label{eq:13:2:151}
\end{gather}
\begin{equation*}
\braOketred{ l^{\prime}}{\bigl(\widehat{\vect{n}} \times\delop_{\Omega}\bigr)}{l}=i \sqrt{l(l+1)} \Fact{l^{\prime}} \delta_{l l^{\prime}}.
\end{equation*}
Equation \eqref{eq:13:2:150} follows from the orthogonality of the vectors in question. Equation \eqref{eq:13:2:151} may also be obtained by taking into account the fact that
\begin{equation*}
\left[\widehat{\vect{n}} \times \delop_{\Omega}\right]=i \widehat{\vect{L}} .
\end{equation*}

Matrix elements of scalar and vector products of the operator \(\widehat{\vect{J}}\) and the operators \(\widehat{\vect{n}}\) and \(\delop_{\Omega}\) may be derived from Eqs. \eqref{eq:13:2:52}-\eqref{eq:13:2:56} and \eqref{eq:13:2:61}-\eqref{eq:13:2:65}
\begin{gather}
\braOket{l^{\prime} s^{\prime} J^{\prime} M^{\prime}}{(\widehat{\vect{n}} \cdot \widehat{\vect{J}})}{ l s J M}=\delta_{s^{\prime} s} \delta_{J^{\prime} J} \delta_{M^{\prime} M}(-1)^{s+l^{\prime}+J+1} \Fact{J l}\sqrt{J(J+1)} 
\clebsch{l}{0}{1}{0}{l^{\prime}}{0}
\sixj{l}{s}{J}{J}{1}{l^{\prime}},  \label{eq:13:2:152}\\
\braOketred{l^{\prime} s^{\prime} J^{\prime}}{(\widehat{\vect{n}} \cdot \widehat{\vect{J}})}{ l s J}=\delta_{s^{\prime} s} \delta_{J^{\prime} J}(-1)^{s+l^{\prime}+J+1} \Fact{J J l}\sqrt{J(J+1)}
\clebsch{l}{0}{1}{0}{l^{\prime}}{0}
\sixj{l}{s}{J}{J}{1}{l^{\prime}} ,\notag
\end{gather}
and
\begin{multline}
\braOket{ l^{\prime} s^{\prime} J^{\prime} M^{\prime}}{[\widehat{\vect{n}} \times \widehat{\vect{J}}]_{\kappa}}{l s J M}=\frac{i}{2}(-1)^{J+s+l^{\prime}}\bigl\{J^{\prime}\bigl(J^{\prime}+1\bigr)-J(J+1)-2\bigr\} \Fact{J l}  \\
\clebsch{l}{0}{1}{0}{l^{\prime}}{0}
\sixj{l}{s}{J}{J^{\prime}}{1}{l^{\prime}} 
\clebsch{J}{M}{1}{\kappa}{J^{\prime}}{M^{\prime}}, \label{eq:13:2:153}
\end{multline}
\begin{equation*}
\braOketred{ l^{\prime} s^{\prime} J^{\prime}}{[\widehat{\vect{n}} \times \widehat{\vect{J}}]}{l s J}=\frac{i}{2}(-1)^{J+s+l^{\prime}}\bigl\{J^{\prime}\bigl(J^{\prime}+1\bigr)-J(J+1)-2\bigr\} \Fact{J l J^{\prime}} 
\clebsch{l}{0}{1}{0}{l^{\prime}}{0}
\sixj{l}{s}{J}{J^{\prime}}{1}{l^{\prime}}.
\end{equation*}
The scalar product of the operators \(\widehat{\vect{n}}\) and \(\widehat{\vect{J}}\) is commutative. At the same time, the vector product of these operators is not commutative, i.e.,
\begin{gather}
\braOket{ l^{\prime} s^{\prime} J^{\prime} M^{\prime}}{\left[ \widehat{\vect{J}} \times \widehat{\vect{n}}\right]_{\kappa}}{l s J M} \notag\\
=\frac{i}{2}(-1)^{J+s+l^{\prime}+1}\bigl\{J^{\prime}\bigl(J^{\prime}+1\bigr)-J(J+1)+2\bigr\} \Fact{J l} 
\clebsch{l}{0}{1}{0}{l^{\prime}}{0}
\sixj{l}{s}{J}{J^{\prime}}{1}{l^{\prime}}
\clebsch{J}{M}{1}{\kappa}{J^{\prime}}{M^{\prime}} , \label{eq:13:2:154}
\end{gather}
\begin{equation*}
\braOketred{ l^{\prime} s^{\prime} J^{\prime}}{\left[ \widehat{\vect{J}} \times \widehat{\vect{n}}\right]}{l s J}
=\frac{i}{2}(-1)^{J+s+l^{\prime}+1}\bigl\{J^{\prime}\bigl(J^{\prime}+1\bigr)-J(J+1)+2\bigr\} \Fact{J l J^{\prime}} 
\clebsch{l}{0}{1}{0}{l^{\prime}}{0}
\sixj{l}{s}{J}{J^{\prime}}{1}{l^{\prime}}.
\end{equation*}
In a similar manner we get
\begin{gather}
\braOket{ l^{\prime} s^{\prime} J^{\prime} M^{\prime}}{\bigl(\delop_{\Omega} \cdot \widehat{\vect{J}}\bigr)}{l s J M}=\delta_{J^{\prime} J} \delta_{s^{\prime} s} \delta_{M^{\prime} M}(-1)^{s+l^{\prime}+J} \Fact{J l^\prime} \sqrt{ J(J+1)}\notag \\
\times\left[l \sqrt{\frac{l+1}{2 l+3}} \delta_{l^{\prime} l+1}+(l+1) \sqrt{\frac{l}{2 l-1}} \delta_{l^{\prime} l-1}\right]
\sixj{l}{s}{J}{J}{1}{l^{\prime}},  \label{eq:13:2:155}\\
\braOket{ l^{\prime} s^{\prime} J^{\prime} M^{\prime}}{\left[\delop_{\Omega} \times \widehat{\vect{J}}\right]_{\kappa}}{l s J M}=\frac{i}{2}(-1)^{J+s+l^{\prime}+1}\Fact{J l^\prime} \bigl\{J^{\prime}\bigl(J^{\prime}+1\bigr)-J(J+1)-2\bigr\} \notag\\
\times\left[l \sqrt{\frac{l+1}{2 l+3}} \delta_{l^{\prime} l+1}+(l+1) \sqrt{\frac{l}{2 l-1}} \delta_{l^{\prime} l-1}\right]
\sixj{l}{s}{J}{J^{\prime}}{1}{l^{\prime}} 
\clebsch{J}{M}{1}{\kappa}{J^{\prime}}{M^{\prime}} , \label{eq:13:2:156}
\end{gather}
\begin{multline*}
\braOketred{ l^{\prime} s^{\prime} J^{\prime}}{\bigl(\delop_{\Omega} \cdot \widehat{\vect{J}}\bigr)}{l s J}=\delta_{J^{\prime} J} \delta_{s^{\prime} s}(-1)^{s+l^{\prime}+J} \Fact{J J l^\prime} \sqrt{ J(J+1)} \\
\times\left[l \sqrt{\frac{l+1}{2 l+3}} \delta_{l^{\prime} l+1}+(l+1) \sqrt{\frac{l}{2 l-1}} \delta_{l^{\prime} l-1}\right]
\sixj{l}{s}{J}{J}{1}{l^{\prime}},
\end{multline*}
\begin{multline*}
\braOketred{ l^{\prime} s^{\prime} J^{\prime}}{\left[\delop_{\Omega} \times \widehat{\vect{J}}\right]}{l s J}=\frac{i}{2}(-1)^{J+s+l^{\prime}+1}\Fact{J l^\prime}\Fact{J^{\prime}} \bigl\{J^{\prime}\bigl(J^{\prime}+1\bigr)-J(J+1)-2\bigr\} \\
\times\left[l \sqrt{\frac{l+1}{2 l+3}} \delta_{l^{\prime} l+1}+(l+1) \sqrt{\frac{l}{2 l-1}} \delta_{l^{\prime} l-1}\right]
\sixj{l}{s}{J}{J^{\prime}}{1}{l^{\prime}}.
\end{multline*}
As in the foregoing case, the scalar product of the vector operators \(\delop_{\Omega}\) and \(\widehat{\vect{J}}\) is commutative, while the vector product is not commutative. Therefore,
\begin{gather}
\braOket{ l^{\prime} s^{\prime} J^{\prime} M^{\prime}}{ \left[\widehat{\vect{J}} \times \delop_{\Omega}\right]_{\kappa}}{l s J M}=\frac{i}{2}(-1)^{J+s+l^{\prime}}\bigl\{J^{\prime}\bigl(J^{\prime}+1\bigr)-J(J+1)+2\bigr\} \Fact{J l^\prime} \notag \\
\times\left[l \sqrt{\frac{l+1}{2 l+3}} \delta_{l^{\prime} l+1}+(l+1) \sqrt{\frac{l}{2 l-1}} \delta_{l^{\prime} l-1}\right]
\sixj{l}{s}{J}{J^{\prime}}{1}{l^{\prime}} 
\clebsch{J}{M}{1}{\kappa}{J^{\prime}}{M^{\prime}}, \label{eq:13:2:157}
\end{gather}
\begin{multline*}
\braOketred{ l^{\prime} s^{\prime} J^{\prime}}{ \left[\widehat{\vect{J}} \times \delop_{\Omega}\right]}{l s J}=\frac{i}{2}(-1)^{J+s+l^{\prime}}\bigl\{J^{\prime}\bigl(J^{\prime}+1\bigr)-J(J+1)+2\bigr\} \Fact{J l^\prime J^{\prime}} \\
\times\left[l \sqrt{\frac{l+1}{2 l+3}} \delta_{l^{\prime} l+1}+(l+1) \sqrt{\frac{l}{2 l-1}} \delta_{l^{\prime} l-1}\right]
\sixj{l}{s}{J}{J^{\prime}}{1}{l^{\prime}}.
\end{multline*}
The matrix elements of scalar and vector products of the operator \(\widehat{\vect{L}}\) and the operators \(\widehat{\vect{n}}\) and \(\delop_{\Omega}\) may be obtained from Eqs. \eqref{eq:13:2:73}-\eqref{eq:13:2:77} and \eqref{eq:13:2:82}-\eqref{eq:13:2:85}. In the case of scalar products we get
\begin{align}
\braOket{ l^{\prime} m^{\prime}}{(\widehat{\vect{n}} \cdot \widehat{\vect{L}})}{l m} & =0, \notag\\
\braOket{ l^{\prime} m^{\prime}}{\bigl(\delop_{\Omega} \cdot \widehat{\vect{L}}\bigr)}{l m} & =0 . \label{eq:13:2:158}
\end{align}
These results follow from the orthogonality of the vectors \(\widehat{\vect{n}}, \delop_{\Omega}\) and \(\widehat{\vect{L}}\). The vector products \([\widehat{\vect{n}} \times \widehat{\vect{L}}]\) and \(\left[\delop_{\Omega} \times \widehat{\vect{L}}\right]\) are not commutative. Hence,
\begin{gather}
\braOket{ l^{\prime} m^{\prime}}{\left[\widehat{\vect{n}} \times \widehat{\vect{L}}\right]_{\kappa}}{l m}=-\frac{i}{2}\left[l^{\prime}\bigl(l^{\prime}+1\bigr)-l(l+1)-2\right] \frac{\Fact{l}}{\Fact{l^{\prime}}} 
\clebsch{l}{0}{1}{0}{l^{\prime}}{0} 
\clebsch{l}{m}{1}{\kappa}{l^{\prime}}{m^{\prime}}=i
\braOket{ l^{\prime} m^{\prime}}{\bigl(\delop_{\Omega}\bigr)_{1 \kappa}}{l m},  \label{eq:13:2:159}\\
\braOket{ l^{\prime} m^{\prime}}{\left[\widehat{\vect{L}} \times \widehat{\vect{n}}\right]_{\kappa}}{l m}=\frac{i}{2}\left[l^{\prime}\bigl(l^{\prime}+1\bigr)-l(l+1)+2\right] \frac{\Fact{l}}{\Fact{l^{\prime}}}
\clebsch{l}{0}{1}{0}{l^{\prime}}{0} 
\clebsch{l}{m}{1}{\kappa}{l^{\prime}}{m^{\prime}}, \label{eq:13:2:160}
\end{gather}
\begin{equation*}
\braOketred{ l^{\prime}}{\left[\widehat{\vect{n}} \times \widehat{\vect{L}}\right]}{l}=-\frac{i}{2}\left[l^{\prime}\bigl(l^{\prime}+1\bigr)-l(l+1)-2\right] \Fact{l}
\clebsch{l}{0}{1}{0}{l^{\prime}}{0},
\end{equation*}
\begin{equation*}
\braOketred{ l^{\prime}}{\left[\widehat{\vect{L}} \times \widehat{\vect{n}}\right]}{l}=\frac{i}{2}\left[l^{\prime}\bigl(l^{\prime}+1\bigr)-l(l+1)+2\right] \Fact{l}
\clebsch{l}{0}{1}{0}{l^{\prime}}{0}
\end{equation*}
and also
\begin{multline}
\braOket{ l^{\prime} m^{\prime}}{\bigl\{\delop_{\Omega} \times \widehat{\vect{L}}\bigr\}_{\kappa}}{l m} \\
=\frac{i}{2}\left[l^{\prime}\bigl(l^{\prime}+1\bigr)-l(l+1)-2\right]\left[l \sqrt{\frac{l+1}{2 l+3}} \delta_{l^{\prime} l+1}+(l+1) \sqrt{\frac{l}{2 l-1}} \delta_{l^{\prime} l-1}\right]
\clebsch{l}{m}{1}{\kappa}{l^{\prime}}{m^{\prime}},  \label{eq:13:2:161}
\end{multline}
\begin{multline}
\braOket{ l^{\prime} m^{\prime}}{\bigl\{\widehat{\vect{L}} \times \delop_{\Omega}\bigr\}_{\kappa}}{l m} \\
=-\frac{i}{2}\left[l^{\prime}\bigl(l^{\prime}+1\bigr)-l(l+1)+2\right]\left[l \sqrt{\frac{l+1}{2 l+3}} \delta_{l^{\prime} l+1}+(l+1) \sqrt{\frac{l}{2 l-1}} \delta_{l^{\prime} l-1}\right]
\clebsch{l}{m}{1}{\kappa}{l^{\prime}}{m^{\prime}} , \label{eq:13:2:162}
\end{multline}
\begin{equation*}
\braOketred{ l^{\prime}}{\bigl\{\delop_{\Omega} \times \widehat{\vect{L}}\bigr\}}{l}
=\frac{i}{2}\left[l^{\prime}\bigl(l^{\prime}+1\bigr)-l(l+1)-2\right]\left[l \sqrt{\frac{l+1}{2 l+3}} \delta_{l^{\prime} l+1}+(l+1) \sqrt{\frac{l}{2 l-1}} \delta_{l^{\prime} l-1}\right] \Fact{l^{\prime}},
\end{equation*}
\begin{equation*}
\braOketred{ l^{\prime}}{\bigl\{\widehat{\vect{L}} \times \delop_{\Omega}\bigr\}}{l}
=-\frac{i}{2}\left[l^{\prime}\bigl(l^{\prime}+1\bigr)-l(l+1)+2\right]\left[l \sqrt{\frac{l+1}{2 l+3}} \delta_{l^{\prime} l+1}+(l+1) \sqrt{\frac{l}{2 l-1}} \delta_{l^{\prime} l-1}\right] \Fact{l^{\prime}}.
\end{equation*}
The matrix elements of scalar and vector products of the operators \(\widehat{\vect{L}}\) and \(\widehat{\vect{J}}\) may be derived from Eqs. \eqref{eq:13:2:86}\eqref{eq:13:2:89}. The scalar product is commutative:
\begin{equation}
\braOket{ l^{\prime} s^{\prime} J^{\prime} M^{\prime}}{(\widehat{\vect{L}} \cdot \widehat{\vect{J}})}{l s J M}=\delta_{l^{\prime} l} \delta_{s^{\prime} s} \delta_{J^{\prime} J} \delta_{M^{\prime} M} \frac{J(J+1)+l(l+1)-s(s+1)}{2}, \label{eq:13:2:163}
\end{equation}
\begin{equation*}
\braOketred{ l^{\prime} s^{\prime} J^{\prime}}{(\widehat{\vect{L}} \cdot \widehat{\vect{J}})}{l s J}=\delta_{l^{\prime} l} \delta_{s^{\prime} s} \delta_{J^{\prime} J} \Fact{J} \frac{J(J+1)+l(l+1)-s(s+1)}{2}.
\end{equation*}
The vector product is not commutative:
\begin{gather}
\braOket{ l^{\prime} s^{\prime} J^{\prime} M^{\prime}}{\bigl\{\widehat{\vect{L}} \times \widehat{\vect{J}}\bigr\}_{\kappa}}{l s J M}=\frac{i}{2}(-1)^{s+l+J+1} \delta_{l^{\prime} l} \delta_{s^{\prime} s} \left[ J(J+1)-J^{\prime}\bigl(J^{\prime}+1\bigr)+2\right] \notag\\
\times \Fact{J l}\sqrt{ l(l+1)}
\sixj{l}{s}{J}{J^{\prime}}{1}{l}
\clebsch{J}{M}{1}{\kappa}{J^{\prime}}{M^{\prime}},  \label{eq:13:2:164}
\end{gather}
\begin{equation*}
\braOketred{ l^{\prime} s^{\prime} J^{\prime}}{\bigl\{\widehat{\vect{L}} \times \widehat{\vect{J}}\bigr\}}{l s J}=\frac{i}{2}(-1)^{s+l+J+1} \delta_{l^{\prime} l} \delta_{s^{\prime} s} \left[ J(J+1)-J^{\prime}\bigl(J^{\prime}+1\bigr)+2\right]  \Fact{J l J^{\prime}}\sqrt{ l(l+1)}
\sixj{l}{s}{J}{J^{\prime}}{1}{l}
\end{equation*}
\begin{gather}
\braOket{ l^{\prime} s^{\prime} J^{\prime} M^{\prime}}{\left[\widehat{\vect{J}} \times \widehat{\vect{L}}\right]_{\kappa}}{l s J M}=\frac{i}{2}(-1)^{s+l+J+1} \delta_{l^{\prime} l} \delta_{s^{\prime} s}\left[J^{\prime}\bigl(J^{\prime}+1\bigr)-J(J+1)+2\right] \notag\\
\times \Fact{J l}\sqrt{ l(l+1)}
\sixj{l}{s}{J}{J^{\prime}}{1}{l^{\prime}}
\clebsch{J}{M}{1}{\kappa}{J^{\prime}}{M^{\prime}}, \label{eq:13:2:165}
\end{gather}
\begin{equation*}
\braOketred{ l^{\prime} s^{\prime} J^{\prime}}{\left[\widehat{\vect{J}} \times \widehat{\vect{L}}\right]}{l s J}=\frac{i}{2}(-1)^{s+l+J+1} \delta_{l^{\prime} l} \delta_{s^{\prime} s}\left[J^{\prime}\bigl(J^{\prime}+1\bigr)-J(J+1)+2\right] \\
\times \Fact{J l J^{\prime}}\sqrt{ l(l+1)}
\sixj{l}{s}{J}{J^{\prime}}{1}{l^{\prime}}.
\end{equation*}
Matrix elements of scalar and vector products of the spin operator \(\widehat{\vect{S}}\) and the operators \(\widehat{\vect{n}}, \delop_{\Omega}\) and \(\widehat{\vect{L}}\) (which commutative with \(\widehat{\vect{S}})\) may be found from Eqs. \eqref{eq:13:2:95}-\eqref{eq:13:2:99}:

\begin{equation}
\braOket{ l^{\prime} s^{\prime} J^{\prime} M^{\prime}}{(\widehat{\vect{n}} \cdot \widehat{\vect{S}})}{l s J M}=(-1)^{l+s+J} \delta_{J J^{\prime}} \delta_{M M^{\prime}} \delta_{s s^{\prime}} \Fact{s l}\sqrt{s(s+1)}
\clebsch{l}{0}{1}{0}{l^{\prime}}{0}
\sixj{1}{s}{s}{J}{l}{l^{\prime}}, \label{eq:13:2:166}
\end{equation}
\begin{equation*}
\braOketred{ l^{\prime} s^{\prime} J^{\prime}}{(\widehat{\vect{n}} \cdot \widehat{\vect{S}})}{l s J}=(-1)^{l+s+J} \delta_{J J^{\prime}} \delta_{s s^{\prime}} \Fact{s l J}\sqrt{s(s+1)}
\clebsch{l}{0}{1}{0}{l^{\prime}}{0}
\sixj{1}{s}{s}{J}{l}{l^{\prime}}.
\end{equation*}
\begin{align}
\braOket{ l^{\prime} s^{\prime} J^{\prime} M^{\prime}}{\bigl(\delop_{\Omega} \cdot \widehat{\vect{S}}\bigr)}{l s J M}&=(-1)^{l+s+J+1} \delta_{J J^{\prime}} \delta_{M M^{\prime}} \delta_{s s^{\prime}} \Fact{s l^\prime}\sqrt{s(s+1)} \notag\\
& \times\left[l \sqrt{\frac{l+1}{2 l+3}} \delta_{l^{\prime} l+1}+(l+1) \sqrt{\frac{l}{2 l-1}} \delta_{l^{\prime} l-1}\right]
\sixj{1}{s}{s}{J}{l}{l^{\prime}},  \label{eq:13:2:167}
\end{align}
\begin{multline*}
\braOketred{ l^{\prime} s^{\prime} J^{\prime}}{\bigl(\delop_{\Omega} \cdot \widehat{\vect{S}}\bigr)}{l s J}=(-1)^{l+s+J+1} \delta_{J J^{\prime}} \delta_{s s^{\prime}} \Fact{s l^\prime J}\sqrt{s(s+1)} \\
\times\left[l \sqrt{\frac{l+1}{2 l+3}} \delta_{l^{\prime} l+1}+(l+1) \sqrt{\frac{l}{2 l-1}} \delta_{l^{\prime} l-1}\right]
\sixj{1}{s}{s}{J}{l}{l^{\prime}}
\end{multline*}\begin{align}
& 
\braOket{ l^{\prime} s^{\prime} J^{\prime} M^{\prime}}{(\widehat{\vect{L}} \cdot \widehat{\vect{S}})}{l s J M}=\delta_{l l^{\prime}} \delta_{s s^{\prime}} \delta_{J J^{\prime}} \delta_{M M^{\prime}} \frac{1}{2}[J(J+1)-l(l+1)-s(s+1)],  \label{eq:13:2:168}
\end{align}
\begin{equation*}
\braOketred{ l^{\prime} s^{\prime} J^{\prime}}{(\widehat{\vect{L}} \cdot \widehat{\vect{S}})}{l s J}=\delta_{l l^{\prime}} \delta_{s s^{\prime}} \delta_{J J^{\prime}} \Fact{J} \frac{1}{2}[J(J+1)-l(l+1)-s(s+1)].
\end{equation*}
\begin{align}
\braOket{l^{\prime} s^{\prime} J^{\prime} M^{\prime}}%
{\left[\widehat{\vect{n}} \times\widehat{\vect{S}}\right]_{\kappa}}%
{l s J M} 
=&i(-1)^{l^{\prime}+J+s} \delta_{ss'} \Fact{J l} 
\clebsch{l}{0}{1}{0}{l^{\prime}}{0} \notag\\
& \times \frac{1}{2}\left[\bigl(J^{\prime}-l^{\prime}\bigr)\bigl(J^{\prime}+l^{\prime}+1\bigr)-(J-l)(J+l+1)\right]
\sixj{J^{\prime}}{l^{\prime}}{s}{l}{J}{1}
\clebsch{J}{M}{1}{\kappa}{J^{\prime}}{M^{\prime}},  \label{eq:13:2:169}
\end{align}
\begin{multline*}
\braOketred{l^{\prime} s^{\prime} J^{\prime}}{\left[\widehat{\vect{n}} \times\widehat{\vect{S}}\right]}{l s J}
=i(-1)^{l^{\prime}+J+s} \delta_{ss'} \Fact{J l J^{\prime}}
\clebsch{l}{0}{1}{0}{l^{\prime}}{0} \\
\times \frac{1}{2}\left[\bigl(J^{\prime}-l^{\prime}\bigr)\bigl(J^{\prime}+l^{\prime}+1\bigr)-(J-l)(J+l+1)\right]
\sixj{J^{\prime}}{l^{\prime}}{s}{l}{J}{1}
\end{multline*}
\begin{align}
& 
\braOket{ l^{\prime} s^{\prime} J^{\prime} M^{\prime}}{\left[\bigl(\delop_{\Omega}\bigr) \times \widehat{\vect{S}}\right]_{\kappa}}{l s J M}=i(-1)^{l^{\prime}+J+s+1} \delta_{s s^{\prime}} \Fact{J l^\prime}
\sixj{J^{\prime}}{l^{\prime}}{s}{l}{J}{1} \notag\\
& \times \frac{1}{2}\left[\bigl(J^{\prime}-l^{\prime}\bigr)\bigl(J^{\prime}+l^{\prime}+1\bigr)-(J-l)(J+l+1)\right]\left[l \sqrt{\frac{l+1}{2 l+3}} \delta_{l^{\prime} l+1}+(l+1) \sqrt{\frac{l}{2 l-1}} \delta_{l^{\prime} l-1}\right]
\clebsch{J}{M}{1}{\kappa}{J^{\prime}}{M^{\prime}},  \label{eq:13:2:170}
\end{align}

\begin{multline*}
\braOketred{ l^{\prime} s^{\prime} J^{\prime}}{\left[\bigl(\delop_{\Omega}\bigr) \times \widehat{\vect{S}}\right]}{l s J}=i(-1)^{l^{\prime}+J+s+1} \delta_{s s^{\prime}} \Fact{J l^\prime J^{\prime}}
\sixj{J^{\prime}}{l^{\prime}}{s}{l}{J}{1} \\
\times \frac{1}{2}\left[\bigl(J^{\prime}-l^{\prime}\bigr)\bigl(J^{\prime}+l^{\prime}+1\bigr)-(J-l)(J+l+1)\right]\left[l \sqrt{\frac{l+1}{2 l+3}} \delta_{l^{\prime} l+1}+(l+1) \sqrt{\frac{l}{2 l-1}} \delta_{l^{\prime} l-1}\right].
\end{multline*}

\begin{align}
\braOket{l^{\prime} s^{\prime} J^{\prime} M^{\prime}}{\left[\widehat{\vect{L}}_{1} \times \widehat{\vect{S}}_{1}\right]_{\kappa}}{l s J M}=&i(-1)^{l+J+s} \Fact{J l} \sqrt{ l(l+1)} \delta_{l l^{\prime}} \delta_{s s^{\prime}} \notag\\
& \times \frac{1}{2}\left[\bigl(J^{\prime}-l\bigr)\bigl(J^{\prime}+l+1\bigr)-(J-l)(J+l+1)\right]
\sixj{J^{\prime}}{l}{s}{l}{J}{1}
\clebsch{J}{M}{1}{\kappa}{J^{\prime}}{M^{\prime}} , \label{eq:13:2:171}
\end{align}
\begin{multline*}
\braOketred{l^{\prime} s^{\prime} J^{\prime}}{\left[\widehat{\vect{L}}_{1} \times \widehat{\vect{S}}_{1}\right]}{l s J}=i(-1)^{l+J+s} \Fact{J l J^{\prime}} \sqrt{ l(l+1)} \delta_{l l^{\prime}} \delta_{s s^{\prime}} \\
\times \frac{1}{2}\left[\bigl(J^{\prime}-l\bigr)\bigl(J^{\prime}+l+1\bigr)-(J-l)(J+l+1)\right]
\sixj{J^{\prime}}{l}{s}{l}{J}{1}.
\end{multline*}
The scalar product of the operators \(\widehat{\vect{S}}\) and \(\widehat{\vect{J}}\) is commutative and its matrix elements are determined by
\begin{equation}
\braOket{l^{\prime} s^{\prime} J^{\prime} M^{\prime}}{(\widehat{\vect{S}} \cdot \widehat{\vect{J}})}{ l s J M}=\delta_{l l^{\prime}} \delta_{s s^{\prime}} \delta_{J J^{\prime}} \delta_{M M^{\prime}} \frac{1}{2}[J(J+1)+s(s+1)-l(l+1)], \label{eq:13:2:172}
\end{equation}
\begin{equation*}
\braOketred{l^{\prime} s^{\prime} J^{\prime}}{(\widehat{\vect{S}} \cdot \widehat{\vect{J}})}{ l s J}=\delta_{l l^{\prime}} \delta_{s s^{\prime}} \delta_{J J^{\prime}} \Fact{J} \frac{1}{2}[J(J+1)+s(s+1)-l(l+1)].
\end{equation*}
The vector product of the same operators is not commutative. The matrix elements of this product are given by
\begin{align}
\braOket{ l^{\prime} s^{\prime} J^{\prime} M^{\prime}}{\left[\widehat{\vect{S}} \times \widehat{\vect{J}}\right]_{\kappa}}{l s J M}=&\frac{i}{2}(-1)^{l+s+J^{\prime}+1} \delta_{s s^{\prime}} \delta_{l l^{\prime}}\left[J(J+1)-J^{\prime}\bigl(J^{\prime}+1\bigr)+2\right] \notag\\
&\times \Fact{J s}\sqrt{ s(s+1)}
\sixj{s}{l}{J}{J^{\prime}}{1}{s} 
\clebsch{J}{M}{1}{\kappa}{J^{\prime}}{M^{\prime}},  \label{eq:13:2:173}\\
\braOket{ l^{\prime} s^{\prime} J^{\prime} M^{\prime} }{ \left[\widehat{\vect{J}} \times \widehat{\vect{S}}\right]_{\kappa}}{l s J M}=&\frac{i}{2}(-1)^{l+s+J^{\prime}+1} \delta_{l l^{\prime}} \delta_{s s^{\prime}}\left[J^{\prime}\bigl(J^{\prime}+1\bigr)-J(J+1)+2\right] \notag\\
&\times \Fact{J s}\sqrt{ s(s+1)}
\sixj{s}{l}{J}{J^{\prime}}{1}{s} 
\clebsch{J}{M}{1}{\kappa}{J^{\prime}}{M^{\prime}}, \label{eq:13:2:174}
\end{align}
\begin{align*}
\braOketred{ l^{\prime} s^{\prime} J^{\prime}}{\left[\widehat{\vect{S}} \times \widehat{\vect{J}}\right]}{l s J}=\frac{i}{2}(-1)^{l+s+J^{\prime}+1} \delta_{s s^{\prime}} \delta_{l l^{\prime}}\left[J(J+1)-J^{\prime}\bigl(J^{\prime}+1\bigr)+2\right]  \Fact{J s J^{\prime}}\sqrt{ s(s+1)}
\sixj{s}{l}{J}{J^{\prime}}{1}{s},\\
\braOketred{ l^{\prime} s^{\prime} J^{\prime} }{ \left[\widehat{\vect{J}} \times \widehat{\vect{S}}\right]}{l s J}=\frac{i}{2}(-1)^{l+s+J^{\prime}+1} \delta_{l l^{\prime}} \delta_{s s^{\prime}}\left[J^{\prime}\bigl(J^{\prime}+1\bigr)-J(J+1)+2\right]  \Fact{J s J^{\prime}}\sqrt{ s(s+1)}
\sixj{s}{l}{J}{J^{\prime}}{1}{s}.
\end{align*}
\chapter{Computer generation of numerical and algebraic expressions}\label{chap:14}\index{computer algebra}\index{computer generation of tables}

The algebraic tables of Secs.~\ref{sec:8.12}, \ref{sec:9.11} and \ref{sec:10.11},
and the numerical tables of Secs.~\ref{sec:8.13} and \ref{sec:10.12}, together
with Tables 9.9--9.11, may be reproduced entry by entry by computer algebra. This
chapter records the method used, so that any individual entry can be
regenerated and checked rather than taken on trust. The programs are listed in
Sec.~\ref{sec:14.5}.

\section{Scope}\label{sec:14.1}\index{computer algebra!scope}
Each algebraic table fixes one small angular momentum and leaves the remaining
arguments symbolic: $b$ in the Clebsch-Gordan tables, $d$ in the $6j$ tables,
and the triad $(\alpha \beta \gamma)$ in the $9j$ tables. The corresponding sum
is then finite with a number of terms bounded by that fixed argument, and is
therefore a finite symbolic computation. Nothing in the method is restricted to
the ranges printed here; the same programs generate entries beyond them, at
greater cost.

\section{Algebraic generation}\label{sec:14.2}\index{computer algebra!algebraic tables}

\subsection{The Clebsch-Gordan coefficients}\label{sec:14.2.1}\index{Clebsch-Gordan coefficient!computer generation}
For fixed $b$ and $\beta$, and with $c=a+k$, the summation index of the Van der
Waerden-Racah sum of Sec.~\ref{sec:8.2.1} runs over at most $2b+1$ values, so
the coefficient is a short symbolic sum. The difficulty is that its terms
contain factorials of $a\pm\alpha$, which are not polynomial in $a$. It is
removed by the change of variable
\begin{equation}
p=a+\alpha, \qquad q=a-\alpha ,\label{eq:14:2:1}
\end{equation}
both of which are non-negative \emph{integers} even when $a$ and $\alpha$ are
half-integral. Every factorial argument is then an explicit integer offset from
$p$, $q$ or $p+q=2a$, and each such factorial may be replaced by a rational
function through
\begin{equation}
\frac{(x+m)!}{x!}=
\begin{cases}
(x+1)(x+2)\cdots(x+m), & m \geq 0,\\[2pt]
\bigl[x(x-1)\cdots(x+m+1)\bigr]^{-1}, & m<0 .
\end{cases}\label{eq:14:2:2}
\end{equation}
The factor $p!\,q!$ produced by the sum cancels exactly against the one under
the square root, and the coefficient takes the form
\begin{equation}
\clebsch{a}{\alpha}{b}{\beta}{c}{\gamma}=T(p,q)\,\bigl[W(p,q)\bigr]^{1/2},
\label{eq:14:2:3}
\end{equation}
with $T$ and $W$ explicit rational functions. No general-purpose simplification
of factorials or gamma functions is required at any stage.

Eq.~\eqref{eq:14:2:3} does not fix the presentation, since any factor may be
moved under the radical as its square. The tables of Sec.~\ref{sec:8.12} absorb
every factor that is of one sign throughout the physical domain
$|\gamma| \leq c$, such as $(c+\gamma)$ or $(c-\gamma+1)$, and leave outside
only those that change sign, such as $\gamma$ itself. Applying that rule
reproduces the printed entries.

\subsection{The $6j$ symbols}\label{sec:14.2.2}\index{6j symbol@$6j$ symbol!computer generation}
The analogous integer variables for the $6j$ symbols are the triangle deficits
\begin{equation}
u=s-2a, \qquad v=s-2b, \qquad w=s-2c, \qquad s=a+b+c=u+v+w .\label{eq:14:2:4}
\end{equation}
Writing the summation index of Eq.~\eqref{eq:9:2:1} as $N=s+t$, and taking
$e=c+n$ and $f=b+m$ as in the tables, the seven factorial arguments of that
equation become
\begin{equation}
\begin{gathered}
N-a-b-c=t, \qquad N-a-e-f=t-n-m, \\
N-b-d-f=v+t-d-m, \qquad N-c-d-e=w+t-d-n, \\
a+b+d+e-N=d+n-t, \qquad a+c+d+f-N=d+m-t, \\
b+c+e+f-N=u+n+m-t ,
\end{gathered}\label{eq:14:2:5}
\end{equation}
each an explicit integer offset from $t$, $u$, $v$ or $w$. The index $t$ runs
over the explicit finite range
\begin{equation}
\max(0,\,n+m) \leq t \leq \min(d+n,\,d+m),\label{eq:14:2:6}
\end{equation}
so the sum has at most $2d+1$ terms. Here the cancellation is complete:
$\Delta(abc)$ contributes $[u!\,v!\,w!/(s+1)!]^{1/2}$, the sum contributes
$(s+1)!/(u!\,v!\,w!)$, and $\Delta(aef)$ contributes the inverse, leaving
\begin{equation}
\sixj{a}{b}{c}{d}{e}{f}=(-1)^{s}\,T(u,v,w)\,\bigl[W(u,v,w)\bigr]^{1/2}
\label{eq:14:2:7}
\end{equation}
with no symbolic factorial remaining. The phase arises because
$(-1)^{N}=(-1)^{s}(-1)^{t}$, which is why every entry of Sec.~\ref{sec:9.11}
carries $(-1)^{s}$ or $(-1)^{s+1}$.

The factors of $T$ and $W$ are linear in $a$, $b$, $c$ and are printed in the
notation of the tables by reading off their coefficients: $(1,1,1)$ gives
$s+k$, $(-1,1,1)$ gives $s-2a+k$, $(0,2,0)$ gives $2b+k$, and so on. The factor
left outside the radical is expressed as a polynomial in $X$ by division, $X$
being quadratic in $a$.

\subsection{The $9j$ symbols}\label{sec:14.2.3}\index{9j symbol@$9j$ symbol!computer generation}
The $9j$ symbols of Sec.~\ref{sec:10.11} are not given by a single sum of the
above type. They are obtained from the sum over three $6j$ symbols,
Eq.~\eqref{eq:10:2:20}. For the tabulated symbols the triad $(\beta a x)$
restricts the intermediate momentum to
\begin{equation}
x=a+\xi, \qquad -\beta \leq \xi \leq \beta ,\label{eq:14:2:8}
\end{equation}
an explicit sum of at most $2\beta+1$ terms. Each of the three $6j$ symbols is
carried by a permutation of its columns together with an interchange of the
upper and lower arguments in two of them into the form
$\bigl\{{A\ B\ C \atop d\ C{+}n\ B{+}m}\bigr\}$ treated in
Sec.~\ref{sec:14.2.2}, so the same machinery evaluates all three; only the
deficits \eqref{eq:14:2:4} need re-expressing in terms of those of $(abc)$.

The phases collapse. Summing the three factors $(-1)^{A+B+C}$ with the
$(-1)^{2x}$ of Eq.~\eqref{eq:10:2:20} gives the exponent
$4a+4\xi+2(a+b+c)+\beta+\mu$, and since $2a$, $2\xi$ and $a+b+c$ are integers
every term is even except the last two:
\begin{equation}
\ninej{a+\lambda}{b+\mu}{c+\nu}{a}{b}{c}{\alpha}{\beta}{\gamma}
=(-1)^{\beta+\mu}\sum_{\xi}(2a+2\xi+1)\,T_1T_2T_3\,[W_1W_2W_3]^{1/2},
\label{eq:14:2:9}
\end{equation}
independent of $\xi$ and of $a$, $b$, $c$. The radicands of the individual
terms differ by perfect squares of rational functions, so a common radical may
be extracted and the whole reduces to a rational multiple of a single square
root. That the ratios are perfect squares is tested, not assumed.

\section{Numerical evaluation}\label{sec:14.3}\index{computer algebra!numerical tables}
Numerical values are obtained in exact rational arithmetic. Since every
coefficient of Secs.~\ref{sec:8.13} and \ref{sec:10.12}, and of
Tables 9.9--9.11, is a signed square root of a rational number, the value $V$ is squared, $V^{2}$ is
confirmed to be rational, and the entry is presented as
\begin{equation}
V=\pm\left(\frac{P}{Q}\right)^{1/2},\label{eq:14:3:1}
\end{equation}
the decimal being derived from the fraction rather than the reverse. The
$12j$ symbols are evaluated from their defining single sums over four $6j$
symbols, Eqs.~\eqref{eq:10:13:6} and \eqref{eq:10:13:26}.

\section{Verification}\label{sec:14.4}\index{computer algebra!verification}
Each generator is checked against an independent implementation of the same
coefficient at a large number of numerical points, at two levels: the sum
itself, and the regrouping into the printed form, so that an error in a phase
or in the choice of factors placed under the radical cannot pass unnoticed.
Points are chosen asymmetrically as well as symmetrically, since a factor such
as $X$ or $Z$ changes sign only when one momentum dominates the other two, and
a symmetric sample would not detect a factor wrongly treated as of one sign.
All entries of the tables cited in Sec.~\ref{sec:14.1} agree. The relations of
Sec.~\ref{sec:10.13} are checked against each other in the same way.

\section{The programs}\label{sec:14.5}\index{computer algebra!programs}
The generators and their checks are written in Python using
\textsf{SymPy}. Each generator produces a single entry, given the fixed
argument and the row and column indices of the table:
\begin{lstlisting}[language=bash]
gen_8_12_cg_tables.py  --b 1 --beta=0 --k=0
gen_9_11_6j_tables.py  --d 1 --m=0 --n=0
gen_10_11_9j_tables.py --alpha 3/2 --beta 3/2 --gamma 0 --lam=1/2 --mu=1/2 --nu=0
\end{lstlisting}
Omitting the row and column indices produces a whole table. The companion
programs \texttt{check\_8\_12.py}, \texttt{check\_9\_11.py},
\texttt{check\_10\_11.py} and \texttt{check\_10\_13.py} perform the
verification of Sec.~\ref{sec:14.4}. The same functions are available
interactively in the two notebooks described in Sec.~\ref{sec:14.6}.

\section{Use of the notebooks}\label{sec:14.6}\index{computer algebra!notebooks}
Two Jupyter notebooks make the generators available without recourse to the
command line. Both require only \textsf{SymPy}; if \textsf{ipywidgets} is
also present the arguments are set from menus, and if it is not, each quantity
is instead obtained by editing the arguments at the head of a cell and running
it. Every result is displayed both as a typeset formula and as the
corresponding \textsf{SymPy} expression, the latter being the form to copy if
the entry is to be used in further computation.

\subsection{Algebraic expressions}\label{sec:14.6.1}
The notebook \texttt{vmk\_algebraic\_tables.ipynb} has one part for each of
the three families of Sec.~\ref{sec:14.2}. In each, one selects the fixed
argument and the row and column of the table:
\begin{lstlisting}[language=Python]
clebsch(b=1, beta=0, k=0)             # <a alpha, b beta | c gamma>, c = a+k
sixj(d=1, m=0, n=0)                   # {a b c; d e f}, e = c+n, f = b+m
ninej(1, 1, 0, lam=1, mu=0, nu=0)     # alpha, beta, gamma, then lambda, mu, nu
\end{lstlisting}
A further cell in each part prints an entire table at once. Setting
\texttt{show\_source=True} adds the \LaTeX{} source of the entry. The
Clebsch-Gordan entries are recomputed as soon as an argument is changed; the
$6j$ and $9j$ entries are not, since a single entry of the larger tables takes
appreciably longer to construct, and are instead computed on request. A closing
part runs the verification programs of Sec.~\ref{sec:14.4} over a reduced
range of arguments.

\subsection{Numerical values}\label{sec:14.6.2}
The notebook \texttt{vmk\_numeric\_tables.ipynb} evaluates the
Clebsch-Gordan coefficients and the $6j$, $9j$ and $12j$ symbols at given
numerical arguments, in the form \eqref{eq:14:3:1}:
\begin{lstlisting}[language=Python]
clebsch(1, 0, 1, 0, 2)                # gamma = alpha + beta is supplied
sixj(1, 1, 1, 1, 1, 1)
ninej(0.5, 0.5, 1, 0.5, 0.5, 1, 1, 1, 2)
twelvej1(1, 1, 1, 1, 1, 1, 1, 1, 1, 1, 1, 1)    # eq (10.13.6)
twelvej2(1, 1, 1, 1, 1, 1, 1, 1, 1, 1, 1, 1)    # eq (10.13.26)
\end{lstlisting}
A coefficient that vanishes is reported together with the condition that fails,
whether a triangular condition on a particular triad or a projection outside
its permitted range, so that a zero is never returned without explanation.
Further cells reproduce the numerical tables themselves under the conditions
stated there, and a final part evaluates the algebraic expressions of
Sec.~\ref{sec:14.2} at numerical arguments and compares them with the values
obtained directly, the two being independent routes to the same quantity.

% Structured bibliography for
% "Quantum Theory of Angular Momentum" (Varshalovich, Moskalev, Khersonskii).
%
% The reference data lives in references.bib.  This file drives biblatex to
% print the four source parts separately while keeping the running 1..145
% numbering of the printed book (defernumbers assigns numbers in print order;
% sorting=none keeps each part in the citation/database order).
%
% Build:  xelatex references  ->  bibtex references  ->  xelatex references x2
% (or:  latexmk -xelatex references.tex)
% The bibtex backend is used because it needs no extra binary; if you have a
% working biber, switch backend=bibtex to backend=biber and run biber instead.
%
% To fold this into the main book instead of building it standalone, move the
% \usepackage[...]{biblatex} and \addbibresource lines into VMK.tex's preamble,
% \include this file's body, and run the bibliography backend on VMK.

% Cite everything so the whole database is printed, in database order.
\nocite{*}

\begin{center}\Large\bfseries REFERENCES\end{center}

\printbibliography[keyword=books,  heading=subbibliography, title={I.~~Books and Reviews}]
\printbibliography[keyword=papers, heading=subbibliography, title={II.~~Papers on Special Problems}]
\printbibliography[keyword=tables, heading=subbibliography, title={III.~~Tables}]
\printbibliography[keyword=added,  heading=subbibliography, title={IV.~~References Added in This Edition}]

\section*{V.~~Additional Papers Since the Book Came Out}
\emph{(none)}

%\include{glossary} no longer needed, as the glossary is now integrated into theindex.
%Neither index is numbered, so the ToC entries have to be added by hand.
\cleardoublepage\addcontentsline{toc}{chapter}{Subject Index}
\printindex
\cleardoublepage\addcontentsline{toc}{chapter}{Index of Symbols}
\printindex[sym]
\end{document}